\documentclass{elsart1p}

\usepackage{graphicx}
\usepackage{epsfig}
\usepackage{amsmath,amssymb,bm}

\usepackage{cite}

\def\subsubsubsection#1{\textbf{#1}}
\input symbols

\begin{document}
\begin{frontmatter}


\begin{flushright}
\small{
BNL-90299-2009-BC,
CERN-PH-TH-2009-112, 
FERMILAB-PUB-09-323-T,
LAL 09-111,
MPP-2009-88, 
MZ-TH/09-22, 
MKPH-T-09-14,
SLAC-R-926,
TUM-HEP-728/09,
WSU-HEP-0902
}
\end{flushright}


\title{Flavor Physics in the Quark Sector}


\author[aff0]{M.  Antonelli}
\author[aff1]{D. M. Asner}
\author[aff2]{D.  Bauer}
\author[aff3]{T.  Becher}
\author[aff4]{M.  Beneke}
\author[aff5]{A. J. Bevan}
\author[aff6,aff7]{M.  Blanke}
\author[aff0]{C.  Bloise}
\author[aff8]{M.  Bona}
\author[aff9]{A.  Bondar}
% ...
\address[aff0]{INFN LNF, Via Enrico Fermi 40, 00044 Frascati, Italy}
\address[aff1]{Carleton University, 1125 Colonel By Drive, Ottawa, ON, Canada K1S 5B6}
\address[aff2]{Imperial College London, London, SW7 2AZ, United Kingdom}
\address[aff3]{Fermi National Accelerator Laboratory, P.O. Box 500 Batavia, IL 60510-5011, USA}
\address[aff4]{Institut f\"ur Theoretische Physik E, RWTH Aachen University, 52056, Germany}
\address[aff5]{Queen Mary, University of London, E1 4NS, United Kingdom}
\address[aff6]{Technische Universit\"{a}t M\"{u}nchen, Excellence Cluster Universe, Boltzmannstra{\ss}e 2, 85748 Garching, Germany}
\address[aff7]{Max-Planck-Institut f\"{u}r Physik, Foehringer Ring 6, 80805 M\"{u}nchen, Germany}
\address[aff8]{CERN CH-1211 Geneve 23, Switzerland}
\address[aff9]{Budker Institute of Nuclear Physics, 11, Prosp. Akademika Lavrentieva Novosibirsk 630090, Russian Federation}
% ...

\begin{abstract}
In the past decade, one of the major challenges of particle physics has been to gain an in-depth understanding of the role of quark flavor.
 In this time frame, measurements and the theoretical interpretation of their results have advanced tremendously. 
 A much broader understanding of flavor particles has been achieved, apart from their masses and quantum numbers, there now exist detailed measurements of 
the characteristics of their interactions allowing stringent tests of Standard Model predictions. 
 Among the most interesting phenomena of flavor physics is the violation of the CP symmetry that has been subtle and difficult to explore.  In the past, 
 observations of CP violation were confined to neutral $K$ mesons, but since the early 1990s, a large number of CP-violating processes have been studied in 
detail in neutral $B$ mesons. In parallel, measurements of the couplings of the heavy 
 quarks and the dynamics for their decays in large samples of $K, D$, and $B$ mesons have been greatly improved in accuracy and the results are being used as 
probes in the search for deviations from the Standard Model.

In the near future, there will be a transition from the current to a new generation of experiments, thus a review of the status of quark flavor physics is 
timely. This report is the result of the work of the physicists attending the $5^{th}$ CKM workshop, hosted by the University of Rome "La Sapienza", 
September 9-13, 2008.  It summarizes the results of the current generation of experiments that is about to be completed and it confronts these results with the 
theoretical understanding of the field which has greatly improved in the past decade.

\end{abstract}

\end{frontmatter}


\vskip 1. cm
\vskip 1. cm
\tableofcontents

\section{Introduction}
\label{sec:intro}
\newenvironment{comment}[1]{}{}

In the past decade, one of the major challenges of particle physics has been to gain an in-depth understanding of the role of quark flavor.
In this time frame, measurements and the theoretical interpretation of their results have advanced tremendously. 
A much broader understanding of flavor particles has been achieved, apart from their masses and quantum numbers, there now exist detailed measurements of the characteristics of their interactions allowing stringent tests of Standard Model predictions. 

Among the most interesting phenomena of flavor physics is the violation of the CP symmetry that has been subtle and difficult to explore.  In the past, observations of CP violation were confined to neutral $K$ mesons, but since the early 1990s, a large number of CP-violating processes have been studied in detail in neutral $B$ mesons. In parallel, measurements of the couplings of the heavy 
quarks and the dynamics for their decays in large samples of $K, D$, and $B$ mesons have been greatly improved in accuracy and the results are being used as probes in the search for deviations from the Standard Model.


In the near future, there will be a transition from the current to a new generation of experiments, thus a review of the status of quark flavor physics is 
timely. This report is the result of the work of the physicists attending the $5^{th}$ CKM workshop, hosted by the University of Rome "La Sapienza", 
September 
9-13, 2008.  It summarizes the results of the current generation of experiments that is about to be completed and it confronts these results with the theoretical understanding of the field which has greatly improved in the past decade.

In this section  the basic formalism of the study of the quark couplings will be introduced and the relationship between CKM matrix elements and observables 
will be discussed. The last paragraph will then detail the plan of the report and the content of the rest of the sections.

\subsection{CKM matrix and the Unitarity Triangle}
\label{sec:ckmgen}
\setcounter{equation}{0}


The unitary CKM matrix~\cite{Cabibbo:1963yz,Kobayashi:1973fv}
connects the {\it weak
eigenstates} $(d^\prime,s^\prime,b^\prime)$ and the corresponding {\it
mass eigenstates} $d,s,b$ (in both basis the up-type mass matrix is diagonal and
the up-type quarks are unaffected by this transformation):
\begin{equation}\label{2.67}
\left(\begin{array}{c}
d^\prime \\ s^\prime \\ b^\prime
\end{array}\right)=
\left(\begin{array}{ccc}
V_{ud}&V_{us}&V_{ub}\\
V_{cd}&V_{cs}&V_{cb}\\
V_{td}&V_{ts}&V_{tb}
\end{array}\right)
\left(\begin{array}{c}
d \\ s \\ b
\end{array}\right)\equiv\hat V_{\rm CKM}\left(\begin{array}{c}
d \\ s \\ b
\end{array}\right).
\end{equation}
The CKM matrix contains all the flavor-changing and CP-violating couplings
of the Standard Model.

Several parameterizations of the CKM matrix have been proposed in the
literature. This report will use the standard parametrization~\cite{Chau:1984fp}
recommended by the Particle Data Group~\cite{Amsler:2008zz}.
We also introduce the generalization of the Wolfenstein 
parametrization~\cite{Wolfenstein:1983yz} presented in~\cite{Buras:1994ec} and
discuss its connection to the Unitarity Triangle parameters.

\subsubsection{Standard parametrization}
            \label{sec:sewm:stdparam}

With $c_{ij}=\cos\theta_{ij}$ and $s_{ij}=\sin\theta_{ij}$
($i,j=1,2,3$), the standard parametrization is given by:
\begin{equation}\label{2.72}
\hat V_{\rm CKM}=
\left(\begin{array}{ccc}
c_{12}c_{13}&s_{12}c_{13}&s_{13}e^{-i\delta}\\ -s_{12}c_{23}
-c_{12}s_{23}s_{13}e^{i\delta}&c_{12}c_{23}-s_{12}s_{23}s_{13}e^{i\delta}&
s_{23}c_{13}\\ s_{12}s_{23}-c_{12}c_{23}s_{13}e^{i\delta}&-s_{23}c_{12}
-s_{12}c_{23}s_{13}e^{i\delta}&c_{23}c_{13}
\end{array}\right)\,,
\end{equation}
where $\delta$ is the phase necessary for {\rm CP} violation. $c_{ij}$
and $s_{ij}$ can all be chosen to be positive and $\delta$ may vary in
the range $0\le\delta\le 2\pi$. However, measurements of CP violation
in $K$ decays force $\delta$ to be in the range $0<\delta<\pi$, as
the sign of the relevant hadronic parameter is fixed.

From phenomenological studies we know that $s_{13}$ and $s_{23}$ are
small numbers: ${\cal O}(10^{-3})$ and ${\cal O}(10^{-2})$, respectively.
Consequently, to a very good accuracy,
\begin{equation}\label{2.73}
s_{12}\simeq|V_{us}|, \quad s_{13}\simeq | V_{ub}|, \quad s_{23}\simeq
| V_{cb}|.
\end{equation}
Thus these three parameters can be extracted from tree level decays
mediated by the transitions $s \to u$, $b \to u$ and $b \to c$, respectively.
The remaining parameter, the phase $\delta$, is responsible for the violation
of the CP symmetry. It can clearly be extracted from CP-violating transitions
but also from CP-conserving ones using three-generation unitarity, through
the construction of the Unitarity Triangle, as discussed below.

\subsubsection{Wolfenstein parametrization and its generalization}
\label{Wolf-Par}

The absolute values of the elements of the CKM matrix show a
hierarchical pattern with the diagonal elements being close to unity,
the elements $\Vus$ and $|V_{cd}|$ being of order $0.2$, the elements
$\Vcb$ and $\Vts$ of order $4\cdot 10^{-2}$ whereas $|V_{ub}|$ and
$\Vtd$ are of order $5\cdot 10^{-3}$. The Wolfenstein 
parametrization~\cite{Wolfenstein:1983yz} exhibits this hierarchy in a 
transparent 
manner. It is an
approximate parametrization of the CKM matrix in which each element is
expanded as a power series in the small parameter $\lambda\sim |V_{us}|
\approx 0.22$,
\begin{equation}\label{2.75} 
\hat V=
\left(\begin{array}{ccc}
1-{\lambda^2\over 2}&\lambda&A\lambda^3(\varrho-i\eta)\\ -\lambda&
1-{\lambda^2\over 2}&A\lambda^2\\ A\lambda^3(1-\varrho-i\eta)&-A\lambda^2&
1\end{array}\right)
+{\cal{O}}(\lambda^4)\,,
\end{equation}
and the set (\ref{2.73}) is replaced by
\begin{equation}\label{2.76}
\lambda, \qquad A, \qquad \varrho, \qquad {\rm and}~~ \eta \, .
\end{equation}
Because of the smallness of $\lambda$ and the fact that for each element 
the expansion parameter is actually $\lambda^2$, this is a rapidly converging
expansion.

The Wolfenstein parametrization is certainly more transparent than
the standard parametrization. However, if one requires sufficient 
level of accuracy, the terms of ${\cal{O}}(\lambda^4)$ and 
${\cal{O}}(\lambda^5)$ 
have to be included in phenomenological applications.
This can be done in many ways~\cite{Buras:1994ec}. The
point is that since (\ref{2.75}) is only an approximation the {\em exact}
definition of the parameters in (\ref{2.76}) is not unique in terms of the 
neglected order ${\cal O}(\lambda^4)$. 
This situation is familiar from any perturbative expansion, where
different definitions of expansion parameters (coupling constants) 
are possible.
This is also the reason why in different papers in the
literature different ${\cal O}(\lambda^4)$ terms in (\ref{2.75})
 can be found. They simply
correspond to different definitions of the parameters in (\ref{2.76}).
Since the physics does not depend on a particular definition, it
is useful to make a choice for which the transparency of the original
Wolfenstein parametrization is not lost.

In this respect
a useful definition adopted by most authors in the literature 
is to go back to the standard parametrization (\ref{2.72}) and to
 {\it define} the parameters $(\lambda,A,\varrho,\eta)$ 
through ~\cite{Buras:1994ec}
\begin{equation}\label{2.77} 
\lambda\equiv s_{12}\,,
\qquad
A \lambda^2\equiv s_{23}\,,
\qquad
A \lambda^3 (\varrho-i \eta)\equiv s_{13} e^{-i\delta}
\end{equation}
to {\it  all orders} in $\lambda$. 
It follows  that
\begin{equation}\label{2.84} 
\varrho=\frac{s_{13}}{s_{12}s_{23}}\cos\delta,
\qquad
\eta=\frac{s_{13}}{s_{12}s_{23}}\sin\delta.
\end{equation}
The expressions (\ref{2.77}) and (\ref{2.84}) represent simply
the change of variables from (\ref{2.73}) to (\ref{2.76}).
Making this change of variables in the standard parametrization 
(\ref{2.72}) we find the CKM matrix as a function of 
$(\lambda,A,\varrho,\eta)$ which satisfies unitarity exactly.
Expanding next each element in powers of $\lambda$ we recover the
matrix in (\ref{2.75}) and in addition find explicit corrections of
${\cal{O}}(\lambda^4)$ and higher order terms. Including 
${\cal{O}}(\lambda^4)$ 
and ${\cal{O}}(\lambda^5)$ terms we find
\begin{equation}\label{2.775} 
\hat V=
\left(\begin{array}{ccc}
1-\frac{1}{2}\lambda^2-\frac{1}{8}\lambda^4               &
\lambda+{\cal{O}}(\lambda^7)                                   & 
A \lambda^3 (\varrho-i \eta)                              \\
-\lambda+\frac{1}{2} A^2\lambda^5 [1-2 (\varrho+i \eta)]  &
1-\frac{1}{2}\lambda^2-\frac{1}{8}\lambda^4(1+4 A^2)     &
A\lambda^2+{\cal{O}}(\lambda^8)                                \\
A\lambda^3(1-\overline\varrho-i\overline\eta)                       &  
-A\lambda^2+\frac{1}{2}A\lambda^4[1-2 (\varrho+i\eta)]   &
1-\frac{1}{2} A^2\lambda^4                           
\end{array}\right)
\end{equation}
where
\begin{equation}\label{2.88d}
\overline\varrho\simeq \varrho (1-\frac{\lambda^2}{2})+{\cal O}(\lambda^4),
\qquad
\overline\eta=\eta (1-\frac{\lambda^2}{2})+{\cal O}(\lambda^4).
\end{equation}
An all-order definition of $\overline\varrho$ and $\overline\eta$ will be
given in the next section. We emphasize here that by definition the expression for
$V_{ub}$ remains unchanged relative to the original Wolfenstein 
parametrization and the corrections to $V_{us}$ and $V_{cb}$ appear only at
${\cal{O}}(\lambda^7)$ and ${\cal{O}}(\lambda^8)$, respectively.
The advantage of this generalization of the Wolfenstein parametrization
is the absence of relevant corrections to $V_{us}$, $V_{cd}$, $V_{ub}$ and 
$V_{cb}$ and an elegant change in $V_{td}$ which allows a simple connection to the 
Unitarity Triangle parameters, as discussed below. 

\subsubsection{Unitarity Triangle}
\label{sec:intro:UT}
The unitarity of the CKM matrix implies various relations between its
elements. In particular, we have
\begin{equation}\label{2.87h}
V_{ud}^{}V_{ub}^* + V_{cd}^{}V_{cb}^* + V_{td}^{}V_{tb}^* =0.
\end{equation}
Phenomenologically this relation is very interesting as it involves
simultaneously the elements $V_{ub}$, $V_{cb}$ and $V_{td}$ which are
under extensive discussion at present. Other relevant unitarity 
relations will be presented as we proceed. 

The relation (\ref{2.87h})  can be represented as a {\it unitarity triangle} in
the complex  plane. 
The invariance of (\ref{2.87h})  under any phase-transformations
implies that the  corresponding triangle
is rotated in the plane under such transformations. 
Since the angles and the sides
(given by the moduli of the elements of the
mixing matrix)  in this triangle remain unchanged, they
 are phase convention independent and are  physical observables.
Consequently they can be measured directly in suitable experiments.  
One can construct  five additional unitarity 
triangles~\cite{Aleksan:1994if,Silva:1996ih}
corresponding to other orthogonality relations, like the one in (\ref{2.87h}).
Some of them should be useful
when the data on rare and CP violating decays improve.
The areas ($A_{\Delta}$) of all unitarity triangles are equal and 
related to the measure of CP violation 
$J_{\rm CP}$~\cite{Jarlskog:1985ht}:
$\mid J_{\rm CP} \mid = 2\cdot A_{\Delta}$.

\begin{figure}[hbt]
\begin{center}
\includegraphics[width=9cm]{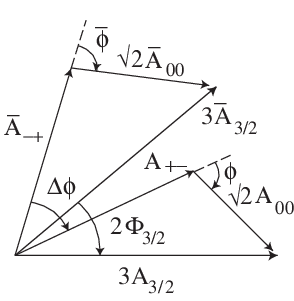}
\end{center}
\vspace{-1cm}
\caption{\it Unitarity Triangle.}
\label{fig:1utriangle}
\end{figure}

The relation (\ref{2.87h}) can be represented as the triangle 
in the complex plane as shown in Fig.~\ref{fig:1utriangle}, where
\begin{equation}\label{2.88}
\overline\varrho+i\overline\eta\equiv\overrightarrow{\mathrm{CA}}=-\frac{V^*_{ub}V_{ud}}{V^*_{cb}V_{cd}}
\end{equation}
and
\begin{eqnarray}
&&\overrightarrow{\mathrm{AB}}=-\frac{V^*_{tb}V_{td}}{V^*_{cb}V_{cd}}=1-\overline\varrho-i\overline\eta\,,\nonumber\\
&&\overrightarrow{\mathrm{CB}}=1\,.
\end{eqnarray}

The parameters $\overline\varrho$ and $\overline\eta$ are the coordinates in the
complex plane of the only non-trivial apex of the Unitarity Triangle. Using their definition
in Eq.~(\ref{2.88}), the exact relation to the parameters $\varrho$ and $\eta$
as given in Eq.~(\ref{2.77}) can be easily found and reads
\begin{equation}\label{2.88b}
\varrho+i\eta=\sqrt{\frac{1-A^2\lambda^4}{1-\lambda^2}} \frac{\overline\varrho+i\overline\eta}
{1-A^2\lambda^4(\overline\varrho+i\overline\eta)}\simeq \left(1+\frac{\lambda^2}{2}\right)
(\overline\varrho+i\overline\eta)+{\cal O}(\lambda^4)\,.
\end{equation}

Phenomenological analyses of the Unitarity Triangles constrain the values of $\overline\varrho$ and
$\overline\eta$. These can be translated to constraints on $\varrho$ and $\eta$ using Eq.~(\ref{2.88b})
and then to the standard parametrization using Eq.~(\ref{2.77}). All recent analyses determine
the $\hat V_{CKM}$ matrix elements in this way, using no expansion whatsoever.

Let us collect useful formulae related to the Unitarity Triangle:
\bi
\item
We can express $\sin(2\alpha_i$), $\alpha_i=
\alpha, \beta, \gamma$, in terms of $(\overline\varrho,\overline\eta)$ using simple
trigonometric formulae:
\begin{equation}\label{2.89}
\sin(2\alpha)=\frac{2\overline\eta(\overline\eta^2+\overline\varrho^2-\overline\varrho)}
  {(\overline\varrho^2+\overline\eta^2)((1-\overline\varrho)^2
  +\overline\eta^2)},  
\end{equation}
\begin{equation}\label{2.90}
\sin(2\beta)=\frac{2\overline\eta(1-\overline\varrho)}{(1-\overline\varrho)^2 + \overline\eta^2},
\end{equation}
 \begin{equation}\label{2.91}
\sin(2\gamma)=\frac{2\overline\varrho\overline\eta}{\overline\varrho^2+\overline\eta^2}\,.
\end{equation}
\item
The lengths of $AC$ and $AB$, denoted by $R_b$ and $R_t$
respectively, are given by
\begin{equation}\label{2.94}
R_b \equiv \frac{\vert V^*_{ub}V_{ud}\vert}{\vert V^*_{cb}V_{cd}\vert}
= \sqrt{\overline\varrho^2 +\overline\eta^2}
\simeq (1-\frac{\lambda^2}{2})\frac{1}{\lambda}
\left| \frac{V_{ub}}{V_{cb}} \right|,
\end{equation}
\begin{equation}\label{2.95}
R_t \equiv \frac{\vert V^*_{tb}V_{td}\vert}{\vert V^*_{cb}V_{cd}\vert} =
 \sqrt{(1-\overline\varrho)^2 +\overline\eta^2}
\simeq\frac{1}{\lambda} \left| \frac{V_{td}}{V_{cb}} \right|\,.
\end{equation}
\item
The unitarity relation (\ref{2.87h}) can be rewritten as
\begin{equation}\label{RbRt}
R_b e^{i\gamma}+R_t e^{-i\beta}=1\,.
\end{equation}
\item
The angle $\alpha$ can be obtained through the relation
\beq\label{e419}
\alpha+\beta+\gamma=\pi\,.
\eeq
\item
In the standard parametrization, the angles $\beta$ and $\gamma$ of the unitarity
triangle are approximately related to the complex phases of the CKM matrix elements
$V_{td}$ and $V_{ub}$ respectively. In particular,
\beq\label{e417}
V_{td}\simeq |V_{td}|e^{-i\beta}\,,\quad V_{ub}\simeq |V_{ub}|e^{-i\gamma}\,.
\eeq
\ei

\subsection{Plan of the report}
The goal of the latest generation of flavor experiments has been not only the measurement of the angles and sides of the unitarity triangles, but the 
measurement of as many redundant observables sensitive to the parameters of the unitarity triangle. On one side in fact the consistency of this plethora of 
measurements is a signal that the CP-violation mechanism is fully understood, on the other side possible deviations from the Standard Model would spoil such 
a consistency. Sensitivity to "New Physics" is therefore proportional to the accuracy we are able to achieve on the Unitarity Triangle. Finally, in case New 
Physics is observed, the Standard Model Unitarity Triangle will have to be measured by means of a subset of observables , those that are not influenced by 
New Physics itself, namely tree dominated processes.

In this report, we first describe general theoretical (Sec.~\ref{sec:theoryPrimers}) and experimental (Sec.~\ref{sec:expPrimers}) tools. Next, the single 
measurements are described and averaged whenever possible. In particular Sec.~\ref{sec:cabibbo} discusses the measurements of the Cabibbo Angle, 
Sec.~\ref{sec:sl} the measurement of  $|V_{cx}|$ and $|V_{ub}|$ in semileptonic decays. Rare decays and measurements of  $|V_{td}|$ and $|V_{ub}|$
are detailed in Sec.~\ref{sec:rare}, while Sec.~\ref{sec:formalism} reports on the mixing and lifetime related measurements, including the time-dependent 
measurements of the phases of the mixing diagram, both for $B_d$ and $B_s$ mesons. All other measurements of angles of the Unitarity 
Triangle are described in Sec.~\ref{sec:tree} and~\ref{sec:angles}: the former shows a large number of measurements of the $\gamma$ angle in tree dominated 
processes, while the latter comprises several techniques to measure $\alpha$, $\beta$, and $\gamma$ in charmless $B$ decays.

These measurements are interpreted altogether in Sec.~\ref{section:globalfits}. First the results of global fits to all observables under the assumption that 
there is no deviation from the Standard Model is presented. This fit returns a very accurate measurement of the position of the apex of the unitarity 
triangle. Next, the redundancy of the measurements is exploited to test the possibility of deviations from the Standard Model both in model independent 
frames and under specific New Physics scenarios.

\section{Theory Primers}
\label{sec:theoryPrimers}
This section contains the description of theretical tools that are common to different fields of flavor physics and that will therefore be used as starting 
point in the subsequent sections.

\subsection{Effective Weak Hamiltonians}
\label{sec:ope}
Flavor-changing hadron transitions are multi-scale processes conveniently
studied using the operator product expansion (OPE)~\cite{Wilson:1969zs,Wilson:1970ag}.
They involve at least two different energy scales: the electroweak scale,
given for instance by the $W$ boson mass $M_W$, relevant for the flavor-changing
weak transition, and the scale of strong interactions $\Lambda_\mathrm{QCD}$,
related to the hadron formation.
Using the OPE, these processes can be described by effective weak Hamiltonians
where the $W$ boson and all heavier particles are eliminated as dynamical degrees
of freedom from the
theory~\cite{Gaillard:1974nj,Altarelli:1974exa,Witten:1976kx,Gilman:1979bc,Guberina:1979ix}.
These Hamiltonians are given by the first term of an expansion in renormalized
local operators of increasing dimensions suppressed by inverse powers of the
heavy scale.

The OPE realizes the scale separation between short-distance (high-energy) and
long-distance (low-energy) physics. The scale $\mu$ at which the local operators
are renormalized sets the threshold between the two regimes.
The effect of particles heavier than $M_W$ enters only through the Wilson
coefficients, namely the effective couplings multiplying the operators of the Hamiltonian.
Short-distance strong-interaction effects are also contained in the Wilson coefficients
and can be computed using renormalization-group improved perturbation theory. Indeed,
Wilson coefficients obey a renormalization group equation (RGE) allowing to resum large
logs of the form $\alpha_s(\mu)^{n+m}\log(M_W/\mu)^n$ to all orders in $n$.
The leading order (LO) resummation corresponds to $m=0$, the next-to-leading order (NLO)
one to $m=1$, and so on. Since the Wilson coefficients depend on short distance
physics only, they behave as effective couplings in the Hamiltonians. They
can be calculated once and for all, i.e. for any external state used to compute
the Hamiltonian matrix elements. Indeed, the complete definition of an effective weak Hamiltonian
requires the choice of the operators and the computation of the corresponding Wilson coefficients.

The dependence on external states, as well as long-distance strong-interaction effects, is
included in the hadronic matrix elements of the local operators and must be evaluated with a
non-perturbative technique (lattice QCD, QCD sum rules, QCDF, SCET, etc.). As non-perturbative
methods can typically compute matrix elements of local operators, this is
a major motivation for using the effective weak Hamiltonians.

We now illustrate the procedure to define the effective weak Hamiltonians and to compute the
Wilson coefficients discussing the case of $\Delta F=1$ transitions, namely processes where 
the quark flavor quantum numbers change by one unit.

The starting point is a generic $S$ matrix element given by the $T$-product of two weak
charged currents computed in the Standard Model (in the following called {\it full} theory
to distinguish it from the {\it effective} theory defined by the effective weak Hamiltonian)
\begin{equation}
\langle F \vert S \vert I \rangle=\int d^4x D^{\mu \nu} \left( x, M_W \right) \langle
F \vert T \Bigl( J_\mu^\mathrm{cc}(x),J_\nu^{cc\,\dagger}(0) \Bigr) \vert I \rangle\,,
\label{eq:Smat}
\end{equation}
where $\langle F \vert$ and $\vert I \rangle$ are the generic final and initial states
and
\begin{equation}
J_\mu^\mathrm{cc}(x)=\frac{g}{\sqrt{2}}\sum_{j=1}^3\left[\left(\sum_{i=1}^2 V_{u^i d^j} \bar u_L^i(x)\gamma_\mu d_L^j(x)\right)
+\bar e^j_L(x) \gamma_\mu \nu_L^j(x)\right]\,,
\end{equation}
where $V$ is the Cabibbo-Kobayashi-Maskawa (CKM) matrix~\cite{Cabibbo:1963yz,Kobayashi:1973fv},
$u^i=\{u,c\}$\footnote{The top
quark is not included as we are building an effective theory valid for energies below $M_W$.},
$d^i=\{d,s,b\}$, $e^i=\{e,\mu,\tau\}$, $\nu^i=\{\nu_e,\nu_\mu,\nu_\tau\}$ and the subscript $L$
denotes the left-handed component of the field.

Given that, using for instance the Feynman gauge,
\begin{equation}
D^{\mu \nu} \left( x, M_W \right)=\int \frac{d^4 q}{(2\pi)^4} e^{-i q\cdot x} \frac{-g^{\mu\nu}}
{q^2-M_W^2+i\varepsilon}=\delta(x) \frac{g^{\mu\nu}}{M_W^2}+\dots\,,
\end{equation}
the two weak currents go at short distances in the large $M_W$ limit. Thus the $S$ matrix
element can be expanded in terms of local operators and gives
\begin{equation}
\langle F  \vert i S \vert I \rangle =4\frac{G_F}{\sqrt{2}}
\sum_i C_i (\mu) \langle F \vert Q_i(\mu) \vert I\rangle + \dots\,, \label{due}
\end{equation}
where $G_F$ is the Fermi constant $G_F/\sqrt{2}=g^2/8 M_W^2$. The dots represent subdominant terms
suppressed by powers of $Q^2/ M_W^2$ where $Q$ is the typical energy scale of the process under
study ($\Lambda_\mathrm{QCD}$ for light hadron decays, $m_b$ for $B$ decays, etc.).

The OPE in Eq.~(\ref{due}) is valid for all possible initial and final states. This allows
for the definition of the effective weak Hamiltonian, given by the operator relation
\begin{equation}
{\cal H}_W^{\Delta F=1}=4\frac{G_F}{\sqrt{2}}  \sum_i C_i (\mu)  Q_i(\mu)=
4\frac{G_F}{\sqrt{2}} \vec Q^T(\mu) \cdot \vec C(\mu) \, .\label{eq:effH}
\end{equation}
The $Q_i(\mu)$ are local, dimension-six operators renormalized at the scale $\mu$ and the $C_i(\mu)$
are the corresponding Wilson coefficients. The set of operators $Q_i(\mu)$ forms a complete basis for
the OPE. This set contains all the linearly-independent, dimension-six operators with the same
quantum numbers of the original weak current product, usually reduced by means of the equations of motion
(although off-shell basis can also be considered). In practice, the operators generated by the
expansion of the {\it full} amplitude (in the so-called ``matching'' procedure described below)
must be complemented by the additional operators generated by the renormalization procedure.
Notice that, in the absence of QCD (and QED) corrections, the effective Hamiltonian in Eq.~(\ref{eq:effH}) reduces
to the Fermi theory of weak interactions. For instance, from the leptonic part of the charged currents, one finds
\begin{equation}
{\cal H}_\mathrm{Fermi}=\frac{G_F}{\sqrt{2}} \bar e\gamma^\mu(1-\gamma_5)\nu_e \bar \nu_\mu\gamma_\mu(1-\gamma_5) \mu\,,
\end{equation}
i.e. the Fermi Hamiltonian describing the muon decay.

For quark transitions, gluonic (and photonic) radiative corrections to amplitudes
computed in terms of local operators produce ultraviolet divergences which are not present in the
full theory. This implies that the local operators $Q_i$ need to be renormalized and depend on the
renormalization scale $\mu$. Therefore $\mu$-dependent Wilson coefficients must be introduced to
cancel this dependence.

Provided that one choses a large enough renormalization scale $\mu \gg \Lambda_\mathrm{QCD}$,
short-distance QCD (and QED) corrections to the Wilson coefficients can be calculated
using a renormalization-group-improved perturbation theory, resumming classes of large logs potentially dangerous
for the perturbative expansion. All non-perturbative effects are confined in the matrix
elements of the local operators. Their calculation requires a non-perturbative technique able
to compute matrix elements of operators renormalized at the scale $\mu$.
In the case of leptonic and semi-leptonic hadron decays, the hadronic effects are confined to the
matrix elements of a single current which can be conveniently written using meson decay constants
(for matrix elements between one hadron and the vacuum) or form factors (for matrix elements between
two hadron states) as for example
\begin{eqnarray}
&&\langle 0 \vert {\bar d}_L \gamma^\mu\gamma_5 u_L \vert \pi^+(q)\rangle = i f_\pi q^\mu\,, \nonumber\\
&&\langle \pi^0(p^\prime) \vert \bar s_L \gamma^\mu d_l \vert K^0(p)\rangle = f^0_+(q^2) (p+p^\prime)^\mu+
f^0_-(q^2)(p-p^\prime)^\mu\,, \quad q^2=(p-p^\prime)^2\,.
\end{eqnarray}
Appearing in different processes, they can be computed using non-perturbative techniques or
measured in one process and used to predict the others.  Predictions for non-leptonic decays, on the other
hand, usually require non-perturbative calculations. Data-driven strategies are possible in
cases where many measurements related by flavor symmetries are available.

The determination of Wilson coefficients at a given order in perturbation theory requires
two steps: ({\it i}) the {\it matching} between the full theory and the effective Hamiltonian 
at a scale $M\sim O(M_W)$ and ({\it ii}) the RGE evolution from the matching scale $M$ down
to the renormalization scale $\mu$.

Let's discuss the second point first. Since ${\cal H} _W^{\Delta F=1}$ in Eq. (\ref{eq:effH}) is
independent of $\mu$, i.e. $\mu^2\frac{d}{d\mu^2}{\cal H} _W^{\Delta F=1}=0$, the Wilson
coefficients $\vec C(\mu)=\left(C_1(\mu), C_2(\mu), \dots\right)$ must satisfy the RGE
\begin{equation}
\mu^2\frac{d}{d\mu^2}\vec C(\mu)=\frac{1}{2}\hat\gamma^T\vec C(\mu)\, ,
\end{equation}
which can be conveniently written as
\begin{equation}
\left(\mu^2\frac{\partial}{\partial\mu^2}+\beta(\alpha_s)
\frac{\partial}{\partial\alpha_s}-\frac{1}{2}\hat\gamma^T(\alpha_s)\right)
\vec C(\mu)=0\, ,
\label{eq:rge}
\end{equation}
where
\begin{equation}
\beta(\alpha_s)=\mu^2\frac{d\alpha_s}{d\mu^2}
\end{equation}
is the QCD $\beta$ function and
\begin{equation}
\hat\gamma(\alpha_s)=2\hat Z^{-1}\mu^2\frac{d}{d\mu^2}\hat Z
\end{equation}
is the operator anomalous dimension matrix. The matrix $\hat Z$
of the renormalization constants is defined by the relation connecting the bare
operators $\vec Q^B$ to the renormalized ones $\vec Q(\mu)$
\begin{equation}
\vec Q(\mu)=\hat Z^{-1}(\mu, \alpha_s) \vec Q^B\,.
\end{equation}

The solution of the system of linear differential equations (\ref{eq:rge}) is found by introducing a
suitable evolution matrix $U(\mu,M_W)$ and by imposing an appropriate set of initial
conditions, usually called matching conditions. The coefficients $\vec C(\mu)$ are given
by\footnote{ The problem of the thresholds due to the presence of heavy quarks with a mass
$M_W \gg m_Q \gg \Lambda_{{\rm QCD}}$ will be discussed below.}
\begin{equation}
\vec C(\mu)=\hat U(\mu,M)\vec C(M)\, ,
\label{eq:cmu}
\end{equation}
with
\begin{equation}
\hat U(m_1,m_2)=T_{\alpha_s}\exp\left(\int_{\alpha_s(m_1)}^{\alpha_s(m_2)}
\frac{d\alpha_s}{\beta(\alpha_s)}\hat\gamma^T(\alpha_s)\right) \, .
\label{eq:solU}
\end{equation}
$T_{\alpha_s}$ is the ordered product with increasing couplings from right to left.

The matching conditions are found by imposing that, at $\mu=M\sim O(M_W)$,
the matrix elements of the original $T$-product of the currents coincide,
up to terms suppressed by inverse powers of $M_W$, with the corresponding
matrix elements of ${\cal H}_W^{\Delta F=1}$. To this end, we introduce the vector
$\vec T$ defined by the relation
\begin{equation}
i\langle \alpha\vert S \vert \beta \rangle = 4\frac{G_F}{\sqrt{2}}\langle \alpha\vert
\vec Q^T \vert \beta \rangle_0 \cdot \vec T(M_W,m_t;\alpha_s) + \dots
\label{eq:matcfull}
\end{equation}
where $\langle \alpha\vert \vec Q^T \vert \beta \rangle _0$ are matrix elements
of the operators computed at the tree level and the dots denote power-suppressed terms.
The vector $\vec T$ contains the dependence on heavy masses and has a perturbative expansion
in $\alpha_s$\footnote{For simplicity, we discuss QCD corrections only. QED corrections
can be considered as well and are included in a similar way.}
On dimensional basis, $\vec T$ can only be a function of $m_t/M_W$ and of $\log(p^2/M_W^2)$
where $p$ generically denotes the external momenta.

We also introduce the matrix $\hat M(\mu)$ such that
\begin{eqnarray}
\langle\alpha\vert{\cal H}_W^{\Delta F=1}\vert\beta\rangle
&=&4\frac{G_F}{\sqrt{2}} \langle \alpha\vert
\vec Q^T(\mu) \vert \beta \rangle \vec C(\mu) \nonumber\\
&=&4\frac{G_F}{\sqrt{2}}\langle \alpha\vert
\vec Q^T \vert \beta \rangle_0\hat M^T(\mu;\alpha_s)\vec C(\mu)\, .
\label{eq:matceff}
\end{eqnarray}
In terms of $\vec T$ and $\hat M$, the matching condition
\begin{equation}
i\langle \alpha\vert S \vert \beta \rangle=
\langle \alpha\vert {\cal H}_W^{\Delta F=1}\vert \beta \rangle \,
\end{equation}
fixes the value of the Wilson coefficients at the scale $M$ as
\begin{equation}
\vec C(M)=[\hat M^T(M;\alpha_s)]^{-1}\vec T(M_W,m_t;\alpha_s)\, .
\label{eq:inicond}
\end{equation}
As the full and the effective theories share the same infrared behavior, the dependence
on the external states on which the matching conditions are imposed drops in
Eq.~(\ref{eq:inicond}), so that any matrix element can be used, even off-shell ones
(with some caution), provided the {\it same} external states are used for computing
matrix elements in both theories.
Notice that the matching can be imposed at any scale $M$ such that large logs do not
appear in the calculation of the Wilson coefficients at that scale, i.e. 
$\alpha_s \log(M/M_W) \ll 1$.

Equation~(\ref{eq:cmu}) is correct if no threshold corresponding to a quark mass
between $\mu$ and $M_W$ is present. Indeed, as $\alpha_s$, $\hat\gamma$ and
$\beta(\alpha_s)$ depend on the number of ``active'' flavors, it is necessary to change
the evolution matrix $\hat U$ defined in Eq.~(\ref{eq:solU}), when passing quark thresholds.
The general case then corresponds to a sequence of effective theories with a decreasing number
of ``active'' flavors. By ``active'' flavor, we mean a dynamical massless ($\mu \gg m_q$)
quark field. The theory with $k$ ``active'' flavors is matched to the one with $k+1$
``active'' flavors at the threshold. This procedure changes the solution for the Wilson
coefficients. For instance, if one starts with five ``active'' flavors at the scale $M_W$ and
chooses $m_c \ll \mu \ll m_b$, the Wilson coefficients become
\begin{equation}
\vec C(\mu)= \W[\mu,M_W] \vec C(M_W)
= \hat U_4(\mu,m_b) \hat T_{45} \hat U_5(m_b,M_W)\vec C(M_W)\, .
\label{ric}
\end{equation}
The matrix $\hat T_{45}$ matches the four and five flavor theories so that the Wilson
coefficients are continuous across the threshold.  The inclusion of the charm threshold proceeds
along the same lines.

So far we have presented the formal solution of the matching and the RGE for the Wilson coefficients.
In practice, we can calculate the relevant functions ($\beta$, $\hat\gamma$, $\hat M$, $\vec T$, etc.)
in perturbation theory only. At the LO, one has
\begin{equation}
\beta(\alpha_s)=-\frac{\alpha_s^2}{4\pi}\beta_0+\dots\,,\quad
\hat\gamma=\frac{\alpha_s}{4\pi}\hat\gamma^{(0)}+\dots \quad 
\vec T=T^{(0)}+\dots\,,\quad \hat M=\hat 1+\dots\,,
\end{equation}
so that the LO Wilson coefficients read
\begin{equation}
\vec C_\mathrm{LO}(\mu)= \left(\frac{\alpha_s(M)}{\alpha_s(\mu)}\right)^{\hat\gamma^{(0)T}/2\beta_0}
\vec T^{0}\,.
\label{eq:WCLO}
\end{equation}
The explicit solution can be found in the basis where the LO anomalous dimension matrix $\hat\gamma^{(0)}$
is diagonal. To go beyond the LO, we have to expand the relevant functions to higher order in $\alpha_s$. 
Discussing the details on higher order calculations goes beyond the purpose of this primer. They can be
found in the original literature cited in the following presentation of the actual effective Hamiltonians
for $\Delta F=1$ and $\Delta F=2$ transitions.

\subsubsection{$\Delta F=1$ effective weak Hamiltionians}
\label{sec:effweak}
\begin{figure}[t]
 \centering
 \includegraphics[width=0.6\textwidth]{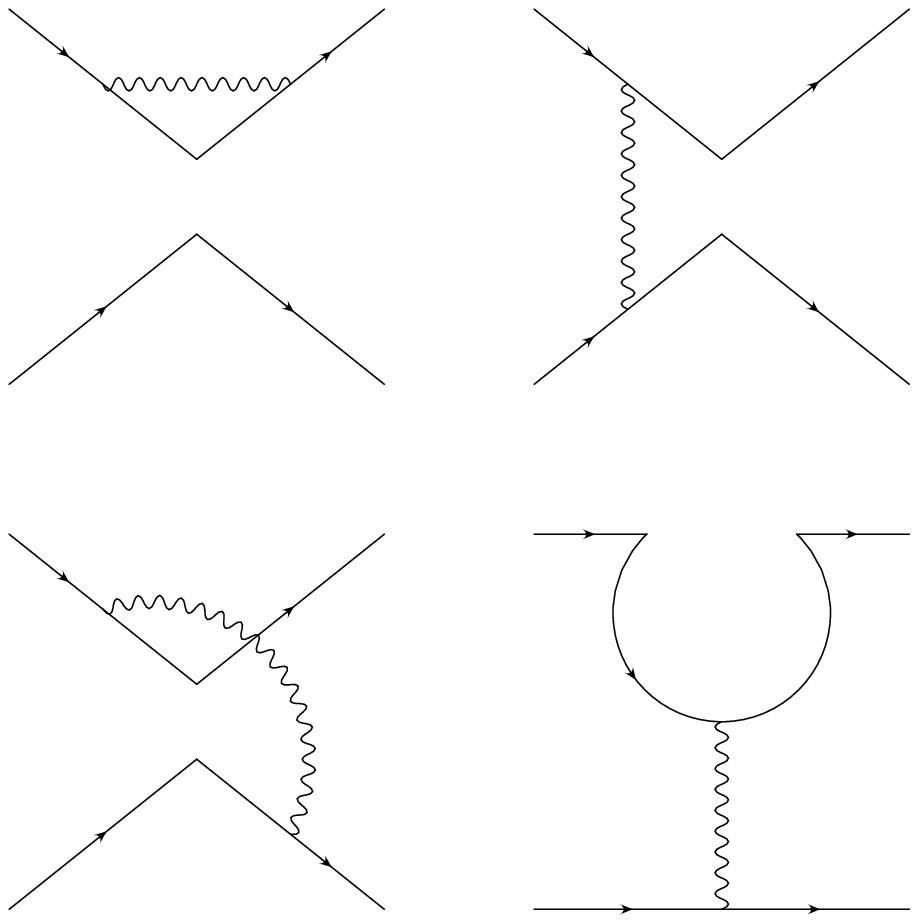}
 \caption{One-loop correction to the $\Delta F=1$ effective weak Hamiltonian.}
 \label{fig:oneloop}
\end{figure}

Even restricting to processes which change each flavor number by no more than one unit, namely $\Delta F=1$ transitions, 
several effective Hamiltonians can be introduced. We start considering the Hamiltonian relevant for transtions
with $\Delta B=1$, $\Delta C=0$, $\Delta S=-1$:
\begin{eqnarray}
{\cal H}_W^{\Delta B=1\,, \Delta C=0\,,  \Delta S=-1}&=&4 \frac{G_F}{\sqrt{2}}\Bigl(\lambda_c^s \bigl(C_1(\mu) Q^c_1(\mu)+
C_2(\mu) Q^c_2(\mu)\bigr) \\
&&+\lambda_u^s \bigl(C_1(\mu) Q^u_1(\mu)+C_2(\mu) Q^u_2(\mu)\bigr)-\lambda_t^s \sum_{i=3}^{10} C_i(\mu) Q_i(\mu)\Bigr)\,,\nonumber
\label{eq:hdb1}
\end{eqnarray}
where the $\lambda^{s}_{q}=V^*_{qb} V_{qs}$ and the operator basis is given by
\begin{equation}
\begin{array}{ll}
Q^q_{1} = {\bar b}_L^\alpha\gamma^\mu q_L^\alpha\, {\bar q}_L^\beta\gamma_\mu s_L^\beta &
Q^q_{2} = {\bar b}_L^\alpha\gamma^\mu q_L^\beta\, {\bar q}_L^\beta\gamma_\mu s_L^\alpha \\
Q_{3} = {\bar b}_L^\alpha \gamma^\mu s_L^\alpha\, \sum_q {\bar q}_L^\beta\gamma_\mu q_L^\beta &
Q_{4} = {\bar b}_L^\alpha \gamma^\mu s_L^\beta\, \sum_q {\bar q}_L^\beta\gamma_\mu q_L^\alpha \\
Q_{5} = {\bar b}_L^\alpha \gamma^\mu s_L^\alpha\, \sum_q {\bar q}_R^\beta\gamma_\mu q_R^\beta~~~~~~~~~&
Q_{6} = {\bar b}_L^\alpha \gamma^\mu s_L^\beta\, \sum_q {\bar q}_R^\beta\gamma_\mu q_R^\alpha \\
Q_{7} = \frac{3}{2}{\bar b}_L^\alpha \gamma^\mu s_L^\alpha\, \sum_q e_q {\bar q}_R^\beta\gamma_\mu q_R^\beta &
Q_{8} = \frac{3}{2}{\bar b}_L^\alpha \gamma^\mu s_L^\beta\, \sum_q e_q {\bar q}_R^\beta\gamma_\mu q_R^\alpha \\
Q_{9} = \frac{3}{2}{\bar b}_L^\alpha \gamma^\mu s_L^\alpha\, \sum_q e_q {\bar q}_L^\beta\gamma_\mu q_L^\beta &
Q_{10} =  \frac{3}{2}{\bar b}_L^\alpha \gamma^\mu s_L^\beta\, \sum_q e_q {\bar q}_L^\beta\gamma_\mu q_L^\alpha\\
\end{array}
\label{eq:basis}
\end{equation}
The sum index $q$ runs over the ``active'' flavors, $\alpha$, $\beta$ are color indices and $e_q$ is the
electric charge of the quark $q$. Besides $Q_1$, which come from the matching, the above operators are
generated by gluon and photon exchanges
in the Feynman diagrams of fig.~\ref{fig:oneloop}. In particular, $Q^q_2$ is generated by current--current
diagrams while $Q_3$--$Q_6$ and $Q_7$--$Q_{10}$ are generated by gluon and photon penguin diagrams
respectively. Notice that the choice of the operator basis in not unique. Different possibilities have
been considered in the literature~\cite{Vainshtein:1975sv,Shifman:1975tn,Shifman:1976ge,Gilman:1982ap,Bijnens:1983ye,Lusignoli:1988fz,Chetyrkin:1996vx}.

The operators basis includes the ten independent operators in Eq.~(\ref{eq:basis}) in the five-flavor effective theory.
Below the bottom threshold, the following relation holds
\begin{equation}
Q_{10}-Q_{9}-Q_4+Q_3=0\, ,
\label{eq:q10}
\end{equation}
so that the independent operators become nine. The basis is further reduced in the
three-flavor theory, i.e. below the charm threshold, due to the additional relations
\begin{equation}
Q_4-Q_3-Q_2+Q_1=0\,,\qquad Q_9-\frac{3}{2}Q_1+\frac{1}{2}Q_3=0\, .
\end{equation}

For $b\to s$ transitions with a photon or a lepton pair in the final state, additional
dimension-six operators must be included in the basis, namely
\begin{eqnarray}
Q_{7\gamma} &=& \frac{e}{16\pi^2} m_b {\bar b}_L^\alpha\sigma^{\mu\nu} F_{\mu\nu} s_L^\alpha
   \nonumber\\
Q_{8g} &=& \frac{g_s}{16\pi^2} m_b {\bar b}_L^\alpha\sigma^{\mu\nu} G_{\mu\nu}^A T^A s_L^\alpha
   \nonumber\\
Q_{9V} &=& \frac{1}{2}{\bar b}_L^\alpha \gamma^\mu s_L^\alpha\, \bar l \gamma_\mu l
   \nonumber\\
Q_{10A} &=& \frac{1}{2}{\bar b}_L^\alpha \gamma^\mu s_L^\alpha\, \bar l \gamma_\mu\gamma_5 l
\label{eq:radbasis}
\end{eqnarray}
where $G^A_{\mu\nu}$ ($F_{\mu\nu}$) is the gluon (photon) field strength tensor and $T^A$ are the 
$SU(3)$ generators.
They contribute an additional term to the Hamiltonian in Eq.~(\ref{eq:hdb1}) so that, up to doubly Cabibbo-suppressed
terms and neglecting the electroweak penguin operators $Q_7$--$Q_{10}$, the effective weak Hamiltonian for these processes
reads
\begin{eqnarray}
{\cal H}_W &=& -4\frac{G_F}{\sqrt{2}} \lambda_t^s \Bigl(\sum_{i=1}^6 C_i(\mu)Q_i(\mu)+C_{7\gamma}(\mu)Q_{7\gamma}(\mu)
+C_{8g}(\mu)Q_{8g}(\mu)\nonumber\\
&&+C_{9V}(\mu)Q_{9V}(\mu)+C_{10A}(\mu)Q_{10A}(\mu)\Bigr)\,,
\label{eq:radh}
\end{eqnarray}
with $Q_{1,2}=Q_{1,2}^c$ defined in Eq.~(\ref{eq:basis}).

At present, the $\Delta F=1$ effective weak Hamiltonian in Eq.~\ref{eq:hdb1}, including electroweak penguin 
operators ($Q_7$--$Q_{10}$ in Eq.~(\ref{eq:basis})), is known at the NNLO in $\alpha_s$~\cite{Gorbahn:2004my}
and at the NLO in $\alpha_e$~\cite{Buras:1992zv,Ciuchini:1993vr}.
The effective Hamiltonian in Eq.~(\ref{eq:radh}) has been fully computed at the NNLO in the strong coupling
constant~\cite{Bobeth:1999mk,Misiak:2004ew,Gorbahn:2005sa,Czakon:2006ss}.

Effective weak Hamiltonians for other transtions can be obtained by trivial changes in the quark fields and
in the CKM matrix elements entering eqs.~(\ref{eq:hdb1}) and (\ref{eq:basis}). In particular
\begin{eqnarray}
 \Delta B=1\,,\Delta C=0\,,\Delta S=0 &:& s \to d \nonumber\\
 \Delta B=0\,,\Delta C=0\,,\Delta S=1 &:& b \to s\,, s \to d \nonumber\\
 \Delta B=0\,,\Delta C=1\,,\Delta S=0 &:& b \to c\,, s \to u\,, c \to s\,, u \to d\,.
\end{eqnarray}

In other cases, for instance $\Delta B=1$, $\Delta C=-1$, $\Delta S=0$ transitions, the Hamiltonian has a
simpler structure, namely
\begin{equation}
 {\cal H}_W^{\Delta B=1,\, \Delta C=-1,\, \Delta S=0}=4 \frac{G_F}{\sqrt{2}}
 V_{cb}^*V_{ud} \Bigl(C_1(\mu) Q'_1(\mu)+C_2(\mu) Q'_2(\mu)\Bigr)
\label{eq:hdb12}
\end{equation}
with
\begin{equation}
 Q'_{1} = {\bar b}_L^\alpha\gamma^\mu c_L^\alpha\, {\bar u}_L^\beta\gamma_\mu d_L^\beta,\quad
 Q'_{2} = {\bar b}_L^\alpha\gamma^\mu c_L^\beta\, {\bar u}_L^\beta\gamma_\mu d_L^\alpha\,.
\label{eq:basis2}
\end{equation}
Only current--current operators enter this Hamiltonian. Penguin operators are not generated as the considered transitions
involve four different flavors. Other Hamiltonians share this feature and can be obtained from eqs.~(\ref{eq:hdb12}) and
(\ref{eq:basis2}) with the following replacements
\begin{eqnarray}
 \Delta B=1\,,\Delta C=1\,,\Delta S=0 &:& c \to u\,, u \to c \nonumber\\
 \Delta B=1\,,\Delta C=-1\,,\Delta S=-1 &:& d \to s \nonumber\\
 \Delta B=1\,,\Delta C=1\,,\Delta S=-1 &:& c \to u\,, u \to c\,, d \to s \nonumber\\
 \Delta B=0\,,\Delta C=-1\,,\Delta S=1 &:& b \to s \nonumber\\
 \Delta B=0\,,\Delta C=1\,,\Delta S=1 &:& b \to s\,, c \to u\,, u \to c\,.
\end{eqnarray}
Clearly the (omitted) Hermitean-conjugate terms in the Hamiltonians mediate transitions with opposite $\Delta F$. 

Notice that physics beyond the SM could change not only the Wilson coefficients through
the matching conditions, but also the operator basis where new spinor and color structures may appear.
Indeed the most general $\Delta F=1$ basis contains a large number of operators making it hardly useful.
On the other hand, a possible definition of the class of new physics models with minimal flavor violation
is that these models produce only real corrections to the SM Wilson coefficients
without changing the operator basis of the effective weak Hamitonian~\cite{Buras:2000dm}.

\subsubsection{$\Delta F=2$ effective weak Hamiltionians}
The $\Delta F=2$ effective weak Hamiltionians are simpler than the $\Delta F=1$ ones. In the SM, the operator
basis includes one operator only. For example, the $\Delta S=2$ effective Hamiltonian is commonly written as
\begin{equation}
{\cal H}_W^{\Delta S=2}=\frac{G_F^2}{4\pi^2}M_W^2\Bigl(\lambda_c^2\eta_1 S_0(x_c)+\lambda_t^2\eta_2 S_0(x_t)+
\lambda_t\lambda_c\eta_3 S_0(x_t,x_c) \Bigr) \hat Q_s
\label{eq:ds2}
\end{equation}
where $\lambda_q=V_{qs}^*V_{qd}$, the functions $S_0$ of $x_q=m_q^2/M_W^2$ come from the LO matching conditions,
the coefficents $\eta_i$ account for the RGE running and NLO effects. Starting from the dimension-six operator
\begin{equation}
Q_s=\bar s_L \gamma_\mu d_L\, \bar s_L \gamma^\mu d_L\,.
\end{equation}
$\hat Q_s$ is defined as $\hat Q_s=K(\mu) Q_s(\mu)$, where $K(\mu)$ is the appropriate
short-distance factor which makes $\hat Q$ independent of $\mu$~\cite{Buras:1983ap}.
The matrix element of this operator between $K^0$ and $\bar K^0$ is
parameterised in terms of the RG-invariant bag parameter ${\hat B}_K$ (see Sec.~\ref{sec:formalism}).

The Hamiltonian in Eq.~(\ref{eq:ds2}) describes only the short-distance part of
the $\Delta S=2$ amplitude. Long-distance contributions generated by the exchange of hadronic states are
also present. These contributions break the OPE producing additional terms which are diffcult to estimate.
This is the case of the $K^0$--$\bar K^0$ mass difference $\Delta M_K$ which therefore cannot be
reliably predicted. On the other hand, the CP-violation parameter $\eps_K$, related to
Im$\langle \bar K^0 \vert{\cal H}_W^{\Delta S=2} \vert K^0\rangle$, is short-distance dominated and thus
calculable.

Concerning $\Delta B=2$ transitions, namely the $B^0_d$--$\bar B^0_d$ and $B^0_s$--$\bar B^0_s$ mixing amplitudes,
virtual top exchange gives the dominant contributions in the SM. Therefore these amplitudes are short-distance
dominated and described by matrix elements of the Hamiltonian
\begin{equation}
{\cal H}_W^{\Delta B=2}=\frac{G_F^2}{4\pi^2}M_W^2 (\lambda_t^q)^2\eta_2 S_0(x_t)\hat Q_b^q
\label{eq:db2}
\end{equation}
where
\begin{equation}
Q_b^q=\bar b_L \gamma_\mu q_L\, \bar b_L \gamma^\mu q_L\,,\qquad q=\{d,s\}\,,
\end{equation}
and $\hat Q_b^q$ is defined similarly to the $\Delta S=2$ case in terms of the bag-parameter
$\hat Q_b^q$ (see Sec.~\ref{sec:formalism}).

At present, $\Delta F=2$ effective Hamiltonians are known at the NLO in the strong coupling
constants~\cite{Buras:1990fn,Herrlich:1993yv,Herrlich:1996vf}.

It is worth noting that, unlike $\Delta F=1$ Hamiltonians, generic new physics contributions to $\Delta F=2$
transitions generate few additional operators allowing for model-independent studies of $\Delta F=2$ 
processes where the Wilson coefficients at the matching scale are used as new physics
parameters~\cite{Bona:2007vi}.

Finally, we mention that the absorbtive part of $\Delta F=2$ amplitudes, related to the neutral mesons width
differences, can also be calculated using an OPE applied to the rates rather than to the amplitudes. We
refer the interested reader to Sec.~\ref{sec:formalism}. for details on this calculation.

\subsection{Factorization}
\label{sec:hqet-scet}

In the previous section it was shown how to integrate out physics at the
electroweak scale, resulting in 10 four-fermion operators $O_{1}$-$O_{10}$. In order to measure 
the decay rates or CP-asymmetries in non-leptonic decays of a $B$ meson to 
two light pseudoscalar mesons (either $\pi$ or $K$), one needs
information about the matrix elements of these operators between the initial 
$B$ meson and the given final state. The nature of the strong interaction implies
that these matrix elements can not be calculated perturbatively, and one either 
has to resort to non-perturbative methods to calculate these matrix elements or
extract them from data. 

In order to determine the required matrix elements from data and still obtain information 
about the electroweak physics requires to have more experimental input than unknown 
matrix elements. It has been known for a long time that in the  $B \to \pi \pi$ system 
there are more measurements than 
non-perturbative parameters, which allows to measure some fundamental parameters
of the CKM matrix~\cite{Gronau:1990ka}. However, of the 8 possible measurements, only 6
have been made to this point, one of which still has very large uncertaities. Thus,
in practice, even in the $\pi \pi$ system some additional information is required
in order to have detailed information about the electroweak phases. The situation is worse 
once we include Kaons in the final state, and without using additional theoretical information, 
there are more unknown parameters than there are measurements. 

Factorization utilizes an expansion in $\Lambda_{\rm QCD}/m_b$ in order to 
simplify the required matrix elements, resulting in new relations in the limit
$\Lambda_{\rm QCD}/m_b \to 0$. 
Theoretically, this limit can be taken using diagrammatic factorization
techniques (QCD factorization)~\cite{Beneke:1999br,Beneke:2000ry,Beneke:2001ev}
 or, equivalently, soft-collinear effective theory (SCET)~\cite{Bauer:2000ew,Bauer:2000yr,Bauer:2001ct,Bauer:2001yt}, 
together with heavy quark effective
theory (HQET). Before detailing how the factorization theorems arise in the
effective field theory approach, we give a simple physical picture of 
factorization, known as color transparency. 

As discussed above, the decay $B \to M_1 M_2$ is described 
by the matrix elements of local four-fermion operators, allowing the  $b$ quark to decay to three
light quarks. Two of these quarks will form the meson $M_1$, while the meson 
$M_2$ is formed from the third light quark together with the spectator quark of the $B$ meson. 
The dominant contribution to a given decay arises from operators for which the two light quark 
forming $M_1$  are in a color singlet configuration.These two quarks in a  color singlet configuration will 
 only interact non-perturbatively
with the remaining system once their separation is of order 
$1/\Lambda_{\rm QCD}$. Due to the large energy $E \sim m_b/2$ of the light mesons, 
this separation only occurs when the two quarks are a distance 
$d \sim E_\pi/\Lambda_{\rm QCD}^2$ from the origin of the decay, and therefore out of the reach 
$d \sim 1/\Lambda_{\rm QCD}$ of the 
non-perturbative physics of the $B$ meson. Thus, the non-perturbative dynamics of 
one of the two mesons is independent of the rest of the system. Since the second light 
meson requires the spectator quark of the $B$ meson, no such factorization should be expected. 

Using effective field theory methods allows to prove this intuitive result rigorously,
while at the same time allowing in principle to go beyond the leading order result in 
$\Lambda_{\rm QCD}/m_b$. The first step in the factorization proof is to separate the different 
energy scales in the system, by constructing the correct effective field theory. In the rest frame of the $B$ meson, the two light mesons 
decay back-to-back with energy $m_B/2$, and we label the directions of the two mesons by four-vectors $n$ and $\bar n$. To describe these two energetic mesons we require 
collinear quark and gluon fields which are labeled by the direction of flight $n$ or $\bar n$ of the meson. We will call the collinear quark fields $\chi_{n/\bar n}$ and $A_{n/\bar n}$, respectively. In order to describe the heavy $B$ meson, we require soft heavy quark and soft light quark and gluon fields, which we call $h_s$, $q_s$ and $A_s$, respectively. Since it is the two light quarks in the $n$ direction that form the meson $M_1$, we will also write
$M_n \equiv M_1$ and $M_{\bar n} \equiv M_2$. 

The important property of SCET/HQET that allows to prove the factorization theorem is that to leading order in $\Lambda_{\rm QCD}/m_b$ the collinear fields in the different directions do not interact with one another. 
Furthermore, all interactions between collinear and soft fields can be removed from the Lagrangian by 
redefining the collinear fields to be multiplied by a soft Wilson line $Y_n$, which depends on the direction $n$ of the collinear field it belongs to. Since all interactions between the different sectors disappear at leading order, the Lagrangian can be written as
\begin{equation}
{\cal L}_{\rm eff} = {\cal L}_n + {\cal L}_{\bar n} + {\cal L}_s + {\cal O}(\Lambda_{\rm QCD}/m_b)\,.
\end{equation}
The 4-quark operators $O_i$ describing the decay of the heavy $b$ quark are matched onto 
operators in the effective field theory, which are constructed out of the collinear and soft fields. This 
allows to write written as
\begin{equation}
O_i = C_i \otimes O_i^{n \bar n} = C_i \otimes  \big[\bar h_s \Gamma_i Y_{\bar n} \chi_{\bar n} \big] \, \big[  \bar \chi_n Y^\dagger_{n} \Gamma_i Y_n \chi_{n} \big]\,.
\end{equation}
Here $C_i$ denotes the Wilson coefficient of the operators and describes the physics occurring at the scale
$m_b$, and the different operators are distinguished by their Dirac and color structure $\Gamma_i$. The symbol $\otimes$ denotes a convolution between the Wilson coefficients and operators, which is due to the fact that 
the Wilson coefficients can depend on the large energies of the light quarks. Note that if the two collinear quarks in the $n$ direction form a color singlet (meaning $\Gamma_i$ is color singlet), then we can use the unitarity of Wilson lines $Y^\dagger_n Y_n = 1$ to write
\begin{equation}
O_i = C_i \otimes  \big[\bar h_s \Gamma_i Y_{\bar n} \chi_{\bar n} \big] \, \big[  \bar \chi_n \Gamma_i \chi_n \big]\,.
\end{equation}
Since the Wilson lines $Y_n$ describe the coupling of the collinear fields $\chi_n$ to the rest of the system, their cancellation is the field theoretical realization of the physical picture given before. 

The absence of interactions between the fields in the $n$ direction from the rest of the system can be used 
to separate the matrix element of the operators $O_i$ as
\begin{eqnarray}
\langle M_n M_{\bar n} | O_i | B_s \rangle &=&  C_i \otimes \langle M_n M_{\bar n} | O_i^{n \bar n} | B_s \rangle = C_i \otimes \langle M_n | \bar \chi_n \Gamma_i \chi_{n}  | 0 \rangle \, \langle M_{\bar n} | h_s \Gamma_iY_{\bar n} \chi_{\bar n} | B_s \rangle \nonumber\\
&=& C_i \otimes \phi_{M_n} \otimes \zeta_{BM_{\bar n}}\,.
\label{firstresult}
\end{eqnarray}
Here $\phi_M$ denotes the light cone distribution function of the meson $M$, while $\zeta_{BM}$ denotes the matrix element describing the $B \to M$ transition. Thus, the matrix element of the required operators factor into a convolution of a perturbatively calculable Wilson coefficient $C_i$, a matrix element describing the $B \to M_2$ transition, as well as the wave function of the meson $M_1$. The wave functions of the light pseudoscalar mesons have been measured in the past and are known relatively well, and some of the  $B \to M_2$ matrix elements can be measured in semileptonic $B$ decays. Thus, much information for the matrix elements of the operators $O_i$ can be measured in other processes, allowing to use the non-leptonic data on to extract information about the weak scale physics. 

 There are several different approaches to understanding factorization and they go by the names QCD Factorization (QCDF)~\cite{Beneke:1999br,Beneke:2000ry,Beneke:2001ev}, 
perturbative QCD (PQCD)~\cite{Li:1994iu,Li:1995jr,Yeh:1997rq,Keum:2000ph,Keum:2000wi,Chen:2001pr} and soft-collinear effective theory (SCET)~\cite{Bauer:2004tj,Bauer:2005kd,Williamson:2006hb} in the literature. All three approaches agree with everything discussed up to this point, and the main differences arises when trying to factorize the matrix elements $\zeta_{BM}$ further. This can be achieved by matching onto a second effective theory which integrates out physics at the scale $\mu_i \sim \sqrt{\Lambda_{\rm QCD} m_b}$, which allows to write 
\begin{equation}
\zeta_{BM} = J \otimes \phi_B \otimes \phi_M\,.
\label{secondfactorization}
\end{equation}
Here $J$ is a matching coefficient that can be calculated perturbatively in an expansion in $\alpha_s(\mu_i)$. A naive calculation of this function $J$ unfortunately leads to a singular convolution with the wave functions $\phi_M$ and $\phi_B$, and it is the resolution of this problem that separates the different approaches. The SCET approach to factorization simply never performs the second step of the factorization theorem and uses directly the results in Eq.~(\ref{firstresult}) but requiring the most experimental information. The PQCD results regulate the singular convolution with an unphysical transverse momentum of the light meson. These results are therefore on less solid theoretical footing, but require the least amount of experimental input. QCDF uses a mixture of both approaches and only uses Eq.~(\ref{secondfactorization}) in cases where no singular convolutions are obtained. Note however, that for power corrections included into QCDF a different logic is used and a new non-perturbative parameter is included to parameterize singular convolutions. 

Besides the differences in the treatment of singular convolutions, there are also differences in how matrix elements of operators containing charm quarks are treated. The theoretical question is whether such contributions can be calculated perturbatively or if they lead to new non-perturbative effects. The SCET approach does not attempt to calculate these matrix elements perturbatively, while QCDF and PQCD do use perturbation theory. The differences between the different approaches are summarized in Tab.~\ref{tab:comparison}.
\begin{table}
\begin{center}
\caption{Comparison of the different approaches to Factorization
\label{tab:comparison}}
\begin{tabular}{l||c|c|c}
&SCET&QCDF&PQCD\\\hline
Expansion in $\alpha_s(\mu_i)$ & No & Yes & Yes \\
Singular convolutions & N/A & New parameters & "Unphysical" $k_T$ \\
Charm Loop & Non-perturbative & Perturbative & Perturbative \\
Number of paramterers & Most & Middle & Least\\ 
\end{tabular}
\end{center}
\end{table}


\subsection{Lattice QCD}
\label{sec:thPrim:LQCD}

The tools explained in the previous two sections are used to separate
the physical scales of flavor physics into the weak scale, the
heavy-quark scale, and the nonperturbative~QCD scale.
At the short distances of the first two, QCD effects can be treated
with perturbation theory, as part of the evaluation of the Wilson
coefficients.
At longer distances, where QCD confines, perturbative QCD breaks down:
to obtain the hadronic matrix elements of the operators, one must 
tackle nonperturbative QCD.

In some cases general features of field theory---symmetry, 
analyticity and unitarity, the renormalization group---are enough.
For example, using the fact that QCD preserves CP one can show that 
the nonperturbative hadronic amplitude drops out of the CP 
asymmetry for a process like $B\to\psi K_S$.
Another set of examples entails using one process to ``measure'' the 
hadronic matrix element, and then using this ``measurement'' in 
other, more intriguing, processes.

In general, however, one would like to compute hadronic matrix 
elements.
The end objective is to see whether new physics lurks at short 
distances, so it is essential that one start with the QCD 
Lagrangian.
Any approach will involve some approximation and compromise---QCD is 
too hard otherwise, so it is just as essential that any uncertainties 
be systematically reducible and under quantifiable control.

One method that has these aims is based on lattice gauge theory,
which provides a mathematically sound definition of the gauge theory.
In QCD, or any quantum field theory, anything of interest can be 
related to a correlation function
\begin{equation}
	\langle O_1(x_1)O_2(x_2)\cdots O_n(x_n) \rangle = \frac{1}{Z}
		\int \prod_{x,\mu}dA_\mu(x) \prod_x d\bar{q}(x)dq(x)\,
		O_1(x_1)O_2(x_2)\cdots O_n(x_n)\, e^{-S},
	\label{eq:lqcd:funcInt}
\end{equation}
where the $O_i(x)$ are local, color singlet operators built out of 
quark fields $q$, antiquark fields $\bar{q}$, and gluon 
fields~$A_\mu$, and $S$ is the classical action.
The normalization factor $Z$ is defined so that $\langle1\rangle=1$.
For brevity, color, flavor, and (for $q$, $\bar{q}$) Dirac 
indices are implied but not written out.
As it stands, Eq.~(\ref{eq:lqcd:funcInt}) requires a definition of the 
products over the continuous spacetime label~$x$.
A~mathematically sound way to do so is to start with a discrete 
spacetime variable, labeling the sites of a four-dimensional 
spacetime lattice.
The idea goes back to Heisenberg, but for QCD and other gauge
theories, the key came when Wilson showed how to incorporate local
gauge invariance with the lattice~\cite{Wilson:1974sk}.
If the lattice has $N_S^3\times L_4$ sites, the spatial size of the 
finite volume is $L=N_Sa$, where $a$ is the lattice spacing, and 
temporal extent $L_4=N_4a$.

The lattice regulates the ultraviolet divergences that appear in
quantum field theory and reduces the mathematical problem to one
similar to statistical mechanics.
Familiar perturbation theory can be derived starting with lattice 
field theory, but many other theoretical tools from condensed matter 
theory are available~\cite{Kogut:1982ds}.
In the years after Wilson's paper there were, for example, many 
attempts to calculate hadron masses with strong coupling expansions.

If the lattice has a finite extent, then the system defined by 
Eq.~(\ref{eq:lqcd:funcInt}) has a finite, albeit large, number of 
degrees of freedom.
That means that the integrals can, in principle, be evaluated on a 
computer.
In the rest of this report all applications of lattice QCD use this 
approach.
In this section we provide a summary of the methods and a guide to 
estimate the inevitable errors that enter when mounting large-scale 
computing.

To start, let us leave the quarks and antiquarks aside and consider
how many gluonic integration variables are needed.
One would like the lattice spacing $a$ to be smaller than a hadron,
and the spatial volume should be large enough to contain at least one
hadron.
A desirable target is then $N_S=L/a=32$, which is typical by now, and 
some groups use even larger lattices.
For reasons explained below, the temporal extent~$N_4$ is often taken 
to be 2 or~3 times larger than~$N_S$.
Taking the gluon's 8 colors and the 4-fold Lorentz index into 
account, the functional integral has 
$8\times4\times32^3\times64\sim10^8$ dimensions.
This is practical with Monte Carlo methods, generating an 
ensemble of random values of the fields and replacing the 
right-hand side of Eq.~(\ref{eq:lqcd:funcInt}) with
\begin{equation}
	\langle O_1(x_1)O_2(x_2)\cdots O_n(x_n) \rangle = \frac{1}{C}
		\sum_{c} w(A^{(c)})\,
		O_1(x_1)O_2(x_2)\cdots O_n(x_n),
	\label{eq:lqcd:funcSum}
\end{equation}
where the weight $w$ for the $c$th configuration is specified below,
and $C$ is chosen so that ${\langle1\rangle=1}$.
If the weight $e^{-S}$ in Eq.~(\ref{eq:lqcd:funcInt}) is real and 
positive, then the random fields can be generated with distribution 
$e^{-S}$, in which case the weights are field independent.
This is called importance sampling, and without it numerical lattice 
field theory is impractical.

In Minkowski space the weight is actually a phase factor $e^{iS_{\rm M}}$.
That means that the weight fluctuates wildly, leading to enormous 
cancellations that are impossible to deal with numerically.
For that reason, numerical LQCD calculations are carried out in 
Euclidean space or, equivalently, with imaginary time.
With this restriction it remains straightforward to compute hadron 
masses and many matrix elements.
If, however, the coordinates $x_i$ in the original correlation 
function must have timelike or lightlike separation, then the 
function lies beyond current computational techniques.

Fermions, such as quarks, are special for several reasons.
To impose the Pauli exclusion principle, the quark fields are 
Grassman numbers, \emph{i.e.}, they anticommute with each other, 
$q_iq_j=-q_jq_i(1-\delta_{ij})$.
The integration is a formal procedure called Berezin integration.
Fortunately, in cases of practical interest, the integration can be 
carried out by hand.
The quark part of the action takes the form
\begin{equation}
	S_{\bar{q}q} = \sum_{ij}\bar{q}_j\,M_{ji}q_i,
\end{equation}
where $i$ and $j$ are multi-indices for spacetime, spin, color, and flavor.
The matrix $M$ is some lattice version of the Dirac operator.
It is easy to show that
\begin{equation}
	\int\prod_{ij}d\bar{q}_jdq_ie^{-S_{\bar{q}q}}=\det M.
\end{equation}
Similarly, if quark fields appear in the operators, each instance of
$q_i\bar{q}_j$ is replaced, using the Wick contraction, by the
quark propagator $M_{ij}^{-1}$.
The determinant and $M^{-1}$ both depend on the gauge field;
we simply carry out the quark and antiquark integration by hand and 
the gluon integration with the Monte Carlo, now with weight
$\det M\,e^{-S_{\rm gauge}}$.
The computation of $M^{-1}_{ij}$ is demanding and the computation of 
$\det M$ is very demanding.

Another peculiar feature of fermions is an obstacle to realizing
chiral symmetry on the lattice \cite{Friedan:1982nk,Nielsen:1980rz},
often called the fermion doubling problem, because a simple
nearest-neighbor version of the Dirac operator leads to a 16-fold
duplication of states.
As a consequence, several formulations of lattice fermions are used 
in numerical lattice QCD.
With staggered fermions~\cite{Banks:1975gq,Susskind:1976jm} some of 
the doubling remains, but a subset of the chiral symmetry is preserved.
With Wilson fermions~\cite{Wilson:1975hf} all doubling is removed, but 
all of the (softly broken) chiral symmetries are explicitly broken.
The Ginsparg-Wilson relation~\cite{Ginsparg:1981bj}, which is derived
from the renormalization group, shows how to preserve a remnant of
chiral symmetry.
Specific solutions are the fixed-point 
action~\cite{Wiese:1993cb,Hasenfratz:1998ri},
domain-wall fermions~%
\cite{Kaplan:1992bt,Shamir:1993zy,Furman:1994ky,Blum:1996jf},
and the overlap~\cite{Neuberger:1997fp,Neuberger:1998wv}.
In the approaches satisfying the Ginsparg-Wilson relation, the chiral 
transformation turns out to depend on the gauge field~\cite{Luscher:1998pqa}.
From a theoretical perspective these are the most attractive, but 
from a practical perspective the staggered and Wilson formulations 
are numerically faster.

To obtain a finite problem, numerical lattice QCD uses a 
finite spacetime volume, so one must specify boundary conditions.
In most cases, one identifies the field with itself, up to a phase:
\begin{equation}
	q(x+L_\mu e_\mu) = e^{i\theta_\mu}q(x),
\end{equation}
where $e_\mu$ is a unit vector and $L_\mu$ is the total extent, both 
in the $\mu$ direction.
If $\theta_\mu=0$ this is called a periodic boundary condition;
if $\theta_\mu=\pi$ this is called an antiperiodic boundary condition;
and otherwise this is called a twisted boundary
condition~\cite{Bedaque:2004kc,deDivitiis:2004kq} (although ``twisted
boundary condition'' has other meanings too~\cite{'tHooft:1979uj}).
In a finite volume, the spectrum is discrete.
The allowed 3-momenta are
\begin{equation}
	\bm{p} = \frac{\bm{\theta}}{L} + \frac{2\pi}{L}\bm{n},
\end{equation}
where $\bm{n}$ is a vector of integers.
One should bear in mind the discrete momentum follows from the finite 
volume, \emph{not} the lattice itself.
For one-particle states finite-volume effects are exponentially 
suppressed in periodic and antiperiodic~\cite{Luscher:1985dn}, 
as well as (partially) twisted~\cite{Sachrajda:2004mi}, boundary conditions.
For multi-particle states the boundary effects are larger and more 
interesting~\cite{Luscher:1991cf}, as discussed for $K\to\pi\pi$ in
Ref.~\cite{Lellouch:2000pv}.

To determine the CKM matrix we need the matrix elements of the 
electroweak Hamiltonian derived in Sec.~\ref{sec:ope}.
In most cases, we are interested in transitions with at most one 
hadron in the initial or final state.
These quantities are determined from 2- and 3-point correlation 
functions, as follows.
A first step is to determine the mass.
Let $O$ be an operator with the quantum numbers ($J^{PC}$, etc.) of
the state of interest.
For large temporal extent $L_4$, and temporal separation $x_4>0$,
the 2-point correlation function
\begin{equation}
	\langle O(x)O^{\dagger}(0)\rangle = 
		\langle 0|\hat{O}(x) \hat{O}^\dagger(0)|0\rangle,
	\label{eq:lqcd:2-pt}
\end{equation}
where $|0\rangle$ is the QCD vacuum state and the hat indicates an 
operator in Hilbert space.
Because these calculations are in Euclidean space, the time 
dependence of the annihilation operator is
\begin{equation}
	O(x) = e^{x_4\hat{H}} \hat{O}e^{-x_4\hat{H}},
\end{equation}
where $\hat{H}$ is the Hamiltonian.
In deriving Eq.~(\ref{eq:lqcd:2-pt}) the eigenvalue of $\hat{H}$ in 
$|0\rangle$ is set to zero.
Inserting a complete set of eigenstates of $\hat{H}$ into 
Eq.~(\ref{eq:lqcd:2-pt}), one has
\begin{equation}
	\langle O(x)O^{\dagger}(0)\rangle = \sum_n
		\langle 0|\hat{O}e^{-x_4\hat{H}}|n\rangle\langle n|\hat{O}^\dagger|0\rangle =
		\sum_n e^{-x_4E_n} |\langle n|\hat{O}^\dagger|0\rangle|^2,
	\label{eq:lqcd:expos}
\end{equation}
where $E_n$ is the energy of the $n$th state.
If $|n\rangle$ is a single-particle state with zero 3-momentum, this
energy is the mass.
Taking $x_4$ large enough the state with the lowest-lying mass 
dominates, and this is how masses are computed in lattice QCD:
evaluate the left-hand side of Eq.~(\ref{eq:lqcd:expos}) with Monte 
Carlo techniques, and fit the right-hand side to a sum of exponentials.

Now suppose that one would like to consider the case where one is 
interested in a simple matrix element, one where an operator from the 
effective Hamiltonian annihilates the hadron.
One can obtain the matrix element by computing another 2-point 
correlation function,
\begin{equation}
	\langle J(x)O^{\dagger}(0)\rangle = 
		\langle 0|\hat{J}(x) \hat{O}^\dagger(0)|0\rangle =
		\sum_n e^{-x_4E_n}
		\langle 0|\hat{J}|n\rangle\langle n|\hat{O}^\dagger|0\rangle.
	\label{eq:lqcd:J-pt}
\end{equation}
With the energies and overlaps $\langle n|\hat{O}^\dagger|0\rangle$ 
from the mass calculation, this calculation yields the transition 
matrix elements $\langle 0|\hat{J}|n\rangle$.

Most of the transitions of interest in flavor physics involve mesons, 
so it is worth illustrating how the quark propagators $M^{-1}$ come in.
For the charged Kaon, for example, we take the operator 
$O=\bar{s}\gamma_5u$, and the 2-point function is computed via
\begin{equation}
	\langle\bar{s}\gamma_5u(x)\bar{u}\gamma_5s(0)\rangle = 
		-\langle \mathop{\rm tr}[G_u(x,0)\gamma_5G_s(0,x)\gamma_5]\rangle_A,
	\label{eq:lqcd:Kaon}
\end{equation}
where the trace is over color and Dirac indices, the average on the 
right-hand side is over gluon fields, and the quark propagator 
$G_f(x,y)$ is the solution of
\begin{equation}
	\sum_xM(w,x)G_f(x,y) = \delta_{wy}
\end{equation}
for flavor $f$, with color and Dirac indices implied.
For the decay of a Kaon to leptons, the transition operator
$J=\bar{s}\gamma_4\gamma_5u$, and the computation of
Eq.~(\ref{eq:lqcd:J-pt}) simply replaces the first $\gamma_5$ on both
sides of Eq.~(\ref{eq:lqcd:Kaon}) with $\gamma_4\gamma_5$.

In neutral meson mixing and in semileptonic and radiative decays one
encounters hadronic matrix elements with one hadron in both the initial
and final states.
For these one computes a 3-point correlation function,
\begin{equation}
	\langle O_f(x)J(y)O_i^\dagger(0)\rangle = \sum_{mn}
		e^{-(x_4-y_4)E_{fm}} e^{-y_4E_{in}}
		\langle0|\hat{O}_f|fm\rangle
		\langle fm|\hat{J}|in\rangle
		\langle in|\hat{O}_i^\dagger|0\rangle.
\end{equation}
The energies $E_{fm}$, $E_{in}$ and amplitudes
$\langle0|\hat{O}_f|fm\rangle$,
$\langle in|\hat{O}_i^\dagger|0\rangle$ are computed from 2-point 
functions, so the 3-point function yields
$\langle fm|\hat{J}|in\rangle$.
As before, for mesons (and baryons) the left-hand side is computed by 
contracting quark and antiquark fields in favor of quark propagators.

Hadron masses and decay amplitudes computed with lattice QCD depend 
on the bare gauge coupling and the bare quark masses, $1+n_f$ free 
parameters, if $n_f$ flavors are relevant to the problem at hand.
The bare gauge coupling is related to the lattice spacing via 
renormalization.
Thus, all dimensional quantities are really ratios of the quantity of 
interest compared to some fiducial quantity with dimensions of mass.
This standard mass should be one that is either not very sensitive to 
the quark masses, such as some of the mass splitting in quarkonium, 
or whose mass dependence is seen to be under good control, such 
as~$f_\pi$.
The bare quark masses are fixed through the simplest hadron masses: 
$m_\pi^2$ and $m_K^2$ for the light and strange quarks, and the $D_s$ 
and $B_s$ or $\eta_c$ and $\Upsilon$ masses for charmed and bottom 
quarks.

In computational physics it is important to know how to estimate 
uncertainties.
In lattice QCD uncertainties arise, in principal, from the nonzero
lattice spacing and the finite volume.
In practice, the algorithms for computing $\det M$ and $M^{-1}$ slow 
down as the quark masses are reduced.
Consequently, the calculations cited elsewhere in this report are
based on simulations with light quark masses that are higher than those of
the up and down quarks in nature.
Also in practice, one must be careful with heavy quarks, because the 
ultraviolet cutoff of currently available lattices, $1/a$ or $\pi/a$, 
is not (much) higher than the $b$-quark mass.

Fortunately, all these uncertainties may be assessed and quantified 
with effective field theories.
(For a review of lattice QCD developed from this perspective, 
see~\cite{Kronfeld:2002pi}.)
For the so-called chiral extrapolation, lattice practitioners use
chiral perturbation theory ($\chi$PT) to extend the reach from
feasible light quark masses down to the physical up- and down-quark
masses.
This is the same $\chi$PT discussed in Sec.~\ref{sec:ChPT}, although 
some practical considerations differ.
Often applications of $\chi$PT to lattice QCD incorporate the leading 
discretization effects of the lattice.
A chiral extrapolation entails a fit to numerical lattice-QCD data,
and the associated uncertainty is estimated from a combination of
quantitative measures, like goodness of fit, and qualitative
considerations, such as the smallness of the quark mass and the effect
of higher-order terms.
In addition, $\chi$PT can be used to estimate finite-size effects, 
because the largest ones typically stem from processes in which a 
virtual pion is emitted, traverses the (periodic) boundary, and is 
then reabsorbed~\cite{Luscher:1985dn,Gasser:1987zq,Colangelo:2006mp}.

Discretization effects can be understood and controlled with the 
Symanzik effective field theory~\cite{Symanzik:1983dc,Symanzik:1983gh}.
The central Ansatz here is that lattice gauge theory is described by 
a continuum effective field theory.
For QCD
\begin{equation}
	\mathcal{L}_{\rm LGT} \doteq \mathcal{L}_{\rm QCD} +
		\sum_i a^{\dim\mathcal{L}_i-4} K_i\mathcal{L}_i,
	\label{eq:lqcd:Sym}
\end{equation}
where the sum runs over operators $\mathcal{L}_i$ of dimension~5 or 
higher, and the power of $a$ follows from dimensional analysis.
The coefficient $K_i$ subsumes short-distance effects, analogously to 
the Wilson coefficients in Sec.~\ref{sec:ope}.
The right-hand side of Eq.~(\ref{eq:lqcd:Sym}) is a tool to analyze 
the left-hand side or, more precisely, numerical data generated with 
the lattice Lagrangian~$\mathcal{L}_{\rm LGT}$.
If $a$ is small enough, the higher dimensional operators may be 
treated as perturbations, leading to two key insights.
The first is to justify an extrapolation in $a$ to the continuum
limit.
More powerfully, if one can show for any (expedient) observable that, 
say, all the dimension-5 $K_i$ vanish, then one knows that they 
vanish for all processes.
The systematic reduction of the first several $K_i$ is known as the 
Symanzik improvement program.
With chirally symmetric actions, the dimension-5~$K_i$ vanish by symmetry,
so these are automatically ${\rm O}(a)$~improved.

For heavy quarks it is often the case that $m_Qa\not\ll1$ and, hence, 
special care is needed.
It is often said that lattice gauge theory breaks down, but it is 
more accurate to say that the most straightforward application of the 
Symanzik effective theory breaks down.
For most calculations relevant to the CKM unitarity triangle, it is 
simpler to use HQET as a theory of cutoff 
effects~\cite{Kronfeld:2000ck,Harada:2001fi,Harada:2001fj}.
This is possible because every (sensible) approach to heavy quarks on 
the lattice enjoys the same static limit and heavy-quark symmetries.
So the same set-up as in Sec.~\ref{sec:hqet-scet} is possible, just 
with different short-distance structure---because the lattice changes 
short distance.
Analogously to Symanzik, one can set up an improvement program.
Now, however, the approach to the continuum limit is not so simple as 
$O(a)$ or $O(a^2)$.
Nevertheless, most serious calculations with heavy-quarks use this
formalism, or something equivalent, to estimate heavy-quark
discretization effects.
For further details on techniques for heavy quarks, 
see~\cite{Kronfeld:2003sd}.
A~more recent development is to map out the $m_Qa$ dependence in finite 
volume~\cite{deDivitiis:2003wy,Lin:2006ur}, 
where both $m_Qa\ll1$ and $m_Qa\approx1$ are feasible
(cf.\ Sec.~\ref{sec:sl}).

One uncertainty that is not amenable to effective field theory (and is, 
therefore, devilish to quantify) stems from the so-called quenched 
approximation~\cite{Marinari:1981qf,Weingarten:1981jy}.
It corresponds to replacing the computationally demanding $\det M$ in 
the weight by $1$ and attempting to compensate by shifts in the bare 
gauge coupling and bare quark masses.
Physically this corresponds to keeping valence quarks but treating 
sea quarks as a dielectric medium.
This approximation is, by now, a historical artifact.
All calculations that aspire to play a role in flavor physics now have 
either $n_f=2$ or $2+1$ flavors of sea quarks.
In both cases the 2 light quarks are taken as light as possible, as a 
basis for chiral extrapolation. 
For $2+1$ the third flavor is tuned to have the mass of the strange 
quark, whereas $n_f=2$ means that the strange quark is quenched.
\begin{figure}
	\centering
	\includegraphics[height=8cm]{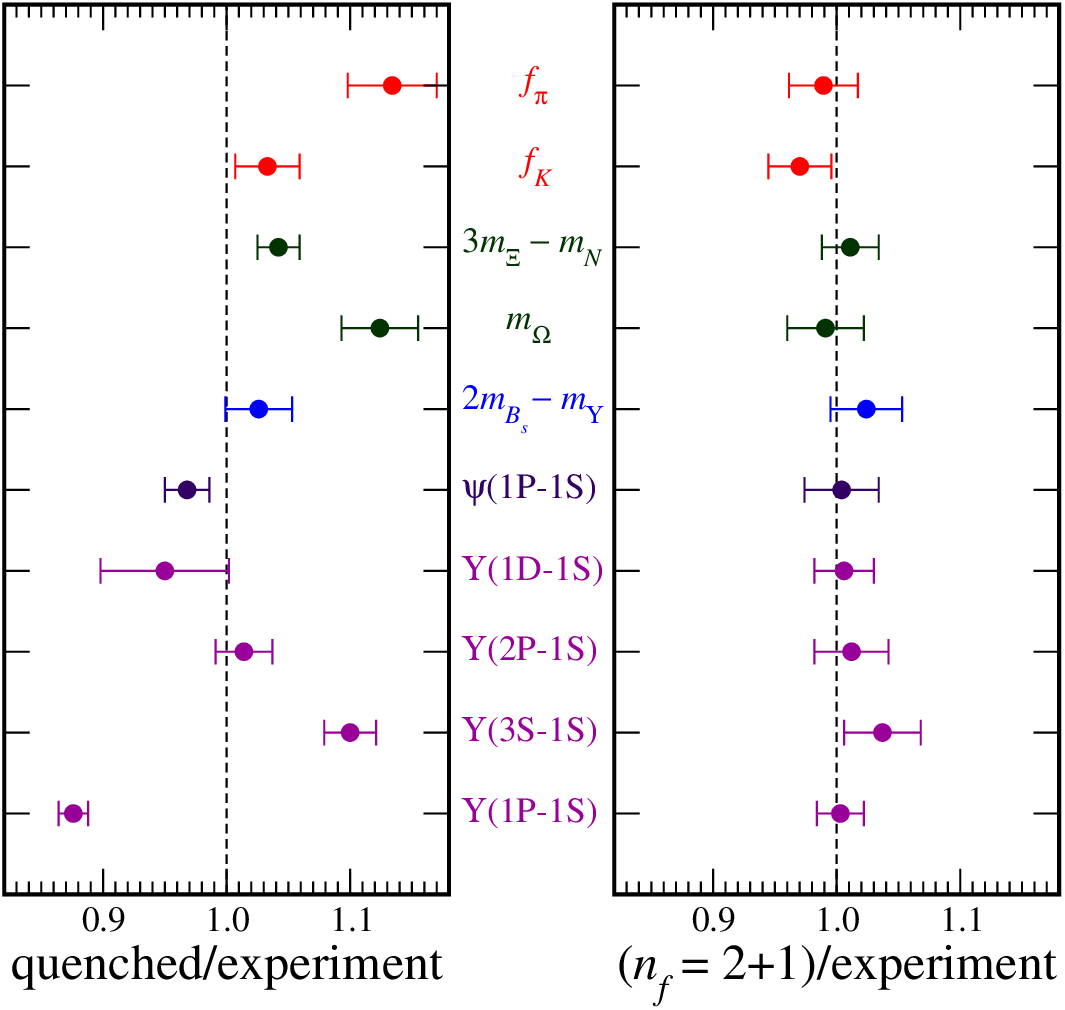}
	\caption[fig:lqcd:lattice]{Comparison of quenched and 2+1 flavor 
	lattice QCD, plotting the ratio of calculated results to 
	laboratory measurements~\cite{Davies:2003ik}.
	The quenched results deviate by as much as 10--15\%---not bad for 
	a strongly-coupled field theory, but not good enough for flavor
	physics.
	With 2+1 flavors of sea quarks, however, the agreement is at the 
	few-percent level.}
	\label{fig:lqcd:ratio}
\end{figure}
A comparison of quenched and 2+1 flavor QCD is shown in
Fig.~\ref{fig:lqcd:ratio}, adapted from Ref.~\cite{Davies:2003ik}.

The results shown in Fig.~\ref{fig:lqcd:ratio}, and many quoted in the 
rest of this report, have been obtained with staggered sea 
quarks~\cite{Bernard:2001av,Aubin:2004wf}, which provide the least 
computationally demanding method for computing $\det M$~\cite{Jansen:2008vs}.
A~drawback in this method is that staggered quarks come in four species, 
and a single quark flavor is simulated with 
$[\det_4M]^{1/4}$~\cite{Hamber:1983kx}, where 
the subscript emphasizes the number of species in the determinant.
There are concerns whether the fourth root really yields QCD in the 
continuum limit, although all published 
criticisms~\cite{Creutz:2007yg,Creutz:2007pr,Creutz:2007rk}
have been refuted~\cite{Bernard:2006vv,Bernard:2007eh}.
The theoretical arguments~\cite{Shamir:2004zc,Shamir:2006nj} in favor of 
this procedure are still being digested, although there is a 
significant body of supporting circumstantial 
evidence~\cite{Durr:2005ax,Sharpe:2006re,Kronfeld:2007ek}.
Whatever one thinks of the rooted staggered sea, it should be clear that 
these calculations should be confirmed.
Other methods for sea quarks are accumulating sufficiently high
statistics, so one can anticipate competitive results not only with 
staggered sea quarks~\cite{Bazavov:2009bb}, but also with
Symanzik-improved Wilson sea quarks~\cite{Aoki:2008sm,Durr:2008zz},
twisted-mass Wilson sea quarks~\cite{Boucaud:2007uk},
domain-wall sea quarks~\cite{Allton:2008pn}, and
overlap sea quarks~\cite{Aoki:2008tq}.

Calculations with 2 flavors of sea quarks have an uncertainty from 
quenching the strange quark.
The error incurred may be as large as 3--5\%, but is again hard to pin down.
In many cases, for example the $\Omega^-$ mass, no significant effect is 
seen.
When using 2-flavor results in this report, we take the original 
authors' estimates of the error for quenching the strange quark.
If they have omitted this line from the error budget, we then
assign a conservative 5\%~error.

Numerical lattice QCD has developed over the past thirty years, and 
much of the literature has aimed to develop numerical methods.
Such work is not limited to algorithm development, but also to 
demonstrate how a phenomenologically relevant calculation could or 
should be carried out.
Inevitably, some papers straddle the middle ground between 
development and mature results, with the consequence that some 
interesting papers have incomplete error budgets.
Where such results are used later in the report, we try to account for 
omitted uncertainties in a rational way.


\subsection{Chiral Perturbation Theory}
\label{sec:ChPT}

Chiral perturbation theory (ChPT) is the effective field theory describing 
strong and electroweak interactions of the light pseudo-scalar mesons
($\pi$, $K$, $\eta$)  at low energy, in a regime where standard 
perturbative methods 
are inapplicable~\cite{Weinberg:1978kz,Gasser:1983yg,Gasser:1984gg}.  
ChPT relies on our understanding of 
the chiral 
symmetry of QCD in the limit of massless light quarks  ($m_u = m_d = m_s = 
0$),    
its spontaneous symmetry breaking according to the pattern 
$SU(3)_L \times  SU(3)_R \to  SU (3)_V$  
and its explicit breaking due to non-vanishing quark masses. 

In the massless limit $m_q=0$, the QCD Lagrangian for light quarks 
($q^\top = (u,d,s)$)
\begin{equation} \label{eq:qcd}
\mathcal{L}_{\rm QCD} =
- \displaystyle\frac{1}{4}\, G^a_{\mu\nu} G^{\mu\nu}_a
 + i \,\bar{q}_L \gamma^\mu D_\mu q_L  
 + i \,\bar{q}_R \gamma^\mu D_\mu q_R 
-   \bar{q}_L \, m_q \, q_R  -  \bar{q}_L \, m_q \, q_L
\end{equation}
is invariant under global independent  $SU(3)_L\times SU(3)_R$ 
transformations of
the left- and right-handed quarks in flavor space: $q_{L,R} \to
g_{L,R} \; q_{L,R} \,$, $g_{L,R} \in SU(3)_{L,R}$. 
The absence of $SU(3)$ multiplets of opposite parity in the hadronic 
spectrum 
suggests that  the chiral group
$G=SU(3)_L\times SU(3)_R$ is spontaneously broken to the diagonal
subgroup $H = SU(3)_V$, i.e. the symmetry is realized \`a la 
Nambu-Goldstone~\cite{Goldstone:1961eq,Nambu:1997wa,Weinberg:1968de}.
According to Goldstone's theorem~\cite{Goldstone:1961eq} then, the 
spectrum of QCD should contain 
an octet of pseudoscalar massless bosons, in one to one correspondence to 
the broken 
symmetry generators. These are identified with  the $\pi$, $K$, and $\eta$ 
mesons,  which
 would be massless 
in the exact chiral limit of $m_{u,d,s} = 0$, but acquire a finite mass  
in the real world due to explicit chiral symmetry breaking induced 
by $m_q \neq 0$.  
Pions, Kaons, and eta remain, however, the lowest lying hadronic 
excitations. The existence of a gap separating $\pi, K, \eta$  from the 
rest of 
the spectrum makes it possible to build an effective theory involving only 
Goldstone modes.  

The basic building blocks of the effective theory are the Goldstone fields 
$\varphi$.
Intuitively, the massless Goldstone modes describe excitations of the 
system along the 
directions in field space that connect degenerate vacuum configurations 
(think about 
the circle of minima in a "Mexican-hat" potential).   
Mathematically, this means that the Goldstone fields parametrize the 
elements $u(\varphi)$  of the coset space 
$SU(3)_L\times SU(3)_R/$ $SU(3)_V$~\cite{Coleman:1969sm,Callan:1969sn}.
The  transformation of $\varphi$  under $G$ is determined by the action of 
$G$ 
on the elements $u(\varphi)$ of the coset space
\begin{equation} \label{eq:uphi}
u(\varphi) \to u(\varphi')=g_R u(\varphi) h(g,\varphi)^{-1} 
= h(g,\varphi) u(\varphi) g_L^{-1}  
\end{equation} 
where $g=(g_L,g_R) \in G$.
The explicit form of 
$h(g,\varphi) \in SU(3)_V$  will not be needed here. 
An explicit parametrization of $u(\varphi)$ is given by 
\begin{equation}
u(\varphi) = \exp{\left\{\frac{i}{\sqrt{2} F} \varphi \right\}} \ , 
\end{equation}
with 
$$
\varphi =  
\left(
\begin{array}{ccc}
 \displaystyle\frac{1}{\sqrt 2}\,\pi^0 + \displaystyle\frac{1}{\sqrt
 6}\,\eta_8 
& \pi^+ & K^+ \\
\pi^- & - \displaystyle\frac{1}{\sqrt 2}\,\pi^0 + 
\displaystyle\frac{1}{\sqrt 6}\,\eta_8 
& K^0 \\
 K^- & \bar{K}^0 & - \displaystyle\frac{2}{\sqrt 6}\,\eta_8 
\end{array}
\right)
\ . $$

The structure of the effective Lagrangian ${\cal L}_{\rm eff}$ 
is determined by chiral symmetry and the discrete  symmetries of QCD. 
${\cal L}_{\rm eff}$ has to be invariant under chiral transformations, 
up to explicit symmetry breaking terms that transform like the quark mass 
term in the 
QCD Lagrangian~(\ref{eq:qcd}).  As a consequence, ${\cal L}_{\rm eff}$   
is organized as an expansion in powers of 
(i) derivatives (momenta) of the Goldstone fields and 
(ii) light quark masses ($m_q$).
Since the meson masses squared  are proportional to the quark masses, 
the two expansions are related 
($m_q \sim  O(M_M^2) \sim O(p^2)$) and the mesonic effective chiral 
Lagrangian takes the form 
\begin{equation}
{\cal L}_{\rm eff}  = \sum_{n \geq 1} \,  {\cal L}_{2n}   \ , \qquad 
\qquad  {\cal L}_{2n} \sim  O(p^{2n})  ~.
\label{eq:LLLeff}
\end{equation}
The power counting parameter is given by  the ratio $p^2 \sim p_{\rm 
ext}^2/\Lambda_\chi^2$ 
of a typical external momentum (or quark mass) over the intrinsic scale  
$\Lambda_\chi$,  
set by the lightest  non-Goldstone states 
($\Lambda_\chi \sim 1\  {\rm GeV}$).
To each order in the expansion, the effective Lagrangian contains a number 
of low-energy constants  (LECs) not fixed by symmetry consideration,
encoding underlying QCD dynamics. \\
The leading order effective Lagrangian reads (in terms of $U(\varphi) = 
u(\varphi)^2$), 
\begin{equation}
{\cal L}_2 = \frac{F^2}{4} \  {\rm Tr} \, \Bigg[  \partial_\mu U \, 
\partial^\mu U     + 
2 B  \, m_q   \left( U  + U^\dagger \right)  \Bigg] 
\end{equation} 
where $m_q = {\rm diag} (m_u, m_d, m_s)$
and the trace is performed over the $SU(3)$ indices. The dimensionful 
constants $F$ and $B$ 
are  related to the pion decay constant and the quark condensate by 
$F_\pi = F \, (1 + O(m_q))$ and $ \langle 0| \bar{u} u | 0 \rangle = - F^2 
\, B \, (1 + O(m_q))$. 
${\cal L}_2$ contains the Gell-Mann-Oaks-Renner~\cite{GellMann:1968rz}  
and Gell-Mann-Okubo~\cite{GellMann:1962xb,Okubo:1962zz}
mass relations and allows one to calculate physical processes, such as 
$\pi \pi$ scattering, 
to $O(p^2)$ in terms of just $F_\pi$ and $M_M^2$  ($M_\pi^2 = B (m_u + 
m_d)$, ...). 

The power of the effective field theory approach 
is that it allows to systematically improve 
the calculations of low-energy processes by considering higher-order 
terms in the momentum/light-quark-mass expansion. 
As shown by Weinberg~\cite{Weinberg:1978kz}, at any given 
order in this expansion only a finite number of couplings in 
(\ref{eq:LLLeff}) appear.
For instance at $O(p^4)$ a given amplitude receives 
contributions only from: (i) tree-level diagrams with one insertion 
from ${\cal L}_4$; (ii) one-loop diagrams with  all vertices 
from ${\cal L}_2$.  The loop diagrams perturbatively unitarize the 
theory and introduce physical infrared singularities due to pseudoscalar 
meson intermediate states 
(the chiral logs,  $\sim m_q \log m_q$).  
However, loops also introduce ultraviolet 
divergences. Using  a regularization compatible with chiral symmetry, the 
counterterms necessary to absorb the divergences must have the same form 
as the terms present in ${\cal L}_4$: thus, one loop divergences 
simply renormalize the LECs of $O(p^4)$.  
This argument generalizes to any order in the low-energy expansion: 
the effective theory is renormalizable order 
by order in the low-energy expansion. 

The finite parts of the LECs can be fitted to experiment or extracted 
by matching to lattice QCD results (or other, less systematic 
approximations to 
non-perturbative QCD dynamics). 
The accuracy of a given  calculation is bounded by the size of  higher 
order terms 
in the low-energy expansion. 
State of the art calculations in the strong sector go up to 
$O(p^6)$~\cite{Bijnens:2004pk}.

To illustrate the general features discussed above, 
we report here the expression of the the pion decay  constant to 
$O(p^4)$~\cite{Gasser:1984gg}
\begin{equation}
F_\pi= F \Bigg[ 1 - 2 \mu_\pi - \mu_K 
%
%
+ \frac{8 B}{F^2} \Big( \hat{m} \, L_5^r(\mu)  +  (m_u + m_d + m_s) L_4^r 
(\mu)  \Big)   
\Bigg]~.
\end{equation}
Here  $\mu_P = M_P^2/(32 \pi^2 F^2)   \log (M_P^2/\mu^2)$, 
$M_\pi^2 = B (m_u + m_d)$,  $M_K = B (m_s +  \hat{m})$, and  $\hat{m} = 
1/2 (m_u + m_d)$. 
Moreover,  $\mu$ is the renormalization scale and $L_{4,5}^{r} (\mu)$ 
are two finite scale-dependent LECs. This expression illustrates the 
appearance of 
calculable chiral logarithms (with unambiguous coefficients) as well as 
polynomial terms in
the quark masses multiplied by  a priori unknown coefficients. 
Expressions of this type are used to extrapolate lattice QCD results 
from unphysical quark masses to the physical point. Nowadays, this is one 
of the most relevant applications of ChPT in CKM physics.
An important recent development in this area is the  use of $SU(2)$
ChPT~\cite{Allton:2008pn,Roessl:1999iu},
in which  Kaons are treated  as external massive fields,
to study the extrapolation of Kaon amplitudes in $m_{u,d}$
(see Sec.~\ref{sect:CabibboLattice} for discussion and applications)

\medskip

The framework presented above describes the strong interactions of 
Goldstone modes. It has been extended in several directions, 
highly relevant to CKM physics,  to include:  
\begin{itemize}
\item  non-leptonic weak interactions of Goldstone modes 
($\Delta 
S=1,2$)~\cite{Cronin:1967jq,Kambor:1989tz,Cirigliano:1999ie,Ecker:2000zr}; 
\item  interactions of soft Goldstone modes with heavy particles 
(heavy mesons \cite{Wise:1992hn,Burdman:1992gh} 
and baryons \cite{Gasser:1987rb,Jenkins:1990jv}); 
\item interaction of Goldstone modes with external electromagnetic fields 
and weak gauge bosons 
(this is achieved by adding external sources that couple to quark 
bilinears in  the QCD 
Lagrangian~\cite{Gasser:1983yg,Gasser:1984gg}); 
\item  other dynamical fields in the low-energy theory, such as 
photons~\cite{Urech:1994hd}
and light leptons~\cite{Knecht:1999ag} 
(the amplitudes are expanded to $O(e^2 p^{2n})$, $e$ being the 
electromagnetic coupling).
\end{itemize}

\subsection {Beyond the Standard Model}
 \label{sect:BSMprimer}

Despite its impressive phenomenological success,
the SM should be regarded as a low-energy effective theory. 
 Viewing the SM as an effective theory
poses two main questions: which is the
{\em energy scale} and which are the 
{\em interactions and symmetries properties} 
of the new degrees of freedom. 
So far we have no unambiguous answer for both
these questions; however, a strong theoretical prejudice 
for new degrees of freedom around the TeV scale follows 
from a natural stabilization of the mechanism of electroweak
symmetry breaking. In this perspective, low-energy 
flavor physics provide a powerful tool to 
address the second question, and in particular to explore the 
symmetries properties of the new degrees of freedom.

In order to describe New Physics (NP) effects in 
flavor physics we can follow two main strategies, 
whose virtues and limitations can be summarised as follows:

\begin{itemize}
\item
{\em Generic Effective Field Theory (EFT) approaches.} \\
Assuming the new degrees to be heavier than SM fields,
we can integrate them out and describe NP
effects by means of a generalization of the Fermi
Theory: the SM Lagrangian becomes the renormalizable part of a more 
general 
local Lagrangian which includes an infinite tower of higher-dimensional 
operators, constructed in terms of SM fields and 
suppressed by inverse powers of an effective scale 
$\Lambda_{\rm NP} > M_W$. \\
This general bottom-up approach allows us to analyse all realistic 
extensions of the SM in terms of a limited number of parameters 
(the coefficients of the higher-dimensional operators). 
The drawback of this method is the impossibility 
to establish correlations 
of NP effects at low and high energies: the scale
$\Lambda_{\rm NP}$ defines the cut-off of the EFT. 
However, correlations among different low-energy processes
can be established implementing specific symmetry properties
on the EFT, such as the Minimal Flavor Violation hypothesis
(see Sec.~\ref{sect:EFT}).
The experimental tests of such correlations allow 
us to test/establish general features of the new theory
which holds independently of the dynamical details of the model. 
In particular, $B$, $D$ and $K$ decays are extremely useful in 
determining the flavor-symmetry breaking pattern of the NP model. 

\item
{\em Explicit Ultraviolet completions.} \\
The generic EFT approach is somehow the 
opposite of the standard top-down strategy, 
where a given NP theory --and a specific set of parameters-- 
are employed to evaluate possible deviations from the SM.  
The top-down approach usually allows us to establish several correlations,
both at low energies and between low- and high-energy observables. 
In the following we will discuss in 
some detail this approach in the case of 
Minimal Supersymmetric extension of the SM (see Sec.~\ref{sect:SUSY}).
The price to pay of this strategy is the loss of generality. 
This is quite a high price given our limited knowledge about 
the physics above the electroweak scale. 

\end{itemize}

\subsubsection{Model-independent approaches and the MFV hypothesis}
\label{sect:EFT}

The NP contributions should naturally induce large effects 
in processes which are severely suppressed in the SM,
such as meson-antimeson mixing ($\Delta F=2$ amplitudes)
or flavor-changing neutral-current (FCNC) rare decays.
Up to now there is no evidence of deviations from the 
SM in these processes and this implies 
severe bounds on the effective scale of various dimension-six
operators in the EFT approach. For instance, the good agreement between SM 
expectations and experimental determinations of $K^0$--${\bar K}^0$ 
mixing leads to bounds above $10^4$~TeV for the effective scale 
of $\Delta S=2$ operators, i.e.~well above the few TeV 
range suggested by a natural stabilization of the  
electroweak-symmetry breaking mechanism. 

The apparent contradiction between these 
two determinations of  $\Lambda$ is a manifestation of what in 
many specific frameworks (supersymmetry, technicolor, etc.)
goes under the name of {\em flavor problem}:
if we insist on the theoretical prejudice that new physics has to 
emerge in the TeV region, we have to conclude that the new theory 
possesses a highly non-generic flavor structure. 
Interestingly enough, this structure has not been clearly identified yet,
mainly because the SM (the low-energy 
limit of the new theory), doesn't possess an exact flavor symmetry.
Within a model-independent approach, we should try to deduce 
this structure from data, using the experimental information on FCNC
transitions to constrain its form. 

\paragraph{Generic bounds on loop-mediated amplitudes.}
In several realistic NP models we can neglect non-standard effects 
in all cases where the corresponding effective operator is generated at 
the
tree-level within the SM. This general assumption implies that the
experimental determination of the CKM matrix via tree-level
processes is free from the contamination of NP contributions.  
Using this determination we can unambiguously predict 
meson-antimeson mixing and FCNC amplitudes within the SM.
Comparing these predictions with data allows to derive 
general constraints on NP which holds in a wide class of models.

The most constrained sector is the one of $\Delta F=2$ transitions, 
where almost all the interesting amplitudes have been measured
with good accuracy. An updated analysis of the present constraints 
from these measurements will be presented in Sec.~\ref{sect:gfitsNP}.
The main conclusions that can be drawn form this analysis 
can be summarized as follows: 
\begin{itemize}
\item 
In all the three accessible short-distance amplitudes       
($K^0$--$\bar K^0$, $B_d$--$\bar B_d$, and $B_s$--$\bar B_s$)
the magnitude of the new-physics amplitude cannot exceed, in size, 
the SM short-distance contribution. The latter is suppressed both 
by the GIM mechanism and by the hierarchical structure of the 
CKM matrix. As a result, new-physics models with TeV-scale flavored 
degrees of freedom and $\mathcal{O}(1)$ flavor-mixing couplings are 
essentially ruled out. For instance, considering 
a generic $\Delta F=2$ effective Lagrangian of the form 
\begin{equation}
\mathcal{L}^{\Delta F=2} = \sum_{i\not=j} \frac{c_{ij}}{\Lambda^2} 
(\bar d_L^i \gamma^\mu d_L^j )^2~,
\end{equation}
where $d^i$ denotes a generic down-type quark $(i=1,2,3)$ 
and $c_{ij}$ are dimensionless couplings, the condition 
$|\mathcal{A}^{\Delta F=2}_{\rm NP}| <  |\mathcal{A}^{\Delta F=2}_{\rm SM} 
|$
implies
\begin{eqnarray}
\Lambda < \frac{ 3.4~{\rm TeV} }{| V_{ti}^* V_{tj}|/|c_{ij}|^{1/2}  }
\approx \left\{ \begin{array}{l}  
9\times 10^3~{\rm TeV} \times |c_{sd}|^{1/2} \!\!\!\!\!\!\! \\ 
4\times 10^2~{\rm TeV} \times |c_{bd}|^{1/2} \!\!\!\!\!\!\! \\
7\times 10^1~{\rm TeV} \times |c_{bs}|^{1/2} \!\!\!\!\!\!\! 
\end{array} 
\right.  
\label{eq:bound}
\end{eqnarray}
\item 
In the case of $B_d$--$\bar B_d$ and $K^0$--$\bar K^0$ mixing,
which are both well measured, there is still room 
for a new-physics contribution comparable to the 
 SM one. However, this is possible only 
if the new-physics contribution is aligned in phase 
with respect to the SM amplitude. The situation is quite different 
in the case of $B_s$--$\bar B_s$ mixing, where present measurements 
allow a large non-standard CP violating phase.
\end{itemize}
As we will discuss in the following, a natural mechanism to
reconcile the stringent bounds  in Eq.~(\ref{eq:bound}) with the 
expectation 
$\Lambda \sim $ few TeV is obtained with the 
Minimal Flavor Violation hypothesis.

\paragraph{Minimal Flavor Violation.}
\label{sect:MFV} 
A very reasonable, although quite pessimistic, solution
to the flavor problem is the so-called 
Minimal Flavor Violation (MFV) hypothesis.
Under this assumption, flavor-violating 
interactions are linked to the
known structure of Yukawa couplings also beyond the SM. 
As a result, non-standard contributions in FCNC 
transitions turn out to be suppressed to a level consistent 
with experiments even for $\Lambda \sim$~few TeV.
One of the most interesting aspects of the MFV hypothesis 
is that it can naturally be implemented within the 
EFT approach to NP.
The effective theories based on this symmetry principle
allow us to establish unambiguous correlations 
among NP effects in various rare decays.
These falsifiable predictions are the key ingredients   
to identify in a model-independent way which are the 
irreducible sources of flavor symmetry breaking.

The MFV hypothesis consists of two ingredients~\cite{D'Ambrosio:2002ex}: 
i)~a {\em flavor symmetry} and ii)~a set of {\em symmetry-breaking 
terms}. 
The symmetry is defined from the SM Lagrangian in absence 
of Yukawa couplings. This is invariant 
under a large gbobal symmetry of flavor transformations: 
${\mathcal G}_{q} \otimes 
{\mathcal G}_{\ell} \otimes U(1)^5$,
where 
\begin{equation}
{\mathcal G}_{q}
= {SU}(3)_{Q_L}\otimes {SU}(3)_{U_R} \otimes {SU}(3)_{D_R}~,
\qquad 
{\mathcal G}_{\ell}
=  {SU}(3)_{L_L} \otimes {SU}(3)_{E_R}~.
\end{equation}
The $SU(3)$ groups refer to a rotation in flavor space 
(or a flavor mixing) among the three families of 
basic SM fields: the quark and lepton doublets, 
$Q_L$ and $L_L$, and the three singlets $U_R$, $D_R$ and 
$E_R$. Two of the five $U(1)$ groups 
can be identified with the total baryon and lepton number
(not broken by the SM Yukawa interaction), while an independent 
$U(1)$ can be associated to the weak hypercharge. 
Since hypercharge is gauged and involves also 
the Higgs field, it is more convenient not to include 
it in the flavour group, which would then be defined as 
${\mathcal G}_{\rm SM} = {\mathcal G}_{\ell} \otimes U(1)^4$~\cite{Feldmann:2009dc}.

Within the SM this large global symmetry, and particularly the ${\rm SU}(3)$ 
subgroups controlling flavor-changing transitions, is 
explicitly broken by the Yukawa interaction
\begin{equation}
\mathcal{L}_Y  =   {\bar Q}_L Y_D D_R  H
+ {\bar Q}_L {Y_U} U_R  H_c
+ {\bar L}_L {Y_E} E_R  H {\rm ~+~h.c.}
\label{eq:LYY}
\end{equation}
The most restrictive hypothesis 
we can make to {\em protect} in a consistent way quark-flavor mixing 
beyond the SM is to assume that $Y_D$ and $Y_U$ are the only 
sources of  ${\mathcal G}_{q}$ breaking also in the NP model.
To implement and interpret this hypothesis in a consistent way, 
we can assume that ${\mathcal G}_{q}$ is a good symmetry, 
promoting $Y_{U,D}$ to be non-dynamical fields (spurions) with 
non-trivial transformation properties under this symmetry
\begin{equation}
Y_U \sim (3, \bar 3,1)_{{\mathcal G}_{q}}~,\qquad
Y_D \sim (3, 1, \bar 3)_{{\mathcal G}_{q}}~.\qquad
\end{equation}
If the breaking of the symmetry occurs at very high energy scales 
at low-energies we would only be sensitive to the background values of 
the $Y$, i.e. to the ordinary SM Yukawa couplings. 
Employing the effective-theory language, 
we then define that an effective theory satisfies the criterion of
Minimal Flavor Violation in the quark sector 
if all higher-dimensional operators,
constructed from SM and $Y$ fields, are invariant under CP and (formally)
under the flavor group ${\mathcal G}_{q}$ \cite{D'Ambrosio:2002ex}.

According to this criterion one should in principle 
consider operators with arbitrary powers of the (dimensionless) 
Yukawa fields. However, a strong simplification arises by the 
observation that all the eigenvalues of the Yukawa matrices 
are small, but for the top one, and that the off-diagonal 
elements of the CKM matrix are very suppressed. 
%
$Y$
%
As a consequence, in the limit where we neglect light quark masses,
the leading $\Delta F=2$ and $\Delta F=1$ FCNC amplitudes get exactly 
the same CKM suppression as in the SM: 
\begin{eqnarray}
  \mathcal{A}(d^i \to d^j)_{\rm MFV} &=&   (V^*_{ti} V_{tj})^{\phantom{a}} 
 \mathcal{A}^{(\Delta F=1)}_{\rm SM}
\left[ 1 + a_1 \frac{ 16 \pi^2 M^2_W }{ \Lambda^2 } \right]~,
\\
  \mathcal{A}(M_{ij}-\bar M_{ij})_{\rm MFV}  &=&  (V^*_{ti} V_{tj})^2  
 \mathcal{A}^{(\Delta F=2)}_{\rm SM}
\left[ 1 + a_2 \frac{ 16 \pi^2 M^2_W }{ \Lambda^2 } \right]~.
\label{eq:FC}
\end{eqnarray}
where the $\mathcal{A}^{(i)}_{\rm SM}$ are the SM loop amplitudes 
and the $a_i$ are $\mathcal{O}(1)$ real parameters. The  $a_i$
depend on the specific operator considered but are flavor 
independent. This implies the same relative correction 
in $s\to d$, $b\to d$, and  $b\to s$ transitions 
of the same type.

As pointed out in Ref.~\cite{Buras:2000dm}, within the MFV
framework several of the constraints used to determine the CKM matrix
(and in particular the unitarity triangle) are not affected by NP.
In this framework, NP effects are negligible not only in tree-level
processes but also in a few clean observables sensitive to loop
effects, such as the time-dependent CPV asymmetry in $B_d \to J/\Psi
K_{L,S}$. Indeed the structure of the basic flavor-changing coupling
in Eq.~(\ref{eq:FC}) implies that the weak CPV phase of $B_d$--$\bar
B_d$ mixing is arg[$(V_{td}V_{tb}^*)^2$], exactly as in the SM.  
This construction provides a natural (a posteriori) justification 
of why no NP effects have been observed in
the quark sector: by construction, most of the clean observables
measured at $B$ factories are insensitive to NP effects in the MFV
framework.

\begin{table}[t]
\begin{minipage}{\textwidth}
\begin{center}
\caption{\label{tab:MFV} Bounds on the scale of new physics 
for some representative $\Delta F=2$~\cite{Bona:2007vi} 
and $\Delta F=1$~\cite{Hurth:2008jc} MFV operators 
(assuming effective coupling $1/\Lambda^2$).}
\begin{tabular}{l|l|l}
Operator & ~$\Lambda_i@95\%$~prob.~[TeV] & ~Observables \\
\hline
$H^\dagger \left( {\bar D}_R  \lambda_d 
\lambda_{\rm FC} \sigma_{\mu\nu} Q_L \right) (e F_{\mu\nu})$
& ~$6.1$ & ~$B\to X_s \gamma$, $B\to X_s \ell^+ \ell^-$\\
$\frac{1}{2} (\bar Q_L  Y_U Y_U^\dagger \gamma_{\mu} Q_L)^2 
\phantom{X^{X^X}_{iii}}$
& ~$5.9$ & ~$\epsilon_K$, $\Delta m_{B_d}$, $\Delta m_{B_s}$ \\   
$\left( {\bar Q}_L \lambda_{\rm FC} \gamma_\mu
Q_L \right) ({\bar E}_R \gamma_\mu E_R)$  
& ~$2.7$ & ~$B\to X_s \ell^+ \ell^-$, $B_s\to\mu^+\mu^-$
\end{tabular}
\end{center}
\end{minipage}
\end{table}

In  Tab.~\ref{tab:MFV} we report a few representative 
examples of the bounds on the higher-dimen\-sio\-nal operators
in the MFV framework.
As can be noted, the built-in CKM suppression
leads to bounds on the effective scale of new physics 
not far from the TeV region. These bounds are very similar to the 
bounds on flavor-conserving operators derived by precision electroweak 
tests. 
This observation reinforces the conclusion that a deeper study of 
rare decays is definitely needed in order to clarify 
the flavor problem: the experimental precision on the clean 
FCNC observables required to obtain bounds more stringent 
than those derived from precision electroweak tests
(and possibly discover new physics) is typically
in the $1\%-10\%$ range.

Although the MFV seems to be a natural solution to the flavor problem, 
it should be stressed that we are still
very far from having proved the validity of this hypothesis from 
data.\footnote{In the EFT language we can say that there is still room for 
sizable new sources of favour symmetry breaking beside the SM 
Yukawa couplings~\cite{Feldmann:2006jk}.} 
A proof of the MFV hypothesis 
can be achieved only with a positive evidence of physics beyond 
the SM exhibiting the flavor-universality pattern
(same relative correction in $s\to d$, $b\to d$, and  $b\to s$ transitions
of the same type) predicted by the MFV assumption.

The idea that the CKM matrix rules the strength of FCNC 
transitions also beyond the SM has become a very popular 
concept in the recent literature and has been implemented 
and discussed in several works. 
It is worth stressing that the CKM matrix 
represents only one part of the problem: a key role in
determining the structure of FCNCs  is also played  by quark masses, 
or by the Yukawa eigenvalues. In this respect, the MFV 
criterion illustrated above provides the maximal protection 
of FCNCs (or the minimal violation of flavor symmetry), 
since the full structure of Yukawa matrices is preserved. 
At the same time, this criterion is based on a 
renormalization-group-invariant 
symmetry argument, which can be implemented 
independently of any specific hypothesis about the 
dynamics of the new-physics framework. 
The only difference between weakly- and strongly-iteracting 
theories at the TeV scale is that in the latter case the 
expansion in powers of the Yukawa spurions cannot be truncated 
to the first non-trivial terms~\cite{Feldmann:2008ja,Kagan:2009bn}
(leaving more freedom for non-negligible effects 
also in up-type FCNC amplitudes~\cite{Kagan:2009bn}).
This model-independent structure does not hold in 
most of the alternative definitions of MFV models 
that can be found in the literature. For instance, 
the definition of Ref.~\cite{Buras:2003jf} 
(denoted constrained MFV, or CMFV)
contains the additional requirement that the 
effective FCNC operators playing a significant 
role within the SM are the only relevant ones 
also beyond the SM. 
This condition is realized only in weakly coupled 
theories at the TeV scale with only one light Higgs doublet, 
such as the MSSM with small $\tan\beta$.
It does not hold in several other frameworks, such as 
Higgsless models, or the MSSM with large $\tan\beta$.

\paragraph{MFV at large $\tan\beta$.}
\label{eq:largetanb}
If the Yukawa Lagrangian contains only one Higgs field,
we can still assume that the Yukawa couplings are the only 
irreducible breaking sources of ${\mathcal G}_{q}$, but 
we can change their overall normalization.

A particularly interesting scenario
is the two-Higgs-doublet model where 
the two Higgses are coupled separately to up-
and down-type quarks:
\begin{equation}
\mathcal{L}^{2HDM}_{Y}  =   {\bar Q}_L Y_D D_R  H_D
+ {\bar Q}_L Y_U U_R  H_U
+ {\bar L}_L Y_E E_R  H_D {\rm ~+~h.c.}
\label{eq:LY2}
\end{equation}
This Lagrangian is invariant under an extra ${\rm U}(1)$
symmetry with respect to the one-Higgs Lagrangian 
in Eq.~(\ref{eq:LYY}): a symmetry under which 
the only charged fields are $D_R$ and $E_R$ 
(charge $+1$) and  $H_D$ (charge $-1$).
This symmetry, denoted ${\rm U}_{\rm PQ}$, 
prevents tree-level FCNCs and implies that $Y_{U,D}$ are the only 
sources of ${\mathcal G}_{q}$ breaking appearing in the Yukawa
interaction (similar to the one-Higgs-doublet
scenario). Coherently with the MFV hypothesis, we can then 
assume that $Y_{U,D}$ are the only relevant 
sources of ${\mathcal G}_{q}$ breaking appearing 
in all the low-energy effective operators. 
This is sufficient to ensure that flavor-mixing 
is still governed by the CKM matrix, and naturally guarantees
a good agreement with present data in the $\Delta F =2$
sector. However, the extra symmetry of the Yukawa interaction allows
us to change the overall normalization of 
$Y_{U,D}$ with interesting phenomenological consequences
in specific rare modes. 

The normalization of the  Yukawa couplings is controlled
by the ratio of the vacuum expectation values (vev) of the two Higgs 
fields, 
or by the parameter $\tan\beta = \langle H_U\rangle/\langle H_D\rangle$.
For $\tan\beta >\!\! > 1 $ the smallness of the $b$ quark 
and $\tau$ lepton masses can be attributed to the smallness 
of $1/\tan\beta$ rather than to the corresponding Yukawa couplings.
As a result, for $\tan\beta >\!\! > 1$ we cannot anymore neglect 
the down-type  Yukawa coupling. 
Moreover, the ${\rm U}(1)_{\rm PQ}$ symmetry cannot be exact:
it has to be broken at least in the scalar potential
in order to avoid the presence of a massless pseudoscalar Higgs.
Even if the breaking of  ${\rm U}(1)_{\rm PQ}$  and 
 ${\mathcal G}_{q}$ are decoupled, the presence of 
${\rm U}(1)_{\rm PQ}$ breaking sources can have important 
implications on the  structure of the Yukawa interaction,
especially if $\tan\beta$ is large~\cite{Hall:1993gn,Blazek:1995nv,
Isidori:2001fv,D'Ambrosio:2002ex}.
We can indeed consider new dimension-four operators such as
\begin{equation}
 \epsilon~ {\bar Q}_L  Y_D D_R  (H_U)^c
\qquad {\rm or} \qquad 
 \epsilon~ {\bar Q}_L  Y_UY_U^\dagger Y_D D_R  (H_U)^c~,
\label{eq:O_PCU}
\end{equation}
where $\epsilon$ denotes a generic MFV-invariant
${\rm U}(1)_{\rm PQ}$-breaking source. Even if $\epsilon \ll 1 $, 
the product $\epsilon \times  \tan\beta$ can be $\mathcal{O}(1)$, 
inducing large corrections to the down-type Yukawa 
sector:
\begin{equation}
 \epsilon~ {\bar Q}_L  Y_D D_R  (H_U)^c 
\ \stackrel{vev}{\longrightarrow}  \
 \epsilon~  {\bar Q}_L  Y_D D_R   \langle H_U \rangle = 
  (\epsilon\times\tan\beta)~  {\bar Q}_L  Y_D D_R   \langle H_D \rangle~.
\end{equation}

Since the $b$-quark Yukawa coupling becomes $\mathcal{O}(1)$, 
the large-$\tan\beta$ regime is particularly interesting 
for helicity-suppressed observables in $B$ physics. 

One of the clearest phenomenological 
consequences is a suppression (typically in the $10-50\%$ range)
of the $B \to \ell \nu$ decay
rate with respect to its SM expectation~\cite{Hou:1992sy,Isidori:2006pk}.
Potentially measurable effects in the $10-30\%$ range
are expected also 
in $B\to X_s \gamma$~\cite{Degrassi:2000qf,Carena:2000uj} and 
$\Delta M_{B_s}$~\cite{Buras:2002vd,Buras:2001mb}. 
The most striking signature could arise from the 
rare decays $B_{s,d}\to \ell^+\ell^-$
whose rates could be enhanced over the SM expectations 
by more than 
one order of 
magnitude~\cite{Hamzaoui:1998nu,Choudhury:1998ze,Babu:1999hn}.
An enhancement of both $B_{s}\to \ell^+\ell^-$ and 
$B_{d}\to \ell^+\ell^-$ respecting the MFV relation
$\Gamma(B_{s}\to \ell^+\ell^-)/\Gamma(B_{d}\to \ell^+\ell^-)
\approx |V_{ts}/V_{td}|^2$ would be an unambiguous signature 
of MFV at large $\tan\beta$~\cite{Hurth:2008jc}.

\subsubsection{The Minimal Supersymmetric extension of the SM  (MSSM)}
\label{sect:SUSY}

The MSSM is one of the most well-motivated and definitely the most 
studied extension of the SM at the TeV scale. For a detailed 
discussion of this model we refer to the specialised literature
(see e.g.~Ref.~\cite{Martin:1997ns}).
Here we limit our self to analyse some properties 
of this model relevant to flavor physics. 

The particle content of the MSSM consist of the SM gauge and 
fermion fields plus a scalar partner for each quark and lepton 
(squarks and sleptons) and a spin-1/2 partner for each 
gauge field (gauginos). The Higgs sector has two Higgs
doublets with the corresponding spin-1/2 partners (higgsinos)
and a Yukawa coupling of the type in Eq.~(\ref{eq:LY2}).
While gauge and Yukawa interactions of the model are 
completely specified in terms of the corresponding SM
couplings, the so-called soft-breaking sector\footnote{~Supersymmetry 
must be broken in order to be consistent with obsevations
(we do not observe degenerate spin partners in nature). The 
soft breaking terms are the most general supersymmetry-breaking 
terms which prserve the nice ultraviolet properties of the 
model. They can be divided into two main classes: 
1) mass terms which break the mass degeneracy of the  
spin partenrs (e.g.~sfermion or gaugino mass terms); 
ii) trilinear couplings among the scalar fields 
of the theory (e.g.~sfermion-sfermion-Higgs couplings).}
of the theory contains several new free parameters, most of which are 
related to flavor-violating observables. For instance the 
$6\times6$ mass matrix of the up-type squarks, 
after the up-type Higgs field gets a vev ($H_U \to \langle H_U \rangle$), 
has the following structure
\begin{equation}
{\tilde M}_U^2 = 
\left(
\begin{array}{cc}
  {\tilde m}_{Q_L}^2   &  A_U \langle H_U \rangle \\
  A_U^\dagger \langle H_U \rangle & {\tilde m}_{U_R}^2 
\end{array}
\right)~  + ~\mathcal{O}\left( m_Z, m_{\rm top} \right)~,
\end{equation}
where ${\tilde m}_{Q_L}$, ${\tilde m}_{U_R}$, and $A_U$ are $3\times3$ 
unknown 
matrices.
Indeed the adjective {\em minimal} in the MSSM acronyms refers 
to the particle content of the model but does not specify 
its flavor structure. 

Because of this large number of free parameters, we cannot 
discuss the implications of the MSSM in flavor physics 
without specifying in more detail the flavor structure of the model. 
The versions of the MSSM analysed in the literature range from 
the so-called Constrained MSSM (CMSSM), where the complete model 
is specified in terms of only four free parameters (in addition to the 
SM couplings), to the MSSSM without $R$ parity and generic flavor 
structure, which contains a few hundreds of new free parameters.

Throughout the large amount of work in the past decades it has 
became clear that the MSSM with generic flavor structure and 
squarks in the TeV range is not compatible with precision tests 
in flavor physics. This is true even if we impose $R$ parity, 
the discrete symmetry which forbids single s-particle production, 
usually advocated to prevent a too fast proton decay. In this 
case we have no tree-level FCNC amplitudes, but the loop-induced
contributions are still too large compared to the SM ones 
unless the squarks are highly degenerate or have very small
intra-generation mixing angles. This is nothing but a 
manifestation in the MSSM context of the general flavor problem
illustrated in Sec.~\ref{sect:EFT}.

The flavor problem of the MSSM is an important clue about the underling 
mechanism of supersymmetry breaking. On general grounds, mechanisms 
of SUSY breaking with flavor universality (such as gauge mediation) 
or with heavy squarks (especially in the case of the 
first two generations) tends to be favoured. However, several options are 
still open.
These range from the very restrictive CMSSM case, which is a 
special case of MSSM with MFV, to more general scenarios with 
new small but non-negligible sources of flavor symmetry breaking.

\paragraph{Flavor Universality, MFV, and RGE in the MSSM.}
\label{sect:MSSM_MFV}
Since the squark fields have well-defined transformation
properties under the SM quark-flavor group ${\mathcal G}_q$,
the MFV hypothesis can easily be implemented in the MSSM 
framework following the general rules outlined in 
Sec.~\ref{sect:MFV}. 

We need to consider all possible interactions compatible 
with i) softly-broken supersymmetry; ii) the breaking of 
${\mathcal G}_q$ via the spurion fields $Y_{U,D}$. 
This allows to express the squark mass terms and 
the trilinear quark-squark-Higgs couplings 
as follows~\cite{Hall:1990ac,D'Ambrosio:2002ex}:
\begin{eqnarray}
{\tilde m}_{Q_L}^2 &=& {\tilde m}^2 \left( a_1 {1 \hspace{-.085cm}{\rm l}} 
+b_1 Y_U Y_U^\dagger +b_2 Y_D Y_D^\dagger 
+b_3 Y_D Y_D^\dagger Y_U Y_U^\dagger
+b_4 Y_U Y_U^\dagger Y_D Y_D^\dagger +\ldots
 \right)~, \nonumber  \\
{\tilde m}_{U_R}^2 &=& {\tilde m}^2 \left( a_2 {1 \hspace{-.085cm}{\rm l}} 
+b_5 Y_U^\dagger Y_U +\ldots \right)~,  \qquad 
A_U ~=~ A\left( a_3 {1 \hspace{-.085cm}{\rm l}} 
+b_6 Y_D Y_D^\dagger +\ldots \right) Y_U~,\qquad 
\label{eq:MSSMMFV}
\end{eqnarray}
and similarly for the down-type terms. 
The dimensionful parameters $\tilde m$ and $A$,
expected to be in the range few 100 GeV -- 1 TeV, 
set the overall scale of the soft-breaking terms.
In Eq.~(\ref{eq:MSSMMFV}) we have explicitly shown  
all independent flavor structures which cannot be absorbed into 
a redefinition of the leading terms (up to tiny contributions 
quadratic in the Yukawas of the first two families),  
when $\tan\beta$ is not too large and the bottom Yukawa coupling 
is small, the terms quadratic in $Y_D$ can be dropped.

In a bottom-up approach, the dimensionless coefficients 
$a_i$ and $b_i$ should be considered as free parameters 
of the model. Note that  this structure is 
renormalization-group invariant: the values of 
$a_i$ and $b_i$ change according to the 
Renormalization Group (RG) flow, but the general structure 
of Eq.~(\ref{eq:MSSMMFV})
is unchanged. This is not the case if the $b_i$ are set to zero,
corresponding to the so-called hypothesis of {\em flavor universality}.
In several explicit mechanism of supersymmetry breaking, 
the condition of flavor universality
holds at some high scale $M$, such as the scale of 
Grand Unification in the CMSSM (see below) or the mass-scale of the
messenger particles in gauge mediation (see Ref.~\cite{Giudice:1998bp}). 
In this case  non-vanishing 
$b_i \sim (1/4\pi)^2 \ln M^2/ {\tilde m}^2$ are 
generated by the RG evolution. 
As recently 
pointed out in Ref.~\cite{Paradisi:2008qh,Colangelo:2008qp}, 
the RG flow in the MSSM-MFV 
framework exhibit quasi infra-red fixed points: even
if we start with all the $b_i =\mathcal{O}(1)$ at some high scale, 
the only non-negligible terms at the TeV scale are those 
associated to the $Y_U Y_U^\dagger$ structures.

If we are interested only in low-energy processes we can integrate 
out the supersymmetric particles at one loop and project this 
theory into the general EFT discussed in the previous sections. 
In this case the coefficients of the dimension-six effective operators 
written in terms of SM and Higgs fields (see Tab.~\ref{tab:MFV}) 
are computable in terms of the supersymmetric soft-breaking parameters.
The typical effective scale suppressing these operators 
(assuming an overall coefficient $1/\Lambda^2$) is
$\Lambda \sim  4 \pi\tilde m$.
Looking at the bounds in Tab.~\ref{tab:MFV}, we then conclude that  
if MFV holds, the present bounds on FCNCs do not exclude squarks in 
the few hundred GeV mass range, i.e.~well within the LHC reach.

\paragraph{The CMSSM framework.}
\label{sec:CMSSM}
The CMSSM, also known as mSUGRA, is the supersymmetric extension 
of the SM with the minimal particle content and the maximal 
number of universality conditions on the soft-breaking terms. 
At the scale of Grand Unification ($M_{\rm GUT} \sim 10^{16}$~GeV) 
it is assumed that there are only three independent 
soft-breaking terms: the universal gaugino mass (${\tilde m}_{1/2}$), 
the universal trilinear term ($A$), and the universal 
sfermion mass ($\tilde m_0$). The model has two additional 
free parameters in  the Higgs sector (the so-called $\mu$ 
and $B$ terms), which control the vacuum expectation 
values of the two Higgs fields (determined also by the 
RG running from the unification scale down to the electroweak scale). 
Imposing the correct $W$- and $Z$-boson masses allow us 
to eliminate one of these Higgs-sector parameters, the remaining 
one is usually chosen to be $\tan\beta$. As a result, the model is fully 
specified in terms of the three high-energy parameters
$\{ {\tilde m}_{1/2}, {\tilde m}_0, A \}$, and the 
low-energy parameter $\tan\beta$.\footnote{More precisely, 
for each choice of $\{ {\tilde m}_{1/2}, {\tilde m}_0, A, \tan\beta\}$
there is a discrete ambiguity related to the sign of the $\mu$ term.}
This constrained version of the MSSM is an example of a SUSY
model with MFV. Note, however, that the model is much more 
constrained than the general MSSM with MFV: in addition to be 
flavor universal, the soft-breaking terms at the unification 
scale obey various additional constraints 
(e.g.~in Eq.~(\ref{eq:MSSMMFV}) we have $a_1=a_2$ and $b_i=0$).

\medskip 

In the MSSM with $R$ parity we can distinguish 
five main classes of one-loop diagrams contributing 
to FCNC and CP violating processes with external down-type 
quarks. They are distinguished according to the virtual particles running 
inside the loops: $W$ and up-quarks (i.e.~the leading SM amplitudes),
charged-Higgs and up-quarks, charginos and up-squarks, 
neutralinos and down-squarks, gluinos and down-squarks. Within the CMSSM, 
the charged-Higgs and chargino exchanges yield the dominant 
non-standard contributions. 

Given the low number of free parameters, the CMSSM is very predictive 
and phenomenologically constrained by the precision measurements 
in flavor physics. The most powerful low-energy constraint comes 
from $B \to X_s \gamma$. For large values of $\tan \beta$, strong 
constraints are also obtained from $B_s \to \mu^+ \mu^-$, 
$\Delta M_s$ and from $B(B \to \tau \nu)$. If these observables 
are within the present experimental bounds, the constrained nature 
of the model implies essentially no observable deviations from the SM 
in other flavor-changing processes. Interestingly enough, the CMSSM 
satisfy at the same time the flavor constraints and those from 
electroweak precision observables for squark masses below 1 TeV
(see e.g.~\cite{Buchmueller:2008qe,Buchmueller:2007zk}).

In principle, within the CMSSM the relative phases of the free parameters
leads to two new observable CP-violating phases (beside the CKM phase).
However, these phases are flavor-blind and turn out to be severely 
constrained by the experimental bounds on the electric dipole moments. 
In particular, the combination of neutron and electron edms forces 
these phases to be at most of $\mathcal{O}(10^{-2})$ for squark masses 
masses below 1 TeV. Once this constraints are  satisfied, 
the effects of these new phases in the $B$, $D$ and $K$ systems 
are negligible.

\paragraph{The Mass Insertion Approximation in the general MSSM.}
\label{sec:genFCNC}
Flavor universality at the GUT scale is not a 
general property of the MSSM, even if the model is embedded 
in a Grand Unified Theory. If this assumption is relaxed, 
new interesting phenomena can occur in flavor physics. 
The most general one is the appearance of gluino-mediated
one-loop contributons to FCNC 
amplitudes~\cite{Ellis:1981ts,Barbieri:1981gn}.

The main problem when going beyond simplifying assumptions, 
such as flavor universality of MFV, is the proliferation in 
the number of free parameters. A useful model-independent 
parameterization to describe the new phenomena occurring 
in the general MSSM with R parity conservation 
is the so-called mass insertion (MI) 
approximation~\cite{Hall:1985dx}. Selecting a flavor basis 
for fermion and sfermion states  where all the couplings 
of these particles to neutral gauginos are flavor diagonal,
the new flavor-violating effects are parametrized in terms 
of the non-diagonal entries of the sfermion mass matrices. 
More precisely, denoting by $\Delta$ the off-diagonal terms in the
sfermion mass matrices (i.e. the mass terms relating sfermions of the
same electric charge, but different flavor), the sfermion propagators
can be expanded in terms of $\delta = \Delta/ \tilde{m}^2$,
where $\tilde{m}$ is the average sfermion mass.  As long as $\Delta$
is significantly smaller than $\tilde{m}^2$ (as suggested by the absence 
of sizable deviations form the SM), one can truncate the series 
to the first term of this expansion and the experimental information
concerning FCNC and CP violating phenomena translates into upper
bounds on these $\delta$'s~\cite{Gabbiani:1996hi}.

The major advantage of the MI method is that it is not necessary to 
perform a full diagonalization of the sfermion mass matrices, 
obtaining a substantial simplification in the comparison of 
flavor-violating effects 
in different processes. There exist four type of mass insertions 
connecting
flavors $i$ and $j$ along a sfermion propagator:
$\left(\Delta_{ij}\right)_{LL}$, $\left(\Delta_{ij}\right)_{RR}$,
$\left(\Delta_{ij}\right)_{LR}$ and
$\left(\Delta_{ij}\right)_{RL}$. The indices $L$ and $R$ refer to the
helicity of the fermion partners.

In most cases the leading non-standard amplitude 
is the gluino-exchange one, which is enhanced by one or two powers of 
the ratio $(\alpha_{\rm strong}/\alpha_{\rm weak})$ with respect 
to neutralino- or chargino-mediated amplitudes. When analysing the 
bounds, it is customary to consider one non-vanishing MI at a time, 
barring accidental cancellations.  This procedure is
justified a posteriori by observing that the MI bounds have
typically a strong hierarchy, making the destructive interference
among different MIs rather unlikely. The bound thus obtained 
from recent measurements in $B$ and $K$ 
physics\footnote{~The bounds on the 1-2 sector  
are obtained from  the measurements of $\Delta M_K$,
  $\varepsilon$ and $\varepsilon^\prime/\varepsilon$.
  In particular $\Delta M_K$ and $\varepsilon$
  bound the real and imaginary part of the product
  $(\delta^d_{12} \delta^d_{12} )$,
  while $\varepsilon^\prime/\varepsilon$ puts a bound on 
Im($\delta^d_{12}$). 
  The bounds on the 1-3 sector
  are obtained from $\Delta M_{B_d}$ (modulus) and
  the CP violating asymmetry in $B\to J/\Psi K$ (phase). 
  The bounds on the $2-3$ sector are derived mainly from 
  $\Delta M_{B_s}$, $B\to X_s \gamma$ and $B \to X_s \ell^+\ell^-$.}
are reported in 
Tab.~\ref{tab:MIquarks}.\footnote{
The leading $\Delta F=1$ and $\Delta F=2$ gluino-mediated 
amplitudes in the MI approximation 
can be found in Ref.\cite{Gabbiani:1996hi}.
In the $\Delta F=2$ case also the NLO QCD 
corrections to effective Hamiltonian are 
known~\cite{Ciuchini:2006dw}. 
A more complete set of supersymmetric amplitudes
in the MI approximation, including chargino-mediated  
relevant in the  large-$\tan \beta$ limit, 
can be found in  Ref.~\cite{Foster:2005wb}.}
\begin{table}
  \centering
\caption{Upper bounds at 95\% C.L.~on the dimensionless 
down-type mass-insertion parameters (see text) 
for squark and gluino masses of 350 GeV (from 
Ref.~\cite{Buchalla:2008jp}). } 
\label{tab:MIquarks}
  \begin{tabular}{llll} 
    $\vert(\delta^d_{12})_{LL,RR}\vert < 1\cdot 10^{-2} $\
    &
    $\vert( \delta^d_{12} )_{LL=RR} \vert < 2 \cdot 10^{-4}$\
    &
    $\vert( \delta^d_{12})_{LR}\vert < 5\cdot 10^{-4} $\
    &
    $\vert( \delta^d_{12})_{RL}\vert < 5 \cdot 10^{-4} $ \    
\\ \hline
    $\vert(\delta^d_{13} )_{LL,RR} \vert < 7\cdot 10^{-2}$\
    &
    $\vert( \delta^d_{13})_{LL=RR}\vert < 5 \cdot 10^{-3}$\
    &
    $\vert( \delta^d_{13} )_{LR}\vert < 1\cdot 10^{-2}$  \
    &
    $\vert( \delta^d_{13} )_{RL}\vert < 1 \cdot 10^{-2}$\
\\ \hline
    $\vert( \delta^d_{23} )_{LL}\vert < 2\cdot 10^{-1}$\
    &
    $\vert(\delta^d_{23} )_{RR}\vert < 7 \cdot 10^{-1}$\
    &
    $\vert( \delta^d_{23} )_{LL=RR} \vert <  5\cdot 10^{-2}$\
    &
    $\vert(  \delta^d_{23} )_{LR,RL}\vert < 5 \cdot 10^{-3}$ \
\\ \\
\end{tabular}
\end{table}
The bounds mainly depend on the gluino and on the average 
squark mass, scaling as the inverse mass (the inverse mass 
square) for bounds derived from $\Delta F=2$ ($\Delta F=1$)
observables. 

The only clear pattern emerging from these bounds 
is that there is no room for sizable 
new sources of flavor-symmetry breaking. 
However, it is too early to draw
definite conclusions since some of the bounds,
especially those in the 2-3 sector,
are still rather weak. As suggested by various authors 
(see e.g. ), the possibility 
of sizable deviations from the SM in the 2-3 
sector could fit well with the large 2-3 mixing 
of light neutrinos, in the context of 
a unification of quark and lepton 
sectors~\cite{Chang:2002mq,Ciuchini:2007ha}.

\subsubsection{Non-supersymmetric extensions of the Standard Model}

We conclude this chapter outlining two of the 
general features of flavor physics appearing 
in non-supersymmetric extensions of the Standard Model,
without entering  the details of specific theories. 

In models with generic flavor structure, the most stringent 
constraints on the new flavor-violating couplings are tipically
derived from Kaon physics (as it also happens for the bounds 
in Tab.~\ref{tab:MIquarks}). This is a consequence of the 
high suppression, within the SM,
of short-distance dominated 
FCNC amplitudes between the first two families:
\begin{equation}
\mathcal{A}(s \to d)_{\rm SM} = \mathcal{O}(\lambda^5)~, 
\qquad  
\mathcal{A}(b \to d)_{\rm SM} = \mathcal{O}(\lambda^3)~, 
\qquad  
\mathcal{A}(b \to s)_{\rm SM} = \mathcal{O}(\lambda^2)~.
\end{equation}
As a result, a natural place to look for sizable deviations
from the SM are rare decays $K\to \pi\nu\bar\nu$ 
and $K_L \to \pi^0\ell^+\ell^-$ (see for instance the 
expectations for these decays in the {\em Littlest Higgs 
model} without \cite{Buras:2006wk} and with \cite{Blanke:2006eb,Blanke:2007wr,Goto:2008fj,Blanke:2009xx}) T-parity.
These decays allow us to explore the 
sector of $\Delta F=1$ $s\to d$ transitions, 
that so far is only loosely tested.

An interesting alternative to MFV, which naturally 
emerges in models with {\em Extra Space-time 
Dimensions} (or models with strongly interacting dynamics 
at the TeV scale), is the hypothesis of 
hierarchical fermion 
profiles~\cite{ArkaniHamed:1999dc,Gherghetta:2000qt,Huber:2000ie,
Agashe:2004cp,Agashe:2005hk} (which is equivalent to the 
hypothesis of hierarchical kinetic terms~\cite{Davidson:2007si}). 
Contrary to MFV, this hypothesis 
(often denoted as NMFV or RS-GIM mechanism) 
is not a symmetry principle but a dynamical argument: 
light fermions are weakly coupled to 
the new TeV dynamics, with a strength inversely 
proportional to their Yukawa coupling (or better the square root
of their SM Yukawa coupling). Also in the case the most significant 
constraints are derived from Kaon physics. However, in this case 
the stringent constraints from $\epsilon_K$ and 
$\epsilon^\prime_K$ generically disfavour visible effects in other observables,
although it is still possible to have some effect, in particular in the phase
of the $B_s$ mixing amplitude~\cite{Blanke:2008yr,Blanke:2008zb}.
In view of the little CP problem in the kaon, several modifications of  
the quark-flavor sector of warped extra-dimensional models have been
proposed. Most of them try to implement the notion of MFV
into the RS framework~\cite{Cacciapaglia:2007fw,Csaki:2008eh,Csaki:2009wc}
by using flavor symmetries. The
downside of these constructions is that they no longer try to explain
the fermion mass hierarchy, but only accommodate it with the least
amount of flavor structure, making this class of models hard to probe
via flavor precision tests.

\section{Experimental Primers}
\label{sec:expPrimers}
This section contains all the relevant information on experiments and experimental techniques which are needed throughout the report. 
\subsection{Overview of experiments}

\subsubsection{Kaon experiments}

In recent years, many experiments have been performed to precisely measure many Kaon decay 
parameters. Branching ratios (BR's) for main, subdominant, and rare decays, lifetimes, parameters of decay densities, 
and charge asymmetries have been measured
with unprecedented accuracy for $K_S$, $K_L$, and $K^\pm$. Different techniques have been used, often allowing 
careful checks of the results from experiments with independent sources of systematic errors. 

In the approach of NA48~\cite{Fanti:2007vi} at the CERN SPS and KTeV~\cite{AlaviHarati:2002ye} at the Fermilab Tevatron,
Kaons were produced by the interactions of intense high-energy 
proton beams on beryllium targets (see Tab.~\ref{tab:na48ktev}). 
Both experiments were designed to measure the direct CP violation parameter 
$\Re(\varepsilon^\prime/\varepsilon)$ via the double ratio of 
branching fractions for $K_S$ and $K_L$ decays to $\pi^+\pi^-$ and $\pi^0\pi^0$ final states.
In order to confirm or disprove the conflicting results of 
the former-generation experiments, NA31~\cite{Barr:1993rx} and  E-731~\cite{Gibbons:1993zq}, the goal was to reach
an uncertainty of a few parts in $10^4$. This not only
requires intense $K_L$ beams,
so as to guarantee the observation of at least $10^8$ decays of the rarest of the four modes, i.e., $K_L\to\pi^0\pi^0$; it
also made it necessary to achieve a high level of cancelation of the systematic uncertainties for $K_L$ and $K_S$ 
detection, separately for neutral and charged decay modes, as well as rejection of the order of $10^6$
for the most frequent $K_L$ backgrounds, $K_L\to 3\pi^0$ and $K_L\to\pi \ell\nu$.

In both setups, the target producing the $K_L$ beam is the origin of coordinates. $K_L$'s 
are transported by a $\sim100$-m long beam line, with magnetic filters to remove unwanted particles and collimators 
to better define the Kaon-beam direction, to a fiducial decay volume (FV). The FV 
is surrounded by veto detectors, for rejecting decay products emitted at large angles and therefore 
with relatively low energy; this is particularly useful for the rejection of $K_L\to 3\pi^0$ background.
The FV is followed by a tracker to measure the charge, multiplicity, and momentum of charged decay products, and
by a fast scintillator hodoscope to provide the first-level trigger and determine the event time. 
The tracking resolution $\sigma_p/p$ is $(4\oplus p[\mathrm{GeV}]/11)\times 10^{-3}$ for NA48 and 
$(1.7\oplus p[\mathrm{GeV}]/14)\times 10^{-3}$ for KTeV.
In the downstream (forward) region, both experiments use 
fine-granularity, high-efficiency calorimeters to accurately measure multiplicity and energy of photons and electrons
for the identification of $K_L\to2\pi^0$. The KTeV calorimeter is made of pure CsI, while the NA48 calorimeter
is made of liquid krypton.
The energy resolution $\sigma_E/E$ is $3.2\%/\sqrt{E[\mathrm{GeV}]}\oplus 9\%/E[\mathrm{GeV}]\oplus0.42\%$ for NA48 and
$2\%\sqrt{E[\mathrm{GeV}]}\oplus0.4\%$ for KTeV.
Behind the calorimeter, the detectors are completed by calorimeters for
muon detection.
Different methods are used for the production of a $K_S$ beam. In NA48, a channeling crystal
bends a small and adjustable fraction of protons that do not interact in the $K_L$ target to a dedicated beam line; 
these protons are then
transported and collimated to interact with a second target located few meters before the FV, thus producing a
$K_S$/$K_L$ beam with momentum and direction close to those of the $K_L$ beam, 
so that most of $K_S$ decays are in the FV.
$K_S$ decays are identified by tagging protons on the secondary beam line
using time of flight.
In KTeV, two $K_L$ beams are produced at the first target, 
with opposite transverse momenta in the horizontal direction, and a thick regenerator is placed in one of the two
beams to produce $K_S$, again a few meters before the FV. $K_S$ and $K_L$ decays are distinguished by their different transverse position on the detector.
In both setups, one measures decays from a $K_L$ beam with $<\sim10^{-6}$ 
contamination from $K_S$, and from an enriched-$K_S$ 
beam contaminated by a $K_L$ component, which is determined very precisely during analysis.

\begin{table}[htb]
\centering
\caption{\label{tab:na48ktev} Typical beam parameters for $K$ production in the NA48, KTeV, ISTRA+, and E787/E949 experiments.}
\begin{tabular}{c|c|c|c|c}
 Experiment & proton energy (GeV) & $K$, spill/cycle         & $K$ momentum & Beam type  \\ \hline
 NA48       & 450 & $1.5\times10^{12}$, 2.4~s/14.4~s & (70--170)~GeV  & $K_S$--$K_L$  \\ 
 NA48/1     & 400 & $5\times10^{10}$, 4.8~s/16.2~s & (70--170)~GeV  & $K_S$  \\ 
 NA48/2     & 400 & $7\times10^{11}$, 4.8~s/16.8~s & 60~GeV  & $K^\pm$  \\ 
 KTeV       & 450--800    & $3\times10^{12}$, 20~s/60~s      & (40--170)~GeV  & $K_S$--$K_L$, $K_L$           \\
 ISTRA+     & 70          & $3\times10^{6}$, 1.9~s/9.7~s     & 25~GeV         & $K^-$                        \\ 
 E787       & 24          & 4--7$\times10^6$, 1.6~s/3.6~s    & 710/730/790~MeV, stopped & $K^+$             \\ 
 E949       & 21.5        & 3.5$\times10^6$, 2.2~s/5.4~s     & 710~MeV, stopped & $K^+$                      \\ \hline 
\end{tabular}
\end{table}

The KTeV experiment at Fermilab underwent different phases. 
The E-799 KTeV phase-I used the apparatus of the E-731 experiment~\cite{Gibbons:1993zq}, upgraded
to handle increased $K_L$ fluxes and to study multibody rare $K_L$ and $\pi^0$ decays. In phase-II of E-799,
a new beam line and a new detector were used, including a new CsI
calorimeter and a new transition radiation detector, thus allowing a sensitivity of $10^{-11}$ on the BR of many $K_L$
decay channels and improving by large factors the accuracy on the ratio of BR's of all of the main $K_L$ channels. 
Finally, using the E-832 experimental configuration $\mathrm{Re}(\varepsilon^\prime/\varepsilon)$ was
measured to few parts in $10^{-4}$~\cite{AlaviHarati:2002ye}. 
The NA48 program involved different setups as well. After operating to
simultaneously produce $K_L$'s and $K_S$'s, the beam parameters were optimized in the NA48/1 phase 
to produce a high-intensity $K_S$ beam
for the study of rare $K_S$ decays, reaching the sensitivity of $10^{-10}$ for some specific channels and 
especially improving knowledge on those with little background from the accompanying $K_L$ decay to the same final
state. 
Subsequent beam  and detector upgrades, including the insertion of a Cerenkov beam counter (``NA48/2 setup'') 
allowed production of simultaneous unseparated charged Kaon 
beams for the measurement of CP violation from the charge asymmetry in the Dalitz densities for three-pion decays~\cite{Batley:2007md}. 
The NA48/2 phase allowed the best present sensitivities for many rare $K^\pm$ decays to be reached, with BR's 
as low as $10^{-8}$ and
improved precision for the ratios of BR's of the main $K^\pm$ channels.
A recent run made in 2007 by the NA62 collaboration using
the NA48/2 setup was dedicated to a precision measurement of the ratio $\Gamma(K_{e2})/\Gamma(K_{\mu2})$.
A future experiment is foreseen at the CERN SPS for the measurement of the ultra-rare decay $K^+\to\pi^+\nu\overline{\nu}$ 
with a 10\% accuracy~\cite{Anelli:2005ju,na62status}.

An unseparated charged Kaon beam was also exploited for study of charged Kaon decay parameters with the ISTRA+ 
detector~\cite{Ajinenko:2001iz} at the
U-70 proton synchrotron in IHEP, Protvino, Russia. A beam (see Tab.~\ref{tab:na48ktev}), 
with $\sim3\%$ $K^-$ abundance is analyzed by a magnetic 
spectrometer with four proportional chambers and a particle identification is provided by three Cerenkov counters.
The detector concept is similar to those presented above, with the tracking of charged decay products provided
by drift chambers, drift tubes, and proportional chambers and with the calorimetry for photon vetoing at large
angle or energy measurement at low angle performed by lead-glass detectors.

A different approach for the study of the ultra-rare $K\to\pi\nu\overline{\nu}$ decay 
and the search for lepton-flavor violating transitions was taken
by the E787~\cite{Adler:1997am,Adler:2000by,Adler:2004hp} and E949~\cite{Adler:2008zza} experiments at the Alternating Gradient Synchrotron (AGS) of the Brookhaven National Laboratory. Charged 
Kaons were
produced by 24-GeV protons interacting on a fixed target. A dedicated beam line transported, purified and momentum selected 
Kaons. The beam (see Tab.~\ref{tab:na48ktev}) had 
adjustable momenta from 670~MeV to 790~MeV and a ratio of Kaons to
pions of $\sim4/1$.

The detector design was optimized to reach sensitivities of the order of $10^{-10}$ on the BR's for decays of $K^\pm$
to charged particles, especially lepton-flavor violating decays, such as $K\to\pi\mu e$: 
for this purpose, redundant and independent measurements for particle identification and
kinematics were provided, as well as efficient vetoing for photons. 
The beam was first analyzed by Cerenkov and wire-chamber detectors, and later slowed down
by a passive BeO degrader and an active lead-glass radiator, the Cerenkov light of which was used to veto pions
and early $K$ decays. Kaons were then stopped inside an active target made of scintillating fibers. 
The charged decay products emitted at large angle
were first analyzed in position, trajectory, and momentum by a drift chamber; their range 
and kinetic energy was then measured in a Range Stack alternating plastic scintillator with passive material. The readout
of the Range Stack photomultipliers was designed to record times and shapes of pulses up to 6.4~$\mu$s after the trigger, thus allowing the
entire chain of $\pi\to\mu\to e$ decays to be detected and allowing clean particle identification. 
The detector was surrounded by electromagnetic calorimeters for 
hermetic photon vetoing: a lead/scintillator barrel and two CsI-crystal endcaps. Two lead/scintillating-fiber 
collars allowed vetoing of charged particles emitted at small angles. Using this setup, the best sensitivity 
to date was obtained for the BR for $K\to\pi\nu\nu$, reaching the $10^{-10}$ level.


Precision studies of $K_S$, $K_L$, and $K^\pm$ main and subdominant decays were performed 
with a different setup using 
the KLOE detector at the DA$\Phi$NE. 
DA$\Phi$NE, the Frascati $\phi$ factory, is an $e^{+}e^{-}$ collider
working at $\sqrt{s}\sim m_{\phi} \sim 1.02$~GeV. $\phi$ mesons are produced
essentially at rest with a visible cross section of $\sim$~3.1~$\mu$b
and decay into $K_SK_L$ and $K^+K^-$ pairs with BR's of $\sim 34\%$ and $\sim49\%$, respectively.
During KLOE data taking, which started in 2001 and concluded in 2006, the peak luminosity of DA$\Phi$NE improved
continuously, reaching $\sim2.5\times10^{32}$~cm$^{-2}$~s$^{-1}$ at the end. The total luminosity integrated
at the $\phi$ peak is $\sim2.2$~fb$^{-1}$, corresponding to $\sim2.2$ ($\sim3.3$) billion $K^0\overline{K^0}$
($K^+K^-$) pairs.

Kaons get a momentum of $\sim$~100~MeV/$c$ which translates into a low speed, $\beta_{K}\sim$~0.2.
$K_S$ and $K_L$ can therefore be distinguished by their mean decay lengths:
$\lambda_{S} \sim $~0.6~cm and $\lambda_{L} \sim $~340~cm.
$K^+$ and $K^-$ decay with a mean length of $\lambda_\pm\sim $~90~cm and can be 
distinguished from their decays in flight to one of the two-body final states 
$\mu\nu$ or $\pi\pi^0$.

The Kaon pairs from $\phi$ decay are produced in a pure $J^{PC}=1^{--}$ quantum state, so that 
observation of a $K_L$ ($K^+$) in an event signals, or tags, the presence of a $K_S$ ($K^-$)
and vice versa; highly pure and nearly monochromatic $K_S$, $K_L$, and $K^\pm$
beams can thus be obtained and exploited to achieve high precision in the measurement of 
absolute BR's.

The analysis of Kaon decays is performed with the KLOE detector, consisting essentially of a drift chamber, DCH, surrounded by an
electromagnetic calorimeter, EMC. A superconducting coil provides a 0.52~T magnetic field.
The DCH~\cite{Adinolfi:2002uk} is a cylinder of 4~m in diameter
and 3.3~m in length, which constitutes a fiducial volume 
for $K_L$ and $K^\pm$ decays extending for $\sim0.5\lambda_L$ and $\sim1\lambda_\pm$. 
The momentum resolution for tracks 
at large polar angle is $\sigma_{p}/p \leq 0.4$\%. 
The invariant mass reconstructed from the momenta of the two pion tracks of a $K_S\to\pi^+\pi^-$ decay peaks
around $m_K$ with a resolution of $\sim$800~keV, thus allowing clean $K_L$ tagging. 
The c.m.\ momenta reconstructed from identification of 1-prong $K^\pm\to\mu\nu,\pi\pi^0$ 
decay vertices in the DC peak around the expected values with a resolution of 1--1.5~MeV, 
thus allowing clean and efficient $K^\mp$ tagging. 

The EMC is a lead/scintillating-fiber sampling calorimeter~\cite{Adinolfi:2002zx}
consisting of a barrel and two endcaps, with good
energy resolution, $\sigma_{E}/E \sim 5.7\%/\sqrt{\rm{E(GeV)}}$, and excellent 
time resolution, $\sigma_{T} =$~54~ps$/\sqrt{\rm{E(GeV)}} \oplus 50$ ps. 
About 50\% of the $K_L$'s produced reach the EMC, where most interact.
A signature of these interactions is the presence of  
an high-energy cluster not connected to any charged track, with a time corresponding to a low velocity: 
the resolution on $\beta_K$ corresponds to a resolution of $\sim1$~MeV
on the $K_L$ momentum. This allows clean $K_S$ tagging.
The timing capabilities of the EMC are exploited to precisely reconstruct 
the position of decay vertices of $K_L$ and $K^\pm$ to $\pi^0$'s from the
cluster times of the emitted photons, thus allowing a precise measurement 
of the $K_L$ and $K^\pm$ lifetimes.

With this setup, KLOE reached the best sensitivity for absolute BR's of the main $K^\pm$, $K_L$, and $K_S$ channels
(dominating world data in the latter case) and improved the knowledge of semileptonic decay rate densities and 
lifetimes for $K^\pm$ and $K_L$.

\subsubsection{\boldmath $B$ Factories}
\label{sec:primers:bfactories}
The high statistics required to perform precise flavor physics with $B$ mesons has been accomplished 
by B-Factories colliding electrons and positrons at the energy of the \FourS\ resonance ($\epem\to\FourS
\B\Bb$): CESR at LEPP (Cornell, USA), PEP-II~\cite{:1993mp} at SLAC (Stanford, USA) and KEK-B~\cite{Bondar:2001ud} at 
KEK (Tsukuba, Japan). 
Measurements that exploit the evolution of the observables with the decay time of the mesons 
also require asymmetric beams in order to ensure a boost to the produced mesons.

To this aim PEP-II (KEK-B) collide 3.1 (3.5) \gev\ positrons on 9.0 (8.0) \gev 
electrons, thus achieving a boost $\beta\gamma=0.56 (0.43)$. The other design parameters of the B-Factories 
are listed in Tab:~\ref{tab:primers:bfactories}. The design instantaneous luminosities were $10^{33}$, $3\times 10^{33}$, and $1\times 10^{34} \cm^{-2}\sec^{-1}$ for CESR, PEP-II and KEK-B, respectively.
 
\begin{table}
\centering
\caption{Accelerator parameters of the B-Factories. The design parameters are given for PEP-II and KEK-B. The final 
running parameters for CESR are given.
\label{tab:primers:bfactories}
}
\begin{tabular}{l||c|c|c|c|c|}
\hline
  & CESR & \multicolumn{2}{|c|}{KEK-B}& \multicolumn{2}{|c|}{PEP-II}\\ 
  &  & LER & HER & LER & HER \\ \hline
Energy (\gev) & 5.29 & 3.5 & 8.0 & 3.1 & 9.0 \\ \hline
Collision mode & 2~mrad & \multicolumn{2}{|c|}{11$mrad$}& \multicolumn{2}{|c|}{Head-on}\\ \hline
Circumference (\m) & 768 & \multicolumn{2}{|c|}{3018}& \multicolumn{2}{|c|}{2199}\\ \hline
$\beta^*_x/\beta^*_y$ (\cm) &100/1.8 & 100/1 & 100/1 & 37.5/1.5& 75/3\\ \hline
$\xi^*_x/\xi^*_y$ &0.03/0.06  & \multicolumn{2}{|c|}{0.05/0.05}& \multicolumn{2}{|c|}{0.03/0.03}\\ \hline
$\epsilon^*_x/\epsilon^*_y~(\pi \mathrm{rad}-\nm)$ & 210/1 & 19/0.19 & 19/0.19 & 64/2.6& 48.2/1.9\\ \hline
relative energy spread ($10^{-4}$)& 6.0 & 7.7 & 7.2 & 9.5 & 6.1 \\ \hline
Total Current (A) & 0.34 & 2.6 & 1.1 & 2.14 & 0.98 \\ \hline
number of bunches & 45 & \multicolumn{2}{|c|}{5120}& \multicolumn{2}{|c|}{1658}\\ \hline
RF Frequency ($MHz$)/ Voltage ($MV$)&500/5 & 508/22 & 508/48 & 476/9.5 & 476/17.5\\ \hline
number of cavities & 4 & 28 & 60 & 10 & 20 \\ \hline
\end{tabular}
\end{table}

The accelerator performances have actually overcome the design: CESR has ceased its operations as B-Factory in 1999 with
a peak luminosity ${\cal{L}}=1.2\times 10^{33}~\cm^{-2}\sec^{-1}$, PEP-II has ended 
its last run in April 2008 with a peak 
luminosity of $12\times 10^{33}\cm^{-2}\sec^{-1}$ and KEK-B, which is still operational and awaits an upgrade (Super-KEK-B),
 has achieved a luminosity as high as $1.7\times 10^{34}~\cm^{-2}\sec^{-1}$. The total collected luminosities are $15.5$, $553$ and $895~\invfb$
for CESR, PEP-II and KEK-B, respectively.

The detectors installed on these accelerators, CLEO-II/II.V/III\footnote{The detector went through several major upgrades during its lifetime. In this section only the final configuration, CLEO-III, is described. The size of the $\Upsilon(4S)$ data-sets collected were $4.7~\invfb$, $9.0~\invfb$, $9.1~\invfb$ with CLEO-II, CLEO-II.V and CLEO-III, respectively.}~\cite{Andrews:1982dp,Kubota:1991ww,Viehhauser:2001ue,Peterson:2002sk,Artuso:2005dc} at CESR, BaBar~\cite{Aubert:2001tu} at PEP-II 
and Belle~\cite{Abashian:2000cg} at KEK-B, are multipurpose
 and require exclusive and hermetic reconstruction of 
the decay products of all generated particles. To this aim the following requirements must be met: (1)
accurate reconstruction of charged-particle trajectories; (2) precise measurement of neutral particle energies; 
and (3) good identification of photons, electrons, muons, charged Kaons, \KS\ mesons and \KL\ mesons. 

The most challenging experimental requirement is the detection of the decay points of the short-lived B mesons. 
CLEO, BaBar and Belle use double-sided silicon-strip detectors allowing full tracking of low-momentum tracks. Four, three and five cylindrical layers are used at CLEO, Belle and BaBar, respectively. To minimize the contribution of multiple scattering, these detectors are located 
at small radii close to the interaction point. For tracking outside the silicon detector, and the measurement of 
momentum, all experiments use conventional drift chambers with a helium-based gas mixture to minimize multiple 
scattering and synchrotron radiation backgrounds.

The other difficult requirement for the detectors is the separation of Kaons from pions. 
At high momentum, this is needed to distinguish topologically identical final states such as $B^{0}\to\pi^{+}\pi^{-}$ and $B^{0}\to K^{+}\pi^{-}$ from one another. At lower momenta, particle 
identification is essential for B flavor tagging.

Three different approaches to high-momentum particle identification have been implemented, 
all of which exploit Cerenkov radiation. At CLEO a proximity focusing RICH with CH$_4$/TEA as the photosensitive medium and LiF as the radiator. The system relies on an expansion gap between the radiator and photon detector to separate the Cherenkov light without the use of additional focusing elements. The RICH has good $K$-$\pi$ separation for charged tracks above $700~\mevc$; below this momenta $\dedx$ measurements in the drift chamber are used for particle identification.  

At Belle, aerogel is used as a radiator. 
Blocks of aerogel are read out directly by fine-mesh phototubes that have high gain and operate reliably in a 
1.5-Tesla magnetic field. Because the threshold momentum for pions in the aerogel is $1.5~\gevc$, below this 
momentum K/$\pi$ separation is carried out using high-precision time-of-flight (TOF) scintillators with a resolution of 
95 ps.
 The aerogel and TOF counter system is complemented by $\dedx$ measurements in the central drift chamber. 
The $\dedx$ system provides $K/\pi$ separation below $0.7~\gevc$ and above $2.5~\gevc$ in the relativistic rise region.

At BaBar, Cerenkov light is produced in quartz bars and then transmitted by total internal reflection to the outside 
of the detector through a water tank to a large array of phototubes where the ring is imaged. The detector is called 
DIRC (Detector of Internally Reflected Cerenkov light). It provides particle identification for particles above $700~\mevc$. 
Additional particle identification is provided by $\dedx$ measurements in the drift chamber and the five-layer silicon 
detector.

To detect photons and electrons, all detectors use large arrays of CsI(Tl) crystals located inside the coil 
of 
the magnet. In BABAR and Belle, another novel feature is the use of resistive plate chambers (RPC) inserted into the 
steel 
return yoke of the magnet. This detector system is used for both muon and $\KL$ detection. At CLEO the iron return yoke of the solenoid is instrumented with plastic streamer counters to identify muons. 

To read out the detectors, BABAR uses electronics based on digital pipelines and incurs little or no dead-time. Belle 
uses 
charge-to-time (Q-to-T) converters that are then read out by multihit time-to-digital counters (TDCs). This allows a 
uniform 
treatment of timing and charge signals. Details of the CLEO data-acquisition system can be found in Ref.~\cite{Viehhauser:2001ue}; the system can handle trigger rates of 1~kHz well above the normal operating conditions (100~Hz).

\subsubsection{\boldmath $\tau$-charm Factories }
\begin{table}
\begin{center}
\caption{Accelerator parameters of $\tau$-charm factories.
\label{tab:primers:charmfactories}
}
\begin{tabular}{lccc}
\hline\hline  
  & BEPC & CESR-c & BEPC-II \\ \hline 
Max. energy (\gev) & 2.2 & 2.08 & 2.3 \\
Collision mode & Head-on & $\pm 3.3$~mrad & 22~mrad \\
Circumference (\m) & 240 & 768 & 240 \\
$\beta^*_x/\beta^*_y$ (\cm) &120/5 & 94/1.2 & 100/1.5 \\ 
$\xi^*_x/\xi^*_y~(10^{-4})$ &350/350  & 420/280 & 400/400 \\
$\epsilon^*_x/\epsilon^*_y (\pi~\mathrm{rad}-\nm)$ & 660/28 & 120/3.5 & 144/2.2 \\
relative energy spread ($10^{-4}$)& 5.8 & 8.2 & 5.2  \\
Total Current (A) & 0.04 & 0.072 & 0.91 \\ 
number of bunches & 1  & 24 & 93 \\
RF Frequency ($MHz$)/ Voltage ($MV$)&200/0.6-1.6 & 500/5 & 500/1.5 \\
number of cavities & 4 & 4 & 2 \\ \hline\hline 
\end{tabular}
\end{center}
\end{table}

Recently there have been two accelerators that have been operating near the $\tau$-charm threshold: BEPC at IHEP (Beijing, China)  and CESR-c \cite{cleocreport} at LEPP (Cornell, USA). The center-of-mass-energy ranges covered are $3.7-5.0~\gev$ and $3.97-4.26~\gev$ by BEPC and CESR-c, respectively. The peak instantaneous luminosities achieved are $12.6\times 10^{30}~\cm^{-2}\sec^{-1}$ and $76~\times 10^{30}~\cm^{-2}\sec^{-1}$. The other parameters of BEPC and CESR-c are given in Tab.~\ref{tab:primers:charmfactories}.

At CESR-c the CLEO-III detector, described in Sec.~\ref{sec:primers:bfactories}, was modified for lower energy data-taking and renamed CLEO-c \cite{cleocreport}. The principal differences were the reduction of the magnetic field from 1.5~T to 1~T and the replacement of the silicon vertex detector by a six-layer inner drift chamber. Both these modifications improved the reconstruction of low momentum tracks. CLEO-c collected 27 million $\psitwos$ events, $818~\invpb$ of integrated luminosity at the $\psiprpr$ and $602~\invpb$ of integrated luminosity at a center-of-mass energy of 4.17~\gev. The latter data set includes a over half a million $D_{s}\Dbar^{*}_{s}$ events. 

The most recent detector installed on BEPC is BES-II \cite{Bai:1994zm,Bai:2001dw}. BES-II collected samples of 58 million $\jpsi$ and 14 million $\psitwos$ events. In addition, an energy scan was performed between center-of-mass energies 3.7 to $5.0~\gev$ to determine both $R$ and the resonances parameters of the higher-mass charmonium states. BES-II tracking was performed by a drift chamber surrounding a straw tube vertex detector.\footnote{The vertex detector was originally operated at Mark III.} A scintillating time-of-flight detector with 180~ps resolution is used for particle identification along with $\dedx$ measurements in the drift chamber. There are sampling electromagnetic-shower counters in the barrel and endcap made from layers of streamer tubes sandwiched between lead absorbers. Outside the 0.4~T solenoid the iron flux return is instrumented with proportional tubes to detect muons.

BEPC and BES-III have recently undergone significant upgrades (see for example \cite{Harris:2008tx}). The BEPC-II accelerator has a design luminosity 100 times greater than BEPC with a peak of $10^{33}~\cm^{-2}\sec^{-1}$. The other parameters of BEPC-II are given in Tab.~\ref{tab:primers:charmfactories}. The BES-III detector has the following components: a He-based drift-chamber, a time-of-flight system with $\sim$100~ps resolution, a CsI(Tl) crystal calorimeter, a 1~T superconducting solenoid and the return yoke is instrumented with RPCs for muon identification. BES-III began taking data in the summer of 2008 and a $\psitwos$ data sample of $10~\invpb$ has already been collected. The collection of unprecedented samples of $\jpsi$, $\psitwos$ and $D$ mesons produced just above open-charm threshold are expected in the coming years.

\subsubsection{Hadron Colliders}
\label{sec:hadrcolliders}

 High energy proton-(anti)proton collisions offer superb opportunities 
for beauty and charm physics due to large production cross section and, 
in contrast to electron-positron colliders running at the \Y4S, the possibility
of studying all species of $b$-mesons and baryons. Present generation experiments,
CDF and \Dzero\, operate at the Fermilab Tevatron providing \ppbar\  collisions at
$\sqrt{s}=1.96$\tev\ in the Run II started in 2002, 
while experiments at the soon to be operated LHC collider
at CERN will study proton-proton collisions at $\sqrt{s}=14$\tev. 
The Tevatron collides {\ppbar} bunches every $396\, \ns$, corresponding to 
an average of $2$ inelastic collisions per crossing at a luminosity of ${\cal L} =
1\times 10^{32}\, \rm cm^{-2}s^{-1}$, typical of the data used to produce the physics results 
discussed here. More recently Tevatron provided peak luminosities in excess 
of $3\times 10^{32}\, \rm cm^{-2}s^{-1}$, and delivered in total 6.5 \ifb as of this writing.

The cross section for centrally produced b-hadrons has been measured with a variety 
of techniques at Tevatron and found to be consistent with NLO theoretical calculations:
an early measurement using inclusive $J/\psi$ down to $P_T=0$ in the rapidity 
range $|y|<0.6$ found $\sigma(\ppbar\rightarrow b + X)= 17.6 \pm 0.4 ({\rm. stat.})
\ase{2.5}{2.3}({\rm sys.})$ $\mu$b\cite{Acosta:2004yw}, while a more recent one using 
fully reconstructed $B^+\rightarrow J/\psi K^+$ measured
$\sigma(\ppbar\rightarrow B^+ + X, P_T>6 \ {\gevc},\ |y|\leq 1)= 2.78 \pm 0.24$ 
$\mu$b\cite{Aaltonen:2009xn} which gives more of an idea of the usable cross-section
for central detectors like CDFII and $\D0$. The fragmentation fraction of b-quarks in 
$B_{u,d}$ and \Bs\ mesons has been measured to be consistent at Tevatron and at 
LEP, with roughly 1 \Bs\ meson produced every 4 $B^+$ or $B^0$, while the rate of 
b-baryons has been reported to be higher at Tevatron with a possible mild $P_T$ 
dependence~\cite{Aaltonen:2008zd}. 
 
The huge production rate for heavy flavored particles has to be contrasted, however, with
the overwhelming inelastic proton-(anti)proton interaction rate which is typically three order 
of magnitudes higher. This poses a fundamental experimental challenge for detectors
ad hadron colliders, which needs to devise trigger strategies in order to be able to record 
as pure a signal as possible while discarding uninteresting events.  

The Tevatron experiments exploit conceptually similar, multi-purpose central detectors with
a cylindrical symmetry around the beam axis, in contrast the dedicated future experiment
at LHC collider (LHCb) employs a radically different forward geometry, in order to 
exploit the rapidly increasing \bbbar\ cross section at high rapidity.

Key elements in the design of detectors for heavy flavor physics at hadron colliders 
are: large magnetic spectrometers for charged particle momentum measurements; 
precision vertex detectors for proper decay time determination and signal separation,
low energy electron and muon identification for triggering, flavor tagging, and 
identification of rare leptonic decays; high rate capability for 
data acquisition and trigger systems. Additionally $\pi$-K identification 
is crucial for flavor tagging and signal separation, and, thus, a significant part of 
the design of the dedicated LHCb detector, while in central muti-purpose 
detectors limited particle id is available with the exception of CDF-II which benefits
from dE/dx and TOF measurements. In the following we will briefly describe the 
CDF~\cite{Acosta:2004yw} and \Dzero~\cite{Abazov:2005pn} 
detectors relevant for the experimental results discussed in this report.

\subsubsubsection{CDF and \Dzero\ detectors}
\label{sec:cdfdetector}

 The CDF-II detector spectrometer is built around an axially symmetric
Central Outer Tracker (COT), a open-cell drift chamber
that provides charged track identification and measurement of
the momentum transverse to the $p\bar{p}$ beams ($p_T$) in the central
region($|\eta| \le 1.2$) for tracks with $p_T> 400\, \rm MeV/{\it c}$.
The active volume of the COT covers extends from a radius of  $40$ to $140\, \rm cm$,
with up to 96 axial and stereo measurement points inside a superconducting solenoid that provides
a $1.4\, \rm T$ axial magnetic field.  \Dzero\ Central Fiber Tracker fills a significant smaller 
space inside a 2 T solenoid, $20$ to $50\, \rm cm$, with 16000 channel organized in 8 alternating
axial and stereo layers each providing a doublet of measurement points.
The $p_T$ resolution is found to be $\delta_{p_T}/p_T \sim 0.001 \cdot p_T (\gevc)$ in 
the CDF tracker. This results in precise invariant mass reconstruction which 
provides excellent signal-to-background ratio for fully reconstructed $B$ and $D$ decay modes.
%
%


Tracks found in the central tracker are extrapolated inward and matched to hits in silicon
microvertex detectors in both CDF and \Dzero.
The CDF detector (SVX II + ISL) uses double sided silicon microstrip technology
providing tracking information in the $r$-$\phi$ and $r$-$z$ planes in the 
pseudo-rapidity range $|\eta|<2$.  
The detector has up to 7 layers of double-sided silicon at radial distances ranging from
$2.5\, \rm cm$ to $28\, \rm cm$ from the beamline.



Within the SVX is the innermost single-sided, radiation hard silicon layer (Layer 00), which is 
mounted directly onto the beam pipe at a radius of $1.35$ to $1.62\, \rm cm$\cite{Hill:2004qb}.   
The impact parameter resolution of the tracking system with, and
without, the inclusion of Layer 00 is shown in Figure~\ref{fig:ip-l00-svt}.
The impact parameter resolution for high $p_T$ charged tracks
is $\sim 25\mu m$ taking in to account the $32\mu m$ contribution from the transverse size 
of the interaction region\cite{Hill:2004qb}.

\Dzero\ silicon microstrip tracker (SMT) is composed of cylindrical barrels with 4 layers
of double-sided detectors interspersed with disks in the central part, and complemented
with large forward disk at both ends, a design optimized for tracking up to
$|\eta|<3$. In addition, in 2006 a new innermost layer (Layer 0) 
was installed inside the existing detector. This has improved the impact parameter
resolution and will prevent the expected performance degradation due to 
radiation damage of the innermost SMT layer during the rest of the Tevatron 
run~\cite{Tsybychev:2007zz}.

\begin{figure}[t]
\center
\epsfig{file=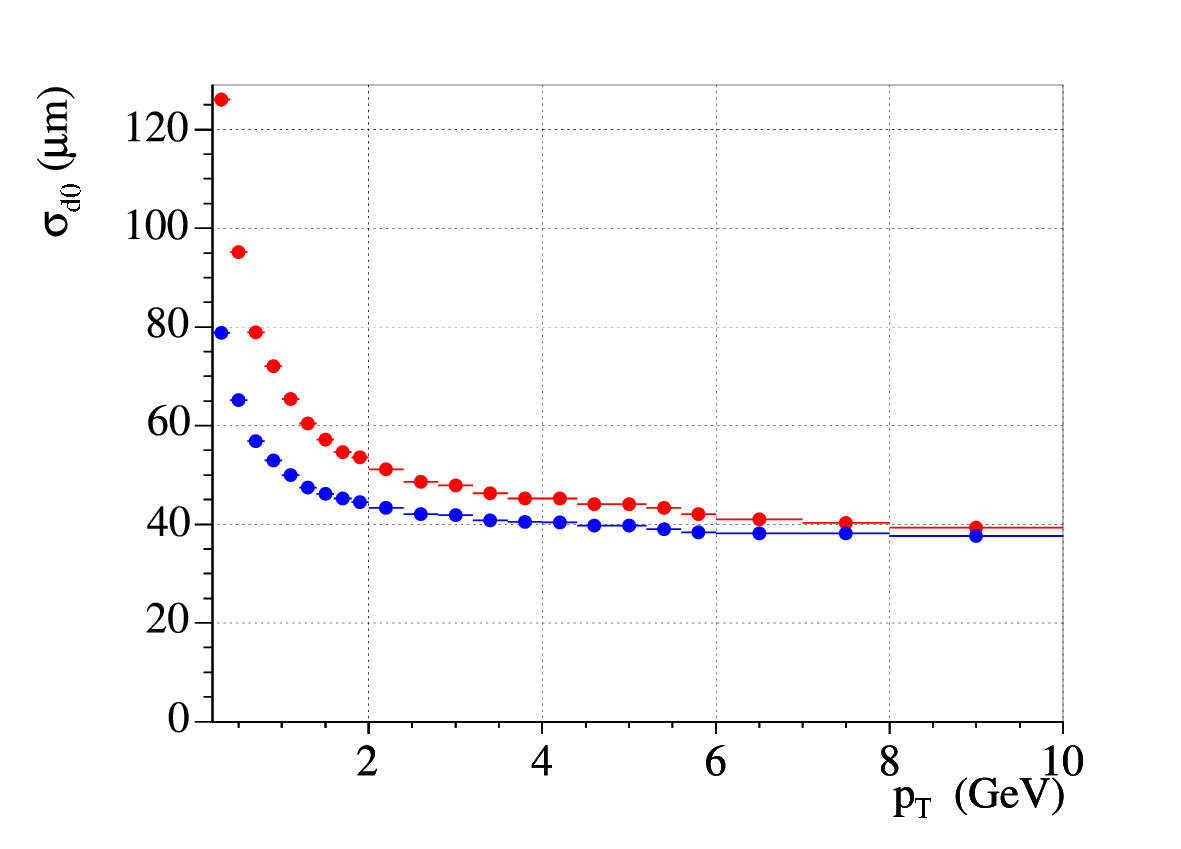,height=6truecm,width=0.48\textwidth}
\psfig{figure=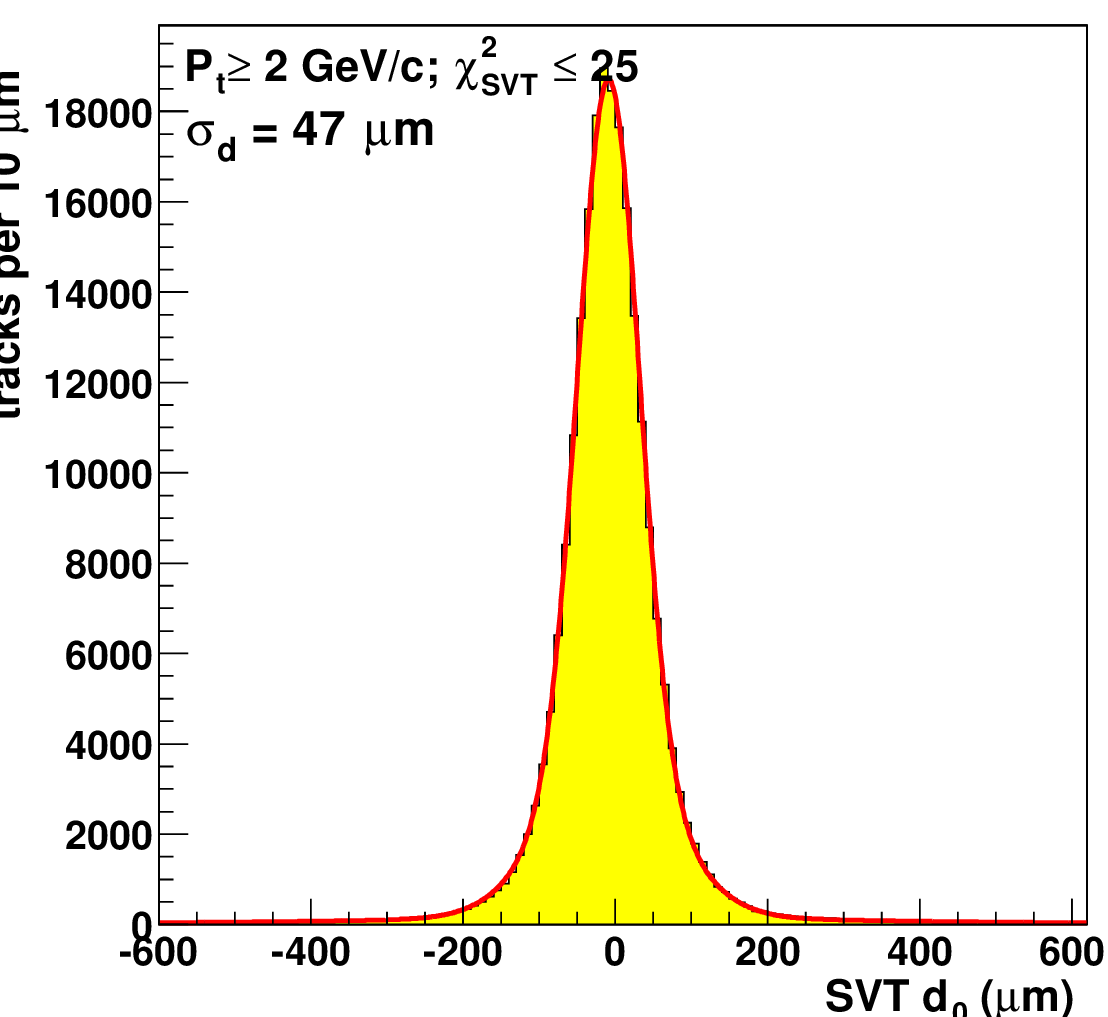,width=0.44\textwidth}
\caption{ CDF impact parameter resolution tracking tracks with Layer 00 hits (blue points) and without 
Layer 00 hits (red points.) a) and 
Silicon Vertex Trigger (SVT) impact parameter distribution for a generic sample of tracks b).  
}
\label{fig:ip-l00-svt}
\end{figure}

The silicon vertex detectors are crucial for precise decay length determination of $b$ decays 
in time dependent measurement. Moreover the 3D vertex reconstruction allowed by the 
combined $r$-$\phi$ and $r$-$z$ measurements provides efficient background rejection 
against the large background of prompt events.


%
%

Particle identification in CDF is provided by $dE/dx$ in the
central drift chamber and a time-of-flight (TOF) system consisting 
of 216 scintillator bars located between the COT and
the solenoid~\cite{Cabrera:2002vp}. The TOF, with a resolution of around 110 ps, 
provides at least $2 \sigma$ $K/\pi$ separation for $p_T<1.5\, \gevc$.
For $p_T > 2.0$ {\gevc}, the separation provided by $dE/dx$ between
pions and Kaons is equivalent to 1.4\,$\sigma$ between two Gaussians
while the separation for pions and electrons is $2.5\,\sigma$ at 
$p_T = 1.5\,{\gevc}$.  

%
%
Outside the solenoid are electromagnetic and hadronic
calorimeters covering the pseudo-rapidity region $|\eta|<3.5$ in CDF and 
up to$|\eta|<4.0$ in \Dzero.

%
%
Muon detectors are located behind the hadron calorimeters,
The CDF muon systems are segmented into four components,
the Central Muon system (CMU) provides coverage
for $|\eta|<0.6$ and $p_T > 1.5\rm \gevc$ and sits
behind  $\sim \! 5.5$ interaction lengths ($\lambda$) of material
primarily consisting of the iron of the hadronic calorimeter.
The Central Muon upgrade (CMP) sits behind an additional $60$ cm, 
$\sim \! 3 \lambda$ of steel, providing
identification for muons with $p_T > \, 3.0 \gevc$ in  $|\eta|<0.6$,
with higher purity than muons identified only in the CMU.
The Central Muon extension (CMX) consists of
eight layers drift chambers arranged in conic sections
and provides coverage for $0.6<|\eta|<1.1$ and $p_T > 2.0\,  \gevc$,
and is located behind absorber material corresponding to $\sim \! 6$ 
up to $\sim \! 10$ interaction lengths.
The \Dzero\ muon system sits outside of a thick absorber ($>10\ \lambda$),
and consists of a layer of tracking detectors and 
scintillation trigger counters inside a $1.8\, \rm T$ iron toroid,
followed by two additional layers outside the toroid. The muon 
coverage extends to $|\eta|=2$. Magnet polarities are regularly
reversed during data collection, thus providing an important way to 
control charge dependent effects in muon reconstruction that might affect 
semileptonic asymmetry measurement.

\subsubsubsection{Triggers}
\label{sec:cdftrigger}

Data acquisition and trigger system for experiments at hadron colliders have 
to sustain an extremely high collision rate, 7.6(40) MHz at Tevatron(LHC),
and reduce it to approximately 100-1000 Hz of interesting events that
can be saved permanently for physics analysis, thus providing rejection 
factor $> 10^4$ against uninteresting proton-(anti)proton collisions.    
The most straightforward way to achieve such a goal is to design electron and muon
based triggers, using single or multi-lepton signatures, that allow to select
significantly pure samples of heavy flavor decays thanks to the 
large semileptonic branching ratios, or by isolating final states containing
e.g. $\jpsi$. Rate is controlled primarily with lepton transverse momentum requirement, 
that has to be kept as low as possible in order to maximize signal efficiency.
Inclusive electron and muon selection with a threshold of 6-8 \gevc are typical at Tevatron. 
Much lower thresholds are possible for events with two leptons, approaching
the minimum detectable transverse momentum in each detector ( $2\ \gevc$ at Tevatron). 

This strategy has been implemented by all the present and forthcoming experiments 
and provided the majority of the result for rare decays and lifetime measurements at Tevatron 
in the last decade. A clear limitation of this approach is that it lacks the ability to select
fully hadronic decays of b-hadrons. In the context of CKM-related physics the latter are 
important for the study of either 2 body charmless decays, or $B\to DK$ decays 
involved in the measurement of the angle $\gamma$ in tree processes, and, most importantly,
for selecting large samples of fully reconstructed $\bsd$ and $\bsdppp$ that
lead to the first observation of $\BsBsbar$ mixing in 2006~\cite{Abulencia:2006ze}. 
To overcome this limitation 
the CDF collaboration pioneered the technique of online reconstruction of charged
tracks originating from decay vertexes far from the collision point due to the 
significant boost and lifetime of B-mesons produced at high energy hadron colliders.
The key innovation introduced for Run II in the CDF trigger was in fact 
the Silicon Vertex Trigger (SVT) \cite{svt} processor. At the second level of the trigger system, 
information from the silicon vertex detector is combined with tracks reconstructed
at the first level trigger in the drift chamber. High resolution SVT-tracks are 
then provided within the latency of $\approx 20 \mus$, and are 
used to select events characterized by two tracks with high 
impact parameter and vertex decay length greater than 200 \mum, thus
providing a rejection factor of 100-1000 while maintaining 
a significant efficiency for B decays. The impact parameter resolution of the SVT, shown
in Figure~\ref{fig:ip-l00-svt}, is approximately $50\, \rm \mu m$, which
includes a contribution of $32\, \rm \mu m$ from the width of the
$p\bar{p}$ interaction region. It has to be noted, however, that selecting events 
based upon decay length information, introduces an important inefficiency at small values 
of proper decay time. We will describe how this bias has been incorporated in the analysis 
in Section~\ref{sec:expPrimers:vtx}.


%
%


\subsection{Common experimental tools}

In the following the most relevant experimental techniques for flavor
physics will be briefly discussed. Time dependent measurements require 
excellent vertexing and flavor tagging capabilities, crucial in the 
latter case is particle identification and $\pi$-K separation. Finally
noise suppression, recoil tagging technique and Dalitz-plot
analysis techniques will be discussed.

\subsubsection{Time-dependent measurements}
\label{sec:tdep}
It is possible to measure phases of the CKM matrix elements, and therefore CP violating quantities, by exploiting the different time evolution of the two 
mass eigenstates of the $B_0$ meson system, $B_L$ and $B_H$. At B-Factories, where a $B_0$ meson is produced coherently with its antiparticle, the 
probability density function of observing a $B$ decay into a flavor eigenstate (called $B_{tag}$) and for whom $\eta=-1(+1)$ if \Bz (\Bzb) and the   
other one, called $B_{reco}$, in a given final 
state $f$ at times that differ by \deltat\ is
\begin{equation}
f_\eta(\deltat) =  
\frac{\Gamma}{4}\,{\rm e}^{ - \Gamma | \deltat | }  
\left\{ 1
+\eta  \left[
S \sin{ \deltam  \deltat } 
- C\cos{ \deltam  \deltat }  
\right]  \right\}, \label{eq:direct}
\end{equation}
where the decay width difference between the two mass eigenstates is neglected, $\deltam$ is the 
mass difference,
\begin{eqnarray}
S=\frac{2 Im\lambda}{1  + |\lambda|^2} \hspace{2cm}
C=\frac{1-|\lambda|^2}{1  + |\lambda|^2},
\end{eqnarray}
and 
\begin{equation}
\lambda=-{\frac{|\matrixelement\Bz{{\cal H}_{\Delta B=2}}\Bzbar|}
                  {\matrixelement\Bz{{\cal H}_{\Delta B=2}}\Bzbar}}
          {\frac{\matrixelement f{{\cal H}_{\Delta B=1}}\Bzbar}
                  {\matrixelement f{{\cal H}_{\Delta B=1}}\Bz}}.\label{eq:lambda}
\end{equation}
Depending on the choice of the final state $f$, S can be related to different phases of the CKM matrix elements. In particular if $f$ is a flavor eigenstate 
then $\lambda=0$ and $C=1$ and $S=0$, no phase can be measured but there is sensitivity to $\deltam$; likewise if $f$ is a CP eigenstate, $\lambda$ is a pure 
phase and this is usually the cleanest configuration to measure CP violation parameters, although all non-zero values of $\lambda$ allow such measurements.

At hadron colliders the same considerations apply, a part from the fact that \deltat\ 
measures the time between the $B$ meson production and its decay and 
that $\eta=-1(+1)$ for an initially produced \Bz(\Bzb). The initial $B$ flavor can be measured 
either by observing the decay products of the other hadron with a $b$ quark in the event, or by utilizing information on the jet of particles the $B$ meson is contained into.

There are therefore three key ingredients in these measurements: the identification of the flavor of the meson produced in association with the one 
reconstructed in the channel $f$ (the so-called $B$-tagging), the measurement of $\deltat$ which requires the reconstruction of the decay vertex of at least 
one $B$ meson (both mesons in the case of $B$-factories), and the reconstruction of the $B$ meson in the final state $f$ with the least possible background.

The experimental uncertainties on these quantities alter the probability  density function of the measured quantities, function which is used in the 
likelihood fits implemented to perform these measurements. Instead of Eq.~\ref{eq:direct} one can then write
\begin{equation}
f_\eta(\deltat) =  
\frac{\Gamma}{4}\,{\rm e}^{ - \Gamma | \Delta t_{true} | }  
\left\{ 1 +
\eta {\cal D} \left[
S \sin{ \deltam  \Delta t _{true}} 
- C\cos{ \deltam  \Delta t _{true} }  
\right]  \right\}\otimes {\cal{R}}(\deltat-\Delta t_{true})  
+  f^{bkg}_\eta(\deltat) , \label{eq:alllike}
\end{equation}
where $\otimes$ indicates the convolution, ${\cal{D}}=1-p_w$ is the tagging dilution ($p_w$ is the probability of incorrectly tagging a meson), {\cal{R}} is 
the vertexing resolution function, and $f^{bkg}_\eta$ is the probability density function for the background.

The next sections describe the techniques adopted for tagging, vertexing reconstruction and background rejection and the means available to estimate the 
quantities that enter into Eq.~\ref{eq:alllike}.

\subsubsection{$B$ Flavor Tagging}
\label{sec:expPrimers:bFlavorTagging}


One of the key components in the measurement of neutral $B$~meson flavor oscillations 
or time dependent CP asymmetries is identifying the flavor of the $B$~meson 
(containing a $\bar{b}$~antiquark) or $\bar{B}$~meson (containing a $b$~quark) 
at production, in the case of incoherent mixing at hadronic colliders, or at the moment 
the other $b$-meson decays in the case of \BzBzb\ from \Y4S. 
We refer to this method of identifying the $B$~hadron flavor as
``$B$~flavor tagging''. The figure
of merit to compare different tagging methods or algorithms is the so-called
effective tagging power $\eD=\varepsilon (1-2\,p_W)^2$, where the
efficiency $\varepsilon$ represents the fraction of events for which a
flavor tag exists and $p_W$ is the mistag probability indicating the
fraction of events with a wrong flavor tag. The mistag probability is
related to the dilution: ${\mathcal D} = 1 - 2\, p_W$. The experimentally observed 
mixing or CP asymmetries are, in fact, proportional to the dilution $\cal D$. 
A flavor tag which always returns the correct tag has a dilution of
1, while a random tag yielding the correct flavor 50\% of the time has a
dilution of zero. 

Several methods to tag the initial $b$~quark flavor have been used both 
at B-factories and hadron collider experiments.
The flavor tagging methods can be divided into two groups, those that 
identify the flavor of the other $b$-hadron  produced in the same event 
(opposite side tag - OST), and those that tag the initial flavor of 
the $B$~candidates itself (same side tag - SST). 
The latter, being based on charge correlation between initial 
$b$~quark and fragmentation particles is only possible at hadron colliders 
or Z-pole experiments.

\begin{figure}
\begin{center}
  \epsfig{file=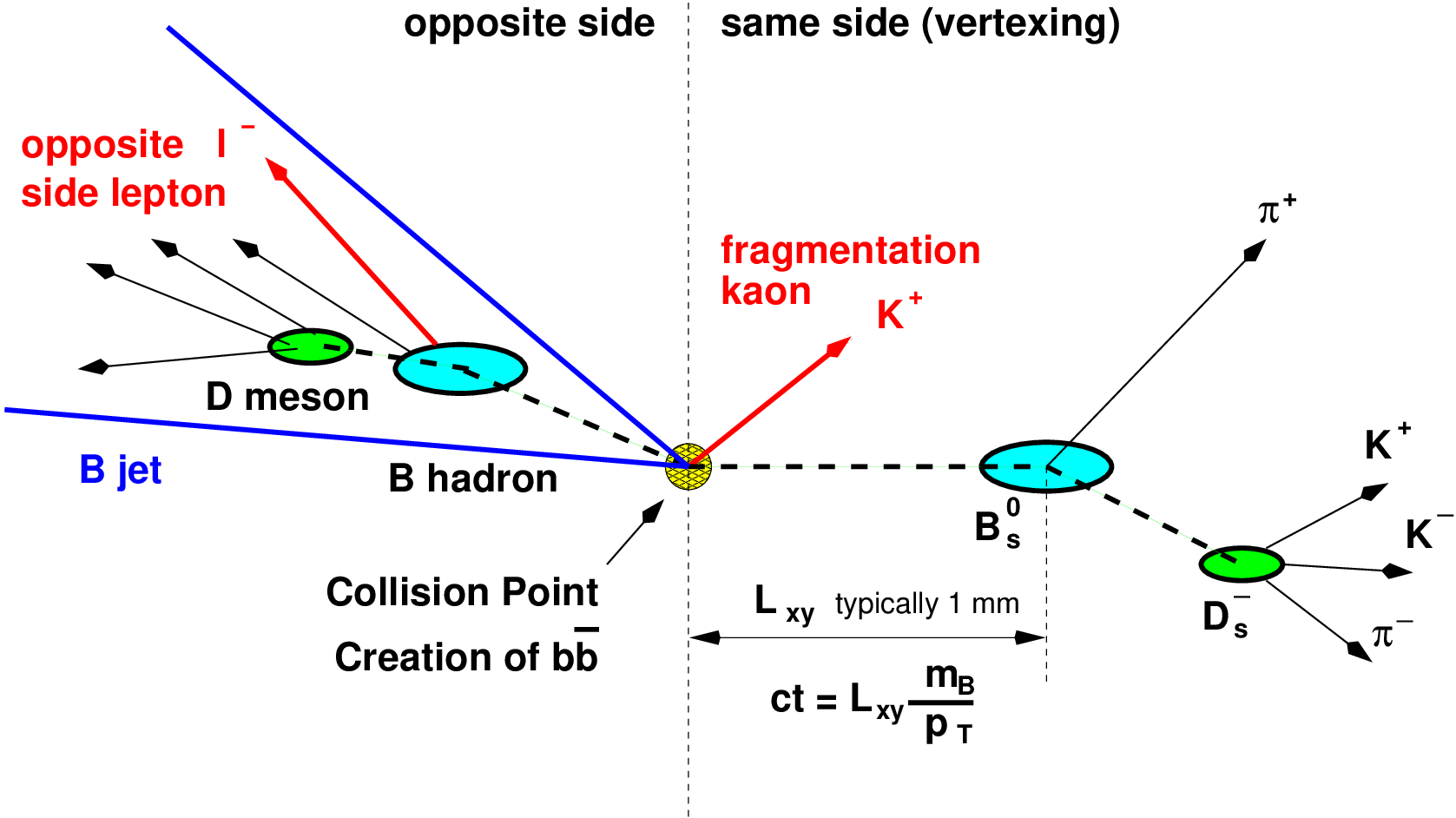,width=0.5\textwidth}
\caption{Sketch of typical $b\bar b$ event indicating several $B$~flavor
  tagging techniques.}
\label{fig:tags}
\end{center}
\end{figure}

Fig.~\ref{fig:tags} is a sketch of a $b\bar b$~event showing the $B$ and
$\bar B$~mesons originating from the primary $p\bar p$ interaction vertex
and decaying at a secondary vertex indicating possible flavor tags on the
decay vertex side (SST) as well as opposite side tags. 

 In the following the main aspects
of the opposite side taggers used at both Tevatron and B-Factories
 and of the SST used for the \BsBsbar\ oscillation 
observation and in the first $\phi_s$ determination at Tevatron will be 
briefly discussed.

\subsubsubsection{Opposite Side Tags}

Both experiments at hadron colliders and B-Factories exploit three feature of $B$ decays 
to estimate the flavour of the opposite $B$ meson.

The ``lepton tagging'' looks for an electron or muon from the semileptonic decay
of the opposite side $B$~hadron in the event. The charge of this lepton is
correlated with the flavor of the $B$~hadron: an $\ell^-$ comes from a $b
\ra c\, \ell^-\bar\nu X$ transition, while an $\ell^+$ originates from a
$\bar b$ quark. Since the semileptonic $B$~branching fraction is small,
${\cal B}(B\ra\ell X)\sim 20$\%, lepton tags are expected to have low
efficiency but high dilution because of the high purity of lepton
identification. 

The strangeness of Kaons or $\Lambda$ from the subsequent charm
decay $c\ra s X$ is also correlated with the $B$~flavor, e.g. a $K^-$ results
from the decay chain $b \ra c \ra s$ while a $K^+$ signals a $\bar b$
flavor. Searching for a charged Kaon from the opposite side $B$ hadron
decay is referred to as ``Kaon tagging''. This method is expected to have
high efficiency but low dilution at hadron colliders since the challenge is 
to first identify Kaons among a vast background of pions through  
particle-id techniques, and then to discriminate the $B$~decay Kaon candidates
from all prompt Kaons produced in the collision by 
relying on Kaon impact parameter and reconstruction of secondary vertexes 
in the opposite side~\cite{Salamanna:2006cd}.

Finally all other information carried by the tracks among the decays of the $B$ mesons constitutes the third large tagging category. 
On average in fact the most energetic charged decay product carries the charge of the original $b$ quark. 
At Tevatron the ``jet charge tagging'' exploits the fact that
the sign of the momentum weighted sum of the particle charges of the
opposite side secondary vertex from $b$ (\Dzero\cite{Abazov:2006qp}) 
or $b$~jet (CDF~\cite{Lecci:2005gf}) is correlated to the charge of the $b$~quark.
Jet charge tags can reach very high efficiency but with low dilution.
Furthermore, more than 20\% of $B$ decays contain 
charged $D^*$ mesons which decay 66\% of the times into a soft pion with the same charge. Soft pions can therefore also have a high charge correlation with 
the original $b$ quark. 
The Belle and BaBar experiments input to multivariate tagging algorithms the charge of all tracks, with special treatment for the softest in the event to 
take into account this effect.

The algorithms to combine all the information use multivariate technique either exploiting 
directly the available output of the various tagging algorithms or starting by assigning each 
track candidate of coming from the "tagging" $B$ meson into one category between
lepton, kaon, soft pion (only for B-Factories) or generic track. Each experiment then has a different approach to exploit the information.

The BaBar experiment uses one Neural Network (NN) per category with different quantities in input depending on the category (see Ref.~\cite{Aubert:2002rg} 
for details): for instance the "Lepton" 
category would contain lepton identification quantities and the momentum. The output of these NNs based on single-particle information are themselves 
combined into several event-by-event NNs, that assess the likelihood of the flavor assignment. The tagging categories are mutually exclusive and for each 
event only one NN is evaluated. The algorithm of the Belle experiment is similar but exploits likelihood instead of NNs and has a single output (called $r$).
In both cases the algorithms are tuned on MC, but the mistag probability is estimated on data control samples.

The experimental sensitivity is maximized upon using the expected dilution
on an event by event basis, employing parameterizations derived by 
a combination of simulation and real data. As an example, the dilution of the lepton 
tagging is parameterized  as a function of the lepton identification quality 
and of the \ptrel\ of the tagging lepton (CDF~\cite{Giurgiu:2005mp,Tiwari:2007he}) or 
of the lepton jet-charge (\Dzero\cite{Abazov:2006qp}). The quantity \ptrel\ is
defined as the magnitude of the component of the tagging-lepton momentum
that is perpendicular to the axis of the jet associated with the lepton
tag. Variation of the dilution as a function of \ptrel\ is shown
left side of Fig.~\ref{fig:tagcdf} for electron tags in CDF. 
The dilution is lower for low $\ptrel$ because fake leptons and leptons from sequential
semileptonic decay ($b \rightarrow c \rightarrow \ell^+$) tend to have
relatively low \ptrel~ values. Also, to maximize the tagging power the dilution of the jet charge tags can be 
calculated separately for different quality of the opposite side secondary
vertex information and parametrized as a linear function in the quantity 
$|Q_{jet}|\cdot{\cal P}_{\rm nn}$, 
where ${\cal P}_{\rm nn}$ expresses a probability for the jet to be a $b$~jet, 
as displayed in Fig.~\ref{fig:tagcdf} for jets containing a 
well separated secondary vertex in the CDF case.
Flavor misidentification can occur because the jet charge does not reflect perfectly the 
true charge of the original $b$~quark, due e.g. to mixing. In addition, the selected 
tagging jet may contain only a few or no tracks from the actual opposite side 
$B$~hadron decay.

\begin{figure}
\centerline{
\includegraphics[height=6cm,width=0.51\textwidth]{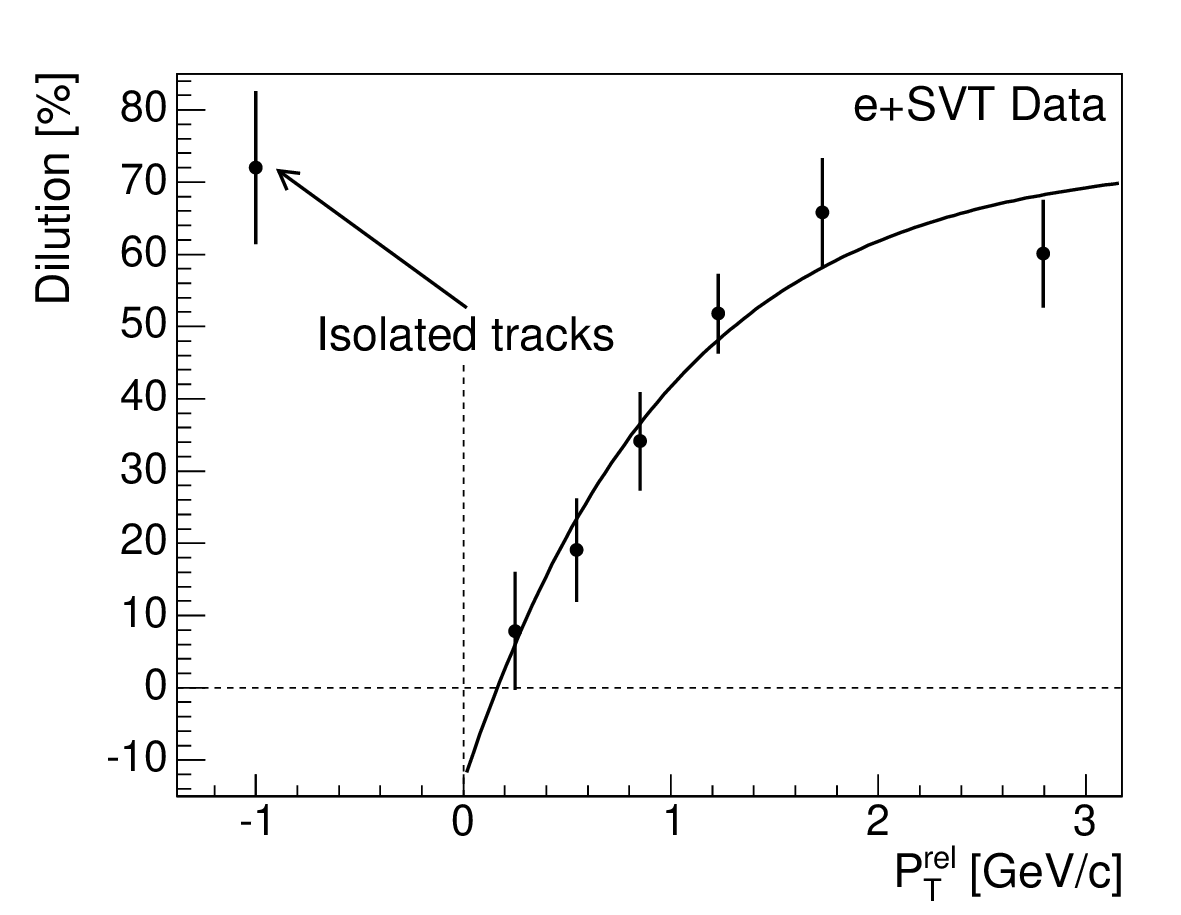}
\includegraphics[width=0.44\textwidth]{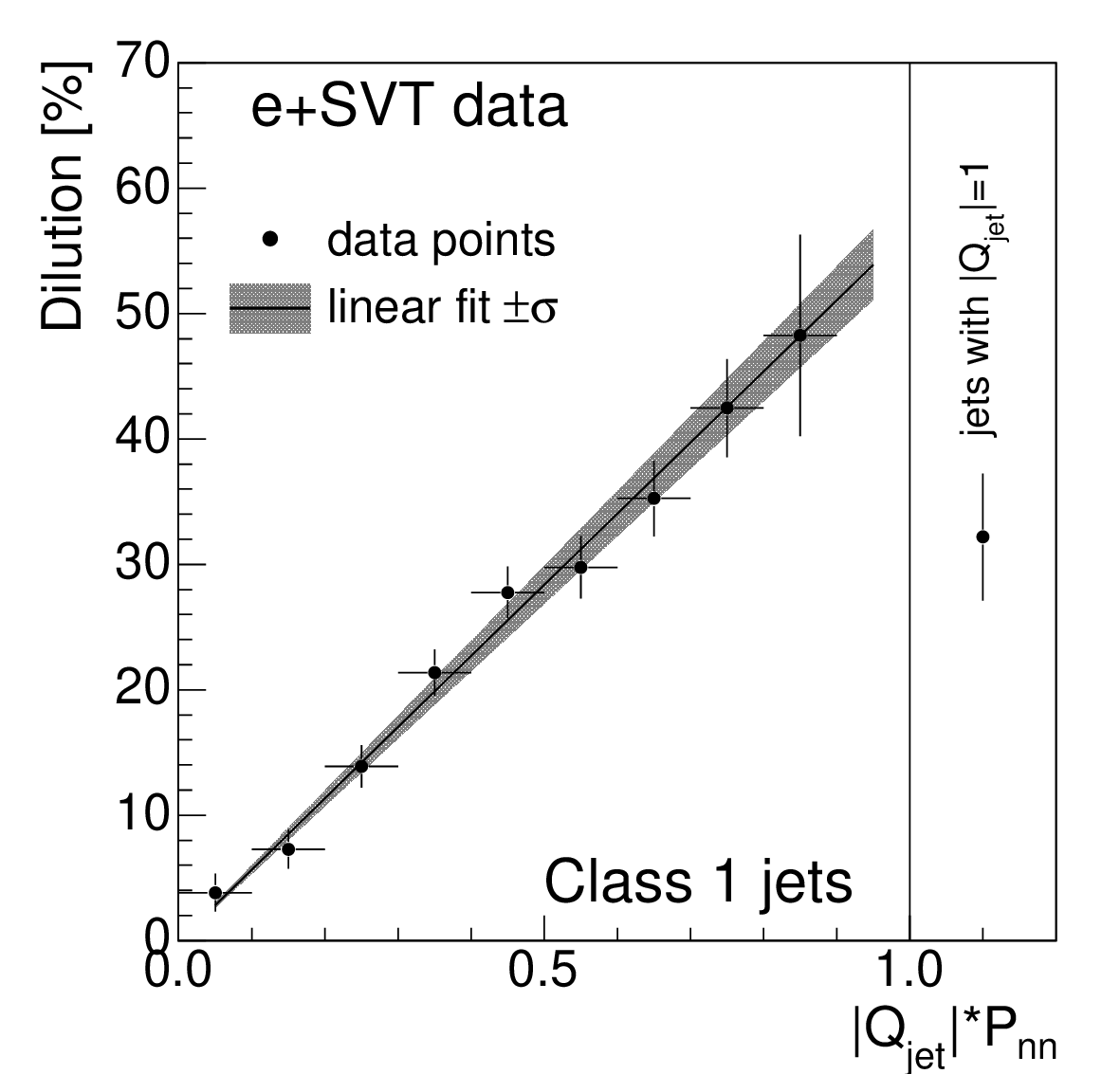}
}
\caption{Variation of dilution of the electron tags with $p_T^{\rm rel}$(left).
Dilution as a function of $|Q_{jet}|\cdot{\cal P}_{\rm nn}$ for ''jet-charge''
tagging algorithm  (right).
\label{fig:tagcdf}
}
\end{figure}

At Tevatron, the typical flavor tagging power of a single tagging algorithm 
is ${\cal O}$(1\%). Limitations in
opposite side tagging algorithms arise because the second bottom hadron is
inside the detector acceptance in less than 40\% of the time or it is
possible that the second $B$~hadron is a neutral $B$~meson that mixed into
its antiparticle.  For example, the low efficiency of an opposite side
lepton tag of $\sim20$\% from the semileptonic $B$~hadron branching fraction
together with a dilution of $\sim30$\% results in an estimated
$\eD\sim0.4\times 0.2\times0.3^2\sim0.01$. At B-factories, better hermeticity
of detectors, enhanced particle identification capability, and the absence of 
incoherent mixing as a source of dilution makes it possible to reach combining all 
the information together en effective tagging power $\eD\sim0.30$ in 
both Belle and Babar. 
As an example of tagging performances for each experiment considered here,
the obtained efficiencies~$\varepsilon$, effective dilutions $\langle {\cal D}\rangle$,
and effective tagging powers \eD\ are shown in
Table~\ref{tab:ostperformance}. 

In the case of opposite side flavor tags, the dilution is expected to be
independent of the type of $B$~meson (\Bz,\Bu,\Bs) under study, hence
can be studied on large inclusive semileptonic samples (CDF)
or on \Bz\ or \Bu\ samples (\Dzero) and then applied in \Bs\ related measurement. 
 The final calibration of the opposite side tagging methods come from a
measurement of the $B^0$~oscillation frequency \deltamd\ in hadronic and
semileptonic samples of $B$~mesons at both B-factories and Tevatron experiments.
A perfectly calibrated tagging method applied to a large sample of \Bzb~mesons should
result in a precise measurement of \deltamd. In turn one can use the well known
world average value of \deltamd to check and re-calibrate the
predicted dilutions of the opposite side tagging algorithms.

\begin{table}
\caption{Tagging performances of the  opposite side tagging
algorithms at BaBar~\cite{:2009yr}, Belle~\cite{Chen:2005dra}, 
\Dzero~\cite{Abazov:2006qp}, and  CDF~\cite{cdfost}. Note that the individual tagger
performance in the latter case are determined in non-exclusive sample so their sum is greater
than the neural network (NN) based combined opposite side tagging for CDF. 
All errors given are statistical.
\label{tab:ostperformance}
}
\begin{center}
\begin{tabular}{|l|l|cc|cc|cc|}
\hline
\multicolumn{2}{|l|}{Category} & \multicolumn{2}{|c|}{Efficiency $\varepsilon$~[\%]}
& \multicolumn{2}{|c|}{Effect.~dilution $\langle {\cal D}\rangle$~[\%]}
& \multicolumn{2}{|c|}{Tagging Power \eD~[\%]} \\ \hline
BaBar & Belle ($r\in$) &BaBar & Belle &BaBar & Belle&BaBar & Belle \\ \hline
Lepton&0.875-1& 8.96$\pm$0.07&14.4$\pm$0.9&99.4$\pm$0.3&97.0$\pm$0.5& 7.98$\pm$0.11&13.5$\pm$0.9\\ 
Kaon I &0.75-0.875& 10.82$\pm$0.07 &9.8$\pm$0.7& 89.4$\pm$0.3&78.2$\pm$0.9&8.65$\pm$0.3&6.0$\pm$0.5\\
Kaon II &0.625-0.75&17.19$\pm$0.09 &10.7$\pm$0.8& 71.0$\pm$0.4&68.4$\pm$1.0& 8.68$\pm$0.17&5.0$\pm$0.5\\
Kaon-Pion&0.5-0.625& 13.67$\pm$0.08 &10.8$\pm$0.8& 53.4$\pm$0.4&55.0$\pm$1.1&  3.91$\pm$0.12 & 3.3$\pm$0.4\\
Pion &0.25-0.5& 14.18$\pm$0.08 &14.6$\pm$0.9& 35.0$\pm$0.4 &36.0$\pm$0.8& 1.73$\pm$0.09& 1.9$\pm$0.2\\
Other &0-0.25& 9.54$\pm$0.07 &39.7$\pm$1.5& 17.0$\pm$ 0.5 &7.2$\pm$0.7& 0.27 $\pm$0.04 &0.2$\pm$0.1\\ \hline
\multicolumn{2}{|l|}{Total Tagging Power} & &&&& 31.2$\pm$ 0.3 & 29.9$\pm$1.2 \\ 
\hline
\multicolumn{2}{|l|}{}& CDF & \Dzero& CDF & \Dzero& CDF & \Dzero\\ \hline
 \multicolumn{2}{|l|}{Muon}        & $ 5.5\pm0.1$ & $ 6.6\pm0.1$   & $35.3\pm1.1$ &$47.3\pm2.7$ & $0.68\pm0.05$& $1.48\pm0.17$  \\
 \multicolumn{2}{|l|}{Electron}    & $ 3.1\pm0.1$  & $ 1.8\pm0.1$ & $30.7\pm1.1$ & $34.1\pm5.8$  & $0.29\pm0.01$ & $0.21\pm0.07$ \\
 \multicolumn{2}{|l|}{Jet Charge}  & $90.5\pm0.1$  & $ 2.8\pm0.1$ &  $9.5\pm0.5$ & $42.4\pm4.8$ & $0.80\pm0.05$ & $0.50\pm0.11$\\
 \multicolumn{2}{|l|}{Kaon}        & $18.1\pm0.1$   &  N/A        & $11.1\pm0.9$ &  N/A         & $0.23\pm0.02$ & N/A \\
 \multicolumn{2}{|l|}{Total Tagging Power} && &&& 1.81$\pm$ 0.10 & 2.19$\pm$ 0.22\\ \hline
\end{tabular}
\end{center}
\end{table}

Ã


\subsubsubsection{Same Side Flavor Tagging}

The initial flavor of a $B$~meson can additionally be tagged 
by exploiting correlations of the $B$~flavor
with the charge of particles produced in association with it (SST). 
Such correlations arise from $b$~quark
hadronization and from $B^{**}$~decays. In the case of a \Bub or
\Bzb~mesons, the fragmentation particles are mainly pions while $\Bsb$~meson
are primarily accompanied by fragmentation Kaons. In the $\Bsb$~meson case
we thus refer to this method as ``same side Kaon tagging'' (SSKT).
In the simplest 
picture, where only pseudo-scalar mesons are produced directly by
the fragmentation process, the following charged stable mesons are
expected: a $\Bzb$ will be produced along with
$\pi^-$, a \Bub will be produced with a $\pi^+$ or a
$K^+$, and a {\Bsb} will be produced with a $K^-$.
Corresponding relations are true for the charge conjugated
$B$~mesons. The idea of the same side tagging algorithm is to
identify the leading fragmentation track charge and to determine the $B$
initial flavor accordingly.


Several advantages compared to the opposite side
tagging algorithms are worth mentioning. The SST shows a
high efficiency since the leading fragmentation track is in the
same detector region as the signal $B$ hadron, thus, within the
detector acceptance, and there are also no limitations due to
branching ratios. The search region for same side
tagging tracks is limited near the signal $B$
direction. Due to this geometrical restriction,
the SST is robust against background from the underlying event or
multiple interactions. Finally neutral meson mixing does not dilute 
the useful charge correlation. These advantages are reflected 
in an higher flavor tagging dilution.

Unlike the opposite side flavor tagging algorithms, the performance of the
same side algorithm cannot easily be quantified using data. Since SST
is based on information from the signal $B$ fragmentation process,
its performance depends on the signal $B$ species. Therefore,
\Bu and \Bz modes can not be used to calibrate the same side
tagging performance for \Bs mesons. Instead, prior to the actual observation 
of \Bs\ mixing, the experiments had to rely upon 
Monte Carlo simulation to quantify the performance of same side
tagging for {\Bs} mesons. High statistics \Bu and $\Bz$ modes 
have been used to verify that specifically tuned Monte
Carlo program accurately model the fragmentation process. 

The CDF algorithm~\cite{Belloni:2007zz} starts selecting charged tracks with
$p_T \geq 450 \mevc$, good momentum and impact parameter resolution as 
potential tagging tracks. 
Fragmentation tracks originate from the primary vertex, therefore 
an impact parameter significance less than 4 is required. 
To reject background from multiple interactions,
the tracks are required to be close to the \Bs\ candidates 
both along the beam direction and in 
$\Delta R = \sqrt{\Delta\eta^2+\Delta\phi^2}\leq$ 0.7. 

About $60$\% of the tagged events have one and only one
tagging track. Of the remaining events approximately one-third have all
tagging tracks with the same charge.  Therefore, the subsequent tagging
algorithm makes a choice between multiple, oppositely
charged tracks in about one-fourth of all tagged events. 
Several variables have been employed. The most sensitive was found to be 
the maximum longitudinal component of the tagging tracks with respect to 
the B momentum, and after that the largest likelihood to be a Kaon based on 
TOF and dE/dx measurements. 
A neural network is finally used to combine the available information.  
Examples of the dependence of the dilution on the variables discussed
above are given in Fig.~\ref{fig:CLL} for the subsample with only one
tagging track. 

\begin{figure}[hbtp]
\includegraphics[width=0.32\textwidth]{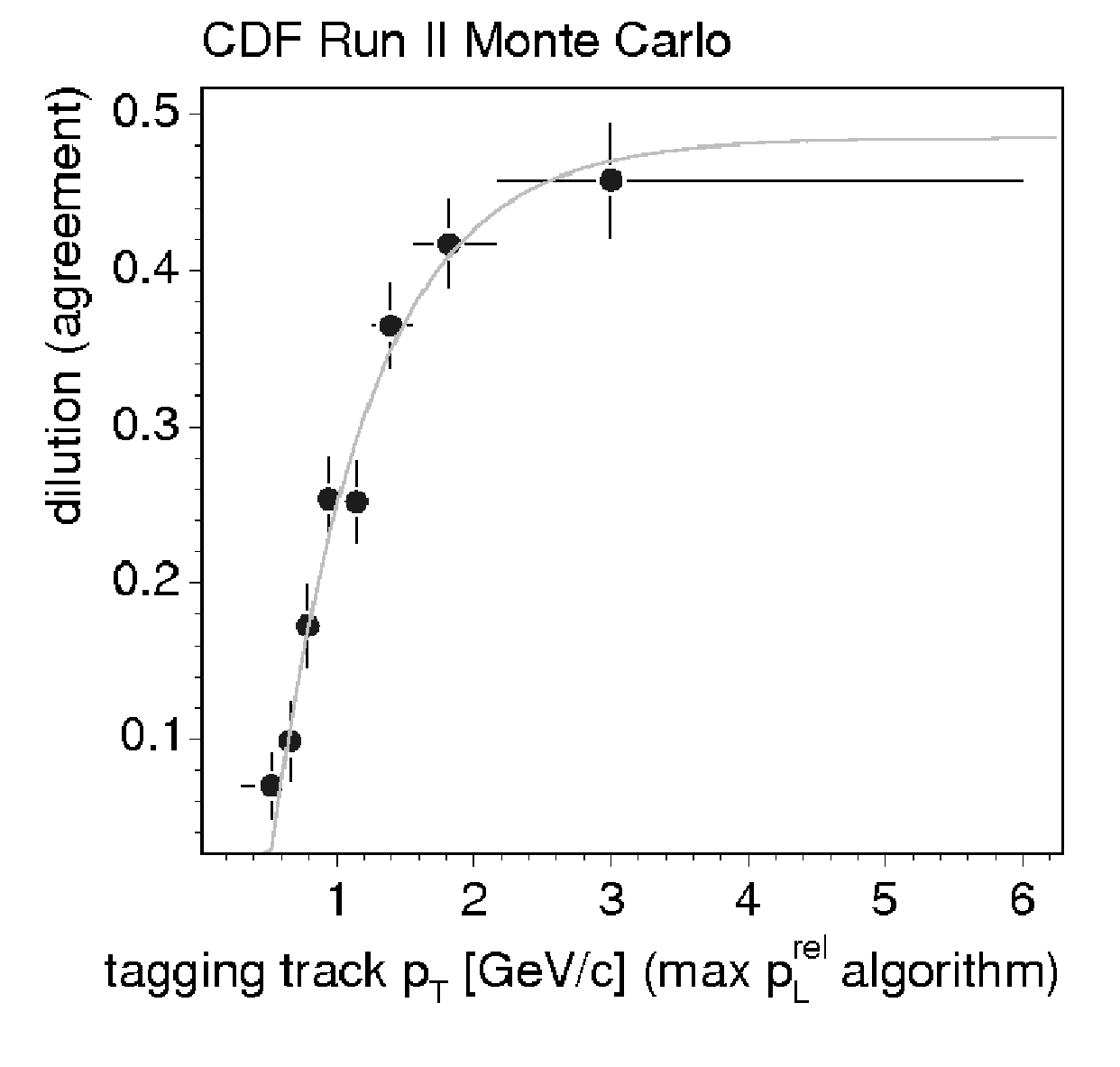}
\includegraphics[width=0.32\textwidth]{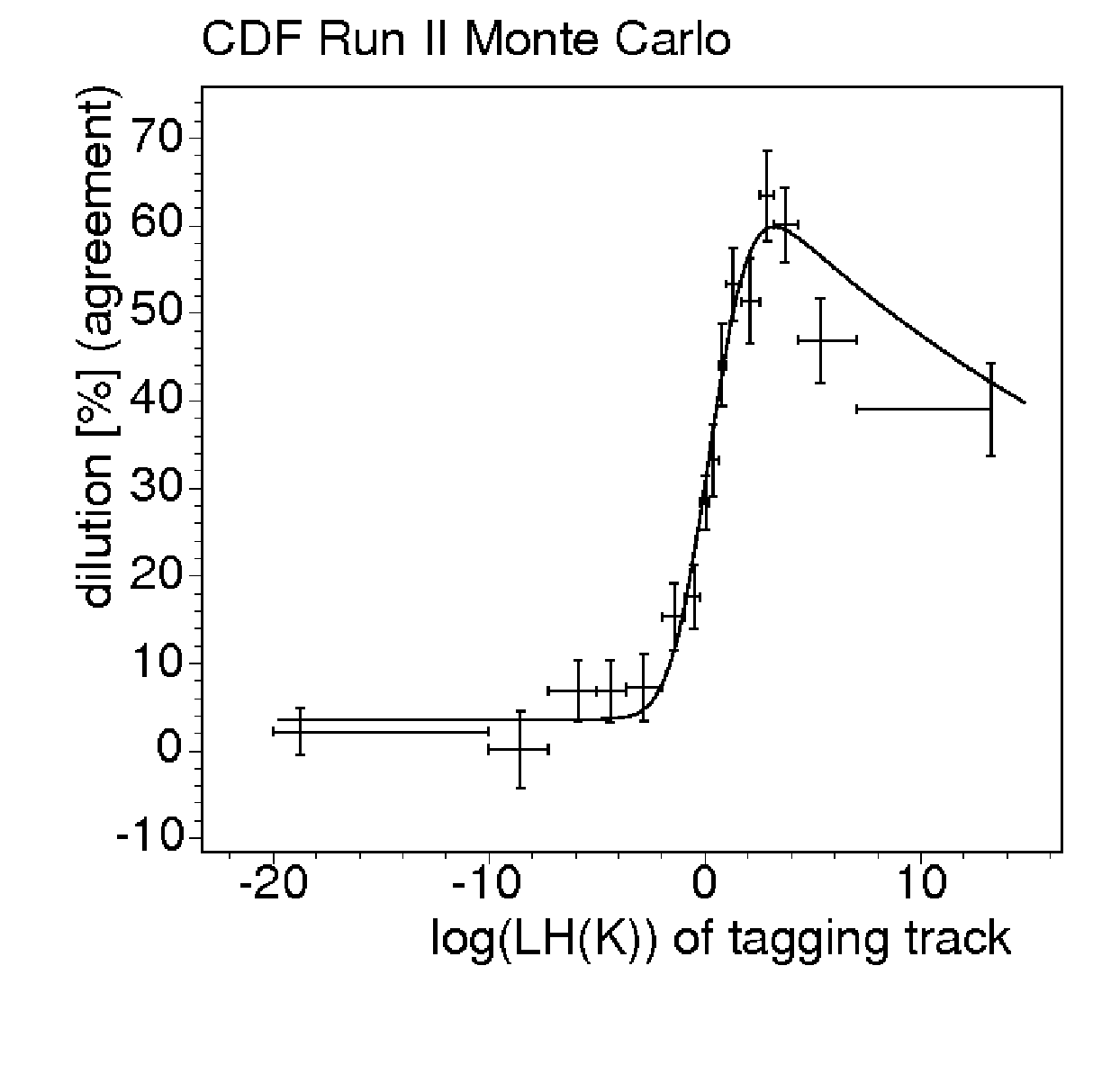}
\includegraphics[width=0.30\textwidth,height=4.cm]{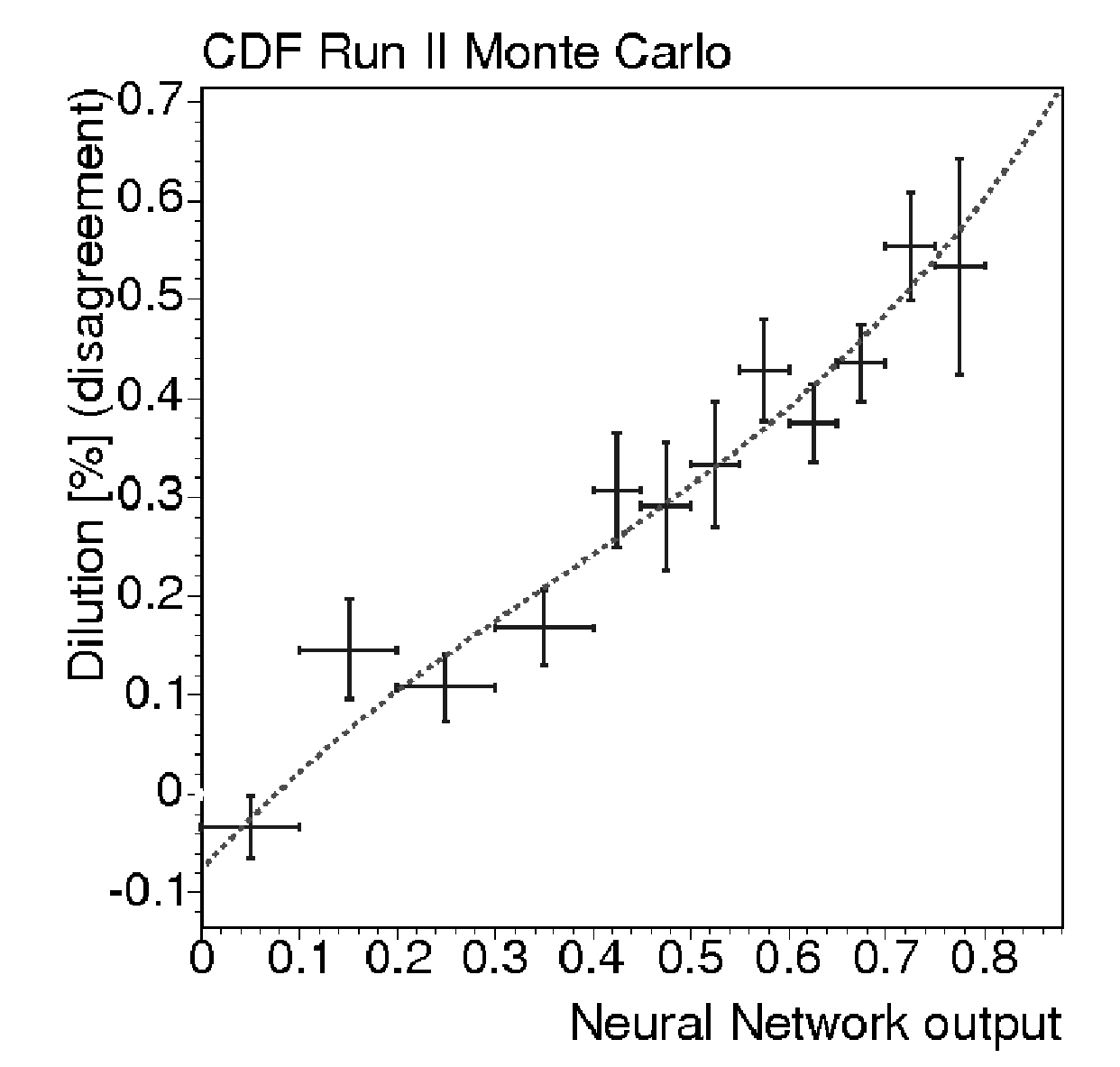}
\caption{(Left panel) Dilution of the maximum
  $p_L^{rel}$ algorithm as a function of tagging track $p_T$. (Middle panel)
Dilution for the Kaon identification based algorithm as
  function of Kaon likelihood. (Right panel)
  Dilution of the NN algorithm as a function of NN variable. The dots represent Monte Carlo data, the line
is the parametrization, which has been used to determine the event-by-event
  dilution. Events with only one tagging track
  candidate around the $B_s$ meson are displayed.}
\label{fig:CLL}
\end{figure}

 The performance of the SSKT algorithm has been evaluated for
\Bu, \Bz and \Bs modes on several decay channels
(see Table~\ref{tab:NNresults}).
The agreement between simulation and data in \Bu and \Bz modes suggests that 
the simulation can predict the tagger performance across all $B$ species. 
The measured differences are 
used to evaluate a systematic uncertainty on SSKT for \Bs mesons. 
Since the algorithm rely on the number of Kaons produced in 
the fragmentation process leading to the production of \Bs mesons 
an additional important uncertainty is derived by the difference in data
and simulation of the number of Kaons around the \Bs direction of flight.
Smaller systematic uncertainties arise considering $b$-quark production mechanism, 
fragmentation models, $B^{**}$ rate and event pile-up.

\begin{table}[htbp]
\caption{Performance of the NN based algorithm in data
  and Monte Carlo. Only statistical uncertainties are quoted.}
\label{tab:NNresults}
\begin{center}
\begin{tabular}{l|l|c|c|c}
\hline
\multicolumn{2}{c|}{[\%]} & $\bud$ & $\bdd$ & $\bsd$ \\
\hline
{MC}
& $\epsilon$  &   55.9  $\pm$ 0.1    &  56.6 $\pm$ 0.1 & 52.1 $\pm$ 0.3  \\
&  $\left<{\cal D}\right>$        &   26.8  $\pm$ 0.2    &  16.1 $\pm$ 0.6 & 29.2 $\pm$ 0.7  \\
\hline
{data (1 fb$^{-1}$)}
& $\epsilon$  &  58.2 $\pm$ 0.3  &   57.2  $\pm$ 0.3   & 49.3 $\pm$ 1.3 \\
&  $\left<{\cal D}\right>$         &  26.4 $\pm$ 0.8  &   15.2  $\pm$ 1.7   & --- \\
\hline
\end{tabular}
\end{center}
\end{table}

The tagging dilution evaluated for the $\bsd$ sample
using the event-by-event predicted dilution derived from Monte Carlo 
yields  $\left<{\cal D}\right> =  24.9-29.3 ^{+3.3}_{-4.3}$\% 
(for comparison using the maximum $p_L^{rel}$ only gives $\left<{\cal
D}\right> = 23.7^{+2.6}_{-4.5}$ \%). The overall SSKT tagging 
figure of merit is $\varepsilon \left<{\cal D}\right>^2
=3.1-4.3 ^{+1.0}_{-1.4}$, including statistical and systematic uncertainties 
(the given range reflect the performance of the CDF TOF system in different data taking
periods).
This result can be compared to the overall OST $\eD = 1.8 \pm 0.1 \%$ for opposite
side tagging on the same channel (note the significant channel dependence of the 
measured \eD, mostly related to the $B$~meson $p_T$ spectrum of the reconstructed decays). 

Also the \Dzero\ experiment recently introduced a same side tagger~\cite{:2008fj}.
The track with $p_T>500$\mevc\ closest in $\Delta R$ to the \Bs\ candidate
flight direction is selected for tagging. Dilution is studied as a function of
the product of the tagging track charge and $\Delta R$, as well as forming 
a same side jet charge from the transverse momentum weighted sum of all tracks
within a narrow cone around the \Bs flight direction. The combined \eD\ from 
OST and SST quoted by the \Dzero\ collaboration is $4.68 \pm 0.54 \%$ to be compared
to $2.48 \pm 0.22 \%$ from the OST alone.

\subsubsection{Vertexing}
\label{sec:expPrimers:vtx}
 For time dependent measurements determining the elapsed times ($\deltat$ in Sec.~\ref{sec:tdep}) is crucial.
This is obtained by first measuring a  length $L$ and then computing $\deltat=L/(c~\beta~\gamma)$. 
The vertexing techniques utilized to measure $L$ are significantly different at B-Factories and hadron 
colliders 
because of the different boost and because time dependent measurements 
have two different needs: measure the difference in time between the two $B$ mesons in an event at the B-Factories and measure the time of flight since the 
production of the $B$ meson of insterest at the hadron colliders. The two approaches are therefore described separatetly in the following.

\subsubsubsection{Vertex reconstruction at B-Factories}

\begin{figure}[ht]
\begin{center}
\epsfig{file=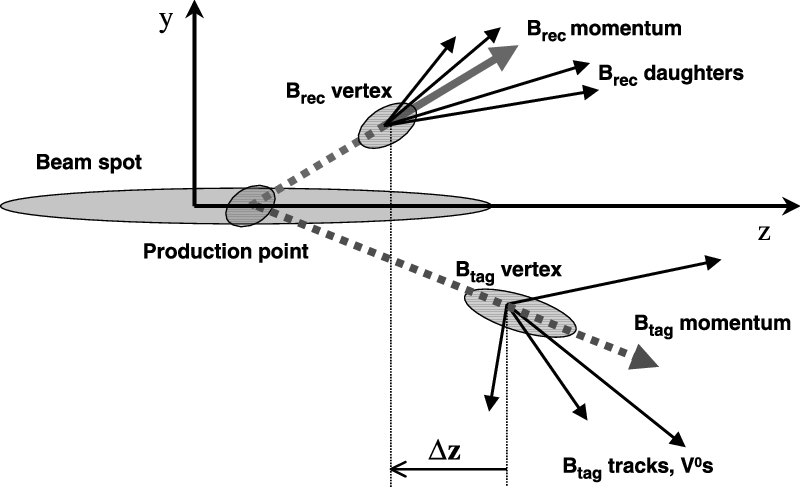,width=6.00truecm}
\caption{Schematic view of the vertexing algorithm at B-Factories.
}
\label{fig:BF_vtx}
\end{center}
\end{figure}

The $B_{reco}$ vertex is reconstructed from charged tracks and photon candidates that are combined to make up intermediate mesons (e.g., $J/\psi$, D, \KS)
and then treated as virtual particles. The trajectory of these virtual particles is computed from those of their decay particles, and,
when appropriate, mass constraints are imposed to improve the knowledge of the kinematics. In the case of charmonium states such as $J/\psi\KS$,
 Belle uses only the dileptons from the $J/\psi$ decay. In Belle, the vertex of the signal candidate is constrained to come from the beam-spot in the x-y
 plane and convolved with the finite B-meson lifetime. BABAR uses the beam-spot information only in the tag vertex reconstruction.
The resulting spatial resolution depends on the final state; it is typically  65~$\mu$m in BABAR and  75$\mu$m in Belle.

BABAR determines the Btag vertex by exploiting the knowledge of the center-of-mass four-momentum and an estimate of the interaction point or beam-spot 
position.   
This information, along with the measured three-momentum of the fully reconstructed Breco candidate,
 its decay vertex, and its error matrix, permits calculation
of the Btag production point and three-momentum, with its associated error matrix (see Fig.~\ref{fig:BF_vtx}).
All tracks that are not associated with the Breco reconstruction are considered; \KS\ and $\Lambda$ candidates are used as input
 to the fit in place of their daughters, but tracks consistent with photon conversions are excluded.
To reduce the bias from charm decay products, the track with the largest $\chi^2$ vertex contribution if greater than 6 is removed
and the fit is iterated until no track fails this requirement.

Belle reconstructs the Btag vertex from well-reconstructed tracks that have hits in the silicon vertex detector and are not assigned to the Breco vertex.
 Tracks from K0S candidates and tracks farther than 1.8 mm in z or 500 $\mu$m in r  from the Breco vertex are excluded.
An iterative fit to these tracks is performed with the constraint that the vertex position be consistent with the beam spot. If the overall
$\chi^2$  is poor, the track with the worst $\chi^2$ contribution is removed, unless it is identified as a high-momentum lepton.
 In this case, the lepton is retained and the track with the second-largest $\chi^2$  is removed.

The resolution on \deltaz\ is dominated by the Btag vertex reconstruction and therefore is nearly independent of the reconstructed CP decay mode.
 Based on Monte Carlo simulation, it is estimated to be 190~$\mu$m. The \deltaz\ measurement is converted to a \deltat\ measurement,
and the corresponding resolution is 1.1 ps in BABAR and 1.43 ps in Belle because of the different center-of-mass boosts.

\subsubsubsection{Decay Length Measurements at Tevatron}
\label{sec:vtxTevatron}
In the Tevatron detectors, with a central geometry,
the decay length is best measured in the transverse plane, the proper time $t$ is computed 
from the flight distance in the transverse plane, $L_{xy}$. Thus, the expression for $t$ and
its resolution are:
\begin{eqnarray}
t &=&  \frac{L}{c \beta\gamma} = L_{xy} \frac{m_B}{c~p_T}\,  
\hspace{0.1\textwidth} ;\hspace{0.1\textwidth}
\sigma_t = \sigma_{L_{xy}} \frac{m_B}{c~p_T} \oplus \frac{\sigma_{p_T}}{p_T} t
\label{eq:t}
\end{eqnarray}
 For fully reconstructed decays, the only significant uncertainty
 is from the decay distance measurement. Partially reconstructed
 decays have an additional term from $p_T$ uncertainty which grows linearly with $t$.


The transverse flight distance of the $B$-meson, $L_{xy}$, is
given by the transverse distance between the location of $p\bar{p}$ interaction,
the Primary Vertex (PV), and the Secondary Vertex (SV), 
i.e. the decay point of the $B$-meson. The position of the PV is determined for each event by 
fitting the tracks in the underlying event to a common origin, exluding the tracks belonging
to the $B$ candidate.


 The secondary vertex is determined by fitting to a common vertex the $B$ dauther 
charged tracks, considering tertiary vertex from charm decay, and mass constraints 
on intermediate resonances where applicable. 
The error estimate on $L_{xy}$ is obtained by combining the PV uncertainty with
that provided by the SV fit. A gaussian resolution function is normally
a good approximation but the error estimate from the vertex fit needs to be multiplied
by a scale factor for a correct measurement. This rescaling is typically calculated from 
the lifetime distribution of prompt background (e.g. from prompt \jpsi\ and 
underlying event tracks for decays involving \jpsi, or from prompt
charm production and fake leptons for semileptonic decays). A peculiar situation
arise when data biased in lifetime, due e.g. to trigger requirements, are used.
In this case special samples can be manufactured combining a prompt charm meson with a
randomly selected charged track, consistent with coming
from the PV. The pseudo-decay length of this events is expected to peak at 0 
and can be used to measure the decay length resolution scale factor.

 The proper decay time resolution for fully reconstructed $\bsd$ and $\bsdppp$
decays with the CDF detector is shown in Fig.~\ref{fig:ctres} (left).
The mean proper decay time uncertainty corresponds to $86 \,\rm fs$, 
which has to be compared with the oscillation period for \Bs\ mesons $\approx 350\,\rm fs$,
and shows the ability of the current Tevatron experiment vertex detectors to resolve
the fast \Bs\ oscillations.

\begin{figure}[bht]
\begin{center}
\epsfig{file=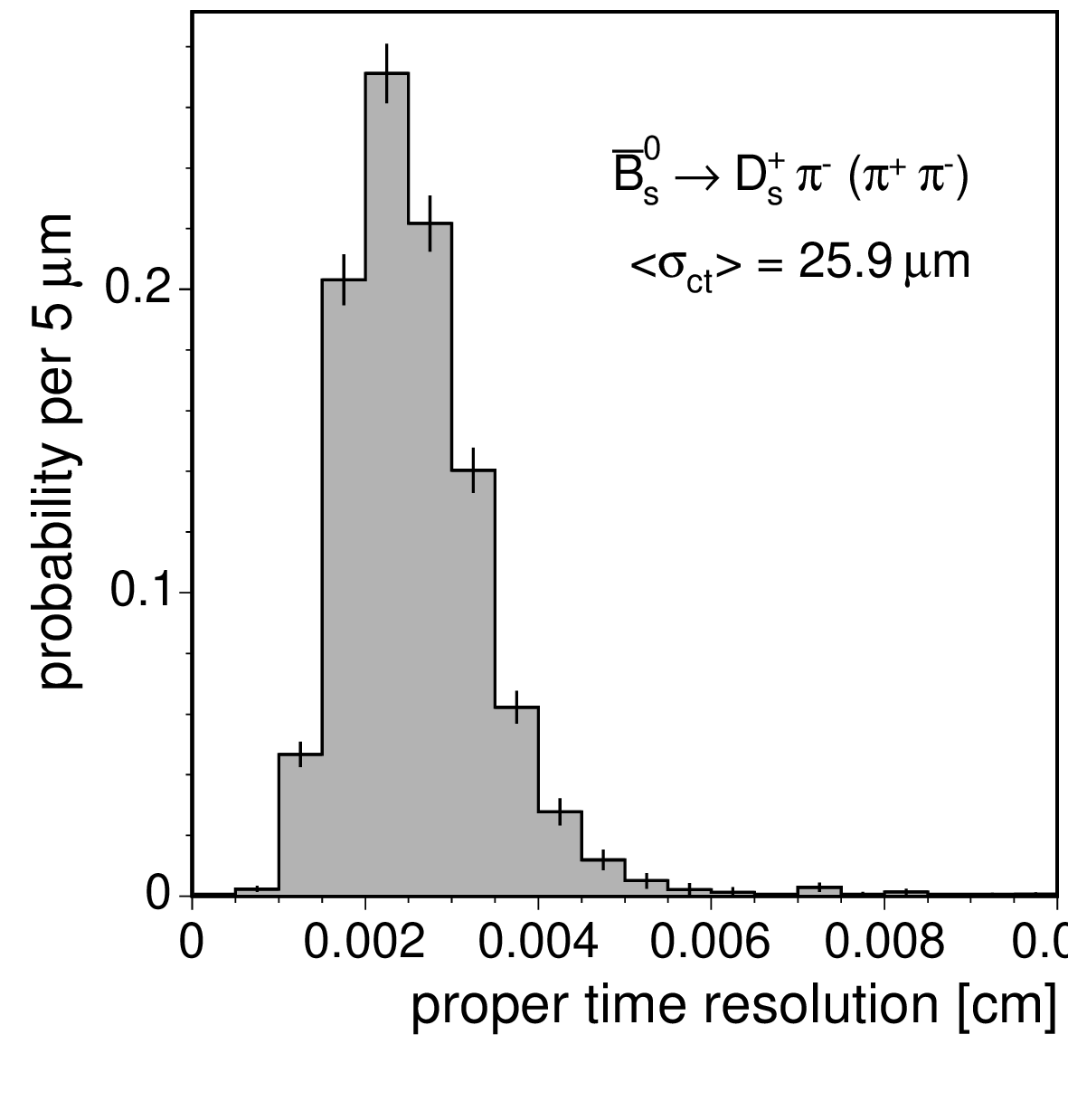,height=6.5cm,width=0.45\textwidth}
\epsfig{file=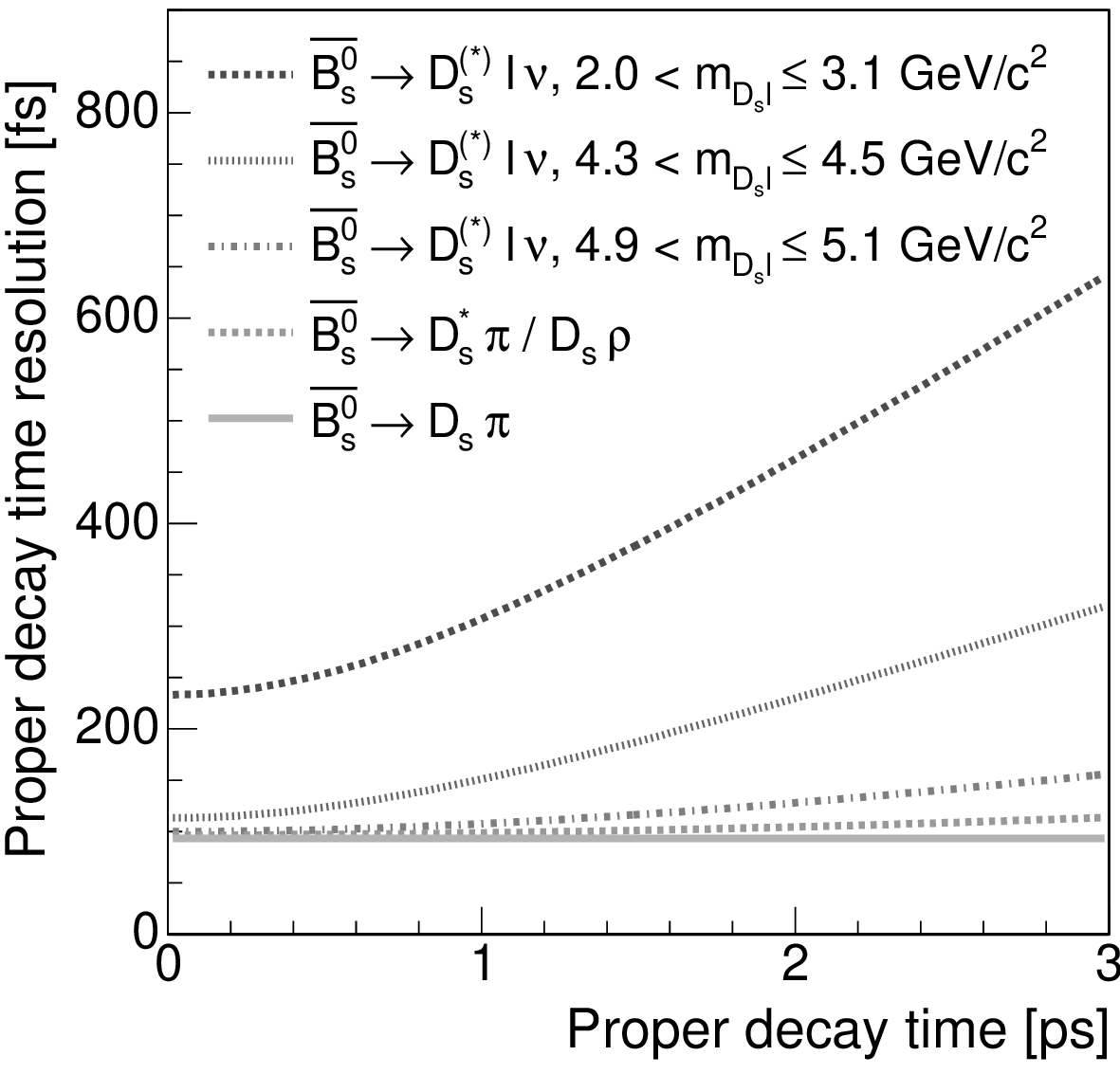,,width=0.48\textwidth}
\caption{The decay time resolution for fully reconstructed \Bs\ decays in CDF (left) 
and the effective resolution for different values of missing (unreconstructed)
semileptonic mass, as a function of the proper decay time (right).}
\label{fig:ctres}
\end{center}
\end{figure}

%
%

 For partially reconstructed decays, like semileptonic decays, there is an important additional
uncertainty in the decay time due to the incompletely measured $p_T$ of the
$B$ meson (Eq.~\ref{eq:t}). 
The distribution $F(k)$ of the fractional missing momentum $k= p_T^{\rm obs}/ p_T(B)$ is
extracted from Monte Carlo simulations and is rather wide with a typical RMS of $10$ to $20\%$.
The gaussian resolution function has to be convoluted with the distribution of this
$k$ factor in any time dependent measurement involving partially reconstructed or 
semi-leptonic decays. In semileptonic decays the missing neutrino momentum is correlated with 
the visible mass $D + \ell$, $M_{D \ell}$, hence it is useful to divide the data
in bins of $M_{D \ell}$ taking advantage of the narrower width of $F(k)$ for 
higher $M_{D \ell}$ as shown in Fig.~\ref{fig:ctres} (right).




 An important complication in time dependent measurement is introduced by reconstruction
or trigger bias on proper time (see e.g. section \ref{sec:hadrcolliders}). To take
in to account this effect a function $\xi(t)$, that describes the acceptance
as a function of proper decay time and is derived from simulations, multiply proper time related
terms in the likelihood fits.
To derive it CDF assumes that for each accepted event
$i$, the expected $ct$ distribution without any bias is an exponential
smeared by the experimental resolution function, where the width is the $ct$
error ($\sigma_{ct_i}$) of that event. The denominator is the sum of the
$N$ expected distributions without any bias,
\begin{eqnarray}
  \label{eqn:eps-b}
  \xi(ct) &\equiv&
  \frac{\mathrm{reconstructed}~ct~\mathrm{after~trigger+selection}}
       {\displaystyle\sum_{i=1}^N
     \frac{1}{\tau}\exp\left(-\frac{ct}{c\tau}\right)
     \otimes G(ct;\sigma_{ct_i})\otimes_k {\cal F}(k)}\,,
\end{eqnarray}
where the smearing with the $k$-factor distribution ${\cal F}(k)$ has to be
included for incompletely reconstructed decays.
The shape of the proper decay length efficiency curve is parametrized using
analytically integrable functions and used to multiply proper time related
terms in the likelihood fits. Fig.~\ref{fig:ctEff} shows a 
representative example of the proper time efficiency from the CDF experiment.
 The rapid turn-on of the curve is due to minimum impact parameter 
and $L_{xy}$ significance requirements at the trigger and reconstruction 
level, while the turn-off at larger
proper decay length is due to upper cut on impact parameter at
the trigger level. Because each $B$ decay
mode has its own kinematic characteristics and selection requirements,
an efficiency curve has to be derived separately for each channel. 

\begin{figure}[t]
\begin{center}
\includegraphics[width=0.45\textwidth]{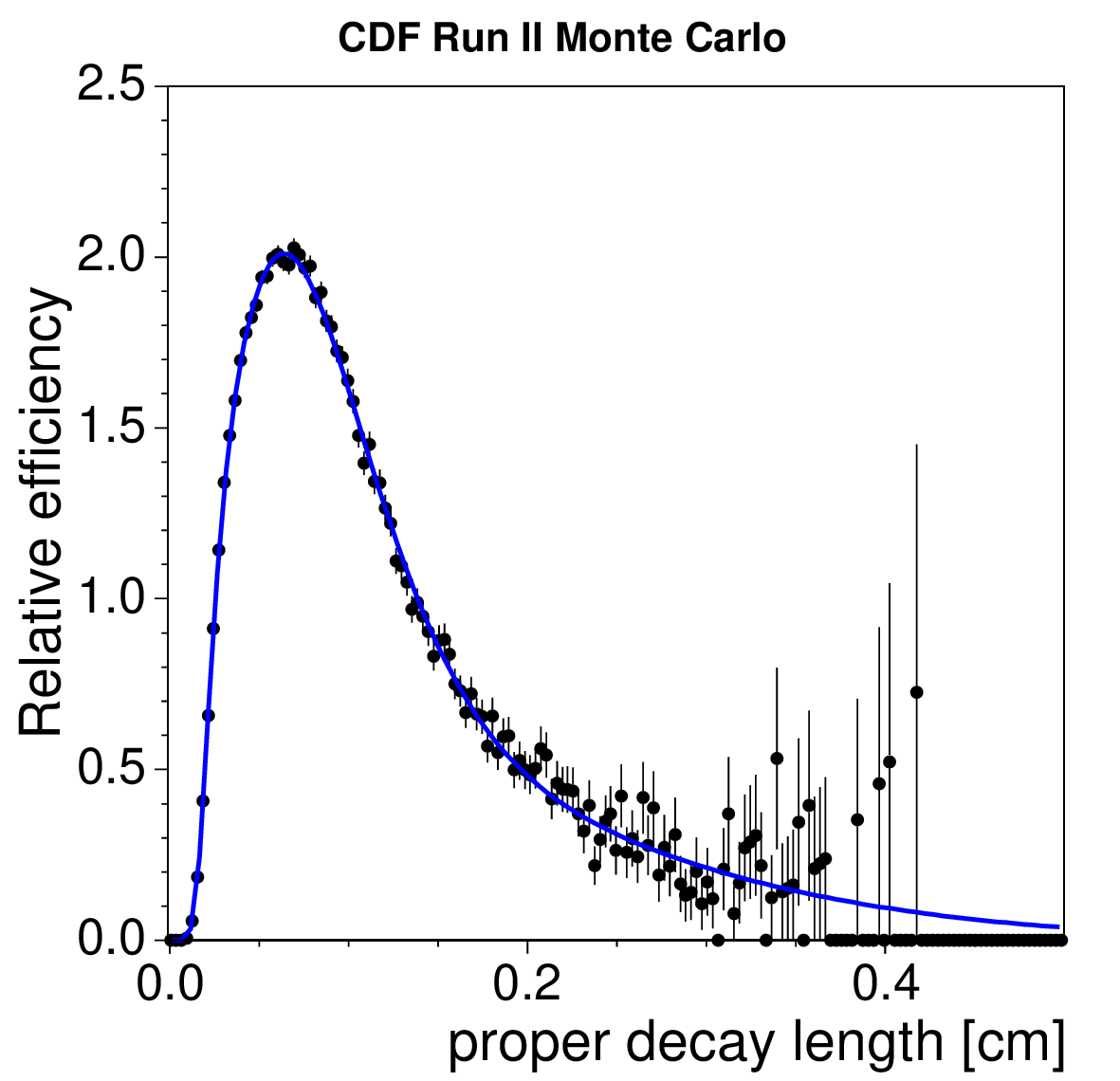}
\includegraphics[width=0.48\textwidth,viewport=0 180 600 810,clip]{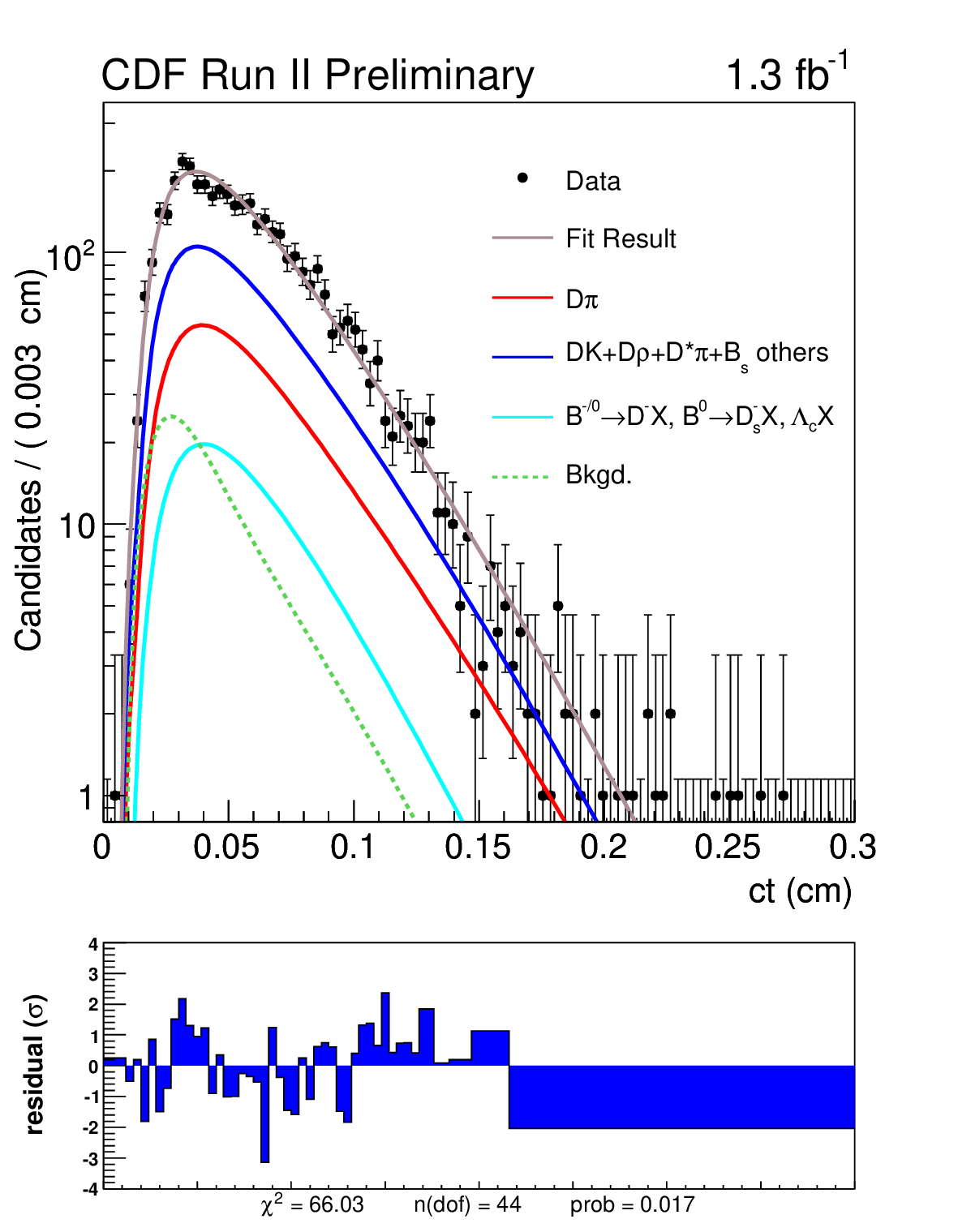}
\caption{A representative example of the dependence of trigger and
selection efficiency on proper decay time from the displaced track trigger
in CDF, vertical scale in arbitrary units (left). Lifetime fit to $\bsd$ sample
from CDF (right)}
\label{fig:ctEff}
\end{center}
\end{figure}

 The method has been extensively validated with Tevatron data,
measuring  \Bz, \Bu and \Bs and $\Lambda_b$ lifetime a variety of  
fully adronic modes. As an example, a recent preliminary determination 
of the \Bs\ lifetime in the $\bsd$ channel is shown 
in Fig.~\ref{fig:ctEff} right~\cite{bslifecdf}, giving 
$c\tau(\Bs) = 455.0 \pm 12.2 \stat \pm 8.2 \syst \mum$, in good agreement with PDG averages.

\subsubsection{Charged Particle Identification }
\label{sec:primers:pid}
%

Identification (ID) of charged particles ($e$, $\mu$, $\pi$, $K$, $p$) plays a crucial role in flavor physics, 
in many cases $\pi/K$ separation being both the most important and experimentally challenging.
Some of the most important PID techniques are sensitive to the particle's
velocity; working in tandem with tracking,
which provides a measurement 
of the particle's momentum, they separate the particles by mass. Other techniques exploit the
unique interaction properties of specific particles.

The purpose of this primer is to describe conceptually the PID techniques 
employed in the detector experiments that provided the results included in this report. 
For a more general discussion, please see Chapters ``Passage of particles through matter''
and ``Particle Detectors'' in Ref.~\cite{Amsler:2008zz}. 

At the track momenta relevant to flavor physics, the rate of \textbf{ionization energy loss}, 
usually denoted \dedx, to a good approximation is a constant for $e^{\pm}$
and a function only of the particle's velocity (but not its type) for the others. 
Measurements of \dedx are naturally provided by nearly all types of tracking detectors.
The type of information provided is either the collected charge or time-over-threshold for
each of the detector elements crossed by the track (which typically number from 8--10 to a few dozen).
The truncated-mean algorithm, which discards a fixed fraction (typically $\sim 30\%$)
of the samples with the highest \dedx values, is usually used to mitigate the effect
of the long tail of the Landau--Vavilov distribution of the individual \dedx samples.

As a function of particle's velocity, the \dedx truncated mean reaches a minimum at 
$\beta\gamma = p/m \approx$ 3.5--4.5 
and rises rapidly as the particle's velocity decreases 
($\dedx \propto 1/\beta^2$ for $\beta\gamma \lesssim 1$). For this reason, \dedx is the most useful 
for $\mu/\pi/K/p$ separation at the momenta where for at least 
one of the particle types being separated $p/m \lesssim 1.4$ (e.g., at $p \lesssim 0.7\, \gevc$ for $K/\pi$ separation). 
At $\beta\gamma \gtrsim 6$, the \dedx
truncated mean experiences a ``relativistic rise'', which is mild in gases, allowing weak (1-2 $\sigma$)
$\pi/K$ separation at $p \gtrsim 1.4\, \gevc$, but nearly non-existent in liquids and solids. 
Depending on the detector and the environment, 
measurements of \dedx can be affected by a large variety of sizable systematic effects, including aging,
and thus development of a \dedx calibration technique that can reliably predict the \dedx mean value
and resolution for a particle of a given type anywhere in the detector can be a great challenge, 
particularly when one wishes to exploit for PID the \dedx ``relativistic rise'' in a gaseous 
tracking system.

Examples of \dedx use in PID include the drift chambers in 
\babar~\cite{Sciolla:1998zs, Aubert:2001tu},
Belle~\cite{Hirano:2000ei},
BESII~\cite{Bai:2001dw},
CDF~\cite{Affolder:2003ep},
CLEO-II~\cite{Kubota:1991ww},
CLEO-III and CLEO-$c$~\cite{Peterson:2002sk}, and
KLOE~\cite{Ambrosino:2004qx}. 
In \babar and Belle, \dedx $K/\pi$ separation at low momenta is very important to 
$B$ flavor tagging, and in CDF the \dedx ``relativistic rise'' is critically
important to the study of $B^0$, $B_s$, $\Lambda_b \to h^+h^{\prime -}$ 
$(h = \pi, K, p)$ decays. The \babar silicon vertex tracker~\cite{Aubert:2001tu}, 
with its 5 double-sided Si layers, is unique among Si vertex detectors at $e^+e^-$
machines in its ability to provide useful \dedx information, which is particularly
valuable for $\pi/e$ separation at $p \lesssim 0.2 \gevc$ (e.g., in charm physics).

\textbf{Time-of-flight (TOF)} PID systems combine knowledge of the particle's creation time 
and trajectory with a high-precision measurement of its arrival time at the TOF detector, 
thus proving a measurement of its velocity. Given the time resolution of the currently
deployed TOF detectors ($\sim 100$-200 ps), they are limited in $\pi/K$ separation of at least $2\sigma$
to $p \lesssim 1.5$ \gevc. Examples include the TOF systems at 
Belle~\cite{Kichimi:2000uu},
BESII~\cite{Bai:2001dw},
CDF~\cite{Cabrera:2002vp}, and
KLOE~\cite{Ambrosino:2004qx}.
Complementarity of TOF and \dedx measurements is evident from the fact that
\dedx separation in gas vanishes for $\pi/K$ at 1.1 \gevc, for $e/\pi$ at 0.16 \gevc,
for $e/K$ at 0.63 \gevc, and for $e/p$ at 1.2 \gevc.

Detectors that exploit the \textbf{Cherenkov--Vavilov} radiation by charged
particles moving faster than $v_{\rm crit} = c/n$, where $n$ is the 
refraction coefficient of a solid, liquid or gaseous radiator, tend to provide the best
velocity-based PID at $p \gtrsim 1 \gevc$. 
The cheapest and most simple are Cherenkov threshold detectors, where the refraction
index of the radiator is chosen in such a way that in the kinematic range of 
interest the lighter of the two particle types being distinguished would 
be superluminal while the other one would not; additional information may be
provided by comparing the observed number of Cherenkov photons with the one expected
for each of the particle types. Belle employs silica aerogel with 
refraction indices varying from 1.01 to 1.03~\cite{Iijima:2000uv}. 

Since Cherenkov radiation is emitted in a
cone with an opening angle $\theta_{\rm C} = \cos^{-1}\frac{1}{n\beta}$, the particle's velocity
can be determined by measuring the cone's opening angle. 
The most common, moderately expensive such technology is RICH (Ring-Imaging CHerenkov), 
where the cone is produced in a transparent solid, liquid or gaseous radiator 
(LiF in CLEO-III and CLEO-$c$,~\cite{Artuso:2005dc}) and projected onto a planar 
photon detector a certain distance away. Another, more expensive but space-saving 
ring-imaging technology is DIRC, used in \babar~\cite{Adam:2004fq},
where the cone of Cherenkov light is produced and captured within a bar of 
synthetic fused silica running the length of the \babar detector.
The $\pi/K$ separation achieved in $B \to X h^{\pm}$ decays in \babar by 
the DIRC (DCH \dedx) varies from $13\sigma$ ($1.0\sigma$) at 1.5 \gevc
to $2.5\sigma$ ($1.9\sigma$) at 4.5 \gevc~\cite{Aubert:2007mj}. However, 
due to the DIRC's mechanical complexity about 18\% of reconstructed high-momentum tracks
in \babar miss the DIRC; similar coverage limitations are usually suffered by RICH and TOF systems
as well.  

For dedicated $e^{\pm}$ ID, 
the most distinctive and frequently used feature of their interactions with matter
is the development of electromagnetic (EM) showers in thick absorbers. 
\textbf{EM calorimeters} seek to contain and measure the total shower energy $E_{\rm cal}$.
For $e^{\pm}$, the ratio $E_{\rm cal}/p$
is close to 1, while for the other charged particles
the $E_{\rm cal}/p$ ratio will be either much smaller than 1 (``minimum-ionizing''), 
have a broad distribution mostly below 1 for those that shower hadronically, 
or have a poorly defined broad distribution for the antiprotons that annihilate 
in the calorimeter. 
Since the shapes of the EM 
showers produced by high-energy $e^{\pm}$ and photons are quite similar, the matching
of calorimeter clusters to tracks extrapolated from the tracking system is of
critical importance. 
The materials used in EM calorimeters the most frequently are blocks of heavy inorganic scintillators
with no longitudinal 
segmentation. 
Thallium-doped CsI is used 
in
\babar~\cite{Aubert:2001tu}, 
Belle~\cite{Abashian:2000cg, Hanagaki:2001fz}, 
CLEO~\cite{Bebek:1987kd}, 
and BESIII.
Even in the absence of longitudinal segmentation, limited information on the longitudinal shower shape
(which is different for $e/\mu/\pi/K/p$) 
can be obtained for particles of sufficiently low momenta (which enter the calorimeter at an angle
sufficiently different from $90^{\circ}$) by combining tracking and lateral cluster-shape information
with a technique recently introduced in \babar~\cite{Brown:2007nf}.
KLOE has a lead--scintillating fiber sampling EM calorimeter~\cite{Adinolfi:2002jk}, 
which, thanks to its longitudinal segmentation, also provides good muon-hadron separation.

Unlike the other long-lived charged particles, muons do not shower.
Hence, dedicated \textbf{muon ID} relies on muons' long path length in absorber thick enough to stop 
hadronic showers (5-8 hadronic interaction lengths is common). Instrumentation of the magnet's iron flux return
with several layers of charged-particle detectors is a good approach since it allows 
monitoring of hadron-shower development (which also enables \KL ID) and
precise matching of tracks with hits in the muon system.
This approach is used in 
\babar~\cite{Aubert:2001tu},
Belle~\cite{Abashian:2002bd},
BESII~\cite{Bai:2001dw}, and
CLEO~\cite{Bortoletto:1991ay, Kubota:1991ww}.

Response of the detector as a whole, and each of the subdetectors individually, 
to the passage of charged
particles of a given type can be studied with high-purity, high-statistics 
calibration samples selected on the basis of the physics and kinematics of certain decays, 
with PID applied to the other particles in the decay to further enhance purity. 
In calibrating the PID response of a given subdetector, PID information 
from the rest of the detector can be used as well. 
Examples of calibration samples used in \epem $B$ factories include
protons from $\Lambda\to\proton\pim$, pions and Kaons from $D^{*+}\to D^0\pi^+\,(\Dz\to\Km\pip)$, 
pions from $\KS \to\pip\pim$, electrons and muons from $\epem \to \ell^+\ell^-\gamma$.

The best PID performance is achieved by combining information from all subdetectors. 
The TOF, \dedx and ring-imaging Cherenkov measurements can be conveniently represented 
in the form of probability-distribution functions (PDFs), which makes likelihood-based hadron 
ID quite close to optimal. On the other hand, the calorimeter and muon-system quantities, 
which are more numerous and can be highly correlated,
are either very difficult or impossible to adequately describe with PDFs.
For this reason, the best PID performance can be achieved by advanced multivariate techniques
such as neural nets and bagged decision trees.

\subsubsection{Background suppression}
\label{sec:primers:bkg}

The isolation of signal events in the presence of significant sources of backgrounds is critical for 
almost all measurements. This usually is achieved by an optimization of the event selection process 
designed to maximize the experimental sensitivity by suppressing the backgrounds effectively while 
retaining
a sizable fraction of the signal. The choice of the method depends on both the nature of the signal and 
background events, and critically on the signal over background ratio which may vary from more than 100 
to $10^{-6}$ or less. 

The separation of signal and background processes relies both on the detector performance as well as 
kinematics of the final state produced. Large acceptance and the high resolution and efficiencies for 
the reconstruction of charged and neutral particles and the identification of leptons and hadrons over 
a wide range of energies are very important.  A low rate of the misidentification of charged hadrons as 
leptons is critical, in particular  for rare processes involving leptons.

Though the cross sections for heavy flavor particle production in hadronic 
interactions exceed the cross sections at \epem\ colliders by several orders of magnitude, their 
fraction of the total interaction rate is small.  Furthermore, the  multiplicity of the final states is 
very large, and thus the combinatorial background to charm and beauty particles is extremely large for 
experiments at hadron colliders and for fixed-target experiments in high momentum hadron beams. 
Typical event triggers rely on the detection of charged hadrons and leptons 
of large transverse momentum and in some cases also on the isolation of decay vertices that are 
displaced from the primary interaction point.  The analyses  often focus on decays involving two- or 
three-body decays to intermediate states of narrow width, for instance \jpsi, $D$ or $D^*$ mesons.
Because of the very large momenta of these intermediate states, the identification of particles that do 
not originate form the primary interaction point is a very powerful tool to suppress backgrounds.  

Background conditions for the detection of charm and beauty particles at \epem\ colliders are markedly 
different.  There are two dominant sources of background, the so-called continuum background and 
combinatorial background from other particles in the final states from decays of resonances under 
study, for instance $\jpsi, \psiprpr,$ or $\Upsilon(nS)$ mesons. 
Two types of processes contribute to continuum background, QED processes, $\epem \to \ellell (\gamma)$ 
with 
$\ell = \electron , \mmu,$ or $\tau$, and quark-pair production, $\epem \to \qqbar $ with $\q = 
u,d,s,(c)$. Both of these processes are impacted by energy losses due to initial state radiation.

At \epem\ colliders operating near kinematic thresholds for pair production of charm or beauty 
particles, for instance the B Factories at \FourS and the Charm Factories at the \psiprpr\ or above, 
the primary particles pairs are produced at very low momenta, leading to event topologies that are 
spherical, not jet-like.

Continuum background is characterized by lower multiplicities and higher momenta of charged and neutral 
particles.   To suppress QED background, selected events are usually required to have at least three 
reconstructed charged particles.  
At sufficiently high c.m. energies, the fragmentation of the light quarks leads to a two-jet topology.  
Such events are characterized by variables that measure the alignment of particles within an event 
along a common axis.  Among the variables that show sharply peaked distributions for jet-like events 
are:
\begin{itemize}
\item thrust, the maximum sum of the longitudinal momenta of all particles
relative to a chosen axis;  the trust distribution peaks at or just below 1.0 for two-body final states 
and two-jet events;
\item $\cos\Delta\theta_{thrust}$, where $\Delta\theta_{thrust}$ is the angle between the thrust axis 
of one or the sum of all particles associated with the signal candidate and the thrust axis of the rest 
of the event;  this distribution is flat for signal events and peaked near 1.0 for continuum 
background;  
\item the energy flow in conical shells centered on the thrust axis, typically nine double cones of 10 
degrees; for continuum events most of the energy is contained in the inner cones, while for the more 
spherical signal events
the energy is shared more uniformly among all cones;     
\item  normalized Legendre moments can be viewed as continuous generalizations of the energy cones, 
typically the first and second of these moments are used, 
$L_j= \sum_i p_i |\cos\theta_i|^j$ with $j=0$ or $j=2$, where $p_i$ and $\theta_i$ are the momentum and 
angle of any particles, except those related to the signal decay, relative to the thrust axis of the 
signal decay.
In many cases these moments provide better discrimination of continuum events than the energy cones.
\item  $R_2=H_2/H_0$, the ratio of second to zeroth Fox-Wolfram movements, 
with $H_2=\sum_{i,j} |p_i| |p_j| \, L_2(\cos\theta_{ij})$, calculated for all particles in the event,  
charged and neutral. The nth Fox-Wolfram moment is the momentum-weighted sum of Legendre polynomial of 
the nth order, computed for the cosine of the angles between all pairs of particles; the ratio $R_2$ 
peaks close to 1.0 for jet-like continuum events.
\end{itemize}
In practice the suppression of the continuum background is achieved by imposing restriction on many of 
these variables, either as sequential individual cuts, or by constructing a multivariable discriminant, 
a decision tree, or employing a neural network.  

Fig.~\ref{fig:primers:jetsupp} shows examples of distributions for two of these variables for 
selected \BB\ events.
\begin{figure}[bht]
\begin{center}
\epsfig{file=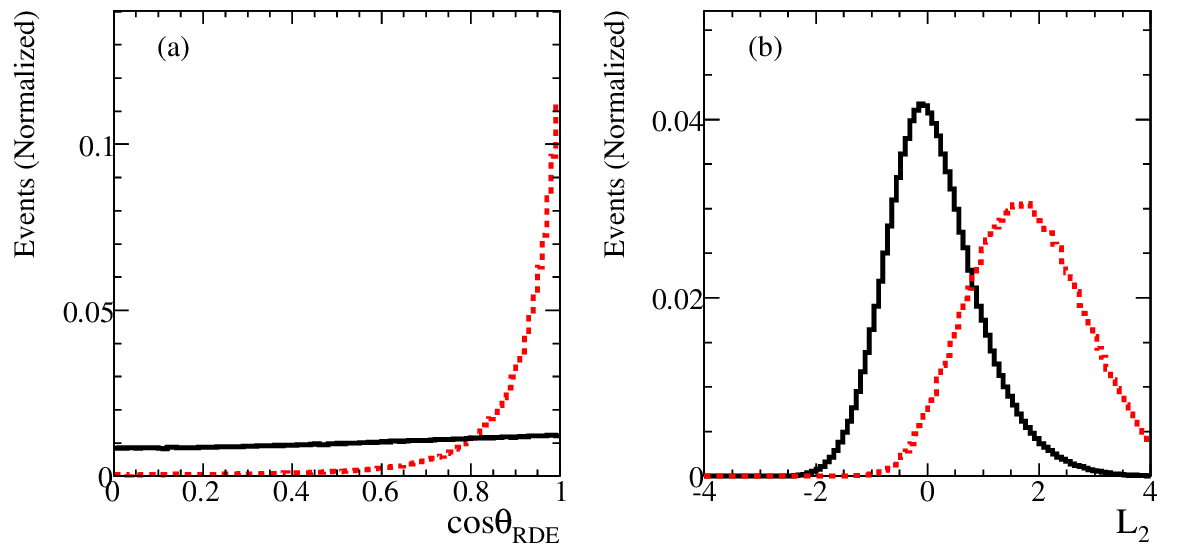,height=5.0cm}
\end{center}
 \caption{\it Distribution of variables used to suppress continuum background in selected candidates 
for $\Bz \to K^- \pi^+ \pi^0$ decays~\cite{Aubert:2008zu} 
a) $\cos\Delta\theta_{thrust}$, b) the normalized Legendre moments $L_2$. 
The solid lines show the expectation for continuum background, the dotted
lines represent the background distributions.
 }
\label{fig:primers:jetsupp}
\end{figure}

For the isolation of exclusive decays of $B$ or $D$ mesons that are pair-produced at Beauty or Charm 
Factories two kinematic variables are commonly used to separate signal from background events.  These 
variables make optimum use of the measured beam energies and are largely uncorrelated. The difference 
of the reconstructed and expected energy for the decay of a meson $M$ is defined as
$\Delta E = (q_M q_0 - s/2)/\sqrt{s}$, 
where $\sqrt{s}=2 E^*_{beam}$ is the total energy of the 
colliding beams in the c.m. frame, and $q_M$ and $q_0$ are the Lorentz vectors representing the 
momentum of the candidate $M$ and of the \epem\ system, $q_0=q_{\en} + q_{\ep}$.  In the c.m. system, 
\begin{equation}
\Delta E = E^*_M - E^*_{beam}, 
\label{eq:deltaE}
\end{equation} 
where $E^*_M$ is the energy of the reconstructed meson $M$.

The second variable is often referred to as the energy-substituted mass, $m_{ES}$.  In the laboratory 
frame, it can be determined from the measured 
three-momentum, $\vec{p}_M$, of the candidate $M$, without explicit knowledge of the masses of the 
decay products,  
$m_{ES}=\sqrt{(s/2 + \vec{p}_M \cdot \vec{p}_0)^2/E_0^2 - \vec{p}^2_M }$.
In the c.m. frame ($\vec{p}_0=0)$, this variable takes the familiar form,
\begin{equation}
m_{ES}=\sqrt{E^{*2}_{beam} - \vec{p}^{*2}_M}, 
\label{eq:mES}
\end{equation} 
where 
$\vec{p}^*_M$ is the c.m. momentum of the meson $M$, derived from the momenta of its decay products, 
and its energy is substituted by $E^*_{beam}$. 

An example of $\Delta E$ and $m_{ES}$ distributions is given in Fig.~
\ref{fig:primers:demes} for a selected sample of rare $B$ decays.
$\Delta E$ is centered on zero and the $m_{ES}$ distribution peaks at the $B$-meson mass. 
While resolution in $\Delta E$ is dominated by detector resolution, the resolution in $m_{ES}$ is 
determined by the spread in the energy of the colliding beams, typically a few \mev.
The flat background is composed of both continuum and \BB\ events, its size depends on the decay mode 
under study and the overall event selection. There is a small component of peaking background due to 
backgrounds with kinematics very similar to the true decays. 

\begin{figure}[bht]
\begin{center}
\epsfig{file=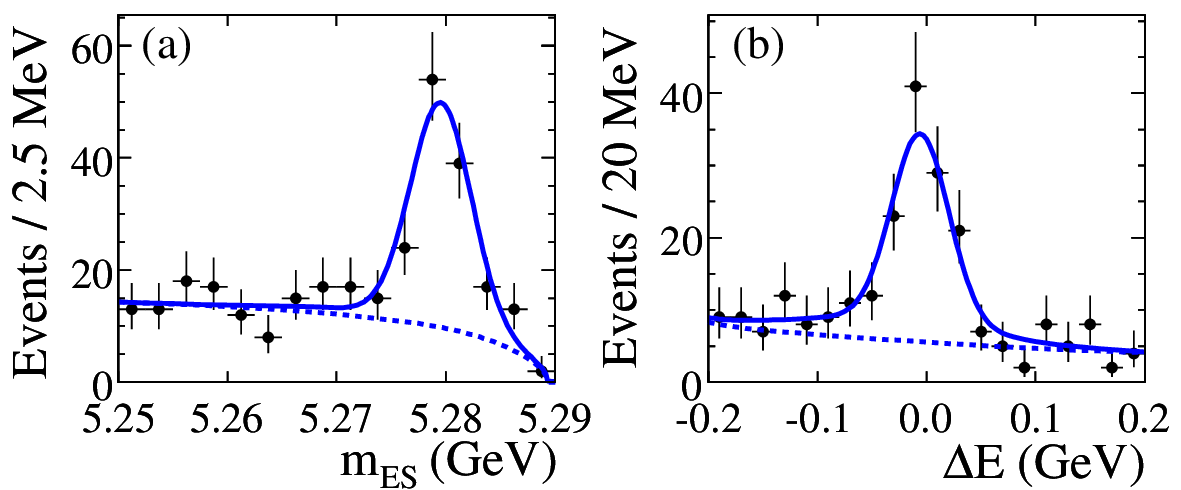,height=4.0cm}
\end{center}
 \caption{\it Distributions of a) $\Delta E$ and b) $m_{ES}$ for a sample of
$\Bz \to \omega \KS$ candidates~\cite{:2008se}.  The solid line represents the result of the fit to the 
data, the dotted line marks the background contributions.
}
\label{fig:primers:demes}
\end{figure}

For decays that cannot be fully reconstructed because of an undetected neutrino or \KL, the separation 
of signal and backgrounds is more challenging. The energy and momentum of the missing particle can be 
inferred from the measurement of all other particles in the event and the total energy and momentum of the colliding beams,
\begin{equation}
(E_{miss}, \vec{p}_{miss}) = (E_0,\vec{p}_0) - (\sum_i E_i, \sum_i \vec{p}_i).  
\label{eq:EPmiss}
\end{equation} 

If the only missing 
particle in the event is a neutrino or \KL,
the missing mass should be close to zero or the Kaon mass and the missing momentum should be non-zero.  
Fig.~\ref{fig:primers:slep}a shows an example of a missing mass squared distribution, $E_{miss}^2- 
|\vec{p}_{miss}|^2$ for $B^- \to D^0 \ell^- \nub$ decays, selected in \BB\ events tagged by a hadronic 
decay of the second $B$ meson in the event.  There is a narrow peak at zero for events in which the 
only missing particle is the neutrino, and a broad enhancement due to $B^- \to D^{*0} \ell^- \nub$ 
decays, in which the low energy pion or photon from the decay $D^{*0} \to D^0 \pi^0$ or $D^{*0} \to D^0 
\gamma$ escaped detection.  Since the second $B$ is fully reconstructed, there is very little 
combinatorial background.

\begin{figure}[bht]
\begin{center}
\epsfig{file=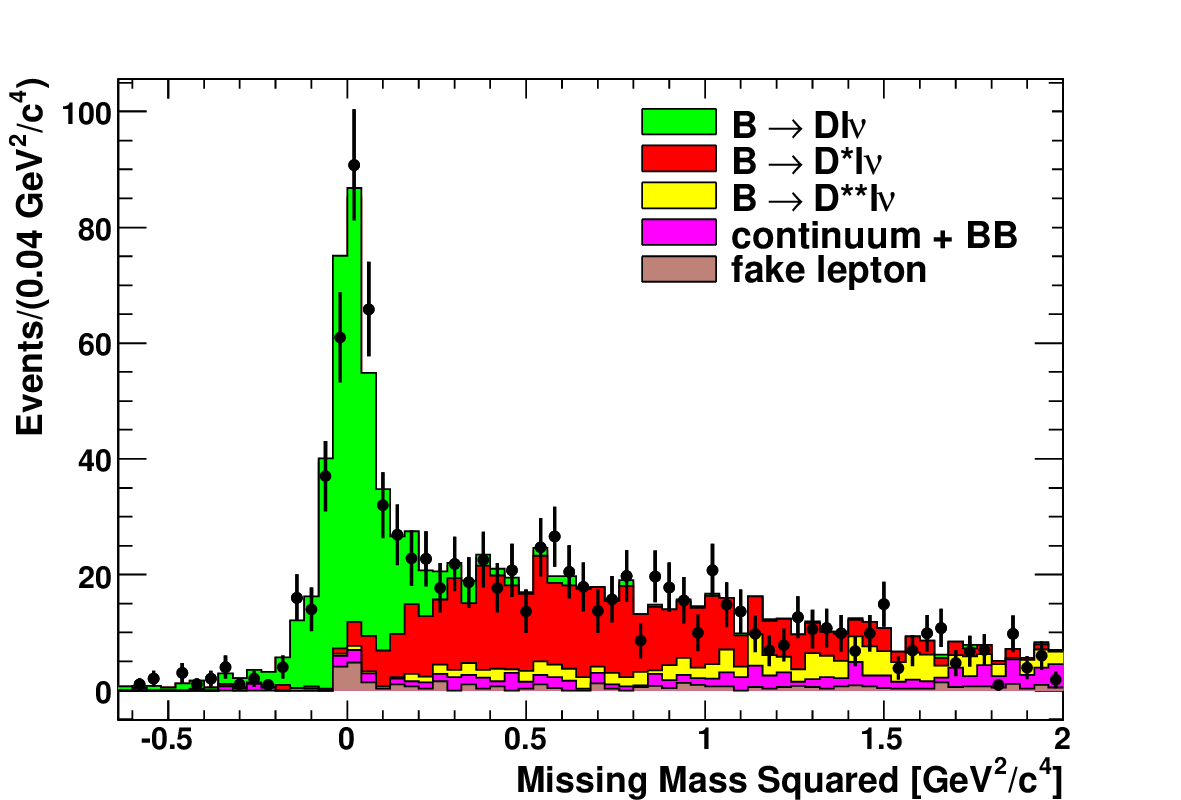,height=5.0cm}~
\epsfig{file=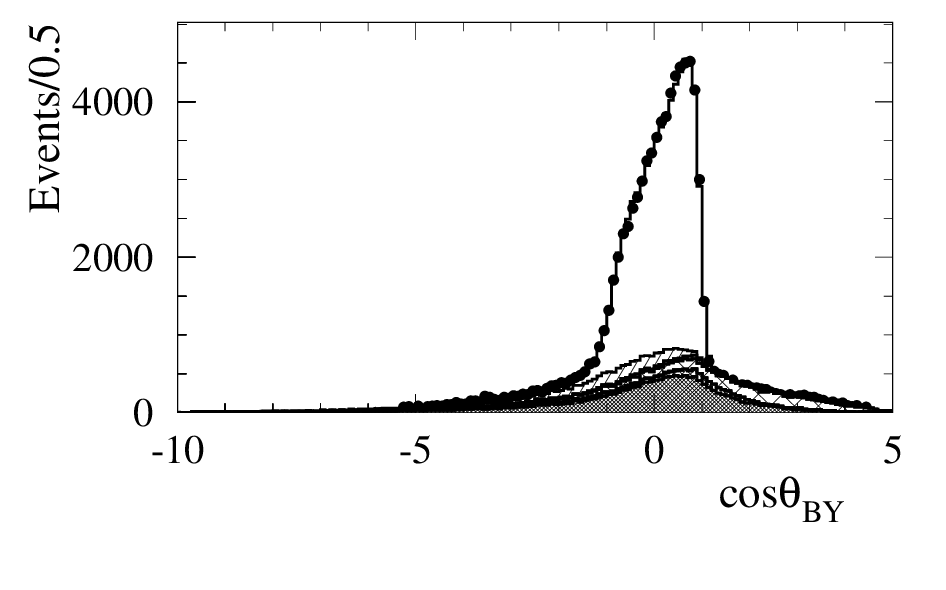,height=4.5cm,width=6.0cm}
\end{center}
 \caption{\it Distributions of the a) the missing mass squared for selected $B \to D \ell \nu$ 
candidates, in \BB\ events tagged by a hadronic decay of the second $B$ meson in the 
event~\cite{:2008ii}, b) $\cos \theta_{BY}$, for a sample of $\Bzb \to \Dstarp \ell^- \nub$ 
candidates~\cite{Aubert:2007rs}
Her the unshaded histrogram indicates the signal distribution, on top of background contributions, 
mostly from other semileptonic $B$ decays.
 }
\label{fig:primers:slep}
\end{figure}

For semileptonic $B$ or $D$ decays, $M \to H \ell \nu$, a variable first introduced by the CLEO 
Collaboration is used to suppress background,
\begin{equation}
\cos \theta_{BY}= \frac{(2 E_B E_Y - M^2_M - M^2_Y)}{2 |\vec{p_M}| |\vec{p}_Y|}.
\label{eq:cosBY}
\end{equation} 

For a true semileptonic decay in which the only missing particle is the neutrino,  $\theta_{BY}$ is the 
angle between the momentum vectors $\vec{p_M}$ and $\vec{p}_Y= \vec{p}_H + \vec{p}_{\ell}$, and the 
condition  $|\cos \theta_{BY}|\leq 1.0$ should be fulfilled, while for background events or 
incompletely reconstructed semileptonic decays the distribution extends to much larger values, thus 
enabling a clear separation from the signal decays (see Fig. \ref{fig:primers:slep}b).


\subsubsection{Recoil Tagging Technique}
\label{sec:recoiltechnique}


At \epem\ colliders charged leptons and heavy flavor particles are produced in pairs,
thus the detection of one member of the pair can be used to tag the presence of the other. 
In particular at Charm and $B$ at B-Factories, operating at or near the threshold 
for charm or beauty particles tagging techniques not only identify the second member of the pair, they also can be used to measure their momentum and energy and uniquely determine their charge and flavor quantum numbers. Furthermore, near threshold, there are no other particle produced, and therefore the combinatorial background is significantly reduced.  In addition, the kinematics of the final state are constrained such that given a fully reconstructed tag of one decays, 
the presence of a missing or undetectable particle like $\nu$ or $\KL$ meson can be identified from the missing momentum and missing energy of the whole event (see for example \cite{He:2007aj}). 

The tagging technique for $\psiprpr\to D\Dbar$ events was first developed by the 
Mark III collaboration \cite{Adler:1989rw} at SLAC, and has since been  exploited in many analyses based on data from by $CLEO_c, BES, KLOE$, and the $B$ Factories. 
For $\psiprpr\to \DzDzb$ events 
there are several tag modes, which can be divided into three categories: 
pure flavor tags such as $\Dz\to\Km\ep\nue$ and $\Dz\to \pim\mup\num$; 
quasi-flavor tags for neutral mesons, such as $\Dz\to \Km\pip$, $\Dz\to \Km\pip\piz$ and $\Dz\to \Km\pip\pip\pim$, for which there is a small doubly-Cabibbo-suppressed contribution, and tags for CP-eigenstates such as $\Dz\to \Kp\Km$ and $\KL\piz$. The quasi-flavor tags can be used to make precision measurements of branching fractions \cite{:2007zt} and partial rates \cite{:2008yi}. The three decays listed correspond to 25\% of the total branching fraction. Since the \psiprpr\ is a $C=-1$ state, the detection of a tag with definite CP means that the other D meson in the event must be of opposite CP. Studies combined flavor and CP-tagged samples of $K\pi$ events \cite{Asner:2005wf} and $\KS\pip\pim$ \cite{Giri:2003ty} have resulted in the determination of the strong-phase parameters in $D$ decay. Using low-multiplicity decays, such as $\Dp\to\Km\pip\pip$ and $\Dp\to \KS\pip$ has resulted in extremely clean samples, even for rare signal decays, and thus precise branching fraction and partial rate measurements. 

Single-tag efficiencies and purities vary considerably depending on the number of tracks and neutrals in the decay. For example, $\Dz\to\Km\pip$ and $\Dp\to \KS\pip\piz$ tags have efficiencies of 65\% and 22\% and sample purities of $\sim 5\%$ and $\sim 50\%$, respectively.  
For fully reconstructed hadronic tags the discriminating variable (shown in 
Fig.~\ref{fig:primers:tags}) is the beam-constrained mass (see  Sec.~\ref{sec:primers:bkg} and Eq.~\ref{eq:mES}).

The recoil technique has also been used successfully in $\epem\to\Dsp\Dssm$ events at CLEO-c to measure branching fractions (\cite{Alexander:2009ux}). Tag decays include $\Dsm\to \Kp\Km\pim$, $\Dsm\to \KS\Km$, $\Dsm\to \Kp\Km\pim\piz$ and $\Dsm\to\pip\pim\pim$ and correspond to approximately 20\% of the total \Dsm\ branching fraction. The $\Dssm\to\Dsm\g/\piz$ candidates are identified with or without the explicit reconstruction of the photon or \piz . 

At the \FourS\ resonance, the higher mass of the $b$ mesons lead to much smaller individual branching fractions for individual decays, which means that the achievable tagging efficiencies are much lower.
Nevertheless, both \babar\ and $Belle$ have developed and employed 
several tagging techniques. 
The cleanest samples are possible for tree-mediated  hadronic decays of the form 
$B \to D^{(*)}X$, where X refers a hadronic state of one or more hadrons, up to five charged mesons (pions or Kaons), up to two neutral pions or a $\KS$, and the  
$D^{0,(*)}$, $D^{+,(*)}$or $D_s^{+,(*)}$ mesons are reconstructed in many different decays modes. The kinematic variables 
\DeltaE\ and \mes, introduced in Sec.~\ref{sec:primers:bkg},
are used to isolate the true tag decays from combinatorial background and  to estimate the purity of the tag samples.
The purity of a given tag mode is used to separate the cleaner samples
from those with high background, the actual choice usually depends on the 
signal mode under study. The tag efficiency is typically  0.3\% and has a 
signal-to-noise ratio of 0.5 (see Fig.~\ref{fig:primers:tags}).

Significantly higher tag efficiencies can be obtained for semileptonic $B$ decays,
for instance $B\to D^{(*)}\ell\nu$ ($\ell=e,\mu$), with a 
branching fraction of more than 7\% for each lepton. For $D$ mesons the same decays listed above are used, are reconstructed and for the $D^*$ mesons the decays are $\Dstarp\to \Dz\pip,\Dp\piz$ and $\Dstarz\to\Dz\piz,\Dz\gamma$. 
Due to the very small mass difference of the $D^*$ and $D$ mesons, the pions and photons from its decay are of low energy, and thus the mass difference 
$\Delta M = m(D\pi) - m(D)$ can be very well measured.   
The presence of a neutrino in the decay can be checked using the variable 
$cos\theta_{BY}$ defined in Eq.~\ref{eq:cosBY}.
As for hadronic tags, tag selection and its efficiency and purity are strongly dependent on the signal decay recoiling against the tag. Typical efficiencies are of order 0.5-1\%.

\begin{figure}[bht]
\begin{center}
\epsfig{file=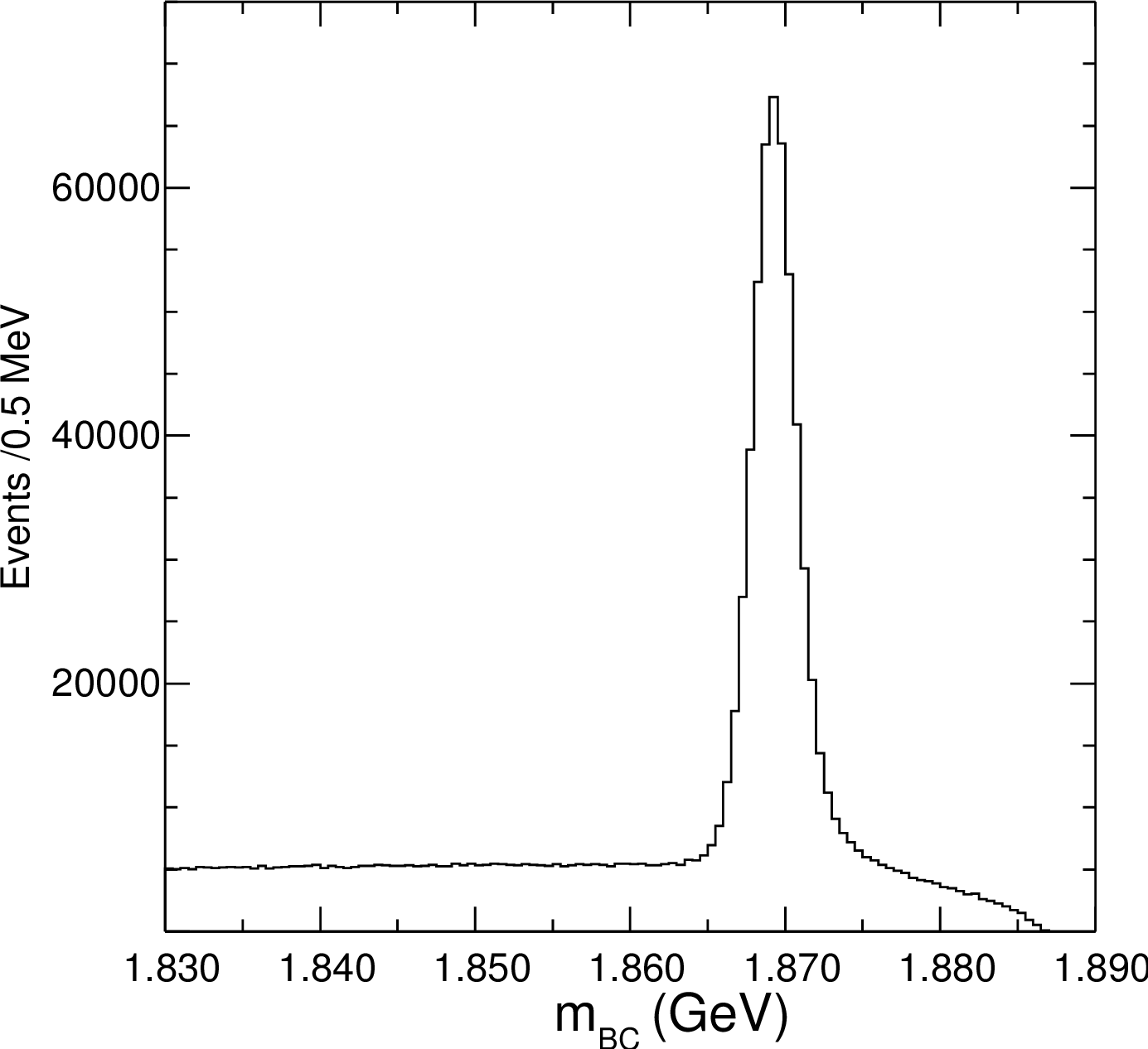,height=5cm}~
\epsfig{file=fig_expPrimters/hadtag_recoil_rw2.eps,height=5cm}~
\end{center}
 \caption{\it Distribution of the energy substituted mass for selected hadronic tag decays
a) for $D$ mesons in \psiprpr\ events at CLEO, and b) for $B$ mesons in \FourS\ decays at \babar.
 }
\label{fig:primers:tags}
\end{figure}

The biggest advantage of the hadronic $B$ tags over the semileptonic $B$ tags 
is the better measurement of the reconstructed $B$ momentum. This permits constraints on the signal decays in the recoil and precise reconstruction of the kinematic variables even in decays with a neutrino or missing neutral Kaon. 
Otherwise the two tags have similar performance. They are completely orthogonal 
samples and thus can be combined .

\subsubsection{Dalitz Plot Analysis}
The partial decay rate of a particle into a multi-body final state  depends on the 
square of a Lorentz invariant matrix element ${\cal M}$. Such matrix element can be independent 
of the specific kinematic configuration of the final state or otherwise  reveal  a non-trivial structure 
in the dynamics of the decay. 
 In the case, for instance, of a three-body decay  $P \to 123$, invariant masses  of pair of particles can be defined as
 $m_{ij}^2 = |p_i + p_j|^2 $  where $p_j$ ($j$=$1$,$2$,$3$) are the four-momenta of the final states particle. A plot of $m_{ij}^2$ versus $m_{ik}^2$ is commonly referred as Dalitz plot \cite{RHDalitz}.

  Dalitz plots distributions  have been used since several decades to study the strong interaction dynamics in particle decays or in scattering experiment. In a three body decay of a meson,    the underlying dynamics can be therefore represented by  intermediate {\it resonances}.  As an example in Fig.\ref{fig:Dalitzplotexample}   a Dalitz plot for the decay $D^0 \to K_S \pi^+ \pi^-$ is shown: there are   several  visible structures  due to competing and interfering resonances. 
  \begin{figure}[!t]
\begin{center}
\includegraphics[ width=.75\textwidth]{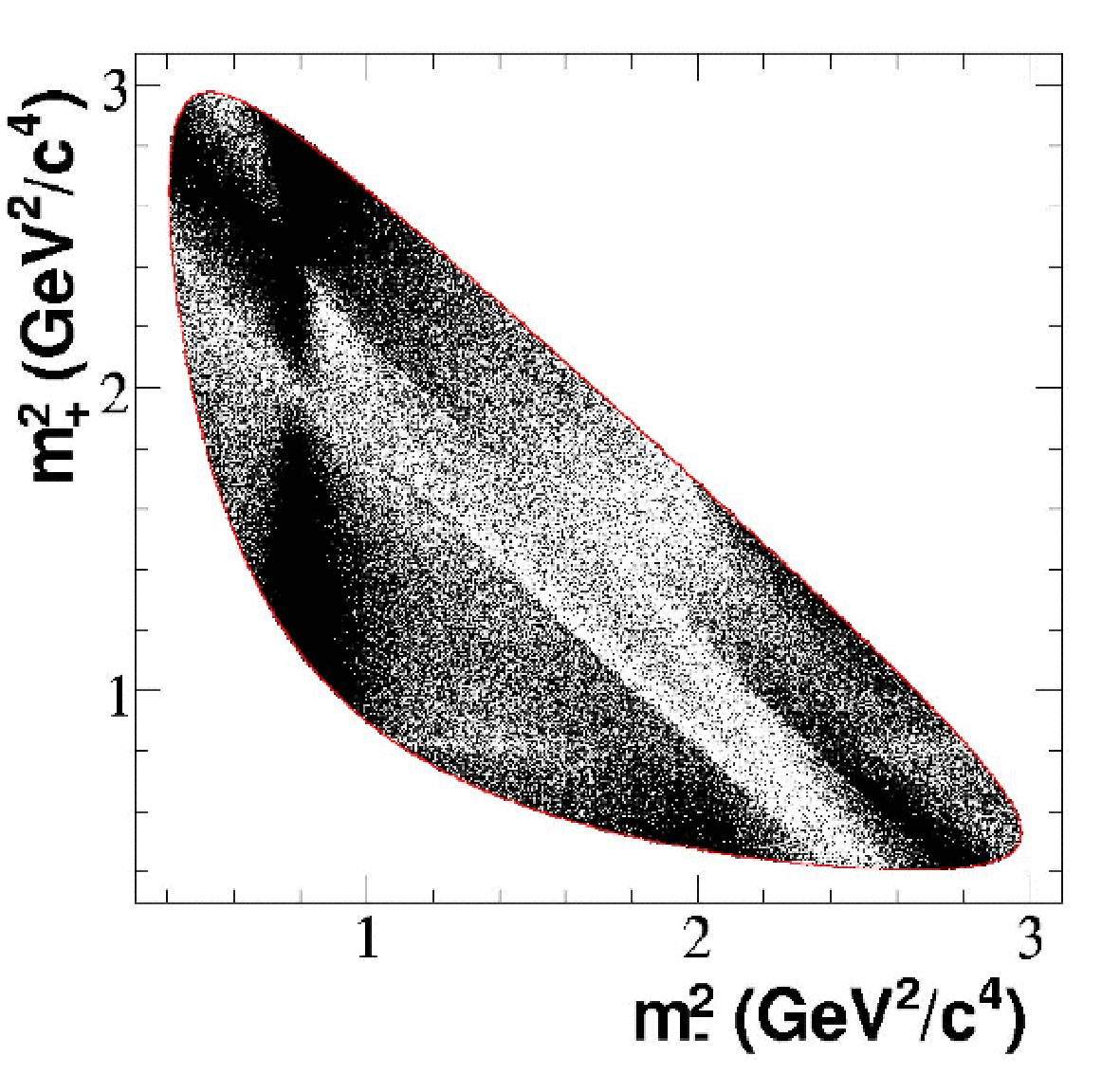}
\end{center}
\caption{Dalitz plot distribution  of a high purity sample of    $D^0 \to K_S \pi^+ \pi^-$, with $m^2_{-} = |p_{K_S}+p_{\pi^-}|^2$ and $m^2_{+} = |p_{K_S}+p_{\pi^+}|^2$ from \cite{Aubert:2008bd}. The most visible  features are described by a $K^{*-}$(892) resonance (vertical band with two lobes) and a $\rho$(770) resonance (diagonal band with two lobes). Interferences between resonances are distorting the distribution.    
The contours (solid red line) represent the kinematic limits of the decay. }
\label{fig:Dalitzplotexample}
\end{figure}

   It is  therefore  a    common practice  to parameterize the matrix element as a coherent sum of two-body amplitudes (subscript $r$) ~\cite{Yao:2006px},

\begin{equation}
{\cal M}   \equiv   \sum_r a_r e^{i \phi_r} {\cal A}_r(m_{13}^2,m_{23}^2) 
\label{eq:genAmp}
\end{equation} 
An additional  constant "non-resonant" term  $a_{\rm NR} e^{i \phi_{\rm NR}}$  is sometimes included. 

The parameters $a_r$  and $\phi_r$  are the magnitude and phase of the  amplitude for the component $r$. 
In the case of a $D^0$ decay the function ${\cal A}_r = F_D \times F_r \times T_r \times W_r$ is a Lorentz-invariant expression where $F_D$ ($F_r$) is the Blatt-Weisskopf centrifugal barrier factor for the $D$  (resonance)  decay vertex~\cite{ref:blatt-weisskopf} 
 $T_r$ is the resonance propagator, and $W_r$ describes the angular distribution in the decay.

For $T_r$  a relativistic Breit-Wigner (BW) parameterization with mass-dependent width is commonly used (for definitions see review in ~\cite{Yao:2006px}). BW mass and width values are usually taken from scattering experiment or world averages  provided by Particle Data Group.

The angular dependence $W_r$ reflects the spin of the resonance and  is described using either Zemach tensors~\cite{Zemach:1963bc,Zemach:1965zz,Filippini:1995yc} where transversality is enforced or the helicity formalism~\cite{ref:jacob-wick,Chung:1993da} when  a 
longitudinal component in the resonance propagator is allowed  (see Ref.~\cite{Yao:2006px} for a comprehensive summary).

   Alternative parameterizations have been used especially to represent spin zero (S-wave) resonances. For this component  the presence of several broad and overlapping  resonances  makes  a  simple BW model  not adequate.For instance,  K-matrix formalism  with the P-vector approximation~\cite{Wigner:1946zz,Aitchison:1972ay} was used for $\pi\pi$ S-wave components.

 In the context of flavor physics  Dalitz model have been used as  effective parameterizations to derive   strong phase  dependence. The knowledge of strong phases is  relevant for analysis where the extraction of weak phases  can be obtained through interferences between different resonances. Moreover, in the case of neutral meson decays the interference between flavor mixing and decay leads  to time-dependent  analyses (either for CP or flavor  mixing measurements). For this reason Dalitz models have been included in such  analyses (that are frequently referred  for short as {\it  time-dependent}  Dalitz analyses).
 
\section{Determination of $\Vud$ and $\Vus$.}
\label{sec:cabibbo}

Unitarity of the bare (unrenormalized) CKM
\cite{Cabibbo:1963yz,Kobayashi:1973fv} $3\times3$ quark mixing
 matrix $V^0_{ij}$, $i=u,c,t$
$j=d,s,b$ implies the orthonormal tree level relations

\begin{equation}
\sum_i V^{0\ast}_{ij} V^0_{ik} = \sum_i V^{0\ast}_{ji} V^0_{ki} =
\delta_{jk} \label{eqone}
\end{equation}

 Standard Model quantum loop effects are important and corrected
for such that Eq.~(\ref{eqone}) continues to hold at the renormalized
level \cite{Marciano:1975cn}. That prescription generally involves
normalization of all charged current semileptonic amplitudes relative
to the Fermi constant

\begin{equation}
G_\mu = 1.166371 (6) \times 10^{-5} {\rm GeV}^{-2} \label{eqtwo}
\end{equation}

 obtained from the precisely measured (recently improved) muon
lifetime \cite{Chitwood:2007pa}

\begin{equation}
\tau_\mu = \Gamma^{-1} (\mu^+\to e^+\nu_e \bar\nu_\mu (\gamma)) =
2.197019 (21)\times 10^{-6} {\rm sec} \label{eqthree}
\end{equation}

 In all processes, Standard Model
SU(3)$_C\times$SU(2)$_L\times$U(1)$_Y$ radiative corrections are
explicitly accounted for \cite{Marciano:1985pd}.

Of particular interest here is the first row constraint

\begin{equation}
|V_{ud}|^2+|V_{us}|^2+|V_{ub}|^2=1 \label{eqfour}
\end{equation}

 An experimental deviation from that prediction would be evidence
for ``new physics'' beyond Standard Model expectations in the form of
tree or loop level contributions to muon decay and/or the semileptonic
processes from which the $V_{ij}$ are extracted. Of course, if
Eq.~(\ref{eqfour}) is respected at a high level of certainty, it implies
useful constraints on various ``new physics'' scenarios.


\subsection{$V_{ud}$ from nuclear decays}

\label{s:Vudnucl}

Nuclear beta decays between $0^+$ states sample only the vector component
of the hadronic weak interaction.  This is important because the
conserved vector current (CVC) hypothesis protects the vector
coupling constant $G_V$ from renormalization by background strong
interactions.  Thus, the $G_V$ that occurs in nuclei should be the same as the
one that operates between free up and down quarks.  In that case, one
can write $G_V = G_F V_{ud}$, which means that a measurement of $G_V$ in nuclei,
when combined with a measurement of  the Fermi constant $G_F$ 
in muon decay, yields the value of the CKM matrix element $V_{ud}$.
To date, precise measurements of the beta decay between isospin
analog states of spin, $J^{\pi} = 0^+$, and isospin, $T = 1$, provide
the most precise value of $V_{ud}$.

A survey of the relevant experimental data has recently been
completed by Hardy and Towner \cite{Hardy:2008gy}.  Compared to the previous
survey \cite{Towner:2005qc} in 2005 there are 27 new publications,
many with unprecedented precision.  In some cases they have improved
the average results by tightening their error assignments and in others by
changing their central values.  Penning-trap measurements of decay energies
have been especially effective in this regard.

For each transition, three experimental quantities have to be
determined:  the decay energy, $Q_{\rm ec}$; the half-life of the
decaying state, $t_{1/2}$; and the branching ratio, $R$, for
the particular transition under study.  The decay energy is used
to calculate the phase space integral, $f$, where it enters as the
fifth power.  Thus, if $f$ is required to have $0.1 \%$ precision then the
decay energy must be known to $0.02 \%$ -- a demand that is currently
being surpassed by Penning-trap devices.  The partial half-life is defined
as $t = t_{1/2}/R$ and the product $ft$ is
\begin{equation}
ft = \frac{K}{G_F^2   V_{ud}^2   \    \langle \tau_+ \rangle^2} ,
\label{ft}
\end{equation}
where $K/(\hbar c)^6 = 2 \pi^3 \hbar \ln 2/(m_e c^2)^5 =
8120.2787(11) \times 10^{-10}$ GeV$^{-4}$ s.  When isospin is an
exact symmetry the initial and final states, being isospin
analogs, are identical except that a proton has switched to a neutron.  
Since the operator describing the transition is simply the isospin ladder
operator, $\tau_+$, its matrix element, $\langle \tau_+ \rangle$, is independent
of nuclear structure and is given by an isospin Clebsch-Gordan coefficient, which for
isospin $T = 1$ states has the value $\sqrt{2}$.  Hence,
\begin{equation}
ft = \frac{K}{2 \, G_F^2  V_{ud}^2} ,
\label{ft1}
\end{equation}
and according to CVC the $ft$ value is a constant independent of
the nucleus under study.  In practice, however, isospin is always a broken
symmetry in nuclei, and beta decay occurs in the presence of
radiative corrections, so a `corrected' $ft$ value is defined by
\begin{equation}
\F t \equiv ft (1 + \delta_R^{\prime}) \left ( 1 -
\left ( \delta_C - \delta_{NS} \right ) \right )
= \frac{K}{2 \, G_F^2  V_{ud}^2 \, (1 + \DRV )} ;
\label{Ft1}
\end{equation}
so it is this corrected $\F t$ that is a constant.  Here the radiative
correction has been separated into three components: (i)   $\DRV $ is a
nucleus-independent part  that includes  the universal short-distance
component $S_{EW}$ affecting all semi-leptonic decays, defined later 
in Eq.~(\ref{eq:SEW}). Being a constant,  $\DRV$  is placed on
the right-hand-side of Eq.~(\ref{Ft1});  (ii)  $\delta_R^{\prime}$ is
transition dependent, but only in a trivial way, since it just depends on the
nuclear charge, $Z$, and the electron energy, $E_e$; while
$\delta_{NS}$ is a \underline{small} nuclear-structure dependent
term that requires a shell-model calculation for its evaluation.
(iii)  Lastly, $\delta_C$ is an isospin-symmetry breaking correction, typically
of order $0.5 \%$, that also requires a shell-model calculation
for its evaluation.

\begin{table}[t]
\caption{Experimental $ft$ values for $0^+ \rightarrow 0^+$ superallowed
Fermi beta decays, the trivial nucleus-dependent component of the
radiative correction, $\delta_R^{\prime}$, the nuclear-structure
dependent isospin-symmetry-breaking and radiative correction taken
together, $\delta_C - \delta_{NS}$, and the corrected $\F t$ values.
The last line gives the average $\F t$ value and the $\chi^2$ of the fit.
\label{t:ftvalues}}
\begin{center}
\begin{tabular}{rrrrr}
& & & & \\[-3mm]
Parent & $ft(s)$ & $\delta_R^{\prime}(\%)$ & $\delta_C - \delta_{NS} (\%)$ &
$\F t (s)$ 
\\[1mm]
\hline
& & & & \\[-3mm]
$^{10}$C  &  $3041.7 \pm 4.3$ & $1.679 \pm 0.004$ & $0.520 \pm 0.039$ &
$3076.7 \pm 4.6$ \\
$^{14}$O  &  $3042.3 \pm 2.7$ & $1.543 \pm 0.008$ & $0.575 \pm 0.056$ &
$3071.5 \pm 3.3$ \\
$^{22}$Mg &  $3052.0 \pm 7.2$ & $1.466 \pm 0.017$ & $0.605 \pm 0.030$ &
$3078.0 \pm 7.4$ \\
$^{26}$Al$^m$&  $3036.9 \pm 0.9$ & $1.478 \pm 0.020$ & $0.305 \pm 0.027$ &
$3072.4 \pm 1.4$ \\
$^{34}$Cl &  $3049.4 \pm 1.2$ & $1.443 \pm 0.032$ & $0.735 \pm 0.048$ &
$3070.6 \pm 2.1$ \\
$^{34}$Ar &  $3052.7 \pm 8.2$ & $1.412 \pm 0.035$ & $0.845 \pm 0.058$ &
$3069.6 \pm 8.5$ \\
$^{38}$K$^m$ &  $3051.9 \pm 1.0$ & $1.440 \pm 0.039$ & $0.755 \pm 0.060$ &
$3072.5 \pm 2.4$ \\
$^{42}$Sc &  $3047.6 \pm 1.4$ & $1.453 \pm 0.047$ & $0.630 \pm 0.059$ &
$3072.4 \pm 2.7$ \\
$^{46}$V  &  $3050.3 \pm 1.0$ & $1.445 \pm 0.054$ & $0.655 \pm 0.063$ &
$3074.1 \pm 2.7$ \\
$^{50}$Mn &  $3048.4 \pm 1.2$ & $1.444 \pm 0.062$ & $0.695 \pm 0.055$ &
$3070.9 \pm 2.8$ \\
$^{54}$Co &  $3050.8 \pm 1.3$ & $1.443 \pm 0.071$ & $0.805 \pm 0.068$ &
$3069.9 \pm 3.3$ \\
$^{62}$Ga & $ 3074.1 \pm 1.5$ & $1.459 \pm 0.087$ & $1.52 \pm 0.21$ &
$3071.5 \pm 7.2$ \\
$^{74}$Rb & $ 3084.9 \pm 7.8$ & $1.50 \pm 0.12$ & $1.71 \pm 0.31$ &
$3078 \pm 13$ \\[4mm]
& & & Average $\overline{\F t}$ & $3072.14 \pm 0.79$ \\
& & & $\chi^2/\nu$ & 0.31 \\
\end{tabular}
\end{center}
\end{table}

In Tab.~\ref{t:ftvalues} are listed the experimental $ft$ values from
the survey of Hardy and Towner \cite{Hardy:2008gy} for 13 transitions, of which 10
have an accuracy at the $0.1 \%$ level, and three at up to the $0.4 \%$ level.
Also listed are the theoretical corrections, $\delta_R^{\prime}$ and
$\delta_C - \delta_{NS}$, taken from Ref.~\cite{Towner:2007np}, and the
corrected $\F t$ values.  This data set is sufficient to provide a
very demanding test of the CVC assertion that the $\F t$ values
should be constant for all nuclear superallowed transitions of this
type.  In Fig.~\ref{fig:vud_towner} the uncorrected $ft$ values in the upper panel 
show considerable scatter, the lowest and
highest points differing by 50 parts in 3000.  This scatter is
completely absent in the corrected $\F t$ values shown in the
lower panel of Fig.~\ref{fig:vud_towner}, an outcome principally
due to the nuclear-structure-dependent corrections, $\delta_C - \delta_{NS}$, 
thus validating the theoretical calculations at the level of current experimental precision. 
The data in Tab.~\ref{t:ftvalues} and Fig.~\ref{fig:vud_towner}
are clearly satisfying the CVC test.
The weighted average of the 13 data is
\begin{equation}
\overline{\F t} = 3072.14 \pm 0.79 ~ {\rm s} ,
\label{Ftavg}
\end{equation}
with a corresponding chi-square per degree of freedom of
$\chi^2 / \nu = 0.31$.   Eq.~(\ref{Ftavg}) confirms
the constancy of $G_V$ -- 
the CVC hypothesis -- at the level of $1.3 \times 10^{-4}$.

Before proceeding to a determination of $V_{ud}$ it has to be noted that
the isospin-symmetry-breaking correction, $\delta_C$, is taken from
Towner and Hardy \cite{Towner:2007np} who calculated proton and neutron
radial functions as eigenfunctions of a Saxon-Woods potential.
An alternative procedure used in the past by
 Ormand and Brown \cite{Ormand:1985zz,Ormand:1989hm,Ormand:1995df}
takes the radial functions as eigenfunctions of a Hartree-Fock
mean-field potential.  The corrections obtained by Ormand and Brown
were consistently smaller than the Saxon-Woods values and this
difference was treated as a systematic error in previous surveys.  In their most recent survey,
though, Hardy and Towner \cite{Hardy:2008gy} repeated the Hartree-Fock calculations, but with
a change in the calculational procedure, and obtained results that were
closer to the Saxon-Woods values.  Even so, when these Hartree-Fock
$\delta_C$ values are used in Eq.~(\ref{Ft1}) the 
$\chi^2$ of the fit  to
$\F t = {\rm constant}$  becomes a factor of three larger.  
This in itself might be sufficient reason to reject the
Hartree-Fock values, but to be safe an average of the Hartree-Fock
and Saxon-Woods $\overline{\F t}$ values was adopted and a systematic
error assigned that is half the spread between the two values.
This leads to
\bea
\overline{\F t} & = & 3071.83 \pm 0.79_{\rm stat} \pm 0.32_{\rm syst}~{\rm s}
\nonumber \\
& = & 3071.83 \pm 0.85 ~{\rm s} .
\label{Ftavg2}
\eea
In the second line the two errors have been combined in quadrature.

Recently, Miller and Schwenk \cite{Miller:2008my} have explored the formally
complete approach to isospin-symmetry breaking, but produced no numerical
results.  The Towner-Hardy \cite{Hardy:2008gy} values quoted here are
based on
a model whose approximations can be tested for  $A = 10$ by comparing with
the
large no-core shell-model calculation of  Caurier {\it et al}
\cite{Caurier:2002hb},
which is as close to an exact calculation as is currently possible.
The agreement between  the two  suggests that any further systematic error
in the
isospin-breaking correction is likely to be small.

The CKM matrix element $V_{ud}$ is then obtained from
\begin{equation}
V_{ud}^2 = \frac{K}{2 G_F^2 (1 + \DRV ) \overline{\F t}} ,
\label{Vud2}
\end{equation}
where $\DRV$ is the nucleus-independent radiative correction taken from Marciano and
Sirlin \cite{Marciano:2005ec}: {\it viz.}
\begin{equation}
\DRV = (2.631 \pm 0.038) \% .
\label{DRV}
\end{equation}
With $\overline{\F t}$ obtained from Eq.~(\ref{Ftavg2}), the value of $V_{ud}$ becomes
\begin{equation}
V_{ud} = 0.97425 \pm 0.00022 .
\label{Vud}
\end{equation}
Compared to the Hardy-Towner survey \cite{Towner:2005qc} of 2005, which
obtained $V_{ud} = 0.97380(40)$, the central value has
shifted by about one standard deviation primarily as a result
of Penning-trap decay-energy measurements and a reevaluation of
the isospin-symmetry breaking correction in 2007 \cite{Towner:2007np}.
The error is dominated by theoretical uncertainties; experiment only
contributes 0.00008 to the error budget.
Currently the largest contribution
to the error budget comes from the nucleus-independent radiative
correction $\DRV $ -- 
recently reduced by a factor of two by Marciano and Sirlin \cite{Marciano:2005ec}.
Further improvements here will need some
theoretical breakthroughs.  Second in order of significance
are the nuclear-structure-dependent corrections $\delta_C$ and
$\delta_{NS}$.  So long as $0^+ \rightarrow 0^+$ nuclear decays
provide the best access to $V_{ud}$, these corrections will
need to be tested and honed.  Here is where nuclear
experiments will continue to play a critical role.

\begin{figure}[thb]
\begin{center}
\resizebox{0.7\textwidth}{!}{\includegraphics{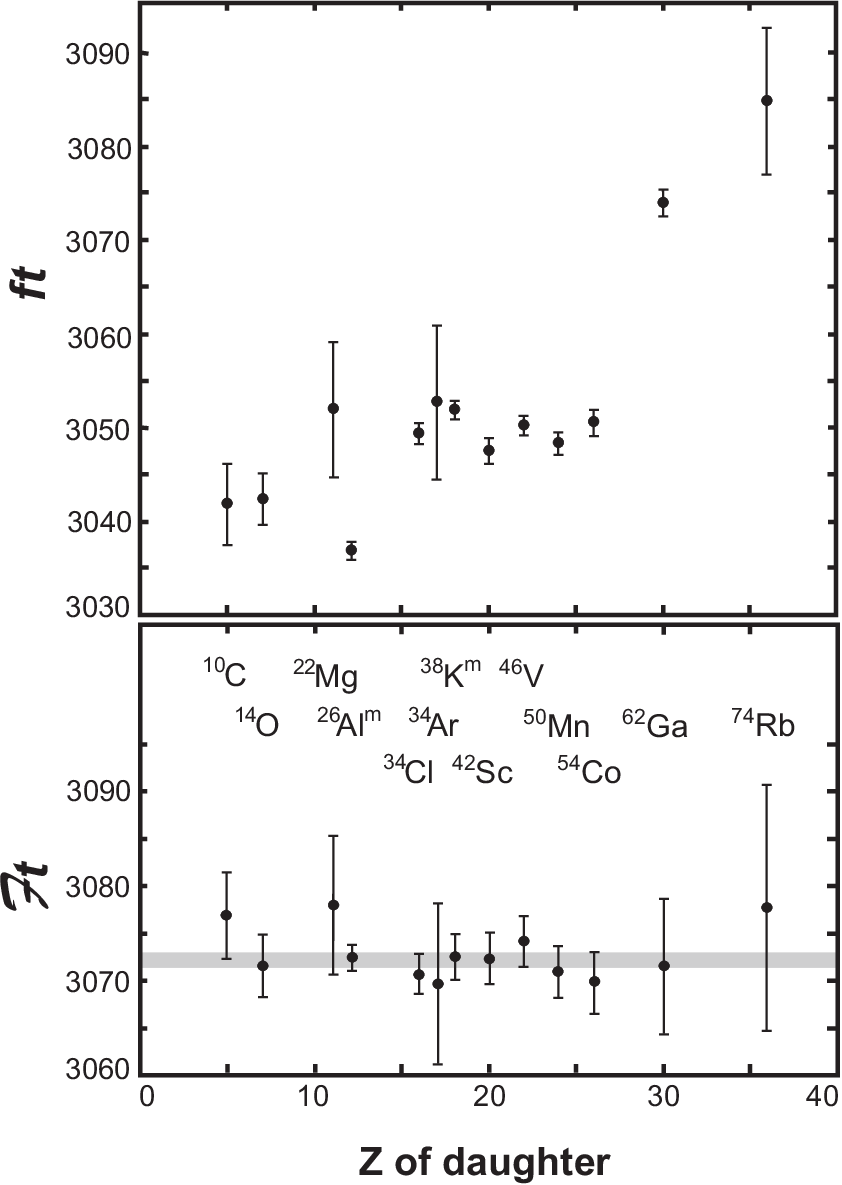}}
\end{center}
\caption{\label{fig:vud_towner}
In the top panel are plotted the uncorrected experimental $ft$
values as a function of the charge on the daughter nucleus.
In the bottom panel, the corresponding $\F t$ values as defined in  
Eq.~(\ref{Ft1}) are given.  The horizontal grey band in the
bottom panel gives one standard deviation around the average
$\overline{\F t}$.}
\end{figure}



\subsection{$V_{ud}$ from neutron  decay}
\label{s:Vudneutron}

Although the result is not yet competitive, to extract $V_{ud}$ from neutron
$\beta$-decay is appealing because it does not
require the application of corrections for isospin-symmetry breaking
effects, $\delta_C$, or nuclear-structure effects, $\delta_{NS}$, as
defined in the previous section on nuclear $\beta$-decay.  However,
it should be noted that the transition-dependent radiative correction,
$\delta'_R$, and the nucleus-independent radiative correction,
$\DRV$, must still be applied to neutron $\beta$-decay
observables; and the latter is, in fact, the largest contributor to the uncertainty
in the nuclear value for $V_{ud}$.

In contrast to nuclear $\beta$-decays between $0^+$ states, which
sample only the weak vector interaction, neutron $\beta$-decay
proceeds via a mixture of the weak vector and axial-vector
interactions.  Consequently, three parameters are required for a
description of neutron $\beta$-decay: $G_F$, the fundamental weak
interaction constant; $\lambda \equiv g_A/g_V$, the ratio of the weak
axial-vector and vector coupling constants; and the parameter of
interest, $V_{ud}$.  Thus, measurements of at least two observables
(treating $G_F$ as an input parameter) are required for a
determination of $V_{ud}$.

A value for $\lambda$ can be extracted from measurements of
correlation coefficients in polarized neutron $\beta$-decay.  Assuming
time-reversal invariance, the differential decay rate distribution of
the electron and neutrino momenta and the electron energy for
polarized $\beta$-decay is of the form \cite{Jackson:1957zz}
\begin{equation}
\frac{dW}{dE_{e}d\Omega_{e}d\Omega_{\nu}}
\propto p_{e}E_{e}(E_{0} - E_{e})^{2}
\left[1 + a \frac{\vec{p}_{e} \cdot \vec{p}_{\nu}}
{E_{e}E_{\nu}} + \langle\vec{\sigma}_{n}\rangle \cdot
\left( A\frac{\vec{p}_{e}}{E_{e}} + B\frac{\vec{p}_{\nu}}{E_{\nu}}
\right) \right],
\label{eq:neutron-correlations}
\end{equation}
where $E_{e}$ ($E_{\nu}$) and $\vec{p}_{e}$ ($\vec{p}_{\nu}$) denote,
respectively, the electron (neutrino) energy and momentum; $E_{0}$
($=782$ keV + $m_e$) denotes the $\beta$-decay endpoint energy, with
$m_e$ the electron mass; and $\langle \vec{\sigma}_n \rangle$ denotes
the neutron polarization.  Neglecting recoil-order corrections, the
correlation coefficients $a$ (the $e$-$\overline{\nu}_e$-asymmetry),
$A$ (the $\beta$-asymmetry), and $B$ (the
$\overline{\nu}_e$-asymmetry) can be expressed in terms of $\lambda$
as \cite{Wilkinson:1982hu,Gardner:2000nk}
\begin{equation}
a = \frac{1 - \lambda^2}{1 + 3\lambda^2},~~~~~
A = -2\frac{\lambda^2 + \lambda}{1 + 3\lambda^2},~~~~~
B = 2\frac{\lambda^2 - \lambda}{1 + 3\lambda^2}.
\label{eq:correlations-lambda}
\end{equation}
At present, these correlation parameters have values $a = -0.103 \pm
0.004$, $A = -0.1173 \pm 0.0013$, and $B = 0.983 \pm 0.004$
\cite{Amsler:2008zzb}.  Although $B$ has been measured to the highest precision
(0.41\%), the sensitivity of $B$ to $\lambda$ is a factor $\sim 10$
less than that of $a$ and $A$.  Thus, the neutron $\beta$-asymmetry
$A$ yields the most precise result for $\lambda$.

A second observable is the neutron lifetime, $\tau_n$, which can be
written in terms of the above parameters as
\cite{Marciano:2005ec,Czarnecki:2004cw,Abele:2008zz}
\begin{equation}
\frac{1}{\tau_n} = \frac{G_F^2 m_e^5}{2\pi^3} |V_{ud}|^2
(1 + 3\lambda^2) f (1 + \mathrm{RC}).
\label{eq:taun}
\end{equation}
Here, $f = 1.6887 \pm 0.00015$ is a phase space factor, which includes
the Fermi function contribution \cite{Wilkinson:1982hu}, and $(1 +
\mathrm{RC}) = 1.03886 \pm 0.00039$ denotes the total effect of all
electroweak radiative corrections
\cite{Marciano:2005ec,Czarnecki:2004cw}.  After insertion of the
numerical factors in Eq.~(\ref{eq:taun}), a value for $V_{ud}$ can be
determined from $\tau_n$ and $\lambda$ according to
\cite{Marciano:2005ec,Czarnecki:2004cw}
\begin{equation}
|V_{ud}|^2 = \frac{4908.7 \pm 1.9~\mathrm{s}}{\tau_n (1 + 3\lambda^2)}.
\label{eq:taun-numerical}
\end{equation}

The current status of a neutron-sector result for $V_{ud}$ is
summarized in Fig.\ \ref{fig:vud_neutron}, where $|\lambda|$ is
plotted on the horizontal axis, and $V_{ud}$ on the vertical axis.  At
present, the Particle Data Group \cite{Amsler:2008zzb} averages the
four most recent measurements of the neutron $\beta$-asymmetry, $A$,
performed with beams of polarized cold neutrons
\cite{Bopp:1986rt,Erozolimsky:1997wi,Liaud:1997vu,Abele:2002wc}, and
one combined measurement of $A$ and $B$ \cite{Mostovoi:2001pv}, to
obtain their recommended value of $\lambda = -1.2695 \pm 0.0029$
(shown as the vertical error band).  It should be noted that the error
on the PDG average for $\lambda$ (0.23\%) is greater than that of the
most precise individual result (0.15\%) \cite{Abele:2002wc}, because
the error on the average has been increased by a $\sqrt{\chi^2/(N-1)}$
scale factor of 2.0 to account for the spread among the individual
data points.  Constraints between the values for $V_{ud}$ and
$\lambda$, computed according to Eq.~(\ref{eq:taun-numerical}) for two
different values for the neutron lifetime, are shown as the angled
error bands.  The band labeled ``PDG 2008'' represents the PDG's
recommended value for $\tau_n = 885.7 \pm 0.8$~s, whereas the other
band relies solely on the most recent result reported for $\tau_n$ of
$878.5 \pm 0.7 \pm 0.3$~s \cite{Serebrov:2004zf}, which disagrees by
$6\sigma$ with the PDG average.  Note that the PDG deliberately chose
not to include this discrepant result in their most recent averaging
procedure.

\begin{figure}[thb]
\begin{center}
\resizebox{0.7\textwidth}{!}{\includegraphics{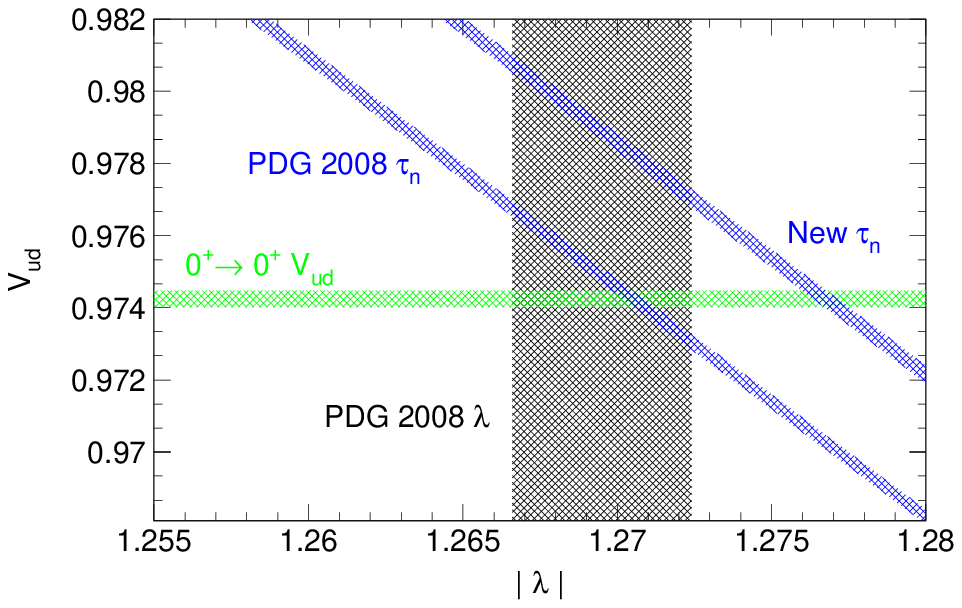}}
\end{center}
\caption{Current status of $V_{ud}$ from neutron $\beta$-decay.  The
vertical error band indicates the current PDG error on $\lambda$.  The
angled error bands show the constraints between $V_{ud}$ and $\lambda$
for two values of the neutron lifetime: the PDG recommended value, and
that from a recent $6\sigma$-discrepant result.  For comparison, the
horizontal error band denotes the value of $V_{ud}$ from $0^+$ nuclear
$\beta$-decays discussed in the previous section.}
\label{fig:vud_neutron}
\end{figure}

The intersection of the error band for $\lambda$ with the error band
defined by the neutron lifetime determines the value for $V_{ud}$.
Assuming the PDG value of $\tau_n = 885.7 \pm 0.8$~s yields
\cite{Amsler:2008zzb}
\begin{equation}
V_{ud} = 0.9746 \pm 0.0004_{\tau_n} \pm 0.0018_{\lambda}
\pm 0.0002_{\mathrm{RC}},
\end{equation}
where the subscripts denote the error sources.  If the discrepant
neutron lifetime result of $878.5 \pm 0.7 \pm 0.3$~s were employed
instead, it would suggest a considerably larger value, $V_{ud} =
0.9786 \pm 0.0004_{\tau_n} \pm 0.0018_{\lambda} \pm
0.0002_{\mathrm{RC}}$.  For comparison, the value for $V_{ud}$ from
nuclear $\beta$-decay discussed in the previous section is shown as
the horizontal band.  The neutron $\beta$-decay result derived from
the PDG's recommended values for $\tau_n$ and $\lambda$ is seen to be
in excellent agreement with that from nuclear $\beta$-decay, albeit
with an error bar that is a factor $\sim 7$--8 larger.

An ongoing series of precision measurements of neutron $\beta$-decay
observables aims to reduce the error on $\lambda$ and resolve the
lifetime discrepancy.  The goal of two currently running experiments,
the PERKEO III experiment at the Institut Laue-Langevin
\cite{Abele:2008zz} (using a beam of cold neutrons) and the UCNA
experiment at Los Alamos National Laboratory \cite{Pattie:2008da}
(using stored ultracold neutrons), are sub-0.5\% measurements of the
neutron $\beta$-asymmetry, $A$.
Since these two experiments employ different experimental approaches,
they are sensitive to different systematic uncertainties. The
combination of their results will reduce the $\lambda$-induced
uncertainty for $V_{ud}$ by up to a factor of $\sim 3$.

Finally, although the error on $\tau_n$ is not the dominant
uncertainty, the $6\sigma$ discrepancy between the PDG average and the
most recent result is clearly unsatisfactory.  Indeed, multiple groups
are now attempting to measure $\tau_n$ to a level of precision ranging
between 1~s and 0.1~s.
Hence, the next round of experiments should reach sufficient
precision to definitively discriminate between the PDG average and the
recent discrepant result.

\subsection{$V_{ud}$ from pionic beta decay} 
\label{sec:pibeta}

$V_{ud}$ can also be obtained from the pion beta decay, $\pi^+
\rightarrow \pi^0 e^+ \nu_e [\gamma]$, which is a pure vector
transition between two spin-zero members of an isospin triplet and is
therefore analogous to the superallowed nuclear decays.  Like neutron
decay, it has the advantage that there are no nuclear-structure
dependent corrections to be applied.  Its major disadvantage, however,
is that it is a very weak branch, $ \mathcal{O}(10^{-8})$, in the
decay of the pion.  The corresponding decay width can be decomposed as
\begin{equation}
\label{pie3}
\Gamma_{\pi_{e 3}} =
\frac{G_F^2 M_{\pi^\pm}^5}{64 \pi^3} 
S_{\rm EW}\, \Big| V_{ud} f_+(0) \Big|^2\,
I_0^{\pi \pi}  \,\left(1 + 
\delta_{\rm EM}
\right)~.
\end{equation}
In the above equation $S_{EW}$ represents the universal short-distance
electroweak correction (Eq.~\ref{eq:SEW}), $f_+(0)$ is the vector
form-factor at zero momentum transfer, $I_0^{\pi \pi}$ the phase space
factor, and $\delta_{\rm EM} $ the long-distance electromagnetic
correction.  As far as the strong interaction is concerned,
the Ademollo-Gatto theorem \cite{Ademollo:1964sr} requires the
deviation of $f_+(0)$ from its value $1$ in the isospin limit to be
quadratic in the quark mass difference $m_d - m_u$. This results in a
very tiny correction $f_+(0) -1 = - 7 \times 10^{-6}$ at one-loop
\cite{Cirigliano:2002ng} and leads to the expectation that higher
order strong interaction corrections will not disturb this nice
picture. The corrections in (\ref{pie3}) are therefore dominated by
electromagnetic contributions.  The long-distance electromagnetic
corrections can be separated into a shift to the phase space integral
$\delta I^{\pi \pi} / I_0^{\pi \pi} = 1.09 \times 10^{-3}$ as well as
a structure dependent term \cite{Cirigliano:2002ng}
\begin{eqnarray}
\label{fpem}
\frac{1}{2} \, \cdot \, \delta_{\rm EM} \big|_{\rm str. dep.}  
&=&
- 4 \pi \alpha \Bigg\{ \frac{2}{3} X_1 + \frac{1}{2} X_6^{\rm phys}(\mu)
+\frac{1}{32 \pi^2} \bigg(3 + \log \frac{m_e^2}{M_{\pi^\pm}^2}
+3 \log \frac{M_{\pi^\pm}^2}{\mu^2} \bigg) \Bigg\} 
\no
\\
&=& (5.11 \pm 0.25) \times 10^{-3}~,
\end{eqnarray}
where we have used the recent results of \cite{DescotesGenon:2005pw}
for the electromagnetic coupling constants $X_{1,6}$ entering in
(\ref{fpem}) (with a fractional uncertainty of $100 \%$) to update the
numerical result of Ref.~\cite{Cirigliano:2002ng}.  Higher order
corrections are expected to be strongly suppressed by $\sim (M_\pi / 4
\pi f_\pi)^2$.  Combining the updated theory with the branching
fraction $BR( \pi^+ \rightarrow \pi^0 e^+\nu_e [\gamma]) = (1.040 \pm
0.004 ({\rm stat} \pm 0.004 ({\rm syst}) ) \times 10^{-8}$ from the
PIBETA experiment~\cite{Pocanic:2003pf}, we find:
\begin{equation}
V_{ud} = 0.9741(2)_{\rm th} (26)_{\rm exp}.
\end{equation}
$V_{ud}$  from pion beta decay is in agreement with the more 
precise result, Eq.~(\ref{Vud}), from nuclear decays. 
A  tenfold improvement on the experimental measurement would 
be needed to make this extraction competitive with nuclear decays.  


\subsection{Determination of $\Vus$ from $K_{\ell 2}$ and $K_{\ell 3}$} 
Here we discuss the determination of $\Vus$
from the combination of leptonic pion and Kaon decay and from
 semileptonic Kaon decay.
We start with the status of the theoretical description of leptonic
pion and Kaon decays and of semileptonic Kaon decays within the SM, and the report
 on the status of the experimental results, particularly
for the semileptonic decay. 


\subsubsection{$P_{\ell 2}$    ($P = \pi, K$)  rates within the SM}
\label{sect:Pl2}
Including all known short- and long-distance electroweak corrections,
and parameterizing the hadronic effects in terms of a few
dimensionless coefficients, the inclusive $P \to \ell \bar{\nu}_\ell
(\gamma)$ decay rate can be written as
\cite{Marciano:1993sh,Cirigliano:2007ga}

\begin{eqnarray}
\Gamma_{P_{\ell 2 (\gamma)}}
\!
&=& 
\Gamma^{(0)}_{P_{\ell 2}} 
S_{\rm EW}
\Bigg\{
1 + \frac{\alpha}{\pi}  \,  F (m_\ell^2/M_P^2) 
\Bigg\}
\Bigg\{  1 - \frac{\alpha}{\pi}  
\bigg[
\frac{3}{2}  \log \frac{M_\rho}{M_P}  
 + 
 c_1^{(P)}  
\nonumber \\
& &
{} \qquad \qquad
+ 
\frac{m_\ell^2}{M_\rho^2}  
\bigg( c_2^{(P)}  \, \log \frac{M_\rho^2}{m_\ell^2}  
+  c_3^{(P)} 
+ c_4^{(P)} (m_\ell/M_P) \bigg) 
\nonumber \\
& &
{} \qquad \qquad
-  \frac{M_P^2}{M_\rho^2} \,  \tilde{c}_{2}^{(P)}  \, 
\log \frac{M_\rho^2}{m_\ell^2} 
\bigg]
\Bigg\}~, 
\end{eqnarray}
where the decay rate in the absence of radiative corrections is given
by
\begin{equation}
\Gamma^{(0)}_{P_{\ell 2}} 
=  
\frac{G_F^2 |V_P|^2  f_P^2 }{4 
\pi} \, 
M_P  \, m_\ell^2  \, \left(1 - \frac{m_\ell^2}{M_P^2} \right)^2 ,
\qquad
V_\pi = V_{ud}, \quad V_K = V_{us}~.
\end{equation}
The factor $S_{\rm EW}$ describes the short-distance electromagnetic
correction \cite{Sirlin:1977sv,Sirlin:1981ie} which is universal for
all semileptonic processes. To leading order it is given by
\begin{equation}
S_{\rm EW} = 1 + \frac{2 \alpha}{\pi} \log \frac{M_Z}{M_\rho}~.
\label{eq:SEW}
\end{equation}

Including also the leading QCD corrections \cite{Marciano:1993sh}, it
assumes the numerical value $S_{\rm EW} = 1.0232$. The first term in
curly brackets is the universal long-distance correction for a
point-like meson. The explicit form of the one-loop function $F(x)$
can be found in \cite{Marciano:1993sh}. The structure dependent
coefficients $c_1^{(P)}$ are independent of the lepton mass $m_{\ell}$
and start at order $e^2 p^2$ in chiral perturbation theory. The other
coefficients appear only at higher orders in the chiral expansion. The
one-loop result (order $e^2 p^2$) for $c_1^{(P)}$ is given by
\cite{Knecht:1999ag},
\begin{eqnarray}
c_1^{(\pi)}
&=&
  - 4 \pi^2 E^r(M_\rho) - \frac{1}{2} + 
\frac{Z}{4} \bigg(3 +2 \log \frac{M_\pi^2}{M_\rho^2} + \log 
\frac{M_K^2}{M_\rho^2} \bigg)~,
\\
c_1^{(K)}
&=&
  - 4 \pi^2 E^r(M_\rho) - \frac{1}{2} + 
\frac{Z}{4} \Big(3 +2 \log \frac{M_K^2}{M_\rho^2} + \log 
\frac{M_\pi^2}{M_\rho^2} \Big)~,
\end{eqnarray}
where the electromagnetic low-energy coupling $Z$ arising at order $e^2 
p^0$ can be expressed through the pion mass difference by the relation 
\begin{equation}
M_{\pi^\pm}^2 - M_{\pi^0}^2 = 8 \pi \alpha Z f_\pi^2 + \ldots~.
\end{equation}
The quantity $E^r(M_\rho)$, being a certain linear combination of 
$e^2 p^2$ low-energy couplings \cite{Knecht:1999ag}, cancels in the ratio 
$\Gamma_{K_{\ell 2 (\gamma)}} / \Gamma_{\pi_{\ell 2 (\gamma)}}$.
As suggested by Marciano \cite{Marciano:2004uf}, a determination of 
$|V_{us} / V_{ud}|$ can be obtained by
combining the experimental values for the decay rates with the  
lattice determination of $f_K / f_\pi$ via
\begin{equation}\label{eq:leprat}
\frac{|V_{us}| f_K}{|V_{ud}| f_\pi} = 0.23872(30) 
\Bigg( \frac{\Gamma_{K_{\ell 2 (\gamma)}}} 
{\Gamma_{\pi_{\ell 2 (\gamma)}}}
\Bigg)^{1/2}~.
\end{equation}
The small error is an estimate of unknown electromagnetic contributions 
arising at order $e^2 p^4$.

In the standard model, the ratios $R^{(P)}_{e/\mu} = 
\Gamma_{P \to e \bar{\nu}_e (\gamma)} /
\Gamma_{P \to \mu \bar{\nu}_\mu (\gamma)}$ 
are helicity suppressed as a consequence of the $V-A$ structure of the 
charged currents, constituting sensitive probes of new physics. 
In a first systematic calculation to order $e^2 p^4$, the radiative 
corrections to $R^{(P)}_{e/\mu}$ have been 
obtained with an unprecedented theoretical accuracy 
\cite{Cirigliano:2007ga,Cirigliano:2007xi}.
The two-loop effective theory results were complemented with a matching 
calculation of an associated counterterm, giving 
\begin{equation}
R^{(\pi)}_{e/\mu} = (1.2352 \pm 0.0001) \times 10^{-4}~, \qquad
R^{(K)}_{e/\mu} = (2.477 \pm 0.001) \times 10^{-5}~.
\end{equation}
The central value of 
$R^{(\pi)}_{e/\mu}$
agrees with the results of a previous calculations 
\cite{Marciano:1993sh,Finkemeier:1995gi}, pushing the theoretical 
uncertainty below the $0.1$ per mille level. The discrepancy with a 
previous determination of 
$R^{(K)}_{e/\mu}$ can be traced back to inconsistencies in the analysis of 
\cite{Finkemeier:1995gi}.

\subsubsection{$K_{\ell3}$ rates within the SM}
\label{sect:Kl3}
The photon-inclusive $K_{\ell3}$ 
decay rates are conveniently decomposed as~\cite{Amsler:2008zzb}
\begin{equation}
\label{eq:Mkl3}
\Gamma_{K_{\ell 3(\gamma)}} =
{ G_F^2 M_K^5 \over 192 \pi^3} C_K^2
  S_{\rm EW}\, \Big| V_{us} f_+^{K^0 \pi^-}(0) \Big|^2\,
I_K^\ell(\lambda_{+,0})\,\left(1 + 
\delta^{K \ell}_{\rm EM}
+
\delta^{K \pi}_{\rm SU(2)}
\right)~,
\end{equation}
where $C_{K}^2=1$ ($1/2$) for the neutral (charged) Kaon decays,
$S_{EW}$ is  the short distance electroweak correction,  
$f_{+}^{K^0 \pi^-} (0)$ is the $K \to \pi$ vector form factor at zero 
momentum transfer, and 
$I_K^\ell(\lambda_{+,0})$ is the phase space integral which  
depends on the (experimentally accessible) slopes of the 
form factors (generically denoted by  $\lambda_{+,\,0}$).
Finally,  $\delta^{K \ell}_{\rm EM}$  represent channel-dependent 
long distance radiative corrections and 
$\delta^{K \pi}_{\rm SU(2)}$ is a correction induced by strong isospin breaking.

\subsubsection*{Electromagnetic effects in $K_{\ell3}$ decays}

\label{sect:KlSM}
The results of the most recent calculation \cite{Cirigliano:2008wn} of the 
four channel-dependent long-distance electromagnetic corrections   
$\delta^{K \ell}_{\rm EM}$ are shown in Tab.~\ref{tab:Kl3radcorr}.
The values given here were obtained to leading nontrivial order in  
chiral effective theory, working with a fully inclusive prescription of 
real photon emission. For the electromagnetic low-energy couplings 
appearing 
in the structure dependent contributions, the recent determinations 
of \cite{Ananthanarayan:2004qk,DescotesGenon:2005pw} were employed.
\begin{table}[ht]
\setlength{\tabcolsep}{3.8pt}
\caption{Summary of the electromagnetic corrections
to the fully-inclusive 
$K_{\ell 3(\gamma)}$ rate~\cite{Cirigliano:2008wn}.} 
\label{tab:Kl3radcorr}
\centering
\begin{tabular}{|c|c|}
\hline
 & $\delta^{K \ell}_{\rm EM}(\%) $   \\
\hline 
$K^{0}_{e 3}$   &  0.99 $\pm$ 0.22 \\
$K^{\pm}_{e3}$    &  0.10 $\pm$ 0.25 \\
$K^{0}_{\mu 3}$ &  1.40 $\pm$ 0.22 \\
$K^{\pm}_{\mu 3}$ &  0.016 $\pm$ 0.25 \\
\hline
\end{tabular}
\end{table}
The errors in Tab.~\ref{tab:Kl3radcorr}
are estimates of (only partially 
known)  higher order contributions. The associated correlation matrix was 
found~\cite{Cirigliano:2008wn}
\be
\left(
\begin{array}{cccc}
  1.0  &  0.081  &  0.685   & -0.147 \\
      &   1.0  & -0.147   &  0.764 \\
      &        &    1.0 &  0.081 \\
      &        &        &  1.0
\end{array}
\right)~.
\end{equation}
It is also useful to record the uncertainties on the linear combinations 
of 
$\delta^{K \ell}_{\rm EM}$ that are relevant for lepton universality and 
strong isospin-breaking tests \cite{Cirigliano:2008wn}:
\begin{eqnarray}
\delta_{\rm EM}^{K^0 e}
- 
\delta_{\rm EM}^{K^0 \mu}
&=& 
(-0.41 \pm 0.17) \% \\
\delta_{\rm EM}^{K^\pm e}
- 
\delta_{\rm EM}^{K^\pm \mu}
&=& 
(0.08 \pm 0.17) \% \\
\delta_{\rm EM}^{K^\pm e}
- 
\delta_{\rm EM}^{K^0 e}
&=& 
(-0.89 \pm 0.32) \% \\
\delta_{\rm EM}^{K^\pm \mu}
- 
\delta_{\rm EM}^{K^0 \mu}
&=& 
(-1.38 \pm 0.32) \% ~.
\end{eqnarray}

The corresponding electromagnetic corrections to the Dalitz plot densities 
can also be found in \cite{Cirigliano:2008wn}. It is important to notice that the 
corrections to the Dalitz distributions can be locally large (up to 
$ \sim 10 \%$) with considerable cancellations in the integrated 
electromagnetic corrections. 

\subsubsection*{Isospin breaking correction in $K_{\ell3}$ decays}
\label{sect:isobreak}

In (\ref{eq:Mkl3}), the same form factor $f_+^{K^0 \pi^-} (0)$ 
(at zero-momentum transfer) is pulled out for all decay channels, where
\begin{equation}
\delta_{\rm SU(2)}^{K^0 \pi^-} = 0~, \quad
\delta_{\rm SU(2)}^{K^\pm \pi^0} = 
\Bigg( 
\frac
{f_+^{K^\pm \pi^0} (0)}
{f_+^{K^0 \pi^-} (0)}
\Bigg)^2 -1~.
\end{equation}
Note that the form factors denote the pure QCD quantities plus the 
electromagnetic contributions to the meson masses and to $\pi^0$-$\eta$ 
mixing. The isospin breaking parameter $\delta_{\rm SU(2)}^{K^\pm \pi^0}$ 
is related to the $\pi^0$-$\eta$ mixing angle via \cite{Cirigliano:2001mk}  
\begin{equation}
\label{isobreak}
\delta_{\rm SU(2)}^{K^\pm \pi^0} = 
2 \sqrt{3} 
\Big( \varepsilon^{(2)} + \varepsilon^{(4)}_{\rm S} + 
\varepsilon^{(4)}_{\rm EM} + 
\ldots \Big) 
\end{equation}
The dominant lowest-order contribution can be expressed in terms of
quark masses \cite{Gasser:1984ux}:
\begin{equation}
\varepsilon^{(2)} = \frac{\sqrt{3}}{4} \frac{m_d - m_u}{m_s - 
\widehat{m}}~,
\qquad
\widehat{m} = \frac{m_u + m_d}{2}~.
\end{equation}
The explicit form of the 
strong and electromagnetic  
higher-order corrections in Eq.~(\ref{isobreak}) can
 be found in~\cite{Cirigliano:2001mk}.  
The required determination of the quark mass ratio
\begin{equation}
R = \frac{m_s - \widehat{m}}{m_d - m_u}
\end{equation}
uses the fact that the double ratio
\begin{equation}
Q^2 = 
\frac{m_s^2 - \widehat{m}^2}{m_d^2 - m_u^2} =
R \,
\frac{m_s / \widehat{m} +1}{2} 
\end{equation}
can be expressed in terms of pseudoscalar masses and a purely 
electromagnetic contribution \cite{Gasser:1984ux}:
\begin{equation}
Q^2 = \frac{\Delta_{K \pi} M_K^2 \big(1 + \cO(m_q^2)\big)}{M_\pi^2 
\big[ \Delta_{K^0 K^+} + \Delta_{\pi^+ \pi^0}
- 
( \Delta_{K^0 K^+} + \Delta_{\pi^+ \pi^0})_{\rm EM}
\big]}~,
\quad
\Delta_{PQ} = M_P^2 -M_Q^2~. 
\end{equation}
Due to Dashen's theorem \cite{Dashen:1969eg}, the electromagnetic term
vanishes to lowest order $e^2 p^0$. At next-to-leading order it is given 
by \cite{Urech:1994hd,Neufeld:1994eg}
\begin{eqnarray} 
(\Delta_{K^0 K^+} + \Delta_{\pi^+ \pi^0})_{\rm EM} &=& 
e^2 M_K^2 \Bigg[ \frac{1}{4 \pi^2} \bigg(3 \ln \frac{M_K^2}{\mu^2}
- 4 + 2 \ln \frac{M_K^2}{\mu^2} \bigg) 
+ \frac{4}{3} (K_5 + K_6)^r (\mu) 
\no \\
& & {} 
- 8 (K_{10} + K_{11})^r (\mu) 
+ 16 Z L_5^r(\mu) \Bigg] + \cO(e^2 M_\pi^2)~.
\label{devDash}
\end{eqnarray}
Based on their estimates for the electromagnetic low-energy 
couplings entering in (\ref{devDash}), Ananthanarayan and Moussallam 
\cite{Ananthanarayan:2004qk} found a rather large deviation from Dashen's limit,
$
(\Delta_{K^0 K^+} + \Delta_{\pi^+ \pi^0})_{\rm EM} =
-1.5 \, \Delta_{\pi^+ \pi^0}~, 
$
which corresponds to \cite{Kastner:2008ch}
$Q = 20.7 \pm 1.2$ (the error accounts for the uncertainty due to higher order 
corrections). Such a small value for $Q$ (compared to $Q_{\rm Dashen} = 
24.2$) is also supported \cite{Donoghue:1996zn,Bijnens:1996kk,Amoros:2001cp}
by previous studies\footnote{Note however that a recent analysis of 
$\eta \to 3 \pi$ at the two-loop level \cite{Bijnens:2007pr} favors the
 value $Q = 23.2$.}.
Together with \cite{Kastner:2008ch} $m_s / \widehat{m} = 24.7 \pm 1.1$ (see also
\cite{Leutwyler:1996qg}) one finds $R = 33.5 \pm 4.3$
and finally, together with a determination of 
$\varepsilon^{(4)}_{\rm S}$ and 
$\varepsilon^{(4)}_{\rm EM}$, the result 
\cite{Kastner:2008ch}
\begin{equation}
\label{eq:iso-brk}
\delta_{\rm SU(2)}^{K^\pm \pi^0} = 0.058(8)~.
\end{equation}


\subsubsection{$K_{\ell 3 }$ form factors}
\label{sec:ffpara}

The hadronic  $K \to \pi$ matrix element of the vector current
is described by two form factors (FFs),  $f_+(t)$ and  $f_-(t)$
\be
\langle\pi^{-}\left(  p_\pi\right)  |\bar{s}\gamma^{\mu}u|K^{0}\left(  p_K\right)
\rangle = (p_K+p_\pi) ^\mu f_+^{}(t) +(p_K-p_\pi) ^\mu f_-^{}(t)
\label{eq:HadronMatrix}
\end{equation}
where $t=(p_K-p_\pi)^2= (p_\ell+p_\nu)^2$. 
The vector form factor $f_+(t)$ represents the P-wave projection of the crossed channel 
matrix element $ \langle 0 |\bar{s}\gamma^{\mu}u| K \pi \rangle $ whereas the S-wave projection 
is described by the scalar form factor defined as
\be
f_0(t)= f_+(t) + \frac{t}{m_K^2-m_\pi^2} f_-(t)~.
\label{eq:f0def}
\end{equation}
By construction,  $f_0(0)=f_+(0)$.

In order to compute the phase space integrals appearing in Eq.~(\ref{eq:Mkl3}) 
we need experimental or theoretical inputs about the  $t$-dependence
of $f_{+,0}(t)$. In principle, chiral perturbation theory (ChPT) 
and lattice QCD are useful tools to set theoretical constraints.
However, in practice the  $t$-dependence of the FFs at present 
is better determined by measurements and by combining measurements 
and dispersion relations. To that aim, we introduce the normalized FFs
\be
\tilde f_{+}(t)=\frac{f_{+}(t)}{f_{+}(0)}~,~\tilde f_{0}(t)=\frac{f_{0}(t)}{f_{0}(0)}~,~\tilde f_+(0)=\tilde f_{0} (0)=1~.
\label{eq:Normff}
\end{equation}

Whereas $\tilde f_{+}(t)$ is accessible in the $K_{e 3}$ and $K_{\mu3}$ decays, $\tilde f_{0}(t)$ 
is more difficult to measure since it is only accessible in $K_{\mu3}$ decays, being
 kinematically suppressed in $K_{e3}$ decays, 
and is strongly correlated with $\tilde f_{+}(t)$. 

Moreover, measuring the scalar form factor is of special interest 
due to the existence of the Callan-Treiman (CT) theorem \cite{Callan:1966hu} 
which predicts the value of the scalar form factor at the so-called CT
point, namely $t\equiv \Delta_{K \pi}= m_K^2-m_\pi^2$,
\be
C \equiv \tilde f_0(\Delta_{K\pi})=\frac{f_K}{f_\pi}\frac{1}{f_+(0)}+ \Delta_{CT},
\label{eq:CTrel}
\end{equation}
where $\Delta_{CT} \sim  {\mathcal{O}} (m_{u,d}/4 \pi F_{\pi})$ is a small correction.  
ChPT at NLO in the isospin limit~\cite{Gasser:1984ux} gives 
\be
\label{eq:DeltaCT}
\Delta_{CT}=(-3.5\pm 8)\times 10^{-3}~,
\end{equation}
where the error is a conservative estimate of the higher order corrections~\cite{Leutwyler}. 
A complete two-loop calculation 
of $\Delta_{CT}$~\cite{Bijnens:2007xa}, as well as a computation 
at $\mathcal{O}(p^4, e^2p^2, (m_d-m_u))$\cite{Kastner:2008ch}, 
consistent with this estimate, have been recently presented. 

The measurement of $C$ 
 provide a powerful consistency check of 
the lattice QCD calculations of  $f_K/f_\pi$ and $f_+(0)$, 
as will be discussed in Sec.~\ref{sec:CTtest}. 

Another motivation to measure the shape of the scalar form factor very accurately is that knowing the 
slope and the curvature of the scalar form factor allows one to perform a matching
 with the 2-loop ChPT  
calculations \cite{Bernard:2007tk} and then determine fundamental constants of QCD
 such as $f_+(0)$ or the low-energy constants (LECs) 
$C_{12}$, $C_{34}$ which appear in many ChPT calculations.

\subsubsection*{Parametrization of the form factors and dispersive approach}

To determine the FF shapes, different experimental analyses of 
$K_{\ell 3}$ data have been performed in the last few years, 
by KTeV, NA48, and KLOE for the neutral mode and by ISTRA+ for the charged mode. 

Among the different parameterizations available, one can distinguish two classes~\cite{Bernard:2009zm}.
The class called class II in this reference contains parameterizations based on mathematical 
rigorous expansions where the slope, the curvature 
and all the higher order terms of the expansion are free parameters of the fit. 
In this class, one finds the Taylor expansion
\be
\tilde{f}_{+,0}^{Tayl}(t) = 1 + \lambda'_{+,0} \frac{t}{M_\pi^2} + \frac{1}{2}\lambda''_{+,0} \left(\frac{t}{m_\pi^2}\right)^2 
+ \frac{1}{6}\lambda'''_{+,0} \left(\frac{t}{m_\pi^2}\right)^3 + \ldots~,
\label{Taylor}
\end{equation}
where $\lambda'_{+,0}$ and $\lambda''_{+,0}$ are the slope and the curvature of the FFs respectively, 
but also the so-called  z-parametrization~\cite{Hill:2006bq}.


As for parameterizations belonging to class I, they correspond to parameterizations for which by using 
physical inputs, specific relations between the slope, the curvature and all the higher order terms of 
the Taylor expansion, Eq.~(\ref{Taylor}) are imposed. 
This allows to reduce the correlations between the fit parameters since 
only one parameter is fitted for each FF. In this class, one finds the pole parametrization 
\be
\tilde{f}_{+,0}^{Pole}(t) = \frac{M_{V,S}^2}{M_{V,S}^2-t}~,
\label{pole}
\end{equation}
in which dominance of a single resonance is assumed and its mass $M_{V,S}$ is the fit parameter. 
Whereas for the vector FF a pole parametrization with the dominance of the $K^*(892)$ ($M_V \sim 892$ MeV) is in 
good agreement with the data, for the scalar FF there is no such obvious dominance. 
One has thus to rely, at least for $\tilde f_0(t)$, on a dispersive parametrization. 
In such a construction, in addition to guarantee the good 
properties of analyticity and unitarity of the FFs, physical inputs such as 
the low energy $K \pi$ data and, in the case of the vector form factor, the 
dominance of K*(892) resonance are used.  

The vector and scalar form factors are analytic functions in the complex $t$-plane, except for
 a cut along the positive real 
axis, starting at the first physical threshold where they develop discontinuities. They are real for $t < t_{ \rm th}=(m_K+m_\pi)^2$. 
Cauchy's theorem 
implies that $\tilde{f}_{+,0}(t)$ can be written as a dispersive integral along the physical cut
\begin{equation}
\label{dis}
\tilde f_{+,0}(t) \; = \; \frac{1}{\pi} \int\limits^\infty_{t_{\rm th}}\!\! ds'\,
\frac{\Im \tilde f_{+,0}(s')}{(s'-t-i0)} + {\rm subtractions} \,,
\end{equation}
where all the possible on-shell states contribute to its imaginary part Im$\tilde f_{+,0}(s')$.
 A number of subtractions is needed to make the integral convergent. 

A particularly appealing dispersive parametrization for the scalar form factor is the one proposed in Ref.~\cite{Bernard:2006gy}. 
Two subtractions are performed, one at $t=0$ where by definition $\tilde f_{0}(0)=1$,
 see Eq.~(\ref{eq:Normff}), and the other one at the CT point. 
%
%
This leads to 
\be
\tilde f_0^{Disp}(t)=\exp\Bigl{[}\frac{t}{\Delta_{K\pi}}(\mathrm{ln}C- G(t))\Bigr{]}~, 
\label{Dispf}
\end{equation}
with
\be 
G(t)=\frac{\Delta_{K\pi}(\Delta_{K\pi}-t)}{\pi}~ \int_{(m_K+m_\pi)^2}^{\infty}
\frac{ds}{s}
\frac{\phi_0(s)}
{(s-\Delta_{K\pi})(s-t-i\epsilon)}~, 
\label{G}
\end{equation}
assuming that the scalar FF has no zero. 
In this case the only free parameter to be determined from a fit to the data is $C$. 
$\phi_0(s)$ represents
the phase of the form factor. According to Watson's theorem
\cite{Watson:1952ji}, this phase can be identified in the elastic region with the S-wave, $I=1/2$ $K\pi$ scattering phase. 
The fact that two subtractions have been made in writing Eq.~(\ref{Dispf}) allows
to minimize the contributions from the unknown high-energy
phase in the dispersive integral. 
The resulting function $G(t)$, Eq.~(\ref{G}),
does not exceed 20\% of the expected value of ln$C$ limiting the
theoretical uncertainties which represent at most 10\% of the value of
$G(t)$ \cite{Bernard:2006gy}.

A dispersive representation for the vector FF has been built in a 
similar way \cite{Passemar:2007qe}. 
Since there is no analog of the CT theorem, in this case, the two subtractions are performed at $t=0$. 
Assuming that the vector FF 
has no zero, one gets
\be
\tilde f_+^{Disp}(t)=\exp\Bigl{[}\frac{t}{m_\pi^2}\left(\Lambda_+ + H(t)\right)\Bigr{]}~,~ 
H(t)=\frac{m_\pi^2t}{\pi} \int_{(m_K+m_\pi)^2}^{\infty}
\frac{ds}{s^2}
\frac{\phi_+ (s)}
{(s-t-i\epsilon)}~. 
\label{Dispfp}
\end{equation}
with $\Lambda_+~\equiv m_\pi^2 d \tilde f_+(t)/dt|_{t=0}$ is the fit parameter and
$\phi_+(s)$ the phase of the vector form factor. Here, in the elastic region, $\phi_+(t)$ equals 
the $I=1/2$, P-wave K$\pi$ scattering phase according to Watson's theorem \cite{Watson:1952ji}.  
Similarly to what happens for $G$, the two subtractions minimize the contribution coming from the 
unknown high energy phase resulting in a relatively small uncertainty on $H(t)$. 
Since the dispersive integral $H(t)$ represents at most 20\% of the expected value of $\Lambda_+$, 
the latter can then be determined with a high precision knowing $H(t)$ much less precisely. 
For more details on the dispersive representations and a 
detailed discussion of the different sources of theoretical uncertainties 
entering the dispersive parametrization via the function $G$ and $H$,
see \cite{Bernard:2006gy} and \cite{Passemar:2007qe}.

Using a class II parametrization  for the FFs in a fit to $K_{\ell 3}$ decay distribution, 
only two parameters ($\lambda_+'$ and $\lambda''_+ $ for a Taylor expansion, 
Eq.~(\ref{Taylor}))
%
%
can be determined for $\tilde f_+(t)$ and only one  parameter
($\lambda'_0$ for a Taylor expansion)
for $\tilde f_0(t)$. 
Moreover these parameters are strongly correlated. 
It has also been shown in Ref. \cite{Bernard:2006gy} that in order to describe the FF shapes accurately in the physical region, 
one has to go at least up to the second order in the Taylor expansion. 
Neglecting the curvature in the parametrization of $\tilde f_0(t)$  
generates a bias in the extraction of $\lambda_0'$ which is then overestimated \cite{Bernard:2006gy}.  
Hence,  using a class II parametrization  for $\tilde f_0(t)$ doesn't allow 
 it to be extrapolated from the physical region ($m_{\ell}^2 < t < t_{0}=(m_K-m_\pi)^2$)
 up to the CT point  with a reliable precision.  
To measure the FF shapes from $K_{\ell3}$ decays with the precision 
demanded in the extraction of $|V_{us}|$, it is preferable to use 
a parametrization in class I.     


\subsubsection{Lattice determinations of $f_+(0)$ and $f_K / f_\pi$}
\label{sect:CabibboLattice}

In this section we summarize the status of results of lattice QCD simulations 
for the semileptonic Kaon decay form factor $f_+(0)$ and for the ratio of 
Kaon and pion leptonic decay constants, $f_K / f_\pi$. For a brief introduction
to lattice QCD we refer the reader to section~\ref{sec:thPrim:LQCD}.

\subsubsection*{Theoretical estimates of $f_+(0)$\label{sec:fplus}}

The vector form factor at zero-momentum transfer, $f_+(0)$, is the key hadronic 
quantity required for the extraction of the CKM matrix element $|V_{us}|$ from 
semileptonic $K_{\ell 3}$ decays (cf. equation (\ref{eq:Mkl3})).
Within SU(3) ChPT one can perform a systematic expansion of $f_+(0)$ of the type 
 \be
    f_+(0) = 1 + f_2 + f_4 + ... ~ , 
    \label{eq:chiralPT}
 \end{equation}
where $f_n = {\O}[M_{K, \pi}^n / (4 \pi f_\pi)^n]$ and the first term is 
equal to unity due to the vector current conservation in the SU(3) limit. 
Because of the Ademollo-Gatto (AG) theorem \cite{Ademollo:1964sr}, 
the first non-trivial term 
$f_2$ does not receive contributions from the local operators of the effective 
theory and can be computed unambiguously in terms of the Kaon and pion masses 
($M_K$ and $M_\pi$) and the pion decay constant $f_\pi$. 
It takes the value $f_2 = -0.023$ at the physical point \cite{Leutwyler:1984je}.
The task is thus reduced to the problem of finding a prediction for the quantity 
$\Delta f$, defined as
 \be
    \Delta f \equiv f_4 + f_6 + ... = f_+(0) - (1 + f_2) ~ ,
    \label{eq:deltaf}
 \end{equation}
which depends on the low-energy constants (LECs) of the effective theory and 
cannot be deduced from other processes.

The original estimate made by Leutwyler and Roos 
\cite{Leutwyler:1984je} was based on the quark model yielding 
$\Delta f = -0.016(8)$.
More recently other analytical approaches have tried to determine the next-to-next-to-leading 
order (NNLO) term $f_4$ by writing it as 
 \be
    f_4 = L_4(\mu) + f_4^{loc}(\mu) ~ ,
    \label{eq:f4}
 \end{equation}
where $\mu$ is the renormalization scale, $L_4(\mu)$ is the loop contribution computed 
in Ref.~\cite{Bijnens:2003uy} 
and $f_4^{loc}(\mu)$ is the ${\O}(p^6)$ local contribution.
For the latter various models have been adopted, namely the quark model in 
Ref.~\cite{Bijnens:2003uy}, 
the dispersion relations in Ref.~\cite{Jamin:2004re} and the $1 / N_c$ 
expansion in Ref.~\cite{Cirigliano:2005xn},
 obtaining $\Delta f = 0.001(10), ~ -0.003(11), ~ 0.007(12)$, respectively.
These values are compatible with zero within the uncertainties and are 
significantly larger than the LR estimate, leading to smaller SU(3)-breaking 
effects on $f_+(0)$.

Notice that in principle the next-to-next-to-leading order (NNLO) term $f_4$ may 
be obtained from the slope and the curvature of the scalar form factor $f_0(q^2)$, 
but present data from $K \to \pi \mu \bar{\nu}_\mu$ decays are not precise enough 
for an accurate determination.

A precise evaluation of $f_+(0)$, or equivalently $\Delta f$, requires the use of 
non-perturbative methods based on the fundamental theory of the strong interaction, 
such us lattice QCD simulations.
Such determinations started recently with the quenched simulations of Ref.~\cite{Becirevic:2004ya}, 
where it was shown that $f_+(0)$ can be determined at the physical point with a 
$\simeq 1 \%$ accuracy.
The findings of Ref.~\cite{Becirevic:2004ya} triggered various unquenched calculations of 
$f_+(0)$, namely those of Refs.~\cite{Tsutsui:2005cj,Dawson:2006qc,Brommel:2007wn}
 with $N_f = 2$ with pion 
masses above $\simeq 500~\mev$ and two very recent ones from Ref.~\cite{Boyle:2007qe} with 
$N_f = 2 + 1$ and Ref.~\cite{Lubicz:2009ht} with $N_f = 2$.  
In the former the simulated pion masses start from $330~\mev$, while in the latter,
 they start from $260~\mev$.
In both cases the error associated with the chiral extrapolation was significantly 
reduced with respect to previous works thanks to the lighter pion masses.

In Ref.~\cite{Lubicz:2009ht} the chiral extrapolation was performed using both SU(3) 
and SU(2) ChPT for $f_2$ (see Ref.~\cite{Flynn:2008tg}).
In the latter case the Kaon field is integrated out and the effects of the strange 
quark are absorbed into the LECs of the new effective theory.
The results obtained using SU(2) and SU(3) ChPT are found to be consistent within 
the uncertainties, giving support to the applicability of chiral
perturbation theory at this order. We note that since no predictions in 
chiral perturbation theory for 
$\Delta f$ as a function of the quark masses exists in a closed form, the
lattice data for $\Delta f$ is currently extrapolated to the physical point
using phenomenologically motivated ans\"atze.

The results for $f_+(0)$ and $\Delta f$ are summarized in Tab. \ref{tab:fplus}, 
together with some relevant details concerning the various lattice set-ups, 
and those of $f_+(0)$ are shown in Fig.~\ref{fig:fplus}.
It can be seen that: 
\begin{itemize}
 \item[i)]  all lattice results suggest a negative, sizable value for 
	$\Delta f$ in agreement with the LR estimate, but at variance with 
	the results of the analytical approaches of Refs.
	\cite{Bijnens:2003uy,Jamin:2004re,Cirigliano:2005xn}, and 
 \item[ii)] the two recent lattice calculations of Refs.\cite{Boyle:2007qe,Lubicz:2009ht} 
	have reached an encouraging precision of $\simeq 0.5 \%$ on the 
	determination of $f_+(0)$.
\end{itemize}

\begin{table}[!htb]
\caption{Summary of model and lattice results for $f_+(0)$ and $\Delta f$.
The lattice errors include both statistical and systematic uncertainties.
\label{tab:fplus}}
\begin{center}
\begin{tabular}{||c|c||c|c||c|c|c|c||}
\hline
 $\mbox{Ref.}$ & $Model/Lattice$ & $f_+(0)$     & $\Delta f$    & $M_\pi~\mbox{(MeV)}$ & $M_\pi L$ & $a~\mbox{(fm)}$ & $N_f$ \\ \hline \hline
 \cite{Leutwyler:1984je}     & $LR$            & $0.961~(~8)$ & $-0.016~(~8)$ & & & & \\ \hline
\cite{Bijnens:2003uy}& $ChPT+LR$       & $0.978~(10)$ & $+0.001~(10)$ & & & & \\ \hline
 \cite{Jamin:2004re}    & $ChPT+disp.$    & $0.974~(11)$ & $-0.003~(11)$ & & & & \\ \hline 
 \cite{Cirigliano:2005xn}    & $ChPT+1/N_c$    & $0.984~(12)$ & $+0.007~(12)$ & & & & \\ \hline \hline
 \cite{Becirevic:2004ya} & $SPQ_{cd}R$     & $0.960~(~9)$ & $-0.017~(~9)$ & $\gtrsim 500$ & $\gtrsim 5$ & $\simeq 0.07$ & $0$ \\ \hline \hline
 \cite{Tsutsui:2005cj}  & $JLQCD$         & $0.967~(~6)$ & $-0.010~(~6)$ & $\gtrsim 550$ & $\gtrsim 5$ & $\simeq 0.09$ & $2$ \\ \hline
 \cite{Dawson:2006qc}  & $RBC$           & $0.968~(12)$ & $-0.009~(12)$ & $\gtrsim 490$ & $\gtrsim 6$ & $\simeq 0.12$ & $2$ \\ \hline
 \cite{Brommel:2007wn}  & $QCDSF$         & $0.965~(~?)$ & $-0.012~(~?)$ & $\gtrsim 590$ & $\gtrsim 6$ & $\simeq 0.08$ & $2$ \\ \hline
 \cite{Lubicz:2009ht} & $ETMC$          & $0.956~(~8)$ & $-0.021~(~8)$ & $\gtrsim 260$ & $\gtrsim 4$ & $\simeq 0.07$ & $2$ \\ \hline \hline
 \cite{Boyle:2007qe}  & $RBC+UKQCD$     & $0.964~(~5)$ & $-0.013~(~5)$ & $\gtrsim 330$ & $\gtrsim 4$ & $\simeq 0.11$ & $2+1$ \\ \hline \hline

\end{tabular}
\end{center}

\end{table}

\begin{figure}[!hbt]
\centering
\resizebox{0.7\textwidth}{!}{\includegraphics[angle=-90]{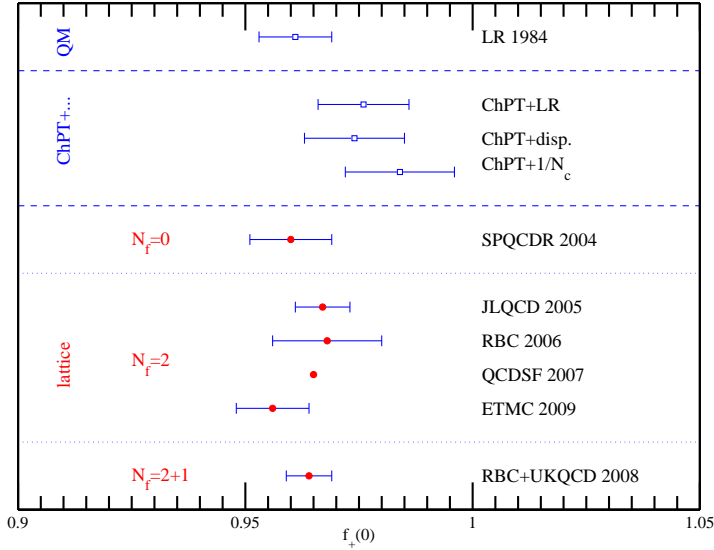}}
\caption{Results of model (squares) and lattice 
(dots) calculations of $f_+(0)$.  }\label{fig:fplus}
\end{figure}

Since simulations of lattice QCD are carried out in a finite volume, the 
momentum transfer $q^2$ for the conventionally used periodic fermion
boundary conditions takes values corresponding to the Fourier modes of the 
Kaon or pion. Using a phenomenological ansatz for the 
$q^2$-dependence of the form factor one interpolates to $q^2=0$ where
$f_+(0)$ is extracted, thereby introducing a major systematic uncertainty.  
A new method based on the use of partially twisted boundary conditions
(cf. section \ref{sec:thPrim:LQCD})
has been developed \cite{Boyle:2007wg} which allows 
this uncertainty to be entirely removed by simulating directly at
 the desired kinematic point $q^2=0$. 

Although the impact of discretization effects is expected to be small\footnote{The analysis from ETM\cite{Lubicz:2009ht}, with fixed simulated 
quark mass, confirms that discretization effects
 are small with respect to present uncertainty.}
, we emphasize that all available lattice calculations have been 
carried out at a single lattice spacing.

A systematic study of the scaling behavior of 
$f_+(0)$, using partially twisted boundary conditions 
 and the extension of the simulations to lighter pion masses in order 
to improve the chiral extrapolation 
will be the priorities for the upcoming lattice studies of $K_{\ell 3}$ decays.

\subsubsection*{Theoretical estimates of $f_K / f_\pi$\label{sec:fKfpi}}

As  was pointed out in Ref.~\cite{Marciano:2004uf}, an  alternative to $K_{\ell 3}$ 
decays for obtaining a precise determination of $|V_{us}|$ is provided by the 
Kaon(pion) leptonic decays $K(\pi) \to \mu \bar{\nu}_\mu (\gamma)$.
In this case, the key hadronic quantity is the ratio of the Kaon and pion decay 
constants, $f_K / f_\pi$.

In contrast to $f_+(0)$, the pseudoscalar decay constants are not protected by 
the AG theorem \cite{Ademollo:1964sr} 
against corrections linear in the SU(3) breaking.
Moreover the first non-trivial term (of order ${\O}(p^4)$) in the chiral 
expansion of $f_K / f_\pi$ depends on the LECs and therefore it cannot be 
predicted unambiguously within ChPT.
This is the reason why the most precise determinations of $f_K / f_\pi$ 
 come from lattice QCD simulations.

During the recent years various collaborations have provided new results for 
$f_K / f_\pi$ using unquenched gauge configurations with both 2 and 2+1 
dynamical flavors.
They are summarized in Tab. 
\ref{tab:fKfPi}, together with some relevant details concerning the various lattice 
set-ups. They are shown graphically in Fig.~\ref{fig:fKfPi}.

\begin{table}[!htb]

\caption{Summary of lattice results for $f_K / f_\pi$.
The errors include both statistical and systematic uncertainties.
\label{tab:fKfPi}}
\begin{center}
\begin{tabular}{||c|c||c||c|c|c|c||}
\hline
 $\mbox{Ref.}$    & $Collaboration$ & $f_K / f_\pi$         & $M_\pi~\mbox{(MeV)}$ & $M_\pi L$ & $a~\mbox{(fm)}$ & $N_f$ \\ \hline \hline
 \cite{Aubin:2004fs,Bazavov:2009bb}      & $MILC$          & $1.197~_{-13}^{+~7}~$ & $\gtrsim 240$ & $\gtrsim 4$ & $\to 0$        & $2+1$ \\ \hline
 \cite{Follana:2007uv}     & $HPQCD$         & $1.189~(~7)$          & $\gtrsim 250$ & $\gtrsim 4$ & $\to 0$        & $2+1$ \\ \hline
 \cite{Durr2008}  & $BMW$           & $1.185~(15)$          & $\gtrsim 190$ & $\gtrsim 5$ & $\to 0$        & $2+1$ \\ \hline 
 \cite{Aubin:2008ie}    & $Aubin\, \emph{et al.}$        & $1.191(23)$ & $\gtrsim 240$ & $\gtrsim 3.8$ & $\to 0$  & $2+1$ \\ \hline
 \cite{Blossier:2009bx}      & $ETMC$          & $1.210~(18)$          & $\gtrsim 260$ & $\gtrsim 4$ & $\to 0$        & $2$   \\ \hline\hline 
 \cite{Beane:2006kx}    & $NPLQCD$        & $1.218~_{-24}^{+11}~$ & $\gtrsim 290$ & $\gtrsim 4$ & $\simeq 0.13$  & $2+1$ \\ \hline
 \cite{Allton:2008pn} & $RBC/UKQCD$     & $1.205~(65)$          & $\gtrsim 330$ & $\gtrsim 4$ & $\simeq 0.11$  & $2+1$ \\ \hline
 \cite{Aoki:2008sm}   & $PACS-CS$       & $1.189~(20)$          & $\gtrsim 160$ & $\gtrsim 2$ & $\simeq 0.09$  & $2+1$ \\ \hline \hline

\end{tabular}
\end{center}

\end{table}

\begin{figure}[ht]
\centering
\resizebox{0.7\textwidth}{!}{\includegraphics[angle=-90]{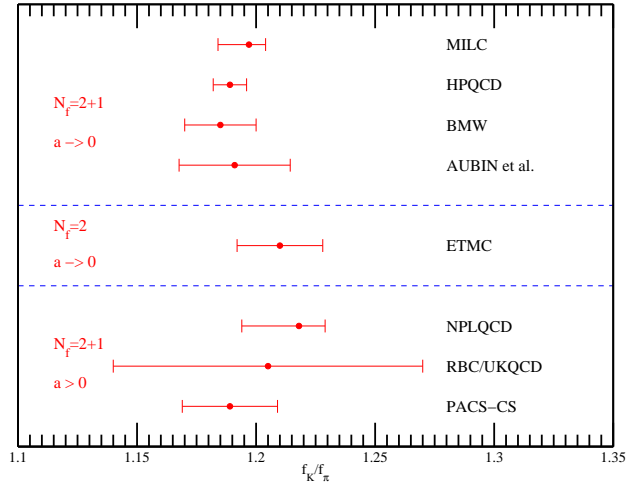}}
\caption{Results of lattice calculations of $f_K / f_\pi$.}
\label{fig:fKfPi}
\end{figure}

A few comments are in order: 
\begin{itemize}
\item[i)] finite size effects are kept under good control by the constraint $M_\pi L 
\gtrsim 4$, which is adopted by all collaborations except Ref.~\cite{Aoki:2008sm}; 
\item[ii)] the continuum extrapolation, which allows  discretization 
effects to be safely removed, has been performed by several collaborations; 
\item[iii)] the convergence of the SU(3) chiral expansion for $f_K / f_\pi$ appears to 
be questionable, mainly because large NLO corrections are already required to 
account for the large difference between the experimental value of $f_\pi$ 
and the value of the decay constant in the massless SU(3) limit;
\item[iv)] the convergence of the SU(2) chiral expansion is much better and thanks 
to the light pion masses reached in the recent lattice calculations, the 
uncertainty related to the chiral extrapolation to the physical point is kept 
to the percent level \cite{Allton:2008pn};
\item[v)] little is known about the details of the chiral and continuum
extrapolation in Ref.~\cite{Follana:2007uv} (HPQCD) which is currently
the most precise lattice prediction for $f_K/f_\pi$; in particular about the
priors on many parameters that have been introduced;
\item[vi)] It is worth repeating (cf. section \ref{sec:thPrim:LQCD}) that there
exist conceptional concerns about the staggered fermion formulation - the
results by MILC, HPQCD, Aubin \emph{et al.} and NPLQCD  use staggered
fermions and need to be confirmed by conceptually clean fermion formulations.
\end{itemize}
\subsubsection*{Summary of lattice results}\label{sec:VudVus:lat_res}

We note that the Flavia Net Lattice Averaging Group (FLAG) has just started
to periodically 
compile and publish (web and journal) lattice QCD results for SM observables and
parameters. In
addition, averages will be computed where feasible and a classification
of the quality of lattice results by means of a simple color coding 
will be provided in order to facilitate understanding of lattice results
for non-experts. For a first status report see \cite{FLAG:Colangelo}.

Hence, no average over lattice results will be provided here. We merely identify
those results that have a good control over systematic uncertainties and
have been published in journals and refer the reader to the forthcoming
FLAG document for averages.

For $f_+(0)$ the  2+1 flavor result by the RBC+UKQCD 
\cite{Boyle:2007qe} collaboration is the most advanced calculation,

\begin{equation}
\ba{llll}
f_+(0)=0.964(5) &\;\;N_f=2+1\,.
    \label{eq:final_fplus} 
\ea
\end{equation} 
while for 2 flavors it is the result by ETM \cite{Lubicz:2009ht},
\begin{equation}
\ba{llll}
f_+(0)=0.956(8) &\;\;N_f=2\,.
\ea
\end{equation}

For $f_K / f_\pi$ with $N_f=2+1$ dynamical quarks, the currently
most precise predictions are by MILC \cite{Bazavov:2009bb}
\begin{equation}
\ba{llll}
f_K/f_\pi=1.197(^{+\,7}_{-13})&\;\;N_f=2+1\,,
\ea
\end{equation}
and HPQCD \cite{Follana:2007uv}
\begin{equation}
\ba{llll}
f_K/f_\pi =1.189(7)&\;\;N_f=2+1\,,
\ea
\end{equation}
both using the same set of staggered sea quark configurations.

For illustrative purposes the latter result will be used later in
section \ref{sec:results}. We also emphasize the   currently 
most precise result with $N_f=2$ dynamical quarks by the ETM
collaboration \textrm{\cite{Blossier:2009bx}}:
\be
\ba{llll}
f_K/f_\pi=1.210(18) &(N_f=2)\,.\\
\ea
\end{equation}
At the current level of precision the comparison of the 
$N_f=2$ and $N_f=2+1$ result indicates a rather
small contribution of the strange sea quarks to the ratio of decay constants.

\subsubsection{Data Analysis}
\label{sec:data}
We perform fits to world data on the BRs and lifetimes for the
$K_L$ and $K^\pm$, with the constraint that BRs add to unity.
This is the correct way of using
 the new measurements. 
 A detailed description of the fit  is given in 
Ref~\cite{Antonelli:2008jg}.  The present version of
our fits uses only published measurements.

\subsubsection*{$K_L$   leading branching ratios and $\tau_L$ }

\label{sec:KL}
Numerous measurements of the principal $K_L$ BRs, or of various ratios
of these BRs, have been published recently. For the purposes of evaluating
\Vusf, these data can be used in a PDG-like fit to the $K_L$ BRs and lifetime,
so all such measurements are interesting.

KTeV has measured five ratios of the six main $K_L$ BRs~\cite{Alexopoulos:2004sx}.
The six channels
involved account for more than 99.9\% of the $K_L$ width and KTeV combines the
five measured ratios to extract the six BRs. We use the five measured ratios
in our analysis:
$\BRT{K_{\mu3}}/\BRT{K_{e3}} = 0.6640(26)$,
$\BRT{\pi^+\pi^-\pi^0}/\BRT{K_{e3}} = 0.3078(18)$,
$\BRT{\pi^+\pi^-}/\BRT{K_{e3}} = 0.004856(28)$,
$\BRT{3\pi^0}/\BRT{K_{e3}} = 0.4782(55)$, and
$\BRT{2\pi^0}/\BRT{3\pi^0} = 0.004446(25)$. The errors on these measurements are
correlated; this is taken into account in our fit.

NA48 has measured the ratio of the BR for $K_{e3}$ decays to the sum of BRs
for all decays to two tracks, giving
$\BRT{K_{e3}}/(1-\BRT{3\pi^0}) = 0.4978(35)$ \cite{Lai:2004bt}.

Using $\phi\to K_L K_S$ decays in which the $K_S$ decays to $\pi^+\pi^-$,
providing normalization, KLOE has directly measured the BRs for the four
main $K_L$ decay channels \cite{Ambrosino:2005ec}.
The errors on the KLOE BR values are dominated
by the uncertainty on the $K_L$ lifetime $\tau_L$; since the dependence of
the geometrical efficiency on $\tau_L$ is known, KLOE can solve for $\tau_L$
by imposing $\sum_x \BRT{K_L\to x} = 1$ (using previous averages for the minor
BRs), thereby greatly reducing the uncertainties on the BR values obtained.
Our fit makes use of the KLOE BR values before application of this constraint:
\BRT{K_{e3}} = 0.4049(21),
\BRT{K_{\mu3}} = 0.2726(16),
\BRT{3\pi^0} = 0.2018(24), and
\BRT{\pi^+\pi^-\pi^0} = 0.1276(15).
The dependence of these values on $\tau_L$ and the correlations between the
errors  are taken into account.
KLOE has also measured $\tau_L$ directly, by fitting the proper decay time
distribution for $K_L\to3\pi^0$ events, for which the reconstruction
efficiency is high and uniform over a fiducial volume of $\sim$$0.4\lambda_L$.
They obtain $\tau_L=50.92(30)$~ns \cite{Ambrosino:2005vx}.

There are also two recent measurements of \BRT{\pi^+\pi^-}/\BRT{K_{\ell3}},
in addition to the KTeV measurement of \BRT{\pi^+\pi^-}/\BRT{K_{e3}} discussed above.
The KLOE collaboration 
obtains \BRT{\pi^+\pi^-}/\BRT{K_{\mu3}} = \SN{7.275(68)}{-3} \cite{Ambrosino:2006up},
while NA48 obtains \BRT{\pi^+\pi^-}/\BRT{K_{e3}} = \SN{4.826(27)}{-3}
\cite{Lai:2006cf}. All measurements are fully inclusive of inner
bremsstrahlung. The KLOE measurement is fully inclusive of the direct-emission
(DE) component, DE contributes negligibly to the KTeV measurement, and a
residual DE contribution of 0.19\% has been subtracted from the NA48 value
to obtain the number quoted above. 

We fit the 13 recent measurements listed above, together with eight
additional ratios of the BRs for subdominant decays. The complete 
list of 21 inputs is given in Table~\ref{tab:KLinputs}. As free parameters, our fit has
the seven largest $K_L$ BRs (those to $K_{e3}$, $K_{\mu3}$, $3\pi^0$,
$\pi^+\pi^-\pi^0$, $\pi^+\pi^-$, $\pi^0$ and $\gamma\gamma$) and the 
$K_L$ lifetime, as well
as two additional parameters necessary for the treatment of the
direct emission (DE) component in the radiation-inclusive $\pi^+\pi^-$
decay width. Our definition of ${\BR}(\pi^+\pi^-)$ is now fully inclusive 
of inner bremsstrahlung (IB), but exclusive of the DE component.
The fit also includes ${\BR}(\pi^+\pi^-\gamma)$ and 
${\BR}(\pi^+\pi^-\gamma_{\rm DE})$, the branching ratios for decays to 
states with a photon with $E^*_\gamma > 20$~MeV, and with a photon
from DE with $E^*_\gamma > 20$~MeV, respectively. Other parameterizations
are possible, but this one most closely represents the input data set
and conforms to recent PDG usage. With 21 input measurements, 10 free
parameters, and the constraint that the sum of the BRs (except for
${\BR}(\pi^+\pi^-\gamma)$, which is entirely included in the sum
of ${\BR}(\pi^+\pi^-)$ and  ${\BR}(\pi^+\pi^-\gamma_{\rm DE})$) equal 
unity, we have 12 degrees of freedom. The fit gives $\chi^2=19.8$
($P=7.1\%$).

\begin{table}
\caption{\label{tab:KLBR}
Results of fit to $K_L$ BRs and lifetime.}
\begin{center}
\begin{tabular}{l|c|r}
Parameter & Value & $S$ \\
\hline
\BRT{K_{e3}} & 0.4056(9) & 1.3 \\
\BRT{K_{\mu3}} & 0.2704(10) & 1.5 \\
\BRT{3\pi^0} & 0.1952(9) & 1.2 \\
\BRT{\pi^+\pi^-\pi^0} & 0.1254(6) & 1.1 \\
\BRT{\pi^+\pi^-} & \SN{1.967(7)}{-3} & 1.1 \\
\BRT{\pi^+\pi^-\gamma}  & \SN{4.15(9)}{-5} & 1.6 \\
\BRT{\pi^+\pi^-\gamma} DE& \SN{2.84(8)}{-5} & 1.3 \\
\BRT{2\pi^0} & \SN{8.65(4)}{-4} & 1.4 \\
\BRT{\gamma\gamma} & \SN{5.47(4)}{-4} & 1.1 \\
$\tau_L$ & 51.16(21)~ns & 1.1 \\
\end{tabular}
\end{center}
\vskip 0.3cm
\end{table}
The evolution of the average values of the BRs for
$K_{L\ell3}$ decays and for
the important normalization channels is shown in \Fig{fig:kpmavg}.

\begin{figure}[ht]
\begin{center}
\resizebox{0.8\textwidth}{!}{\includegraphics{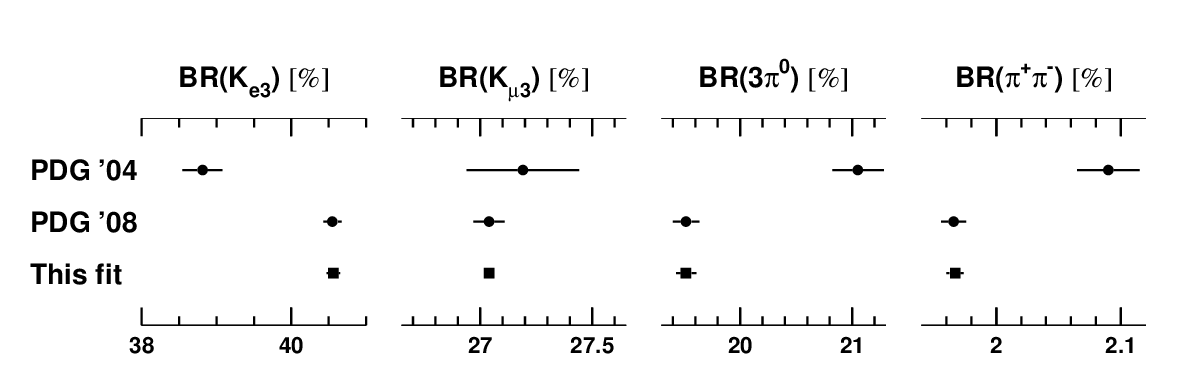}}
\end{center}
\caption{\label{fig:klavg}
Evolution of average values for main $K_L$ BRs.}
\end{figure}

\begin{table}
\caption{Input data used for the fit to $K_L$ BRs and lifetime (all the references refer to PDG08~\cite{Amsler:2008zzb}).}
\label{tab:KLinputs}
\center
\begin{tabular}{lcl}
\hline\hline
 Parameter                          & Value         & Source        \\[0.5ex]\hline
 $\tau_{K_L}$                       & 50.92(30) ns  & Ambrosino 05C      \\
 $\tau_{K_L}$                       & 51.54(44) ns  & Vosburgh 72  \\
 $\BR{K_{e3}}$                      & 0.4049(21)    & Ambrosino 06      \\
 $\BR{K_{\mu3}}$                    & 0.2726(16)    & Ambrosino 06      \\
 $\BR{K_{\mu3}}/\BR{K_{e3}}$        & 0.6640(26)    & Alexopoulos 04      \\
 $\BR{3\pi^0}$                      & 0.2018(24)    & Ambrosino 06      \\
 $\BR{3\pi^0}/\BR{K_{e3}}$          & 0.4782(55)    & Alexopoulos 04   \\
 $\BR{\pi^+\pi^-\pi^0}$             & 0.1276(15)    & Ambrosino 06      \\
 $\BR{\pi^+\pi^-\pi^0}/\BR{K_{e3}}$ & 0.3078(18)    & Alexopoulos 04   \\
 $\BR{\pi^+\pi^-}/\BR{K_{e3}}$      & 0.004856(29)  & Alexopoulos 04    \\
 $\BR{\pi^+\pi^-}/\BR{K_{e3}}$      & 0.004826(27)  & Lai 07      \\
 $\BR{\pi^+\pi^-}/\BR{K_{\mu3}}$    & 0.007275(68)  & Ambrosino 06F   \\
 $\BR{K_{e3}}/\BR{\mbox{2 tracks}}$ & 0.4978(35)    & Lai 04B      \\
 $\BR{\pi^0\pi^0}/\BR{3\pi^0}$      & 0.004446(25)  & Alexopoulos 04  \\
 $\BR{\pi^0\pi^0}/\BR{\pi^+\pi^-} $ & 0.4391(13)    & PDG etafit~\cite{Amsler:2008zzb}   \\
 $\BR{\gamma\gamma}/\BR{3\pi^0}$    & 0.00279(3)    & Adinolfi 03      \\
 $\BR{\gamma\gamma}/\BR{3\pi^0}$    & 0.00281(2)    & Lai 03      \\
 $\BR{\pi^+\pi^-}/\BR{\pi^+\pi^-(\gamma)}$  & 0.0208(3)  & Alavi-Harati 01B    \\
 $\BR{\pi^+\pi^-\gamma_{DE}}/\BR{\pi^+\pi^-\gamma}$  & 0.689(21)  & Abouzaid 06A      \\
 $\BR{\pi^+\pi^-\gamma_{DE}}/\BR{\pi^+\pi^-\gamma}$  & 0.683(11)  & Alavi-Harati 01B     \\
 $\BR{\pi^+\pi^-\gamma_{DE}}/\BR{\pi^+\pi^-\gamma}$  & 0.685(41)  & Ramberg 93      \\
\hline\hline
\end{tabular}
\end{table}

\subsubsection*{$K_S$  leading branching ratios and $\tau_S$}

\label{sec:KS}

KLOE has measured the ratio ${\rm BR}(K_S\to\pi e \nu)/{\rm BR}
(K_S\to\pi^+\pi^-)$ with 1.3\% precision~\cite{Ambrosino:2006si}, making
possible an independent determination of $|V_{us}|\,f_+(0)$ to better 
than 0.7\%. In \cite{Ambrosino:2006sh}, KLOE combines
the above measurement with their measurement
${rm \BR}(K_S\to\pi^+\pi^-)/{rm \BR}(K_S\to\pi^0\pi^0) = 2.2459(54)$.
Using the constraint that the $K_S$ BRs sum to unity and assuming
the universality of lepton couplings, they determine the BRs for 
$\pi^+\pi^-$, $\pi^0\pi^0$, $K_{e3}$, and $K_{\mu3}$ decays.

Our fit is an extension of the analysis in \cite{Ambrosino:2006sh}.
We perform a fit to the data on the $K_S$ BRs to $\pi^+\pi^-$, 
$\pi^0\pi^0$, and $K_{e3}$ that uses, in addition to the above 
two measurements:
\begin{itemize}
\item the measurement from NA48, 
$\Gamma{K_S\to\pi e\nu}/\Gamma{K_L\to\pi e\nu}$~ \cite{Batley:2007zzb}, where the denominator
is obtained from the results of our $K_L$ fit;
\item the measurement of $\tau_S$ (not assuming $CPT$) from NA48 
\cite{Amsler:2008zzb}, 89.589(70)~ps; 
\item the measurement of $\tau_S$ (not assuming $CPT$) from KTeV 
\cite{Amsler:2008zzb}, 89.58(13)~ps;
\item the result ${\rm BR}{K_{\mu3}}/{\rm BR}{K_{\e3}} = 0.66100(214)$,
obtained from the assumption of universal lepton couplings, the
values of the quadratic (vector) and linear (scalar) form-factor
parameters from our fit to form-factor data, and the long-distance 
electromagnetic corrections discussed in Sec.~\ref{sect:Kl3}.
\end{itemize} 
The free parameters are the four BRs listed above plus $\tau_S$.
With six inputs and one constraint (on the sum of the BRs), the fit
has one degree of freedom and gives $\chi^2=0.0038$ ($P=95\%$).
The results of the fit are listed in Table~\ref{tab:KSfit}.

 \begin{table}
 \caption{Results of fit to $K_S$ BRs and lifetime}
 \label{tab:KSfit}
\begin{center}
 \begin{tabular}{lcl}
 \hline\hline
 Parameter & Value & $S$ \\
 \hline
 $\mathcal{B}{\pi^+\pi^-}$ & 0.6920(5) & 1.0 \\
 $\mathcal{B}{\pi^0\pi^0}$ & 0.3069(5) & 1.0 \\
 $\mathcal{B}{K_{e3}}$ & $7.05(8)\times10^{-4}$ & 1.0 \\
 $\mathcal{B}{K_{\mu3}}$ & $4.66(6)\times10^{-4}$ & 1.0 \\
 $\tau_S$ & $4.66(6)\times10^{-4}$ & 1.0 \\
 \hline
 \end{tabular}
\end{center}
 \end{table}

\subsubsection*{$K^\pm$ leading branching ratios and $\tau^\pm$}

There are several new results providing information on $K^\pm_{\ell3}$
rates. 
The NA48/2 collaboration has  published measurements of the three ratios
\BRT{K_{e3}/\pi\pi^0}, \BRT{K_{\mu3}/\pi\pi^0}, and
\BRT{K_{\mu3}/K_{e3}} \cite{Batley:2006cj}.
These measurements are not independent; in our fit, we use the values
$\BRT{K_{e3}/\pi\pi^0} = 0.2470(10)$ and
$\BRT{K_{\mu3}/\pi\pi^0} = 0.1637(7)$ and take their correlation
into account.

KLOE has measured the absolute BRs for the
$K_{e3}$ and $K_{\mu3}$ decays
\cite{:2007xm}.
In $\phi\to K^+ K^-$ events, $K^+$ decays into $\mu\nu$ or $\pi\pi^0$
are used to tag a $K^-$ beam, and vice versa. KLOE performs four
separate measurements for each $K_{\ell3}$ BR, corresponding to the
different combinations of Kaon charge and tagging decay.
The final averages are $\BRT{K_{e3}} = 4.965(53)(38)\%$ and
$\BRT{K_{\mu3}} = 3.233(29)(26)\%$.
KLOE has also measured the absolute
branching ratio for the $\pi\pi^0$\cite{:2007xs} and $\mu \nu$ decay\cite{Ambrosino:2005fw}.

 Our fit takes into account the correlation between these values, as
 well as their dependence on the $K^\pm$ lifetime.
 The world average value for $\tau_\pm$ is nominally quite precise.
 However, the PDG error is scaled by 2.1; the confidence level for the
 average is 0.17\%. It is important to confirm the value of $\tau_\pm$.
 The  new measurement from KLOE,
 $\tau_\pm = 12.347(30)$~ns,  agrees with the PDG average.

Our fit for the six largest $K^\pm$ branching ratios and lifetime
uses the measurements in Table~\ref{tab:KPMinputs}, including the six 
measurements noted above. We have recently carried out a comprehensive 
survey of the $K^\pm$ data set, which led to the elimination of 11
measurements currently in the 2008 PDG fit. 
 Finally, we note that after the elimination
of the 1970 measurement of $\Gamma(\pi^\pm\pi^\pm\pi^\mp)$ from Ford et al.(
Ford70 in Ref.~\cite{Amsler:2008zzb}), the input data set provides no strong
 constraint on the $\pi^\pm\pi^\pm\pi^\mp$ branching ratio, which increases the uncertainties
on the resulting BR values. The fit uses 17 input measurements, seven
free parameters, and one constraint, giving 11 degrees of freedom.
We obtain the results in Table~\ref{tab:KpmBR}. 
The fit gives $\chi^2=25.8$ ($P=0.69\%$). The comparatively low $P$-value
reflects some tension between the KLOE and NA48/2 measurements of the 
$K_{\ell3}$ branching ratios.

\begin{table}
\caption{\label{tab:KpmBR}
Results of fit to $K^\pm$ BRs and lifetime.}
\begin{center}
\begin{tabular}{l|c|r}
Parameter & Value & $S$ \\
\hline
\BRT{K_{\mu2}}      & 63.47(18)\%   & 1.3 \\
\BRT{\pi\pi^0}      & 20.61(8)\%   & 1.1 \\
\BRT{\pi\pi\pi}     &  5.573(16)\% & 1.2 \\
\BRT{K_{e3}}        &  5.078(31)\%  & 1.3 \\
\BRT{K_{\mu3}}      &  3.359(32)\%  & 1.9 \\
\BRT{\pi\pi^0\pi^0} &  1.757(24)\%  & 1.0 \\
$\tau_\pm$         & 12.384(15)~ns & 1.2 \\
\end{tabular}
\end{center}
\vskip 0.3cm
\end{table}

\begin{figure}[ht]
\begin{center}
\resizebox{0.7\textwidth}{!}{\includegraphics{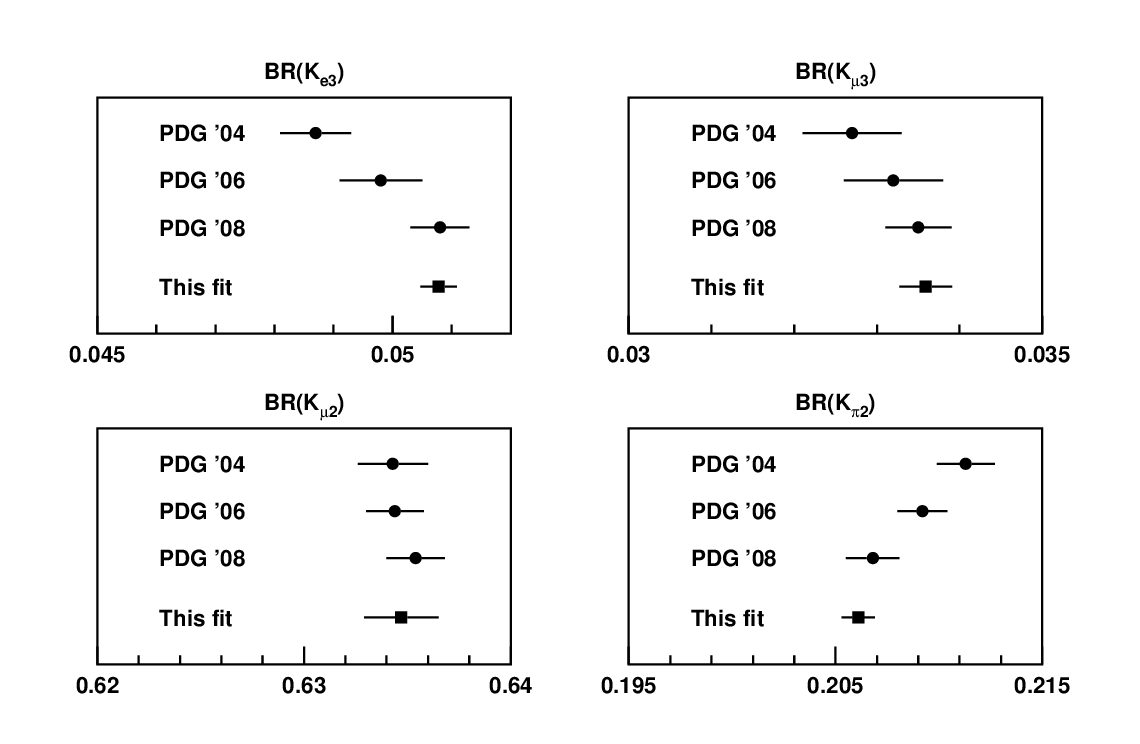}}
\end{center}
\caption{\label{fig:kpmavg}
Evolution of average values for main $K^\pm$ BRs.}
\end{figure}

Both the significant evolution of the average values of the $K_{\ell3}$
BRs and the effect of the correlations with \BRT{\pi\pi^0} are evident
in \Fig{fig:kpmavg}.

\begin{table}
\caption{Input data used for the fit to $K^\pm$ BRs and lifetime (all the references refer to PDG08~\cite{Amsler:2008zzb}). 
The two 1995 values of the
$K^\pm$ lifetime from Koptev et al. are averaged with $S=1.6$ before being
included in the fit as a single value. 
}
\center
\begin{tabular}{lcl}
\hline\hline
  Parameter                                          & Value         & Source           \\[0.5ex]\hline
  $\tau_{K^\pm}$                                     & 12.368(41) ns & Koptev 95 (*)    \\
  $\tau_{K^\pm}$                                     & 12.380(16) ns & Ott 71           \\ 
  $\tau_{K^\pm}$                                     & 12.443(38) ns & Fitch 65B        \\ 
  $\tau_{K^\pm}$                                     & 12.347(30) ns & Ambrosino 08     \\ 
  $\BR{K_{\mu2}}$                                    & 0.6366(17)    & Ambrosino 06A    \\ 
  $\BR{\pi\pi^0}$                                    & 0.2066(11)    & \cite{:2007xs}\\
  $\BR{\pi\pi^0}/\BR{K_{\mu2}}$                      & 0.3329(48)    & Usher 92         \\ 
  $\BR{\pi\pi^0}/\BR{K_{\mu2}}$                      & 0.3355(57)    & Weissenberg 76   \\
  $\BR{\pi\pi^0}/\BR{K_{\mu2}}$                      & 0.3277(65)    & Auerbach 67      \\
  $\BR{K_{e3}}$                                      & 0.04965(53)   & Ambrosino 08A    \\
  $\BR{K_{e3}}/\BR{\pi\pi^0\!+K_{\mu3}+\!\pi2\pi^0}$ & 0.1962(36)    & Sher 03          \\ 
  $\BR{K_{e3}}/\BR{\pi\pi^0}$                        & 0.2470(10)    & Batley 07A       \\
  $\BR{K_{\mu3}}$                                    & 0.03233(39)   & Ambrosino 08A    \\
  $\BR{K_{\mu3}}/\BR{\pi\pi^0}$                      & 0.1636(7)     & Batley 07A       \\  
  $\BR{K_{\mu3}}/\BR{K_{e3}}$                        & 0.671(11)     & Horie 01         \\ 
  $\BR{\pi\pi^0\pi^0}$                               & 0.01763(26)   & Aloisio 04A      \\ 
  $\BR{\pi\pi^0\pi^0}/\BR{\pi\pi\pi}$                & 0.303(9)      & Bisi 65          \\
\hline\hline
\end{tabular}
\label{tab:KPMinputs}
\end{table}

\subsubsection*{Measurement of BR($K_{e2})$/BR($K_{\mu2}$)}

Experimental knowledge of $K_{e2}/K_{\mu2}$ was poor until recently.
 The current world average
 $R_K = \BRT{K_{e2}}/\BRT{K_{\mu2}}= (2.45 \pm 0.11) \times 10^{-5}$ dates back to
 three experiments
of the 1970s~\cite{Amsler:2008zzb} and has a precision of about 5\%.
Two new  measurements were reported recently by NA62 and KLOE
 (see Tab.~\ref{tab:ke2kmu2}).
A preliminary result based on about 14,000 $K_{e2}$ events, was presented at the 2009 winter
conferences by the KLOE collaboration~\cite{spadaroke2}.
Preliminary result from NA62, based on about 50,000 $K_{e2}$ events from the 2008
 data set was presented in at KAON 2009~\cite{Gudzoke2}.
Both the KLOE and the NA62 measurements are inclusive with respect to final state radiation
  contribution due to bremsstrahlung.
The small contribution of $K_{l2\gamma}$ events from direct photon emission from the decay vertex was subtracted by each of the experiments.
Combining these new results with the current PDG value yields a current world average of
\begin{equation}
R_K  = ( 2.498 \pm 0.014 ) \times 10^{-5},
\label{eqn:ke2kmu2}
\end{equation}
in good agreement with the SM expectation~\cite{Cirigliano:2007ga} and,
 with a relative error of $0.56\%$, an order of magnitude more precise than the previous
 world average.

\begin{table}[t]
      \caption{Results and prediction for $R_K =  \BRT{K_{e2}}/\BRT{K_{\mu2}}$.}
  \begin{center}
      \begin{tabular}{lc}
        \hline \hline
        & $R_K$ $[10^{-5}]$  \\ \hline
        PDG                    & $2.45 \pm 0.11$ \\
        NA48/2    & $2.500 \pm 0.016$ \\
        KLOE             & $2.493 \pm 0.031 $ \\ \hline
        SM prediction & $2.477 \pm 0.001$ \\
        \hline \hline
                                                  & \\*[-3mm]
      \end{tabular}
      \label{tab:ke2kmu2}
  \end{center}
\end{table}

\subsubsection*{\mathversion{bold}Measurements of $K_{\ell3}$ slopes}

\label{sect:explsl}

For $K_{e3}$ decays, recent measurements of the quadratic slope parameters
of the vector form factor (${\lambda_+',\lambda_+''}$), see Eq.~\ref{Taylor} are available from
KTeV \cite{Alexopoulos:2004sy},
KLOE \cite{Ambrosino:2006gn}, ISTRA+ \cite{Yushchenko:2004zs}, and
NA48 \cite{Lai:2004kb}.

We show the results of a fit to the $K_L$ and $K^-$ data in the
first column of \Tab{tab:e3ff}, and to only the $K_L$ data in the
second column. With correlations correctly
taken into account, both fits give good
values of $\chi^2/{\rm ndf}$. The significance of the quadratic
term is $4.2\sigma$ from the fit to all data, and $3.5\sigma$ from
the fit to $K_L$ data only.
\begin{table}
\caption{Average of quadratic fit results for $K_{e3}$ slopes.}
\center
\label{tab:e3ff}
\begin{tabular}{lcc}
\hline\hline
& $K_L$ and $K^-$ data & $K_L$ data only \\
& 4 measurements & 3 measurements \\
& $\chi^2/{\rm ndf} = 5.3/6$ (51\%) &
  $\chi^2/{\rm ndf} = 4.7/4$ (32\%)\\
\hline
\SN{\lambda_+'}{3}
     & $25.2\pm0.9$ & $24.9\pm1.1$ \\
\SN{\lambda_+''}{3}
     & $1.6\pm0.4$ & $1.6\pm0.5$ \\
$\rho(\lambda_+',\lambda+'')$
     & $-0.94$ & $-0.95$ \\
$I(K^0_{e3})$
     & 0.15463(21) & 0.15454(29) \\
$I(K^\pm_{e3})$
     & 0.15900(22) & 0.15890(30) \\
\hline\hline
\end{tabular}
\end{table}

Including or excluding the $K^-$ slopes
has little impact on the values of $\lambda_+'$ and $\lambda_+''$;
in particular, the values of the phase-space integrals change by just
0.07\%. The errors on the phase-space integrals are significantly
smaller when the $K^-$ data are included in the average. 

KLOE, KTeV, and NA48 also quote the values shown in \Tab{tab:pole}
for $M_V$ from pole (see Eq.~\ref{pole}) fits to $K_{L\:e3}$ data. The average value of
$M_V$ from all three experiments is
$M_V = 875\pm5$~MeV with $\chi^2/{\rm ndf} = 1.8/2$.
The three values are quite compatible with each other and
reasonably close to the known value of the $K^{\pm*}(892)$
mass ($891.66\pm0.26$~MeV). The values for $\lambda_+'$ and $\lambda_+''$
from expansion of the pole parametrization are qualitatively in
agreement with the average of the quadratic fit results.
More importantly, for the evaluation of the phase-space
integrals, using the average of quadratic or pole fit results gives
values of $I(K^0_{e3})$ that differ by just 0.03\%.
%
%

\begin{table}
\caption{Pole fit results for $K^0_{e3}$ slopes.}
\center
\begin{tabular}{lc|c}
\hline\hline
Experiment & $M_V$ (MeV) & $\left<M_V\right> = 875\pm5$ MeV \\
KLOE & $870\pm6\pm7$ &  $\chi^2/{\rm ndf} = 1.8/2$ \\
KTeV & $881.03\pm7.11$ & \SN{\lambda_+'}{3} = 25.42(31) \\
NA48 & $859\pm18$ & $\lambda_+''=2\times\lambda_+'^{\,2}$ \\
& &  $I(K^0_{e3})$ = 0.15470(19) \\
\hline\hline
\end{tabular}
\label{tab:pole}
\end{table}

 For $K_{\mu3}$ decays, recent measurements of the
 slope parameters $({\lambda_+',\lambda_+'',\lambda_0})$ are
 available from  KTeV \cite{Alexopoulos:2004sy},  KLOE \cite{:2007yza},
 ISTRA+ \cite{Yushchenko:2003xz}, and NA48 \cite{Lai:2007dx}.
 We will not use the  ISTRA+ result for the average because systematic
 errors have not been provided.
 We use the $K_{e3}-K_{\mu3}$ averages provided by the experiments
 for KTeV and KLOE. NA48 does not provide such an average, so we
 calculate it for inclusion in the fit.

 We have studied the statistical sensitivity of the
 form-factor slope measurements using Monte Carlo techniques.
 The conclusions of this study are a) 
 that neglecting a quadratic term in the parametrization of the 
 scalar form factor when fitting results  leads to  a shift of the value
 of the linear term by about 3.5 times the value of the quadratic
 term; and b)
 that because of correlations, it is impossible to measure the 
 quadratic slope parameter from quadratic fits to the data at 
 any plausible level of statistics.
 The use of the linear representation of the scalar form factor 
 is thus inherently unsatisfactory. The effect is relevant
 when testing the CT theorem Eq.~(\ref{eq:CTrel}) discussed
 in section~\ref{sec:CTtest}.

 The results of the combination are listed in \Tab{tab:l3ff}.

\begin{figure}[ht]
\begin{center}
\resizebox{0.95\textwidth}{!}{\includegraphics{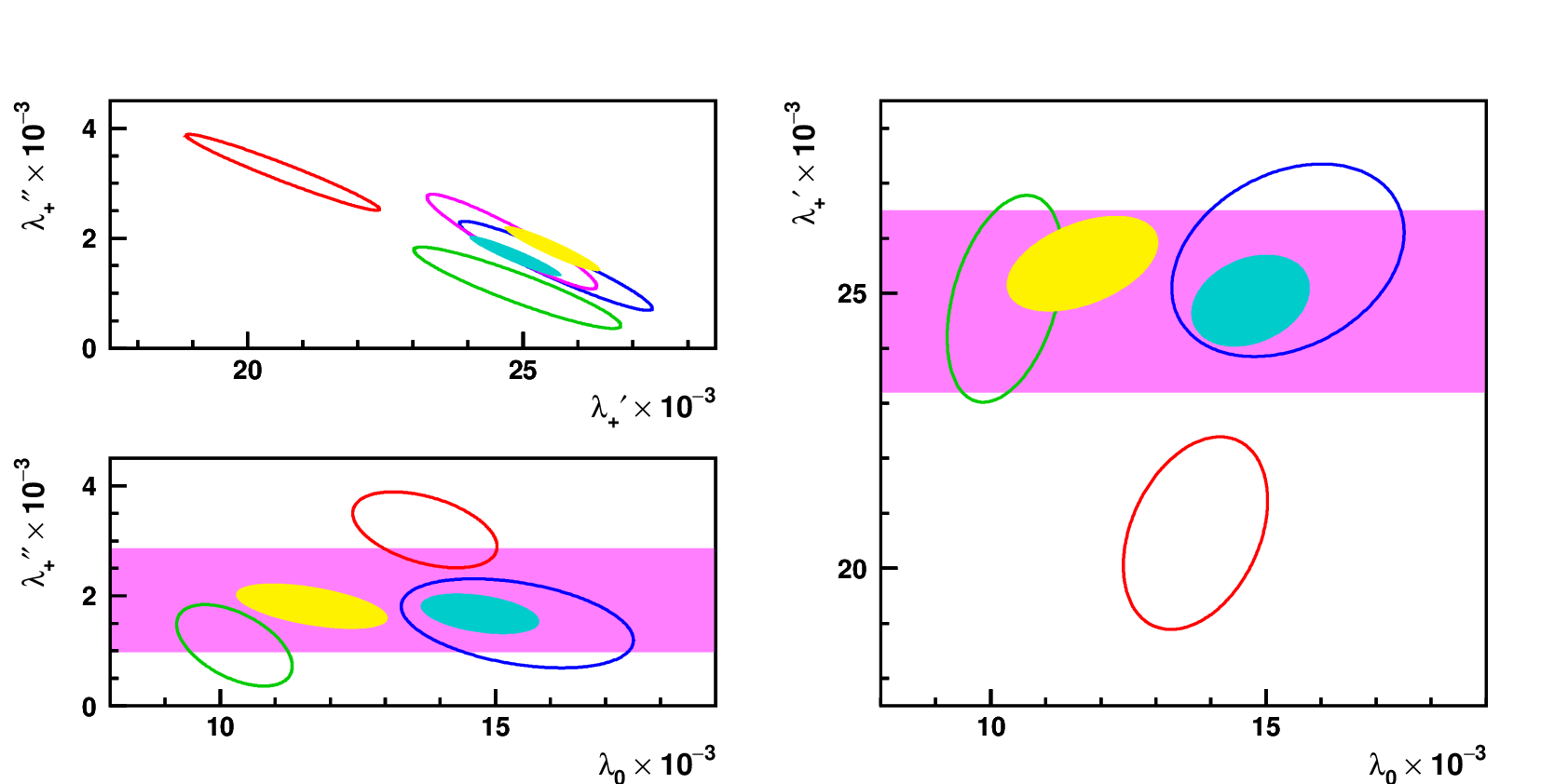}}
\end{center}
\caption{\label{fig:FFm3} 1-$\sigma$ contours for $\lambda_+'$, $\lambda_+''$,
  $\lambda_0$ determinations from  KLOE(blue ellipse),
  KTeV(red ellipse), NA48(green ellipse), and world 
 average with(filled yellow ellipse) and without(filled cyan ellipse) 
 the NA48 $K_{\mu3}$ result.}
\end{figure}

\begin{table}
\caption{Averages of quadratic fit results for $K_{e3}$ and $K_{\mu3}$ slopes.}
\center
\begin{tabular}{lc}
\hline\hline
$\chi^2/{\rm ndf}$               & 29/8 $(3\times 10^{-4})$  \\
$\lambda_+'\times 10^3 $         & $24.5\pm0.9$ ($S=1.1$)  \\
$\lambda_+'' \times 10^3 $       & $1.8\pm0.4$ ($S=1.3$)   \\
$\lambda_0\times 10^3 $          & $11.7\pm1.4$ ($S=1.9$)  \\
$\rho(\lambda_+',\lambda_+'')$   & $-0.94$                          \\
$\rho(\lambda_+',\lambda_0)$     & $+0.44$                          \\
$\rho(\lambda_+'',\lambda_0)$    & $-0.52$                          \\
$I(K^0_{e3})$                    & 0.15449(20)	        \\
$I(K^\pm_{e3})$                  & 0.15885(21)	        \\
$I(K^0_{\mu3})$                  & 0.10171(32)	        \\
$I(K^\pm_{\mu3})$                & 0.10467(33)	        \\
$\rho(I_{e3},I_{\mu3})$          & $+0.53$               \\
\hline\hline
\end{tabular}
\label{tab:l3ff}
\end{table}

The value of $\chi^2/{\rm ndf}$ for all measurements is terrible;
we quote the results with scaled errors. This leads to
errors on the phase-space integrals that are $\sim$60\% larger after inclusion
of the new  $K_{\mu3}$  NA48 data.

The evaluations of the phase-space integrals for all four modes are listed in each case.
Correlations are fully accounted for, both in the fits and in the evaluation
of the integrals. The correlation matrices for the integrals are of the
form
\begin{displaymath}
\begin{array}{cccc}
+1 & +1 & \rho & \rho \\
+1 & +1 & \rho & \rho \\
\rho & \rho & +1 & +1 \\
\rho & \rho & +1 & +1 \\
\end{array}
\end{displaymath}
where the order of the rows and columns is $K^0_{e3}$, $K^\pm_{e3}$,
$K^0_{\mu3}$, $K^\pm_{\mu3}$,
and $\rho = \rho(I_{e3},I_{\mu3})$ as listed in the table.

Adding the $K_{\mu3}$ data to the fit does not cause drastic changes
to the values of the phase-space integrals for the $K_{e3}$ modes:
the values for $I(K^0_{e3})$ and $I(K^\pm_{e3})$ in \Tab{tab:l3ff}
are qualitatively in agreement with those in \Tab{tab:e3ff}.
As in the case of the fits to the $K_{e3}$ data only, the significance of the
quadratic term in the vector form factor is strong ($3.6\sigma$ from the
fit to all data).


\subsection{$\Vus$ determination from tau decays}
\label{sec:cab:tau}

A very precise determination of $V_{us}$ can be obtained from
the semi-inclusive hadronic decay width of the $\tau$ lepton into final
states with strangeness \cite{Gamiz:2004ar,Gamiz:2007qs}. 
The ratio of the Cabibbo-suppressed and Cabibbo-allowed
$\tau$ decay widths directly measures $(V_{us}/V_{ud})^2$, up to very small
SU(3)-breaking corrections which can be theoretically estimated with the
needed accuracy.

The inclusive character of the total $\tau$ hadronic width renders
possible an accurate calculation of the ratio
\cite{Braaten:1988hc,Narison:1988ni,Braaten:1991qm,LeDiberder:1992te,Pich:1994zx}
\begin{equation}\label{eq:cab:r_tau_def}
 R_\tau \,\equiv\, { \Gamma [\tau^- \to \nu_\tau
 \,\mathrm{hadrons}\, (\gamma)] \over \Gamma [\tau^- \to \nu_\tau e^-
 {\bar \nu}_e (\gamma)] }\, = \, R_{\tau,V} + R_{\tau,A} + R_{\tau,S}\, ,
\end{equation}
using analyticity constraints and the operator product expansion.
One can separately compute the contributions associated with
specific quark currents:
$R_{\tau,V}$ and $R_{\tau,A}$ correspond to the Cabibbo-allowed
decays through the vector and axial-vector currents, while
$R_{\tau,S}$ contains the remaining Cabibbo-suppressed
contributions.

To a first approximation the Cabibbo mixing can be directly obtained
from experimental measurements, without any theoretical input.
Neglecting the small SU(3)-breaking corrections from the $m_s-m_d$
quark-mass difference, one gets:
\begin{equation}
 |V_{us}|^{\mathrm{SU(3)}} =\: |V_{ud}| \:\left(\frac{R_{\tau,S}}{R_{\tau,V+A}}\right)^{1/2}
 =\: 0.210\pm 0.003\, .
\end{equation}
We have used $|V_{ud}| = 0.97425\pm 0.00022$ (cf. Eq.~(\ref{Vud})),    
$R_\tau = 3.640\pm 0.010$       
and the value $R_{\tau,S}=0.1617\pm 0.0040$ \cite{Gamiz:2007qs}, which results from the
recent BaBar \cite{Aubert:2007mh} and Belle \cite{:2007rf} measurements of Cabibbo-suppressed
tau decays \cite{Banerjee:2007nk}.
The new branching ratios measured by BaBar and Belle are all smaller than the previous
world averages, which translates into a smaller value of $R_{\tau,S}$ and $|V_{us}|$.
For comparison, the previous value $R_{\tau,S}=0.1686\pm 0.0047$ \cite{Davier:2005xq} 
resulted in $|V_{us}|^{\mathrm{SU(3)}}=0.215\pm 0.003$.

This rather remarkable determination is only slightly shifted by
the small SU(3)-breaking contributions induced by the strange quark mass.
These corrections can be theoretically
estimated through a QCD analysis of the difference
\cite{Gamiz:2004ar,Gamiz:2007qs,Pich:1999hc,Chen:2001qf,Chetyrkin:1998ej,Korner:2000wd,Maltman:2006st,Kambor:2000dj,Maltman:1998qz,Baikov:2004tk}
\begin{equation}
 \delta R_\tau  \,\equiv\,
 {R_{\tau,V+A}\over |V_{ud}|^2} - {R_{\tau,S}\over |V_{us}|^2}\, .
\end{equation}
Since the strong interactions are flavor blind, this quantity vanishes in the SU(3) limit.
The only non-zero contributions are proportional 
to the mass-squared difference $m_s^2-m_d^2$ or to vacuum expectation
values of SU(3)-breaking operators such as $\delta O_4
\equiv \langle 0|m_s\bar s s - m_d\bar d d|0\rangle = (-1.4\pm 0.4)
\cdot 10^{-3}\; \mathrm{GeV}^4$ \cite{Gamiz:2004ar,Pich:1999hc}. The dimensions of these operators
are compensated by corresponding powers of $m_\tau^2$, which implies a strong
suppression of $\delta R_\tau$ \cite{Pich:1999hc}:
\begin{equation}\label{eq:cab:dRtau}
 \delta R_\tau \,\approx\,   24\, S_{\mathrm{EW}}\; \left\{ {m_s^2(m_\tau^2)\over m_\tau^2} \,
 \left( 1-\epsilon_d^2\right)\,\Delta(\alpha_s)
 - 2\pi^2\, {\delta O_4\over m_\tau^4} \, Q(\alpha_s)\right\}\, ,
\end{equation}
where $\epsilon_d\equiv m_d/m_s = 0.053\pm 0.002$ \cite{Leutwyler:1996qg}.
The perturbative QCD
corrections $\Delta(\alpha_s)$ and
$Q(\alpha_s)$ are known to $O(\alpha_s^3)$ and $O(\alpha_s^2)$,
respectively \cite{Pich:1999hc,Baikov:2004tk}.

The theoretical analysis of $\delta R_\tau$
involves the two-point vector and axial-vector correlators,
which have transverse ($J=1$) and longitudinal ($J=0$) components.
The $J=0$ contribution to $\Delta(\alpha_s)$ shows a rather
pathological behavior, with clear signs of being a non-convergent perturbative
series. Fortunately, the corresponding longitudinal contribution to
$\delta R_\tau$ can be estimated phenomenologically with a much better
accuracy, $\delta R_\tau|^{L}\, =\, 0.1544\pm 0.0037$ 
\cite{Gamiz:2004ar,Jamin:2006tj},
because it is dominated by far by the well-known $\tau\to\nu_\tau\pi$
and $\tau\to\nu_\tau K$ contributions \cite{Maltman:2001gc}. 
To estimate the remaining  $L+T$ 
component, one needs an input value for the strange quark mass. Taking the
range
$m_s(m_\tau) = (100\pm 10)\:\mathrm{MeV}$ \
[$m_s(2\:\mathrm{GeV}) = (96\pm 10)\:\mathrm{MeV}$],
which includes the most recent determinations of $m_s$ from QCD sum rules
and lattice QCD \cite{Jamin:2006tj},
one gets finally $\delta R_{\tau,th} = \delta R_\tau|^{L} + \delta R_\tau|^{L+T}=
0.216\pm 0.016$, which implies \cite{Gamiz:2007qs}
\begin{equation}\label{eq:cab:Vus_det}
 |V_{us}| \, =\, \left(\frac{R_{\tau,S}}{\frac{R_{\tau,V+A}}{|V_{ud}|^2}-\delta
 R_{\tau,\mathrm{th}}}\right)^{1/2} \; =\;
 0.2165\pm 0.0026_{\mathrm{\, exp}}\pm 0.0005_{\mathrm{\, th}}\, .
\end{equation}
A larger central value, $|V_{us}| = 0.2212\pm 0.0031$, 
is obtained with the old world average for $R_{\tau,S}$.


Notice that the theoretical input only appears through the quantity
$\delta R_{\tau,th}$,
which is one order of magnitude smaller than the ratio
$R_{\tau,V+A}/|V_{ud}|^2 = 3.665\pm 0.012$. Theoretical uncertainties are
thus very suppressed,
although a number of issues deserve further investigation.
These include (i) an assessment of the uncertainty due
to different prescriptions (Contour Improved Perturbation Theory versus
Fixed Order Perturbation Theory)
for the slow-converging $D=2$, L+T correlator series,
which could shift $\vert V_{us}\vert$ by up to $\sim
0.0020$~\cite{Maltman:2008na};
(ii) addressing the stability of the extracted $\vert V_{us}\vert$
by using  alternate sum rules that involve different weights, $w(s)$,
and/or spectral
integral endpoints $s_0<m_\tau2$~\cite{Maltman:2006st,Maltman:2009bh}.
With theory errors at the level of Eq. (\ref{eq:cab:Vus_det}), 
experimental errors would dominate, in contrast to the
situation encountered in $K_{\ell 3}$ decays.

The phenomenological determination of $\delta R_\tau|^{L}$ contains a hidden
dependence on $V_{us}$ through the input value of the Kaon decay constant
$f_K$. Although the numerical impact of this dependence is negligible,
it can be taken explicitly into account. 
Using the measured $K^-/\pi^-\to\bar\nu_\mu \mu^-$ decay widths and the $\tau$
lifetime \cite{Amsler:2008zzb}, one can determine the Kaon and pion contributions to $R_\tau$ 
with better accuracy than the direct $\tau$ decay measurements, with the results
$R_\tau|^{\tau^-\to\nu_\tau K^-} = (0.04014\pm 0.00021)$ and
$R_\tau|^{\tau^-\to\nu_\tau \pi^-} = (0.6123\pm 0.0025)$.
The corresponding longitudinal contributions are just given by
$R_\tau|^{\tau^-\to\nu_\tau P^-}_{_L} \equiv 
R_\tau|^{\tau^-\to\nu_\tau P^-} - R_\tau|^{\tau^-\to\nu_\tau P^-}_{_{L+T}}=
-2 (m_P^2/m_\tau^2) R_\tau|^{\tau^-\to\nu_\tau P^-}$ 
($P=K,\pi$).

Subtracting the longitudinal contributions from Eq.~(\ref{eq:cab:Vus_det}),
one gets an improved formula to determine $V_{us}$ with the best possible
accuracy \cite{Gamiz:2007qs}:
\begin{equation}\label{eq:cab:Vus_improvedForm}
 |V_{us}|^2 \, =\, \frac{\tilde R_{\tau,S}}{\frac{\tilde R_{\tau,V+A}}{|V_{ud}|^2}-\delta
 \tilde R_{\tau,\mathrm{th}}} \,\equiv\,
 \frac{R_{\tau,S} - R_\tau|^{\tau^-\to\nu_\tau K^-}_{_L}}{\frac{R_{\tau,V+A}
 -R_\tau|^{\tau^-\to\nu_\tau \pi^-}_{_L}}{|V_{ud}|^2}-
 \delta\tilde R_{\tau,\mathrm{th}}}
 \, ,
\end{equation}
where
$\delta\tilde R_{\tau,\mathrm{th}}\,\equiv\, \delta\tilde R_\tau|^{L}
+ \delta R_{\tau,\mathrm{th}}|^{L+T} =
(0.033\pm 0.003) + (0.062\pm 0.015) = 0.095\pm 0.015$.
The subtracted longitudinal correction $\delta\tilde R_\tau|^{L}$ is now much
smaller because it does not contain any pion or Kaon contribution.
Using the same input values for $R_{\tau,S}$ and $R_{\tau,V+A}$, one recovers
the $V_{us}$ determination obtained before in Eq.~(\ref{eq:cab:Vus_det}), with 
an error of $\pm\, 0.0030$.

Sizable changes on the experimental determination of $R_{\tau,S}$ are to be expected from
the full analysis of  the huge BaBar and Belle data samples. In particular, the high-multiplicity
decay modes are not well known at present
and their effect has been just roughly estimated or simply ignored.
Thus, the result (\ref{eq:cab:Vus_det}) could easily fluctuate in the near future.
However, it is important to realize that the final error of the $V_{us}$ determination from
$\tau$ decay is likely to remain
 dominated by the experimental uncertainties. If $R_{\tau,S}$
is measured with a 1\% precision, the resulting $V_{us}$ uncertainty will
get reduced to around 0.6\%, i.e. $\pm 0.0013$, making $\tau$ decay the competitive source of
information about $V_{us}$.

An accurate measurement of the invariant-mass distribution of the final ha\-drons
in Cabibbo-suppressed $\tau$ decays could make possible a simultaneous determination
of $V_{us}$ and the strange quark mass, through a correlated analysis of
several SU(3)-breaking observables constructed with weighted moments of the
hadronic distribution \cite{Gamiz:2004ar,Pich:1999hc,Chen:2001qf}.
However, the extraction of $m_s$ suffers from
theoretical uncertainties related to the convergence of the associated perturbative QCD series.
A better understanding of these QCD corrections is needed in order to improve the present
determination of $m_s$ 
\cite{Gamiz:2004ar,Pich:1999hc,Maltman:2006st,Kambor:2000dj,Maltman:1998qz,Baikov:2004tk}.

\subsection{Physics Results}
\label{sec:results}
In this section we summarize the results for $\Vus$ 
discussed in the previous sections and based on these results we 
give constraints on physics beyond the SM. Instead of averages for
lattice results for $f_K/f_\pi$ we use  $f_K/f_\pi=1.189(7)$ by HPQCD 
\cite{Follana:2007uv} for illustrative purposes (cf. the discussion at the 
end of section \ref{sect:CabibboLattice}).

\subsubsection{Determination of $\vert V_{us}\vert\times f_{+}(0)$ and
 $\vert V_{us}\vert/\vert V_{ud}\vert\times f_K/f_\pi$}

 This section describes the results that are independent of the 
 theoretical parameters   $f_{+}(0)$ and $f_K/f_\pi$.

\subsubsection*{Determination of $\vert V_{us}\vert\times f_{+}(0)$ }
 The value of $\vert V_{us}\vert\x f_{+}(0)$ has been determined
 from (\ref{eq:Mkl3}) using
 the world average values reported in section~\ref{sec:data}
 for lifetimes, branching ratios and phase space integrals, 
 and the radiative and $SU(2)$ breaking corrections
 discussed in section~\ref{sect:Kl3}.

\begin{table}[t]
\caption{\label{tab:Vusf0}
 Summary of $\vert V_{us}\vert\x f_{+}(0)$ determination from all channels.}
\begin{center}
\begin{tabular}{l|c|c|c|c|c|c}
mode               & $\vert V_{us}\vert\x f_{+}(0)$  & \% err & BR   & $\tau$ & $\Delta$& Int \\
\hline
$K_L \to \pi e \nu$     & 0.2165(5)    & 0.26   & 0.09 & 0.20   & 0.11  & 0.06\\
$K_L \to \pi \mu \nu$   & 0.2175(6)    & 0.32   & 0.15 & 0.18   & 0.15  & 0.16\\
$K_S \to \pi e \nu$     & 0.2157(13)   & 0.61   & 0.60 & 0.03   & 0.11  & 0.06\\
$K^\pm \to \pi e \nu$   & 0.2162(11)   & 0.52   & 0.31 & 0.09   & 0.41  & 0.06\\
$K^\pm \to \pi \mu \nu$ & 0.2168(14)   & 0.65   & 0.47 & 0.08   & 0.42  & 0.16\\
 average & 0.2166(5)  &    &  &   &   & 
\end{tabular}
\end{center}
\end{table}

\begin{figure}[t]
\begin{center}
\resizebox{0.6\textwidth}{!}{\includegraphics{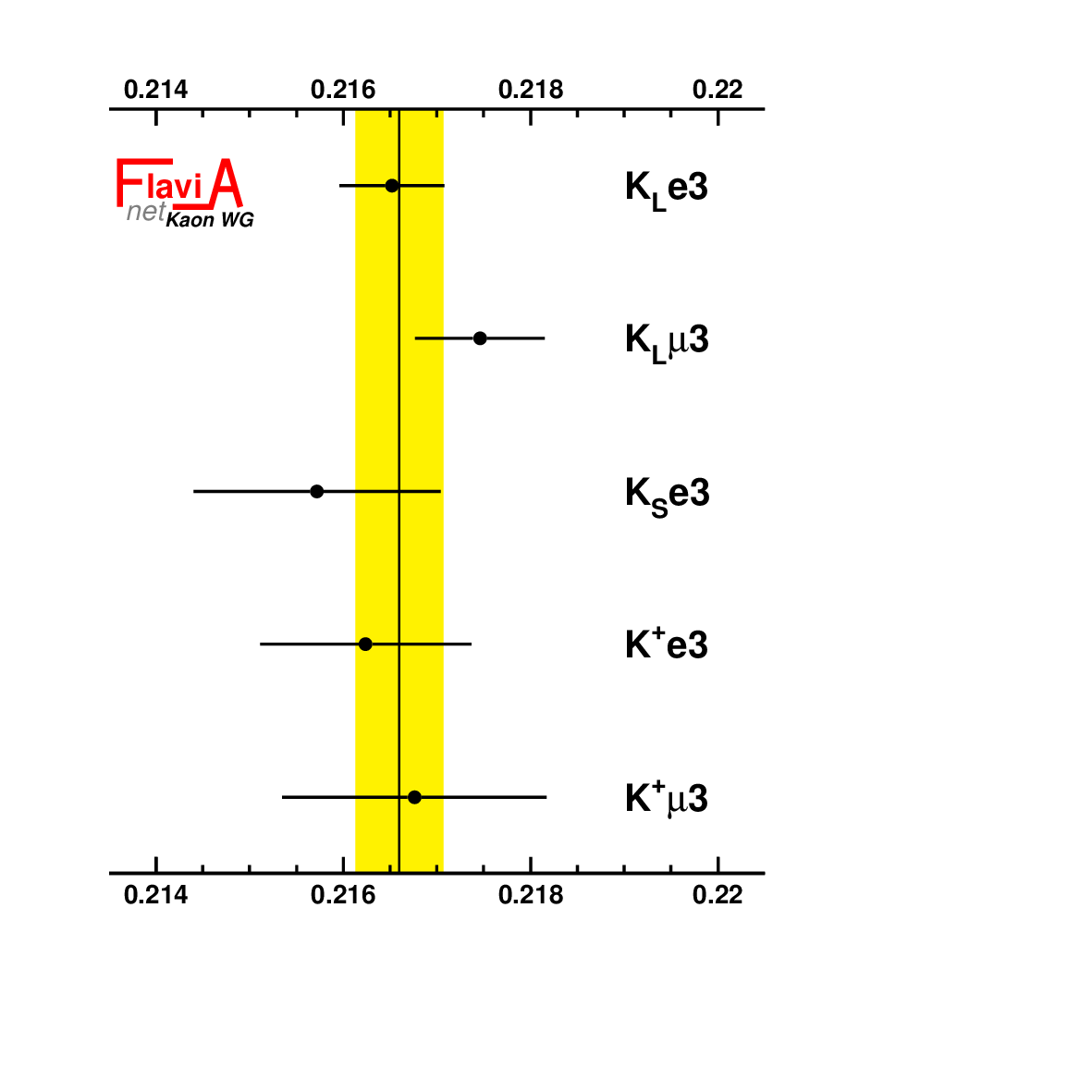}}
\end{center}
\caption{ Display of $\vert V_{us}\vert\x f_{+}(0) $ for all channels. 
\label{fig:Vusf0} }
\end{figure}

The results  are given in Tab.~\ref{tab:Vusf0},
and are shown in \Fig{fig:Vusf0} for  $K_L \to \pi e \nu$, $K_L \to \pi \mu \nu$,
$K_S \to \pi e \nu$, $K^\pm \to \pi e \nu$, $K^\pm \to \pi \mu \nu$, 
and for the combination.
The average,   
\be
\vert V_{us}\vert\x f_+(0)=0.2166(5),
\label{eq:vusfres}
\end{equation}
 has an uncertainty of about of $0.2\%$.  
The results from the five modes are in good agreement, the
fit probability is 55\%. 
In particular, comparing the values of $\vert V_{us}\vert\x f_{+}(0)$
obtained from $K^0_{\ell3}$ and $K^\pm_{\ell3}$ we obtain
a value of the SU(2) breaking correction 
$$
\delta^K_{SU(2)_{exp.}}= 5.4(8)\%
$$
in agreement with the CHPT calculation reported in Eq.~\ref{eq:iso-brk}:
$\delta^K_{SU(2)}= 5.8(8)\%$.

\subsubsection{A test of lattice calculation: the Callan-Treiman relation}
\label{sec:CTtest}
 As described in Sec.~\ref{sec:ffpara} the Callan-Treiman relation fixes the
 value of scalar form factor at $t=m_K^2-m_\pi^2$ 
(the so-called Callan-Treiman point) to the
 ratio $(f_K/f_\pi)/f_+(0)$. The dispersive parametrization for the scalar
 form factor proposed in~\cite{Bernard:2006gy} and discussed in Sec.~\ref{sec:ffpara}
 allows the available measurements of the scalar form factor to be  transformed
 into a precise information on $(f_K/f_\pi)/f_+(0)$, completely independent of
 the lattice estimates.

 Very recently  KLOE~\cite{Testa:2008xz}, KTeV~\cite{KTeVprivate}, ISTRA+~\cite{ISTRAprivate},  
 and NA48\cite{Lai:2007dx} have produced results on the
 scalar FF behavior using the dispersive parametrization.
 The results are given in Tab.~\ref{tab:ffdis} for all four experiments.

\begin{table}[ht]
\caption{\label{tab:ffdis}
Experimental results for log(C).}
\centering
\begin{tabular}{l|c|c}
Experiment & $\log(C)$& mode \\
\hline
KTeV   & 0.195(14)  & $K_{L\mu3}$   \\
KLOE   & 0.217(16)  & $K_{L\mu3}$ and $K_{Le3}$   \\
NA48   & 0.144(14)  & $K_{L\mu3}$   \\
ISTRA+ & 0.211(13)  & $K^-_{\mu3}$    \\
\hline
\end{tabular}\\[1mm]
\end{table}


\begin{figure}[t]
\begin{center}
\resizebox{0.6\textwidth}{!}{\includegraphics{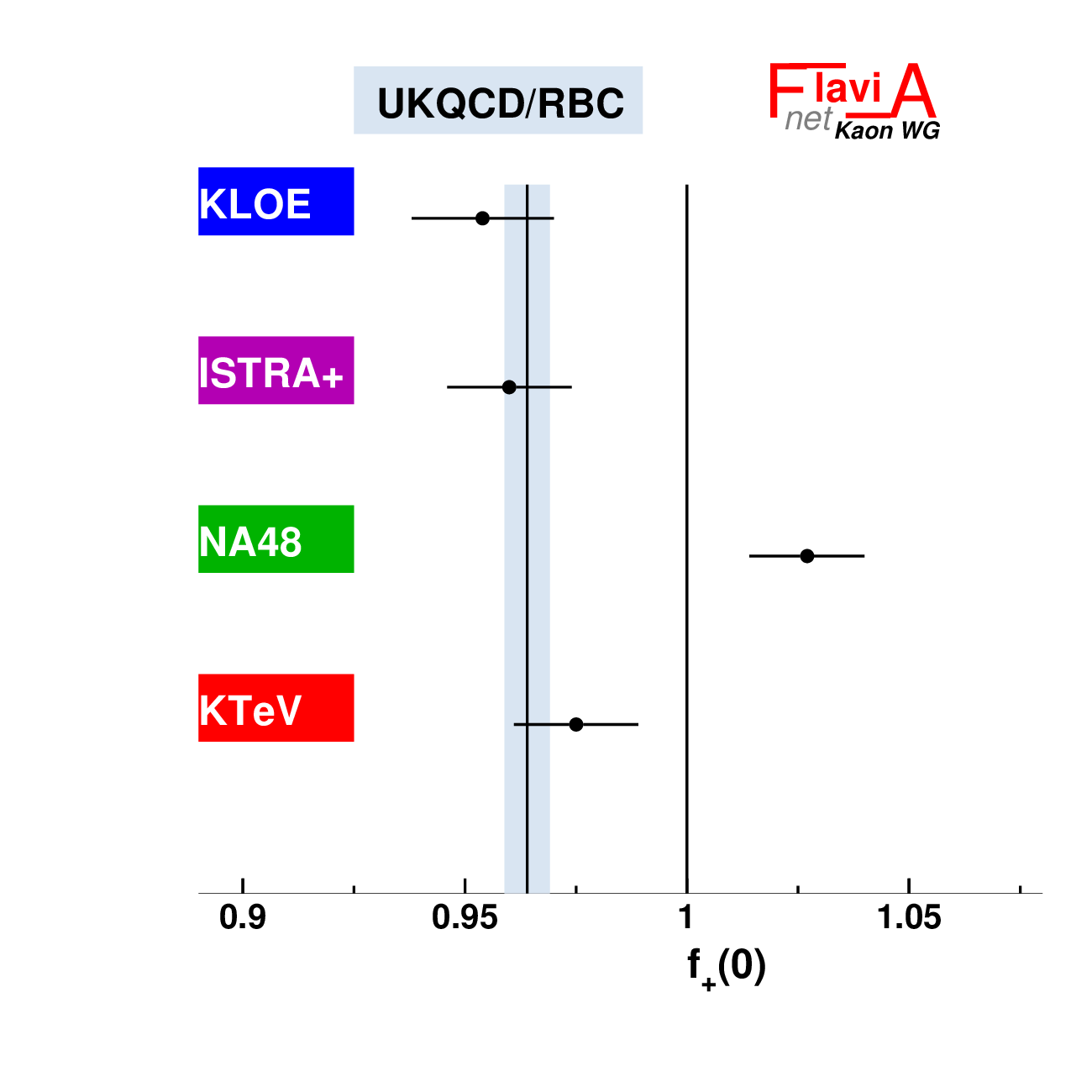}}
\end{center}
\caption{ Values for $f_+(0)$ determined from the scalar form factor slope using the Callan-Treiman 
relation and  $f_K/f_\pi=1.189(7)$. The UKQCD/RBC result $f_+(0)=0.964(5)$ is also
shown.\label{fig:CTtest} }
\end{figure}

Fig. \ref{fig:CTtest} shows the values for $f_+(0)$ 
determined from the scalar form factor slope
measurements obtained using the Callan-Treiman relation and
$f_K/f_\pi=1.189(7)$. The value of $f_+(0)=0.964(5)$ from UKQCD/RBC is also shown.
As already noted in Sec.~\ref{sec:data}, the NA48 result is 
difficult to accommodate. Here one can see that 
this results is also  inconsistent with the theoretical 
estimates of $f_+(0)$. In particular, it violates the 
Fubini-Furlan bound $f_+(0)<1$~\cite{Fubini:1970ss}. For this
reason, the NA48 result will be excluded when using the
Callan-Treiman constraint.

 We combine the average of the above results, $\log C =0.207\pm0.008$,  
 with the lattice determinations of  $f_K/f_\pi=1.189(7)$ and 
 $f_+(0)=0.964(5)$ using the constraint given by the Callan-Treiman
 relation. The results of the combination are given in Tab.~\ref{tab:ffsct}.
\begin{table}[ht]
\caption{\label{tab:ffsct}Results from the form factor fit.}
\begin{center}
\begin{tabular}{c|c|c}
 $\log C$  & $f_+(0)$ & $f_K/f_\pi$ \\
\hline
  0.204(6) & 0.964(4) & 1.187(6)\\
\hline
\multicolumn{3}{c}{correlation matrix}\\
\hline
          1.    & -0.44  &  0.52 \\
                &  1.    & 0.28 \\
                &        &  1.   \\

\end{tabular}
\end{center}
\end{table}
 The fit probability is 99\%, confirming the agreement between
 experimental measurements and lattice determinations.
 The accuracies of $f_K/f_\pi$ and $f_+(0)$ are also slightly improved, and 
 this effect can be better seen in the ratio  $f_+(0)/(f_K/f_\pi)$,
 which is directly related to the Callan-Treiman constraint.

\subsubsection*{Determination of 
$\vert V_{us}\vert/\vert V_{ud}\vert\times f_K/f_\pi$ }
\label{sec:fkfpvusvud}

An independent  determination of \Vus\ is obtained from $K_{\ell2}$ decays. 
The most important mode is  $K^+\to\mu^+\nu$, which has been  measured 
 by KLOE with a relative uncertainty of about $0.3\%$.  
Hadronic uncertainties are minimized by making use of the ratio 
$\Gamma(K^+\to\mu^+\nu)/\Gamma(\pi^+\to\mu^+\nu)$.

Using the world average values
of BR($K^\pm\to\mu^\pm\nu $) and of $\tau^\pm$ given in Sec.~\ref{sec:data}
and the value of $\Gamma(\pi^\pm\to\mu^\pm\nu)=38.408(7)~\mu s^{-1}$
from \cite{Amsler:2008zzb} we obtain:
\beq
\vert V_{us}\vert/\vert V_{ud}\vert\x f_K/f_\pi = 0.2758 \pm  0.0007~. 
\label{eq:vusvudres}
\eeq

\begin{figure}[t]
\begin{center}
\resizebox{0.6\textwidth}{!}{\includegraphics{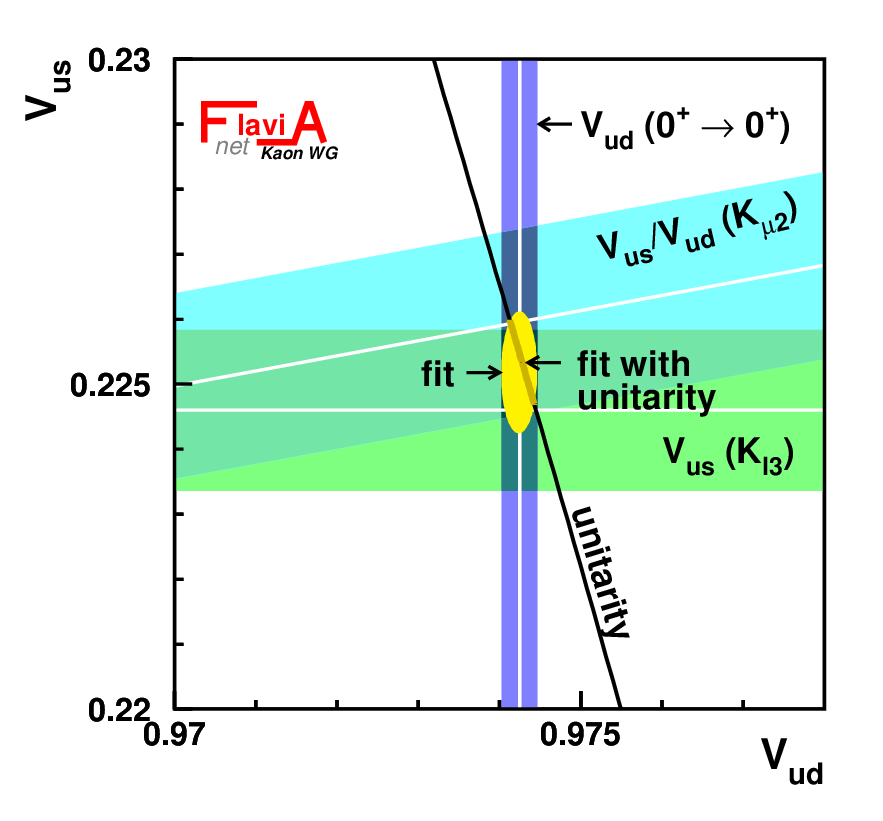}}
\end{center}
\caption{\label{fig:vusuni} Results of fits to \Vud, \Vus, and $\Vus/\Vud$.}
\end{figure}

\subsubsection{Test of Cabibbo Universality or CKM unitarity}

 To determine $|V_{us}|$ and   $|V_{ud}|$ we use the value
 $\vert V_{us}\vert\x f_{+}(0)=0.2166(5)$ reported in 
 Tab.~\ref{tab:Vusf0}, the result
 $\vert V_{us}\vert/\vert V_{ud}\vert f_K/f_\pi = 0.2758(7)$ discussed in
 Sec.~\ref{sec:fkfpvusvud}, $f_+(0) = 0.964(5)$, and $f_K/f_\pi = 1.189(7)$. 
 From the above we find:
\bea
 \vert V_{us}\vert&=& 0.2246\pm  0.0012 \qquad [K_{\ell 3}~{\rm only}]~, \\
 \vert V_{us}\vert/\vert V_{ud}\vert&=& 0.2319\pm  0.0015 \qquad [K_{\ell 2}~{\rm only}]~.
\eea
 A slightly less precise determination of $\vert V_{us}\vert/\vert V_{ud}\vert=0.2304(^{+0.0026}_{-0.0015})$ is obtained using the value of $f_K/f_\pi$ from MILC~\cite{Bazavov:2009bb}.
 These determinations can be used in a fit together with the 
 the  evaluation of \Vud\ from
 $0^+\to0^+$ nuclear beta decays quoted in section~\ref{s:Vudnucl}:
 $|V_{ud}|$=0.97425\plm0.00022.
The global fit gives 
\be
\Vud = 0.97425(22) \qquad \Vus = 0.2252(9) \qquad [K_{\ell 3,\ell 2}~+~ 0^+\to0^+]~,
\label{eq:VusVud}
\end{equation}
with $\chi^2/{\rm ndf} = 0.52/1$ (47\%). This result does not make use 
of CKM unitarity. If the  unitarity constraint is included, 
the fit gives 
\beq
\Vus=\sin\,\theta_C=\lambda=0.2253(6) \qquad [{\rm with~unitarity}]
\eeq
Both results are illustrated in \Fig{fig:vusuni}.



Using the (rather negligible) $|V_{ub}|^2\simeq 1.5\times10^{-5}$ 
 in conjunction with the above results
 leads to 

\be
|V_{ud}|^2+|V_{us}|^2+|V_{ub}|^2 = 0.9999 (4)_{V_{ud}} (4)_{V_{us}} = 0.9999 (6) \label{eqtwelve}
\eeq

 The outstanding agreement with unitarity provides an impressive
confirmation of Standard Model radiative corrections
 \cite{Czarnecki:2004cw,Marciano:2005ec}(at about the 60
sigma level!). It can be used to constrain ``new physics'' effects
which, if present, would manifest themselves as a deviation from 1, \emph{i.e.}\
what would appear to be a breakdown of unitarity.

We will give several examples of the utility Eq.~(\ref{eqtwelve}) provides
for constraining ``new physics''. Each case is considered in
isolation, \emph{i.e.}\ it is assumed that there are no accidental
cancellations. 

 \subsubsection*{Exotic Muon Decays}

If the muon can undergo decay modes beyond the Standard Model
$\mu^+\to e^+\nu_e\bar\nu_\mu$ and its radiative extensions, those
exotic decays will contribute to the muon lifetime. That would mean
that the ``real'' Fermi constant, $G_F$, is actually smaller than the
value in Eq.~(\ref{eqtwo}) and we should be finding

\be
|V_{ud}|^2+|V_{us}|^2+|V_{ub}|^2 = 1-BR (\rm exotic~muon~decays) \label{eqthirteen}
\eeq

A unitarity sum below 1 could be interpreted as possible evidence
for such decays. Alternatively, Eq.~(\ref{eqtwelve}) provides at
(one-sided) 95\% CL

\be
BR(\rm exotic~muon~decays) < 0.001 \label{eqfourteen}
\eeq

That is, of course, not competitive with, for example, the direct
bound $BR(\mu^+\to e^+\gamma)< 1\times 10^{-11}$ \cite{Amsler:2008zzb}. However, for
decays such as $\mu^+\to e^+\bar\nu_e\nu_\mu$ (wrong neutrinos),
Eq.~(\ref{eqfourteen}) is about a factor of 10 better than the direct
constraint \cite{Amsler:2008zzb} $BR(\mu^+\to e^+\bar\nu_e\nu_\mu)<0.012$.
 That constraint is useful for possible future neutrino factories where
 the neutrino beams originate from muon decays. If such a decay were to exist,
 it would provide a background to neutrino oscillations.

Another way to illustrate the above constraint is to extract the Fermi
constant from nuclear, $K$ and $B$ decays assuming the validity of CKM
 unitarity without employing muon decay. Values in Eq.~\ref{eq:VusVud} give

\be
G_F^{\rm CKM} = 1.166279 (261)\times 10^{-5}{\rm GeV}^{-2} \quad {\rm CKM~Unitarity}
\label{eqfifteen} 
\eeq

which is in fact the second best determination of $G_F$, after
Eq.~(\ref{eqtwo}). The comparison between $G_\mu$ in Eq.~(\ref{eqtwo})
and $G^{\rm CKM}_F$ in Eq.~(\ref{eqfifteen})  is providing the constraints on
``new physics'', if it affects them differently. So far, they are equal to
 within errors.

 \subsubsection*{Heavy Quarks and Leptons}

As a second example, consider the case of new heavy quarks or leptons
that couple to the ordinary 3 generations of fermions via mixing
\cite{Marciano:1975cn}. For a generic heavy charge $-1/3$ $D$ quark from a 4th
generation, mirror fermions, SU(2)$_L$ singlets etc., one finds at the
one-sided 95\% CL

\be
|V_{uD}|\le 0.03 \label{eqsixteen}
\eeq

Considering that $|V_{ub}|\simeq 0.004$, such an indirect
constraint appears not  to be very stringent but it can be useful in some
 models to rule out large loop induced effects from mixing. In the case of heavy
neutrinos with $m_N>m_\mu$, one finds similarly

\be
|V_{\ell N}|< 0.03 \quad , \quad \ell=e,\mu \label{eqseventeen}
\eeq

 \subsubsection*{Four Fermion Operators}

If there are induced dim. 6 four fermion operators of the form 

\be
\mp i\frac{2\pi}{\Lambda^2} \bar u \gamma_\mu d\bar e_L \gamma^\mu\nu_e
\label{eqeighteen} 
\eeq
where $\Lambda$ is a high effective mass scale due to
compositeness, leptoquarks, excited $W^\ast$ bosons (\emph{e.g.}\
extra dimensions) or even heavy loop effects, they will interfere with
  the Standard Model beta decay amplitudes and
 give $G^{\rm CKM}_F = G_\mu\left(1\pm \frac{\sqrt{2}\pi}{G_\mu\Lambda^2}\right)$.
  One finds at  90\%CL

\be
\Lambda > 30 {\rm ~TeV}  \label{eqnineteen}
\eeq

Similar constraints apply to new 4 fermion lepton
operators that contribute to $\mu^+\to e^+\nu_e\bar\nu_\mu$. Of
course, in some cases there can be a cancellation between semileptonic
and purely leptonic effects and no bound results.

The high scale bounds in Eq.~(\ref{eqnineteen}) apply most directly
to compositeness because no coupling suppression was assumed. For
leptoquarks, $W^\ast$ bosons etc.\ the bounds should be about an order
of magnitude smaller due to weak couplings. A $m_{W^*}$ bound of about
 4$\sim$6 TeV results if we assume it affects leptonic and semileptonic
 decays very differently; but that assumption may not be valid and may
 need to be relaxed (see below). In the case of new loop
effects, those bounds should be further reduced by another order of
magnitude. For example, we next consider the effect of heavy $Z^\prime$
bosons in loops that enter muon and charged current semileptonic decays
differently where a bound of about 400 GeV is obtained.

 \subsubsection*{Additional $Z^\prime$ Gauge Bosons}

As next example, we consider the existence of additional $Z^\prime$
bosons that influence unitarity at the loop level by affecting muon
and semi-leptonic beta decays differently\cite{Marciano:1987ja}. 
 In general, we found that the unitarity sum was predicted
to be greater than one in most scenarios. In fact, one expects

\bea
|V_{ud}|^2+|V_{us}|^2+|V_{ub}|^2 &=& 1 + 0.01\lambda\ell n\; X/(X-1) \nonumber \\
X &=& m^2_{Z^\prime}/m^2_W \label{eqtwenty}
\eea

where $\lambda$ is a model dependent quantity of $O(1)$. It
can have either sign, but generally $\lambda>0$.

In the case of SO(10) grand unification $Z^\prime=Z_\chi$ with
$\lambda\simeq0.5$, one finds at one-sided 90\% CL

\be
m_{Z_\chi} > 400 {\rm GeV} \label{eqtwentyone}
\eeq

That bound is somewhat smaller than tree level bounds on $Z^\prime$ bosons from
 atomic parity violation and polarized Moller scattering \cite{Anthony:2005pm,Czarnecki:2005pe} as well
 as the direct collider search bounds \cite{Amsler:2008zzb} $m_{Z_\chi}>720$ GeV\null.

 \subsubsection*{Charged Higgs Bosons}

A particularly interesting test is the comparison of the $\vert V_{us}\vert$ 
value extracted from the helicity-suppressed $K_{\ell 2}$ decays
with respect to the value extracted from the  helicity-allowed $K_{\ell 3}$
  modes.
To reduce theoretical uncertainties from $f_K$ and electromagnetic 
corrections in $K_{\ell 2}$,
we exploit the ratio $Br(K_{\ell2})/Br(\pi_{\ell2})$ and 
we study the quantity
\be
R_{l23}=\left\vert \frac{V_{us}(K_{\ell 2})}{V_{us}(K_{\ell 3})}\x 
\frac{V_{ud}(0^+\to 0^+)}{V_{ud}(\pi_{\ell 2})}\right\vert\,.
\end{equation}
Within the SM, 
 $R_{l23}=1$, while deviation from 1 can be induced by 
 non-vanishing scalar- or  right-handed currents.
 Notice that in $R_{l23}$ the  hadronic uncertainties
enter through  $(f_K/f_\pi)/f_+(0)$.

Effects of scalar currents due to a charged Higgs 
 give~\cite{Antonelli:2008jg}
 \be
 R_{l23} = \left\vert 1 - \frac{m^2_{K^+}}{M^2_{H^+}}\left(1-\frac{m_d}{m_s} \right)
\frac{\tan^2\beta}{1+\epsilon_0\tan\beta}\right\vert~,
\end{equation}
whereas  for right-handed currents we have
 \be
 R_{l23} = 1 -2\,\left(\epsilon_s-\epsilon_{ns}\right)~.
\end{equation}

In the case of scalar densities (MSSM),
the unitarity relation between $\vert V_{ud}\vert$ extracted from 
  $0^+\to0^+$ nuclear beta decays and $\vert V_{us}\vert$ extracted from 
$K_{\ell3}$ remains valid as soon as form factors are experimentally determined.
This constrain  together with the experimental
information of $\log C^{MSSM}$ 
can be used in the global fit to improve the accuracy of the determination 
of $R_{l23}$, which in this scenario turns to be 
\be
\left. R_{l23} \right|^{\rm exp}_{\rm scalar}  =  1.004 \pm  0.007~.
\end{equation}
Here $(f_K/f_\pi)/f_+(0)$ has been fixed from lattice. This ratio
is the key quantity to be improved in order to reduce 
present uncertainty on $R_{l23}$. 

The  measurement of $R_{l23}$ above can be used to set bounds
on the charged Higgs mass and $\tan\beta$.
Fig. \ref{fig:higgskmunu} shows the excluded region at 95\%
 CL in the $M_H$--$\tan\beta$ plane (setting $\epsilon_0=0.01$).
 The measurement of  BR($B \to \tau \nu$)~\cite{Isidori:2006pk,Hou:1992sy,
Akeroyd:2003zr}
 can be also used to set a similar  bound
in the  $M_H$--$\tan\beta$  plane. While $B\to\tau \nu$
 can exclude quite an extensive region of this plane,
 there is an uncovered region in the exclusion corresponding 
 to a destructive interference between the charged-Higgs 
 and the SM amplitude.
 This region is fully covered by the $K\to \mu \nu$ result.

\begin{figure}[t]
\centering
\resizebox{0.7\textwidth}{!}{\includegraphics{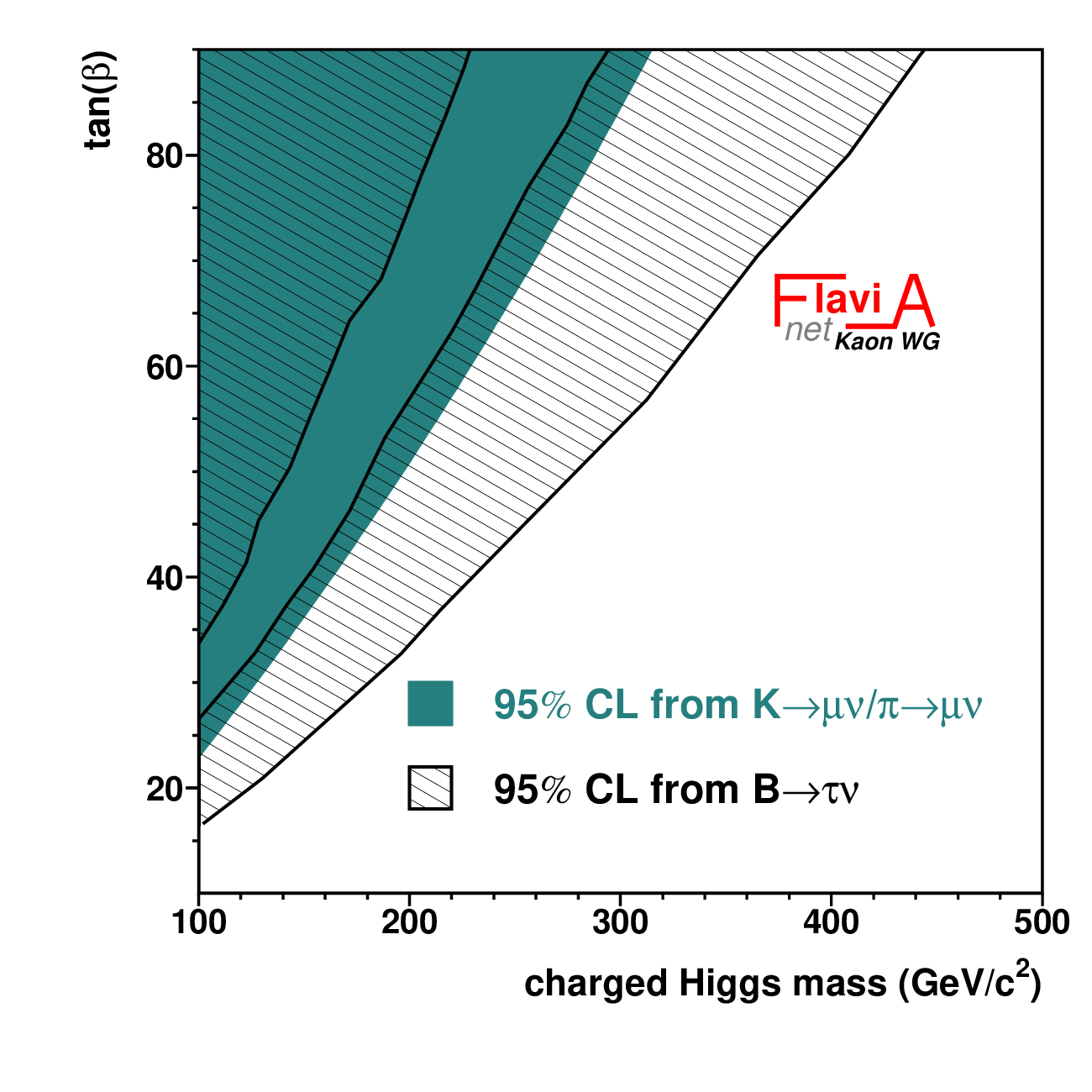}}
\caption{\label{fig:higgskmunu}
Excluded region in the charged Higgs mass-$\tan\beta$ plane.
The region excluded by  $B\to \tau \nu $ is also indicated.}
\end{figure}

In the case of right-handed currents~\cite{Bernard:2006gy},
$R_{l23}$ can be obtained from a global fit to the values
of eqs.~(\ref{eq:vusfres}) and (\ref{eq:vusvudres}). 
Here $\log C^{\rm exp}$ is free of new physics effects and 
can be also used to constrain $(f_K/f_\pi)/f_+(0)$ 
together with lattice results (namely the values in Tab.~\ref{tab:ffsct}).
The result is
\be
\left. R_{l23} \right|^{\rm exp}_{\rm RH curr.} =  1.004 \pm  0.006~. 
\end{equation}

In addition, interesting unitarity constraints can be placed on
supersymmetry \cite{Barbieri:1985ff,Hagiwara:1995fx,Kurylov:2001zx} where SUSY loops affect muon and
semileptonic decays differently. Again, one expects constraints up to 
mass scales of $O$(500 GeV), depending on the degree of
cancellation between squark and slepton effects.

In the future, the unitarity constraint could improve from $\pm
0.0006$ to $\pm0.0004$ if $f_+(0)$  and $f_K/f_\pi$ errors as well as
 uncertainties from radiative corrections can be reduced.
 Such an improvement will be difficult, but particularly well motivated
 if an apparent violation starts to emerge or the LHC makes a
 relevant ``new physics'' discovery.

As an added comment, we again mention that eqs.~(\ref{eqtwo}) and
 (\ref{eqfifteen}) represent our two best measurements of the Fermi constant.
 Their agreement reinforces the validity of using $G_\mu$ to normalize
 electroweak charged and neutral current amplitudes in other precision
 searches for ``new physics''. In fact, either $G_\mu$ or $G^{\rm CKM}_F$ could
 be used without much loss of sensitivity, since all other experiments are
 currently less precise than both. For example, one of the next best
 determinations of the Fermi constant  (which is insensitive to $m_t$)
 comes from \cite{Czarnecki:2004cw}

\be
G^{(2)}_F = \frac{\pi\alpha}{\sqrt{2} m^2_W \sin^2\theta_W(m_Z)_{\overline{MS}} (1-\Delta r (m_Z)_{\overline{MS}})} \label{eqtwentyfive}
\eeq

where

\alpheqn

\bea
\alpha^{-1} & = & 137.035999084 (51) \label{eqtwentysix} \\
m_W & = & 80.398 (25) \rm{~GeV} \\
\sin^2\theta_W (m_Z)_{\overline{MS}} & = & 0.23125 (16) \\
\Delta r (m_Z)_{\overline{MS}} & = & 0.0696 (2)
\eea

\reseteqn

One finds

\be
G^{(2)}_F = 1.165629 (1100) \times 10^{-5} \rm{~GeV}^{-2} \label{eqtwentyseven}
\eeq
with an uncertainty about 180 times larger than $G_\mu$ and about 4 times
 larger than $G^{\rm CKM}_F$. The value in Eq.~(\ref{eqtwentyseven}) is,
 nevertheless, very useful for constraining ``new physics'' that affects
 it differently than $G_\mu$ or $G^{\rm CKM}_F$. Perhaps the two best examples
 are the $S$ parameter \cite{Peskin:1990zt,Marciano:1990dp}

\be
S\simeq \frac{1}{6\pi} N_D \label{eqtwentyeight}
\eeq
which depends on the number of new heavy SU(2)$_L$ doublets 
(e.g. $N_D=4$ in the case of a 4th generation) and a generic $W^*$
 Kaluza-Klein excitation associated with extra dimensions \cite{Czarnecki:2004cw} that
 has the same quark and lepton couplings. Either would contribute to $G_\mu$ or
 $G^{\rm CKM}_F$ but not to $G^{(2)}_F$. Therefore, one has the relation

\be
G_\mu \simeq G^{\rm CKM}_F \simeq G^{(2)}_F (1+0.0085S + {\cal O}(1) \frac{m^2_W}{m^2_{W^*}}) \label{eqtwentynine}
\eeq
The good agreement among all three Fermi constants then suggests $m_{W^*}>
2$ \linebreak $\sim3$ TeV and $S\simeq 0.1\pm0.1$ (consistent with zero).
 Those constraints are similar to what is obtained from global fits to all
 electroweak data. Taken at face value they suggest any ``new physics'' near
 the TeV scale that we hope to unveil at the LHC is hiding itself quite well
 from us in precision low energy data. It will be interesting to see what the
 LHC finds.

\subsubsection{Tests of Lepton Flavor Universality in $K_{\ell 2}$ decays}

The ratio $R_K = \Gamma({K_{\mu2}})/\Gamma({K_{e2}})$ can be precisely calculated 
within the
 Standard Model.
 Neglecting radiative corrections, it is given by 
\be
R_K^{(0)} = \frac{m_e^2}{m_\mu^2} \: \frac{(m_K^2 - m_e^2)^2}{(m_K^2 - m_\mu^2)^2} = 2.569 \times 10^{-5},
\end{equation}
and reflects the strong helicity suppression of the electron channel.
Radiative corrections have been computed with effective
 theories~\cite{Cirigliano:2007ga},
yielding the final SM prediction
\bea
R^{\rm SM}_K &=& R_K^{(0)} ( 1 + \delta R_K^{\text{rad.corr.}})  \nn \\
&=& 2.569 \times 10^{-5} \times ( 0.9622 \pm 0.0004 ) =(2.477 \pm 0.001) \times 10^{-5}~. 
\eea

Because of the helicity suppression within then SM, the 
 $K_{e2}$  amplitude is a prominent candidate
 for possible sizable contributions from physics beyond the SM. Moreover,
 when normalizing to the $K_{\mu2}$ rate, we obtain an extremely precise 
 prediction of the $K_{e2}$ width within the SM. In order to be visible 
 in the $K_{e2}/K_{\mu2}$ ratio, the new physics must  violate lepton 
 flavor universality.

 Recently it has been pointed out that in a supersymmetric framework 
sizable violations of  lepton universality can be expected
 in $K_{l2}$ decays~\cite{Masiero:2005wr}. At the tree level, 
 lepton flavor violating terms are forbidden in the MSSM. 
 However, these appear at the one-loop level, where an effective 
 $H^+ l \nu_\tau$ Yukawa interaction is generated. Following the
 notation of Ref.~\cite{Masiero:2005wr},
 the non-SM contribution to $R_K$ can be written as 
\be
R_K^{\text{LFV}} \approx R_K^{\text{SM}} \left[ 1 + \left(
\frac{m_K^4}{M_{H^\pm}^4} \right) \left( \frac{m_\tau^2}{m_e^2} \right) |\Delta_{13}|^2 \tan^6 \beta \right]~.
\label{eqn:susy}
\end{equation}
 The lepton flavor violating coupling $\Delta_{13}$, being generated at the 
loop level, could reach values of $\cO(10^{-3})$.
 For moderately large $\tan \beta$ values, 
 this contribution may therefore
 enhance $R_K$ by up to a few percent.
 Since the additional term in Eq.~\ref{eqn:susy} goes with the fourth power of the
 meson mass, no similar effect is expected in $\pi_{l2}$ decays.

 The world average result for $R_K$ presented in Sec. \ref{sec:data} 
 gives strong constraints 
 for $\tan \beta$ and $M_{H^\pm}$, as shown in Fig.~\ref{fig:susylimit}.
 For values of $\Delta_{13} \approx   10^{-3}$
 and  $\tan \beta > 50$ the charged Higgs masses is pushed 
 above 1000~GeV/$c^2$ at 95\% CL.

\begin{figure}[t]
\centering
\resizebox{0.7\textwidth}{!}{\includegraphics{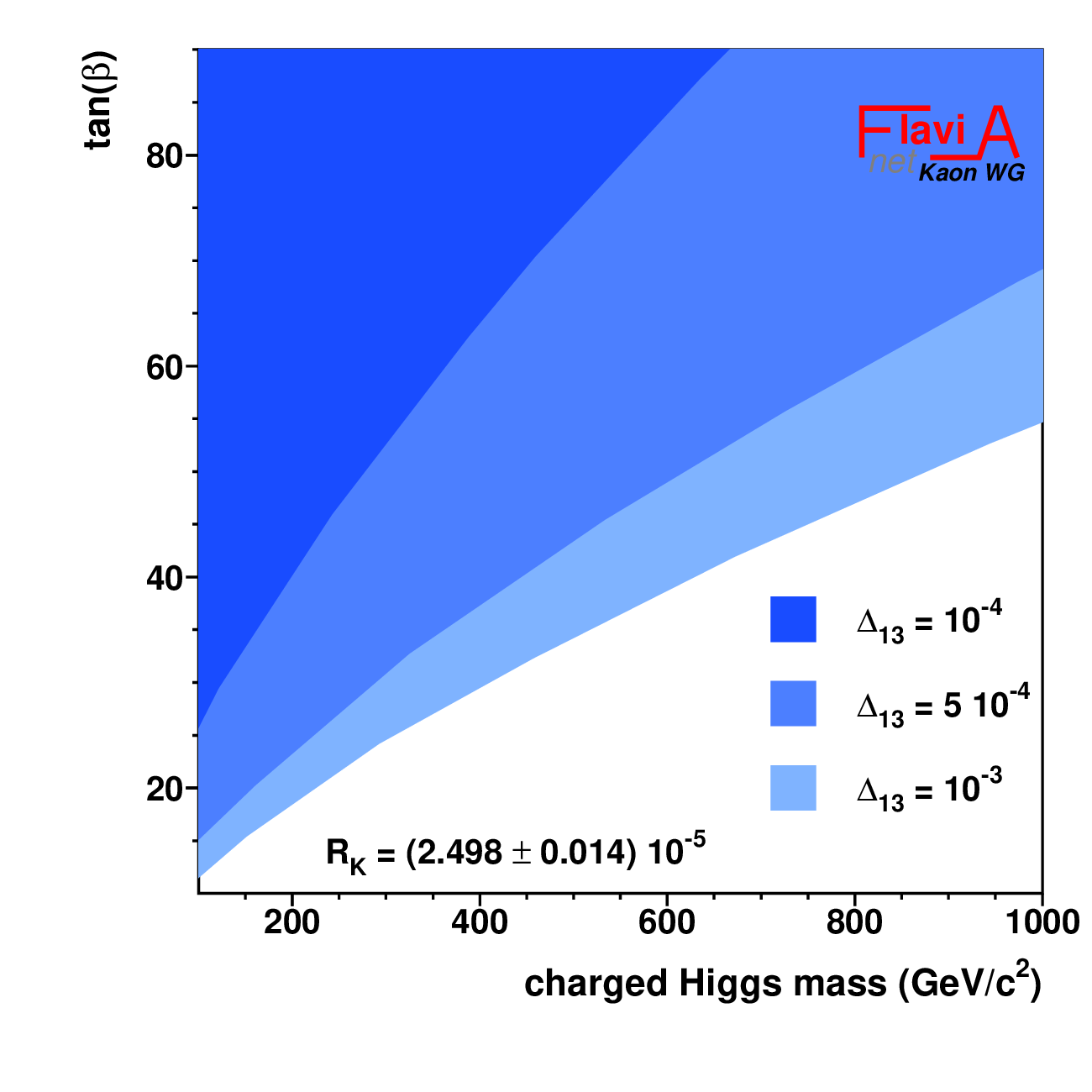}}
\caption{Exclusion limits at $95\%$ CL on $\tan \beta$ and the charged
Higgs mass $M_{H^\pm}$ 
from $R_K$ for different
values of $\Delta_{13}$. }
\label{fig:susylimit}
\end{figure}

\newcommand{\el}{$E_\ell$}
\newcommand{\mx}{$M_X$}
\newcommand{\mxqsq}{$(M_X, q^2)$}
\newcommand{\elsmax}{$(E_\ell, s^{\rm max})$}
\newcommand{\smax}{$s^{\rm max}$}
\newcommand{\pplus}{$P^+$}
\newcommand{\qq}{q^2}
\newcommand{\qz}{q_0}
\newcommand{\as}{\alpha_s}
\newcommand{\mupi}{\mu_\pi^2}
\newcommand{\mug}{\mu_G^2}
\newcommand{\rd}{\rho_D^3}
\newcommand{\rls}{\rho_{LS}^3}
\newcommand{\muwa}{\mu_{\scriptscriptstyle\rm WA}}

\section[Semileptonic $B$ and $D$ decays: $|V_{cx}|$ and $|V_{ub}|$]{\boldmath
Semileptonic $B$ and $D$ decays: $|V_{cx}|$ and $|V_{ub}|$} 
\label{sec:sl}

In this section, we address semileptonic decays that proceed at the tree 
level of the weak interaction.
We focus on decays of the lightest pseudoscalar mesons, $D$ for charm 
and $B$ for bottom, because higher excitations decay hadronically (or, 
in case of the $B^*$, radiatively) to the $D$ and $B$ and thus have
negligibly small semileptonic partial widths.
The amplitude for quark flavor change in these processes is proportional
to a CKM matrix element, providing a direct way to ``measure'' 
the CKM matrix.

Purely leptonic decays of pseudoscalars are, of course, also directly 
sensitive to the CKM matrix, but they require a spin flip.
Their rate is, hence, helicity suppressed  by a factor $(m_\ell/m_P)^2$, 
where $m_P$ is the pseudoscalar meson mass and $m_\ell$ the mass of 
the daughter lepton.
This suppression makes purely leptonic decays more sensitive to
non-Standard processes, and therefore less reliable channels for 
the determination of CKM matrix elements than semileptonic decays.

As with the determination of \Vus in the semileptonic decay 
$K\to\pi\ell\nu$, discussed in Sec.~\ref{sec:cabibbo}, one can determine 
\Vcs from $D\to K\ell\nu$,
\Vcd from $D\to\pi\ell\nu$,
\Vub from $B\to\pi\ell\nu$, and
\Vcb from $B\to D^{(*)}\ell\nu$,
by combining measurements of the differential decay rate with lattice-QCD 
calculations for the hadronic part of the transition, commonly described 
with form factors.
This section starts with the three heavy-to-light decays, and then 
proceeds to heavy-to-heavy decays for which heavy-quark symmetry plays a 
crucial role.
\Vcb and \Vub can also be determined from \emph{inclusive} semileptonic 
$B$ decays, because the large energy scale \mb and the inclusion of all 
final-state hadrons makes these processes amenable to the 
operator-product expansion (OPE).
Within the OPE the short-distance QCD can be calculated in perturbation 
theory, and the long-distance QCD can be measured from kinematic 
distributions.
While this is rather straightforward for \Vcb\, it is more subtle \Vub\, 
so these two topics are treated in separate subsections.

\subsection{Exclusive semileptonic $B$ and $D$ decays to light 
mesons~$\pi$ and~$K$}

\subsubsection{Theoretical Background} 

Heavy-to-light semileptonic decays, in which a $B$ or $D$ meson decays
into a light pseudoscalar or vector meson (such as a pion or
$\rho$ meson), are sensitive probes of quark flavor-changing
interactions.
The decay rate for $H \to P \ell \nu$ semileptonic decay is given by
\begin{eqnarray}
    \hspace*{-1em}
    \frac{d\Gamma}{dq^2} = \frac{G_F^2|V_{qQ}|^2}{24\pi^3}
        \frac{(q^2-m_\ell ^2)^2 \sqrt{E_P^2-m_P^2}}{q^4 m_H^2}
	\left\{\left(1+\frac{m_\ell ^2}{2q^2}\right) m_H^2(E_P^2-m_P^2)
	\left[f_+(q^2)\right]^2 \right. \nonumber \\ \left.
        + \frac{3m_\ell^2}{8q^2}(m_H^2-m_P^2)^2\left[f_0(q^2)\right]^2\right\},
    \label{slep:eq:B2pi}
\end{eqnarray}
where $q \equiv p_H - p_P$ is the momentum transferred to the lepton
pair and $|V_{qQ}|$ is the relevant CKM matrix element.  The form
factors, $f_+(q^2)$ and $f_0(q^2)$, parametrize the hadronic matrix
element of the heavy-to-light vector current,
$V^\mu\equiv i{\bar q}\gamma^\mu Q$:
\begin{equation}
    \langle P | V^\mu | H \rangle  = f_+(q^2)
	\left(p^\mu_H + p^\mu_P - \frac{m_H^2 - m_P^2}{q^2}\,q^\mu \right) +
	f_0(q^2) \frac{m_H^2 - m_P^2}{q^2}\,q^\mu ,
\end{equation}
where $E_P= (m_H^2+m_P^2 -q^2)/2m_H$ is the energy of the light meson in
the heavy meson's rest frame.
The kinematics of semileptonic decay require that the form factors are
equal at zero momentum-transfer, $f_+(0) = f_0(0)$.
In the limit $m_\ell\to0$, which is a good approximation for 
$\ell=e$, $\mu$, the form factor $f_0(q^2)$ drops out and the
expression for the decay rate simplifies to
\begin{equation}
    \frac{d\Gamma}{dq^2} = \frac{G_F^2 |V_{qQ}|^2}{192 \pi^3 m_H^3}
	\left[
(m_H^2 + m_P^2 - q^2)^2 - 4 m_H^2 m_P^2 \right]^{3/2} |f_+(q^2)|^2 .
\end{equation}
Using the above expression, a precise experimental measurement of the
decay rate, in combination with a controlled theoretical calculation of
the form factor, allows for a clean determination of the CKM matrix
element $|V_{qQ}|$.

\subsubsubsection{Analyticity and unitarity}

It is well-established that the general properties of analyticity and
unitarity largely constrain the shapes of heavy-to-light semileptonic
form factors~\cite{Bourrely:1980gp,Boyd:1994tt,%
Lellouch:1995yv,Boyd:1997qw,Arnesen:2005ez}.
All form factors are analytic in $q^2$ except at physical poles and
threshold branch points.
Because analytic functions can always be expressed as convergent power
series, this allows the form factors to be written in a particularly
useful manner.

Consider a change of variables that maps $q^2$ in the semileptonic
region onto a unit circle:
\begin{equation}
    z(q^2, t_0) = \frac{\sqrt{1 - q^2/t_+}-\sqrt{1-t_0/t_+}}%
        {\sqrt{1-q^2/t_+}+\sqrt{1-t_0/t_+}} ,
    \label{eq:slep:z_trans}
\end{equation}
where $t_+\equiv(m_H + m_P )^2$, $t_-\equiv(m_H - m_P )^2$, and $t_0$ is
a constant to be discussed later.
In terms of this new variable, $z$, the form factors have a simple form:
\begin{equation}
    P(q^2) \phi(q^2,t_0) f(q^2) = \sum_{k=0}^{\infty} a_k(t_0) z(q^2,t_0)^k .
    \label{eq:slep:z_exp}
\end{equation}
In order to preserve the analytic structure of $f(q^2)$, the function
$P(q^2)$ vanishes at poles below the $H$-$P$ pair-production threshold
that contribute to $H$-$P$ pair-production as virtual intermediate
states.
For example, in the case of $B\to\pi\ell\nu$ decay,  $P(q^2)$
incorporates the location of the $B^*$ pole:
\begin{equation}
    P^{B\to\pi\ell\nu}_+(q^2)  = z(q^2, m_{B^*}) .
\end{equation}
For the case of $D$~meson semileptonic decays, the mass of the
$D^*$~meson is above the $D$-$\pi$ production threshold, 
but the $D^*_s$ is below $D$-$K$ production threshold.
Hence
\begin{eqnarray}
    P^{D\to\pi\ell\nu}_+(q^2)  &=& 1 ,\\
    P^{D\to K \ell\nu}_+(q^2)  &=& z(q^2, m_{D^*_s}) .
\end{eqnarray}
In the expression for $f(q^2)$, Eq.~(\ref{eq:slep:z_exp}), $\phi(q^2,
t_0)$ is any analytic function.
It can be chosen, however, to make the unitarity constraint on the
series coefficients have a simple form.
The standard choice for $\phi_+(q^2,t_0)$, which enters the expression
for $f_+(q^2)$, is~\cite{Arnesen:2005ez}:
\begin{eqnarray}
    \phi_+(q^2, t_0) & = & \sqrt{\frac{3}{96 \pi \chi^{(0)}_J}}
        \left( \sqrt{t_+ - q^2} + \sqrt{t_+ - t_0}  \right)
	\left( \sqrt{t_+ - q^2} + \sqrt{t_+ - t_-} \right)^{3/2}  \nonumber\\
        & \times & \left( \sqrt{t_+ - q^2} + \sqrt{t_+} \right)^{-5}
	\frac{(t_+ - q^2)}{(t_+ - t_0)^{1/4}} ,
    \label{eq:slep:phi}
\end{eqnarray}
where $\chi^{(0)}_J$ is a numerical factor that can be calculated using
perturbation theory and the operator product expansion.
A~similar function can be derived for the irrelevant form factor $f_0(q^2)$.


Given the above choices for $P(q^2)$ and $\phi(q^2, t_0)$, unitarity
constrains the size of the series coefficients:
\begin{equation}
    \sum_{k=0}^{N} a_k^2 \lesssim 1,
    \label{eq:slep:a_const}
\end{equation}
where this holds for any value of $N$.
In the case of the $B\to\pi\ell\nu$ form factor, the sizes of the series
coefficients ($a_k$s) turn out to be much less than~1~\cite{Bailey:2008wp}.
Becher and Hill recently pointed out that this is due to the fact that
the $b$-quark mass is so large, and used heavy-quark power-counting to
derive a tighter constraint on the $a_k$s:
\begin{equation}
    \sum_{k=0}^{N} a_k^2 \leq \left( \frac{\Lambda}{m_Q}\right)^3 
    \label{eq:slep:hq_const} ,
\end{equation}
where $\Lambda$ is a typical hadronic scale~\cite{Becher:2005bg}.
The above expression suggests that the series coefficients should be
larger for $D$-meson form factors than for $B$-meson form factors.
This, however, has not been tested.

In order to accelerate the convergence of the power-series in $z$, the
free parameter $t_0$ in Eq.~(\ref{eq:slep:z_trans}) can be chosen to
make the range of $|z|$ as small as possible.
For the value $t_0=0.65t_-$ used in Ref.~\cite{Arnesen:2005ez}, the
ranges of $|z|$ for some typical heavy-to-light semileptonic decays are
given in Tab.~\ref{tab:slep:z_regions}.
The tight heavy-quark constraint on the size of the coefficients in the
$z$-expansion, in conjunction with the small value of $|z|$, ensures
that only the first few terms in the series are needed to describe
heavy-to-light semileptonic form factors to a high accuracy.
\begin{table}
    \caption{Physical region in terms of the variable $z$ for various
    semileptonic decays given the choice $t_0=0.65t_-$.}
\centering
    \begin{tabular}{cc}
    \hline \hline
	$B\to\pi l \nu$  & $- 0.34 <  z < 0.22 $ \\
	$D\to \pi l \nu$ & $- 0.17 <  z < 0.16 $ \\
	$D\to K l \nu$  & $- 0.04 <  z < 0.06 $ \\
    \hline \hline
    \end{tabular}
    \label{tab:slep:z_regions}
\end{table}

Other model-independent parameterizations of heavy-to-light semileptonic
form factors base on analyticity and unitarity have been proposed and
applied to the case of $B\to\pi\ell\nu$ decay by Bourrely, Caprini, and
Lellouch~\cite{Bourrely:2008za} and by Flynn and
Nieves~\cite{Flynn:2006vr,Flynn:2007ii}.
Bourrely \emph{et al.}\ use the series expansion in $z$ described above,
but choose simpler outer function, $\phi(q^2, t_0) = 1$.
This leads, however, to a more complicated constraint on the series
coefficients, which is no longer diagonal in the series index $k$.
Flynn and Nieves use multiply-subtracted Omn\`es dispersion relations to
parametrize the form factor shape in terms of the elastic $B$-$\pi$
scattering phase shift and the value of $f_+(q^2)$ at a few subtraction
points below the $B$-$\pi$ production threshold.

\subsubsubsection{Lattice QCD} 

In lattice-QCD calculations and in heavy-quark effective theory (HQET), it
is easier to work with a different linear combination of the form
factors:
\begin{equation}
    \langle P | V^\mu | H \rangle  = \sqrt{2m_H} \left[
        v^\mu f_\parallel(E_P) + p_\perp^\mu f_\perp(E_P) \right] ,
\end{equation}
where $v^\mu = p^\mu_H / m_H$ is the velocity of the heavy meson,
$p_\perp^\mu = p_P^\mu - (p_P \cdot v)v^\mu$ is the component of the
light meson momentum perpendicular to $v$, and $E_P = p_P \cdot v =
(m_H^2 + m_P^2 - q^2)/(2 m_H)$ is the energy of the light meson in the
heavy meson's rest frame.
In the heavy meson's rest frame, the form factors $f_\parallel(E_P)$ and
$f_\perp(E_P)$ are directly proportional to the hadronic matrix elements
of the temporal and spatial vector current:
\begin{eqnarray}
    f_\parallel(E_P) & = & \frac{\langle P | V^0 | H \rangle}{\sqrt{2 m_H}}
    \label{eq:slep:fpara} \\
    f_\perp(E_P) & = & \frac{\langle P | V^i | H \rangle}{\sqrt{2 m_H}}
        \frac{1}{p_P^i} .
    \label{eq:slep:fperp}
\end{eqnarray}
Lattice QCD simulations therefore typically determine $f_\parallel(E_P)$
and $f_\perp(E_P)$, and then calculate the form factors that appear in
the heavy-to-light decay width using the following equations:
\begin{eqnarray}
    f_0 (q^2) & = & \frac{\sqrt{2 m_H}}{m_H^2 - m_P^2} \left[
        (m_H - E_P) f_\parallel(E_P) + (E_P^2 - m_P^2) f_\perp(E_P) \right], \\
    f_+ (q^2) & = & \frac{1}{\sqrt{2 m_H}} \left[
        f_\parallel (E_P) + (m_H - E_P) f_\perp (E_P) \right].
\end{eqnarray}
These expressions automatically satisfy the kinematic constraint
$f_+(0)=f_0(0)$.

The goal is to evaluate the hadronic matrix elements on the right-hand
side of Eqs.(\ref{eq:slep:fpara}) and (\ref{eq:slep:fperp}) via
numerical simulations in lattice QCD.
Such simulations are carried out with  operators, $V_\mu^L$, written in
terms of the lattice heavy and light quark fields appearing in the
lattice actions.
Hence, an important step in any lattice determination of hadronic matrix
elements is the matching between continuum operators such as $V_\mu$ and
their lattice counterparts.
The matching takes the form
\begin{equation}
    \langle P | V_\mu | H \rangle = Z^{Qq}_{V_\mu}
        \langle P | V_\mu^L | H \rangle .
    \label{eq:slep:V_renorm}
\end{equation}
For heavy-light currents with dynamical (as opposed to static) heavy
quarks, the matching factors $Z^{Qq}_{V_\mu}$ have been obtained to date
either through a combination of perturbative and nonperturbative methods
or via straight one-loop perturbation theory.
Uncertainties in $Z^{Qq}_{V_\mu}$ can be a major source of systematic
error in semileptonic form factor calculations and methods are being
developed for complete nonperturbative determinations in order to reduce
such errors in the future.

Another important feature of lattice simulations is that calculations
are carried out at nonzero lattice spacings and with \emph{up}- and
\emph{down}-quark masses $m_q$ that are larger than in the real world.
Results are obtained for several lattice spacings and for a sequence of
$m_q$ values and one must then extrapolate to both the continuum and the
physical quark mass limits.
These two limits are intimately connected to each other, and it is now
standard to use chiral perturbation theory ($\chi$PT) that has been
adapted to include discretization effects~\cite{Lee:1999zxa,Aubin:2003mg,%
Bar:2003mh,Sharpe:2004ny,Bar:2005tu,Aubin:2007mc}. 

The initial pioneering work on $B$ and $D$ meson semileptonic decays on
the lattice were all carried out in the quenched
approximation~\cite{Bowler:1999xn,Abada:2000ty,Aoki:2001rd,%
ElKhadra:2001rv,Shigemitsu:2002wh}.
This approximation which ignores effects of sea quark-antiquark pairs
has now been overcome and most recent lattice calculations include
vacuum polarization from $N_f=2+1$ or $N_f=2$ dynamical light quark
flavors.
Unquenched calculations of $B \to \pi\ell\nu$ semileptonic decays have 
been carried out by the Fermilab/MILC and the HPQCD collaborations using 
the MILC collaboration $N_f=2+1$ 
configurations~\cite{Okamoto:2004xg,Dalgic:2006dt,Bailey:2008wp}.
Both collaborations use improved staggered (AsqTad) quarks for light
valence and sea quarks.
They differ, however, in their treatment of the heavy $b$ quark.
Fermilab/MILC employs the heavy clover action and HPQCD the
nonrelativistic NRQCD action.
The dominant errors in both calculations are due to statistics and the
chiral extrapolation.
The next most important error stems from discretization corrections for
the Fermilab/MILC and operator matching for the HPQCD collaborations,
respectively.
It is important that simulations based on other light quark lattice
actions be pursued in the future as a cross check.


In the case of $D \rightarrow K$ and $D \rightarrow \pi$ semileptonic
decays, there exists to-date only one $N_f=2+1$ calculation, again based 
on AsqTad light and clover heavy quarks, by the
Fermilab Lattice and MILC collaborations~\cite{Okamoto:2004xg}.
Recently two groups have initiated $N_f=2$ calculations, and
their results are still at a preliminary stage. 
The ETM collaboration uses ``twisted mass'' light and charm quarks at
maximal twist~\cite{Blossier:2008dj}, whereas Be\'cirevi\'c, Haas and
Mescia use improved Wilson quarks and configurations created by the
QCDSF collaboration~\cite{Becirevic:2007cr,Haas:2008fw}.
The latter group employs double ratio methods and twisted boundary
conditions to allow more flexibility in picking out many values of
$q^2$.
There has also been a recent exploratory study with improved Wilson
quark action which, although still quenched, is at a very small lattice
spacing of around 0.04~fm~\cite{Khan:2007hn}.
These authors have considered both $B$ and $D$ decays.

\subsubsubsection{Light-cone QCD Sum Rules} 

Light-cone sum rules (LCSR)~\cite{Balitsky:1989ry,Braun:1988qv,Chernyak:1990ag}
combine the idea of the original QCD sum rules
\cite{Shifman:1978bx,Shifman:1978by} with the elements of the theory of
hard exclusive processes.
LCSR are  used in a wide array of applications (for a review, see
\cite{Colangelo:2000dp}), in particular, for calculating
$B\to\pi,K,\eta,\rho,K^*$ and $D\to \pi,K$ form factors~\cite{Belyaev:1993wp,%
Belyaev:1994zk,Khodjamirian:1997ub,Bagan:1997bp,Khodjamirian:2000ds,Ball:2004ye,%
Ball:2006yd,Ball:2006eu,Ball:2007hb,Duplancic:2008ix,Duplancic:2008tk}.
The starting point is a specially designed correlation function 
where the  product of two currents is 
sandwiched between the vacuum and an on-shell state. 
In the case of $\bar{B}^0\to\pi^+$ form factor
\begin{eqnarray}
    F_\mu (p,q) & = & i\int d^4xe^{iqx}
        \langle \pi^+(p)\mid T\{\bar{u}\gamma_\mu b(x),
        m_b\bar{b}i\gamma_5 d(0)\}\mid 0\rangle\nonumber \\
        & = & \left(\frac{2f_B f^+_{B\pi}(q^2)m_B^2}{m_B^2-(p+q)^2} +
        \sum\limits_{B_h}\frac{2f_{B_h}
	f^+_{B_h\pi}(q^2)m_{B_h}^2}{m_{B_h}^2-
	(p+q)^2}\right)p_\mu+ O(q_\mu)\,,
    \label{eq:slep:corr}
\end{eqnarray}
where the factor proportional to $p_\mu$ is transformed into a hadronic 
sum by inserting a complete set of hadronic states between the currents. 
This sum also represents, schematically, a dispersion integral
over the hadronic spectral density.
The lowest-lying $B$-state contribution contains the desired $B\to \pi$ form
factor multiplied by the $B$ decay constant.

At spacelike $(p+q)^2\ll m_b^2$ and at small and
intermediate $q^2\ll m_b^2$, the time ordered product in Eq.~(\ref{eq:slep:corr})
may also be expanded near the light-cone $x^2\sim 0$, thereby resumming local
operators into distribution amplitudes:
\begin{equation}
    F((p+q)^2,q^2) = \sum\limits_{t=2,3,4}
	\int Du_i \sum\limits_{k=0,1}\left(\frac{\alpha_s}{\pi}\right)^k
	\; T^{(t)}_k((p+q)^2,q^2,u_i,m_b,\mu)
	\varphi^{(t)}_\pi(u_i,\mu)\,. 
    \label{eq:slep:ope}
\end{equation}
This generic expression is a convolution (at the factorization scale
$\mu$) of calculable short-distance coefficient functions $T^{(t)}_k$
and universal pion light-cone distribution amplitudes (DA's)
$\varphi^{(t)}_\pi(u_i,\mu)$ of twist $t\geq 2$. The integration goes
over the pion momentum fractions $u_i={u_1,u_2,...}$ distributed among
quarks and gluons.  Importantly, the contributions to Eq.~(\ref{eq:slep:ope})
corresponding to higher twist and/or higher multiplicity pion DA's are
suppressed by inverse powers of the $b$-quark virtuality $
((p+q)^2-m_b^2)$, allowing one to retain a few low twist contributions
in this expansion. 
Currently, analyses of Eq.~(\ref{eq:slep:corr}) can include all LO 
contributions of twist 2,3,4 quark-antiquark and quark-antiquark-gluon
DA's of the pion and the $O(\alpha_s)$ NLO corrections to the twist 2
and 3 two-particle coefficient functions.

Furthermore, one uses quark-hadron duality to 
approximate the sum over excited $B_h$ states 
in Eq.~(\ref{eq:slep:corr}) by the result from the 
perturbative QCD calculation introducing 
the effective threshold parameter $s_0^B$. 
The final step involves  a Borel transformation 
$(p+q)^2 \to M^2$, where  the scale of  the Borel parameter 
$M^2$ reflects  the characteristic virtuality 
at which the correlation function is calculated. 
The resulting LCSR for the $B\to \pi$ form factor  
has the following form
\begin{equation}
    f_{B\pi}^+(q^2)\:=\:\dfrac{e^{m_B^2/M^2}}{2m_B^2f_B}
        \dfrac{1}{\pi}\int_{m_b^2}^{s_0^B}ds\,\mbox{Im}\,F^{(OPE)}(s,q^2)
	e^{-s/M^2},
    \label{eq:slep:lcsr}
\end{equation}
where $\mbox{Im}F^{(OPE)}$ is directly calculated 
from the double expansion (\ref{eq:slep:ope}). 
The intrinsic uncertainty 
introduced by the quark-hadron duality approximation
is minimized by calculating the $B$ meson mass 
using the derivative of the same sum rule.   
The main input parameters,
apart from $\alpha_s$ and $b$ quark mass (taken in the $\overline{\rm MS}$
scheme), include the nonperturbative 
normalization constants and nonasymptotic coefficients
for each given twist component, e.g., for the twist-2 
pion DA $\varphi_\pi$ these are 
$f_\pi$ and the Gegenbauer moments $ a_i$. For twist-3,4 
the recent analysis can be found in Ref.~\cite{Ball:2007zt}.  
For the $B$-meson decay constant entering LCSR (\ref{eq:slep:lcsr})
one usually employs the conventional
QCD sum rule  for the two-point correlator
of $\bar{b}i\gamma_5 q$ currents with $O(\alpha_s)$ 
accuracy (the most complete sum rule in $\overline{\rm MS}$-scheme
is presented  in \cite{Jamin:2001fw}).
More details on the numerical results, sources of uncertainties
and their estimates
can be found in the recent update \cite{Duplancic:2008ix}.
Further improvement of the LCSR calculation of heavy-to-light
form factors is possible,
if one gets a better understanding of the quark-hadron 
duality approximation in $B$ channel, and a 
more accurate estimation of nonperturbative 
parameters of pion DA's.

Despite their intrinsically approximate nature, LCSRs represent a useful
analytic method providing a unique possibility to calculate both hard
and soft contributions to the transition form factors.  Different
versions of LCSR employing $B$-meson distribution amplitudes
\cite{Khodjamirian:2006st} as well as the framework of SCET
\cite{DeFazio:2005dx, DeFazio:2007hw} have also been introduced.

\subsubsection{Measurements of $D$ Branching Fractions and $q^2$ Dependence}

In the last few years, a new level of precision has been achieved in 
measurements of branching fractions and hadronic form factors 
for exclusive semileptonic $D$ decays by the Belle, BaBar, and CLEO 
collaborations. In this section, we focus on semileptonic decays, 
$D \to P \ell \nu_{\ell}$, where $D$ represents a $D^0$ or $D^+$, 
$P$ a pseudoscalar meson, charged or neutral, either $\pi$ or $K$, and 
$\ell$ a muon or electron.  
In addition, we also present a BaBar analysis of $D_s^+ \to K^+ K^- 
\ell^+ \nu_{\ell}$,  which provided first evidence of an S-wave contribution.

The results from the $B$-Factories (Babar and Belle) are based on very large
samples of $D$ mesons produced via the process $\epem \to c{\bar c}$ recorded 
at about 10.58 \gev\ c.m. energy. 
CLEO-c experiment relies on a sample of $\psiprpr\to D\Dbar$ events,
which is smaller, but allows for very clean tags and excellent $q^2$ 
resolution. 
Two of the four recent analyses tag events by reconstructing a hadronic 
decay of one of the $D$ mesons in the event, in addition to the semileptonic
decay of the other. The total number of tagged events serves as a 
measure of the total sample of $D$ mesons and thus provides the absolute 
normalization for the determination of the semileptonic branching 
fractions. 
Untagged analyses typically rely on the relative normalization to a 
sample of $D$ decays with a well measured branching fraction.
The analyses use sophisticated techniques for background suppression 
(Fisher discriminants) and resolution enhancement (kinematic fits).
The neutrino momentum and energy is equated with the reconstructed 
missing momentum and energy relying on energy-momentum conservation.
The detailed implementation and resolution varies significantly among
the measurements and cannot be presented here in detail. 

The BaBar Collaboration reports a study of $D^{0} \to K^{-} e^{+}\nu_e$ 
based on a luminosity of $75~\textrm{fb}^{-1}$~\cite{Aubert:2006mc}. 
They analyze $D^{*+} \to D^{0} \pi^{+}$ decays, with $D^{0} \to K^{-} 
e^{+} \nu_e$.
The analysis exploits the two-jet topology of $\epem \to c{\bar c}$ events.
The events are divided by the plane perpendicular to the event thrust 
axis into two halves, each equivalent to a jet produced by $c$- or 
$\cbar$-quark fragmentation.
The energy of each jet is estimated from its measured mass and the total 
c.m. energy.
To determine the momentum of the $D$ and the energy of the neutrino a 
kinematic fit is performed to the total event, constraining the 
invariant mass of the $K^{-} e^{+} \nu_e$ candidate  to the $D^{0}$ mass.
The $D$ direction is approximated by the direction opposite the vector 
sum of the momenta of all other particles in the event, except the Kaon 
and lepton associated with the signal candidate.
The neutrino energy is estimated as the difference between the total 
energy of the jet containing the Kaon and charged lepton and the sum 
of the particle energies in that jet.
To suppress combinatorial background each $D^{0}$ candidate is combined 
with a \pip of the same charge as the lepton and the mass difference is 
required to be small,  $\delta M = M(D^{0}\pi^+)-M(D^{0}) < 0.160~\gev$.
The background-subtracted $q^2$ distribution is corrected for efficiency 
and detector resolution effects.

For BaBar's analysis~\cite{Aubert:2006mc},
the normalization of the form factor at $q^{2}=0$ is 
$f_{+}^{K}(0)=0.727\pm0.007\pm0.005\pm0.007$, where the 
first error is statistical, the second systematic, and the third
due to uncertainties of external input parameters. 
In addition to the traditional parametrization of the form factors 
as a function of $q^2$ using pole approximations, BaBar also performed a 
fit in terms of the expansion in the parameter $z$. The results are 
presented in Fig.~\ref{fig:dslep:babar_zfit}.
A fit to a polynomial shows that data are compatible with a linear 
dependence, which is fully consistent with the modified pole ansatz for 
$f_+(q^2)$.
\begin{figure}[bp]
\centering
    \includegraphics[width=0.6\textwidth]{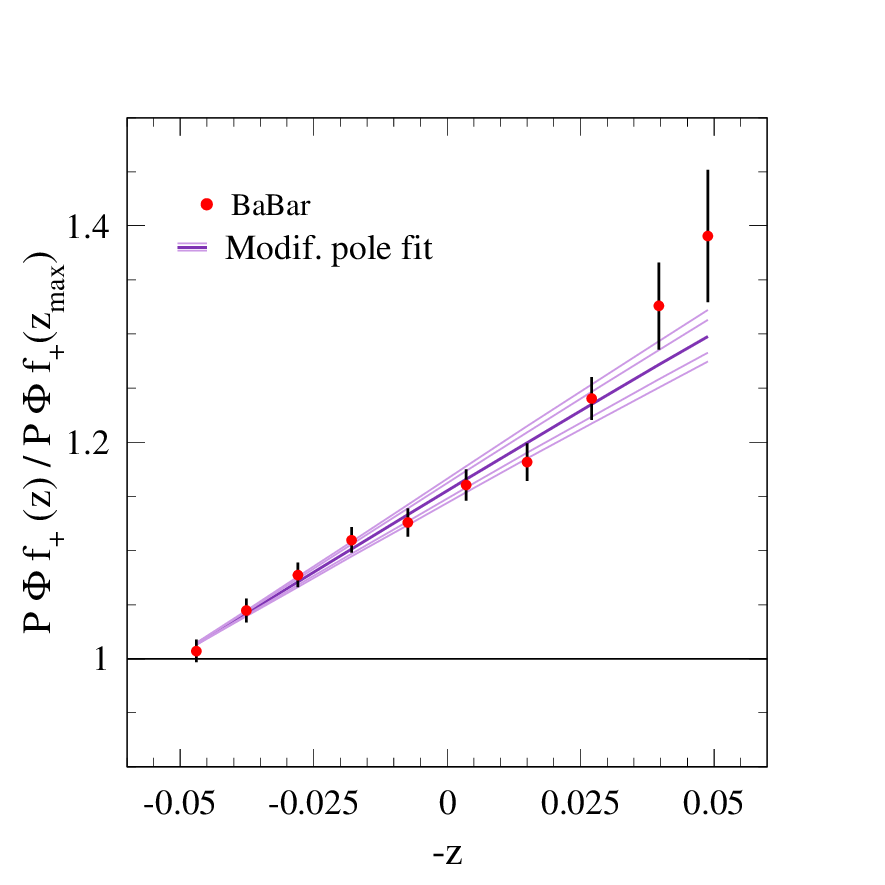}
    \caption{Babar analysis of 
        $D^0\to K^-e^{+}\nu_e$~\cite{Aubert:2006mc}: Measured values 
        for $P\times\Phi\times f_+$ versus $-z$, normalized to 1.0 at 
        $z=z_{max}$. The straight lines represent the expectation from the 
        fit to the modified pole ansatz, the  result in the center, as 
        well as the statistical and total uncertainties on either side.}
    \label{fig:dslep:babar_zfit}
\end{figure}

BaBar also reports the branching fraction for $D^0\to K^-e^+\nu_e$.
To obtain the normalization for the signal sample, they perform a 
largely identical analysis to isolate a sample of 
$D^0\to K^-\pi^{+}$ decays, and combine it with the world average 
${\cal B}(D^{0}\to K^{-} \pi^{+})=(3.80 \pm 0.07)\%$.
The result, the ratio of branching fractions,
$R_{D}={\cal B}(D^{0} \to K^{-} e^{+} \nu_e)/%
{\cal B}(D^{0}\to K^{-} \pi^{+}) = 0.927\pm0.007\pm0.012$,  translates to 
${\cal B}(D^{0} \to K^{-} e^{+} \nu_e)=%
(3.522 \pm 0.027 \pm 0.045 \pm 0.065)\%$, 
where the last error represents the uncertainty of 
${\cal B}(D^{0}\to K^{-}\pi^+)$. 

The Belle Collaboration has analyzed a sample of $282~\textrm{fb}^{-1}$, 
recorded at or just below the \FourS\ resonance~\cite{Widhalm:2006wz}. 
They search for the process, 
$\epem \to c{\bar c} \to D_{\rm tag}^{(*)}D^{*+}_{{\rm sig}}X$, 
with $D^{*+}_{{\rm sig}}\to D^{0}\pi^{+}_{\rm soft}$~\cite{Widhalm:2006wz}.
Here $X$ represents additional particles from $c$-quark fragmentation.
The $D_{\rm tag}$ is reconstructed as a $D^{0}$ or $D^{+}$, 
in decay modes $D \to K(n\pi)$ with $n=1,2,3$. 
In events that contain a $D^{*+}_{{\rm sig}}$,  the recoil of the 
$D_{\rm tag}^{(*)}X\pi^{+}_{\rm soft}$ provides an estimate of the signal
$D^{0}$-meson energy and momentum vector.
Figure~\ref{fig:dslep:belleinvmass} shows the invariant mass spectrum as 
derived from the $D_{\rm tag}^{(*)}X\pi^{+}_{\rm soft}$ system. 
\begin{figure}[bp]
\centering
    \includegraphics[width=0.70\textwidth]{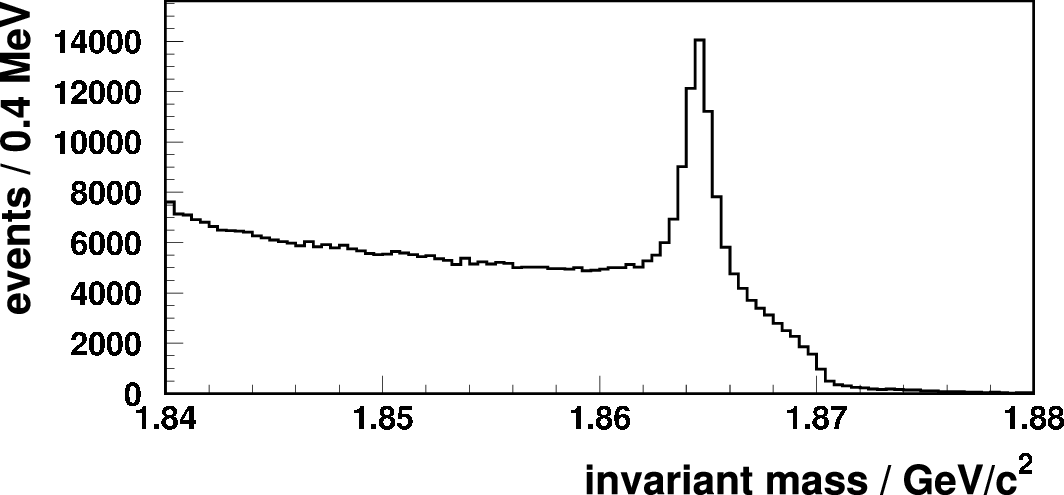}
    \caption{Belle experiment~\cite{Widhalm:2006wz}: 
    Invariant mass distribution for $D^0_{\rm sig}$ candidates.}
    \label{fig:dslep:belleinvmass}
\end{figure}
This distribution determines the number of $D^{0}$'s in the candidate 
sample and provides an absolute normalization. In this sample a 
search for semileptonic decays $D^{0} \to \pi^{-}\ell^{+} \nu_{\ell}$ or 
$D^{0} \to K^{-} \ell^{+} \nu_{\ell}$ is performed; here the charged
lepton is either an electron or muon. Pairs of a hadron and a lepton of 
opposite sign are identified  and the neutrino four-momentum is obtained 
from energy-momentum conservation.  Fig.~\ref{fig:dslep:bellemissmass} 
shows the distribution for the missing mass squared, $M_{\nu}^2$, 
which for signal events is required to be consistent with zero, 
$<0.05~\textrm{GeV}^2/c^{4}$.
\begin{figure}[bp]
\centering
    \includegraphics[width=0.47\textwidth]{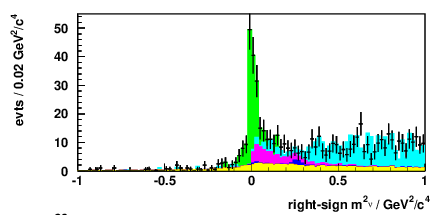}\hfill
    \includegraphics[width=0.47\textwidth]{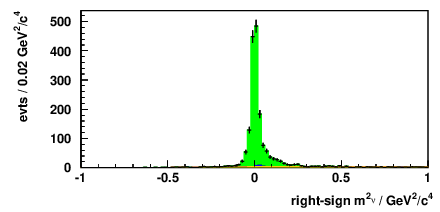}
    \caption{Belle experiment~\cite{Widhalm:2006wz,Widhalm:2007mi}:
    Missing mass squared distribution for $D^0_{\rm sig}$ candidates.
    Left: $D^{0} \to \pi^{-}\ell^{+} \nu_{\ell}$;
    right:  $D^{0} \to K^{-}\ell^{+} \nu_{\ell}$. The $D^{0} \to K^{-}
    \ell^{+} \nu_{\ell}$  and fake $D^0$ backgrounds are derived from
    data and are shown in magenta and yellow respectively. The cyan
    histogram shows the contribution from $D^{0} \to K^{*}/\rho \ell^{+}
    \nu_{\ell}$ as determined from simulation.}
    \label{fig:dslep:bellemissmass}
\end{figure}
The resulting branching fractions are 
${\cal B}(D^{0} \to K^{-} \ell^{+} \nu_{\ell})=(3.45\pm0.07\pm0.20)\%$ and
${\cal B}(D^{0} \to \pi^{-} \ell^{+} \nu_{\ell})=(0.255\pm0.019\pm0.016)\%$.
The measured form factors as a function of $q^{2}$ are also included 
in Fig.~\ref{fig:dslep:allfpq2k} for both decay modes. 
The normalization of the form factors at $q^{2}=0$ are 
$f_{+}^{K}(0)=0.695\pm0.007\pm0.022$ and  
$f_{+}^{\pi}(0)=0.624\pm0.020\pm0.030$. 

The CLEO Collaboration analyzed data recorded at the mass at the 
$\psi(3770)$ resonance, which decays exclusively to $D\bar{D}$ pairs.
They report measurements of semileptonic decays of both $D^{0}$ and 
$D^{+}$, for both untagged and tagged events.
For the untagged analysis~\cite{:2007sm} the normalization of $D\bar{D}$
pairs is based  on a separate analysis~\cite{:2007zt}. 
Individual hadrons, $\pi^{-}$, $\pi^{0}$, $K^{-}$, or $K_{\rm S}$, are 
paired with an electron and the missing momentum and energy 
of the entire event are used to estimate the neutrino four-momentum.
The missing mass squared is required to be consistent with zero.
Additionally, the four-momentum of the signal candidates, i.e., the sum of the 
hadron, lepton and neutrino energies  must be consistent
with the known energy and mass of the $D$ meson.
The yield of $D$ mesons is extracted in five $q^2$ bins.  
The CLEO Collaboration reports the branching fractions,
${\cal B}(D^{0} \to K^{-} \e^{+} \nu_e)       =(3.56 \pm 0.03  \pm 0.09)\%$,
${\cal B}(D^{0} \to \pi^{-} \e^{+} \nu_e)     =(0.299\pm 0.011 \pm 0.09)\%$,
${\cal B}(D^{+} \to {\bar K^{0}} \e^{+} \nu_e)=(8.53 \pm 0.13 \pm 0.23)\%$,
and ${\cal B}(D^{+} \to \pi^{0} \e^{+} \nu_e) =(0.373\pm 0.022 \pm 0.013)\%$. 
Figure~\ref{fig:dslep:allfpq2k}
includes the CLEO-c untagged results for $f_{+}$($q^{2})$ versus $q^2$.

Recent results of the CLEO-c tagged analysis~\cite{:2008yi}   
were reported for the first time at this workshop.
This analysis is based on a luminosity of 281~$\textrm{pb}^{-1}$.   
To tag events, all events are required to 
have a hadronic $D$ decay, fully reconstructed in one of eight channels for 
$D^{0}$ and one of six channels for $D^{+}$. Since the $D {\bar D}$ system 
is produced nearly at rest, the $D$ candidate should have an energy 
consistent with the beam energy. The beam-energy substituted mass, $m_{ES}$, 
is required to be consistent with the known $D$ mass.
For this sample of events, an electron is paired with a hadron, 
$\pi^{-}$, $\pi^{0}$, $K^{-}$, or $K_{\rm S}$. 
In $D {\bar D}$ events with a signal semileptonic decay, the only 
unidentified particle is the neutrino.
Its energy and momentum are derived from the missing energy and 
momentum.
The measured difference of these two quantities, $U=E_{\nu}-P_{\nu}$, is 
used to discriminate signal from background. 
Fig.~\ref{fig:dslep:cleocu} shows the $U$ distribution for the four 
semileptonic decay modes.
The requirement of a hadronic tag results in extremely pure samples.
For the decay $D^{0} \to K^{-} \e^{+} \nu_e$ the signal-to-noise ratio is 
about 300.
Based on these selected samples CLEO-c reports the branching fractions,  
${\cal B}(D^{0} \to K^{-} \e^{+} \nu_e)       =(3.61  \pm 0.05  \pm 0.05)\%$,
${\cal B}(D^{0} \to \pi^{-} \e^{+} \nu_e)     =(0.314 \pm 0.013 \pm 0.004)\%$,
${\cal B}(D^{+} \to {\bar K^{0}} \e^{+} \nu_e)=(8.90  \pm 0.17 \pm 0.21)\%$,
and ${\cal B}(D^{+} \to \pi^{0} \e^{+} \nu_e) =(0.384 \pm 0.027 \pm 0.023)\%$.
Figure~\ref{fig:dslep:allfpq2k}
shows the CLEO-c results for $f_{+}$($q^{2}$) versus $q^2$.

The CLEO Collaboration has computed the average of the untagged and tagged 
results, taking into account all correlations. The results for the
branching fractions are shown in Tab.~\ref{tab:cleocavgbr}.
The untagged analysis contains about 2.5 times more events but
has larger backgrounds and different systematic uncertainties.
The product of the form factor $f_+(0)$ and the CKM matrix element is 
extracted from the combined measurements, 
$f_{+}^{K}$($0$)$|V_{cs}|=0.744\pm0.007\pm 0.005$ and
$f_{+}^{\pi}$($0$)$|V_{cd}|=0.143\pm 0.005 \pm 0.002$. 

\begin{table}[tp]
    \caption{CLEO-c: Absolute branching fractions 
        for tagged, untagged and averaged results.}
    \label{tab:cleocavgbr}
    \centering
    {\small
    \begin{tabular}{lccc}
    \hline\hline
      & Tagged  & Untagged   & Average   \\
    \hline
    $\pi^-e^+\nu_e$&$0.308\pm 0.013\pm 0.004 $&$0.299\pm 0.011\pm 0.008 $&$ 0.304 \pm 0.011\pm 0.005 $ \\
    $\pi^0e^+\nu_e$&$0.379\pm 0.027\pm 0.002 $&$0.373\pm 0.022\pm 0.013$&$ 0.378 \pm 0.020\pm 0.012$ \\
    $K^-  e^+\nu_e$&$3.60 \pm 0.05 \pm 0.05 $& $3.56 \pm 0.03 \pm 0.09 $&$3.60  \pm 0.03 \pm 0.06 $ \\
    $\bar{K}^0e^+\nu_e$&$8.87 \pm 0.17\pm 0.21 $&$ 8.53 \pm 0.13\pm 0.23$&$8.69 \pm 0.12\pm 0.19 $ \\
    \hline\hline
    \end{tabular} }
\end{table}

\begin{figure}[bp]
\centering
    \includegraphics[width=0.70\textwidth]{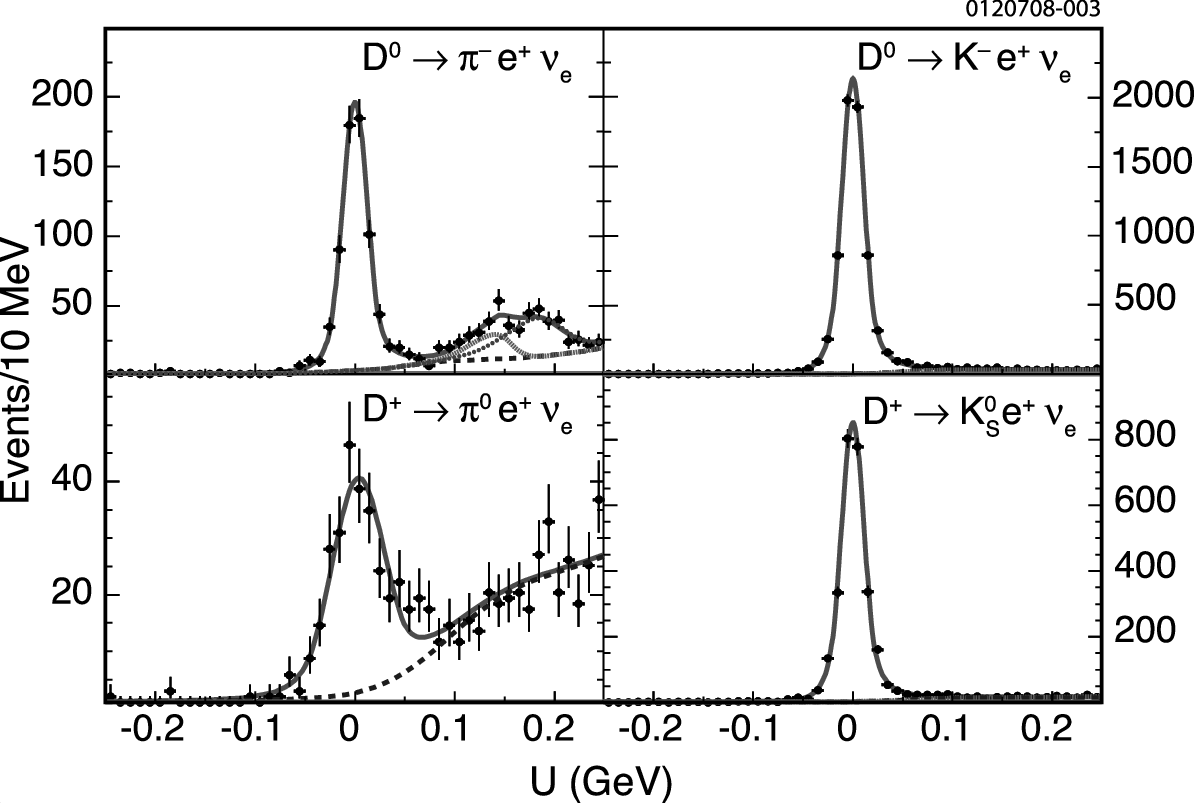}
    \caption{CLEO-c tagged analysis~\cite{:2008yi}:  Signal distributions ($U=E_{\nu}-P_{\nu}$) 
     for the four semileptonic $D$ decay channels.}
    \label{fig:dslep:cleocu}
\end{figure}
%
%

Since the time that the above results were reported at CKM2008,
CLEO collaboration has completed a new tagged analysis which is based on 
the entire 818~$\textrm{pb}^{-1}$ of data recorded at the $\psi(3770)$
resonance~\cite{Besson:2009uv}.
The results for the most recent branching fraction measurements are,  
${\cal B}(D^{0} \to K^{-} \e^{+} \nu_e)       =(3.50  \pm 0.03  \pm 0.04)\%$,
${\cal B}(D^{0} \to \pi^{-} \e^{+} \nu_e)     =(0.288 \pm 0.008 \pm 0.003)\%$,
${\cal B}(D^{+} \to {\bar K^{0}} \e^{+} \nu_e)=(8.83  \pm 0.10 \pm 0.20)\%$,
and ${\cal B}(D^{+} \to \pi^{0} \e^{+} \nu_e) =(0.405 \pm 0.016 \pm
0.009)\%$. The measured form factors as a function of $q^{2}$ for this
analysis are shown at the bottom of Fig.~\ref{fig:dslep:allfpq2k}. 
The product of the form factor $f_+(0)$ and the CKM matrix element is 
extracted from an isospin-combined fit which yields  
$f_{+}^{K}$($0$)$|V_{cs}|=0.719\pm0.006\pm 0.005$ and
$f_{+}^{\pi}$($0$)$|V_{cd}|=0.150\pm 0.004 \pm 0.001$. The new
CLEO-c results are consistent with the previous CLEO-c measurements and
supersede those measurements.  

At this conference BaBar reported a measurement of  $D_s^+ \to K^+ K^-
\ell^+ \nu_{\ell}$ decays~\cite{Aubert:2008rs}. Events with a $K^+K^-$ mass in 
the range $1.01-1.03\gevcc$ are selected, corresponding to $\phi \to 
K^+K^-$ decays, except for a small S-wave contribution which is observed 
for the first time.
Since the final state meson is a vector, the decay rate depends on 
five variables, the mass squared of the $K^+K^-$ pair, $q^2$ and three 
decay angles, 
and on three form factors, $A_1$, $A_2$ and $V$, for which  the $q^2$
dependence is assumed to be dominated by a single pole,
\begin{equation}
    V(q^2)       = \frac{V(0)}{1-q^2/m_V^2}, \quad
    A_{1,2}(q^2) = \frac{A_{1,2}(0)}{1-q^2/m_{A}^2}, \\
\end{equation}
with a total of five parameters, the normalizations $V(0), A_1(0), A2(0)$ and
the pole masses $m_V$ and $m_A$.  In a data sample of $214 \invfb$, 
the BaBar Collaboration selects about 25,000 
signal decays, about 50 times more than the earlier analysis
by FOCUS \cite{Link:2004qt}. The signal yield and the form factor ratios are 
extracted from a binned maximum likelihood fit to the four-dimensional decay 
distribution,
$r_2=A_2(0)/A_1(0)=0.763 \pm 0.071 \pm 0.065$ and 
$r_V=V(0)/A_1(0)  =1.849 \pm 0.060 \pm 0.095$, as well as the pole mass 
$m_A=2.28 ^{+0.23}_{-0.18} \pm 0.18 \gevcc$.
The sensitivity to $m_V$ is weak and therefore this parameter is fixed to 
$2.1\gevcc$.
The result of the fit is shown in Fig.~\ref{fig:dslep:babar_KKenu}.
The small S-wave contribution, which can be associated with $f_0 \to
K^+K^-$ decays, corresponds to $(0.22^{+0.12}_{-0.08}\pm0.03)\%$ of the
$K^+K^- e^+\nu_e$ decay rate.
\begin{figure}[bp]
\centering
    \includegraphics[width=0.6\textwidth]{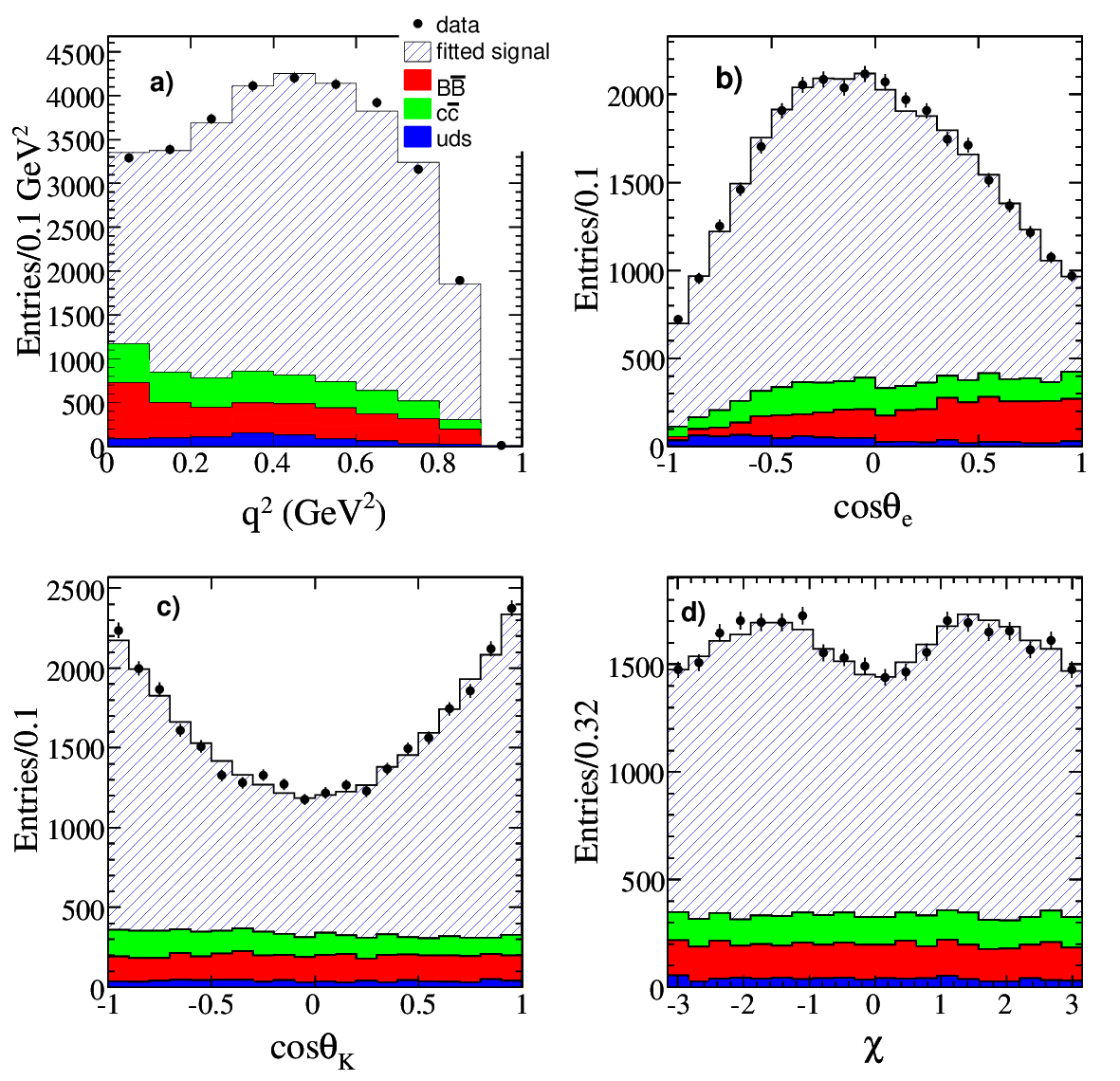}
    \caption{BaBar~\cite{Aubert:2008rs}: Projected distributions of                  the four kinematic variables. The data (points with statistical
errors) are compared to the sum of four contributions:
the fitted signal (hatched histograms) and the estimated background
contributions (different colored histograms) from \BB, \ccbar, and the 
sum of \uubar, \ddbar, and \ssbar events.}
    \label{fig:dslep:babar_KKenu}
\end{figure}
The $D_s^+ \rightarrow K^+ K^- e^+ \nu_e$ branching fraction is measured
relative to the decay $D_s^+ \rightarrow  K^+ K^- \pi^+$, resulting in 
${\cal B}(D_s^{+} \to K^+K^- e^{+} \nu_e)/%
{\cal B}(D_s^{+}\to K^+K^- \pi^{+}) = 0.558\pm0.007\pm0.016$, from which 
the absolute total branching fraction 
${\cal B}(D_s^{+} \to \phi e^{+} \nu_e)=%
(2.61 \pm 0.03 \pm 0.08 \pm 0.15)\%$ is obtained.   
By comparing this quantity with the predicted decay rate, using the 
fitted parameters for the form factors, the absolute 
normalization $A_1(0)=0.607 \pm 0.011 \pm 0.019 \pm 0.018$ was 
determined for the first time.
The third error stated here refers to the combined uncertainties from 
various external inputs, namely branching fractions for $D^+_s$, and 
$\phi$, the  $D^+_s$ lifetime and $|V_{cs}|$.
Lattice QCD calculations for this decay have been performed only in the
quenched approximation. They agree with the  experimental results for
$A_1(0)$, $r_2$ and $m_A$, but are lower than the measured value 
of $r_V$. It would be interesting to see if unquenched calculations are
in better agreement with experimental results.

In summary, BaBar, Belle and CLEO-c have measured $D$ meson semileptonic
branching fractions and hadronic form factors in a variety of decay modes,
using complementary experimental approaches. 
The results from the experiments are highly consistent. 
With lattice QCD prediction for the form factors, these results will allow
a precise determination of $V_{cs}$ and $V_{cd}$. 
Fig.~\ref{fig:dslep:allfpq2k} shows a compilation of all form factor 
measurements, $f_{+}(q^{2})$ versus~$q^2$.
\begin{figure}[bp]
\centering
    \includegraphics[width=0.47\textwidth]{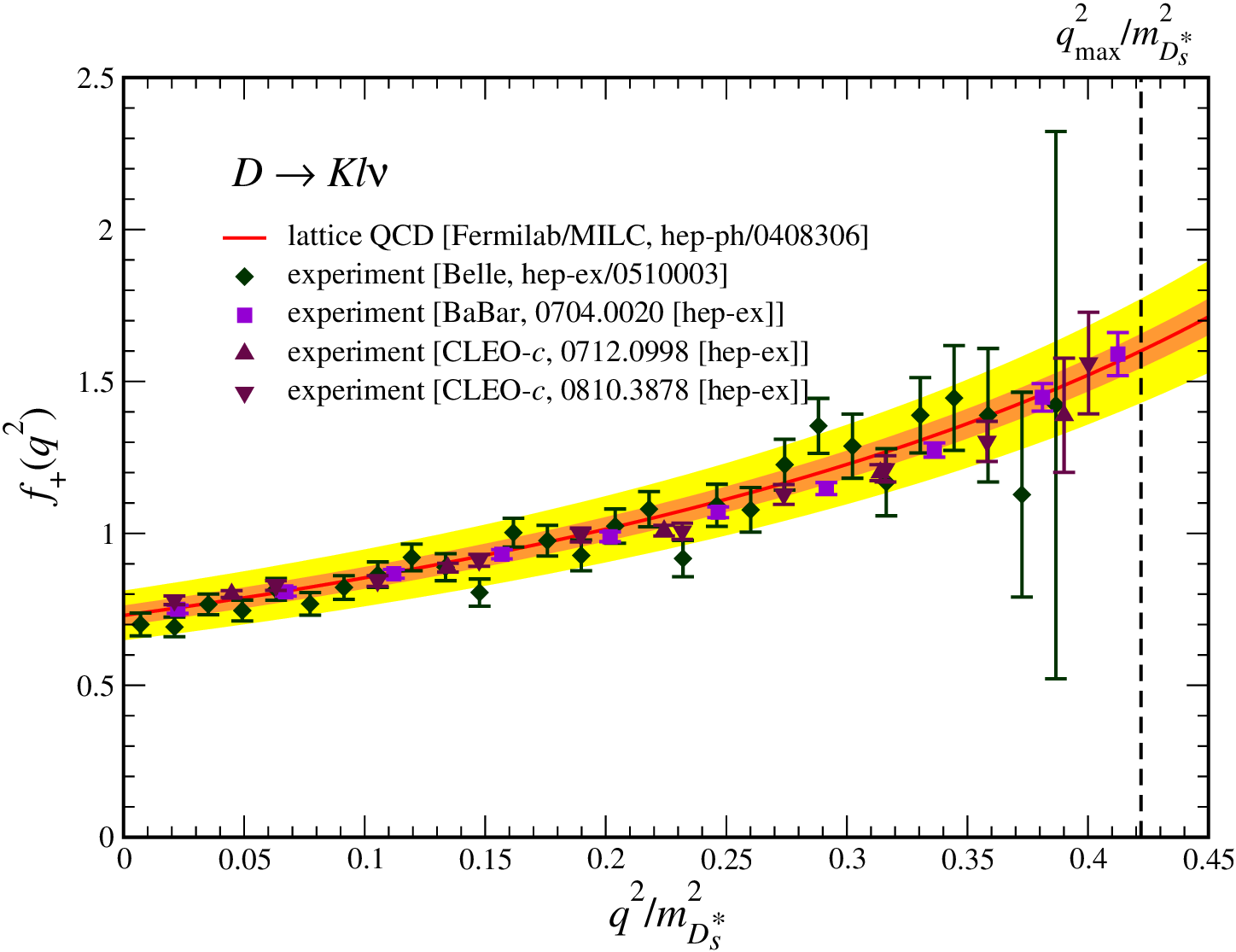}\hfill
    \includegraphics[width=0.47\textwidth]{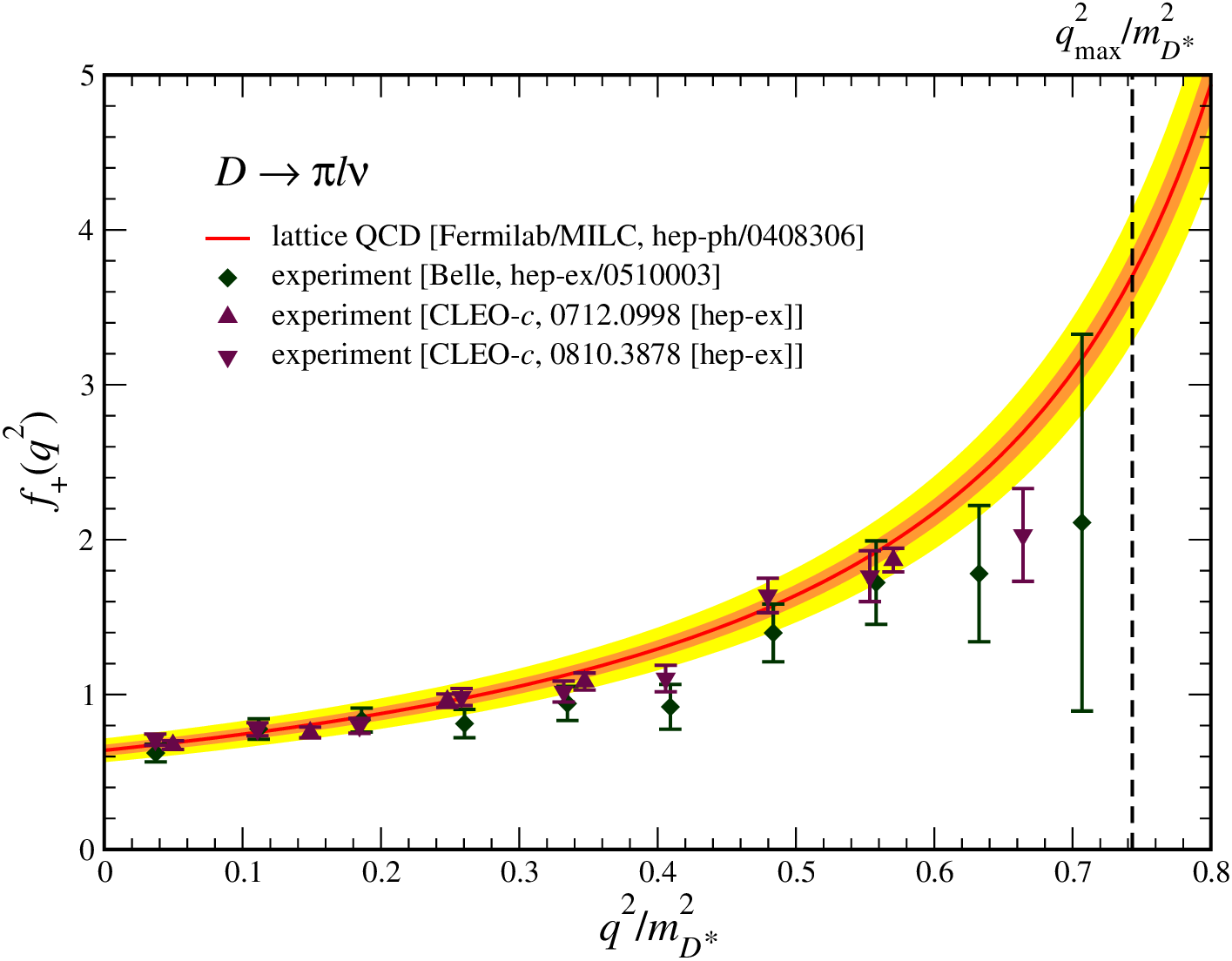}\\
    \includegraphics[width=0.47\textwidth]{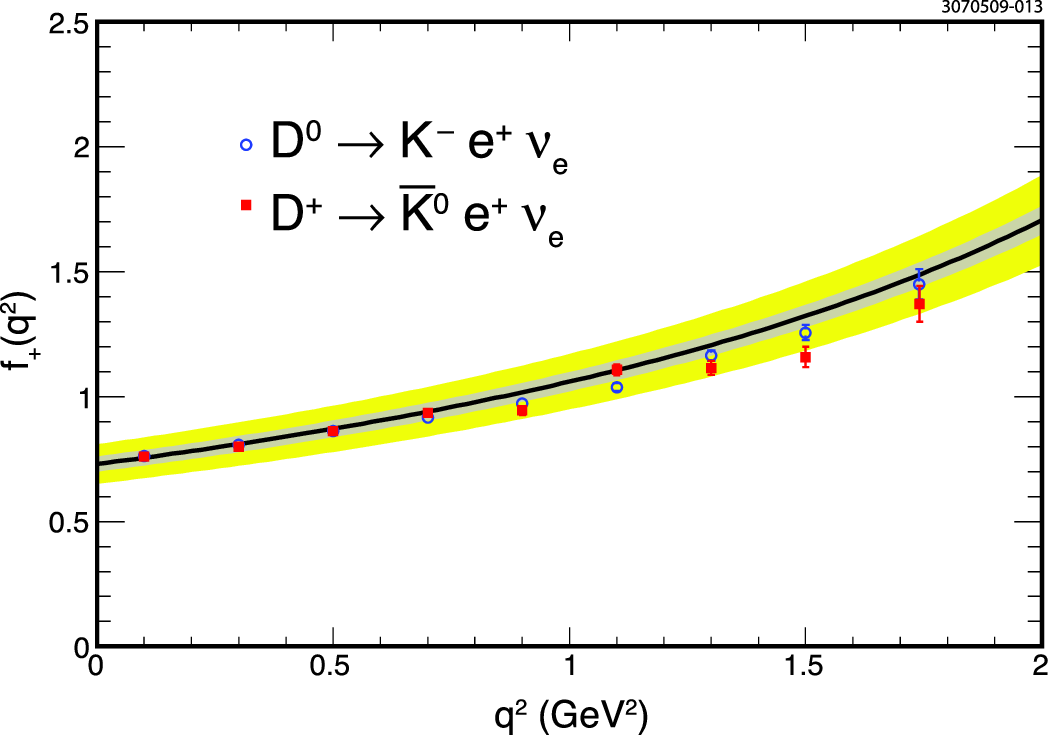}\hfill
    \includegraphics[width=0.47\textwidth]{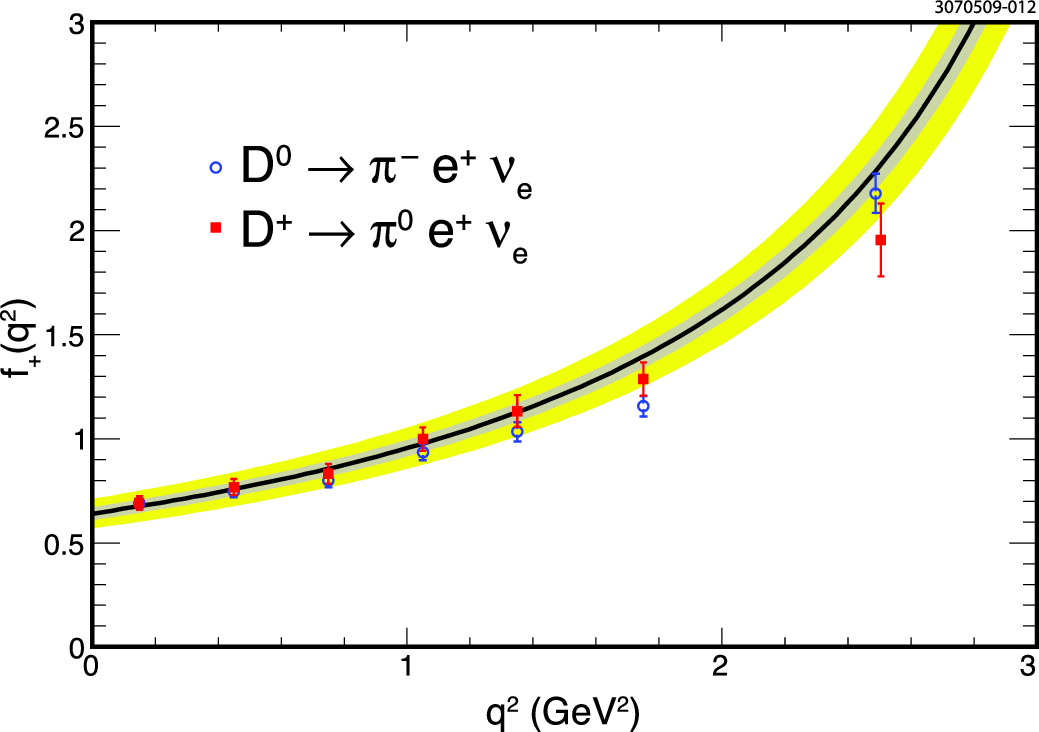}
     \caption{Compilation of the form factor $f_{+}(q^{2})$  versus 
        $q^2$ for the semileptonic $D$ decays  with a Kaon (left) and 
        pion (right).
        Top plots, adapted from Ref.~\cite{Bernard:2009ke}, include  
        measurements through the end of 2008.
        Bottom plots show results of a recent CLEO-c 
        analysis~\cite{Besson:2009uv}.
        In all plots the lines are the LQCD results of Ref.~\cite{Aubin:2004ej};
        the inner band represents statistical uncertainty and the outer 
        band includes the systematic uncertainty.}
    \label{fig:dslep:allfpq2k}
\end{figure}
\begin{table}[tp]
\caption{Summary of the form factors parameters obtained by the different 
experiments for $D \to K$ semileptonic decays. The first column gives 
the simple pole mass, the second the parameter $\alpha$ used in the 
modified pole model, and the third the normalization.}
\label{tab:FFDtoKsumary}
\centering
\begin{tabular}{l@{\quad}l@{\quad}l@{\quad}l}
\hline\hline
& \multicolumn{1}{c}{$M_{\rm pole} [\gevcc]$} & \multicolumn{1}{c}{$\alpha$}
& \multicolumn{1}{c}{$f_+(0)$} \\
\hline
Belle~\cite{Widhalm:2006wz} &$1.82\pm 0.04\pm 0.03 $&$0.52\pm 0.08\pm 0.06 $&$ 0.695 \pm 0.007\pm 0.022 $ \\
BaBar~\cite{Aubert:2006mc} &$1.884\pm0.012\pm0.015 $&$0.38\pm 0.02\pm 0.03$&$ 0.727 \pm 0.007\pm 0.005\pm 0.007$ \\ 
CLEO-c~\cite{Besson:2009uv} & $1.93 \pm 0.02 \pm 0.01$ & $0.30 \pm 0.03 \pm
0.01 $ & $0.739 \pm 0.007 \pm 0.005 $\\
LQCD~\cite{Aubin:2004ej}   &                        &$0.50\pm 0.04\pm 0.07$&$ 0.73 \pm 0.03\pm 0.07 $ \\
\hline\hline
\end{tabular}
\end{table}
\begin{table}[tp]
\caption{Summary of the form factors parameters obtained by the 
different experiments for $D \to \pi$ semileptonic decays. The first 
column gives the simple pole mass, the second the parameter $\alpha$  
used in the modified pole model, and the third the normalization.}
\label{tab:FFDtopisumary}
\centering
\begin{tabular}{l@{\quad}l@{\quad}l@{\quad}l}
\hline\hline
& \multicolumn{1}{c}{$M_{\rm pole} [\gevcc]$} & \multicolumn{1}{c}{$\alpha$}
& \multicolumn{1}{c}{$f_+(0)$} \\
\hline
Belle~\cite{Widhalm:2006wz} &$1.97\pm 0.08\pm 0.04 $&$0.10\pm 0.21\pm 0.10 $&$ 0.624 \pm
0.020\pm 0.030 $ \\
CLEO-c~\cite{Besson:2009uv} & $1.91 \pm 0.02 \pm 0.01$ & $0.21 \pm 0.07 \pm
0.02 $ & $0.666 \pm 0.019 \pm 0.004 \pm 0.003$\\
LQCD~\cite{Aubin:2004ej} &   &$0.44\pm 0.04\pm 0.07 $&$ 0.64 \pm 0.03\pm 0.06 $ \\
 \hline\hline
\end{tabular}
\end{table}
All analyses presented here have performed studies 
of the $q^2$ parameterizations and extractions of the associated parameters. 
A summary of these measurements is given in Tabs.~\ref{tab:FFDtoKsumary} 
and~\ref{tab:FFDtopisumary}, as well as the values obtained by lattice 
QCD computation~\cite{Aubin:2004ej}. 
The reader is referred to the references for more details.  

Measurements of $D \to \pi \ell \nu_{\ell}$ and 
$D \to V \ell \nu_{\ell}$ will benefit
from the increased data samples expected in the near future. Of
particular interest is the anticipated $\psi(3770)$ running of BES-III.
The BES-III Collaboration began data accumulation in July of 2008. 
The experiment is comparable to CLEO-c in detector design but has superior muon
identification performance, but worse performance for hadron identification, and is expected to accumulate at least an
order of magnitude more data. The muon identification will allow access
to all the semileptonic modes covered in this section from a single
experiment.

\subsubsection{Measurements of $B$ branching fractions and $q^2$ dependence}

Exclusive semileptonic decays $\B\to\X_u \ell\nu$, where $X_u$ denotes a
charmless hadronic final state, have been reported by the CLEO, BaBar,
and Belle collaborations~\cite{Adam:2007pv,Athar:2003yg,Aubert:2006px,%
:2008gka,Aubert:2006ry,Aubert:2008ct,Aubert:2005cd,Hokuue:2006nr,%
:2008kn,Abe:2003rh}.
The specification of the final state provides good kinematic
constraints and an effective background rejection, but results in lower
signal yields compared with inclusive measurements.  
Three experimental techniques that differ in the way the
second \B meson in the \BB event is treated have been employed in these
measurements.  The second \B meson is either fully reconstructed in a
hadronic decay mode (``hadronic tags''), partially reconstructed in a
semileptonic decay mode (``semileptonic tags'') or not reconstructed at
all (``untagged''). The tagged and untagged methods differ greatly in
terms of signal efficiency and purity.

\subsubsubsection{\boldmath$\B \to \pi \ell \nu$} 

The $\B \to \pi \ell \nu$ decay is the most promising decay mode for a
precise determination of \Vub, both for experiment and for theory.  A
number of measurements with different tagging techniques exist, but at
present the untagged analyses, which were first performed by the CLEO
collaboration~\cite{Athar:2003yg}, still provide the most precise
results.  In untagged analyses, the momentum of the neutrino is inferred
from the missing energy and momentum in the whole event. The neutrino is
combined with a charged lepton and a pion to form a $\B \to \pi \ell
\nu$ candidate.  The biggest experimental challenge is the suppression
of the $\B\to\X_c\ell\nu$ background.  Additional background sources are
$\epem \to \qqbar$ $(q=u,d,s,c)$ continuum events, which dominate at
low $q^2$, and feed-down from other $\B\to\X_u\ell\nu$ decays, which
dominate at high $q^2$.

The BaBar experiment has measured the $\B \to \pi \ell \nu$ branching
fraction and $q^2$ spectrum with a good accuracy~\cite{Aubert:2006px}.
In this analysis, the signal yields are extracted from a
maximum-likelihood fit to the two-dimensional \DeltaE vs. \mes
distribution of the signal $\B$ meson in twelve bins of $q^2$ (see
Fig.~\ref{fig:slep:piBaBarnotag}). This fit allows for an extraction of
the $q^2$ dependence of the form factor $f_+(q^2)$. The shape of the
measured spectrum is compatible with the ones predicted from
LQCD~\cite{Dalgic:2006dt,Okamoto:2004xg} and LCSR~\cite{Ball:2004ye}
calculations, but incompatible with the ISGW2 quark
model~\cite{Scora:1995ty}.  A fit to the $q^2$ spectrum using the
Becirevic-Kaidalov (BK) parametrization yields a shape parameter $\alpha
= 0.52 \pm 0.05 \pm 0.03$ with a goodness-of-fit of $P(\chi^2)=0.65$.
Other parameterizations, e.g. the $z$-expansion, have been used in a
simultaneous fit of the BaBar data and LQCD
calculations~\cite{Bailey:2008wp}.  The measured partial branching
fractions are extrapolated to the full decay rate and, in combination
with recent form-factor calculations, used to determine \Vub

The leading experimental systematic uncertainties are associated with
the reconstruction of charged and neutral particles, which impact the
modeling of the missing momentum reconstruction, and with backgrounds
from continuum events at low $q^2$ and from $\B\to\X_u\ell\nu$ decays at
high $q^2$.  Due to the feed-down from $\B\to\rho\ell\nu$ decays, 
the uncertainties on the branching fraction and form factors for this
decay mode contribute to the systematic uncertainty.  A simultaneous
measurement of $\B\to\pi\ell\nu$ and $\B\to\rho\ell\nu$ decays can
reduce this uncertainty.

\begin{figure}[bp]
\centering
    \includegraphics[width=0.47\textwidth]{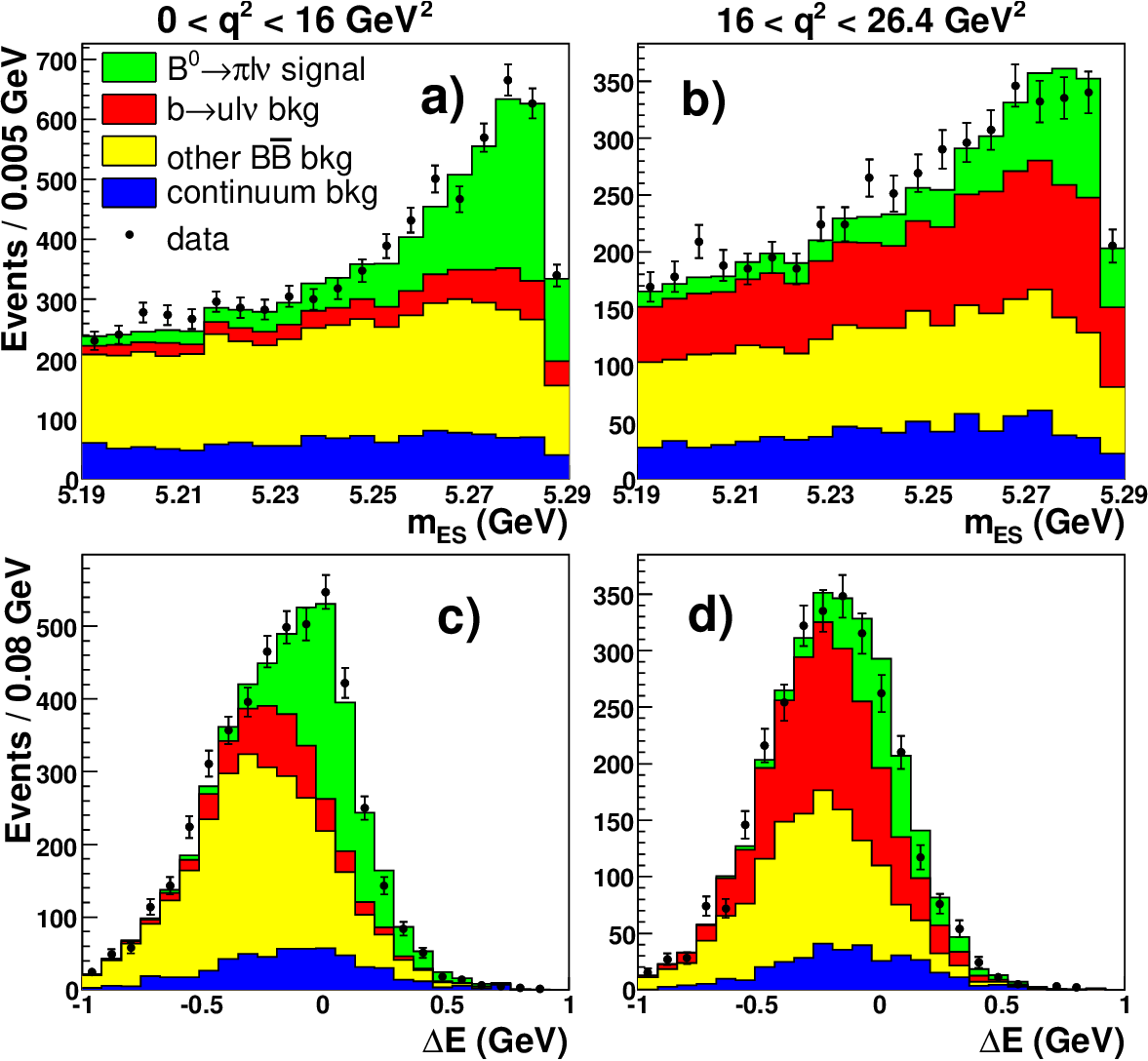}
    \includegraphics[width=0.47\textwidth]{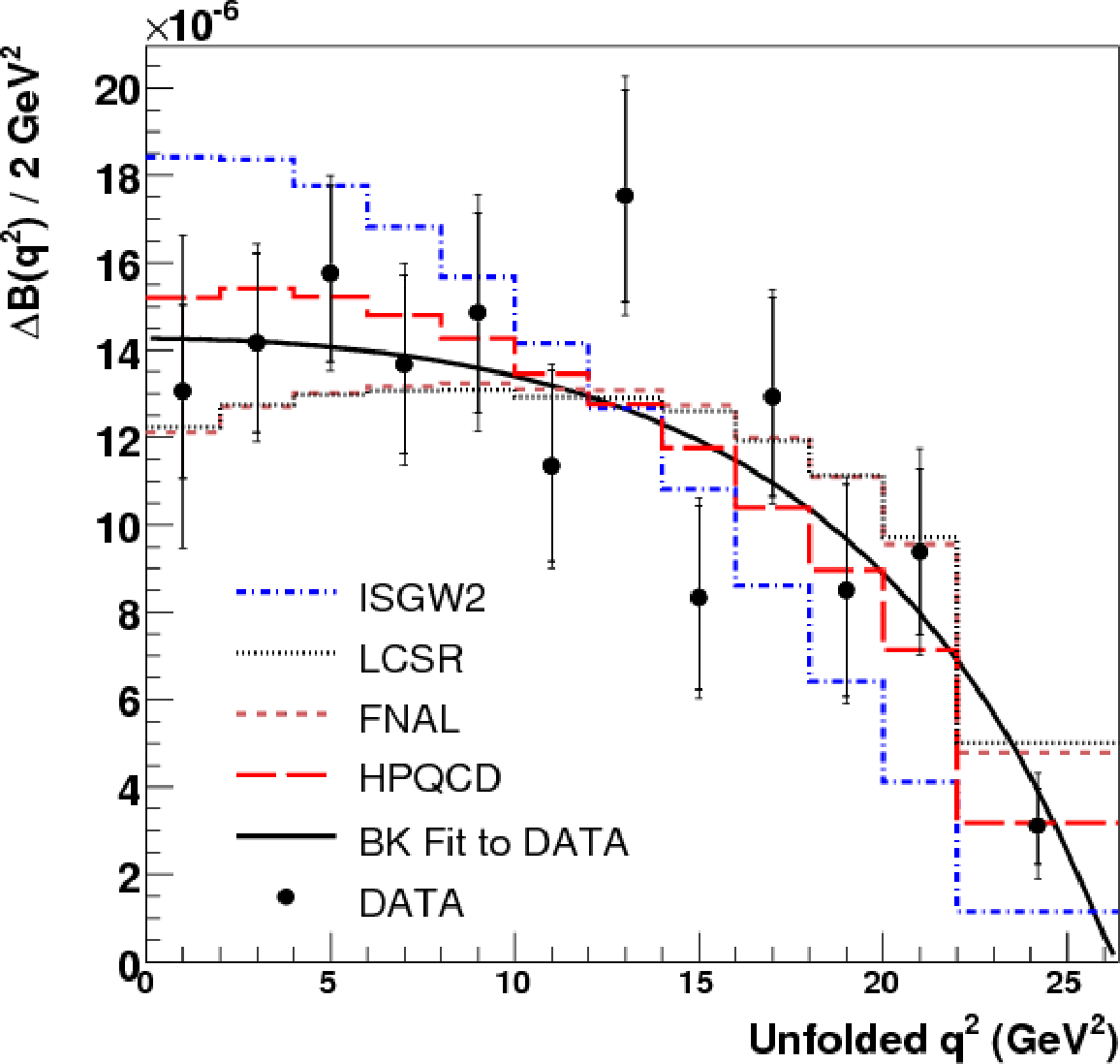}
    \caption{Untagged $\B \to \pi \ell \nu$ measurement from
	BaBar~\cite{Aubert:2006px}.
	Left: \DeltaE and \mes projections for $q^2 < 16~\gev^2$ and
	$q^2 > 16~\gev^2$.  Right: Measured $q^2$ spectrum compared
	with a fit of the BK parametrization and with theory
	predictions from LQCD~\cite{Dalgic:2006dt,Okamoto:2004xg},
	LCSR~\cite{Ball:2004ye} and the ISGW2 quark
	model~\cite{Scora:1995ty}.}
    \label{fig:slep:piBaBarnotag}
\end{figure}

Recently several tagged measurements have
appeared~\cite{Aubert:2006ry,:2008gka,Hokuue:2006nr,:2008kn}.
They have led to a simpler and more precise reconstruction of the
neutrino momentum and have low backgrounds and a uniform acceptance in
$q^2$.  This is achieved, however, at the expense of much smaller signal
samples which limit the statistical precision of the form-factor
measurement.  Semileptonic-tag measurements have a signal-to-background
ratio of around 1--2 and yield $\sim 0.5$ signal decays per $\invfb$. The
signal is extracted from the distribution of events in $\cos^2\phi_B$, 
where $\phi_B$ is
the angle between the direction of either \B meson and the plane
containing the momentum vectors of the tag-side $\Dstar\ell$ system and
the signal-side $\pi\ell$ system~\cite{:2008gka}.  Hadronic-tag
measurements reach signal-to-background ratios of up to $\sim 10$ and
yield $\sim 0.1$ signal decays per $\invfb$. Here the signal is
extracted from the missing-mass squared distribution (see
Fig.~\ref{fig:slep:hadtag}).

Tab.~\ref{tab:slep:BFpi} summarizes all $\B \to \pi \ell \nu$
branching-fraction measurements; shown are the total branching fraction
as well as the partial branching fractions for $q^2<16~\gev^2$ and
$q^2>16~\gev^2$ with statistical and systematic uncertainties.  The
measurements agree well among each other. A combination of all
measurements results in an average branching fraction of $1.34\times
10^{-4}$ with a precision of $6\%$ ($4\%$ statistical and $4\%$
systematic).

\begin{table}[tp]
    \centering
\caption{Total and partial branching fractions for $\B^0 \to \pim \ell^+ \nu$ 
with statistical and systematic uncertainties. 
Measurements of ${\cal B} (\B^+ \to \piz \ell^+ \nu)$
have been multiplied by a factor $2\tau_{B^0}/\tau_{B^+}$.}
\label{tab:slep:BFpi}
\renewcommand{\arraystretch}{1.1}
\begin{tabular}{lcccc} \hline\hline
           &${\cal L} (\invfb)$ & ${\cal B} \times 10^4$   & $\Delta {\cal B}(q^2<16) \times 10^4$ & $\Delta {\cal B}(q^2>16) \times 10^4$ \\ \hline
                  
BaBar no tag (\pim)~\cite{Aubert:2006px}& $206$ & $1.45 \pm 0.07 \pm 0.11$ & $1.08 \pm 0.06 \pm 0.09$ & $0.38 \pm 0.04 \pm 0.05$ \\
CLEO no tag (\pim,\piz)~\cite{Adam:2007pv}  & $ 16$ & $1.38 \pm 0.15 \pm 0.11$ & $0.97 \pm 0.13 \pm 0.09$ & $0.41 \pm 0.08 \pm 0.04$ \\ \hline
BaBar sl.\ tag (\pim)~\cite{:2008gka}    & $348$ & $1.39 \pm 0.21 \pm 0.08$ & $0.92 \pm 0.16 \pm 0.05$ & $0.46 \pm 0.13 \pm 0.03$ \\ 
Belle sl.\ tag (\pim)~\cite{Hokuue:2006nr}    & $253$ & $1.38 \pm 0.19 \pm 0.15$ & $1.02 \pm 0.16 \pm 0.11$ & $0.36 \pm 0.10 \pm 0.04$ \\
BaBar sl.\ tag (\piz)~\cite{:2008gka}    & $348$ & $1.80 \pm 0.28 \pm 0.15$ & $1.38 \pm 0.23 \pm 0.11$ & $0.45 \pm 0.17 \pm 0.06$ \\
Belle sl.\ tag (\piz)~\cite{Hokuue:2006nr}    & $253$ & $1.43 \pm 0.26 \pm 0.15$ & $1.05 \pm 0.23 \pm 0.12$ & $0.37 \pm 0.15 \pm 0.04$ \\ \hline
BaBar had.\ tag (\pim)~\cite{Aubert:2006ry}  & $211$ & $1.07 \pm 0.27 \pm 0.19$ & $0.42 \pm 0.18 \pm 0.06$ & $0.65 \pm 0.20 \pm 0.13$ \\
Belle had.\ tag (\pim)~\cite{:2008kn}  & $605$ & $1.12 \pm 0.18 \pm 0.05$ & $0.85 \pm 0.16 \pm 0.04$ & $0.26 \pm 0.08 \pm 0.01$ \\
BaBar had.\ tag (\piz)~\cite{Aubert:2006ry}  & $211$ & $1.54 \pm 0.41 \pm 0.30$ & $1.05 \pm 0.36 \pm 0.19$ & $0.49 \pm 0.23 \pm 0.12$ \\
Belle had.\ tag (\piz)~\cite{:2008kn} & $605$ & $1.24 \pm 0.23 \pm 0.05$ & $0.85 \pm 0.16 \pm 0.04$ & $0.41 \pm 0.11 \pm 0.02$ \\ \hline
Average                         &    & $1.36 \pm 0.05 \pm 0.05$ & $0.94 \pm 0.05 \pm 0.04$ & $0.37 \pm 0.03 \pm 0.02$ \\
\hline\hline
\end{tabular}
\end{table}

\subsubsubsection{\boldmath$\B \to \eta /\eta'/\rho/\omega \ell \nu$}

In addition to $\B \to \pi \ell \nu$, the experiments have measured other
semileptonic final states with a pseudoscalar meson, 
$\eta$~\cite{Athar:2003yg,Aubert:2006gba,:2008gka,Aubert:2008ct} 
or $\eta'$~\cite{Adam:2007pv,Aubert:2006gba,:2008gka}, 
or a vector meson, 
$\rho$~\cite{Athar:2003yg,Adam:2007pv,Hokuue:2006nr,:2008kn,Aubert:2005cd}
or $\omega$~\cite{Abe:2003rh,Aubert:2008ct}. 
They are important ingredients to the determination of the composition
of the inclusive $\B\to\X_u\ell\nu$ rate.  They may also help to further
constrain theoretical form-factor calculations and provide valuable
cross-checks for the determination of \Vub from $\B \to \pi \ell \nu$.
The LQCD calculations for these final states are challenging. For the
flavor-neutral final-state mesons, $\eta$, $\eta'$ and $\omega$, the
matrix element contains contributions from quark-disconnected diagrams.
For the $\rho$ final state, the large width of the $\rho$ resonance
complicates the calculations.

The $\eta$ and $\eta'$ modes have been measured by the CLEO and BaBar
collaboration.  The limit on ${\cal B} (\B\to\eta'\ell\nu)$ published by
BaBar~\cite{:2008gka} agrees only marginally with the CLEO
result~\cite{Adam:2007pv} (at the $2.6 \sigma$ level).  Further
measurements are needed to resolve this discrepancy.  In the future, a
measurement of the ratio $R_{\eta'\eta} = {\cal
B}(\B\to\eta'\ell\nu)/{\cal B}(\B\to\eta\ell\nu)$ would be interesting
to constrain the gluonic singlet contribution to the $\B\to\eta^{(\prime)}$ form
factor, as proposed in~\cite{Ball:2007hb}.

The $\B \to \rho \ell \nu$ decay has a larger rate than charmless
semileptonic decays into pseudoscalar mesons, but one must deal with the
non-resonant $\pi\pi$ contribution, which leads to a sizable systematic
uncertainty. The kinematics of decays with vector mesons are described
by three form factors. The statistical precision in current analyses is
still too low to measure these form factors. As an example,
Fig.~\ref{fig:slep:hadtag} shows the missing-mass and $q^2$ spectra of
$\B \to \rho \ell \nu$ and $\B \to \omega \ell \nu$ decays measured by
the Belle collaboration in a hadronic-tag analysis~\cite{:2008kn}.
Tab.~\ref{tab:slep:BFhigher} summarizes the most precise
branching fraction results for semileptonic $\B$ decays to low-mass
charmless hadrons heavier than the pion.

\begin{table}[tp]
\caption{Total branching fractions for exclusive $\B \to X_u\ell\nu$ 
decays with $X_u = \eta,\eta',\rho, \rm or~\omega$. 
$^\dagger$The BaBar collaboration reports an upper   
limit of ${\cal B}(\B^+\to\eta'\ell^+\nu)<0.47$ at $90\%$ CL~\cite{:2008gka}.}
\label{tab:slep:BFhigher}
\begin{center}
  \begin{tabular}{lccc} \hline
   Decay mode            & ${\cal B} \times 10^4$ & $\sigma_{stat} \times 10^4$ & $\sigma_{syst} \times 10^4$ \\ \hline
   $\B^+\to\eta\ell^+\nu$   (BaBar average)~\cite{Aubert:2008ct}       & 0.37 & 0.06 & 0.07 \\
   $\B^+\to\eta'\ell^+\nu$  (CLEO no tag)~\cite{Adam:2007pv}$^\dagger$ & 2.66 & 0.80 & 0.56 \\
   $\B^0\to\rho^-\ell^+\nu$ (average)         & 2.80 & 0.18 & 0.16 \\
   $\B^+\to\omega\ell^+\nu$ (BaBar no tag)~\cite{Aubert:2008ct}        & 1.14 & 0.16 & 0.08 \\ \hline
  \end{tabular}
\end{center}
\end{table}

\begin{figure}[bp]
    \centering
    \includegraphics[width=0.25\linewidth]{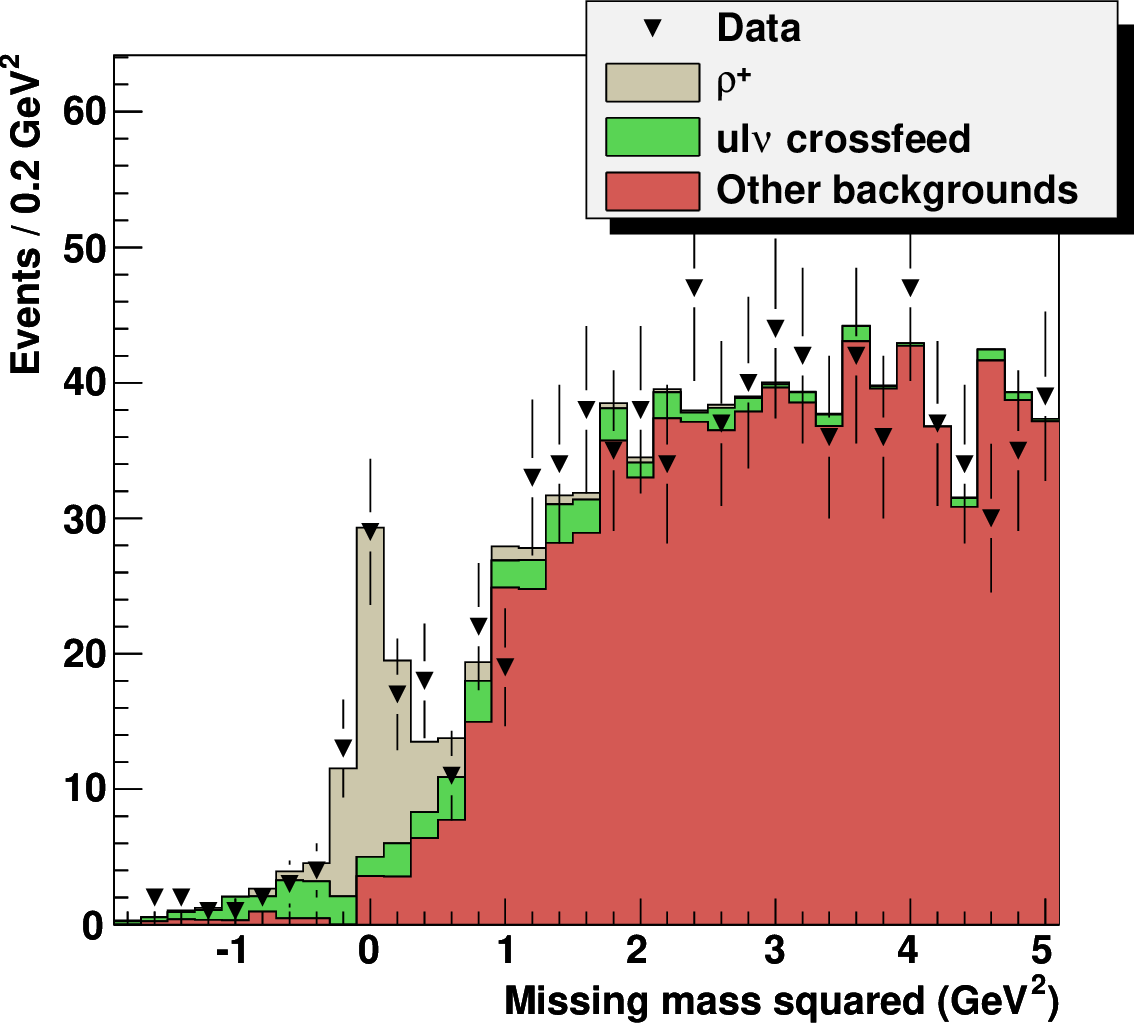}\hfill
    \includegraphics[width=0.25\linewidth]{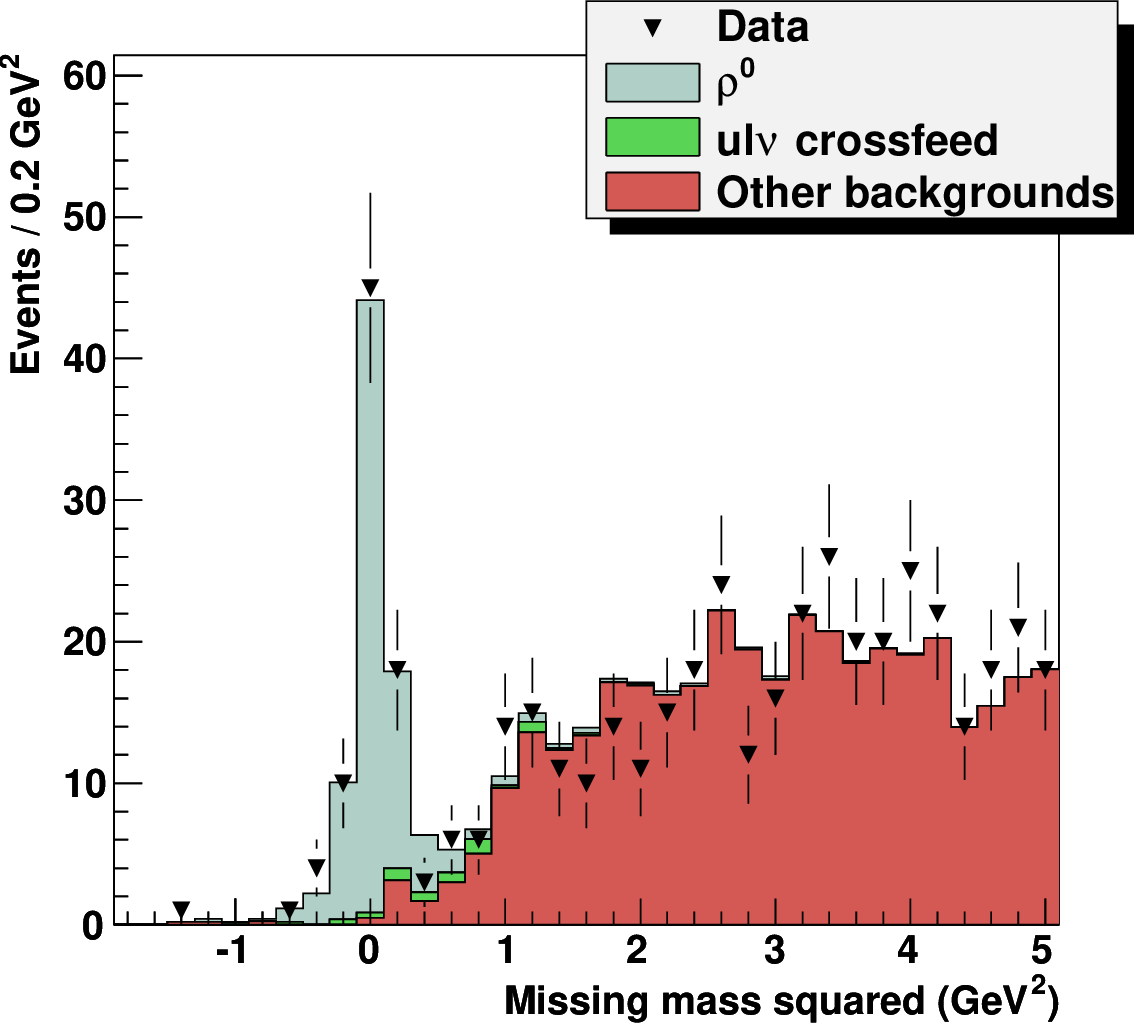}\hfill
    \includegraphics[width=0.25\linewidth]{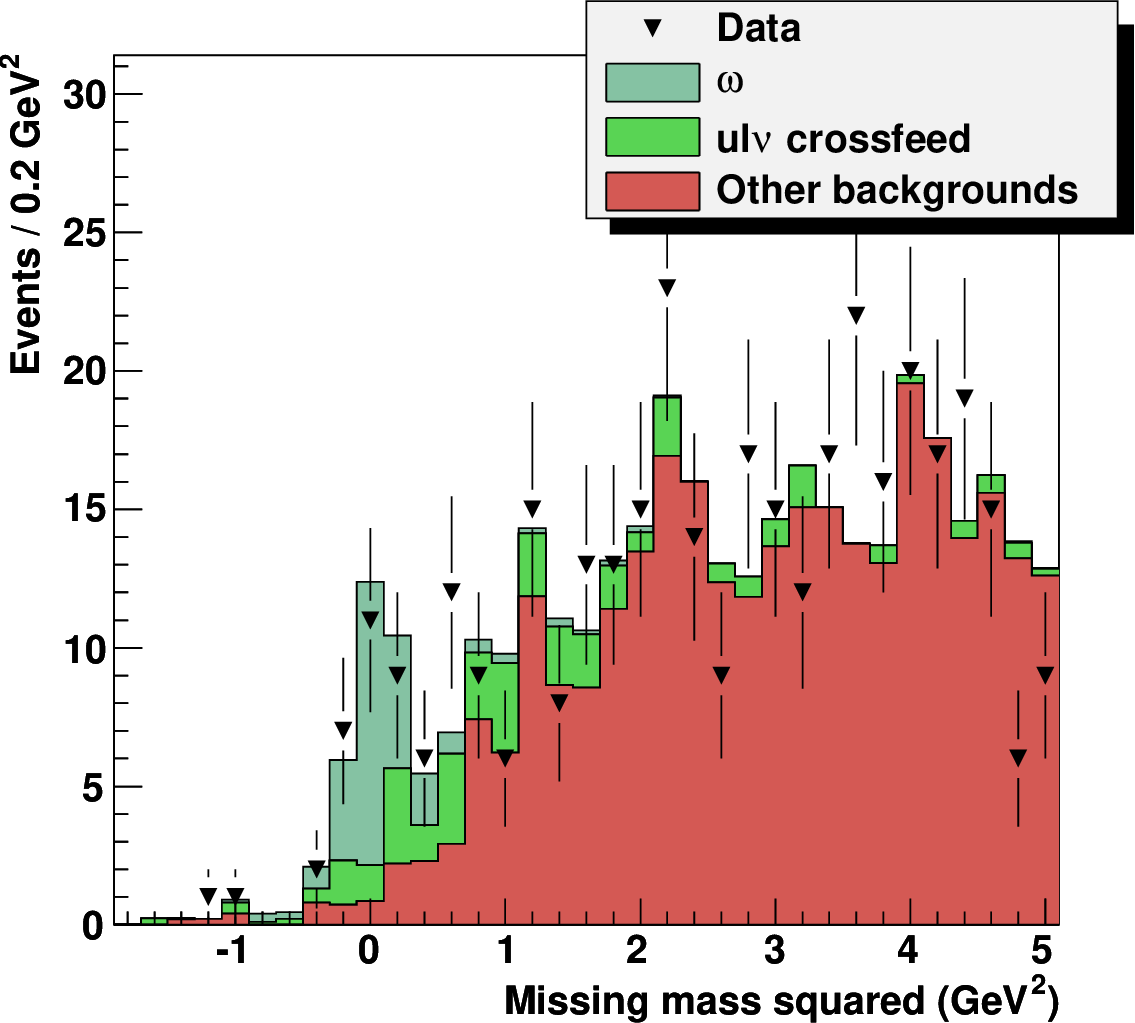}\\[0.3cm]
    \includegraphics[width=0.25\linewidth]{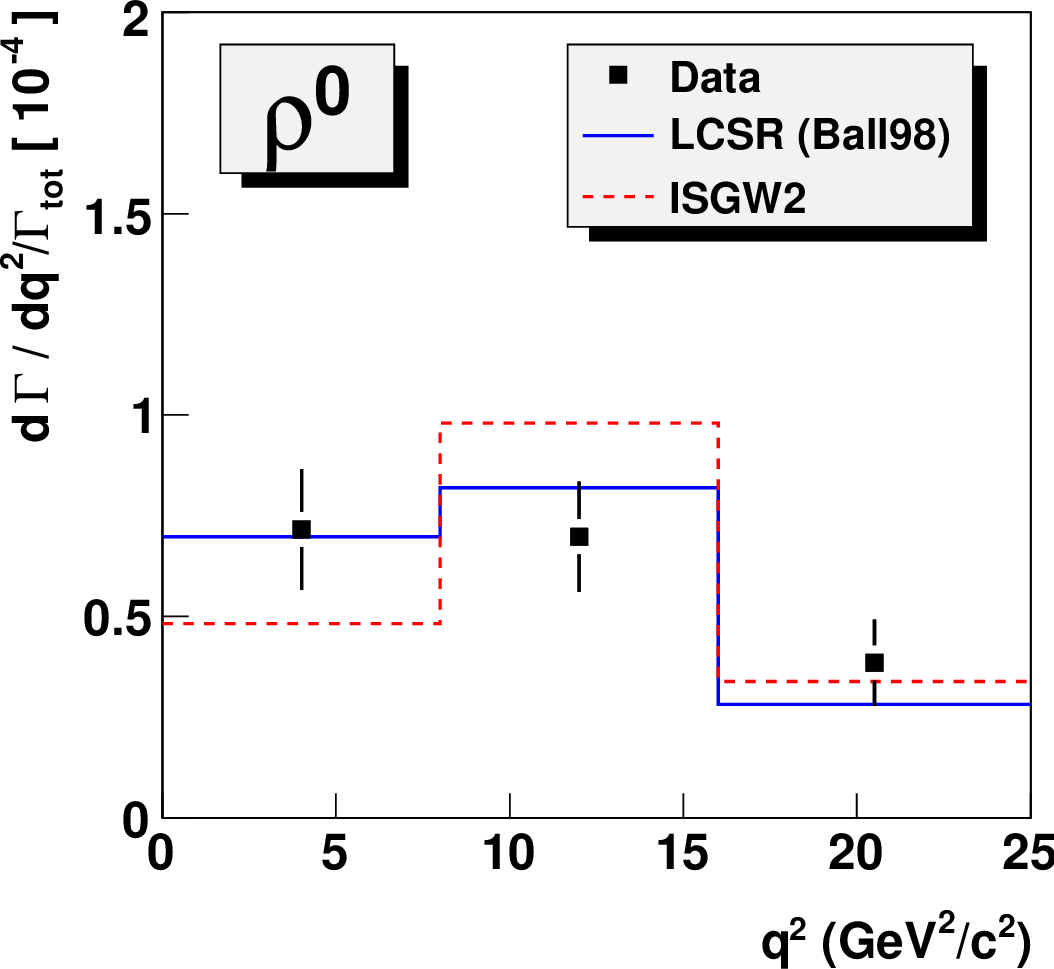}\hfill
    \includegraphics[width=0.25\linewidth]{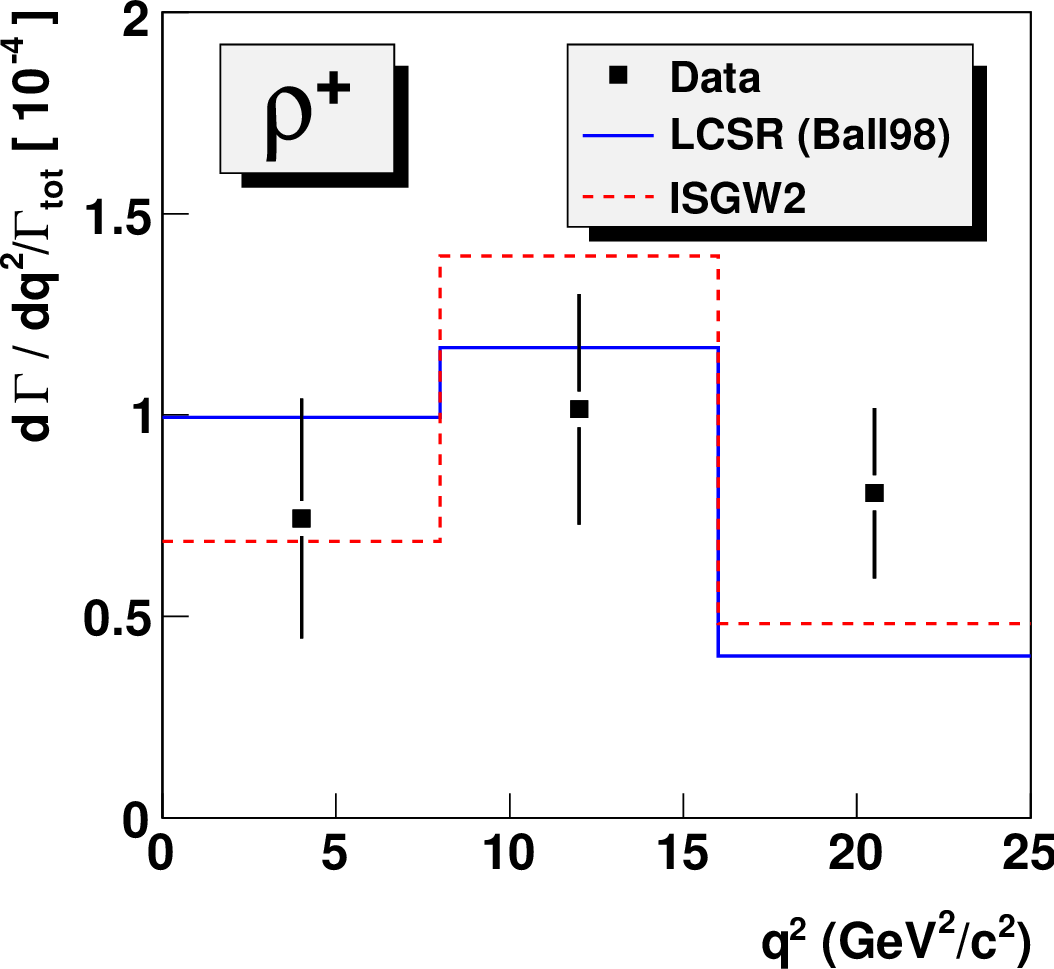}\hfill
    \includegraphics[width=0.25\linewidth]{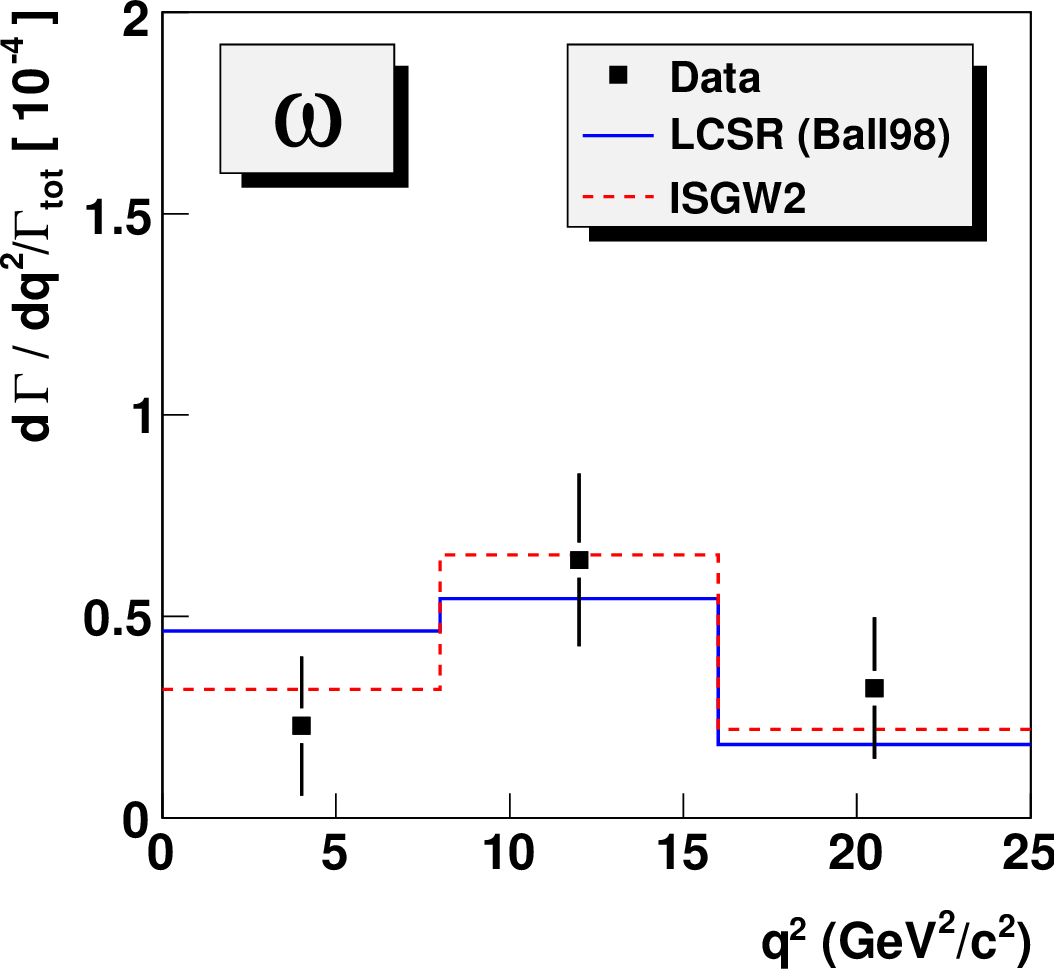}\hfill
    \caption{Belle hadronic-tag measurements~\cite{:2008kn}:
     Missing-mass squared distributions and $q^2$ spectra for 
     $\B \to \rho\ell\nu$ and $\B \to \omega\ell\nu$ decays.}
    \label{fig:slep:hadtag}
\end{figure}

\subsubsubsection{Prospects for exclusive charmless decays}

The outlook for further improvements in these measurements for the full
\B-factory datasets and for a Super \B factory is good.  It can be
expected that for $\B\to\pi\ell\nu$ the untagged measurements will
remain the most precise up to integrated luminosities of several \invab.
To reduce the systematic uncertainties of untagged measurements, a
better knowledge of inclusive $\B\to X_u\ell\nu$ decays is important,
since they are the biggest limitation in the high-$q^2$ region where
LQCD calculations exist.  In addition, a significant fraction of the \BB
background comes from events, where the signal \B meson has been wrongly
reconstructed by assigning one or more particles from the decay of the
other \B meson to the signal decay. To reduce this uncertainty, much
effort is needed to improve the simulation of generic \B-meson decays.
With the full \B-factory dataset, a precision of about 4-5\% should 
be achievable for the total $\B\to\pi\ell\nu$ branching fraction.

The tagged measurements in particular will improve with larger data
samples.  The systematic uncertainties in these measurements have a
significant statistical component and thus the total experimental error
is expected to fall as $1/\sqrt{N}$.  For the higher-mass states, the
tagged measurements should soon give the most precise branching-fraction
results.  However, the larger data samples from untagged analyses will
be needed to extract information on the three form factors involved in
decays with a vector meson.  For an integrated luminosity of
1--2~$\invab$, several thousand $\B\to\rho\ell\nu$ and
$\B\to\omega\ell\nu$ decays can be expected.  These signal samples will
allow us to obtain some information on the form factors or ratios of
form-factors through a simultaneous fit of the $q^2$ spectrum and
decay-angle distributions, similar to the study of $\B\to\Dstar\ell\nu$
decays. A measurement of all three form factors will most likely not be
feasible with the current \B-factory data samples.

\subsubsection{Determination of $|V_{cs}|$, $|V_{cd}|$, $|V_{ub}|$}

Once both the form factor $|f_+(q^2)|^2$ and the experimental decay
width $\Gamma (q_\textrm{min})$ are known, the
CKM matrix element $|V_{qQ}|$ can be determined in several ways.   We
briefly describe the two most common methods below.

Until recently the standard procedure used to extract CKM matrix
elements from exclusive semileptonic decays has been to integrate the
theoretically determined form factor over a region of $q^2$ and then
combine it with the experimentally measured decay rate in this region:
\begin{equation}
    \frac{\Gamma (q_\textrm{min})}{|V_{qQ}|^2} =
        \frac{G_F^2}{192 \pi^3 m_H^3}
        \int_{q^2_\textrm{min}}^{q^2_\textrm{max}} dq^2 \left[
        (m_H^2 + m_P^2 - q^2)^2 - 4 m_H^2 m_P^2 \right]^{3/2}
        |f_+(q^2)|^2 .
    \label{eq:slep:VqQ_standard}
\end{equation}
The integration requires a continuous parametrization of the form
factor between $q^2_\textrm{min}$ and $q^2_\textrm{max}$ that is
typically obtained by fitting the theoretical form factor result to a
model function such as the Be\'cirevi\'c-Kaidalov
(BK)~\cite{Becirevic:1999kt} or Ball-Zwicky (BZ)
parametrization~\cite{Ball:2004ye}.
The three-parameter BK Ansatz,
\begin{eqnarray}
    f_+(q^2) & = & \frac{f_+(0)}{\left(1-q^2/m_{B^*}^2\right)
        \left(1-\alpha \, q^2/m_{B^*}^2 \right)},
    \label{eq:slep:BK_f+}\\
    f_0(q^2) & = &
        \frac{f_+(0)}{\left(1-q^2/\beta m_{B^*}^2 \right)},
    \label{eq:slep:BK_f0}
\end{eqnarray}
incorporates many essential features of the form factor shape such as
the kinematic constraint at $q^2=0$, heavy-quark scaling, and the
location of the $B^*$ pole.  The four-parameter BZ Ansatz extends the BK
expression for $f_+(q^2)$ by including an additional pole to capture the
effects of multiparticle states.

In general, the use of a model function to parametrize the form factor
introduces assumptions that make it difficult to quantify the agreement
between theory and experiment and gives rise to a systematic uncertainty
in the CKM matrix element $|V_{qQ}|$ that is hard to estimate.  It is
likely that this error can be safely neglected when interpolating
between data points. Thus the choice of fit function should have only a
slight impact on the exclusive determinations of \Vcs and \Vcd because
lattice-QCD calculations and experimental measurements possess a
large region of overlap in~$q^2$.
It is less clear, however, how well the BK and BZ Ans\"atze can be 
trusted to extrapolate the form factor shape beyond the reach of 
the numerical lattice-QCD data or the experimental data.
Thus one should be cautious in using them for the exclusive
determination of \Vub via Eq.~(\ref{eq:slep:VqQ_standard}), since an
extrapolation in $q^2$ is necessary both for lattice QCD,
which is most accurate at high $q^2$, and for
experimental measurements, which are most precise at low values of $q^2$.
In particular, comparisons of lattice and experimental determinations of 
BK or BZ fit parameters are potentially misleading, because apparent 
inconsistencies could simply be due to the inadequacy of the
parametrization.

Recently, several groups have begun to use model-independent
parameterizations for the exclusive determination of
\Vub~\cite{Arnesen:2005ez,Becher:2005bg,VandeWater:2006zz,Flynn:2006vr,%
Flynn:2007ii,Bourrely:2008za,Bailey:2008wp}.
This avoids the concerns about the BK and BZ Ans\"atze outlined above, and
should become the standard method for determining \Vub and other CKM
matrix elements from semileptonic decays in the near future.  For
concreteness, here we focus on the $z$-expansion given in
Eq.~(\ref{eq:slep:z_exp}), but the procedure for determining \Vub
outlined here should apply to other model-independent parameterizations.
Because the $z$-expansion relies only on analyticity and unitarity, it
can be trusted to extrapolate the form factor shape in $q^2$ beyond the
reach of the data.  One can easily check for consistency between theory
and experiment using this parametrization by fitting the data
separately and comparing the slope ($a_1/a_0$), curvature ($a_2/a_0$),
and so forth.  Finally, because as many terms can be added to the
convergent series as are needed to describe the $B\to\pi\ell\nu$ form
factor to the desired accuracy, the parametrization can be
systematically improved as theoretical and experimental data get better.

Once the shapes of the theoretical and experimental form factor data are
determined to be consistent, the CKM matrix element \Vub is given simply
by the ratio of the normalizations,
$\Vub=a_0^\textrm{exp.}/a_0^\textrm{theo.}$.
The total uncertainty in \Vub can be reduced, however, by fitting the
theoretical and experimental data simultaneously, leaving the relative
normalization as a free parameter to be determined~\cite{Bailey:2008wp}.
The combined fit incorporates all of the available data, thereby
allowing the numerical lattice QCD data primarily to dictate the shape
at high $q^2$ and the experimental data largely to determine the shape
at low $q^2$.  Although the theoretical and experimental data are
uncorrelated, it is important to include the correlations between
experiments or between theoretical calculations, despite the fact that
they can be difficult to ascertain.  Fig.~\ref{fig:slep:z_fit} shows an
example combined fit to the model-independent $z$-parametrization that
uses 2+1 flavor lattice QCD results from
Fermilab/MILC~\cite{Bailey:2008wp} and experimental data from
BABAR~\cite{Aubert:2006px}.
\begin{figure} \begin{center}
    \includegraphics[width=3in]{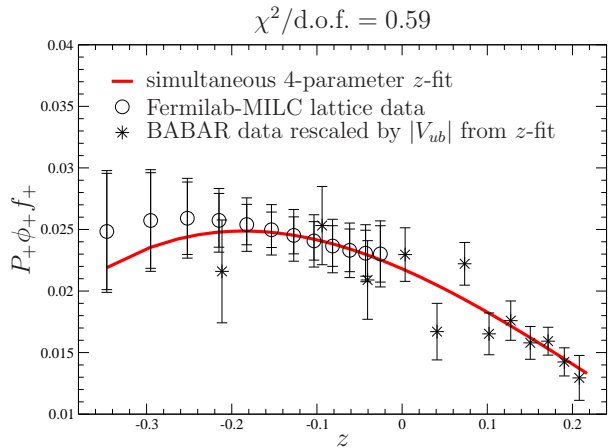}
    \caption{Model-independent determination of $|V_{ub}|$ from a
    simultaneous fit of lattice and experimental $B\to\pi\ell\nu$
    semileptonic form factor data to the
    $z$-parametrization~\cite{Bailey:2008wp}.  Inclusion of terms in
    the power-series through $z^3$ yields the maximum uncertainty in
    $|V_{ub}|$;  the corresponding 4-parameter $z$-fit is given by
    the red curve in both plots. The circles denote the
    Fermilab-MILC lattice-QCD calculation~\cite{Bailey:2008wp}, while the 
    stars indicate the 12-bin Babar data~\cite{Aubert:2006px}, 
    rescaled by the value of $|V_{ub}|$
    determined in the simultaneous $z$-fit.}
    \label{fig:slep:z_fit}
\end{center} \end{figure}

Combining the most recent experimental measurements of $D\to K\ell\nu$ and 
$D\to\pi\ell\nu$ form factors with the 2+1 flavor lattice QCD 
calculations from the Fermilab/MILC 
collaboration~\cite{Aubin:2004ej,Amsler:2008zzb}, 
CLEO finds~\cite{Besson:2009uv}
\begin{eqnarray}
    \Vcd &=& 0.234 \pm 0.007 \pm 0.025, \label{eq:slep:Vcd} \\
    \Vcs &=& 0.985 \pm 0.012 \pm 0.103, \label{eq:slep:Vcs}
\end{eqnarray}
where the errors are experimental and theoretical, respectively.
These determinations rely upon the BK parametrization, both to
parametrize the theoretical $D\to\pi\ell\nu$ and $D\to K\ell\nu$ form
factor shapes for use in Eq.~(\ref{eq:slep:VqQ_standard}) and within the
lattice QCD calculation itself.  Although this is unlikely to introduce
a significant systematic error, use of one of the many model-independent
functional forms available would be preferable.  The largest
uncertainties in both \Vcs and \Vcd are from discretization errors in the
lattice QCD calculation, and can be reduced by simulating at a finer
lattice spacing.  
Because the lattice calculations of the $D\to\pi\ell\nu$ and $D\to
K\ell\nu$ form factors can improved in a straightforward manner, without
requiring new techniques, we expect the errors in both \Vcd and \Vcs to
decrease significantly in the near future.

Most recent exclusive determinations of \Vub rely upon the 2+1 flavor
lattice QCD calculations of the $B\to\pi\ell\nu$ form factor of the
HPQCD and Fermilab/MILC
collaborations~\cite{Okamoto:2004xg,Dalgic:2006dt,Amsler:2008zzb}.
Those which use model-independent parameterizations of the form factor
shape often incorporate additional theoretical points from light cone
sum rules,  soft collinear effective theory, and chiral perturbation
theory~\cite{Arnesen:2005ez,Becher:2005bg,Flynn:2007ii,Bourrely:2008za}.
All of the results for \Vub are consistent within uncertainties.  We
show a representative sample of these results, along with two
model-dependent determinations that rely on the BK and BZ
parameterizations for comparison, in Fig.~\ref{fig:slep:Vub}.  Below we
quote the most recent calculation by Fermilab/MILC because this is the
only one to use a model-independent parametrization along with the full
correlation matrices, derived directly from the data, for both theory
and experiment~\cite{Bailey:2008wp}:
\begin{eqnarray}
    \Vub = (3.38 \pm 36)\times 10^{-3},
    \label{eq:slep:Vub} 
\end{eqnarray}
where the total uncertainty is the sum of statistical, systematic, and
experimental errors added in quadrature.  The dominant theoretical
uncertainty in \Vub comes from statistics and the extrapolation to the
physical up and down quark masses and to the continuum.  The
sub-dominant uncertainties, which are of comparable size, are due to the
perturbative renormalization of the heavy-light vector current and
heavy-quark discretization errors in the action and current.  All of
these errors can be reduced by increasing statistics and simulating at a
finer lattice spacing.  
We therefore expect the total uncertainty in \Vub determined from
$B\to\pi\ell\nu$ semileptonic decay to decrease in the next few years.
\begin{figure} 
    \centering
    \includegraphics[width=3in]{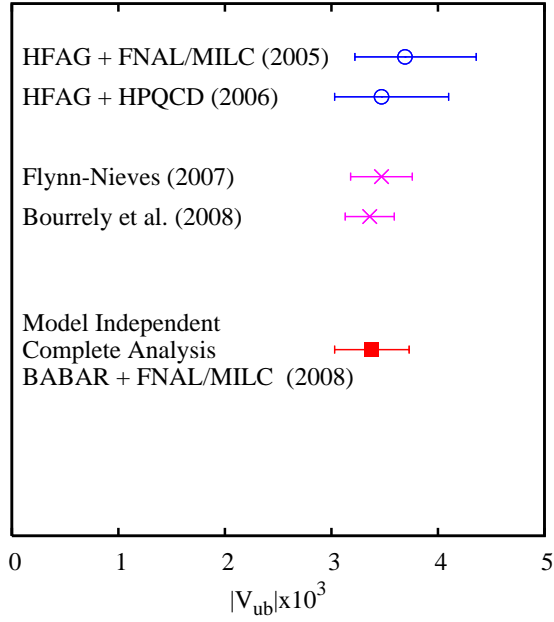}
    \caption{Determinations of \Vub that rely upon 2+1 flavor
        lattice QCD calculations.  The upper two results use the BK and
        BZ  parameterizations, respectively, to describe the the
        $B\to\pi\ell\nu$ form factor, while the lower three results use
        different model-independent parameterizations.}
    \label{fig:slep:Vub}
\end{figure}


%

\subsection{$B\to D^{(*)}\ell\nu$ decays for \Vcb}

\subsubsection{Theoretical background: HQS and HQET}
\label{sec:slep:exclusiveBDtheory}

The matrix elements of semileptonic decays can be related to a set of
form factors.  In the conventions of refs.~\cite{deDivitiis:2007ui,%
deDivitiis:2007uk,deDivitiis:2008df}, the matrix elements relevant for
$B\to D^{(*)} \ell \nu$ decays are
\begin{eqnarray}
    \frac{\langle D| \mathcal{V}^\mu |B \rangle}{\sqrt{m_B m_{D}}} & = &
        (v_B + v_D)^\mu h_+ + (v_B-v_D)^\mu h_-, \\
    \frac{\langle D^*_\alpha |\mathcal{V}^\mu | B \rangle}{\sqrt{m_Bm_{D^*}}} & = &
        \varepsilon^{\mu \nu\rho\sigma} v^\nu_B v^\rho_{D^*}
        \epsilon^{*\sigma}_\alpha h_V, \\
    \frac{\langle D^*_\alpha |\mathcal{A}^\mu | B \rangle}{\sqrt{m_Bm_{D^*}}} & = &
    i \epsilon^{* \nu}_\alpha [h_{A_1}(1+w)g^{\mu \nu} - 
    (h_{A_2}v^\mu_B + h_{A_3}v^\mu_{D^*})v^\nu_B],
\end{eqnarray}
where $m_B$ and $m_{D^{(*)}}$ are the masses of the $B$ and $D^{(*)}$
mesons, respectively, $v_{B,D^{(*)}} = p_{B,D^{(*)}}/m_{B,D^{(*)}}$ is
the 4-velocity of the mesons, $\varepsilon^{\mu \nu\rho\sigma}$ is the
totally antisymmetric tensor in 4 dimensions, and $\epsilon^\mu_\alpha$
is the polarization vector of $D^*_\alpha$, with
\begin{equation}
    \sum_{\alpha=1}^3 \epsilon^{* \mu}_\alpha \epsilon^\nu_\alpha = 
        -g^{\mu \nu} +v^\mu_{D^*} v^{\nu}_{D^*}.
\end{equation}

The form factors depend on the heavy-light meson masses, and on the
velocity transfer from initial to final state $w=v_B \cdot v_{D^{(*)}}$.
The values of $w$ are constrained by kinematics to fall in the range
\begin{equation}
    1 \leq w \leq \frac{m^2_B + m^2_{D^{(*)}}}{2 m_B m_{D^{(*)}}},
\end{equation}
with the largest value of $w$ around~1.5.
The usual invariant $q^2=m^2_B + m^2_{D^{(*)}}-2wm_B m_{D^{(*)}}$.

The differential rate for the decay $B\to D \ell \nu$ is
\begin{equation}
    \frac{d\Gamma(B\rightarrow D \ell \nu)}{dw} =
        \frac{G^2_F}{48\pi^3}m^3_{D}(m_B+m_{D})^2(w^2-1)^{3/2} 
        |V_{cb}|^2 |\mathcal{G}(w)|^2,
    \label{slep:eq:BtoD}
\end{equation}
with
\begin{equation}
    \mathcal{G}(w) = 
    h^{B\rightarrow D}_+(w) - \frac{m_B-m_D}{m_B+m_D} h^{B \rightarrow D}_-(w).
\end{equation}
The differential rate for the semileptonic decay 
$\overline{B}\to D^*\ell\overline{\nu}_\ell$ is
\begin{equation}
    \frac{d\Gamma(B\rightarrow D^* \ell \nu)}{dw} =
        \frac{G^2_F}{4\pi^3}m^3_{D^*}(m_B-m_{D^*})^2\sqrt{w^2-1} 
        |V_{cb}|^2\chi(w)|\mathcal{F}(w)|^2,
    \label{slep:eq:BtoDstar}
\end{equation}
where $\chi(w)|\mathcal{F}_{B \rightarrow D^*}(w)|^2$ contains a
combination of four form factors that must be calculated
nonperturbatively.
At zero recoil $\chi(1)=1$, and $\mathcal{F}(1)$
reduces to a single form factor, $h_{A_1}(1)$.  At non-zero recoil, all
four form factors contribute, yielding
\begin{eqnarray}
    \chi(w) & = & \frac{w+1}{12}
        \left(5w+1-\frac{8w(w-1)m_B m_{D^*}}{(m_B-m_{D^*})^2}\right), \\
    \mathcal{F}(w) & = & h_{A_1}(w) \frac{1+w}{2}
        \sqrt{\frac{H^2_0(w)+H^2_+(w)+H^2_-(w)}{3\chi(w)}},
\end{eqnarray}
with
\begin{eqnarray}
    H_0(w) & = & 
        \frac{w-m_{D^*}/m_B-(w-1)R_2(w)}{1-m_{D^*}/m_B}, \\
    H_\pm(w) & = & t(w)\left[1\mp \sqrt{\frac{w-1}{w+1}}R_1(w)\right],  \\
    t^2(w) & = & \frac{m^2_B-2 w m_B m_{D^*}+m^2_{D^*}}{(m_B-m_{D^*})^2}, \\
    R_1(w) & = &\frac{h_V(w)}{h_{A_1}(w)},  \\
    R_2(w) & = & \frac{h_{A_3}(w)+ (m_{D^*}/m_B)h_{A_2}(w)}{h_{A_1}(w)}.
\end{eqnarray}
Eqs.~(\ref{slep:eq:BtoD}) and~(\ref{slep:eq:BtoDstar}) hold for 
vanishing lepton mass, and there are corrections analogous to those in 
Eq.~(\ref{slep:eq:B2pi}).
For semimuonic decays, these effects are included in recent experimental 
analyses.

In the limit of infinite heavy-quark mass, 
all heavy quarks interact in the same way in heavy light mesons.
This phenomenon is known as heavy quark symmetry (HQS). 
For example, given that a heavy quark has spin quantum number
$1/2$, the quark has a chromomagnetic moment $g/(2 m_Q)$, which vanishes 
as the heavy quark $m_Q$ goes to
infinity.  Thus, in a meson, the interaction between the spin of the
heavy quark and the light degrees of freedom is suppressed.  The
heavy-light meson is then symmetric under a change in the $z$-component
of the heavy-quark spin, and this is known as heavy-quark spin symmetry.

In the heavy-quark limit we have that the velocity of the heavy quark is
conserved in soft processes.  Thus, the mass-dependent piece of the
momentum operator can be removed by a field redefinition,
\begin{eqnarray}
    h_Q(v,x) = \frac{1\, + \not \! v}{2}e^{im_Q v \cdot x} Q(x),
\end{eqnarray}
where $(1 +\! \not \! v)/2$ is a projection operator, 
and $Q(x)$ is the conventional quark field in QCD.  
If the quark has a total momentum $P^\alpha$, the new field carries a
residual momentum $k^\alpha = P^\alpha -m_Q v^\alpha$.  In the limit
$m_Q \rightarrow \infty$, the effective Lagrangian for heavy quarks
interacting via QCD becomes
\begin{eqnarray}
    \mathcal{L}_{\rm HQET} = \overline{h}_Q i v\cdot D h_Q,
    \label{eq:hqetlagrangian}
\end{eqnarray}
where $D^\alpha = \partial^\alpha - i g_s t_a A^\alpha_a$ is the
covariant derivative.  For large but finite $m_Q$, this Lagrangian
receives corrections from terms of higher-dimension proportional to
inverse powers of $m_Q$.  These corrections break the HQS of the leading
order Lagrangian, but are well-defined at each order of the expansion,
and can be included in a systematic way.  The resulting Lagrangian is
known as the Heavy-Quark Effective Theory (HQET).  The higher-dimension
operators in the HQET come with coefficients that are determined by 
matching to the underlying, fundamental theory, namely~QCD.

In lattice simulations, it is not possible to treat quarks where the
mass in lattice units $am_Q$ is large compared to 1 using conventional
light-quark methods.  All lattice heavy-quark methods make use of HQET
in order to avoid the large discretization effects that would result
from such a naive treatment.  For lattices currently in use, $am_c\sim
0.5-1.0$ and $am_b\sim 2-3$, so HQET methods are essential for precision
calculations.
For a technical review of these methods, see Ref.~\cite{Kronfeld:2003sd}.

One approach consists in simulating a discrete version
of the HQET action, introduced in Ref.~\cite{Eichten:1989zv}, by
treating the sub-leading operators as insertions in correlation
functions.  The matching procedure is particularly complicated on the
lattice because of the presence of power divergences that arise as a
consequence of the mixing of operators of lower dimensions with the
observable of interest, but it can be carried out with non-perturbative
accuracy~\cite{Heitger:2003nj} by means of a finite volume technique
(see also~\cite{Sommer:2006sj} for a review of the subject).

The Fermilab approach makes use of the fact that the Wilson fermion
action reproduces the static quark action in the infinite mass limit.
Higher dimension operators can then be adjusted in a systematic way.  Each
higher dimension operator has a counterpart in HQET, and once the
coefficients of the new operators are tuned to the appropriate values,
the lattice action gives the continuum result, to a given order in HQET.
To order $\Lambda_{\rm QCD}/2m_Q$, the only new operator is a single
dimension 5 term, and this is the same term that is added to the Wilson
fermion action to improve it in the light quark sector.  (The power of 2
is a combinatoric factor appropriate to the HQET expansion.)  This
improved action is known as the Sheikholeslami- Wohlert action
\cite{Sheikholeslami:1985ij}, and the tunings of the parameters in this
action appropriate to heavy quarks is the Fermilab method now in common
use \cite{ElKhadra:1996mp,Kronfeld:2000ck}.  Higher order improvement to
the Fermilab method, including operators of even higher dimension, has
been proposed in Ref.~\cite{Oktay:2008ex}.

Another approach to handle with heavy quarks on the lattice is the
so-called ``step-scaling method''~\cite{Guagnelli:2002jd}.  Within the
step-scaling method the dynamics of the heavy quarks is resolved by 
making simulations
on small volumes ($L\simeq 0.5$~\fm) without recurring to any
approximation but introducing, at intermediate stages, finite volume
effects.  These are subsequently accounted for by performing simulations
on progressively larger volumes and by relying on the observation that
sub-leading operators enter the HQET expansion of finite volume effects
multiplied by inverse powers of $L m_{Q}$.  The success of this approach
depends on the possibility of computing the finite volume observable,
finite volume effects and their product with smaller errors and
systematics with respect to the ones that would be obtained by a direct
calculation.  The strength of the method is a great freedom in the
definition of the observable on finite volumes provided that its
physical value is recovered at the end of the procedure.

The Fermilab Lattice Collaboration introduced a double ratio in order to
compute $h_+$ at zero-recoil \cite{Hashimoto:1999yp}
\begin{equation}
    \frac{\langle D|\overline{c}\gamma_4 b|\overline{B}\rangle
        \langle\overline{B}|\overline{b}\gamma_4c|D\rangle}{%
        \langle D|\overline{c}\gamma_4 c|D\rangle
        \langle\overline{B}|\overline{b}\gamma_4 b|\overline{B}\rangle} =
    \left|h_{+}(1)\right|^2.
    \label{eq:hpl}
\end{equation}
This double ratio has the advantage that the statistical errors and many
of the systematic errors cancel.  The discretization errors are
suppressed by inverse powers of heavy-quark mass as
$\alpha_s(\Lambda_{\rm QCD}/2m_Q)^2$ and $(\Lambda_{\rm QCD}/2m_Q)^3$
\cite{Kronfeld:2000ck}, and much of the current renormalization cancels,
leaving only a small correction that can be computed perturbatively
\cite{Harada:2001fj}.  The extra suppression of discretization errors by
a factor of $\Lambda_{\rm QCD}/2m_Q$ occurs at zero-recoil for
heavy-to-heavy transitions, and is a consequence of Luke's Theorem
\cite{Luke:1990eg}.

In order to obtain $h_-$, it is necessary to consider non-zero recoil
momenta.  In this case, Luke's theorem does not apply, and the HQET
power counting leads to larger heavy-quark discretization errors.
However, this is mitigated by the small contribution of $h_-$ to the
branching fraction.  The form factor $h_-$ can be determined from the
double ratio \cite{Hashimoto:1999yp}
\begin{equation}
    \frac{\langle D|\overline{c}\gamma_jb|\overline{B}\rangle
        \langle D|\overline{c}\gamma_4  c|D\rangle}{%
        \langle D|\overline{c}\gamma_4 b|\overline{B}\rangle
        \langle D|\overline{c}\gamma_j b|D\rangle} =
        \left[1-\frac{h_-(w)}{h_+(w)}\right]
        \left[1+\frac{h_-(w)}{2h_+(w)}(w-1)\right],
    \label{eq:hmi}
\end{equation}
which can be extrapolated to the zero-recoil point $w=1$.
Using the double ratios of Eqs.~(\ref{eq:hpl}) and~(\ref{eq:hmi}) the
latest (preliminary) unquenched determinations of $h_+(1)$ and $h_-(1)$
from the Fermilab Lattice and MILC Collaborations combine to
give~\cite{Okamoto:2005zg}
\begin{equation}
    \mathcal{G}(1)=1.074(18)(16),
   \label{slep:eq:G(1)}
\end{equation}
where the first error is statistical and the second is the sum of all
systematic errors in quadrature.

The form factor at zero-recoil needed for $B\to D^*\ell\nu$ has been
computed by the Fermilab Lattice and MILC Collaborations using the
double ratio \cite{Bernard:2008dn}
\begin{equation}
    \frac{\langle D^*|\overline{c}\gamma_j \gamma_5 b|\overline{B}\rangle
        \langle\overline{B}|\overline{b}\gamma_j\gamma_5c|D^*\rangle}{%
        \langle D^*|\overline{c}\gamma_4 c|D^*\rangle
        \langle\overline{B}|\overline{b}\gamma_4 b|\overline{B}\rangle} =
        \left|h_{A_1}(1)\right|^2,
    \label{eq:hA1}
\end{equation}
where again, the discretization errors are suppressed by inverse powers
of heavy-quark mass as $\alpha_s(\Lambda_{\rm QCD}/2m_Q)^2$ and
$(\Lambda_{\rm QCD}/2m_Q)^3$, and much of the current renormalization
cancels, leaving only a small correction that can be computed
perturbatively \cite{Harada:2001fj}.  They extrapolate to physical light
quark masses using the appropriate rooted staggered chiral perturbation
theory \cite{Laiho:2005ue}.  Including a QED correction of $0.7\%$
\cite{Sirlin:1981ie}, they obtain~\cite{Bernard:2008dn}
\begin{equation}
    \mathcal{F}(1)=0.927(13)(20),
   \label{slep:eq:F(1)}
\end{equation}
where the first error is statistical and the second is the sum of
systematic errors in quadrature.


Because of the kinematic suppression factors $(w^2-1)^{3/2}$ and
$(w^2-1)^{1/2}$ appearing in Eqs.~(\ref{slep:eq:BtoD}) and
(\ref{slep:eq:BtoDstar}), respectively, the experimental decay rates at
zero recoil must be obtained by extrapolation.  The extrapolation is
guided by theory, where Ref.~\cite{Caprini:1997mu} have used dispersive
constraints on the form factor shapes, together with heavy-quark
symmetry to provide simple, few parameter, extrapolation formulas
expanded about the zero-recoil point,
\begin{eqnarray}
    h_{A_1}(w) = h_{A_1}(1) \left[1 - 8\rho_{D^*}^2z + 
        (53\rho_{D^*}^2-15)z^2 - (231\rho^2_{D^*}-91)z^3\right], \\
    R_{1}(w) = R_{1}(1)-0.12(w-1)+0.05(w-1)^2, \\
    R_{2}(w) = R_{2}(1)+0.11(w-1)-0.06(w-1)^2, \\
    {\cal G}(w) = {\cal G}(1) \left[1 - 8\rho_{D}^2z + 
        (51\rho_{D}^2-10)z^2 -(252\rho^2_{D}-84)z^3\right],
\end{eqnarray}
with
\begin{equation}
    z = \frac{\sqrt{w+1}-\sqrt{2}}{\sqrt{w+1}+\sqrt{2}}.
\end{equation}
This approach is employed below to determine $\mathcal{G}(1)\Vcb$ 
and~$\mathcal{F}(1)\Vcb$.

These extrapolations
introduce a systematic error into the extraction of \Vcb that, although
mild for $B\to D^*\ell\nu$, can be eliminated by calculating the form
factors at non zero recoil.
A~first step on this route has been done by
applying the step scaling method to calculate, in the \emph{quenched}
approximation, $\mathcal{G}(w)$ and $\mathcal{F}(w)$ for values of
$w$ where experimental data are directly available.  The form
factors have been defined on the lattice entirely in terms of ratios of
three-point correlation functions, analogously to the double ratios
discussed above, obtaining in such a way a remarkable statistical and (a
part from quenching) systematic accuracy.  All the details of the
calculations, including chiral and continuum extrapolations and
discussions on the sensitiveness of finite volume effects on the heavy
quark masses, can be found in refs.~\cite{deDivitiis:2007ui,de
Divitiis:2007uk,deDivitiis:2008df}.  The results are shown in
Tab.~\ref{tab:gwssm} and Tab.~\ref{tab:fwssm}.
\begin{table}[tp]
    \centering
    \caption{Quenched results for $\mathcal{G}(w)$ and
    $\Delta(w)$ at non zero
    recoil~\cite{deDivitiis:2007ui,deDivitiis:2007uk}.  The notation ``(q)''
    stays for the unknown systematics coming from the quenching
    approximation.  QED corrections not included.}
    \label{tab:gwssm}
    \begin{tabular}{ccccccc}
        \hline\hline
        $w$ & $\mathcal{G}(w)$ & $\Delta(w)$ \\ 
        \hline
         1.000          & 1.026(17)(q)  & 0.466(26)(q)\\[-5pt]
         1.030          & 1.001(19)(q)  & 0.465(25)(q)\\[-5pt]
         1.050          & 0.987(15)(q)  & 0.464(24)(q)\\[-5pt]
         1.100          & 0.943(11)(q)  & 0.463(24)(q)\\[-5pt]
         1.200          & 0.853(21)(q)  & 0.463(23)(q)\\
        \hline\hline
    \end{tabular}
\end{table}
\begin{table}[tp]
    \centering
    \caption{Quenched results for $\mathcal{F}(w)$ and
    $\mathcal{F}(w)/\mathcal{G}(w)$ at non zero
    recoil~\cite{deDivitiis:2008df}.  The notation ``(q)'' stands for the
    unknown systematics coming from the quenching approximation.  QED
    corrections not included.}
    \label{tab:fwssm}
    \begin{tabular}{ccc}
        \hline\hline
        $w$ & $\mathcal{F}(w)$ & $\mathcal{F}(w)/\mathcal{G}(w)$ \\
        \hline
        1.000 & 0.917(08)(05)(q) & 0.878(10)(04)(q) \\[-5pt]
        1.010 & 0.913(09)(05)(q) & 0.883(09)(04)(q) \\[-5pt]
        1.025 & 0.905(10)(05)(q) & 0.891(09)(04)(q) \\[-5pt]
        1.050 & 0.892(13)(04)(q) & 0.905(10)(04)(q) \\[-5pt]
        1.070 & 0.880(17)(04)(q) & 0.914(12)(05)(q) \\[-5pt]
        1.075 & 0.877(18)(04)(q) & 0.916(12)(05)(q) \\[-5pt]
        1.100 & 0.861(23)(04)(q) & 0.923(16)(05)(q) \\
        \hline\hline
    \end{tabular}
\end{table}
The quantity $\Delta(w)$ appearing in Tab.~\ref{tab:gwssm} is
required to parametrize the decay rate $B\to D \tau \nu_{\tau}$ and its
knowledge with non perturbative accuracy opens the possibility to
perform lepton-flavor universality checks on the extraction of \Vcb
from this channel.  On the one hand, the phenomenological relevance of
the results of Tab.~\ref{tab:gwssm} and Tab.~\ref{tab:fwssm} is limited by
the quenching uncertainty that cannot be reliably quantified.  On the
other hand, these results shed light on the systematics on \Vcb
coming from the extrapolation of the experimental decay rates at zero
recoil.
The agreement at zero recoil with the full QCD results, 
Eqs.~(\ref{slep:eq:G(1)}) and (\ref{slep:eq:F(1)}), suggests that the 
unestimated quenching error may be comparable to the present statistical 
error.

\subsubsection{Measurements and Tests}

Measurements of the partial decay widths 
$d\Gamma/dw$ for the decays $B\rightarrow D^{(*)} \ell \nu$
have been performed for more than fifteen years on data recorded at the 
\FourS\ resonance (CLEO, Babar, Belle), and at LEP. 
Though this review will cover only the most recent
measurements, it will offer an almost complete overview 
of the analysis techniques employed so far. 

A semileptonic decay is reconstructed by combining a charged lepton, $\ell$,
either an electron or a muon, and a charm meson of the appropriate
charge and flavor.
To reject non-\BB\ background, only leptons with momentum $p_\ell <
2.3$ \gevc\ are accepted. To suppress fake leptons and leptons from 
secondary decays, a lower bound $p_\ell $ is usually applied,  
in the range from 0.6 to 1.2 \gevc, depending on the analysis.
$D$ mesons are fully reconstructed in several hadronic decay channels. 
Charged and neutral $D^{*}$ are identified by their decays to $D \pi$. 
In \FourS\ decays, the energy and momentum of the $B$ mesons, $E_B$ and
$|\vec{p}_B|$, are well known\footnote{In LEP experiments the
direction of the $B$ meson is obtained from the vector joining the
primary vertex to the $B$ decay vertex, the neutrino energy is computed
from the missing energy in the event. A missing energy technique is
also applied by \FourS\ experiments to improve background
rejection in $B\rightarrow D \ell \nu$ measurements.}. 
Since the neutrino escapes detection, the $B$ decay usually is not 
completely reconstructed.
However, kinematic constraints can be applied to reject background.  
In particular, if the massless neutrino is the only unobserved particle, 
the $B$-meson direction is constrained to lie on a cone centered along 
the $D^{(*)}\ell$ momentum vector, $\vec{p}_{D^{(*)}\ell}$, with an 
opening angle $\theta_{BY}$ bounded by the condition 
$|\cos\theta_{BY}|\le1$
(see Eq.~\ref{eq:cosBY} for the exact definition).
Background events from random $D^{(*)}\ell$ combinations are spread over 
a much larger range in $\cos\theta_{BY}$ and decays of the type 
$B\rightarrow D^{(*)} \pi\pi \ell \nu $, where the additional pions are 
not reconstructed, accumulate mainly below $\cos\theta_{BY}= -1$.
 
The differential decay rate 
$d^4\Gamma/dw d\cos\theta_\ell d\cos\theta_V d\chi$ depends on four 
variables: $w=v_B \cdot v_{D^{(*)}}$, $\theta_\ell$, the angle between 
the lepton direction in the virtual $W$ rest frame and the $W$ direction 
in the $B$ rest frame, $\theta_V$, the angle between the $D$-meson 
direction in the $D^*$ rest frame and the $D^*$ direction in the $B$ 
rest frame, and $\chi$,  the angle between the plane determined from the 
$D^*$ decay products and the plane defined by the two leptons.
In HQET, the decay rate is parametrized in term of four quantities:
the normalization $\mathcal{F}(1)\Vcb$, the slope $\rho^2_{D^*}$, and 
the form-factor ratios $R_1(1)$ and $R_2(1)$.  
Many measurements of $\mathcal{F}(1)\Vcb$ and $\rho^2_{D^*}$ rely on the
differential decay rates, integrated over the three angles,
$d\Gamma(B\to D^*\ell\nu)/dw$ and thus require external knowledge of 
$R_1(1)$ and $R_2(1)$.

Following the first measurement by CLEO~\cite{Duboscq:1995mv}, the 
Babar~\cite{Aubert:2007rs}, and 
Belle~\cite{Adachi:2008nd} Collaborations have employed much larger 
samples of reconstructed neutral $B$ mesons to determine $R_1(1)$ and 
$R_2(1)$ from a fit to the four-dimensional differential decay rate.
Figure~\ref{slep:f:Belleplots} shows a comparison of the 
data and the fit results from the recent Belle analysis, for the 
projections of the four kinematic variables.
\begin{figure}[bp]
    \centering
    \includegraphics[width=0.4\textwidth]{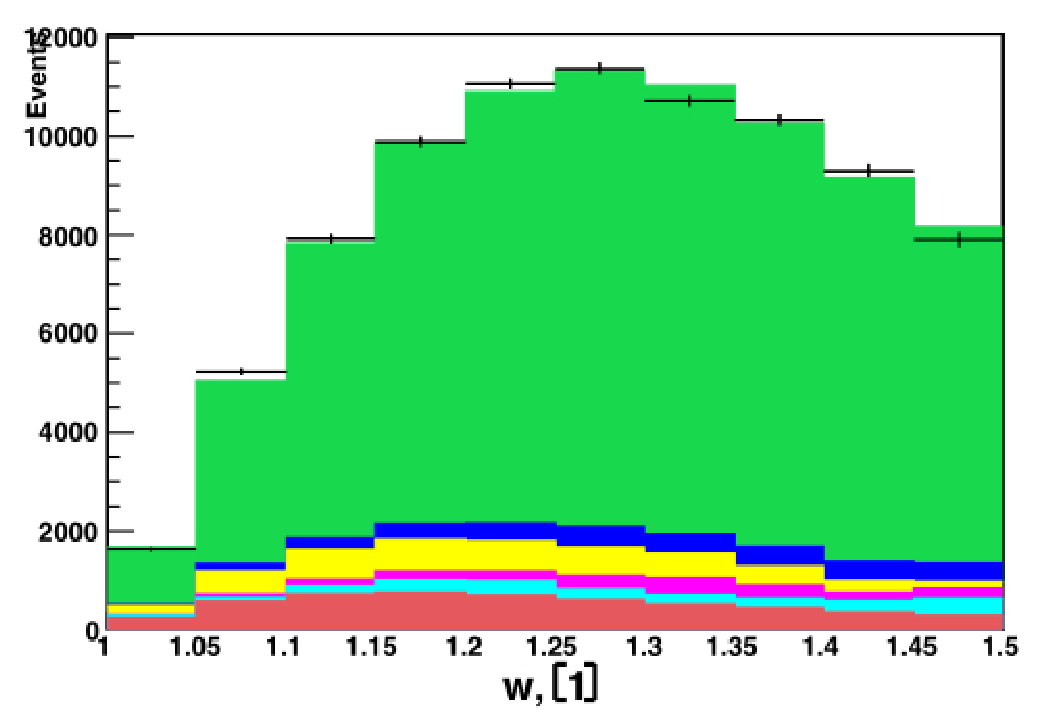} \hfill
    \includegraphics[width=0.4\textwidth]{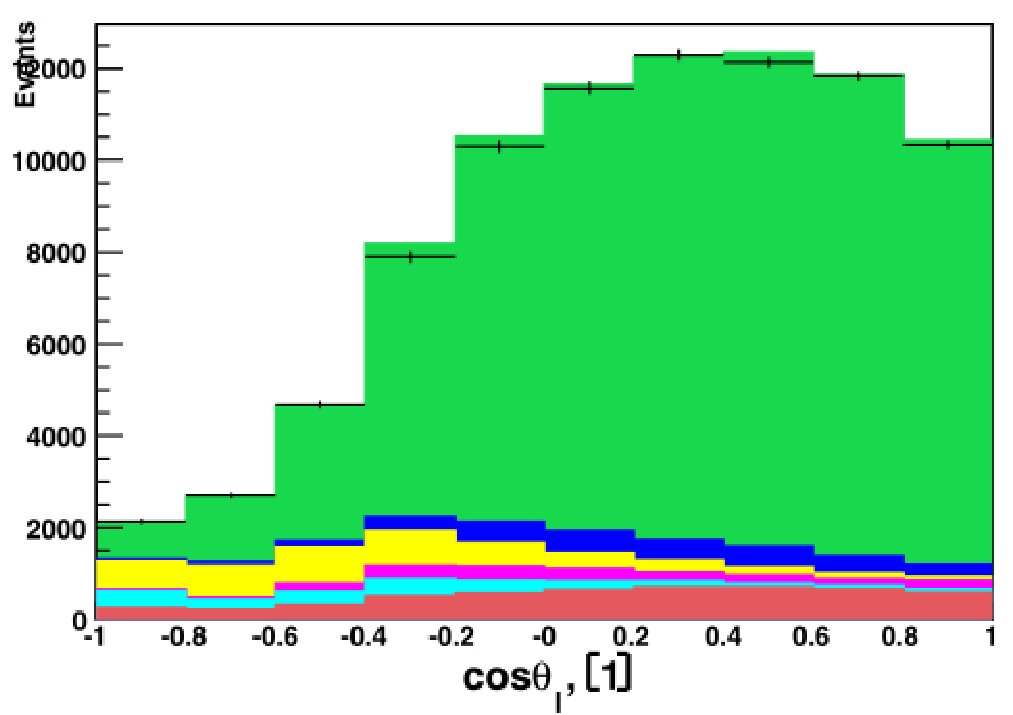} \\
    \includegraphics[width=0.4\textwidth]{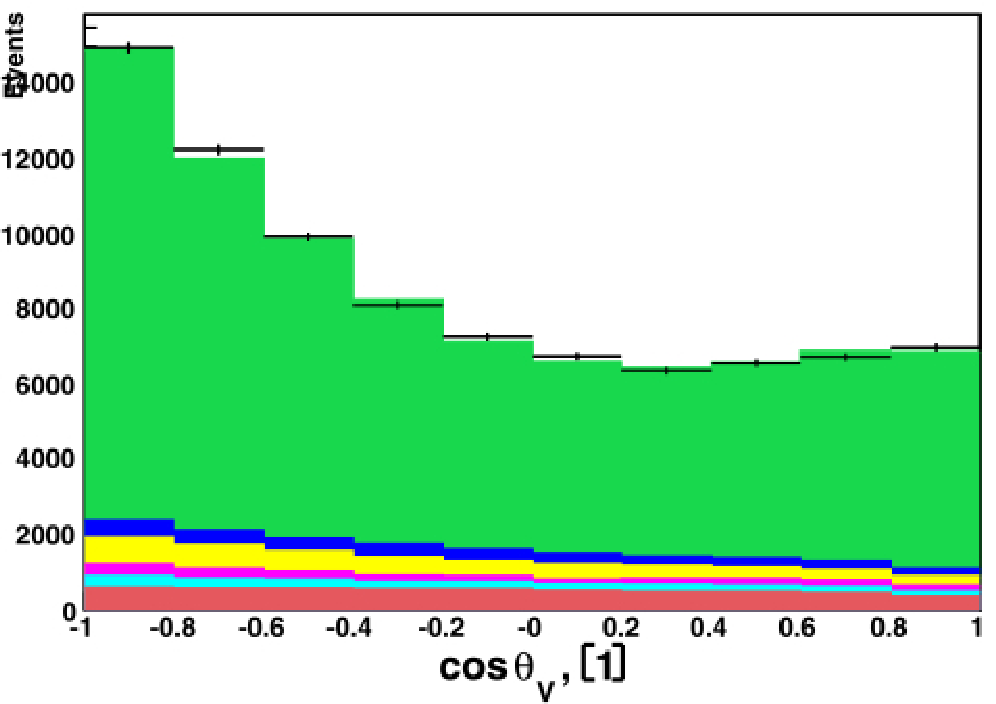} \hfill
    \includegraphics[width=0.4\textwidth]{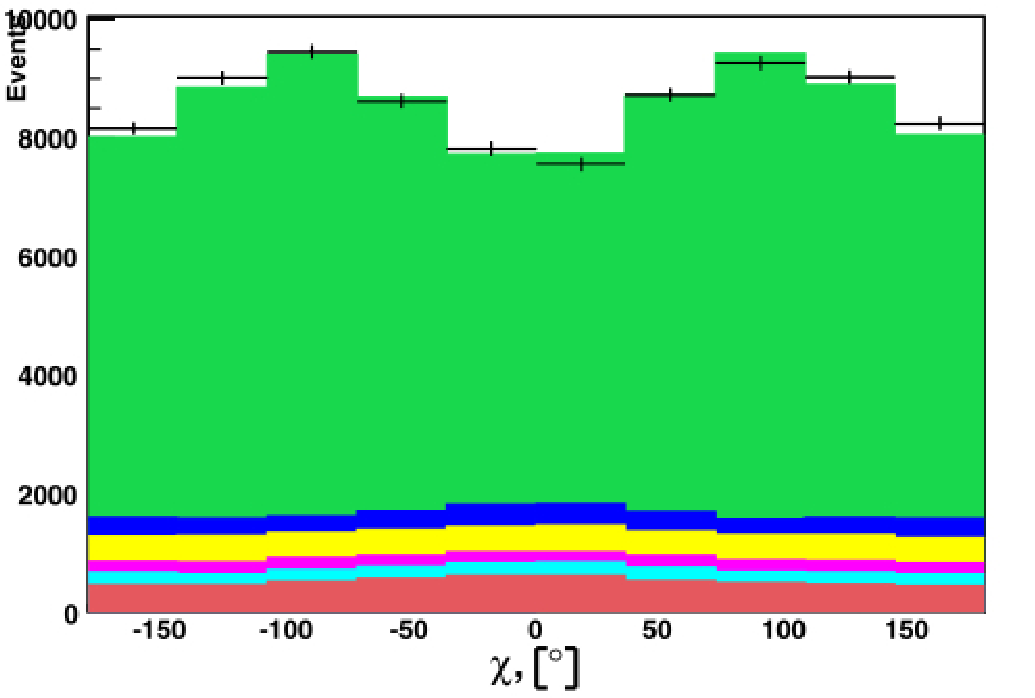} 
    \caption{Belle~\cite{Adachi:2008nd}: Results of the four-dimensional 
        fit to the $B^0 \to D^{*+} \ell \nu$ decay rate in terms one 
        one-dimensional projections: $w$ (top-left), 
        $\cos\theta_\ell$ (top-right),
        $\cos\theta_V$ (bottom left) and $\chi$ (bottom right). 
        The data (points) are compared to the sum of the fitted
        contribution, signal (green) and several background
        sources (in different colors).}
    \label{slep:f:Belleplots} 
\end{figure}
Tab.~\ref{slep:t:FF} lists the results of the fully-differential 
measurements from Babar and Belle.
\begin{table}[tp]
    \centering
    \caption{Summary of the $B$-factories results on form factors and 
        \Vcb from semileptonic $B$ decays.} 
    \label{slep:t:FF}
    \begin{tabular}{cccccc}
    \hline\hline
    Mode & Ref. & $\mathcal{F}(1)\Vcb \cdot 10^3$ & $\rho^2_{D^*}$ & $R_1$ & $R_2$ \\
    \hline
    $B^{-}\to D^{*0}\ell^-\bar\nu$       & \cite{Aubert:2007qs} &
        $35.9\pm 0.6\pm 1.4$ & $1.16\pm 0.06\pm 0.08$ & - & - \\
    $\bar{B^{0}}\to D^{*+}\ell^-\bar\nu$ & \cite{Adachi:2008nd} &
        $34.4\pm 0.2\pm 1.0$ & $1.29\pm 0.05\pm 0.03$ &
            $1.50\pm 0.05\pm 0.06$ & $0.84\pm 0.03\pm 0.03$ \\
    $\bar{B^{0}}\to D^{*+}\ell^-\bar\nu$ & \cite{Aubert:2007rs} &
        $34.4\pm 0.3\pm 1.1$ & $1.19\pm 0.05\pm 0.03$ &
            $1.43\pm 0.06\pm 0.04$ & $0.83\pm 0.04\pm 0.02$ \\
    $B\to D^{(*)}\ell^-\bar\nu$          & \cite{Aubert:2008yv} &
        $35.9\pm 0.2\pm 1.2$ & $1.22\pm 0.02\pm 0.07$ & - & - \\
    \hline\hline
    \end{tabular}
    \begin{tabular}{cccc}
    \hline\hline
    Mode & Ref. & $\mathcal{G}(1) \Vcb \cdot 10^3$& $\rho^2_{D}$ \\ 
    \hline
    $B \to D \ell^- \bar\nu$ & \cite{Aubert:2008yv} &
        $43.1\pm 0.8\pm 2.3$ & $1.20\pm 0.04\pm 0.07$ \\
    $B \to D \ell^- \bar\nu$ & \cite{:2008ii} & 
        $43.0\pm 1.9\pm 1.4$ & $1.20\pm 0.09\pm 0.04$ \\
    \hline\hline
    \end{tabular}
\end{table}
%

In a recent Babar analysis~\cite{Aubert:2007qs} a sample of about 23,500 
$B^-\rightarrow D^{*0} \ell^- \bar{\nu}$ decays has been 
selected from about $2 \times 10^7$ $\FourS \rightarrow B\bar{B}$ events.
The signal yield is determined in ten bins in $w$ to measure
$d\Gamma(B^- \rightarrow D^{*0} \ell^- \bar{\nu})/dw$ 
with minimal model dependence.  
The fitted values
of $\mathcal{F}(1)\Vcb$ and $\rho^2_{D^*}$ are given in Tab.~\ref{slep:t:FF}.

The large integrated luminosities and the deeper understanding of $B$ mesons 
properties accumulated in recent years have allowed $B$-factories to
perform new measurements of semileptonic decays based on innovative approaches.
Babar has recently published results on
$\mathcal{G} (w)\Vcb$ and $\mathcal{F}(w)\Vcb$ ,
based on an inclusive selection of $B\rightarrow D \ell \nu X$
decays, where only the $D$ meson and the charged lepton are
reconstructed~\cite{Aubert:2008yv}.
To reduce background from $D^{**} \ell \nu$ decays and other background 
sources, the lepton momentum is restricted $p_\ell > 1.2 \gevc$, and the 
$D$ mesons are reconstructed only in the two simplest and cleanest decay 
modes, $D^0 \rightarrow K^- \pi^+$ and  $D^+ \rightarrow K^- \pi^+\pi^+$.

Signal decays with $D$ and  $D^*$ mesons in the final
states are separated from background processes (mainly semileptonic decays 
involving higher mass charm  mesons, $D^{**}$) on a 
statistical basis. The $V-A$ structure of the weak decays favors larger 
values of $p_\ell$ for the vector meson $D^*$ than for the scalar $D$. 
using the three-dimensional distributions of the lepton momentum
$p_\ell$, the $D$ momentum $p_D$, and $\cos\theta_{BY}$. 

The signal and background yields, the values of $\rho^2_D,~\rho^2_{D^*}$, 
$\mathcal{G}(1)\Vcb$ and $\mathcal{F}(1)\Vcb$ are
obtained from a binned $\chi^2$ fit to the three-dimensional 
distributions of the lepton momentum
$p_\ell$, the $D$ momentum $p_D$, and $\cos\theta_{BY}$, separately for 
the $D^0 \ell$ and $D^+ \ell$ samples. 
The contribution from neutral and charged $B$ decays in each
sample are obtained from the ratio of measured branching fractions of
$\FourS\rightarrow B^+B^-$, $\FourS\rightarrow B^0\bar{B}^0$, the 
branching fractions for charged and neutral $D^{*}$ mesons to $D$ mesons, 
and by imposing equal
semileptonic decay rates for charged and neutral $B$ mesons. 
As an example, Fig.~\ref{slep:f:Incl} shows the
results of the fit in one-dimensional projections for the $D^0 e^- \bar{\nu}_e X$ 
sample. 
An alternative fit with R$_1$(1) and R$_2$(1) as free parameters
gives results consistent with the fully differential measurements 
cited above, albeit with larger statistical and systematic errors.
\begin{figure}
    \centering
    \includegraphics[height=5.4cm,trim = 0 125mm 0 0,clip]{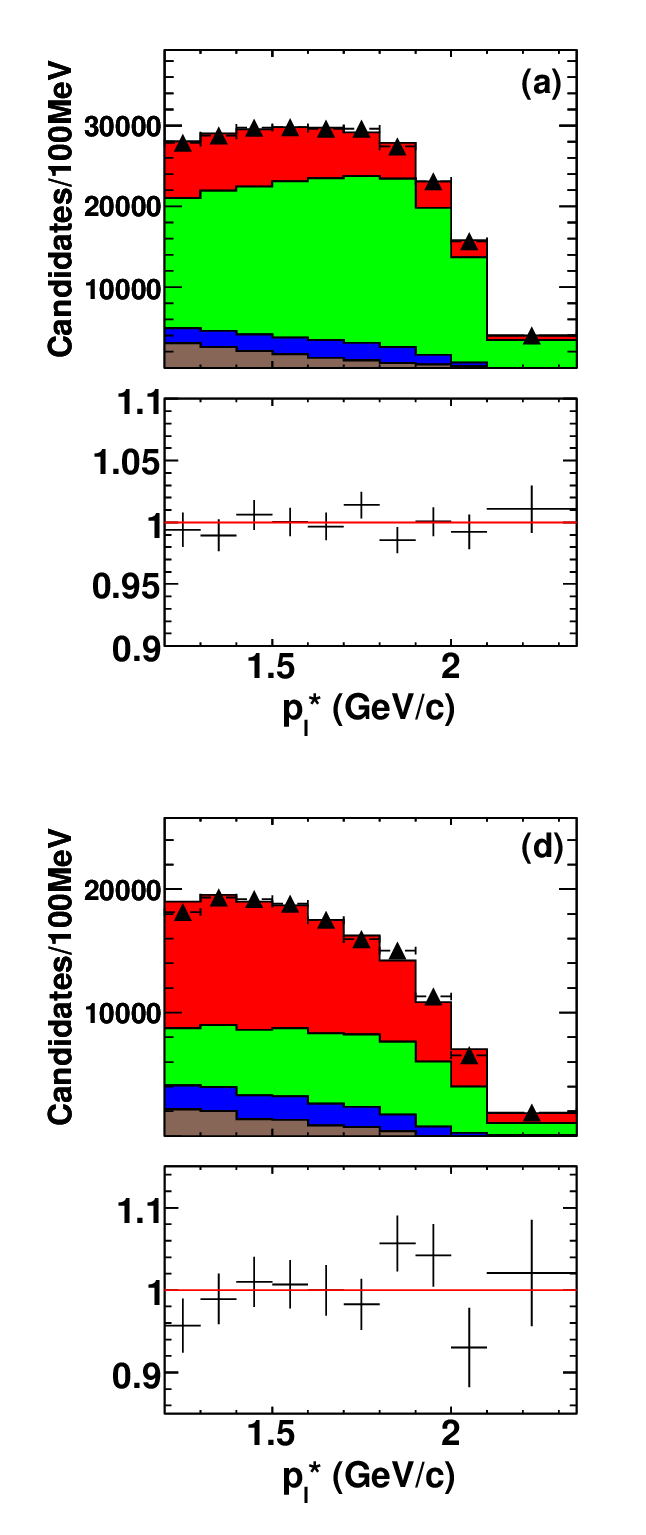}
    \includegraphics[height=5.4cm,trim = 0 125mm 0 0,clip]{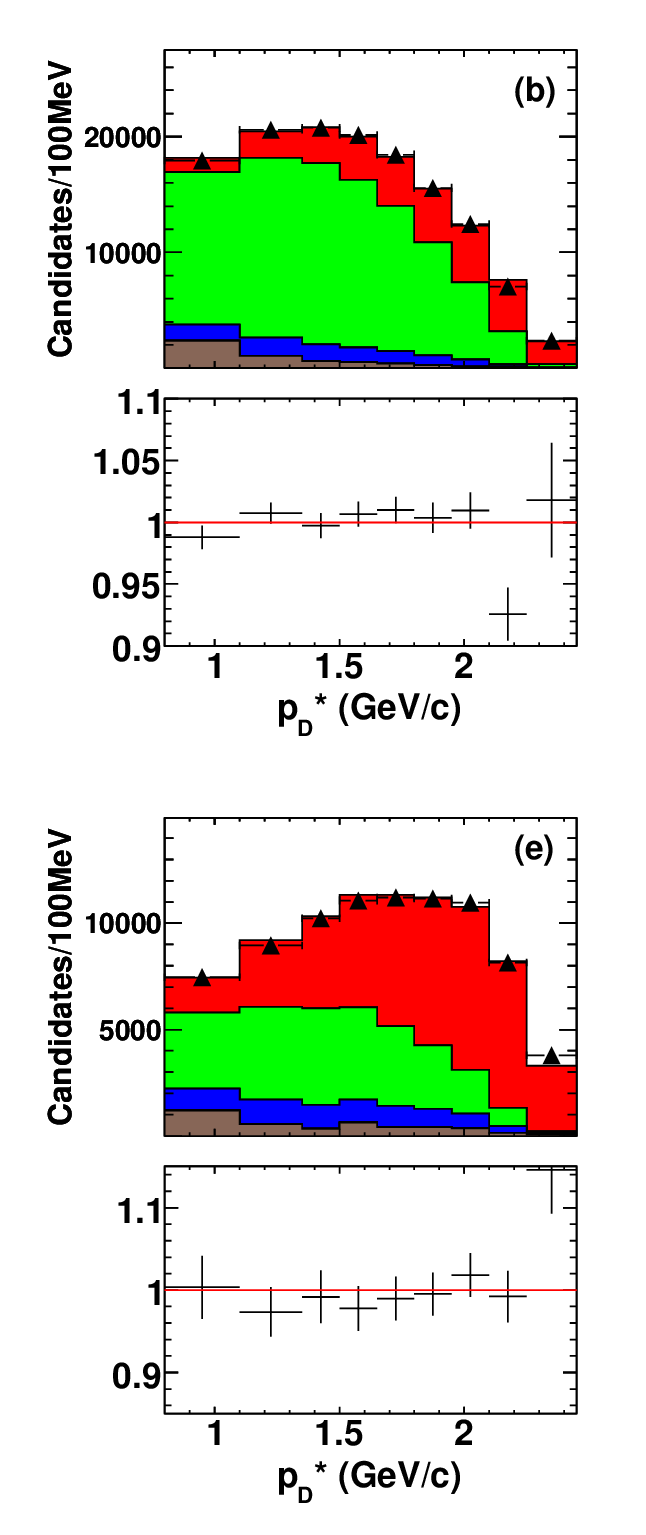}
    \includegraphics[height=5.4cm,trim = 0 125mm 0 0,clip]{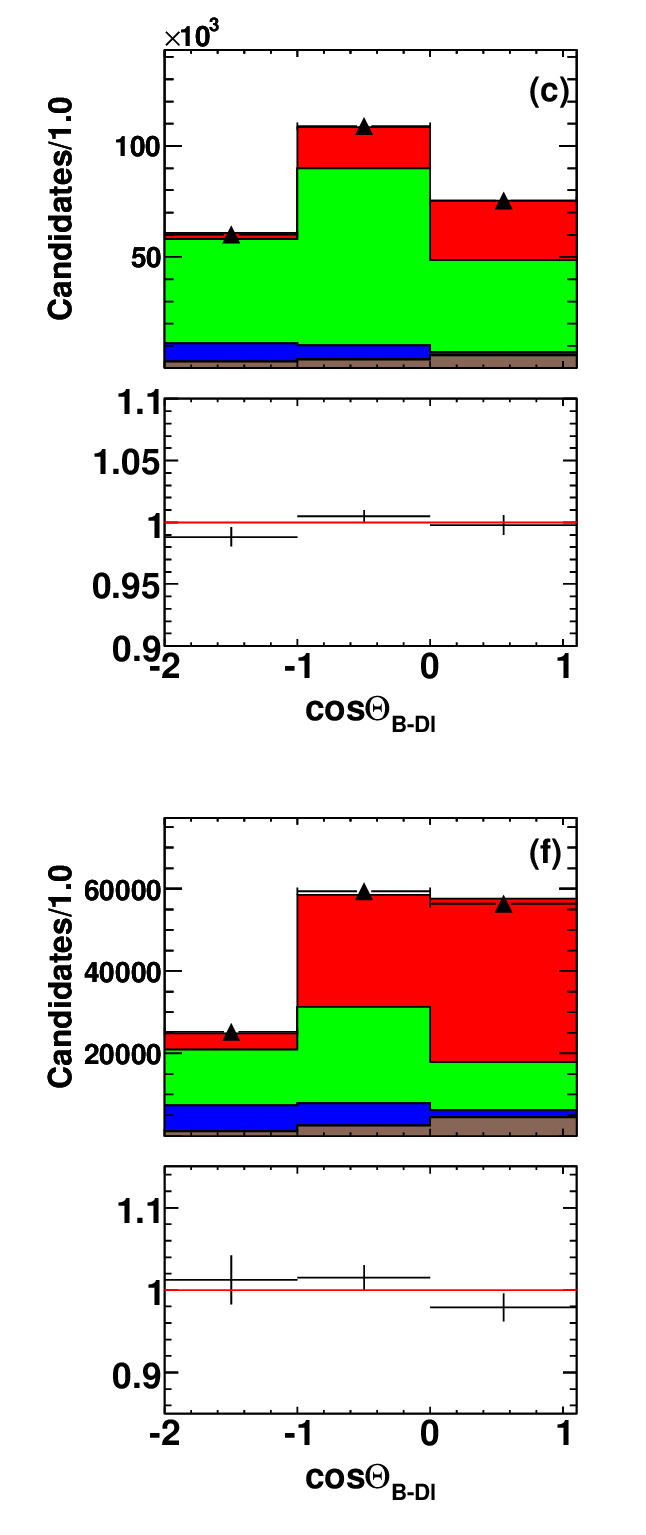}
    \caption{BaBar~\cite{Aubert:2008yv}: 
	Projected distributions for selected $B\rightarrow D^0 e^- \bar{\nu}_e X$ 
      events a) $p_\ell$, (b) $p_D$,  and (c) $\cos\theta_{BY}$. The data (points) 
      are compared to the fit result, showing 
      contributions from $D\ell\nu$ (red), $D^*\ell\nu$ (green),
     $D^{**}\ell\nu$ (blue) decays, and residual background (taupe).}
    \label{slep:f:Incl}
\end{figure}

Since the $D\ell\nu$ and $D^*\ell\nu$ decays are measured simultaneously, the
comparison of their form factors to validate the QCD predictions is
straightforward. The measured form factor ratio at zero recoil
$\mathcal{G}(1)/\mathcal{F}(1)=1.23 \pm 0.09 $ 
confirms the lattice QCD prediction of $1.16 \pm 0.04$. 
The difference of the slope parameters 
$\rho^2_D - \rho^2_{D^*} = 0.01 \pm 0.04$ is consistent with
zero, as predicted~\cite{Grinstein:2001yg}.   

The large luminosity accumulated in the $B$-factories permits the use of
tagged event samples, for which one of the two $B$ mesons is fully 
reconstructed in an hadronic final state (more than 1000 modes are considered) 
and a semileptonic decay of the other $B$ is reconstructed from the remaining particles in the event. Since the momentum of the tagged $B$ 
is measured, the kinematic properties of the semileptonic $B$ are 
fully determined. This technique results in a sizable background reduction
and thus a much lower bound on the lepton momentum ($p_\ell >0.6~\gevc)$, 
a much more precise determination of $w$, and therefore a remarkable
reduction of the systematic error, at the cost of an increase in the
statistical error (the tagging efficiency does not exceed 0.5\%).
While several measurements of semileptonic branching fractions exist to
date, only BaBar has presented a form factor determination,  
$\mathcal{G}(1)\Vcb$ and $\rho^2_{D}$,  with a tagged
sample of $B \to D \ell \nu$ decays~\cite{:2008ii}.

The yield of signal events in ten equal size $w$ bins is obtained from a fit to
the distribution of the  missing mass squared, ${\cal M}^2_\nu = (P_B - P_D - P_\ell)^2$.  An example is shown in Fig.~\ref{slep:f:Tag}.
A fit to the background-subtracted and efficiency-corrected signal yield, summed over charged and neutral $B$ decays, is used to extract the form-factor parameters, the normalization $\mathcal{G}\Vcb$, and the slope, $\rho^2_D$.  The signal yield and the fitted form factor as a function  of $w$ are shown in Fig.~\ref{slep:f:Tag}.
The results of this measurement, and of all the others
discussed so far, are reported in Tab.~\ref{slep:t:FF}.  There is very good consistency among all of the most recent measurements.
\begin{figure}[bp]
    \centering
    \includegraphics[width=0.3\textwidth]{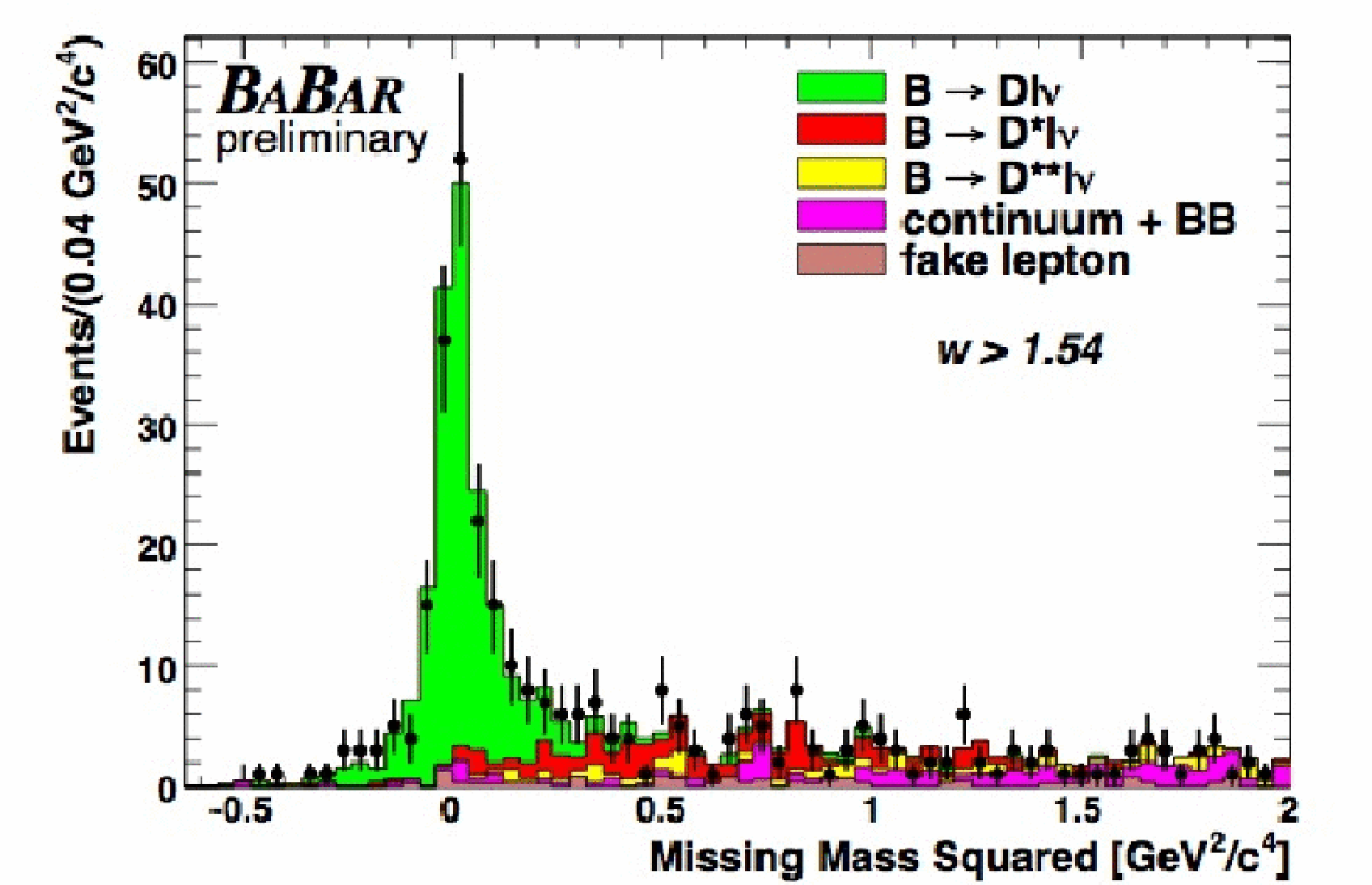} \hfill
    \includegraphics[width=0.64\textwidth]{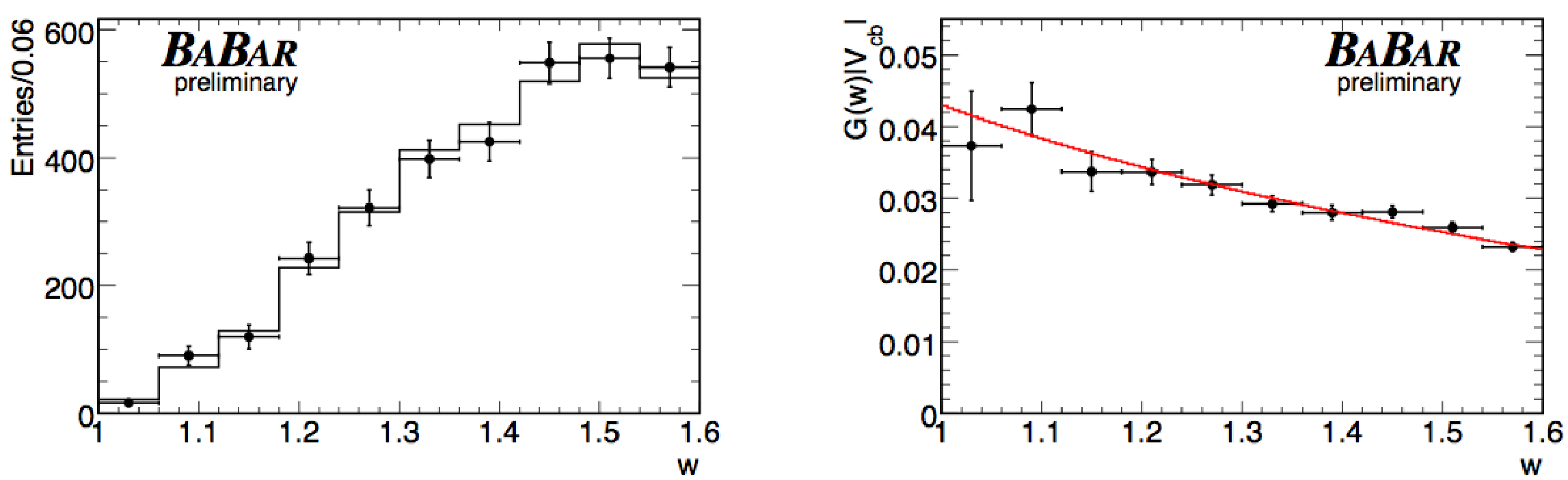}
    \caption{BaBar analysis of tagged $B \to D \ell \mu$ decays~\cite{:2008ii}  
     Data (points) compared to fit results,
     left: ${\cal M}^2_\nu$ for $w>1.54$,  center: signal event
     yield for the sum of charged and neutral $B$ decays,
	right: $\mathcal{G}(w)$ vs $w$, as obtained from 
      efficiency-corrected yields (data points) and the result of the 
      form factor fit (solid line).
    }
    \label{slep:f:Tag}
\end{figure}

By integrating the differential decays rates the branching fractions for
$B\rightarrow D^ \ell \nu$ and $B\rightarrow D^{*} \ell \nu$ decays 
can be determined with good precision.
However, there has been a long standing problem with the measured semileptonic branching fractions. 
The sums of the branching fractions for $B \rightarrow
D \ell^- \bar\nu$, $B \rightarrow D^* \ell^- \bar\nu$ and
$B \rightarrow D^{(*)} \pi  \ell^- \bar\nu $ decays\cite{:2007rb,Aubert:2007qw},
$9.5\pm 0.3\%$ for $\Bp$ and $8.9\pm0.2)\%$ for $\Bz$,
are significantly smaller than the measured inclusive $B \rightarrow X_c \ell \nu$ branching fractions of $10.89 \pm 0.16 \%$ and $10.15 \pm 0.16\%$ for $\Bp$ and $\Bz$, respectively.
Branching fractions for  $B \to D^{**} \ell \nu$ decay are still not well known,
and furthermore, the assumption that the four $D^{**}$ mesons decay exclusively to 
$D\pi$ and $D^*\pi$ final states is largely untested experimentally. And even among the measured values for the single largest $B$ branching fraction, ${\cal B}(B\to D^* \ell\nu)$, there is a spread that exceeds the stated errors significantly. 
 
\subsubsection{Determination of Form Factors and $|V_{cb}|$}


Fig.~\ref{slep:f:cont} shows the one sigma contour plots for all the
measurements of $\mathcal{G}(w)\Vcb$ and $\mathcal{F}(w)\Vcb$
performed so far.
\begin{figure}[bp]
    \centering
    \includegraphics[height=4.5cm]{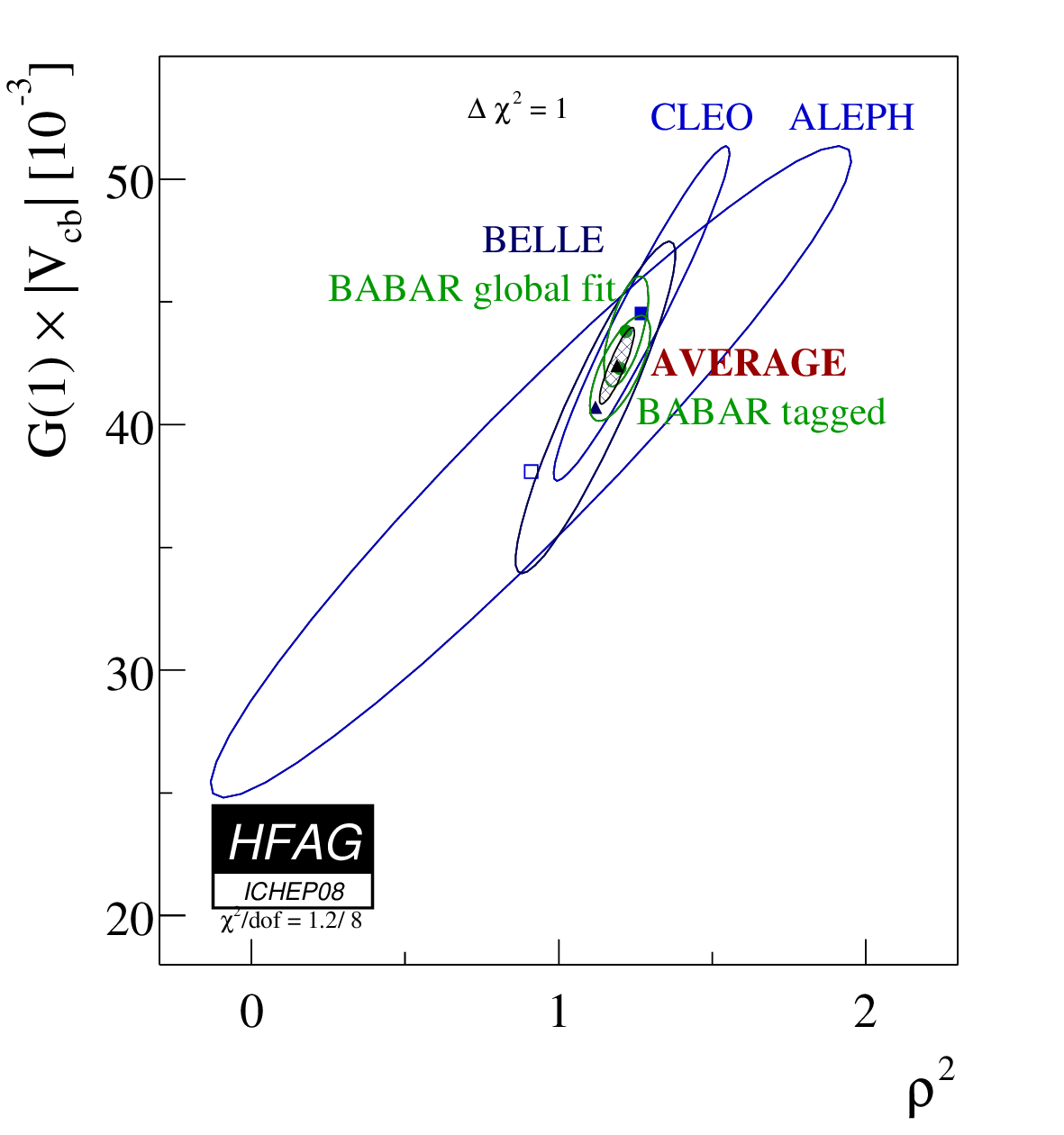} \hfill
    \includegraphics[height=4.5cm]{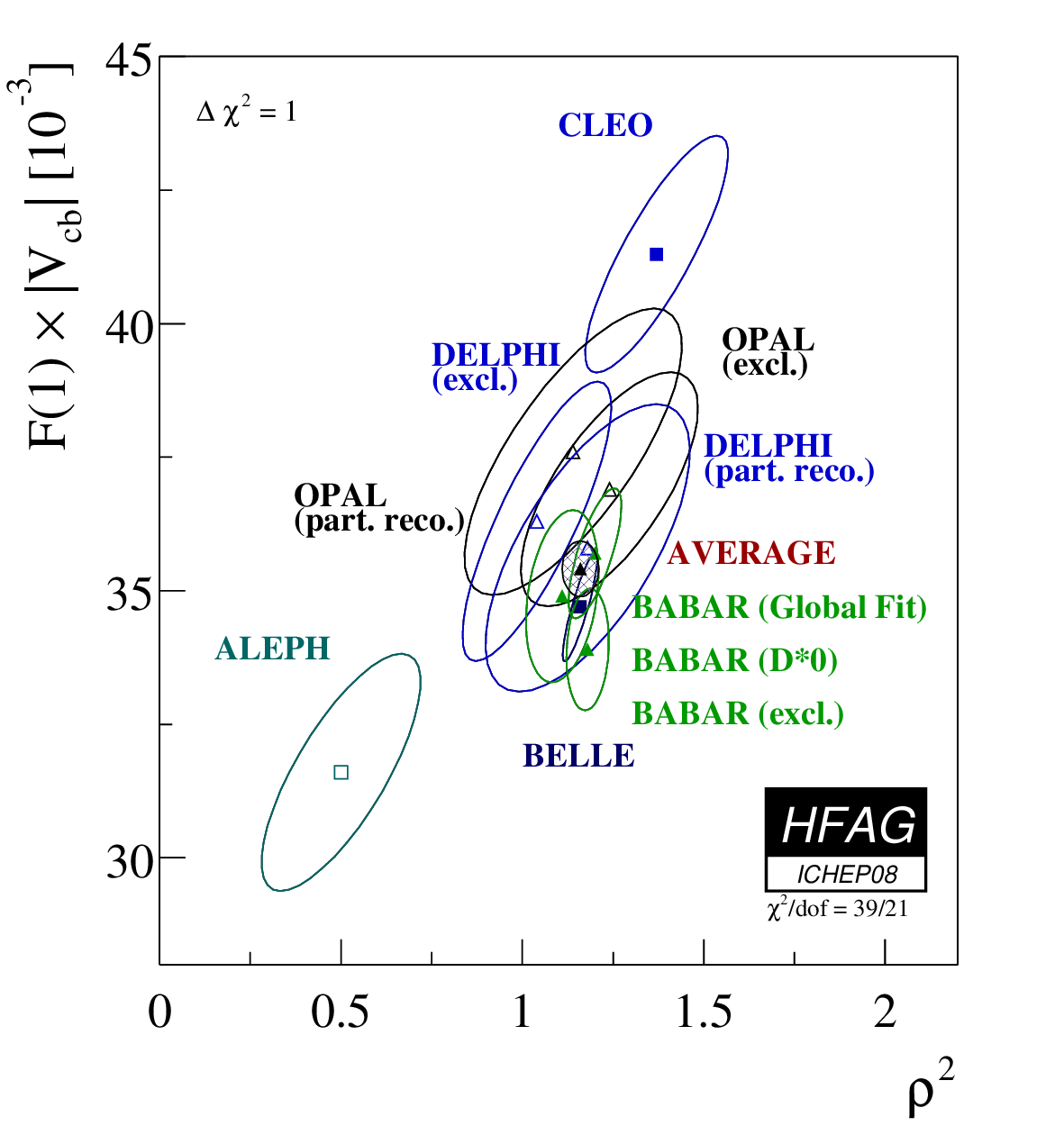} \hfill
    \includegraphics[height=4.5cm]{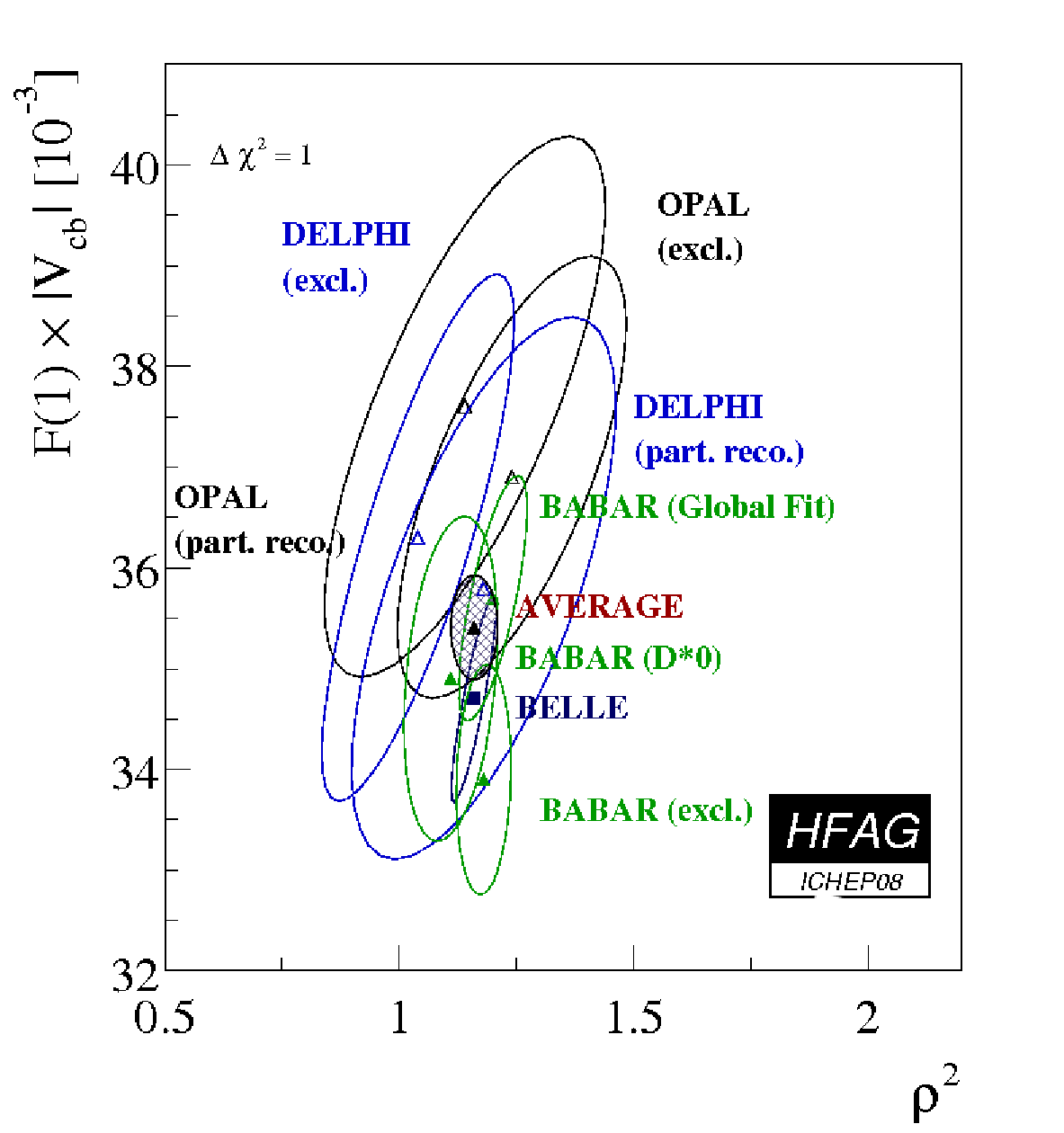}
    \caption{HFAG: One sigma contour plots for all measurements
         of $\mathcal{G}(1)\Vcb$ (left), $\mathcal{F}(1)\Vcb$
        (center), and  $\mathcal{F}(1)\Vcb$ with the two
         measurements that are least consistent with the 
        average removed (right).}
    \label{slep:f:cont}
\end{figure}
While there is a good agreement among the five
measurements of $B\rightarrow D\ell \nu$ decays, 
there is less consistency among the ten D$^*$ results, specifically 
two of the older measurements differ significantly from the recent, more 
precise measurements. 
Tab.~\ref{slep:t:avr} shows the averages of form factor measurements.
\begin{table}[tp]
    \centering
    \caption{Averages for form factors extrapolations and slopes.}
    \label{slep:t:avr}
    \begin{tabular}{ccc}
        \hline\hline
        Process & $\mathcal{G}(1)\Vcb$, $\mathcal{F}(1)\Vcb$ & $\rho^2_{D^{(*)}}$ \\
        \hline
        $B\rightarrow D\ell \nu$ & $42.4\pm 1.6 $& $1.19\pm 0.05$  \\
        $B\rightarrow D^*\ell \nu$ & $35.41 \pm 0.52$    & $1.16 \pm 0.05$  \\
        \hline\hline
    \end{tabular}
\end{table}
Using the values of $\mathcal{G}(1)$ and $\mathcal{F}(1)$ reported in 
Eqs.~(\ref{slep:eq:G(1)}) and~(\ref{slep:eq:F(1)}) we obtain
\begin{eqnarray}
    \Vcb &=& (39.4\pm1.4\pm0.9)\times10^{-3}~\mathrm{from}~B\to D\ell\nu, \\
    \Vcb &=& (38.28\pm0.71\pm0.99)\times10^{-3}~\mathrm{from}~B\to D^*\ell\nu,
\end{eqnarray}
where the first error is from experiment and the second from unquenched
lattice QCD.
The two results agree well. 
It is not straightforward for combine these two results, because the 
correlations between the two sets of measurements and two calculations 
have not been analyzed.
Assuming a correlation of 50\% for both, we obtain the average value 
from exclusive decays 
\begin{equation}
    \Vcb = (38.6\pm 1.1)\times10^{-3},
    \label{slep:eq:exVcb}
\end{equation}
where experimental and lattice-QCD errors have been added in quadrature.

\subsection{Inclusive CKM-favored $B$ decays}
\label{sec:inclusiveVcb}

\subsubsection{Theoretical Background}

The inclusive $\bar B\rightarrow X_c\ell\bar\nu$ decay rate can be
calculated using the operator product expansion (OPE). Applied to heavy
quark decays, the OPE amounts to an expansion in inverse powers of the
heavy quark mass and is often referred to as heavy-quark expansion
(HQE). Using this technique, the non-perturbative input needed to
predict the rate is reduced to a few
matrix elements of local operators in HQET. Together with \Vcb, \mb, and
\mc, these heavy-quark parameters can be extracted from a moment
analysis, i.e. by fitting the theoretical predictions for the decay rate
and moments of decay spectra to the available experimental results.

The application of the OPE to semileptonic heavy hadron decays was
developed quite some time ago
\cite{Chay:1990da,Bigi:1993fe,Blok:1993va,Manohar:1993qn}. A detailed
discussion of the technique can, for example, be found in the textbook
\cite{Manohar:2000dt}. For a review focusing on the extraction of \Vcb
and the heavy quark parameters, see \cite{Benson:2003kp} and the PDG
review \cite{Amsler:2008zzb}. In the following, we briefly recall some
of the basic concepts, review recent progress in evaluating higher-order
perturbative corrections, and briefly discuss possible limitations of
the approach. After this, we review the available experimental data and
the results of the moment analysis.
                                            
The $\bar B\rightarrow X_c\ell\bar\nu$ decay is mediated by the
effective Hamiltonian
\begin{equation}
{\cal H}_{\rm eff}=\frac{G_F}{\sqrt{2}}V_{cb}\,\, {J}^\mu\, J^{\ell}_\mu=  
\frac{G_F}{\sqrt{2}}V_{cb}\, {\bar c}\,\gamma^\mu\,(1-\gamma_5)\,b \; 
\bar\ell\,\gamma_\mu\,(1-\gamma_5)\,\nu \,.
\end{equation}
Neglecting electromagnetic corrections, the decay rate factors into a
product of a leptonic tensor $L_{\mu\nu}$ and a hadronic tensor
$W_{\mu\nu}$, which are given by the matrix elements of two leptonic and
two hadronic currents.
Using the optical theorem, the hadronic tensor can be obtained from the
imaginary part of the forward matrix element of the product ${\bf
T}_{\mu\nu}$ of two weak currents, $2 M_B\,W_{\mu\nu}= -2\,{\rm Im}
\langle B(p_B)|\, {\bf T}_{\mu\nu}\, |B(p_B)\rangle$.
The OPE expands the time-ordered product ${\bf T}_{\mu\nu}$ into a sum of 
local HQET operators $O_i$ of increasing dimension
\begin{equation}\label{eq:HQEandOPE:ope}
{\bf T}_{\mu\nu} = -i\int\!d^4x e^{-iq x} {\bf T}\left[ J_\mu^\dagger(x)\, 
J_\nu (0)\right]  
= \sum_i C^i_{\mu\nu}(v\cdot q,q^2,m_b,m_c)\, O_i(0)\,.
\end{equation}
In order to perform the expansion, a velocity vector $v^\mu$, with
$v^2=1$, is introduced to split the $b$-quark momentum into $p_b^\mu=\mb
v^\mu+ r^\mu$, where the components of the residual momentum $r^\mu$ are
independent of the $b$-quark mass. It is usually chosen to be the meson
velocity, $v^\mu=p_B^\mu/M_\B$. Because Eq.~(\ref{eq:HQEandOPE:ope}) is
an operator relation, it holds for arbitrary matrix elements. To
determine the Wilson coefficients $C_{\mu\nu}^i$ one considers partonic
matrix elements of Eq.~(\ref{eq:HQEandOPE:ope})  in perturbation theory.

The OPE separates the physics associated with large scales such as \mb,
which enter the Wilson coefficients $C_{\mu\nu}^i$, from the
non-perturbative dynamics entering the matrix elements of the operators
$O_i$. In this context, it is important that the operators on the
right-hand side of Eq.~(\ref{eq:HQEandOPE:ope}) are defined in HQET so
that their matrix elements are independent of \mb up to power
corrections and are governed by non-perturbative dynamics associated
with the scale $\Lambda_{\mathrm{QCD}}$. Since the Wilson
coefficients of higher dimensional operators in
Eq.~(\ref{eq:HQEandOPE:ope}) contain inverse powers of \mb, their
contributions to the rate are suppressed by powers of
$\Lambda_{\mbox{\tiny QCD}}/m_b$. The leading operator in
Eq.~(\ref{eq:HQEandOPE:ope}) has dimension three and is given by a
product of two HQET heavy quark fields $O_3=\bar h_v\, h_v$. Up to power
corrections, its $B$-meson matrix element is one. Dimension four
operators can be eliminated using the equation of motion and the leading
power corrections arise from two dimension five operators: the kinetic
operator $O_{\rm kin}$ and the chromomagnetic operator $O_{\rm mag}$,
whose $B$-meson matrix elements are denoted by $\lambda_1$ and
$\lambda_2$ \cite{Falk:1992wt} or $\mu_\pi^2$ and $\mu_G^2$
\cite{Blok:1993va}. Different schemes are used to define these
parameters, but to leading order and leading power they are given by
\begin{eqnarray}
\langle O_{\rm kin} \rangle&\equiv& \frac{1}{2M_B} \langle \bar B(p_B)|\, \bar{h}_v (i D)^2 {h}_v\, |\bar B(p_B)\rangle =  - \mu_{\pi}^2= \lambda_1 \, , \label{matrixelements} \\
\langle O_{\rm mag}\rangle &\equiv& \frac{1}{2M_B}\langle \bar B(p_B)|\,\frac {g }{2}\, \bar{h}_v  \sigma_{\mu\nu}G^{\mu\nu} {h}_v\, |\bar B(p_B)\rangle  =\mu_{G}^2=  3 \lambda_2 \,.\nonumber 
\end{eqnarray}
In order for the OPE to converge, it is necessary that the scales
entering the Wilson coefficients are all larger than
$\Lambda_{\mbox{\tiny QCD}}$. This condition is violated in certain
regions of phase-space. In order to get reliable predictions, one needs
to consider sufficiently inclusive quantities such as the total rate,
which takes the form
\cite{Chay:1990da,Bigi:1993fe,Blok:1993va,Manohar:1993qn}
\begin{equation}\label{eq:HQEandOPE:rate}
\Gamma(\bar B\rightarrow X_c \ell \bar\nu) = \frac{G_F^2 |V_{cb}|^2 m_b^5}{192\pi^3} 
\left\{f(\rho) + k(\rho) \frac{\mu_\pi^2}{2m_b^2} + g(\rho) \frac{\mu_G^2}{2m_b^2}
 \right \} \, ,
\end{equation}
up to corrections suppressed by $(\Lambda_{\mbox{\tiny QCD}}/m_b)^3$,
and with $\rho=m_c^2/m_b^2$. The Wilson coefficients $f(\rho)$,
$k(\rho)$ and $g(\rho)$ can be calculated in perturbation theory. They
are obtained by taking the imaginary part of $C_{\mu\nu}^i$, contracting
with the lepton tensor $L^{\mu\nu}$ and integrating over the leptonic
phase space. We have written the expansion in inverse powers of \mb, but
it is the energy release $\Delta E \sim m_b-m_c$ which dictates the size
of higher order corrections. Other suitable inclusive observables
include the spectral moments
\begin{equation}\label{eq:HQEandOPE:momentsdef}
\left\langle E_\ell^n E_X^m (M_X^2)^l\right\rangle =
\frac{1}{\Gamma_{0}}\int_{E_0}^{E_{\rm max}}\!\!\!\! 
dE_\ell \int\!\!dE_X\,\int \!\!dM_X^2 
\frac{d\Gamma}{dE_X\, dM_X^2\, dE_\ell } E_\ell^n E_X^m (M_X^2)^l,
\end{equation}
with $\Gamma_0=\Gamma(E_\ell > E_0)$ for low values of $n$, $m$, and
$l$, with a moderate lepton energy cut $E_0$. The OPE for the moments
Eq.~(\ref{eq:HQEandOPE:momentsdef}) depends on the {\it same} operator
matrix elements as the rate Eq.~(\ref{eq:HQEandOPE:rate}), but the
calculable Wilson coefficients $f(\rho)$, $g(\rho)$, and $k(\rho)$ will
be different for each moment. Note that the coefficient $k(\rho)$ of the
kinetic operator is linked to the leading power coefficient $f(\rho)$,
for example $k(\rho)=-f(\rho)$ for the total rate. The corresponding
relations for the moments are given in \cite{Becher:2007tk}.

By measuring the rate and several spectral moments
Eq.~(\ref{eq:HQEandOPE:momentsdef}), and fitting the theoretical
expressions to the data, one can simultaneously extract \Vcb, the quark
masses  \mb and \mc, as well as the heavy quark parameters such as
$\mu_\pi$ and $\mu_G$. Two independent implementations of this moment
analysis are currently used \cite{Bauer:2002sh,Bauer:2004ve} and
\cite{Buchmuller:2005zv} (based on \cite{Benson:2003kp,Gambino:2004qm}).
Both groups include terms up to third order in $\Lambda_{\rm QCD}/m_b$
\cite{Gremm:1996df} and evaluate leading order Wilson coefficients to
one-loop accuracy
\cite{Jezabek:1988iv,Jezabek:1988ja,Czarnecki:1989bz,Czarnecki:1994pu,Voloshin:1994cy,Trott:2004xc,Falk:1995me}.
In addition, they also include the part of the two-loop corrections
which is proportional $\beta_0$
\cite{Luke:1994du,Ball:1995wa,Gremm:1996gg,Falk:1997jq,Uraltsev:2004in,Gambino:2004qm,Aquila:2005hq}.
However, the two fits use different schemes for the masses and heavy
quark parameters. The analysis of \cite{Buchmuller:2005zv} is performed
in the kinetic scheme \cite{Uraltsev:1996rd}, while
\cite{Bauer:2002sh,Bauer:2004ve} adopt the $1S$-scheme
\cite{Hoang:1998hm} as their default choice. Both schemes, as well as
others, such as the potential-subtracted  \cite{Beneke:1998rk} and the
shape-function scheme \cite{Bosch:2004th}, are designed to improve the
perturbative behavior by reducing the large infrared sensitivity
inherent in the pole scheme. Two-loop formulae for the conversion among
the different schemes can be found in  \cite{Neubert:2004sp}.

It has been noticed that the two-loop terms appearing in the conversion
of \mb among schemes were in some cases larger than the uncertainties
quoted after fitting in a given scheme \cite{Neubert:2008cp}. This
indicates that higher-order corrections to the Wilson coefficients can
no longer be neglected. Recently, a number of new perturbative results
for the Wilson coefficients have become available, however, they have
not yet been implemented into the moment analysis. The Wilson
coefficient of the leading order operator $O_3$ has been evaluated to
two-loop accuracy \cite{Melnikov:2008qs,Pak:2008qt}. The numerical
technique used in \cite{Melnikov:2008qs} allows for the calculation of
arbitrary moments and its results are confirmed by an independent
analytical calculation of the rate and the first few $E_\ell$ and $E_X$
moments \cite{Pak:2008qt}. An earlier estimate of the two-loop
corrections \cite{Czarnecki:1998kt} needed to be revised in view of the
new results \cite{Dowling:2008ap}. At the same accuracy, one should also
include the one-loop corrections to the coefficients of the kinetic and
chromomagnetic operators. So far, only the corrections for the kinetic
operator are available \cite{Becher:2007tk}. Furthermore, the tree-level
OPE has been extended to fourth order in ${\Lambda_{\rm QCD}}/{m_b}$
\cite{Dassinger:2006md}.

In addition to perturbative and non-perturbative corrections, the
hadronic decay rates will contain terms which are not captured by the
OPE. While such terms are exponentially suppressed in completely
Euclidean situations, they are not guaranteed to be negligible for the
semileptonic rate and its moments \cite{Shifman:2000jv}. These
violations of quark-hadron duality are difficult to quantify. Model
estimates seem to indicate that the effects on the rate are safely below
the 1\% level for the total rate \cite{Bigi:2001ys,Bigi:2002fj}, but
they could be larger for the spectral moments. Other issues studied in
the recent literature concern the role of the charm quarks
\cite{Bigi:2005bh,Breidenbach:2008ua}  and potential new physics effects
\cite{Dassinger:2007pj,Dassinger:2008as}.

\subsubsection{Measurements of Moments}

Measurements of the semileptonic \B~branching fraction and inclusive
observables in $\B\to X_c\ell\nu$~decays relevant to the determination
of the heavy quark parameters in the OPE have been obtained by the
BaBar~\cite{Aubert:2004td,Aubert:2004tea,Aubert:2007yaa},
Belle~\cite{Urquijo:2006wd,Schwanda:2006nf}, CDF~\cite{Acosta:2005qh},
CLEO~\cite{Csorna:2004kp} and DELPHI~\cite{Abdallah:2005cx} Collaborations.
The photon-energy spectrum in $\B\to X_s\g$~decays,
which is particularly sensitive to the \b-quark mass, \mb, has been
studied by BaBar~\cite{Aubert:2006gg,Aubert:2005cua},
Belle~\cite{Schwanda:2008kw,Abe:2008sxa} and
CLEO~\cite{Chen:2001fja}. In this section, we briefly review new or
updated measurements of $\B\to X_c\ell\nu$~decays.

BaBar has updated their previous measurement of the hadronic mass moments
$\langle M^{2n}_X\rangle$ \cite{Aubert:2004tea} and obtained preliminary
results based on a dataset of 210~\invfb taken at the
\FourS~resonance~\cite{Aubert:2007yaa}. In this analysis, the hadronic
decay of one \B~meson in $\FourS\to\B\Bbar$ is fully reconstructed
($\B_\mathrm{tag}$) and the semileptonic decay of the second \B is
inferred from the presence of an identified lepton (\electron or \mmu)
among the remaining particles in the event ($\B_\mathrm{sig}$). This
fully reconstructed tag provides a significant reduction in
combinatorial backgrounds and results in a sample of semileptonic decays 
with a purity
of about 80\%. Particles that are not used in the reconstruction of $\B_\mathrm{tag}$ and are not identified as the charged lepton 
are assigned to the $X_c$~system, and its mass $M_X$ is
calculated using some kinematic constraints for the whole event.

From the $M_X$ spectrum, BaBar calculates the hadronic mass moments
$\langle M^n_X\rangle$, $n=1,\dots,6$ as a function of a lower limit on 
the lepton momenta in
the center-of-mass (c.m.) frame ranging from 0.8 to 1.9~\gevc. These
moments are distorted by acceptance and finite resolution
effects and an event-by-event correction is derived from Monte Carlo
(MC) simulated events. These corrections are approximated as linear functions 
of the observed mass with
coefficients that depend on the lepton momentum, the multiplicity of the
$X_c$~system and $E_\mathrm{miss}-c|\vec p_\mathrm{miss}|$, where
$E_\mathrm{miss}$ and $\vec p_\mathrm{miss}$ are the missing energy
and 3-momentum in the event, respectively. Note that in this analysis
mixed mass and c.m.\ energy moments $\langle N^{2n}_X\rangle$,
$n=1,2,3$, with $N_X=M^2_Xc^4-2\tilde\Lambda E_X+\tilde\Lambda^2$ and
$\tilde\Lambda=0.65$~\gev are measured in addition to ordinary hadronic mass
moments. These mixed moments are expected to better constrain some
heavy quark parameters, though  they are not yet used in global fit
analyses.

Belle has recently measured the c.m.\ electron energy~\cite{Urquijo:2006wd}
and the hadronic mass~\cite{Schwanda:2006nf} spectra in $\B\to
X_c\ell\nu$~decays, based on 140~\invfb of \FourS~data. The
experimental procedure is very similar to the BaBar analysis, {\it
i.e.}, the hadronic decay of one \B~meson in the event is fully reconstructed. 
The main difference to the  BaBar analysis is that
detector effects in the spectra are removed by unfolding using the
Singular Value Decomposition (SVD) algorithm~\cite{Hocker:1995kb} with
a detector response matrix determined by MC simulation. The moments are
calculated from the unfolded spectra. Belle measures the partial semileptonic
branching fraction and the c.m.\ electron energy moments $\langle
E^n_e\rangle$, $n=1,\dots,4$, for minimum c.m.\ electron energies
ranging from 0.4 to 2.0~\gev. In the hadronic mass
analysis~\cite{Schwanda:2006nf}  the first and second moments
of $M^2_X$ are measured for minimum c.m.\ lepton energies between 
0.7 and 1.9~\gev.

Another interesting analysis of inclusive $\B\to X_c\ell\nu$~decays comes
from the DELPHI experiment~\cite{Abdallah:2005cx} operating at LEP. 
In this study,
the \b-frame lepton energy $\langle E^n_l\rangle$, $n=1,2,3$, and the
hadronic mass $M^{2n}_X$, $n=1,\dots,5$, moments are measured without
applying any selection on the lepton energy in the \b-frame. This is
possible because DELPHI measures decays of \b-hadrons in
$\Z\to\b\bbar$~events. \b-hadrons are produced with significant
kinetic energy in the laboratory frame, so that charged leptons produced
at rest in the \b-frame can be observed in the detector.

\subsubsection{Global Fits for \Vcb and \mb}

The OPE calculation of the $\B\to X_c\ell\nu$ weak decay rate depends
on a set of heavy quark parameters that contain the soft QCD
contributions. These parameters can be determined from other inclusive
observables in \B~decays, namely the lepton energy~$\langle
E^n_\ell\rangle$ and hadronic mass moments~$\langle M^{2n}_X\rangle$
in $\B\to X_c\ell\nu$ and the photon energy moments~$\langle
E^n_\g\rangle$ in $\B\to X_s\g$. Once these parameters are known, \Vcb
can be determined from measurements of the semileptonic \B~branching
fraction. This is the principle of the global fit analyses for
\Vcb. On the theory side, these analyses require OPE predictions of
the aforementioned inclusive observables, in addition to a calculation
of the semileptonic width. At present, two independent sets of
theoretical formulae have been derived including non-perturbative
corrections up to $\mathcal{O}(1/\mb^3)$, referred to as the
kinetic~\cite{Benson:2003kp,Gambino:2004qm,Benson:2004sg} and the 1S
scheme~\cite{Bauer:2004ve}, according to the definition of
the \b-quark mass used.

Tab.~\ref{slep:incVcb:tab:1} summarizes the results of the global fit analyses
performed by BaBar~\cite{Aubert:2007yaa} and Belle~\cite{Schwanda:2008kw}
in terms of \Vcb and the \b-quark mass \mb. 
\begin{table}[tp]
    \centering
    \caption{Results of the global fit analyses by BaBar and Belle in
        terms of \Vcb and the \b-quark mass~\mb, including 
        the $\chi^2$ of the fit over the number of degrees of
        freedom. Note that the fit results for \mb 
        in the kinetic and
        1S~schemes can be compared only after scheme translation.} 
    \label{slep:incVcb:tab:1}
    \begin{tabular}{l|ccc}
    \hline \hline
      & \Vcb (10$^{-3}$) & \mb (\gev) & $\chi^2/\mathrm{ndf.}$\\
    \hline
    BaBar kinetic~\cite{Aubert:2007yaa} & $41.88\pm 0.81$ &
    $4.552\pm 0.055$ & $8/(27-7)$\\
    Belle kinetic~\cite{Schwanda:2008kw} & $41.58\pm 0.90$ &
    $4.543\pm 0.075$ & $4.7/(25-7)$\\
    Belle 1S~\cite{Schwanda:2008kw} & $41.56\pm 0.68$ & $4.723\pm
    0.055$ & $7.3/(25-7)$\\
    \hline \hline
    \end{tabular}
\end{table}
BaBar uses 27 and Belle 25
measurements of the partial $\B\to X_c\ell\nu$~branching fraction and
the moments in $B\to X_c\ell\nu$ and $B\to X_s\g$~decays. Measurements at
different thresholds in the lepton or photon energy are highly correlated. Correlations between measurements
and between their theoretical predictions must therefore be accounted
for in the definition of the $\chi^2$ of the fit. The BaBar analysis performs a
fit in the kinetic mass scheme only. In this framework, the free
parameters are: \Vcb, $\mb(\mu)$, $\mc(\mu)$, $\mu_\pi^2(\mu)$,
$\mu^2_G(\mu)$, $\rho^3_D(\mu)$ and $\rho^3_{LS}(\mu)$, where $\mu$ is
the scale taken to be 1~\gev. In addition, Belle fits their data also
in the 1S~scheme. Here, the free parameters are: \Vcb, \mb,
$\lambda_1$, $\rho_1$, $\tau_1$, $\tau_2$ and $\tau_3$. The only
external input in these analyses is the average $B$~lifetime
$\tau_B=(1.585\pm 0.006)$~ps~\cite{Barberio:2008fa}.

HFAG has combined the available $\B\to X_c\ell\nu$ and $\B\to X_s\g$
data from different experiments to extract \Vcb and~\mb.
Using 64 measurements in total
(Tab.~\ref{slep:incVcb:tab:2}), the analysis is carried out in the kinetic
scheme. 
\begin{table}[tp]
    \caption{Measurements of the lepton energy~$\langle E^n_\ell\rangle$
        and hadronic mass moments~$\langle M^{2n}_X\rangle$ in $\B\to
        X_c\ell\nu$ and the photon energy moments~$\langle E^n_\g\rangle$ in
        $\B\to X_s\g$ used in the combined HFAG fit.} 
    \label{slep:incVcb:tab:2}
    \centering
    \begin{tabular}{l|ccc}
        \hline \hline
        Experiment &  $\langle E^n_\ell\rangle$ & $\langle
        M^{2n}_X\rangle$ & $\langle E^n_\g\rangle$\\
        \hline
        BaBar & $n=0,1,2,3$~\cite{Aubert:2004td} &
        $n=1,2$~\cite{Aubert:2007yaa} & 
        $n=1,2$~\cite{Aubert:2006gg,Aubert:2005cua}\\
        Belle & $n=0,1,2,3$~\cite{Urquijo:2006wd} &
        $n=1,2$~\cite{Schwanda:2006nf} & $n=1,2$~\cite{Abe:2008sxa}\\
        CDF & & $n=1,2$~\cite{Acosta:2005qh} & \\
        CLEO & & $n=1,2$~\cite{Csorna:2004kp} &
        $n=1$~\cite{Chen:2001fja}\\
        DELPHI & $n=1,2,3$~\cite{Abdallah:2005cx} &
        $n=1,2$~\cite{Abdallah:2005cx} & \\
        \hline \hline
    \end{tabular}
\end{table}
The procedure is very similar to the analyses of the $B$-factory
datasets described above. The results for \Vcb, \mb and
$\mu^2_\pi$ are quoted in Tab.~\ref{slep:incVcb:tab:3} 
\begin{table}
    \centering
    \caption{Combined HFAG fit to all experimental data
        (Tab.~\ref{slep:incVcb:tab:2}). In the last column we quote
        the $\chi^2$ of the fit over the number of degrees of
        freedom.}
    \label{slep:incVcb:tab:3}
    \begin{tabular}{l|cccc}
        \hline\hline
        Dataset & \Vcb (10$^{-3}$) & \mb (\gev) & $\mu^2_\pi$ ($\gev^2$) &
        $\chi^2/\mathrm{ndf.}$\\
        \hline
        $X_c\ell\nu$ and $X_s\g$ & $41.54\pm 0.73$ & $4.620\pm 0.035$ &
        $0.424\pm 0.042$ & $26.4/(64-7)$\\
        $X_c\ell\nu$ only & $41.31\pm 0.76$ &  $4.678\pm 0.051$ &
        $0.410\pm 0.046$ & $20.3/(53-7)$\\
        \hline \hline
    \end{tabular}
\end{table}
and Fig.~\ref{slep:incVcb:fig:1}.
\begin{figure}
    \centering
    \includegraphics[width=0.45\columnwidth]{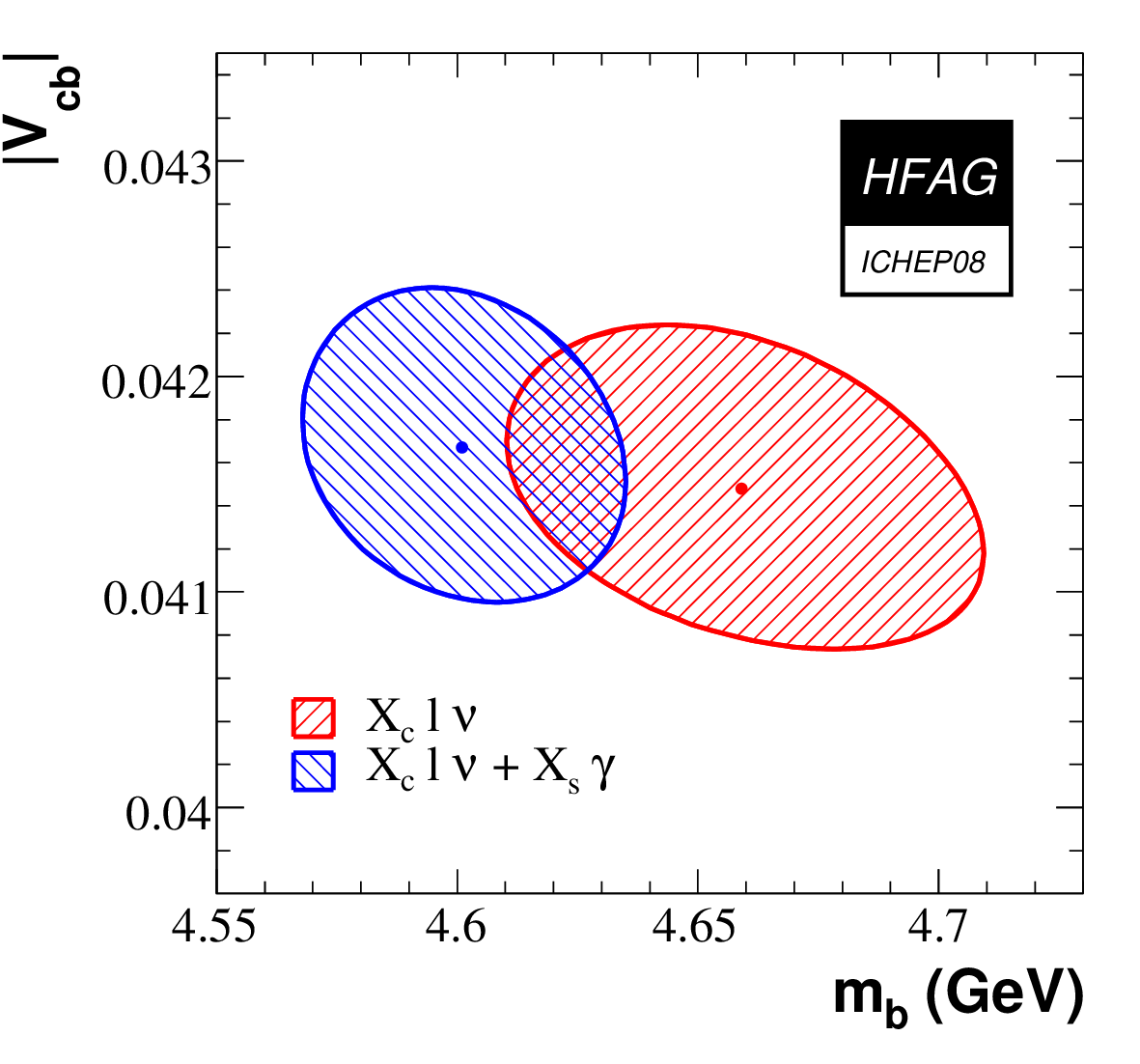}
    \includegraphics[width=0.45\columnwidth]{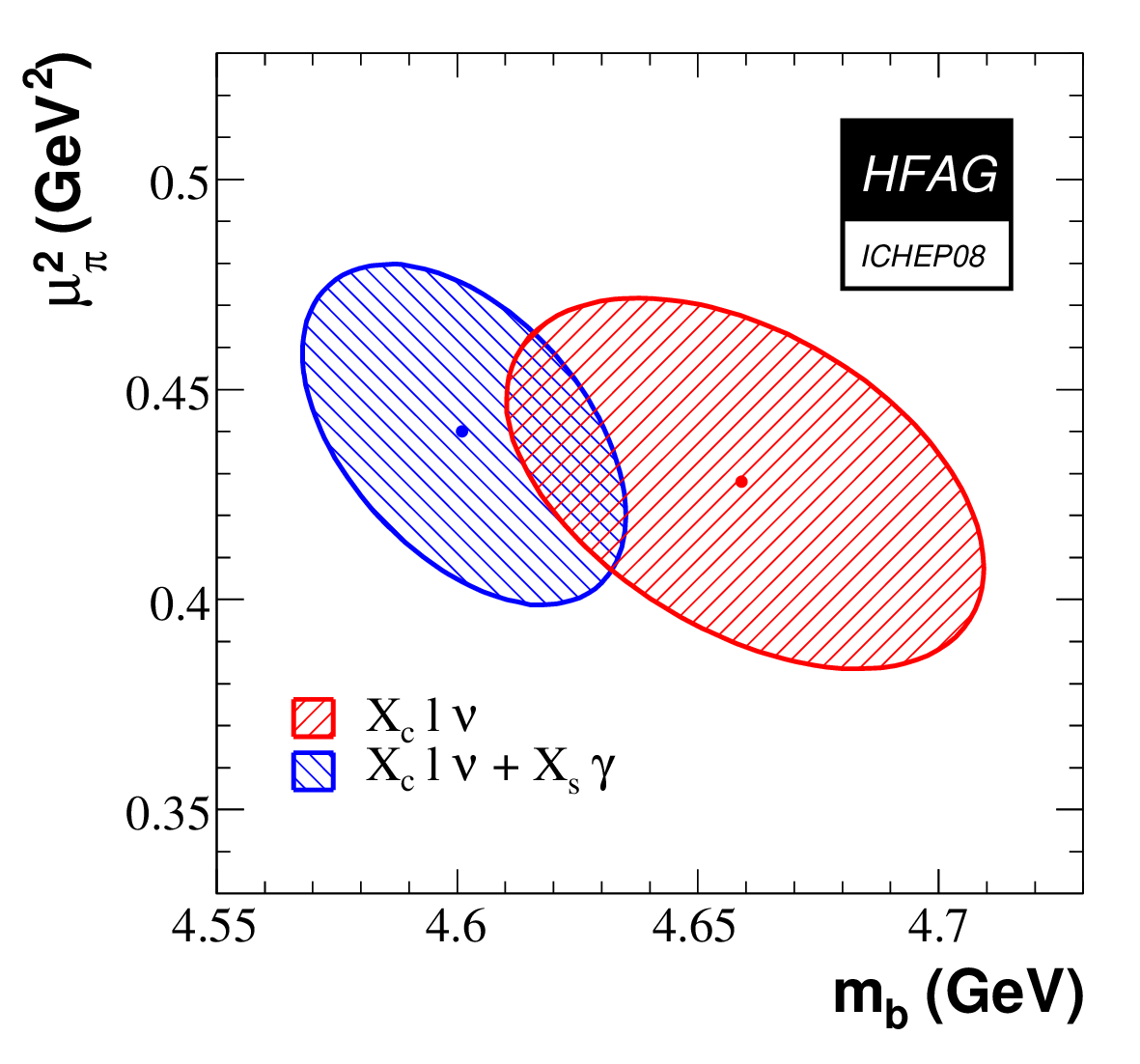}
    \caption{$\Delta\chi^2=1$~contours for the HFAG fit for \Vcb
        and \mb in the $(\mb,\Vcb)$ and $(\mb,\mu^2_\pi)$~planes, with and
        without $B\to X_s\g$~data.}
    \label{slep:incVcb:fig:1}
\end{figure}
Recently, concerns have been raised about the
inclusion of $\B\to X_s\g$~moments, because their prediction is not based
on pure OPE but involves modeling of non-OPE contributions using a
shape function. We therefore also quote the results of a fit without the
$\B\to X_s\g$ data (53 measurements).

The current result for \Vcb\ based on fits to lepton-energy, hadronic-mass, and photon-energy moments by HFAG is
\begin{equation}
    \Vcb = (41.54\pm0.73)\times10^{-3},
    \label{slep:incVcb:eq:Vcb}
\end{equation}
where theoretical and experimental uncertainties have been combined.
This value differs from the exclusive determination of \Vcb, 
Eq.~(\ref{slep:eq:exVcb}), at the $2\sigma$ level.
Note that the inclusive fits lead to values $\chi^2$ that are 
substantially smaller than should be expected, which may point to a 
problem with the input errors or correlations.
The determination of \mb and \mc will be further discussed in
Sec.~\ref{sec:mbdetermination}.

\subsection{Inclusive CKM-suppressed $B$ decays}
\subsubsection{Theoretical Overview}
\label{sec:inclusivevubtheory}


The inclusive semileptonic B decays into charmless final states are
described by the same local OPE we have considered above for the CKM
favored ones.  The relevant non-perturbative matrix elements are those
measured in the fit to the moments discussed in Sec.~\ref{sec:inclusiveVcb}.
In the total
width there is one additional contribution from a four-quark operator
related to  the Weak Annihilation (WA) between the $b$ quark and a
spectator~\cite{Bigi:1993bh}, whose analogue in the CKM favored decay is
suppressed by the large charm mass. In an arbitrary, properly
defined scheme the total semileptonic width is through $O(1/m_b^3, \as^2)$
\begin{eqnarray}
\label{widthb2u}
\Gamma[\bar{B} \to X_u e \bar{\nu}] &=& \frac{G_F^2 \,m_b^5}{192\pi^3} |V_{ub}|^2
\left[ 1 +  \frac{\alpha_s}{\pi} p_u^{(1)}
              + \frac{\alpha_s^2}{\pi^2} p_u^{(2)} 
              - \frac{\mupi}{2 m_b^2} - \frac{3\mug}{2m_b^2} 
 \right. \nonumber\\ && \hspace{1cm} \left.    
 +  \left(\frac{ 77}{6} + 8 \ln\frac{\muwa^2}{m_b^2}\right)
\frac{\rd}{ m_b^3} 
+\frac{3\rho _{LS}^3}{2 m_b^3} + \frac{32\pi^2}{m_b^3} 
B_{\scriptscriptstyle\rm WA}(\mu_{\scriptscriptstyle\rm WA})  
\right],
\end{eqnarray}
where $B_{\scriptscriptstyle\rm WA}$ is the \B meson matrix element of
the WA operator evaluated at the scale $\muwa$. Since
$B_{\scriptscriptstyle\rm WA}$ vanishes in the factorization
approximation, WA is phenomenologically important only to the extent
factorization is violated at $\muwa$.  We therefore expect it to
contribute less than 2-3\% to the difference between \Bz and \Bu widths
and, due to its isosinglet component, to the total width  of both
neutral and charged \B \cite{Uraltsev:1999rr,Voloshin:2001xi}.  The
latter and the $\ln \muwa$ in the coefficient of $\rd$ originate in
the 
mixing between WA and  Darwin operators \cite{Gambino:2005tp}.  The
dominant parametric uncertainty on the total width currently comes from
\mb, due to the $m_b^5$ dependence.  The theoretical uncertainty from
missing higher order corrections has been estimated to be at most $2\%$
 in the kinetic scheme \cite{Uraltsev:1999rr}.  Assuming  35~MeV
precision on $m_b$,  $|V_{ub}|$ could presently be extracted from the
total decay rate with a theoretical error smaller than 2.5\%.

Unfortunately, most experimental analyses apply severe cuts to avoid the
charm background. The cuts limit the invariant mass of the hadronic
final state, $X$,  and destroy the convergence of the local OPE
introducing a sensitivity to the effects of Fermi motion of the heavy
quark inside the $B$ meson.  These effects are not suppressed by powers of
$1/m_b$ in the restricted kinematic regions. The Fermi motion is
inherently non-perturbative; within the OPE it can be described by a
nonlocal distribution function, called the shape function (SF)
\cite{Neubert:1993ch,Bigi:1993ex}, whose
lowest integer moments are given by the same expectation values of local
operators appearing in Eq.(\ref{widthb2u}). In terms of light-cone
momenta $P^\pm = E_X \mp p_X$,  a typical event in the SF region has
$P^+\ll P^-= O(m_b)$, with $P^+$ not far above the QCD scale.  The
emergence of the SF is also evident in perturbation theory: soft-gluon
resummation gives rise to a $b$ {\it quark} SF when supplemented by an
internal resummation  of running coupling corrections,
see e.g.~\cite{Andersen:2005bj,Andersen:2005mj,Andersen:2006hr,Gardi:2007jx}.
This SF has the required support properties,
namely it extends the kinematic ranges by energies of
$O(\Lambda_{QCD})$, and it is stable under higher order corrections. The
{\it quark} SF can therefore be predicted under a few assumptions, as we
will see below.  The inclusion of power corrections related to the
difference between $b$ quark and $B$ meson SFs and the proper matching
to the OPE are important issues in this context.  An alternative
possibility is to give up predicting the  SF. Since the OPE  fixes 
the first few moments of the SF, one can parametrize it in terms of the
known non-perturbative quantities employing an ansatz for its functional
form. The uncertainty due to the functional form can be evaluated
by varying it, a process that has been recently systematized
\cite{Ligeti:2008ac}.  Finally, one can exploit the universality of the
SF, up to $1/m_b$ corrections, and extract it from the photon spectrum
of $B\to X_s \gamma$, which is governed by the same dynamics as
inclusive semileptonic decays
\cite{Neubert:1993ch,Bigi:1993ex,Lange:2005qn,Lange:2005xz}.
Notice that rates in restricted phase space regions 
always show increased sensitivity to \mb, 
up to twice that in Eq.(\ref{widthb2u}).

Subleading contributions in $1/m_b$ are an important issue, and
acquire a different character depending on the framework in which they are
discussed.  For instance,  if one expands in powers of the heavy quark
mass the non-local OPE that gives rise to the SF, the first subleading
order sees the emergence 
of many largely unconstrained {\it subleading SFs} \cite{Leibovich:2002ys,%
Bauer:2002yu,Bosch:2004cb} that
break the universality noted above.  An alternative procedure has been
developed in \cite{Gambino:2007rp}, where the only expansion in $1/m_b$
is at the level of local OPE. A single {\it finite $m_b$} distribution
function has been introduced for each of the three relevant structure
functions at fixed $q^2$.  All power-suppressed terms are taken into
account in the OPE relations for the integer moments of the SFs, which
are computed, like Eq.(\ref{widthb2u}) through $O(\Lambda^3)$. Finally,
in the context of resummed perturbation theory, power corrections appear
in moment space and can be parametrized.

Perturbative corrections also modify the physical spectra: the complete
$O(\as)$ and $O(\beta_0\as^2)$ corrections to the triple differential
spectrum \cite{{DeFazio:1999sv,Gambino:2006wk}} are available,
while the $O(\as^2)$ have been recently computed in the SF region
only \cite{Bonciani:2008wf,Asatrian:2008uk,Beneke:2008ei,Bell:2008ws}.
There is a clear interplay between perturbative corrections and the
proper definition of the SF beyond lowest order, a problem that has been
addressed in different ways, see below.

The experimental cuts can aggravate the uncertainty due to WA. Indeed,
WA effects are expected to manifest themselves only at maximal $q^2$ and
lead to an uncertainty that depends strongly on the cuts employed.  In
the experimental analyses the high-$q^2$ region could therefore either
be excluded or used to put additional
constraints on the WA matrix element
\cite{Gambino:2005tp,Rosner:2006zz,Aubert:2007tw}.  Moreover, the
high-$q^2$ spectrum is not properly described by the OPE (see
\cite{Gambino:2007rp} and references therein) and should be modeled,
while  its contribution to the integrated rate can be parametrized by
the WA matrix element $B_{\rm WA}$.  In particular, at $\muwa=1\rm GeV$,
the positivity of the $q^2$ spectrum implies a positive value of $B_{\rm
WA}(1 \rm GeV)$, leading to a decrease in the extracted $|V_{ub}|$
\cite{Gambino:2007rp}.

All the problems outlined above have been extensively discussed in the
literature. We will now consider four practical implementations, briefly
discussing  their basic features.

\subsubsubsection{DGE}
The approach of Refs.~\cite{Andersen:2005bj,%
Andersen:2005mj,Andersen:2006hr,Gardi:2007jx} uses resummed
perturbation theory in moment--space to compute the on-shell decay
spectrum in the entire phase space; non-perturbative effects are taken
into account as power corrections in moment space.  Resummation is
applied to both the `jet' and the `soft' (quark distribution or SF)
subprocesses at NNLL\footnote{The `jet' logarithms are similar to those
resummed in the approach of Ref.~\cite{Lange:2005yw}; there however
`soft' logarithms are not resummed.}, dealing directly with the double
hierarchy of scales ($\Lambda\ll \sqrt{\Lambda \mb}\ll \mb$)
characterizing the decay process.  Consequently, the shape of the
spectrum in the kinematic region where the final state is jet-like is
largely determined by a calculation, and less by parametrization.  The
resummation method employed, Dressed Gluon Exponentiation (DGE), is a
general resummation formalism for inclusive distributions near a
threshold~\cite{Gardi:2006jc}.
It goes beyond the standard Sudakov resummation 
by incorporating an internal resummation of running--coupling
corrections (renormalons) in the exponent and has proved effective
in a range of applications~\cite{Gardi:2006jc}. 
DGE adopts the Principal Value procedure to regularize the Sudakov
exponent and thus \emph{define} the non-perturbative parameters. In
particular, this definition applies to
the would-be $1/m_b$ ambiguity of the `soft' Sudakov factor, which
cancels exactly~\cite{Gardi:2004ia} against the pole--mass renormalon
when considering the spectrum in physical hadronic variables. The same
regularization used in the Sudakov exponent must be applied in the
computation of the regularized $b$ pole mass from  the input
$m_b^{\overline {\rm MS}}$.
This makes DGE calculation consistent with the local OPE up to
${\cal O}(\Lambda^2/m_b^2)$. 

\subsubsubsection{ADFR}
In this model based on perturbative resummation \cite{Aglietti:2007ik}
the integral in the Sudakov exponent is regulated by the use of the
{\it analytic coupling} \cite{Shirkov:1997wi}, which
is finite in the infrared and is meant to account for all
non-perturbative effects.  The resummation is performed at NNLL, while
the non-logarithmic part of the spectra is computed at $O(\as)$ in the
on-shell scheme, setting the pole $b$ mass numerically equal to $M_B$.
In contrast with DGE, this procedure does not enforce the cancellation
of the renormalon ambiguity associated with \mb, and thus it violates
the local OPE at $O(\Lambda/\mb)$, resulting  in an uncontrolled ${
O}(\Lambda)$ shift of the $P^+$ spectrum.  The model reproduces  $b$
fragmentation data and the photon spectrum in $B\to X_s\gamma$, but does
not account for $O(\Lambda/\mb)$ power corrections relating different
processes. The normalization (total rate) is fixed by the total width of
$B\to X_c \ell \nu$, avoiding the $\mb^5$ dependence, but introducing a
dependence on \mc.

\subsubsubsection{BLNP}
The SF approach of Ref.~\cite{Lange:2005yw} employs
a modified expansion in inverse powers of \mb, where at each order the
dynamical effects associated with soft gluons, $k^+ \sim P^+\sim
\Lambda$ are summed into non-perturbative shape functions.  As mentioned
above, at leading power there is one such function; beyond this order
there are several different functions. To extend the calculation beyond
this particular region, the expansion is designed to match the local OPE
when integrated over a significant part of the phase space. In this way
two systematic expansions in inverse powers of
the mass are used together. In this {\it multiscale} OPE, developed following 
SCET methodology (cf.\ Sec.~\ref{sec:hqet-scet}), the differential width 
is given by  
\begin{equation}
\frac{d\Gamma}{dP^+\,dP^-\,dE_l}= H J \otimes S + \frac1{\mb} H'_i J_i \otimes S_i' + \dots
\end{equation}
where soft (S), jet (J), and hard (H) functions depend on momenta $\sim
\Lambda, \sqrt{\Lambda\mb}, \mb$, respectively. The jet and hard
functions are computed perturbatively at $O(\as)$ in the {\it shape
function scheme}, resumming Sudakov logs at NNLL, while the soft
functions are parametrized at an intermediate scale, $\mu\sim 1.5\gev$,
using the local OPE constraints on their first moments computed at
$O(1/m_b^2)$ and a set of functional forms. Although the subleading SFs
are largely unconstrained, BLNP find that the experimentally-relevant
partial branching fractions remain under good control: the largest
uncertainty in the determination of \Vub is due to \mb.

\subsubsubsection{GGOU}
The kinetic scheme used in Sec.~\ref{sec:inclusiveVcb} to define the 
OPE parameters, is employed in \cite{Gambino:2007rp}  to  introduce the
distribution functions through a factorization formula for the structure
functions $W_i$,
\begin{equation}\label{eq:conv2}
 W_i(q_0,q^2) \propto   
 \int dk_+ \ F_i(k_+,q^2,\mu)  W_i^{\rm pert}
\left[ q_0 - \frac{k_+}{2} \left( 1 - \frac{q^2}{m_b M_B} \right), q^2,\mu \right],
\end{equation}
where the distribution functions $F_i(k_+,q^2,\mu)$  depend on the
light-cone momentum $k_+$, on $q^2$ (through subleading effects) and on
the infrared cutoff $\mu$ \cite{Gambino:2007rp}.  As the latter inhibits soft
gluon emission, the spectrum has only collinear singularities whose
resummation  is numerically irrelevant.  The perturbative corrections in
$W_i^{\rm pert}$ include  $O(\as^2\beta_0)$ contributions, which alone
decrease the value of \Vub by about 5\%.  The functions
$F_i(k_+,q^2,\mu)$ are constrained by the local OPE expressions for
their first moments at fixed $q^2$ and $\mu=1\gev$, computed including
all $1/m_b^3$ corrections. A vast range of functional forms is explored,
leading to a 1-2\% uncertainty on \Vub \cite{Gambino:2007rp}.

Although conceptually quite different,  the above approaches generally
lead to roughly consistent results when the same inputs are used and the
theoretical errors are taken into account.
\begin{figure}
    \centering
    \includegraphics[width=0.47\textwidth]{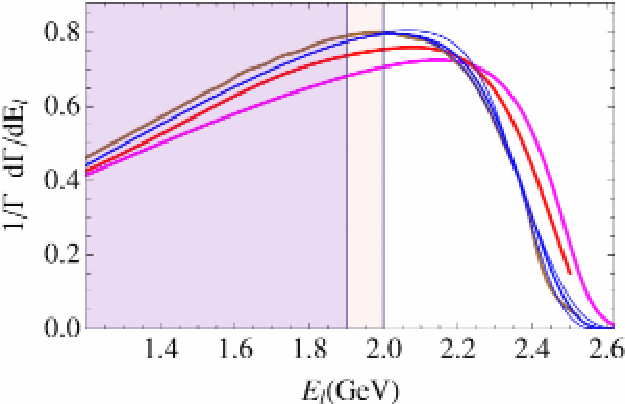}
    \includegraphics[width=0.47\textwidth]{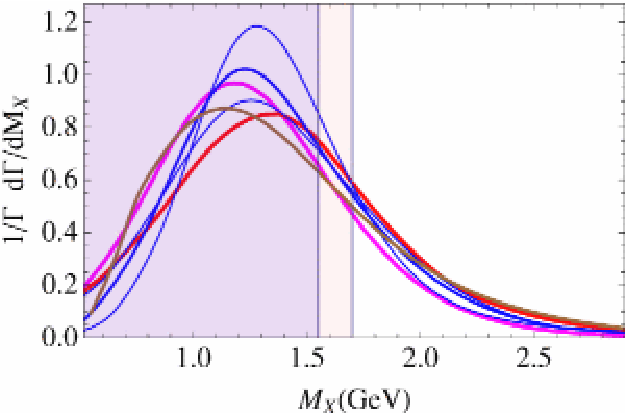}
    \caption{Comparison of different theoretical treatments of
    inclusive $b\to u$ transitions: (a) $E_l$ spectrum; (b) $M_X$ spectrum.
    Red, magenta, brown and blue lines refer, respectively, to DGE,
    ADFR, BLNP, GGOU with a sample of three different functional forms. The actual experimental 
cuts at $E_{l}=1.9,2.0$\gev and $M_X=1.55,1.7$\gev are also indicated.
}
    \label{fig:slep:incVub}
\end{figure}
In Fig.~\ref{fig:slep:incVub}(a)
we show the normalized electron energy spectrum computed in the
latest implementations of the four approaches.
Except in ADFR, the spectrum depends sensitively on \mb.
An accurate measurement of the electron spectrum can discriminate
between at least some of the methods. The same applies to the $M_X$ spectrum,
which is shown in Fig.~\ref{fig:slep:incVub}(b)  for
$E_l>1\gev$.

\subsubsection{Review of $m_b$ determinations}
\label{sec:mbdetermination}


As we have just seen, 
theoretical predictions of inclusive B~decays can depend strongly on
$m_b$. Thus, uncertainties in the knowledge of $m_b$ can affect the
determination of other parameters. To achieve the high precision in the
theoretical predictions required by experimental data it 
is important to avoid the ${\rm O}(\Lambda_{\rm QCD})$ renormalon ambiguities
related to the pole mass parameter and to consider the quark masses as
renormalization scheme dependent couplings of the Standard Model Lagrangian
that have to be determined from processes that depend on them. Thus
having precise control over the scheme-dependence of the bottom quark mass
parameters is as important as reducing their numerical uncertainty.
 
Predictions for $B$ meson decays also suffer from renormalon ambiguities
of order $\Lambda_{\rm QCD}^2/m_b$ or smaller. These ambiguities cannot
in general be removed solely by a particular choice of a bottom mass
scheme. Additional subtractions in connection with fixing specific
schemes for higher order non-perturbative matrix element in the
framework of the OPE are required to remove these ambiguities. Some
short-distance bottom mass schemes have been proposed together with
additional subtractions concerning the kinetic operator $\lambda_1$ or
$\mu_\pi^2$. 
In the following we briefly review  the prevalent perturbative bottom mass 
definitions which were employed in recent analyses of inclusive $B$ decays.
A more detailed review on quark mass definitions including analytic formulae
has been given in the CKM~2003 Report~\cite{Battaglia:2003in}.


\subsubsubsection{$\overline{\rm MS}$ mass}:
The most common short-distance mass parameter it the $\overline{\rm MS}$ mass
$\overline m_b(\mu)$, which is defined by regularizing QCD with
dimensional regularization and subtracting the UV divergences in the 
$\overline{\rm MS}$ scheme.
As a consequence the $\overline{\rm MS}$ mass depends on the
renormalization scale $\mu$. 
Since the UV subtractions do not contain any infrared sensitive terms, 
the $\overline{\rm MS}$ mass is only sensitive to scales of order or
larger than $m_b$. The $\overline{\rm MS}$ mass is therefore disfavored for
direct use in the theoretical description of inclusive $B$ decays. 
However, it is still useful as a reference mass.
The relation between the pole mass and the $\overline{\rm MS}$ mass is known to 
${\rm O}(\alpha_s^3)$~\cite{Gray:1990yh,Melnikov:2000qh,Chetyrkin:1999qi}, 
see Sec.~2.1 of the 2003
report~\cite{Battaglia:2003in} for analytic formulae.

\subsubsubsection{Threshold masses}
The shortcomings of the $\overline{\rm MS}$ masses in describing
inclusive $B$ decays can be resolved by so-called threshold 
masses~\cite{Hoang:2000yr}. The prevalent threshold mass definitions are the kinetic,
the shape function and the 1S mass schemes. They are free of an ambiguity of
order $\Lambda_{\rm QCD}$ through in general scale-dependent subtractions. 

The \emph{kinetic mass} is defined as~\cite{Bigi:1996si,Bigi:1994ga} 
\begin{eqnarray}
m_{b,{\rm kin}}^{}(\mu_{\rm kin}^{}) 
& = &
m_{b,{\rm pole}} - \left[\bar\Lambda(\mu_{\rm kin}^{})\right]_{\rm pert} 
- \left[\frac{\mu^2_\pi(\mu_{\rm kin}^{})}
{2 m_{b,{\rm pole}}}\right]_{\rm pert}
\,,
\label{kindef}
\end{eqnarray}
where $\left[\bar\Lambda(\mu^{}_{\rm kin})\right]_{\rm pert}$ and
$\left[\mu_\pi^2(\mu^{}_{\rm kin})\right]_{\rm pert}$ are perturbative
evaluations of HQET matrix elements that describe the difference
between the pole and the $B$ meson mass. The term $\mu_{\rm kin}$ is the
subtraction scale. To avoid the appearance of large logarithmic terms
it should be chosen somewhat close to the typical momentum
fluctuations within the $B$ meson.
The relation between the kinetic mass and the pole mass
is known to ${\rm O}(\alpha_s^2)$ and ${\rm O}(\alpha_s^3
\beta_0^2)$~\cite{Czarnecki:1997sz,Melnikov:1998ug}, see the 2002
report~\cite{Battaglia:2003in} for analytic formulae.

The \emph{shape function mass}~\cite{Bosch:2004th,Neubert:2004sp} is defined
from the condition that the OPE for the first moment of the leading order shape
function for the $B\to X_u\ell\nu$ and $B\to X_s\gamma$ decays in the
endpoint regions vanishes identically. The relation between
the shape function mass and the pole mass is known at ${\cal
  O}(\alpha_s^2)$ and reads
\begin{eqnarray}
   m_b^{\rm SF}(\mu_{\rm SF},\mu) 
   &=& m_{b,\rm pole} - \mu_{\rm SF}\,\frac{C_F\alpha_s(\mu)}{\pi}
    \left[ 1 - 2\ln\frac{\mu_{\rm SF}}{\mu} + \frac{\alpha_s(\mu)}{\pi}\,
    k_1(\mu_{\rm SF},\mu) \right]  \nonumber\\
 &&\mbox{}
- \frac{\mu_\pi^2(\mu_{\rm SF},\mu)}{3\mu_{\rm SF}}\,
    \frac{C_F\alpha_s(\mu)}{\pi} \left[ 2\ln\frac{\mu_{\rm SF}}{\mu}
    + \frac{\alpha_s(\mu)}{\pi}\,k_2(\mu_{\rm SF},\mu) \right] ,
\end{eqnarray}
where
\begin{eqnarray}
   k_1(\mu_{\rm SF},\mu)
   &=& \frac{47}{36}\,\beta_0
    + \left( \frac{10}{9} - \frac{\pi^2}{12} - \frac94\,\zeta_3
    + \frac{\kappa}{8} \right) C_A
    + \left( - 8 + \frac{\pi^2}{3} + 4\zeta_3 \right) C_F \nonumber\\
\lefteqn{
+ \left[ - \frac43\,\beta_0
    + \left( - \frac23 + \frac{\pi^2}{6} \right) C_A
    + \left( 8 - \frac{2\pi^2}{3} \right) C_F \right] \ln\frac{\mu_{\rm
      SF}}{\mu}
}    \nonumber\\
\lefteqn{
    + \left( \frac12\,\beta_0 + 2C_F \right) \ln^2\frac{\mu_{\rm SF}}{\mu} \,,
}   \\
   k_2(\mu_{\rm SF},\mu)
   &=& - k_1(\mu_{\rm SF},\mu) 
    + \frac76\,\beta_0 + \left( \frac13 - \frac{\pi^2}{12} \right) C_A
    + \left( - 5 + \frac{\pi^2}{3} \right) C_F
 \nonumber \\
\lefteqn{
    + \left( - \frac12\,\beta_0 - C_F \right) \ln\frac{\mu_{\rm SF}}{\mu} \,.
}
\end{eqnarray}
The relation depends on the momentum cutoff $\mu_{\rm SF}$ which
enters the definition of the first moment and on the
(non-perturbative and infrared subtracted) kinetic energy matrix
element $\mu^2_\pi$ defined from the ratio of the second and zeroth
moment of the shape function. Since the shape function is
renormalization scale dependent, the shape function mass depends on
also on the renormalization scale $\mu$. In practical applications
the SF mass has been considered for $\mu=\mu_{\rm SF}$, 
$m_b^{\rm SF}(\mu_{\rm SF})\equiv  m_b^{\rm SF}(\mu_{\rm SF},\mu_{\rm SF})$.

The \emph{1S mass}~\cite{Hoang:1998ng,Hoang:1998hm,Hoang:1999zc} is 
defined as one half of the perturbative series for the mass of the $n=1$,
${}^3S_1$ bottomonium bound state in the limit 
$m_b\gg m_b v\gg m_b v^2\gg \Lambda_{\rm QCD}$. 
In contrast to the kinetic and shape-function masses,
the subtraction scale involved in the 1S mass
is tied dynamically to the inverse Bohr radius $\sim m_b\alpha_s$ of
the bottomonium ground state and therefore does not appear as an
explicit parameter.
  The 1S mass scheme is known completely to
${\rm O}(\alpha_s^3)$,  see the 2003
report~\cite{Battaglia:2003in} for analytic formulae.

In Tab.~\ref{tabmasses} the numerical values of the bottom quark kinetic,
shape function and 1S masses are provided for different values for the strong
coupling taking the $\overline{\rm MS}$ mass $\overline m_b(\overline m_b)$ as
a the reference input. Each entry corresponds to the mass using the
respective 1-loop/2-loop/3-loop relations as far as they are available and 
employing a common renormalization scale for the strong coupling when the pole
mass is eliminated. As the 
renormalization scale we employed $\mu=\overline m_b(\overline m_b)$ to
minimize the impact of logarithmic terms involving the cutoff scales 
$\mu_{\rm kin,SF} $ and the scale  $\overline m_b(\overline
m_b)$~\cite{Hoang:2008yj}.  Numerical 
approximations for the conversion formulae at the respective 
highest available order accounting in particular
for the dependence on $\alpha_s^{(n_f=5)}(M_Z)$ and the renormalization scale 
$\mu$ read: 
\begin{eqnarray}
m_b^{\rm 1S} & = &
1.032\, m_b^{\rm kin}(1~\mbox{GeV}) 
  \, +\, 1.9\,\Delta\alpha_s
  \, - \,0.003 \,\Delta\mu
\,,
\\[4mm]
m_b^{\rm SF}(1.5~\mbox{GeV}) & = &
1.005\,m_b^{\rm kin}(1~\mbox{GeV})
  \, +\, 0.9\, \Delta\alpha_s
  \, - \,0.006 \, \Delta\mu
  \, - \,0.003 \, \Delta\mu_\pi^2
\,,\quad
\\[4mm]
m_b^{\rm SF}(1.5~\mbox{GeV}) & = &
0.976\,m_b^{\rm 1S}
  \, -\, 0.9\, \Delta\alpha_s
  \, + \,0.001 \, \Delta\mu
  \, - \,0.003 \, \Delta\mu_\pi^2
\,,\quad
\\[4mm]
\overline m_b(\overline m_b) & = &
0.917\, m_b^{\rm kin}(1~\mbox{GeV})
  \, -\, 8.2\, \Delta\alpha_s
  \, + \,0.005 \,\Delta\mu
\,,
\\[4mm]
\overline m_b(\overline m_b) & = &
0.888\, m_b^{\rm 1S} 
  \, -\, 9.9\, \Delta\alpha_s
  \, + \,0.006 \, \Delta\mu
\,,
\\[4mm]
\overline m_b(\overline m_b) & = &
0.916\,m_b^{\rm SF}(1.5~\mbox{GeV})
  \, -\, 8.0\, \Delta\alpha_s
  \, + \,0.017 \, \Delta\mu
  \, + \,0.003 \, \Delta\mu_\pi^2
\,,
\end{eqnarray}
where 
$\Delta\alpha_s = [\alpha_s^{(5)}(M_Z)-0.118]~\textrm{GeV}$,
$\Delta\mu  =  (\mu - 4.2~\mbox{GeV})$,
$\Delta\mu_\pi^2 =
[\mu_\pi^2(1.5~\textrm{GeV}) - 0.15~\textrm{GeV}^2]~\textrm{GeV}^{-1}$.
The formulae agree with the respective exact relations to better 
than $10$~MeV (for $3.7~\mbox{GeV}<\mu<4.7$~GeV). The theoretical
uncertainties from missing higher order terms are reflected in the
renormalization scale dependence of the conversion formulae.
\begin{table}[t] 
\centering
    \caption{Numerical values of the bottom quark kinetic, 1S and shape 
function masses in 
units of GeV for a given $\overline{\rm MS}$ value $\overline m_b(\overline
m_b)$ using  $\mu=\overline 
m_b(\overline m_b)$, $n_l=4$ and three values of 
$\alpha_s^{(5)}(m_Z)$. Flavor matching was carried out at 
$\mu=\overline m_b(\overline m_b)$. For the shape function mass
$\mu_\pi^2(1.5~\mbox{GeV}) = 0.15~\mbox{GeV}^2$ was adopted. 
Numbers with an asterisk are given in the large-$\beta_0$
approximation.}
\label{tabmasses}
\begin{tabular}{|c|c|c|c|} \hline
$\overline m_b(\overline m_b)$ & 
$m_{b,\rm kin}(1\,\mbox{GeV})$ &
$m_{b,{\rm 1S}}$ & 
$m_{b,\rm SF}(1.5\,\mbox{GeV})$
\\[0.05cm]
\hline
\multicolumn{4}{|c|}{{\small $\alpha_s^{(5)}(m_Z)=0.116$ 
}}\\
\hline
{\small 4.10} & 
{\small 4.36/4.42/4.45$^*$} & 
{\small 4.44/4.56/4.60}& 
{\small 4.34/4.44/-}\\
\hline
{\small 4.15} & 
{\small 4.41/4.48/4.50$^*$} & 
{\small 4.49/4.61/4.65}& 
{\small 4.39/4.50/-}\\
\hline
{\small 4.20} & 
{\small 4.46/4.53/4.56$^*$} & 
{\small 4.54/4.66/4.71}& 
{\small 4.45/4.55/-}\\
\hline
{\small 4.25} & 
{\small 4.52/4.59/4.61$^*$} & 
{\small 4.60/4.72/4.76}& 
{\small 4.50/4.61/-}\\
\hline
{\small 4.30} & 
{\small 4.57/4.64/4.67$^*$} & 
{\small 4.65/4.77/4.81}& 
{\small 4.56/4.66/-}\\
\hline
\multicolumn{4}{|c|}{{\small $\alpha_s^{(5)}(m_Z)=0.118$ 
}}\\
\hline
{\small 4.10} & 
{\small 4.37/4.44/4.46$^*$} & 
{\small 4.45/4.57/4.62}& 
{\small 4.35/4.46/-}\\
\hline
{\small 4.15} & 
{\small 4.42/4.49/4.52$^*$} & 
{\small 4.50/4.63/4.67}& 
{\small 4.40/4.51/-}\\
\hline
{\small 4.20} & 
{\small 4.47/4.55/4.57$^*$} & 
{\small 4.55/4.68/4.73}& 
{\small 4.46/4.57/-}\\
\hline
{\small 4.25} & 
{\small 4.52/4.60/4.63$^*$} & 
{\small 4.61/4.73/4.78}& 
{\small 4.51/4.62/-}\\
\hline
{\small 4.30} & 
{\small 4.58/4.66/4.69$^*$} & 
{\small 4.66/4.79/4.84}& 
{\small 4.56/4.68/-}\\
\hline
\multicolumn{4}{|c|}{{\small $\alpha_s^{(5)}(m_Z)=0.120$ 
}}\\
\hline
{\small 4.10} & 
{\small 4.37/4.45/4.48$^*$} & 
{\small 4.46/4.59/4.64}& 
{\small 4.36/4.48/-}\\
\hline
{\small 4.15} & 
{\small 4.43/4.51/4.54$^*$} & 
{\small 4.51/4.64/4.70}& 
{\small 4.41/4.53/-}\\
\hline
{\small 4.20} & 
{\small 4.48/4.56/4.59$^*$} & 
{\small 4.56/4.70/4.75}& 
{\small 4.47/4.59/-}\\
\hline
{\small 4.25} & 
{\small 4.54/4.62/4.65$^*$} & 
{\small 4.62/4.75/4.80}& 
{\small 4.52/4.64/-}\\
\hline
{\small 4.30} & 
{\small 4.59/4.67/4.71$^*$} & 
{\small 4.67/4.81/4.86}& 
{\small 4.57/4.70/-}\\
\hline
\end{tabular}
\end{table}

\subsubsubsection{Bottom quark mass determinations}

There are two major methods to determine the bottom mass with high
precision: spectral sum rules using data for the bottom production
rate in $e^+e^-$ collisions, and fits to moments obtained from
distributions of semileptonic $B\to X_c\ell\nu$ and radiative 
$B\to X_s\gamma$ decays. Both rely on the validity of the
operator product expansion and the input of higher order perturbative
corrections. The results obtained from both
methods are compatible.
Lattice determinations still have larger uncertainties and 
suffer from systematic effects, which need to be better understood to
be competitive to the previous two methods. A summary of recent bottom
mass determinations is given in Tab.~\ref{tabcollection}. 

\subsubsubsection{Spectral $e^+e^-$ sum rules}

The spectral sum rules start from the
correlator $\Pi(q^2)$ of two electromagnetic bottom quark currents and
are based on the fact that
derivatives of $\Pi$ at $q^2=0$ are related to moments of
the total cross section $\sigma(e^+e^-\to b\bar b)$,
\begin{equation}
{\cal M}_n 
\, = \,
\frac{12\,\pi^2\,Q_b^2}{n!}\,
\bigg(\frac{d}{d q^2}\bigg)^n\,\Pi(q^2)\bigg|_{q^2=0}
\, = \,
\int \frac{d s}{s^{n+1}}\,R(s)
\,,
\label{Mdef}
\end{equation}
where $R=\sigma(e^+e^-\to b\bar b)/\sigma(e^+e^-\to\mu^+\mu^-)$.  From
Eq.\,(\ref{Mdef}) it is possible to determine the bottom quark 
mass using an operator product expansion~\cite{Novikov:1976tn,Reinders:1984sr}. 
One has to restrict the moments to
$n\lesssim 10$ such that the momentum range contributing to the
moment is sufficient larger than $\Lambda_{\rm QCD}$ and the
perturbative contributions dominate.
Here the most important non-perturbative matrix element is the gluon
condensate, but its contribution is very small.

\begin{table}[t!]  
    \centering
\caption{Collection in historical order in units of GeV of recent bottom
quark mass determinations from spectral sum rules and 
the $\Upsilon(\mbox{1S})$ mass.
Only results where $\alpha_s$ was taken as an input are shown. The
uncertainties quoted in the respective references have been added
quadratically. All numbers have been taken from the respective
publications.}
\label{tabcollection}
\begin{small}
\begin{tabular}{llrl} \hline
 author & $\overline m_b(\overline m_b)$ & other mass\mbox{\hspace{5mm}}  & 
  \mbox{\hspace{5mm}}comments, Ref.
\\ \hline\hline
\multicolumn{4}{|c|}{ nonrelativistic spectral sum rules }\\
\hline\hline
  Melnikov \hfill  98
    & $4.20\pm 0.10$
    & $M_{\rm kin}^{1 \rm GeV}=4.56\pm 0.06$
    & NNLO, $m_c=0$~\protect\cite{Melnikov:1998ug} 
\\ \hline
  Hoang  \hfill 99
    & $4.20\pm 0.06$ 
    & $M^{}_{\rm 1S}=4.71\pm 0.03$
    & NNLO, $m_c=0$~\protect\cite{Hoang:1999ye} 
\\ \hline
  Beneke  \hfill 99
    & $4.26\pm 0.09$ 
    & $M_{\rm PS}^{2 \rm GeV}=4.60\pm 0.11$
    & NNLO, $m_c=0$~\protect\cite{Beneke:1999fe}
\\ \hline
  Hoang \hfill  00
    & $4.17\pm 0.05$ 
    & $M^{}_{\rm 1S}=4.69\pm 0.03$
    & NNLO, $m_c\neq 0$~\protect\cite{Hoang:2000fm}
\\ \hline
  Eidem\"uller \hfill  02
    & $4.24\pm 0.10$ 
    & $M_{\rm PS}^{2 \rm GeV}=4.56\pm 0.11$
    & NNLO + ${\rm O}(\alpha_s^2)$, $m_c= 0$~\protect\cite{Eidemuller:2002wk}
\\ \hline
  Pineda \hfill  06
    & $4.19\pm 0.06$ 
    & $M_{\rm PS}^{2 \rm GeV}=4.52\pm 0.06$
    & NNLL partial, $m_c= 0$~\protect\cite{Pineda:2006gx}
\\ \hline\hline
\multicolumn{4}{|c|}{ relativistic spectral sum rules }\\
\hline\hline
  K\"uhn  \hfill 01  
    & $4.19\pm 0.05$ 
    & 
    & ${\rm O}(\alpha_s^2)$~\protect\cite{Kuhn:2001dm} 
\\ \hline
  Bordes  \hfill 02  
    & $4.19\pm 0.05$ 
    & 
    & ${\rm O}(\alpha_s^2)$, \mbox{finite energ. s.r.}~\protect\cite{Bordes:2002ng} 
\\ \hline
  Corcella  \hfill 02  
    & $4.20\pm 0.09$ 
    & 
    & ${\rm O}(\alpha_s^2)$, continuum err.incl.~\protect\cite{Corcella:2002uu} 
\\ \hline
  Hoang  \hfill 04
    & $4.22\pm 0.11$ 
    & 
    & ${\rm O}(\alpha_s^2)$, contour improved~\protect\cite{Hoang:2004xm} 
\\ \hline
  Boughezal  \hfill 06
    & $4.21\pm 0.06$ 
    & 
    & ${\rm O}(\alpha_s^3)$~\protect\cite{Boughezal:2006px} 
\\ \hline
  K\"uhn  \hfill 07
    & $4.16\pm 0.03$ 
    & 
    & ${\rm O}(\alpha_s^3)$~\protect\cite{Kuhn:2007vp} 
\\ \hline\hline
\multicolumn{4}{|c|}{ moments from $B\to X_c\ell\nu$ and $B\to
  X_s\gamma$ distributions }\\
\hline\hline
  HFAG (ICHEP 08) \hfill 08
    & $4.28\pm 0.07$ 
    & $M_{\rm kin}^{1 \rm GeV}=4.66\pm 0.05$
    & $B\to X_c\ell\nu$, ${\rm O}(\alpha_s^2\beta_0)$~\protect\cite{Barberio:2008fa} 
\\
    & $4.23\pm 0.05$ 
    & $M_{\rm kin}^{1 \rm GeV}=4.60\pm0.03$
    & $B\to X_c\ell\nu$, $B\to X_s\gamma$, ${\rm O}(\alpha_s^2\beta_0)$~\protect\cite{Barberio:2008fa} 
\\
    & $4.17\pm 0.04$ 
    & $M_{\rm 1S}=4.70\pm 0.03$
    & $B\to X_c\ell\nu$ \& $B\to X_s\gamma$, ${\rm O}(\alpha_s^2\beta_0)$~\protect\cite{Barberio:2008fa} 
\\
    & $4.22\pm 0.07$ 
    & $M_{\rm 1S}=4.75\pm 0.06$
    & $B\to X_c\ell\nu$, ${\rm O}(\alpha_s^2\beta_0)$~\protect\cite{Barberio:2008fa} 
\\ \hline
\end{tabular}
\end{small}
\end{table}

{\it Nonrelativistic $e^+e^-$ sum rules:}
For the large $n$, $4\lesssim n\lesssim 10$, the moments are dominated by the
bottomonium bound states region and the experimentally
unknown parts of the $b\bar b$ continuum cross section are suppressed.
Depending on the moment the overall experimental uncertainties in the b quark
mass are between $15$ and $20$~MeV. Sum rule analyses using threshold masses and based
on NNLO fixed order computations in the framework of
NRQCD~\cite{Melnikov:1998ug,Hoang:1999ye,Beneke:1999fe,Hoang:2000fm} yield consistent
results but suffer from relatively large NNLO corrections to the
normalization of the moments ${\cal M}_n$. Uncertainties in the bottom
mass at the level of below $50$ to $100$~MeV were achieved by making
assumptions on the behavior of higher order corrections. The use of
renormalization group improved NRQCD computations in Ref.~\cite{Pineda:2006gx}
yields an uncertainty of $60$~MeV without making such assumptions. However,
the analysis of Ref.~\cite{Pineda:2006gx} neglects known large NNLL order
contributions to the anomalous dimension of the quark pair production
currents~\cite{Hoang:2003ns}.

{\it Relativistic sum rules:}
For small $n$, $1\le n\lesssim 4$, the experimentally unmeasured parts of 
the $b\bar b$ continuum cross section above the $\Upsilon$ resonance
region constitute a substantial contribution to the spectral moments
and uncertainties below the $100$~MeV level are only
possible using theory to predict the continuum
contributions~\cite{Corcella:2002uu}.
For the theoretical determination of the moments usual fixed order
perturbation theory can be employed. The most recent bottom quark mass
determinations~\cite{Boughezal:2006px,Kuhn:2007vp} use perturbation
theory at ${\rm O}(\alpha_s^3)$ and  obtain $\overline m_b(\overline
m_b)$ with an uncertainty between $25$ and $58$~MeV.

\subsubsubsection{Inclusive $B$ decay moments}

As already discussed in Sec.~\ref{sec:inclusiveVcb} the analysis of 
moments of lepton energy and hadron invariant mass
moments obtained from spectra in the semileptonic decay $B\to X_c\ell\nu$
and of radiative photon energy moments from $B\to X_s\gamma$ allows to
determine the bottom quark threshold masses. 
Currently the theoretical input for the
moment computations includes ${\rm O}(\alpha_s)$  and ${\cal
  O}(\alpha_s^2\beta_0)$ corrections for the partonic contribution
and tree-level Wilson coefficients for the power
corrections~\cite{Gambino:2004qm,Bauer:2004ve,Aquila:2005hq}; 
it would be desirable to include  the known full 
${\rm O}(\alpha_s^2)$ corrections into the analysis.
For what concerns the determination of \mb, the results based on
combined $B\to X_c\ell\nu$
and  $B\to X_s\gamma$ data are in agreement with the $e^+e^-$ sum rule
determinations, while using only  $B\to X_c\ell\nu$ data leads to slightly larger
\mb with larger error, which are, however, still compatible with the other
determinations. In fact, the semileptonic moments are mostly sensitive to a
combination of \mb and \mc, as apparent from Fig.\ref{fig:slep:mbmc}, where
various determinations of \mc and \mb are compared.    
\begin{figure}[bp]
    \centering
    \includegraphics[width=0.7\textwidth]{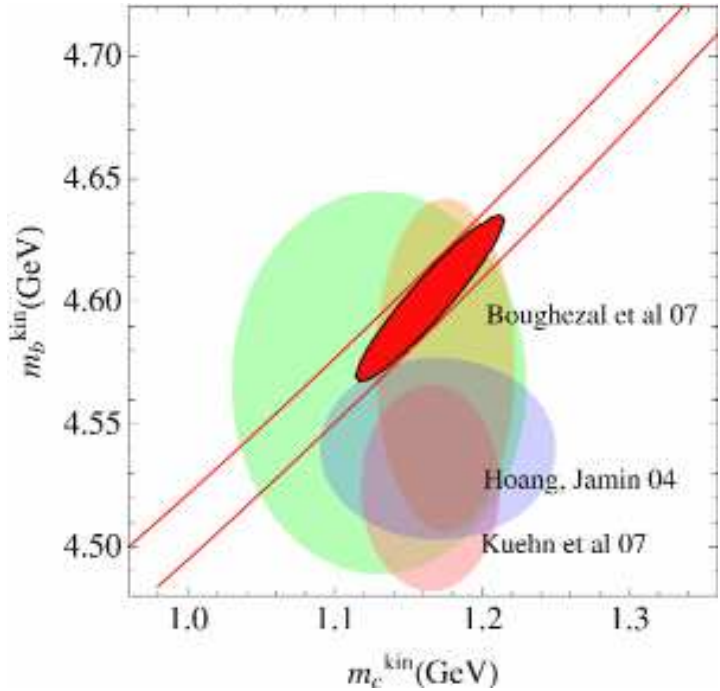}
    \caption{Comparison of different determinations of \mc and \mb in
        the kinetic scheme. The red ellipse refers to the semileptonic
        fit discussed in Sec.~\ref{sec:inclusiveVcb}, the large green 
        ellipse to the 2007 PDG 
        values, and the others to various  $e^+e^-$ sum rule
        determinations listed in Table \ref{tabcollection}, taking
        into account the sizable theoretical error in the change of
        scheme. Figure updated from \cite{Gambino:2008fj}.}
    \label{fig:slep:mbmc}
\end{figure}

\subsubsubsection{Lattice QCD}

In principle, lattice QCD should provide sound ways of determining the 
quark masses: each bare mass is adjusted until one particular 
hadron mass agrees with experiment.
In practice, there are several approaches.
One is to convert the bare mass of the lattice action to a more familiar 
renormalization scheme. 
Another is to define the mass via ratios of matrix elements derived from 
the CVC or PCAC relations.
Finally, one can compute short-distance objects that are sensitive to 
the (heavy) quark masses, for which continuum perturbation theory can be 
used~\cite{Bochkarev:1995ai}.
For the first two methods, a matching procedure is needed to relate the 
bare lattice mass, or currents, to a continuum scheme, such as 
$\overline{\rm MS}$.
The matching can be done in perturbation theory---the state of 
the art for light quarks is two-loop~\cite{Mason:2005bj}---or 
via nonperturbative matching~\cite{Sommer:2006sj}.
When considering nonperturbatively matched results from lattice QCD, one 
should bear in mind that the match is to an RI-MOM scheme or to the 
renormalization-group independent (RGI) mass.
Final conversion to the $\overline{\rm MS}$ scheme always entails 
perturbation theory, because dimensional regulators, and hence their 
minimal subtractions, are defined only in perturbation theory.

For bottom quarks, light-quark methods do not carry over 
straightforwardly~\cite{Kronfeld:2003sd}.
Consequently, unquenched determinations of \mb have been limited to 
one-loop accuracy~\cite{Gray:2005ur} while nonperturbatively matched 
determinations remain quenched~\cite{Heitger:2009qf}.
They are, thus, not competitive with the other determinations of \mb 
discussed here.
For charm the situation is almost the same, except on the finest lattices 
with the most-improved actions.
Then, as discussed below, it is possible to use moments of the 
charmonium correlator and continuum O($\alpha_s^3$) perturbation 
theory~\cite{Allison:2008xk}, or to employ two-loop matching, which is 
still in progress~\cite{Allison:2008ri}.

\subsubsubsection{Charm mass determinations}

Due to the increased precision in the data and in the theoretical description  
the charm quark mass is also an important input parameter in the analysis of
inclusive $B$ decays.
Due to its low mass the use of threshold masses is not imperative for the
charm quark, and the most common scheme is the $\overline{\rm MS}$ mass. 
The most precise measurements are obtained from
$e^+e^-$ sum rules. More recently, charm mass measurements with small
uncertainties are also obtained from inclusive $B$ decays.
In the $e^+e^-$ sum rule analyses of
Refs.~\cite{Boughezal:2006px,Kuhn:2007vp} based on 
fixed order perturbation theory at ${\rm O}(\alpha_s^3)$
the results $\overline m_c(\overline m_c)= 1.295\pm 0.015$~GeV and
$1.286\pm 0.013$~GeV, respectively, were obtained. In was, however, pointed
out in Ref.~\cite{Hoang:2004xm} based on an ${\rm O}(\alpha_s^2)$ analysis
that carrying out the analysis in fixed-order 
perturbation theory might underestimate the theory error due to a discrepancy of
the predictions in fixed-order and in contour-improved perturbation theory.
In the analysis of Ref.~\cite{Allison:2008xk} lattice calculations of moments
of different current-current correlators, defined in analogy to
Eq.~(\ref{Mdef}), and ${\rm O}(\alpha_s^3)$ fixed-order computations of these
moments were combined and the result $\overline m_c(\overline m_c)= 1.268\pm
0.009$~GeV was obtained. This analysis avoids the usually large conversion
uncertainties when lattice masses are converted to the $\overline{\rm MS}$
continuum mass, however, it might also suffer from the theory issue pointed
out in Ref.~\cite{Hoang:2004xm}. Thus this issue certainly deserves further
investigation. More recently, measurement of the charm mass with small
uncertainties were also obtained from fits to inclusive $B$ decay spectra.
In Refs.~\cite{Hoang:2005zw} and \cite{Buchmuller:2005zv} the results
$\overline m_c(\overline m_c)= 1.22\pm 0.06$~GeV and
$1.24\pm 0.09$~GeV, respectively, were obtained. These results are compatible
with the $e^+e^-$ sum rule analyses. 

\subsubsection{Measurements and tests}

The experimental measurements of inclusive charmless semileptonic $B$
decays are dominated by measurements at the $\FourS$ resonance.  They
fall into two broad categories: so-called ``tagged'' measurements, in
which the companion $B$ meson is fully reconstructed in a hadronic
decay mode (see Sec.~\ref{sec:recoiltechnique}), 
which allows an unambiguous association of particles with the
semileptonic $B$ decay and the determination of the $B$ decay rest frame;
and untagged measurements, in which only a charged lepton and, in some cases,
the missing momentum vector for the event are measured.

The untagged measurements tend
to have high efficiency but poor signal to noise, and are sensitive to
$\epem\to\qqbar$ continuum background.  
The main source of background is from $b\to c\ell\nub$ decays.
Existing measurements all require the lepton momentum to exceed
$1.9~\gev$ in the $\FourS$ rest frame.
Those analyses that utilize the missing momentum
vector generally have improved background rejection, but also have
additional uncertainties due to the modeling of sources of missing
momentum, such as imperfect track and cluster reconstruction, the
response to neutral hadrons and the presence of additional neutrinos.
The partial branching fraction in a specified kinematic region is
determined in some analyses by a cut-and-count method, and in others
by a fit of the measured spectrum to the predicted shapes of the
signal and background components.  In all cases the fits use coarse
binning in regions where the differential distributions are highly sensitive
to details of the shape function.

The tagged measurements require the presence of an electron or muon
with $E_\ell>1.0~\gev$ amongst the particles not used in the reconstruction
of the hadronic $\B$ decay.
These analyses provide measurements of the kinematic variables
of the hadronic system associated with the semileptonic decay,
such as $m_X$ and $P_+$, as well as of $q^2$.
They also provide additional handles for suppressing background,
which comes predominantly from the Cabibbo-favored decays $b\to
c\ell\nub$; these include charge correlations between the
fully-reconstructed $B$ meson and the lepton, the veto of Kaons from
the semileptonically-decaying $\Bb$, and constraints on the charge sum
of reconstructed tracks and on the reconstructed missing mass-squared
in the event.  This power has a cost; the net selection efficiency is
$<1\%$ relative to an untagged analysis, and is not well
understood in absolute terms due to incomplete knowledge of the
decay modes that contribute to the fully-reconstructed $B$ meson
sample.  As a result, these analyses measure
ratios of branching fractions, usually relative to the inclusive
semileptonic partial branching fraction for $E_e>1.0~\gev$.
Examples of measurements from these two categories are shown in
Figs.~\ref{fig:inclVubEeMeas} and ~\ref{fig:inclVubMxMeas}.

\begin{figure}
\centering
    \includegraphics[width=0.57\textwidth]{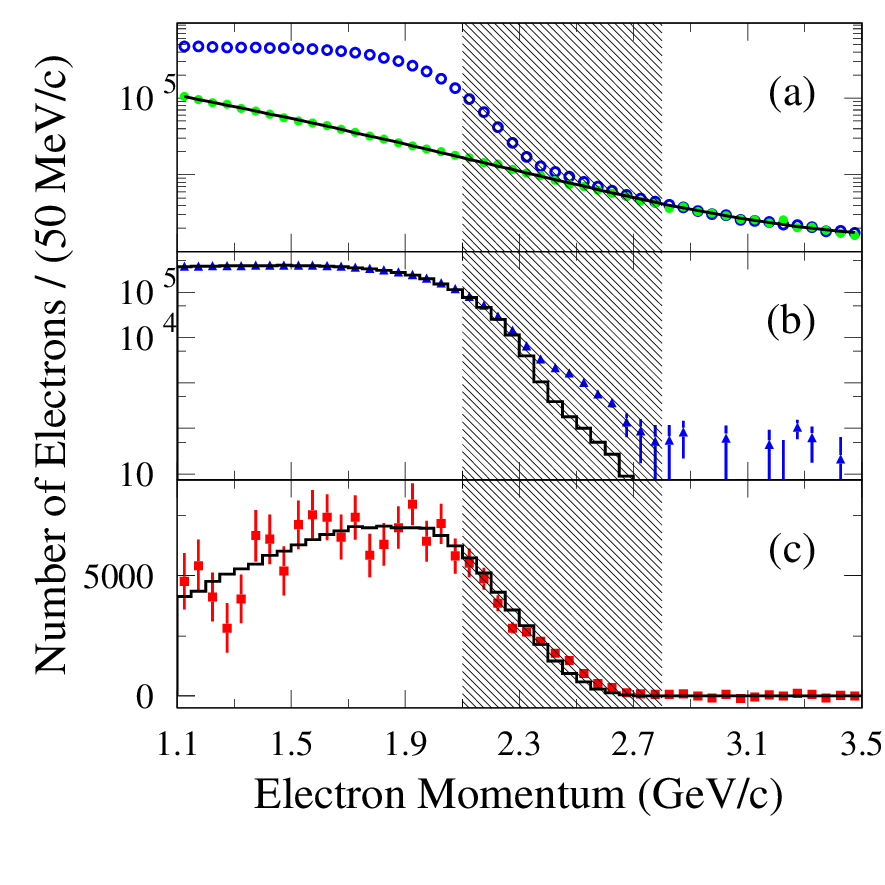}
\caption{The inclusive electron energy spectrum\cite{Aubert:2005mg} 
from BaBar is shown for (a) on-peak data and $\qq$ continuum (histogram);
(b) data subtracted for non-$\BB$ contributions (points) and the simulated
contribution from $\B$ decays other than $b\to u\ell\nu$ (histogram); 
and (c) background-subtracted data (points) with a model of the
$b\to u\ell\nu$ spectrum (histogram).
}
    \label{fig:inclVubEeMeas}
\end{figure}

\begin{figure}
\centering
\includegraphics[width=0.41\textwidth]{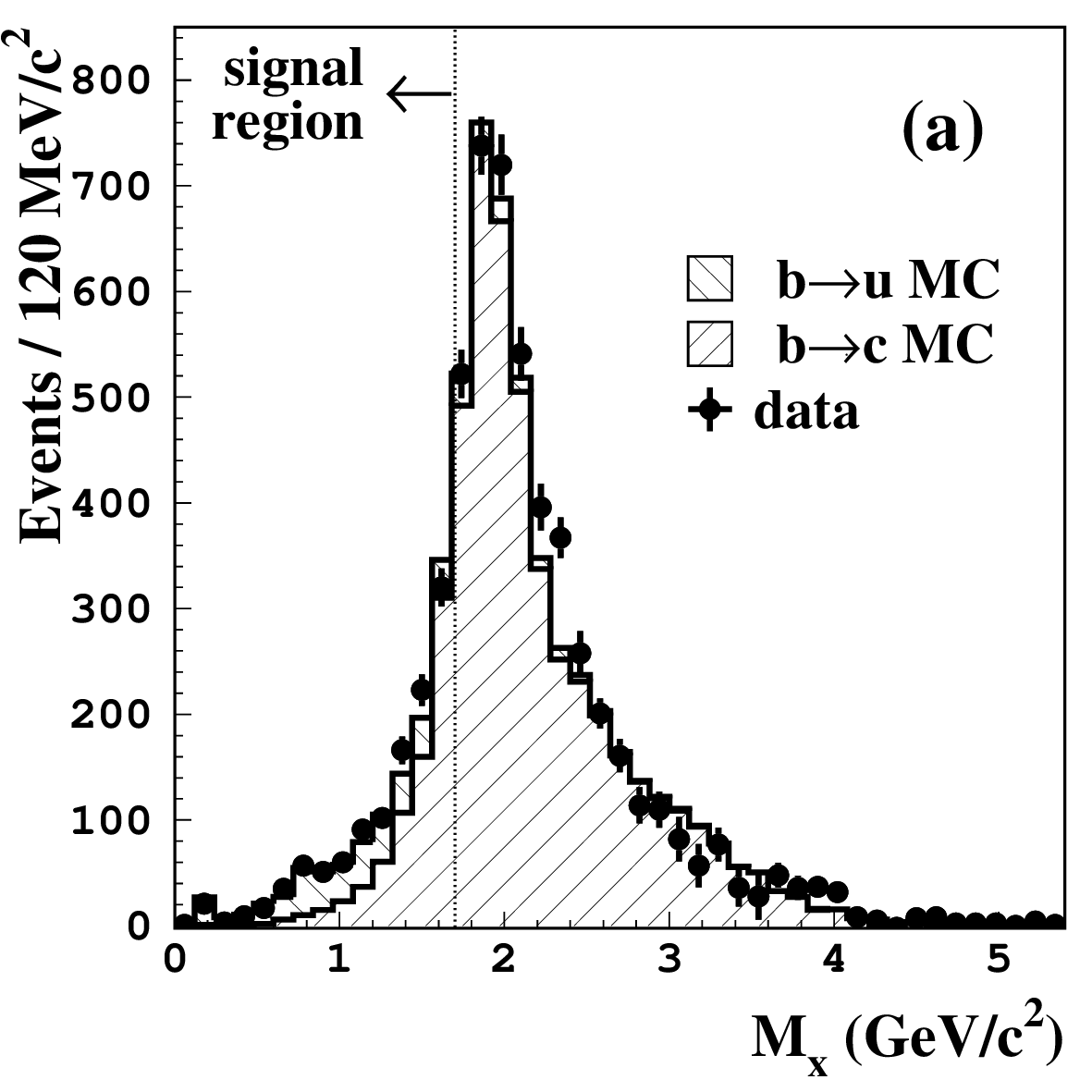}
\includegraphics[width=0.41\textwidth]{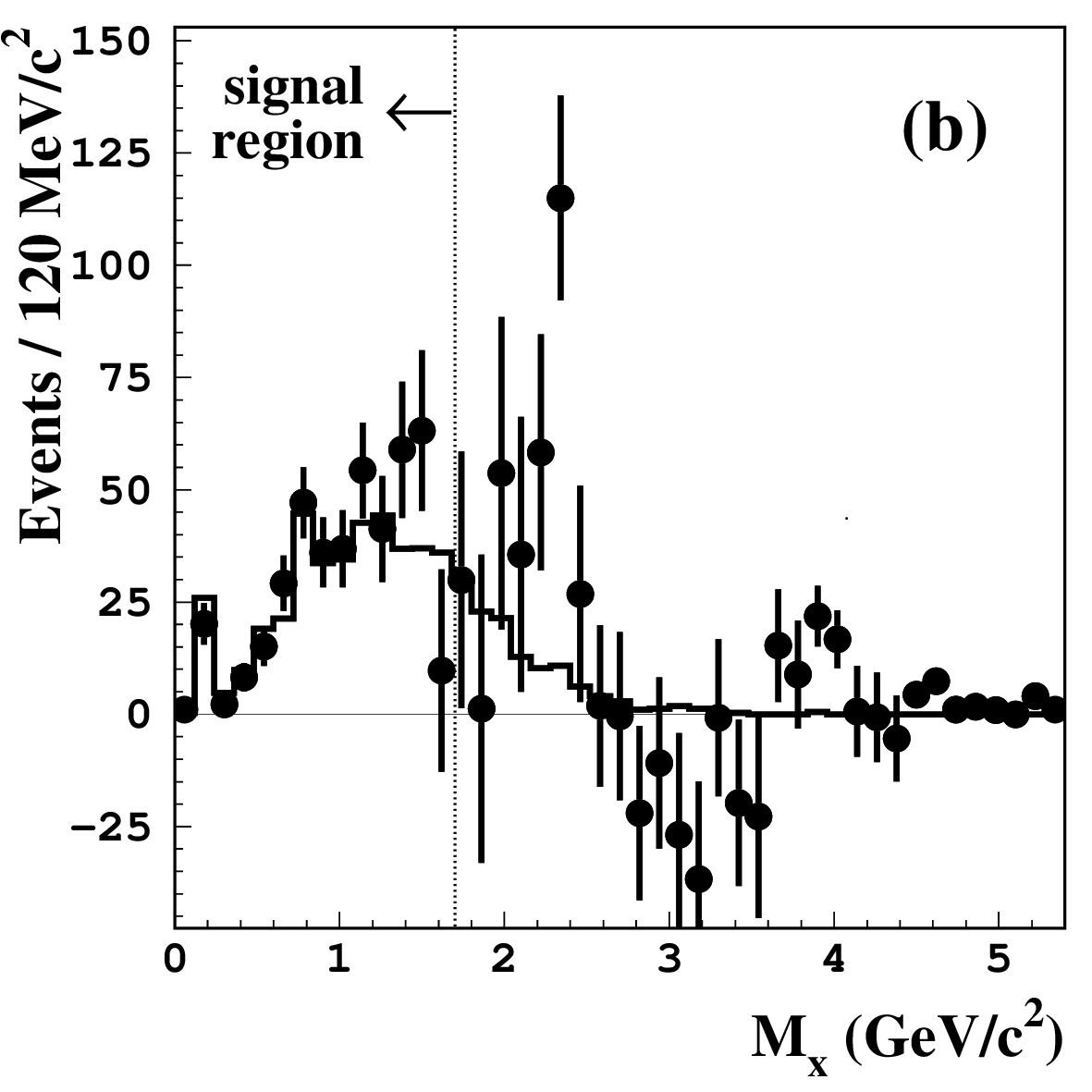}
\caption{The hadronic invariant mass spectrum\cite{Bizjak:2005hn}
in Belle data (points) is shown in (a) with histograms corresponding
to the fitted contributions from $b\to c\ell\nu$ and $b\to u\ell\nu$.
After subtracting the expected contribution from $b\to c\ell\nu$,
the data (points) are compared to a model $b\to u\ell\nu$ spectrum
(histogram) in (b).
}
    \label{fig:inclVubMxMeas}
\end{figure}

The large $b\to c\ell\nub$ background is reduced in most analyses by
making restrictive kinematic cuts.  The measured quantity is then a
(sometimes small) fraction of the full $b\to u\ell\nub$ rate.  
As discussed in section~\ref{sec:inclusivevubtheory},
these restrictions introduce sensitivity to the non-perturbative shape
function, and significantly increase the sensitivity to $m_b$.  The
choice of kinematic cuts is a balance between statistical and systematic
uncertainties, which increase as kinematic cuts are relaxed, 
and theoretical and parametric uncertainties, which decrease under these
conditions.

The best determinations to date of various $b\to u\ell\nub$ partial
rates are given in Tab.~\ref{tab:inclusivevub}.  The experimental
systematic uncertainties affecting all analyses are due to track
reconstruction and electron identification.  Untagged analyses are
relatively more sensitive to bremsstrahlung and radiative corrections.
The tagged analyses have additional uncertainties due to the
determination of event yields via fits to the invariant mass spectra of
fully-reconstructed $B$ candidates.  Uncertainties due to the modeling
of $b\to c\ell\nub$ decays are correlated between measurements, but
their magnitude varies depending on the cuts applied and the analysis
strategy; most analyses include some data-based evaluation of the
level of this background.  The leading sources of uncertainty arise
from uncertainties in the form factors for $\Bb\to D^*\ell\nub$ decays
and limited knowledge of semileptonic decays to higher mass charm
states.  The modeling of $b\to u\ell\nub$ decays is relevant to all
analyses to the extent that the precise mix of exclusive states, which
is not well measured, affects the acceptance and reconstruction
efficiency.  In some analyses an additional sensitivity arises due to
the use of a $b\to u\ell\nub$ component in a fit to the measured
kinematic observable; the shape of this component then affects the
result.  The sensitivity of each analysis to weak annihilation varies
as a function of the acceptance cuts used.

The larger data sets now available allow less restrictive kinematic
cuts that encompass up to $90\%$ or more of the total $b\to u\ell\nub$ rate,
which significantly reduces the impact of theoretical uncertainties.
A preliminary result from Belle~\cite{newBELLEbulnu} uses a multivariate
analysis on a tagged
sample to measure the full $b\to u\ell\nub$ rate for $E_\ell>1.0~\gev$,
and quotes an experimental uncertainty of $6\%$ on $\Vub$ and smaller
theoretical uncertainties than measurements made in more restrictive
kinematic regions. Given the challenging nature of the
measurements that include large regions dominated by $b\to c\ell\nub$
decays, it is valuable to have results based on complementary
techniques.  Untagged measurements of the fully inclusive electron
spectrum can also be pushed further into the region dominated by $b\to
c\ell\nub$ decays; there are prospects for pushing down to
$E_e>1.6~\gev$ while maintaining experimental errors at the
$<5\%$ level on $\Vub$.  

The availability of measured partial rates in different kinematic
regions allows a test of the theoretical predictions, as ratios of
partial rates are independent of $\Vub$.  One gauge of the consistency
of the measured and predicted partial rates is the $\chi^2$ of the
$\Vub$ average within each theoretical framework.  These are given in
Tab.~\ref{tab:inclusivevub}
In each case a reasonable $\chi^2$
probability is obtained.  One can also probe directly the ratios of particular
partial rates.  

\subsubsection{Determination of \Vub}

As described in the previous section, the large background 
from the $b\to c \ell \nu$ decays is the chief experimental limitation
in the determination of the total branching fraction for 
$b\to u \ell \nu$ decays.
The different analyses
are characterized by kinematic cuts applied on: the lepton energy (\el),
the invariant mass of the hadron final state (\mx), the light-cone
component of the hadronic final state momentum along the jet direction
(\pplus), the two dimensional distributions \mx-$q^2$ and \el-\smax ,
where $q^2$ is the squared transferred momentum to the lepton pair and
\smax\ is the maximal \mx$^2$ at fixed $q^2$ and \el. 
Given the large variety of analyses performed, and the differences
in background rejection cuts used in the different experimental techniques, 
each analysis measures a partial rate in a different phase-space region.
The differential
rates needed from the theory to extract \Vub\ from the experimental
results have been calculated using each theoretical approach.  
The challenge of averaging the \Vub\ measurements from the 
different analyses is due mainly to the complexity of combining
measurements performed with different systematic assumptions
and with potentially-correlated systematic uncertainties.
Different analyses often use a different decomposition of their
systematic uncertainties, so achieving consistent definitions for any
potentially correlated contributions requires close coordination
with the experiments. 
Also, some tagged analyses produce partial rates in several kinematic 
variables, like \mx, \mx -$q^2$ and \pplus, based on the same data 
sample, so the
statistical correlation among the analyses needs to be accounted for.
As a result, only those analyses for which the statistical 
correlation is provided are included in the average.
Systematic uncertainties that are uncorrelated with any other sources
of uncertainty appearing in an average are lumped with the statistical 
error.  Those systematic errors correlated with at least one other measurement
are treated explicitly.  Examples of correlated systematic errors 
include uncertainties in the branching fractions for exclusive 
$ b \to c \ell \nu$ and $b \to u \ell \nu$ decay modes,
the tracking, particle identification and luminosity uncertainties for
analyses performed in the same experiment, etc.

The theoretical errors for a given calculation are considered completely
correlated among all the analyses. 
No uncertainty is assigned for possible duality violations.

For BLNP, we have considered theoretical errors due to the HQE
parameters $m_b$ and $\mu_\pi^2$, the functional form of the shape
function, the subleading shape functions, the variation of the matching
scales, and weak annihilation.

For DGE, the theoretical errors are due to the effect of the $\alpha_s$
and $m_b$ uncertainties on the prediction of the event fraction and the
total rate, weak annihilation and the change and variation of the scale
of the matching scheme.

The theoretical errors for GGOU are from the value of $\alpha_s$, $m_b$
and non-perturbative parameters, higher order perturbative and
non-perturbative corrections, the modeling of the $q^2$ tail, the weak
annihilation matrix element and the functional form of the distribution
functions  at fixed  $q^2$ and $\mu=1\gev$.

Finally, the theoretical errors considered for ADFR are related to the
uncertainties on $\alpha_s$, \Vcb, 
$m_c$, and the semileptonic
branching fraction.  In addition, a different method to extract
\Vub\ from the semileptonic rate is used, which does not depend on the
inclusive semileptonic charm rate, and pole quark masses are employed
instead of the $\overline{\rm MS}$ ones.

The theoretical errors are all characterized by uncertainties whose size
and derivative as a function of the rate are different, affecting in
different ways the \Vub\ averages.

The methodology and the results provided by the Heavy Flavor Averaging
Group (HFAG) are presented in this section.  To meaningfully combine the
different analyses, the central values and errors are rescaled to a
common set of input parameters.  Specifically for the $b\to u \ell \nu$
analyses, the average $B$ lifetime used for the measurements is
$(1.573\pm 0.009)$~ps.  Moreover, a rescaling factor to account for
final state radiation is applied to the partial branching fractions used
for the CLEO and Belle endpoint measurements.  


The fit performed to obtain the value of the $b$ quark mass is described
in Sec.~\ref{sec:inclusiveVcb}.  The  value of \mb from the 
global fit in the kinetic scheme is used for all
the four frameworks for consistency, translated to the different mass
schemes as needed.  Note that the models depend strongly
on the $b$ quark mass, except for ADFR, so it is very important to use a
precise determination of the $b$ quark mass.  The results obtained by
these methods and the corresponding averages are shown in
Tab.~\ref{tab:inclusivevub}. 
\begin{table}[tb]
    \centering
    \caption{Partial branching fraction and $\Vub$ from inclusive 
        $b\to u\ell\nub$ measurements. The values determined using 
        different theoretical calculations are given along with the 
        corresponding theory uncertainty; the experimental error on 
        $\Vub$ is quoted separately.
        The $f_u$ values are from BLNP.
        The ADFR values for the endpoint analyses refer to 
        $E_e>2.3$\gev.} 
    \label{tab:inclusivevub}
    \begin{tabular}{lcccccc}\hline\hline
    Method & $\Delta$BF$\times 10^5$ & $f_u^{BLNP}$ 
    & \multicolumn{4}{c|}{$(\Vub \times 10^5)$} \\
    (GeV)  &       &                         
    &      BLNP &      GGOU &      DGE   & ADFR\\ \hline

    $E_e>2.1$\cite{Bornheim:2002du} 
         
            & $33\pm 2\pm 7$ & 0.20 
            & $383 \pm  45^{+32}_{-33}$
            & $368 \pm  43^{+24}_{-38}$
            & $358 \pm  42^{+28}_{-25}$ 
            & $349 \pm  20^{+24}_{-24}$ \\ 
    $E_e$-$q^2$\cite{Aubert:2005im} 
         
            & $44\pm 4\pm 4$ & 0.20 
            & $428 \pm  29^{+36}_{-37}$
            & not~avail.
            & $404 \pm  27^{+28}_{-30}$
            & $390 \pm  26^{+23}_{-24}$ \\ 
    $m_X$-$q^2$\cite{Kakuno:2003fk} 
         
            & $74\pm 9\pm 13$ & 0.35 
            & $423 \pm  45^{+29}_{-30}$
            & $414 \pm  44^{+33}_{-34}$
            & $420 \pm  44^{+23}_{-18}$
            & $397 \pm  42^{+23}_{-23}$ \\ 
    $E_e>1.9$\cite{Limosani:2005pi} 
         
            & $85\pm 4\pm 15$ & 0.36 
            & $464 \pm  43^{+29}_{-31}$
            & $453 \pm  42^{+22}_{-30}$
            & $456 \pm  42^{+28}_{-24}$
            & $326 \pm  17^{+22}_{-22}$ \\ 
    $E_e>2.0$\cite{Aubert:2005mg} 
         
           & $57\pm 4\pm 5$ & 0.28 
           & $418 \pm  24^{+29}_{-31}$ 
           & $405 \pm  23^{+22}_{-32}$
           & $406 \pm  27^{+27}_{-26}$ 
           & $346 \pm  14^{+24}_{-23}$ \\ 
    $m_X<1.7$\cite{Bizjak:2005hn} 
         
            & $123\pm 11\pm 12$ & 0.69 
            & $390 \pm  26^{+24}_{-26}$ 
            & $386 \pm  26^{+18}_{-21}$
            & $403 \pm  27^{+26}_{-20}$
            & $393 \pm  26^{+24}_{-24}$ \\ 

    $m_X<1.55$\cite{Aubert:2007rb} 
         
            & $117\pm 9\pm 7$ & 0.61 
            & $402 \pm  19^{+27}_{-29}$
            & $398 \pm  19^{+26}_{-28}$
            & $423 \pm  20^{+21}_{-16}$
            & $404 \pm  19^{+25}_{-26}$ \\ 
    $m_X$-$q^2$\cite{Aubert:2007rb} 
         
            & $77\pm 8\pm 7$ & 0.35 
            & $432 \pm  28^{+29}_{-31}$
            & $422 \pm  28^{+33}_{-35}$
            & $426 \pm  28^{+23}_{-19}$
            & $415 \pm  27^{+24}_{-24}$ \\ 
    $P^+<0.66$\cite{Aubert:2007rb} 
         
            & $94\pm 9\pm 8$ & 0.60 
            & $365 \pm  24^{+25}_{-27}$
            & $343 \pm  22^{+28}_{-27}$
            & $370 \pm  24^{+31}_{-24}$
            & $356 \pm  23^{+23}_{-23}$ \\ 

    Average & & 
    &   $406 \pm  15^{+25}_{-27}$ 
    &   $403 \pm  15^{+20}_{-25}$ 
    &   $425 \pm  15^{+21}_{-17}$ 
    &   $384 \pm  13^{+23}_{-20}$ \\
    $\chi^2/\mathrm{d.f.}$ & & 
    & $13.9/8$ & $9.4/7$ & $7.1/8$ & $16.1/8$ \\ 
    \hline\hline
    \end{tabular}
\end{table}

All the methods are consistent with the current data.
Fig.~\ref{fig:slep:compare-inc-Vub} compares \Vub\ extracted in each 
experimental analysis using different frameworks.
\begin{figure}[bp]
    \centering
    \includegraphics[width=0.57\textwidth]{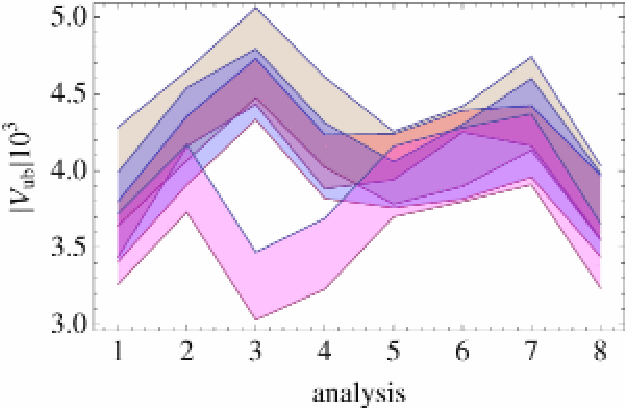}
    \caption{Comparison of \Vub\ extracted from experiment  as in 
        Tab.~\ref{tab:inclusivevub} (with the exception of the second 
        line) using the color code introduced in 
        Fig.~\ref{fig:slep:incVub} for the four frameworks.
        The bands correspond to theory errors depurated of common 
        parametric errors.}
    \label{fig:slep:compare-inc-Vub}
\end{figure}
The results of DGE, BLNP, GGOU  agree in all cases within theoretical 
non-parametric errors.  We take as our evaluation of \Vub from inclusive 
semileptonic decays the arithmetic average of the values and errors of
these three determinations to find
\begin{equation}
    \Vub = (411^{+27}_{-28})\times10^{-5}.
    \label{eq:slep:incVub}
\end{equation}

Although in these three cases the $\chi^2/{\rm d.f.}$ reported in 
Tab.~\ref{tab:inclusivevub} is good, a small WA contribution can 
marginally improve it. 
Differences among these theory approaches can be uncovered by additional 
experimental information on the physical spectra.
For instance, the endpoint analyses of \babar\ and Belle already  
allow us to extract \Vub\ at values of $E_{\rm cut}$ ranging from 1.9 to 
2.3~\gev. 
The two plots in Fig~\ref{fig:slep:compare-El-Vub} compare \Vub\ 
extracted in the four theory frameworks at various $E_{\rm cut}$. 
\begin{figure}[bp]
    \includegraphics[width=0.47\textwidth]{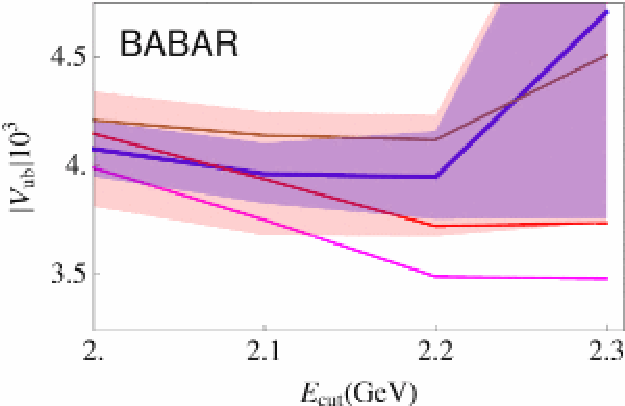}\hfil
    \includegraphics[width=0.47\textwidth]{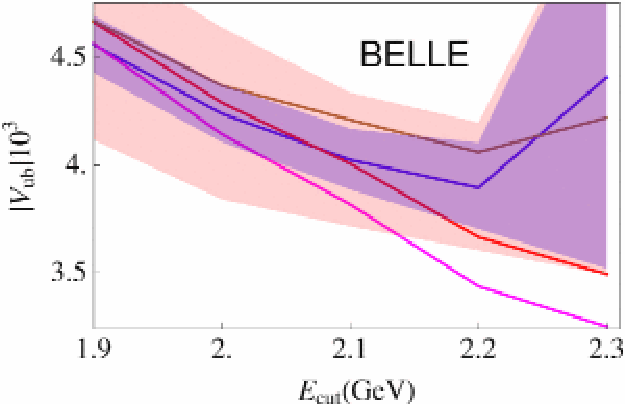}
    \caption{\Vub\ extracted from the lepton endpoint by \babar\ and 
        Belle as a function of the cut on the lepton energy. 
        The bands correspond to GGOU theory errors depurated of the 
        parametric and WA error and to the latter combined with the 
        experimental one.}
    \label{fig:slep:compare-El-Vub}
\end{figure}
\babar's  more precise results lead to stable values of \Vub\ for $E_{\rm cut}\le 
2.2$\gev in BLNP and GGOU, but it must be stressed that the shape of the 
spectrum strongly depends on \mb and no conclusion can presently be drawn. 


As mentioned above, the leading shape function can also be measured in
$b\to s \gamma$ decays, and there are prescriptions that relate directly
the partial rates for $b\to s \gamma$ and $b\to u \ell\nu$
decays~\cite{Neubert:1993um,Leibovich:1999xf,
Lange:2005qn,Lange:2005xz}, thus avoiding any parametrization of the
shape function.  However, uncertainties due to the sub-leading shape
function remain.  
The \babar\ measurement in Ref.~\cite{Aubert:2005mg} has
been analyzed by in Ref.~\cite{Golubev:2007cs} to obtain
\Vub $= (4.28 \pm 0.29 \pm 0.29 \pm 0.26 \pm
0.28)\times 10^{-3}$ and 
\Vub $= (4.40 \pm 0.30 \pm 0.41 \pm 0.23)\times
10^{-3}$ using calculations from Refs.~\cite{Leibovich:1999xf} and
\cite{Lange:2005qn,Lange:2005xz}, respectively.  These results are
consistent with the inclusive \Vub\ average.

Another approach is to measure $b\to u\ell\nu$ transitions over the
full phase space, thereby reducing theoretical uncertainties.  In
the first measurement of this type, \babar~\cite{Aubert:2006qi} found
\Vub $= (4.43 \pm 0.45 \pm 0.29)\times 10^{-3}$.
A preliminary BELLE measurement of $90\%$ of the full
$b\to u\ell\nub$ rate quotes $\Vub\times 10^{5}$ as follows:~\cite{newBELLEbulnu}
$437\pm 26{}^{+23}_{-21}$ (BLNP),
$446\pm 26{}^{+15}_{-16}$ (DGE),
$441\pm 26{}^{+12}_{-22}$ (GGOU).  The last error in each case combines
uncertainties from theory and $m_b$, and is smaller than in
less-inclusive measurements.

The inclusive determinations of \Vub\ are about $\sim 2\sigma$ larger
than those obtained from exclusive $B\to\pi\ell\nu$.  
The estimated uncertainty on \Vub\ from inclusive decays is presently smaller
than from exclusive decays.
%
The value of \Vub\ predicted from the measured $\sin2\beta$ value is
closer to the exclusive result~\cite{Bona:2006ah}.  

The experimental results and theoretical computations presented in this
chapter represent an
enormous effort, and their distillation into determinations of \Vub
and \Vcb have required close communication among the participants.

\section{Rare decays and measurements of  $|V_{td}/V_{ts}|$}
\label{sec:rare}
\subsection{Introduction}

In this chapter we will discuss a particular subclass of $B$, $K$, and
$D$ meson decays, so-called rare decays. These transitions have been
the subject of a considerable number of experimental and theoretical
investigations. Being rare processes mediated by loop diagrams in the
SM, they all test the flavor structure of the underlying theory at
the level of quantum corrections and provide information on the
couplings and masses of heavy virtual particles appearing as
intermediate states. The resulting sensitivity to non-standard
contributions, such as charged Higgs bosons, SUSY particles,
Kaluza-Klein (KK) excitations or other exotics arising in extensions
of the SM, allows for an indirect observation of NP, a strategy
complementary to the direct production of new particles. Whereas the
latter option is reserved to the Tevatron and the LHC, the indirect
searches performed by CLEO, \babar, Belle, and other low-energy
experiments already impose severe restrictions on the parameter space
of a plethora of NP scenarios, while they do not exclude the
possibility that CDF, D0, or LHC$b$ may find significant deviations
from the SM expectations in certain rare processes, and thus evidence
for NP, prior to a direct discovery of the associated new states by
the high-$p_T$ experiments ATLAS and CMS.

Among the rare decays, the radiative $b \to (s,d) \gamma$ transitions
play a special role. Proceeding at rates of order $G_F^2 \alpha$, they
are parametrically enhanced over all other loop-induced, non-radiative
rare decays that are proportional to $G_F^2 \alpha^2$.  The helicity
violating $b \to (s,d) \gamma$ amplitudes are dominated by
perturbative QCD effects which replace the quadratic GIM suppression
present in the electroweak vertex by a logarithmic one. This mild
suppression of the QCD corrected amplitudes reduces the sensitivity of
these processes to high-scale physics, but makes them wonderful
laboratories to study both perturbative and non-perturbative
strong-interaction phenomena. Since the $b \to (s,d) \gamma$
transitions receive sizable contributions from top-quark loops
involving the couplings $|V_{ts}|$ or $|V_{td}|$, radiative $B$-meson
decays may be in addition used to test the unitarity of the CKM matrix
and to over-constrain the Wolfenstein parameters $\bar \rho$ and $\bar
\eta$. The theoretical and experimental status of both the inclusive
$B \to X_{s,d} \gamma$ and exclusive $B \to (K^\ast, \rho, \omega)
\gamma$ modes is reviewed in Sec.s 6.2 and 6.3.

Useful complementary information on the chiral nature of the flavor
structure of possible non-standard interactions can be obtained from
the studies of purely leptonic and semileptonic rare decays.
Tree-level processes like $B \to \tau \nu$ or $B \to D \tau \nu$
provide a unique window on scalar interactions induced by charged
Higgs bosons exchange, while loop-induced decays such as $B_{s, d} \to
\mu^+ \mu^-$ and $B \to (X_s, K, K^\ast) \ell^+ \ell^-$ also probe the
magnitude and phase of $SU(2)$ breaking effects arising from
$Z$-penguin and electroweak box amplitudes. The latter contributions
lead to a quadratic GIM mechanism in the corresponding decay
amplitudes and therefore to an enhanced sensitivity to the scale of
possible non-standard interactions. In contrast to the two-body decay
modes $B \to \tau \nu$ and $B_{s,d} \to \mu^+ \mu^-$, the three-body
decays $B \to D \tau \nu$ and $B \to (X_s, K, K^\ast) \ell^+ \ell^-$
allow one to study non-trivial observables beyond the branching
fraction by kinematic measurements of the decay products. In the
presence of large statistics, expected from the LHC and a future super
flavor factor, angular analyses of the $b \to c \tau \nu$ and $b \to
s \ell^+ \ell^-$ channels will admit model-independent extractions of
the coupling constants multiplying the effective interaction
vertices. The recent progress achieved in the field of purely leptonic
and semileptonic rare decays is summarized in Sec.s 6.4 to 6.6.

Our survey is rounded off in Sec.s 6.7 and 6.8 with concise
discussions of various rare $K$ and $D$ meson decays. In the former
case, the special role of the $K \to \pi \nu \bar \nu$ and $K_L \to
\pi^0 \ell^+ \ell^-$ modes is emphasized, which due to their
theoretical cleanness and their enhanced sensitivity to both
non-standard flavor and CP violation, are unique tools to discover
or, if no deviation is found, to set severe constraints on non-MFV
physics where the hard GIM cancellation present in the SM and MFV is
not active.


\subsection{Inclusive $B\to X_{s,d}\gamma$}

\subsubsection{Theory of inclusive $B\to X_{s,d}\gamma$}
\label{sec:6:bsgamma_incl_theo}

The inclusive decay $B\to X_{s} \gamma$ is mediated by a FCNC and is
loop suppressed within the SM. Comparing the experimentally measured
branching fraction with that obtained in the SM puts constraints on
all NP models which alter the strength of FCNCs. These constraints are
quite stringent, because theory and experiment show good agreement
within errors that amount to roughly $10\%$ on each side. To reach
this accuracy on the theory prediction requires to include QCD
corrections to NNLO in perturbation theory. In this section we
describe the SM calculation of the branching fraction to this order,
elaborate on some theoretical subtleties related to experimental cuts
on the photon energy, and give examples of the implications for NP
models. We also summarize the status of $B\to X_d\gamma$ decays, for
which experimental results have recently become available.

The calculation of QCD corrections to the $B\to X_s \gamma$ branching
fraction is complicated by the presence of widely separated mass
scales, ranging from the mass of the top quark and the electroweak
gauge bosons to those of the bottom and charm quarks. A
straightforward expansion in powers of $\alpha_s$ leads to terms of
the form $\alpha_s \ln (M_W/m_b) \sim 1$ at each order in perturbation
theory, so fixed-order perturbation theory is inappropriate. One uses
instead the EFT techniques discussed in Sec.~\ref{sec:ope}, to set
up an expansion in RG improved perturbation theory. After integrating
out the top quark and the electroweak gauge bosons, the leading-power
effective Lagrangian reads
\begin{equation} \label{eq:rare:Leff}
  {\cal L}_{\rm eff}={\cal L}_{{\rm QCD}\times{\rm QED}}+
  \frac{G_F}{\sqrt{2}}\sum_{q=u,c} V^\ast_{qs}V_{qb} \left[
    C_1(\mu) \hspace{0.25mm} Q_1^q +
    C_2(\mu) \hspace{0.25mm} Q_2^q +
    \sum_{i=3}^8 C_i(\mu) \hspace{0.25mm} Q_i\right]\,.
\end{equation}
The Wilson coefficients $C_i$ are obtained at a high scale $\mu_0\sim
M_W$ as a series in $\alpha_s$ by matching Green's functions in the SM
with those in the EFT. They are then evolved down to a low scale
$\mu\sim m_b$ by means of the RG. Solving the RG equations requires
the knowledge of the anomalous dimensions of the operators, and the counting in
RG-improved perturbation theory is such that the anomalous dimensions
must be known to one order higher in $\alpha_s$ than the matching
coefficients themselves. The Wilson coefficients and anomalous
dimensions to the accuracy needed for the NNLO calculation were
obtained in \cite{Bobeth:1999mk, Misiak:2004ew} and
\cite{Gorbahn:2004my, Gorbahn:2005sa, Czakon:2006ss}, respectively.

The final step in the calculation consists in the evaluation of the
decay rate $\Gamma(B\to X_s\gamma)_{E_\gamma>E_0}$ using the effective
Lagrangian (\ref{eq:rare:Leff}). The cut on the photon energy is
required to suppress background in the experimental measurements. The
rate is calculated in the heavy-quark expansion, which uses that
$\Lambda_{\rm QCD}\ll m_b, m_c$. The leading-order result can be
written as
\begin{equation} \label{eq:rare:gbs}
  \Gamma(B \to X_s\gamma)_{E_\gamma>E_0}=
  \frac{G_F^2\alpha m_b^5}{32\pi^4}|V^\ast_{ts}V_{tb}|^2 \sum_{i,j=1}^8
  C_i (\mu) C_j (\mu) \hspace{0.25mm} G_{ij}(E_0) \,,
\end{equation}
where we have neglected contributions from $Q_{1,2}^u$, which are CKM
suppressed. The functions $G_{ij}$ can be calculated in fixed-order
perturbation theory as long as $\Lambda_{\rm QCD}\ll m_b-2E_0 =
\Delta$. In that case, they are obtained from the partonic matrix
elements of the $b\to X_s \gamma$ decay. Results at NLO in $\alpha_s$
are known completely \cite{Buras:2002tp}. At NNLO, exact results are
available only for $G_{77}$ \cite{Melnikov:2005bx, Blokland:2005uk,
  Asatrian:2006ph}. Concerning the NNLO corrections to the other
elements $G_{ij}$, it is reasonable to focus on terms where $i,j\in
\{1,2,7,8\}$, since the Wilson coefficients $C_{3-6}$ are small. For
those terms, the set of NNLO diagrams generated by inserting a bottom,
charm, or light-quark loop into the gluon lines of the NLO diagrams
are also known \cite{Ligeti:1999ea, Bieri:2003ue, Boughezal:2007ny,
  Ewerth:2008nv}, with the exception of $G_{18}$ and $G_{28}$.  An
estimate of the remaining NNLO corrections was performed in
\cite{Misiak:2006ab}, by calculating the full corrections to the
elements $G_{ij}$ in the asymptotic limit $m_c \gg m_b/2$, and then
interpolating them to three different boundary conditions at $m_c=0$
to find results at the physical value $m_c\approx m_b/4$.

The results of the various NNLO corrections discussed above lead to
the numerical analysis of \cite{Misiak:2006zs}, which found
\begin{equation} \label{eq:rare:brsg}
  \BR(B \to X_s \gamma)_{E_\gamma>1.6\gev}
  =(3.15\pm 0.23)\times 10^{-4} \,.
\end{equation}
The total error was obtained by adding in quadrature the uncertainties
from hadronic power corrections ($5\%$), parametric dependences
($3\%$), and the interpolation in the charm quark mass ($3\%$). The
most significant unknown stems from hadronic power corrections scaling
as $\alpha_s \Lambda_{\rm QCD}/m_b$ \cite{Lee:2006wn}. For the $(Q_7,
Q_8)$ interference, this involve hadronic matrix elements of
four-quark operators with trilocal light-cone structure, which were
estimated in the vacuum insertion approximation to change the
branching fraction by about $-[0.3,3.0]\%$. Corrections of similar or
larger size may arise from non-local $\alpha_s\Lambda_{\rm QCD}/m_b$
corrections due to the $(Q_{1,2}, Q_7)$ interference, but these have
not yet been estimated.

The fixed-order calculation relies on the parametric counting $\Delta
\sim m_b$. However, measurements of the branching fractions are
limited to values above a photon energy cut $E_0=1.6\gev$,
corresponding to $\Delta \sim 1.4\gev$, so it can be argued the
counting $\Lambda_{\rm QCD }\ll \Delta \ll m_b$ is more
appropriate. In that case, to properly account for the photon energy
cut requires to separate contributions from a hard scale $\mu_h\sim
m_b$, the soft scale $\mu_s\sim \Delta$, and an intermediate scale
$\mu_i \sim \sqrt{m_b\Delta}$. An EFT approach able to separate these
scales and to resum large logarithms of their ratios was developed in
\cite{Neubert:2004dd}, and extended to NNLO in RG-improved
perturbation theory in \cite{Becher:2005pd, Becher:2006qw,
  Becher:2006pu}. An approach which used the same factorization of
scales, but a different approach to resummation, called dressed gluon
exponentiation (DGE), was pursued in \cite{Andersen:2005bj,
  Andersen:2006hr}. Compared to \cite{Becher:2006pu}, the DGE approach
includes additional effects arising from the resummation of
running-coupling corrections in the power-suppressed $\Delta/m_b$
contributions.

The consistency between the SM prediction (\ref{eq:rare:brsg}) and the
experimental world average as given in
Tab.~\ref{tab:6:bsgamma-exp-incl-bf}, provides strong constraints on
many extensions of the SM. The prime example is
the bound on the mass of the charged Higgs boson in the 2HDM of type
II (2HDM-II) \cite{Ciuchini:1997xe, Borzumati:1998tg, Bobeth:1999ww}
that amounts to $M_{H^+} > 295\gev$ at $95\%$ CL
\cite{Misiak:2006zs}, essentially independent of $\tan \beta$. This is
much stronger than other available direct and indirect constraints on
$M_{H^+}$.

The inclusive $b \to s \gamma$ transition has also received a lot of
attention in SUSY extensions of the SM \cite{Bobeth:1999ww,
  Ciuchini:1998xy, Degrassi:2000qf, Carena:2000uj, Degrassi:2006eh}.
In the limit of $M_{\rm SUSY} \gg M_W$, SUSY effects can be absorbed
into the coupling constants of local operators in an EFT
\cite{D'Ambrosio:2002ex}. The Higgs sector of the MSSM is modified by
these non-decoupling corrections and can differ notably from the
native 2HDM-II model. Some of the corrections to $B \to X_s \gamma$ in
the EFT are enhanced by $\tan \beta$, as $\alpha_s \tan \beta
\sim 1$ for $\tan \beta \gg 1$, and need to be resummed if
applicable. In the large $\tan \beta$ regime the relative sign of the
chargino contribution is given by $-{\rm sgn}(A_t \mu)$. For ${\rm
  sgn} (A_t \mu) > 0$, the chargino and charged Higgs boson
contributions interfere constructively with the SM amplitude and this
tends to rule out large positive values of the product of the
trilinear soft SUSY breaking coupling $A_t$ and the Higgsino parameter
$\mu$. In the MSSM with generic sources of flavor violation, $B \to
X_s \gamma$ implies stringent bounds on the flavor-violating entries
in the down-squark mass matrix. In particular, for small and moderate
values of $\tan \beta$ all four mass insertions $(\delta_{23}^d)_{AB}$
with $A, B = L, R$ except for $(\delta_{23}^d)_{RR}$ are determined
entirely by $B \to X_s \gamma$~\ref{sec:genFCNC}. The bounds on $|(\delta_{23}^d)_{AB}|$
amount to $4 \times 10^{-1}$, $6 \times 10^{-2}$, and $2 \times
10^{-2}$ for the $LL$, $LR$, and $RL$ insertion.

In the portion of the SUSY parameter space with inverted scalar mass
hierarchy, realized in the class of SUSY GUT scenarios, chargino
contributions to $b \to s \gamma$ are strongly enhanced. As a result,
SUSY GUT models with third generation Yukawa unification and universal
squark and gaugino masses at the GUT scale are unable to accommodate
the value of the bottom-quark mass without violating either the
constraint from $B \to X_s \gamma$ or $B_s \to \mu^+ \mu^-$, unless
the scalar masses are pushed into the few TeV range
\cite{Albrecht:2007ii}. A potential remedy consists in relaxing Yukawa
to $b$--$\tau$ unification, but even then the predictions for ${\cal
  B} (B \to X_s \gamma)$ tend to be at the lower end of the range
favored by experiment \cite{Altmannshofer:2008vr}.

In non-SUSY extensions of the SM, contributions due to Kaluza-Klein~(KK)
excitations in models with universal extra dimensions (UEDs) interfere
destructively with the SM amplitude, $B \to X_s \gamma$ leads to
powerful bounds on the inverse compactification radius $1/R$~\cite{Agashe:2001xt,Buras:2003mk}.
Exclusion limits have been obtained in the five- and
six-dimensional case and amount to $1/R > 600\gev$
\cite{Haisch:2007vb} and $1/R > 650\gev$~\cite{Freitas:2008vh} at $95
\%$ CL. These bounds exceed the limits that can be derived from any
other direct measurement.

The discussion so far dealt with $B\to X_s\gamma$. Recently, a first
measurement of the $B\to X_d\gamma$ branching fraction has been
presented \cite{:2008ig}. Compared to $B\to X_s\gamma$, the nominal
theoretical difference is to replace $s\to d$ in the effective
Lagrangian (\ref{eq:rare:Leff}), in which case the terms proportional
to $Q_{1,2}^u$ are no longer CKM suppressed. The implications of this
have been studied in \cite{Ali:1998rr}, where it was pointed out that
the ratio $\BR(B \to X_s\gamma)/\BR(B \to X_d \gamma)$ can be
calculated with reduced theoretical uncertainty. This was used along
with the experimental results to determine $|V_{td}/V_{ts}|$ in
\cite{:2008ig}. A possible subtlety is that in \cite{Ali:1998rr} the
total branching fraction has been calculated, whereas the experimental
measurements are limited to the region $M_{X_s}<1.8\gev$ of hadronic
invariant masses, where ``shape-function'' effects are expected to be
important.

\subsubsection{Experimental methods and status of $B \to X_{s,d}
  \gamma$}

\label{sec:6:bsgamma_incl_exp}

The analysis of the inclusive $B \to X_s\g$ decay at the $B$ factories
is rather complicated. The quantities to be measured are the
differential decay rate, i.e., the photon energy spectrum as well as
the total branching fraction. There are three methods for the
inclusive analyses: fully inclusive, semi-inclusive, and $B$ recoil.

The idea of the fully inclusive method is to subtract the photon
energy spectrum of the on-resonant $e^+e^- \to \Upsilon(4S) \to \BB$
events by that of the continuum $e^+e^- \to \qqbar$ events. This
method is free from the uncertainty of the final state, and can
exploit the whole available statistics. However, the signal purity is
very low, and the background suppression is a key issue. The photon
energy is obtained in the $\Upsilon(4S)$ rest frame and not in the $B$
rest frame, since the momentum of the $B$ is unknown.

Panel (a) of Fig.~\ref{fig:6:bsgamma-exp-incl-photon_spectrum} shows
the photon energy spectrum after suppressing the continuum background,
using event topology, and vetoing high energy photons from $\pi^0$ or
$\eta$ using the invariant mass of the candidate high energy photon,
and of any other photons in the event. The largest background is from the
continuum events, and is subtracted using the continuum data. This
subtraction requires 
correction due to small center-of-mass energy difference
for the event selection
efficiency, photon energy, and photon multiplicity between the
on-resonant and continuum sample. As shown in panel (b) of
Fig.~\ref{fig:6:bsgamma-exp-incl-photon_spectrum}, the subtracted
spectrum still suffers from huge backgrounds from $B$ decays, which
are subtracted using the MC sample. Here, the MC sample needs to be
calibrated with data using control samples to reproduce the yields of
$\pi^0$, $\eta$, etc.
The final photon spectrum, obtained with the prescribed procedure, 
for $b \to s\g$ events, is shown in panel (c) of
Fig.~\ref{fig:6:bsgamma-exp-incl-photon_spectrum}. It can be seen that
the errors increase rapidly for photon energies below 2\gev due to the
very large continuum background in that region. For this reason all
measurements of the branching ratio introduce a cutoff
$E_\gamma^\mathrm{min}$ and then extrapolate to get
$\BR(E_\gamma>1.6\gev)$ which is compared to the theory
prediction. Measurements by CLEO, \babar, and Belle using the fully
inclusive method are listed in
Tab.~\ref{tab:6:bsgamma-exp-incl-bf}. The results are consistent with
the SM expectation (\ref{eq:rare:brsg}).
\begin{figure}
 \begin{center}
   \caption{\label{fig:6:bsgamma-exp-incl-photon_spectrum}%
     (a) On-resonant data (open circle), scaled continuum data (open
     square) and continuum background subtracted (filled circle)
     photon energy spectrum.  (b) The spectra of photons from $B$
     decays (MC).  (c) The extracted photon spectrum for $B \to
     X_s\gamma$. The plots are taken from \cite{Abe:2008sx}.}
  \begin{tabular}{ccc}
   (a) & (b) & (c) \\
   \includegraphics[scale=0.25,clip]{%
   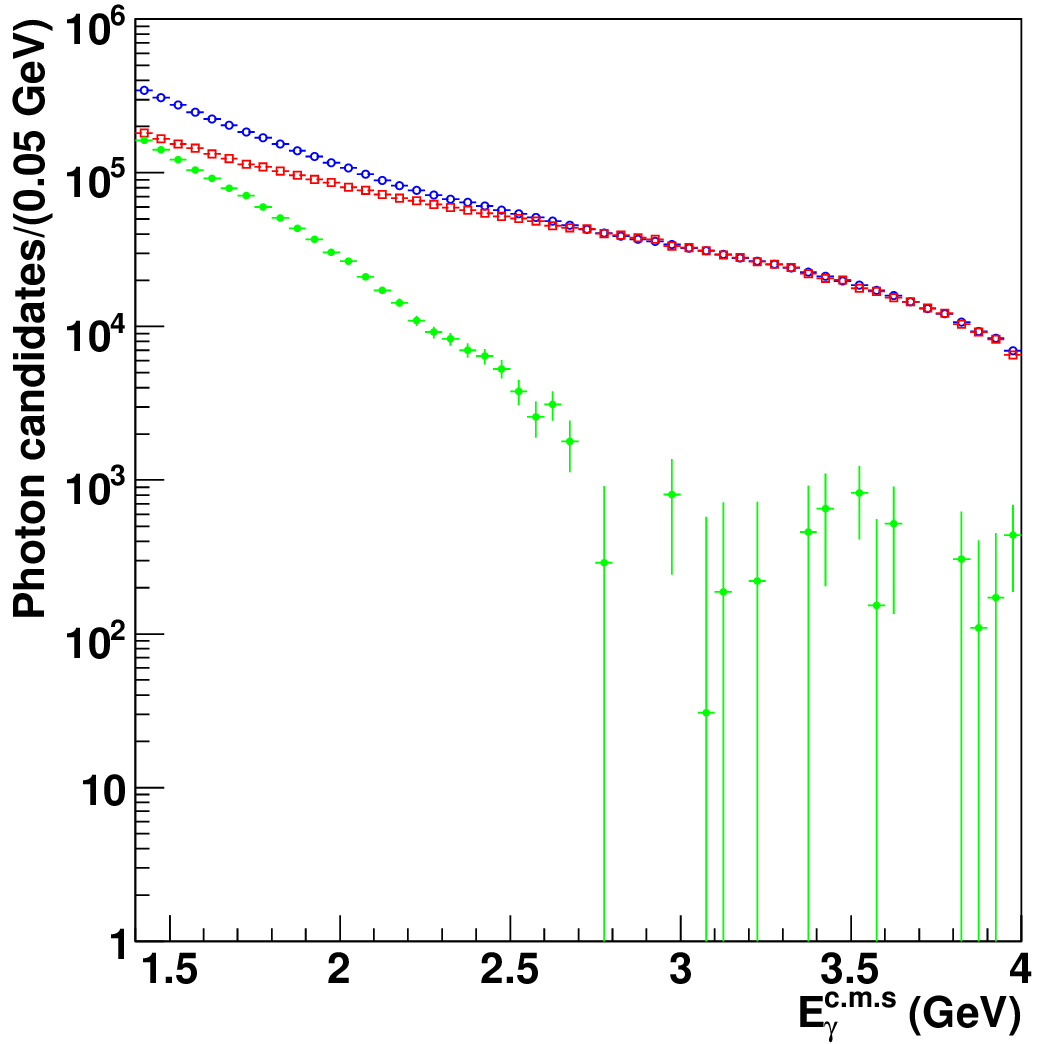} &
   \includegraphics[scale=0.25,clip]{%
   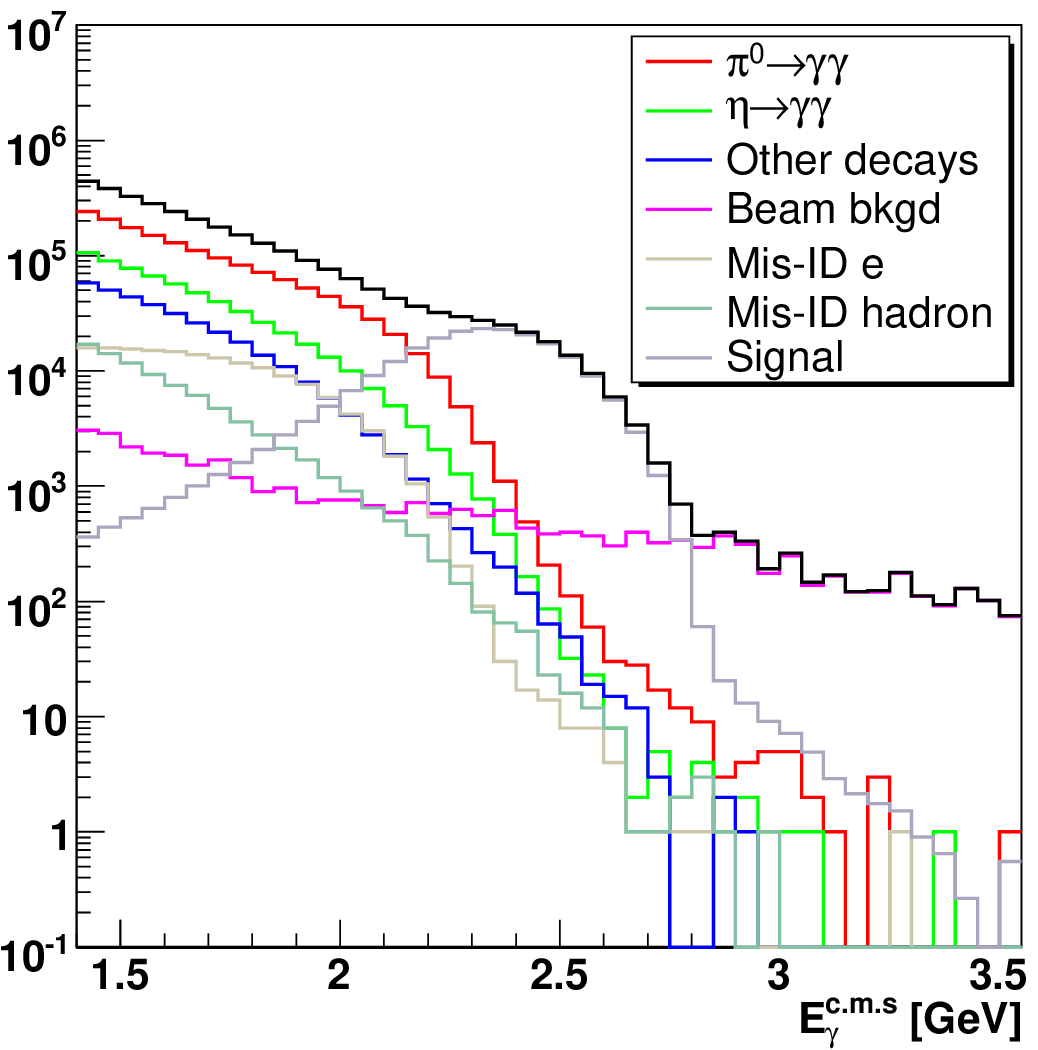} &
   \includegraphics[scale=0.25,clip]{%
   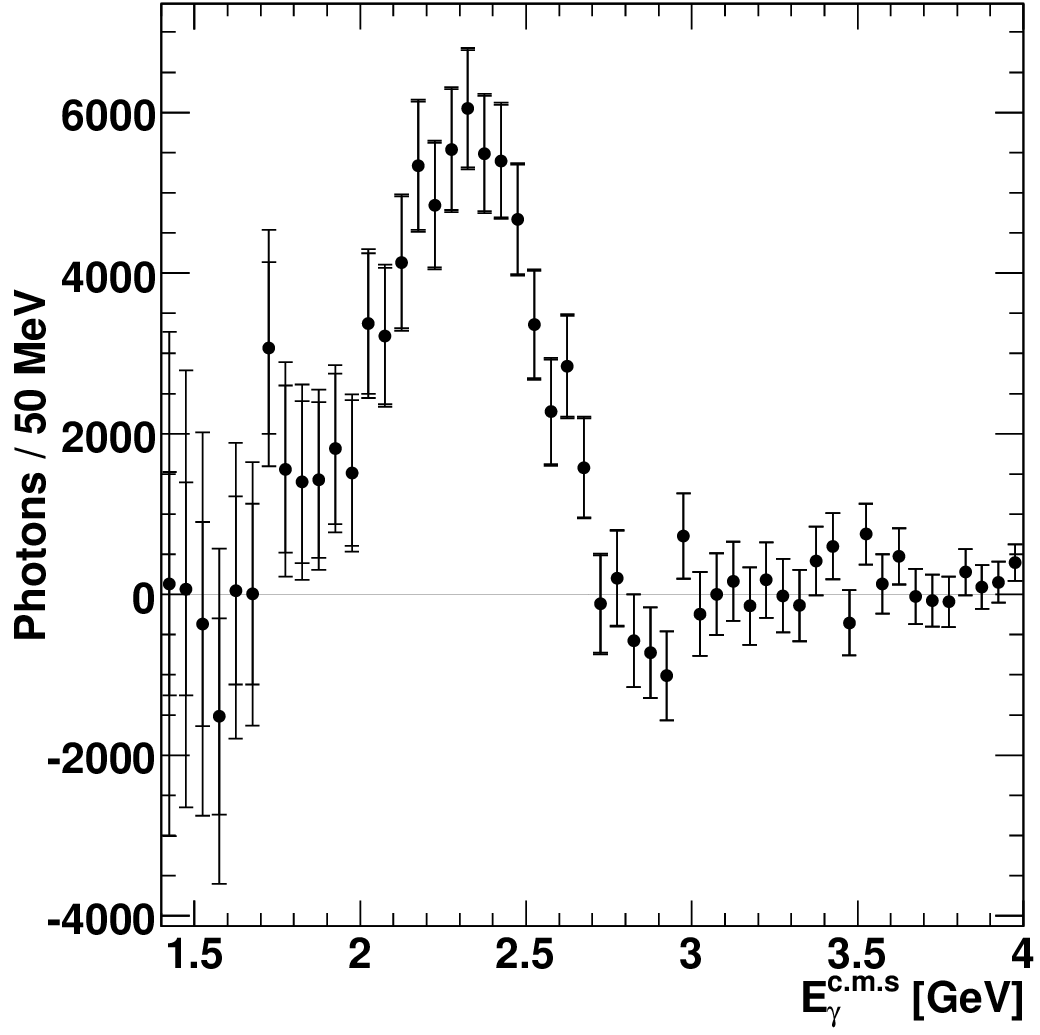}
  \end{tabular}
 \end{center}
\end{figure}

In the semi-inclusive method, also called ``sum-of-exclusive'' method,
the reconstruction of the $B \to X_s\g$ signal is performed by the sum
of certain hadronic final states $X_s$ that are exclusively
reconstructed. Typically, $X_s$ is reconstructed from one Kaon plus
up to
four pions including up to one or two neutral pions, but also modes
with three Kaons or an $\eta$ are used. The advantage of this method
is a better signal purity compared to the fully inclusive method. The
background suppression is still important, but the detailed correction
of the MC samples and the precise determination of the luminosity of
the off-resonance sample, used in the fully inclusive method, are not
necessary. Another advantage is that the photon energy in the $B$ rest
frame can be measured from the mass of the $X_s$ system. However, this
method can reconstruct only a part of the $X_s$ system, and suffers
from the large uncertainty in the fraction of the total width present
in the exclusive modes that are reconstructed. The measurements from
\babar and Belle are listed in Tab.~\ref{tab:6:bsgamma-exp-incl-bf}.

It is also possible to measure the CP asymmetry, $A_{\rm CP}$, of $B
\to X_s\gamma$ with the semi-inclusive method, since most of the final
states provide flavor information. In the SM, $A_{\rm CP}$ is
predicted to be less than $1\%$~\cite{Kagan:1998bh,Hurth:2003dk}, but some models beyond the SM
predict much larger values of $A_{\rm CP}$~\cite{Kagan:1998bh,Baek:1998yn,Kiers:2000xy,Hurth:2003dk}. The measurement of
\babar leads to $A_{\rm CP} = -0.010 \pm 0.030_{\rm stat} \pm
0.014_{\rm syst}$ for $M_{X_s} < 2.8\gev$~\cite{:2008gvb} while Belle
finds $A_{\rm CP} = 0.002 \pm 0.050_{\rm stat} \pm 0.030_{\rm syst}$
for $M_{X_s} < 2.1\gev$~\cite{Nishida:2003yw}.

\babar recently reported a first measurement of $B \to X_d\g$ using
the semi-inclusive approach~\cite{:2008ig}. In this analysis, seven
exclusive final states in the range $0.6\gev < M_{X_d} < 1.8\gev$ are
reconstructed.  Although the analysis suffers from a large background
from continuum events, mis-reconstructed $B \to X_s\g$ events, and an
large uncertainty in missing modes, \babar obtained the branching
fraction in this mass range to be $(7.2 \pm 2.7_{\rm stat} \pm
2.3_{\rm syst}) \times 10^{-6}$.

In the $B$ recoil method, one of the two produced $B$ mesons is fully
reconstructed in a hadronic mode, and an isolated photon is identified
in the rest of the event.  This method provides a very clean signal,
and one obtains simultaneously the flavor, charge, and momentum of the $B$ meson.
The drawback is a very low efficiency.  In the analysis
with $210\, \invfb$ by \babar~\cite{Aubert:2007my}, $6.8 \times 10^5$
$B$ mesons are tagged and $119 \pm 27$ signal events are found.  The
result is limited by statistics and is not competitive with the other
methods listed in Tab.~\ref{tab:6:bsgamma-exp-incl-bf}.  However, this
method is promising for a future super flavor factory.
\begin{table}
 \begin{center}
   \caption{\label{tab:6:bsgamma-exp-incl-bf}%
     Inclusive branching fractions of radiative $B$ decays.
     $E_\gamma^\mathrm{min}$ and $\mathcal{B}(E_\gamma >
     E_\gamma^\mathrm{min})$ are the minimum energy and branching
     fraction reported in the paper, while $\mathcal{B}(E_\gamma >
     1.6\gev)$ is the rescaled branching fraction.  The size of the
     data sets is given in units of $\invfb$ and the branching
     fractions are in units of $10^{-6}$.}
   \catcode`;=\active \def;{\phantom{0}}
  \begin{tabular}{lccccc}
   \hline\hline
   \multicolumn{1}{c}{Method} & Data set & $E_\gamma^\mathrm{min}$
   & $\mathcal{B}(E_\gamma > E_\gamma^\mathrm{min})$
   & $\mathcal{B}(E_\gamma > 1.6\gev)$ & Ref. \\ \hline
   CLEO fully inclusive & $;;9$ & $2.0$ & $305 \pm 41 \pm 26$
	       & $329 \pm 53$ & \cite{Chen:2001fja} \\
   \babar fully inclusive & $;82$ & $1.9$ & $367 \pm 29 \pm 34 \pm 29$
	       & $392 \pm 56$ & \cite{Aubert:2006gg} \\
   \babar semi-inclusive & $;82$ & $1.9$
	   & $327 \pm 18  \,^{+55}_{-40} \,^{+4}_{-9}$
	       & $349 \pm 57$ & \cite{Aubert:2005cua} \\
   \babar $B$-recoil & $210$ & $1.9$ & $366 \pm 85 \pm 60$
	       & $391 \pm 111$ & \cite{Aubert:2007my} \\
   Belle semi-inclusive & $;;6$ & $2.24$ & ---
	       & $369 \pm 94$ & \cite{Abe:2001hk} \\
   Belle fully inclusive & $605$ & $1.7$ & $332 \pm 16 \pm 37 \pm 1$
	       & $337 \pm 43$ & \cite{Abe:2008sx} \\ \hline
   Average     & -- & -- & -- & $352 \pm 23 \pm 9$ & \\ \hline
   Theory prediction & -- & -- & -- & $315 \pm 23$ & \cite{Misiak:2006zs} \\
   \hline\hline
  \end{tabular}
 \end{center}
\end{table}

\subsubsection{Theory of photon energy spectrum and moments}

The basic motivation to study the photon energy spectrum in $B \to X_s
\gamma$ is the fact that backgrounds prohibit a measurement of the
branching fraction for non-hard photons. Despite significant progress,
the current measurements still have sizable errors below $E_{\gamma}
\sim 2\gev$. Raising the photon energy cut $E_\gamma > E_0$
significantly increases the accuracy of the measurements, but requires
an larger extrapolation to the ``total'' width, thereby introducing
some model dependence.

In contrast to the branching ratio, the photon energy spectrum is
largely insensitive to NP \cite{Kagan:1998ym}. It can thus be used for
precision studies of perturbative and non-perturbative
strong-interaction effects. In particular, the measured spectrum
allows to extract the value of the mass of the bottom quark from its
first moment $\left\langle E_{\gamma}\right \rangle \sim m_b/2$, its
average kinetic energy $\mu_{\pi}^2$ from its second moment, and gives
direct information on the importance of the $B$ meson ``shape
function'' for different values of $E_0$. The measurements of $m_b$
and $\mu_{\pi}^2$ using $B\to X_s \gamma$ are complementary to the
determinations using the inclusive moments of $B\to X_c
\ell{\bar{\nu}}$. Fits to the measured moments
\cite{Buchmuller:2005zv, Schwanda:2008kw} based on the
1S~\cite{Bauer:2004ve} and the kinetic scheme \cite{Bigi:1993ex,
  Gambino:2004qm} have been very useful, and constitute today an
important input to the determination of
$|V_{ub}|$~\cite{Barberio:2008fa}.

The calculation of the $B \to X_s \gamma$ photon spectrum is a complex
theoretical problem. First of all, there is no unique way to define
the total $B \to X_s \gamma$ width owing to both the soft divergence
of the $(Q_7, Q_8)$ interference term and the possibility of secondary
photon emission in non-radiative $b\to s$ decays. Furthermore, the
local OPE in $B \to X_s \gamma$ does not apply to contributions from
operators other than $Q_7$, in which the photon couples to light
quarks \cite{Kapustin:1995fk, Ligeti:1997tc}. 
In the case of the $(Q_7, Q_8)$ interference, the resulting ${\cal O} (\alpha_s \Lambda_{\rm
  QCD}/m_b)$ corrections to the total rate have been estimated in
\cite{Lee:2006wn}. A detailed study of the impact of these and similar
enhanced non-local power-corrections on the photon energy spectrum is
in progress~\cite{}.

Even in the case of the $Q_7$ self-interference, where an local OPE
for the total width exists, the calculation of the spectrum is highly
non-trivial. The main difficulty arises due to the jet-like structure
of the decay process, where an energetic hadronic system, $E_{X_s}\sim
E_{\gamma} \sim {\cal O}(m_b/2)$, with a relatively small mass, ${\cal
  O}(\sqrt{m_b \Lambda_{\rm QCD}})$, recoils against the photon.

In the endpoint region, the $B \to X_s \gamma$ spectrum can be
computed as a convolution between a perturbatively calculable
coefficient function and the quark distribution function
$S(l^+)$. While the non-perturbative content of $S(l^+)$ is in
principle clear, a calculation by existing non-perturbative methods is
not possible, so that in practice one must model the function $S(l^+)$
using a suitable parametrization. Information on the corresponding
model parameters can be obtained from the experimental measurements of
the first few moments of inclusive decay spectra, which in turn
determine the moments of the ``shape function''.

Significant progress has been made since the first dedicated
calculation of the spectrum \cite{Kagan:1998ym}. The state-of-the-art
calculations are based on the factorization picture of inclusive
decays near the endpoint \cite{Korchemsky:1994jb}. Consider for
example the measurement of the partial $B \to X_s \gamma$ width with
$E_{\gamma}>E_0$. The key observation is that near the endpoint, i.e.,
for $\Delta \ll m_b$, and to leading order in $\Lambda_{\rm QCD}/m_b$
there are three separate dynamical processes which are quantum
mechanically incoherent. A soft subprocess, $S$, which is characterized by soft gluons
with momenta of order $\Delta=m_b-2E_{\gamma}$, a jet subprocess, $J$,
summing up collinear hard radiation with virtualities of order of
$\sqrt{m_b\Delta}$, and a hard function, $H$, associated with virtual
gluons with momenta of order $m_b$. The factorized decay width takes
the form
\begin{equation}
\label{eq:soft_jet_hard_factorization}
\Gamma(\Delta)=H(m_b) J(\sqrt{m_b\Delta}) \otimes S(\Delta) 
\,+\,{\cal O}(\Lambda_{\rm QCD}/m_b) \,.
\end{equation}
This factorization formula, originally proposed in
\cite{Korchemsky:1994jb}, was rederived in the framework of
SCET~\cite{Bauer:2000ew, Bosch:2004th}. It serves as a basis for a
range of approaches, facilitating the resummation of large Sudakov
logarithms associated with the double hierarchy of scales $m_b\gg
\sqrt{m_b\Delta}\gg \Delta$. This includes DGE \cite{Gardi:2004ia,
  Gardi:2005yi, Andersen:2005mj, Andersen:2005bj, Andersen:2006hr} and
a multi-scale OPE (MSOPE)~\cite{Neubert:2005nt, Becher:2005pd,
  Becher:2006qw, Becher:2006pu}.

Using SCET, it is possible to systematically define additional
non-local operators that contribute to the decay spectra at subleading
powers of $\Lambda_{\rm QCD}/m_b$~\cite{Bauer:2001mh, Lee:2004ja,
  Beneke:2004in}. Unfortunately, the corresponding subleading ``shape
functions'' are not well constrained since starting at ${\cal
  O}(\Lambda_{\rm QCD}/m_b)$ the number of functions exceeds the
number of observables. Thus, estimating non-perturbative corrections
to (\ref{eq:soft_jet_hard_factorization}) remains a notoriously
difficult task.

While (\ref{eq:soft_jet_hard_factorization}) only holds for $\Delta
\ll m_b$, its use may vary depending on the extent at which effects on
the lowest scale are described by perturbation theory. If $\Delta \gg
\Lambda_{\rm QCD}$ it is useful to compute the quark distribution
function perturbatively~\cite{Gardi:2004ia, Becher:2005pd} rather than to
parametrize it. In contrast, when $\Delta \sim \Lambda_{\rm QCD}$ this
function becomes non-perturbative. Two different approaches based on
SCET have been developed to deal with these two regimes, MSOPE for the
former and a formalism based on parametrization of the shape functions
for the latter \cite{Bosch:2004th}. In contrast, DGE, which is at the
outset derived in the regime $\Delta \gg \Lambda_{\rm QCD}$, has been
extended to the regime $\Delta \sim \Lambda_{\rm QCD}$ by constraining
the Borel transform of $S(l^+)$ and then parametrizing
non-perturbative corrections depending on the soft scale in moment
space.

Beyond the conceptual issues discussed so far, much progress has been
made on the calculation side. In particular, the $Q_7$
self-interference part of the spectrum has been computed to NNLO
accuracy~\cite{Melnikov:2005bx, Asatrian:2006sm}. In addition, all the
necessary ingredients for Sudakov resummation at NNLO of both the soft
and the jet function are in place \cite{Korchemsky:1992xv,
  Gardi:2004ia, Moch:2004pa, Becher:2005pd, Becher:2006qw}. Some
additional higher-order corrections are known, in particular,
running-coupling corrections \cite{Ligeti:1999ea, Gardi:2005yi,
  Andersen:2006hr}, but unfortunately complete NNLO calculations of
the $(Q_7, Q_{1,2})$ and $(Q_{1,2}, Q_{1,2})$ interference terms are
not available at present.

Systematic NNLO analysis of the $B \to X_s \gamma$ branching fraction
and spectrum have been performed by three groups \cite{Misiak:2006zs,
  Andersen:2006hr, Becher:2006pu}. While the first analysis works at
fixed-order in perturbation theory, the latter two articles are based
on Sudakov resummation utilizing
(\ref{eq:soft_jet_hard_factorization}). The MSOPE result has been
combined with the fixed-order predictions by computing the fraction of
events $1-T$ that lies in the range $E_0 = [1.0, 1.6]\gev$. 
The analysis \cite{Becher:2006pu} finds $1- T = 0.07
{^{+0.03}_{-0.05}}_{\rm pert} \pm 0.02_{\rm hard} \pm 0.02_{\rm
  pars}$, where the individual errors are perturbative, hadronic, and
parametric. The quoted value is almost twice as large as the estimate
$1-T = 0.04 \pm 0.01_{\rm pert}$ obtained in fixed-order perturbation
theory \cite{Misiak:2006zs}. In contrast, in the DGE approach
\cite{Andersen:2006hr} one finds a much thinner tail of the photon
energy spectrum at NNLO, $1- T = 0.016 \pm 0.003_{\rm pert}$, which is
consistent with the result obtained in fixed order perturbation
theory. 

Given the common theoretical basis for the resummation, the opposite
conclusions drawn in \cite{Becher:2006pu} and \cite{Andersen:2006hr}
may look surprising. The main qualitative differences between the two
calculations are as follows. First, the result \cite{Becher:2006pu} is
plagued by a significant additional theoretical error related to
low-scale, $\mu \sim \Delta$, perturbative corrections, indicating the
presence of large subleading logarithmic corrections to the soft
function. In contrast, the DGE approach \cite{Andersen:2006hr}
supplements Sudakov resummation with internal resummation of
running-coupling corrections, which is necessary to cure the endpoint
divergence of the fixed-logarithmic-accuracy expansion. Second, the
MSOPE approach \cite{Becher:2006pu} identified a high sensitivity to
the matching procedure, dealing with terms that are suppressed by
powers of $\Delta/m_b$ for $E_\gamma \sim m_b/2$, but are not small
away from the endpoint \cite{Andersen:2006hr, Misiak:2008ss}. The
analysis \cite{Andersen:2006hr}, on the other hand, has used
additional information on the small $E_{\gamma}$ behavior of the
different interference terms, which is known to all orders in
perturbation theory, to extend the range of applicability of
resummation to the tail region.

In conclusion, progress on the theory front, in particular in
factorization and resummation of perturbation theory, and in explicit
higher order calculations, significantly improved our knowledge of the
photon energy spectrum in $B \to X_s \gamma$. Nevertheless,
uncertainties of both perturbative and non-perturbative origin remain,
which deserve further theoretical investigations.

\subsubsection{Experimental results of photon energy spectrum and moments}
\label{sec:6:bsgamma_photon_energy}

In the case of the semi-inclusive and $B$ recoil methods, the photon
energy spectrum can be measured directly in the $B$ meson rest frame.
The semi-inclusive method suffers from large uncertainty from
the hadronic system, while the $B$ recoil method requires much more
statistics. Presently, precise measurements of the photon energy
spectrum are therefore provided only with the fully inclusive method.

In the fully inclusive method, it is not possible to know the momentum
of the $B$ meson for each photon, so only the photon energy
distribution in the $\Upsilon(4S)$ rest frame is directly
measurable. As a result the raw photon energy spectrum has to be
corrected not only for the photon detection efficiency of the
calorimeter and other selection efficiency, but also for the smearing
effect between the $B$ meson and the $\Upsilon(4S)$ frame. The energy
spectrum is also smeared by the response of the calorimeter.  The
correction due to smearing, which is also referred to as
``unfolding'', depends on the signal models and its parameters.
CLEO~\cite{Chen:2001fja} uses a signal model by Ali and
Greub~\cite{Ali:1995bi}, while a model by Kagan and
Neubert~\cite{Kagan:1998ym} is used by both
\babar~\cite{Aubert:2006gg} and Belle~\cite{Abe:2008sx}. The latter
collaboration also considers other models~\cite{Lange:2005qn,
  Lange:2005yw, Andersen:2006hr, Benson:2004sg}.

The photon energy spectrum is often represented in terms of the first
two moments, i.e., mean and variance, above a certain energy
threshold.  For example, \babar~\cite{Aubert:2006gg} obtains $\langle
E_\gamma \rangle = (2.346 \pm 0.032_{\rm stat} \pm 0.011_{\rm syst}
)\gev$ and $\langle E^2_\gamma \rangle - \langle E_\gamma \rangle^2 =
(0.0226 \pm 0.0066_{\rm stat} \pm 0.0020_{\rm syst})\gev^2$ for
$E_\gamma > 2.0\gev$, while Belle~\cite{Abe:2008sx} $\langle E_\gamma
\rangle = (2.281 \pm 0.032_{\rm stat} \pm 0.053_{\rm syst} \pm
0.002_{\rm boos}) \, {\rm GeV}$ and $\langle E^2_\gamma \rangle -
\langle E_\gamma \rangle^2 = (0.0396 \pm 0.0156_{\rm stat} \pm
0.0214_{\rm syst} \pm 0.0012_{\rm boos}) \, {\rm GeV}^2$ for $E_\gamma
> 1.7 \, {\rm GeV}$, where the last errors in the Belle measurements
are from the boost correction. In these measurements, branching
fractions and moments with different photon energy thresholds are also
obtained. Parameters useful for the $\Vcb$ and $\Vub$ determinations
such as the bottom-quark mass $m_b$ or its mean momentum squared
$\mu_\pi^2$ can be obtained by fitting the theoretical predictions to
the measured moments. This is discussed in detail in 
Sec.~\ref{sec:inclusiveVcb}. 

One of the challenges in the measurement is to lower the energy
threshold of the photon. The contamination of the background from \B
decays becomes more severe rapidly, as the threshold is lowered,
especially in the region below $2\gev$.  So far,
with growing data sets, measurements with lower photon energy
threshold have been performed. However, the results with lower cut on
the photon energy tend to give larger systematic, and
model errors. This raises the question which threshold value is
optimal to determine $\Vcb$ and $\Vub$. Another issue is that the
uncertainty of $m_b$ is included in the signal model error for the
measurements of the moments, but the measured moments themselves are
used to determined $m_b$. The latter issue could probably be avoided
by performing a simultaneous determination of the parameters in
question from the raw photon energy spectrum.

\subsection{Exclusive $B\to V \gamma$ decays}

\subsubsection{Theory of exclusive $B\to V \gamma$ decays}
\label{sec:rare:theory_excl}

The exclusive decays $B_{(s)}\to V\gamma$, with $V \in
\{K^\ast,\rho,\omega,\phi\}$, are mediated by FCNCs and thus test the
flavor sector in and beyond the SM. After matching onto the effective
Lagrangian (\ref{eq:rare:Leff}), the main theoretical challenge is to
evaluate the hadronic matrix elements of the operators $Q_{1-8}$. QCDF
is a model-independent approach based on the heavy-quark expansion
\cite{Beneke:2001at, Bosch:2001gv, Ali:2001ez}, and the bulk of this
section is devoted to describing this formalism.  At the end of the
section we briefly mention the ``perturbative QCD'' (pQCD) approach
\cite{Keum:2004is, Lu:2005yz, Matsumori:2005ax}. Although the hadronic
uncertainties inherent to the exclusive decay modes are a barrier to
precise predictions, we shall see that the exclusive decays
nonetheless provide valuable information on the CKM elements
$|V_{td}/V_{ts}|$ and allow to put constraints on the chiral structure
of possible non-standard interactions.

QCDF is the statement that in the heavy-quark limit the hadronic
matrix element of each operator in the effective Lagrangian can be
written in the form
\begin{equation} \label{eq:rare:ff}
  \left \langle V \gamma \left | Q_i
    \right | \bar B \right \rangle = T_i^{\rm I}  \, F^{B \to V_\perp} +
  \int_0^{\infty}\frac{d\omega}{\omega}\phi^B_+(\omega)
  \int_0^1 du \, \phi_{\perp}^V(u)
  T^{\rm II}_i(\omega,u)
  \,
  +{\cal O}\left(\frac{\Lambda_{\rm QCD}}{m_b}\right) \,.
\end{equation}
The form factor $F^{B\to V_\perp}$ and the light-cone distribution
amplitudes (LCDAs) $\phi^B_+, \, \phi^V_\perp$ are non-perturbative,
universal objects. The hard-scattering kernels $T_i^{\rm I, II}$ can
be calculated as a perturbative series in $\alpha_s$. The elements
$T_i^{\rm I}$ ($T_i^{\rm II}$) are referred to as ``vertex
corrections'' (``spectator corrections''). The hard-scattering kernels
have been known completely at order $\alpha_s$ (NLO) for some time
\cite{Beneke:2001at, Bosch:2001gv, Ali:2001ez}, and recently some of
the $\alpha_s^2$ (NNLO) corrections have also been calculated
\cite{Ali:2007sj}.

An all orders proof of the QCDF formula (\ref{eq:rare:ff}) was
performed in \cite{Becher:2005fg}, using the technology of SCET. The
EFT approach also allows to separate physics from the two perturbative
scales $m_b$ and $\sqrt{m_b\Lambda_{\rm QCD}}$, and to resum
perturbative logarithms of their ratio using the RG. The numerical
impact of this resummation has been investigated in \cite{Ali:2007sj,
  Becher:2005fg}.

The predictive power of QCDF is limited by hadronic uncertainties
related to the LCDAs and QCD form factors, as well as by power
corrections in $\Lambda_{\rm QCD}/m_b$. For instance, the form factors
$F^{B\to V_\perp}$ can be calculated with QCD sum rules to an accuracy
of about $15\%$, which implies an uncertainty of roughly $30\%$ on the
$B\to V\gamma$ branching fractions. More troublesome is the issue of
power corrections. A naive dimensional estimate indicates that these
should be on the order of $10\%$, but this statement is hard to
quantify. Since SCET is an effective theory which sets up a systematic
expansion in $\alpha_s$ and $\Lambda_{\rm QCD}/m_b$, it has the
potential to extend the QCDF formalism to subleading order in
$\Lambda_{\rm QCD}/m_b$.  However, in cases where power corrections
have been calculated, the convolution integrals over momentum
fractions do not always converge \cite{Kagan:2001zk}. These ``endpoint
divergences'' are at present a principle limitation on the entire
formalism.

Although a comprehensive theory of power corrections is lacking, it is
nonetheless possible to estimate some of the corrections which are
believed to be large, or which play an important role in
phenomenological applications. One such correction stems from the
annihilation topology, which has been shown to factorize at leading
order in $\alpha_s$ \cite{Bosch:2001gv}. Annihilation gives the
leading contribution to isospin asymmetries, and is also important for
$B^{\pm}\to \rho^{\pm}\gamma$ branching fractions, where it is
enhanced by a factor of $C_{1,2}/C_7$. The $\Lambda_{\rm QCD}/m_b$
corrections from annihilation have been included in all recent
numerical studies \cite{Ball:2006eu, Bosch:2004nd, Beneke:2004dp,
  Ali:2006fa}, and part of the $\Lambda_{\rm QCD}^2/m_b^2$ correction,
so-called ``long-distance photon emission'', has been calculated in
\cite{Ball:2006eu}. Some additional $\alpha_s \Lambda_{\rm QCD}/m_b$
corrections from annihilation and spectator scattering needed to
calculate isospin asymmetries were dealt with in
\cite{Kagan:2001zk}. Corrections from three-particle Fock states in
the $B$ and $V$ mesons, most significant for indirect CP asymmetries,
were estimated in \cite{Ball:2006eu}.

We now give numerical results for some key observables in $B\to V
\gamma$ decays, and compare them with experiment. The ratio of $B\to
K^\ast \gamma$ and $B\to \rho \gamma$ branching fractions is useful
for the determination of $|V_{td}/V_{ts}|$. To understand why this is
the case, consider the expression
\begin{equation} \label{eq:rare:BrRat}
  \frac{\BR(B^0\to \rho^0\gamma)}{\BR(B^0\to
    K^{\ast0}\gamma)}
  =\frac{1}{2\xi^2}\left|\frac{V_{td}}{V_{ts}}\right|^2
  \bigg[1-2 R_{ut} \epsilon_0  \cos \alpha \cos\theta_0 +R_{ut}^2 
  \epsilon_0^2\bigg] \,.
\end{equation}
Analogous expressions hold for charged decays and $B\to \omega\gamma$.
The quantities $\epsilon_0$ and $\cos \theta_0$ can be calculated in
QCDF, and vanish at leading order in $\Lambda_{\rm QCD}/m_b$ and
$\alpha_s$. Beyond leading order they are approximately $10\%$, but
the factor inside the brackets remains close to unity, due to a
additional suppression from the CKM factors $\cos \alpha \sim 0.1$ and
$R_{ut}=|(V_{ud} V_{ub})/(V_{td}V_{tb})| \sim 0.5$. Therefore, by far,
the dominant theoretical uncertainty is related to the form factor
ratio $\xi = F^{B\to K^*}/F^{B\to \rho}$. The ratio of form factors
can be calculated with better accuracy than the form factors
themselves and has been estimated using light-cone sum rules to be
$1.17\pm 0.09$ \cite{Ball:2006eu}. Extracting $|V_{td}/V_{ts}|$ from
(\ref{eq:rare:BrRat}) and averaging with determinations from the
charged mode and the $B\to \omega\gamma$ decay yields the results
given in Sec.~\ref{sec:6:VtdVts}.

Direct and isospin CP asymmetries, $A_{\rm CP}$ and $A_{\rm I}$,
provide useful tests of the SM and the QCDF approach. In QCDF, direct
CP asymmetries in $B\to V\gamma$ decays are suppressed by at least one
power of $\alpha_s$ and isospin asymmetries by at least one power of
$\Lambda_{\rm QCD}/m_b$, so both of these are predicted to be
small. We first consider $B\to \rho\gamma$ decays. In that case the
QCDF prediction for the direct CP asymmetry is about $-10\%$
\cite{Beneke:2004dp, Ali:2006fa} and agrees well with the recent
experimental results quoted in
Sec.~\ref{sec:rare_exclusive_bsgamma_exp}. The QCDF result for
$A_{\rm I}$ depends strongly on $\cos \alpha$, but in the preferred
range of $\alpha$ near $90^{\circ}$ is roughly between zero and
$-10\%$ \cite{Beneke:2004dp, Ali:2006fa}.  Values closer to the
central experimental value can be generated if one assumes a large
contribution from non-perturbative charming penguins
\cite{Kim:2008rz}, which would be in contradiction with the power
counting of QCDF.  Given the large experimental errors it is not yet
possible to draw a definite conclusion. For $B\to K^\ast \gamma$
decays, the direct CP asymmetries are strongly suppressed due to the
CKM structure of the decay amplitude.  The isospin asymmetry comes out
to be $(3\pm 4)\%$ \cite{Barberio:2008fa}, which is compatible with
predictions from QCDF \cite{Ball:2006eu, Kagan:2001zk, Bosch:2004nd,
  Beneke:2004dp}.  This isospin asymmetry is very sensitive to the
magnitude and sign of the ratio $C_6/C_7$.

Finally, we consider indirect CP asymmetries. In the SM, these are
suppressed by powers of $m_{s,d}/m_b$ or arise from the presence of
three-particle Fock states in the $B$ and $V$ mesons, which are
$\Lambda_{\rm QCD}/m_b$ corrections to the leading order factorization
formula \cite{Grinstein:2004uu}. A calculation performed in
\cite{Ball:2006eu} indicates that the corrections from three-particle
Fock states are much smaller than the generic size of a $\Lambda_{\rm
  QCD}/m_b$ power correction, so that the indirect CP asymmetries are
estimated to be below the $3\%$ level for all decay modes. The
asymmetries could be much larger in extensions of the SM with altered
chiral structure such as left-right symmetric models
\cite{Ball:2006eu}. The current experimental results are within their
large errors consistent with zero \cite{Ushiroda:2006fi,
  Aubert:2008gy}.

A modified implementation of the heavy-quark expansion is provided by
the pQCD approach \cite{Keum:2004is, Lu:2005yz, Matsumori:2005ax}.
The main difference compared to QCDF is that pQCD attempts to
calculate the QCD form factors perturbatively. The assumptions
required for such a treatment have been questioned in
\cite{DescotesGenon:2001hm}. However, numerical results for most
observables are in rough agreement with those from QCDF. A recent
comparison between the branching fractions, isospin and CP asymmetries
obtained within the two theoretical setups can be found in
\cite{Ball:2006eu}.

\subsubsection{Experimental results for exclusive $B \to V \gamma$ decays}
\label{sec:rare_exclusive_bsgamma_exp}

The exclusive reconstruction of radiative $B \to V\g$ decays or other
multi-body decays such as $B \to K\pi\gamma$ is usually
straightforward. The dominant background originates from the continuum
process $e^+e^- \to \qqbar$, which is experimentally suppressed by
means of event shape variables.

Vetoing high energetic photons from $\pi^0$ or $\eta$ is also useful.
The background from $\B$ decays is small in the low hadronic mass
region, but becomes larger for higher hadronic mass, i.e., lower
photon energy. Therefore, in the analysis of the exclusive final
states with more than two particles, it is necessary to apply a cut on
the hadronic mass, which is typically around $2$ to $2.5 \, {\rm
  GeV}$. The contribution of the cross-feed from radiative $\B$ decays
to other final states also becomes a significant background in some
modes.

The first observation of radiative $\B$ decays has been established in
1993 by CLEO \cite{Ammar:1993sh} by a measurement of the $B \to
\Kstar\g$ mode. They found 13 events in the signal region in a data
sample of $1.4\, \invfb$, and measured the branching fraction
$\mathcal{B}(\B \to \Kstar\g) = (45 \pm 15_{\rm stat} \pm 3_{\rm
  syst}) \times 10^{-6}$. Now, the measurements by \babar and Belle are
based on data set that are more than 100 times larger and start to be
dominated by systematics, as can be seen from
Tab.~\ref{tab:6:bsgamma-exp-excl-bf}. Unfortunately, it is not easy
to predict the branching fractions of exclusive modes precisely, and
hence it is difficult to compare the results with theory.

\begin{table}
 \begin{center}
   \caption{\label{tab:6:bsgamma-exp-excl-bf}%
     Measured branching fractions of radiative $B$ decays.  Only modes
     with evidence are listed.  The size of the data sets is given in
     the units of $\invfb$.  }
  \catcode`;=\active \def;{\phantom{0}}
  \begin{tabular}{c|ccc|ccc}
    \hline \hline
    & \multicolumn{3}{c}{Belle} & \multicolumn{3}{c}{\babar} \\
    Mode & $\BR$ ($10^{-6}$) & Data set & Ref.
    & $\BR$ ($10^{-6}$) & Data set & Ref. \\ \hline
    $\Bz \to \Kstarz\g$
    & $40.1 \pm 2.1 \pm 1.7$ & $;78$ & \cite{Nakao:2004th}
    & $45.8 \pm 1.0 \pm 1.6$ & $347$ & \cite{:2008cy} \\
    $\Bz \to \Kstarp\g$
    & $42.5 \pm 3.1 \pm 2.4$ & $;78$ & \cite{Nakao:2004th}
    & $47.3 \pm 1.5 \pm 1.7$ & $347$ & \cite{:2008cy} \\
    $\Bp \to K_1(1270)^+\g$
    & $43 \pm 9 \pm 9$ & $140$ & \cite{Yang:2004as} & -- & -- & -- \\
    $\Bz \to K_2^*(1430)^0\g$
    & $13 \pm 5 \pm 1$ & $;29$ & \cite{Nishida:2002me}
    & $12.2 \pm 2.5 \pm 1.0$ & $;81$ & \cite{Aubert:2003zs} \\
    $\Bp \to K_2^*(1430)^+\g$ & -- & -- & --
    & $14.5 \pm 4.0 \pm 1.5$ & $;81$ & \cite{Aubert:2003zs} \\
   $\Bp \to \Kp\eta\g$
   & $8.4 \pm 1.5 \,^{+1.2}_{-0.9}$ & $253$ & \cite{Nishida:2004fk}
   & $7.7 \pm 1.0 \pm 0.4$ & $423$ & \cite{Aubert:2008js} \\
   $\Bz \to \Kz\eta\g$
   & $8.7 \,^{+3.1}_{-2.7} \,^{+1.9}_{-1.6}$ & $253$ & \cite{Nishida:2004fk}
   & $7.1 \,^{+2.1}_{-2.0} \pm 0.4$ & $423$ & \cite{Aubert:2008js} \\
   $\Bp \to \Kp\etapr\g$
   & $3.2 \,^{+1.2}_{-1.1} \pm 0.3$ & $605$ & \cite{:2008ru}
   & --& -- & -- \\
   $\Bp \to \Kp\phi\g$
   & $3.4 \pm 0.9 \pm 0.4$ & $;90$ & \cite{Drutskoy:2003xh}
   & $3.5 \pm 0.6 \pm 0.4$ & $211$ & \cite{Aubert:2006he} \\
   $\Bp \to \proton\Lbar\g$
   & $2.45 \,^{+0.44}_{-0.38} \pm 0.22$ & $414$ & \cite{Wang:2007as}
   & -- & -- & -- \\
   $\Bp \to \Kp\pim\pip\g$
   & $25.0 \pm 1.8 \pm 2.2$ & $140$ & \cite{Yang:2004as}
   & $29.5 \pm 1.3 \pm 2.0$ & $211$ & \cite{Aubert:2005xk} \\
   $\Bp \to \Kz\pip\piz\g$ & -- & -- & --
   & $45.6 \pm 4.2 \pm 3.1$& $211$ & \cite{Aubert:2005xk} \\
   $\Bz \to \Kz\pip\pim\g$
   & $24.0 \pm 4.0 \pm 3.0$ & $140$ & \cite{Yang:2004as}
   & $18.5 \pm 2.1 \pm 1.2$ & $211$ & \cite{Aubert:2005xk} \\
   $\Bz \to \Kp\pim\piz\g$ & -- & -- & --
   & $40.7 \pm 2.2 \pm 3.1$ & $211$ & \cite{Aubert:2005xk} \\
   $B_s^0 \to \phi\g$ & $57 \,^{+18}_{-15} \,^{+12}_{-11}$
       & $;24$ &\cite{:2007ni}
   & --  & --  & -- \\ \hline
   $\Bp \to \rho^+\g$ & $0.87 \,^{+0.29}_{-0.27} \,^{+0.09}_{-0.11}$
       & $605$ & \cite{Taniguchi:2008ty}
   & $1.20 \,^{+0.42}_{-0.37} \pm 0.20$ & $423$ & \cite{:2008gf} \\
   $\Bz \to \rho^0\g$ & $0.78 \,^{+0.17}_{-0.16} \,^{+0.09}_{-0.10}$
       & $605$ & \cite{Taniguchi:2008ty}
   & $0.97 \,^{+0.24}_{-0.22} \pm 0.06$ & $423$ & \cite{:2008gf} \\
   $\Bz \to \omega\g$ & $0.40 \,^{+0.19}_{-0.17} \pm 0.13$
       & $605$ & \cite{Taniguchi:2008ty}
   & $0.50 \,^{+0.27}_{-0.23} \pm 0.09$ & $423$ & \cite{:2008gf} \\
   \hline\hline
  \end{tabular}
 \end{center}
\end{table}

What can be predicted more precisely are the direct CP or charge
asymmetry $A_{\rm CP}$ and the isospin asymmetry $A_\mathrm{I}$. They are
defined as
\begin{equation} \label{eq:rare:ACPAI}
\begin{split}
  A_{\rm CP} &= \frac{\Gamma(\Bbar \to \Kstarb\g) - \Gamma(\B \to
    \Kstar\g)}{%
    \Gamma(\Bbar \to \Kstarb\g) + \Gamma(\B \to \Kstar\g)} \,, \\[2mm]
  A_\mathrm{I} &= \frac{\Gamma(\Bz \to \Kstarz\g) - \Gamma(\Bp \to
    \Kstarp\g)}{%
    \Gamma(\Bz \to \Kstarz\g) + \Gamma(\Bp \to \Kstarp\g)} \,,
\end{split}
\end{equation}
and similarly for the other decay modes. In the case of $B \to
\Kstar\g$, \babar obtained $A_{\rm CP} = -0.009 \pm 0.017_{\rm stat}
\pm 0.011_{\rm syst}$ and $A_\mathrm{I} = 0.029 \pm 0.019_{\rm stat}
\pm 0.016_{\rm syst} \pm 0.018_{\rm prod}$~\cite{:2008cy} while the
results of Belle read $A_{\rm CP} = -0.015 \pm 0.044_{\rm stat} \pm
0.012_{\rm syst}$ and $A_\mathrm{I} = 0.034 \pm 0.044_{\rm stat} \pm
0.026_{\rm syst} \pm 0.025_{\rm prod}$~\cite{Nakao:2004th}. The last
errors in $A_{\rm I}$ arise from the production ratio of $B^0$ and
$B^+$ for which \babar and Belle assume the values $1.044 \pm 0.050$
and $1.020 \pm 0.034$, respectively. The direct CP asymmetry has also
been measured in the $B \to \rho \gamma$ system by Belle which finds
$A_{\rm CP}=-0.11\pm0.32_{\rm stat}\pm 0.09_{\rm syst}$
\cite{Taniguchi:2008ty}. The corresponding experimental results for
the isospin asymmetry read $A_{\rm I}=-0.43{^{+0.25}_{-0.22}}_{\rm
  stat}\,\pm 0.10_{\rm syst}$ from \babar \cite{:2008gf} and $A_{\rm I}
=-0.48{^{+0.21}_{-0.19}}_{\rm stat}\,{^{+0.08}_{-0.09}}_{\rm syst}$
from Belle \cite{Taniguchi:2008ty}. Within errors, the measured values
of $A_{\rm CP}$ and $A_{\rm I}$ are consistent with the SM predictions
discussed in Sec.~\ref{sec:rare:theory_excl}.

Another important variable is the time-dependent CP asymmetry.  In the
SM, the photon from the $b \to s\g$ process is almost polarized.
Photons from $\Bz$ are right-handed, while photons from $\Bzb$ are
left-handed.  So if the photon is completely polarized, $\Bz$ and
$\Bzb$ cannot decay into a common final state, and mixing-induced CP
violation does not happen. Indeed, the time-dependent CP violation
(tCPV) in radiative $\B$ decays $\B \to f_{\rm CP}\g$, where $f_{\rm
  CP}$ denotes a CP eigenstate, is expected to be within a few percent
even when we consider the possible enhancement due to the strong
interaction. Therefore, the measurement of tCPV for $b \to s\g$ is a
probe of the photon polarization, and large values of tCPV would be a
signal of the presence of non-standard right-handed interactions.

The final state in $\Kstarz \to \KS\piz$ is a CP eigenstate, but it is
not essential whether the decay goes through $\Kstarz$ or not.
Actually, final states can be any of the type $P_1P_2\g$, where $P_1$
and $P_2$ are pseudoscalar mesons~\cite{Atwood:2004jj}. In
consequence, the measurements have been performed not only for $\B \to
\Kstarz\g \to \KS\piz\g$ but also for the non-resonant mode $\B \to
\KS\piz\g$. In Tab.~\ref{tab:6:bsgamma-exp-excl-tcpv} we list the
measured $S$ terms of the various tCPV. Since the final state
$\KS\piz\g$ does not include charged tracks that come from the $\B$
vertex, the $\B$ decay vertex has to be calculated using the $\KS$
trajectory, which causes lower efficiency.  Although the error is
still large, the result is consistent with vanishing CP asymmetry.

Many other exclusive final states have also been found by \babar and
Belle. Tab.~\ref{tab:6:bsgamma-exp-excl-bf} shows the decays with
experimental evidence and their branching fractions. Radiative decays
through kaonic resonances are observed for $\B \to K_2^*(1430)\g$ and
$\B \to K_1(1270)\g$, in addition to $\B \to \Kstar\g$.  The other
listed modes are three- or four-body decays.  Measurements of these
branching ratios provide a better understanding of the composition of
$b \to s\g$ final states, and potentially reduce the systematic errors
due to hadronization in the inclusive analysis with the sum of
exclusive method.  Some exclusive modes can also be used to study the
tCPV.  As shown in Tab.~\ref{tab:6:bsgamma-exp-excl-tcpv}, \babar has
performed the first measurement of tCPV for $\Bz \to \KS\eta\g$, while
Belle has reported the first evidence of $\Bp \to \Kp\etapr\g$, whose
neutral mode is also usable for an tCPV analysis.

\begin{table}
 \begin{center}
   \caption{\label{tab:6:bsgamma-exp-excl-tcpv}%
     Measurements of tCPV of radiative $B$ decays. Only the $S$ terms
     are shown. The size of the data sets is given in units of
     $\invfb$.  }
  \begin{tabular}{c|ccc|ccc}
    \hline\hline
    & \multicolumn{3}{c}{Belle} & \multicolumn{3}{c}{\babar} \\
    Mode & $S$ & Data set & Ref.  & $S$ & Data set & Ref. \\ \hline
    $\Bz \to \Kstarz\g$
    & $-0.32\,^{+0.36}_{-0.33} \pm 0.05$ & $492$ & \cite{Ushiroda:2006fi}
    & $-0.03 \pm 0.29 \pm 0.03$ & $423$ & \cite{Aubert:2008gy} \\
    $\Bz \to \KS\piz\g^\dag$
    & $-0.10 \pm 0.31 \pm 0.07$ & $492$ & \cite{Ushiroda:2006fi}
    & -- & -- & -- \\
    $\Bz \to \KS\piz\g^\ddag$ & -- & -- & --
    & $-0.78 \pm 0.59 \pm 0.09$ & $423$ & \cite{Aubert:2008gy} \\
    $\Bz \to \KS\eta\g$ & -- & -- & --
    & $-0.18\,^{+0.49}_{-0.46} \pm 0.12$ & $423$ & \cite{Aubert:2008js} \\
    $\Bz \to \KS\rho^0\g$
    & $0.11 \pm 0.33\,^{+0.05}_{-0.09}$ & $605$ & \cite{Li:2008qm}
    & -- & -- & -- \\
    \hline\hline
    \multicolumn{7}{l}{${}^\dag$ $M_{K\pi} < 1.8 \, {\rm GeV}$~~
      ${}^\ddag$ $1.1 \, {\rm GeV} < M_{K\pi} < 1.8 \, {\rm GeV}$} \\
  \end{tabular}
 \end{center}
\end{table}

Belle has recently reported the measurement of tCPV in $\Bz \to
\KS\rho^0\g \to \KS\pi^+\pi^-\g$~\cite{Li:2008qm}. The advantage of
this mode is that the $B$ decay vertex can be determined from two
charged pions.  On the other hand, there exists a contamination from
other decays with the same final state such as $\Bz \to
\Kstarp\pim\g$.  Since $K_1(1270)$ and $K^*(1680)$ have significant
branching fractions to $K\rho$, it is necessary to estimate the
fraction of $\B \to K_1(1270)\g$ and $\B \to K^*(1680)\g$ in the
entire $\B \to K\pi\pi\g$ decay. Belle uses the charged mode $\Bp \to
\Kp\pi^+\pi^-\g$ in order to disentangle the composition, and,
assuming the isospin relation, estimates the dilution factor to the
effective $S$ in the $\rho^0$ mass window. The result listed in
Tab.~\ref{tab:6:bsgamma-exp-excl-tcpv} shows that the size of the
error is competitive to those for $\Bz \to \Kstarz\g$.

Radiative decays of the $B_s$ meson have been studied by Belle using
the data taken at the $\Upsilon(5S)$ center-of-mass energy, and the
decay $B_s \to \phi\gamma$ has been observed as shown in
Tab.~\ref{tab:6:bsgamma-exp-excl-bf}. \lhcb is expected to perform the
study of the time-dependent asymmetry of this
mode~\cite{Muheim:2008vu}.  With respect to the $B_d$ system, there is
an additional observable $A^{\Delta}$ in the formula of the asymmetry:
\begin{equation}
  A_{\rm CP}(t) = \frac{ S \sin(\Delta m_s t) - C \cos(\Delta m_s t)}{%
    \cosh(\Delta\Gamma_s t / 2) - A^{\Delta} \sinh(\Delta\Gamma_s /2)}\,.
\end{equation}
The extra contribution $A^{\Delta}$ parametrizes the fraction of
wrongly polarized photons, and is sensitive to NP as well as the $S$
term.  According to the MC simulation, \lhcb is expected to reach
sensitivities of $\sigma(A^{\Delta}) \sim 0.22$ and $\sigma(S) \sim
0.11$ for $2\, \invfb$, which demonstrates that the prospects for a
measurement of the photon polarization at \lhcb are promising.

\subsubsection{Determinations of $|V_{td}/V_{ts}|$ from $b \to (s,d)
  \gamma$}
\label{sec:6:VtdVts}

Since the $b \to d\gamma$ process is suppressed by a factor of
$|V_{td}/V_{ts}|$ compared to $b \to s\gamma$, its branching fraction
is useful to extract the ratio $|V_{td}/V_{ts}|$ by means of
(\ref{eq:rare:BrRat}). The exclusive modes to be studied in the case
of $b \to d\gamma$ are $B \to (\rho,\omega)\gamma$.  Due to their
small branching fractions, the continuum background suppression is a
key issue in the analysis.  In addition, the good particle
identification of the \babar and Belle detectors is essential to
separate $B \to \rho\g$ from $B \to K^*\g$. Both \babar and Belle have
observed clear signals of these modes. The current values of the
branching fractions are given in
Tab.~\ref{tab:6:bsgamma-exp-excl-bf}.

The input value for the extraction of $|V_{td}/V_{ts}|$ is the
branching ratio of $B \to (\rho,\omega)\gamma$ and $B \to K^*\gamma$.
One can perform a simultaneous fit to $B \to (\rho,\omega)\gamma$ and
$B \to K^*\gamma$ or calculate the ratio from the individual fits to
$B \to (\rho,\omega)\gamma$ and $B \to K^*\gamma$, so as to cancel
common systematic errors. In order to obtain the combined branching
fraction of $B \to (\rho,\omega)\gamma$, one assumes the isospin
relation $\mathcal{B}(B \to (\rho,\omega)\gamma) = \mathcal{B}(B^+ \to
\rho^+\gamma) = 2\, (\tau_{B^+}/\tau_{B^0}) \, \mathcal{B}(B^0 \to
\rho^0\gamma) = 2\, (\tau_{B^+}/\tau_{B^0}) \, \mathcal{B}(B^0 \to
\omega\gamma)$.  From the combined branching fraction of $B \to
\rho^+\gamma$, $B \to \rho^0\gamma$, and $B \to \omega\gamma$, \babar
and Belle have extracted the values $0.039 \pm 0.008$ and $0.0284 \pm
0.0050_{\rm stat} {\,^{+0.0027}_{-0.0029}}_{\rm syst}$ for
$\mathcal{B}(B \to (\rho,\omega)\gamma)/\mathcal{B}(B \to K^*\gamma)$,
respectively. These measurements translate into $|V_{td}/V_{ts}| =
0.233 {\,^{+0.025}_{-0.024}}_{\rm expr} \, \pm 0.021_{\rm theo}$ for
\babar \cite{:2008gf} and $0.195 {\,^{+0.020}_{-0.019}}_{\rm expr} \pm
0.015_{\rm theo}$ for Belle \cite{Taniguchi:2008ty}, where the first
(second) error in $|V_{td}/V_{ts}|$ is of experimental (theoretical)
nature. The values extracted from the individual decay modes can also
be found in the latter references.

Future precise measurements of $B \to X_d\gamma$ also provide a
promising way to determine the ratio $|V_{td}/V_{ts}|$. Using the
value of $\mathcal{B}(B \to X_d\gamma)$ as given in
Sec.~\ref{sec:6:bsgamma_incl_exp} leads to $|V_{td}/V_{ts}| = 0.177
\pm 0.043_{\rm expr} \pm 0.001_{\rm theo}$ \cite{:2008ig}. Although
the given theory error is likely to be underestimated, as it does not
take into account an uncertainty due to the experimental cut on
$M_{X_d}$, the quoted numbers make clear that determinations of
$|V_{td}/V_{ts}|$ from $B \to X_d\gamma$ are at the moment essentially
only limited by experiment.

So far, the central values of $|V_{td}/V_{ts}|$ extracted from $b \to
(s,d) \gamma$ are compatible with the ones following from $B_{d,s}$
mixing \cite{Abulencia:2006ze}, although both the experimental and
theoretical uncertainties are significantly larger in the former
case. While thus not suitable for a precise determination of
$|V_{td}/V_{ts}|$, the $b \to (s,d) \gamma$ results are complementary
to those from neutral meson mixing, since they could be affected
differently by NP. It is therefore worthwhile to try to improve the
measurements of $b \to (s,d) \gamma$ with one order of magnitude
larger luminosities.


\subsection{Purely leptonic rare decays}

\subsubsection{Theory of purely leptonic rare decays}

The charged-current processes $P\to\ell\nu$ are the simplest
flavor-violating helicity suppressed observables. Both in the SM and
models of NP with a extended Higgs sector these modes appear already
at the tree level. The charged Higgs contribution is proportional to
the Yukawa couplings of quarks and leptons, but it can compete with
the contribution arising form $W^{\pm}$-boson exchange due to the
helicity suppression of $P\to\ell\nu$~\cite{Hou:1992sy}. Taking into
account the resummation of the leading $\tan \beta = v_u/v_d$
corrections to all orders, the $H^\pm$ contributions to the $P \to
\ell \nu $ amplitude within a MFV supersymmetric framework leads to
the following ratio~\cite{Akeroyd:2003zr, Isidori:2006pk}
\begin{equation}
 R_{P\ell\nu} = \frac{\BR^{\rm SM}(P \to \ell\nu)}{\BR^{\rm SUSY}(P
\to \ell\nu)} = \left[1-\left(\frac{m^{2}_P}{m^{2}_{H^\pm}}\right)
\frac{\tan^2\beta}{1+\epsilon_0\tan\beta} \right]^2 \,,
\label{eq:Btn}
\end{equation}
where $\epsilon_0$ denotes the effective coupling which parametrizes
the non-holomorphic corrections to the down-type Yukawa interaction.
One typically has $\epsilon_0\sim 10^{-2}$. For a natural choice of
the MSSM parameters, the relation (\ref{eq:Btn}) implies a suppression
with respect to the SM in the $B \to \tau \nu$ decay of ${\cal
  O}(10\%)$, but an enhancement is also possible for very light
$M_{H^\pm}$.

Performing a global fit of the unitarity triangle, one obtains the following SM prediction
$\BR(B \to \tau \nu)_{\rm SM} = (0.87 \pm 0.19) \times 10^{âˆ}4]$.
The major part of the total error stems from the
uncertainty due to the $B$-meson decay constant $f_B$. The latter prediction is 1.7$\sigma$
below the current world average $\BR(B \to \tau \nu)_{\rm exp} = (1.51 \pm 0.33) \times 10^{−4}$.
However, systematic errors in the lattice determinations of $f_B$ in conjunction with the limited experimental
statistics  do not allow to draw a clear-cut conclusion about the presence of
beyond the SM physics in $B \to \tau \nu$ at the moment.


The expression for $R_{K\mu \nu}$ is obtained from (\ref{eq:Btn}) by
replacing $m_B^2$ with $m_K^2$. Although the charged Higgs
contributions are now suppressed by a factor $m_K^2/m_B^2\sim 1/100$,
$K \to \ell \nu$ is competitive with $B \to \tau \nu$ due to the
excellent experimental resolution~\cite{Antonelli:2008jg} and the good
theoretical control of the former. The best strategy to fully exploit
the NP sensitivity of the $K_{l2}$ system is to consider the
observable $R_{K\mu\nu}/R_{\pi\mu\nu}$~\cite{Isidori:2006pk,
  Antonelli:2008jg} that is proportional to $(f_K/f_{\pi})^2$. Once a
well established unquenched lattice calculations of $f_K/f_{\pi}$ will
be available, $R_{K\mu\nu}/R_{\pi\mu\nu}$ will play a relevant role in
both constraining and probing scenarios with a extended Higgs sector.

The SM prediction for the $B_s \to \mu^+ \mu^-$ branching fraction is
$\BR(B_s\to\mu^{+}\mu^{-})_{{\rm SM}}=(3.37\pm 0.31)\times
10^{-9}$~\cite{Buras:2003td} while the current 95\% CL upper bound
from CDF reads $\BR(B_s\to \mu^{+} \mu^{-})_{\rm exp} < 5.8
\times 10^{-8}$ \cite{:2007kv}, which still leaves room for
enhancements of the branching fraction relative to the SM of more than
factor of 10. In particular, the MSSM with large $\tan\beta$ provides
a natural framework where large departures from the SM expectations of
$\BR(\Bs \to \mumu)$ are
allowed~\cite{Babu:1999hn}.

The important role of $\BR(B_{s,d} \to \ell^+ \ell^-)$ in the large
$\tan\beta$ regime of the MSSM has been widely discussed in the
literature. The leading non-SM contribution to $B\to \ell^+\ell^-$
decays is generated by a single tree-level amplitude, i.e., the
neutral Higgs exchange $B\to A^0,H^0 \to \ell^+\ell^-$. Since the
effective FCNC coupling of the neutral Higgs bosons appears only at
the quantum level, in this case the amplitude has a strong dependence
on other MSSM parameters of the soft sector in addition to $M_{A^0}
\sim M_{H^0}$ and $\tan\beta$. In particular, a key role is played by
the $\mu$ term and the up-type trilinear soft-breaking term, $A_U$,
which control the strength of the non-holomorphic terms. The leading
parametric dependence of the scalar FCNC amplitude from these
parameters is given by
\begin{equation}
{\cal A} (B_s \to \mu^+ \mu^-) \propto \frac{m_b m_\mu}{M_{A^0}^2}
 \frac{\mu A_U}{M^2_{\tilde q}} \tan^3\beta \; m_b (\bar b_R s_L)
 (\bar \mu_L \mu_R) \,.
\end{equation}

More quantitatively, the pure SUSY contributions can be summarized by
the approximate formula
\begin{equation}
\BR(B_s\to\mu^+\mu^-)\simeq \frac{5\times10^{-8}}{\left(1+0.5
\displaystyle \, \frac{\tan \beta}{50}\right)^4}
\Bigg(\frac{\tan\beta}{50}\Bigg)^6 \, \left(\frac{500
\rm{GeV}}{M_{A^0}}\right)^4 \left(\frac{\epsilon_{Y}}{3\times
10^{-3}}\right)^{2} \,,
\label{bsmumuSUSY}
\end{equation}
where $\epsilon_{Y} \sim 3\times 10^{-3}$ holds in the limit of all
the SUSY masses and $A_U$ equal. The approximation (\ref{bsmumuSUSY})
shows that $\BR(B_s\to\mu^+\mu^-)$ already poses interesting
constraints on the MSSM parameter space, especially for light
$M_{A^0}$ and large values of $\tan\beta$. However, given the specific
dependence on $\mu$ and $A_U$, the present $\BR(B_s \to \mu^+\mu^-)$
bound does not exclude the large $\tan\beta$ effects in $P\to\ell\nu$
already discussed.

\subsubsection{Experimental results on purely leptonic rare decays}

To measure the branching fraction for $B \to \tau \nu$ is a big
challenge as there are at least three neutrinos in the final state. To
get a sufficiently pure signal sample the recoil technique discussed
in Sec.~\ref{sec:recoiltechnique} is used. On the tagging side a
semi-leptonic or a fully reconstructed hadronic state is required, and
on the signal side the visible particles from the $\tau$ decay. On top
of this the most powerful discriminating variable is excess energy in
the calorimeter.

The first Belle analysis used fully hadronic tag decays and had a
$3.5\,\sigma$ signal with $449\times 10^6$ \BB
pairs~\cite{Ikado:2006un}. \babar used both hadronic and semileptonic
tag decays and had a $2.6\,\sigma$ signal with $383\times 10^6$ \BB
pairs~\cite{Aubert:2007bx,Aubert:2007xj}. The latest Belle analysis
uses semileptonic tag decays with one prong $\tau$ decays and
$657\times 10^6$ \BB pairs. In this sample they find 154 signal events
with a significance of $3.8\,\sigma$. This results in a branching
fraction of $(1.65{^{+0.38}_{-0.37}}_{\rm stat}{^{+0.35}_{-0.37}}_{\rm
  syst})\times 10^{-4}$. All the results are summarized in
Tab.~\ref{tab:BtoEllNuBF}. Searches have also been made for the decay
$\Bp \to \mup \nu_\mu$ where \babar has set a 90\% CL upper limit of
$1.3 \times 10^{-6}$~\cite{Aubert:2008ri} and Belle at $1.7 \times
10^{-6}$~\cite{Satoyama:2006xn}.
\begin{table}
  \centering
  \caption{Summary of the $\B \to \tau \nu_\tau$ measurements.}
  \begin{tabular}{llcccc}
    \hline
    \hline
    Experiment & 
            Tagging method & 
                          Data set & Significance & $\BR(10^{-4})$ & Ref.\\
    \hline
    Belle & Hadronic & 449M   & $3.5\,\sigma$ & 
          $1.79^{+0.56}_{-0.49}{}^{+0.46}_{-0.51}$ &
                                        \cite{Ikado:2006un} \\ 
    \babar & Semileptonic & 383M   & - & 
       $0.9\pm 0.6 \pm 0.1$ &
                                        \cite{Aubert:2007bx} \\
    \babar & Hadronic    & 383M   & $2.2\,\sigma$ & 
       $1.8^{+0.9}_{-0.8}\pm 0.4$ &
                                        \cite{Aubert:2007xj} \\
    Belle & Semileptonic & 657M   & $3.8\,\sigma$ & 
          $1.65^{+0.38}_{-0.37}{}^{+0.35}_{-0.37}$ &
                                        \cite{:2008ch} \\ 
    \hline
    Average &    &        &               & 
                                   $1.51\pm 0.33$ & \\
    \hline
    \hline
  \end{tabular}
  \label{tab:BtoEllNuBF}
\end{table}

Searches for $\Bs \to \mumu$ are only carried out at hadron machines,
whereas $B_d \to \mumu$ is being searched for at the \B-factories as
well, even if the measurements are no longer competitive with the
Tevatron results. CDF and D0 build multivariate discriminants that
combine muon identification with kinematics and lifetime information.
This keeps signal efficiency high while rejecting $\mathcal{O}(10^6)$
larger backgrounds including Drell-Yan continuum, sequential $b \to c
\to s$ decays, $b\bar{b} \to \mumu + X$ decays, and hadrons faking
muons.  Background estimates are checked in multiple control regions,
and then the signal-like region of the discriminant output is
inspected for excess of events clustering at the \B mass. The overlap
between \Bs and $B_d$ search regions, due to limited mass resolution,
is smaller at CDF allowing independent results on each mode. There is
no evidence of a signal and the best limit at 90\% CL is $\BR(\Bs \to
\mumu) < 4.7 \times 10^{-8}$~\cite{:2007kv}.

In the near future it is expected that both CDF and D0 will reach a limit of
$\BR(\Bs \to \mumu)$ at $2 \times 10^{-8}$ with 8\invfb of data. This is just
a factor six above the SM expectation and will set serious constraints on NP as
outlined in the previous section. Assuming no signal, \lhcb will be able to
exclude $\BR(\Bs \to \mumu)$ to be above the SM level with just 2\invfb of
data corresponding to one nominal year of data taking. A $5\,\sigma$ discovery
at the SM level will require several years of data taking and all three LHC
experiments are competitive for this~\cite{Martinez:2007mi,Aad:2009wy}.

Other rare leptonic decay modes have been searched for including rare
$\Dz$ decays and the LFV decay $\B \to e \mu$. All of these results
are summarized in Tab.~\ref{tab:BtoEllEll}.
\begin{table}
 \centering
 \caption{An overview of the limits set on the decays of the type
   $\B \to \ellell$.}
 \begin{tabular}{llccc}
   \hline
   \hline
   Experiment &
           Decay &
                         Data set & 90\% CL Limit ($\times 108$) & Ref.\\
   \hline
   D0    &    $\Bs \to \mumu$            &    $1.3\invfb$    & 9.4    &    \cite{Abazov:2007iy} \\
   CDF    &     $\Bs \to \mumu$              &     $2.0\invfb$       & 4.7      &    \cite{:2007kv} \\
   CDF    &     $\Bs \to e^\pm\mu^\mp$      &     $2.0\invfb$       & 20      &     \cite{Aaltonen:2009vr} \\
   CDF    &    $\Bs \to \epem$            &    $2.0\invfb$       & 28      &    \cite{Aaltonen:2009vr} \\
   CDF    &     $B_d \to \mumu$           &     $2.0\invfb$       & 1.5      &    \cite{:2007kv} \\
   CDF    &    $B_d \to e^\pm\mu^\mp$     &    $2.0\invfb$     & 6.4      &     \cite{Aaltonen:2009vr} \\
   CDF    &    $B_d \to \epem$           &    $2.0\invfb$     & 8.3     &    \cite{Aaltonen:2009vr} \\
   \babar &    $B_d \to \mumu$            &    384M            &11.3     &     \cite{Aubert:2007hb} \\
   \babar &    $B_d \to \epem$            &     384M            & 5.2      &     \cite{Aubert:2007hb} \\
   \babar &     $B_d \to e^\pm\mu^\mp$     &     384M           & 9.2      &     \cite{Aubert:2007hb} \\
   Belle  &    $B_d \to \mumu$           &     85M             & 16       &     \cite{Chang:2003yy} \\
   Belle  &    $B_d \to \epem$            &    85M             & 19       &     \cite{Chang:2003yy} \\
   Belle  &     $B_d \to e^\pm\mu^\mp$     &     85M            & 17       &     \cite{Chang:2003yy} \\
   \hline
   \hline
 \end{tabular}
 \label{tab:BtoEllEll}
\end{table}

\subsection{Semileptonic modes}

\subsubsection{$B\to D\tau\nu$ modes}

In the framework of the 2HDM-II, charged Higgs boson exchange
contributes significantly not only to $B\to\tau\nu$ but also to $B\to
D\tau\nu$ decays already at tree level, if $\tan\beta =
\mathcal{O}(50)$.  Due to the recent data accumulated at the $B$
factories, these channels become a standard tool to constrain the
effective coupling $g_S$ of a charged Higgs boson to right-handed
down-type fermions \cite{Nierste:2008qe, Kamenik:2008tj,
Eriksson:2008cx}.

While $\BR(B\to\tau\nu)$ is more sensitive to charged-Higgs
effects than $\BR(B\to D\tau\nu)$, the latter branching
fraction has a much smaller theoretical uncertainty. The prediction
for $\BR(B\to\tau\nu)$ involves the $B$-meson decay constant
$f_B$, which is obtained from lattice calculations, and the CKM
element $|V_{ub}|$, both suffering from large errors,
$\delta(|V_{ub}|f_B)\sim 20\%$. In contrast, the vector and scalar
form factors $F_V$ and $F_S$ in $B\to D\tau\nu$ are well under
control, $\delta(|V_{cb}|F_V)<4\%$ and $\delta(|V_{cb}|F_S)<7\%$.
First, $|V_{cb}|F_V(q^2)$ is extracted from the measured $q^2$
spectrum in $B\to D\ell\nu$ \cite{Barberio:2008fa}. Second, $F_S(q^2)$
is constrained by $F_V$ at $q^2= (p_B-p_D)^2=0$ and by heavy-quark
symmetry at maximal $q^2$. Since two parameters are sufficient to
describe the $B\to D$ form factors, $F_S(q^2)$ is thus fixed
\cite{Nierste:2008qe, Trine:2008qv}. Thanks to this good precision,
present data on $\BR(B\to D\tau\nu)$ can almost completely
exclude the window around $g_S=2$ left by $\BR(B\to\tau\nu)$
at $95\%$ CL \cite{Trine:2008qv}.

Since charged-Higgs effects exhibit a $q^2$ dependence distinct from
longitudinal $W^\pm$-boson exchange, the differential distribution
$d\Gamma(B\to D\tau\nu)/dq^2$ is more sensitive than the branching
ratio $\BR(B\to D\tau\nu)$ \cite{Kiers:1997zt}. Notice that in
the differential distribution charged-Higgs effects can be detected
not only from the normalization of the decay mode, but also from the
shape of the spectrum.

To further increase the sensitivity to charged Higgs boson exchange,
one can include information on the polarization of the $\tau$ lepton.
Though the latter is not directly accessible at the $B$ factories, in
the decay chain $B\to D\nu [\tau\to\pi\nu]$ the $\tau$ spin is
directly correlated with the direction of the pion in the final state.
To combine this correlation with the sensitivity from the $q^2$
distribution, an unbinned fit to the triple-differential distribution
$d\Gamma{(B\to
  D\bar{\nu}[\tau^-\to\pi^-\bar{\nu}])}/(dE_D\,dE_{\pi}\,d\cos\theta_{D\pi})$
should be performed \cite{Nierste:2008qe}. Here $E_D,\ E_{\pi},$
denote the energies of the mesons and $\theta_{D\pi}$ is the angle
between $D$ and $\pi^-$ in the $B$ rest frame.  The exploration of
both differential distributions in a comprehensive experimental
analysis makes the $B\to D\tau\nu$ mode particularly well-suited to
detect charged-Higgs effects and to distinguish them from other
possible NP contributions.


\subsection{Semileptonic neutral currents decays}
\subsubsection{Theory of inclusive $B \to X_s \ellp \ellm$}
\label{sec:rare:bslltheory}

The study of $b \to s \ell^+ \ell^-$ transitions can yield useful
complementary information, when confronted with the less rare $b \to s
\gamma$ decays, in testing the flavor sector of the SM. In
particular, a precise measurement of the inclusive $B \to X_s \ell^+
\ell^-$ decay distributions would be welcome in view of NP searches,
because they are amenable to clean theoretical descriptions for
dilepton invariant masses in the ranges $q^2 \in [1, 6] \, {\rm
  GeV}^2$ and $q^2 > 14.4 \, {\rm GeV}^2$. The inclusive $B \to X_s
\ell^+\ell^-$ rate can be written as follows
\begin{equation}
  \frac{d^2\Gamma}{dq^2 \, d\cos\theta_l} =
  \frac38 \left[ (1+\cos^2 \theta_l) \, {H_T(q^2)} + 2 \, \cos\theta_l \, 
    {H_A(q^2)}
    + 2 \, (1-\cos^2\theta_l) \, {H_L(q^2)} \right] \,,
\label{eq:rare:semileptonicrate}
\end{equation}
where $q^2 = (p_\ellp + p_\ellm)^2$ and $\theta_l$ is the angle between
the negatively charged lepton and the $\Bbar$ meson in the center-of-mass
frame of the lepton pair. At leading order and up to an overall
$(m_b^2 - q^2)^2$ factor one has 
\begin{equation}
\begin{split}
  H_T (q^2) &\propto 2 q^2 \left [ \left (C_9+ 2C_7 \,
      \frac{m_b^2}{q^2} \right )^2 + C_{10}^2 \right ]
  \,, \\[1mm]
  H_A (q^2) &\propto -4 q^2 C_{10} \left (C_9+ 2C_7 \,
    \frac{m_b^2}{q^2} \right )
  \,, \\[2mm]
  H_L (q^2) &\propto \left [ \left (C_9+ 2C_7 \right )^2 + C_{10}^2
  \right ] \,.
\end{split}
\nonumber
\end{equation}
The coefficients $H_i (q^2)$ are three independent functions of the
Wilson coefficients of the effective Hamiltonian (\ref{eq:rare:Leff}).
Hence separate measurements of these three quantities lead to better
constraints on the coefficients $C_7$, $C_9$, and $C_{10}$. In terms
of the functions $H_i (q^2)$ the total rate and the forward-backward
asymmetry (FBA) are given by $d\Gamma/dq^2 = H_T (q^2)+H_L(q^2)$ and
$dA_{\rm{FB}}/dq^2 = 3/4 \, H_A(q^2)$. The double differential rate
(\ref{eq:rare:semileptonicrate}) is known at NNLO in
QCD~\cite{Bobeth:1999mk, Asatryan:2001zw, Asatryan:2002iy,
  Ghinculov:2002pe, Asatrian:2002va, Ghinculov:2003bx,
  Ghinculov:2003qd, Bobeth:2003at, Asatrian:2003yk, Gorbahn:2004my}
and at NLO in QED~\cite{Bobeth:2003at, Huber:2005ig, Huber:2007vv}. In
addition non-perturbative corrections scaling as $\Lambda_{\rm
  QCD}^2/m_b^2$, $\Lambda_{\rm QCD}^3/m_b^3$, or $\Lambda_{\rm
  QCD}^2/m_c^2$~\cite{Falk:1993dh, Ali:1996bm, Chen:1997dj,
  Buchalla:1997ky, Buchalla:1998mt, Bauer:1999kf, Ligeti:2007sn} have
been calculated.

A comment on QED corrections is necessary. After inclusion of the NLO
QED matrix elements, the electron and muon channels receive
contributions proportional to $\ln(m_b^2/m_\ell^2)$. These results
correspond to the process $B \to X_s \ellp \ellm$ in which QED
radiation is included in the $X_s$ system and the dilepton invariant
mass does not contain any photon. In the \babar and Belle experiments
the inclusive decay is measured as a sum over exclusive states. As a
consequence the log-enhanced QED corrections are not directly
applicable to the present experimental results and have to be modified
\cite{Huber:2008ak}. We also add that potentially large corrections to
$R_K = \Gamma (B \to X_s \mu^+ \mu^-)_{q^2 \in [q_0^2, q_1^2]}/\Gamma
(B \to X_s e^+ e^-)_{q^2 \in [q_0^2, q_1^2]}$, which in the SM is to
an excellent approximation equal to 1, can arise from collinear photon
emission. Since the actual net effect of these corrections depends on
the experimental cuts, an good understanding of this issue is crucial
to put reliable bounds on possible NP effects from a measurement of
$R_K$.

Cuts on the dilepton and hadronic invariant masses are necessary to
reject backgrounds from resonant charmonium production, $B\to X_s
\psi(c\bar c)\to X_s \ellp \ellm$, and double semileptonic decays,
$B\to X_c \ellm \bar\nu \to X_s \ellp \ellm \nu\bar\nu$,
respectively. The first cut, in particular, forces us to consider
separately the low- and high-$q^2$ regions corresponding to dilepton
invariant masses of $q^2\in [1,6] \, \gev^2$ and $q^2 > 14.4 \,
\gev^2$, respectively. In the low-$q^2$ region the OPE is well behaved
and power corrections are small, but the effect of the $M_{X_s}$ cut
is quite important. The present experimental analyses correct for this
effect utilizing a Fermi motion model~\cite{Ali:1998nq}. In the
high-$q^2$ region $M_{X_s}$ cuts are irrelevant but the OPE itself
breaks down, resulting in large $\Lambda_{\rm QCD}/m_b$ power
corrections. Both these problems can be addressed as discussed at the
very end of this subsection.

The most up-to-date SM predictions in the case of muons in the final
state read
\begin{equation} \label{eq:rare:SMbsll}
\begin{split}
  & {\cal B}_{q^2 \in [1,6] \gev^2} =
  ( 1.59  \pm 0.11 ) \times 10^{-6} \,, \\[1mm]
  & {\cal B}_{q^2 > 14 \, \gev^2} = (2.42 \pm 0.66) \times 10^{-7}  \,, \\[1mm]
  & q_0^2 = ( 3.50 \pm 0.12) \, {\rm GeV}^2 \,, \\[1mm]
  & \bar{\cal A}_{q^2 \in [1,3.5] \gev^2}  =  ( -9.09 \pm 0.91 )\% \,, \\[1mm]
  & \bar{\cal A}_{q^2 \in [3.5,6] \gev^2} = ( 7.80 \pm 0.76 )\% \,,
\end{split}
\end{equation}
where $q_0^2$ denotes the location of the zero in the FBA spectrum and
$\bar {\cal A}_{\rm bin}$ are the integrated FBA in the $q^2 \in
[1,3.5] \gev^2$ and $q^2 \in [3.5,6] \gev^2$ bins. We emphasize that
the quoted errors do not take into account uncertainties related to
the presence of enhanced local power corrections scaling as $\alpha_s
\Lambda_{\rm QCD}/m_b$. Based on simple dimensional reasons these
unknown corrections can be estimated to induce errors at the order of
$5\%$.

Finally, let us mention three possible improvements in the
experimental analyses. First, a measurement of the low-$q^2$ rate
normalized to the semileptonic $B\to X_u \ell \nu$ rate with the same
$M_{X_s}$ cut would have a much reduced sensitivity to the actual
$M_{X_s}$ cut employed~\cite{Lee:2005pwa}. Second, the convergence of
the OPE is greatly enhanced for the high-$q^2$ rate normalized to the
semileptonic $B\to X_u \ell \nu$ rate with the same $q^2$
cut~\cite{Ligeti:2007sn}, as can be seen by comparing the relative
error in (\ref{eq:rare:SMbsll}) with the SM prediction for this new
ratio which reads ${\cal R}_{q^2 > 14 \, \gev^2} = (2.29 \pm 0.30)
\times 10^{-3}$~\cite{Huber:2007vv}. Third, the angular decomposition
of the rate and the separate extraction of $H_T(q^2)$ and $H_A(q^2)$
would result in much stronger constraints on the Wilson
coefficients~\cite{Lee:2006gs}.

\subsubsection{Experimental results on inclusive $\B \to X_s   \ellp \ellm$}

In a fully inclusive analysis of the rare electroweak penguin decay $B
\to X_s \ell^+\ell^-$, where \ellell is either $e^+ e^-$ or $\mu^+
\mu^-$, some difficulties arise, since an abundant source of leptons
is produced in semileptonic $B$ and $D$ decays. For example, the
branching fraction for two semileptonic $B$ decays, ${\cal B}(B \to
X_c \ell \nu)=(10.64 \pm 0.11)\%$ \cite{Barberio:2008fa}, is about
four orders of magnitude larger than that of the signal. Since
standard kinematic constraints like the the beam-energy-substituted
mass, \mes, or the difference between the reconstructed $B$ meson
energy in the center-of-mass frame and its known value, \DeltaE,
cannot be used here, one needs to develop other analysis
strategies. So far two alternative methods were developed that allows
one to reduce these backgrounds. The first so-called recoil method is
based on kinematic constraints of the $\Upsilon(4S) \ra B \bar B$
decays. By performing a complete reconstruction of the other $B$ meson
in a hadronic final state plus requiring a lepton pair the residual
background consists of two consecutive semileptonic decays of the
signal $B$ candidate. This is reduced by requirements on missing
energy in the whole events, event shapes, and vertex
information. Since the $B$ reconstruction efficiency is of the order
of $0.1\%$, the present $B \bar B$ sample are not sufficiently large
to use this method. The second so-called semi-inclusive method
consists of summing up exclusive final states.

Both \babar and Belle focused on the second method. Using 89 (152)
million $B \bar B$ events \babar (Belle) reconstructed final states
from a $K^+$ or a $K^0_S$ and up to two (four) pions recoiling against
the lepton pair, where at most one $\pi^0$ was
accepted~\cite{Aubert:2004it, Iwasaki:2005sy}. In both analyses, event
shape variables, kinematic variables, and vertex information are
combined into likelihood functions for signal, $B \bar B$ backgrounds,
and $e^+ e^- \ra q\bar q$ continuum backgrounds. The likelihood ratios
are optimized to enhance signal-like events. The signal is extracted
from an extended maximum likelihood fit to the \mes distribution after
selecting a signal-like region in \DeltaE. Both analyses found
significant event yields, measuring branching fractions of
\begin{equation}
\begin{split}
  {\cal B}(B \ra X_s \ell^+ \ell^-) & = (5.6 \pm 1.5_{\rm stat}\pm
  0.6_{\rm syst} \pm 1.1_{\rm mode})
  \times 10^{-6} \,, \\[1mm]
  {\cal B}(B \ra X_s \ell^+ \ell^-)& = \left (4.11\pm 0.83_{\rm
      stat} {^{+0.85}_{-0.81}}_{\rm syst} \right ) \times 10^{-6} \,,
\end{split}
\end{equation}
\noindent
where the $J/\psi$ and $\psi(2S)$ veto regions have been excluded and
the third error of the \babar number corresponds to the uncertainty
induced by the Fermi motion model~\cite{Ali:1998nq}. The partial
branching fractions in bins of $q^2$ as measured by \babar and Belle
are summarized in Tab.~\ref{tab:xsll}. \babar also measured the direct
CP asymmetry $(N_{\bar B} -N_B)/(N_{\bar B} + N_B) =-0.22\pm 0.26 \pm
0.02$, where $N_{B (\bar B)}$ are the signal yields for $B ({\bar B})
\ra X_s \ell^+ \ell^-$. All results are consistent with the SM
predictions discussed in Sec.~\ref{sec:rare:bslltheory}.

\begin{table}
\centering
\caption{\babar and Belle measurements of the partial branching fractions  
for the $B \ra X_s \ell^+ \ell^-$ decay in different bins of $q^2$. 
The $J/\psi$ and $\psi(2S)$ veto regions differ 
for the $e^+ e^-$ and $\mu^+ \mu^-$ modes. The latter are shown in parentheses.}
{\footnotesize
\begin{tabular}{|c|c|c|}
  \hline 
  Experiment & $q^2 \, \rm [GeV^2]$ & ${\cal B} \, [10^{-7}] $ \\ \hline
  BaBar~\cite{Aubert:2004it}  & 0.04--1.0& $0.8 \pm 3.6^{+0.7}_{-0.4}$ \\
  & 1.0--4.0 & $16\pm 6 \pm 5$ \\
  & 4.0--7.29 (7.84) & $18 \pm 8 \pm 4 $\\
  & 10.56 (10.24)--11.90 (12.60) & $10 \pm 8 \pm 2$ \\
  & 14.44--25.0 & $6.4 \pm 3.2^{+1.2}_{-0.9}$ \\
  \hline \hline
  Belle~\cite{Iwasaki:2005sy} & 0.04--1.0 & $11.34\pm 4.83^{+4.60}_{-2.71}$ \\
  & 1.0--6.0 & $14.93\pm 5.04^{+4.11}_{-3.21}$ \\
  & 6.0--7.27 (7.55) \& 10.54 (10.22)-- 11.81 (12.50) \& 14.33 (14.33)--14.4 & $7.32\pm 6.14^{+1.84}_{-1.91}$ \\
  & 14.4--25.0&  $4.18\pm 1.17^{+0.61}_{-0.68}$ \\
  \hline
\end{tabular}
}
\label{tab:xsll}
\end{table}

\subsubsection{Theory of exclusive $\b \to \s  \ellp \ellm$ modes}
\label{sll-exclusive}
The theoretical calculation of exclusive $\b \to s \ellell$ amplitudes
is complicated by the fact that one encounters non-factorizable QCD
dynamics. Some of these effects can be estimated using perturbative
methods based on the heavy-quark expansion. To be concrete, we focus
on the decays $\B \to {\Kstar}\ellell$ and comment on other decay
modes at the end of this section.

Assuming the $K^*$ to be on the mass shell, the decay $\bar {B^0}\to
\bar K^{*0}(\to K^- \pi ^+) \ell^+ \ell^-$ is completely described by
four independent kinematic variables; namely, the lepton-pair
invariant mass, $q^2$, and the three angles $\theta_l$,
$\theta_{K^*}$, $\phi$. The sign of the angles for the $\Bbar$ decay
show great variation in the literature. Therefore we present here an
explicit definition.  $\vec{p}$ denote three momentum vectors in the
$\Bbar$ rest frame, $\vec{q}$ the same in the di-muon rest frame, and
$\vec{r}$ in the \Kstarzb rest frame, the $z$-axis is defined as as
the direction of the \Kstarzb in the $\Bbar$ rest frame. Three unit
vectors are given in the following way: the first one is in the
direction of the $z$-axis where the $\theta$ angles are measured with
respect to, and the other two are perpendicular to the di-muon and
\Kstarzb decay planes.
\begin{equation}
  \label{eq:unitvecBdb}
  \vec{e}_z =  
  \frac{\vec{p}_{\Km}+\vec{p}_{\pip}}{|\vec{p}_{\Km}+\vec{p}_{\pip}|}\, ,
  \qquad
  \vec{e}_l=
  \frac{\vec{p}_{\mun}\times\vec{p}_{\mup}}
       {|\vec{p}_{\mun}\times\vec{p}_{\mup}|}\, , 
  \qquad 
  \vec{e}_K=
  \frac{\vec{p}_{\Km}\times\vec{p}_{\pip}}
       {|\vec{p}_{\Km}\times\vec{p}_{\pip}|}\, .
\end{equation}
It follows  for the $\Bbar$ 
\begin{equation}
  \label{eq:AngleDefBdb}
  \cos\theta_l = \frac{\vec{q}_{\mup}\cdot\vec{e}_z}{|\vec{q}_{\mup}|}\, ,
  \qquad
  \cos\theta_{K} = \frac{\vec{r}_{\Km}\cdot\vec{e}_z}{|\vec{r}_{\Km}|}
\end{equation}
and
\begin{equation}
  \label{eq:angledef2Bdb}
  \sin\phi= (\vec{e}_l\times \vec{e}_K)\cdot \vec{e}_z\, , \qquad \cos\phi=
  \vec{e}_K\cdot \vec{e}_l\, .
\end{equation}

The angles are defined in the intervals
\begin{equation}
  -1\leqslant\cos\theta_l\leqslant 1\, ,\qquad
  -1\leqslant\cos\theta_{K}\leqslant 1\, , \qquad
  -\pi\leqslant\phi < \pi\, ,
\end{equation}
where in particular it should be noted that the $\phi$ angle is signed.

In words, for the $\Bbar$ the angle $\theta_l$ is measured as the
angle between the $\ell^+$ and the $z$-axis in the dimuon rest
frame. As the $\Bbar$ flies in the direction of the $z$-axis in the
dimuon rest frame this is equivalent to measuring $\theta_l$ as the
angle between the $\ell^+$ and the $\Bbar$ in the di-lepton rest
frame. The angle $\theta_K$ is measured as the angle between the Kaon
and the $z$-axis measured in the \Kstarzb rest frame. Finally $\phi$
is the angle between the normals to the planes defined by the $K\pi$
system and the $\mu^+\mu^-$ system in the rest frame of the $\Bbar$
meson.

For the $\B$ the definition is such that the angular distributions
will stay the same as for the $\Bbar$ in the absence of \CP
violation. This means that for all the definitions above, $\ell^-$ is
interchanged with $\ell^+$, $K^+$ with $K^-$ and and \pip with \pim.

Following \cite{Lee:2006gs}, the doubly differential decay rate for
$\Bbar \to {\bar \Kstar}\ellell$ can be decomposed as in the inclusive case
(\ref{eq:rare:semileptonicrate}). Here the helicity amplitude
$H_T(q^2)$ determines the rate for transversely polarized $K^*$
mesons, $H_L(q^2)$ the longitudinal rate, and $H_A(q^2)$ is
responsible for the lepton FBA. In terms of transversity amplitudes,
which are relevant for the angular analysis of $\B \to
{\Kstar}(\Kaon\pi)\ellell$, these functions read \cite{Kruger:2005ep}
\begin{equation}
\begin{split}
 H_T(q^2) &= |A_{\perp\, L}|^2 + |A_{\perp\, R}|^2 +
             |A_{\parallel\, L}|^2 + |A_{\parallel\, R}|^2 \,,\\[1mm]
 H_L(q^2) &= |A_{0\, L}|^2 + |A_{0\, R}|^2\,,\\[1mm]
 H_A(q^2) &= 2 \, {\rm Re} \left[ A_{\parallel\, R} \, A_{\perp\, R}^*
     - A_{\parallel\, L} \, A_{\perp\, L}^* \right] \,. 
\label{eq:rare:Hfunc}
\end{split}
\end{equation}
The transversity amplitudes themselves can be written as
\cite{Beneke:2001at, Beneke:2004dp, Kruger:2005ep}
\begin{equation}
\begin{split}
 A_{\perp\, L,R} &  
  \propto  \left[ { (C_9 \mp C_{10})} \, \frac{{ V(q^2)}}{m_\B + m_{\Kstar}}
       + \frac{2 m_\b}{q^2} \, { {\cal T}_1(q^2)} \right] \,,
\\[1mm]
 A_{\parallel\, L,R} &  
  \propto  \left[ {(C_9 \mp C_{10})} \, \frac{{ A_1(q^2)}}{m_\B - m_{\Kstar}}
       + \frac{2 m_\b}{q^2} \, { {\cal T}_2(q^2)} \right] \,,
\\[1mm]
 A_{0\, L,R} & 
\propto 
\left[ { (C_9 \mp C_{10})} \left\{ 
   \frac{{ A_1(q^2)}}{m_\B - m_{\Kstar}}
- \frac{m_\B^2-q^2}{m_\B^2}\, \frac{{A_2(q^2)}}{m_\B + m_{\Kstar}} \right\} 
\right. \cr  & \left. \qquad
       + \frac{2 m_\b}{m_\B^2}  \left\{ { {\cal T}_2(q^2)}
       - \frac{m_\B^2-q^2}{m_\B^2} \, { {\cal T}_3(q^2)} \right\}
 \right] \,.
\label{eq:rare:Adecomp}
\end{split}
\end{equation}
Here we neglected some terms of order $m_{\Kstar}^2/m_\B^2$, and did
not show the kinematic normalization factors which can be found in
\cite{Kruger:2005ep}.  The ingredients in (\ref{eq:rare:Adecomp}) are:
first, the SM short-distance Wilson coefficients $C_{9,10}$ of the $\b
\to \s\ellell$ operators in the weak effective Lagrangian
(\ref{eq:rare:Leff}), which are to be tested against NP.\footnote{NP
  contributions to the operators $Q_{7-10}'$, that are obtained from
  $Q_{7-10}$ by exchanging left- by right-handed fields everywhere,
  can easily be included \cite{Kruger:2005ep}.}  Second, the vector-
and axial-vector $\B \to {\Kstar}$ transition form factors $V,A_{1,2}$
which have to be estimated by non-perturbative methods. Third, the
$q^2$-dependent functions ${\cal T}_i(q^2)$ that contain factorizable
and non-factorizable effects from virtual photons via the operators
$Q_{1-8}$ in (\ref{eq:rare:Leff}).  In the ``naive factorization
approximation'', the functions ${\cal T}_i(q^2)$ are again expressed
in terms of short-distance Wilson coefficients, $\B \to {\Kstar}$
transition form factors, and quark-loop functions, which are
perturbative if $q^2$ lies outside the vector-resonance
region. Corrections to ``naive factorization'' can and should be
systematically computed in the $m_\b \to \infty$ limit, if we restrict
ourselves\footnote{In principle, the region $4m_\c^2 \ll q^2 \leq
  m_\b^2$ can be treated in heavy hadron chiral perturbation theory
  \cite{Grinstein:2004vb}.}  to the window $q^2 \in [1 ,6]
\gev^2$. The QCDF theorem \cite{Beneke:2001at,Beneke:2004dp} which can
be further justified in SCET, takes the schematic form
\begin{equation}
\begin{split}
  {
\begin{array}{l} 
{\cal T}_1(q^2)\simeq \frac{m_\B^2}{m_\B^2-q^2} \, {\cal T}_2(q^2) 
\\
{\cal T}_3(q^2) - \frac{m_\B^2}{m_\B^2-q^2} \, {\cal T}_2(q^2) 
\end{array}
}& \, \simeq \
\left\{ \begin{array}{c} 
{ \xi_\perp(q^2)} \, { C_\perp(q^2)} 
+ 
 { \phi_\B^\pm(\omega)} \otimes { \phi_{\Kstar}^\perp(u)} \otimes
  { T_\perp(\omega,u)} \,,
\\[0.4em]
{ \xi_\parallel(q^2)} \, { C_\parallel(q^2)} 
+ 
 { \phi_\B^\pm(\omega)} \otimes { \phi_{\Kstar}^\parallel(u)} \otimes
  { T_\parallel(\omega,u)} \,,
\end{array}
\right.
\label{eq:Tfact}
\end{split}
\end{equation}
where $\xi_{\perp,\parallel}$ are universal form factors arising in
the combined heavy-quark-mass and large-recoil-energy limit
\cite{Charles:1998dr, Beneke:2000wa}, $C_{\perp,\parallel}$ and
$T_{\perp,\parallel}$ are perturbative coefficient functions including
vertex corrections and spectator effects, respectively, and $\phi_\B$
and $\phi_{\Kstar}$ denote hadronic LCDAs which again have to be
estimated from non-perturbative methods.  On the one hand, the
reduction of form factors in the symmetry limit is a crucial
ingredient to obtain a precise estimate of the FBA
\cite{Burdman:1998mk, Ali:1999mm, Beneke:2000wa}.  On the other hand,
observables like the isospin asymmetry between charged and neutral
decays are sensitive to $\Lambda_{\rm QCD}/m_\b$ corrections to
(\ref{eq:Tfact}), which generally are small but difficult to estimate
very precisely \cite{Feldmann:2002iw, Beneke:2004dp,
  Altmannshofer:2008dz}.

To be concrete, let us quote some theoretical predictions for
individual SM rates and asymmetries in the low-$q^2$ region, following
the numerical analysis in \cite{Beneke:2004dp} but using updated
values for the $B$ lifetimes. We first note that the hadronic
uncertainties for the partial rates in that region are dominated by
the form factor uncertainties, and therefore should be considered as
less useful for precision tests of the SM.  These uncertainties drop
out to a large extent in the prediction for the FBA in particular in
the vicinity of the zero of the FBA\footnote{The form factor
  dependence could be further reduced by normalizing the FBA to the
  transverse rate, instead of the full rate.}. This is illustrated in
panel (a) of Fig.~\ref{fig:sll-exclusive:asymm}. For the zero of the
FBA one obtains
\begin{equation}
\begin{split}
  q_0^2(\B^0 \to \Kstar{}^{0} \ellell) & =  
  \left (4.36^{+0.33}_{-0.31}\right )\gev^2 \,, \\[1mm] 
  q_0^2(\B^\pm \to \Kstar{}^{\pm}\ellell)&=  
  \left ( 4.15^{+0.27}_{-0.27} \right)\gev^2 \,.
\end{split}
\end{equation}
Considering the FBA for the partially integrated rates
\begin{equation}
A_{\rm FB} = 
\frac{\int_0^1 \frac{d\Gamma}{d \cos\theta_l} d \theta_l - 
      \int_{-1}^0 \frac{d\Gamma}{d \cos\theta_l} d \theta_l}
     {\int_0^1 \frac{d\Gamma}{d \cos\theta_l} d \theta_l +
      \int_{-1}^0 \frac{d\Gamma}{d \cos\theta_l} d \theta_l}
\end{equation} 
one obtains
\begin{equation}
A_{\rm FB}^{\text{low-}q^2}= 
\left\{ 
\begin{array}{ll}
  -0.033^{+0.014}_{-0.016} &\,, \quad \text{for} 
  ~\B^0 \to \Kstar{}^{0} \ellell \,,  \\[0.2em] 
  -0.062^{+0.018}_{-0.023} &\,, \quad \text{for} 
  ~\B^\pm \to \Kstar{}^{\pm}\ellell \,.
\end{array}
\right.
\end{equation}

\begin{figure}[t!]
 \begin{center}
  (a) ~\includegraphics[width=0.4\textwidth]{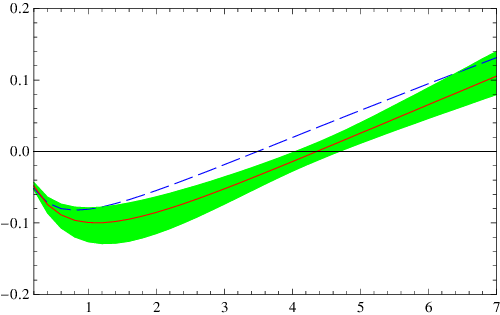} \
  ~~~(b) ~\includegraphics[width=0.4\textwidth]{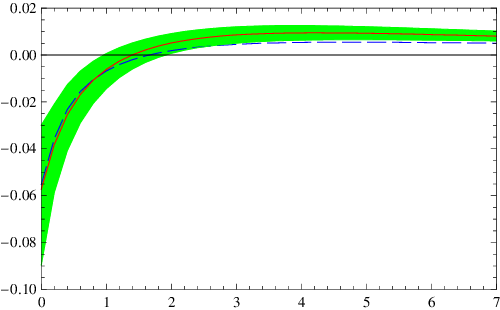}
 \end{center}
 \caption{\label{fig:sll-exclusive:asymm} (a) Theoretical estimate for
   differential FBA in $\B^0 \to \Kstar{}^{0} \ellell$. (b) Estimate
   for differential isospin asymmetry. The dashed line denotes the LO
   result. The solid line with the error band the NLO prediction with
   parametric uncertainties.}
\end{figure}

The corresponding predictions for the isospin asymmetry are shown in
panel (a) of Fig.~\ref{fig:sll-exclusive:asymm}, and the partially
integrated isospin asymmetry is estimated as
\begin{equation}
  A_{\rm I}^{\text{low-}q^2} 
  = \frac{\int d\Gamma^0 - \int d\Gamma^\pm}{\int d\Gamma^0 + \int d\Gamma^\pm}
  =  0.007^{+0.003}_{-0.003}
  \,.
\end{equation}
Notice that the perturbative errors can be reduced by resummation of
large logarithms in SCET \cite{Ali:2006ew} or the computation of
higher-order corrections, but irreducible systematic uncertainties
from both higher-order $\Lambda_{\rm QCD}/m_b$ corrections, and the
restricted precision of the form factor estimates from LCSR or LQCD
remain.

Let us finally consider further exclusive decay modes that can be used
to test the $\b \to \s \ellell$ transition. The decay into a
pseudoscalar Kaon, $\B \to K \ellell$, is similar to the decay into a
longitudinal vector meson \cite{Beneke:2001at, Bobeth:2007dw}. An
interesting observable for the identification of NP is the ratio $R_K$
already mentioned in Sec. \ref{sec:rare:bslltheory}.  One should
also mention the decay $B_s \to \phi\ell^+\ell^-$, where a recent
model-independent analysis of NP effects based on ``naive''
factorization has been given \cite{Yilmaz:2008pa}. A SM analysis
including NLO effects is straightforward and will be discussed
elsewhere \cite{Beneke:2009new}.

A related process is $\B \to \rho\ellell$ which probes the $\b\to
\d\ellell$ transition in and beyond the SM. Due to the different CKM
hierarchy it may show potentially larger isospin and \CP-violating
effects than its counterparts in $b\to s \ell^+ \ell^-$
\cite{Beneke:2001at}.  It is also useful as a cross-check for the
factorization approach.

\subsubsection{Angular observables in $\B \to K^*  \ellp \ellm$}
\label{sll-angular}
Besides the branching fractions, the FBA and CP-violating observables,
the exclusive decay $\Bzb \to \Kstarzb \ellell$ with an angular
analysis of the subsequent $\Kstarzb \to \Km \pip$ decay offers the
possibility to further constrain NP \cite{Kruger:2004, Kruger:2005ep,
  Lunghi:2006hc, Bobeth:2008ij, Egede:2008uy, Altmannshofer:2008dz}.
The decay is described by 4 independent kinematic variables: the
lepton-pair invariant mass squared, $q^2$, and the three angles
$\theta_l$, $\theta_{K}$, $\phi$. Summing over final-state spins, the
differential decay distribution can be expressed in terms of 9
independent functions \cite{Kruger:1999xa, Melikhov:1998cd,
  Kim:2000dq, Kim:2001xua, Faessler:2002ut}, which are related to the
transversity amplitudes\footnote{Another transversity amplitude $A_t$
  does not contribute for massless leptons.} discussed around
(\ref{eq:rare:Hfunc}) and (\ref{eq:rare:Adecomp}), and which are
invariant under the following symmetry
transformations \cite{Egede:2008uy}
\begin{equation}
\begin{split}
  A_{i\, L} & \to \cos\theta \, e^{+ i \phi_L} \, A_{i\, L} -
  \sin\theta \, e^{- i \phi_R} \, A^*_{i\, R}
  \,, \\
  A_{i\, R} & \to \sin\theta \, e^{- i \phi_L} \, A^*_{i \, L} +
  \cos\theta \, e^{+ i \phi_R} \, A_{i\, R} \,,
  \\[0.2em]
  A_{\perp L} & \to + \cos\theta \, e^{+ i \phi_L} \, A_{\perp L}
  +\sin\theta \, e^{- i \phi_R} \, A^*_{\perp R}
  \,, \\
  A_{\perp R} & \to - \sin\theta \, e^{- i \phi_L} \, A^*_{\perp L} +
  \cos\theta \, e^{+ i \phi_R} \, A_{\perp R} \,.
\end{split}
\end{equation}
Here $i=\|,0$. Any experimental observable constructed from the
transversity amplitudes thus has to be invariant under these
symmetries or would require to measure the helicity of the decay
products which is not possible at \lhcb or a super flavor factory. For
instance, this excludes the asymmetry $A_T^{(1)}$ defined in
\cite{Melikhov:1998cd}, despite its very attractive NP sensitivity
\cite{Kruger:2005ep, Lunghi:2006hc}.

As it has been emphasized in \cite{Egede:2008uy}, one can construct
angular observables which simultaneously fulfill a number of
requirements, namely: i) small theoretical uncertainties due to
cancellations of form-factor dependencies, ii) good experimental
resolution at \lhcb and/or super flavor factory, iii) high sensitivity
to NP effects, including contributions from new operators in the weak
effective Hamiltonian. Focusing on the sensitivity to right-handed
operator $Q_7'$, where one would encounter the combination of Wilson
coefficients $(C_7+C_7')$ in $A_{\perp L,R}$ and $(C_7-C_7')$ in
$A_{\parallel L,R}$ and $A_{0L,R}$, the authors of \cite{Egede:2008uy}
identify the following three observables to satisfy the above criteria
\begin{equation}
 \label{eq:rare:ATDef} 
 A_T^{(2)} =\frac{|A_\bot|^2 - |A_\||^2}{|A_\bot|^2 + |A_\||^2} 
\,, \quad
 A_T^{(3)} =
  \frac{|A_{0} A_{\|}^*|}{ |A_{0}| \,  |A_\bot|} \,,
\quad 
 A_T^{(4)} = \frac{|A_{0L} A_{\bot L}^* - A_{0R}^* A_{\bot R}| }{|
    A_{0} A_{\|}^*|} \,,
\end{equation}
where $ A_i A^*_j = A^{}_{i L} A^*_{jL} + A^{}_{iR} A^*_{jR} $.  In
particular, the dependence on the form factors $\xi_{\bot,\|}$ drops
out to first approximation if one neglects $\alpha_s$ and
$\Lambda_{\rm QCD}/m_b$ corrections.

\begin{figure}
\begin{center}
\includegraphics[width=0.4\textwidth]{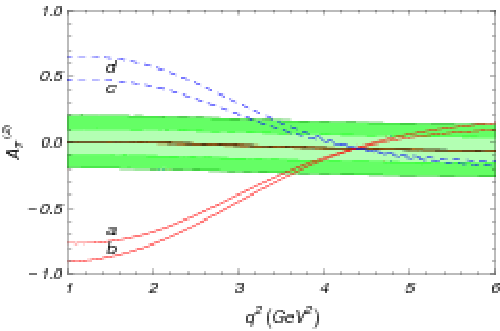}
\includegraphics[width=0.4\textwidth]{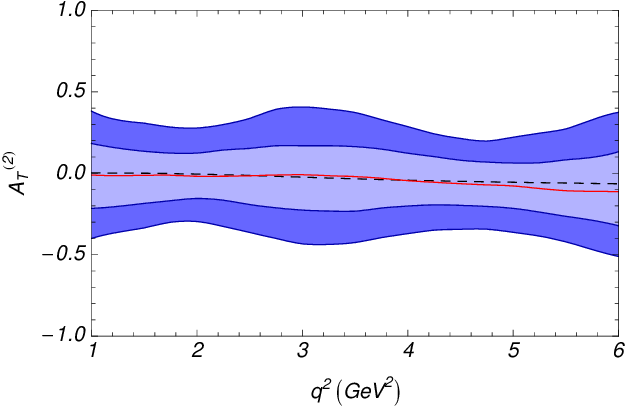}

\vspace{1mm}

\includegraphics[width=0.4\textwidth]{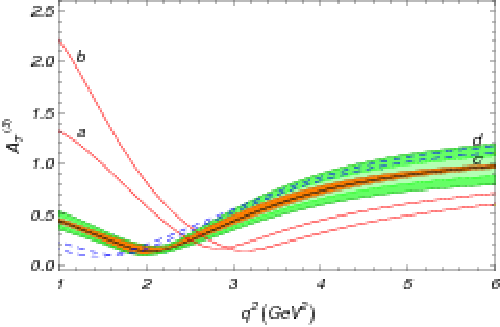}
\includegraphics[width=0.4\textwidth]{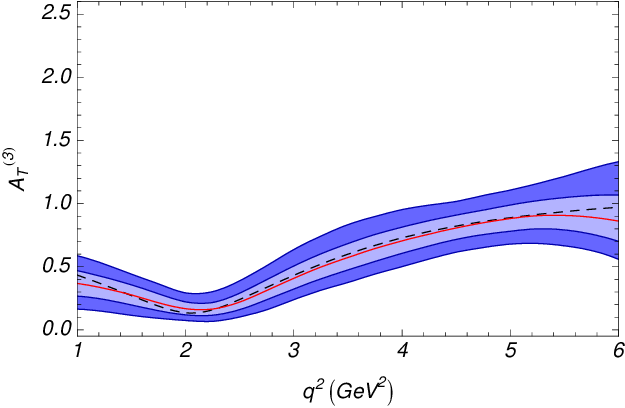}

\vspace{1mm}

\includegraphics[width=0.4\textwidth]{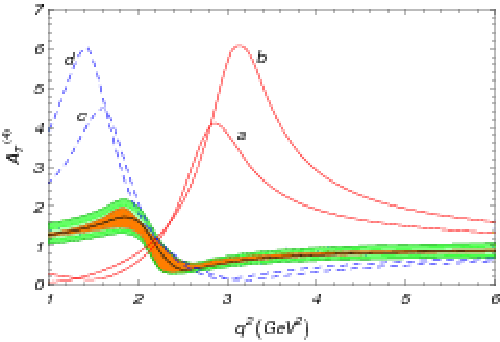}
\includegraphics[width=0.4\textwidth]{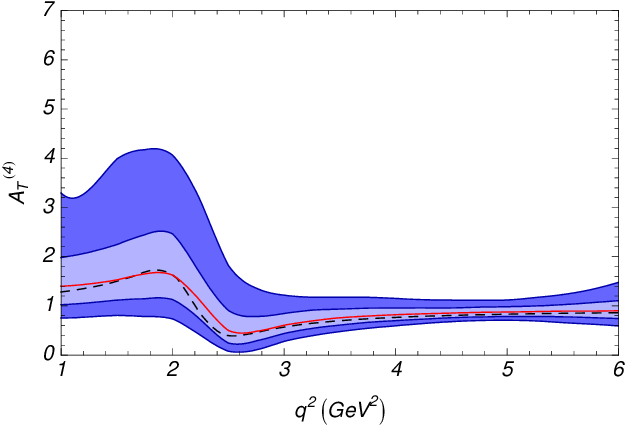}

\end{center}
\caption{The asymmetries $A_T^{(2)}$, $A_T^{(3)}$, and $A_T^{(4)}$ as
  a function of $q^2$, with theoretical errors (left panels), and
  experimental errors (right panels). See text for details. Fig.s
  taken from \cite{Egede:2008uy}.}
  \label{fig:rare:AT24}
\end{figure}

In Fig.~\ref{fig:rare:AT24}, the theoretical estimates and
experimental sensitivity for $A_T^{(2)}$, $A_T^{(3)}$, and $A_T^{(4)}$
are plotted as a function of $q^2$.  In each theoretical plot on the
left-hand side the thin dark line is the central NLO result for the SM
and the narrow inner dark (orange) band corresponds to the related
uncertainties due to both input parameters and perturbative scale
dependence.  Light gray (green) bands refer to $\Lambda_{\rm QCD}/m_b
= \pm 5\%$ corrections considered for each spin amplitude, while for
the darker gray (green) one considers $\Lambda_{\rm QCD}/m_b = \pm
10\%$ corrections.  The curves labeled (a) to (d) correspond to four
different benchmark points in the MSSM. For more details we refer to
\cite{Egede:2008uy}.  The experimental sensitivity for a data set
corresponding to $10~{\rm fb}^{-1}$ of integrated luminosity at \lhcb
is given in each figure on the right, assuming SM rates. Here the
solid (red) line shows the median extracted from the fit to the
ensemble of data, and the dashed (black) line shows the theoretical
input distribution.  The inner and outer bands correspond to 1$\sigma$
and 2$\sigma$ experimental errors.

The observables $A_T^{(3)}$ and $A_T^{(4)}$ offer sensitivity to the
longitudinal spin amplitude $A_{0 L,R}$ in a controlled way, i.e., the
theoretical uncertainties from NLO corrections turn out to be very
small. Concerning the sensitivity to right-handed currents, one
observes sizable deviations from the SM for $A_T^{(2)}$, $A_T^{(3)}$,
and $A_T^{(4)}$ in the 4 SUSY benchmark scenarios studied in
\cite{Egede:2008uy}. For a recent discussion of other NP scenarios we
refer to \cite{Altmannshofer:2008dz}.  Comparing the theoretical and
experimental figures, it can be seen that in particular $A_T^{(3)}$
offers great promise to distinguish between such NP models.

\subsubsection{Experimental results on exclusive $b \to (s,d)  \ellp \ellm$}
The exclusive electroweak decay $B \ra K \ell^+ \ell^-$ is a $b \ra s$
transition that was first observed by Belle~\cite{Abe:2001dh} in a
sample of 31 million $B \bar B$ events. Using 123 million $B \bar B$
events \babar confirmed the observation and reported first evidence for
$B \ra K^* \ell^+ \ell^-$ ~\cite{Aubert:2003cm} which was confirmed
later by Belle~\cite{Ishikawa:2003cp}. In the most recent studies
\babar and Belle have reconstructed ten final states consisting of
$K^\mp$, $K^0_S (\ra \pi^+ \pi^-)$, $K^\mp \pi^\pm $, $K^\mp \pi^0$ or
$K^0_S (\ra \pi^+ \pi^-) \pi^\mp$ besides the lepton pair using 384
million and 657 million $B \bar B$ events, respectively
\cite{:2008ju, Aubert:2008ps,:2009zv}. The signal yields in
individual final states are extracted from the \mes and \DeltaE
distributions. The main background arises from random combinations of
leptons from $B$ and $D$ decays. As in the semi-inclusive analysis
this combinatorial background is suppressed by using event shape
variables, kinematic variables, and vertex information that are
combined into a neural network (\babar) or a likelihood ratio
(Belle). The multivariate observables are optimized separately for
each mode, for each type of background, $B \bar B$ or $e^+ e^- \ra q
\bar q$, and each $q^2$ region.

Total branching fractions measured by \babar, Belle, and CDF are in
agreement with each other and the SM predictions
~\cite{Ali:1999mm,Ali:2002jg}. The interest, however, has shifted towards
rate asymmetries, since many uncertainties in both predictions and
measurements cancel as explained in
Sec.~\ref{sll-exclusive}. \babar and Belle so far studied isospin
asymmetries, $A^{K^{(*)}}_{\rm I}$, direct CP asymmetries,
$A^{K^{(*)}}_{\rm CP}$, and lepton forward-backward asymmetries,
$A_{\rm FB}$, as well as the $K^*$ longitudinal polarization, $F_L$,
and the ratio of rates to $\mu^+ \mu^-$ and $e^+ e^-$ final states,
$R_{K^{(*)}}$. With increased statistics both experiments started to
explore the $q^2$ dependence of these observables.

The CP-averaged isospin asymmetry and direct CP asymmetry are defined
by
\begin{equation}
  \begin{split}
    A^{K^{(*)}}_{\rm I} & = \frac {\BR(\Bz \to K^{(*)0}\ellell) -
      (\tau_0/\tau_+)\BR(\B^{\pm} \to K^{(*)\pm}\ellell)} {{\cal
        B}(\Bz \to K^{(*)0}\ellell) +
      (\tau_0/\tau_+)\BR(\B^{\pm} \to K^{(*)\pm}\ellell)}, \\
    A_{\rm CP}^{K^{(*)}} & = \frac {\BR(\overline{B} \rightarrow
      \overline{K}^{(*)}\ellell) - \BR(B \rightarrow
      K^{(*)}\ellell)} {\BR(\overline{B} \rightarrow
      \overline{K}^{(*)}\ellell) + \BR(B \rightarrow
      K^{(*)}\ellell)},
\end{split}
\end{equation}
where $\tau_0$ and $\tau_+$ are the $B^0$ and $B^+$ lifetimes,
respectively.  $A_{\rm CP}$ is predicted to be ${\cal O}(10^{-3})$ in
the SM. NP at the electroweak scale, however, could produce a
significant enhancement~\cite{Bobeth:2008ij}. The ratios $R_{K^{(*)}}$
are sensitive to the presence of a neutral SUSY Higgs
boson~\cite{Yan:2000dc}. In the SM, $R_{K}$ is expected to be unity
modulo a small correction accounting for differences in phase space
~\cite{Hiller:2003js} and possibly QED radiation. For
$m_{\ell^+\ell^-} \geq 2m_{\mu}$, $R_{K^*}$ should be also close to
unity. Due to the $1/q^2$ dependence of the photon penguin
contribution, however, there is a significant rate enhancement in the
$B \to K^* e^+ e^-$ mode for $m_{\epem}<2m_{\mu}$ decreasing the SM
expectation of $R_{K^*}$ to 0.75. New scalar and pseudoscalar
contributions may modify this prediction. The possible size of these
effects is however already bounded severely by the Tevatron limits on
$B_s \to \mu^+ \mu^-$.

Present results of branching fractions, rate-based asymmetries, and
lepton-flavor ratios are summarized in Tab.s~\ref{tab:K-stuff} and
\ref{tab:rates}. At the present level of
precision branching fractions, $R_{K^{(*)}}$, and $A_{\rm CP}$ are in
good agreement with the SM. While $A_{\rm I}$ agrees with the SM for
large values of $q^2$, the \babar measurement of $A_{\rm I}$ in the
low-$q^2$ region deviates from the SM expectation
\cite{Feldmann:2002iw} by almost $4~\sigma$ for the combination of the
$B \to K \ell^+ \ell^-$ and $B \to K^* \ell^+ \ell^-$ modes. Though
consistent with the SM expectation the Belle results support the \babar
observations at low $q^2$.

\begin{table}
\centering
\caption{Measurements of the partial branching fractions 
and isospin asymmetries for the $B \ra K \ell^+ \ell^-$ and 
$B \ra K^* \ell^+ \ell^-$ decays in different bins of $q^2$.}
{\footnotesize
\begin{tabular}{|c|c|c|c|c|}
\hline 
Experiment & Mode &$q^2 \, \rm [GeV^2]$ & $\BR \, [10^{-7}] $ & $A_{\rm I}$ \\ \hline
\babar~\cite{Aubert:2008ps} & $B \to K \ell^+ \ell^-$& $\ 0.1$--7.02 & $0.181^{+0.39}_{-0.36}\pm 0.008$ & $-1.43^{+0.56}_{-0.85} \pm 0.05$  \\
& & 10.24--12.96 or $ >$14.06 & $ 0.135^{+0.040}_{-0.037}\pm 0.007$ & $\ 0.28^{+0.24}_{-0.30} \pm 0.03$  \\
&  $B \to K^* \ell^+ \ell^-$ &  $\ 0.1$--7.02 & $0.43^{+0.11}_{-0.10}\pm 0.03$ & $-0.56^{+0.17}_{-0.15} \pm 0.03$ \\
&  & 10.24--12.96 or $>$14.06 & $0.42^{+0.10}_{-0.10}\pm 0.03$ & $\ 0.18^{+0.36}_{-0.28} \pm 0.04$  \\
\hline \hline
Belle~\cite{:2009zv}  &  $B \to K \ell^+ \ell^-$ & $\ \ 0.0$--2.0$\ $ & $0.81^{+0.18}_{-0.16} \pm 0.05$ & $-0.33^{+0.33}_{-0.25}\pm 0.05$  \\
& & $\ \ 2.0$--5.0$\ \ $ & $0.58^{+0.16}_{-0.14}\pm 0.04$ & $-0.49^{+0.45}_{-0.34}\pm 0.04$  \\
& & $\ \ 5.0$--8.86 & $0.86^{+0.18}_{-0.16}\pm 0.05$ & $ -0.19^{+0.26}_{-0.22}\pm 0.05$  \\
& & 10.09--12.86 & $0.55^{+0.16}_{-0.14}\pm 0.03 $ & $ -0.29^{+0.37}_{-0.29}\pm 0.05$  \\
& & 14.18--16.0 $\ $& $0.38^{+0.19}_{-0.12}\pm 0.02$ & $-0.40^{+0.61}_{-0.69}\pm 0.04$  \\
& & $>16.0$ & $0.98^{+0.20}_{-0.18}\pm 0.06$ & $ 0.11^{+0.24}_{-0.21}\pm 0.05$   \\ \hline
& &$\ \ 1.0$--6.0$\ \ $ & $1.36^{+0.23}_{-0.21}\pm 0.08$ & $-0.41^{+0.25}_{-0.20}\pm 0.04$   \\ \hline
& $B \to K^* \ell^+ \ell^-$ & $\ \ 0.0$--2.0 & $1.46^{+0.40}_{-0.35} \pm 0.12$ & $-0.67^{+0.18}_{-0.16}\pm 0.03$  \\
& & $\ \ 2.0$--5.0$\ \ $ & $1.29^{+0.38}_{-0.34}\pm 0.10$ & $\ 1.17^{+0.72}_{-0.82}\pm 0.02$   \\
& & $\ \ 5.0$--8.86 & $0.99^{+0.41}_{-0.36}\pm 0.08$ & $ -0.47^{+0.31}_{-0.29}\pm 0.04$   \\
& & 10.09--12.86 & $2.24^{+0.44}_{-0.40}\pm 0.18 $ & $\ 0.00^{+0.20}_{-0.21}\pm 0.05$  \\
& & 14.18--16.0$ \ $ & $1.05^{+0.29}_{-0.26}\pm 0.08$ & $\ 0.16^{+0.30}_{-0.35}\pm 0.05$   \\
& & $>16.0$ & $2.04^{+0.27}_{-0.24}\pm 0.16$ & $ -0.02^{+0.20}_{-0.21}\pm 0.05$  \\ \hline
 & & $\ \ 1.0$--6.0$\ $ & $1.49^{+0.45}_{-0.40}\pm 0.12$ &  \\
\hline
\hline
CDF~\cite{Aaltonen:2008xf}  &  $B^+ \to K^+ \ell^+ \ell^-$ & $<$8.4 or 10.2--13.0 or $>$14.1  & $5.9\pm1.5\pm0.4$  & $-0.33^{+0.33}_{-0.25}\pm 0.05$  \\
CDF~\cite{Aaltonen:2008xf}  &  $B^0 \to K^{*+} \ell^+ \ell^-$ &  $<$8.4 or 10.2--13.0 or $>$14.1  & $8.1\pm3.0\pm1.0$  & \\

\hline
\end{tabular}
}
\label{tab:K-stuff}
\end{table}

\begin{table}
\centering
\caption{\babar and Belle measurements of total branching fractions, 
CP asymmetries, and lepton flavor ratios for the $ B \ra K \ell^+ \ell^-$ 
and $B \ra K^* \ell^+ \ell^-$ decays. For $B \to K^* \ell^+ \ell^-$ 
the pole region, $q^2 < m_\mu^2$, is included in $R_{K^*}$. 
The CP asymmetries are given for $B \to K^+ \ell^+ \ell^-$ and 
the combined $B \ra K^*\ell^+ \ell^-$ modes.}
{\footnotesize
\begin{tabular}{|c|c|c|c|c|}
\hline 
Experiment &Mode & $\BR \, [10^{-7}] $ & $ A_{\rm CP}$ & $ R_{K^{(*)}}$ \\ \hline
\babar~\cite{Aubert:2008ps} & $B \to K \ell^+ \ell^-$ & $\ 3.9^{+0.7}_{-0.7}\pm 0.2 $ & $-0.18^{+0.18}_{-0.18}\pm 0.01$ & $0.96^{+0.44}_{-0.34} \pm 0.05$ \\
& $B \to K^* \ell^+ \ell^-$ & $ 11.1^{+1.9}_{-1.8} \pm 0.7$ &$\ \ 0.01^{+0.16}_{-0.15}\pm 0.01$ &$1.1^{+0.42}_{-0.32}\pm 0.07$  \\
\hline \hline
Belle~\cite{:2009zv}  & $B \to K\ell^+ \ell^-$ & $\ 4.8^{+0.5}_{-0.4}\pm 0.3$ & $-0.04^{+0.1}_{-0.1}\pm 0.02 $  & $1.03^{+0.19}_{-0.19}\pm 0.06$ \\
& $B \to K^* \ell^+ \ell^-$ & $10.8^{+1.1}_{-1.0} \pm 0.9$ & $-0.10^{+0.10}_{-0.10}\pm 0.01$& $0.83^{+0.17}_{-0.17}\pm 0.08$ \\
\hline
\end{tabular}
}
\label{tab:rates}
\end{table}

The angular distribution of $B \ra K^* \ell^+ \ell^-$ depends on the
three angles defined in eqs.~\ref{eq:AngleDefBdb} and
\ref{eq:angledef2Bdb}. The one-dimensional angular distributions in
$\cos \theta_K$ and $\cos\theta_\ell$ simply are
\begin{equation}
\begin{split}
  W(\theta_K) &= \frac{3}{2} F_L \cos^2 \theta_K +
  \frac{3}{4}(1- F_L) (1-\cos^2 \theta_K) \,, \\
  W(\theta_\ell)& = \frac{3}{4} F_L (1- \cos^2 \theta_\ell) +
  \frac{3}{8}(1-F_L )(1+ \cos^2 \theta_\ell)+ A_{\rm FB} \cos
  \theta_\ell \,.
\end{split}
\end{equation}
\noindent 
While $W(\theta_K)$ depends only on $F_L$, $W(\theta_\ell)$ depends
both on $F_L$ and $A_{\rm FB}$. The FBA is proportional to the
difference of two interference terms that include products of the
Wilson coefficients $C_9 C_{10}$ and $C_7 C_{10}$. In the first term
the main $q^2$ dependence originates from the $q^2$ dependence of
$C_9$ while in the second term it results from the $1/q^2$ dependence
of the photon penguin contribution.
 
\babar and Belle measured $F_L$ and $A_{\rm FB}$ in different bins of
$q^2$. After extracting the event yield from the $m_{ES}$
distribution, $F_L$ is determined first from a fit to $W(\theta_K)$.
Then $A_{\rm FB}$ is determined from a fit to $W(\theta_\ell)$ for
fixed signal yields and fixed $F_L$. The results are summarized in
Tab.~\ref{tab:afb}. The \babar and Belle results for $F_L$ and $A_{\rm
  FB}$ in comparison to their SM predictions and three scenarios, that
result from changing the sign of the Wilson coefficients $C_7$, or
$C_9 C_{10}$, or both combinations with respect to the SM values are
shown in Fig.~\ref{fig:sll-afb}. At the present level of precision
both $F_L$ and $A_{\rm FB}$ are consistent with the SM
expectations. For $B \ra K \ell^+ \ell^-$, the measurement of $A_{\rm
  FB}$ is consistent with zero as expected in the SM. It is important
to emphasize, that models in which the sign of $C_7$ is reversed while
$C_{9,10}$ receive only small non-standard corrections are disfavored
at the $3 \sigma$ level by the combination of the $\BR (B \to X_s
\gamma)$ and $\BR (B \to X_s \ell^+ \ell^-)$ measurements
\cite{Gambino:2004mv}. The hypothetical NP scenario corresponding to
the green dashed curves in Fig.~\ref{fig:sll-afb} is thus in variance
with the available data on the inclusive $b \to s \gamma, \ell^+
\ell^-$ transitions. This observation makes clear that to bound the
values of the various Wilson coefficients one should exploit all the
experimental information in the $b \to s \gamma$ and $b \to s \ell^+
\ell^-$ sector combining both inclusive and exclusive channels.

\begin{table}
\centering
\caption{\babar and Belle measurements of the 
$K^*$ longitudinal polarizations and the lepton FBAs for the 
$B \ra K^* \ell^+ \ell^-$ decays in different bins of $q^2$.}
{\footnotesize
\begin{tabular}{|c|c|c|c|}
  \hline \hline 
  Experiment & $q^2 \, \rm [GeV^2]$ & $F_L$ & $A_{\rm FB}$ \\ \hline
  \babar~\cite{:2008ju} & $\ 0.1$--6.25 & $\ 0.35^{+0.16}_{-0.16}\pm 0.04$ & $0.24^{+0.18}_{-0.23}\pm 0.05$ \\
  & 10.24--12.96 or $>$14.06 & $\ 0.71^{+0.20}_{-0.22}\pm 0.04$ &$0.76^{+0.52}_{-0.32}\pm 0.07$  \\
  \hline \hline
  Belle~\cite{:2009zv} &$\ \ 0.0$--2.0  & $\ 0.29^{+0.21}_{-0.18}\pm 0.02$ & $0.47^{+0.26}_{-0.32}\pm 0.03$ \\
  & $ \ \ 2.0$--5.0 & $\ 0.75^{+0.21}_{-0.22}\pm 0.05$ & $0.14^{+0.20}_{-0.26}\pm 0.07 $ \\
  &$ \ \ 5.0$--8.86 & $\  0.65^{+0.26}_{-0.27}\pm 0.06 $ & $0.47^{+0.16}_{-0.25}\pm 0.14$ \\
  & $10.09$--12.86 & $\ 0.17^{+0.17}_{-0.15}\pm 0.03$ & $0.43^{+0.18}_{-0.20}\pm 0.03$ \\
  & $14.18$--16.0 & $ -0.15^{+0.27}_{-0.23}\pm 0.07 $ & $0.70^{+0.16}_{-0.22}\pm 0.10$ \\
  & $>16.0$  & $\ 0.12^{+0.15}_{-0.13}\pm 0.02$ & $ 0.66^{+0.11}_{-0.16}\pm 0.04$ \\ \hline
  & $\ \ 1.0$--6.0 & $\ 0.67^{+0.23}_{-0.23}\pm 0.05$ & $ 0.26^{+0.27}_{-0.30}\pm 0.07$ \\
  \hline
\end{tabular}
}
\label{tab:afb}
\end{table}

\begin{figure}[t!]
 \begin{center}
 \mbox{
\includegraphics[height=0.2\textheight]{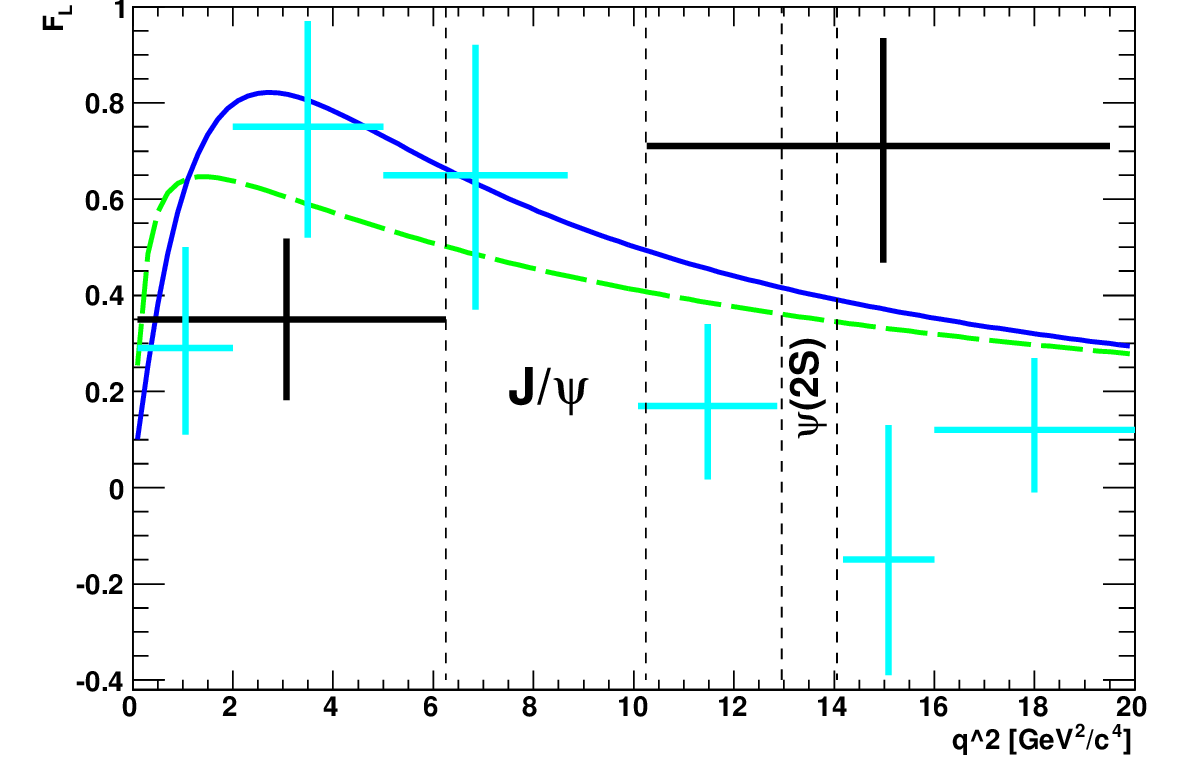} 

\vspace{5mm}

\includegraphics[height=0.2\textheight]{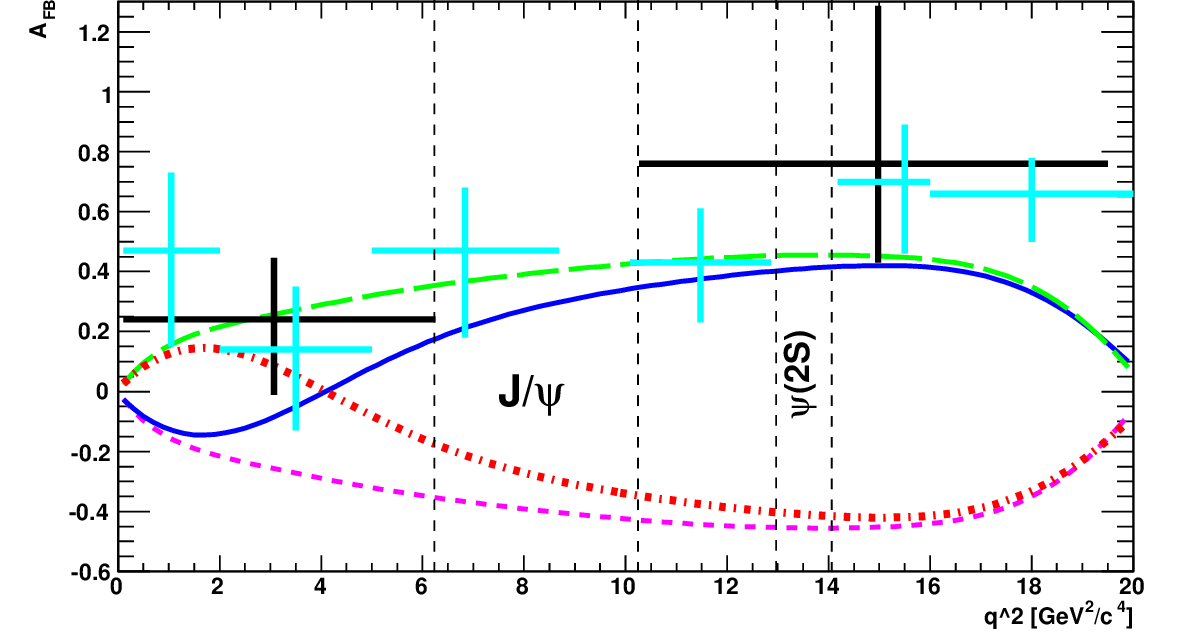}
}
 \end{center}
 \caption{\label{fig:sll-afb} Measurements of $F_L$ (left) and of
   $A_{\rm FB}$ (right) as a function of $q^2$ for $B \ra K^* \ell^+
   \ell^-$ from \babar (black points) and Belle (cyan points). The
   curves show predictions for four cases, the SM (blue solid curve),
   the flipped-sign $C_7$ case (green dashed curve), the case of
   flipped-sign $C_9 C_{10}$ (magenta dotted curve), and the case with
   both flipped-sign $C_7$ and $C_9 C_{10}$ (red dash-dotted curve).
 }
\end{figure}

The exclusive decays $B \ra \pi (\rho) \ell^+ \ell^-$ are the
corresponding $b \ra d$ transitions that are suppressed with respect
to the $b \ra s$ transitions by $|V_{td}/V_{ts}|^2$. \babar
\cite{Aubert:2007mm} searched for $B \ra \pi \ell^+ \ell^-$ events
using 230 million $B \bar B$ events while Belle \cite{Wei:2008nv}
searched for $B \ra (\pi, \rho, \omega) \ell^+ \ell^-$ modes using 657
million $B \bar B$ events. The lowest branching fraction upper limit
is set for the $B \ra \pi \ell^+ \ell^-$ mode by Belle yielding ${\cal
  B}(B \ra \pi \ell^+ \ell^-) < 4.9 \times 10^{-8}$ at 90\% CL which
just lies a factor of around $1.5$ above the central value of the SM
prediction which reads $\BR(B \ra \pi \ell^+ \ell^-) =(3.3\pm
1.0) \times 10^{-8}$~ \cite{Aliev:1998sk}.

The \lhcb experiment will collect $\sim7\times10^3$ fully reconstructed
$\Bz \to \Kstarz \mup\mun$ events per 2\invfb integrated
luminosity~\cite{Dickens:1028123}. At the LHC design luminosity, such an
data-set will be acquired in a single year of data-taking. Before such
conditions are achieved, with even the data from the LHC pilot run, a
0.1\invfb integrated luminosity would therefore give a
comparable number of events to the final dataset expected from the
B-factory experiments~\cite{:2008ju,:2009zv}
and CDF~\cite{Aaltonen:2008xf}. The branching ratio for this decay
is already measured with a precision comparable to the level of
theoretical uncertainties. \lhcb's experimental exploration will
therefore focus on angular observables such as the forward-backward
asymmetry $A_{\rm FB}$.

Given the background expected from simulation
studies~\cite{Dickens:1028123}, \lhcb will be able to determine the
zero-crossing point of this asymmetry by counting forward- and
backward-events with a statistical precision of $0.5\gev^2$ with
2\invfb integrated luminosity~\cite{Dickens:1045395}. Additional
systematic contributions to this precision from e.g.\ the
determination of acceptance and trigger efficiencies are under
study. A measurement with a statistical uncertainty at the level of
present theoretical uncertainties on the zero-crossing point (see
Sec.~\ref{sll-exclusive}) will therefore require a 10\invfb integrated
luminosity. However, using information from the $\theta_K$ angular
distribution, in particular, by making a simultaneous fit for both the
$A_{\rm FB}$ and $F_L$ observables, a factor $\sim2$ increase in the
statistical precision can be obtained~\cite{Reece:1103766}. Adding the
information from the angle $\phi$, a full angular fit will give a
further $\sim30\%$ increase in the
precision~\cite{Egede:1142152}. More significantly, such a full
angular fit will give access to the underlying amplitudes, from which
any observable can then be formed. As detailed in
Sec.~\ref{sll-angular}, \lhcb will be able to measure, with good
precision, other theoretically well controlled observables such as
$A_T^{(2,3,4)}$~(see Eq~\ref{eq:rare:ATDef}), which will give very
different new physics sensitivity to $A_{\rm FB}$. Studies indicate
that full angular fits can be made to converge with data-sets in
excess of the expectation from 2\invfb integrated luminosity. In
practise, performing such a fit will require excellent understanding
of the trigger and detector efficiencies and will be a later \lhcb
measurement.


\subsubsection{Rare $K \to \pi\nu\bar{\nu}, \ell^+\ell^-$ decays in
  and beyond the SM}
\label{sec:rare-Kaon}
The rare decays $K_{L}\rightarrow \pi^{0}\nu\bar{\nu}$,
$K^{+}\rightarrow\pi^{+}\nu\bar{\nu}$, and
$K_{L}\rightarrow\pi^{0}\ell^{+}\ell^{-}$ proceed dominantly through
heavy-quark induced FCNC. Since their rates in the SM are predicted
with high precision, they offer the cleanest and clearest window into
the sector of $s\rightarrow d$ transitions. Their study is thus
complementary to $B$ physics in searching for NP, and constraining the
possible models.

\paragraph{Prediction within the SM.} The electroweak processes
inducing the rare $K$ decays are of three types: $Z$ penguin and
$W^\pm$ boxes, single- and double-photon penguin. The former as well
as the CP-violating single-photon penguin, are always dominated by
short-distance physics, i.e., the top- and charm-quark
contribution. On the other hand, the CP-conserving photon penguins are
fully dominated by the long-distance up-quark contribution, in which
case they get further enhanced by the $\Delta I=1/2$ rule. These
contributions are to be evaluated in $\chi$PT, by relating them to
other, well-measured observables.

For $K_{L}\rightarrow\pi^{0}\nu\bar{\nu}$ and
$K^{+}\rightarrow\pi^{+}\nu\bar{\nu}$, short-distance physics
dominates because of the absence of photon penguins and the quadratic
GIM breaking exhibited in the $Z$ penguin. The calculation of the
branching ratios can be split into several pieces. First, the top
quark contribution $X_{t}$ is known including NLO QCD
effects \cite{Buchalla:1993bv,Misiak:1999yg}. While NLO electroweak effects have been
estimated in the large top-quark mass limit \cite{Buchalla:1997kz}. In
the charm-quark sector, the NNLO QCD \cite{Buras:2005gr, Buras:2006gb}
and electroweak \cite{Brod:2008ss} corrections have been computed,
significantly reducing the scheme and scale ambiguities in the
corresponding quantity $P_{c}$. For both these contributions, the
matrix elements of the resulting dimension-six operator, encoded in
$\kappa_{L}$ and $\kappa_{+}$, are obtained from the full set of
$K_{\ell3}$ data, including isospin-breaking and long-distance QED
corrections \cite{Mescia:2007kn}. Higher-dimensional contributions for
the charm quark, which are negligible in the case of the top quark
since they are suppressed by $m_{K}^{2}/m_t^{2}$, as well as the
residual up-quark contributions are parametrized by $\delta P_{u,c}$,
which has been estimated using $\chi$PT \cite{Isidori:2005xm}. The
error on $\delta P_{u,c}$ may be reduced through LQCD studies
\cite{Isidori:2005tv}. Finally, the rate for $K_{1}\approx K_{S}$ and
$K_{2}\approx K_{L}$ are similar, and thus indirect CP-violation,
$K_{L}\rightarrow\varepsilon K_{2}\rightarrow\pi^{0}\nu\bar{\nu}$, is
below the percent level since the smallness of $\varepsilon_{K} \sim
10^{-3}$ cannot be compensated \cite{Buchalla:1996fp}. Putting all
these pieces together, the $K\rightarrow\pi\nu\bar{\nu}$ rates are
predicted with a high level of accuracy in the SM
\begin{equation}
\begin{split}
  \mathcal{B}\left(  K_{L}\rightarrow\pi^{0}\nu\bar{\nu}\right)   
  &=\left(2.54\pm0.35\right) \times 10^{-11} \,, \\
  \mathcal{B}\left(  K^{+}\rightarrow\pi^{+}\nu\bar{\nu}\right)   
  &=\left(8.51\pm0.73\right) \times 10^{-11} \,.
\end{split}
\end{equation}
The composition of the quoted errors is as follows
$69\%_{\text{CKM}}$, $12\%_{\rm para}$, $15\%_{X_{t}}$,
$4\%_{\kappa_{L}}$ and $52\%_{\text{CKM}}$, $17\%_{\rm para}$,
$12\%_{X_{t}}$, $12\%_{\delta P_{u,c}}$, $5\%_{P_{c}}$, 
$2\%_{\kappa_{+}}$, where the parametric uncertainty combines the
errors on $m_{t}$, $m_{c}$, and $\alpha_{s}$.

For the $K\rightarrow\pi\ell^{+}\ell^{-}$ modes, besides the
short-distance top- and charm-quark contributions, some long-distance
effects arise due to the photon penguins. For the CP-odd $K_{2}$, the
single-photon penguin is CP-violating, hence still short-distance
dominated, and is known precisely \cite{Buchalla:1995vs}. On the other
hand, the double-photon penguin is a purely long-distance
CP-conserving contribution. It has been evaluated from
$K_{L}\rightarrow\pi^{0}\gamma\gamma$ data, and turns out to be
competitive in the muon case\cite{Buchalla:2003sj,
  Isidori:2004rb}. For the $K^{+}$ and the CP-even $K_{1}$, the
CP-conserving single-photon penguin completely dominates, hence these
modes do not give us access to the short-distance physics. Further,
this photon penguin is large enough to compensate for
$\varepsilon_{K}\sim10^{-3}$ in the indirect CP-violating
$K_{L}\rightarrow\varepsilon
K_{1}\rightarrow\pi^{0}\gamma^{\ast}\rightarrow\pi^{0}\ell^{+}\ell^{-}$
contribution \cite{D'Ambrosio:1998yj}. This piece can be brought under
control thanks to the
$\mathcal{B}(K_{S}\rightarrow\pi^{0}\ell^{+}\ell^{-})$ measurements,
up to its interference sign \cite{Bruno:1992za, Friot:2004yr,
  Buchalla:2003sj, Mescia:2006jd}. Nevertheless, the current
experimental accuracy for $\mathcal{B}%
(K_{S}\rightarrow\pi^{0}\ell^{+}\ell^{-})$ still represents the
largest source of uncertainty in the
$\mathcal{B}(K_{L}\rightarrow\pi^{0}\ell^{+}\ell^{-})$ predictions,
which are
\begin{equation}
\begin{split}
  \mathcal{B}(K_{L}\rightarrow\pi^{0}e^{+}e^{-}) &
  =3.54_{-0.85}^{+0.98}\;
  \left (1.56_{-0.49}^{+0.62} \right )\times 10^{-11} \,, \\[1mm]
  \mathcal{B}(K_{L}\rightarrow\pi^{0}\mu^{+}\mu^{-})&
  =1.41_{-0.26}^{+0.28}\; \left (0.95_{-0.21}^{+0.22} \right )\times
  10^{-11}\,,
\end{split}
\end{equation}
for constructive (destructive) interference.

For the $K_{L}\rightarrow\ell^{+}\ell^{-}$ modes, though the
short-distance top- and charm-quark contributions are predicted with
excellent accuracy \cite{Gorbahn:2006bm}, it is the long-distance
two-photon penguin which dominates. Its theoretical estimation is
problematic because, contrary to
$K_{L}\rightarrow\pi^{0}\gamma\gamma\rightarrow\pi^{0}\ell^{+}\ell^{-}$,
it i) diverges in $\chi$PT \cite{Isidori:2003ts} and ii) produces the
final lepton pair in the same state as the short-distance processes,
and hence interferes with them with an unknown sign. Better
measurements of $K_S \to \pi^0 \gamma \gamma$ and $K^+ \to \pi^+
\gamma \gamma$ could settle this issue \cite{Gerard:2005yk}. These two
problems have, up to now, upset attempts to extract the subleading
short-distance top- and charm-quark components from the well-measured
$\mathcal{B}(K_{L}\rightarrow\mu^{+}\mu^{-})$.

\paragraph{Sensitivity to NP effects.}

Rare $K$ decays are ideally suited to search for NP effects. Indeed,
besides the loop suppression of the underlying FCNC processes, they
are significantly CKM suppressed. Compared to ${\cal A}(b\rightarrow
s,d)$, the amplitudes in the $s \to d$ sector scale as 
\begin{equation} 
\begin{split}
  {\cal A} (s&\rightarrow d)\sim|V_{td}^\ast V_{ts}|\sim\lambda^{5}\,, \\
  {\cal A} (b&\rightarrow d)\sim|V_{td}^\ast V_{tb}|\sim\lambda^{3}\,, \\
  {\cal A} (b&\rightarrow s)\sim|V_{ts}^\ast
  V_{tb}|\sim\lambda^{2}\,,\label{Eq:rare:rareK3}
\end{split}
\end{equation}
with $\lambda\sim0.22$. If NP is generic, i.e., it does not follow the
CKM scaling (\ref{Eq:rare:rareK3}), it is clear that the constraints
from rare $K$ decays are typically the most stringent. Stated
differently, a measurement of $K_L\rightarrow\pi^0\nu\bar{\nu}$ close
to its SM prediction is the most difficult to reconcile with the
existence of generic NP at a reasonably low scale around a TeV.

NP models in which the CKM scalings (\ref{Eq:rare:rareK3}) are
preserved are referred to as of MFV type
\cite{D'Ambrosio:2002ex}. When this is the case, NP can show up at a
low scale without violating experimental bounds, including those from
rare $K$ decays. In addition, when MFV is enforced within a particular
model like the MSSM, the effects are expected to be rather small,
often beyond the experimental sensitivity. This has been analyzed at
moderate \cite{Isidori:2006qy} or large
$\tan\beta$\cite{Isidori:2001fv, Isidori:2002qe, Isidori:2006jh},
without $R$-parity \cite{Nikolidakis:2007fc}, or with MFV imposed at the
GUT scale \cite{Paradisi:2008qh, Colangelo:2008qp}. Turning this
around, the rare $K$ decays emerge as one of the best places to look
for deviations of the MFV hypothesis \cite{Bobeth:2005ck,
  Haisch:2007ia, Hurth:2008jc}. If the flavor-breaking transitions
induced by the NP particles are not precisely aligned with those of
the SM, large effects can show up. This is true even given the current
measurement of the $K^{+}\rightarrow\pi^{+}\nu\bar{\nu}$ mode. The
model-independent bound it implies on the
$K_{L}\rightarrow\pi^{0}\nu\bar{\nu}$ mode is still about 30 times
higher that the SM prediction \cite{Grossman:1997sk}.

Each NP model affects the basic electroweak FCNC differently. If it
enters into the $Z$ penguin, the two $K\rightarrow\pi\nu\bar{\nu}$
modes exhibit the best sensitivity. This happens for example in the
MSSM from chargino-squark loops at moderate $\tan\beta
$\cite{Nir:1997tf, Buras:1997ij, Colangelo:1998pm, Buras:2000qz,
  Buras:2004qb} or charged-Higgs-quark loops at large $\tan\beta$
\cite{Isidori:2006jh}, with $R$-parity violation \cite{Grossman:2003rw,
  Deshpande:2004xc,D eandrea:2004ae}, in little Higgs models  without \cite{Buras:2006wk} and with 
\cite{Blanke:2006eb,Blanke:2007wr,Goto:2008fj,Blanke:2009am} T-parity, and in the presence of
extra-dimensions \cite{Buras:2002ej, Freitas:2008vh,
  Blanke:2008yr}. In most of these models, correlated changes to the
short-distance photon penguin are induced
\cite{Cho:1996we,Bobeth:2001jm}, and these could then be probed and
disentangled using the $K_{L}\rightarrow\pi^{0}\ell^{+}\ell^{-}$
modes. Combined measurements of all the rare $K$ decay modes can serve
as a powerful discriminator among models \cite{Isidori:2004rb,
  Mescia:2006jd}. Further, purely electromagnetic effects could also
be present, as in the electromagnetic operators, for which the
$K_{L}\rightarrow\pi^{0}\ell^{+}\ell^{-}$ modes are clean probes while
$\varepsilon^{\prime}$ is problematic \cite{Buras:1999da}.

In addition, NP could occur with helicity-suppressed couplings
proportional to the fermion mass. Typical examples are the neutral
Higgs-induced FCNC, as generated in the MSSM at large $\tan\beta$
\cite{Hall:1993gn, Babu:1999hn, Isidori:2001fv,
  Isidori:2002qe}. Obviously, the
$K_{L}\rightarrow\pi^{0}\mu^{+}\mu^{-}$ and
$K_{L}\rightarrow\mu^{+}\mu^{-}$ modes are the only available windows
for such helicity-suppressed effects in the $s\rightarrow d$
sector. Therefore, these effects can in principle be disentangled from
NP in the $Z$ or photon penguins by a combined analysis of all the
rare $K$ decay modes \cite{Mescia:2006jd}.

In conclusion, the $K\rightarrow\pi\nu\bar{\nu}$ modes offer one of
the best opportunities to find a irrefutable signal of NP in the field
of flavor physics. Furthermore, combining information on the
different $K\rightarrow\pi\nu\bar{\nu}$ and
$K_{L}\rightarrow\pi^{0}\ell^{+}\ell^{-}$ channels allows one to probe
and disentangle NP effects in most of the different types of FCNC
interactions. Being either free of hadronic uncertainties, or these
being under sufficiently good theoretical control, the stage is set
for a complete and detailed study of $s\rightarrow d$ transitions.

\subsubsection{Experimental status of $K \to \pi\nu\bar{\nu}$ and $K_L\to\pi\ell^+\ell^-$}

The E787 and E949 experiments have established the feasibility of
observing the $K^+\to\pi^+\nu\bar\nu$ decay using a stopped Kaon
beam~\cite{Adler:2008zza}. Observation of seven candidate events by
E787 and E949 yields ${\cal B}(K^+\to\pi^+\nu\bar\nu) = \left
  (1.73^{+1.15}_{-1.05} \right )\times10^{-10}$ when the relative
acceptance and measured background are taken into account with a
likelihood method~\cite{Artamonov:2008qb}.  It has been estimated
that, assuming the SM decay rate, a stopped $K^+$ experiment could
accumulate hundreds of $K^+\to\pi^+\nu\bar\nu$ events, using a copious
proton source such as Project-X at FNAL~\cite{ProjectX_GoldBook}.  The
NA62 experiment at CERN seeks to observe on the order of a hundred
$K^+\to\pi^+\nu\bar\nu$ decays using a decay-in-flight technique in an
unseparated 75 GeV beam.

The experiment E391a has set a limit of ${\cal
  B}(K_L\to\pi^0\nu\bar\nu) < 670\times10^{-10}$ at 90\% CL in a
sample of $5.1\times10^9$ $K_L$ decays~\cite{Ahn:2007cd}. The
experimental result is still larger than the model-independent
limit~\cite{Grossman:1997sk} of ${\cal B}(K_L\to\pi^0\nu\bar\nu) <
14.6\times10^{-10}$ at 90\% CL implied by the $K^+\to\pi^+\nu\bar\nu$
results. E391a is currently analyzing an additional $3.6\times10^9$
$K_L$ decays and plans to implement an upgraded detector in the
experiment E14 at JPARC that would have a sensitivity comparable to
the expected SM $K_L\to\pi^0\nu\bar\nu$ decay rate.

The experimental limits on $K_L\to\pi^0 e^+e^-$ and $K_L\to\pi^0
\mu^+\mu^-$ are $2.8\times10^{-10}$ and $3.8\times10^{-10}$ at 90\% CL
by the KTeV collaboration~\cite{Amsler:2008zz}.  The $K_L\to\pi^0
e^+e^-$ mode suffers from an irreducible background from
$K_L\to\gamma\gamma e^+e^-$ decays, ${\cal B}( K_L\to\gamma\gamma
e^+e^-) = (5.95\pm0.33)\times10^{-7}$, that can be suppressed by a
precise diphoton mass resolution. There are currently no experiments
planned to continue the search for these decays.


\subsection{Rare D meson decays}

\subsubsection{Rare leptonic decays}

In the Standard Model (SM) flavor-changing neutral current (FCNC) decays of 
charm hadrons are highly suppressed by the GIM mechanism
\cite{Glashow:1970gm}.  In the process $\Dz \to X_u\ellell$
this leads to branching fractions of $\order(10^{-8})$~\cite{Burdman:2001tf}. However, this
process can be enhanced by the presence of long-distance
contributions, increasing the branching fractions by several orders of
magnitude~\cite{Burdman:2001tf}. The effect of these long distance contributions
from intermediate resonances can be separated by examining the invariant
mass of the lepton pair (e.g. $\phi\to\ellell$). In radiative charm
decays (e.g. $c\to u\gamma$), the long distance contributions are not
so easily determined, making it increasingly difficult to study the
short-distance effects. 
The branching fractions of the $\Dz \to \ellell$ final state are
predicted to be $\order(10^{-13})$~\cite{Burdman:2001tf}, including contributions from long
distance processes.

Lepton family-number violating (LFV), and lepton-number violating (LV) decays are strictly
forbidden in the SM. The processes are allowed
in extensions to the SM with non-zero neutrino mass but at a very low level~\cite{Burdman:2001tf}.
A large impact is expected to come from R-parity violating
super-symmetry. Depending on the size of the R-parity violating
couplings, branching fractions for these processes can be enhanced up to
the  $\order(10^{-6})$ level for differing $c\to u\ellell$ processes.

The search for FCNC processes in charm decays has not received the attention that 
the $K$ and $\B$ meson sectors have attracted. The current measurements
of these decays (Tab.~\ref{t1}-\ref{t4}) agree with SM predictions, and there are ongoing efforts to improve
both theoretical predictions and experimental limits. There is also ongoing
effort to measure new effects such as \emph{CP} violation in these processes.

\begin{table}[!h]
\centering
\caption{90\% confidence limits on Flavor-changing neutral current, (FCNC), lepton family-number
(LFV) violating, or lepton-number (LV) violating decay modes of the \Dp~(left) and the \Ds~(right)~\cite{Amsler:2008zzb}.}
  \begin{minipage}[t]{0.45\linewidth}
\begin{tabular}{ lllll }
\hline \hline
Process & Decay type & \multicolumn{2}{c}{Upper limit} & Reference \\
\hline
\pip\epem & FCNC & $<$ 7.4 & $\times 10^{-6}$ &  \cite{He:2005iz}\\
\pip\mumu & FCNC & $<$ 3.9 & $\times 10^{-6}$ &  \cite{Abazov:2007kg}\\
$\rho^+$\mumu & FCNC & $<$ 5.6 & $\times 10^{-4}$ &  \cite{Kodama:1995ia}\\
\Kp\epem & N/A\footnote{These modes are not a useful test for FCNC, because both quarks must change flavor.} & $<$ 6.2 & $\times 10^{-6}$ &  \cite{He:2005iz}\\
\Kp\mumu & N/A$^{\rm{\footnotesize{a}}}$ & $<$ 9.2 & $\times 10^{-6}$ &  \cite{Link:2003qp}\\
\pip\epm\ensuremath{\mu^\mp} & LFV & $<$ 3.4 & $\times 10^{-5}$ &  \cite{Aitala:1999db}\\
\Kp\epm\ensuremath{\mu^\mp} & LFV & $<$ 6.8 & $\times 10^{-5}$ &  \cite{Aitala:1999db}\\
\pim\ep\ep & LV & $<$ 3.6 & $\times 10^{-6}$ &  \cite{He:2005iz}\\
\pim\mup\mup & LV & $<$ 4.8 & $\times 10^{-6}$ &  \cite{Link:2003qp}\\
\pim\ep\mup & LV & $<$ 5.0 & $\times 10^{-5}$ &  \cite{Aitala:1999db}\\
$\rho^-$\mup\mup & LV & $<$ 5.6 & $\times 10^{-4}$ &  \cite{Kodama:1995ia}\\
\Km\ep\ep & LV & $<$ 4.5 & $\times 10^{-6}$ &  \cite{He:2005iz}\\
\Km\mup\mup & LV & $<$ 1.3 & $\times 10^{-5}$ &  \cite{Link:2003qp}\\
\Km\ep\mup & LV & $<$ 1.3 & $\times 10^{-4}$ &  \cite{Frabetti:1997wp}\\
\Kstarm\mup\mup & LV & $<$ 8.5 & $\times 10^{-4}$ &  \cite{Kodama:1995ia} \\
\hline \hline
\end{tabular}
  \end{minipage}
  \hspace*{\fill}
  \begin{minipage}[t]{0.45\linewidth}

\begin{tabular}{ lllll }
\hline \hline
Process & Decay type & \multicolumn{2}{c}{Upper limit} & Reference \\
\hline
\pip\epem         & N/A$^{\rm{\footnotesize{a}}}$ & $<$ 2.7 & $\times 10^{-4}$ &  \cite{Aitala:1999db}\\
\pip\mumu         & N/A$^{\rm{\footnotesize{a}}}$  & $<$ 2.6 & $\times 10^{-5}$ &  \cite{Link:2003qp}\\
\Kp\epem      & FCNC & $<$ 1.6 & $\times 10^{-3}$ &  \cite{Aitala:1999db}\\
\Kp\mumu      & FCNC & $<$ 3.6 & $\times 10^{-5}$ &  \cite{Link:2003qp}\\
\Kstarm\mumu & FCNC & $<$ 1.4 & $\times 10^{-3}$ &  \cite{Kodama:1995ia}\\
\pip\epm\ensuremath{\mu^\mp} & LFV & $<$ 6.1 & $\times 10^{-4}$ &  \cite{Aitala:1999db}\\
\Kp\epm\ensuremath{\mu^\mp}  & LFV & $<$ 6.3 & $\times 10^{-4}$ &  \cite{Aitala:1999db}\\
\pim\ep\ep    & LV & $<$ 6.9 & $\times 10^{-4}$ &  \cite{Aitala:1999db}\\
\pim\mup\mup  & LV & $<$ 2.9 & $\times 10^{-5}$ &  \cite{Link:2003qp}\\
\pim\ep\mup   & LV & $<$ 7.3 & $\times 10^{-4}$ &  \cite{Aitala:1999db}\\
\Km\ep\ep     & LV & $<$ 6.3 & $\times 10^{-4}$ &  \cite{Aitala:1999db}\\
\Km\mup\mup   & LV & $<$ 1.3 & $\times 10^{-5}$ &  \cite{Link:2003qp}\\
\Km\ep\mup    & LV & $<$ 6.8 & $\times 10^{-4}$ &  \cite{Aitala:1999db}\\
\Kstarm\mup\mup & LV & $<$ 1.4 & $\times 10^{-3}$ &  \cite{Kodama:1995ia}\\
\hline \hline
\end{tabular}
  \end{minipage}
\label{t1}
\end{table}

\begin{table}[!h]
\centering
\caption{90\% confidence limits on flavor-changing neutral current (FCNC), or lepton-number
  (LV) violating decay modes of the
  \textbf{$\Lambda_c$}~\cite{Amsler:2008zzb}.} 
\begin{tabular}{ lllll }
\hline \hline
Process & Decay type & \multicolumn{2}{c}{Upper limit} & Reference \\
\hline
$p$\mumu &  FCNC & $<$ 3.4 & $\times 10^{-4}$ &  \cite{Kodama:1995ia}\\
$\Sigma^-$\mup\mup & LV & $<$ 7.0 & $\times 10^{-4}$ &  \cite{Kodama:1995ia}\\
\hline \hline
\end{tabular}
\label{t3}
\end{table}

\begin{table}[!h]
\centering
\caption{90\% confidence limits on flavor-changing neutral current (FCNC), lepton family-number
(LFV) violating, or lepton-number (LV) violating decay modes of the \textbf{\Dz}~\cite{Amsler:2008zzb}.}
\label{t4}
  \begin{minipage}[t]{0.45\linewidth}
\begin{tabular}{ lllll }
\hline \hline
Process & Decay type & \multicolumn{2}{c}{Upper limit} & Reference \\
\hline
$\gaga$           & FCNC & $<$ 2.7 & $\times 10^{-5}$ &  \cite{Coan:2002te}\\
\epem         & FCNC & $<$ 1.2 & $\times 10^{-6}$ &  \cite{Aubert:2004bs}\\
\mumu         & FCNC & $<$ 1.3 & $\times 10^{-6}$ &  \cite{Aubert:2004bs}\\
\piz\epem     & FCNC & $<$ 4.5 & $\times 10^{-5}$ &  \cite{Freyberger:1996it}\\
\piz\mumu     & FCNC & $<$ 1.8 & $\times 10^{-4}$ &  \cite{Kodama:1995ia}\\
$\eta$\epem     & FCNC & $<$ 1.1 & $\times 10^{-4}$ &  \cite{Freyberger:1996it}\\
$\eta$\mumu     & FCNC & $<$ 5.3 & $\times 10^{-4}$ &  \cite{Freyberger:1996it}\\
\pipi\epem    & FCNC & $<$ 3.73 & $\times 10^{-4}$ &  \cite{Aitala:2000kk}\\
$\rho^0$\epem   & FCNC & $<$ 1.0 & $\times 10^{-4}$ &  \cite{Freyberger:1996it}\\
\pipi\mumu    & FCNC & $<$ 3.0 & $\times 10^{-5}$ &  \cite{Aitala:2000kk}\\
$\rho^0$\mumu   & FCNC & $<$ 2.2 & $\times 10^{-5}$ &  \cite{Aitala:2000kk}\\
$\omega$\epem   & FCNC & $<$ 1.8 & $\times 10^{-4}$ &  \cite{Freyberger:1996it}\\
$\omega$\mumu   & FCNC & $<$ 8.3 & $\times 10^{-4}$ &  \cite{Freyberger:1996it}\\
\KpKm\epem    & FCNC & $<$ 3.15 & $\times 10^{-4}$ &  \cite{Aitala:2000kk}\\
$\phi$\epem     & FCNC & $<$ 5.2 & $\times 10^{-5}$ &  \cite{Freyberger:1996it}\\
\KpKm\mumu    & FCNC & $<$ 3.3 & $\times 10^{-5}$ &  \cite{Aitala:2000kk}\\
$\phi$\mumu     & FCNC & $<$ 3.1 & $\times 10^{-5}$ &  \cite{Aitala:2000kk}\\
\Kzb\epem     & N/A\footnote{These modes are not a useful test for FCNC, because both quarks must change flavor.}      & $<$ 1.1 & $\times 10^{-4}$ &  \cite{Freyberger:1996it}\\
\Kzb\mumu     &  N/A$^{\rm{\footnotesize{a}}}$     & $<$ 2.6 & $\times 10^{-4}$ &  \cite{Kodama:1995ia}\\
\Km\pip\epem  & FCNC & $<$ 3.85 & $\times 10^{-4}$ &  \cite{Aitala:2000kk}\\
\Kstarzb\epem &   N/A$^{\rm{\footnotesize{a}}}$   & $<$ 4.7 & $\times 10^{-5}$ &  \cite{Aitala:2000kk}\\
\Kp\pip\mumu  & FCNC & $<$ 3.59 & $\times 10^{-4}$ &  \cite{Aitala:2000kk}\\
\hline \hline
\end{tabular}
  \end{minipage}
  \hspace*{\fill}
  \begin{minipage}[t]{0.45\linewidth}
\begin{tabular}{ lllll }
\hline \hline
Process & Decay type & \multicolumn{2}{c}{Upper limit} & Reference \\
\hline

\Kstarzb\mumu &   N/A$^{\rm{\footnotesize{a}}}$  & $<$ 2.4 & $\times 10^{-5}$ &  \cite{Aitala:2000kk}\\
\pipi\piz\mumu& FCNC & $<$ 8.1 & $\times 10^{-4}$ &  \cite{Kodama:1995ia}\\
\epm\ensuremath{\mu^\mp}         & LFV     & $<$ 8.1 & $\times 10^{-7}$ &  \cite{Aubert:2004bs}\\
\piz\epm\ensuremath{\mu^\mp}     & LFV     & $<$ 8.6 & $\times 10^{-5}$ &  \cite{Freyberger:1996it}\\
$\eta$\epm\ensuremath{\mu^\mp}     & LFV     & $<$ 1.0 & $\times 10^{-4}$ &  \cite{Freyberger:1996it}\\
\pipi\epm\ensuremath{\mu^\mp}    & LFV     & $<$ 1.5 & $\times 10^{-5}$ &  \cite{Aitala:2000kk}\\
$\rho^0$\epm\ensuremath{\mu^\mp}   & LFV     & $<$ 4.9 & $\times 10^{-5}$ &  \cite{Freyberger:1996it}\\
$\omega$\epm\ensuremath{\mu^\mp}   & LFV     & $<$ 1.2 & $\times 10^{-4}$ &  \cite{Freyberger:1996it}\\
\KmKp\epm\ensuremath{\mu^\mp}    & LFV     & $<$ 1.8 & $\times 10^{-4}$ &  \cite{Aitala:2000kk}\\
$\phi$\epm\ensuremath{\mu^\mp}     & LFV     & $<$ 3.4 & $\times 10^{-5}$ &  \cite{Freyberger:1996it}\\
\Kzb\epm\ensuremath{\mu^\mp}     & LFV     & $<$ 1.0 & $\times 10^{-4}$ &  \cite{Freyberger:1996it}\\
\Km\pip\epm\ensuremath{\mu^\mp}  & LFV     & $<$ 5.53 & $\times 10^{-4}$ &  \cite{Freyberger:1996it}\\
\Kstarzb\epm\ensuremath{\mu^\mp} & LFV     & $<$ 8.3 & $\times 10^{-5}$ &  \cite{Aitala:2000kk}\\
\pim\pim\ep\ep + c.c   & LV & $<$ 1.12 & $\times 10^{-4}$ &  \cite{Aitala:2000kk}\\
\pim\pim\mup\mup + c.c & LV & $<$ 2.9 & $\times 10^{-5}$ &  \cite{Aitala:2000kk}\\
\Km\pim\ep\ep + c.c    & LV & $<$ 2.06 & $\times 10^{-4}$ &  \cite{Aitala:2000kk}\\
\Km\pim\mup\mup + c.c  & LV & $<$ 3.9 & $\times 10^{-4}$ &  \cite{Aitala:2000kk}\\
\Km\Km\ep\ep + c.c     & LV & $<$ 1.52 & $\times 10^{-4}$ &  \cite{Aitala:2000kk}\\
\Km\Km\mup\mup + c.c   & LV & $<$ 9.4 & $\times 10^{-5}$ &  \cite{Aitala:2000kk}\\
\pim\pim\ep\mup + c.c  & LV & $<$ 7.9 & $\times 10^{-5}$ &  \cite{Aitala:2000kk}\\
\Km\pim\ep\mup + c.c   & LV & $<$ 2.18 & $\times 10^{-4}$ &  \cite{Aitala:2000kk}\\
\Km\Km\ep\mup + c.c    & LV & $<$ 5.7 & $\times 10^{-5}$ &  \cite{Aitala:2000kk}\\
\hline \hline
\end{tabular}
  \end{minipage}

\end{table}

Therefore, searching for FCNC, LFV, or LV modes in the charm sector
is a relatively inviting place to investigate new physics in the SM.
Similar arguments hold for rare decays in the $K$ and $B$
sector. However, the charm system is unique in that it couples an
up-type quark to new physics. 

It is clear that due to the relatively little experimental progress in
this area within the last decade and the large data sets from the
flavor factories, that there is a several orders of magnitude 
in precision to be gained from re-reanalyzing these measurements with meaningful
limits to be derived which may have the potential to constrain
parameter space for many new physics models. At present the upper
limits for branching fractions for those modes more recently
measured~\cite{Aubert:2004bs, He:2005iz,Abazov:2007kg} are 
starting to confine the allowed parameter space of R-parity violating
super-symmetric models.

\subsubsection{ $D$ and $D_s$ decay constants from lattice QCD}

Quark confinement inside hadrons makes the direct experimental 
determination of how quarks change from one flavor to another 
via the weak interactions impossible. 
Instead we must study experimentally the decay of a hadron,
calculate the effect of the strong force 
on the quarks in the hadron and then correct 
for this to expose the quark interaction with the 
W boson. The simplest such hadron decay is annihilation 
of a charged pseudoscalar into a W and thence into 
a lepton and an antineutrino. 
The leptonic width of such a pseudoscalar meson, $P$, of quark content 
$a\overline{b}$ (or $\overline{a}b$) is given
by:
\begin{equation}
\Gamma(P \rightarrow l \nu_l (\gamma)) = \frac{G_F^2 
|V_{ab}|^2}{8\pi}f_{P}^2m_l^2m_{P}\left( 1-\frac{m_l^2}{m_{P}^2}\right)^2.
\label{eq:gamma}
\end{equation}
$V_{ab}$ is from the Cabibbo-Kobayashi-Maskawa (CKM) 
matrix element which encapsulates the Standard Model 
description of quark coupling to the $W$. 
$f_{P}$, the decay constant, parametrizes the amplitude for 
the meson annihilation to a $W$
and is basically the probability for the 
quark and antiquark to be in the same place.  
It is defined by:
\begin{equation}
f_Pm_P = <0|\overline{\psi}(x) \gamma_0 \gamma_5 \psi(x) |P(\vec{p}=0)>
\label{eq:fdef}
\end{equation}
Note that $f_P$ is a property of the meson in pure QCD. 
In the real world there is also electromagnetism 
and so the experimental 
rate must be corrected for this. It is a small (1-2\%) effect, 
except for very heavy mesons ($B$s) decaying to very light 
leptons (electrons)~\cite{Yao:2006px}.
If $V_{ab}$ is known from elsewhere
an 
experimental value for $\Gamma$ gives $f_{P}$, 
to be compared to theory. 
If not, an accurate theoretical value for $f_{P}$, 
combined with experiment, can yield a value 
of $V_{ab}$. 

Accuracy in both experiment and theory is important 
for useful tests of the Standard Model.  Here the numerical 
techniques of lattice QCD come to the fore for the 
theoretical calculation because it is now possible 
to do such calculations accurately~\cite{Davies:2003ik} and the pseudoscalar 
decay constant is one of the simplest quantities to 
calculate in lattice QCD. 

A lattice QCD calculation proceeds by splitting space-time up into 
a lattice of points (with spacing $a$) and generating sets of gluon fields 
on 
the lattice that are `typical snapshots of the vacuum'. For 
accurate calculations these snapshots need to include the 
effect of quark-antiquark pairs, known as `sea' quarks, 
generated by energy fluctuations in the vacuum. The important 
sea quarks are those which cost little energy to make i.e. 
the light $u$, $d$ and $s$. Unfortunately in lattice QCD 
it is numerically expensive to work with sea $u$ and $d$ masses 
that are close to their physical values and we have to 
extrapolate to the physical point from heavier values using 
chiral perturbation theory.  Valence quarks that make up a 
hadron are propagated through these gluon fields, allowing 
any number of interactions. We tie together appropriate valence quark 
and antiquark propagators 
to make, for example, a meson correlator which is then printed out as a 
function 
of lattice time, $t$ (we sum on spatial lattice sites to project 
on to zero spatial momentum). We fit as a function of $t$ to  
a multi-exponential form: 
\begin{equation}
<0|H^{\dag}(0)H(t)|0> = \sum_i A_i (e^{-E_it} + e^{-E_i(T-t)})
\end{equation}
where $T$ is the time extent of the lattice. The smallest value of 
$E_i$ (corresponding to the state that survives to large $t$) is 
the ground state mass in that channel, and $A_i$ is the square 
of the matrix element between the vacuum and $P$ of the operator $H$ used 
to create and 
destroy the hadron. If $H$ is the local temporal axial current of 
equation~\ref{eq:fdef}
(and this is the operator used if the valence quark and antiquark 
are simply tied together at the same start and end points matching colors 
and spins)
then $A_0$ will be directly related 
to the decay constant of the ground state pseudoscalar.

For $K$ and $\pi$ mesons several very accurate decay constant 
determinations have been done now in lattice QCD including 
the full effect of $u$, $d$ and $s$ quarks in the sea, 
and at several values of the lattice spacing. 
Extrapolations 
to the physical point in the $u/d$ mass and $a = 0$ have 
been done with a full error budget. The lattice 
value of $f_K/f_{\pi}$ can be used to determine $V_{us}$ to 
1\% accuracy. (* This is 
presumably described in the subsection on strange physics, so you only 
need a reference to that here *).  

\begin{figure}
\begin{center}
\includegraphics[width=6.5cm,angle=-90]{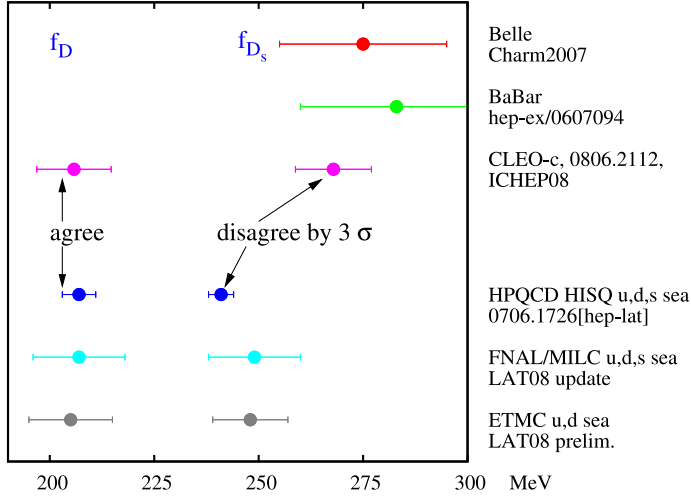}
\end{center}
\caption{A comparison of lattice results for the $D$ and $D_s$ decay 
constant~\cite{Follana:2007uv, Aubin:2005ar, Blossier:2008dj} and 
experimental results obtained from the leptonic decay rate using CKM 
elements $V_{cs}$ and 
$V_{cd}$ from elsewhere~\cite{:2007zm, :2008sq, 
Zhang:2008fe, Aubert:2006sd,Widhalm:2007mi}. There is agreement between 
lattice and experiment for $f_D$, but not for $f_{D_s}$. }
\label{fig:fd}
\end{figure}

For the charged charmed mesons $D_d$ and $D_s$ the lattice determination 
of 
the decay constant can be compared to an experimental one derived 
from the leptonic decay rate if values of
$V_{cd}$ and $V_{cs}$ are assumed (usually $V_{us} = V_{cd}$ and $V_{ud} = 
V_{cs}$).
This is an important test of modern lattice techniques, that can be 
used to calibrate lattice errors. The errors expected in the $D/D_s$ 
case are similar to those for $K/\pi$. The statistical errors on the 
raw lattice numbers are similar, and the extrapolations that must 
be done in the $u/d$ mass are less of an issue than for $\pi$. 
Indeed for the $D_s$, which has no valence $u/d$ quark, there
is very little dependence on the 
(sea) $u/d$ mass and so very little extrapolation. 

The extrapolations to $a=0$ are worse for $D/D_s$ than for 
$K/\pi$ and the reason is that the charm quark mass in lattice units, 
$m_ca$, is relatively large. Typically a lattice result at non-zero 
lattice spacing will have a power series dependence on the lattice 
spacing with the scale of the $a$-dependent terms set by a typical 
momentum inside the bound state. Extrapolations to $a=0$ can then 
be done using this functional form, and the resulting error will 
depend on the size of the extrapolation. For charm physics the 
scale of the $a$-dependent terms is set by $m_c$ and we expect
\begin{equation}
m = m_{a=0}(1+A(m_ca)^2 + B(m_ca)^4 + \ldots).
\end{equation}
The extrapolation can then be quite severe, and will determine the final 
error, if we do not take steps 
to control or eliminate terms in this series by improving the action. 

The Highly Improved Staggered Quark (HISQ) action for charm 
quarks~\cite{Follana:2006rc} 
eliminates the $(m_ca)^2$ term and results at three values of 
the lattice spacing then give an accurate extrapolation to $a=0$ with 
a 2\% final error~\cite{Follana:2007uv}. 
Alternatives to this are the `Fermilab interpretation' of improved Wilson 
quarks~\cite{Aubin:2005ar} and the twisted mass formalism~\cite{Blossier:2008dj}. Both 
have larger errors than for 
HISQ at present. Improved Wilson quarks have discretization 
errors at $\alpha_s(m_ca)$ in principle but the Fermilab interpretation 
removes the leading errors that come from the kinetic energy, and 
experience 
has shown that $a$-dependence is small in this formalism. However, 
relativity 
 is given up and this means, for example, that the masses of mesons 
cannot be as accurately tuned and a renormalisation factor is needed to 
relate the decay constant on the lattice to a result appropriate to the 
real world (at $a=0$). The twisted mass formalism uses a relativistic 
framework with errors appearing first at $m_ca)^2$. It has so far been 
applied at two values of the lattice spacing for gluon field 
configurations 
that do not include $s$ sea quarks, so are not completely realistic. 

Fig.~\ref{fig:fd} shows a comparison of these lattice calculations 
to new experimental results from CLEO-c for both $f_D$ and $f_{D_s}$ 
(which appeared after~\cite{Follana:2007uv, Aubin:2005ar}) and older results 
for $f_{D_s}$ from BaBar~\cite{Aubert:2006sd}. The 
good 
agreement between 
lattice and experiment for $f_D$ (and $f_K$ and $f_{\pi}$) 
contrasts with the $3\sigma$ disagreement for $f_{D_s}$ and 
it has been suggested that this is a harbinger of new 
physics~\cite{Dobrescu:2008er}. 
Improved experimental errors for $f_{D_s}$ will shed light on 
this. Meanwhile, lattice calculations and their systematic errors are also 
being tested against other 
quantities in charm physics~\cite{Davies:2008nq}. 

\subsubsection{Experimental results on $f_D$ }

Fully leptonic decays of $D^{+}_{(s)}$ mesons depend
upon both the weak and strong interactions. The weak
part is straightforward to describe in terms of the
annihilation of the quark antiquark pair to a $W^{+}$
boson. The strong interaction is required to describe
the gluon exchange between the quark and antiquark. The
strong interaction effects are parametrized by the
decay constant, $f_{D^{+}_{(s)}}$, such that the total
decay rate is given by 
\begin{displaymath}
 \Gamma(D^{+}_{(s)}\to l^{+}\nu) =
\frac{G_{F}^{2}}{8\pi}f^{2}_{D^{+}_{(s)}}m^{2}_{l}M_{D^{
+}_{(s)}}[1-\frac{m_{l}^{2}}{M^{2}_{D^{+}_
{(s)}}}]^{2}|V_{cd(s)}|^{2} \;, 
\end{displaymath}
where $G_{F}$ is the Fermi coupling constant,
$M_{D^{+}_{(s)}}$ and $m_{l}$ are the $D^{+}_{(s)}$
meson and final state lepton masses, respectively, and
$V_{cd(s)}$ is a Cabibbo-Kobayashi-Maskawa (CKM) matrix
element. The values of $V_{cd}$ and $V_{cs}$ can be
equated $V_{us}$ and $V_{ud}$, which are well known.
Therefore, within the standard model, measurements of
the fully leptonic decay rates allow a determination of
$f_{D^{+}_{(s)}}$.

Measurements of $f_{D^{+}_{(s)}}$ can be compared to
calculations from theories of QCD, the most precise of
which use unquenched lattice techniques (see for example
Ref. \cite{Follana:2007uv}.) Similar calculations of strong
parameters in $B$ meson decay 
are relied upon to extract CKM matrix elements, such as
$|V_{td}|/|V_{ts}|$ from the rates of $B$ mixing.
Therefore, comparing predictions for $f_{D^{+}_{(s)}}$
to measurements is important for validating the QCD
calculation techniques. Deviations of
experimental measurements from theoretical predictions
may be a consequence of non-SM physics (see for example
Ref.~\cite{Dobrescu:2008er}).

CLEO-c provides the most precise experimental
determinations of $f_{D^{+}}$ \cite{:2008sq} and
$f_{D^{+}_{s}}$ \cite{Alexander:2009ux,Onyisi:2009th} to date.
All measurements at CLEO-c exploit the recoil technique
described in Sec.~\ref{sec:recoiltechnique}.

The determination of $f_{D^{+}}$ uses six hadronic
decays of the $D^{-}$ as tags: $K^{+}\pi^{-}\pi^{-}$,
$K^{+}\pi^{-}\pi^{-}\pi^{0}$, $K^{0}_{S}\pi^{-}$,
$K^{0}_{S}\pi^{-}\pi^{-}\pi^{+}$,
$K^{0}_{S}\pi^{-}\pi^{0}$ and $K^{+}K^{-}\pi^{-}$. 
The analysis is performed on $818~\mathrm{pb}^{-1}$ of
$e^{+}e^{-}\to \psi(3770) \to D\bar{D}$ data.
460,000 tagged events are reconstructed. The fully leptonic decay 
reconstructed is $D^{+}\to \mu^{+}\nu_{\mu}$, Events are
considered as signal if they contain a single additional
charged track of opposite charge to the fully
reconstructed tag decay. Events with additional
neutral energy deposits in the calorimeter are vetoed.
The beam-energy constrained missing-mass squared, $MM^2$
is computed:
\begin{displaymath}
 MM^{2}=(E_{beam}-E_{\mu^{+}})^{2}-(-\mathbf{p}_{D^{-}}-
\mathbf{p}_{\mu^{+}})^{2} \;,
\end{displaymath}
where $E_{beam}$ is the beam energy,
$\mathbf{p}_{D^{-}}$ is the three-momentum of the fully
reconstructed $D^{-}$ decay and
$E_{\mu^{+}}~(\mathbf{p}_{\mu^{+}})$ is the energy
(three-momentum) of the $\mu^{+}$ candidate. For signal events 
the measured $MM^{2}$ will be close to
zero (the $\nu$ mass). 

The sample of events is then divided depending upon
whether the energy the $\mu^{+}$ candidate deposits in
the electromagnetic calorimeter is more or less than
300\mev; 98.8\% of $\mu^{+}$ deposit less than 300~MeV.
The yield of $D^{+}\to\mu^{+}\nu$ events is extracted by
a fit to the $MM^{2}$ distribution of $\mu$ candidates
depositing less than 300\mev.

 The fit to data produces the following results:
\begin{displaymath}
 \mathcal{B}(D^{+}\to \mu^{+}\nu) = (3.82\pm 0.32 \pm
0.09)\times 10^{-4} \; ,
\end{displaymath}
and
\begin{displaymath}
 f_{D^{+}} = (205.8\pm 8.5 \pm 2.5) \mathrm{MeV} \; ,
\end{displaymath}
where first uncertainty is statistical and the second
uncertainty is systematic. Furthermore, the ratio
between $\mu\nu$ and the small
$D^{+}\to\tau^{+}(\pi^{+}\nu)\nu$ contribution has been
fixed to the SM expectation. The systematic uncertainty
contains significant contributions from radiative
corrections, particle identification efficiency and
background assumptions. The measurement is in good
agreement with the theoretical prediction of Follana
{\it et al.} \cite{Follana:2007uv} of
\mbox{$f_{D^{+}}=(207\pm 4)~\mathrm{MeV}$}.

The CLEO-c measurements of $f_{D^{+}_{s}}$ are made with
a data set corresponding to $600~\mathrm{pb}^{-1}$ of
integrated luminosity collected at a center-of-mass
energy of 4.170\gev, which is close to the maximum of
the $D^{+}_{s}D^{*-}_{s}$ production cross-section. One
analysis reconstructs $D^{+}_{s}\to\mu^{+}\nu$ and
$D^{+}_{s}\to\tau^{+}(\pi^{+}\bar{\nu})\nu$ events
\cite{Alexander:2009ux} and the other reconstructs 
$D^{+}_{s}\to\tau^{+}(e^{+}\nu\bar{\nu})\nu$ events
\cite{Onyisi:2009th}. These are briefly reviewed in
turn.

The analysis of $D^{+}_{s}\to\mu^{+}\nu$ and
$D^{+}_{s}\to\tau^{+}(\pi^{+}\bar{\nu})\nu$ reconstructs
$D_{s}^{-*}\to D_{s}^{-}\gamma$ tags in nine hadronic
$D^{+}_{s}$ decay modes. The number of tags
reconstructed is approximately 44,000. Signal candidates
are reconstructed in an almost identical fashion to the
measurements of $f_{D^{+}}$ and the resulting $MM^{2}$
distribution and fit are shown in Fig.~\ref{fig:fDs}
(a). The principal results from this analysis are: 
\begin{eqnarray*}
  \mathcal{B}(D_{s}^{+}\to\mu^{+}\nu) & = & (0.591\pm
0.037\pm 0.018)\% \; ,\\
  \mathcal{B}(D_{s}^{+}\to\tau^{+}\nu) & = & (6.42\pm
0.81 \pm 0.18)\% \; ,
\end{eqnarray*}
and 
\begin{displaymath}
 f_{D^{+}_{s}} = (263.3 \pm 8.2 \pm 3.9)~\mathrm{MeV}. 
\end{displaymath}
The main systematic uncertainty is from the $D_{s}^{*+}$
tag yields.
 
\begin{figure}[tb]
 \begin{center}
 \begin{tabular}{cc}
 \epsfig{file=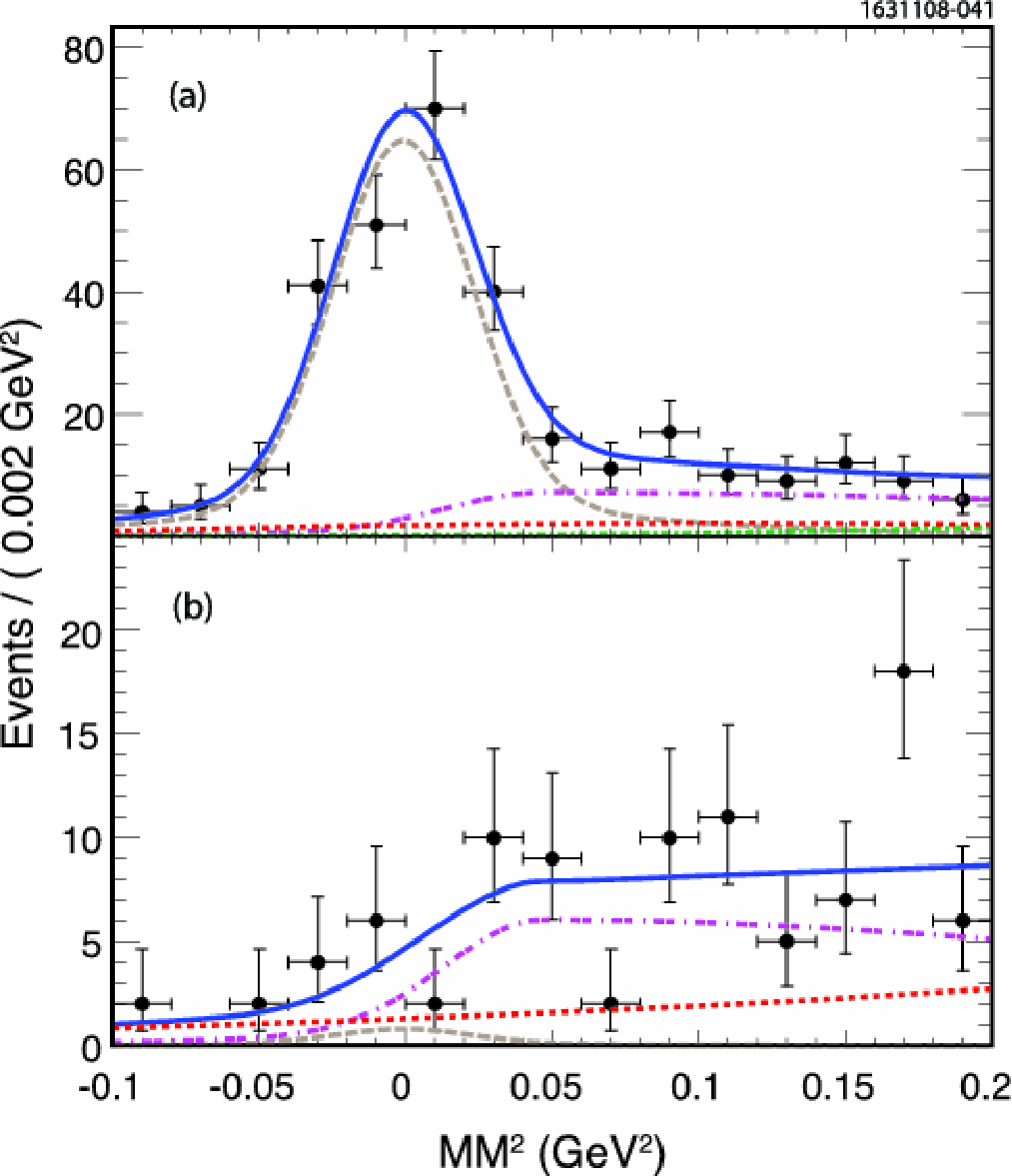,width=0.45\textwidth} &
 \epsfig{file=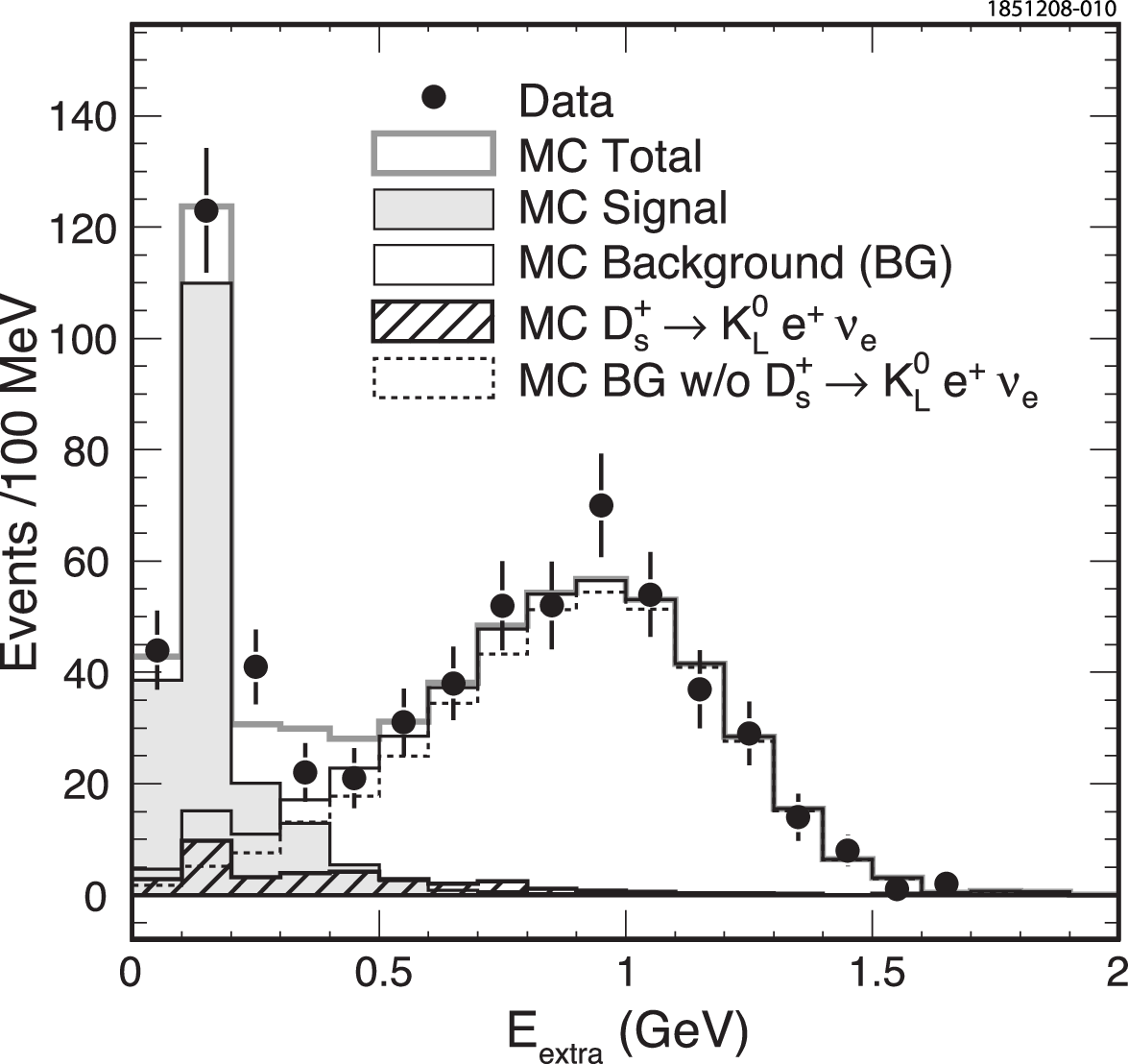,width=0.45\textwidth}  \\
 (a) & (b)
 \end{tabular}
 \caption{\label{fig:fDs}Fits to the (a) $MM^{2}$ and
(b) $E_{\mathrm{extra}}$ distributions used to extract
the $D_{s}^{+}\to l\nu$ yields at CLEO-c. In (a) the
solid line is the total fit result, the dashed line is
the $\mu\nu$ component, the dot-dashed line is the
$\tau\nu$ component and the dotted line is background.}
\end{center}
\end{figure}

The selection of
$D^{+}_{s}\to\tau^{+}(e^{+}\nu\bar{\nu})\nu$ only uses
three of the purest $D_{s}^{-}$ tags: $\phi\pi^{-}$,
$K^{*0}K^{-}$ and $K^{-}K^{0}_{S}$. There are 26,300
tagged events reconstructed. Events with a single
charged track of opposite sign, which is compatible with
being an electron, are selected as signal candidates.
The distribution of extra energy, $E_{\mathrm{extra}}$,
in these events, with the background evaluated in the
tag $D_{s}^{-}$ mass sidebands subtracted, is shown in
Fig.~\ref{fig:fDs} (a). The signal peaks close to
150\mev, which is the energy of the photon in
$D_{s}^{*+}\to D_{s}^{+}\gamma$ decays. A binned fit to
this distribution gives the following results:
\begin{eqnarray*}
  \mathcal{B}(D_{s}^{+}\to\tau^{+}\nu) & = & (0.530\pm
0.47\pm 0.22)\% \; 
\end{eqnarray*}
and 
\begin{displaymath}
 f_{D^{+}_{s}} = (252.5 \pm 11.1 \pm 5.2)~\mathrm{MeV}. 
\end{displaymath}
The main systematic uncertainty is from the estimation
of the $D_{s}^{+}\to K^{0}_{L}e^{+}\nu$ peaking
backgrounds. 

The two results for $f_{D_{s}^{+}}$ give an average
value of 
\begin{displaymath}
 f_{D^{+}_{s}} = (259.5 \pm 6.6 \pm 3.1)~\mathrm{MeV} \;
,
\end{displaymath}
which is $2.3\sigma$ larger than the recent lattice
calculation $f_{D^{+}_{s}}=(241\pm 3)~\mathrm{MeV}$
\cite{Follana:2007uv}. 

\babar~\cite{Aubert:2006sd} and Belle~\cite{Widhalm:2006wz} have also
measured $f_{D^{+}_{s}}$, but the results are much less precise than
those from CLEO-c. However, these results were made with a fraction of
their data sets and will be updated.

be\section{Measurements of $\Gamma$, $\Delta \Gamma$, $\Delta m$ and mixing-phases in $K$, $B$, and $D$ meson decays }
\label{sec:formalism}

The phenomenon of meson-antimeson oscillation, being a flavor changing neutral current (FCNC) process, is very sensitive to heavy
degrees of freedom propagating in the mixing amplitude and, therefore, it
represents one of the most powerful probes of New Physics (NP).
In $K$ and $B_{d,s}$ systems the comparison of observed meson mixing with the
Standard Model (SM) prediction has achieved a good accuracy and plays a
fundamental role in constraining not only the Unitarity Triangle (UT) but also
possible extensions of the SM.
Very recently the evidence for flavor oscillation in the $D$ system has been also revealed, providing complementary
information with respect to the $K$ and $B_{d,s}$ systems, since it involves mesons with up-type quarks.

We recall here the basic formalism of meson-antimeson mixing, starting from the $K$ system.
In principle, one could describe neutral meson mixing with a unique formalism. However, we
present different formalisms for $K$, $B$, and $D$ mixing to make contact with previous literature,
considering also that different approximations are used.

The neutral Kaons $K^0=(\bar s d)$ and $\bar K^0=(s \bar d)$ are flavor eigenstates which in the SM can mix via weak interactions through the box diagrams shown in Fig.~\ref{fig:box}.
In the presence of flavor mixing the time evolution of the $K^0$-$\bar K^0$ system is described by
\begin{equation}\label{eq:evol}
i \frac{d}{dt}\left(\begin{array}{c} K^0(t) \cr \bar K^0(t) \end{array}\right)=\hat H  \left(\begin{array}{c} K^0(t) \cr \bar K^0(t) \end{array}\right)\,,
\end{equation}
where the the Hamiltonian $\hat H$ is a $2\times 2$ non-hermitian matrix which can be decomposed as $\hat H = \hat M -i \hat \Gamma/2$.
The matrices $\hat M$ and $\hat \Gamma$ are hermitian and their elements respectively 
describe the dispersive and absorptive part of the time evolution of the Kaon states.

We note that, in terms of $K^0$ and $\bar K^0$, the CP eigenstates are given by\footnote{The
phase convention is chosen so that
$CP |K^0 \rangle=|\bar K^0\rangle$ and  $CP |\bar K^0 \rangle=|K^0\rangle$.}
\begin{eqnarray}\label{eq:CPeig}
K_\pm=\frac{1}{\sqrt{2}}(K^0\pm\bar K^0)\,,\qquad & CP |K_\pm \rangle=\pm|K_\pm \rangle\,.
\end{eqnarray} 

The Hamiltonian eigenstates, called {\it short} and {\it long} due to the significant
difference between their decay time, can be written as
\begin{equation}\label{eq:KLS}
K_S=\frac{K_++\bar \epsilon\, K_-}{\sqrt{(1+|\bar \epsilon|^2)}}\,, \qquad
K_L=\frac{K_-+\bar \epsilon\, K_+}{\sqrt{(1+|\bar \epsilon|^2)}}\,.
\end{equation}
They coincide with CP eigenstates but for a small admixture governed by a small complex parameter $\bar \epsilon$, defined as
\begin{equation}\label{eq:epsbar}
\frac{1-\bar \epsilon}{1+\bar \epsilon}=-\sqrt{\frac{M^*_{12}-i\Gamma^*_{12}/2}{M_{12}
-i\Gamma_{12}/2}}\,.
\end{equation}
\begin{figure}[t]
\centering
\includegraphics[angle=0, width=.8\textwidth]{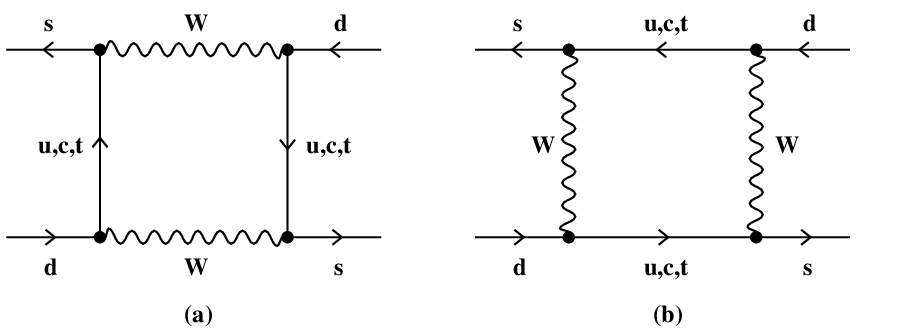}
\caption{\it Box diagrams contributing to $K^0-\bar K^0$ mixing
in the SM.}
\label{fig:box}
\end{figure}

In the $K$ system, the smallness of $\bar \epsilon \simeq \mathcal{O}(10^{-3})$ implies
Im$M_{12} \ll$~Re$M_{12}$ and Im$\Gamma_{12} \ll$~Re$\Gamma_{12}$.
Consequently, the mass difference between the mass eigenstates $K_L$ and $K_S$
can be well approximated by the simple expression:
\begin{equation}\label{eq:DMK}
\Delta M_K \equiv M_{K_L}-M_{K_S}=-2 Re M_{12}\,,
\end{equation}
where the off-diagonal element $M_{12}$ is given by
\begin{equation}\label{eq:M12K}
M_{12} = \langle K^0 | {\cal H}_{eff}^{\Delta S=2} | \bar K^0 \rangle\,,
\end{equation}
with ${\cal H}_{eff}^{\Delta S=2}$ being the effective Hamiltonian that describes $\Delta S=2$
transitions, defined in Sec.~\ref{sec:ope}. Notice that Eq.~(\ref{eq:DMK}) only gives the
short-distance part of $\Delta M_K$. Long-distance contributions, however, are present
due to the exchange of light meson states and are difficult to estimate. On the other hand,
the imaginary part of the amplitudes discussed below are not affected by these contributions,
which have Cabibbo-suppressed imaginary part.

We now discuss the parameters $\epsK$ and $\epsp/\epsK$. which are
used to measure indirect and direct CP violatons in the $K$ system.
Since a two pion final state is CP even while a three pion final state
is CP odd, $K_{\rm S}$ and $K_{\rm L}$ preferably decay to $2\pi$ and
$3\pi$ respectively, via the following CP conserving decay modes:
$K_{\rm L}\to 3\pi$ (via $K_2$), $K_{\rm S}\to 2 \pi$ (via $K_1$).
Given how $K_{\rm L}$ and $K_{\rm S}$ are not CP eigenstates,
they can also decay, with small branching fractions, as follows: $K_{\rm
  L}\to 2\pi$ (via $K_1$), $K_{\rm S}\to 3 \pi$ (via $K_2$).  This CP
violation is called indirect as it does not proceed via explicit breaking
of the CP symmetry in the decay itself, but via the mixing of the CP
state with opposite CP parity to the dominant one.  We note that
$\bar\epsilon$ depends on the phase convention of the $K$ meson states
and hence it is not measurable by itself. A phase-independent
parameter which provides a measure of the indirect CP violation is
\begin{equation}\label{ek}
\epsK
={{A(K_{\rm L}\rightarrow(\pi\pi)_{I=0}})\over{A(K_{\rm 
S}\rightarrow(\pi\pi)_{I=0})}},
\end{equation}
which, using the already-mentioned approximations as well as $\Delta M\simeq-\Delta\Gamma$ and
$\Gamma_{12}\simeq (A_0^*)^2$, is related to $\bar\epsilon$ by
\begin{equation}\label{eq:basic2}
\epsK=\bar\epsilon+i \xi=\left(\frac{\mathrm{Im}M_{12}}{2 \mathrm{Re}M_{12}}+\xi\right)
\,\exp({\rm i}\phi_\epsilon)\sin \phi_\epsilon\,,\quad\textrm{with}\quad\xi=\frac{\mathrm{Im}
A_0}{\mathrm{Re} A_0}\,,
\end{equation}
where the phase $\phi_\epsilon$ is measured to be $(43.51 \pm 0.05)^\circ$~\cite{Amsler:2008zz}.
The amplitude $A_0$ appearing in Eq.~(\ref{eq:basic2}) is defined through
\begin{eqnarray} 
&A(K^0\rightarrow\pi^+\pi^-)=\sqrt{\frac{2}{3}} A_0 e^{i\delta_0}+ \sqrt{\frac{1}
{3}} A_2 e^{i\delta_2}\,,&\nonumber\\
&A(K^0\rightarrow\pi^0\pi^0)=\sqrt{\frac{2}{3}} A_0 e^{i\delta_0}-2\sqrt{\frac{1}{3}} A_2 e^{i\delta_2}\,,&
\label{eq:ai}
\end{eqnarray} 
where the subscript $I=0,2$ denotes states with isospin $0,2$. Hence, $\delta_{0,2}$ are the corresponding strong phases, and
the weak CKM phases are contained in $A_0$ and $A_2$.
Indirect CP violation reflects the fact that the mass
eigenstates are not CP eigenstates. Direct
CP violation, on the other hand, is realized via a 
direct transition of a CP odd to a CP even state or vice-versa. 
A measure of the direct CP violation in $K_L\to \pi\pi$ is characterized
by a complex parameter $\epsp$ defined as\footnote{Actually direct CP violation is accounted for
by Re$(\epsp)$.}
\begin{equation}\label{eprime}
\epsp={\frac{1}{{\sqrt 2}}}\mathrm{Im}
\left(\frac{A_2}{A_0}\right) e^{i\Phi},\quad\quad
\Phi=\pi/2+\delta_2-\delta_0,
\end{equation}
with amplitudes $A_{0,2}$ defined in Eq.~(\ref{eq:ai}).
Extracting the strong phases $\delta_{0,2}$ from $\pi\pi$ scattering
yields $\Phi\approx \pi/4$.
Experimentally, the ratio $\epsp/\epsK$
can be determined by measuring the ratios
\begin{equation}\label{eq:etasK}
\eta_{00}={\frac{A(K_{\rm L}\to\pi^0\pi^0)}{A(K_{\rm S}\to\pi^0\pi^0)}},
            \qquad
  \eta_{+-}={\frac{A(K_{\rm L}\to\pi^+\pi^-)}{A(K_{\rm S}\to\pi^+\pi^-)}}.
\end{equation}
In fact, from Eqs.~(\ref{eq:etasK}) and (\ref{eq:ai}), one finds
\begin{equation}
\eta_{00} \simeq \epsK-2\epsp,~~~~
\eta_{+-} \simeq \epsK+ \epsp\,,
\end{equation}
by exploiting the smallness of $\epsK$ and $\epsp$, using Im$A_i\ll$Re$A_i$ and
$\omega=\mathrm{Re} A_2/Re A_0=0.045\ll 1$, which corresponds to
the $\Delta I =1/2$ rule.
The ratio ${\epsp}/{\epsK}$  can then be measured from
\begin{equation}
\label{eq:eta00}
\left|{{\eta_{00}}\over{\eta_{+-}}}\right|^2\simeq 1 -6\; 
\mathrm{Re} \left(\frac{\epsp}{\epsK} \right)\,.
\end{equation}

The formalism recalled in the case of the $K$ system is basically the same for
the $B_d$ and $B_s$ systems.
There is however a notation difference for the neutral mass eigenstates, which are denoted
{\it heavy} and {\it light} and are expressed in terms of the flavor eigenstates as
\begin{equation}\label{eq:BLH}
B_q^{L,H}=\frac{1}{\sqrt{(1+|(q/p)_q|^2)}}(B_q \pm (q/p)_q \bar B_q)\,,\qquad (q=d,s)
\end{equation}
with $(q/p)_q$ parameterizing indirect CP violation. This parameter is similar to
$\bar \epsilon$ for Kaons. Comparing Eqs.~(\ref{eq:KLS}) and (\ref{eq:BLH}), one finds
$q/p=(1-\bar\epsilon)/(1+\bar\epsilon)$.
  
Similar to the Kaon case, the phase of $(q/p)_q$ depends on the phase convention of the
$B$ meson states, but the absolute value $|(q/p)_q|$ can be measured.
Further interesting experimental observables in the $B_d$ and $B_s$ systems are the mass and width differences:
$\Delta M_{B_q}\equiv M_{B_H}-M_{B_L}$ and $\Delta \Gamma_{B_q}\equiv \Gamma_{B_L}-\Gamma_{B_H}$.
They can be written in terms of the dispersive, $M^q_{12}$, and absorptive, $\Gamma^q_{12}$, matrix elements as
\begin{eqnarray}\label{eq:obsBnoappr}
&(\Delta M_{B_q})^2-\frac{1}{4} (\Delta \Gamma_{B_q})^2 = 4  |M^{q}_{12}|^2- |\Gamma^{q}_{12}|^2\,,&\nonumber\\
&\Delta M_{B_q} \Delta \Gamma_{B_q}=-4 Re (M^q_{12} \Gamma^{q*}_{12})\,, \quad 
|(q/p)_q|=\left| \sqrt{\frac{2 M^{q*}_{12}-i \Gamma^{q*}_{12}}{2 M^{q}_{12}-i \Gamma^{q}_{12}}} \right |\,.&
\end{eqnarray}
The dispersive element $M^q_{12}$ is related to the matrix element of the effective $\Delta B=2$ Hamiltonian, defined in
Sec.~\ref{sec:ope}, as it can be straightforwardly derived from Eq.~(\ref{eq:M12K}).
The absorptive matrix element $\Gamma^q_{12}$ can be written as
\begin{equation}\label{eq:Gamma12}
\Gamma^q_{12}=\frac{1}{2 M_{B_q}}\,\mathrm{Disc}
\langle B_q | i \int d^4x {\cal T}\, ({\cal H}_{eff}^{\Delta B=1}(x) \,{\cal H}_{eff}^{\Delta B=1}(0)) | \bar B_q \rangle\,,
\end{equation} 
where ``Disc'' picks up the discontinuities across the physical cut in the time-ordered product of the $\Delta B=1$ Hamiltonians,
defined in Sec.~\ref{sec:ope}.
The relations in Eq.~(\ref{eq:obsBnoappr}) can be simplified by exploiting the smallness of the ratio
$\Gamma^q_{12}/M^q_{12} \sim {\cal O}(m_b^2/m_t^2) \sim 10^{-3}$, which allows neglecting ${\cal O}(m_b^4/m_t^4)$ terms,
so that one can write
\begin{equation}\label{eq:obsB}
\Delta M_{B_q}= 2  |M^{q}_{12}|\,,\quad \Delta \Gamma_{B_q} = -\Delta M_{B_q} Re \left (\frac{\Gamma^q_{12}}{M^q_{12}} \right)\,,
\quad |(q/p)_q|=1-\frac{1}{2} Im \left( \frac{\Gamma^q_{12}}{M^q_{12}} \right)\,.
\end{equation}

Other important CP-violating observables, associated in the CKM phase convention to the phases of the $B_q$ mixing amplitudes,
are the CKM angles
\begin{equation}
\bb=\mathrm{arg}\left(-\frac{V_{cb}^*V_{cd}}{V_{tb}^*V_{td}}\right)\,,\quad
\bb_s=\mathrm{arg}\left(-\frac{V_{tb}^*V_{ts}}{V_{cb}^*V_{cs}}\right)\,\,.
\end{equation}
Note that $V_{cb}^*V_{cd}$ and $V_{cb}^*V_{cs}$ are approximately real in the CKM phase convention so that $M_{12}^d\simeq\vert
M^d_{12}\vert e^{2\mathrm{i}\bb}$ and $M_{12}^s\simeq\vert M_{12}^s\vert e^{-2\mathrm{i}\bb_s}$. Moreover, the two
angles have different size: $\bb\sim 1$ and $\bb_s\sim\lambda^2\sim {\cal O}(10^{-2})$.

The angle $\bb$ can be measured, for instance, in the time-dependent CP asymmetry of $b\to c\bar c s$ transitions. In particular,
for the golden channel $B_d\to J/\psi K_S$ and neglecting doubly Cabibbo-suppressed contributions to the decay amplitudes,
 one obtains
\begin{equation}
a_{CP}^{B\to J/\psi\, K_S}(t)=\sin(2\bb)\sin(\Delta M_{B_d} t)\,.
\end{equation}
We refer the reader to Sec.~\ref{sec:lifemix:beta} for more details. The extraction of the angle $\bb_s$ is more problematic.
In analogy to the previous case, the decay $B_s\to J/\psi\,\phi$ is sensitive to the phase of the $B_s$ mixing amplitude.
In this case, a time-dependent angular analysis is required to separate CP-odd and
CP-even contributions. This analysis provides a joint measurement of $\Gamma_s$, $\Delta\Gamma_s$ and $\phi_s$, where
$\phi_s=\arg(\lambda_{J/\psi\,\phi})$ with $\lambda_{J/\psi\,\phi}=(q/p)_s 
A(\bar B_s\to (J/\psi\,\phi)_f)/A(B_s\to (J/\psi\,\phi)_f)$ with $f=\{0,\parallel,\perp\}$ (details on the experimental
measurements can be found in Sec.~\ref{sec:lifemix:blife:phi-s}). Discarding doubly Cabibbo-suppressed terms, the ratios
$\bar A/A$ are real, so that $\phi_s$ could give access to the mixing phase. In the same approximation,
the SM mixing phase $\bb_s$ vanishes. Therefore, a measurement of the $B_s$ mixing phase as small as the SM one requires
controlling the Cabibbo-suppressed terms in the decay amplitudes. For an attempt see Ref.~\cite{Faller:2008gt}.
On the other hand, an unsuppressed $B_s$ mixing phase generated by NP can be measured using $B_s\to J/\psi\,\phi$
with good accuracy.
Another angle related to the $B_s$ mixing phase is $\phi_s^\prime=\arg(-M_{12}^s/\Gamma_{12}^s)$, from which one can write
\begin{equation}
\Delta\Gamma_{B_s} =2\vert \Gamma_{12}^s\vert \cos\phi_s^\prime,\quad a^s_\mathrm{fs}=\frac{\vert \Gamma_{12}^s \vert}
{\vert M_{12}^s \vert}\sin\phi_s^\prime\,.
\end{equation}
Similar to $\phi_s$, $\phi_s^\prime$ coincides with the phase of the $B_s$ mixing amplitude up to doubly
Cabibbo-suppressed terms. Therefore, $\phi_s^\prime\simeq \phi_s$ is a sensitive probe of large NP contributions to the
$B_s$ mixing phase, but cannot be used to determine $\bb_s$. Indeed, $\phi_s^\prime$ turns out to be very small
in the SM, as the leading term in $\Gamma_{12}^s$ has the same phase of $M_{12}^s$ while the corrections are both
Cabibbo and GIM suppressed.

Finally, the study of mixing and CP violation in the $D$ system is based on the same formalism as
for $B$ mesons.
A peculiarity of $D$ mesons is that CP violation in mixing is strongly suppressed
within the SM by the CKM combination $V_{cb} V^*_{ub}$, 
so that the matrix elements $M^D_{12}$ and $\Gamma^D_{12}$ 
of the $\Delta C=2$ effective Hamiltonian (see Sec.~\ref{sec:ope})
are real to a good approximation.
Long-distance contributions which cannot be accounted
for by the effective Hamiltonian plague computations of $D$-$\bar D$ mixing observables, even more than in the case of $\Delta M_K$.
The short-distance (SD) part of the mass and width differences can be computed. They are given by:
\begin{eqnarray}
\label{eq:DMD}
\Delta M_D^{\rm SD}&=&M_{D_H}-M_{D_L}=2{\rm Re}\sqrt{(M^D_{12}-\frac{i}{2}\Gamma^D_{12})(M^{D*}_{12}-
\frac{i}{2}\Gamma^{D*}_{12})}\,,\nonumber\\
\Delta \Gamma_D^\mathrm{SD}&=&\Gamma_{D_H}-\Gamma_{D_L}= -4\mathrm{Im}\sqrt{(M^D_{12}-\frac{i}{2}\Gamma^D_{12})(M^{D*}_{12} - 
\frac{i}{2}\Gamma^{D*}_{12})}\,.
\end{eqnarray}
For the $D$ system, the experimental information on the mass and width differences
is provided by the time-integrated observables
\begin{equation}\label{eq:xDyD}
x_D=\frac{\Delta M_D}{\Gamma_D}\,, \qquad y_D=\frac{\Delta \Gamma_D}{2 \Gamma_D}\,.
\end{equation}

\subsection{The $K$-meson system}
\subsubsection{ Theoretical prediction for $\Delta M_K$, $\epsK$, and $\epsp/\epsK$ }
\label{sec:KmixingTh}

The expression for the mass difference $\deltamK$ has been given in Eq.~(\ref{eq:DMK}).
The short-distance contributions, which are represented by the real parts of the box diagrams (see Fig.~\ref{fig:box}) with charm quark and top quark exchanges, are known at NLO in QCD~\cite{Herrlich:1993yv}.
The dominant contribution is represented by charm exchanges, due to the smallness of the real parts of the CKM top quark couplings, which is not compensated by the effect of having heavier quarks running in the loop.
Non-negligible contribution comes from the box diagrams with simultaneous charm and top exchanges.
In spite of the accuracy achieved in the short-distance part, a reliable theoretical prediction of $\deltamK$ is prevented by relevant long-distance contributions which are difficult to estimate. The calculated short-distance part leads in fact to a value of  $\sim 80$\%~\cite{Bijnens:1990mz} of the experimentally observed mass difference between the neutral Kaon states of $\deltamK=(3.483\pm0.006)\cdot10^{-12}\mev$~\cite{Amsler:2008zz}.  

Theoretical predictions for the parameter of indirect CP violation $\epsK$ have been sofar obtained from the expression~(\ref{eq:basic2}) by neglecting the term $\xi$, which constitutes a small contribution, and approximating the phase $\phi_\epsilon$ to $\pi/4$, so that $|\epsK|$ can be written as: 
\begin{equation}\label{eq:epsKmodulus}
|\epsK|=\C_{\varepsilon} \hat B_K A^2
\lambda^6 \etabar \{-\eta_1 S_0(x_c) (1 - \frac{\lambda^2}{2}) + \eta_3
S_0(x_c, x_t) + \eta_2 S_0(x_t) A^2 \lambda^4 (1 -\rhobar) \}\,,
\end{equation}
where $C_\varepsilon=\frac{G_F^2 f_K^2 M_K M_W^2}{6\sqrt{2}\pi^2\deltamK}$.
However, it has been recently pointed out~\cite{Buras:2008nn} that the adopted approximations might no longer be justified,
due to the improved theoretical accuracy in both perturbative and non-perturbative contributions.
In Eq.~(\ref{eq:epsKmodulus}) the Inami-Lim functions $S_0(x_{c,t})$ and  $S_0(x_c,x_t)$~\cite{Inami:1980fz} contain the box-contributions from the charm and top-quark exchange with $x_i=m_i^2/M_W^2$, while $\eta_{i}$ ($i=1,2,3$) describe (perturbative) short-distance QCD-corrections~\cite{Herrlich:1993yv,Buras:1990fn,Herrlich:1996vf}. The Kaon bag parameter \rgbk measures the deviation of the $\Delta S=2$ hadronic matrix element from its value in the vacuum saturation approach:
\begin{equation}\label{eq:bk}
\rgbk\;=\;\frac{\langle\Kzb| \hat{Q}^{\Delta S=2}|\Kz\rangle}{\frac83 f_K^2m_K^2}\,.
\end{equation}
Therefore, \rgbk contains all the non-perturbative QCD contributions for $\epsK$. Currently the best determination of this parameter is available from lattice simulations of QCD with either 2+1 or 2 dynamical quark flavors, which avoid the systematic uncertainty due to ``quenching'' in earlier lattice studies done without dynamical quarks (cf.\ Sec.~\ref{sec:thPrim:LQCD}). At this time, the most accurate results (obtained independently with 2+1 dynamical quark flavors) by RBC/UKQCD \cite{Antonio:2007pb} and Aubin et al.\ \cite{Aubin:2009jh} quote a combined statistical and systematic uncertainty of 5.4 and 4.0 per cent for \rgbk, respectively. That means that the contribution from \rgbk to the  total uncertainty in $\epsK$ is now comparable to the second biggest contribution, which originates from \Vcb. This CKM-matrix element is nowadays known with 2.3 per cent accuracy~\cite{Amsler:2008zz} but enters $\epsK$ in the fourth power.

 In most current lattice calculations, due to algorithmic and computational limitations, the simulated up- and down-quark masses are heavier than their 
physical values, thus requiring an extrapolation for \rgbk to those light quark masses guided by chiral perturbation theory (ChPT, see 
Sec.~\ref{sec:ChPT}). %
Fig.~\ref{fig:bk} summarizes the currently available lattice results with either $N_f=2+1$ or $2$ dynamical quarks. The RBC/UKQCD~\cite{Antonio:2007pb,Allton:2008pn} and the HPQCD~\cite{Gamiz:2006sq} results both were obtained with $N_f=2+1$, where the former used the domain-wall and the latter the staggered fermion formulation. The work by Aubin et al.\ \cite{Aubin:2009jh} used a mixed action approach, where domain-wall valence fermions have been calculated on a 2+1 flavor background of dynamical staggered quarks. In case of the RBC/UKQCD result, crucial ingredients for obtaining a small uncertainty in the final number were the use of Kaon SU(2)-ChPT to extrapolate to light physical quark masses and the use of a non-perturbative renormalization technique. The HPQCD result was obtained using degenerate Kaons (made of two quarks of mass $m_s/2$) and used a perturbative renormalization technique. Results with $N_f=2$ are available from JLQCD~\cite{Aoki:2008ss} using dynamical overlap fermions and from ETMC~\cite{Dimopoulos:2008hb} using twisted mass fermions\footnote{The ETMC result is still preliminary. No systematic errors are included yet.},
both with non-perturbative renormalization. While ETMC also used Kaon  SU(2)-ChPT, the JLQCD result was extrapolated using NLO SU(3)-ChPT with analytic NNLO-terms added. All these dynamical results except the one from Aubin et al.\ were obtained at a single value for the finite lattice spacing $a$ (values indicated in Fig.~\ref{fig:bk}), meaning that a continuum extrapolation is still missing. But the RBC/UKQCD and the HPQCD results account for this fact in the systematic error estimate. For the near future, one should expect updates to these results, containing, e.g.\ simulations at finer lattice spacings and lighter dynamical quark masses and therefore further increasing the accuracy of the value \rgbk obtained from lattice calculations.

Estimates of the Kaon bag parameter in the chiral limit ($m_u,m_d, m_s \to 0$) are available from lattice simulations~\cite{Allton:2008pn}, large $N_C$ approximation~\cite{Bardeen:1987vg,Bijnens:1995br,Bijnens:2006mr} or the QCD-hadronic duality~\cite{Prades:1991sa}.
\begin{figure}[!t]
\begin{center}
\includegraphics[angle=-90, width=.9\textwidth]{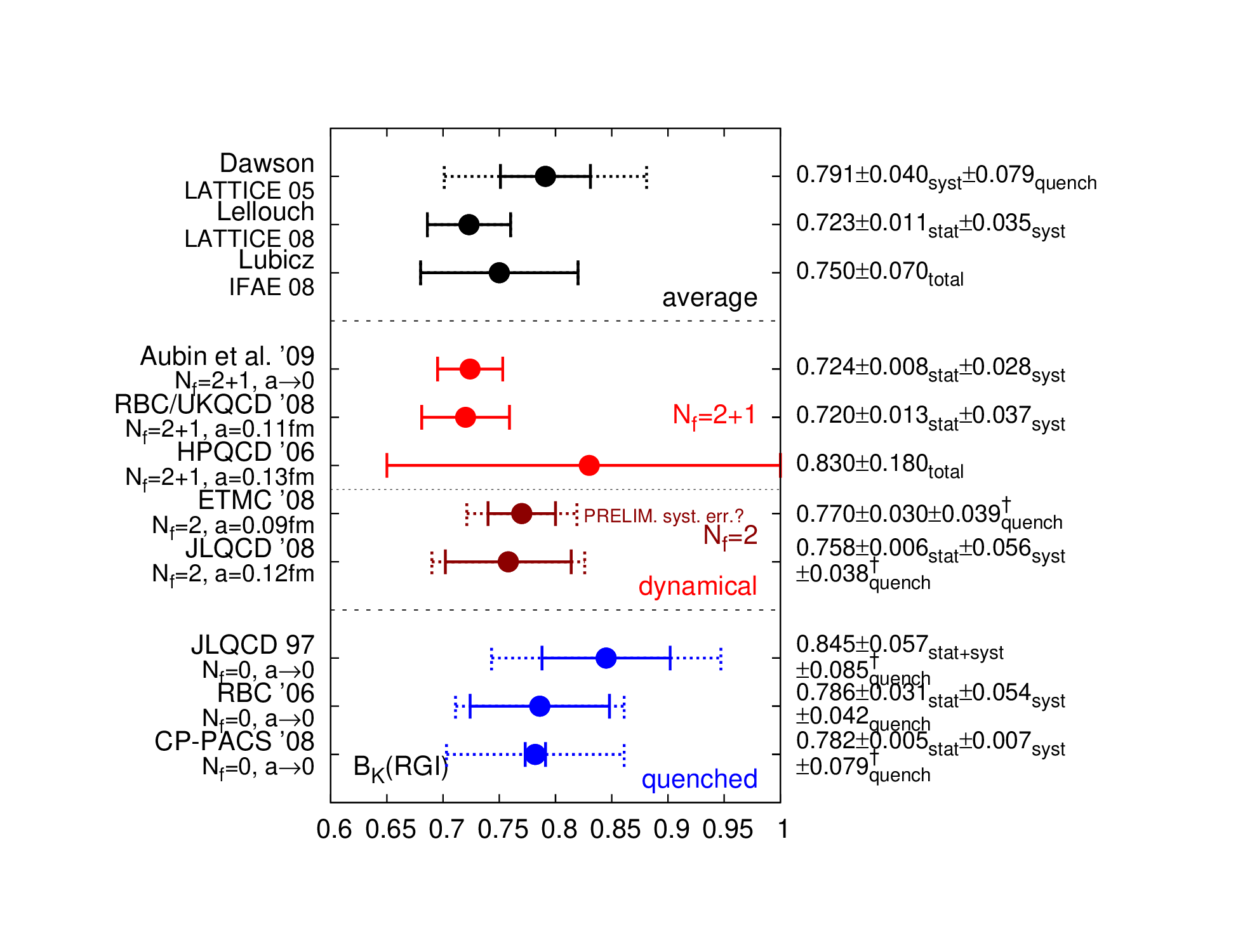}
\end{center}
\caption{Summary of lattice results for \rgbk: included are recent results from dynamical $N_f=2+1$ \cite{Aubin:2009jh,Antonio:2007pb,Allton:2008pn,Gamiz:2006sq} and $N_f=2$ \cite{Aoki:2008ss,Dimopoulos:2008hb} simulations. Also shown are quenched ($N_f=0$) results from  JLQCD \cite{Aoki:1997nr}, RBC \cite{Aoki:2005ga}, and CP-PACS \cite{Nakamura:2008xz}.  For comparison, the figure includes the old lattice average from the Lattice 2005 conference \cite{Dawson:2005za} and two recently published averages \cite{Lubicz:2008am,Lellouch:2009fg}, too. {\it Solid errorbars} do not include the error due to quenching, which is added in the {\it dashed errorbars}. \newline$^\dagger$ A (conservative) quenching error of 5\% or 10\% has been assigned to $N_f=2$ or quenched results, respectively, where no estimate for this systematic error has been provided (see Sec.~\ref{sec:thPrim:LQCD}).}
\label{fig:bk}
\end{figure}

As far as indirect CP violation is concerned, the parameter $\epsp/\epsK$ can be written by using the operator-product expansion (OPE) as an expression involving the hadronic $K\to(\pi\pi)$ operators $Q_6$ and $Q_8$ given in Eq.~(\ref{eq:basis}) of Sec.~\ref{sec:ope}, see e.g.~\cite{Buras:1998raa}.
Here $\epsp$ can be expressed in terms of isospin amplitudes and is an experimentally measurable (complex) parameter. The hadronic matrix elements $Q_6$, $Q_8$ contribute most to the uncertainties in the theoretical prediction for $\epsp/\epsK$. On the lattice, usually an indirect approach~\cite{Bernard:1985wf} is pursued, which measures $K\to\pi$ and $K\to{\rm vacuum}$ operators instead and relates those via chiral perturbation to the wanted $K\to(\pi\pi)$ operators. For quenched studies see~\cite{Noaki:2001un,Blum:2001xb}, for recent work using dynamical domain-wall fermions see~\cite{Li:2008kc}. The latter work raises some doubts about the applicability of this approach, which is based on SU(3)-ChPT to be valid around the strange quark mass. A different technique, based on the calculation of finite volume correlation functions, is described in~\cite{Lellouch:2000pv}, which might turn out to be more successful in the future. See also~\cite{Giusti:2006mh} for an alternative approach, i.e. to study $\epsp/\epsK$ in the small box approach ($\eps$-regime).


\subsubsection{Experimental methods and results}
The Kaon system was the first playground for the understanding of the 
violation of the CP symmetry. In the years before the $B$ system was 
investigated,
all forms of CP violation had been observed in the Kaon system. These are CP violation
in the Kaon mixing, with the
measurement of \RepsK, 
in the direct decay, providing a non-null
result on \Repsp\ and 
in the interference between mixing and decay, through the 
determination of the $\eta_{+-}$ phase, $\phi_{+-}$. 
CP violation  has been observed in the 
\kl\ decay to the CP-even eigenstate of two pions, 
in the time-integrated charge asymmetry of the \kl\ semileptonic decay rates,  
in the \kl\ $\to \pi^+ \pi^- \gamma$ channel, and 
in the angular asymmetry of the  
\kl\ $\to e^+ e^- \pi^+ \pi^-$ decays.

Direct CP violation through $\Delta S$= 1 processes has been  
measured as a tiny difference in the ratios of the  
branching ratios of the \kl\ to the CP-even eigenstates, 
\kl\ $\to \pi^+ \pi^-$ and \kl\ $\to \pi^0 \pi^0$, normalized to the 
\ks\ branching ratio for the same final state (Eq.~\ref{eq:eta00}).
Precise measurements in the Kaon sector have been obtained with 
different techniques  by 
the present generation of experiments (KTeV, NA48 and KLOE).
Results on all of the major branching ratios, lifetimes 
and the \kl\ mass are summarized in Sec.~\ref{sec:data}, 
where analyses of interest for the \Vus\ determination are discussed. 
Several of these new measurements are not in good agreement with 
the average of 
older data.
This 
is the case of the CP-violating decays, \kl\ $\to \pi^+ \pi^-$,  
\kl\ $\to \pi^0 \pi^0$, 
whose branching ratios as measured by KTeV~\cite{Alexopoulos:2004sx} 
in year 2004 are 
5\% and 8\%  lower, respectively, than previous world averages. 
KLOE and NA48 recently confirmed\cite{Ambrosino:2006up,Lai:2006cf} the KTeV 
result on the BR(\kl\ $\to \pi^+ \pi^-$).

KTeV has measured the BR(\kl\ $\to  \pi^+ \pi^-$) and the 
 BR(\kl\ $\to  \pi^0 \pi^0$)\cite{Alexopoulos:2004sx} from the
analysis of all of the main \kl\ decay channels, as described in Sec.~\ref{sec:data}. 
The CP-violating charged channel was selected among events not satisfying 
the criteria for semileptonic and $\pi^+ \pi^- \pi^0$ decays, by
imposing cuts 
on the m$_{\pi \pi}$ invariant mass   
and on the two-track transverse momentum-squared.
The \kl\ $\to \pi^0 \pi^0$ events are identified by the reconstruction 
of exactly four clusters in the calorimeter. The photons are paired to have 
two pions pointing to a single decay vertex and the pion invariant mass 
is required to be consistent with the Kaon mass. 
Major systematic uncertainties come from the precision on the knowledge of the 
efficiency reconstruction 
and to a less extent from radiative corrections to charged modes,  
Monte Carlo statistics and background subtraction.

The KLOE measurement\cite{Ambrosino:2006up} 
has been obtained from the relative 
ratio of \kl\ $\to \pi^+ \pi^-$ and \kl\ $\to \pi \mu \nu$ decays,  
the absolute semileptonic branching ratio BR(\kl\ $\to \pi \mu \nu$) 
being previously 
determined\cite{Ambrosino:2005ec} from the
measurement of all the 
major decay modes, using the tagging technique to obtain the absolute  
branching fractions and imposing the constraint on the sum of the 
branching fractions to solve the dependence on the \kl\ lifetime.
The \kl\ sample at KLOE, operating at the Frascati $\phi$ factory, is 
tagged by the reconstruction of the 
\ks\ $\to \pi^+ \pi^-$ decays, 
giving a precise determination of the \kl\ momentum. In order to minimize the 
difference on trigger efficiency between the two selected channels, the pions 
from \ks\ decay are requested to release in the calorimeter energy enough 
to trigger the data acquisition system.     

The CP-violating channel was selected by a fit with a linear combination 
of Monte Carlo shapes for signal and background to the 
$\sqrt{E^{2}_{miss}+|p^{2}_{miss}|}$
distribution, where
$E_{miss}$ is the missing energy in the hypothesis of the two charged 
pion decay.
The precision is dominated by the accuracy on tagging and tracking 
efficiency, which depend on 
corrections applied to the Monte Carlo sample, necessary to resolve 
small discrepancies between the Monte Carlo-predicted 
distributions and those obtained from data control samples.   

NA48 measured the relative decay widths 
$\Gamma(\kl\ \to \pi^+ \pi^-)$/$\Gamma(\kl\ \to \pi e \nu)$\cite{Lai:2006cf}
from a sample of two-track events selected for the analysis, which results in 
the semileptonic Ke3 branching ratio normalized to all of the two-track 
modes\cite{Lai:2004bt}.
The CP-violating channel was selected by analysis requirements on the m$_{\pi \pi}$ 
invariant mass, 
 the Kaon transverse momentum-squared,  the ratio  
of the reconstructed energy and the particle momentum, E/P ( which is very effective in 
separating electrons from $\mu$ and $\pi$ ), and finally using the muon veto 
system for K$\mu$3 rejection.    
Systematic uncertainties are due to the knowledge of  Kaon 
spectrum, background contamination from K$\mu$3 decays, and to a less extent  
radiative corrections, trigger efficiencies and Monte Carlo statistics.
The obtained branching fractions are summarized in Tab.~\ref{tab:pipi}, 
together with the CP-violation parameter $|\eta_{+-}|$,
defined in Eq.~\ref{eq:etasK}, and \RepsK. 

\begin{table}[h!t!b!]
\caption{\kl\ $\to  \pi^+ \pi^-$ branching ratios as measured by 
KTeV\cite{Alexopoulos:2004sx}, KLOE\cite{Ambrosino:2006up} and 
NA48\cite{Lai:2006cf}. 
compared with previous world average\cite{Hagiwara:2002fs}.
 For \RepsK , the average value 
of $\phi_{+-} = (43.4 \pm 0.7)^{\circ}$ and \Repsp\ = (16.5$\pm$2.6)$ \times 10^{-4}$ have been used. }
\label{tab:pipi}
\begin{center}
\begin{tabular}{|c|c|c|c|} 
\hline
\hline
Source& BR$(\kl \to \pi^+ \pi^-)$& $|\eta_{+ -}|$ & \RepsK  \\
\hline
\hline
PDG  04  & ($20.90 \pm 0.25$)$10^{-4}$\cite{Hagiwara:2002fs} &($22.88 \pm 0.14$)$10^{-4}$
&($16.6 \pm 0.2$)$10^{-4}$       \\
KTeV 04  & ($19.75 \pm 0.12$)$10^{-4}$\cite{Alexopoulos:2004sx} &($22.28 \pm 0.10$)$10^{-4}$
&($16.1 \pm 0.2$)$10^{-4}$   \\
KLOE 06  & ($19.63 \pm 0.21$)$10^{-4}$\cite{Ambrosino:2006up} & ($22.19 \pm 0.13$)$10^{-4}$
&($16.1 \pm 0.2$)$10^{-4}$   \\
NA48 07  & ($19.69 \pm 0.19$)$10^{-4}$\cite{Lai:2006cf} & ($22.23 \pm 0.12$)$10^{-4}$
&($16.1 \pm 0.2$)$10^{-4}$   \\
\hline
\hline
\end{tabular}
\end{center}
\end{table}

Direct CP violation has been established by 
precision measurements from NA48\cite{Batley:2002gn} 
and KTeV\cite{AlaviHarati:2002ye}, 
giving 
$\Repsp = (14.7\pm2.2) \times 10^{-4}$
 and $\Repsp = (20.7\pm2.8) \times 10^{-4}$, 
respectively. 
%
The measurements hitherto summarized are used in the fit 
procedure described  in the 
``CP violation in Klong decays'' 
Review\cite{Amsler:2008zzb}, which  obtains  
the value of 
\begin{equation}
\label{eq:resultepsK}
|\epsilon_K|  = (2\dot |\eta_{+-}|+ | \eta_{00}|)/3 = (2.229\pm0.012) \times 10^{-3}
\end{equation}
The KTeV collaboration recently announced the final result on 
\Repsp, obtained with an improved analysis of the entire data 
set\cite{Yamanaka:2008hk,Worcester:2007zz}. 
The systematic error 
was reduced from 2.4 $\times 10^{-4}$ to 1.8 $\times 10^{-4}$ 
and the statistical 
uncertainty from 1.5 $\times 10^{-4}$ 
to 1.1 $\times 10^{-4}$, giving 
\Repsp\ = (19.2$\pm$1.1$_{stat} \pm$1.8$_{syst}) \times 10^{-4}$.
The results from the two experiments are consistent within 1.7 $\sigma$.
The new average, after scaling uncertainties to take into account the 
consistency level of the measurements, and including contributions 
from $\Delta$I=3/2 amplitudes 
\cite{Sozzi:2004py},   
not present in Eq.323,  
is \Repsp\ = (16.4$\pm$1.9)$ \times 10^{-4}$.

The KTeV experiment also measured the phase of the CP-violating decays.
They used 
simultaneous measurements of events from two nearly parallel 
Kaon beams, 
with one of the beams passing through a
thick regenerator, for precise determination of 
acceptances and contamination for both the charged ($\pi^+ \pi^-$) and 
 neutral ($\pi^0 \pi^0$) modes.  
To reduce systematic uncertainties, the regenerator
positions were alternated between the two beams once per
minute.
\Repsp\ has been obtained by a fit to the vacuum-to-regenerator ratio for 
charged and neutral modes, taking into account the  
\kl-\ks\ interference pattern in the 
regenerator sample\cite{AlaviHarati:2002ye}.
Together with  \Repsp, the fit provides the best results on 
$\phi_{00} - \phi_{+-}  = (0.29 \pm 0.31)^{\circ}$, 
$\Delta M_K = M_{\kl}-M_{\ks} = 3.465(7) 10^{-12}$ MeV  
and the \ks\ lifetime, $\tau_S = 89.62(5) 10^{-12}$~s.

The unitarity relation applied to time evolution of the neutral Kaon 
state leads to the Bell-Steinberger 
relation expressing CP and CPT violation parameters in terms of Kaon decay 
widths. 
 \begin{eqnarray}
   && \left[ {{\Gamma_S + \Gamma_L}\over{\Gamma_S-\Gamma_L}}+
i\tan\phi_{\rm SW}\right]
    \times 
    \left[\frac{\Re(\epsilon)}{1+|\epsilon|^2 }
-i\Im(\delta) \right] \qquad \nonumber \\ 
   &&=   {1\over{\Gamma_S-\Gamma_L}} \sum_f \cA_L(f) \cA^*_S(f),
\label{eq:b-s}
\end{eqnarray}
where 
\begin{equation}
\phi_{\rm SW} = \arctan\left( \frac{ 2 \Delta M_K}{\Gamma_S-\Gamma_L} \right)~.
\end{equation}

Besides testing CPT symmetry, the unitarity constraint for the neutral 
Kaon system, which receives relevant 
contributions only from few final states, 
really improves the precision on the \RepsK~parameter.  
The measurements of the \kl\ , \ks\ branching fractions and lifetimes, 
together with the KLOE 
upper limits on the \ks\ $\to  \pi^0 \pi^0 \pi^0$ mode\cite{Ambrosino:2005iw},  
on the time-integrated charge asymmetry 
of the \ks\ semileptonic decay\cite{Ambrosino:2006si}, 
and the new result on $\phi_{+-}$ announced by 
KTeV\cite{Yamanaka:2008hk,Worcester:2007zz}, 
have improved the accuracy on both 
CP- and CPT-violation parameters,
\RepsK\  and  \Imdelta.
The results published in Ref.\cite{Ambrosino:2006ek} were 
mostly based on the KLOE measurements. These 
have been revised for the ``CPT invariance tests in neutral Kaon decay'' 
Review\cite{Amsler:2008zzb} 
using the entire set of published data from KTeV and NA48, and then  
updated by the FlaviaNet Kaon Working Group to include the 
preliminary results on the $\eta_{\pi \pi}$ 
phases from the KTeV experiment\cite{Yamanaka:2008hk,Worcester:2007zz}.
The results are summarized in Tab.~\ref{tab:reps}.

\begin{table}[h!t!b!]
\caption{CP and CPT violation parameters from the unitarity constraint 
(Bell-Steinberger relation).}
\label{tab:reps}
\begin{center}
\begin{tabular}{|c|c|c|c|} 
\hline
\hline
Source& \RepsK & \Imdelta\ & Ref. \\
\hline
\hline
KLOE 06  & ($15.96 \pm 0.13$)$10^{-4}$ & ($0.04 \pm 0.21$)$10^{-4}$ &
\cite{Ambrosino:2006ek} \\
PDG 08  & ($16.12 \pm 0.06$)$10^{-4}$ &($-0.06 \pm 0.19$)$10^{-4}$ & 
\cite{Amsler:2008zzb}  \\
FlaviaNet 08  & ($16.12 \pm 0.06$)$10^{-4}$ &($-0.01 \pm 0.14$)$10^{-4}$ & 
  \\
\hline
\hline
\end{tabular}
\end{center}
\end{table}


Overall, the   
measurements in the Kaon sector to date constitute a  
precise data set consistent with CPT symmetry and unitarity.
The comparison with CP-violation parameters in the B sector confirms  
that the CKM mechanism is the major source of CP-violation in meson
decays.
Still, Kaon physics has to meet the challenging experimental program
on CP violation in very rare, and especially \kl\ $\to \pi \nu \nu$,
decays which is extremely promising for constraining 
the CKM parameters and the physics beyond the SM as discussed in Sec.~\ref{sec:rare-Kaon}.


\subsection{The B-meson system}
\subsubsection{Lifetimes, $\Delta \Gamma_{B_q}$, $A^q_{SL}$ and $\Delta M_{B_q}$}
\label{sec:BmixingTh}


Heavy meson mixing plays a particularly important role in placing constraints on NP, since this loop process can be computed quite reliably using the heavy-quark expansion (HQE). 
Similarly, the hierarchy of lifetimes of heavy hadrons can be understood 
in the HQE, which makes use of the disparity of scales present in the
decays of hadrons containing b-quarks. HQE predicts the ratios of lifetimes of beauty
mesons\cite{Bigi:1994wa,Voloshin:2000zc,Neubert:1996we,Rosner:1996fy}, which now agree with the 
experimental observations within experimental and theoretical 
uncertainties. The most recent theoretical predictions show evidence of excellent agreement 
of theoretical and experimental results~\cite{Ciuchini:2001vx,Franco:2002fc,Beneke:2002rj,Gabbiani:2003pq,Gabbiani:2004tp,Badin:2007bv}. 
This agreement also provides us with some confidence that quark-hadron
duality, which states that smeared partonic amplitudes can be replaced by the hadronic
ones, is expected to hold in inclusive decays of heavy flavors. It should be pointed out 
that the low experimental value of the ratio $\tau(\Lambda_b)/\tau(B_d)$ has long been a 
puzzle for the theory. Only recent next-to-leading order (NLO) calculations of
perturbative QCD~\cite{Ciuchini:2001vx,Franco:2002fc,Beneke:2002rj} and $1/m_b$ corrections~\cite{Gabbiani:2003pq,Gabbiani:2004tp,Badin:2007bv}
to spectator effects as well as recent Tevatron measurements practically eliminated this discrepancy. 

The inclusive decay rate of a heavy hadron $H_b$ and $B$-meson mixing parameters can be  most 
conveniently computed by employing the optical theorem to relate the decay width to the imaginary part 
of the forward matrix element of the transition operator:
\begin{equation} \label{rate}
\Gamma(H_b)=\frac{1}{2 M_{H_b}}\, Disc \langle H_b | i \int d^4x {\cal T}\, ({\cal H}_{eff}^{\Delta B=1}(x) \,{\cal H}_{eff}^{\Delta B=1}(0)) |  H_b \rangle\,,
\end{equation}
where $H_{\mbox{\scriptsize eff}}^{\Delta B=1}$ represents the effective $\Delta B=1$ Hamiltonian, given in Sec.~\ref{sec:ope}.

In the heavy-quark limit, the energy release is large, so that the correlator in
Eq.~(\ref{rate}) is dominated by short-distance physics.
The OPE can be applied as explained in Sec.~\ref{sec:ope}, leading to a prediction for the decay widths of Eq.~(\ref{rate}) as
a series of local operators of increasing dimension suppressed by powers
of $1/m_b$:
\begin{equation}\label{expan}
\Gamma(H_b)= \frac{1}{2 M_{H_b}} \sum_k \langle H_b |{\cal T}_k | H_b \rangle
=\sum_{k} \frac{C_k(\mu)}{m_b^{k}}
\langle H_b |{\cal O}_k^{\Delta B=0}(\mu) | H_b \rangle,
\end{equation}
with the scale dependence of the Wilson coefficients compensated by the scale dependence of the matrix elements.

It is customary to make predictions for the ratios of lifetimes (widths),
as many theoretical uncertainties cancel out in the ratio. Since the differences of lifetimes 
should come from the differences in the light sectors of heavy hadrons, at the leading order in 
HQE all beauty hadrons with light spectators have the same lifetime.
The difference between meson and baryon lifetimes first occurs at order
$1/m_b^2$ and is essentially due to the different structure of mesons and baryons, amounting to at most 
$1-2\%$~\cite{Neubert:1996we}.

The main effect appears at the $1/m_b^3$ level and comes from
dimension-six four-quark operators, whose contribution is
enhanced due to the phase-space factor $16 \pi^2$. They are thus capable of inducing
corrections of order $16 \pi^2 (\Lambda_{QCD}/m_b)^3$ = ${\cal O}(5-10\%)$.
These operators introduce through the so-called Weak Annihilation (WA) and Pauli Interference
(PI) diagrams, a difference in lifetimes for both
heavy mesons and baryons. Their effects have been computed~\cite{Neubert:1996we,Shifman:1984wx,Bigi:1993fe,Guberina:1979xw,Bilic:1984nq,Guberina:1986gd,Guberina:2000de}
 including NLO perturbative
QCD corrections~\cite{Ciuchini:2001vx,Franco:2002fc,Beneke:2002rj} and $1/m_b$ corrections~\cite{Gabbiani:2003pq,Gabbiani:2004tp,Badin:2007bv}.
The non-perturbative contribution is enclosed in the matrix elements of the mentioned operators, which are the following four
\begin{eqnarray}
O_1^q = \bar b_i\gamma^{\mu}(1-\gamma_5)b_i\bar q_j\gamma_{\mu}(1-\gamma_5)q_j, &&
O_2^q = \bar b_i\gamma^{\mu}\gamma_5b_i\bar
q_j\gamma_{\mu}(1-\gamma_5)q_j,\nonumber \\
\widetilde{O}_1^q = \bar b_i\gamma^{\mu}(1-\gamma_5)b_j
\bar q_i\gamma_{\mu}(1-\gamma_5)q_j, &&
\widetilde{O}_2^q = \bar b_i\gamma^{\mu}\gamma_5 b_j
\bar q_i\gamma_{\mu}(1-\gamma_5)q_j.
\end{eqnarray}
The matrix elements of these operators are parameterized in a different way depending on whether or not the light quark $q$ of the operator enters as a valence quark in the external hadronic state~\cite{Ciuchini:2001vx}.
In this way one can distinguish the contribution of the contraction of the light quark in the operator with the light quark in the hadron, which is the only one calculated in lattice QCD~\cite{DiPierro:1998ty,DiPierro:1999tb,DiPierro:1998cj,Becirevic:2001fy}.
Computing the contribution of the contraction of two light quarks in the operator, which vanishes in the vacuum saturation approximation, has been so far prevented by the difficult problem of subtracting power divergences.
By combining the results for the perturbative and non-perturbative contributions discussed above, the theoretical predictions for the lifetime ratios read
\begin{eqnarray}
\tau(B^+)/\tau(B^0) = 1.06\pm0.02\,, \,
\tau(B_s)/\tau(B^0) = 1.00\pm0.01\,, \,
\tau(\Lambda_b)/\tau(B^0) = 0.91\pm0.04\,.\nonumber\\  
\end{eqnarray}

Similar calculations yield B-mixing parameters presented in the form of expansion in 
$1/m_b^n$.
The width difference is related the matrix elements $M^q_{12}$ and $\Gamma^q_{12}$ as in Eq.~(\ref{eq:obsB}) and by using the HQE it can be written as
\begin{eqnarray}\label{corr}
\Delta \Gamma_{B_q} 
&=&\frac{G^2_F m^2_b}{6\pi (2 M_{B_q})}(V^*_{cb}V^{\phantom{*}}_{cq})^2\cdot\\
&&\left\{\left[F(z)+P(z)\right] \langle Q \rangle + 
\left[F_S(z)+P_S(z)\right] \langle Q_S \rangle
+\delta_{1/m} + \delta_{1/m^2}\right\}\,,
\nonumber
\end{eqnarray}
where $z=m_c^2/m_b^2$ and the two $\Delta B=2$ operators are defined as
\begin{equation}\label{qqs}
Q = (\bar b_iq_i)_{V-A}(\bar b_jq_j)_{V-A},\qquad
Q_S= (\bar b_iq_i)_{S-P}(\bar b_jq_j)_{S-P} ~.
\end{equation}
The matrix elements for $Q$ and $Q_S$ are known to
be
\begin{eqnarray}\label{meqs}
\langle Q \rangle &\equiv& \langle \overline B_q\vert Q\vert B_q\rangle 
= f^2_{B_q}M^2_{B_q}2\left(1+\frac{1}{N_c}\right)B_{B_q},
\nonumber \\[0.1cm]
\langle Q_S \rangle &\equiv& \langle \overline B_q\vert Q_S \vert B_q\rangle 
= -f^2_{B_q}M^2_{B_q}
\frac{M^2_{B_q}}{(m_b+m_s)^2}\left(2-\frac{1}{N_c}\right)B^S_{B_q},
\nonumber
\end{eqnarray}
A theoretical prediction for the $B_{d,s}$ width differences then requires to calculate non-perturbatively the decay constants $f_{B_{d,s}}$ and the bag parameters $B_{B_{d,s}}$ and $B^S_{B_{d,s}}$.
Several unquenched lattice calculations of the decay constants have been performed with $N_f=2$ or $N_f=2+1$ dynamical fermions~\cite{AliKhan:2000eg,AliKhan:2001jg,Bernard:2002pc,Aoki:2003xb,Wingate:2003gm,Gray:2005ad,Gamiz:2009ku,Bernard:2007zz}.
They have been obtained by treating the $b$ quark on the lattice with two different approaches, either FNAL~\cite{ElKhadra:1996mp} or non-relativistic QCD. A collection of these results is provided in Ref.~\cite{Lubicz:2008am}, where the following averages are estimated
\begin{equation}\label{eq:fBds}
f_{B_d}=(200 \pm 20) \mev\,, \qquad f_{B_s}=(245 \pm 25)\mev\,.
\end{equation}
The average for $f_{B_s}$ takes into account all the existing $N_f=2$ and $N_f=2+1$ results.
For $f_{B_d}$ the lattice determination is more delicate, because its value is enhanced by chiral logs effects relevant at low quark masses. In order to properly account for these effects, simulations at light values of the quark mass (typically $m_{ud} < m_s/2$) are required.
For this reason, the $f_{B_d}$ average provided in Ref.~\cite{Lubicz:2008am} and given in Eq.~(\ref{eq:fBds}) is derived by taking into account only the results obtained by the HPQCD~\cite{Gray:2005ad} and FNAL/MILC~\cite{Bernard:2007zz} collaborations, by using the MILC gauge field configurations generated at light quark masses as low as $m_s/8$.
A more recent HPQCD calculation~\cite{Gamiz:2009ku} of $f_{B_d}$ and $f_{B_s}$, as well as of the bag parameters $B_{B_d}$ and $B_{B_s}$, came out after the averages in Ref.~\cite{Lubicz:2008am} were performed.
Since the new results are consistent with the old ones, the averages~\cite{Lubicz:2008am} can be considered up to date.

Also for the bag parameters, a collection of quenched~\cite{Lellouch:2000tw,Becirevic:2001xt,Aoki:2002bh} and unquenched ($N_f=2$ and $N_f=2+1$)~\cite{Aoki:2003xb,Gamiz:2009ku,Dalgic:2006gp,Albertus:2007zz} results can be found in Ref.~\cite{Lubicz:2008am}.
A first observation is that the dependence on the light quark mass, that should allow to distinguish between $B_d$ and $B_s$ mesons, is practically invisible.
For the $B_{B_{d,s}}$ bag parameters, the unquenched results tend to be slightly lower than the quenched determinations, though still well compatible within the errors, and lead to the averages~\cite{Lubicz:2008am}
\begin{equation}\label{eq:Bds}
B_{B_d}^{\overline{\rm{MS}}}(m_b)=B_{B_s}^{\overline{\rm{MS}}}(m_b)=0.80 \pm 0.08\,,
\end{equation}
in the $\overline{\rm{MS}}$ scheme at the renormalization scale $\mu=m_b$, which correspond to the renormalization group invariant parameters
\begin{equation}\label{eq:BdsRGI}
\hat B_{B_d}=\hat B_{B_s}=1.22 \pm 0.12\,.
\end{equation}
The bag parameters $B^S_{B_{d,s}}$ have been recently calculated without the unquenched approximation only by one lattice collaboration~\cite{Dalgic:2006gp}, finding no evidence of quenching effects.
The averages given in Ref.~\cite{Lubicz:2008am}, include also previous quenched lattice results, and in the $\overline{\rm{MS}}$ scheme at the renormalization scale $\mu=m_b$ they read
\begin{equation}\label{eq:BSds}
B^S_{B_d}=B^S_{B_s}=0.85 \pm 0.10\,.
\end{equation}

The Wilson coefficients of these operators have been computed at NLO in QCD~\cite{Beneke:1998sy,Beneke:2003az,Ciuchini:2003ww}
and, together with $1/m_b$-suppressed effects~\cite{Beneke:1996gn}, lead to the theoretical predictions~\cite{Lenz:2006hd}
\begin{equation}
\frac{\Delta \Gamma_{B_d}}{\Gamma_{B_d}}= (4.1 \pm 0.9 \pm 1.2) \cdot 10^{-3}\,,\qquad \qquad \frac{\Delta \Gamma_{B_s}}{\Gamma_{B_s}}=(13 \pm 2 \pm 4) \cdot 10^{-2}\,.
\label{eq:DGammath}
\end{equation}
We observe that the theoretical predictions above are obtained by expressing the ratio $\Delta \Gamma/\Gamma$ as $(\Delta \Gamma/\Delta M)_{th.} /(\Delta M/\Gamma)_{exp.}$, i.e. by using the available accurate experimental measurements for the lifetimes and the mass differences. 
Moreover, they are obtained in a 
different operator basis $\{Q,\tilde Q_s\}$,  with $\tilde Q_S=(\bar b_iq_j)_{S-P}(\bar b_jq_i)_{S-P}$, where there do not appear strong cancelations due to NLO and $1/m_b$-suppressed contributions~\cite{Lenz:2006hd}. 
The differences between the central values of 
$\Delta\Gamma_{B_q}/\Gamma_{B_q}$ computed in the ``old'' and ``new'' bases, which
come from uncalculated $\alpha_s/m_b$ and $\alpha_s^2$ corrections, turn out to be quite large. A conservative $30$\% uncertainty is taken into account by the second errors in Eq.~(\ref{eq:DGammath}).

The experimental observable $|(q/p)_q|$, whose
deviation from unity describes CP violation due
to mixing, is related to $M^q_{12}$ and $\Gamma^q_{12}$, through
Eq.~(\ref{eq:obsB}). The theoretical prediction of $|(q/p)_q|$
is therefore based on the same perturbative and
non-perturbative calculation discussed for the width differences,
while the CKM contribution to $|(q/p)_q|$ is different
from that in $\Delta \Gamma_{B_q}/ \Gamma_{B_q}$.
The updated theoretical predictions are~\cite{Beneke:1998sy,Beneke:2003az,Ciuchini:2003ww}
\begin{equation}\label{eq:qp}
|(q/p)_d| - 1 = (2.96 \pm 0.67)\cdot 10^{-4}\,,\quad
|(q/p)_s| - 1 = −(1.28 \pm 0.28)\cdot 10^{-5}\,.
\end{equation}
Experimentally, information on the CP violation parameter $|(q/p)_q|$ is provided by the measurement of the semileptonic CP asymmetry, defined as
\begin{equation}\label{eq:ASL}
A^q_{SL}=\frac{\Gamma(B^0(t) \to l^- \bar \nu X) -\Gamma(\bar B^0(t) \to l^+ \nu X)}
{\Gamma(B^0(t) \to l^- \bar \nu X) -\Gamma(\bar B^0(t) \to l^+ \nu X)}
\end{equation}
which is related to $|(q/p)_q|$ through
\begin{equation}\label{eq:ASLqp}
A^q_{SL}=2(1- |(q/p)_q|)\,.
\end{equation}

We conclude this section on $B$ mesons by discussing the mass difference.
In contrast to $\Delta M_K$, in this case the long-distance contributions are estimated to be very small and $\Delta M_{B_{d,s}}$ is very well approximated by the relevant box diagrams, which are analogous to those shown in Fig.~\ref{fig:box} for Kaons.
Moreover, due to $m_{u,c} \ll m_t$, only the top sector can contribute significantly, whereas the charm sector and the mixed top-charm contributions are entirely negligible.
Thus, the theoretical expression for $M_{12}^{q}$, to which the mass difference is related through Eq.~(\ref{eq:obsB}), can be written as
\begin{eqnarray}
 M^q_{12}=\frac{G_F^2 M_{B_q} M_W^2}{12 \pi^2} \left(V_{tb} V_{tq}^*\right)^2 \eta_B
S_0(x_t) f_{B_q}^2 {\hat B}_{B_q},\qquad (q=d,s)\,, 
\end{eqnarray}
where $S_0(x_t)$ is the Inami-Lim function and $\hat\eta_B\approx 0.551$ represents the NLO QCD correction~\cite{Buras:1990fn}. 

The mass differences in the $B_d$ and  $B_s$ systems are proportional
to $|V_{td}|^2$ and $|V_{ts}|^2$, respectively, thus representing
important constraints on the UT,
provided that the multiplied hadronic matrix elements
are calculated. 
In order to involve reduced hadronic uncertainties, it is convenient to use as experimental constraints the ratio $\Delta M_{B_s}/\Delta M_{B_d}$ and $\Delta M_{B_s}$, since the strange-bottom sector is not affected by the uncertainty due to the chiral extrapolation.
On the other hand the UT analysis, being overconstrained, can be performed without using some inputs. In this way, the mass difference $\Delta M_{B_s}$ can be predicted with an accuracy of approximately $10$\%, as shown in Sec.~\ref{section:globalfits} where the whole UT analysis is discussed.

\label{sec:lifemix:blife:exp}

%
%
%
%
Experiments have published measurements of all flavors of $B$ hadrons. The B factories
have produced precision measurements of the $B^+$ and $B^0$ lifetimes. In addition
to the light mesons, the Tevatron experiments have also measured $B_s^0$, $\Lambda_b$ and $B_c$ lifetimes.
Selected measurements and the agreement of the world average fit with theoretical prediction are listed
in Table \ref{tab:lifemix:blifes}.

A typical lifetime measurement consists of two steps: signal isolation
and lifetime fitting.  In the signal isolation step one tries to
obtain the cleanest signal without cutting on lifetime-related
variables (impact parameter, vertex displacement etc). Fitting the
lifetime distribution is usually done with a log likelihood fit
(binned or unbinned) in which the signal lifetime distribution is a
convolution of an exponential and the detector lifetime resolution
function, while the background lifetime distribution is typically an
ad-hoc parametrization based on a sample in which the signal was
anti-selected (e.g. mass sidebands).

The B factory experiments have measured $B^+$ and $B^0$ lifetimes in
fully reconstructed hadronic decays as well as semileptonic $B^0
\rightarrow l D^* \nu$ and partially reconstructed hadronic $B^0
\rightarrow D^{*-} \pi^+$ and $B^0 \rightarrow D^{*-} \rho^+$
decays. The common element of all B factory measurements is the
reconstruction of the lifetime exploiting the boost in the $z$
direction from asymmetric colliding beams. As discussed in Section
\ref{sec:expPrimers:vtx}, the time measurement comes from correcting
the displacement, $L$ by the boost factor: $\Delta t = L/c \beta
\gamma$. The decay time $\Delta t$ is the decay time
between two vertices in the $z$ direction. The $B$ mesons are always
pair produced in the B factory environment, and the decays of each of
the $B$ mesons is a completely independent event from the other.
Therefore, one can measure the lifetime of a $B$ meson by starting the
time interval $\Delta t$ with the decay time of one of the $B$ mesons
and ending it with the decay of the other $B$ meson.
For measurements utilizing partially reconstructed (e.g. $D^*+$) decays,
many handles specific to the B factory environment are used.  In the particular 
example of semileptonic decays involving the, $D^{*+}$,  the 
$D^{*+} \rightarrow \pi^+ D^0$ the $D^0$ decay is not explicitely reconstructed.
Instead, the small phase space for the $D^*$ decay is exploited to infer the 
direction and energy of the $D^*$ from that of the soft pion.
The momenta of the $D^{+}$, the lepton, and the center of mass energy are used
to infer the momentum of the neutrino. The invariant mass of the
neutrino computed from the same variables provides a powerful variable which
eliminates a large fraction of the $B\overline B$ combinatorial background.

Tevatron experiments have access to $B^0$, $B^+$ as well as $B_s^0$,
$\Lambda_b$ and $B_c$ decays and have reported lifetime measurements
for all of them. The $B^0$ and $B^+$ lifetimes are measured both in
semileptonic and fully reconstructed hadronic decays. The corrections
for semileptonic decays are different than in B factory
experiments. Samples are gathered by triggering on a lepton; CDF also
requires a displaced track to further increase the $B$ fraction of the
sample. Lacking the initial energy of the $B$ meson, inclusive
reconstruction is not possible. The $D / D^{*}$ meson has to be
reconstructed explicitely, incurring a large branching ratio
penalty. As discussed in more detail in Section
\ref{sec:expPrimers:vtx}, the energy of the neutrino is also unknown,
so the computation of the decay proper time incurres a $K$-factor
correction which has to be derived from Monte Carlo simulation of $B$
meson decays. The \D0~experiment directly measures the ratio of $B^+$ and
$B^0$ lifetimes by fitting the ratio of observed $B^+$ and $B^0$ decay 
times (no $K$ factor correction is applied) to a predicted distribution which
depends on the lifetime ratio. 
 $B^0, B^+, B_s^0$, and $\Lambda_b$ lifetimes are also
measured with fully reconstructed hadronic final states. For the $B^0,
B^+$ and $B_s^0$ mesons the most common final state is $B \rightarrow
D \pi$. Data samples for these analyses are gathered by triggering on
displaced vertices, as described in Section
\ref{sec:vtxTevatron}. This trigger sculpts the lifetime
distribution. This effect and the correction techniques used in the
analyses are discussed in detail in Section \ref{sec:expPrimers:vtx}.  The
lifetime ratio of $\tau(\Lambda_b)/\tau(B^0)$ has been of considerable
interest. There has been a long standing disagreement between
theoretical predictions and experimental results.  The $\Lambda_b$
lifetime has been measured both in the $\Lambda_b \rightarrow J/\psi
\Lambda$ and $\Lambda_b \rightarrow \Lambda_c \pi$ decays. Data for
the first decay channel is gathered with dimuon triggers which are not
lifetime biased.  Data for the second channel is gathered through the
displaced track trigger and requires similar corrections to those
applied to fully hadronic $B$ meson decays. 
%
The $B_c$ meson occupies a special place amongst the $B$-hadrons as it
can decay weakly via the $b$ or $c$ quark, making its lifetime considerably
shorter than those of light $B$ mesons. Due to its relatively large branching
fraction, Tevatron experiments measure the $B_c$ lifetime in semileptonic
decays ($B_c \rightarrow J/\Psi l \nu$). The $B_c$ mass is measured in
hadronic decays, where lifetime cuts are used to reject background. Similar to
semileptonic light $B$ decays, the $B_c$ momentum cannot be fully reconstructed; 
$K$-factor corrections based on Monte Carlo simulation are necessary.

Table \ref{tab:lifemix:blifes} lists representative measurements from the different
experiments and current world average results. Most of the $B$ hadron lifetimes are now in good
agreement with theoretical predictions\cite{Tarantino:2005zi} \cite{Kiselev:2003mp}, 
except for the $B_s^0$ lifetime which  is currently significantly lower than the predicted value.

%
%
%
\begin{table}
\caption{Representative measurements from different experiments and world average
lifetimes. The ratios of world average hadron lifetimes to the $B^0$ lifetime are compared
to theoretical predictions.\label{tab:lifemix:blifes}}
\begin{center}
\begin{tabular}{
c @{\hspace{0.2 cm}\vline\hspace{0.2 cm}}
c @{\hspace{0.2 cm}\vline\hspace{0.2 cm}}
c @{\hspace{0.2 cm}\vline\hspace{0.2 cm}}
c @{\hspace{0.2 cm}\vline\hspace{0.2 cm}}
c @{\hspace{0.2 cm}\hspace{0.2 cm}}
c}
 Hadron & Source & Result [ps] & Ratio $\tau(B_x)/\tau(B^0) $ & Theory Pred. Ratio \\
\hline
$B^0$ & & & &\\ 
      & \babar  & $1.504 \pm 0.013 ^{+ 0.018}_{-0.013}$ & & \\
      & Belle   & $1.534 \pm 0.008 \pm 0.010$ & &\\
      & CDF     & $1.524 \pm 0.030 \pm 0.016$ & &\\
      & \D0     & $1.414 \pm 0.018 \pm 0.034$ & &\\
      & PDG     & $1.525 \pm 0.009$ & &   \\
\hline
$B^+$ & & & & $  1.04 - 1.08        $  \\
      & Belle   &  $1.635 \pm 0.011 \pm 0.011 $ & &             \\
      & \babar  &  $1.673 \pm 0.032 \pm 0.023 $ & &             \\
      & CDF     &  $1.630 \pm 0.016 \pm 0.011 $ & &             \\
      & \D0     & $---$ &  $1.080 +- 0.016 +- 0.014   $ &              \\
      & PDG     &  $1.638 \pm 0.011$ & $1.07 \pm 0.01$   & \\
\hline
$B_s^0$ & & & &  $0.99 - 1.01 $  \\
      & CDF   &  $1.36 \pm 0.09^{+0.06}_{-0.05}$ & & \\
      & \D0   &  $1.398 ± 0.044^{+0.028}_{-0.025}$ & &  \\
      & PDG   &  $ 1.425 \pm 0.041 $ & $0.934 \pm 0.027$ \\
\hline
$\Lambda_b$ & & & &  $0.87 - 0.95 $  \\
      & \D0   & $1.218^{+0.130}_{-0.115} \pm 0.0427$   &  &\\
      & CDF   & $1.593^{+0.083}_{-0.078} \pm 0.0337  $ &  &\\
      & PDG   & $1.383 ^{+0.049}_{-0.048}$             & $0.91 \pm 0.03$ &  \\
\hline
$B_c^+$ & & & &  $ 0.31 - 0.36 $ \\
        & CDF   & $ 0.475^{+0.053}_{-0.049} \pm 0.018 $  & & \\
        & \D0   & $ 0.448^{+0.038}_{-0.036} \pm 0.032 $  & & \\
        & PDG   & $ 0.453 \pm 0.041 $ & $0.29 \pm 0.03 $ & & \\
\end{tabular}
\end{center}
\end{table}
%
As discussed in Section \ref{sec:BmixingTh}, additional information can be
extracted from the measurement of lifetime differences between the heavy and light eigenstates. 
At the Tevatron the lifetime difference in the $B_s$ system is accessible in 
the decay $B_s \rightarrow J/\psi \phi$ which gives rise to both CP-even and 
CP-odd final states. It is possible to separate the two CP components of the 
decay and measure the lifetime difference through a simultaneous 
fit to the time evolution and angular distributions of the decay products
of the $J/\psi$ and $\phi$ mesons. Fig. \ref{fig:lifetimeproj} shows the
lifetime projections for the  $\Delta \Gamma$  measurements at CDF and 
\Dzero~with the CP even and CP odd components fitted separatly. Both 
experiments have
so far analysed 2.8 fb$^{-1}$ of data. The results \cite{dgcdf, :2008fj} are still
compatible with a $\Delta \Gamma$ of zero. 
It should be noted that if $\Delta \Gamma$ is not zero, the flavor specific
(equal mix of $B_s^H$ and $B_s^L$ at t=0) and CP specific $B_s$ lifetimes will
be distinct. 
\begin{figure}[ht]
\centering
    \includegraphics[width=2.2in]{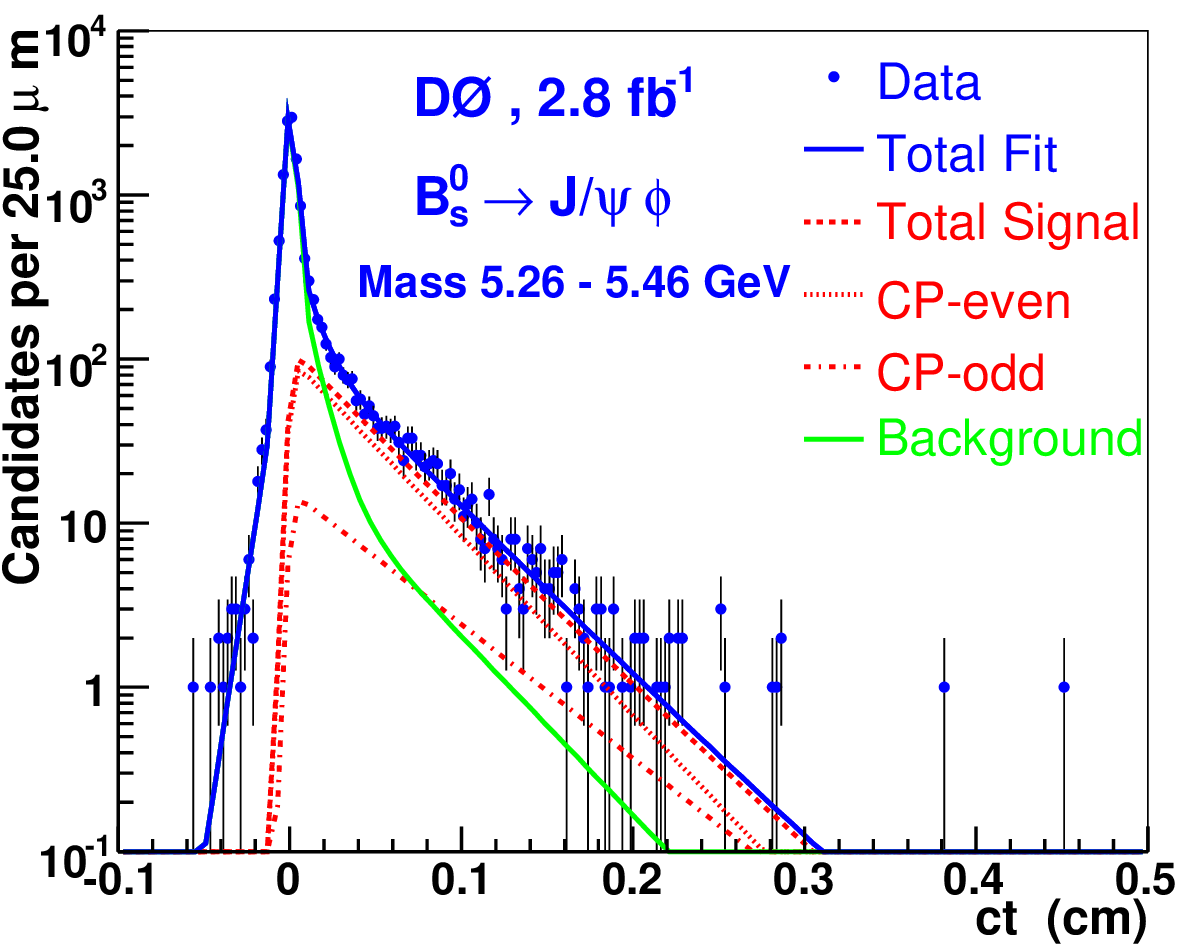}
%
    \includegraphics[width=2.6in]{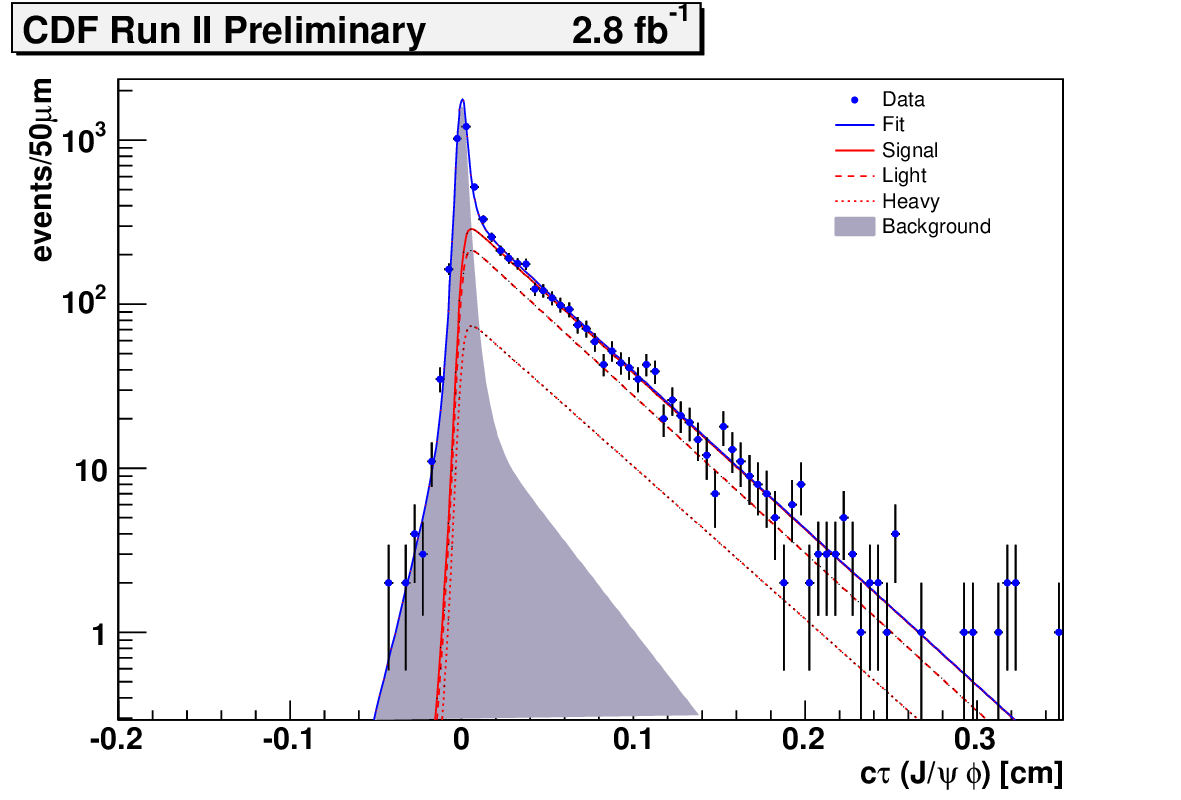}
\caption[$B_s$ lifetime projection]
{Lifetime projection for $B_s^0\rightarrow J/\psi \phi$ decay candiates in the signal region. 
In the left panel, the projection of the \D0~fit. In the right panel, the corresponding 
projection by the CDF collaboration.     \label{fig:deltagamma_Tevatron
}}
\label{fig:lifetimeproj}
\end{figure}
%



\subsubsection{$B$ meson mixing }
\label{sec:lifemix:blife:mix}
\label{sec:bmixingExp}
Mixing measurements utilize lifetime and flavor tagging information,
as described in Section \ref{sec:expPrimers:bFlavorTagging}. As
discussed in Section \ref{sec:tdep}, the time evolution of the
probability density function for a $B$ meson tagged with flavor $\eta$
to decay with flavor $f$ is given by Equation \ref{eq:alllike}. The
relevant experimental parameters that come into play are the effective
tagging power ($\epsilon D^2$), proper time resolution and
signal. These differ significantly between the $B$ factory experiments
and hadron colliders. Opposite side flavor tagger properties for the
different experiments are compared in Table \ref{tab:ostperformance}.
The main difficulty with flavor tagging in hadron colliders is that
the opposite side $B$ meson is not in the detector acceptance most of
the time.  This is not the case with $B$ factories due to the coherent
production and controlled boost of the $B\overline{B}$ meson pair.

Mixing in the $B_d$ meson sector was established 25 years ago by the Argus
Collaboration~\cite{Albrecht:1987dr} and precision measurements where
available since the beginning of the asymmetric B-Factory program,
since they can exploit both the large luminosity and the boost of the
center-of-mass frame. A compilation of measurements from all contributing
experiments is shown in Figure Fig.~\ref{fig:deltamd}.
The most accurate measurements come from Belle and \babar , where
one of the two $B$ mesons is fully via in their semileptonic decay and
the other $B$ tagged as $B^0$ or $\bar{B}^0$.

\begin{figure}[htb]
\begin{center}
\epsfig{file=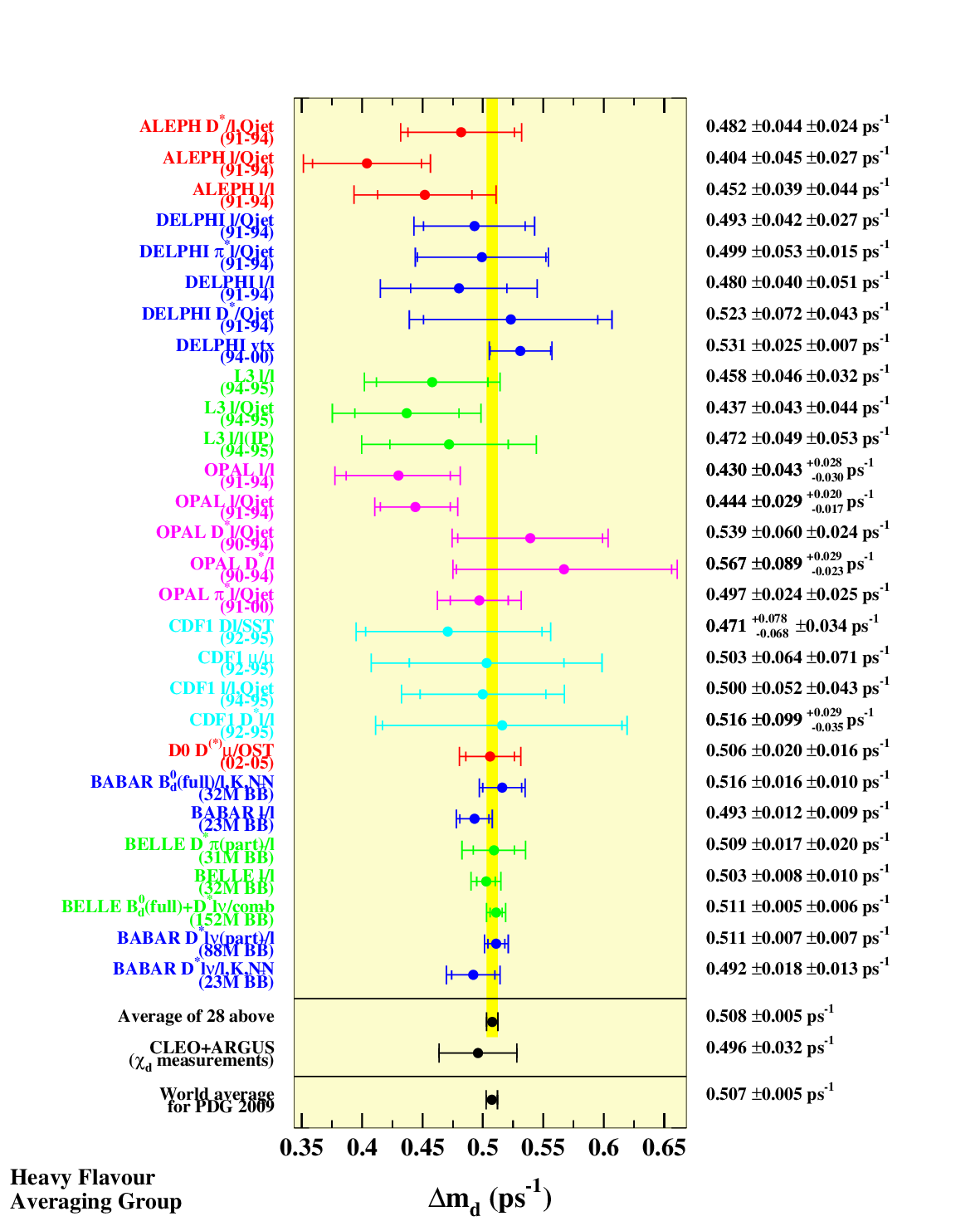,height=0.9 \textheight}
\caption{
Summary and average of  $\Delta M_{B_d}$ measurements. See~\cite{Barberio:2008fa} for the full list of references. \label{fig:deltamd}}
\end{center}
\end{figure}

The most significant contribution of the Tevatron experiments to $B$
mixing is the observation of $B_s^0$ oscillations. The analysis layout
is very similar to that for $B^0$ mixing. Clean signal reconstruction,
proper time resolution and flavor tagger dilution are essential. The
probability density function for tagged decays also follows the
formalism of Equation \ref{eq:alllike}. The significant difference
with respect to $B^0$ mixing is the high expected oscillation
frequency ($\sim 18~{\rm ps^{-1}}$).  The high frequency makes
excellent proper time reconstruction essential; in order to resolve
the oscillation frequency, the detector resolution has to be better
than one oscillation period. The now-standard way of interpreting and
combining results was first proposed by Moser and
Roussarie\cite{Moser:1996xf} and is called an amplitude scan.
Mathematically very similar to a Fourier transformation of the tagged
lifetime distribution, this method involves re-fitting the data with
different probe frequencies, while floating the oscillation
amplitude. The amplitudes are then reported as a function of probe
frequency.  An amplitude significantly different from zero indicates
the presence of an oscillation signal.  Figure
\ref{fig:lifemix:ampliScans} shows the most recent results from CDF
and D0.  The CDF result\cite{Abulencia:2006ze} is consistent with
oscillations at $\Delta M_{B_s}$ = $17.77 \pm 0.10$ (stat) $\pm$ 0.07
(syst) ps$^{-1}$. The signal significance is found to be
$5.4\sigma$. The \D0 result\cite{d0Update} shows consistency with an
oscillation signal at $\Delta M_{B_s} =18.53 \pm 0.93$ (stat) $\pm$
0.30 (sys) ps$^{-1}$. The significance of the signal corresponds to
$2.9\sigma$ and supersedes the original two-sided bound $ 17  <
\Delta M_{B_s} < 21 \rm{ps}^{-1}$ at 90$\%$ CL\cite{Abazov:2006dm}.

\begin{figure}[htb]
\begin{center}
\epsfig{file=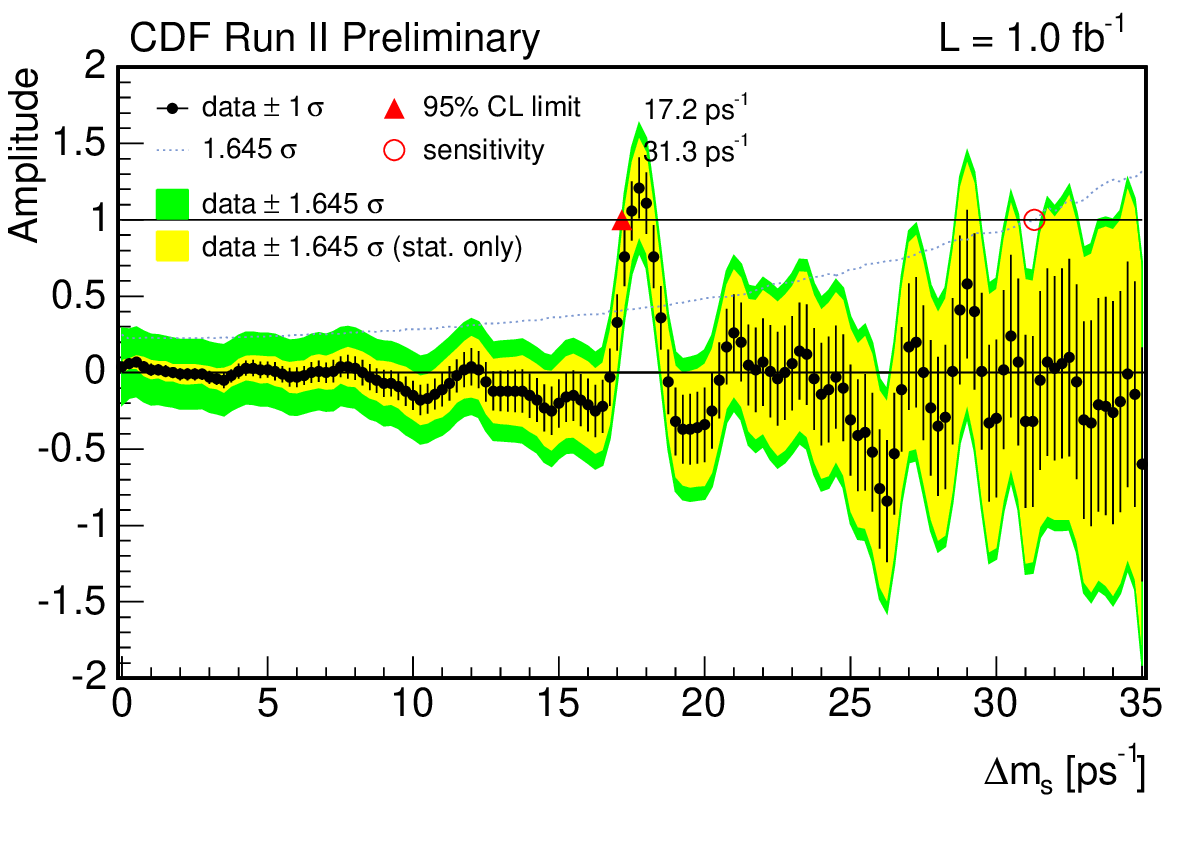, width = 0.47\textwidth}
\epsfig{file=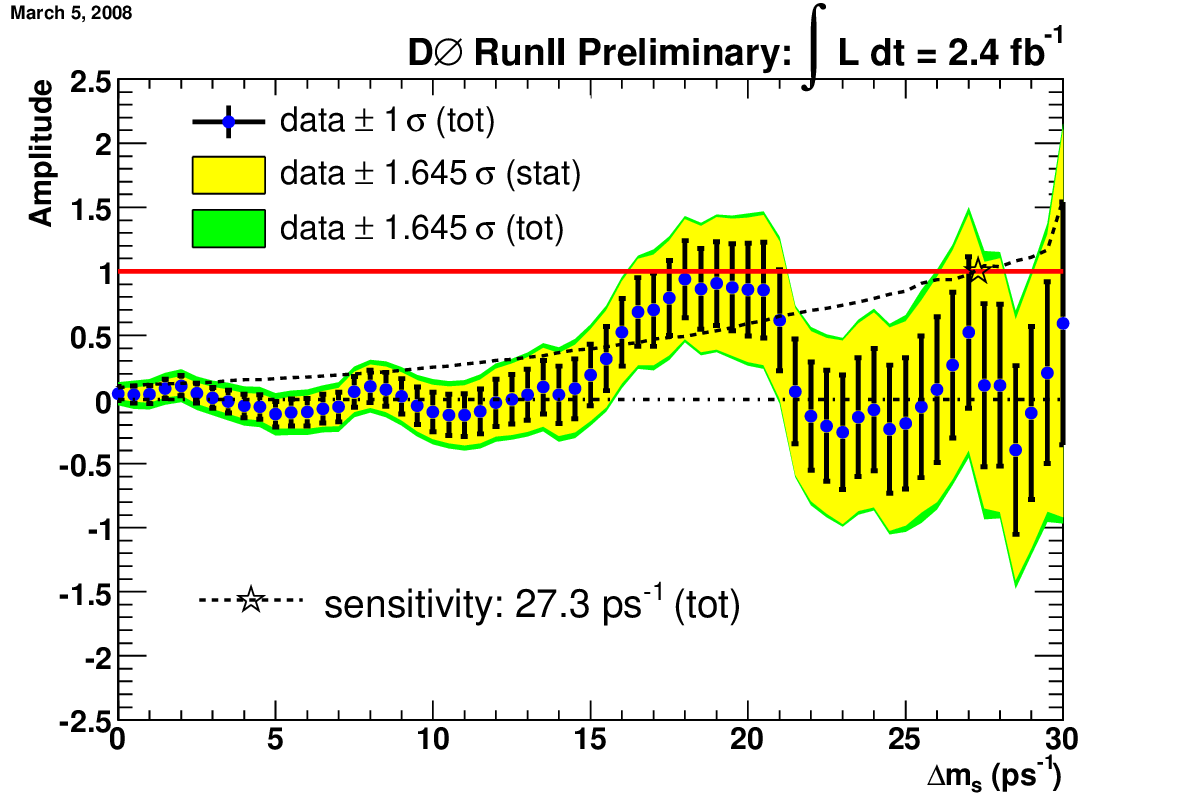, width = 0.48 \textwidth}
\end{center}
\caption{Amplitude scans for the $B_s$ oscillation fits. CDF and \D0
results are shown in the left and right panel, respectively.
\label{fig:lifemix:ampliScans}
}
\end{figure}


\subsubsection{Measurements of the angle \bb in tree dominated processes}
\label{sec:lifemix:beta}

The Standard Model (SM) of electroweak interactions describes charge 
conjugation-parity (\CP) violation as a consequence of an irreducible 
complex phase in the three-generation Cabibbo-Kobayashi-Maskawa (CKM) 
quark-mixing matrix~\cite{Cabibbo:1963yz,Kobayashi:1973fv}. 
In this framework, neutral $B$ decays to 
\CP\ eigenstates containing a charmonium and $K^{0}$ meson provide a 
direct measurement of $\stwob$~\cite{Bigi:1981qs,Carter:1980tk}. 
The unitarity triangle angle $\beta$ (or $\Phi_1$) is 
$\arg \left[\, -V_{\rm cd}^{}V_{\rm cb}^* / V_{\rm td}^{}V_{\rm tb}^*\, \right]$ where
the $V_{ij}$ are CKM matrix elements. 

$\Bz \rightarrow \jpsi \piz$ proceeds instead via a Cabibbo-suppressed  
$\b \rightarrow c \bar c d$ transition: the tree amplitude has
the same weak phase as the $\btoccbars$ transition, therefore we expect 
the corresponding values of $S$ and $C$ to be $-\stwob$ and $0$ respectively, 
unless penguin 
amplitudes or other contributions are significant.

The current status of measurements of $\stwob$ 
from charmonium decays are presented in what follows and cover 
$\btoccbars$ and $\b \rightarrow c \bar c d$ transitions. 
Additional results on determining the sign of $\beta$ are also 
mentioned using the measurement of $\ctwob$ in \btoccbars decays.

Most of the measurements presented here are based on data collected
by the \babar\ and the Belle experiments.
The difference between the proper decay times of the signal 
$B$ meson ($B_{rec}$) and of the other $B$ meson ($B_{tag}$) 
is used to measure the time-dependent $\CP$-asymmmetries, $\Acp$.
The initial flavor of $B_{rec}$ is identified by using information from 
$B_{tag}$.
$\Acp$ is defined as
\begin{equation}
\Acp(t) \equiv \frac{N(\Bzb(t)\to f) - N(\Bz(t)\to f)} {N(\Bzb(t)\to f) + N(\Bz(t)\to f)} = S \sin(\deltamd{t}) - C \cos(\deltamd{t}),
\label{eq:timedependence}
\end{equation}
where $N(\Bzb(t)\to f)$ is the number of \Bzb\ that decay into the 
CP-eigenstate $f$ after a time $t$
and $\deltamd$ is the difference between the \B\ mass eigenstates.
Belle reports results using the variable $A \equiv -\C$.

In the SM, direct CP violation in $\btoccbars$ decays is
negligible. Under this assumption, the \CP\ violation parameters $S$ and $C$ 
are given by $S_{\btoccbars} = -\eta_{f}\stwob$ and $C_{\btoccbars}$ = 0, 
where $\eta_{f}$ is $-$1 for ($\ccbar$)$\KS$ decays (e.g. $\jpsi\KS$, 
$\psitwos\KS$, $\chicone\KS$, $\eta_c \KS$
~\footnote{Charge-conjugate reactions are included implicitly unless otherwise specified.}) and $\eta_{f}$ is $+$1 for the ($\ccbar$)$\KL$ 
(e.g. $\jpsi\KL$) state. 
The $\jpsi\Kstarz (\Kstarz \to \KS\piz)$ final 
state is an admixture of \CP\ even and \CP\ odd amplitudes 
for which we use $\eta_f = \effectiveeta$.
To be consistent with other time-dependent CP measurements, 
we show the results in terms of $C_f = \eta_f C$ and $S_f = \eta_f S$.
Using $425.7 \ifb$ of integrated luminosity, the $\babar$ experiment
measured the time-dependent CP asymmetry parameters for the $\jpsi\KS$, 
$\psitwos\KS$, $\chicone\KS$, $\eta_c \KS$ and $\jpsi\KL$ modes combined~\cite{:2008cp}~\footnote{Unless otherwise stated, all results are quoted with the first error being statistical and the second systematic.}:
\begin{eqnarray}
C_f = 0.026 \pm 0.020 (stat) \pm 0.016 (syst), ~~~
S_f = 0.691 \pm 0.029 (stat) \pm 0.014 (syst). \nonumber
\end{eqnarray}
$C_f$ and $S_f$ for each of the decay modes within the \CP\ sample and
of the $\jpsi\Kz (\KS+\KL)$ sample were also measured ~\cite{:2008cp}. 
These results are preliminary.
The Belle experiment measured these parameters from  
$\jpsi\KS$ and  $\jpsi\KL$  decays using a data sample of $492 \ifb$ 
and found~\cite{Chen:2006nk}:
\begin{eqnarray}
C_f = -0.018 \pm 0.021 (stat) \pm 0.014 (syst), ~~~
S_f = 0.642 \pm 0.031 (stat) \pm 0.017 (syst) .\nonumber
\end{eqnarray}
Belle also reported results from the $\psitwos\KS$ decay using  $605 \ifb$
~\cite{Abe:2007gj}:
\begin{eqnarray}
C_f = -0.039 \pm 0.069 (stat) \pm 0.049 (syst), ~~~
S_f =  0.718 \pm 0.090 (stat) \pm 0.033 (syst). \nonumber
\end{eqnarray}
The analysis of $\btoccbars$ decay modes imposes a constraint on $\stwob$ only, but a four-fold
ambiguity in the determination of the angle $\beta$ remains.  
It is possible to reduce this ambiguity by measuring $\ctwob$
using the angular and time-dependent asymmetry in 
$\Bz\to\jpsi\Kstarz(\Kstarz\to\KS\piz)$ decays.
The results of the fit treating
$\stwob$ and $\ctwob$ as independent variables give
$\ctwob =+3.32_{-0.96}^{+0.76} \pm 0.27$ 
~\cite{Aubert:2004cp} for \babar. 
Using the outcome of fits to simulated samples,
the sign of $\cos2\beta$ is determined to be positive 
at the 86\% confidence level. 
Belle reported $\ctwob =+0.56 \pm 0.11 \pm 0.27$ 
~\cite{Itoh:2005ks}. 
These results are compatible with the Standard Model expectations.
Other measurements also contribute to reduce the ambiguity. 

The time-dependent CP asymmetry parameters were measured in the
$\Bz \rightarrow \jpsi \piz$ decay and are also consistent with the SM.
Using a dataset of $425 \ifb$, the \babar\ experiment 
measured~\cite{Aubert:2008bs}:  
\begin{eqnarray}
C_f = -0.20 \pm 0.19 (stat) \pm 0.03 (syst), ~~~
S_f = -1.23 \pm 0.21 (stat) \pm 0.04 (syst).\nonumber
\end{eqnarray}
This is evidence for CP violation as $S$ and $C$ are measured to have
 non-zero values at a $4\sigma$ confidence level.
The results reported by the Belle experiment using $492 \ifb$  
are~\cite{:2007wd}:
 \begin{eqnarray}
C_f = -0.08 \pm 0.16 (stat) \pm 0.05 (syst), ~~~
S_f = -0.65 \pm 0.21 (stat) \pm 0.05 (syst).\nonumber
\end{eqnarray}
The measurements of $\stwob$ in charmonium decays are in excellent 
agreement with the SM expectations~\cite{Ciuchini:1995cd}. 
The results presented above are summarized in
Fig.~\ref{fig:btoccsS_CP}.
High precision measurements 
using larger datasets are anticipated in the next few years.

\begin{figure}[!ht]
\begin{center}
\includegraphics[height = 0.45 \textwidth]{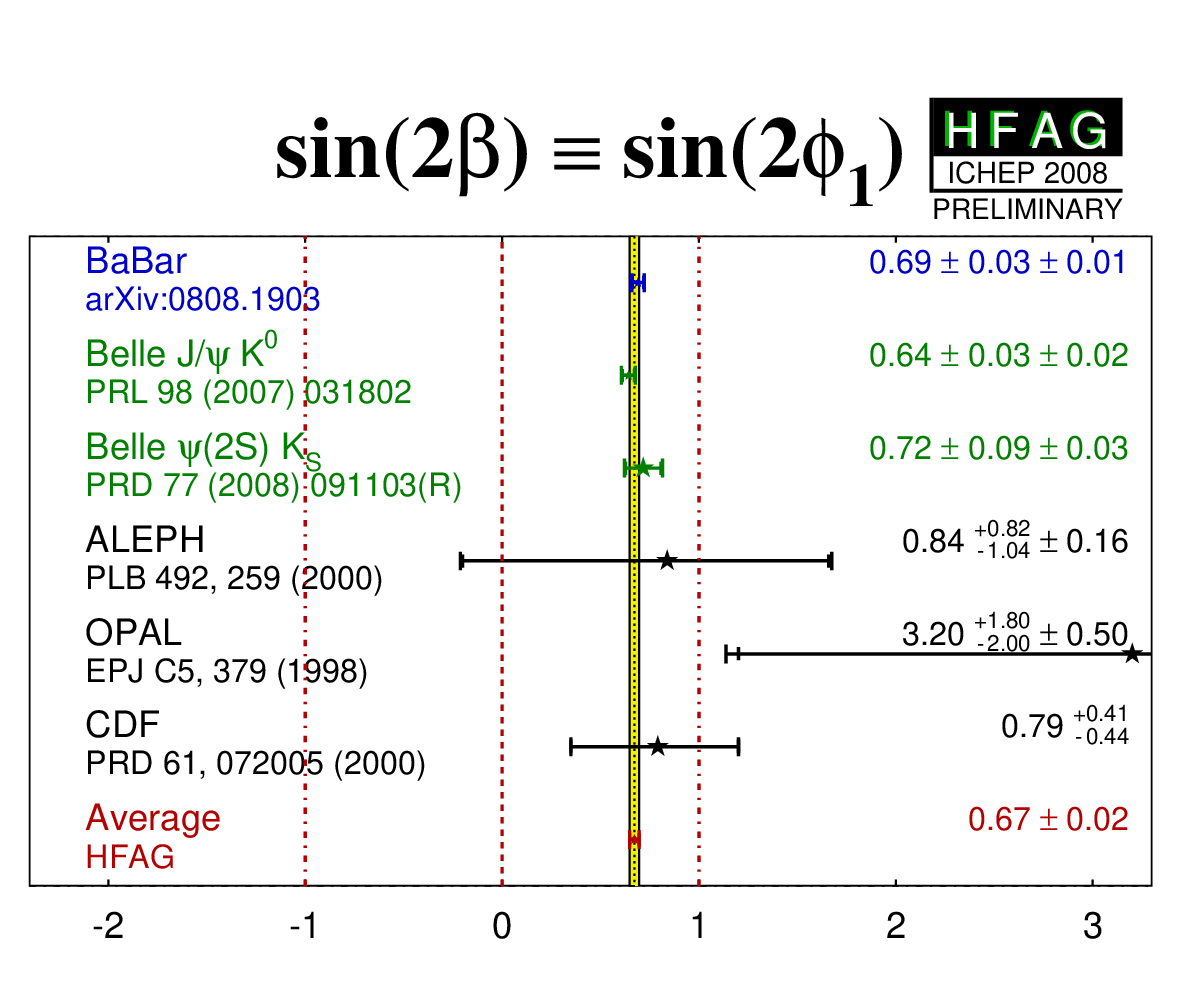}
\includegraphics[height = 0.43 \textwidth]{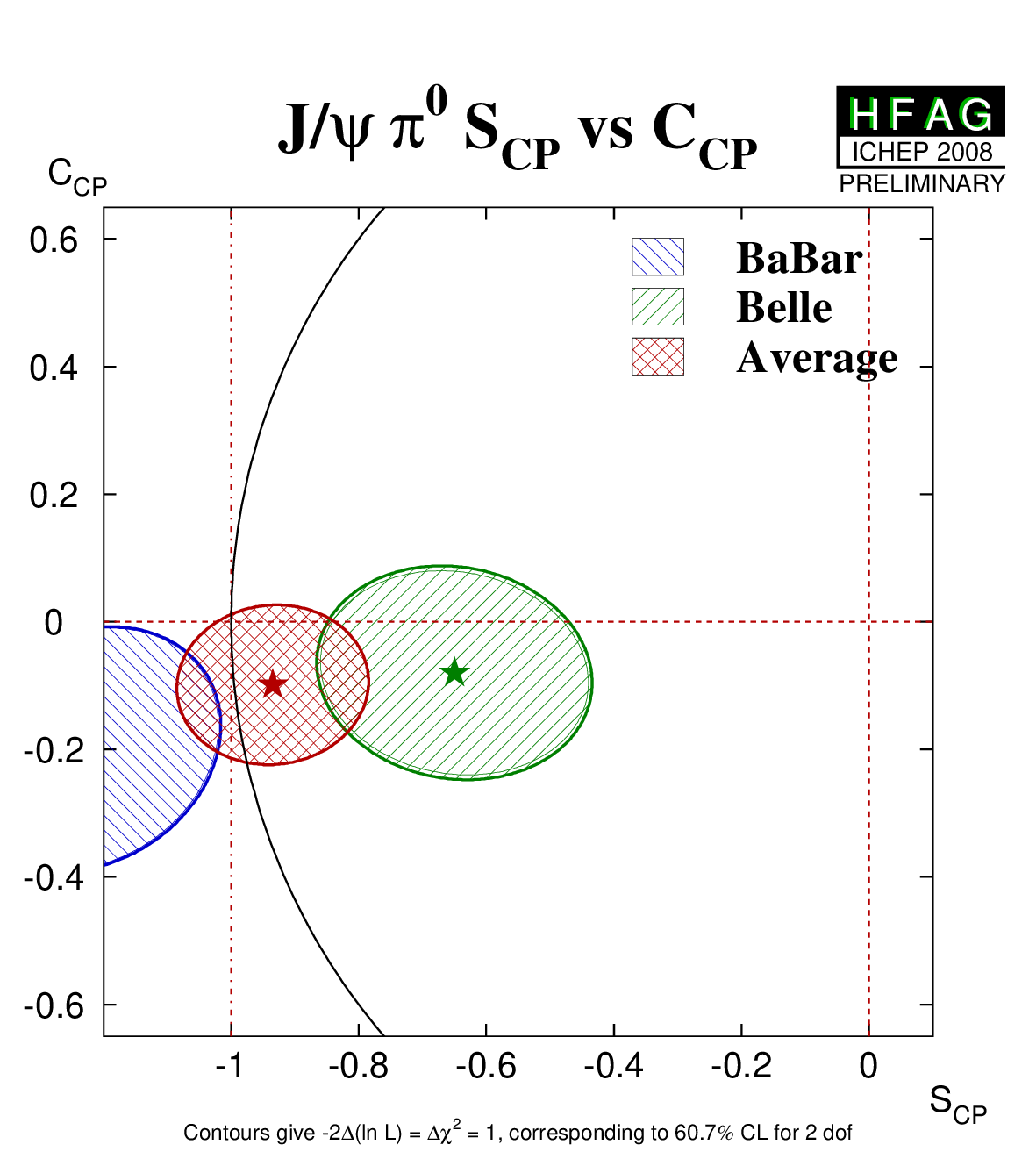}
\end{center}
\caption{HFAG averages. In the left panel, the average of $\stwob$ from all experiments.
In the right panel, the summary plot of $S$ versus $C$ for $\Bz \rightarrow \jpsi \piz$.
\label{fig:btoccsS_CP}}
\end{figure}


\subsubsection{Measurement of the $B_s$ meson mixing phase }
\label{sec:lifemix:blife:phi-s}

The $\Bs$ mixing phase is accessible through the time-evolution of
$\BsJphi$ decays, which is sensitive to the relative phase between the
mixing and the $\bar{b}\to \bar{c}c\bar{s}$ quark-level transition,
$\betas = \betasSM+\betasNP$. This phase is responsible for
\CP-violation. In the Standard Model it equals to
$\betasSM=\arg(-V_{ts}V_{tb}^{*}/V_{cs}V_{cb}^{*}) \approx 0.02$  \cite{Charles:2004jd,Bona:2006ah}. 
Any sizeable deviation from this value would be unambiguous evidence of new physics \cite{Dunietz:2000cr}. 
If new physics contributes a phase
($\betasNP$), this would also enter $\phis = \phisSM - 2\betasNP$,
which is the phase difference between mixing and decay into final
states common to $\Bs$ and \Bsb, and is tiny in the SM: $\phisSM =
\arg(-M_{12}/\Gamma_{12}) \approx 0.004$ \cite{Lenz:2006hd}. The phase
\phis\ enters the decay-width difference between light and heavy
states, $\Delta\Gamma=\Gamma_L-\Gamma_H=2|\Gamma_{12}|\cos(\phis)$ and equals
$\DGSM \approx 2|\Gamma_{12}| = 0.096 \pm 0.036$ ps$^{-1}$ in
the Standard Model \cite{Lenz:2006hd}, thus playing a r\^{o}le in $\BsJphi$
decays. Since the SM values for \betas\ and \phis\ cannot be resolved
with the resolution of current experiments, the following
approximation is used: $\phis \approx -2\betasNP \approx -2\betas$,
which holds in case of sizable NP contributions. \par 

The measurement of \betas\ is analogous to the determination of the phase
$\beta=\arg(-V_{cd}V_{cb}^{*}/V_{td}V_{tb}^{*})$ in $\Bz \to \jpsi\KS$
decays, except for a few additional complications. The oscillation
frequency in the $\Bs$ system is about 35 times higher than in \Bz\ mesons,
requiring excellent decay-time resolution. The decay of a pseudoscalar
meson (\Bs) into two vector mesons ($\jpsi$ and $\phi$) produces two
\CP-even states (orbital angular momentum $L = 0,2$), and one \CP-odd
state ($L=1$), which need to be separated for maximum sensitivity. Finally,
the value of the SM expectation for $\betas$ is approximately $30$
times smaller \cite{Bigi:1981qs} than  $\beta$.\par

Both Tevatron experiments have performed
measurements of the time-evolution of flavor-tagged $\Bs \to
\jpsi(\to\mu^+\mu^-) \phi(\to K^+K^-)$ decays
\cite{Aaltonen:2007he}.
The CDF analysis is described in the following, a similar analysis is performed by \Dzero. 
Events enriched in \jpsi\ decays are
selected by a trigger that requires the spatial matching between a
pair of two-dimensional, oppositely-curved, tracks in the multi-wire
drift chamber (coverage $|\eta|<1$) and their extrapolation outward to
track-segments reconstructed in the muon detectors (drift chambers and
scintillating fibers). In the offline analysis, a kinematic fit to a
common space-point is applied between the candidate \jpsi\ and another
pair of tracks consistent with being Kaons originated from a $\phi$
meson decay. 
An artificial neural network trained on simulated events (to
identify signal, $S$) and $\Bs$ mass sidebands (for background, $B$)
is used for an unbiased optimization of the selection. The quantity
$S/\sqrt{S+B}$ is maximized using kinematic and particle
identification (PID) information. Discriminating observables include
Kaon-likelihood from the combination of \dedx\ and TOF information,
transverse momenta of the $\Bs$ and $\phi$ mesons, the $K^+K^-$ mass,
and the quality of the vertex fit.

The sensitivity to the mixing phase is enhanced if the evolution of
\CP-even eigenstates, \CP-odd eigenstates, and their interference is
separated. This is done by using the angular distributions of final state particles
to statistically determine the \CP-composition of the signal. The
angular distributions are studied in the transversity basis, which
allows a convenient separation between \CP-odd and \CP-even terms in
the equations of the time-evolution.
Sensitivity to the phase increases if the evolution of bottom-strange
mesons produced as $\Bs$ or \Bsb\ are studied independently.  The time
development of flavor-tagged decays contains terms proportional to
$\sin(2\betas)$, reducing the ambiguity with respect to the untagged
case ($\propto |\sin(2\betas)|$). Building on techniques used in the
$\Bs$ mixing frequency measurement \cite{Abulencia:2006ze}, the production
flavor is inferred using flavor tagging techniques discussed in Sec.
\ref{sec:expPrimers:bFlavorTagging}

The tagging
power, $\epsilon D^2 \approx 4.5\%$, is the product of an efficiency
$\epsilon$, the fraction of candidates with a flavor tag, and the
square of the dilution $D=1-2w$, where $w$ is the mistag
probability. 
The proper time of the decay and its resolution are known on a
per-candidate basis from the position of the decay vertex, which is
determined with an average resolution of approximately 27 \mum\ (90
fs$^{-1}$) in $\BsJphi$ decays.
Information on $\Bs$ candidate mass and its uncertainty, angles
between final state particles' trajectories (to extract the
\CP-composition), production flavor, and decay length and its
resolution are used as observables in a multivariate unbinned
maximum likelihood fit of the time evolution. The fit accounts for direct decay
amplitude, mixing followed by the decay, and their
interference. Direct \CP-violation is expected to be small and is not
considered. The outputs of the fit are the phase $\betas$, the decay-width
difference $\Delta\Gamma$, and 25 other ``nuisance'' parameters
($\vec{\nu}$). These include the mean $\Bs$ decay-width ($\Gamma =
(\Gamma_L + \Gamma_H)/2$), the squared magnitudes of linear
polarization amplitudes ($|A_0|^2$, $|A_{\parallel}|^2$,
$|A_{\perp}^2|$), the \CP-conserving (``strong'') phases
($\delta_{\parallel} = \arg(A_{\parallel} A_{0}^{*})$, $\delta_{\perp}
= \arg(A_{\perp}A_{0}^{*})$), and others.  

The acceptance of the
detector is calculated from a Monte Carlo simulation and found to be
consistent with observed angular distributions of random combinations
of four tracks in data. CDF also validated the angular-mass-lifetime model 
by measuring lifetime and polarization amplitudes in 7800 \BdJKst\
decays, which show angular features similar to the $\Bs$ sample:
$c\tau(\Bz) = 456 \pm 6 (stat) \pm 6 (syst) \mum$, $|A_0|^2 = 0.569
\pm 0.009 (stat) \pm 0.009 (syst)$, $|A_\parallel|^2 = 0.211 \pm 0.012
(stat) \pm 0.006 (syst)$, $\delta_\parallel = -2.96 \pm 0.08 (stat)
\pm 0.03 (syst)$, and $\delta_\perp = 2.97 \pm 0.06 (stat) \pm 0.01
(syst)$. The results, consistent and competitive with most recent
$B$--factories' results \cite{Aubert:2007hz}, support the reliability of
the model. Additional confidence is provided by the precise
measurement of lifetime and width-difference in untagged $\BsJphi$
decays \cite{Aaltonen:2007gf}.\par

Tests of the fit on simulated samples show biased, non-Gaussian
distributions of estimates and multiple maxima, because the likelihood
is invariant under the transformation $\mathcal{T} =
(2\betas\to\pi-2\betas, \Delta\Gamma\to-\Delta\Gamma,
\delta_\parallel\to2\pi-\delta_\parallel,\delta_\perp\to\pi-\delta_\perp)$,
and the resolution on \betas\ was found to depend crucially on the
true values of \betas\ and $\Delta\Gamma$.  CDF quotes therefore a
frequentist confidence region in the ($\betas, \Delta\Gamma$) plane
rather than point-estimates for these parameters. Obtaining a correct
and meaningful region requires projecting the full 27-dimensional region
into the ($\betas,\Delta\Gamma$) plane.
A common approximate method is the profile likelihood approach.
For every point in the ($\betas,\Delta\Gamma$) plane, $\hat{\vec{\nu}}$ are the
values of nuisance parameters that maximize the likelihood. Then
$-2\Delta\ln(L_p)$ is typically used as a $\chi^2$ variable to derive
confidence regions in the two-dimensional space
($\betas,\Delta\Gamma$). Simulations show that in the
present case the approximation fails. The resulting regions contain
the true values with lower probability than the nominal confidence
level (C.L.) because the $-2\Delta\ln(L_p)$ distribution has longer
tails than a $\chi^2$. 
A full confidence
region construction is therefore needed, using simulation of a large
number of pseudo-experiments to derive the actual distribution of
$-2\Delta\ln(L_p)$, with a potential for an excessive weakening of the
results from systematic uncertainties. However, in a full confidence
limit construction, the use of $-2\Delta\ln(L_p)$ as ordering function
is close to optimal for limiting the impact of systematic
uncertainties \cite{Cranmer:2003vt,Punzi:2005yq}. With this method, it is possible to
account for the effect of systematic uncertainties just by randomly
sampling a limited number of points in the space of all nuisance
parameters: a specific value $(\betas, \Delta\Gamma)$ is excluded only
if it can be excluded for any assumed value of the nuisance parameters
within $5\sigma$ of their estimate on data. Fig. \ref{fig:contours} shows
the confidence regions obtained by the two experiments with 2.8 \invfb\ of Tevatron data.
\begin{figure}
\centering
\includegraphics[width=0.49\textwidth, angle=0]{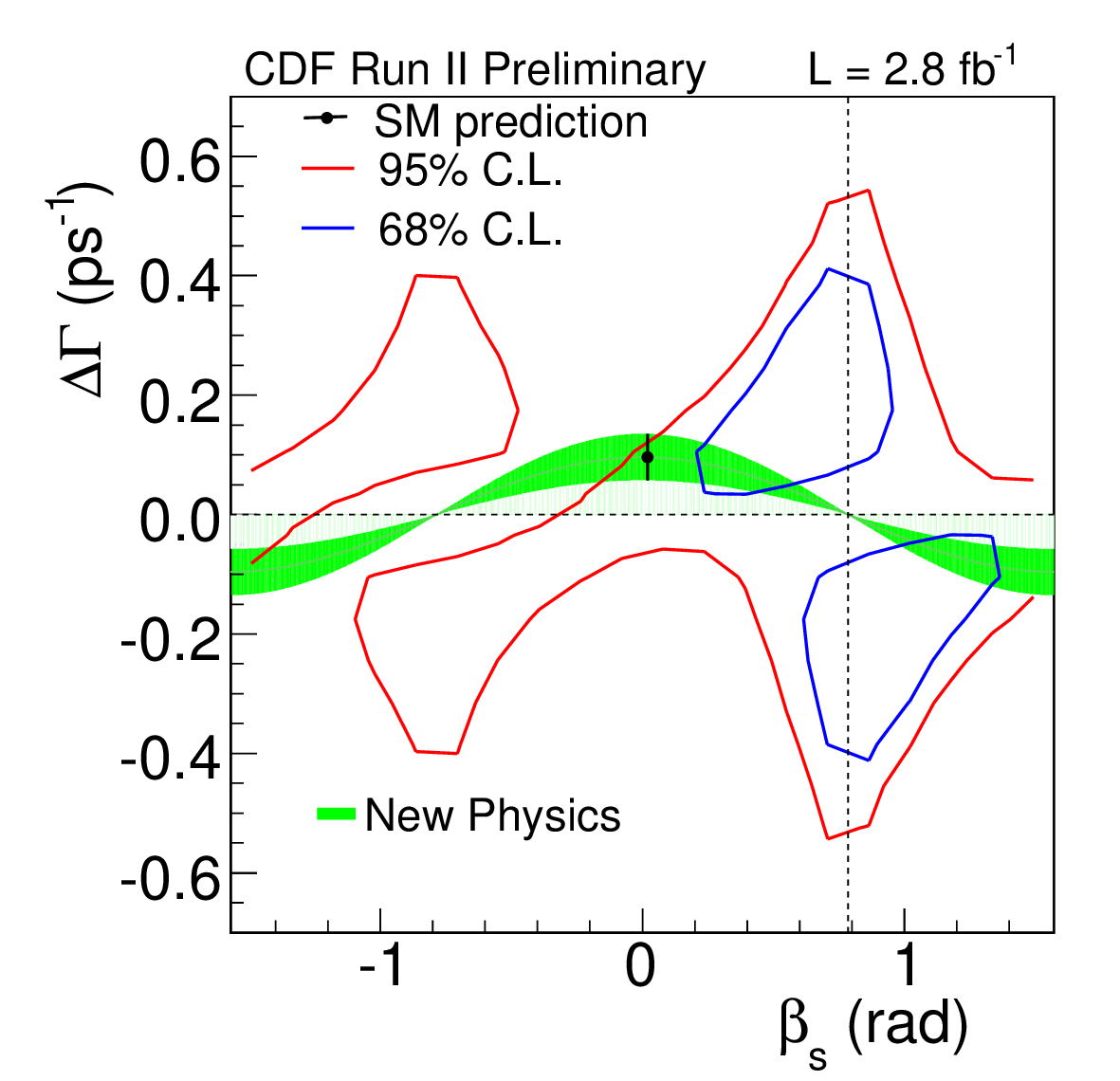}
\includegraphics[width=0.49\textwidth, angle=0]{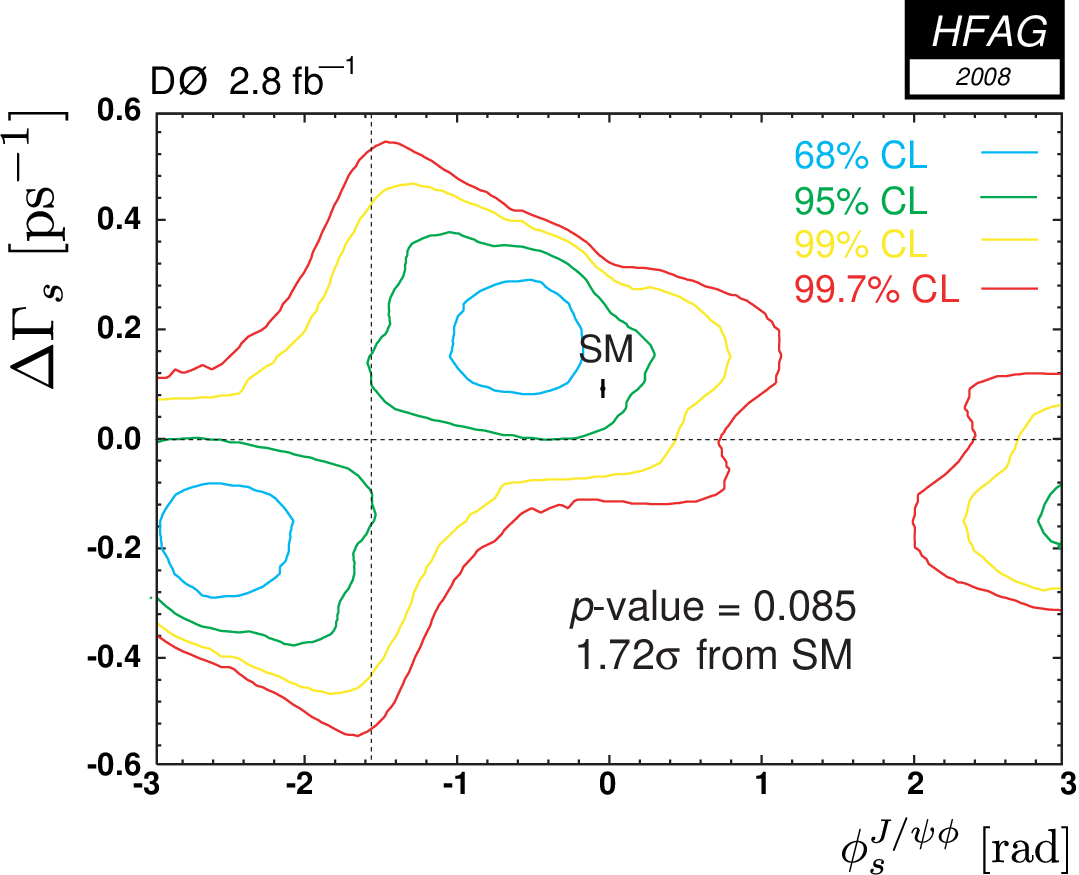}

\caption{\label{fig:contours}Confidence region in the $(\betas,\Delta\Gamma)$ plane obtained with 2.8 \invfb\ of CDF (left panel) and \Dzero (right panel) data. The green band is the region allowed by any NP contribution not entering $|\Gamma_{12}|$, and assuming $2|\Gamma_{12}|= 0.096 \pm 0.036$ ps$^{-1}$ \cite{Lenz:2006hd}.}
\end{figure}

A separate handle on CP violation is available through semileptonic $\Bs$ decays and has been performed
by the D0 collaboration on 2.8 \ifb\ of Tevatron data. The flavor of the $\Bs$ meson in 
the final state is determined by the muon charge in the decay 
$\Bs \rightarrow D^-_s\mu^+\nu X$ with $D^-_s\rightarrow\phi\pi^-$ and $\phi\rightarrow K^+K^-$.
A combined tagging method is then used to determine the initial state flavor. A time-dependent
fit to $\Bs$ candidate distributions yields the CP violation parameter
\begin{equation}
  A_{sl}^s = -0.0024\pm0.0117~(\mbox{stat}) ^{+0.0015}_{-0.0024}~(\mbox{syst}) .
\end{equation}

This is the first direct measurement~\cite{Aubert:2008bs} of the time integrated flavor untagged charge asymmetry in
semileptonic $B^0_s$ decays. $A_{SL}^{s,\mbox{\scriptsize unt.}}$ has also been obtained from a data sample
corresponding to an integrated luminosity of 1.3~fb$^{-1}$ in comparing the decay rate
$\Bs \rightarrow \mu^+ D^-_s \nu X$, $D^-_s\rightarrow \phi\pi^-$, $\phi\rightarrow K^+K^-$ with its charge conjugated
decay rate. The asymmetry amounts to
\begin{equation}
  A^{s,\mbox{\scriptsize unt.}}_{SL} = [1.23\pm0.97~(\mbox{stat}) \pm0.17~(\mbox{syst})] \times 10^{-2} , 
\end{equation}
assuming that $\Delta m_s / \overline{\Gamma_s}\gg 1$. The result can be further related to the
CP-violating phase in $B^0_s$ mixing via
\begin{equation}
  \frac{\Delta\Gamma_s}{\Delta m_s}\tan\phi_s = [2.45\pm1.93~(\mbox{stat}) \pm0.35~(\mbox{syst})] \times 1
0^{-2} .
\end{equation}




\begin{figure}[h]
\begin{center}
\includegraphics[width=0.49 \textwidth, angle = 0]{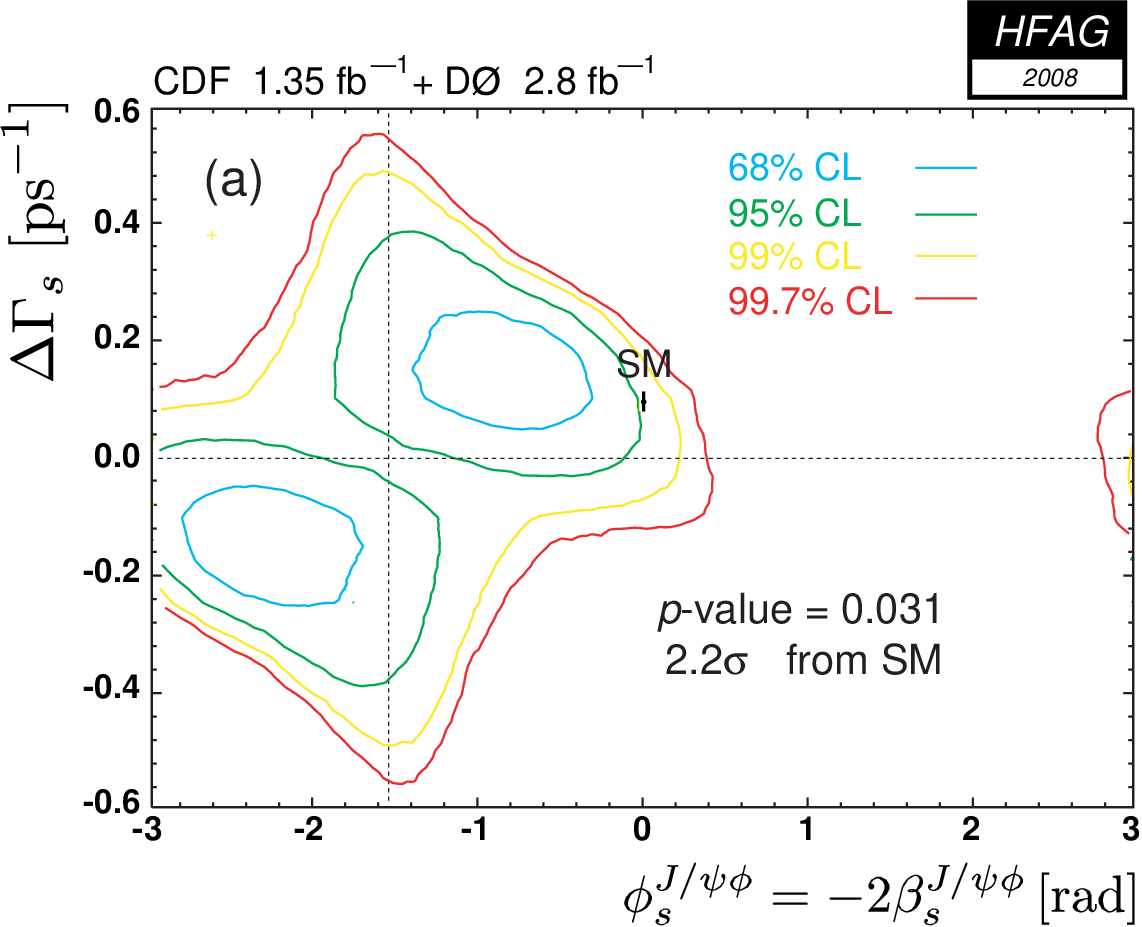}
\includegraphics[width=0.49 \textwidth, angle = 0]{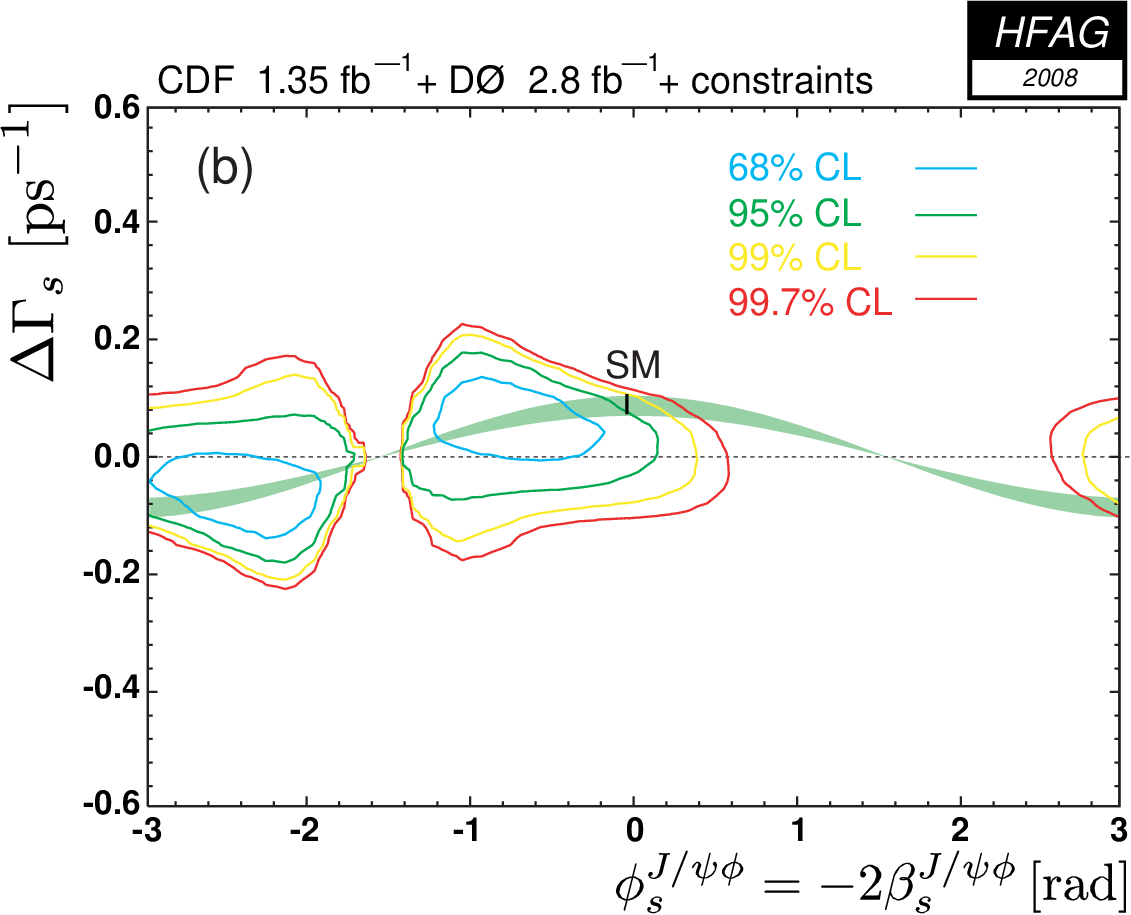}
\end{center}
\caption{\label{fig:HFAG_DO_CDF_phi_s}
Contour plots of \Dzero and CDF combined results in the ($\Delta\Gamma_s, 
\phi_s$) plane for different confidence levels. In the left panel, the combination of two results without constraints. In the right panel, constraints of the measured charge asymmetry $A^s_{SL}$ and $B^0_s$ lifetime are taken into account.
}
\end{figure}

The world best knowledge on the $\Bs$ mixing phase at this time comes from
combining the two Tevatron results and applying all available
constraints to the calculation. Fig. \ref{fig:HFAG_DO_CDF_phi_s}
shows the combination of the results from the two Tevatron
experiments. The current combined result is based
on the 1.35 \ifb\ dataset from CDF and the 2.8 \ifb\ dataset from
D0. The unconstrained result is consistent with the Standard Model
prediction within $2.2\sigma$. Adding constraints from $A_{SL}^s$ and
the \Bs\ flavor specific lifetime measurements, the discrepancy betwen
the Standard Model prediction and the combined result increases to
$2.3 \sigma$.



\subsection{The D-meson system}
\subsubsection{Theoretical prediction for $\Delta M_D$ and CP violation within the SM and beyond }
\label{sec:DmixingTh}

The SM calculation of $\Delta M_D$ is plagued by long-distance
contributions, responsible for very large theoretical uncertainties. 
The short-distance contribution in $\Delta M_D$~\cite{Cheng:1982hq,Datta:1984jx}, indeed, is highly suppressed
�both by a factor $(m_s^2-m_d^2)/M_W^2$ generated by the GIM mechanism and by a
further factor $(m_s^2-m_d^2)/m_c^2$ due to the fact that the external
momentum, of the order of $m_c$, is communicated to the internal light quarks
in box-diagrams. 
These factors explain why the box-diagrams are so small for $D$ mesons
relative to $K$ and $B_{d,s}$ mesons where the GIM mechanism enters as
$m_c^2/M_W^2$ and $m_t^2/M_W^2$ and external momenta can be neglected.

Theoretical estimates of charm mixing in the SM have been performed using 
either quarks or hadrons as basic degrees of freedom.  
The former method, like that used in $B_{d,s}$ mixing, consists in analyzing 
the mixing by using a sum of local operators 
ordered by dimension according to OPE~\cite{Georgi:1992as}.  
Roughly speaking, the result at the leading order in the OPE (where operators of dimension $D=6$ contribute) and in QCD  
from the $\s\sbar$ intermediate state yields  
the result~\cite{Golowich:2005pt}  
$y_{\rm D} \sim F(z) (\Vus/\Vcs)^2 \sim 0.01$ where 
$F(z) = 1/2 + {\cal O}(z)$ with 
$z \equiv (m_s/m_c)^2 \simeq 0.006$.  
This seems to reproduce the correct magnitude.  
Such is, however, not the case, as severe flavor cancellations 
with the $\d\bar d$, $\s\bar d$, $\d\bar s$ 
intermediate states occur (the leading terms in 
the $z$-expansion for $x_{\rm D}$ and $y_{\rm D}$ respectively become 
$z^2$ and $z^3$ at order $\alpha_s^0$ and just $z^2$ 
at order $\alpha_s^1$).  
The result through ${\cal O}(\alpha_s)$ 
is tiny, $x_{\rm D} \simeq y_{\rm D} \sim 10^{-6}$~\cite{Golowich:2005pt}.  
Evidently the OPE for charm is slowly convergent, although  
higher orders of the OPE do contain terms in which the $z$-suppression 
is less severe~\cite{Ohl:1992sr,Bigi:2000wn}.
The problem is that the number of local operators increases sharply 
with the operator dimension $D$ (e.g. $D=6$ has two operators, 
$D=9$ has fifteen, and so on).  To make matters worse, 
the matrix elements of the various local operators are unknown 
and can be only roughly approximated in model calculations.  
QCD lattice determinations would be of great use, but are currently 
unavailable.  

The other method, which considers hadronic degrees of freedom, 
is based on the following relation between 
the width difference and the absorptive matrix element given in Eq.~(\ref{eq:DMD}), with
\begin{equation}\label{eq:Gamma12bis}
\Gamma^D_{12}=\frac{1}{2 M_D}\, Disc \langle D^0 | i \int d^4x {\cal T}\, ({\cal H}_{eff}^{\Delta C=1}(x) \,{\cal H}_{eff}^{\Delta C=1}(0)) | \bar D^0 \rangle\,,
\end{equation} 
To get $y_{\rm D}$, defined in Eq.~(\ref{eq:xDyD}), one inserts intermediate states between the 
$\Delta C=1$ effective Hamiltonians (see Sec.~\ref{sec:ope}).  
This method yielded an early estimate for 
$y_{{\rm B}_s}$ (where the dominant contributions are few in 
number~\cite{Aleksan:1993qp}), but for charm mixing many matrix 
elements contribute.  The result of using a 
theoretical model~\cite{Buccella:1994nf}  
gives $y_{\rm D} \sim 10^{-3}$, which is too small.  
This shows how delicate the sum over many 
contributions seems to be.  

Another approach is to rely more on charm decay data and less on the 
underlying theory~\cite{Donoghue:1985hh,Wolfenstein:1985ft}.  Given that SU(3)
breaking occurs at second order in charm mixing~\cite{Falk:2001hx}, 
perhaps all two-particle and three-particle sectors contribute very 
little.  However, this cannot happen for the four-particle
intermediate states because the decay of $\Dz$ into four-Kaon states is 
kinematically forbidden.  In fact, Ref.~\cite{Falk:2004wg} claims 
that these multiparticle sectors can generate $y_D\sim 10^{-2}$.  
The complicate picture is worsened by the fact that a dispersion relation calculation~\cite{Falk:2004wg}
using charm decay widths as input predicts a negative value for $x_{\rm D}$, i.e. of opposite sign with respect to the experimental measurement.  
The  determination of $x_{\rm D}$ in the SM is certainly subtle enough to 
deserve further study and, at the same time, to strengthen the 
motivation for studying NP models of $\Dz$ mixing. 

NP contributions to charm mixing can affect $y_{\rm D}$ as well as 
$x_{\rm D}$.  We do not consider the former here, instead 
refering the reader to refs.~\cite{Golowich:2006gq,Petrov:2007gp}.
The study of $x_{\rm D}$ in Ref.~\cite{Golowich:2007ka} considers 
$21$ New Physics models, arranged in terms of extra gauge bosons 
(LR models, etc), extra scalars (multi-Higgs models, etc), 
extra fermions (little Higgs models, etc), 
extra dimensions (split fermion models, etc), and 
extra global symmetries (SUSY, etc).
The strategy for calculating the effect of NP on $\Dz$ mixing 
is, for the most part, straightforward.  One considers a 
particular NP model and calculates the mixing amplitude 
as a function of the model parameters.  If the mixing signal 
is sufficiently large, constraints on the parameters are obtained.  
Of these 21 NP models, only four (split SUSY, universal extra
dimensions, left-right symmetric and flavor-changing two-higgs 
doublet) are ineffective in producing charm mixing at the observed 
level.  This has several causes, e.g. the NP mass scale is too 
large, severe cancellations occur in the mixing signal, etc.  
This means that 17 of the NP models can produce charm mixing.  
We refer to Ref.~\cite{Golowich:2007ka} for details.

Finally, we observe that for a deeper understanding of $D^0-\bar D^0$ mixing, there remain additional avenues to explore, among them correlating 
NP contributions between charm mixing and rare charm decays 
and providing a comprehensive account of CP violations 
(both SM and NP) in $D^0-\bar D^0$ mixing.

\subsubsection{Experimental results } 

 Recent studies have shown evidence for mixing in the
\Dz-\Dzb{} system at the 1\%  
level~\cite{Aubert:2007wf, Staric:2007dt, Aubert:2007en,
Aubert:2008zh}. The measured values can be accommodated by the
Standard Model (SM)~\cite{Wolfenstein:1985ft, Donoghue:1985hh,
Bigi:2000wn, Falk:2001hx, Falk:2004wg} where the largest predictions
for $x_D$ and $y_D$ are of $\order(10^{-2})$.  These measurements
provide strong constraints on new physics models~\cite{Burdman:2003rs,
Petrov:2006nc, Golowich:2007ka}. An observation of
\CP violation in \Dz-\Dzb{} mixing with the present experimental
sensitivity would provide evidence for physics 
beyond the SM~\cite{Blaylock:1995ay}, and a
search for this effect in the charm system is considered 
elsewhere~\cite{charmCPV:CC}. Presented here is an overview of 
recent mixing measurements.



The first evidence analysis studies right-sign (RS), Cabibbo-favored (CF)
decay $\Dztokpi$ and the wrong-sign (WS) decay $\DztokpiWS$. The
latter can be produced via the doubly Cabibbo-suppressed (DCS) decay
$\DztokpiWS$ or via mixing followed by a CF decay $\Dzbtokpi$.  The
DCS decay has a small rate \Rdcs of order $\tan^4\theta_C \approx
0.3\%$ relative to CF decay.
\Dz and \Dzb are distinguished by their production in the 
decay $\Dstp\to\pisoft^+\Dz$ where the $\pisoft^+$ is referred to as
the ``slow pion''. In RS decays the $\pisoft^+$ and Kaon have opposite
charges, while in WS decays the charges are the same.

The time dependence of the WS decay rate is used to separate the
contributions of DCS decays from \Dz-\Dzb mixing.  For the WS decay of
a meson produced as a \Dz at time~$t = 0$ in the limit of small mixing
($|x_D|$, $|y_D| \ll 1$) and \CP conservation this is approximated as
\begin{equation}
  \frac{\Tws}{e^{-\Gamma t}} \propto
      \Rdcs + 
      \sqrt{\Rdcs}\yPrime_f\; \Gamma t + 
      \frac{\xPrimeSq_f + \yPrimeSq_f}{4} (\Gamma t)^2\,,
\label{eq:Tws}
\end{equation}
where $\xPrime_f = x_D\cos\delta_{f} + y_D \sin\delta_{f}$,
$\yPrime_f = -x_D\sin\delta_{f} + y_D \cos\delta_{f}$, where $f$ is the
final state accessible to both \Dz and \Dzb decays, and 
$\delta_{f}$ is the relative strong phase between the DCS and CF amplitudes.
This makes it possible to measure the quantities $x_D$ and $y_D$, if the
strong phase difference $\delta_{f}$ is known. To search for \CP violation,
Eq.~(\ref{eq:Tws}) is applied  to \Dz and \Dzb samples separately.

Evidence for $\Dz$-$\Dzb$ mixing in \DztokpiWS decays has been
reported by the \babar\ collaboration~\cite{Aubert:2007wf}. The mixing
parameters were found to be $\xPrimeSq_{K\pi} = [ -0.22 \pm 0.30
\hbox{ (stat.)}\pm 0.21 \hbox{ (syst.)}]\times 10^{-3}$ and
$\yPrime_{K\pi} = [9.7 \pm 4.4 \hbox{ (stat.)}\pm 3.1 \hbox{ (syst.)}] 
\times 10^{-3}$, and a correlation between them of $-0.94$. This
result is inconsistent with the no-mixing hypothesis with a
significance of 3.9~$\sigma$, with no evidence for \CP violation.

The quantum coherence between pair-produced \Dz and \Dzb in
$\psi(3770)$ decays can be used to study charm mixing and to make a
determination of the relative strong phase
$\delta_{K\pi}$~\cite{Asner:2008ft}. Using data collected with the
CLEO-c detector at $E_{cm} =~3.77~\gev$, as well as branching fraction
input from other experiments a value of $\cos\delta_{K\pi} =
1.03^{+0.31}_{-0.17} \pm 0.06$ was found, where the uncertainties are
statistical and systematic, respectively. In addition, by further
including external measurements of charm mixing parameters, another
measurement of $\cos\delta_{K\pi} = 1.10 \pm 0.35
\pm 0.07$, as well as $x_D\sin\delta_{K\pi} = (4.4^{+2.7}_{-1.8} \pm
2.9) \times 10^{-3}$ and $(\delta_{K\pi} = 22^{+11\ +9}_{-12\
-11})$~\degrees, was made.

The initial evidence by the \babar experiment was first confirmed by
the CDF collaboration~\cite{:2007uc, CDF:CC}.  The CDF analysis was
performed on a signal sample of $12.7 \times 10^3$ $D^0\rightarrow
K^+\pi^-$ decays gathered with the displaced track trigger. This
corresponds to an integrated luminosity of 1.5 $fb^{-1,}$. The
analysis considers $D^0$ decays with proper decay times between 0.75
and 10 mean $D^0$ lifetimes.  The mixing parameters are measured to be
$R_D = 3.04 \pm 0.55(\times 10^{-3})$, $\yPrime = 8.54 \pm 7.55
(\times 10^{-3})$, $\xPrimeSq = -0.12 \pm 0.35 (\times 10^{-3})$.  The
data are inconsistent with the no mixing hypothesis ($\yPrime =
\xPrimeSq = 0$) with a probability equivalent to 3.8 Gaussian standard
deviations.


Further evidence for mixing was reported by the \babar\ collaboration
using a time-dependent Dalitz plot analysis of the WS $\Dz\to
\Kp\pim\piz$ decays~\cite{Aubert:2008zh}. The advantage of an
amplitude analysis across the Dalitz plot is that the interference
term in Eq.~\ref{eq:Tws}, produces a variation in average decay time
as a function of position in the Dalitz plot that is sensitive to the
complex amplitudes of the resonant isobars as well as the mixing
parameters. In this study, the change in the average decay time and
the interference between the $\Dz \to \Kstarp \pim$ and $\Dz \to
\rho^-\Kp$ amplitudes are the origin of the sensitivity to mixing.
Assuming \CP conservation, the mixing parameters $\xPrime_{K\pi\piz}
=$ [2.61 $\mbox{}^{\rm +0.57}_{\rm -0.68}$\,(stat.) $\pm$
0.39\,(syst.)]\%, and $\yPrime_{K\pi\piz}=$ [-0.06 $\mbox{}^{\rm
+0.55}_{\rm -0.64}$\,(stat.) $\pm$ 0.34\,(syst.)]\% were
extracted. This result is inconsistent with the no-mixing hypothesis
with a significance of $3.2$~$\sigma$. No evidence of \CP violation in
mixing was observed.


The CLEO collaboration pioneered an analysis of $\Dz\to\KS\pippim$
decays using a time-dependent Dalitz plot
analysis~\cite{Asner:2005sz}, allowing for a direct determination of
$x_D$ and $y_D$. Due to the presence of \CP-eigenstates in the final
state, the amplitudes of \Dz and \Dzb are entangled, so that the
analysis is free of unknown phases.
The Belle collaboration has repeated this analysis~\cite{Abe:2007rd},
first assuming \CP conservation and subsequently allowing for \CP
violation.  Assuming negligible \CP violation, the mixing parameters
$x_D=(0.80\pm0.29^{+0.09 +0.10}_{-0.07 -0.14})\% $ and $y_D=(0.33\pm
0.24^{+0.08 +0.06}_{-0.12 -0.08})\% $ were measured, where the errors
are statistical, experimental systematic, and systematic due to the
amplitude model uncertainties, respectively. This corresponds to a
deviation of 2.4~$\sigma$ significance from the no-mixing
hypothesis. Allowing for CP violation, the $CPV$ parameters
$|q/p|=0.86^{+0.30 +0.06}_{-0.29 -0.03}\pm0.08$ and
$\arg(q/p)=(-14^{+16 +5 +2}_{-18 -3 -4})^\circ$ have been obtained.


One consequence of \Dz-\Dzb{} mixing is that the \Dz\ decay time
distribution can be different for decays to different \CP
eigenstates~\cite{Liu:1994ea}. Using the ratios of lifetimes extracted
from a sample of \Dz mesons produced through the process
$\Dstp\to\Dz\pip$, that decay to \Kmpip, \KmKp, or \pipi, the
lifetimes of the CP-even, Cabibbo-suppressed modes
\KmKp and \pipi are compared to that of the CP-mixed, Cabibbo-favored 
mode \Kmpip to obtain a measurement of \yCP, which in the limit of \CP
conservation corresponds to the mixing parameter $y_D$.  Both
Belle~\cite{Staric:2007dt} and \babar~\cite{Aubert:2007en} have
produced measurements of $\Dz$-$\Dzb$ mixing parameters, at 3.2 and
3.0~$\sigma$ from the no mixing expectation, respectively.  All
current results are shown in Fig.~\ref{fig:YcpAvg}.  No evidence for
a $\CP$ asymmetry between $\Dz$ and $\Dzb$ decays has been found.

\begin{figure}[!ht]
\begin{center}
 	\includegraphics[width=0.75\linewidth]{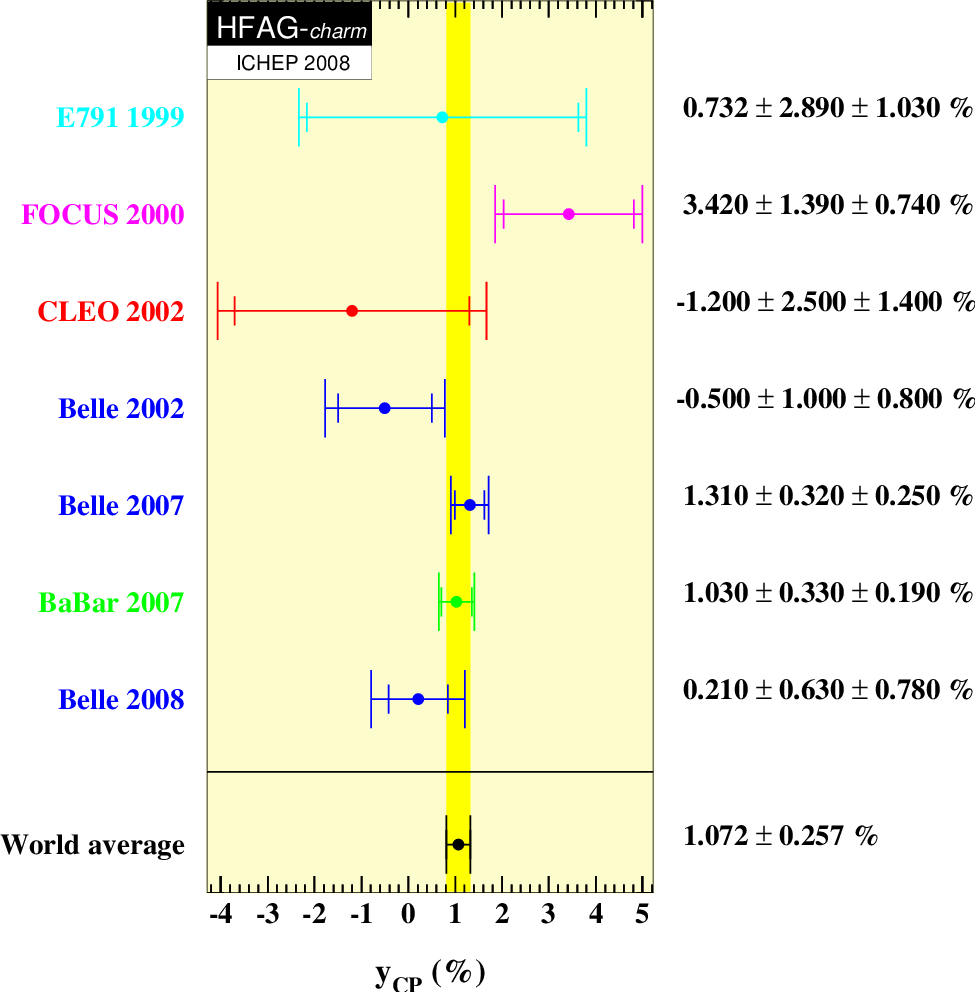}
\end{center}
\caption{\label{fig:YcpAvg}Current measurements of ${y_{CP}}$. The mean $\yCP\approx 1~\%$
  differs significantly from zero~\cite{Barberio:20 08fa}.}
\end{figure}


The mixing parameter $y_{CP}$ has also been measured by the Belle
Collaboration, using a flavor-untagged sample of $D^0\to K_S^0K^+K^-$
decays~\cite{:2008tp}.  By measuring the difference in lifetimes
between \Dz\ mesons decaying to \KS \KpKm in two different m(\KpKm)
regions with different contributions of \CP even and odd eigenstates
they determine $\yCP = (0.21 \pm 0.63\pm 0.78 \pm 0.01(\rm model))\%$.
This result, is also included in Fig.~\ref{fig:YcpAvg}.

\begin{figure}[ht]
\hbox to \hsize{
 \includegraphics[width=0.5\linewidth]{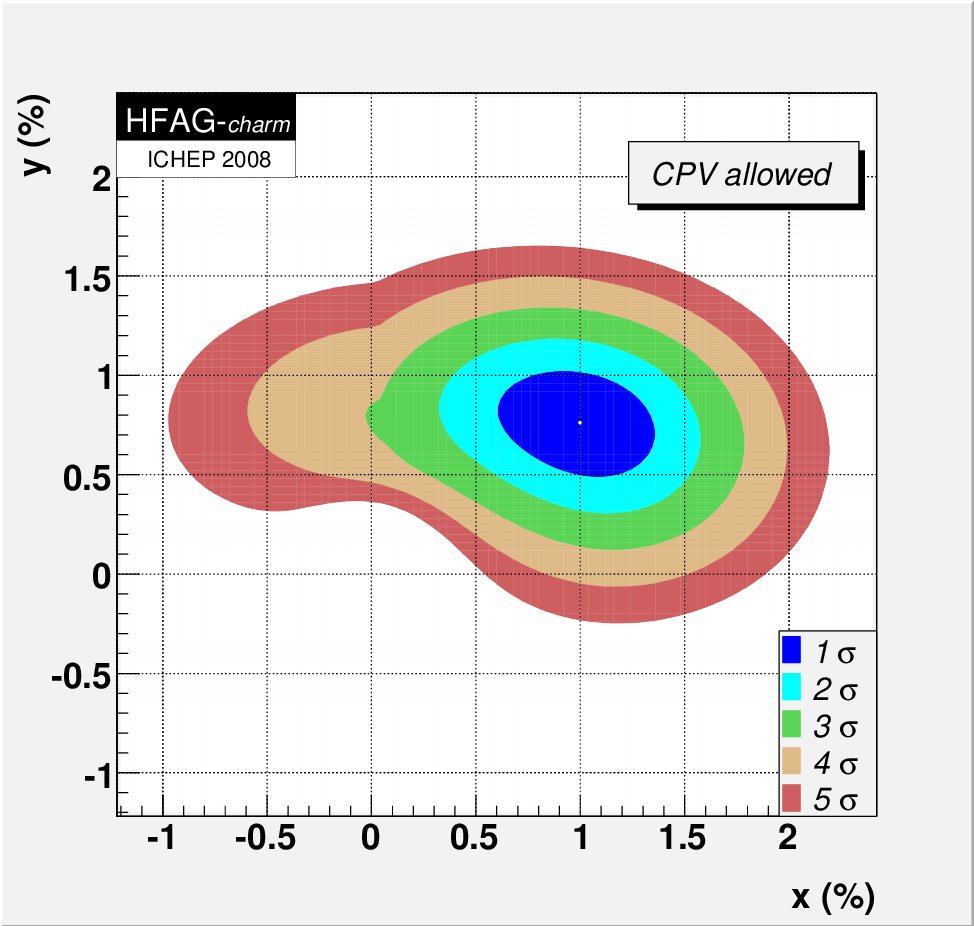}
 \includegraphics[width=0.5\linewidth]{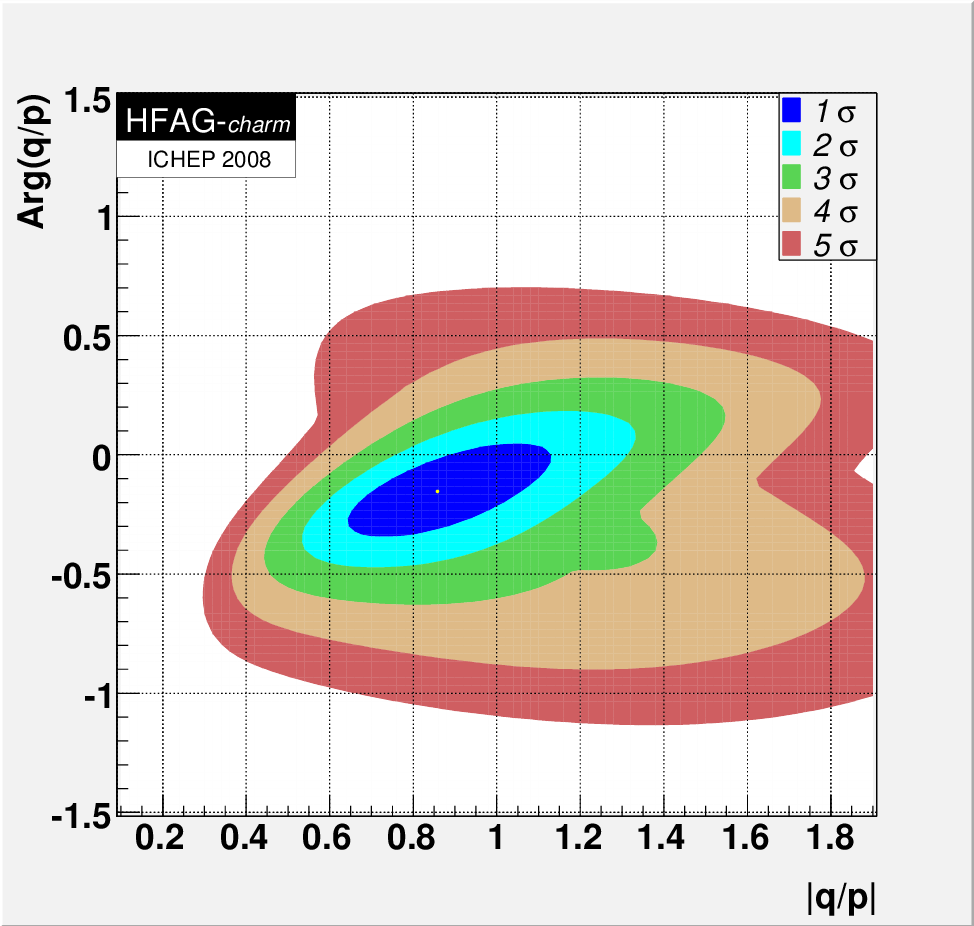}\hfill
}
\caption{\label{fig:hfagAvg}World averages from the Heavy Flavor
  Averaging Group (HFAG): \chisq contours for $x_D$~vs.~$y_D$, and
  $|q/p|$~vs.~Arg$(q/p)$.}
\end{figure}


A global average has been constructed from 28 mixing variables
(including those mentioned above), by the Heavy Flavor Averaging Group
(HFAG)~\cite{Barberio:2008fa}, as shown in Fig.~\ref{fig:hfagAvg}. The
no-mixing point $x_D = y = 0$ is excluded at 9.8~$\sigma$, and the
values $x_D\approx y \approx 1~\%$ are favored, but to date no single
measurement exceeds 5~$\sigma$.

\subsection{Future Outlook}

With planned data taking coming to an end for some of the main
experiments contributing to lifetime and mixing results, attention is
now turning to the flavor program of the Large Hadron Collider. With a
dedicated heavy flavor experiment (LHCb), two powerful multi-purpose
detectors (ATLAS and CMS), and plans for tremendous integrated
luminosity samples, expectations of precision results are very high.
In this chapter we review expected performance for some of the most 
interesting results that are expected to come from the LHC.

\subsubsection{$B$ meson mixing and lifetimes }

The first $B$ meson lifetime measurements at LHC experiments will be
used as calibration measurements to understand detector effects on
time-dependant analyses. Very large samples of fully reconstructed
$B^+$ and $B^0$ candidates will be available very early after the LHC
starts, and will allow comparison with existing precise lifetime
measurements. For example, at ATLAS, 1024 reconstructed $B^0\to J/\psi
K^{*0}$ are expected after 10 $\rm{pb}^{-1}$ of data, which will allow
a lifetime measurement with 10\% precision after approximatively one
month of data taking. Similarly, the LHCb experiment will reconstruct
$1.735$ million $B^+\to J/\psi K^+$ candidates for 2 $\rm{fb}^{-1}$ of
data, with a small background over signal ratio, allowing not to use
any lifetime selection criteria and thus to determine lifetime
resolution functions. Hadronic decay modes will also be reconstructed
with large samples. The LHCb experiment will reconstruct $1.34$
million $B^0 \to D^- (K^-\pi^+\pi^-) \pi^+$ decays in 2 $\rm{fb}^{-1}$
of data. The expected proper time resolution of 33.9 fs will allow
LHCb to reach the current $B^0$ lifetime precision (0.009 ps) with
60000 events, considering only statistical errors.

Measurements of $\Lambda_b$ lifetimes are expected to improve
significantly with LHC results. LHCb expects to reconstruct $2.3\times
10^4$ events for 2 $\rm{fb}^{-1}$ of data in the decay mode $\Lambda_b 
\to J\psi(\mu^+ \mu^-) \Lambda(p\pi)$. The anticipated proper
time resolution for these decays is 41.5 fs, yielding to a lifetime measurement with
a statistical precision of 0.027 ps. The ATLAS experiment will
reconstruct 4500 events in the same decay mode with 10 $\rm{fb}^{-1}$
of data.

LHC experiments plan on precisely measuring the $B_c^+$ lifetime.  The
$B_c^+$ production cross-section is roughly 20 times larger at the LHC
than at the Tevatron. About $10^9$ $B_c^\pm$ will be produced per
year in LHCb. Measurement of the $B_c^+$ lifetime will be an
interesting window on the proportions of its three decay mechanisms:
$b$ decay, $c$ decay and anihilation. The most promising decay chanel
that will be used for the analysis is $B_c^+\to J/\psi
\pi^+$. Assuming a $B_c^+$ production cross-section of 0.4 $\mu$b and
a branching fraction for $B_c^+\to J/\psi \pi^+$ equal to $1.3\times
10^{-3}$, 700 events are expected for 2 $\rm{fb}^{-1}$ of data at
LHCb, and 80 events for 10 $\rm{fb}^{-1}$ at CMS, leading to a
statistical precision on the lifetime measurement of 0.026 ps at LHCb
and 0.055 ps at CMS.


The reconstruction of the flavor specific decay mode $B_s^0 \to D_s^+
\pi^-$ with $D_s^+ \to K^+ K^- \pi^+$ will allow the measurement of
the $B_s^0$ mixing frequency $\Delta m_s$ together with the $B_s^0$
width difference, $\Delta \Gamma_s$. 155000 reconstructed candidates
are expected at LHCb in 2 $\rm{fb}^{-1}$ of data, with a small
background over signal ratio $\frac{B}{S} \in [0.06;0.4]$ at 90\%
confidence level. The mass resolution is expected to be 17 ${\rm
MeV}/c^2$ and a proper time resolution of 33 fs is anticipated. This
implies measurements of the $B_s^0$ lifetime to a precision of 0.013 ps. 
The expected uncertainty on $\Delta m_s$ is 0.008
$\rm{ps}^{-1}$.  $\Delta \Gamma_s$ will be measured to 0.03
$\rm{ps}^{-1}$ precision, assuming a central value of $\Delta
\Gamma_s$ equal to 0.068 $\rm{ps}^{-1}$. A more precise $\Delta
\Gamma_s$ determination is expected to be obtained from the
time-dependant angular analysis of the decay mode $B_s^0 \to J/\psi
\phi$. Preliminary studies show that a precision of 0.021
$\rm{ps}^{-1}$ can be reached at ATLAS with 10 $\rm{fb}^{-1}$ of
data, 0.010 $\rm{ps}^{-1}$ at CMS with 10 $\rm{fb}^{-1}$ of data
assuming perfect tagging, and 0.008 $\rm{ps}^{-1}$ at LHCb with 2
$\rm{ps}^{-1}$ of data.

In summary, very precise lifetime measurements of the $B^0$ and $B^+$
mesons will be available very soon after LHC starts and will be used
to calibrate LHC detectors for further lifetime
measurements. Parameters of the $B_s^0$ hadron ($\tau_s$, $\Delta
m_s$) will reach similar precisions to those currently available for
$B^0$ and $B^+$. These measurements are expected to rapidly become
limited by systematics uncertainties. Precision studies of other $B$
hadrons, such as the lifetime of the $B_c^+$ and $\Lambda_b$, will be 
conducted at the CMS, ATLAS, and LHCb experiments.

\begin{figure}
\begin{center}
\epsfig{file=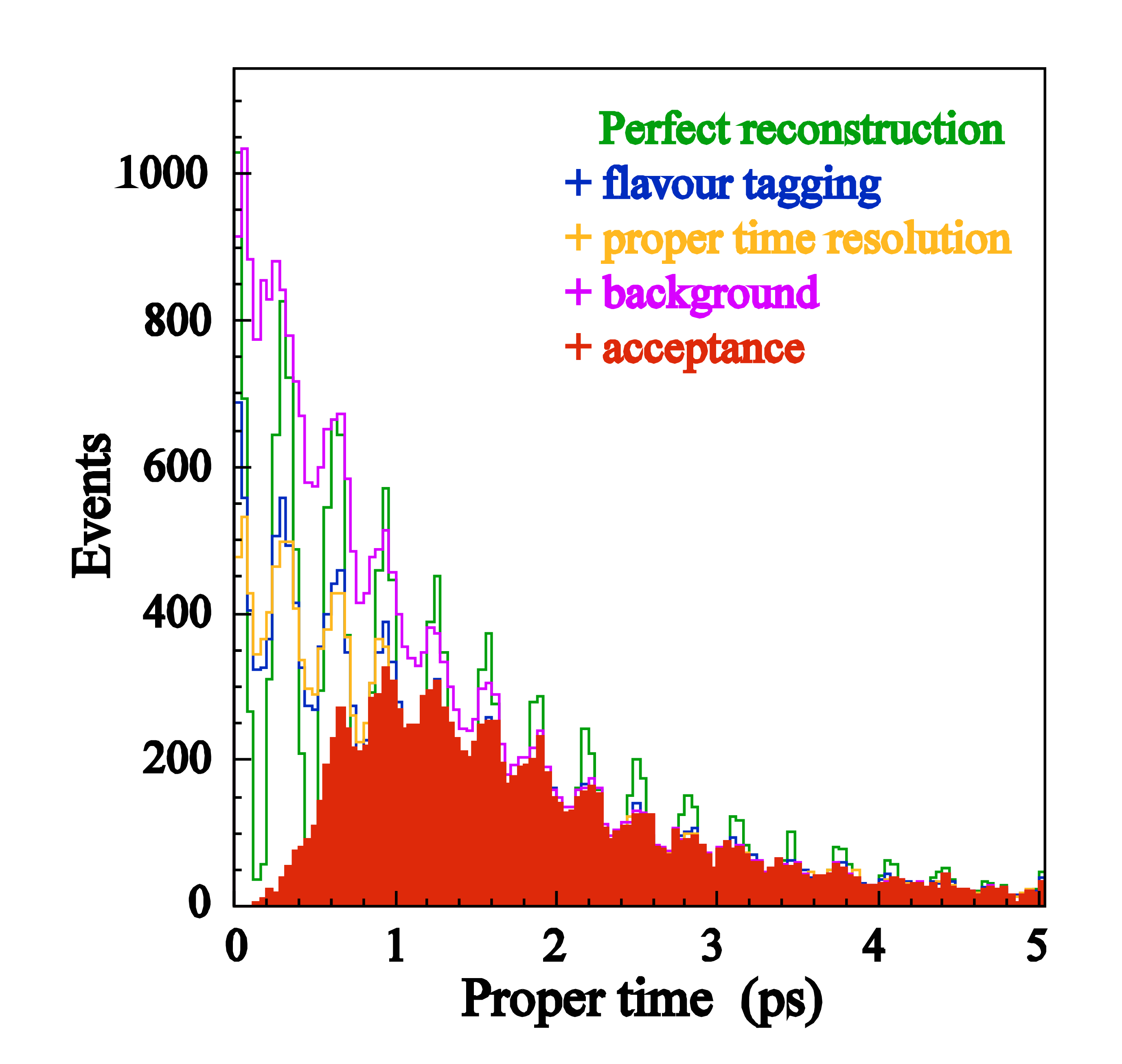, width = 0.5 \textwidth}
\end{center}
\caption{LHCb toy Monte Carlo simulation of the proper time distribution of right-sign tagged
$B_s^0 \to D_s^+ \pi^- $ decays. The toy simulation is based on resolutions, efficiencies, and tagging power
estimated from full detector simulation.
\label{fig:robbe_ct}}
\end{figure}


\subsubsection{Measurements of the $\Bs$ meson mixing phase }



As discussed in Sec. \ref{sec:lifemix:blife:phi-s} and
\cite{Dighe:1998vk}, the most precise measurement of $\beta_s$ can be
obtained via a tagged time-dependent angular analysis of the $\BsJphi$
decay mode. In order to disentangle the two CP eigenstates, the three
amplitudes are statistically separated through an angular analysis.
The oscillation amplitude of the time-dependent angular distributions
is proportional to the CP-violation phase $\beta_s$.  In the following
text we compare the key performance parameters for this measurement
between the three experiments.

\newcommand{\BuJK}{\Bu to \jpsi \Kp} 
Offline selections for the
three experiments are based on basic quantities like particle
identification, $p_T$ of the decay products, vertex quality and, only
for ATLAS and CMS, $b$-vertex displacement.  The MuonID capability is
similar for the three experiments (muon efficiency of $\sim$ 90 $\%$
for a misidentification rate of $\sim 1 \%$, but dependent with $p_T,
\eta$ for central detectors). Hadron identification capability is
higher for LHCb due to the powerful RICH system \cite{rich_lhcb}
which allows to Kaon identification with an efficiency of $\sim 88\%$.
The pion misidentification rate of $\sim 3 \%$.

The expected momentum resolution is $\sigma_p/p =(0.3-0.5)\%$ for LHCb and
$\sigma_{p_{T}}/p_{T} = 1-2 \%$ for ATLAS/CMS.  This provides a 
$B_s$ mass resolution of $\sim$ 17 MeV/c$^2$ for LHCb without use of a
$J/\Psi$ mass constraint in the fit. CMS and ATLAS predict $B_s$ mass resolutions of $\sim 14-16$ MeV /c$^2$, 
using a $J/\Psi$ mass constraint in the fit. LHCb does
not make use of the $J/\Psi$ mass constraint because this requirement
modifies the proper time acceptance of the decaying $B_s$.

ATLAS/CMS use an offline selection with $B_s$ lifetime selection cuts.
This selection gets rid of most of the prompt combinatorial background
but also modifies heavily the proper time acceptance that must be
corrected afterwards.  LHCb will optimize the $B_s$ signal selection by
minimizing the bias on the proper time and angular acceptances.
 

For the time being, LHCb and ATLAS are developing tagged analyses, 
while CMS is currently reporting an untagged one.
ATLAS will use several taggers mainly based on leptons and vertex
charge. The combined tag gives an effective tagging power of
$\epsilon_{eff} = \epsilon_{tag} (1-2 \omega)^2 = 4.6 \%$.
LHCb expects excellent hadron identification and therefore can profit also from
both same side and opposite side Kaon taggers.  The combined tag
is expected to have an effective tagging power of $\epsilon_{eff} =
6.2 \%$.  Tagging calibration will be performed at LHCb using
flavor specific decays, namely \BdJKst\ and \BuJK for calibration of
OS taggers, and $B_s \to D_s \pi$ for calibration of the same side
tagger.
The last key ingredient is the proper time resolution,
$\sigma_\tau$. Expected average proper time resolutions are 83
fs, 77 fs and 40 fs, for ATLAS, CMS and LHCb, respectively.
At the time of this report, Monte Carlo samples with full simulation
which were available for studies have limited statistics: $\sim$ 7
pb$^{-1}$ of inclusive $J/\Psi \to \mu^+ \m u^-$ were available for
the LHCb studies, and 20 - 50 pb$^{-1}$ of $b \to J/\Psi(\mu \mu) X$
for ATLAS/CMS. The Monte Carlo with full detector simulation cannot be
used to perform a full analysis evaluation. However, these samples can
be used to estimate yield, background fractions, mass, proper time and
angle distributions, resolutions, and acceptances.  The extracted
quantities are then used in toy Monte Carlo ensembles in order to
estimate the sensitivity to $2 \beta_s$ (and other parameters) via
results of unbinned maximum likelihood fits.

Tab. \ref{tab:results} summarizes the expected precision for $2
\beta_s$ and $\Delta \Gamma_s$ after 1/4 of a nominal year of
running. The estimated event yield, background contamination,
effective tagging efficiency $\epsilon D^2$ and proper time
resolutions $\sigma(\tau)$ are also listed per experiment.  These
studies assumed values of $2 \beta_s \sim 0.04$ for $\beta_s$ and
$\Delta \Gamma_s$ and $\Delta \Gamma_s/\overline{\Gamma}_s \sim 0.1$.
ATLAS, CMS and LHCb have a strong potential to increase the precision
of the measurements of the $B_s$ CP violating phase well beyond the
present CDF and \D0 results. These precision measurements will open
opportunities to probe for effects beyond the Standard Model.

{\footnotesize
\begin{table}[htb]
\caption{\small Summary table for ATLAS, CMS and LHCb. 
We show the untagged signal yield for a luminosity corresponding to 
a 1/4 year of running at nominal luminosity, the B/S ratio, the effective tagging efficienc
y, 
the proper time resolution and the sensitivity 
on $2\beta_s$ and $\Delta \Gamma_s / \overline{\Gamma}_s$.}
\label{tab:results}

\begin{center}
\begin{tabular}{|c|c|c|c|}\hline
                        & ATLAS  & CMS   & LHCb \\ 
\hline
${\cal L}$[fb$^{-1}]$    & 2.5    & 2.5   & 0.5 \\
signal yield [untagged] & 22.5 k & 27 k   & 28.5 k \\
B/S                     & 0.18   & 0.25   & 2      \\
dominant background     & long-lived & long-lived & prompt \\
$\epsilon D^2$ & 4.6 $\%$ & N/A & 6.2 $\%$ \\
$\sigma(\tau)$ & 83 fs & 77 fs & 40 fs \\ \hline
$\sigma(2 \beta_s)$ & 0.16 & N/A & 0.06 \\
$\sigma(\Delta \Gamma_s /\overline{\Gamma}_s)/(\Delta \Gamma_s /\overline{\Gamma}_s)$ & 0.45 & 0.28 & 0.17 \\ \hline
\end{tabular}\end{center}
\end{table}
}


\subsubsection{$D^0$ mixing and CP violation }
\label{sec:spradlin:lhcb}

As the dedicated flavor experiment at CERN's Large Hadron Collider
(LHC), LHCb is the only LHC experiment currently planning measurements
of $D^0$-$\overline{D^0}$ mixing and charm CP violation.  The following studies
document the expected performance of the LHCb experiment.

Many of the features that make LHCb an excellent $B$ physics
laboratory also make LHCb well-suited for many charm physics studies
at unprecedented levels of precision~\cite{Alves:2008zz}.  The silicon
Vertex Locator (VELO) will provide the excellent vertex resolutions
necessary for time dependent measurements: an estimated
\mbox{$45\,\mathrm{fs}$} proper time resolution is expected for \mbox{$D^0
\rightarrow K^- \pi^+$} decays where the $D^0$ mesons are produced in
$b$-hadron decays.  The LHCb tracking system will supply precise
momentum measurements The projected mass resolution for two body
decays of $D^0$ mesons is estimated to be \mbox{$6\,\mathrm{MeV}/c^2$}
The LHCb Ring Imaging Cherenkov (RICH) detectors will provide
excellent $K$--$\pi$ discrimination over a wide momentum range from
\mbox{$2\,\mathrm{GeV/c}$} to \mbox{$100\,\mathrm{GeV/c}$}.  Finally,
the LHCb trigger system will have a high statistics charm stream, so
that the large charm production in LHC collisions can be exploited for
precision measurements.

\label{sec:spradlin:cpv}

LHCb will perform both time-dependent and time-integrated CP violation
searches.  Each time-dependent $D^0$-$\overline{D^0}$ mixing
measurement will be analyzed in charge conjugate subsets to measure
possible CP violating effects.  Measurements with promptly produced
charm mesons and with charm mesons produced in $b$-hadron decays will
be pursued. Analysis methods for both sources are under
development. Preliminary studies for measurements with secondary charm
are currently more complete.  Initial studies have focused on
$D^{\ast+}$--tagged two-body \mbox{$D^0 \rightarrow h^- h'^+$} decays.
Multi-body decays to charged products and up to one $K_S^0$ are
suitable for precision measurements at LHCb and will be investigated.
In four body hadronic decays, plans for CP violation searches include
complete amplitude analyses and analyses of quantities that are odd
under time reversal.

Simulated events from a full interaction and LHCb detector simulation
have been used to estimate LHCb's potential performance in charm mixing
analyses.  Preliminary event selection studies on these simulated
events indicate a yield of approximately 8~million $D^{\ast+}$--tagged
\mbox{$D^0 \rightarrow K^- K^+$} decays in $10 \ifb$ of
collisions. The $D^{\ast+}$ was produced in a $b$-hadron decay in
these studies.~\cite{Spradlin:2007zz}.  This yield estimate includes
the expected effects of both the L0 and the HLT triggers. This
corresponds to a statistical precision of approximately \mbox{$4
\times 10^{-4}$} for the CP asymmetry search.  The selection used in
the study was optimized for the wrong sign (WS) \mbox{$D^0 \rightarrow
K \pi$} decays. Reoptimizing for \mbox{$D^0 \rightarrow K K$} is expected to
result in even higher yields.
Similar studies predict approximately 1.2~billion $D^{\ast+}$--tagged
\mbox{$D^0 \rightarrow K K$} decays in $10\,\mathrm{fb}^{-1}$ after
the L0 trigger, before the HLT trigger.  Efficient strategies to
select these events in the HLT are under investigation.
  

\label{sec:spradlin:mix}


LHCb will measure {$D^0$-$\overline{D^0}$} mixing in as many channels
as it can efficiently reconstruct.  Initial studies have focused on
the two main mixing measurements possible with $D^{\ast+}$--tagged
two-body \mbox{$D^0 \rightarrow h^- h'^+$} decays---mixing from
analysis of WS $K\pi$ decays, and the ratio of lifetimes of singly
Cabibbo suppressed (SCS) and right sign (RS) decays. 

\label{sec:spradlin:mix:vtx}

Time-dependent analyses require precise measurements of the creation
and decay vertices of the $D^0$ mesons.  The scale of the required
precision is set by the approximately \mbox{$4\ \mathrm{mm}$} mean
laboratory flight distance for a $60\,\mathrm{GeV/c}$ $D^0$ (the mean
momentum of secondary $D^{\ast+}$--tagged $D^0$ decays).  The decay
vertex of a two-body $D^0$ decay can be determined precisely from its
products with a resolution of {$\sim 260\ \mu\mathrm{m}$} along the
beam axis.  
For promptly produced $D^0$ decays, the precisely measured primary interaction
vertex (resolution {$\sim 60\ \mu\mathrm{m}$} along the beam
axis~\cite{Alves:2008zz}) is the creation vertex.
    
For secondary charm decays, the additional charged tracks must come
from the $b$-hadron decay that produced the $D^{\ast+}$.  LHCb has
been developing techniques to partially reconstruct the parent
$b$-hadron that produced the $D^{\ast+}$~\cite{Spradlin:2007zz}.
Initial results from these developments are promising.  As shown in
the $B_{\mathrm{part}}$ column of Tab.~\ref{tab:mix:vtx:res}, using a
partial reconstruction dramatically improves the precision of the
estimated $D^0$ creation vertex and, consequently, the measured $D^0$
proper time.  Fig.~\ref{fig:spradlin:mix:vtx:lt} shows that this
process produces precisely measured proper times that closely
reproduce the generated proper time distribution.  The $b$-hadron
partial reconstruction is approximately 60\% efficient with respect to
all selected secondary $D^{\ast+}$--tagged \mbox{$D^0 \rightarrow h^-
h'^+$} decays.

    \begin{table}
    \caption[ Resolutions of $D^0$, $D^{\ast+}$, and $B_{\mathrm{part}}$
      vertices, and of $D^0$ proper time ]{
	Estimated resolutions of $D^0$, $D^{\ast+}$, and $B_{\mathrm{part}}$
	vertices, and of $D^0$ proper time in simulated LHCb data.
	The $D^0$ proper time, $\tau_{D^0}$, is estimated both using the
	$D^{\ast+}$ vertex as the creation vertex in the first column, and
	using the $B_{\mathrm{part}}$ vertex as the creation vertex in the last
	column.
     \label{tab:mix:vtx:res}}      
	\begin{center}
	\begin{tabular}{r@{\hspace{0.3 cm}\vline\hspace{0.3 cm}}r@{\hspace{0.6 cm}}r@{\hspace{0.3 cm}\vline\hspace{0.3 cm}}r} \hline\hline
        \hspace{3em} &
        $D^{0}$ &
        $D^{*+}$ &
        $B_{\mathrm{part}}$ \\ \hline

        $x$ &
        $22\ \mu\mathrm{m}$ &
        $190\ \mu\mathrm{m}$ &
        $18\ \mu\mathrm{m}$ \\

        $y$ &
        $17\ \mu\mathrm{m}$ &
        $140\ \mu\mathrm{m}$ &
        $18\ \mu\mathrm{m}$ \\

        {$z$} &
        {$260\ \mu\mathrm{m}$} &
        {$4200\ \mu\mathrm{m}$} &
        {$240\ \mu\mathrm{m}$} \\

        {$\tau_{D^0}$} &
        \multicolumn{2}{c@{\hspace{0.3 cm}\vline\hspace{0.3 cm}}}{{$0.47\ \mathrm{ps}$}} &
        {$0.045\,\mathrm{ps}$} \\ \hline\hline
      \end{tabular}
	\end{center}
    \end{table}

    \begin{figure}[ht]
      \begin{center}
        \includegraphics[width=0.49\textwidth]{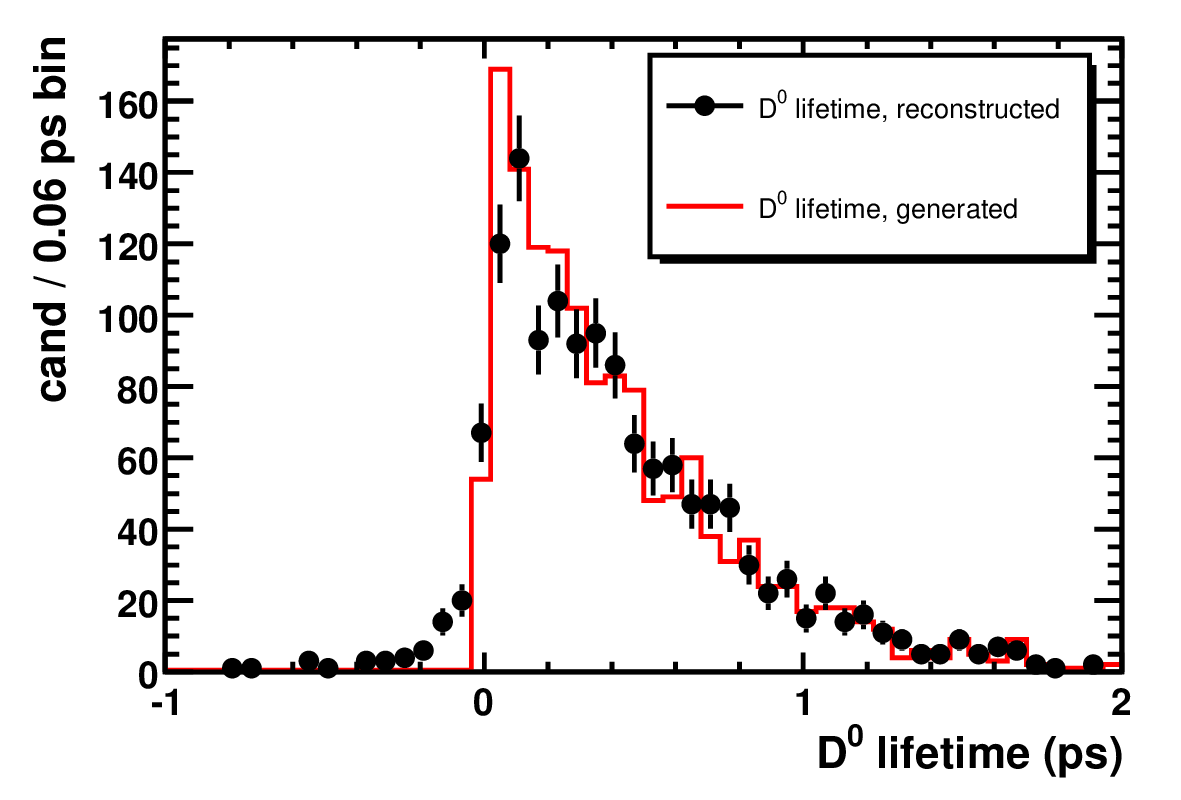}
        \includegraphics[width=0.49\textwidth]{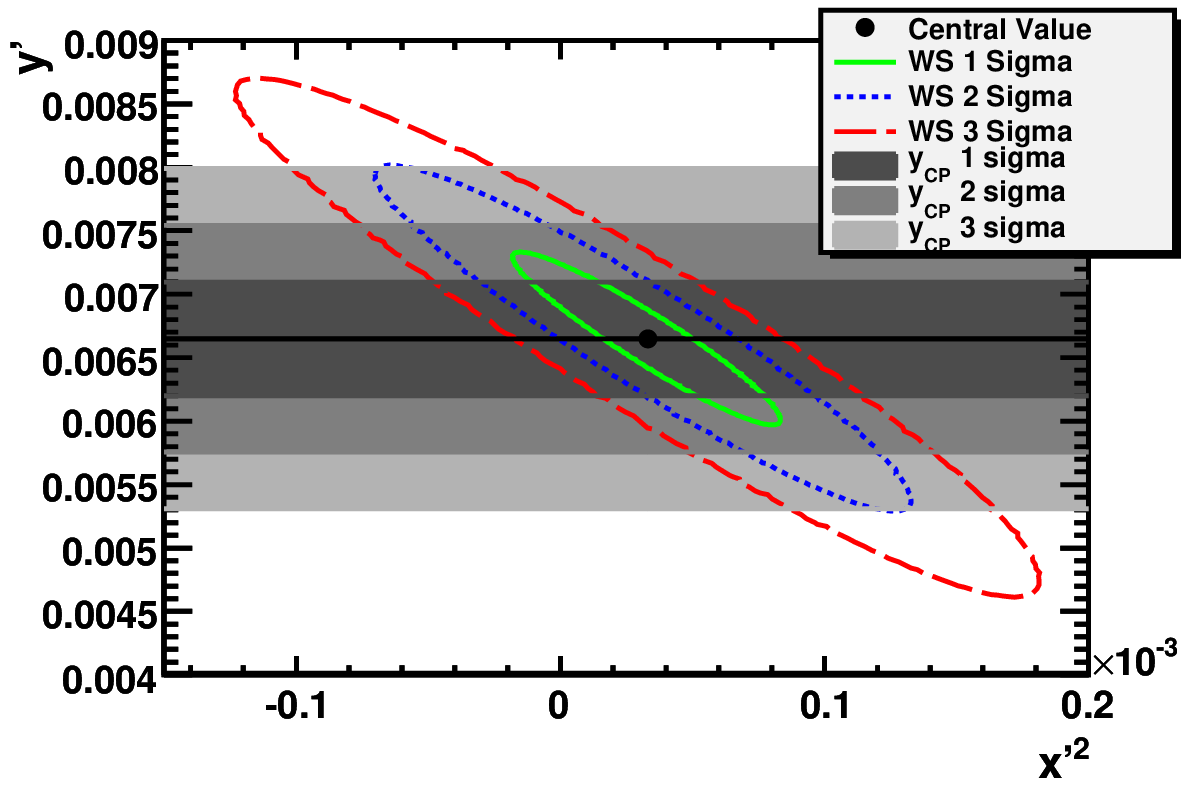}
      \end{center}

    \caption{ In the left panel, the distribution of the proper times
    for simulated $D^0$ mesons from \mbox{$B \rightarrow D^{\ast+} X$}
    decays.  The solid lines are the generated proper times and the
    points are the estimated $D^0$ proper times using the estimated
    parent $B$ decay vertex as the $D^0$ production vertex. In the
    right panel, the sensitivities in $10\,\mathrm{fb}^{-1}$ from the
    WS study and the $y_\mathrm{CP}$ study.  Contours correspond to
    $1\sigma$, $2\sigma$, and $3\sigma$ confidence levels from the WS
    study.  Horizontal bands correspond to $1\sigma$, $2\sigma$, and
    $3\sigma$ confidence levels from the $y_\mathrm{CP}$ study.
    \label{fig:spradlin:mix:vtx:lt}}
\end{figure}

\label{sec:spradlin:mix:sense}

Toy Monte Carlo studies have been used to estimate LHCb's statistical
sensitivities to the mixing parameters $x'^2$ and $y'$ in a two-body
WS mixing study and to the mixing parameter $y_{\mathrm{CP}}$ in a
two-body lifetime ratio study.

Selection studies in fully simulated LHCb events predict
a yield of roughly $230,000$ $D^{\ast+}$--tagged WS decays  \mbox{$10\,\mathrm{fb}^{-1}$} of LHCb data. Again, the $D^{\ast+}$ mesons originate in the decays of $b$-hadrons in this study.
 The \mbox{$10\,\mathrm{fb}^{-1}$}
signal and background yields, proper time resolution, and
proper time acceptance of this selection were used in a toy Monte
Carlo study to estimate the LHCb statistical sensitivity to ${x'}^2$
and $y'$:
\begin{eqnarray*}
  \sigma_{\mathrm{stat}}({x'}^2) = \pm 0.064 \times 10^{-3};~~~ 
  \sigma_{\mathrm{stat}}(y') = \pm 0.87 \times 10^{-3}\ \mbox{\cite{Spradlin:2007zz}}.
\end{eqnarray*}

The same selection studies referred to in
Sec.~\ref{sec:spradlin:cpv} estimate that a lifetime ratio analysis
on \mbox{$10\,\mathrm{fb}^{-1}$} of LHCb data would incorporate
approximately 8 million $D^{\ast+}$--tagged \mbox{$D^0 \rightarrow
  K^-K^+$} decays from $b$-hadron decays.  The
\mbox{$10\,\mathrm{fb}^{-1}$} signal and background yields, the proper
time resolution, and the proper time acceptance of this selection were
used in a toy Monte Carlo study to estimate the LHCb statistical
sensitivity to $y_{\mathrm{CP}}$:

\begin{eqnarray*}
  \sigma_{\mathrm{stat}}(y_{\mathrm{CP}}) = \pm 0.5 \times 10^{-3}\ \mbox{\cite{Spradlin:2007zz}}.
\end{eqnarray*}

\noindent Strategies to reduce the systematic uncertainties to commensurate
precision are in development.  While systematic uncertainties are still
under study, but LHCb will certainly have the statistical power to
make precision measurements in charm CP violation and
$D^0$-$\overline{D^0}$ mixing.




%



\section{Measurement of the angle \gm in tree dominated processes }
\label{sec:tree}
\subsection{Overview of Theoretically Pristine Approaches to Measure $\gamma$}
\label{subsec:DK-intro}

Among the fundamental parameters of the Standard Model of particle physics,
the angle $\gamma = {\rm arg}\left(-V_{ud} V^*_{ub}/V_{cd}V^*_{cb}\right)$ of 
the Unitarity Triangle formed from elements of the Cabibbo-Kobayashi-Maskawa 
quark mixing matrix~\cite{Cabibbo:1963yz,Kobayashi:1973fv}
has a particular importance.
It is the only CP violating parameter that can be measured using 
only tree-level decays,
and thus it provides an essential benchmark in any effort to understand
the baryon asymmetry of the Universe.
Strategies to measure fundamental parameters of the Standard Model and to
search for New Physics by overconstraining the Unitarity Triangle inevitably
require a precise measurement of $\gamma$.

Fortunately, there is a theoretically pristine approach to measure $\gamma$
using tree-dominated $B \to DK$ decays~\cite{Bigi:1988ym,Gronau:1990ra,Gronau:1991dp}. 
The approach exploits the interference between $D^0$ and $\bar{D}^0$
amplitudes that occurs when the neutral $D$ meson is reconstructed in decay
that is accessible to both flavor states.
Feynman diagrams for the relevant $B$ decays are shown in
Fig.~\ref{fig:dk-diags}. 
The original approach  uses $D$ decays to CP
eigenstates~\cite{Gronau:1990ra,Gronau:1991dp}, but variants using
doubly-Cabibbo-suppressed decays~\cite{Atwood:1996ci,Atwood:2000ck},
singly-Cabibbo-suppressed decays~\cite{Grossman:2002aq} and multibody final
states such as $\KS\pip\pim$~\cite{Bondar:BINP,Giri:2003ty,Poluektov:2004mf},
and many others besides, have been proposed.

\begin{figure}
  \centering
  \includegraphics[width=0.4\textwidth,bb=135 575 462 763,clip=true]{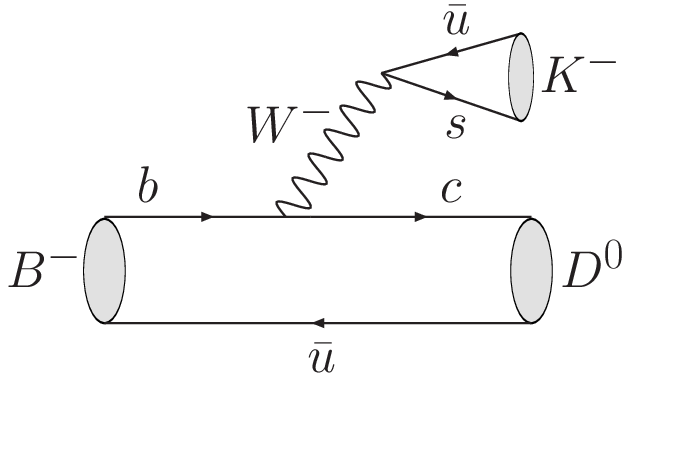}
  \hspace{0.02\textwidth}
  \includegraphics[width=0.4\textwidth,bb=92 516 398 695,clip=true]{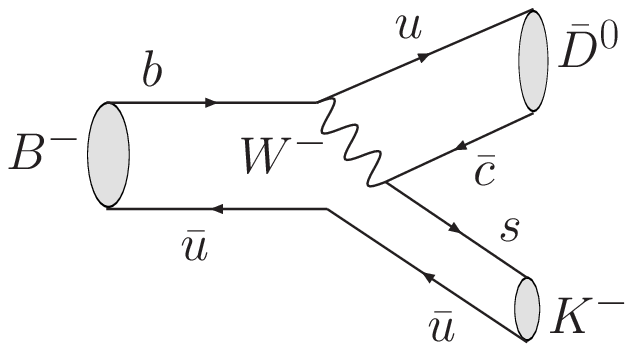}
  \caption{
    Leading Feynman diagrams contributing to the $B^+ \to DK^+$ decay.
    From~\cite{Aubert:2008bd}.
  }
  \label{fig:dk-diags}
\end{figure}

Considering $D$ decays to CP eigenstates (CP even and odd denoted by $D_1$
and $D_2$ respectively), and defining
\begin{equation}
  r_B e^{i \delta_B} = \frac{
    {\cal A}\left(B^+ \to D^0 K^+\right)
  }{
    {\cal A}\left(B^+ \to \overline{D}^0 K^+\right)
  } \, ,
\end{equation}
the dependence on $\gamma$ of the decay rates is found to be as follows
(as illustrated in Fig.~\ref{fig:glw-triangle}).
\begin{eqnarray}
  {\cal A}\left(B^{-}\to D_{1}K^{-}\right) 
  \propto 
  \frac{1}{2} \left(1+r_{B}e^{i \left(\delta_B - \gamma\right)}\right) 
  & \longrightarrow & \\
  & & \hspace{-25mm} \nonumber
  \Gamma\left(B^{-}\to D_{1}K^{-}\right) 
  \propto 
  1+r_{B}^{2}+2r_{B}\cos\left(\delta_B - \gamma\right) 
  \\
  {\cal A}\left(B^{-}\to D_{2}K^{-}\right) 
  \propto 
  \frac{1}{2} \left(1-r_{B}e^{i \left(\delta_B - \gamma\right)}\right) 
  & \longrightarrow & \\
  & & \hspace{-25mm} \nonumber
  \Gamma\left(B^{-}\to D_{2}K^{-}\right) 
  \propto 
  1+r_{B}^{2}-2r_{B}\cos\left(\delta_B - \gamma\right)
  \\
  {\cal A}\left(B^{+}\to D_{1}K^{+}\right) 
  \propto 
  \frac{1}{2} \left(1+r_{B}e^{i \left(\delta_B + \gamma\right)}\right) 
  & \longrightarrow & \\
  & & \hspace{-25mm} \nonumber 
  \Gamma\left(B^{+}\to D_{1}K^{+}\right) 
  \propto 
  1+r_{B}^{2}+2r_{B}\cos\left(\delta_B + \gamma\right)
  \\
  {\cal A}\left(B^{+}\to D_{2}K^{+}\right) 
  \propto 
  \frac{1}{2} \left(1-r_{B}e^{i \left(\delta_B + \gamma\right)}\right) 
  & \longrightarrow & \\
  & & \hspace{-25mm} \nonumber
  \Gamma\left(B^{+}\to D_{2}K^{+}\right) 
  \propto 
  1+r_{B}^{2}-2r_{B}\cos\left(\delta_B + \gamma\right)
\end{eqnarray}

\begin{figure}
  \centering
  \includegraphics[width=0.6\textwidth]{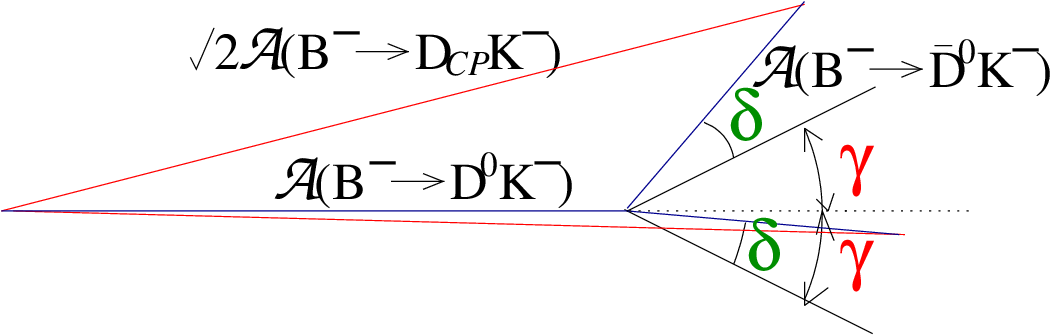}
  \caption{
    Illustration of the sensitivity to $\gamma$ that arises from the
    interference of $B^+ \to D^0K^+$ and $B^+ \to \overline{D}^0K^+$ decay
    amplitudes. 
  }
  \label{fig:glw-triangle}
\end{figure}

From the above expressions it is clear that CP violation effects will be
enhanced for values of $r_B$ close to unity.
It can also be seen that measurements of rates (and rate asymmetries)
alone yield information on $x_\pm = r_B \cos(\delta_B \pm \gamma)$.
This leads to ambiguities in the extraction of $\gamma$.
These can be resolved, and the overall precision improved, when information on 
$y_\pm = r_B \sin(\delta_B \pm \gamma)$ is obtained, 
as can be achieved from Dalitz plot analyses, for example.

To avoid relying on theoretical estimates of the hadronic parameters $r_B$ and
$\delta_B$, these parameters must also be determined from the data.
Once that is done, the underlying method has essentially zero theoretical
uncertainty.  
The largest effects are due to charm mixing and possible CP violation
effects in the $D$ decays~\cite{Grossman:2005rp}.
However, once measured it is possible to take these effects into account in
the analysis.
Similarly, when decays of neutral $B$ mesons are used, there is a potential
systematic effect if the possible
$B^0_{(s)}\textnormal{--}\overline{B}^0_{(s)}$ width difference is
neglected~\cite{Gronau:2004gt,Gronau:2007bh}. 

As already mentioned above, many different decays in the ``$B \to DK$'' family
can be used to gain sensitivity to $\gamma$.
Not only charged but also neutral $B$ decays can be used.
Any decay of the neutral $D$ meson that is accessible to both $D^0$ and
$\overline{D}^0$ can be used.
Furthermore decays with excited $D$ and/or $K$ states not only provide
additional statistics.
In the former case there is an effective strong phase difference of $\pi$
between the cases that the $D^*$ is reconstructed as $D\pi^0$ and
$D\gamma$ that is particularly beneficial when $D$ decays to
doubly-Cabibbo-suppressed final states are analyzed~\cite{Bondar:2004bi}.  
When $K^*$ mesons are used, their natural width can be handled by the
introduction of effective hadronic parameters~\cite{Gronau:2002mu};
alternatively a Dalitz plot analysis of the $B \to DK\pi$ decay removes this
problem and maximizes the sensitivity to $\gamma$~\cite{Gershon:2008pe}.
Ultimately it is clear that the best sensitivity to $\gamma$ will be obtained
by combining as many statistically independent measurements as possible.

\newcommand{\bdk}{$B^{\pm}\to DK^{\pm}$}
\newcommand{\bdkm}{$B^{-}\to DK^{-}$}
\newcommand{\bdkp}{$B^{+}\to DK^{+}$}
\newcommand{\bdtk}{$B^{\pm}\to \tilde{D}K^{\pm}$}
\newcommand{\bdtkp}{$B^{+}\to \tilde{D}_+K^{+}$}
\newcommand{\bdtkm}{$B^{-}\to \tilde{D}_-K^{-}$}

\newcommand{\bdsk}{$B^{\pm}\to D^{*}K^{\pm}$}
\newcommand{\bdskm}{$B^{-}\to D^{*}K^{-}$}
\newcommand{\bdskp}{$B^{+}\to D^{*}K^{+}$}
\newcommand{\bdstk}{$B^{\pm}\to \tilde{D}^{*}K^{\pm}$}
\newcommand{\bdstkp}{$B^{+}\to \tilde{D}^{*}_+K^{+}$}
\newcommand{\bdstkm}{$B^{-}\to \tilde{D}^{*}_-K^{-}$}

\newcommand{\bdks}{$B^{\pm}\to DK^{*\pm}$}
\newcommand{\bdksm}{$B^{-}\to DK^{*-}$}
\newcommand{\bdksp}{$B^{+}\to DK^{*+}$}
\newcommand{\bdtks}{$B^{\pm}\to \tilde{D}K^{*\pm}$}
\newcommand{\bdtksp}{$B^{+}\to \tilde{D}_+K^{*+}$}
\newcommand{\bdtksm}{$B^{-}\to \tilde{D}_-K^{*-}$}

\newcommand{\bddsk}{$B^{\pm}\to D^{(*)}K^{\pm}$}
\newcommand{\bddskm}{$B^{-}\to D^{(*)}K^{-}$}
\newcommand{\bddskp}{$B^{+}\to D^{(*)}K^{+}$}
\newcommand{\bddstk}{$B^{\pm}\to \tilde{D}^{(*)}K^{\pm}$}
\newcommand{\bddstkp}{$B^{+}\to \tilde{D}^{(*)}_+K^{+}$}
\newcommand{\bddstkm}{$B^{-}\to \tilde{D}^{(*)}_-K^{-}$}

\newcommand{\bddsks}{$B^{\pm}\to D^{(*)}K^{(*)\pm}$}
\newcommand{\bddsksm}{$B^{-}\to D^{(*)}K^{(*)-}$}
\newcommand{\bddsksp}{$B^{+}\to D^{(*)}K^{(*)+}$}
\newcommand{\bddstks}{$B^{\pm}\to \tilde{D}^{(*)}K^{(*)\pm}$}
\newcommand{\bddstksp}{$B^{+}\to \tilde{D}^{(*)}_+K^{(*)+}$}
\newcommand{\bddstksm}{$B^{-}\to \tilde{D}^{(*)}_-K^{(*)-}$}

\newcommand{\bdksnr}{$B^{\pm}\to DK^0_S\pi^{\pm}$}
\newcommand{\bdkspnr}{$B^{+}\to DK^0_S\pi^{+}$}
\newcommand{\bdksmnr}{$B^{-}\to DK^0_S\pi^{-}$}
\newcommand{\bdtksnr}{$B^{\pm}\to \tilde{D}K^0_S\pi^{\pm}$}
\newcommand{\bdtkspnr}{$B^{+}\to \tilde{D}_+K^0_S\pi^{+}$}
\newcommand{\bdtksmnr}{$B^{-}\to \tilde{D}_-K^0_S\pi^{-}$}

\newcommand{\bdpi}{$B^{\pm}\to D\pi^{\pm}$}
\newcommand{\bdpip}{$B^{+}\to D\pi^{+}$}
\newcommand{\bdpim}{$B^{-}\to D\pi^{-}$}
\newcommand{\bdtpi}{$B^{\pm}\to \tilde{D}\pi^{\pm}$}
\newcommand{\bdtpip}{$B^{+}\to \tilde{D}_+\pi^{+}$}
\newcommand{\bdtpim}{$B^{-}\to \tilde{D}_-\pi^{-}$}

\newcommand{\bdstpi}{$B^{\pm}\to \tilde{D}^{*}\pi^{\pm}$}
\newcommand{\bdstpip}{$B^{+}\to \tilde{D}^{*}_+\pi^{+}$}
\newcommand{\bdstpim}{$B^{-}\to \tilde{D}^{*}_-\pi^{-}$}
\newcommand{\bdspi}{$B^{\pm}\to D^{*}\pi^{\pm}$}
\newcommand{\bdspim}{$B^{-}\to D^{*}\pi^{-}$}
\newcommand{\bdspip}{$B^{+}\to D^{*}\pi^{+}$}

\newcommand{\bndspi}{$B\to D^{*\pm}\pi^{\mp}$}
\newcommand{\bndspim}{$B\to D^{*+}\pi^{-}$}
\newcommand{\bndspip}{$B\to D^{*-}\pi^{+}$}

\newcommand{\bddspi}{$B^{\pm}\to D^{(*)}\pi^{\pm}$}
\newcommand{\bddspip}{$B^{+}\to D^{(*)}\pi^{+}$}
\newcommand{\bddspik}{$B^{-}\to D^{(*)}\pi^{-}$}
\newcommand{\bddstpi}{$B^{\pm}\to \tilde{D}^{(*)}\pi^{\pm}$}
\newcommand{\bddstpip}{$B^{+}\to \tilde{D}^{(*)}_+\pi^{+}$}
\newcommand{\bddstpim}{$B^{-}\to \tilde{D}^{(*)}_-\pi^{-}$}

\newcommand{\dsdpi}{$D^{*\pm}\to D\pi^{\pm}$}
\newcommand{\dsdpis}{$D^{*\pm}\to D\pi_s^{\pm}$}
\newcommand{\dsdpip}{$D^{*+}\to D^0\pi^{+}$}
\newcommand{\dsdpips}{$D^{*+}\to D^0\pi_s^{+}$}
\newcommand{\dsdpim}{$D^{*-}\to \overline{D}{}^0\pi^{-}$}
\newcommand{\dsdpims}{$D^{*-}\to \overline{D}{}^0\pi_s^{-}$}

\newcommand{\dkpp}{$\overline{D}{}^0\to K^0_S\pi^+\pi^-$}
\newcommand{\dkkk}{$\overline{D}{}^0\to K^0_S K^+ K^-$}
\newcommand{\dtkpp}{$\tilde{D}\to K^0_S\pi^+\pi^-$}

\newcommand{\phm}{\ensuremath{\phantom{-}}}
\newcommand{\phl}{\ensuremath{\phantom{<}}}
\newcommand{\phz}{\ensuremath{\phantom{0}}}
\newcommand{\real}{\Re}
\newcommand{\imag}{\Im}

\def\kskk{\ensuremath{\KS \Kp \Km}\xspace}
\def\Dztokskk{\ensuremath{\Dz \to \kskk}\xspace}
\def\K  {\ensuremath{K}\xspace}

\def\Dztilde   {\ensuremath {\tilde{D}^0}\xspace}
\def\Dstarztilde   {\ensuremath {\tilde{D}^{\ast 0}}\xspace}

\subsection{Experimental results on \gm from $B \to DK$ decays}

\subsubsection{GLW analyses}
\label{subsubsec:glw}


The technique of measuring $\gamma$ proposed by Gronau, London and Wyler (and
called GLW) \cite{Gronau:1990ra,Gronau:1991dp} makes use of $D^0$ decays to
\CP\ eigenstates, such as $K^+K^-$, $\pi^+\pi^-$ (\CP-even) or
$K^0_S\pi^0$, $K^0_S\phi$ (\CP-odd). Since both $D^0$ and
$\bar{D}^{0}$ can decay into the same CP eigenstate ($D_{CP}$, or
$D_1$ for a CP-even state and $D_2$ for a CP-odd state), the
$b\rightarrow c$ and $b\rightarrow u$ processes shown in
Fig.~\ref{fig:dk-diags} interfere in the $B^{\pm} \to D_{CP} K^{\pm}$
decay channel. This interference may lead to direct CP violation.
To measure $D$ meson decays to CP eigenstates a large number of
$B$ meson decays is required since the branching fractions to these
modes are of order 1\%. To extract $\gamma$ using the GLW method,
the following observables sensitive to CP violation are used: the
asymmetries
\begin{equation}
\label{glw_eq1}
\begin{split}
{\cal{A}}_{1,2} & \equiv \frac{{\cal B}(B^- \rightarrow D_{1,2}K^-) -
{\cal B}(B^+ \rightarrow D_{1,2}K^+) }{{\cal B}(B^- \rightarrow
D_{1,2}K^-) + {\cal B}(B^+ \rightarrow D_{1,2}K^+) }\\
& = \frac{2 r_B \sin \delta ' \sin \gamma}{1 + r_B^2 + 2 r_B \cos
\delta ' \cos \gamma}
\end{split}
\end{equation}
and the double ratios
\begin{equation}
\label{glw_eq2}
\begin{split}
{\cal{R}}_{1,2} &\equiv \frac{{\cal B}(B^- \rightarrow D_{1,2}K^-) +
{\cal B}(B^+ \rightarrow D_{1,2}K^+) }{{\cal B}(B^- \rightarrow
D^0 K^-) + {\cal B}(B^+ \rightarrow D^0 K^+) }\\
 &= 1 + r_B^2 + 2 r_B
\cos \delta ' \cos \gamma,
\end{split}
\end{equation}
where
\begin{equation}
\label{glw_eq3}
\delta ' = \left\{
             \begin{array}{ll}
              \delta_B & \mbox {{\rm  for }$D_1$}\\
              \delta_B + \pi&  \mbox{{\rm for }$D_2$}\\
             \end{array}
             \right. ,
\end{equation}
and $r_B$ and $\delta_B$ were defined in the previous section.
The value of $r_B$ is given by the ratio of the CKM matrix elements
$|V_{ub}^*V_{cs\vphantom{b}}^{\vphantom{*}}|/
 |V_{cb}^*V_{us\vphantom{b}}^{\vphantom{*}}|\sim 0.38$
times a color suppression factor.
Here we assume that mixing and CP violation in the neutral $D$ meson
system can be neglected.

Instead of four observables $\mathcal{R}_{1,2}$ and $\mathcal{A}_{1,2}$, only
three of which are independent
(since $\mathcal{A}_1\mathcal{R}_1 = -\mathcal{A}_2\mathcal{R}_2$),
an alternative set of three parameters can be used:
\begin{equation}
  \begin{split}
  x_{\pm} &= r_B\cos(\delta_B\pm \gamma)
          = \frac{\mathcal{R}_1(1\mp\mathcal{A}_1)-
                   \mathcal{R}_2(1\mp\mathcal{A}_2)}{4},
  \end{split}
\end{equation}
and
\begin{equation}
  r_B^2 = \frac{\mathcal{R}_1+\mathcal{R}_2-2}{2}.
\end{equation}
The use of these observables allows for a direct comparison with the
methods involving analyses of the Dalitz plot distributions of multibody $D^0$
decays (see Sec.~\ref{subsubsec:dalitz}),
where the same parameters $x_{\pm}$ are obtained.


Measurements of $B\to D_{\rm CP}K$ decays have been performed by both the
BaBar~\cite{:2008yk} and Belle~\cite{Abe:2006hc} collaborations,
while CDF has recently made measurements using CP-even decays only~\cite{Gibson:2008gk}.
The most recent update is BaBar's analysis using a data sample of 382M
$B\overline{B}$ pairs~\cite{:2008yk}.
The analysis uses $D^0$ decays to $K^+K^-$ and $\pi^+\pi^-$ as CP-even
modes, $K^0_S\pi^0$ and $K^0_S\omega$ as CP-odd modes.

\begin{table}
  \caption{Results of the GLW analysis by BaBar~\cite{:2008yk}.}
\centering
  \label{glw_table}
  \begin{tabular}{|l|l|}
    \hline
    $\mathcal{R}_{1}$ & $\phantom{-}1.06\pm 0.10 \pm 0.05$ \\
    $\mathcal{R}_{2}$ & $\phantom{-}1.03\pm 0.10 \pm 0.05$ \\
    $\mathcal{A}_{1}$ & $+0.27\pm 0.09 \pm 0.04$ \\
    $\mathcal{A}_{2}$ & $-0.09\pm 0.09 \pm 0.02$ \\
    \hline
    $x_+$ & $-0.09\pm 0.05\pm 0.02$ \\
    $x_-$ & $+0.10\pm 0.05\pm 0.03$ \\
    $r_B^2$ & $0.05\pm 0.07\pm 0.03$ \\
    \hline
  \end{tabular}
\end{table}

The results of the analysis (both in terms of asymmetries and double
ratios, and the alternative $x_{\pm}, r^2_B$ set of parameters) are shown in
Tab.~\ref{glw_table}.
As follows from (\ref{glw_eq1}) and (\ref{glw_eq3}), the signs of the
$\mathcal{A}_{1}$ and $\mathcal{A}_{2}$ asymmetries should be opposite,
which is confirmed by the experiment. The $x_{\pm}$ values are in
good agreement with those obtained by the Dalitz plot analysis technique (see \ref{subsubsec:dalitz}).
Note that the measurement of $\mathcal{A}_{1}$ deviates from zero by 2.8
standard deviations.

A summary of measurements of observables with the GLW method is given in
Fig.~\ref{fig:hfag-glw}.  As well as the results using $B\to D_{\rm CP}K$
decays, this compilation also includes measurements from the decay channels
$B\to D^{*}_{\rm CP}K$ and $B\to D_{\rm CP}K^*$.

\begin{figure}
  \centering
  \includegraphics[width=0.44\textwidth]{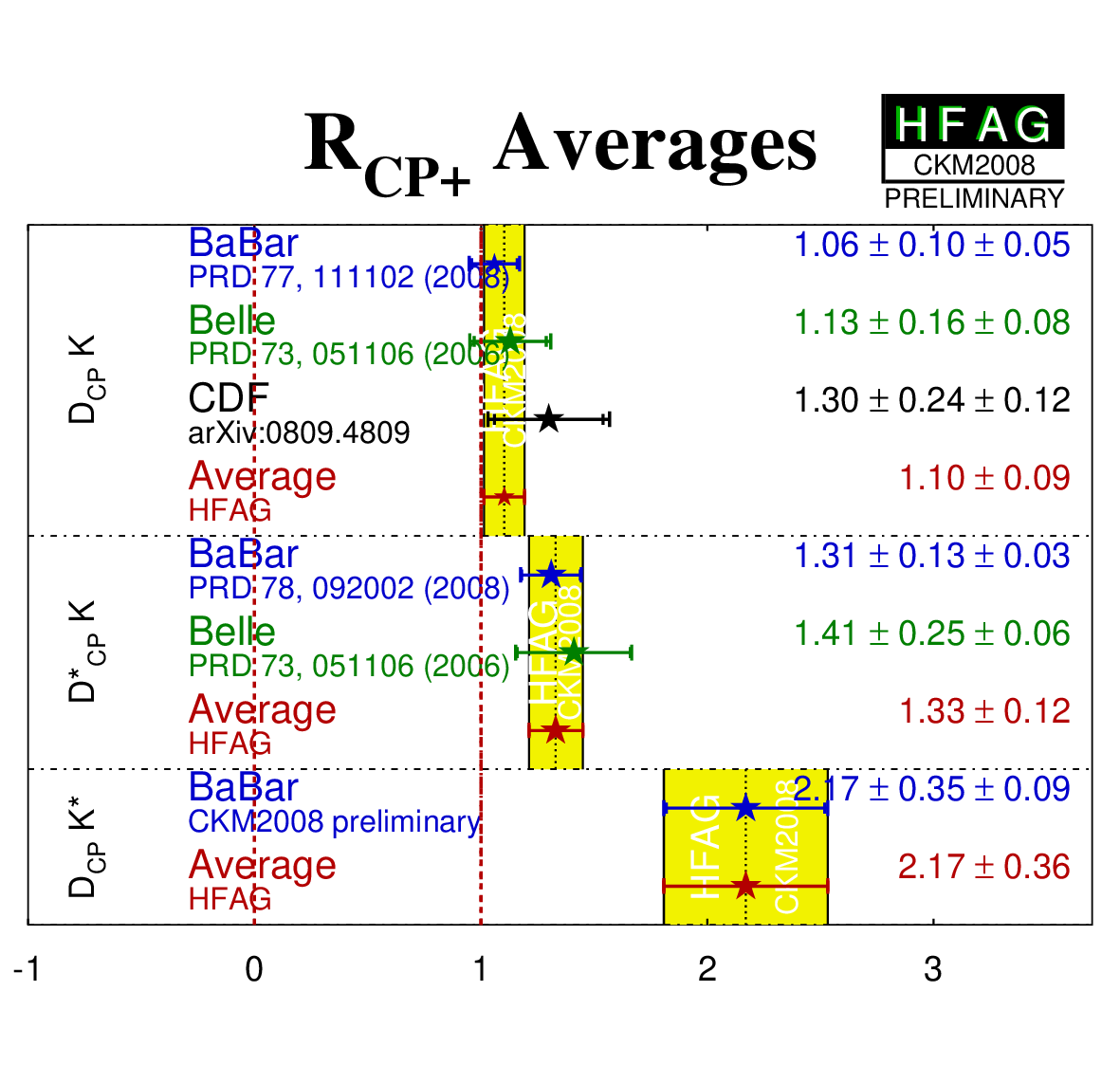}
  \hfill
  \includegraphics[width=0.44\textwidth]{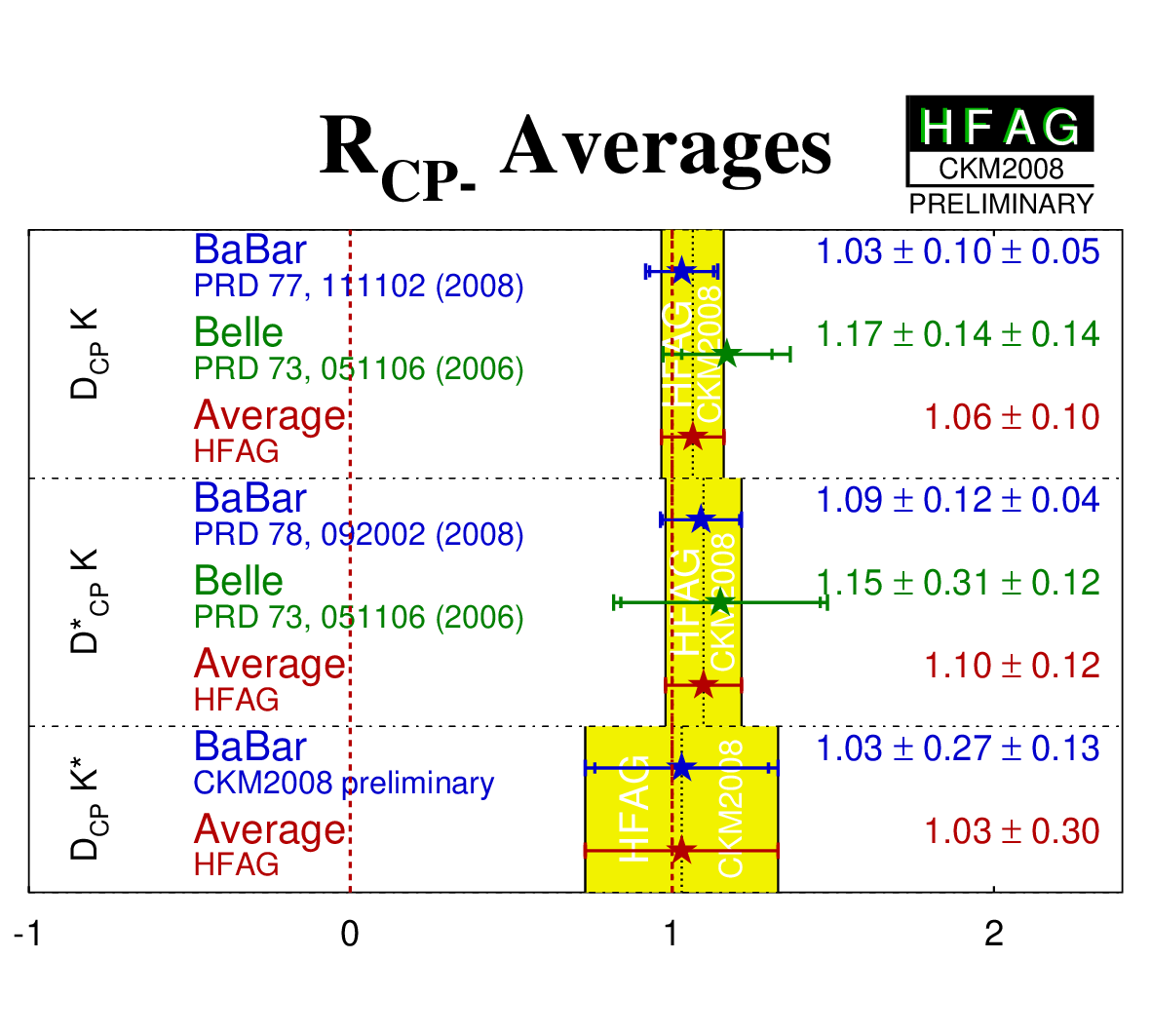}

  \includegraphics[width=0.44\textwidth]{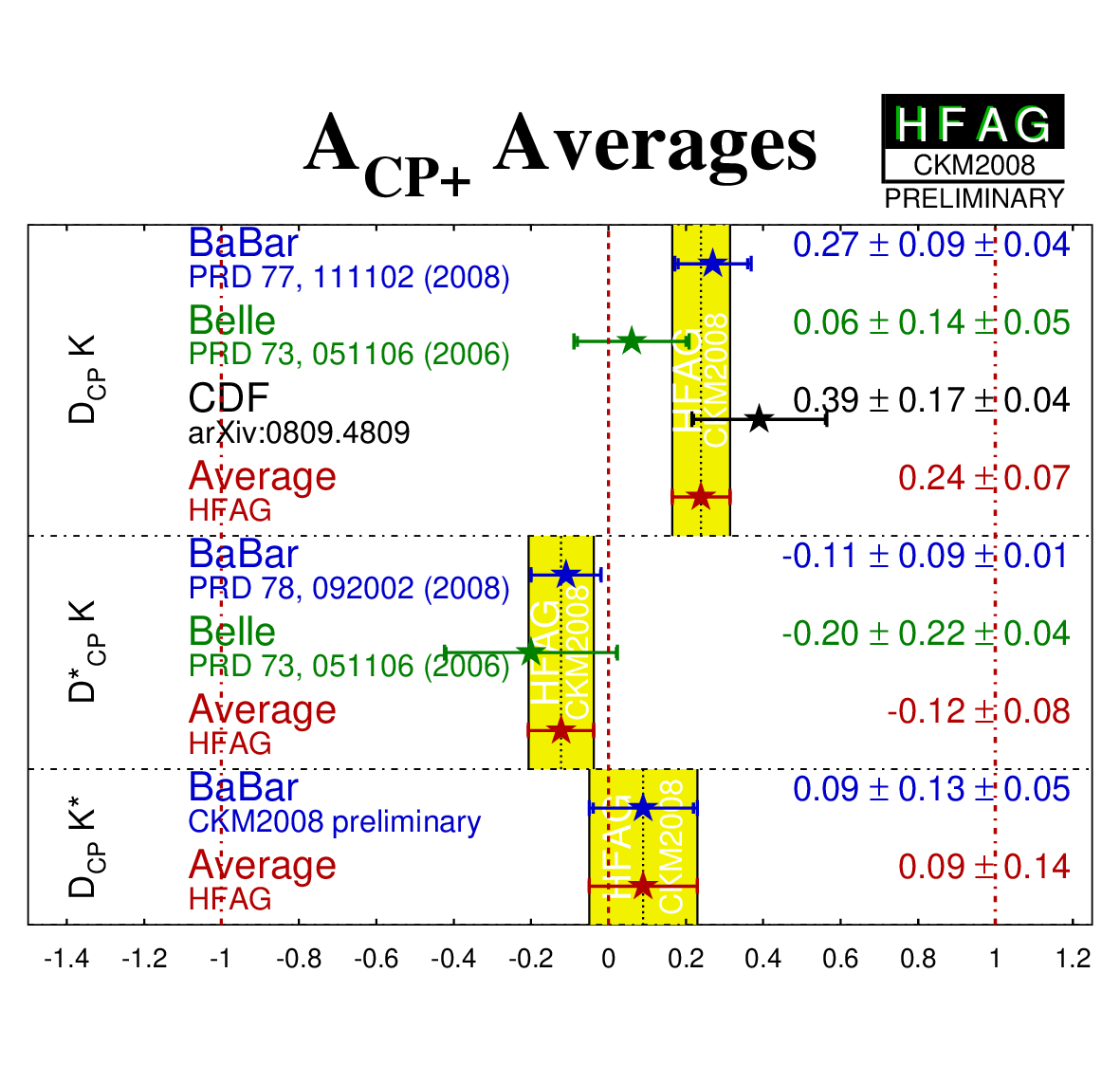}
  \hfill
  \includegraphics[width=0.44\textwidth]{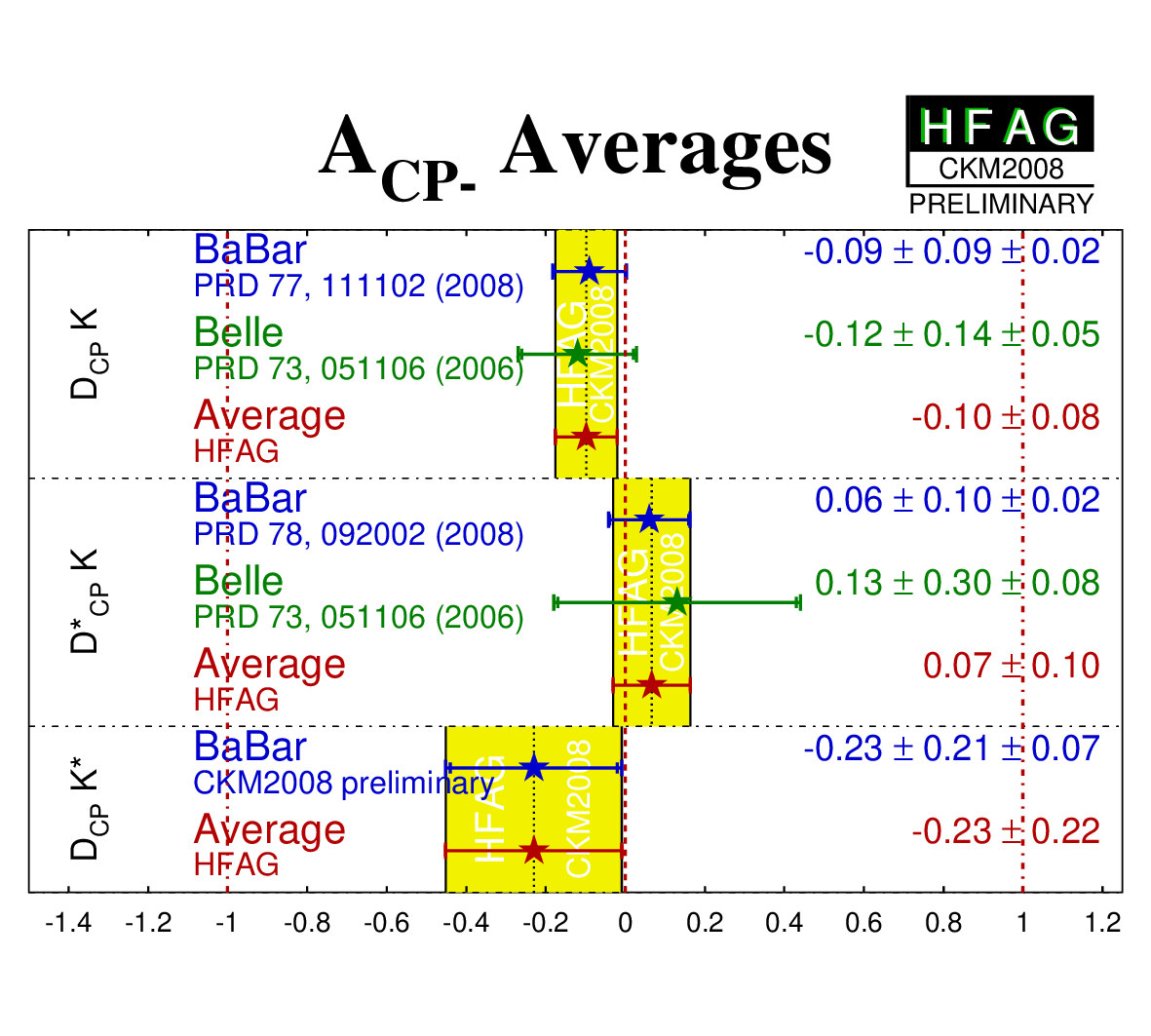}
  \caption{
    Compilations and world averages of measurements of observables using the
    GLW method.
    Top left: $\mathcal{R}_{1}$; top right: $\mathcal{R}_{2}$;
    bottom left: $\mathcal{A}_{1}$; bottom right: $\mathcal{A}_{2}$.
  }
  \label{fig:hfag-glw}
\end{figure}

\subsubsection{ADS analyses}
\label{subsubsec:ads}


The difficulties in the application of the GLW methods are primarily
due to the small magnitude of the CP asymmetry of the
$B^{\pm}\to D_{CP}K^{\pm}$ decay probabilities, which may lead
to significant systematic uncertainties in the measurement of
CP violation. An alternative approach was proposed by Atwood,
Dunietz and Soni \cite{Atwood:1996ci,Atwood:2000ck}.
Instead of using $D^0$ decays to CP eigenstates, the ADS method uses
Cabibbo-favored and doubly Cabibbo-suppressed decays:
$\overline{D}^0\to K^-\pi^+$ and $D^0\to K^-\pi^+$.
In the decays $B^+\to [K^-\pi^+]_D K^+$ and $B^-\to [K^+\pi^-]_D K^-$,
the suppressed $B$ decay corresponds to the Cabibbo-allowed $D^0$ decay,
and vice versa. Therefore, the interfering amplitudes are of similar
magnitudes, and one can expect significant CP asymmetry.

The observable that is measured in the ADS method is
the fraction of the suppressed and allowed branching ratios:
\begin{equation}
  \begin{split}
  \mathcal{R}_{ADS}&=\frac{{\cal B}(B^{\pm}\to [K^{\mp}\pi^{\pm}]_DK^{\pm})}
                   {{\cal B}(B^{\pm}\to [K^{\pm}\pi^{\mp}]_DK^{\pm})}\\
             &=r_B^2+r_D^2+2r_Br_D\cos\gamma\cos\delta,
  \end{split}
\end{equation}
where $r_D$ is the ratio of the doubly Cabibbo-suppressed and
Cabibbo-allowed $D^0$ decay amplitudes~\cite{Barberio:2008fa}:
\begin{equation}
  r_D=\left|\frac{A(D^0\to K^+\pi^-)}{A(D^0\to K^-\pi^+)}\right| =
  0.058\pm 0.001,
\end{equation}
and $\delta$ is the sum of strong phase differences in $B$ and $D$ decays:
$\delta=\delta_B+\delta_D$.
Once a significant signal is seen, the direct CP asymmetry must be measured,
\begin{equation}
  \begin{split}
    \mathcal{A}_{ADS} &=
    \frac{
      {\cal B}(B^{-}\to [K^{+}\pi^{-}]_DK^{-}) -
      {\cal B}(B^{+}\to [K^{-}\pi^{+}]_DK^{+})
    }{
      {\cal B}(B^{-}\to [K^{+}\pi^{-}]_DK^{-}) +
      {\cal B}(B^{+}\to [K^{-}\pi^{+}]_DK^{+})
    } \\
    &=
    \frac{2r_Br_D\sin\gamma\sin\delta}{r_B^2+r_D^2+2r_Br_D\cos\gamma\cos\delta}.
  \end{split}
\end{equation}


Studies of ADS channels have been performed by both BaBar~\cite{Aubert:2005pj}
and Belle~\cite{:2008as}.
Unfortunately, the product branching ratios into the final states of interest
are so small that they cannot be observed using the current experimental
statistics.
The most recent update of the ADS analysis is that from Belle using 657M
$B\overline{B}$ pairs~\cite{:2008as}. The analysis uses \bdk\ decays
with $D^0$ decaying to $K^+\pi^-$ and $K^-\pi^+$ modes (and their
charge-conjugated partners). The ratio of suppressed and allowed modes is
found to be
\begin{equation}
  \mathcal{R}_{ADS}=(8.0^{+6.3}_{-5.7}{}^{+2.0}_{-2.8})\times 10^{-3}.
\end{equation}
Since the signal in  the suppressed modes is not significant, the CP
asymmetry is inevitably consistent with zero:
\begin{equation}
  \mathcal{A}_{ADS}=-0.13^{+0.98}_{-0.88}\pm 0.26.
\end{equation}

A summary of measurements of observables with the ADS method is given in
Fig.~\ref{fig:hfag-ads}.
As well as the results using the decays $B\to D K$ with $D \to K\pi$,
this compilation also includes measurements from the decay channels
$B\to D^{*}K$ with $D \to K\pi$ and the decays $D^* \to D\pi^0$ and $D^* \to
D\gamma$ treated distinctly~\cite{Bondar:2004bi},
$B\to DK^*$ with $D \to K\pi$ and $B\to D K$ with $D \to K\pi\pi^0$.

\begin{figure}
  \centering
  \includegraphics[width=0.44\textwidth]{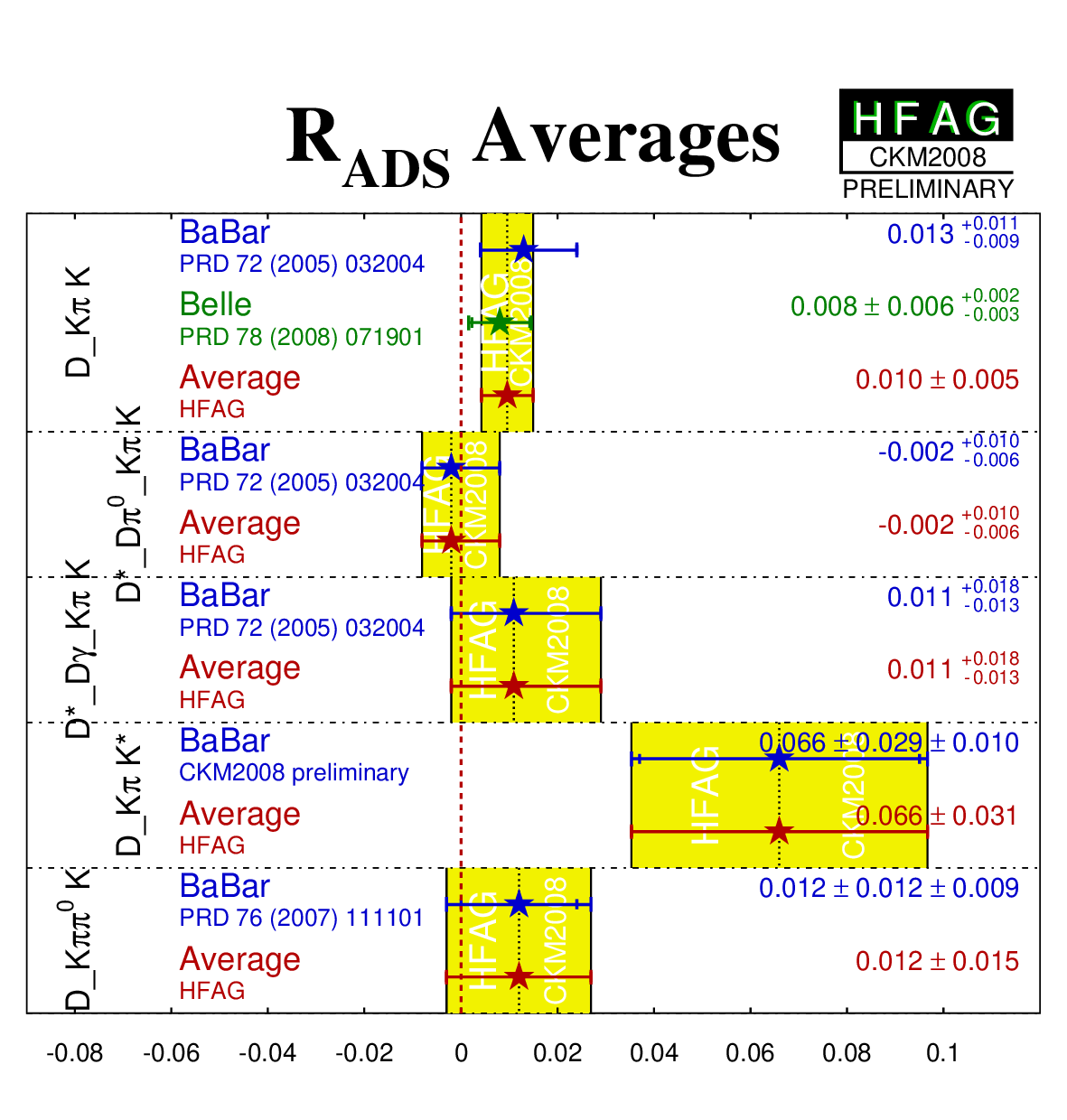}
  \caption{
    Compilations and world averages of measurements of observables using the
    ADS method.
  }
  \label{fig:hfag-ads}
\end{figure}

The ADS analysis currently does not give a significant constraint on
$\gamma$, but it provides important information on the value of $r_B$.
Using the conservative assumption $\cos{\gamma}\cos{\delta}=-1$
one obtains the upper limit $r_B<0.19$ at 90\% CL.
A somewhat tighter constraint can be obtained by using the
$\gamma$ and $\delta_B$ measurements from the Dalitz plot analyses (see
Sec.~\ref{subsubsec:dalitz}), and the recent CLEO-c measurement of the
strong phase
$\delta_D=(22^{+11}_{-12}{}^{+9}_{-11})^{\circ}$~\cite{Rosner:2008fq,Asner:2008ft}.

\subsubsection{Dalitz plot analyses}
\label{subsubsec:dalitz}


A Dalitz plot analysis of a three-body final state of the $D$ meson
allows one to obtain all the information required for determination
of $\gamma$ in a single decay mode. The use of a Dalitz plot
analysis for the extraction of $\gamma$ was first discussed in the context of
the ADS method~\cite{Atwood:1996ci,Atwood:2000ck}. This technique uses the
interference of Cabibbo-favored $D^0\to K^-\pi^+\pi^0$ and doubly
Cabibbo-suppressed $\overline{D}{}^0\to K^-\pi^+\pi^0$ decays. However, the
small rate for the doubly Cabibbo-suppressed decay limits the sensitivity of
this technique.

Three body final states such as
$K^0_S\pi^+\pi^-$~\cite{Giri:2003ty,Bondar:BINP} have been suggested as
promising modes for the extraction of $\gamma$. Like in the GLW or ADS method,
the two amplitudes interfere as the $D^0$ and $\overline{D}{}^0$ mesons
decay into the same final state $K^0_S \pi^+ \pi^-$; we denote the
admixed state as $\tilde{D}_+$. Assuming no CP asymmetry in
neutral $D$ decays, the amplitude of the $\tilde{D}_+$ decay as a
function of Dalitz plot variables $m^2_+=m^2_{K^0_S\pi^+}$ and
$m^2_-=m^2_{K^0_S\pi^-}$ is
\begin{equation}
  f_{B^+}=f_D(m^2_+, m^2_-)+r_Be^{i(\delta_B+\gamma)}f_D(m^2_-, m^2_+)\,,
\end{equation}
where $f_D(m^2_+, m^2_-)$ is the amplitude of the \dkpp\ decay.

Similarly, the amplitude of the $\tilde{D}_-$ decay from $B^{-}\to DK^{-}$ process is
\begin{equation}
  f_{B^-}=f_D(m^2_-, m^2_+)+r_Be^{i(\delta_B-\gamma)}f_D(m^2_+, m^2_-)\,.
\end{equation}
The \dkpp\ decay amplitude can be determined at the $B$ factories from
the large samples of flavor-tagged \dkpp\ decays produced in continuum
$e^+e^-$ annihilation.
[In fact, only $|f_D|^2$ can be determined from flavor tagged data, but a
model assumption can be made to describe the variation of the strong phase
across the Dalitz plot.  Approaches to avoid such model-dependence are
discussed in more detail below.]
Once $f_D$ is known, a simultaneous fit of $B^+$ and $B^-$ data allows the
contributions of $r_B$, $\gamma$ and $\delta_B$ to be separated. The method
has only a two-fold ambiguity: the solutions at $(\gamma,\delta_B)$ and
$(\gamma+180^{\circ}, \delta_B+180^{\circ})$ cannot be distinguished.
References~\cite{Giri:2003ty} and~\cite{Poluektov:2006ia} give more detailed
descriptions of the technique.


Both Belle and BaBar collaborations recently reported updates of their
$\gamma$ measurements using Dalitz plot analysis. The preliminary result
from Belle~\cite{:2008wya} uses a data sample of 657M $B\overline{B}$ pairs
and two modes, \bdk\ and \bdsk\ with $D^{*}\to D\pi^0$. The neutral $D$ meson
is reconstructed in the $K^0_S\pi^+\pi^-$ final state in both cases.


To determine the decay amplitude, $D^{*\pm}$ mesons
produced via the $e^+ e^-\to c\bar{c}$ continuum process are used, which
then decay to a neutral $D$ meson and a charged pion.
The flavor of the neutral $D$ meson is tagged by the charge of the pion
in the decay \dsdpim. $B$ factories offer large sets of such charm data:
$290.9\times 10^3$ events are used in the Belle analysis with only 1.0\%
background.

The description of the \dkpp\ decay amplitude is based on the isobar model.
The amplitude $f_D$ is represented by a coherent sum of two-body decay
amplitudes and one nonresonant decay amplitude.
%
%
The model includes a set of 18 two-body amplitudes:
five Cabibbo-allowed amplitudes: $K^*(892)^+\pi^-$,
$K^*(1410)^+\pi^-$, $K_0^*(1430)^+\pi^-$,
$K_2^*(1430)^+\pi^-$ and $K^*(1680)^+\pi^-$;
their doubly Cabibbo-suppressed partners; eight amplitudes with
$K^0_S$ and a $\pi\pi$ resonance:
$K^0_S\rho$, $K^0_S\omega$, $K^0_Sf_0(980)$, $K^0_Sf_2(1270)$,
$K^0_Sf_0(1370)$, $K^0_S\rho(1450)$, $K^0_S\sigma_1$ and $K^0_S\sigma_2$; and a
flat nonresonant term.

The selection of \bddsk\ decays is based on the CM energy difference
$\Delta E = \sum E_i - E_{\rm beam}$ and the beam-constrained $B$ meson mass
$M_{\rm bc} = \sqrt{E_{\rm beam}^2 - (\sum \vec{p}_i)^2}$, where $E_{\rm beam}$
is the CM beam
energy, and $E_i$ and $\vec{p}_i$ are the CM energies and momenta of the
$B$ candidate decay products. To suppress background from
$e^+e^-\to q\bar{q}$ ($q=u, d, s, c$) continuum events,
variables that characterize the event shape are used.
At the first stage of the analysis, when the $(M_{\rm bc}, \Delta E)$
distribution is fitted in order to obtain the fractions of the background
components, a requirement on the event shape is imposed to suppress the
continuum events. The number of such ``clean'' events is 756 for \bdk\ mode
with 29\% background, and 149 events for \bdsk\ mode with 20\% background.
In the Dalitz plot fit, events are not rejected based on event shape
variables, these are used in the likelihood function to better separate signal
and background events.


The Dalitz distributions of the $B^+$ and $B^-$ samples are fitted
separately, using Cartesian parameters
$x_{\pm}=r_{\pm}\cos(\delta_B\pm\gamma)$ and
$y_{\pm}=r_{\pm}\sin(\delta_B\pm\gamma)$, where the indices ``$+$'' and
``$-$'' correspond to $B^+$ and $B^-$ decays, respectively. In this approach
the amplitude ratios ($r_+$ and $r_-$) are not constrained to be equal for the
$B^+$ and $B^-$ samples. Confidence intervals in $r_B$, $\gamma$ and
$\delta_B$ are then obtained from the $(x_{\pm},y_{\pm})$ using a frequentist
technique. 


The values of the parameters $r_B$, $\gamma$ and $\delta_B$ obtained from the
combination of \bdk\ and \bdsk\ modes are presented in
Tab.~\ref{fc_comb_table}. Note that in addition to the detector-related
systematic error which is caused by the uncertainties of the background
description, imperfect simulation, {\it etc.}, the result suffers from the
uncertainty of the $D$ decay amplitude description. The statistical confidence
level of CP violation for the combined result is $(1-5.5\times 10^{-4})$,
corresponding to 3.5 standard deviations.


\begin{table*}
  \caption{Results of the combination of \bdkp\ and \bdskp\ modes by Belle~\cite{:2008wya}. }
  \label{fc_comb_table}
  \begin{tabular}{|l|c|c|c|c|c|} \hline
  Parameter & $1\sigma$ interval & $2\sigma$ interval &
              Systematic error & Model uncertainty \\ \hline
  $\phi_3$  & $76^{\circ}\;^{+12^{\circ}}_{-13^{\circ}}$
            & $49^{\circ}<\phi_3<99^{\circ}$
            & $4^{\circ}$ & $9^{\circ}$ \\
  $r_{DK}$  & $0.16\pm 0.04$
            & $0.08<r_{DK}<0.24$
            & $0.01$ & $0.05$ \\
  $r_{D^*K}$  & $0.21\pm 0.08$
            & $0.05<r_{D^*K}<0.39$
            & $0.02$ & $0.05$ \\
  $\delta_{DK}$  & $136^{\circ}\;^{+14^{\circ}}_{-16^{\circ}}$
            & $100^{\circ}<\delta_{DK}<163^{\circ}$
            & $4^{\circ}$ & $23^{\circ}$ \\
  $\delta_{D^*K}$  & $343^{\circ}\;^{+20^{\circ}}_{-22^{\circ}}$
            & $293^{\circ}<\delta_{DK}<389^{\circ}$
            & $4^{\circ}$ & $23^{\circ}$ \\
  \hline
  \end{tabular}
\end{table*}

In contrast to the Belle analysis, the BaBar analysis based on a data sample
of 383M $B\overline{B}$ pairs~\cite{Aubert:2008bd} includes seven different
decay modes: \bdk, \bdsk\ with $D^0\to D\pi^0$ and $D\gamma$, and \bdks,
where the neutral $D$ meson is reconstructed in $K^0_S\pi^+\pi^-$
and $K^0_SK^+K^-$ (except for \bdks\ mode) final states. The signal
yields for these modes are shown in Tab.~\ref{babar_yield}.

\begin{table}
  \caption{
    Signal yields of different modes used for Dalitz analysis by
    BaBar collaboration~\cite{Aubert:2008bd}.
  }
  \label{babar_yield}
  \begin{tabular}{|l|l|l|}
    \hline
    $B$ decay & $D$ decay & Yield \\
    \hline
    \bdk\                             & \dkpp    & $600\pm 31$ \\
                                       & \dkkk    & $112\pm 13$ \\
    $B^{\pm}\to [D\pi^0]_{D^{*}}K^{\pm}$ & \dkpp   & $133\pm 15$ \\
                                       & \dkkk    & $32\pm 7$ \\
    $B^{\pm}\to [D\gamma]_{D^{*}}K^{\pm}$ & \dkpp   & $129\pm 16$ \\
                                        & \dkkk    & $21\pm 7$ \\
    $B^{\pm}\to DK^{*\pm}$              & \dkpp   & $118\pm 18$ \\
    \hline
  \end{tabular}
\end{table}


The differences from the Belle model of \dkpp\ decay are as follows:
the K-matrix formalism~\cite{Wigner:1946zz,Aitchison:1972ay,Anisovich:2002ij}
is used to describe the $\pi\pi$ $S$-wave, while the $K\pi$ $S$-wave is
parametrized using $K^*_0(1430)$ resonances and an effective range nonresonant
component with a phase shift~\cite{Aston:1987ir}.
The description of \dkkk\ decay amplitude uses an isobar model that includes
eight two-body decays: $K^0_S a_0(980)^0$, $K^0_S \phi(1020)$,
$K^0_S f_0(1370)$, $K^0_S f_2(1270)^0$, $K^0_S a_0(1450)^0$,
$K^- a_0(980)^+$, $K^+ a_0(980)^-$, and $K^- a_0(1450)^+$. 



The fit to signal samples is performed in a similar way to the Belle analysis,
using an unbinned likelihood function that includes Dalitz plot variables and in addition  $B$ meson invariant mass and event-shape variables to better separate signal and background events.
From the combination of all modes, BaBar obtains
$\gamma=(76^{+23}_{-24}\pm 5\pm 5)^{\circ}$ (mod 180$^{\circ}$),
where the first error is statistical, the second is experimental systematic,
and the third is the $D^0$ model uncertainty.
The values of the amplitude ratios are $r_B=0.086\pm 0.035\pm 0.010\pm 0.011$
for \bdk, $r^*_B=0.135\pm 0.051\pm 0.011\pm 0.005$ for \bdsk, and
$\kappa r_s=0.163^{+0.088}_{-0.105}\pm 0.037\pm 0.021$
for \bdks\ (here $\kappa$ accounts for possible nonresonant \bdksnr\
contribution). The combined significance of direct CP violation is
99.7\%, or 3.0 standard deviations.
%

Summaries of measurements of observables with the Dalitz plot method are given
in Figs.~\ref{fig:hfag-dalitz} and~\ref{fig:hfag-dalitz2}.

\begin{figure}
  \centering
  \includegraphics[width=0.32\textwidth]{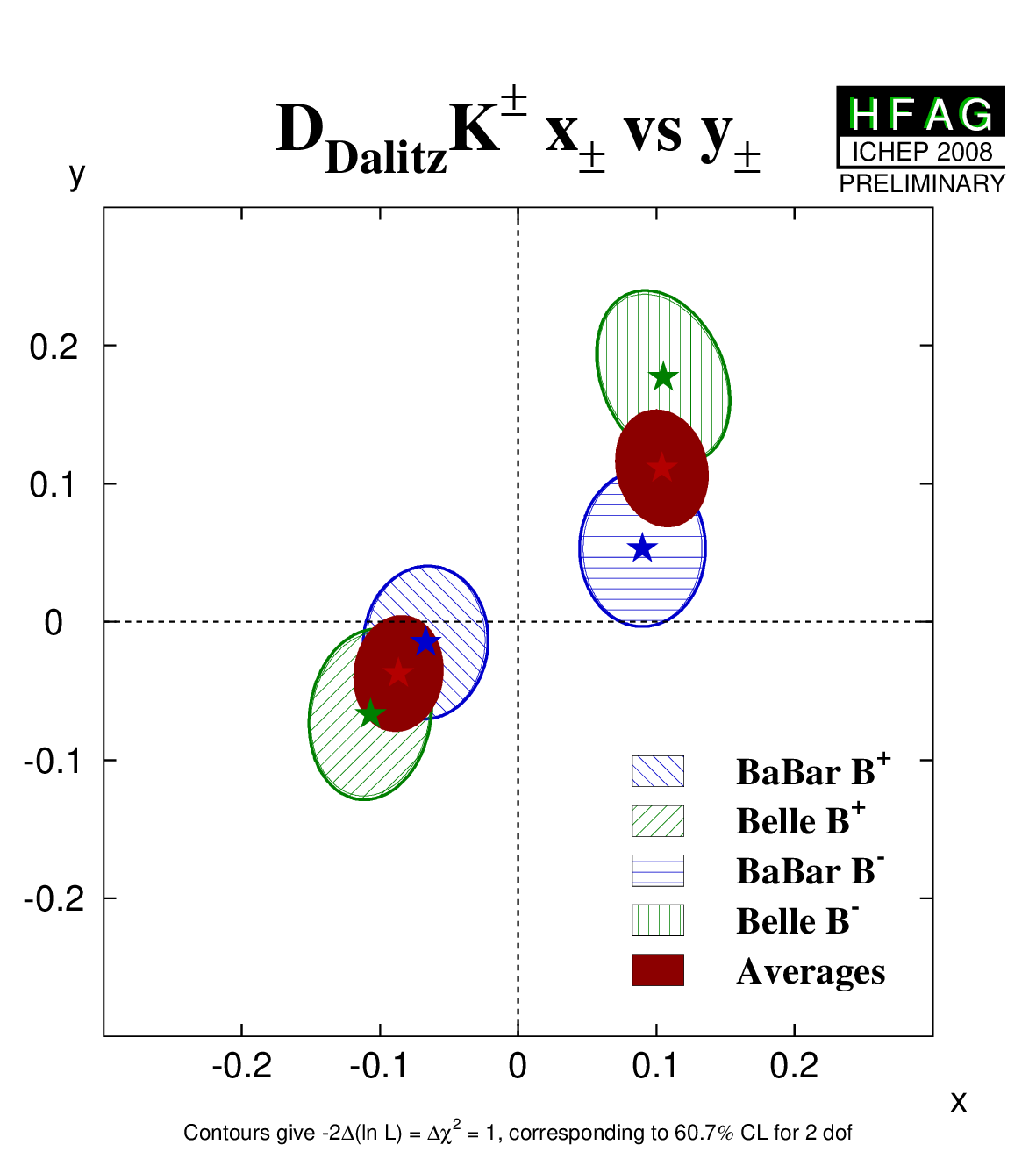}
  \hfill
  \includegraphics[width=0.32\textwidth]{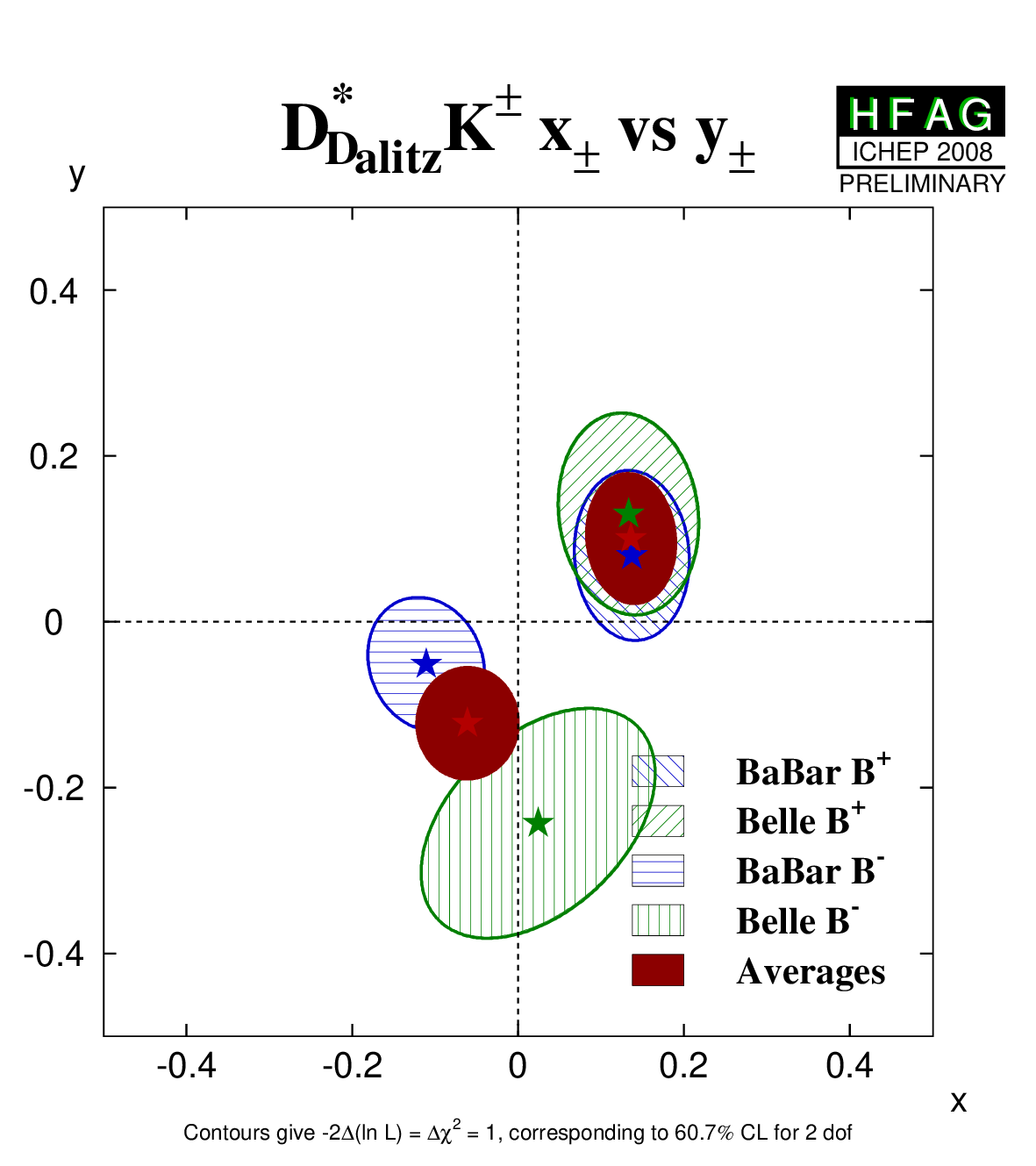}
  \hfill
  \includegraphics[width=0.32\textwidth]{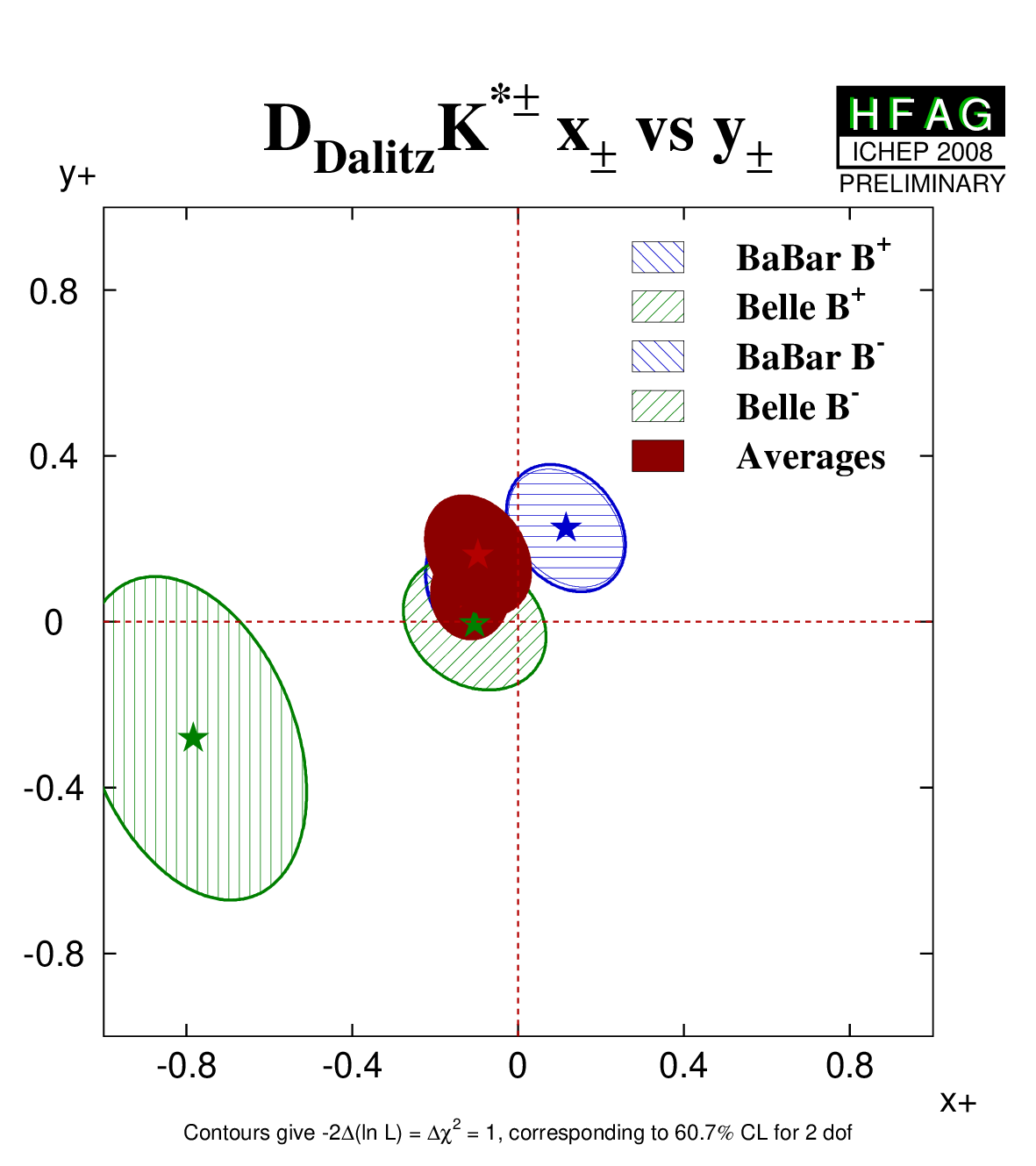}
  \caption{
    World averages of measurements of observables in the Cartesian
    parametrization of the Dalitz method.
    Left: $(x_\pm, y_\pm)$ for $B \to DK$;
    (middle): $(x_\pm, y_\pm)$ for $B \to D^*K$
    ($D^*\to D\pi^0$ and $D^*\to D\gamma$ combined);
    (right): $(x_\pm, y_\pm)$ for $B \to DK^*$.
    The Belle results use only $D \to K_S^0\pi^+\pi^-$, while the BaBar
    results include also $D \to K_S^0K^+K^-$.
    The averages do not include model uncertainities.
  }
  \label{fig:hfag-dalitz}
\end{figure}

\begin{figure}
  \centering
  \includegraphics[width=0.44\textwidth]{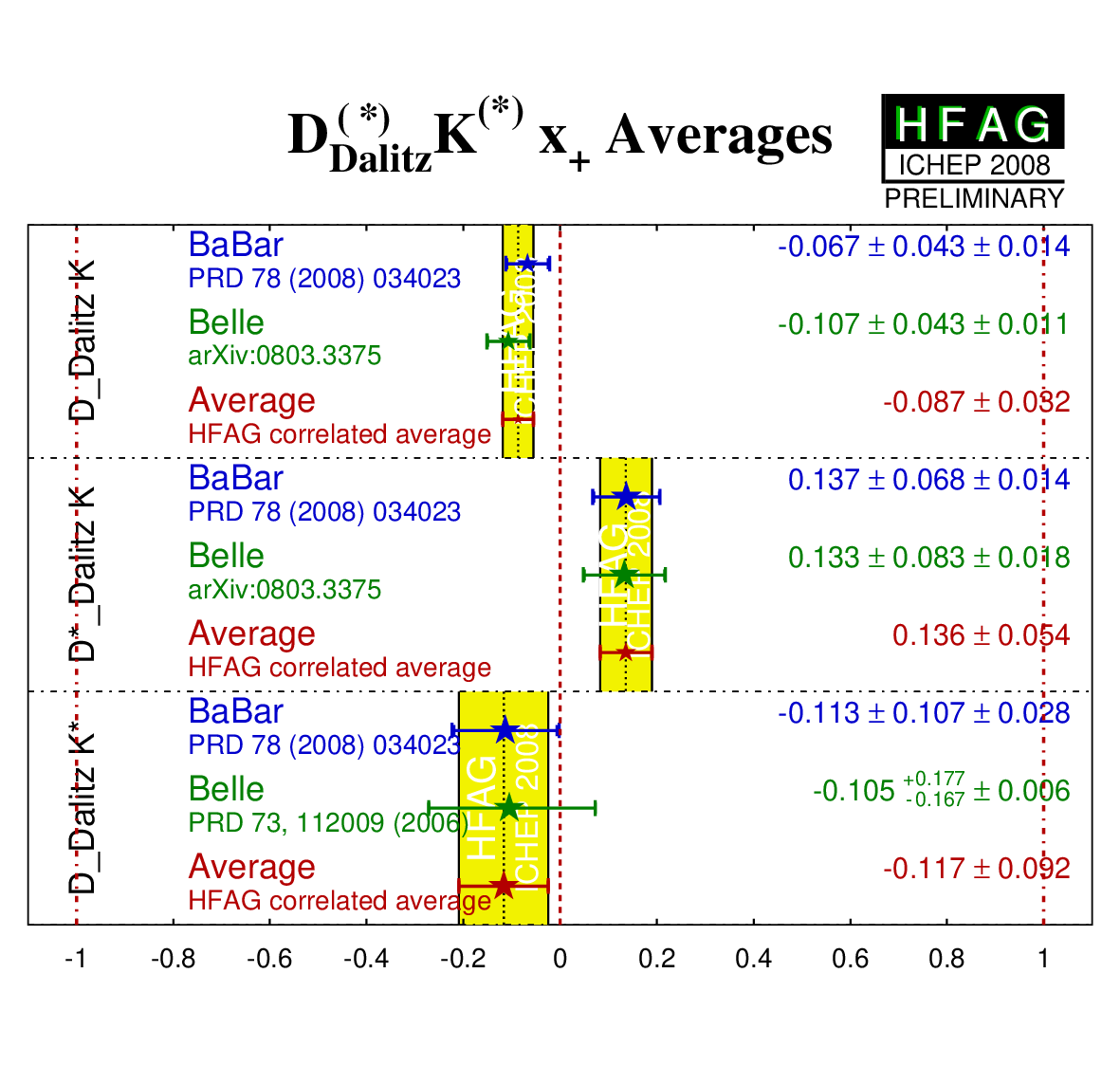}
  \hfill
  \includegraphics[width=0.44\textwidth]{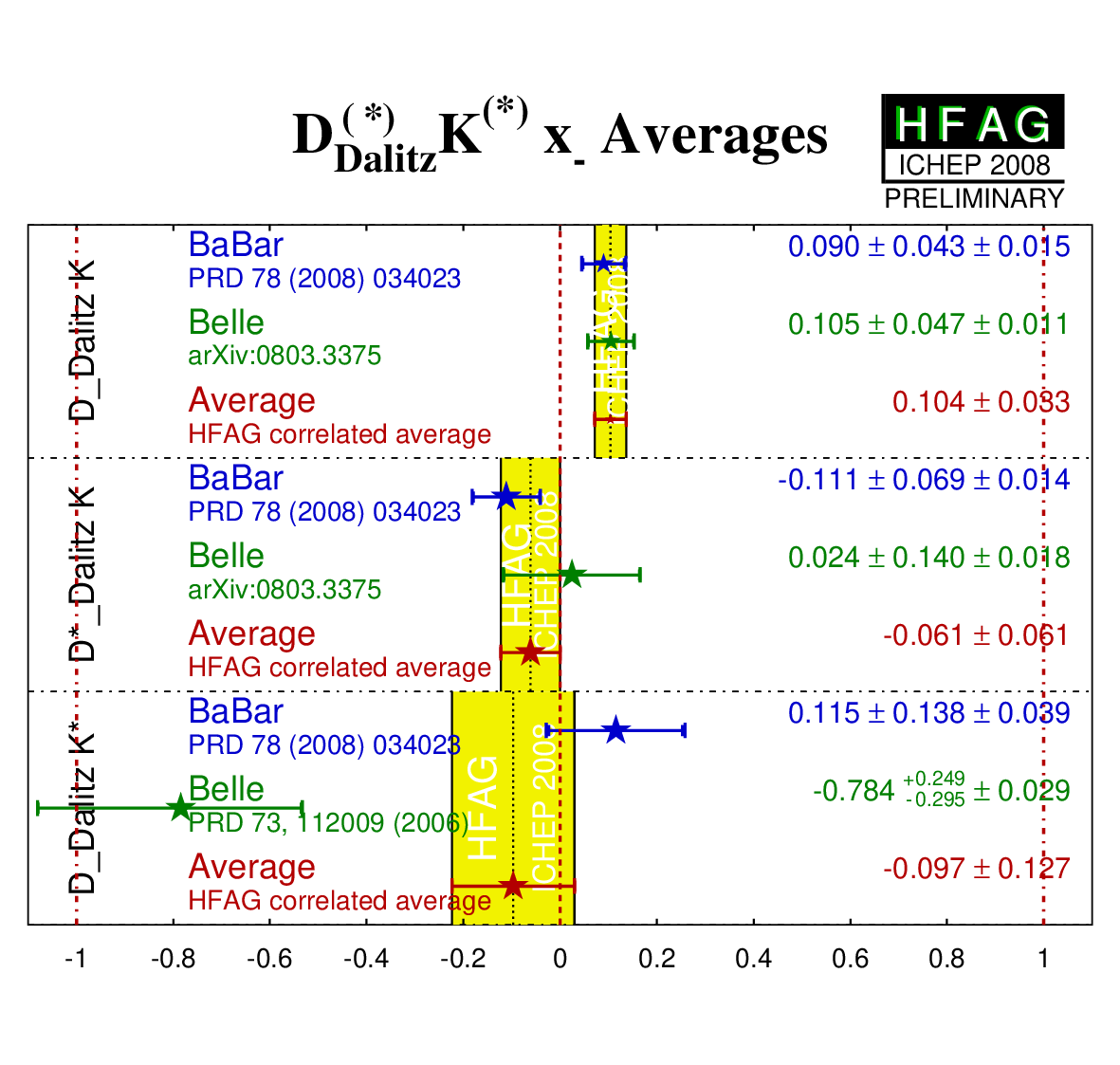}

  \includegraphics[width=0.44\textwidth]{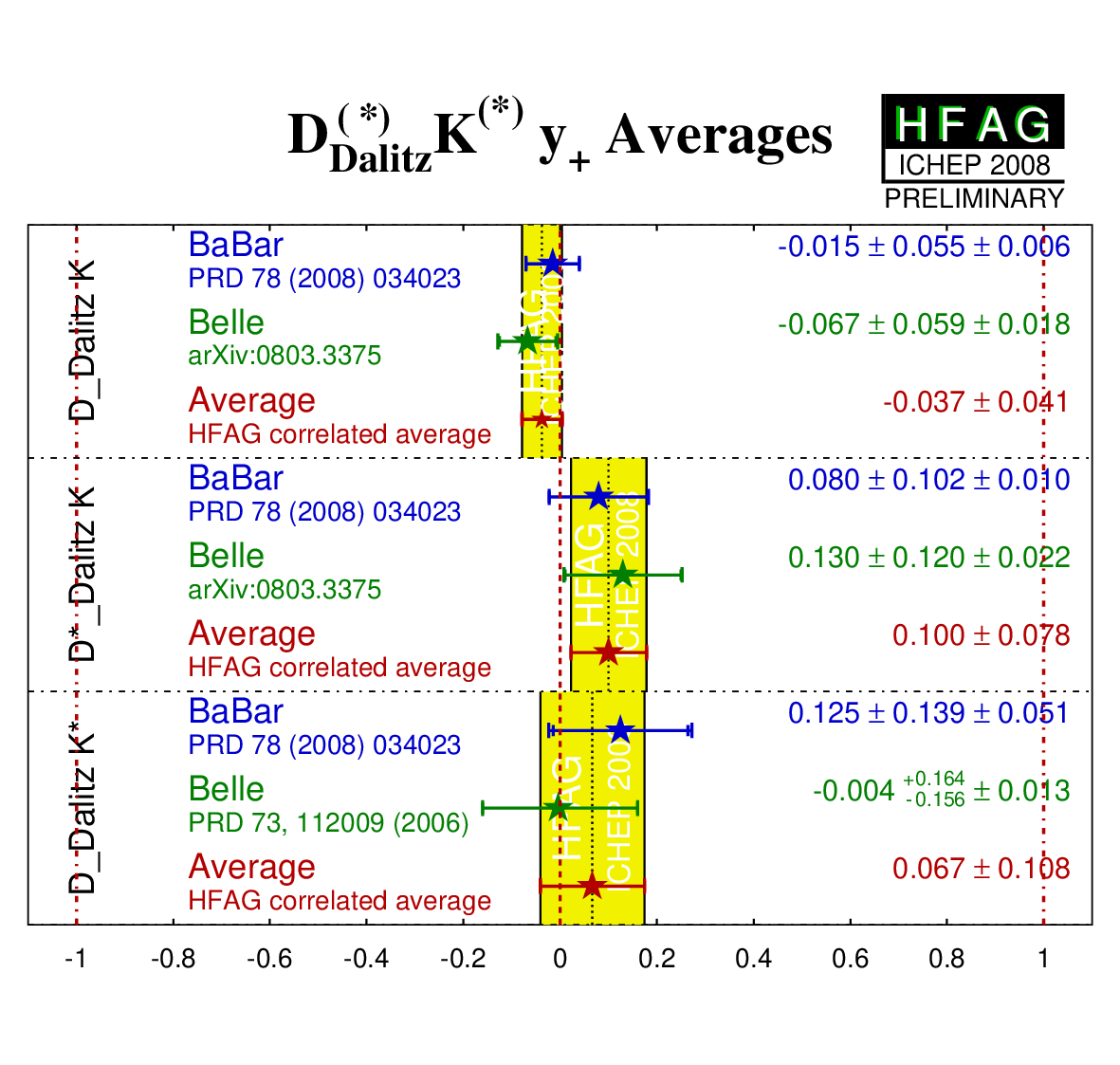}
  \hfill
  \includegraphics[width=0.44\textwidth]{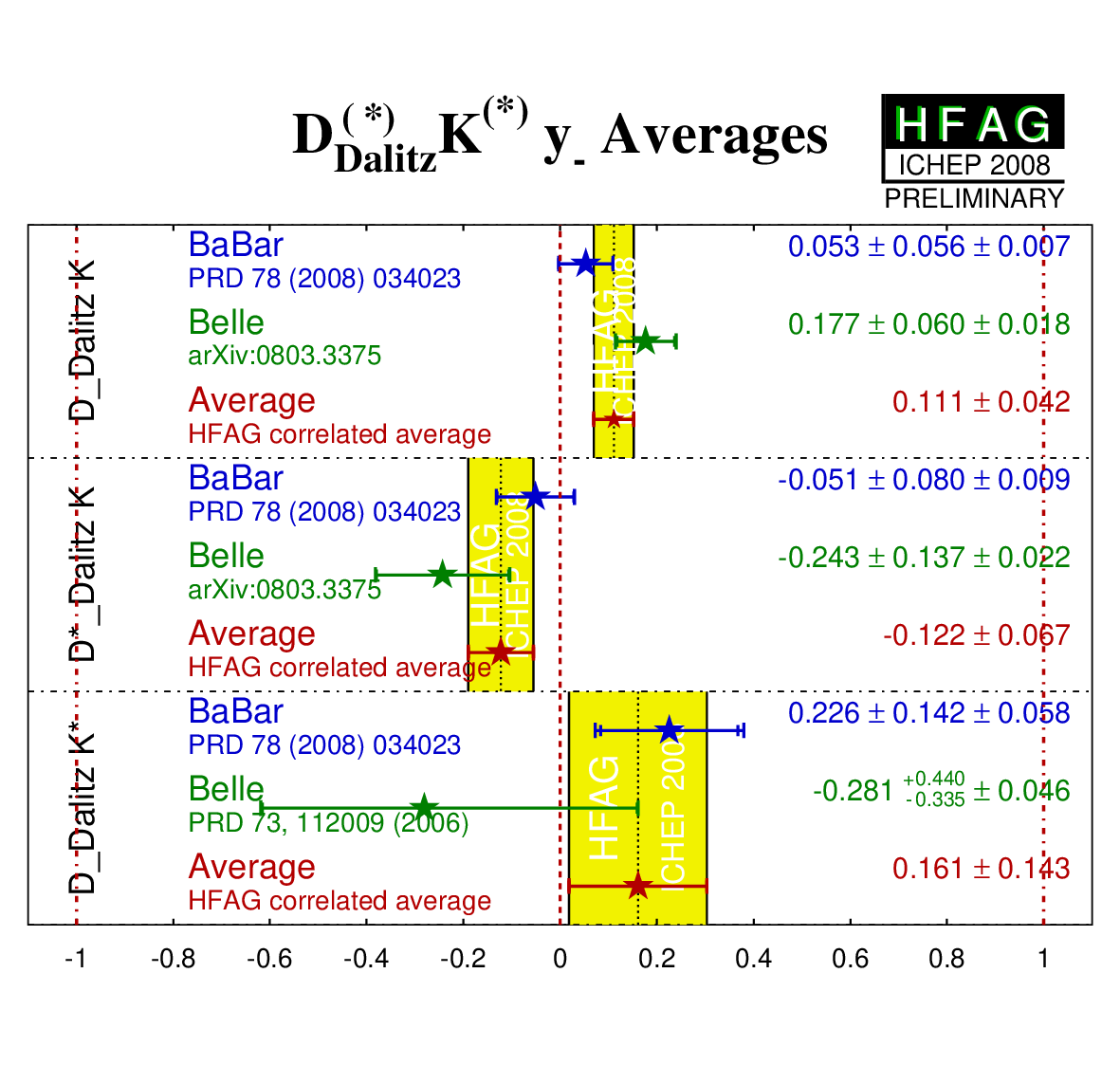}
  \caption{
    World averages of measurements of observables in the Cartesian
    parametrization of the Dalitz method from HFAG~\cite{Barberio:2008fa}.
    Top left: $x_+$; top right: $x_-$;
    bottom left: $y_+$; bottom right: $y_-$.
    The data is described in the caption to Fig.~\ref{fig:hfag-dalitz}.
  }
  \label{fig:hfag-dalitz2}
\end{figure}

\subsubsection{Other techniques}
\label{subsubsec:other}


In decays of neutral $B$ mesons to final states such as $DK$ both amplitudes
involving $D^0$ and $\overline{D}{}^0$ are color-suppressed.
Consequently, the value of $r_B$ is larger, with na\"ive estimates giving
$r_B\sim 0.4$.
In the decay $B^0\to DK^*(892)^0$ the flavor of the $B$ meson is tagged by the
charge of the Kaon produced in the $K^*(892)^0$ decay ($K^+\pi^-$ or
$K^-\pi^+$)~\cite{Dunietz:1991yd}, so that a time-dependent analysis is
not necessary.

Searches for doubly Cabibbo-suppressed decays have not yet yielded a
significant signal, but allow limits to be put on $r_B$.  The most recent
results are from BaBar using a data sample of 465M $B\overline{B}$
pairs~\cite{:2009au}. 
BaBar has studied $D\to K\pi$, $D\to K\pi\pi^0$ and $D\to K\pi\pi\pi$, and has found
${\mathcal R}_{\rm ADS}(K\pi) < 0.244$ at the 95\% confidence level.
The results can be combined using external information from
CLEO-c~\cite{Rosner:2008fq,Asner:2008ft,Lowery:2009id} to obtain 
$r_S \in \left[ 0.07, 0.41 \right]$ at the 95\% confidence level,
where $r_S$ is the equivalent of the parameter $r_B$ modified due to the
finite width of the $K^{*0}$ resonance~\cite{Gronau:2002mu}.

BaBar have also performed a Dalitz plot analysis of the three-body decay
\dkpp\ decay in $B^0\to DK^*(892)^0$~\cite{Aubert:2008yn}.  The technique, and
the decay model are similar to that used for \bdks\ decays (see
Sec.~\ref{subsubsec:dalitz}). The analysis is based on 371M $B\overline{B}$
pairs, and yields the following
constraints: $\gamma=(162\pm 56)^{\circ}$, $r_B<0.55$ with 90\% CL.


It is also possible to measure $\gamma$ by exploiting the interference between
$b \to c$ and $\overline{b} \to \overline{u}$ decays that occurs due to
$B^0$--$\overline{B}{}^{0}$ mixing using a time-dependent analysis.
Since the interference occurs via oscillations, the mixing phase is also
involved and the analysis is sensitive to the combination of angles
$\sin(2\beta+\gamma)$.
In this approach, the abundant decays such as $B \to D\pi$ and $B \to D^*\pi$
can be used; however the size of the CP violation effect depends on the
magnitude of the ratio of the $b \to u $ over $b \to c$ amplitudes, usually denoted $R$, which is
na\"ively expected to take values $R \sim 0.02$ for these decays.
Consequently these measurements are still statistics limited, as well as being
potentially sensitive to systematics caused by any mismodelling of the large
CP-conserving component.
The statistical precision can be improved by using partial reconstruction for
$B \to D^*\pi$ decays as well as the more conventional ``full'' reconstruction.
A summary of measurements of these modes from
BaBar~\cite{Aubert:2005yf,Aubert:2006tw} and Belle~\cite{:2008kr,Ronga:2006hv}
is given in Fig.~\ref{fig:hfag-dpi}.

\begin{figure}
  \centering
  \includegraphics[width=0.44\textwidth]{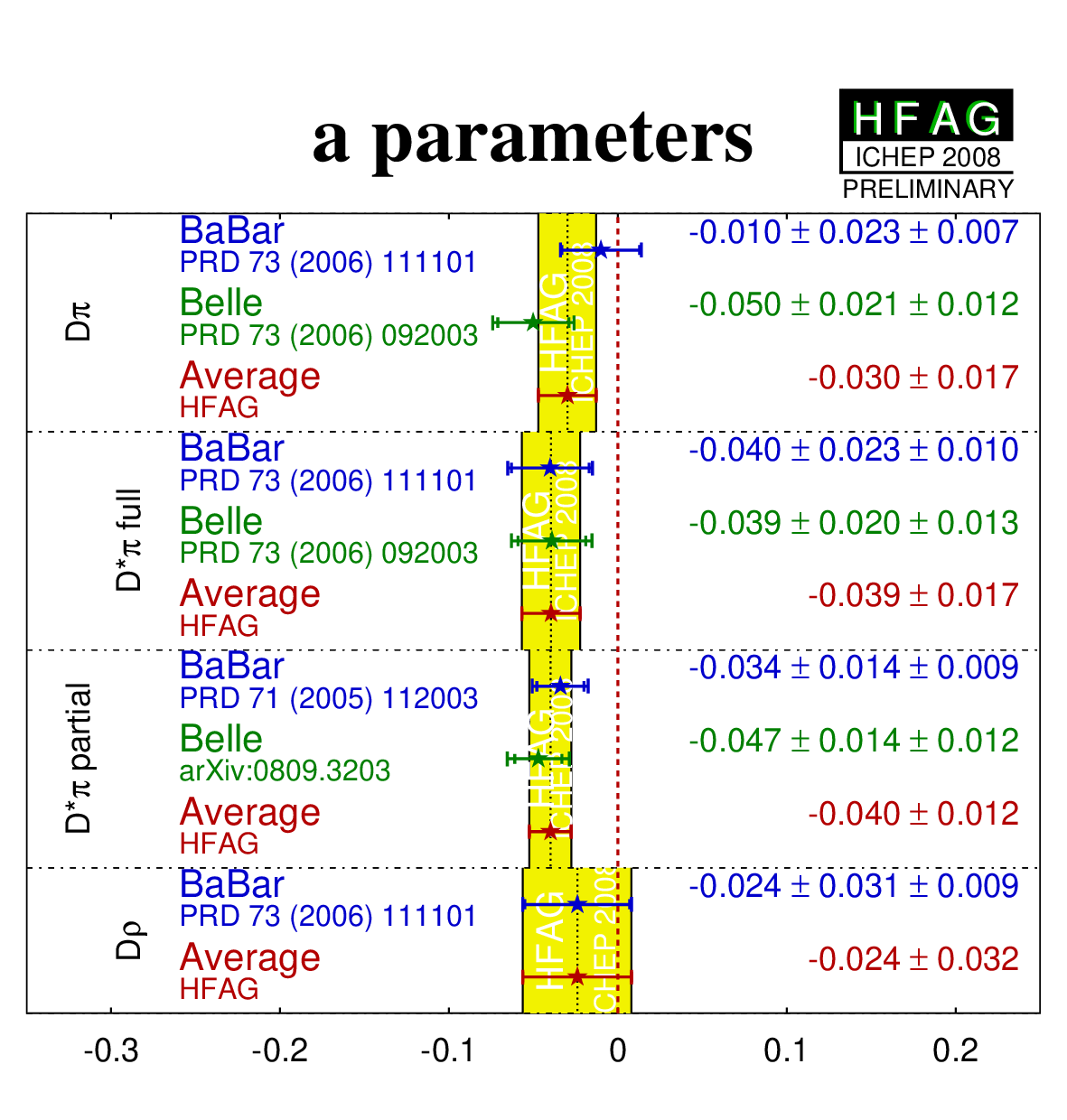}
  \hfill
  \includegraphics[width=0.44\textwidth]{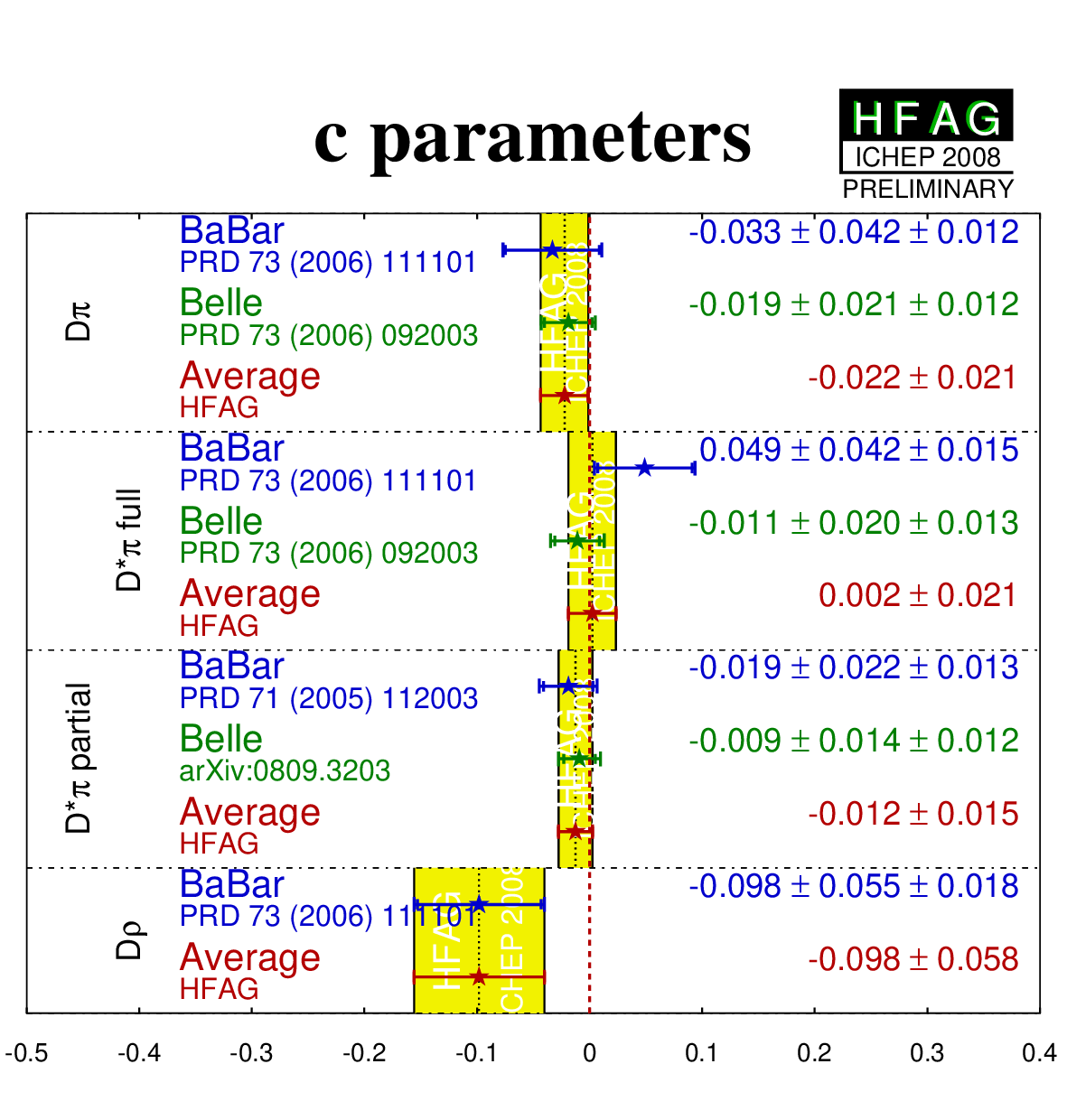}
  \caption{
    Measurements of observables in $B \to D\pi$ and similar final states.
    The parameters used in these compilations are
    $a = (-1)^{L+1} 2 R \sin(2\beta+\gamma)\cos(\delta)$ and
    $c = (-1)^{L+1} 2 R \cos(2\beta+\gamma)\sin(\delta)$,
    where $L$ is the angular momentum in the decay
    ($+1$ for $D\pi$ and $-1$ for $D^*\pi$ and $D\rho$),
    $R$ is the magnitude of the ratio of the $\overline{b} \to \overline{u}$
    and $b \to c$ amplitudes and $\delta$ is their relative phase.
  }
  \label{fig:hfag-dpi}
\end{figure}


Another similar neutral $B$ decay mode is $B^0\to D^{\mp}K^0\pi^{\pm}$,
where time-dependent Dalitz plot analysis is sensitive to
$2\beta+\gamma$~\cite{Aleksan:2002mh,Polci:2006aw}.
One advantage of this technique compared to the methods based
on $B^0\to D^{(*)}\pi$ decays is that, since both
$b\to c$ and $b\to u$ diagrams involved in this decay are color-suppressed,
the expected value of the ratio of their magnitudes $R$ is larger.
Secondly, $2\beta+\gamma$ is measured with only a two-fold ambiguity
(compared to four-fold in $B^0\to D^{(*)}\pi$ decays). In addition,
all strong amplitudes and phases can be, in principle, measured in the
same data sample.

The BaBar collaboration has performed the analysis based on 347M
$B\overline{B}$ pairs data sample~\cite{Aubert:2007qe}.
The $B^0\to D^{\mp}K^0\pi^{\pm}$ Dalitz plot is found to be dominated by
$B^0\to D^{**0}K^0_S$ (both $b\to u$ and $b\to c$ transitions) and
$B^0\to D^-K^{*+}$ ($b\to c$ only) states.
From an unbinned maximum likelihood fit to the time-dependent Dalitz
distribution, the value of $2\beta+\gamma$ as a function of $R$ is
obtained. The value of $R$ cannot be determined with the current data sample,
therefore, the value $R=0.3$ is used, and its uncertainty is taken into
account in the systematic error.
This results in the value $2\beta+\gamma=(83\pm 53\pm 20)^{\circ}$ or
$(263\pm 53\pm 20)^{\circ}$.




\subsection{Outlook on the $\gamma$ measurement }
\label{subsec:outlookgamma}
The world average values that include the latest measurements
presented in 2008 are reported in Sec.~\ref{section:globalfits}. 


For an evaluation of the prospect of $\gamma$ measurement,
it is essential to note the fact that for the first time the value of $r_B$ is
shown to be significantly non-zero. In previous measurements, poor
constraints on $r_B$ caused significantly non-gaussian errors for
$\gamma$, and made it difficult to predict the future sensitivity of
this parameter. Now that $r_B$ is constrained to be of the order
0.1, one can confidently extrapolate the current precision to future
measurements at LHCb and Super-B facilities.

%

The $\gamma$ precision is mainly dominated by Dalitz analyses. These
analyses currently suffer from a hard-to-control uncertainty due to the $D^0$
decay amplitude description, which is estimated to be
5--10$^{\circ}$. At the current level of statistical precision this
error starts to influence the total $\gamma$ uncertainty. A solution
to this problem can be the use of quantum-correlated $D\overline{D}$
decays at $\psi(3770)$ resonance available currently at CLEO-c
experiment, where the missing information about the strong phase in
$D^0$ decay can be obtained experimentally.

\subsubsection{Model-independent Method}
\label{subsubsec:modelindependent}

Giri {\it et al.} proposed \cite{Giri:2003ty} a model-independent
procedure for obtaining $\gamma$, as follows. The Dalitz plot is
divided into $2\mathcal N$ bins, symmetrically about the line
$m^2_+=m^2_-$. The bins are indexed from $-i$ to $i$, excluding
zero. The coordinate transformation $m^2_+\leftrightarrow m^2_-$
thus corresponds to the exchange of bins $i\leftrightarrow -i$. The
number of events in the $i$-th bin of a flavor-tagged $D^0$ decay
$K_S^0\pi^+\pi^-$ Dalitz plot is then expressed as:

\begin{equation}
K_i = A_D\int_{i}|f_D(m^2_+,m^2_-)|^2dm^2_+dm^2_- = A_DF_i,
\label{eq:ki}
\end{equation}

\noindent where $A_D$ is a normalization factor. The coefficients $K_i$ can be
obtained precisely from a very large sample of $D^0$ decays reconstructed in
flavor eigenstate, which is accessible at $B$-factories, for example. The
interference between the $D^0$ and $\bar D^0$ amplitudes is parametrized by
the quantities $c_i$ and $s_i$:

\begin{eqnarray}
c_i &\equiv& \frac{1}{\sqrt{F_iF_{-i}}}
\int_i|f_D(m^2_+,m^2_-)||f_D(m^2_-,m^2_+)|\cos[\Delta\delta_D(m^2_+,m^2_-)]dm^2_+dm^2_-,
\label{eq:ci}\\
s_i &\equiv& \frac{1}{\sqrt{F_iF_{-i}}} \int_i
|f_D(m^2_+,m^2_-)||f_D(m^2_+,m^2_-)|\sin[\Delta\delta_D(m^2_+,m^2_-)]dm^2_+dm^2_-,
\label{eq:si}
\end{eqnarray}

\noindent where the integral is performed over a single bin. The
quantities $c_i$ and $s_i$ are the amplitude-weighted averages of
$\cos{\Delta\delta_{D}}$ and $\sin{\Delta\delta_{D}}$ over each
Dalitz-plot bin. The expected number of events in the bins of the
Dalitz plot of the $D$ decay from \bdk\ is
\begin{equation}
  \langle N_i\rangle = 
  A_B[K_i + r_B^2K_{-i} + 2\sqrt{K_iK_{-i}}(x_\pm c_i+y\pm s_i)],
  \label{n_b}
\end{equation}
where $A_B$ is the normalization constant. As soon as the $c_i$ and
$s_i$ coefficients are known, one can obtain $x_\pm$ and $y_\pm$ values (and
hence $\gamma$ and other related quantities) by a maximum likelihood
fit using equation (\ref{n_b}). In principle, $c_i$ and $s_i$ can be
left as free parameters in a $\tilde D^0\to K_S^0\pi^+\pi^-$
Dalitz-plot analysis from $B^{\pm}$ decays. However, it has been
shown \cite{Bondar:2005ki} that almost infinite statistics of $B$
decays is necessary in that case.

It is important to note that $c_i$ and $s_i$ depend only on the
$D^0$ decay, not the $B$ decay, and therefore these quantities can
be measured using the quantum-correlated $D\overline{D}$ decays of
the $\psi(3770)$ resonance. For example, the expected number of
events in a bin of the Dalitz plot of $D_{CP}$ tagged decays equals
\begin{equation}
  \langle M_i\rangle^\pm = A_{CP}^\pm [K_i + K_{-i} \pm 2\sqrt{K_iK_{-i}}c_i],
\end{equation}
where the $\pm$ indicates  whether the CP tag is CP-even or CP-odd.
This relation can be used to obtain the $c_i$ coefficients, but obtaining
$s_i$ remains a problem. If the binning is fine enough, so
that both the phase difference $\Delta\delta_D$ and the amplitude
$|f_D|$ remain constant across the area of each bin, the expressions
(\ref{eq:ci},\ref{eq:si}) reduce to $c_i=\cos(\Delta\delta_D)$ and
$s_i=\sin(\Delta\delta_D)$. The $s_i$ coefficients can be obtained
as $s_i=\pm\sqrt{1-c_i^2}$. Using this equality if the amplitude
varies across a bin will lead to bias in the $x_\pm,y_\pm$ fit results.
Since $c_i$ is obtained directly, and the absolute value of $s_i$ is
overestimated, the bias will mainly affect $y_\pm$ determination,
resulting in lower absolute values of $y_\pm$.

A unique possibility to find $s_i$ independent of $c_i$ is available
in a sample where both $D$ mesons from the $\psi(3770)$ decay into
the $K_S^0\pi^+\pi^-$ state~\cite{Bondar:2008hh}. Since the
$\psi(3770)$ is a vector, the two $D$ mesons are produced in a $P$-wave,
and the wave function of the two mesons is antisymmetric. Then the
four-dimensional density of the two correlated Dalitz plots is given
by:
\begin{equation}
\begin{split}
  \langle M\rangle_{ij} = A_{\rm corr}[&K_i K_{-j} + K_{-i} K_j - \\
    &2\sqrt{K_iK_{-i}K_jK_{-j}}(c_i c_j + s_i s_j)].
\end{split}
\end{equation}
The indices $i,j$ correspond to the two $D$ mesons from $\psi(3770)$
decay. This decay is sensitive to both $c_i$ and $s_i$ for the price
of having to deal with the four-dimensional phase space.

The original idea of Giri {\it et al.} was to divide the Dalitz plot
into square bins \cite{Giri:2003ty}. In case of limited statistics
unavoidably the number of the bins could be relatively small.
Consequently, a large loss of sensitivity can be expected due to
variation of amplitude and phase over the bin. Bondar {\it et al.}
noted \cite{Bondar:2008hh} that increased sensitivity can be
obtained if the bins are chosen to minimize the variation in
$\Delta\delta_D$ over each bin. One can divide the Dalitz phase
space into ${\mathcal N}$ bins of equal size with respect to
$\Delta\delta_{D}$ as predicted, for example, by the BaBar isobar
model \cite{Aubert:2008bd}. In the half of the Dalitz plot
$m^2_+<m^2_-$, the $i^{th}$ bin is defined by the condition

\begin{equation}
2\pi(i-3/2)/{\mathcal N} < \Delta \delta_D(m^2_+,m^2_-) <
2\pi(i-1/2)/{\mathcal N},
\end{equation}

\noindent The $-i^{th}$ bin is defined symmetrically in the lower
portion of the Dalitz plot. Such a binning with $\mathcal N=8$ is
shown in Fig.~\ref{fig:phasebin}. One might suspect that, since we
are using a model to determine our bins, we are not free of model
dependence. In fact {\it any} binning is acceptable in that it will
give a correct, unbiased answer for $\gamma$, at the cost of larger
uncertainties compared to an optimal binning with respect to
$\Delta\delta_{D}$.

\begin{figure}
\centering
\includegraphics[width=8cm]{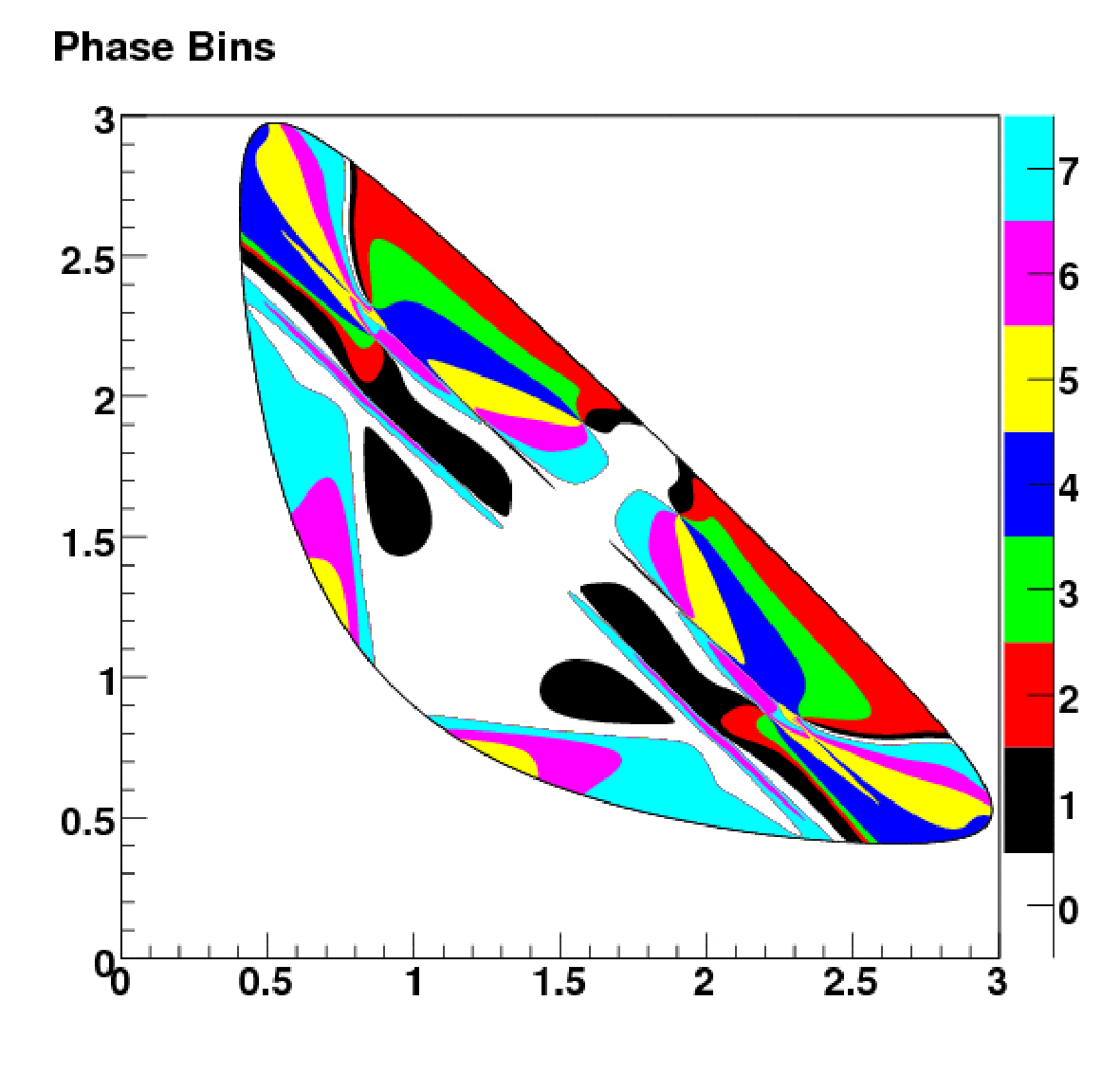}
\caption{Phase binning of the $D^0 \to \bar K_S^0 \pi^+\pi^-$ Dalitz
plot.} \label{fig:phasebin}
\end{figure}

\begin{table}[t]
\centering \caption{Fit results for $c_i$ and $s_i$. The first error
is statistical, the second error is the systematic uncertainty, the
third error is the model uncertainty due to including
$K^0_L\pi^+\pi^-$ events in the analysis.}
\begin{tabular}{ccc}        \hline\hline
 $i$     & $c_i$  & $s_i$  \\\hline
0 & $\hphantom{-}0.743 \pm 0.037 \pm 0.022 \pm 0.013$ & $\hphantom{-}0.014 \pm 0.160 \pm 0.077 \pm 0.045$ \\
1 & $\hphantom{-}0.611 \pm 0.071 \pm 0.037 \pm 0.009$ & $\hphantom{-}0.014 \pm 0.215 \pm 0.055 \pm 0.017$ \\
2 & $\hphantom{-}0.059 \pm 0.063 \pm 0.031 \pm 0.057$ & $\hphantom{-}0.609 \pm 0.190 \pm 0.076 \pm 0.037$ \\
3 & $-0.495 \pm 0.101 \pm 0.052 \pm 0.045$ & $\hphantom{-}0.151 \pm 0.217 \pm 0.069 \pm 0.048$ \\
4 & $-0.911 \pm 0.049 \pm 0.032 \pm 0.021$ & $-0.050 \pm 0.183 \pm 0.045 \pm 0.036$ \\
5 & $-0.736 \pm 0.066 \pm 0.030 \pm 0.018$ & $-0.340 \pm 0.187 \pm 0.052 \pm 0.047$ \\
6 & $\hphantom{-}0.157 \pm 0.074 \pm 0.042 \pm 0.051$ & $-0.827 \pm 0.185 \pm 0.060 \pm 0.036$ \\
7 & $\hphantom{-}0.403 \pm 0.046 \pm 0.021 \pm 0.002$ & $-0.409 \pm 0.158 \pm 0.050 \pm 0.002$ \\
\hline\hline \label{table:ci_si-final}
\end{tabular}
\end{table}

Using 818 pb$^{-1}$ of $e^+e^-$ collisions produced at the $\psi(3770)$, the
CLEO-c collaboration has made a first determination~\cite{Briere:2009aa} of
the strong phase parameters, $c_i$ and $s_i$, 
which are listed in Tab.~\ref{table:ci_si-final}. From a toy Monte
Carlo study with a large sample of $B^{\pm}\to \tilde D^0 K^{\pm}$
data generated with $\gamma=60^\circ$, $\delta_B=130^\circ$ and
$r_B=0.1$, CLEO found that the decay model uncertainty on $\gamma$
is reduced to about $1.7^\circ$ due to these new measurements. As a result,
the precision of the $\gamma$ measurement using $B^+\to\tilde D^0 K^+$ decays
will not be limited by model-dependent assumptions on strong phase behavior
in the $\tilde D^0\to K^0_S\pi^+\pi^-$ decay.


\subsubsection{Prospects for LHCb}
\label{subsec:lhcbgamma}

The measurement of the CKM angle $\gamma$ in tree dominated processes is one
of the principal goals of LHCb.  Extensive simulation studies have been
conducted in a variety of channels.  The results summarized here derive
from~\cite{Akiba:2008zza} and references therein.

LHCb will measure $\gamma$ in tree dominated processes using two main
approaches: 
\begin{enumerate}
\item{{\bf Time-dependent measurements}\hspace*{0.1cm}  The extraction of $\gamma$ has been
studied using  both $B^0 \to D^\mp \pi^\pm$ and $B_s \to D_s^\mp K^\pm$.
Although the CP-asymmetries in these modes involve a contribution
arising from the mixing diagram, this contribution can be subtracted using the result
from complementary measurements in other processes, allowing for a pure tree-level $\gamma$
determination.}
\item{{\boldmath $B \to DK$ \bf strategies} \hspace*{0.1cm}  The modes that have so far been investigated 
which have significant weight in the $\gamma$ measurement include $B^\mp \to DK^\mp$, with the
neutral $D$ reconstructed in the $K^+K^-$, $\pi^+\pi^-$,  $K^\mp \pi^\pm$, $K^\mp \pi^\pm \pi^+ \pi^-$ and
$K^0_S \pi^+ \pi^-$ final states,
and $B^0 \to D(K^\pm \pi^\pm, K^+K^-, \pi^+\pi^-) K^{\ast 0} (K^-\pi^+)$ (+c.c.).
The fact that no initial-state flavor tagging is required means that the relative
sensitivity of the  $B \to DK$ method is particularly high at LHCb compared with
time-dependent measurements, in which the tagging power is in general lower than
is the case at  $\Upsilon(4S)$ experiments.}
\end{enumerate}
The expected yields in 2~$\rm fb^{-1}$ of data taking in these channels are given
in Tab.~\ref{tab:lhcb_yields}.  Note that the goal of the baseline LHCb experiment
is to accumulate around 10~$\rm fb^{-1}$ of integrated luminosity.  In all modes
the selection benefits from the good performance of the $\pi-K$ separation provided by 
the LHCb RICH system.

\begin{table}
\begin{center}
\caption[]{Summary of expected LHCb signal and background yields for  2~$\rm fb^{-1}$.
In those rows where more than one channel is specified 
(eg. $B^\pm \to D(K^\pm \pi^\mp) K^\pm$
or $B^+ \to D(K^+K^- + \pi^+\pi^-) K^+$),
the yields correspond to the {\it sum} over all indicated modes. 
The physics parameters assumed in calculating these numbers can be found in~\cite{Akiba:2008zza}.}
\label{tab:lhcb_yields}
\vspace{0.2cm}
\begin{tabular}{lrr}\hline \hline
Channel & Signal & Background \\ \hline
$B^\pm \to D(K^\pm \pi^\mp) K^\pm$ & 56k  & 35k \\
$B^+ \to D(K^- \pi^+) K^+$ &  680 & 780 \\
$B^- \to D(K^+ \pi^-) K^-$ &  400 & 780 \\
$B^+ \to D(K^+K^- + \pi^+\pi^-) K^+$ & 3.3k & 7.2k \\
$B^- \to D(K^+K^-  + \pi^+\pi^-) K^-$ &  4.4k &  7.2k \\
$B^\pm \to D(K^\pm \pi^\mp \pi^+ \pi^-) K^\pm$ & 61k & 40k \\
$B^+ \to D(K^- \pi^+ \pi^+ \pi^-) K^+$ & 470 & 1.2k \\
$B^- \to D(K^+ \pi^- \pi^+ \pi^-) K^-$ & 350 & 1.2k \\
$B^0 \to D(K^+\pi^-) K^{*0}$, $\bar{B^0} \to D(K^- \pi^+) \bar{K}^{*0}$   &  3.4k &  1.7k\\
$B^0 \to D(K^- \pi^+) K^{*0}$ &  350 &  850 \\
$\bar{B^0} \to D(K^+ \pi^-) \bar{K}^{*0}$ &   230 &  850 \\
$B^0 \to D(K^+K^- + \pi^+\pi^-) K^{*0}$ &  150 &  500 \\
$\bar{B^0} \to D(K^+K^- + \pi^+\pi^-) \bar{K}^{*0}$ &  550  & 500 \\
$B^\pm \to D(K^0_S \pi^+\pi^-) K^\pm$ & 5k & 4.7k \\
$B_s, \bar{B_s} \to D_s^\mp K^\pm$ & 6.2k & 4.3k \\ 
$B^0, \bar{B^0} \to D^\mp \pi^\pm$ & 1,300k & 290k \\ \hline \hline
\end{tabular}
\end{center}
\end{table}

The physics processes underlying the event rates and kinematic distributions 
in the $B \to DK$ channels have many parameters in common.  This 
means that the observables for these channels may be combined
in a global fit to achieve the best possible sensitivity to these
parameters, most notably $\gamma$ itself.  The power of such a fit has
been investigated in a toy Monte Carlo study, taking as input the
expected sensitivities on the observables arising from the full simulation.
For the two and four body $D$ decay modes the observables are the event rates
in each mode;  for the $D \to K^0_S \pi^+\pi^-$ decay they are the
populations of bins in Dalitz space, as defined by the expected strong-phase
difference.  Important components
of this fit are the external constraints which come from the $D$ decay
properties from the quantum-correlated measurements at CLEO-c.  
These are the measured strong phase difference in $D \to K\pi$
decays~\cite{Rosner:2008fq,Asner:2008ft}, 
the measured coherence factor~\cite{Atwood:2003mj} and average strong phase
difference in $D \to K\pi\pi\pi$ decays~\cite{Lowery:2009id}, 
and the expected sensitivity on the cosine and sine of the strong phase
differences in the $D \to K^0_S \pi^+\pi^-$ Dalitz plot
bins~\cite{Briere:2009aa}.\footnote{
  Note that the results shown here take as input preliminary estimates of the
  CLEO-c sensitivity to the $D$-meson decay properties for both $D \to
  K\pi\pi\pi$ and $D \to K^0_S \pi^+\pi^-$.
}
The results of this fit have a dependence on the assumed values of the physics
parameters; the least well known of these is $\delta_{B^0}$, the strong phase
difference between the interfering diagrams in $B^0 \to D K^{\ast 0}$ decays,
and so in Tab.~\ref{tab:lhcb_btodk} the expected sensitivity on $\gamma$ is
shown as a function of this phase.  The CLEO-c inputs allow for a significant
improvement on the overall precision.

\begin{table}
\begin{center}
\caption[]{Expected LHCb sensitivity to $\gamma$ from $B\to DK$ strategies 
for data sets corresponding to integrated 
luminosities of 0.5, 2 and 10~$\mathrm{fb}^{-1}$, with and 
without CLEO-c constraints.}
\label{tab:lhcb_btodk}
\vspace{0.2cm}
\begin{tabular}{lccccc}\hline\hline 
$\delta_{B^{0}}~(^{\circ})$ & 0 & 45 & 90 & 135 & 180 \\ \hline
\multicolumn{6}{c}{$0.5~\mathrm{fb}^{-1}$} \\ \hline
$\sigma_{\gamma}$ without CLEO-c constraints  $(^{\circ})$  &  11.5 &	12.9 & 	13.1 & 	12.5 &	9.7 
\\
$\sigma_{\gamma}$ with CLEO-c constraints  $(^{\circ})$  &  9.0 &	12.0 & 	10.7 & 	11.1 &	8.6 \\ 
\hline
\multicolumn{6}{c}{$2~\mathrm{fb}^{-1}$} \\ \hline
$\sigma_{\gamma}$ without CLEO-c constraints $(^{\circ})$   &  5.8 &	8.3 & 	7.8 & 	8.4 &	5.0 
\\
$\sigma_{\gamma}$ with CLEO-c constraints $(^{\circ})$   &  4.6 &	6.1 & 	5.7 & 	6.0 &	4.3 \\ 
\hline
\multicolumn{6}{c}{$10~\mathrm{fb}^{-1}$} \\ \hline
$\sigma_{\gamma}$ without CLEO-c constraints $(^{\circ})$   &  2.6 &	5.4 & 	3.5 & 	4.8 & 2.4 
\\
$\sigma_{\gamma}$ with CLEO-c constraints $(^{\circ})$   &  2.3 &	3.5 & 	2.9 & 	3.2 &	2.2 \\
\hline\hline
\end{tabular}
\end{center}
\end{table}

The results from the global $B\to DK$ fit may be combined with the expected uncertainty on $\gamma$
from the time-dependent measurements, the most important of which is the analysis of 
$B_s \to D_s^\mp K^\pm$ decays.   The expected precision on $\gamma$ from all of these 
measurements is shown in Tab.~\ref{tab:lhcb_gamglobal}.  It can be seen that with the modes
under consideration a sensitivity of $2-3^\circ$ is expected in the lifetime of the experiment.

\begin{table}
\begin{center}
\caption[]{Expected LHCb combined sensitivity to $\gamma$ from $B\to DK$ and time-dependent 
measurements for data sets corresponding to integrated luminosities of 0.5, 2 and 
10~$\mathrm{fb}^{-1}$.}
\label{tab:lhcb_gamglobal}
\vspace{0.2cm}
\begin{tabular}{lccccc}\hline\hline 
$\delta_{B^{0}}~(^{\circ})$ & 0 & 45 & 90 & 135 & 180 \\ \hline
$\sigma_{\gamma}$ for $0.5~\mathrm{fb}^{-1}$  $(^{\circ})$&  8.1 &	10.1 & 	9.3 & 	9.5 &	7.8 \\
$\sigma_{\gamma}$ for $2~\mathrm{fb}^{-1}$  $(^{\circ})$&  4.1 &	5.1 & 	4.8 & 	5.1 &	3.9 \\
$\sigma_{\gamma}$ for $10~\mathrm{fb}^{-1}$  $(^{\circ})$&  2.0 &	2.7 & 	2.4 & 	2.6 &	1.9 \\
\hline\hline
\end{tabular}
\end{center}
\end{table}


\section{Measurements of the angles of the unitarity triangle in charmless hadronic $B$ decays}
\label{sec:angles}
\subsection{Theory estimates for hadronic amplitudes }

\subsubsection{Angles, physical amplitudes, topological amplitudes}
\label{sssec:angles-theory-intro}
Any standard-model (SM) amplitude for a
decay $\B \to f$ can be written, by integrating out the weak
interactions to lowest order in $G_F$ (Sec. \ref{sec:ope}),
as a linear combination
\begin{equation}
  {\cal A}(\B \to f) = \sum_{i,U} C_i V_{UD} V_{Ub}^*
          \langle f | Q_i | \B \rangle
\end{equation}
of matrix elements of local operators $Q_i$ in QCD $\times$ QED. Here
$D=d,s$ and $U=u,c,t$. By CKM unitarity, one term in the sum over $U$
can be removed. This gives a
decomposition into (physical) tree and penguin amplitudes
(the names are motivated by Wick contractions of the operators $Q_i$
contributing to them),
\begin{eqnarray}
  {\cal A}(\B \to f) &=& T_f e^{i \theta_T} + P_f e^{i \theta_P} ,  \nonumber \\
  {\cal A}(\Bb \to f) &=& T_f e^{-i \theta_T} + P_f e^{-i \theta_P} ,
\end{eqnarray}
where $T_f$ and $P_f$ (``strong amplitudes'') and $\theta_T$ and
$\theta_P$ (``weak phases'') have definite CP transformation properties.
For decays into two light mesons, conventionally $U=c$ (or $U=t$) is
eliminated, giving $\theta_P = \beta$ ($\theta_P=0$), and $\theta_T = \gamma$.
For decays involving charmonium, the tree is associated with $U=c$
($\theta_T=0$), and one of $U=u,t$ is eliminated (both are expected to be
negligible). The prototypical angle measurement derives from the
time-dependent CP asymmetry
\begin{equation}
  A_{\rm CP}(f; t) \equiv \frac{\Gamma(\Bb(t) \to f) -
    \Gamma(\B(t)\to f) }{ \Gamma(\Bb(t) \to f) + \Gamma(\B(t) \to f)}
 \equiv - C_f \cos \Delta m\, t + S_f \sin \Delta m\, t ,
\end{equation}
where $f$ is a CP eigenstate of eigenvalue $\eta_{\rm CP}(f)$,
$\Delta m$ is the absolute value of the mass difference between the two mass
eigenstates in the $\Bz$--$\Bzb$ system, and
\begin{equation}
   C_f = \frac{1-|\xi|^2}{1+|\xi|^2}, \qquad
   S_f = \frac{2 {\rm Im} \xi}{1+|\xi|^2} , \qquad
   \xi = e^{-i 2 \beta} \frac{{\cal A}(\bar B
     \to f)}{{\cal A}(B \to f)} .
\end{equation}
(We assume $CPT$ conservation, and neglect lifetime differences and CP
violation in mixing throughout.) 
If $P_f$ can be neglected, $|\xi|=1$, $C_f=0$, and
$S_f$ gives a clean measurement of $\sin 2
(\beta + \theta_T)$. This is true to very good approximation for
decays into final states containing charmonium such as $B \to J/\psi K_S$
($\theta_T=0$, $- \eta_{\rm CP}(f) S_f = \sin 2\beta$).
It holds less accurately for $b \to \d$ transitions like
$\B \to (\pipi, \rhopi, \rhorho)$, where the
CKM hierarchy is $[ P_f /T_f ]_{\rm CKM} = {\cal O}(1)$, but some
suppression of penguin amplitudes follows from theoretical arguments
reviewed below. In these modes, one has approximately
 $- \eta_{\rm CP}(f) S_f \approx \sin 2 (\beta + \gamma) = - \sin 2 \alpha$.
Conversely, penguin-dominated $b \to s$ modes 
 $\B \to (\pi K, \phi K, \eta^{(\prime)} K, \dots)$,
where $[ T_f /P_f ]_{\rm CKM} = {\cal O}(\lambda^2)$, probe $\stwob$.

In view of these considerations, it is clear that the interpretation
of the time-dependent CP asymmetries (and more generally, the
many charmless $\B$ and $\Bb$ decay rates) in terms of CKM parameters
and possible new-physics contributions requires some information on
at least the amplitude ratios $P/T$, hence on the hadronic matrix
elements $\langle f | Q_i | \B \rangle$.
In principle, the latter are determined by the QCD and electromagnetic
coupling and quark masses via (for the case of a two-particle
final state) four-point correlation functions involving three operators
destroying the $B$-meson and creating the final-state mesons, as well
as one insertion of the operator $Q_i$. Formally, they are expressible
in terms of a path integral
\begin{equation}
 \langle M_1 M_2 | Q_i | B \rangle \sim
   \int \! d A \int \! d \bar \psi\, d \psi\, j_B^\mu(x) j_{M_1}^\nu(y) j_{M_2}^\rho(z)
   Q_i(w) e^{i (S_{QCD + QED})} .
\end{equation}
The currents $j_B$, $j_{M_1}$, $j_{M_2}$ must have the correct quantum
numbers to create/destroy the initial- and final-state particles,
for instance $j_B^\mu = \bar b \gamma^\mu \gamma_5 d$ for a $B^0$
decay, but are otherwise arbitrary. 
In practice, this path integral cannot be evaluated; however, the inner
(fermionic) path integral can be represented as a sum of Wick
contractions which provide a nonperturbative definition of ``topological''
amplitudes (Fig.\ \ref{fig:topo-amps}). We stress that no expansion of
any kind has been made; the lines represent the full inverse Dirac operators,
rather than perturbative (``free'')  propagators, averaged over arbitrary gluon
backgrounds by the outer (gluonic) path integral.
\begin{figure}[tb]
\centering
\parbox{0.4\textwidth}{\vskip-0cm
\includegraphics[width=0.4\textwidth,angle=0]{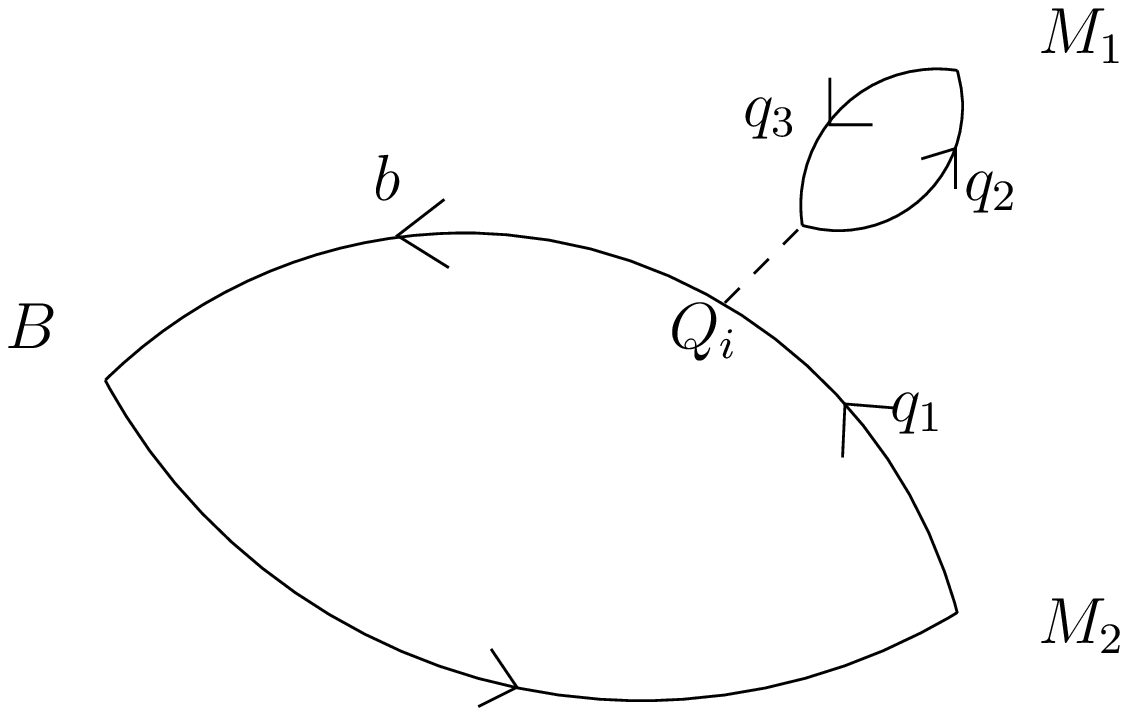}
}
\parbox{0.4\textwidth}{\vskip-0cm
\includegraphics[width=0.4\textwidth,angle=0]{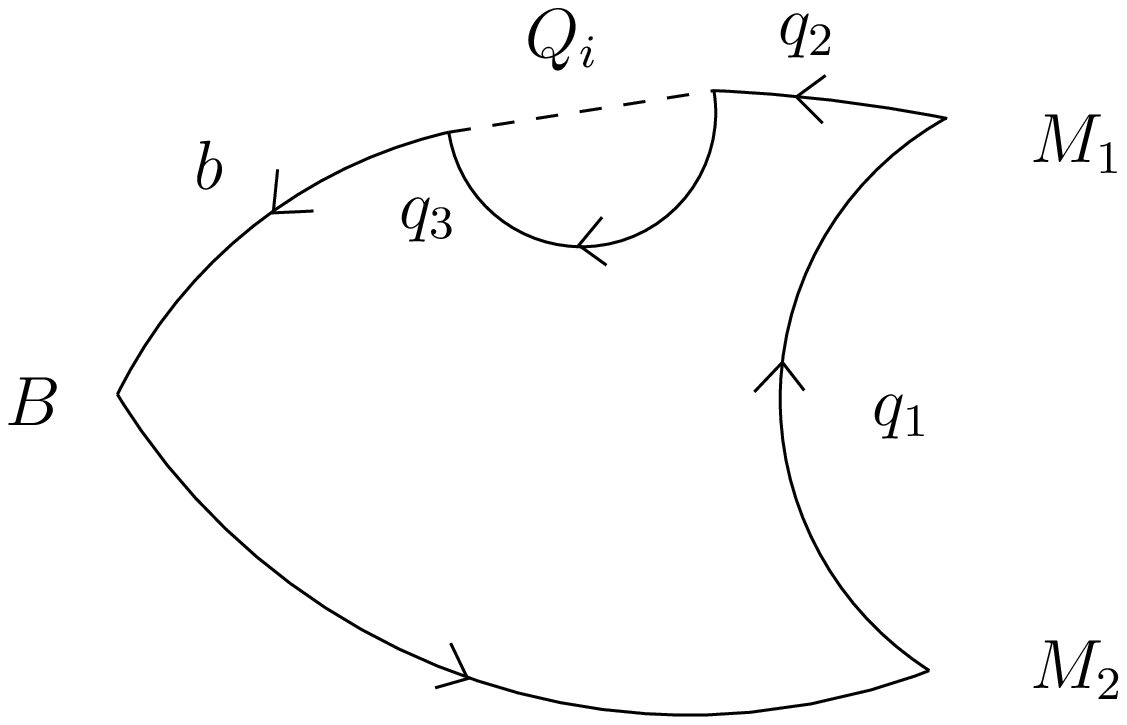}
}
\vspace*{-2.2cm}
\caption{Examples of  Wick
contractions.
Left: Tree contraction.
Right: Penguin contraction.
The scheme-independent topological amplitudes correspond to certain
sums of contractions of several operators in the weak Hamiltonian.
The lines are ``dressed'' propagators, depending on the gluonic background.
Arbitrarily many gluons not shown. From \cite{Buras:1998ra}.
\label{fig:topo-amps}
}
\end{figure}
A complete list has been given in
\cite{Buras:1998ra}. Topological amplitudes can also be defined equivalently
(and were originally) as matrix elements of the $SU(3)$ decomposition
of the weak Hamiltonian \cite{Zeppenfeld:1980ex,Gronau:1994rj}.

Each physical amplitude decomposes into several topological ones. For
a tree, in the notation of \cite{Beneke:2003zv},
\begin{eqnarray}
  T_{M_1 M_2} &=& |V_{ub} V_{uD}| [ A_{M_1 M_2} (
\alpha_1(M_1 M_2) + \alpha_2(M_1 M_2) + \alpha_4^u(M_1 M_2) )
\nonumber \\ &&
\qquad + B_{M_1 M_2} (b_1(M_1 M_2) + b_2(M_1 M_2) + b_3^u(M_1 M_2) +
b_4^u(M_1 M_2) ) 
\nonumber \\ &&
\qquad + {\cal O}(\alpha) ] \; + (M_1 \leftrightarrow M_2) \; .
\end{eqnarray}
The first two terms on the first line are known as the color-allowed
and color-suppressed trees, while the third term is due to a set
of penguin contractions. The terms on the second lines are due to
annihilation topologies, where both fields in the current $j_B$ are
contracted with the weak vertex. We have not spelled out ${\cal O}(\alpha)$
terms, which include the electroweak penguin terms, as well as
long-distance QED effects.\footnote{
These effects include emissions of soft photons from the final-state
particles \cite{Baracchini:2005wp} and are modeled in
extracting the two-body rates and asymmetries (which are not
infrared safe if soft photons are included) from data.
}
Not all topological amplitudes are present for every final state.\footnote{
More precisely, one would write
$\alpha_i(M_1 M_2) \to c_i(M_1 M_2) \alpha_i(M_1 M_2)$ where $c_i(M_1
M_2)$ = 0 if the amplitude is not present and a Clebsch-Gordan
coefficient otherwise \cite{Beneke:2003zv}.}
On the other
hand, if both $a_i(M_1 M_2)$ and $a_i(M_2 M_1)$ are present, they must
be summed.  For instance \cite{Beneke:2003zv},
\begin{eqnarray}
T_{\pi^0 \rho^0} &=&  \frac{i}{2} |V_{ub}V_{ud}| \frac{G_F}{\sqrt{2}} m_B^2
  \Bigg[ f_+^{B \pi}(0) f_\rho (\alpha_2(\pi^0 \rho^0) - \alpha_4^u(\pi^0 \rho^0))
   - f_B f_\pi f_\rho b_1(\pi^0 \rho^0) \nonumber \\
&& \qquad  + A_0^{B \rho}(0) f_\pi
      (\alpha_2(\rho^0 \pi^0) - \alpha_4^u(\rho^0 \pi^0) )
   - f_B f_\pi f_\rho  b_1(\rho^0 \pi^0)
 + {\cal O}(\alpha) \Bigg] ,
\end{eqnarray}
where we have also spelled out the normalization factors
$A_{M_1 M_2}$, which like $B_{M_1M_2}$
consists of form factors,
decay constants, $G_F$, etc. as a convention (and in anticipation of the
heavy-quark expansion), neglecting terms ${\cal O}(m_{\pi}/m_B, m_\rho/m_B)$.
Moreover, for flavor-singlet mesons $M_1$ or $M_2$ there are
additional amplitudes.

Similarly, for a penguin, we have the decomposition
\begin{eqnarray}
  P_{M_1 M_1} &=& |V_{cb} V_{cD}| [ A_{M_1 M_2} \alpha_4^c(M_1 M_2) 
+ B_{M_1 M_2} (b_3^c(M_1 M_2) + b_4^c(M_1 M_2) ) ] 
\nonumber \\ && \; + (M_1 \leftrightarrow M_2) \; .
\end{eqnarray}
The parametrization are general, but
we have now fixed a convention where $V_{tb}^* V_{tD}$ has been eliminated.

Present theoretical knowledge on the topological amplitudes
derives from expansions 
(i) in the Wolfenstein parameter $\lambda$ (see above discussion), 
(ii) around the $SU(3)$ flavor symmetry limit (i.e., in $m_s/\Lambda$),
(iii) in the inverse number of colors $1/N_c$, and
(iv) the heavy-quark expansion in $\Lambda/m_b$ and $\alpha_s$,
where $\Lambda \equiv \Lambda_{\rm QCD}$ is the QCD scale parameter.
The counting for the various topological amplitudes is shown in
Tab. \ref{tab:topo-hierarchies}.
\begin{table}[tb]
\centering
\caption{Hierarchies among topological amplitudes from expansions in
  the Cabibbo angle $\lambda$, in $1/N_c$, and in $\Lambda_{\rm
  QCD}/m_b$. Some multiply suppressed amplitudes (e.g. EW penguin
  amplitudes that are CKM suppressed in $b \to s$ transitions) are omitted.
\label{tab:topo-hierarchies}}
\begin{tabular}{c|cccccccccc}
& $T$/$a_1$ & $C$/$a_2$ & $P_{ut}$/$\alpha_4^u$ & $P_{ct}$/$\alpha_4^c$ &
$P_{EW}$/$\alpha_{3EW}$ & $P_{EW}^{\rm C}$/$\alpha_{4EW}$ &
$b_3^c$ & $b_4^c$ & $E$/$b_1$ & $A$/$b_2$ \\
\hline
Cabibbo $(b\to d)$ & \multicolumn{10}{c}{all amplitudes are
${\cal O}(\lambda^3)$ } \\
Cabibbo $(b\to s)$ & $\lambda^4$ & $\lambda^4$ &
$\lambda^4$ & $\lambda^2$ & $\lambda^2$ & $\lambda^2$ & $\lambda^2$ &
$\lambda^2$ & $\lambda^4$ & $\lambda^4$  \\
$1/N$ & $1$ & $\frac{1}{N}$ & $\frac{1}{N}$  & $\frac{1}{N}$ &
$1$ & $\frac{1}{N}$ & $\frac{1}{N}$ & $\frac{1}{N}$ & $\frac{1}{N}$ & $1$ \\ 
$\Lambda/m_b$ & $1$ & $1$ & $1$ & $1$ & $1$ & $1$ & $\Lambda/m_b$ & $\Lambda/m_b$ & $\Lambda/m_b$ & $\Lambda/m_b$ \\
\end{tabular}
\end{table}
The $\lambda$ and $1/N_c$ counting
provide only (rough) hierarchies. Existing $SU(3)$ analyses work at zeroth
order, providing relations between topological amplitudes for
different final states. (In the case of $\pipi$, $\rhorho$,
isospin alone provides useful relations. This is the basis for the $\alpha$
determinations reviewed in Sec. \ref{sec:chmls-angle-alpha} below.)
The virtue is the possibility to completely eliminate some
of the theoretically difficult amplitudes from the analysis, removing
the need for their theoretical computation.
This comes of the expense of eliminating some of the experimental
information that is in
principle sensitive to short-distance physics (SM and beyond) from
the analysis, as well. For instance, in the $\alpha$ determinations,
six observables are needed to determine one parameter.

Both the $1/N_c$ expansion and the heavy-quark expansion rely
on an expansion in Feynman
diagrams. The virtue of the heavy-quark expansion is that, to lowest
order in the expansion parameter $\Lambda/m_b$, and in some cases
to subleading order, the amplitudes themselves are calculable in
perturbation theory. More precisely, they factorize into products of
form factors and of convolution of a perturbative expression
with non-perturbative meson wave functions.
Moreover, all the $b_i$ (annihilation) amplitudes
are power-suppressed.
The theoretical basis of the $1/m_b$ expansion
is discussed in Sec. \ref{sec:hqet-scet}. The rest of this section
is devoted to quantitative results and phenomenology of the
topological (and physical) amplitudes.

\subsubsection{Tree amplitudes: results}
\label{sssec:angles-theory-tree}
\begin{figure}[tb]
\centering
\parbox{0.29\textwidth}{\vskip-1.5cm
\includegraphics[width=0.36\textwidth,angle=0,trim=1.5cm 0 0 0]{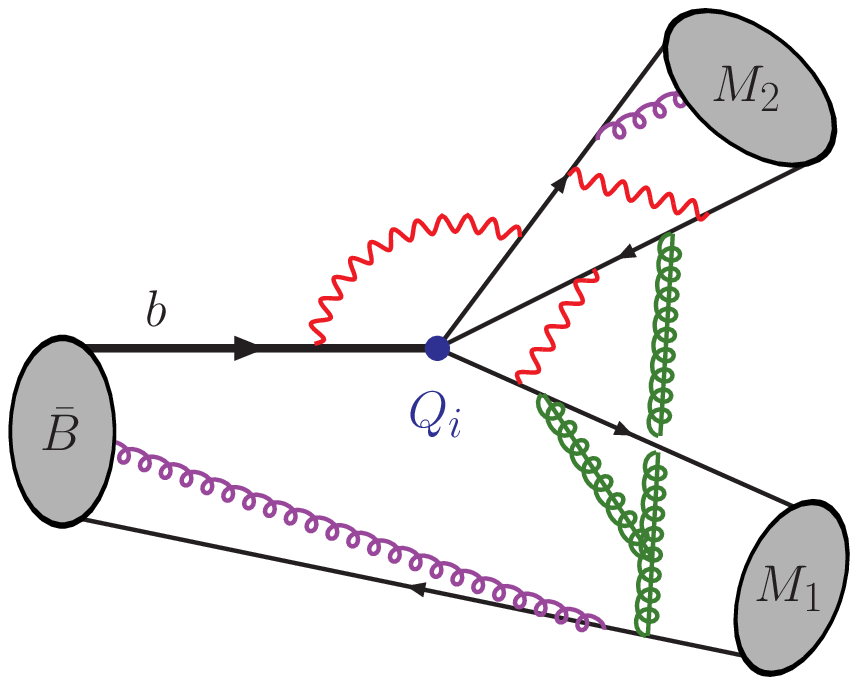}
}
\hspace*{0.5cm}
\parbox{0.32\textwidth}{\vskip-0cm
\includegraphics[width=0.32\textwidth,angle=180]{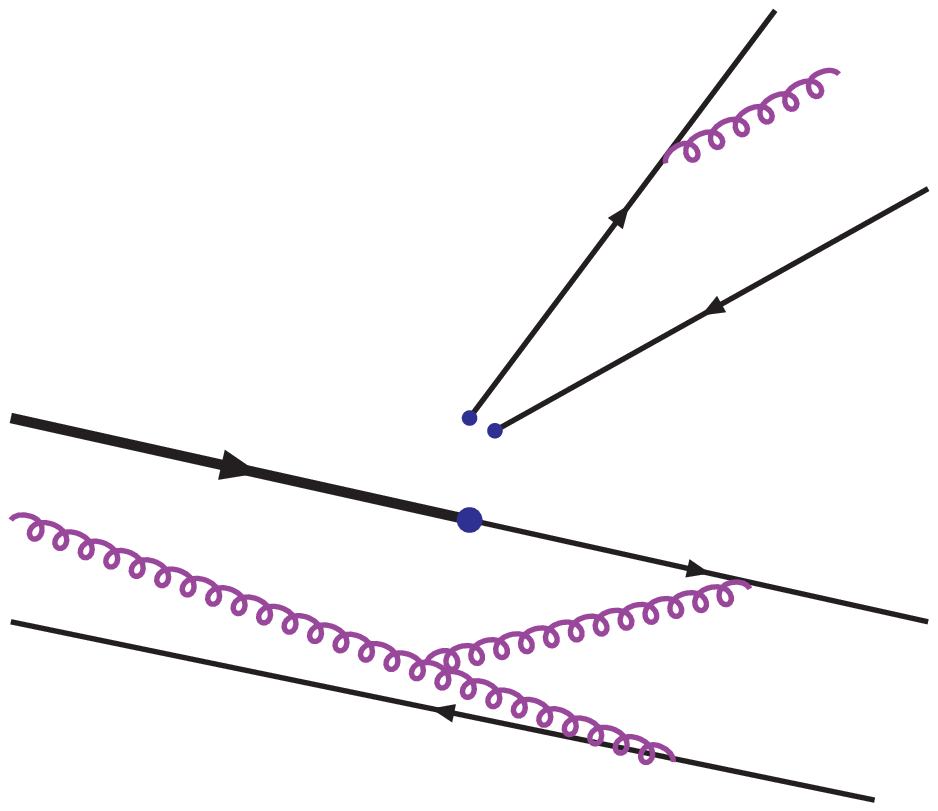}
}
\parbox{0.32\textwidth}{\vskip-0cm
\includegraphics[width=0.32\textwidth,angle=180]{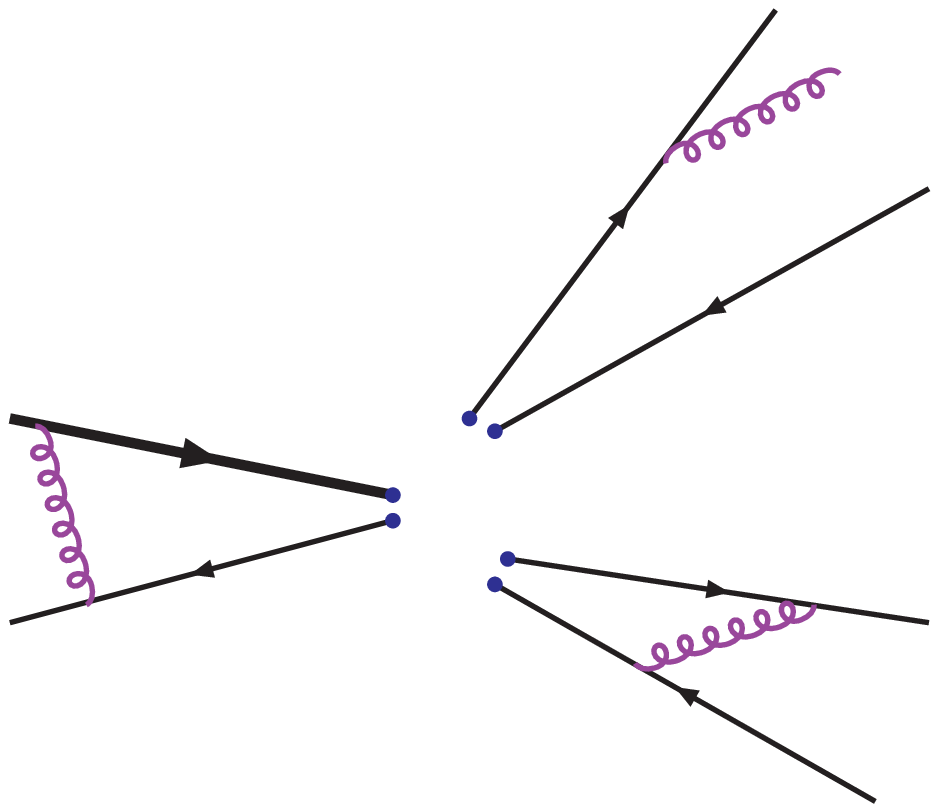}
}
\vspace*{-2.5cm}
\caption{
Factorization of the tree amplitudes. Left: Matrix element of
a weak Hamiltonian current-current operator $Q_{1,2}$ in the effective
$5$-flavor QCD$\times$QED theory. The red, wavy lines close to the vertex
have virtualities of order $m_b^2$; the system of green `cut-spring' lines
connecting to the spectator, of order $\Lambda m_b$. The purple
`spring' lines entering the mesons indicate the soft gluon
background in which the hard subprocess takes place.
Middle: Factorization into a product of a wave function and a form
factor (to be convoluted with a hard kernel $H^{\rm I}$ or $H^{\rm II}$).
Right: The $B$-type bilocal form factor (convoluted with $H^{\rm II}$)
factorizes further into wave functions. (According to the pQCD
framework, this is also true for the soft ($A$-type) form factor.)
\label{fig:fact_tree}
}
\end{figure}

The most complete results are available for the topological tree amplitudes,
whose factorization at leading power in the $1/m_b$ expansion
is pictured in Fig.\ \ref{fig:fact_tree}. The gray blobs and the 
violet `spring' lines contain the soft and collinear gluon degrees of freedom
(virtualities $< \sqrt{\Lambda m_b}$). The hard subgraph, formed by the
remaining gluon lines and the pieces of quark lines between
their attachments and the weak vertex, connects the quark legs of the
weak operators $Q_1$ and $Q_2$ with valence quark lines for the
external states but
not with each other, hence the name "tree". For either operator,
the hard scale $m_b$ (wavy lines) can be
matched onto two operators ${\cal O}^{\rm I,II}$ in SCET$_{\rm I}$
(see Sec.~\ref{sec:hqet-scet}).
At leading power, all terms at ${\cal O}(\alpha_s)$ (NLO)
\cite{Beneke:1999br,Beneke:2000ry,Beneke:2001ev} 
and ${\cal O}(\alpha_s^2)$ (NNLO) 
\cite{Beneke:2005vv,Kivel:2006xc,Pilipp:2007mg,Bell:2007tv,Bell:2009nk}
have been computed. In particular, these results establish the validity of
factorization and the good behavior of the perturbation expansion up to
NNLO.  The hard-collinear scale $\sqrt{\Lambda m_b}$ can also be
factorized. This has been performed for the operators of type II in
\cite{Becher:2004kk,Beneke:2005gs,Kirilin:2005xz,Pilipp:2007mg}.
Again, a stable perturbation expansion is observed.
Depending on the flavor
of the valence quark lines, color factors differ, giving rise to
a ``color-allowed'' amplitude $\alpha_1$ and a ``color-suppressed''
one $\alpha_2$.
The type-II (hard-spectator-scattering) contributions then take the form
\begin{equation}
    A_{M_1 M_2} \alpha_{1,2}^{\rm II} \propto [H^{\rm II} * \phi_{M_2}] *
    [\phi_{B} * J * \phi_{M_1}] 
\end{equation}
of a convolution of hard and
hard-collinear scattering kernels $H^{\rm II}$ and $J$ with meson
wave functions. An alternative is not to perform the hard-collinear
factorization and define a non-local
form factor $\zeta_J = \phi_{B} * J * \phi_{M}$, information on which
has to be extracted from experiment. This works in practice to zeroth
order in $\alpha_s(m_b)$ \cite{Bauer:2004tj}. At higher orders, the kernel
$H^{\rm II}$ acquires a dependence on how the momentum is shared between the
$M_1$ valence quarks, i.e. the convolution $H^{\rm II} * \zeta_J$
becomes nontrivial. No higher-order analyses have been
performed.\footnote{
Strictly speaking, the convolution of the $\zeta_J$ factor with $H^{\rm
II}$ might diverge at the endpoint. Correspondingly, to such a convolution
 in general a non-perturbative soft rescattering phase should be associated.
An endpoint divergence indeed appears in the attempt to perturbatively
factorize $\zeta_J$ at first subleading power, see below.}

For the type-I operators, in the collinear expansion one
encounters divergent convolutions in factorizing the hard-collinear
scale already at the leading power, indicating a soft overlap breaking
(perturbative) factorization of soft and collinear physics.
In this case, however, not performing this
factorization is more feasible, as it leaves a single form
factor (which can be taken to be an ordinary QCD form factor or
the SCET soft form factor) multiplying a convolution of a
hard-scattering kernel with one light-meson wave function,
\begin{equation}
  A_{M_1 M_2} \alpha_{1,2}^{\rm I} \propto f^{B M_1}(0) H^{\rm I} * \phi_{M_2}.
\end{equation}
[By convention, the form factor is factored out into $A_{M_1 M_2}$.]
An alternative treatment is $k_T$ factorization (``pQCD'')~\cite{Keum:2000ph},
where a transverse-momentum-dependent $B$-meson wave function is
introduced, which regularizes the endpoint divergence. In this
case, a convergent convolution arises (at lowest order),
and within the uncertainties on the wave function it is generally
possible to accommodate the observed data.\footnote{
Independently of the convergence issue, a perturbative
calculation in the $k_T$ (or any other) factorization scheme must
demonstrate that the result is dominated by modes which are
perturbative.}

Finally, certain power corrections were identified as potentially large
in \cite{Beneke:2001ev}. 
One class, which is only relevant for final states containing pseudoscalars,
consists of ``chirally enhanced'' terms, which are proportional to the ratio
$r_\chi^P = m_P^2/(m_b m_q)$, where $P$ is a pion or Kaon and $m_q$ an average
of light quark masses; 
another class of certain annihilation topologies with large color factors
(first pointed out in \cite{Keum:2000ph}) is discussed in
Sec.~\ref{sssec:angles-theory-penguin} below.
Power corrections are of phenomenological relevance in $\alpha_2^{\rm II}$,
which contains a chirally-enhanced power correction involving the
large Wilson coefficient $C_1$, where the convolution of
$H^{\rm II}$ with the power-suppressed analogue $\zeta_J^{\rm tw 3}$
of $\zeta_J$ is divergent. These power corrections are therefore
not dominated by perturbative gluon exchange. They have been
modeled in \cite{Beneke:2001ev} by introducing an IR cutoff ${\cal
  O}(500\, {\rm MeV})$ on
the convolutions and associating an arbitrary rescattering phase
with the soft dynamics.

Quantitatively, combining the phenomenological analysis in
\cite{Beneke:2006mk} (where values for hadronic parameters are
specified) with the results of \cite{Bell:2007tv,Bell:2009nk} gives
\begin{eqnarray}
\alpha_1(\pi\pi) &=& 1.015 + [0.025 + 0.012i]_V  + [0.024 +  0.026 i]_{VV}
   \nonumber \\
   &&     -\,\left[\frac{r_{\rm sp}}{0.485} \right]
   \Big\{ [0.020]_{\rm LO} + [0.034 + 0.029i]_{HV} + [0.012]_{\rm tw3} \Big\}
   \nonumber \\
   &=& 0.999^{+0.034}_{-0.072} + (0.009^{+0.024}_{-0.051})i,  \label{eq:a1}
\\[0.2cm]
\alpha_2(\pi\pi) &=& 0.184 - [0.153 + 0.077i]_V - [0.030 + 0.042 i]_{VV}
   \nonumber \\
   && + \,\left[ \frac{r_{\rm sp}}{0.485} \right]
   \Big\{ [0.122]_{\rm LO} + [0.050 +0.053i]_{HV} + [0.071]_{\rm tw3} \Big\}
   \nonumber \\
   &=& 0.245^{+0.228}_{-0.135} + (-0.066^{+0.115}_{-0.081})i.  \label{eq:a2}
\end{eqnarray}
In each amplitude, the terms on the first and second lines
correspond to the type-I and type-II contributions. These are
further split into terms ${\cal O}(1)$, ${\cal O}(\alpha_s)$ (V, LO),
and ${\cal O}(\alpha_s^2)$ (VV, HV), and an estimate of a chirally
enhanced power correction following a model defined in
\cite{Beneke:2001ev} (``tw3'').
The relative normalization factor of the spectator-scattering
contributions, $r_{\rm sp} = (9 f_\pi f_B)/(m_b f_+^{B \pi}(0)
\lambda_B)$, contains the bulk of the parametric uncertainty of that
term.
We observe that the color-allowed tree is perturbatively stable and has small
uncertainties resulting from the poor knowledge of hadronic
input parameters. Moreover, the spectator-scattering contribution
is small, including a weak dependence on endpoint-divergent power
corrections (labeled ``tw3'').

Conversely, the color-suppressed tree amplitude is dominated by
the type-II contribution, and it exhibits large sensitivity
to a chirally enhanced, non-factorizable power correction.
It is important to keep
in mind that the estimate for the latter, unlike all other pieces,
is based on a model. Several phenomenological
analyses of the $B \to \pi \pi$ data
favor large values $\alpha_2(\pi \pi) = {\cal O}(1)$, which is
sometimes called a puzzle.
In the Standard Model, a large value can come from a large $r_{\rm
  sp}$, for instance through the moment
$\lambda_B^{-1} \equiv \int {\rm d} \omega \phi_{B+}(\omega)/\omega$
of the relevant $B$-meson wave function. 
Information on $\lambda_B$ can be obtained by operator product expansions
in HQET \cite{Lee:2005gza,Kawamura:2008vq} and from QCD sum rules
\cite{Ball:2003fq,Braun:2003wx}, but with considerable uncertainties.
 An interesting possibility is to determine
$\lambda_B$ more directly from radiative semileptonic decay,
discussed in Sec.~\ref{sssec:angles-theory-prospects} below.
Second, is not inconceivable that a large value originates from
the presently incalculable twist-three spectator scattering.
Such an interpretation would be consistent with the fact
that data suggest a small value of $\alpha_2(\rho \rho)$, which is not
sensitive to chiral enhancement.

In the treatment advocated in \cite{Bauer:2004tj}, $\zeta_J$ is not
factorized. The generic prediction is
$\arg \alpha_2/\alpha_1 = {\cal O}(\alpha_s)$ (this is set to zero
in the analysis). A prediction on the magnitude
requires knowledge on $\zeta$ and $\zeta_J$ from outside sources,
in analogy with the results described above. For more details,
see \cite{Bauer:2004tj,Williamson:2006hb,Jain:2007dy}.

In the pQCD approach, the issue of a large $\alpha_2/\alpha_1$
(possibly with a large phase) has been addressed in
\cite{Li:2005kt} and again in \cite{Li:2009wb}. The latter
paper augments the structure in the original approach by an extra
soft rescattering factor which represents an
additional non-perturbative parameter that has to be adjusted to
experimental data.
We note that the computation in \cite{Li:2005kt} uses
the hard (type-I) vertex from \cite{Beneke:1999br,Beneke:2000ry,Beneke:2001ev}
as a building block to estimate
NLO effects in the pQCD approach. Taking, for the sake of the
argument, the asymptotic form of the
distribution amplitude $\phi_{M_2}$,  the contribution is proportional
to
\begin{equation}
C_2(\mu)
+ \frac{C_1(\mu)}{N_c}
\left[ 1 +
\frac{ \alpha_s(\mu) \, C_F}{4\pi}  \left(
  - \frac{37}{2} - 3 \, i  \, \pi + 12 \, \ln \frac{m_b}{\mu}
\right)
\right] ,
\end{equation}
where $C_1(\mu)$ is the large current-current Wilson coefficient.
In order to obtain both a large magnitude and phase, one would need
to evaluate this expression at a low scale $\mu \ll m_b$, where
perturbation theory is questionable.\footnote{More precisely, the
apparent $\mu$-dependence is formally
a NNLO effect.} In the pQCD approach
the above expression appears inside a convolution integral, where
the scale $\mu$ is fixed by the internal kinematics of the spectator
scattering.
The enhancement and the large phase of the color-suppressed tree
amplitude found in \cite{Li:2005kt}
therefore has to be associated to a rather low effective renormalization
scale
(see also the discussion in \cite{Beneke:2007zz}).
Correspondingly, scale variations or alternative scale-setting
procedures in the pQCD approach represent an additional source of
potentially large theoretical uncertainties associated to this kind of
NLO effects.

Finally, the (physical) tree amplitudes receive contributions from
penguin and annihilation contractions as discussed above. The
factorization properties of the former are very similar to those
of the penguin amplitudes discussed in the following section and give rise
to corrections that are subleading with respect to $\alpha_1$,
$\alpha_2$. For the annihilation amplitudes $b_1$ and $b_2$ there
is no factorization in the collinear expansion. Both the model
of \cite{Beneke:2001ev} and the $k_T$ factorization of
\cite{Keum:2000ph} result in small numerical
values.\footnote{In \cite{Buras:1998ra}
it has been noted that $b_2$ is leading in the $1/N_c$ counting.
On the other hand, the diagrams that are leading in the $1/N_c$
counting combine
to the product of a decay constant and the matrix element of the
divergence of a current that is conserved in the limit $m_{s,d,u} \to 0$.
A hard-scattering approach then implies
$b_2^{N=\infty} = {\cal O}(m_{s,d,u}/m_b)$. This suppression is also found
in the QCD light-cone sum rules treatment in \cite{Khodjamirian:2005wn}.
Nevertheless we do not know a rigorous argument why
this amplitude could not be as large as ${\cal O}(m_s/\Lambda)$
in $B\to \pi K$ decays.
}

\subsubsection{Penguin amplitudes: results}
\label{sssec:angles-theory-penguin}
The penguin contraction $\alpha_4^c(M_1 M_2)$ entering
the physical penguin amplitude $P_{M_1 M_2}$ decomposes in the
heavy-quark expansion as
\begin{equation}
\alpha_4^c = a_4^c + r_\chi a_6 + \mbox{higher powers and terms not
chirally enhanced} .
\end{equation}
Factorization of $a_4^c$ (as defined here) to leading power has been
argued (to one loop) and the hard kernels computed
in \cite{Beneke:1999br,Beneke:2000ry,Beneke:2001ev}
but  has been the subject of some controversy
over the existence of an extra leading-power long-distance
nonrelativistic ``charming-penguin'' 
contribution \cite{Bauer:2004tj,Beneke:2004bn,Bauer:2005wb}. Such an
incalculable extra term would, in practice, imply that no prediction
for penguin amplitudes can be made.
It appears that this theoretical issue has recently been
resolved~\cite{Beneke:2009az} (in favor of calculability in the sense
of~\cite{Beneke:1999br,Beneke:2000ry,Beneke:2001ev}).
Again, there are two contributions, labeled I and II as in the case of the
trees. The computation of the spectator scattering term $a_4^{\rm II}$
has been performed to  ${\cal O}(\alpha_s^2)$ \cite{Beneke:2006mk,Jain:2007dy}.
The ``scalar-penguin'' term $r_\chi a_6^c$ is again a chirally
enhanced power correction,
which however is calculable. At ${\cal O}(\alpha_s)$ 
\cite{Beneke:1999br,Beneke:2000ry,Beneke:2001ev} it dominates over
$a_4^c$ when $M_2$ is a pseudoscalar (otherwise it is negligible).
Finally, the physical penguin amplitudes contain a penguin annihilation term
with a large color factor that is not chirally enhanced.
Twist-three spectator scattering is unlikely to be very important, as the
type-I contributions are significant (similarly to the color-allowed
tree).
Because the perturbative results for the penguin amplitudes,
in comparison to the tree amplitudes, are rather incomplete at this
time (only one piece at ${\cal O}(\alpha_s^2)$ has been computed, as discussed
above), we refrain from giving numerical results; for an exhaustive compendium
of ${\cal O}(\alpha_s)$ results we refer to \cite{Beneke:2003zv}. 
Rather, we recall the following ``anatomy''. As just mentioned, physical
penguin amplitudes are approximately described in terms of a leading-power
piece, a chirally enhanced power correction, and an annihilation term:
\begin{equation}
  P_{M_1 M_2} \propto a_4^c(M_1 M_2) \pm r_\chi^{M_2} a_6^c(M_1 M_2)
    + \frac{B_{M_1 M_2}}{A_{M_1 M_2}} b_3^c(M_1 M_2) .
\end{equation}
The sign in front of $a_6$ provides for constructive interference
in the case of a $PP$ final state and destructive one for a $VP$ final
state; moreover the enhancement factor $r_\chi$ is absent (or small)
for $PV$ final states. This results in a particular pattern for
the magnitudes and phases of penguin amplitudes (and $P/T$ ratios)
that can be compared to data in $\Delta B = \Delta S = 1$ decays
\cite{Beneke:2003zv,Beneke:2006mk}, with a reasonable agreement within
uncertainties. The comparison also indicates the presence of
substantial annihilation contributions (as included in the `S4'
scenario favored in \cite{Beneke:2003zv}). For instance, a complex
annihilation term is essential to account for the observed
sign of CP asymmetries in
$B^0 \to K^+ \pi^-$ and $B^0 \to \pi^+ \pi^-$. (A caveat to this is
that the ${\cal O}(\alpha_s^2)$ contribution to $a_6$ is currently not
known; as it involves the large coefficient $C_1$ it might make
a non-negligible contribution to the phase of $P/T$.)
As with the endpoint divergent twist-three spectator scattering
(and with the same caveats) the annihilation term is rendered finite
in pQCD ($k_T$ factorization) and one can obtain the ``correct'' sign
of the penguin amplitudes through the annihilation amplitude.
A treatment based on the approach of \cite{Bauer:2004tj}, but extended
by an $a_6$ term and a real annihilation amplitude, can be found in
\cite{Jain:2007dy}. The phenomenologically required phase is assigned there to
a nonperturbative charming-penguin parameter.

\subsubsection{Application to angle measurements}
\label{sssec:angles-theory-angles}
As explained in Sec.~\ref{sssec:angles-theory-intro}, various time-dependent
CP asymmetries measure CKM angles via their $S$-parameter in the limit of
vanishing $T$ or $P$. The predictions obtained from the heavy-quark expansion
can be directly applied to correct for non-vanishing subleading amplitudes. For
the case of the angle $\beta$ in $b \to s$ penguin transition, where
\begin{equation}
  \Delta S_f = - \eta_{\rm CP}(f) S_f - \sin(2 \beta)
    \approx 2 \cos(2 \beta) \sin \gamma\, {\rm Re}\frac{T_f}{P_f} ,
\end{equation}
such analyses have been performed in
\cite{Beneke:2005pu,Cheng:2005bg,Williamson:2006hb,Wang:2008rk,Li:2006jv},
following the different
treatment of hadronic inputs and (divergent) power corrections
outlined above.
Results are compared in Tab. \ref{tab:DeltaS}.
\begin{table}[htb]
\centering
\caption{Predictions for $\Delta S$ defined in the text for several
penguin-dominated modes. 
{\em Note:} For the QCDF results, we quote the result of a scan over
input parameters (conservative). For the SCET results, double results
correspond to two solutions of a fit of hadronic parameters,
and errors are combined in quadrature. Results for $PP$ final states
are from \cite{Williamson:2006hb}, for $PV$ from \cite{Wang:2008rk};
both papers assume $SU(3)$ to reduce the number of theory papers but
differ over the inclusion of certain chirally enhanced terms.
\label{tab:DeltaS}}
\begin{tabular}{c||c|c|c|c}
mode & QCDF/BBNS \cite{Beneke:2005pu} & SCET/BPRS \cite{Williamson:2006hb,Wang:2008rk} & pQCD \cite{Li:2006jv} & experiment \\
\hline
$\phi K_S$ & $0.01$ \dots $0.05$ & $0$ / $0$  & $0.01$ \dots $0.03$ & $-0.23 \pm 0.18$ \\
$\omega K_S$ & $0.01$ \dots $0.21$ & $-0.25$ \dots $-0.14$ / $0.09$ \dots $0.13$ & $0.08$ \dots $0.18$ & $-0.22 \pm 0.24$ \\
$\rho^0 K_S$ & $-0.29$ \dots $0.02$ & $0.11$ \dots $0.20$ / $-0.16$ \dots $-0.11$ & $-0.25$ \dots $-0.09$  & $-0.13 \pm 0.20$ \\
$\eta K_S$ & $-1.67$ \dots $0.27$ & $-0.20$ \dots $0.13$ / $-0.07$ \dots $0.21$ &  &  \\
$\eta' K_S$ & $0.00$ \dots $0.03$ & $-0.06$ \dots $0.10$ / $-0.09$ \dots $0.11$ & & $-0.08 \pm 0.07$ \\
$\pi^0 K_S$ & $0.02$ \dots $0.15$ & $0.04$ \dots $0.10$  & & $-0.10 \pm 0.17$ \\
\end{tabular}
\end{table}

Analogous expressions hold for $b \to d$ transitions. This allows
a measurement of $S_{\pi^+ \pi^-, \pi^+ \rho^-, \rho^+ \rho^-}$ to be
directly turned into one of $\gamma$. These determinations
are competitive with the average of isospin-triangle ``$\alpha$''
determinations, and in fact even of the global unitarity triangle fit:
$\gamma_{\pi \pi} = (70^{+13}_{-10})^\circ$,
$\gamma_{\pi\rho}=(69 \pm 7)^\circ$ \cite{Beneke:2007zz}, and
$\gamma_{\rho_L \rho_L} = (73.2^{+7.6}_{-7.7})^\circ$
\cite{Beneke:2006hg}. (These involve QCDF calculations of $P/T$; we
have not updated experimental inputs.)
For a combination of heavy-quark expansion and $SU(3)$ flavor arguments, see
\cite{Beneke:2006rb}.

\subsubsection{Prospects}
\label{sssec:angles-theory-prospects}
The discovery that predictions for hadronic two-body decay amplitudes
can be made in perturbation theory in an expansion in $\Lambda/m_b$
has led to a lot of activity at the conceptual, technical, and
phenomenological level. At the former, it provides a highly nontrivial
application of soft-collinear effective theory, while at the latter
it bore the promise to discuss many more observables separately than
is possible based on isospin and flavor-$SU(3)$ arguments alone.
So far, the available technical results are between the NLO and NNLO
stage, where they show a good behavior of the perturbation series.
The NNLO computations should be completed
also for the (topological) penguins, including chirally
enhanced power corrections. This means one-loop corrections to $a_6^{\rm II}$
and two-loop corrections to $a_4^{\rm I}$ and $a_6^{\rm II}$,
and analogous electroweak amplitudes.
Not before then will it be really possible to compare to data
(preferably from new-physics-insensitive channels) to assess the importance
of certain incalculable power corrections, which will then likely dominate
the uncertainties on all amplitudes. A related issue is the status
of required nonperturbative inputs -- foremost, form factors and moments of the
$B$-meson wave functions. While some progress on the former is
expected from improved lattice results, the latter has to be obtained
in other ways, such as from QCD light-cone sum rules or from data
itself. Most important are the first inverse moments $\lambda_B^{-1}$
and $\lambda_{B_s}^{-1}$. They are intimately related to the size
of spectator-scattering terms, hence to the color-suppressed tree
(and electroweak-penguin) amplitudes. Interestingly, in the case of
$B_d$ mesons this parameter
can already be constrained from the search for the radiative
semileptonic decay $B^+ \to \gamma \ell^+ \nu$ \cite{Aubert:2007yh}.
Here, a more sophisticated theoretical analysis taking into account
known higher-order and power corrections in that mode would be interesting.

For the non-factorizable power corrections themselves, significant
conceptual progress would be necessary before one might gain
quantitative control.
The fate of soft-collinear factorization
is a hard problem but is important. Meanwhile, a comparison of data
with refined theory predictions may give us more (or less) confidence
in present models of the power corrections.

\subsection{Measurement of \bb }

\subsubsection{Theoretical aspects}

Measurements of time-dependent \CP\ violation in hadronic $b \to s$ penguin
dominated decay modes provide an interesting method to test the SM.
Naively, decays to \CP\ eigenstate final states $f$ (with \CP\ eigenvalues
$\eta_f$) which are dominated by $V_{tb}V_{ts}^{*}$
amplitudes should have small values of 
$\Delta S_f \equiv -\eta_f S_f - S_{\jpsi \KS}$
since, in the SM, 
${\rm arg}\left( V_{tb}V_{ts}^{*} \right) \approx 
{\rm arg}\left( V_{cb}V_{cs}^{*} \right)$.
Although one expects hadronic corrections in these modes to be only 
of ${\cal O}(\lambda^2) \approx 5\%$~\cite{Grossman:1996ke,London:1997zk},
this is difficult to confirm rigorously. 
In fact in the past few years many theoretical
studies~\cite{Beneke:2005pu,Buchalla:2005us,Cheng:2005ug,Cheng:2005bg,Li:2006jv} of the ``pollution'' from the amplitude proportional to 
$V_{ub}V_{us}^{*}$ to these modes have been undertaken. 
Recall that the amplitude can be written as
\begin{equation}
\label{amplitudebtos}
A(\bar B\to f) = V_{cb} V_{cs}^* \,a_f^c+ V_{ub} V_{us}^* \,a_f^u 
\propto 1 + e^{-i\gamma} \,d_f,
\end{equation}
where schematically the hadronic amplitude ratio is given by 
\begin{equation}
\label{schematicdf}
d_f \sim  \left|\frac{V_{ub} V_{us}^*}{V_{cb} V_{cs}^*}\right| 
\,\frac{\{P^u,C,\ldots\}}{P^c+\ldots}.
\end{equation}
Since for small $d_f$, the correction $\Delta S_f \approx 
2 \,\mbox{Re}(d_f) \cos(2\beta)\sin\gamma$, these contributions have to 
be negligibly small for time-dependent CP asymmetry measurements in $b\to s$ 
transitions to provide a clean and viable test of the SM, or $d_f$ 
has to be under very good theoretical control. The problem is that precise 
model independent estimates are rather difficult to make. Most theoretical 
calculations suggest that the two penguin amplitudes $P^c$, $P^u$ 
are similar resulting in a universal positive contribution 0.03 
to $S_f$, while the final-state dependence results mainly from 
the interference of the color-suppressed tree amplitude $C$ with 
the dominant penguin amplitude, $\mbox{Re}(C/P^c)$.
For more detailed reviews, see Refs.~\cite{Browder:2008em,Silvestrini:2007yf}.

In fact, it is important to note that there are actually (at least) three ways
to determine $\stwob$ in the SM:

\begin{itemize}    

\item First, the gold-plated method via $\Bz \ra \jpsi \KS$,

\item Via the $b\to s$ penguin-dominated decay modes,

\item From the ``predicted'' value of $\stwob$, based on the SM CKM Unitarity
  Triangle fit.
  Unlike the previous two, which are directly measured values of $\stwob$,
  the predicted value is typically obtained by using hadronic matrix elements,
  primarily from lattice calculations, along with experimental information on
  \CP\  violating and \CP\ conserving parameters $\epsilon_K$,
  $\Delta m_s/\Delta m_d$ and $V_{ub}/V_{cb}$.
  In fact, recently it has been shown that the precision in one hadronic
  matrix element ($B_K$) has improved so that even without using
  $V_{ub}/V_{cb}$ a non-trivial constraint can be obtained for the predicted
  value of $\stwob$ in the SM~\cite{Lunghi:2008aa}. This is important since
  there is an appreciable disagreement between inclusive and exclusive
  determinations of $V_{ub}$~\cite{Yao:2006px}.

\end{itemize}  
Differences in the resulting three values of $\stwob$ may imply new physics and
need to be carefully understood.   

In the discussion of experimental results below, we see that
ten $b \to s$ penguin dominated decay modes have been identified so far. 
Several theoretical studies find that three of the modes: $\phi \KS$,
$\etapr \KS$ and $\KS\KS\KS$ are the cleanest with SM predictions of 
$\Delta S_f \lesssim 0.05$, since either there is no pollution from 
the color-suppressed tree amplitude, or the penguin amplitude is 
large, in which case $d_f$ is estimated to be only a few percent;
this also generally means that the uncertainties on these estimates are
small. On the other hand, theoretical calculations find appreciably larger tree
contributions (with large uncertainties) in several of the other modes, such
as $\eta \KS$, $\rho \KS$, $\omega \KS$. It therefore no longer seems useful
to average the \CP\ asymmetry over all of the penguin modes. 
Factorization-based calculations suggest that the uncertainty in the 
case of $\pi^0 \KS$ is intermediate between the two sets of final 
states above. However, for $\pi^0 \KS$ additional information is 
available: a general amplitude parametrization of the entire 
set of $\pi K$ final states together with SU(3) flavor 
symmetry allows to constrain $S_{\pi^0\KS}$ by other $\pi K$ and 
$\pi\pi$ observables~\cite{Ciuchini:2008eh,Fleischer:2008wb,Gronau:2008gu}. 
At present this method yields $S_{\pi^0\KS} \simeq 0.8 \textnormal{--} 1$, 
if one allows for an anomalously large color-suppressed 
tree amplitude that is suggested by the current $\pi K$ branching 
fractions and direct CP asymmetries. 
Hence improved measurements of the direct and time-dependent asymmetries may
still provide useful tests of the SM.

Finally, we note that the current
experimental errors of 0.07 ($\etapr\KS$) and 0.17 ($\phi \KS$ and
$\KS\KS\KS$), as shown in Fig.~\ref{fig_chmls}, are statistics dominated 
and are also still large compared to the expected theory uncertainties. 
At a Super Flavor Factory ($\approx 50$--$75 \ {\rm ab}^{-1}$ of data) 
the experimental errors will get significantly reduced down to around
$0.01$--$0.03$~\cite{Hashimoto:2004sm,:2007zzg,Browder:2007gg,Buchalla:2008jp,Browder:2008em}. 
Looking to the future, another interesting channel is
$\Bs\to\phi\phi$~\cite{Grossman:1996ke,Raidal:2002ph}, 
where the na\"{i}ve Standard Model expectation for $S_f$ is zero, 
and which will be measured by LHCb. 
As mentioned above, in the SM it is theoretically quite difficult to explain
$\Delta S_f$ larger than $0.05$ in these modes. 
Therefore if improved experimental measurements show $\Delta S \gtrsim 0.1$
then that would be an unambiguous sign of a \CP-odd phase beyond the
SM-CKM paradigm.

\subsubsection{Experimental results}

\noindent
\underline{$\Bz \ra \etapr \Kz$ and $\Bz \ra \omega \KS$}

Both the BaBar and Belle experiments reconstruct seven decay channels of 
$\Bz \ra \etapr \Kz$,
\begin{itemize}
\item[]$\Bz \ra \etapr(\rho \g, \eta_{\g\g} \pip \pim, \eta_{3\pi} \pip \pim) \KS(\pip \pim)$,
\item[]$\Bz \ra \etapr(\rho \g, \eta_{\g\g} \pip \pim) \KS(\piz \piz)$ and
\item[]$\Bz \ra \etapr(\eta_{\g\g} \pip \pim, \eta_{3\pi} \pip \pim) \KL$.
\end{itemize}
BaBar identifies the decays with a \KS\ using \mes, \DeltaE\ and a Fisher discriminant
which separates continuum from \BB\ events~\cite{:2008se}.
Similarly, Belle uses \Mbc, \DeltaE\ and a likelihood ratio, ${\cal R}_{S/B}$,
which performs the same task of \qqbar\ discrimination~\cite{Chen:2006nk}. 
For \KL\ modes, only the \KL\ direction is measured, so either \mes\ or \DeltaE\ is calculated. BaBar uses \DeltaE\ while Belle chooses \Mbc.
Fig.~\ref{fig_etapk0} shows \deltat\ and asymmetry projections for 
$\Bz \ra \etapr \Kz$. 

\begin{figure}[htb]
  \centering
  \includegraphics[width=110pt,height=!]{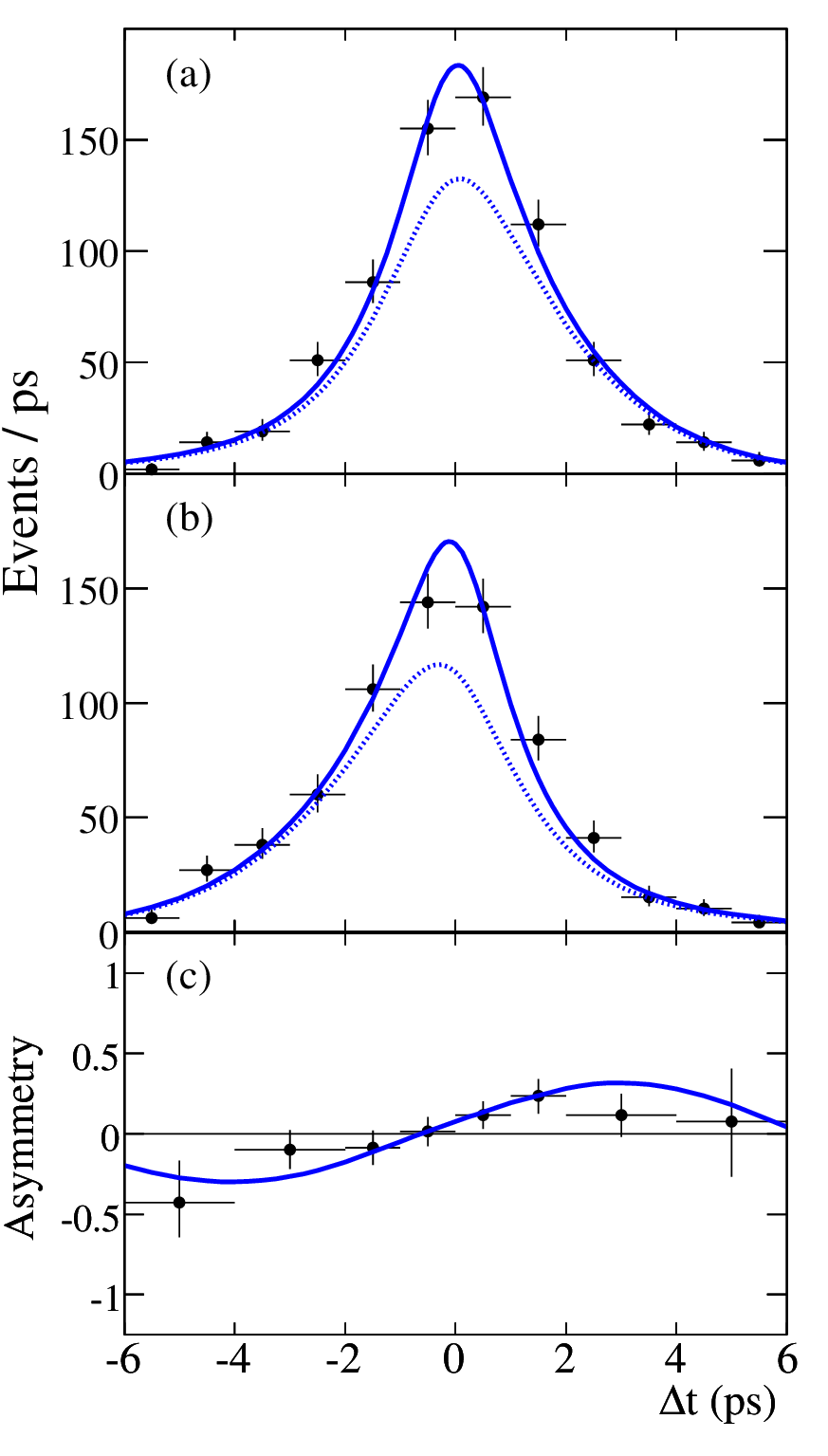}
  \hfill
  \includegraphics[width=110pt,height=!]{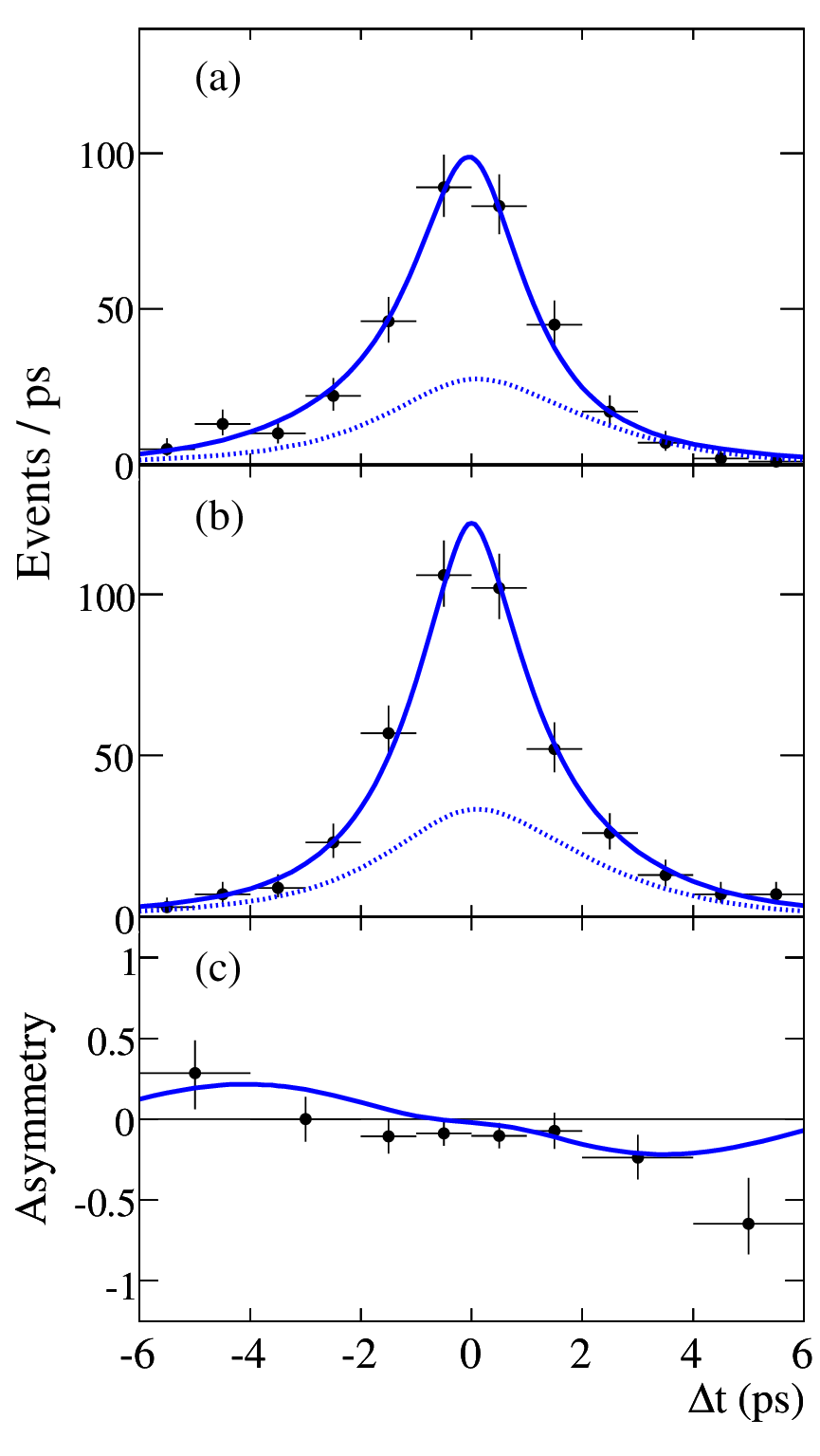}
  \hfill
  \includegraphics[width=150pt,height=160pt]{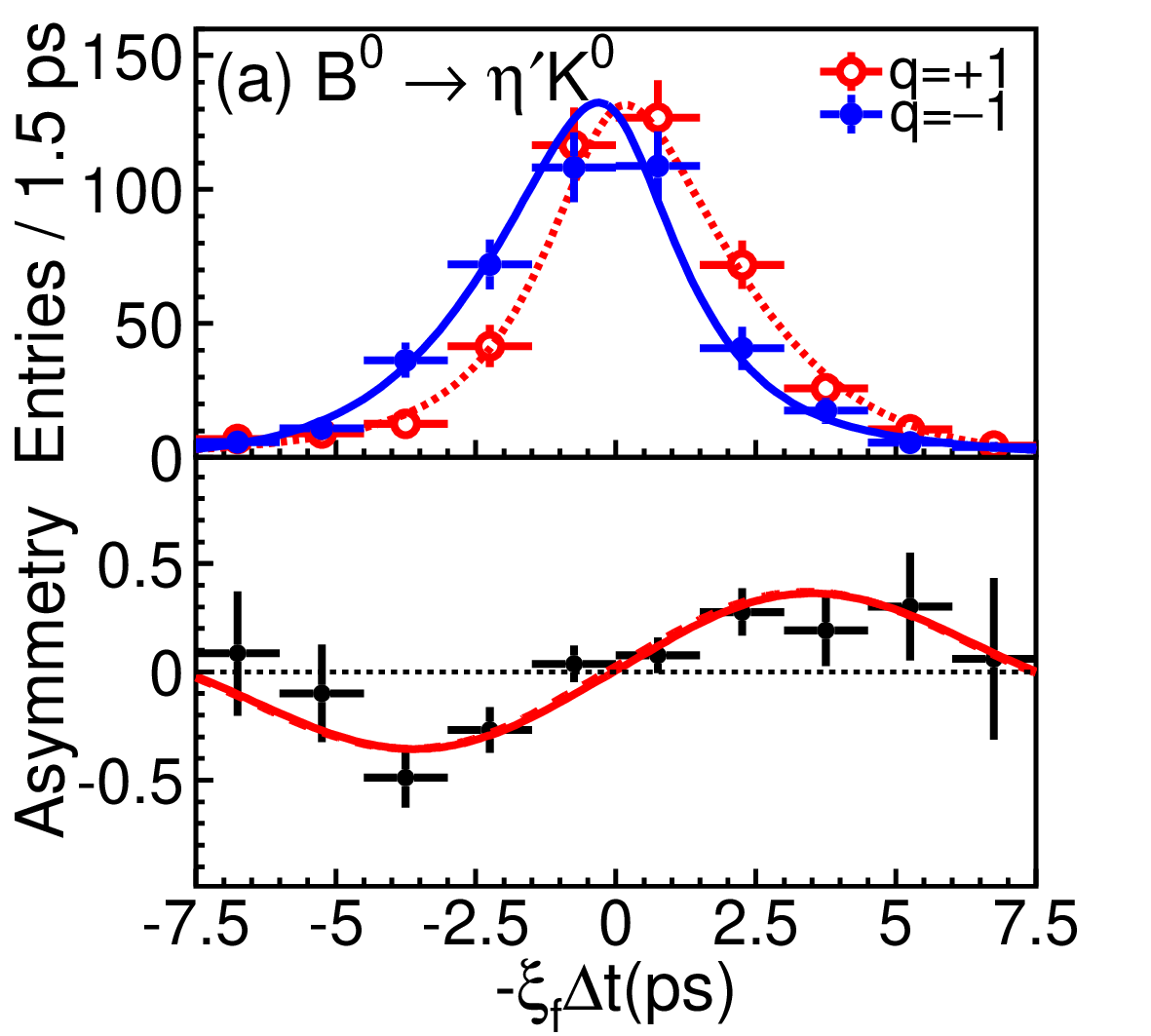}
  \put(-316,178){\Bz\ tags}
  \put(-198,178){\Bz\ tags}
  \put(-316,118){\Bzb\ tags}
  \put(-198,118){\Bzb\ tags}
  \caption{
    Signal enhanced \deltat\ projections and asymmetry plots for $\Bz \ra
    \etapr \Kz$. The left (middle) plot shows BaBar's fit results for $\Bz \ra
    \etapr \KS$ ($\Bz \ra \etapr \KL$) and the right plot shows Belle's
    combined fit result.
  }
  \label{fig_etapk0}
\end{figure}

For $\Bz \ra \omega \KS$, the only useful decay channel is, $\Bz \ra
\omega(\pip \pim \piz) \KS(\pip \pim)$. BaBar uses
\mes, \DeltaE, a Fisher discriminant, the $\omega$ mass and its
helicity to
discriminate between signal and background~\cite{:2008se} while Belle uses \Mbc, \DeltaE, ${\cal R}_{S/B}$ and
the $\omega$ mass~\cite{Abe:2006gy}. The fit results are summarized in
Tab.~\ref{tab_etapk0_wks} and
Fig.~\ref{fig_chmls}.

\begin{table}[htb]
  \caption{Summary of $\Bz \ra \etapr \Kz$ and $\Bz \ra \omega \KS$.}
  \label{tab_etapk0_wks}
  \begin{tabular}
    {@{\hspace{0.5cm}}c@{\hspace{0.5cm}}||@{\hspace{0.5cm}}c@{\hspace{0.25cm}}  @{\hspace{0.25cm}}c@{\hspace{0.25cm}}}
    \hline \hline
    & BaBar & Belle\\
    \hline
    \multicolumn{3}{c}{$\Bz \ra \etapr \Kz$} \\
    Yield ($N(\BB) \times 10^{6}$) & $2515 \pm 69$ \ ($467$) & $1875 \pm 60$ \
    ($535$) \\
    \hline
    \multicolumn{3}{c}{$\Bz \ra \omega \Kz$} \\
    Yield ($N(\BB) \times 10^{6}$) & $163 \pm 18$ \ ($467$) & $118 \pm 18$ \ ($535$) \\
    \hline \hline
  \end{tabular}
\end{table}

In these modes there is no evidence for direct \CP\ violation while
mixing-induced \CP\ violation is consistent with charmonium.
The significance of the mixing-induced \CP\ violation effect in 
$\Bz \ra \etapr \Kz$ is greater than $5\sigma$ in both BaBar and Belle analyses.

\vspace{10pt}
\noindent
\underline{$\Bz \ra \Kz \piz$, $\Bz \ra \KS \KS \KS$ and $\Bz \ra \KS \piz \piz$}

These modes are distinguished by the lack of a primary track coming from the
reconstructed \B\ vertex. In such cases, the \B\ vertex is determined by
extrapolating the \KS pseudo-track back to the interaction point. However, due
to the relatively long lifetime of the \KS\ meson, the vertex reconstruction
efficiency is less than 100\% as the charged pion daughters may not be able to
register hits in the innermost sub-detector. 

For $\Bz \ra \Kz \piz$, BaBar describes signal events with the reconstructed \B\
mass and the mass of the tag-side \B\ calculated from the known beam energy
and reconstructed \B\ momentum constrained with the nominal \B\ mass. In
addition, the cosine of the polar angle of the \B\ candidate in the \FourS\
frame and ratio of angular moments, $L_{2}/L_{0}$, which discriminate against
continuum are also used~\cite{:2008se}. 
Belle uses \Mbc, \DeltaE, and ${\cal R}_{S/B}$ to describe signal events and
additionally considers the $\Bz \ra \KL \piz$ channel for which \DeltaE\ cannot
be calculated~\cite{Fujikawa:2008pk}. Fig.~\ref{fig_k0pi0} shows 
\deltat\ and asymmetry projections for $\Bz \ra \Kz \piz$. 

\begin{figure}[htb]
  \centering
  \includegraphics[width=135pt,height=!]{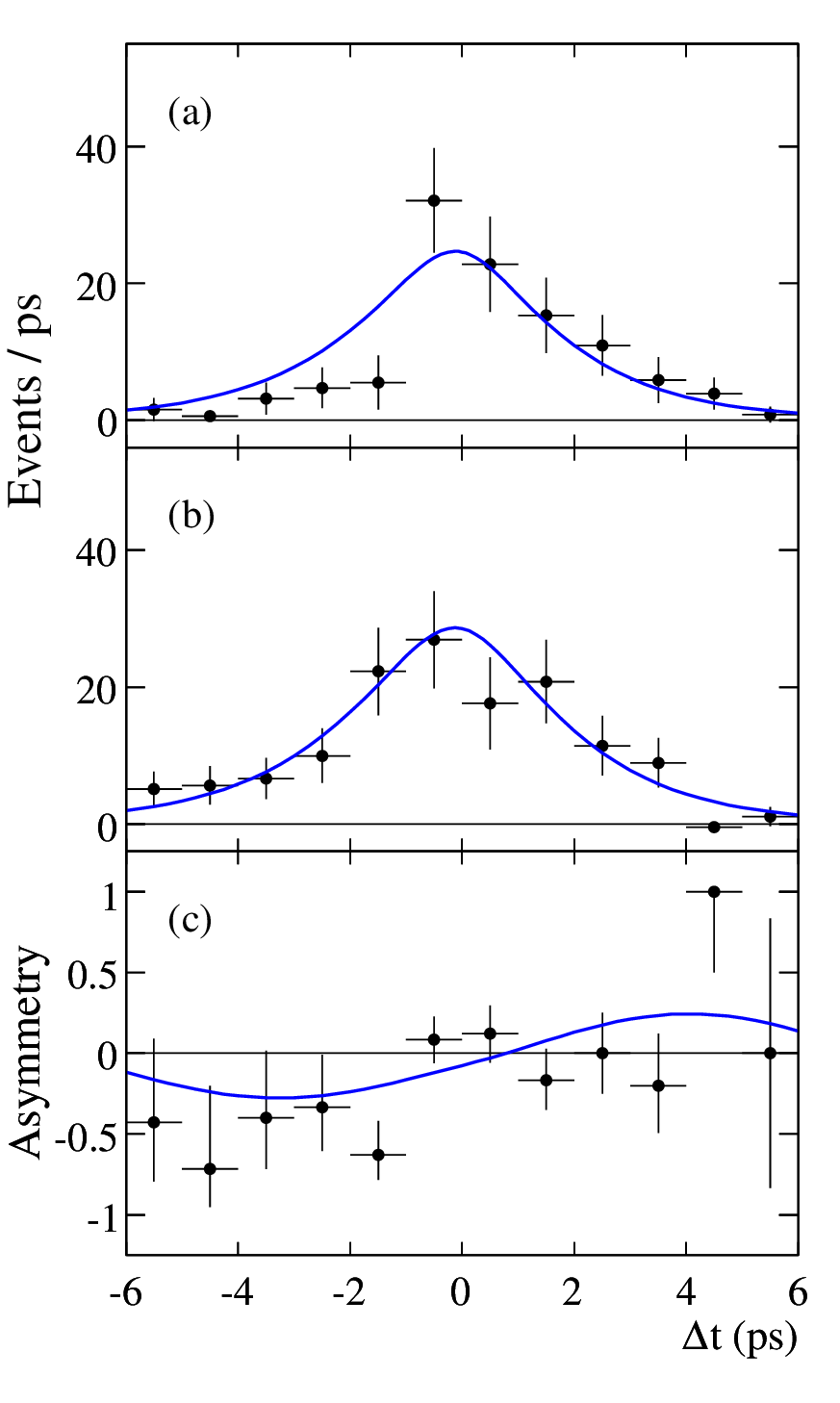}
  \hfill
  \includegraphics[width=175pt,height=!]{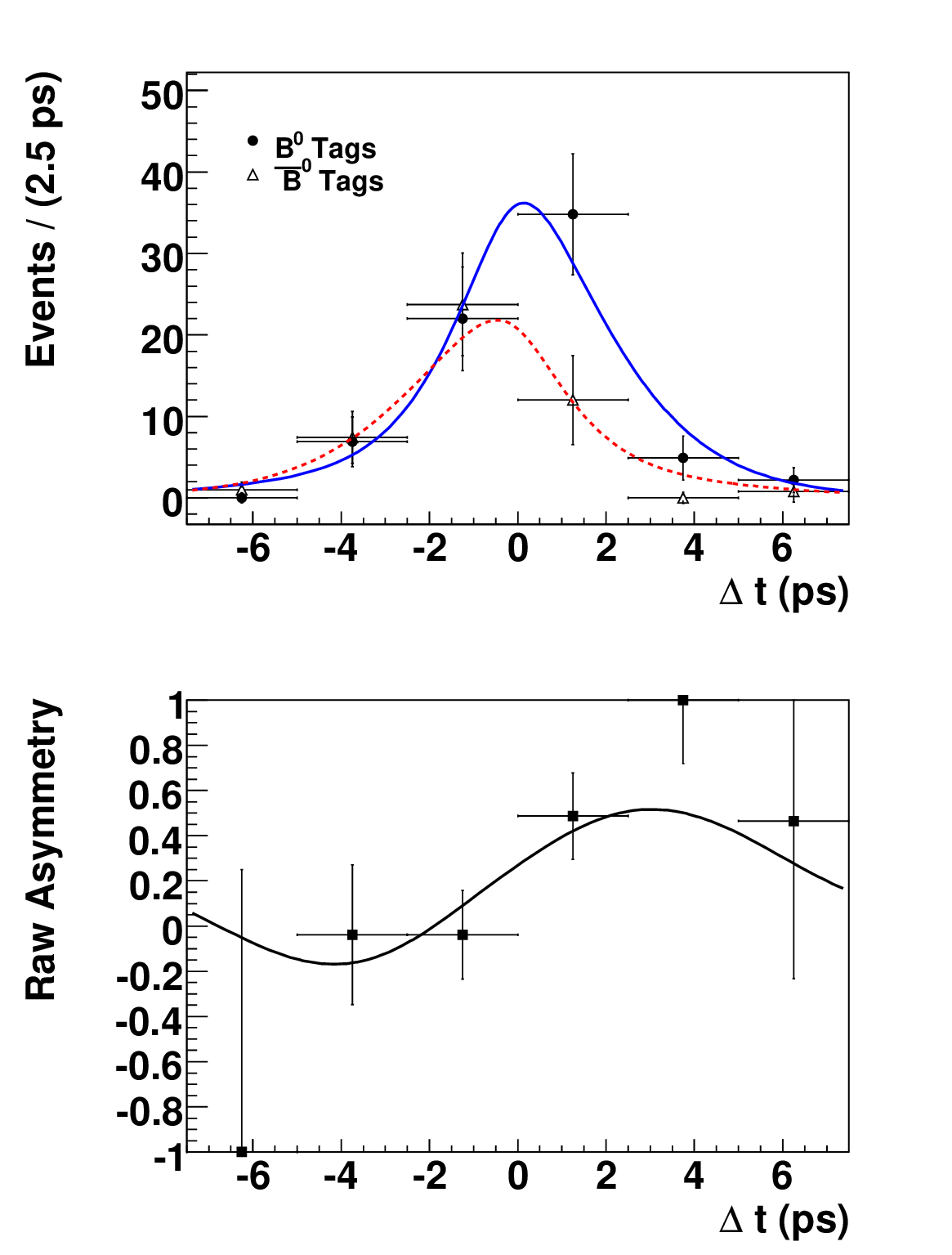}
  \put(-298,205){\Bz\ tags}
  \put(-298,140){\Bzb\ tags}
  \caption{
    Signal enhanced and background subtracted \deltat\ projections and
    asymmetry plots for $\Bz \ra \Kz \piz$. The left plot shows BaBar's fit
    result and the right plot shows Belle fit result.
  }
  \label{fig_k0pi0}
\end{figure}

For $\Bz \ra \KS \KS \KS$ and $\Bz \ra \KS \piz \piz$~\cite{Gershon:2004tk},
BaBar uses \mes, \DeltaE and a neural network (NN) which distinguishes 
\BB\ from \qqbar\ events to describe signal~\cite{ksksks_BaBar, Aubert:2007ub} 
and similarly, Belle uses \Mbc, \DeltaE and 
${\cal R}_{S/B}$~\cite{Chen:2006nk,:2007xd}.  
For $\Bz \ra \KS \KS \KS$, both experiments include the case where one 
\KS\ decays to a neutral pion pair. The fit results are summarized in
Tab.~\ref{tab_k0pi0_ksksks_kspi0pi0} and Fig.~\ref{fig_chmls}.

\begin{table}[htb]
  \caption{Summary of \Bz \ra \Kz \piz, \Bz \ra \KS \KS \KS\ and \Bz \ra \KS \piz \piz.}
  \label{tab_k0pi0_ksksks_kspi0pi0}
  \begin{tabular}
    {@{\hspace{0.5cm}}c@{\hspace{0.5cm}}||@{\hspace{0.5cm}}c@{\hspace{0.25cm}}  @{\hspace{0.25cm}}c@{\hspace{0.25cm}}}
    \hline \hline
    & BaBar & Belle\\
    \hline
    \multicolumn{3}{c}{$\Bz \ra \Kz \piz$} \\
    Yield ($N(\BB) \times 10^{6}$) & $556 \pm 32$ \ ($467$) & $657 \pm 37$ \ ($657$) \\
    \hline
    \multicolumn{3}{c}{$\Bz \ra \KS \KS \KS$} \\
    Yield ($N(\BB) \times 10^{6}$) & $274 \pm 20$ \ ($467$) & $185 \pm 17$ \ ($535$) \\
    \hline
    \multicolumn{3}{c}{$\Bz \ra \KS \piz \piz$} \\
    Yield ($N(\BB) \times 10^{6}$) & $117 \pm 27$ \ ($227$) & $307 \pm 32$ \ ($657$) \\
    \hline \hline
  \end{tabular}
\end{table}

In these modes the direct \CP\ components are all consistent with Standard
Model expectations and the mixing-induced parameters are consistent with
charmonium with current statistics.
The largest discrepancy, which is not statistically significant, is in the
mixing-induced \CP\ violation parameter in $\Bz \ra \KS \piz \piz$, which
appears to have the wrong sign. 

\vspace{10pt}
\noindent
\underline{$\Bz \ra \KS \pip \pim$ and $\Bz \ra \KS \Kp \Km$}
\label{sec:angles:beta-tddp}

To extract \CP\ violation parameters of modes such as $\Bz \ra \KS \rho^0$
($\rho^0 \ra \pi^+ \pi^-$) or $\Bz \ra \KS \phi$ ($\phi \ra K^+ K^-$), it is
necessary to perform a time-dependent Dalitz plot analysis as interfering
resonances in the three-body final states make the results of quasi-two-body
analyses difficult to interpret. As the relative amplitudes and phases of each 
decay channel in the Dalitz plot are determined in such an analysis, the angle
\phioneeff\ can be directly obtained, rather than measuring \Scpeff.

For $\Bz \ra \KS \pip \pim$, the signal model contains the $K^{*+}(892)$,
$K^{*+}_{0}(1430)$, $\rho^{0}(770)$, $f_{0}(980)$, $f_{2}(1270)$,
$f_{\rm X}(1300)$ states and a nonresonant component. BaBar describes signal
events with \mes, \DeltaE and the output of a neutral
network~\cite{Aubert:2007vi} while Belle just uses
\DeltaE~\cite{:2008wwa}. Belle finds two solutions given in
Tab.~\ref{tab_kspipi_belle} with consistent \CP\ parameters but different
$K^{*+}_{0}(1430) \pim$ relative fractions due to the interference between
$K^{*+}_{0}(1430)$ and the non-resonant component. The high $\Kstarp \pim$
fraction of Solution 1 is in agreement with some phenomenological
estimates~\cite{Chernyak:2001hs} and may also be qualitatively favored by the
total $K$--$\pi$ S-wave phase shift as a function of $m(K\pi)$ when compared
with that measured by the LASS collaboration~\cite{Aston:1987ir}. The fit
results for both experiments are summarized in Tab.~\ref{tab_kspipi_kskk} and
Fig.~\ref{fig_kspipi_kskk}, which includes the preferred solution from Belle.  

\begin{table}[htb]
  \caption{Multiple solutions in $\Bz \ra \KS \pip \pim$ at Belle where the
    first error is statistical, the second systematic and the third is the
    model uncertainty.} 
  \label{tab_kspipi_belle}
  \begin{tabular}
    {@{\hspace{0.5cm}}c@{\hspace{0.5cm}}||@{\hspace{0.5cm}}c@{\hspace{0.25cm}}  @{\hspace{0.25cm}}c@{\hspace{0.25cm}}}
    \hline \hline
    & Sol. 1 & Sol. 2 \\
    \hline
    $\phioneeff(\rho^{0}(770) \KS)$ & $(20.0^{+8.6}_{-8.5} \pm 3.2 \pm 3.5)^{\circ}$ & $(22.8 \pm 7.5 \pm 3.3 \pm 3.5)^{\circ}$\\
    $\phioneeff(f_{0}(980) \KS)$ & $(12.7^{+6.9}_{-6.5} \pm 2.8 \pm 3.3)^{\circ}$ & $(14.8^{+7.3}_{-6.7} \pm 2.7 \pm 3.3)^{\circ}$\\
    \hline \hline
  \end{tabular}
\end{table}

The decay $\Bz \ra \KS \Kp \Km$ is also studied with a time-dependent Dalitz
plot analysis. The signal model contains the $f_{0}(980)$, $\phi(1020)$,
$f_{X}(1500)$ and $\chi_{c0}$ states and a nonresonant component. The BaBar
collaboration additionally uses the \KS\ decay channel to neutral pions and
describes signal events with \mes and \DeltaE~\cite{:2008gv}. Similarly, Belle
uses \Mbc\ and \DeltaE~\cite{Dalseno:2008ms}.

\begin{figure}[htb]
  \centering
  \includegraphics[width=0.42\textwidth]{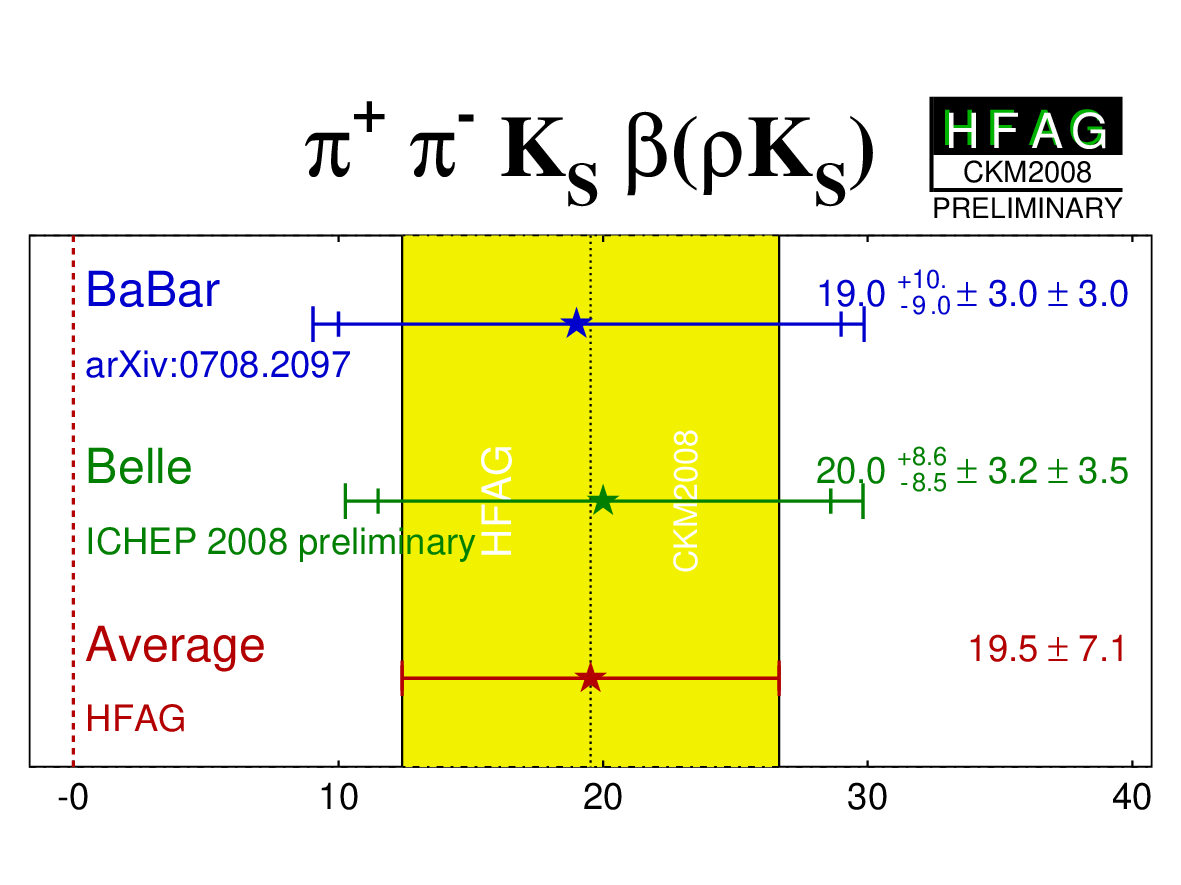}
  \hspace{0.03\textwidth}
  \includegraphics[width=0.42\textwidth]{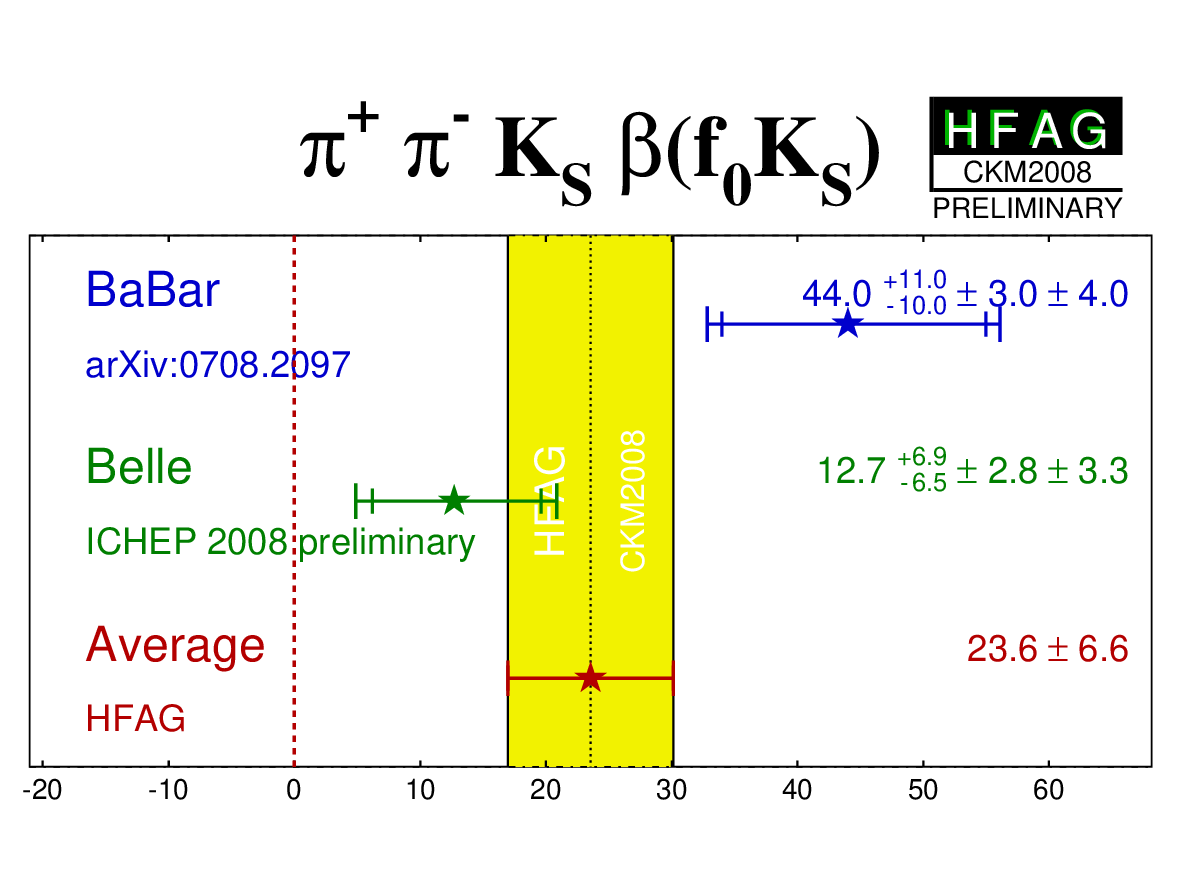}\\
  \includegraphics[width=0.42\textwidth]{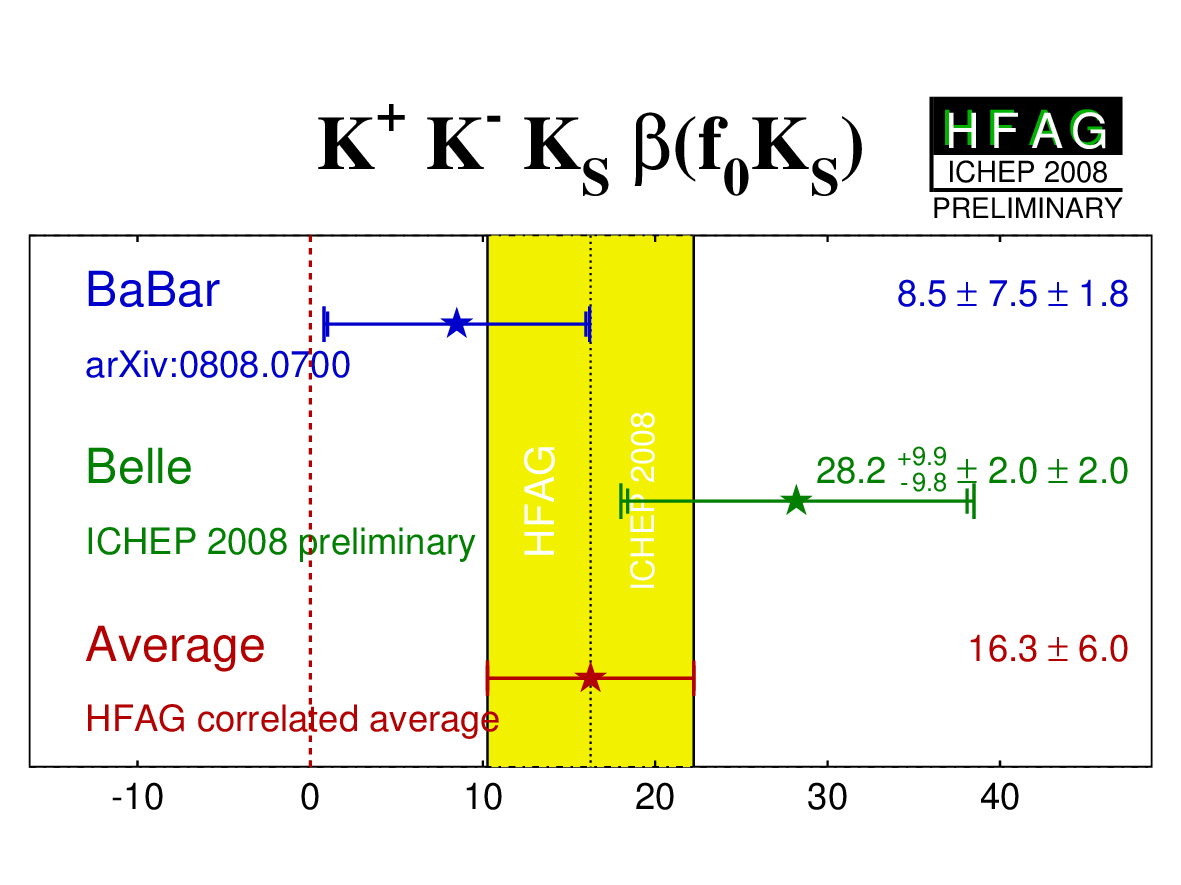}
  \hspace{0.03\textwidth}
  \includegraphics[width=0.42\textwidth]{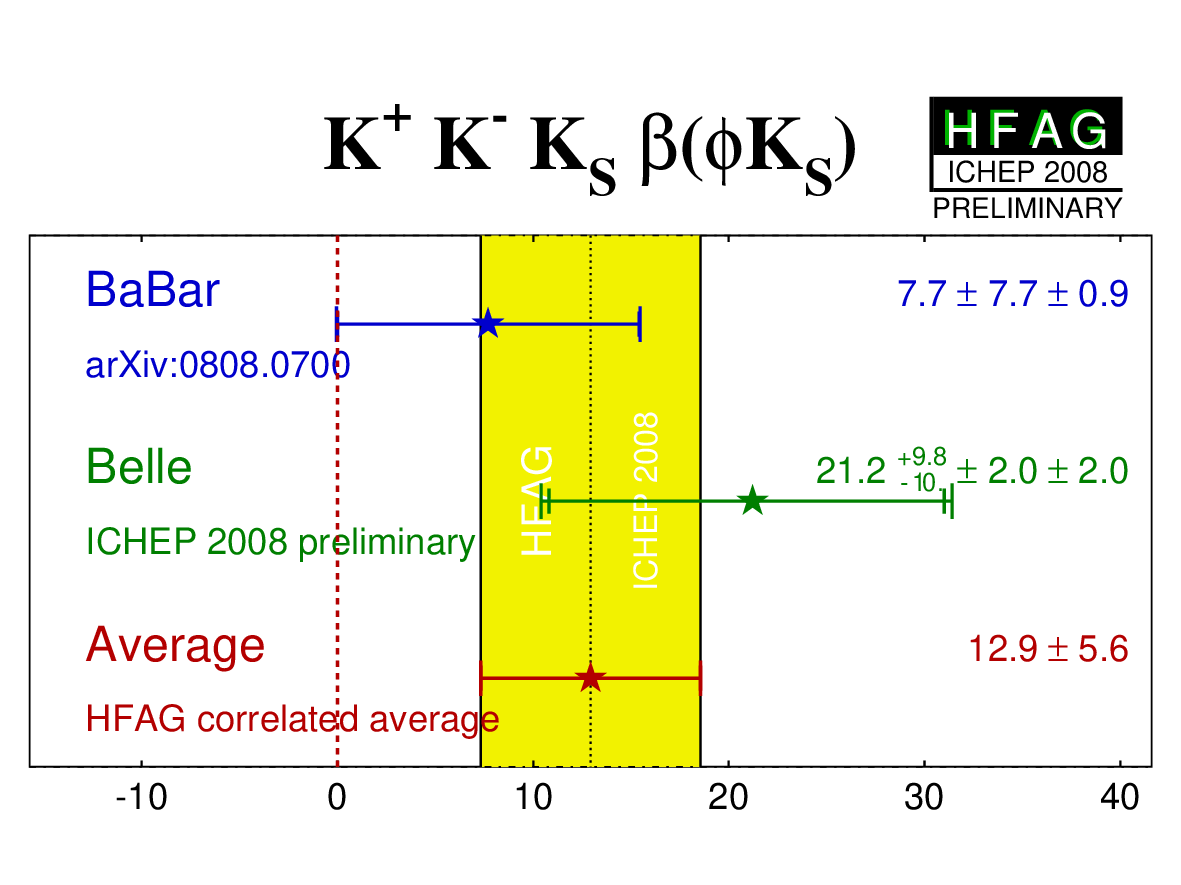}
  \caption{
    \CP\ parameters of $\Bz \ra \KS \pip \pim$ and $\Bz \ra \KS \Kp \Km$.
  }
  \label{fig_kspipi_kskk}
\end{figure}

Belle finds four solutions as shown in Tab.~\ref{tab_kskk_belle} due to the
interference between $f_{0}(980)$, $f_{X}(1500)$ and the non-resonant
component. 
Using external information from $\Bz \ra \KS \pip \pim$, 
if the $f_{X}(1500)$ is the $f_{0}(1500)$ for both $\Bz \ra \KS \pip \pim$ and 
$\Bz \ra \KS \Kp \Km$, the ratio of branching fractions, 
${\cal B}(f_{0}(1500) \ra \pip \pim) / {\cal B}(f_{0}(1500) \ra \Kp \Km)$,
prefers the solution with the low $f_{X}(1500) \KS$ fraction. 
Similarly other measurements of the relative magnitudes of the 
$f_{0}(980) \ra \pip \pi$ and $f_{0}(980) \ra \Kp \Km$ widths
prefer the solution with the low $f_{0}(980) \KS$ fraction. 
The fit results for both experiments are summarized
in Tab.~\ref{tab_kspipi_kskk} and Fig.~\ref{fig_kspipi_kskk}, 
which includes the preferred solution from Belle, while Fig.~\ref{fig_chmls}
gives a summary of these results together with those from other charmless
hadronic $B$ decays. 
In the time-dependent \CP\ violation analyses, there is no evidence for 
direct \CP\ violation and \phioneeff\ is consistent with charmonium. 
\begin{table}[htb]
  \caption{Multiple solutions in $\Bz \ra \KS \Kp \Km$ at Belle where the error is statistical only.}
  \label{tab_kskk_belle}
  \begin{tabular}
    {@{\hspace{0.5cm}}c@{\hspace{0.5cm}}||@{\hspace{0.5cm}}c@{\hspace{0.25cm}}  @{\hspace{0.25cm}}c@{\hspace{0.25cm}}  @{\hspace{0.25cm}}c@{\hspace{0.25cm}}  @{\hspace{0.25cm}}c@{\hspace{0.25cm}}}
    \hline \hline
    & Sol. 1 & Sol. 2 & Sol. 3 & Sol. 4 \\
    \hline
    $\phioneeff(f_{0}(980) \KS)$ & $(28.2^{+9.8}_{-9.9})^{\circ}$ & $(64.1^{+7.6}_{-8.0})^{\circ}$ & $(61.5^{+6.5}_{-6.5})^{\circ}$ & $(36.9^{+10.9}_{-9.6})^{\circ}$\\
    $\phioneeff(\phi(1020) \KS)$ & $(21.2^{+9.8}_{-10.4})^{\circ}$ & $(62.1^{+8.3}_{-8.8})^{\circ}$ & $(65.1^{+8.7}_{-8.7})^{\circ}$ & $(44.9^{+13.2}_{-13.6})^{\circ}$\\
    \hline \hline
  \end{tabular}
\end{table}

\begin{table}[htb]
  \caption{Summary of $\Bz \ra \KS \pip \pim$ and $\Bz \ra \KS \Kp \Km$.}
  \label{tab_kspipi_kskk}
  \begin{tabular}
    {@{\hspace{0.5cm}}c@{\hspace{0.5cm}}||@{\hspace{0.5cm}}c@{\hspace{0.25cm}}  @{\hspace{0.25cm}}c@{\hspace{0.25cm}}}
    \hline \hline
    & BaBar & Belle\\
    \hline
    \multicolumn{3}{c}{$\Bz \ra \KS \pip \pim$} \\
    Yield ($N(\BB) \times 10^{6}$) & $2172 \pm 70$ \ ($383$) & $1944 \pm 98$ \
    ($657$) \\
    \hline
    \multicolumn{3}{c}{$\Bz \ra \KS \Kp \Km$} \\
    Yield ($N(\BB) \times 10^{6}$) & $1428 \pm 47$ \ ($467$) & $1269 \pm 51$ \
    ($657$) \\
    \hline \hline
  \end{tabular}
\end{table}

\begin{figure}[htb]
  \centering
  \includegraphics[height=0.46\textheight]{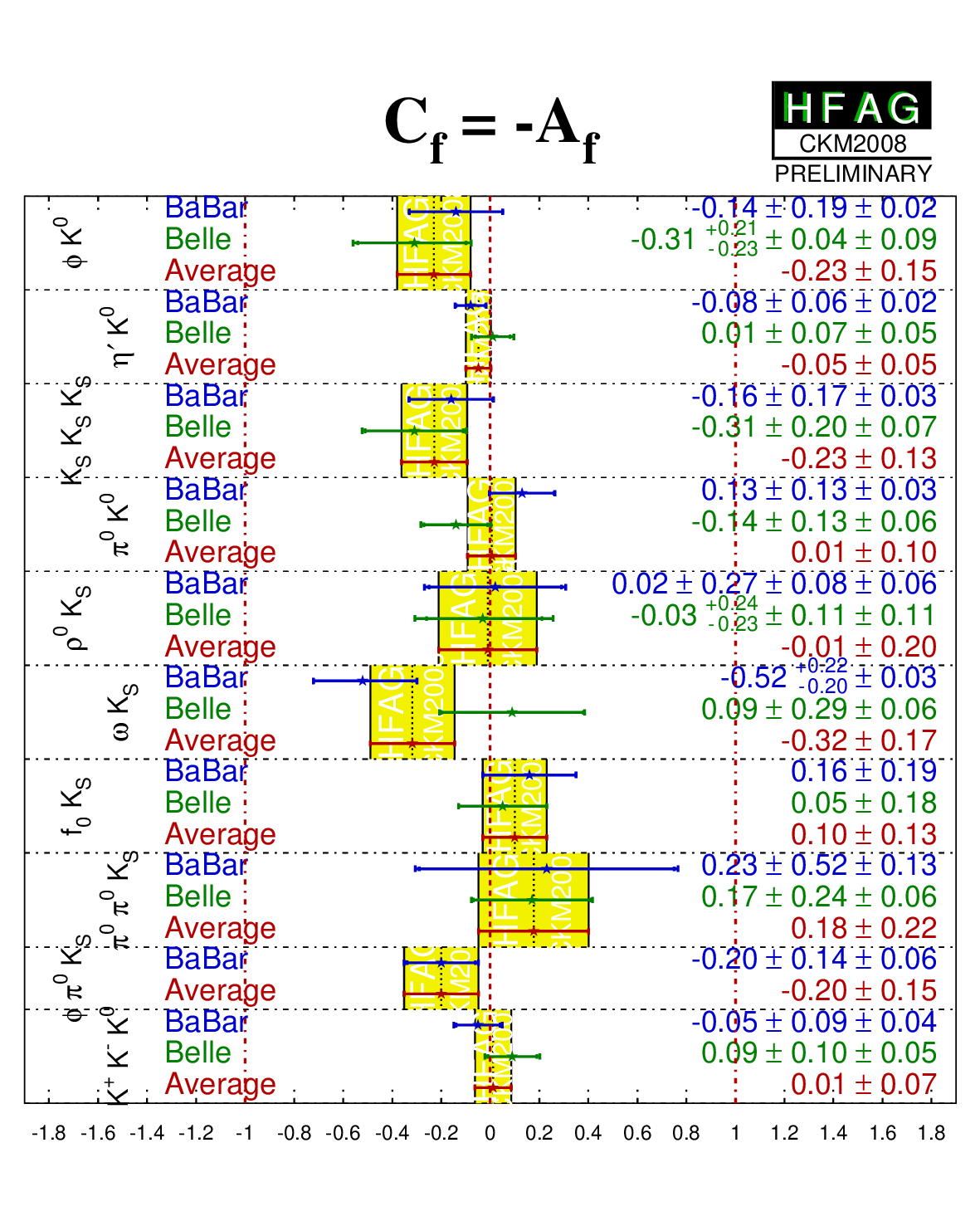}
  \includegraphics[height=0.46\textheight]{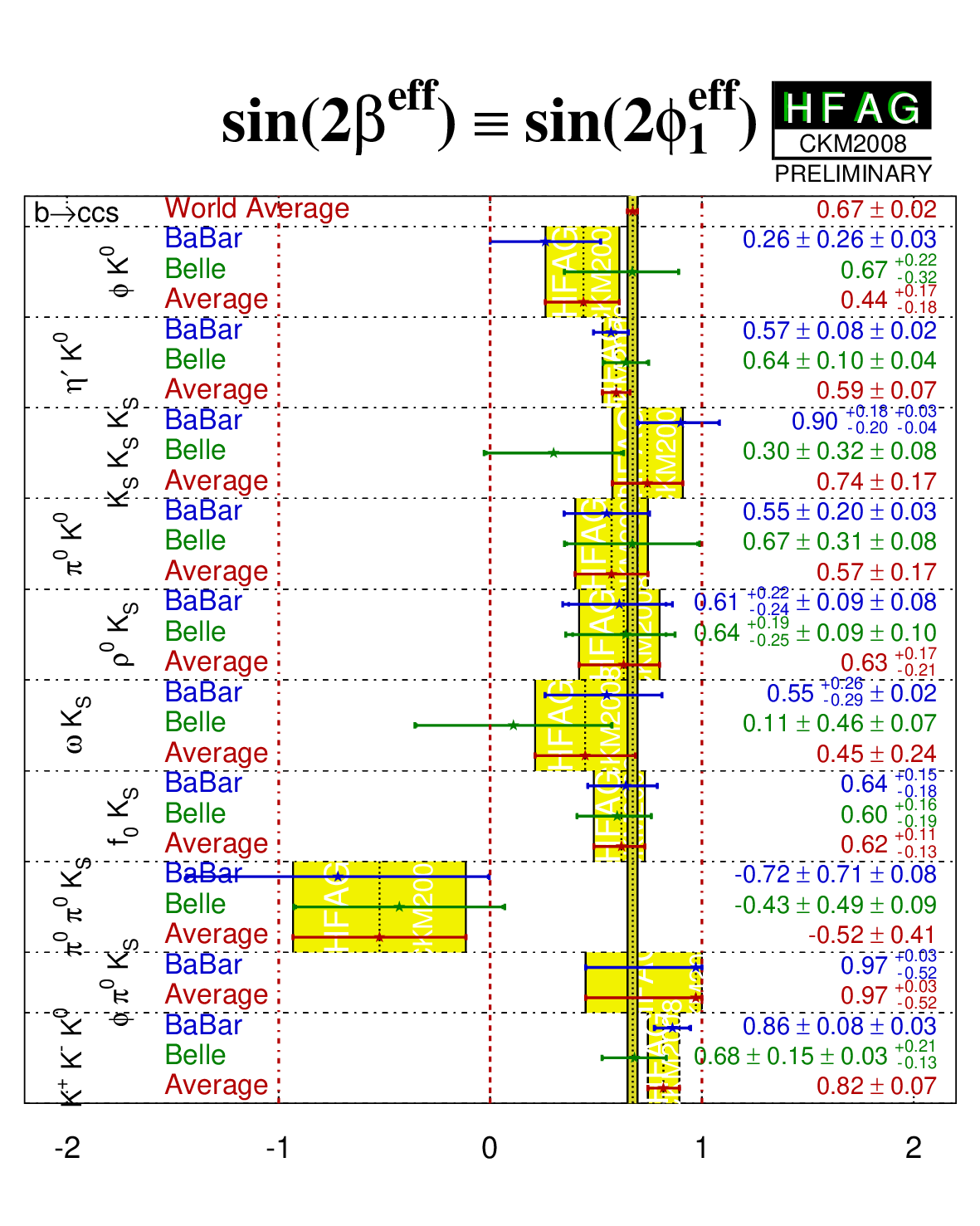}
  \caption{
    \CP\ parameters of charmless hadronic \B\ decays.
  }
  \label{fig_chmls}
\end{figure}

%

\subsection{Measurements of \aa}
\label{sec:chmls-angle-alpha}


\subsubsection{Theoretical aspects}

The $b \to u$ tree amplitude (Fig.~\ref{fig:angles:alpha_diags}(a)) is
proportional to $V_{ub}$ and, in the usual convention, carries the weak
phase~$\gamma$. Since \BzBzb mixing carries the weak phase $2\beta$, at the
tree level the time-dependent \CP-violation measurements in the $\Bz \to
\pip\pim$ and $\Bz \to \rho^+\rho^-$ decays 
are sensitive to $2\beta + 2\gamma = 2\pi - 2\alpha$.

The decay-time distribution for
$\Bz\to\pip\pim$ is given by
\begin{equation}
\frac{dN}{d\deltat} = 
\frac{e^{-\left|\deltat\right|/\tau}}{4\tau}\times
\Bigl\{1-q^{}_{\rm tag}[ 
C^{}_{\pi\pi} \cos(\deltamd\deltat) - S^{}_{\pi\pi} \sin(\deltamd\deltat)]\Bigr\}\,,
\label{eqn:angles:alpha_master}
\end{equation}
where $\tau$ is the neutral $B$\/ lifetime, $\deltamd$ is the \Bz--\Bzb mixing
frequency, $\deltat$ is the difference in decay times $t^{}_{\pi\pi}-t^{}_{\rm tag}$,
and the parameter $q^{}_{\rm tag}$ equals $+1\,(-1)$ when the
tag-side $B$ decays as a $B^0 (\overline{B}{}^0)$. 
The parameter $C^{}_{\pi\pi}$ characterizes direct \CP violation and is also
referred to in the literature as $-{\cal A}^{}_{\pi\pi}$.
At the tree level, the \CP-violating asymmetries $S_{\pi\pi} = \sin{2\alpha}$ 
($\alpha \equiv \arg\left[-V_{td}^{}V_{tb}^{*}/V_{ud}^{}V_{ub}^{*}\right]$)
and $C_{\pi\pi} \equiv -{\cal A}^{}_{\pi\pi} =0$.
However, since the leading higher-order $b \to d$ contribution to the 
$\Bz \to \pip\pim$ 
decay amplitude (Fig.~\ref{fig:angles:alpha_diags}(b))
is sizable and carries the weak phase $-\beta$, direct \CP violation 
$C_{\pi\pi} \neq 0$ becomes possible and 
$S_{\pi\pi} = \sin{2\alpha_{\rm eff}} \sqrt{1-C^2_{\pi\pi}}$, 
where, in general, the phase difference 
$\alpha - \alpha_{\rm eff} = \Delta\alpha \equiv \delta \neq 0$. 
Contributions from physics beyond the Standard Model could affect the
\CP-violating asymmetries $S_{\pi\pi}$ and $C_{\pi\pi}$ primarily through
additional penguin amplitudes.

The value of $\delta$ can be extracted through a model-independent analysis 
that uses the $SU(2)$ isospin-related decays $B^{\pm}\to\pipm\piz$ and
$\Bz\to\piz\piz$~\cite{Gronau:1990ka}.
Let us denote the $B^{ij} \to \pi^i \pi^j$ and $\Bbar^{ij} \to \pi^i \pi^j$
decay amplitudes $A^{ij}$ and $\Abar^{ij}$, respectively. 
Assuming isospin symmetry, these amplitudes are related by the equations
\begin{equation}
A^{+-}/\sqrt{2} + A^{00} = A^{+0},~~~~~~~~\Abar^{+-}/\sqrt{2} + \Abar^{00} = \Abar^{-0},
\end{equation}
\noindent 
which can be represented graphically in the form of ``isospin triangles'' 
(Fig.~\ref{fig:angles:alpha_diags}(c)). 
Neglecting electroweak penguins, $|A^{+0}|=|\Abar^{-0}|$
(evidence of direct \CP\ violation in $B^{\pm}\to\pipm\piz$ would show that
such contributions cannot be neglected, and would be a signal for new physics
contributions).
If the (arbitrary) global phase of all $A^{ij}$ amplitudes is chosen
such that $A^{+0}=\Abar^{-0}$, it can be shown that the phase difference 
between $A^{+-}$ and $\Abar^{+-}$ is $2\delta$.
Note that the value of $\delta$ extracted in this manner carries an eightfold
ambiguity.
Moreover, the value of $\aa$ that is obtained is insensitive to new physics
effects, unless they violate isospin.
In the $B \to \pi \pi$ system (as in the $B \to \rho \rho$ case, discussed
below), knowledge of $A^{00}$ and $\Abar^{00}$ is the limiting factor in 
the extraction of $\delta$.

\begin{figure}[!bp]
\begin{center}
\begin{tabular}{c c c}
  \includegraphics[height=0.70in]{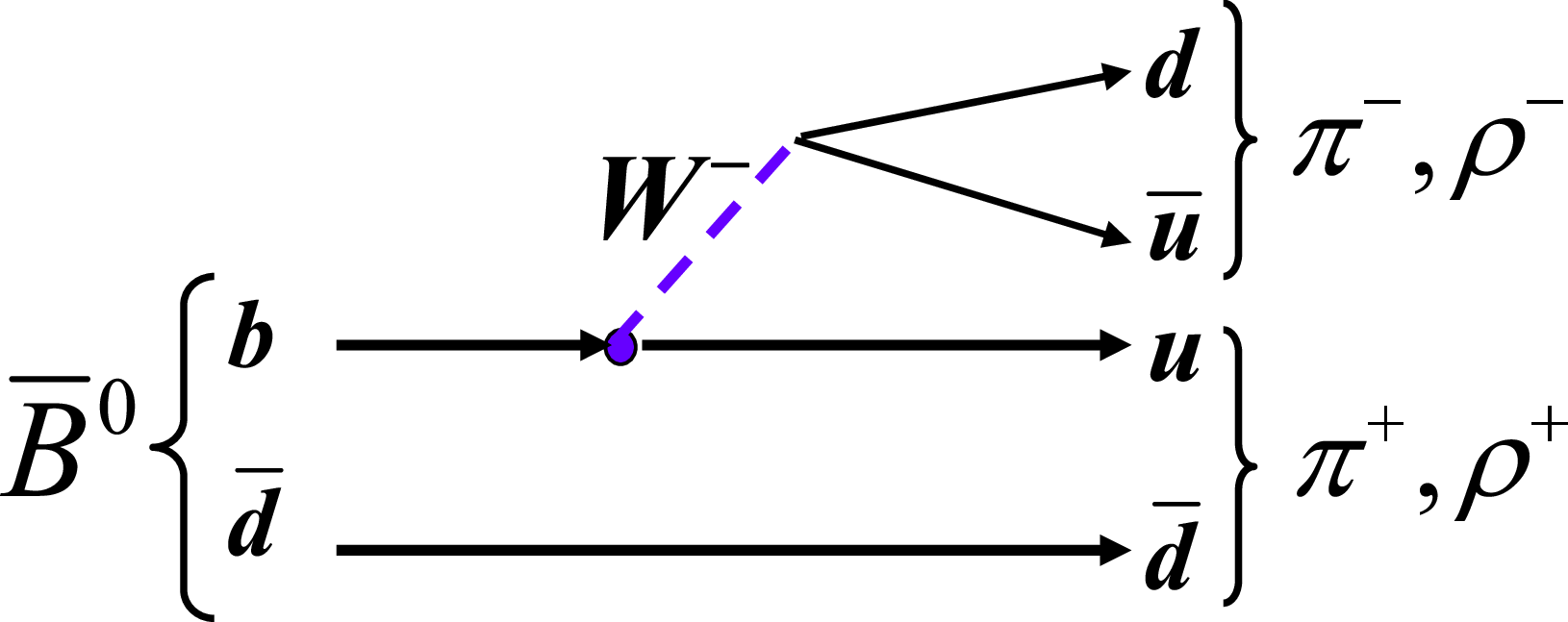} & ~
  \includegraphics[height=0.70in]{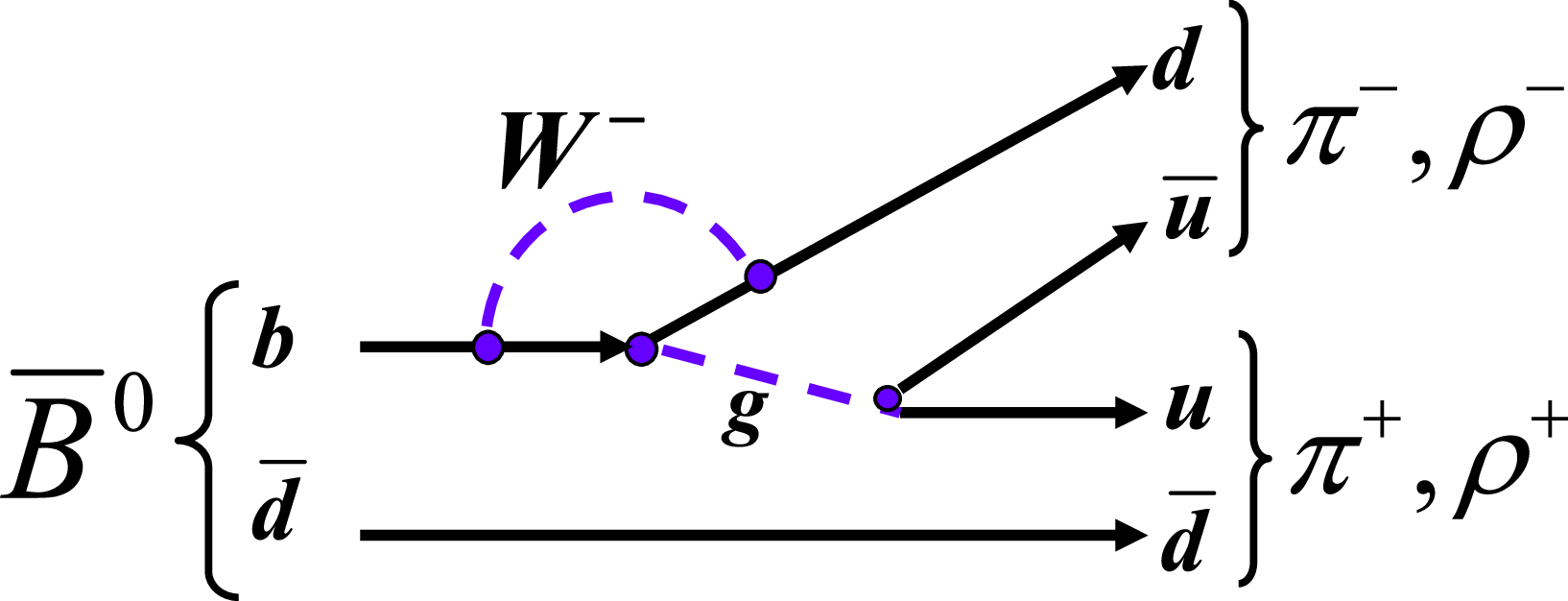} & ~
  \includegraphics[height=0.70in]{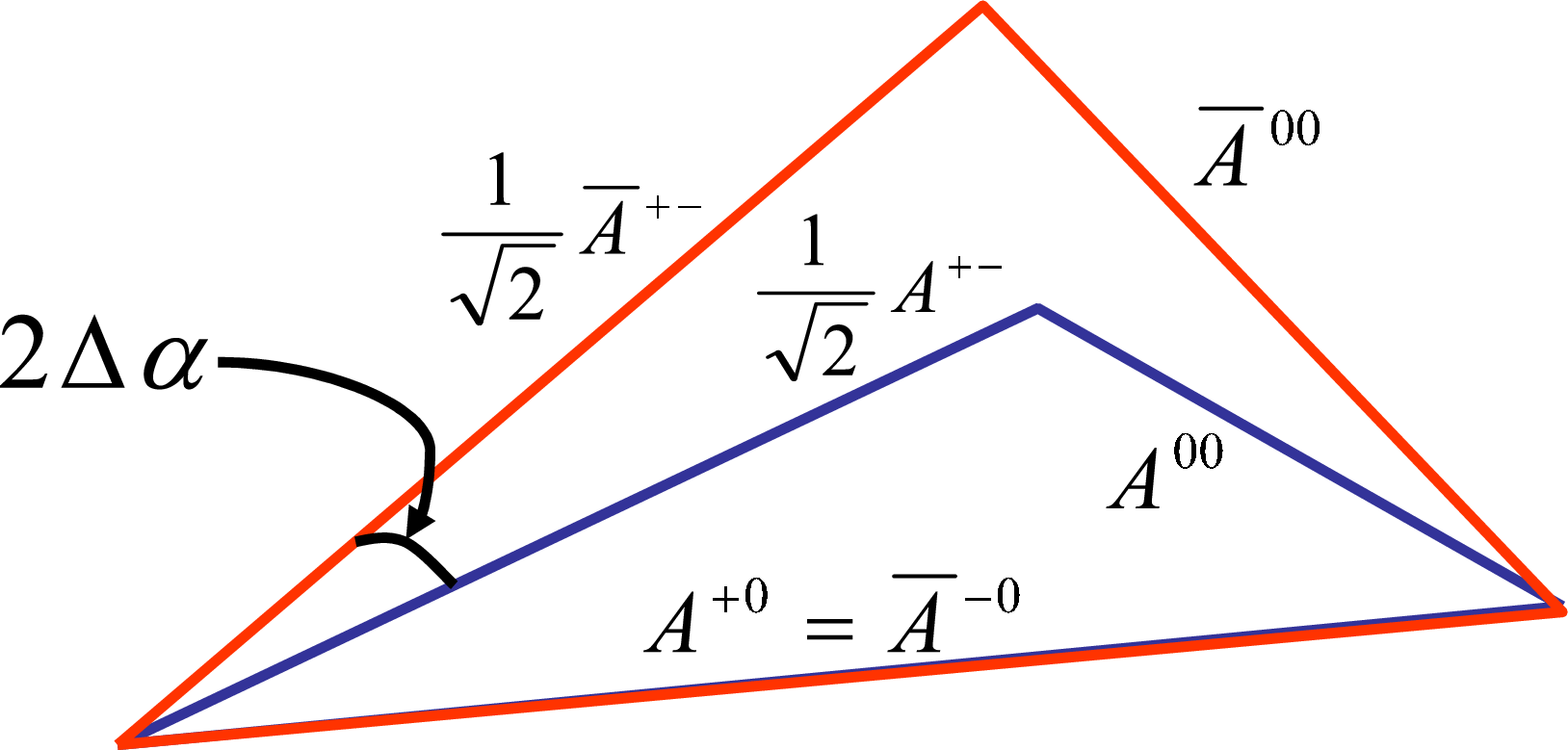} \\  
\end{tabular}
  \put(-375,21){\bfseries \small {(a)}}
  \put(-235,21){\bfseries \small {(b)}}
  \put(-95,21){\bfseries \small {(c)}}
  \caption{
    (a) Tree and (b) gluonic-penguin contributions to 
    $B^0 \to (\pi/\rho)^+ (\pi/\rho)^-$. 
    (c) London--Gronau isospin triangles for $B \to \pi\pi$, 
    $B \to \rho\rho$~\cite{Gronau:1990ka}.
  }
  \label{fig:angles:alpha_diags}
\end{center}
\end{figure}

For $B \to \rho \rho$ decays, the same formalism applies separately to each
helicity amplitude (where $\CP=+1$ ($L=0,2$) and $\CP=-1$ ($L=1$)).
Thus, the extraction of $\aa$ requires knowledge of the polarization. 
In practise, the fraction of longitudinal polarization ($f^{}_L$) is measured
by fitting the $\rho$ helicity angle distribution. The probability 
density function (PDF) used is
\begin{eqnarray}
  \frac{d^2N}{d\cos\theta^{}_1\,d\cos\theta^{}_2} & \ =\  & 
  4f^{}_L\cos^2\theta^{}_1\cos^2\theta^{}_2 +
  (1-f^{}_L)\sin^2\theta^{}_1\sin^2\theta^{}_2\,,
\end{eqnarray}
where $\theta^{}_1$ ($\theta^{}_2$) is the angle between the daughter $\pi^0$
and direction opposite the $\rho^-$ ($\rho^+$) direction in the 
$\rho^+$ ($\rho^-$) rest frame (see Fig.~\ref{fig:angles:helicity}). 
$B^0\ra\rho^+\rho^-$ is found to be almost purely $f^{}_L\!=1$, which implies
that the \CP-odd $L\!=\!1$ component is negligible. 
This high polarization is fortunate, as it gives 
a larger \CP\ asymmetry and thus greater sensitivity to~\aa.
(Conversely, the possibility to resolve some of the ambiguities in the
solution for $\aa$ from the interference between different helicity amplitudes
is precluded.)
Moreover, the contributions from penguin amplitudes
(Fig.~\ref{fig:angles:alpha_diags}b) are found to be small for $B\ra\rho\rho$, 
allowing a determination of \aa\ with small theoretical uncertainty.

\begin{figure}[!htb]
\begin{center}
\epsfig{file=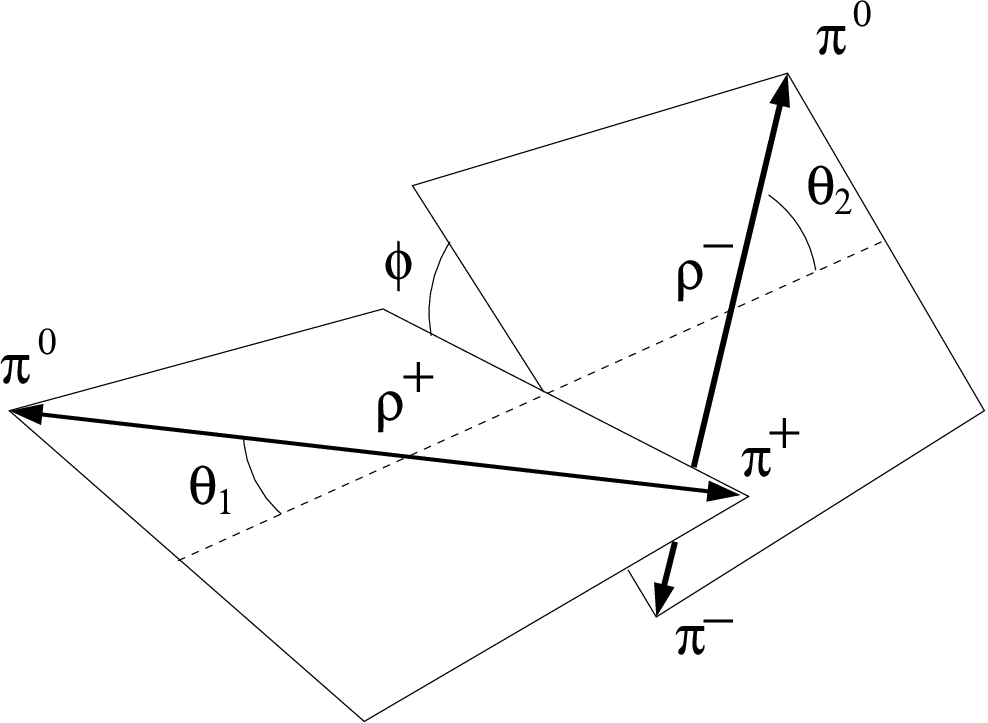,width=3.0in}
\end{center}
\caption{Definition of helicity angles $\theta^{}_1$ and $\theta^{}_2$
used to fit for $f^{}_L$, the fraction of longitudinal polarization. 
\label{fig:angles:helicity}
}
\end{figure}

A second complication in $B \to \rho \rho$ decays is that the final state
$\rho$ mesons have non-zero decay width, and thus their masses are not
necessarily equal. As a consequence, Bose-Einstein symmetry no longer holds, 
and the $I\!=\!1$ isospin state is allowed~\cite{Falk:2003uq}. 
In this case the isospin relations needed to extract \aa\ 
(Fig.~\ref{fig:angles:alpha_diags}c) do not hold. 
The problem can be studied by restricting the $\pi\pi$ 
invariant mass window used to select $\rho\ra\pi\pi$ candidates to a 
narrow range and checking whether the fitted value of \stwoa\ shifts. 
No such shift has been observed, and hence possible isospin violation is below
the sensitivity of current  measurements.

The decays $B^0\ra\rho^+\pi^-$, $B^0\ra\rho^-\pi^+$, and $B^0\ra\rho^0\pi^0$
(collectively referred to as $B^0\ra\rho\pi$) are also mediated by the 
$b\ra u\bar{u}d$ transition, and thus the interference between $B^0\ra\rho\pi$
and $\overline{B}{}^0\ra\rho\pi$ is also sensitive to~\aa. However, these
modes have an advantage over $B\ra\pi\pi$ and $B\ra\rho\rho$ decays, as
pointed out in Ref.~\cite{Snyder:1993mx}: the three-body $\pi^+\pi^-\pi^0$
final state yields a Dalitz plot that can be analyzed to measure all three
$B^0\ra(\rho\pi)^0$ modes simultaneously. 
The decay-time distributions of these three states allows one to resolve the
penguin contribution and determine \aa\ with very little theoretical
uncertainty and only a single unresolvable ambiguity ($\aa \longrightarrow \aa
+ \pi$). In addition, one can use the branching fractions for the charged
modes $B^+\ra\rho^+\pi^0$ and $B^+\ra\rho^0\pi^+$ along with isospin relations
to improve the determination of~\aa~\cite{Lipkin:1991st,Gronau:1991dq}.

The Dalitz plot has a time dependence
\begin{eqnarray}
|A(t,s^{}_+,s^{}_-)|^2 & \ \propto\  & e^{-\Gamma |t|}\Biggl\{
(|A^{}_{3\pi}|^2 + |\overline{A}^{}_{3\pi}|^2) \ -\ \nonumber \\
 & & \hskip0.50in
q^{}_{\rm tag}\cdot (|A^{}_{3\pi}|^2 - |\overline{A}^{}_{3\pi}|^2)\cos(\Delta m\,\Delta t)
\ +\ \nonumber \\
 & & \hskip0.75in
q^{}_{\rm tag}\cdot 2\cdot {\rm Im}\left(\frac{q}{p}A^*_{3\pi}\overline{A}^{}_{3\pi}\right)
\sin(\Delta m\,\Delta t)\Biggr\}\,,
\label{eqn:angles:rhopi_timedep}
\end{eqnarray}
where 
$A_{3\pi}={\cal A}(B^0\ra \pi\pi\pi)$, 
$\overline{A}_{3\pi}={\cal A}(\overline{B}{}^0\ra \pi\pi\pi)$, 
$s^{}_+ = (p_++p_0)^2$, $s^{}_- = (p_-+p_0)^2$, 
and $p^{}_+,\,p^{}_-$, and $p^{}_0$ are the four-momenta of
the $\pi^+\!$, $\pi^-\!$, and $\pi^0$, respectively.
The parameter $q^{}_{\rm tag}$ equals $+1\,(-1)$ when the
tag-side $B$ decays as a $B^0 (\overline{B}{}^0)$, and $q/p$ 
is the ratio of complex coefficients relating the $B^0$ and 
$\overline{B}{}^0$ flavor eigenstates to the mass eigenstates.

The amplitudes $A^{}_{3\pi}$ and $\overline{A}^{}_{3\pi}$ are
further decomposed into
\begin{eqnarray}
A^{}_{3\pi}(s^{}_+, s^{}_-) & \ =\  & 
f^{}_+(s^{}_+, s^{}_-)\,A_+ + f^{}_-(s^{}_+, s^{}_-)\,A_-  + f^{}_0(s^{}_+, s^{}_-)\,A_0 
\label{eqn:angles:rhopi_decomp_a}  \\
\left(\frac{q}{p}\right)\overline{A}^{}_{3\pi}(s^{}_+, s^{}_-) & \ =\  & 
\bar{f}^{}_+(s^{}_+, s^{}_-)\,\overline{A}_+ +
\bar{f}^{}_-(s^{}_+, s^{}_-)\,\overline{A}_-  + 
\bar{f}^{}_0(s^{}_+, s^{}_-)\,\overline{A}_0\,,
\label{eqn:angles:rhopi_decomp_b}
\end{eqnarray}
where the subscript ``+'' represents $\rho^+\pi^-$, 
``$-$'' is for $\rho^-\pi^+$, and ``0'' is for $\rho^0\pi^0$. 
The kinematic functions $f^{}_i$ and $\bar{f}^{}_i$ are the products
of Breit-Wigner functions to describe the $\pi\pi$ lineshape and an
angular function to describe the helicity distribution. The
goal of the analysis is to fit the time-dependence of the Dalitz plot 
to determine the six complex amplitudes $A^{}_i$ and $\overline{A}^{}_i$; 
from these one determines \aa\ via the relationship
\begin{eqnarray}
e^{i 2\aa} & = & \frac{\overline{A}^{}_+
 + \overline{A}^{}_- + 2\overline{A}^{}_0}
{A^{}_+ + A^{}_- + 2A^{}_0}\,.
\label{eqn:angles:rhopi_isospin}
\end{eqnarray}

Note that the description of the $\pi\pi$ lineshape introduces some systematic
error in the Dalitz plot analysis. 
This can be checked by changing the lineshape in within a reasonable range or
by using an alternative SU(3)-based method to extract $\alpha$ that does not
use the tails of $\pi\pi$ lineshapes~\cite{Gronau:2004tm}.

All the above methods use isospin to estimate the penguin pollution.
They are thus theoretically limited by isospin breaking. While hard to compute
these corrections are expected to be at the degree level, with the smallest
impact expected in the $B\to \rho \pi$
extraction~\cite{Gardner:1998gz,Gronau:2005pq,Zupan:2007fq}.

\subsubsection{Experimental measurements}

\underline{$B \to \pi\pi$}

High-quality separation of charged Kaons and pions is a distinctive
experimental challenge in the $\Bz \to \pip\pim$ and $\Bpm \to \pipm\piz$
analyses. Indeed, $\BR(\Bz \to \Kp\pim)/\BR(\Bz \to \pip\pim) \approx 3.8$ and
$\BR(\Bpm \to \Kpm\piz)/\BR(\Bpm \to \pipm\piz) \approx
2.3$~\cite{Barberio:2008fa}, and the separation between the $K\pi$ and
$\pi\pi$ candidates in the kinematic quantity \DeltaE at $e^+e^-$ $B$-meson
factories is only about $1.5 \sigma$. 
Both Belle and \babar\ employ sophisticated likelihood-based pion-Kaon
separation in the branching-fraction and \CP-violation analyses in these
modes. In addition to the $B$ factories, the CDF experiment, thanks to its
$1.4\sigma$ \dedx-based Kaon-pion separation, aided by the invariant-mass
separation of the $\Kpm\pimp$ and $\pip\pim$ candidates, is able to provide a
competitive measurement of the $\BR(\Bz \to \Kp\pim)/\BR(\Bz \to \pip\pim)$
ratio, and thus of the less-well-known $\BR(\Bz \to \pip\pim)$.

\begin{table}[!tp]
  \caption{
    Branching fractions and \CP asymmetries in $B \to \pi\pi$. 
    First error is statistical and second systematic. 
    Please note that Belle quotes ${\cal A} \equiv -C$. 
    The April 2008 online update of the preliminary CDF result is 
    $\BR(\pip\pim) = (5.02 \pm 0.33 \pm 0.35) \times
    10^{-6}$~\cite{Morello:2006pv}.
    Values given in parentheses are the numbers of \BB\ pairs in the datasets
    used in the analyses, where appropriate.
  }
\begin{tabular}{c||r@{~}l|r@{~}l|c}
\hline\hline
& \babar   &                                    & Belle                          &         			& HFAG avg. \\
\hline  
$S_{\pi\pi}$                &  $-0.68 \pm 0.10 \pm 0.03$~\cite{Aubert:2008sb}
&(467M) & $-0.61 \pm 0.10 \pm 0.04$~\cite{Ishino:2006if} &(535M) 	& $ -0.65 \pm 0.07$ \\
$C_{\pi\pi}$                &  $-0.25 \pm 0.08 \pm 0.02$~\cite{Aubert:2008sb} &(467M) & $-0.55 \pm 0.08 \pm 0.05$~\cite{Ishino:2006if} &(535M) 	& $ -0.38 \pm 0.06 $ \\
$\BR(\pip\pim) \times 10^6$ &  $5.5 \pm 0.4 \pm 0.3$~\cite{Aubert:2006fha} &(227M) & $5.1 \pm 0.2 \pm 0.2$~\cite{Abe:2006qx} &(449M)		& $5.16 \pm 0.22$ \\
$\BR(\pip\piz) \times 10^6$ &  $5.02 \pm 0.46 \pm 0.29 $~\cite{Aubert:2007hh} &(383M) & $ 6.5 \pm 0.4^{+0.4}_{-0.5} $~\cite{Abe:2006qx} &(449M)	& $5.59^{+0.41}_{-0.40} $ \\
${\cal A}(\pip\piz)$        &  $0.030\pm 0.039 \pm 0.010 $~\cite{Aubert:2007hh} &(383M) & $ 0.07 \pm 0.03 \pm 0.01 $~\cite{:2008zza} &(535M)	& $ 0.050 \pm 0.025 $ \\
$\BR(\piz\piz) \times 10^6$ &  $1.83 \pm 0.21 \pm 0.13$~\cite{Aubert:2008sb} &(467M)& $1.1 \pm 0.3 \pm 0.1$~\cite{Abe:2006cx} &(535M)		& $1.55 \pm 0.19$ \\
$C_{\piz\piz}$              &  $-0.43 \pm 0.26 \pm 0.05$~\cite{Aubert:2008sb} &(467M)& $-0.44^{+0.62}_{-0.73}~^{+0.06}_{-0.04}$~\cite{Abe:2006cx} &(535M)& $ -0.43^{+0.24}_{-0.25} $ \\
\hline\hline
\end{tabular}
\label{tab:angles:pipi}
\end{table} 

\begin{figure}[!bp]
\begin{center}
\begin{tabular}{c c c c}
  \includegraphics[height=0.93in]{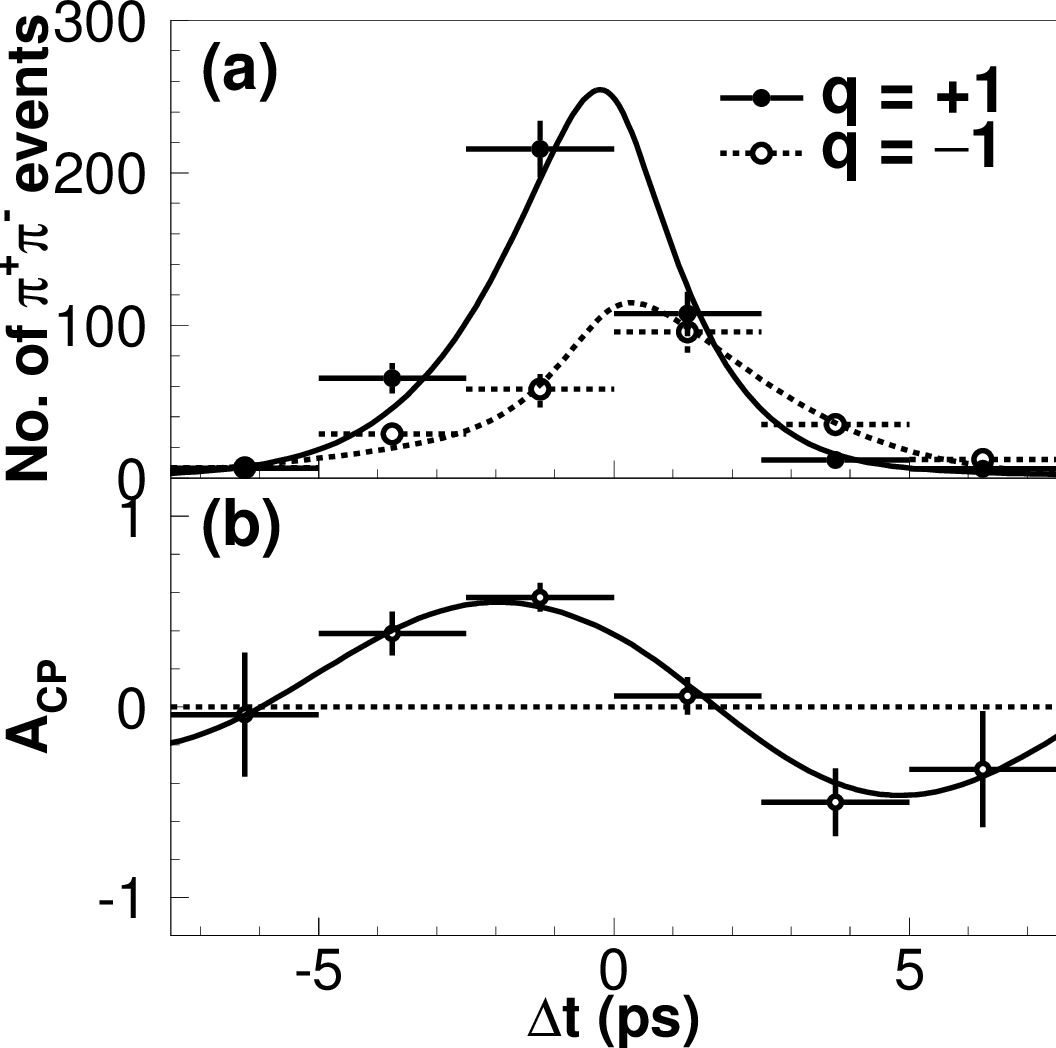} & ~
  \includegraphics[height=0.93in]{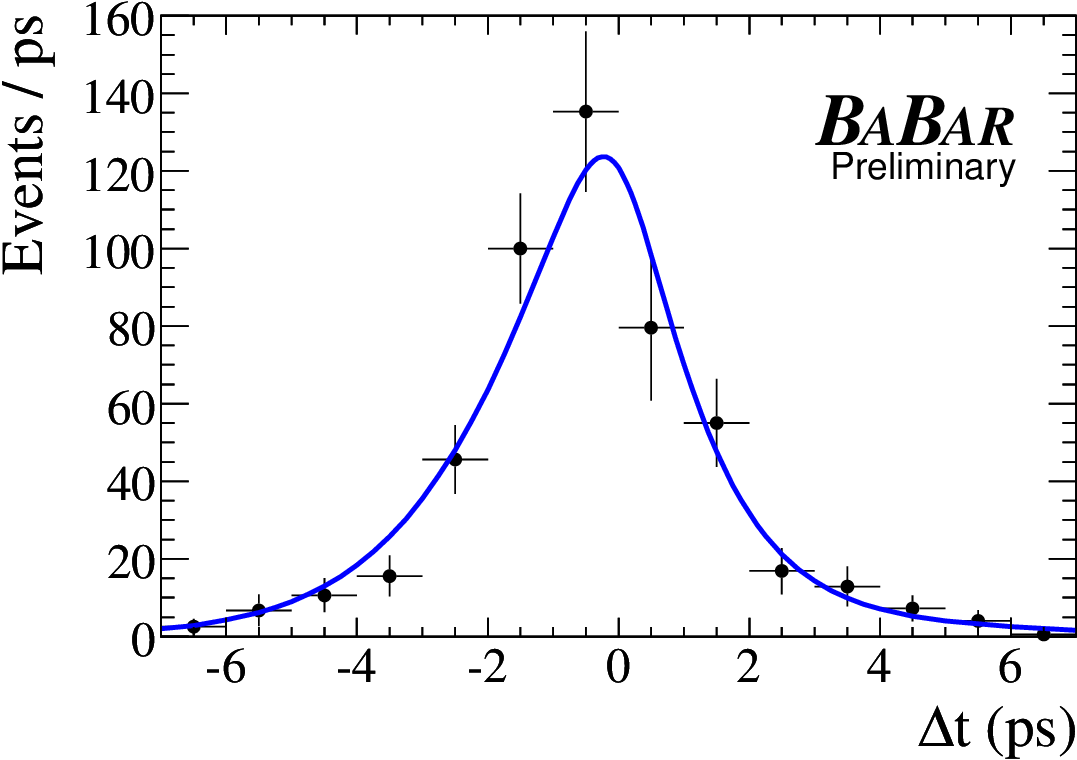} & ~
  \includegraphics[height=0.93in]{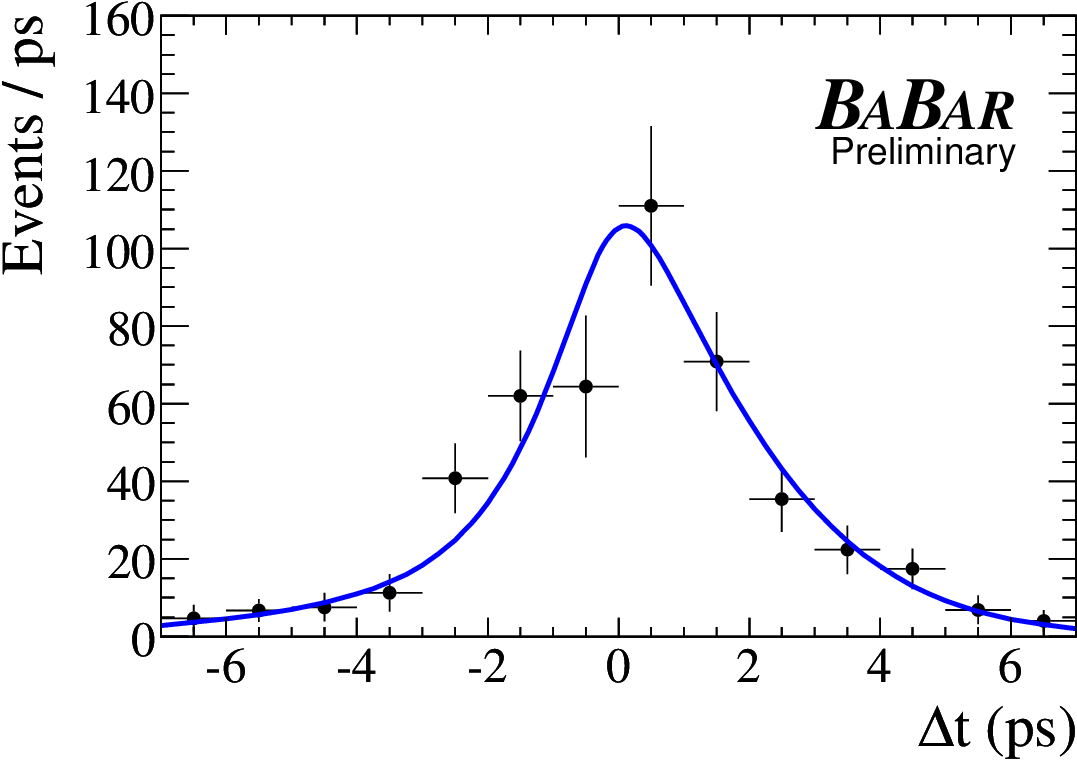} & ~
  \includegraphics[height=0.93in]{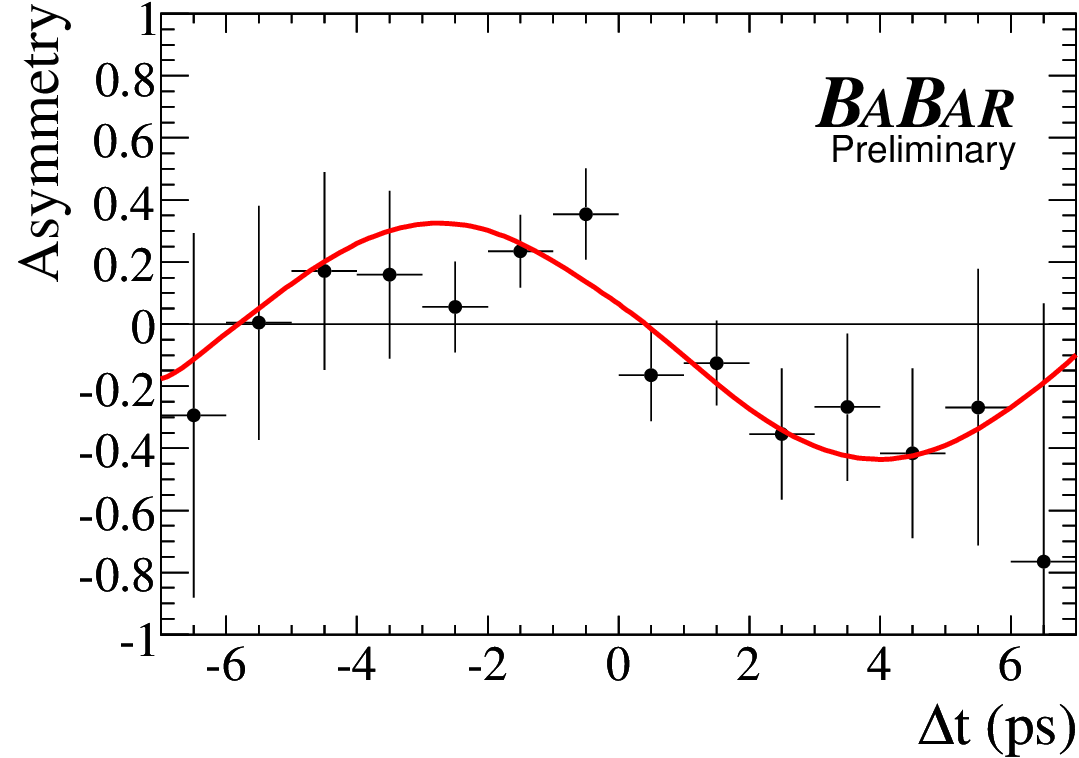} \\ 
\end{tabular}
  \put(-283,27){\bfseries {\small (c)}}
  \put(-182,27){\bfseries {\small (d)}}
  \put(-78,27){\bfseries {\small (e)}}
  \caption{
    (a) Distributions of $\Delta t$ for \Bz $(q=+1)$ and \Bzb $(q=-1)$ tags
    and (b) their \CP-violating asymmetry in $\Bz \to \pip\pim$ signal events
    reported by Belle~\cite{Ishino:2006if}. 
    Distributions of $\Delta t$ for (c) \Bz and (d) \Bzb tags and (e) their
    \CP-violating asymmetry in $\Bz \to \pip\pim$ signal events reported by
    \babar~\cite{Aubert:2008sb}.
}
  \label{fig:angles:dt_pipi}
\end{center}
\end{figure}

\begin{figure}[!bt]
\begin{center}
\vspace{0.2cm}
\begin{tabular}{c c c}
  \includegraphics[height=1.7in]{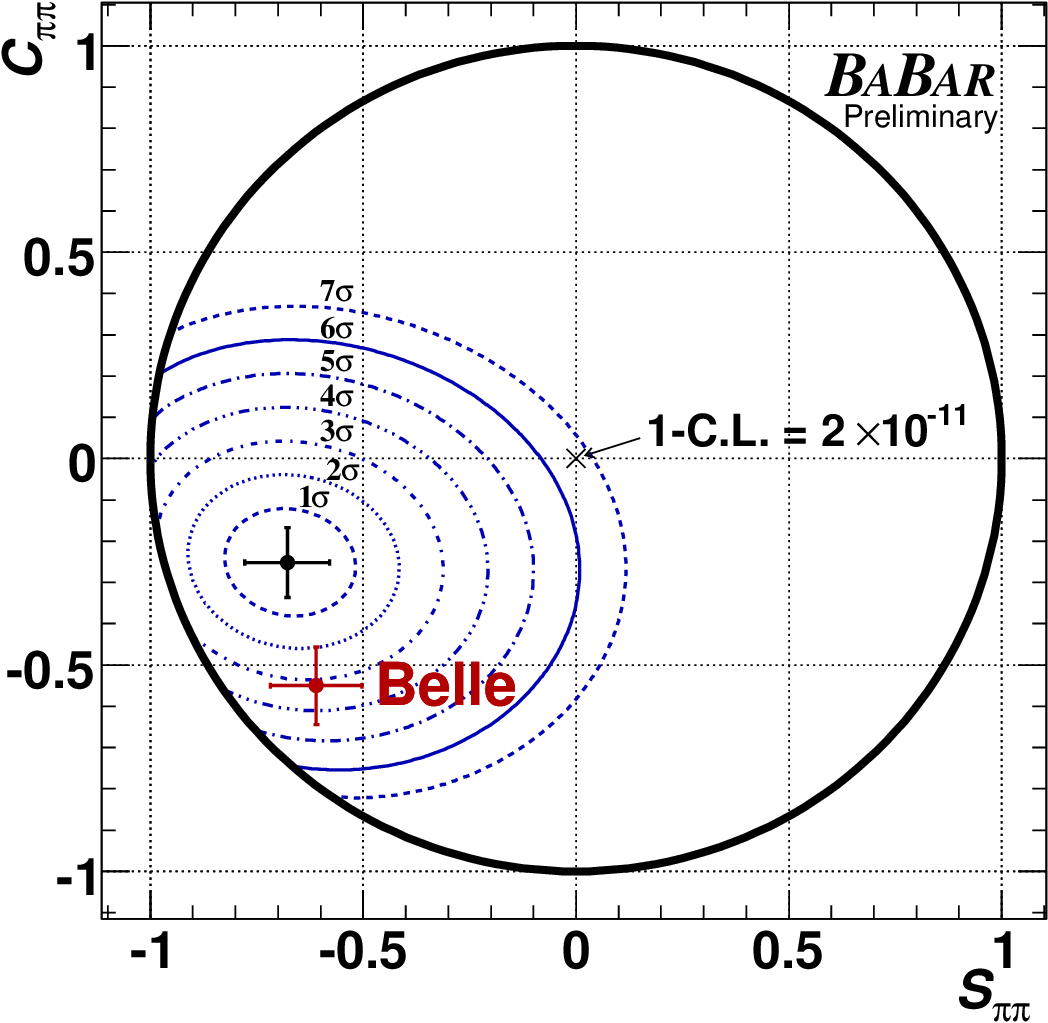}~~ & 
  \includegraphics[height=1.7in]{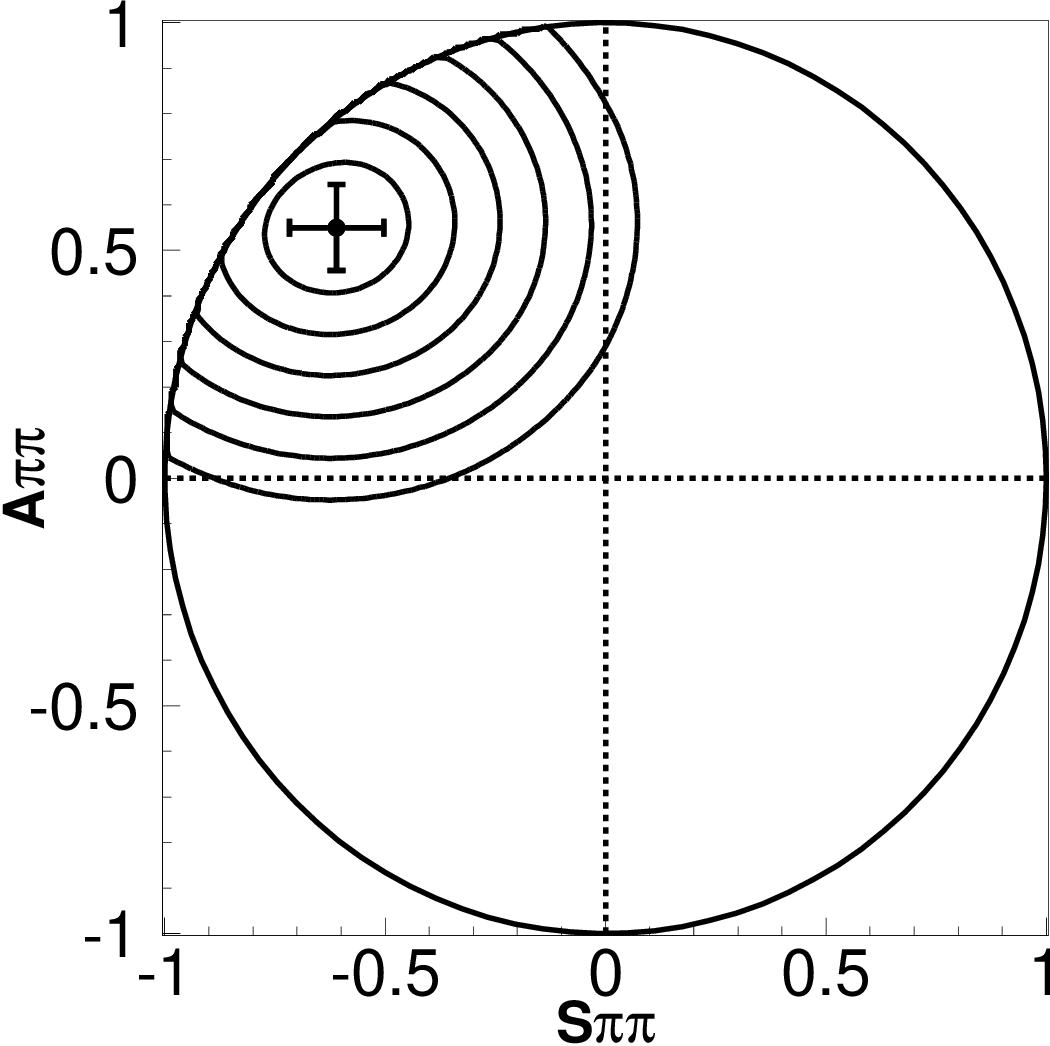}~~ & 
  \includegraphics[height=1.7in]{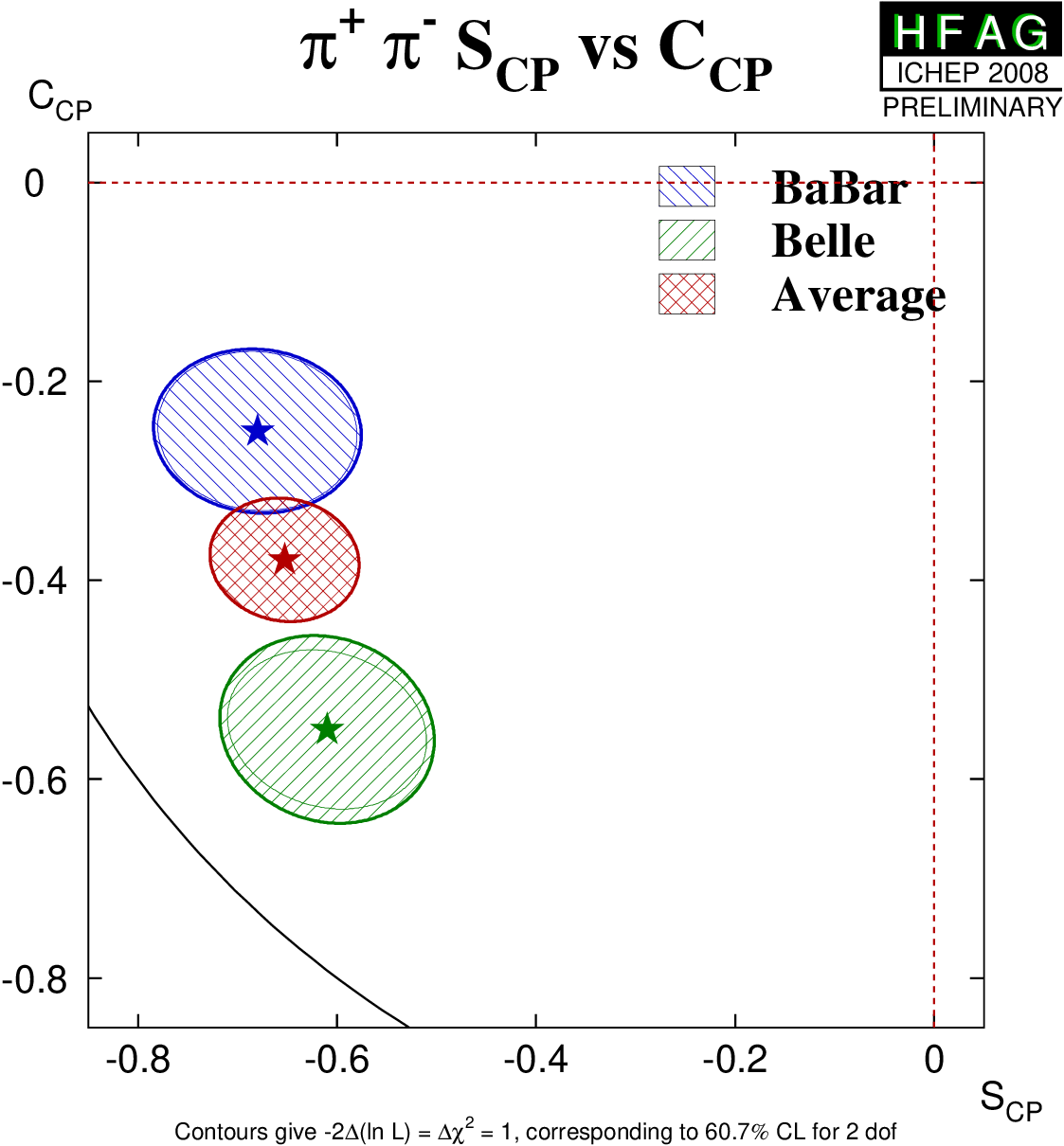} \\ 
\end{tabular}
  \put(-363,54){\bfseries {\small (a)}} 
  \put(-224,54){\bfseries {\small (b)}}
  \put(-100,39){\bfseries {\small (c)}}
  \caption{
    $S_{\pi\pi}$ and $C_{\pi\pi} \equiv -{\cal A}_{\pi\pi}$ in $\Bz\to\pip\pim$: 
    central values, uncertainties, and confidence-level (C.L.) contours for 
    $1-\mathrm{C.L.} = 0.317$ $(1\sigma)$, $4.55 \times 10^{-2}$ $(2\sigma)$,
    $2.70 \times 10^{-3}$ $(3\sigma)$, $6.33 \times 10^{-5}$ $(4\sigma)$,
    $5.73 \times 10^{-7}$ $(5\sigma)$, $1.97 \times 10^{-9}$ $(6\sigma)$ and
    $2.56 \times 10^{-12}$ $(7\sigma)$: (a) \babar~\cite{Aubert:2008sb}, (b)
    Belle~\cite{Ishino:2006if}. 
    (c) \babar and Belle $\Delta \chi^2 = 1$ ($S_{\pi\pi}$, $C_{\pi\pi}$)
    contours, corresponding to 60.7\% C.L., and their HFAG correlated
    average. \babar and Belle results are consistent at 0.055 ($1.9\sigma$)
    C.L.}
  \label{fig:angles:pipi_SC_contours}
\end{center}
\end{figure}

\begin{figure}[!bt]
\begin{center}
\begin{tabular}{c c}
  \includegraphics[height=1.6in]{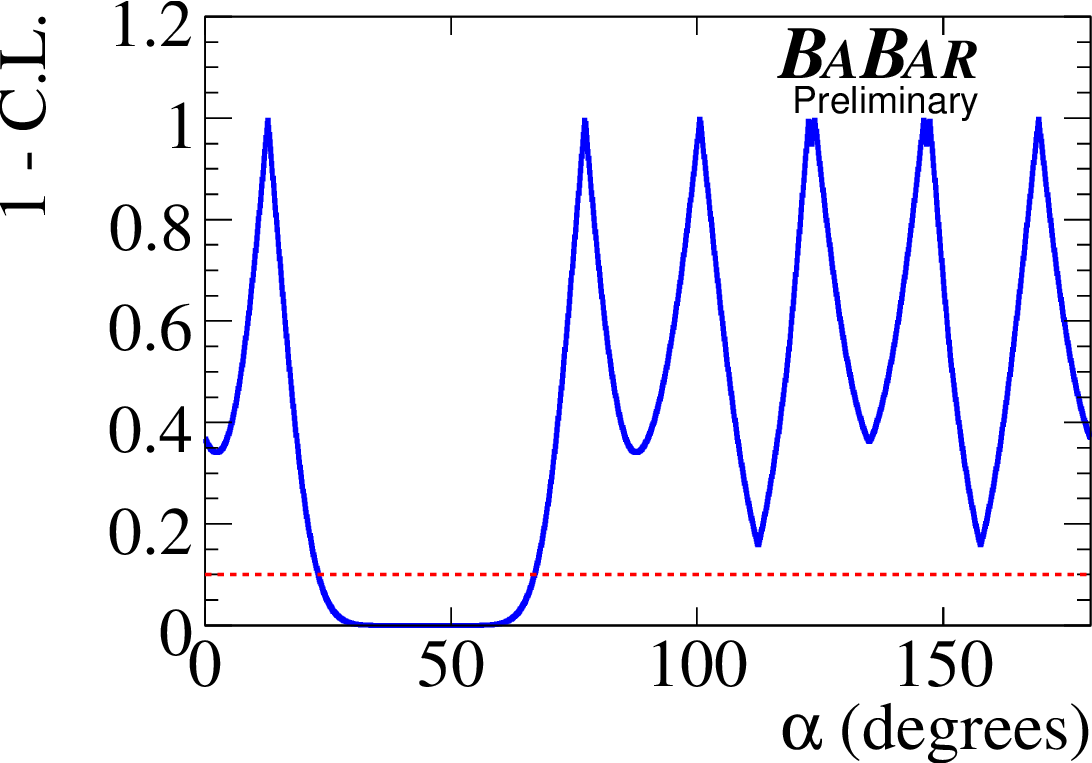}~~~~~~~ & 
  \includegraphics[height=1.6in]{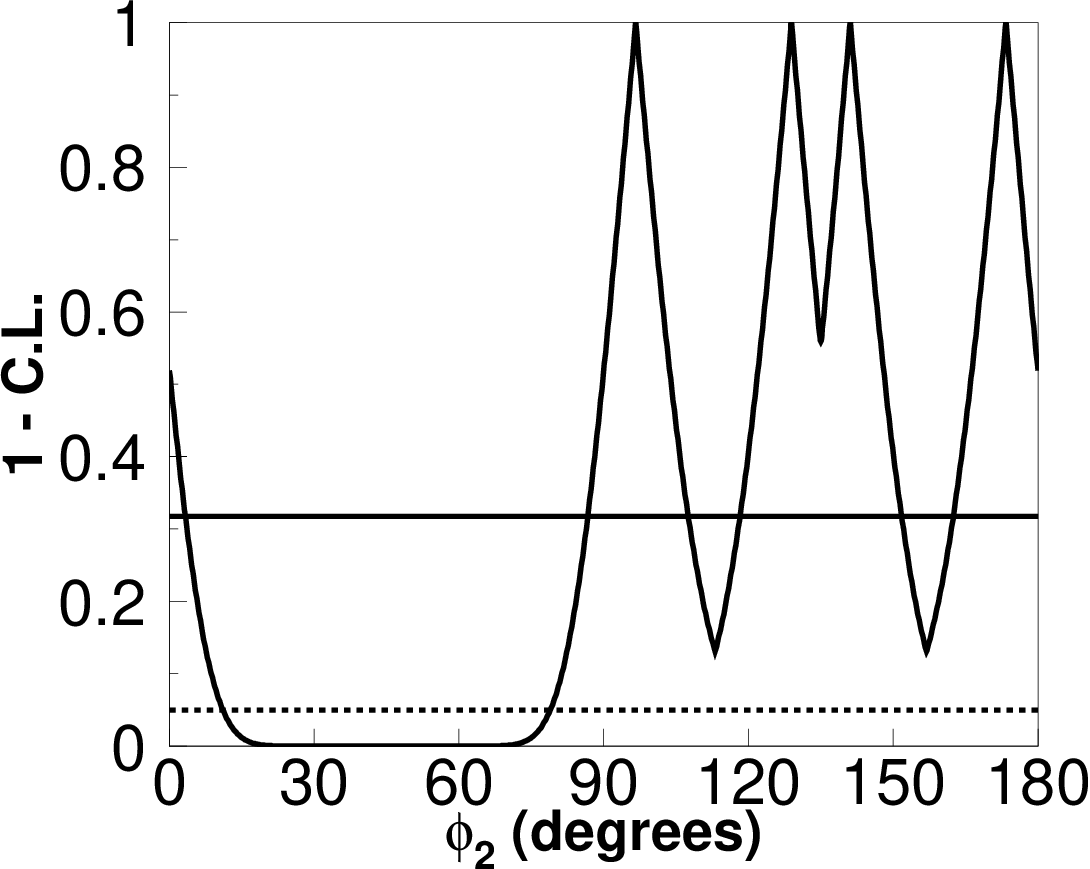} \\ 
\end{tabular}
  \put(-300,50){\bfseries {\small (a)}} 
  \put(-120,50){\bfseries {\small (b)}}
  \caption{
    Constraints on the CKM angle $\alpha$:
    (a) from \babar~\cite{Aubert:2008sb} using only the $B \to \pi\pi$ results
    from \babar; 
    (b) from Belle~\cite{Ishino:2006if}, 
    using Belle's measurements of $S_{\pipi}$ and $C_{\pipi}$ and the Summer
    2006 HFAG world averages for the branching fractions and \CP-violating
    asymmetries in $\Bp \to \pip\piz$ and $\Bz \to \piz\piz$. 
 }
  \label{fig:angles:pipi_alpha}
\end{center}
\end{figure}

The most up-to-date measurements in the $B \to \pi\pi$ modes, along with the 
September 2008 HFAG averages, are quoted in Tab.~\ref{tab:angles:pipi}. 
With the exception of $C_{\piz\piz}$, the sensitivities of the \babar and
Belle measurements are very similar.  
Plots of $\Bz \to \pip\pim$ $\Delta t$ distributions for the 
\Bz and \Bzb tags and their \CP-violating asymmetries are shown in 
Fig.~\ref{fig:angles:dt_pipi}, and the 
($S_{\pi\pi}$, $C_{\pi\pi}$) confidence-level contours are shown in 
Fig.~\ref{fig:angles:pipi_SC_contours}. 
Interpretation of the latest \babar and Belle $B \to \pi\pi$ results in terms
of constraints on the angle $\alpha$ is shown in
Fig.~\ref{fig:angles:pipi_alpha}.  
Only the isospin-triangle relations are used in these constraints. 
Values of $\alpha$ near 0 or $\pi$ can be excluded with additional physics
input~\cite{Aubert:2007hh,Bona:2007qta}. 
The key point is that the isospin analysis requires no knowledge about either
the magnitude or phase of the penguin contribution.
However, using CKM unitarity the relative phase between penguin and tree can
be chosen to be \aa, so that the direct CP violation parameter $C_{\pi\pi}
\propto \aa$.  Consequently, the observation $C_{\pi\pi} \neq 0$ requires $\aa
\neq 0$ (or alternatively hadronic parameters must unphysically tend to
infinity).

Both Belle and \babar observe a non-zero \CP-violating asymmetry $S_{\pi\pi}$
in the time distribution of $\Bz \to \pip\pim$ decays, with significances of
$5.3\sigma$ and $6.3\sigma$, respectively. Belle observes, with a significance
of $5.5\sigma$, direct \CP violation ($C_{\pi\pi} \neq 0$) in $\Bz \to
\pip\pim$; \babar sees $3.0\sigma$ evidence of $C_{\pi\pi} \neq 0$.

\vspace{10pt}
\noindent
\underline{$B\ra\rho\rho$}

The decay $B^0\ra\rho^+\rho^-$ has been measured by Belle and \babar several
times with increasingly larger data samples. Both experiments measure the
branching fraction, $f^{}_L$, and the \CP-violating parameters
$A^{}_{\rho\rho}$ and $S^{}_{\rho\rho}$. 
The most recent results are listed in Tab.~\ref{tab:angles:results_rprm}. 
The measured values of $A^{}_{\rho\rho}$ and $S^{}_{\rho\rho}$ are consistent with zero, i.e., there is no evidence for \CP\ violation. 
The decay-time distributions and \CP\ asymmetry distribution 
($A^{}_{CP}$ in bins of $\Delta t$) are shown in 
Figs.~\ref{fig:angles:brprm_belle} and \ref{fig:angles:brprm_babar}.
From the same analysis, Belle has also set a limit on the nonresonant
$B^0\ra\rho^0\pi^+\pi^-$ contribution at
$\Gamma(\rho^\pm\pi^\mp\pi^0)/\Gamma(\rho^+\rho^-)= 0.063\,\pm0.067$.

\begin{figure}[bth]
\begin{center}
\epsfig{file=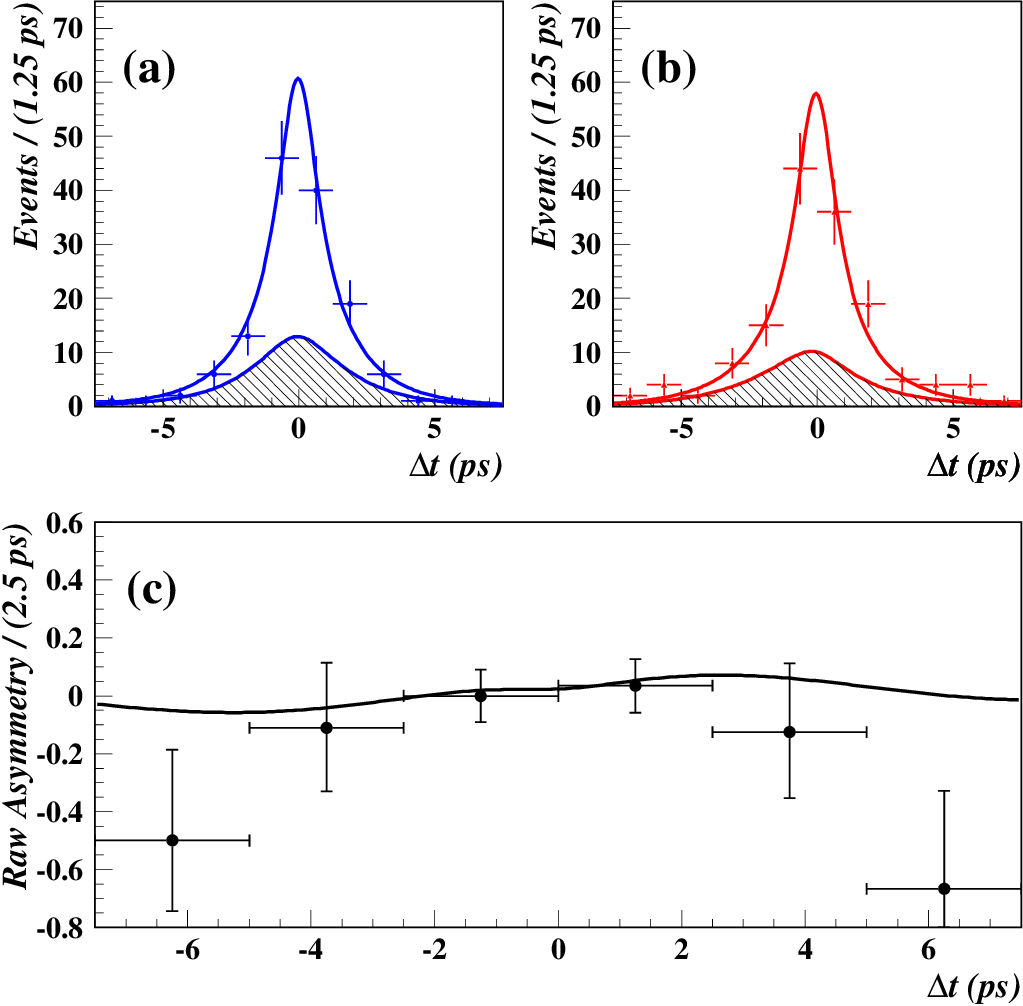,width=3.4in}
\end{center}
\caption{Decay time distributions from Belle~\cite{Abe:2007ez}.
{\it (a)\/} $\Bzb\ra\rho^+\rho^-$ decays 
{\it (b)\/} $B^0\ra\rho^+\rho^-$ decays, and 
{\it (c)\/} the raw asymmetry $(\overline{N}-N)/(\overline{N}+N)$, 
where $\overline{N}$ ($N$) is the number of $\Bzb$ ($B^0$) candidates
including background. The hatched region shows the fit result for 
the signal component, and the solid curve shows the fit result for 
the total.
\label{fig:angles:brprm_belle}}
\end{figure}

\begin{figure}[bth]
\begin{center}
\epsfig{file=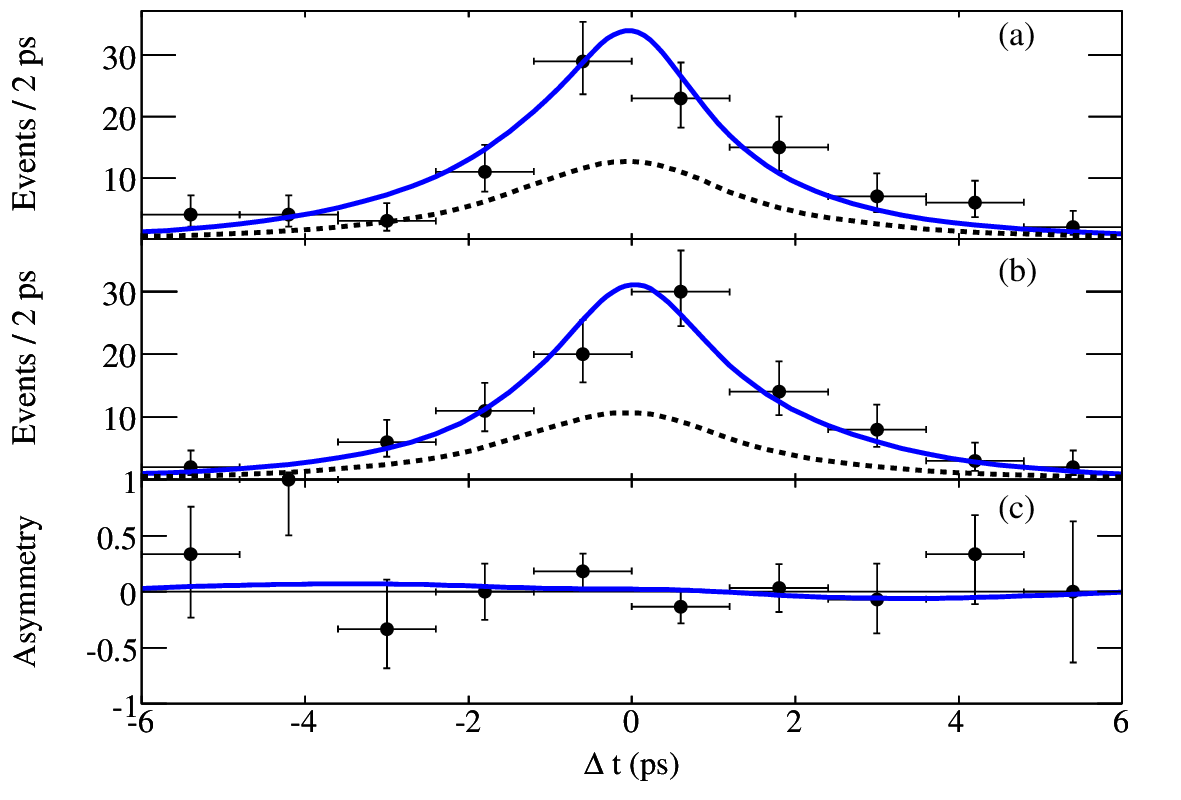,width=3.90in}
\end{center}
\caption{Decay time distributions from \babar~\cite{Aubert:2007nua}.
{\it (a)\/} $\Bzb\ra\rho^+\rho^-$ decays 
{\it (b)\/} $B^0\ra\rho^+\rho^-$ decays, and 
{\it (c)\/} the asymmetry $(\overline{N}-N)/(\overline{N}+N)$, 
where $\overline{N}$ ($N$) is the number of signal
$\Bzb\ra\rho^+\rho^-$ ($B^0\ra\rho^+\rho^-$) decays.
The dashed curve shows the fit result for all backgrounds, 
and the solid curve shows the fit result for the total.
\label{fig:angles:brprm_babar}}
\end{figure}


\begin{table}[htb]
\caption{Belle and \babar results for $B^0\ra\rho^+\rho^-$
  decays~\cite{Aubert:2007nua,Somov:2006sg,Abe:2007ez}. 
  \label{tab:angles:results_rprm}}
\begin{tabular}{l||c|c|c|c|c}
\hline\hline
  & Data     & Branching & $f^{}_L$ & $A^{}_{\rho\rho}$ & $S^{}_{\rho\rho}$ \\
  & (\invfb) & fraction $\times 10^{-6}$ & & &  \\
\hline
Belle &  253/492 & $22.8\,\pm3.8\,^{+2.3}_{-2.6}$ & 
$0.941\,^{+0.034}_{-0.040}\,\pm0.030$ & $0.16\,\pm0.21\,\pm0.08$ & $0.19\,\pm0.30\,\pm0.08$ \\
\babar & 349 & $25.5\,\pm2.1\,^{+3.6}_{-3.9}$ & 
$0.992\,\pm0.024\,^{+0.026}_{-0.013}$ & $-0.01\,\pm0.15\,\pm0.06$ & 
  $-0.17\,\pm0.20\,^{+0.05}_{-0.06}$ \\
\hline\hline
\end{tabular}
\end{table}



The most recent results from  Belle~\cite{Zhang:2003up} and
\babar~\cite{Aubert:2006sb} on the decay $B^+\ra\rho^+\rho^0$ are listed in
Tab.~\ref{tab:angles:results_rprz}. Both measured values of $A^{}_{CP}$ are
consistent with zero, implying that a possible electroweak penguin
contribution is small.  
Belle has also set a limit on the nonresonant $B^+\ra(\rho\pi\pi)^+$
contribution of 
$\Gamma[(\rho\pi\pi)^+]/\Gamma(\rho^+\rho^0)< 0.17$ at 90\% C.L.


\begin{table}[htb]
\caption{Belle and \babar results for $B^+\ra\rho^+\rho^0$ decays, from
Refs.~\cite{Zhang:2003up,Aubert:2006sb}.
\label{tab:angles:results_rprz}}
\begin{tabular}{l||c|c|c|c}
\hline\hline
  & Data     & Branching & $f^{}_L$ & $A^{}_{CP}$ \\
  & (\invfb) & fraction $\times 10^{-6}$  & &   \\
\hline
Belle &  78 & $31.7\,\pm7.1\,^{+3.8}_{-6.7}$ & 
               $0.95\,\pm0.11\,\pm0.02$ & $-0.12\,\pm0.13\,\pm0.10$ \\
\babar & 211 & $16.8\,\pm2.2\,\pm2.3$ & 
               $0.905\,\pm0.042\,^{+0.023}_{-0.027}$ & $0.00\,\pm0.22\,\pm0.03$ \\
\hline\hline
\end{tabular}
\end{table}



The decay $B^0\ra\rho^0\rho^0$ has proved difficult to measure due to its small
branching fraction, and has only recently been observed.
Measurements from \babar~\cite{Aubert:2008iha} and Belle~\cite{:2008et} are
listed in Tab.~\ref{tab:angles:results_rzrz}.
Both experiments obtain the signal yield from unbinned maximum likelihood
fits to $M^{}_{bc}$ (or $\mes \equiv M^{}_{bc}$), \DeltaE, and
$M^{}_{\pi\pi}$. The fit is complicated by possible contributions from
$\rho^0f_0(980)$, $f_0f_0$, $f_0\pi^+\pi^-$, and $a^{}_1\pi$ final states, as
well as from $B^0\ra\rho^0\pi^+\pi^-$ and $B^0\ra\pi^+\pi^-\pi^+\pi^-$. 

The \babar experiment requires that $M^{}_{\pi\pi}\in (0.50,1.05)$\gevcc; 
they subsequently fit to variables \mes, \DeltaE,
helicity angles $\cos\theta^{}_1$, $\cos\theta^{}_2$, and the decay time
difference $\Delta t$. Including the helicity angles in the fit yields
a measurement of $f^{}_L$, and including $\Delta t$ yields a measurement
of $A^{}_{\rho\rho}$ and $S^{}_{\rho\rho}$. 
\babar observes an excess of signal events with $3.1\sigma$
significance, and no significant nonresonant contributions.
The measured values of $A^{}_{\rho\rho}$ and $S^{}_{\rho\rho}$ are consistent with 
zero, i.e., there is no evidence for \CP\ violation. 


The Belle experiment requires $M^{}_{\pi\pi}\in (0.55,1.70)$\gevcc -- a wider
window than that used by \babar (see Fig.~\ref{fig:angles:brzrz_belle}).
Belle observes a higher rate of nonresonant $\rho\pi\pi$ and $4\pi$ components
than \babar does, and the significance of Belle's $\rho^0\rho^0$ signal is 
only $1.0\sigma$.

\begin{figure}[bth]
\vspace*{0.30in}
\begin{center}
\epsfig{file=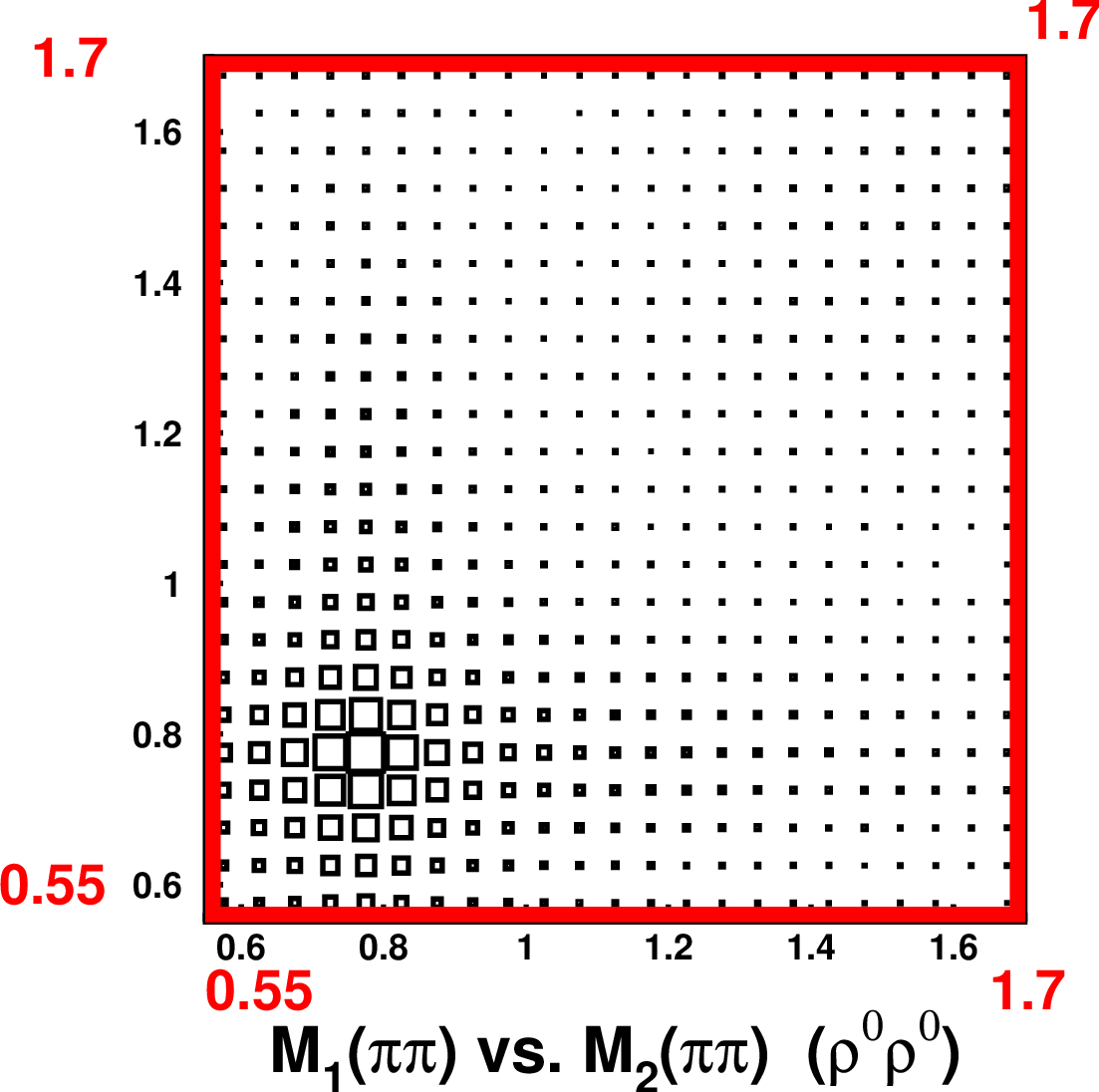,width=1.7in}
\hspace*{0.02in}
\epsfig{file=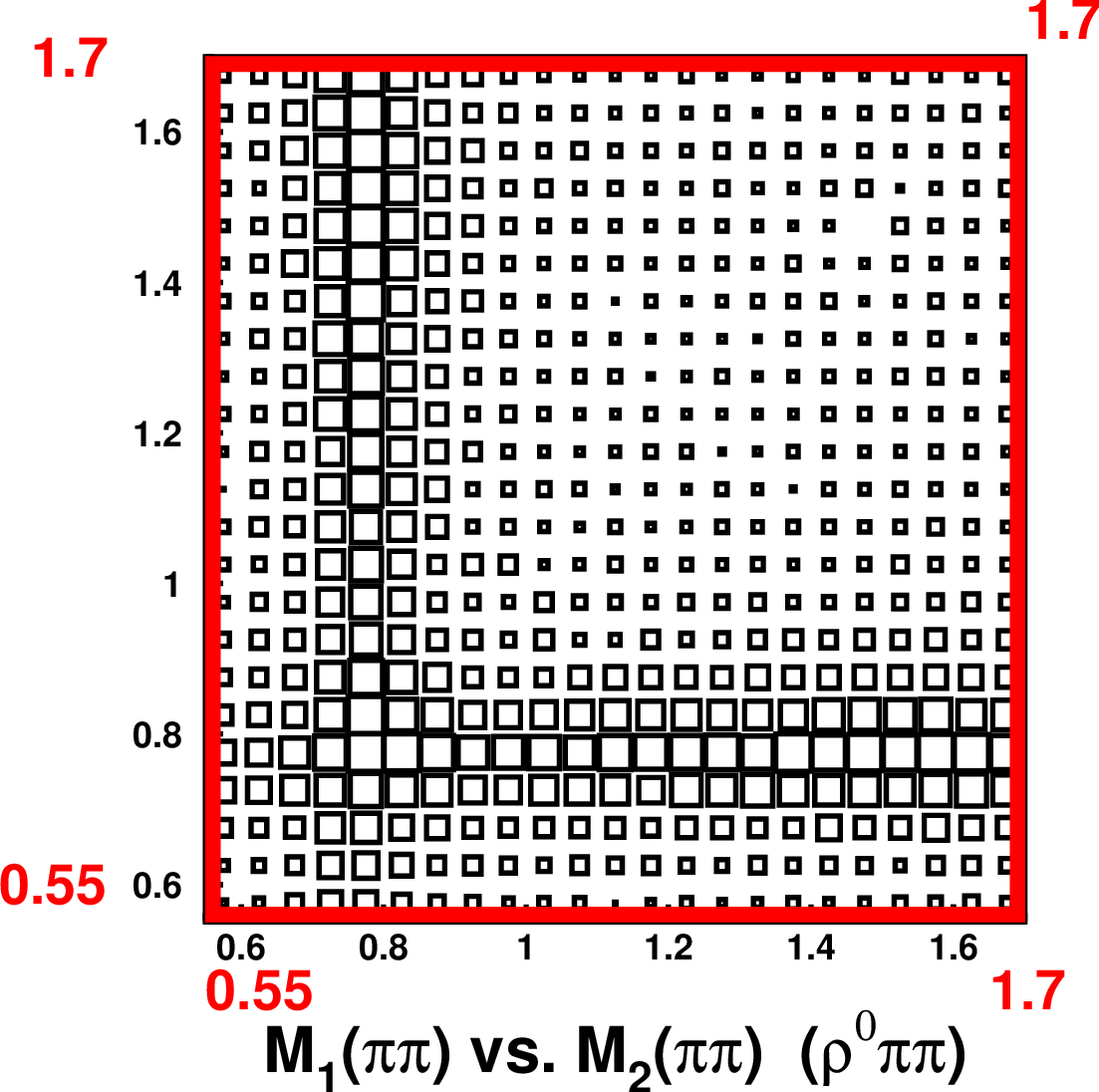,width=1.7in}
\hspace*{0.02in}
\epsfig{file=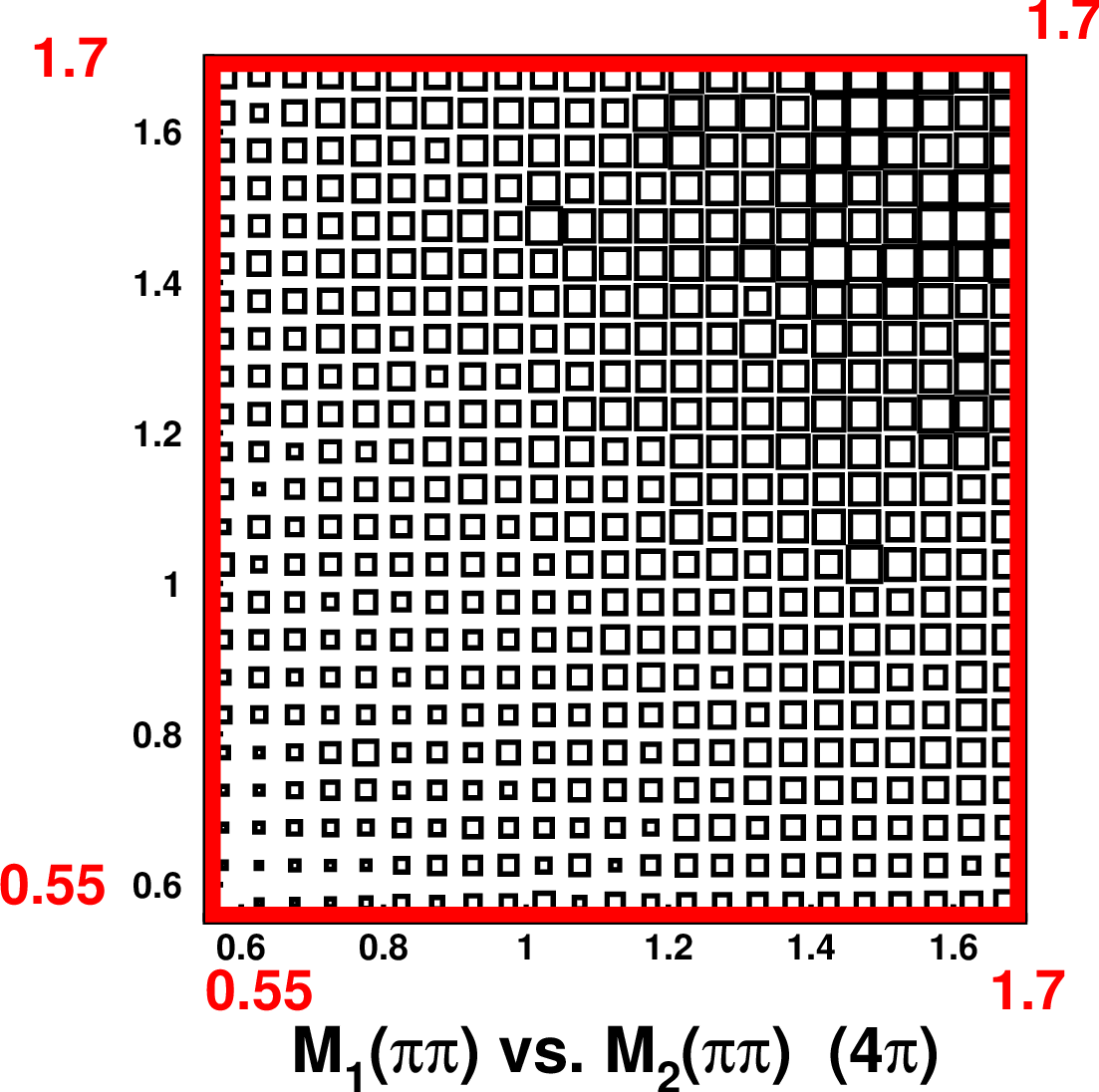,width=1.7in}
\end{center}
\caption{Monte Carlo simulated $M^{}_{\pi\pi}$ distributions
for {\it (a)\/} $B^0\ra\rho^0\rho^0$, {\it (b)\/} $B^0\ra\rho^0\pi^+\pi^-$, and
{\it (b)\/} $B^0\ra\pi^+\pi^-\pi^+\pi^-$ decays, from Belle. The plots are
symmetrized by randomly choosing the mass combination plotted against the 
horizontal axis. The fitted region for Belle is 
$M^{}_{\pi\pi}\in (0.55,1.70)$\gevcc, whereas that 
for \babar is $M^{}_{\pi\pi}\in (0.55,1.05)$\gevcc.
\label{fig:angles:brzrz_belle}}
\end{figure}

\begin{table}[htb]
\caption{Belle and \babar results for $B^0\ra\rho^0\rho^0$ decays~\cite{Aubert:2008iha,:2008et}.
\label{tab:angles:results_rzrz}}
\begin{tabular}{l|c|c|c|c}
\hline\hline
Mode & Branching & $f^{}_L$ & $A^{}_{\rho\rho}$ & $S^{}_{\rho\rho}$ \\
 & fraction $(10^{-6})$ & & &  \\
\hline
\multicolumn{5}{c}{Belle ($605 \invfb$)} \\
$\rho^0 \rho^0$ & $0.4\,\pm0.4\,^{+0.2}_{-0.3}$ & $-$ & $-$ & $-$ \\
$\rho^0\pi^+\pi^-$ & $5.9\,^{+3.5}_{-3.4}\,\pm 2.7$ & $-$ & $-$ & $-$ \\
$\pi^+\pi^-\pi^+\pi^-$ & $12.4\,^{+4.7}_{-4.6}\,^{+2.1}_{-1.9}$ & $-$ & $-$ & $-$ \\
\hline
\multicolumn{5}{c}{\babar($423 \ \invfb$)} \\
$\rho^0 \rho^0$ & \ $0.92\,\pm 0.32\,\pm 0.14$ \ & 
\ $0.75\,^{+0.11}_{-0.14}\,\pm 0.05$ \ & \ $-0.2\,\pm0.8\,\pm0.3$ \ & 
\ $0.3\,\pm0.7\,\pm0.2$ \ \\
$\rho^0\pi^+\pi^-$ & $-1.6\,^{+5.0}_{-4.5}\,\pm 2.2$ 
& $-$ & $-$ & $-$ \\
$\pi^+\pi^-\pi^+\pi^-$ & $3.0\,^{+11.6}_{-9.9}\,\pm 4.1$ 
& $-$ & $-$ & $-$ \\
\hline\hline
\end{tabular}
\end{table}



Both Belle and \babar constrain \aa\ using isospin analysis~\cite{Gronau:1990ka}. 
The fitted observables are the branching fractions and fractions of
longitudinal polarization for $B^+\ra\rho^+\rho^0$, $B^0\ra\rho^+\rho^-$, and
$B^0\ra\rho^0\rho^0$, the coefficients $A^{}_{\rho\rho}$ and $S^{}_{\rho\rho}$
for $B^0\ra\rho^+\rho^-$ decays, 
and $A^{}_{\rho\rho}$ for $B^0\ra\rho^0\rho^0$ decays. 
The fitted parameters are the magnitudes $|{\cal A}(B^0\ra\rho^+\rho^0)|$, 
$|{\cal A}(B^0\ra\rho^+\rho^-)|$, and $|{\cal A}(B^0\ra\rho^0\rho^0)|$, 
the average phase of, and phase difference between, amplitudes 
${\cal A}(B^0\ra\rho^+\rho^-)$ and ${\cal A}(\Bzb\ra\rho^+\rho^-)$, and \aa. 
To obtain a confidence interval for \aa,
the experiments scan values of \aa\ and, for each value, fit the 
measured observables. The resulting $\chi^2$ is input into the cumulative 
$\chi^2$ distribution to obtain a confidence level ($p$-value) for that 
value of \aa. Plotting this confidence level (C.L.) versus \aa\ allows one to
read off a confidence interval. 

The most recent Belle result~\cite{:2008et},
obtained using world average values~\cite{Barberio:2008fa} for all observables
except $B(B^0\ra\rho^0\rho^0)$ for which only the Belle result is used,
is shown in Fig.~\ref{fig:angles:phitwo}\,(top). 
The ``flat-top'' region results from the fact that no measurement of
$A^{}_{\rho\rho}$ for $B^0\ra\rho^0\rho^0$ decays is used.
From the plot one reads off three disjoint 68.3\% C.L. intervals;
the interval consistent with unitarity ($\aa + \bb + \gm = 180^\circ$)
is $(75.8,106.2)^\circ$. 
Requiring symmetric errors gives $\aa=(91.7\pm 14.9)^\circ$.

The most recent \babar result~\cite{Aubert:2008iha}, made using \babar results
exclusively, is shown in Fig.~\ref{fig:angles:phitwo}\,(bottom).  
The dotted contour is the nominal solution; however, including in the fit 
the parameter $S^{}_{\rho\rho}$ from $B^0\ra\rho^0\rho^0$ decays reduces the four-fold 
ambiguity for \aa\ to three solutions (solid contour). 
The final result is expressed in terms of the shift 
$\delta\equiv\aa-\aa_{\rm eff}$ that results from the 
penguin contribution (recall that 
$S^{}_{\rho\rho}=-\sqrt{1-A^2_{\rho\rho}}\,\sin2\aa_{\rm eff}$,
see Eq.~\ref{eqn:angles:alpha_master}). 
The upper limit is $|\delta|< 17.6^\circ$ at 90\% C.L.

\begin{figure}[bth]
\begin{center}
\vbox{
\epsfig{file=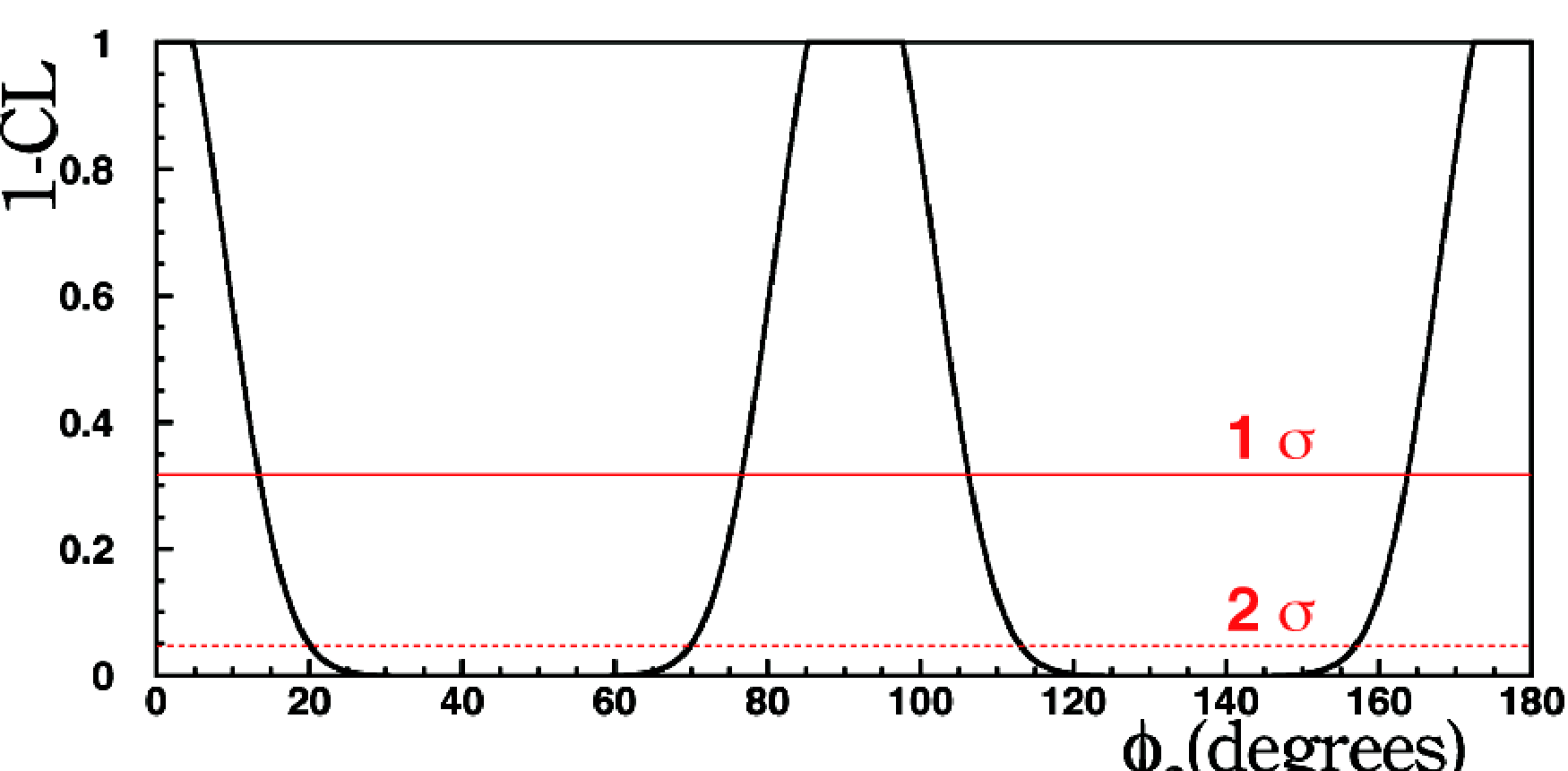,width=3.5in}
\vskip0.20in
\epsfig{file=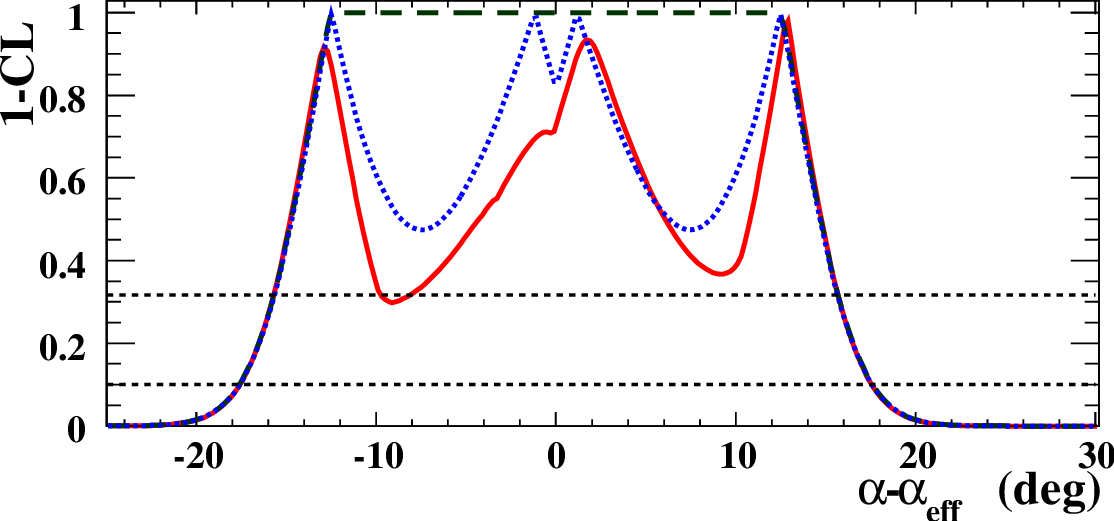,width=3.5in}
}
\end{center}
\caption{Plot of $1\!-\!CL$ versus \aa\ from Belle~\cite{:2008et} (top), 
and $1\!-\!CL$ versus $\aa-\aa_{\rm eff}$ from \babar~\cite{Aubert:2008iha} (bottom). 
From these plots one reads off confidence intervals.
In the top plot, the flat-top region results from not using $A^{}_{\rho\rho}$ 
from $B^0\ra\rho^0\rho^0$ in the fit; in the bottom plot, the solid curve 
results from using both $A^{}_{\rho\rho}$ and $S^{}_{\rho\rho}$ from $B^0\ra\rho^0\rho^0$.
\label{fig:angles:phitwo}}
\end{figure}


\vspace{10pt}
\noindent
\underline{$B^0\ra\rho\pi$}

The time-dependent Dalitz plot analysis of $B^0 \to \pi^+\pi^-\pi^0$ has been
performed by \babar using 346\invfb\ of data~\cite{Aubert:2007jn} and by
Belle using 414\invfb~\cite{Kusaka:2007mj}. 
In principle, one inserts the parametrization
(\ref{eqn:angles:rhopi_decomp_a}) and (\ref{eqn:angles:rhopi_decomp_b}) into
(\ref{eqn:angles:rhopi_timedep}) to obtain the PDF for fitting. However, the
resulting PDF is nonlinear in the amplitudes $A^{}_i$ and $\overline{A}^{}_i$,
and the fit is not well-behaved for current statistics. To stabilize the fit,
one defines new fitting parameters~\cite{Quinn:2000by}
\begin{eqnarray}
U^{\pm}_i & \ =\  & |A^{}_i|^2\ \pm\ |\overline{A}^{}_i|^2 
\label{eqn:angles:rhopi_first} \\
U^\pm_{ij} & \ =\  & A_i\,A^*_j \ \pm\ \overline{A}_i\,\overline{A}^*_j 
\label{eqn:angles:rhopi_second} \\
I_i & \ =\  & {\rm Im}(\overline{A}_i\,A^*_i) \\
{\rm Re} (I^{}_{ij}) & \ =\  & {\rm Re} (\overline{A}_i\,A^*_j \ -\ \overline{A}_j\,A^*_i) 
\label{eqn:angles:rhopi_third} \\
{\rm Im} (I^{}_{ij}) & \ =\  & {\rm Im} (\overline{A}_i\,A^*_j \ +\ \overline{A}_j\,A^*_i)\,.
\label{eqn:angles:rhopi_fourth}
\end{eqnarray}
Eqs. (\ref{eqn:angles:rhopi_first})-(\ref{eqn:angles:rhopi_fourth}) 
define 27 real parameters from six complex amplitudes, and thus these
parameters are not all independent. 
The overall normalization is fixed by setting $U^+_+=1$,
and then there are 26 free parameters in the fit. The fit results for 
\babar and Belle are listed in Tab.~\ref{tab:angles:rhopi_results}.

\begin{table}[htb]
\caption{Fit results for the $U$ and $I$ coefficients from 
Refs.~\cite{Aubert:2007jn} (\babar) and \cite{Kusaka:2007mj} (Belle).
The first error listed is statistical, and the second is systematic.}
\label{tab:angles:rhopi_results}
\begin{tabular}{l||c|c}
\hline\hline
Parameter   & \babar & Belle \\
\hline
$U_+^+$
& $\phantom{-}1.0$ (fixed)   &  $+1$ (fixed)             \\
$U_-^+$
& $\phantom{-}1.32\pm0.12\pm0.05$  & $+1.27 \pm 0.13\pm 0.09$ \\
$U_0^+$
& $\phantom{-}0.28\pm0.07\pm0.04$ & $+0.29 \pm 0.05\pm 0.04$ \\
$U_{+-}^{+,{\rm Re}}$
& $\phantom{-}0.17\pm0.49\pm0.31$ & $+0.49 \pm 0.86\pm 0.52$ \\
$U_{+0}^{+,{\rm Re}}$
& $-1.08\pm0.48\pm0.20$ & $+0.29 \pm 0.50\pm 0.35$ \\
$U_{-0}^{+,{\rm Re}}$
& $-0.36\pm0.38\pm0.08$ & $+0.25 \pm 0.60\pm 0.33$ \\
$U_{+-}^{+,{\rm Im}}$
& $-0.07\pm0.71\pm0.73$ & $+1.18 \pm 0.86\pm 0.34$ \\
$U_{+0}^{+,{\rm Im}}$
& $-0.16\pm0.57\pm0.14$ & $-0.57 \pm 0.35\pm 0.51$ \\
$U_{-0}^{+,{\rm Im}}$
& $-0.17\pm0.50\pm0.23$ & $-1.34 \pm 0.60\pm 0.47$ \\
\hline
$U_+^-$
& $\phantom{-}0.54\pm0.15\pm0.05$ & $+0.23 \pm 0.15\pm 0.07$ \\
$U_-^-$
& $-0.32\pm0.14\pm0.05$ & $-0.62 \pm 0.16\pm 0.08$ \\
$U_0^-$
& $-0.03\pm0.11\pm0.09$ & $+0.15 \pm 0.11\pm 0.08$ \\
$U_{+-}^{-,{\rm Re}}$
& $\phantom{-}2.23\pm1.00\pm0.43$ & $-1.18 \pm 1.61\pm 0.72$ \\
$U_{+0}^{-,{\rm Re}}$
& $-0.18\pm0.88\pm0.35$ & $-2.37 \pm 1.36\pm 0.60$ \\
$U_{-0}^{-,{\rm Re}}$
& $-0.63\pm0.72\pm0.32$ & $-0.53 \pm 1.44\pm 0.65$ \\
$U_{+-}^{-,{\rm Im}}$
& $-0.38\pm1.06\pm0.36$ & $-2.32 \pm 1.74\pm 0.91$ \\
$U_{+0}^{-,{\rm Im}}$
& $-1.66\pm0.94\pm0.25$ & $-0.41 \pm 1.00\pm 0.47$ \\
$U_{-0}^{-,{\rm Im}}$
& $\phantom{-}0.12\pm0.75\pm0.22$ & $-0.02 \pm 1.31\pm 0.83$ \\
\hline
$I_+$
& $- 0.02\pm0.10\pm0.03$ & $-0.01 \pm 0.11\pm 0.04$ \\
$I_-$
& $- 0.01\pm0.10\pm0.02$ & $+0.09 \pm 0.10\pm 0.04$ \\
$I_0$
& $\phantom{-} 0.01\pm0.06\pm0.01$ & $+0.02 \pm 0.09\pm 0.05$ \\
$I_{+-}^{{\rm Re}}$
& $\phantom{-}1.90\pm2.03\pm0.65$ & $+1.21 \pm 2.59\pm 0.98$ \\
$I_{+0}^{{\rm Re}}$
& $\phantom{-}0.41\pm1.30\pm0.41$ & $+1.15 \pm 2.26\pm 0.92$ \\
$I_{-0}^{{\rm Re}}$
& $\phantom{-}0.41\pm1.30\pm0.21$ & $-0.92 \pm 1.34\pm 0.80$ \\
$I_{+-}^{{\rm Im}}$
& $-1.99\pm1.25\pm0.34$ & $-1.93 \pm 2.39\pm 0.89$ \\
$I_{+0}^{{\rm Im}}$
& $-0.21\pm1.06\pm0.25$ & $-0.40 \pm 1.86\pm 0.85$ \\
$I_{-0}^{{\rm Im}}$
& $\phantom{-}1.23\pm1.07\pm0.29$ & $-2.03 \pm 1.62\pm 0.81$ \\
\hline\hline
\end{tabular}
\end{table}


To constrain~\aa, a $\chi^2$ fit is performed to the 27 measured
observables listed in Tab.~\ref{tab:angles:rhopi_results}. The 
$\chi^2$ statistic takes into account all correlations between
the observables. There are in principle 12 free parameters in the
fit, corresponding to the six complex amplitudes 
$A^{}_i$ and $\overline{A}^{}_i$. However, the additional
parameter \aa\ is included along with the (complex) isospin 
relation~(\ref{eqn:angles:rhopi_isospin}); together these reduce the 
number of free parameters to 11. The constraint $U^+_+=1$ fixes the
overall normalization, and a global phase factor can be neglected;
thus the final number of free parameters is nine. 
A scan is performed over values of \aa,
where for each value the other eight parameters are floated in
order to minimize the $\chi^2$. The resulting change in the
$\chi^2$ from the minimum value is converted into a confidence 
level ($CL$) either by using the cumulative $\chi^2$ distribution 
for one degree of freedom, or by finding the $p$-value from an
ensemble of toy MC experiments.

The resulting plots of $1-CL$ versus \aa\ for \babar and Belle are 
shown in Fig.~\ref{fig:angles:rhopi_phi2}. The values of \aa\ that 
have $(1-CL)>0.317$ determine $1\sigma$ confidence intervals for~\aa.
As can be seen from the plots, the $1-CL$ contour has large
variations that result in multiple regions, i.e., non-simply-connected 
intervals. Typically, the experiments quote only the interval 
consistent with unitarity. Belle obtains a second $1-CL$ contour
by including additional observables: the branching fractions
for $B^0\ra\rho^+\pi^-,\,\rho^-\pi^+,\,\rho^0\pi^0$ obtained from 
their analysis, and world average values~\cite{Barberio:2008fa}
for the branching fractions and \CP\ asymmetries measured for the charged
modes $B^\pm\ra\rho^\pm\pi^0$ and $B^\pm\ra\rho^0\pi^\pm$. With these four
new observables, two additional isospin relations are used; the 
final number of parameters floated in the fit is~12. The resulting 
$1-CL$ contour is also shown in Fig.~\ref{fig:angles:rhopi_phi2}. 
The final result from \babar is $\aa = (87\,^{+45}_{-13})^\circ$, 
whereas the final result from Belle is $\aa\in(68^\circ, 95^\circ)$ 
at 68.3\% CL.

\begin{figure}[bth]
\begin{center}
\hbox{
\epsfig{file=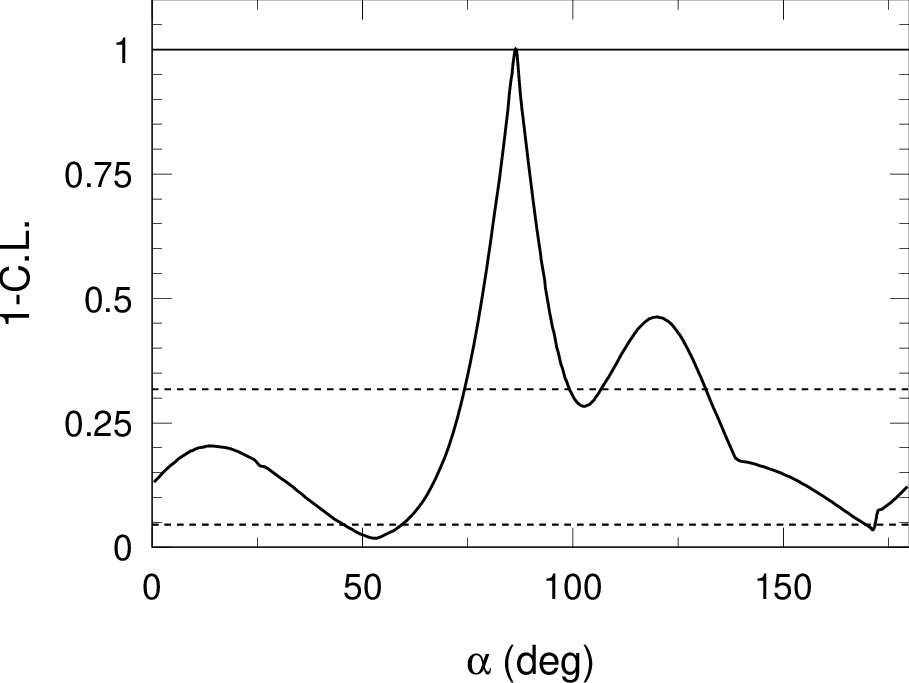,width=2.45in}
\hskip0.10in
\epsfig{file=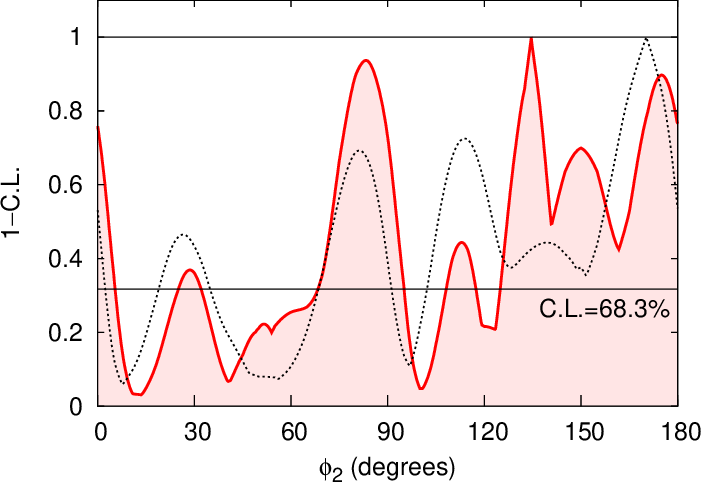,width=2.7in}
}
\end{center}
\caption{$1-CL$ versus \aa\ resulting from $\chi^2$ fits to the 27 
observables listed in Tab.~\ref{tab:angles:rhopi_results}. The left 
plot is from \babar~\cite{Aubert:2007jn}, and the right plot is from 
Belle~\cite{Kusaka:2007mj}. The horizontal line at $1-CL=0.317$ corresponds 
to 68.3\% CL and is used to determine $1\sigma$ confidence intervals 
for~\aa. For Belle, the dashed contour corresponds to a nine-parameter 
fit, and the solid contour corresponds to a twelve parameter fit (see text).
\label{fig:angles:rhopi_phi2}}
\end{figure}

\vspace{10pt}
\noindent
\underline{$B^0\ra a^\pm_1\,\pi^\mp$}

As proposed by Gronau and Zupan~\cite{Gronau:2005kw}, the $\Delta t$ 
distribution for $B^0\ra a^\pm_1\,\pi^\mp$ decays can be fit to determine \aa. 
However, there can be a penguin amplitude that substantially shifts the
measured \aa\ value from the true value, as found for $B\ra\pi\pi$
decays. Thus, to determine \aa\ from $B^0\ra a^\pm_1\,\pi^\mp$ requires
external input, e.g., assuming $SU(3)$ symmetry and using measurements of
$B\ra a^{}_1K$~\cite{Aubert:2007ds} and 
$B^0\ra K^{}_{1A}\pi$~\cite{Aubert:2008ee} decays. 
This method has uncertainties arising from $SU(3)$-breaking corrections and
unknown decay constants $f^{}_{a^{}_1}$ and $f^{}_{K_1}$. 

Experimentally, one simultaneously fits the four distributions
$B^0\ra a^\pm_1\pi^\mp$ and $\Bzb\ra a^\pm_1\pi^\mp$ to the PDF
\begin{eqnarray}
\frac{dN(a^\pm_1\pi^\mp)}{d\Delta t} & = & 
(1\pm{\cal A}^{}_{CP})\frac{e^{-|\Delta t|/\tau}}{8\tau}\times \nonumber \\
 & & \hskip-0.30in 
\Biggl\{1-q^{}_{\rm tag}\Bigl[(C\pm\Delta C)\cos(\Delta m\,\Delta t) -
(S\pm\Delta S)\sin(\Delta m\,\Delta t)\Bigr]\Biggr\},
\end{eqnarray}
where $q^{}_{\rm tag}\!=\!+1\,(-1)$ when the tag-side $B$ decays as a
$B^0\,(\overline{B}{}^0)$. The parameters
${\cal A}^{}_{CP}$, $C$, and $S$ are \CP-violating, and the
parameters $\Delta C$ and $\Delta S$ are \CP-conserving. 
\babar has performed this fit using 349\invfb\ of data~\cite{Aubert:2006gb}; 
the results are listed in Tab.~\ref{tab:angles:results_aonepi}. 
The values obtained are subsequently used to determine 
$\aa_{\rm eff}=\aa - \delta$ within a four-fold ambiguity. 
The solution closest to the \aa\ value favored by $B\ra\rho\rho$ 
and $B\ra\pi\pi$ decays is $\aa_{\rm eff}\!=\!(78.6\pm 7.3)^\circ$.
This result differs from $\aa$ by the unknown penguin 
contribution~$\delta$.

\begin{table}[htb]
\caption{\babar results for $B^0\ra a^\pm_1\,\pi^\mp$ decays, from Ref.~\cite{Aubert:2006gb}. 
\label{tab:angles:results_aonepi}}
\begin{tabular}{l||c}
\hline\hline
Parameter  & Value \\
\hline
${\cal A}^{}_{CP}$ & $-0.07\,\pm0.07\,\pm0.02$ \\
$C$ & $-0.10\,\pm0.15\,\pm0.09$ \\
$S$ & $0.37\,\pm0.21\,\pm0.07$ \\
\hline
$\Delta C$ & $0.26\,\pm0.15\,\pm0.07$ \\
$\Delta S$ & $-0.14\,\pm0.21\,\pm0.06$ \\
\hline\hline
\end{tabular}
\end{table}

\subsection{Measurements of \gm\ in charmless hadronic \B\ decays }
\label{sec:angles:gamma}

In this subsection we review the methods to determine the weak phase in the
CKM matrix using $\Delta S=1$ charmless hadronic \B\ decays: $\B\to K\pi$,
$\B\to K\pi\pi$. Conventionally this is rewritten as a constraint on \gm,
but in many instances it also involves the knowledge of the \BzBzb\ mixing
phase. This is taken to be well known as it is measured precisely in
$\Bz \to \jpsi \KS$. The $\B\to K\pi$ and $\B\to K\pi\pi$ decays are
dominated by QCD penguins and are as such sensitive to new physics effects
from virtual corrections entering at 1-loop. Comparing the extracted value
of \gm\ with that from a tree level determination using
$\B\to DK$ constitutes a test of Standard Model. 

\subsubsection{Constraints from $\B_{(s)} \to hh$}
\label{sec:angles:gamma-hh}

We can write any amplitude as a sum of two terms
\begin{equation}
A(B\to f)=Pe^{i\delta_P} +T e^{i\gamma}e^{i\delta_T},
\end{equation}
where the ``penguin'' $P$ carries only a strong phase $\delta_P$, while the
``tree'' $T$ has both strong phase $\delta_T$ and a weak phase \gm.
The latter flips signs for the \CP\ conjugated amplitude $A(\Bb\to \bar f)$).
The sensitivity to \gm\ comes from the interference of the two
contributions.
In $\Delta S=1$ decays the tree contribution is doubly CKM suppressed --- it
carries the CKM factor $|V_{ub}^*V_{us}|$ -- while the penguin contribution
has a CKM factor $ |V_{cb}^*V_{cs}|$ that is $\sim 1/\lambda^2$ times larger.
We can thus expand in $T/P$, which gives for the direct \CP\ asymmetry and
branching fraction respectively
\begin{equation}
\begin{split}
A_{\CP}&=2\frac{T}{P}\sin(\delta_P-\delta_T)\sin\gm 
+O\Big(\Big(\frac{T}{P}\Big)^2\Big),\\
{\cal B}&=P^2\Big[1+2\frac{T}{P}\cos(\delta_P-\delta_T)\cos\gm 
+O\Big(\Big(\frac{T}{P}\Big)^2\Big)\Big].
\end{split}
\end{equation}
Using the above expression for ${\cal B}(\Bz\to\Kp\pim)$ one can get a
very simple geometric bound on \gm, if $P$ is known. 
Obtaining $P$ from ${\cal B}(\Bp\to\pip\Kz)$ --- neglecting very small color
suppressed electroweak penguins --- one has $|\cos\gm|>\sqrt{1-R}$ valid 
for
$R<1$~\cite{Fleischer:1997um,Neubert:1998pt}
($R$ is defined in Eq.~\ref{eq:rdefs} below). 
At present this gives $\gm<77^\circ$ at $1\sigma$.

The extraction of \gm\ requires more theoretical input. One needs
to determine the strong phase difference $\delta_P-\delta_T$ and the ratio
$T/P$. This can be achieved either by relating $T/P$ to $\Delta S=0$ decays
using SU(3)~\cite{Neubert:1998jq,Buras:1998rb,Buras:2000gc,Buras:2004ub,Gronau:2007af,Gronau:2000md,Chiang:2004nm,Chiang:2003pm,Ciuchini:2008eh,Fleischer:2007hj,Fleischer:1999pa}
or by using the $1/m_b$ expansion and factorization theorems to calculate
the $T/P$ ratio~\cite{Beneke:2003zv,Beneke:2001ev,Bauer:2005kd,Bauer:2004dg,Bauer:2004tj,Li:2005kt}. 

The methods that use SU(3) flavor symmetries exploit the fact that $\Delta
S=0$ decays such as $\B\to \pi\pi$ are tree dominated. The CKM factors
multiplying the ``tree'' ($V_{ub}^*V_{ud}$) and ``penguin'' terms
($V_{cb}^*V_{cd}$) are of comparable size (unlike $\Delta S=1$ decays where
the ``tree'' is CKM suppressed). From these decays one can then determine
the size of $T/P$ and feed it into $\Delta S=1$ decays to extract \gm.
In doing this quite often some $1/m_b$ suppressed annihilation or exchange
amplitudes need to be neglected. These methods are hard to improve
systematically, while already at present the determined value of \gm\ is
dominated by theoretical errors due to SU(3) breaking and the neglected
amplitudes. These were estimated to be of order $8^\circ-10^\circ$
in~\cite{Gronau:2007af} for the extraction of \gm\ from $\B\to \pi\pi$ and
$\B\to \pi K$. Some improvement can be expected, if one does not need to
neglect annihilation amplitudes but rely only on flavor symmetry.  
One interesting method of this type uses $\Bs\to \Kp \Km$ and $\B\to \pip\pim$
decays~\cite{Dunietz:1993rm,Fleischer:1999pa}.
In this analysis, the theoretical error on the extracted value of \gm\ due to
SU(3) breaking was estimated to be of the order of
$5^\circ$~\cite{Fleischer:2007hj}.  
This is a promising avenue of investigation for the LHCb experiment.

If instead of extracting \gm\ the goal is to make a precision test of the
Standard Model, one can rather take as an input the value of \gm\ determined
from $\B\to DK$ or from global fits. 
A theoretically clean prediction  of $S_{\KS\piz}$ is then possible using
isospin relations, while theoretical calculations based on the $1/m_b$
expansion are used only for SU(3) breaking terms~\cite{Fleischer:2008wb} (see
also~\cite{Gronau:2008gu}).


If the $1/m_b$ expansion is used to determine \gm, a number of different
observables can be used, since in principle all the observables are now
calculable. At present in the $1/m_b$ expansion calculations \gm\ is
taken as an input, but it could of course be extracted from data instead.
Different groups  treat differently various terms in the expansion, for
instance expanding or not expanding in $\alpha_S(\sqrt{\Lambda m_b})$,
including different $1/m_b$ suppressed terms in the expansion, etc.,  and
this may lead to slightly different extracted values of \gm\ (but the
estimated theoretical errors should account for the differences). The
important point is that the expansion is systematically improvable so that
the errors could at least in principle be reduced in the future. For
instance, the theoretical errors on the value of \gm\ extracted from
$S_{\rho\pi}$ are about $5^\circ$  and about $10^\circ$ if extracted from
$S_{\pi\pi}$~\cite{Beneke:2003zv}. Much larger errors can be expected for
\gm\ extracted from $\Delta S=1$ decays, since the interference is CKM
suppressed. 


As an example let us consider the ratios
\begin{equation}
R   = \frac{{\cal B}(\Bz\to\pimp\Kpm)\tau_{\Bp}}{{\cal B}(\Bpm\to\pipm\Kz)\tau_{\Bz}},\quad 
R_c = \frac{2{\cal B}(\Bpm\to\piz\Kpm)}{{\cal B}(\Bpm\to\pipm\Kz)}, \quad
R_n = \frac{{\cal B}(\Bz\to\pimp\Kpm)}{2{\cal B}(\Bz\to\piz\Kz)},
\label{eq:rdefs}
\end{equation}
for which part of the theoretical and experimental uncertainties
cancel~\cite{Buras:2004th}.  Tab.~\ref{tab:angles:gamma-kpi} summarizes
the current experimental measurements of the $\B \to K\pi$ branching
fractions and \CP\ asymmetries~\cite{Aubert:2006fha,Aubert:2006gm,Abe:2006xs,Abe:2006qx,Aubert:2007hh,Aubert:2008sb,Barberio:2008fa},
while $\tau_{\Bp}/\tau_{\Bz} = 1.073 \pm 0.008$~\cite{Barberio:2008fa}.
In Tab.~\ref{tab:angles:gamma-kpi} we also quote the resulting world
averages for the ratios, ignoring the correlations between the individual
branching fraction measurements.  These translate into the following bounds on
\gm\ at $68\%$ confidence level~\cite{Li:2005kt}
\begin{equation}
R   \Rightarrow 55^{\circ} < \gm < 95^{\circ}, \qquad
R_c \Rightarrow 55^{\circ} < \gm < 80^{\circ}, \qquad
R_n \Rightarrow 40^{\circ} < \gm < 75^{\circ}.
\end{equation}

\begin{table}
\centering
\caption{Summary of $\B \to K\pi$ experimental measurements.}
\label{tab:angles:gamma-kpi}
\begin{tabular}{lccc}
\hline
\hline
Quantity                    & BaBar Value                           & Belle Value                                    & World Average Value                  \\
\hline
${\cal B}(\Bpm\to\pipm\Kz)$ & $(23.9 \pm 1.1 \pm 1.0)\times10^{-6}$ & $(22.8 \,^{+0.8}_{-0.7} \pm 1.3)\times10^{-6}$ & $(23.1 \pm 1.0)\times10^{-6}$        \\
${\cal B}(\Bpm\to\piz\Kpm)$ & $(13.6 \pm 0.6 \pm 0.7)\times10^{-6}$ & $(12.4 \pm 0.5 \pm 0.6)\times10^{-6}$          & $(12.9 \pm 0.6)\times10^{-6}$        \\
${\cal B}(\Bz\to\pimp\Kpm)$ & $(19.1 \pm 0.6 \pm 0.6)\times10^{-6}$ & $(19.9 \pm 0.4 \pm 0.8)\times10^{-6}$          & $(19.4 \pm 0.6)\times10^{-6}$        \\
${\cal B}(\Bz\to\piz\Kz)$   & $(10.1 \pm 0.6 \pm 0.4)\times10^{-6}$ & $( 9.7 \pm 0.7 \,^{+0.6}_{-0.7})\times10^{-6}$ & $( 9.8 \pm 0.6)\times10^{-6}$        \\
\hline                                                                                     
$A_{\CP}(\Bpm\to\pipm\Kz)$  & $(- 2.9 \pm 3.9 \pm 1.0)\%$           & $(+3   \pm 3   \pm 1)\%$                       & $(+0.9 \pm 2.5)\%$                   \\
$A_{\CP}(\Bpm\to\piz\Kpm)$  & $(+ 3.0 \pm 3.9 \pm 1.0)\%$           & $(+7   \pm 3   \pm 1)\%$                       & $(+5.0 \pm 2.5)\%$                   \\
$A_{\CP}(\Bz\to\pimp\Kpm)$  & $(-10.7 \pm 1.6 \,^{+0.6}_{-0.4})\%$  & $(-9.4 \pm 1.8 \pm 0.8)\%$                     & $(-9.8 \, ^{+1.2}_{-1.1})\%$         \\
$A_{\CP}(\Bz\to\piz\Kz)$    & $(-13   \pm 13  \pm 3)\%$             & $(+14  \pm 13  \pm 6)\%$                       & $(-1 \pm 10)\%$                      \\
\hline
$R$                         & $\cdots$                              & $\cdots$                                       & $0.90 \pm 0.05$                      \\
$R_c$                       & $\cdots$                              & $\cdots$                                       & $1.12 \pm 0.07$                      \\
$R_n$                       & $\cdots$                              & $\cdots$                                       & $0.99 \pm 0.07$                      \\
\hline
\hline
\end{tabular}
\end{table}


The measurements of \Bs\ decays to two light hadrons can provide further
constraints on \gm~\cite{Gronau:2000md}.  
Following its earlier discovery of $\Bs\to\Kp\Km$~\cite{Abulencia:2006psa},
the CDF collaboration has recently produced updated measurements of the
branching fraction and \CP\ asymmetry of the decay
$\Bs\to\Km\pip$~\cite{Aaltonen:2008hg,Morello:2008gy}:
\begin{eqnarray}
{\cal B}(\Bs\to\Km\pip) &=& (5.0 \pm 0.7 \pm 0.8)\times10^{-6} \, , \\
A_{\CP}(\Bs\to\Km\pip) &=& (39 \pm 15 \pm 8)\% \, .
\end{eqnarray}
It has been recently pointed out that these results have implications for
SU(3) and QCD factorization~\cite{Chiang:2008vc}, which prefer a larger
value of the branching fraction for the Standard Model value of \gm.

\subsubsection{Constraints from $\B \to K\pi\pi$ Dalitz-plot analyses}
\label{sec:angles:gamma-exp-kpipi}

Three-body decays have an added benefit that quasi-two-body decays such
as $\B\to \Kstar\pi$ and $\B\to K\rho$ can interfere through the same final
$K\pi\pi$ state.  Measuring the interference pattern in the Dalitz plot
then allows to determine not only the magnitudes of the amplitudes as in
the two body decays, but also the relative phases between the amplitudes.
This can then be used either to check $1/m_b$ predictions or as an
additional input for the determination of the CKM weak phase using flavor
symmetries. We will review such a method
below~\cite{Ciuchini:2006kv,Gronau:2006qn}. The cleanest method requires
measurements from the $\Bz\to\Kp\pim\piz$ and $\Bz\to\KS\pipi$ Dalitz
plots~\cite{Ciuchini:2006kv,Gronau:2006qn,Gronau:2007vr}.
Other methods also use $\Bp\to\KS\pip\piz$~\cite{Ciuchini:2006kv},
$\Bp\to\Kp\pip\pim$ and $\Bz\to\KS\pipi$~\cite{Bediaga:2007zz}, and
$\Bs\to\Kp\pim\piz$ and $\Bs\to\KS\pipi$~\cite{Ciuchini:2006st}.

The main idea of the method~\cite{Ciuchini:2006kv,Gronau:2006qn} is that by
using isospin decomposition one can cancel the QCD penguin contributions
($\Delta I=0$ reduced amplitudes) in $B\to\Kstar\pi$ decays. 
The $I=3/2$ ($\Delta I=1$) final state, is for instance given by
\begin{equation}
\label{eq:angles:A32}
3A_{3/2} =  A(\Bz\to\Kstarp\pim) + \sqrt{2}A(\Bz\to\Kstarz\piz)~,
\end{equation}
with an equivalent definition for the amplitude for charge-conjugated states,
$\bar A_{3/2}$. 
Since both magnitudes and relative phases of amplitudes are measurables,
this is now an observable quantity --- up to an overall phase. In the absence
of electroweak penguin (EWP) terms $A_{3/2}$ carries a weak phase \gm, so
that in this limit 
\begin{equation}
\gm = \Phi_{3/2}\equiv -\frac{1}{2}\mbox{arg}\left(R_{3/2}\right )~,
~~{\rm where}~~
R_{3/2}\equiv \frac{\bar A_{3/2}}{A_{3/2}}~.
\end{equation}


The constraint in $\rhobar-\etabar$ plane in the absence of EWP is a
straight line, $\etabar = \rhobar\,\tan\Phi_{3/2}$. The inclusion of EWP
shifts this constraint to~\cite{Gronau:2006qn}
\begin{equation}
\etabar = \tan\Phi_{3/2}\left [\rhobar + C[1- 2{\rm Re}(r_{3/2})] + {\cal O}(r^2_{3/2})\right]~,
\end{equation}
where $C$ is a quantity that depends only on electroweak physics and is
well known, with a theoretical error below $1\%$ ($\lambda=0.227$)
\begin{equation}
C  \equiv \frac{3}{2}\frac{C_9+C_{10}}{C_1+C_2}\frac{1-\lambda^2/2}
{\lambda^2} = -0.27~,
\end{equation}
while the nonperturbative QCD effects enter only through a complex parameter
\begin{equation}
r_{3/2}  \equiv  \frac{(C_1-C_2)\langle (K^*\pi)_{I=3/2}|{\cal O}_1-{\cal O}_2|B^0\rangle}
{(C_1+C_2)\langle (K^*\pi)_{I=3/2}|{\cal O}_1+{\cal O}_2|B^0\rangle}~.
\end{equation}
Here ${\cal O}_1\equiv (\bar bs)_{\rm V-A}(\bar uu)_{\rm V-A}$ and ${\cal
O}_2\equiv (\bar bu)_{\rm V-A}(\bar us)_{\rm V-A}$ are the V-A
current-current operators. In naive factorization $r_{3/2}$ is found to be
real and small, $r_{3/2} \leq 0.05$~\cite{Ciuchini:2006kv}. This is in
agreement with the estimate using flavor SU(3) from $\B\to\rho\pi$,
$r_{3/2}=0.054 \pm 0.045 \pm 0.023$~\cite{Gronau:2006qn}, where the first
error is experimental and the second an estimate of theoretical errors.
This then gives the constraint 
\begin{equation}
\label{eq:angles:constraint}
\etabar = \tan\Phi_{3/2}\left[\rhobar -0.24\pm0.03\right]\, .
\end{equation}
%

\begin{figure}
  \centering
  \includegraphics{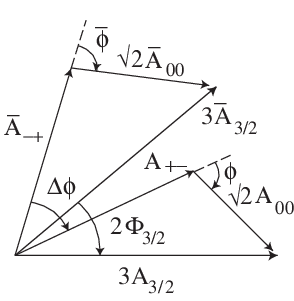}
  \caption{
    Geometry for Eq.~(\ref{eq:angles:A32}) and its charge-conjugate,
    using notations 
    $A_{+-} \equiv A(\Bz\to\Kstarp\pim)$, $A_{00} = A(\Bz\to\Kstarz\piz)$ 
    and similar notations for charge-conjugated modes~\cite{Gronau:2006qn}.
  }
  \label{fig:angles:triangle}
\end{figure}

The phase $\Phi_{3/2}$ can be determined by measuring the magnitudes and
relative phases of the $\Bz\to\Kstarp\pim,~\Bz\to\Kstarz\piz$ amplitudes
and their charge-conjugates.
A graphical representation of the triangle relation
Eq.~(\ref{eq:angles:A32}) and its charge conjugate is given in
Fig.~\ref{fig:angles:triangle}. The above four magnitudes of amplitudes and
the two relative phases,
$\phi\equiv {\rm arg}[A(\Bz\to\Kstarz\piz)/A(\Bz\to\Kstarp\pim)]$
and
$\bar\phi\equiv {\rm arg}[A(\Bzb\to\Kstarzb\piz)/A(\Bzb\to\Kstarm\pip)]$,
determine the two triangles separately.
Their relative orientation is fixed by the phase difference
$\Delta\phi\equiv {\rm arg}[A(\Bz\to\Kstarp\pim)/A(\Bzb\to\Kstarm\pip)]$.

A similar analysis is possible using $B\to\rho K$ decays.  Although each
$\rho$ meson has only a single dipion decay, the relative phase between the
amplitudes in Eq.~(\ref{eq:angles:A32}) can be determined exploiting the fact
that the $\Kstar\pi$ amplitudes appear in both $K\pi\pi$ Dalitz plots and
therefore can be used as a common reference.
The same approach could also be applied to $B\to\Kstar\rho$ decays.

The $\Bp\to\Kp\pip\pim$ Dalitz plot provides the highest signal event yield
of the $K\pi\pi$ Dalitz plots and so can be used to establish a working
isobar model.  This information can be used by the other analyses, leading
to smaller systematic uncertainties.  The $\Kp\pip\pim$ Dalitz plot also
contains the intermediate state $\rho^0(770)\Kp$, which is predicted to
have a large direct \CP\ asymmetry $\sim40\%$.  Measuring this asymmetry,
interesting in its own right, tells us that the tree and penguin
contributions are of similar order and that we do indeed have sensitivity
to \gm\ in these decays.  BaBar~\cite{Aubert:2008bj} and
Belle~\cite{Belle:Kppippim} have recently updated their analyses of this
Dalitz plot and both see strong evidence of direct \CP\ violation in
$\Bp\to\rho^0(770)\Kp$.  The results are in excellent agreement and are
summarized in Tab.~\ref{tab:angles:gamma-kppippim}.
The signal Dalitz-plot model used in these analyses contains contributions
from $\Kstarz(892)\pip$, $\Kstarz_0(1430)\pip$, $\rho^0(770)\Kp$,
$\omega(782)\Kp$, $f_0(980)\Kp$, $f_2(1270)\Kp$, $f_{\rm X}(1300)\Kp$,
$\chiczero\Kp$, and a nonresonant component; the BaBar model also contains
$\Kstarz_2(1430)\pip$.
The main difference between the approaches of the two experiments concerns
the nonresonant model.
Belle uses two $e^{-\alpha s}$ distributions, where $\alpha$ is a free
parameter, one with $s = m^2_{\Kp\pim}$ and one with $s = m^2_{\pipi}$.
BaBar uses a phase-space component in addition to a parametrization of the
low-mass $\Kp\pim$ S-wave that follows that of the LASS
experiment~\cite{Aston:1987ir}.

\begin{table}[htb]
\centering
\caption{Summary of results for $A_{\CP}$ of $\Bp\to\rho^0(770)\Kp$.
         The uncertainties are statistical, systematic and
	 model-dependent respectively.}
\label{tab:angles:gamma-kppippim}
\begin{tabular}{lc}
\hline
\hline
Experiment   & $A_{\CP}(\rho^0(770)\Kp)$            \\
\hline
BaBar        & $(+44 \pm 10 \pm 4 \,^{+5}_{-13})\%$ \\
Belle        & $(+41 \pm 10 \pm 3 \,^{+3}_{-7})\%$  \\
\hline
HFAG Average & $(+42 \,^{+8}_{-10})\%$              \\
\hline
\hline
\end{tabular}
\end{table}

The $\Bz\to\KS\pipi$ Dalitz plot is an extremely rich physics environment.
As well as providing measurements of the \BzBzb\ mixing phase $2\bb$,
discussed in Sec.~\ref{sec:angles:beta-tddp}, it is possible to measure the
phase difference $\Delta\phi$ between $\Bz\to\Kstarp\pim$ and
$\Bzb\to\Kstarm\pip$, one of the crucial ingredients in the determination of
\gm\ with the method of
Refs.~\cite{Ciuchini:2006kv,Gronau:2006qn,Gronau:2007vr}.
Both BaBar~\cite{Aubert:2007vi} and Belle~\cite{Dalseno:2008ms} have
performed time-dependent Dalitz-plot analyses of this mode.  Details of the
analyses are discussed in Sec.~\ref{sec:angles:beta-tddp}.
Belle find two fit solutions that correspond to different interference
between the $\Kstarp_0(1430)$ and nonresonant components.  These two
solutions prefer different values of $\Delta\phi$.
The results for $\Delta\phi$ are illustrated in
Fig.~\ref{fig:angles:gamma-kspipi} and summarized in
Tab.~\ref{tab:angles:gamma-kspipi}.
There is some disagreement between the BaBar and Belle results.
The experimentally measured values of $\Delta\phi$ include the
\BzBzb\ mixing phase since they come from time-dependent analyses.
This has to be removed before the values can be used in the extraction of
\gm.

\begin{figure}[htb]
\centering
\includegraphics[width=0.32\textwidth]{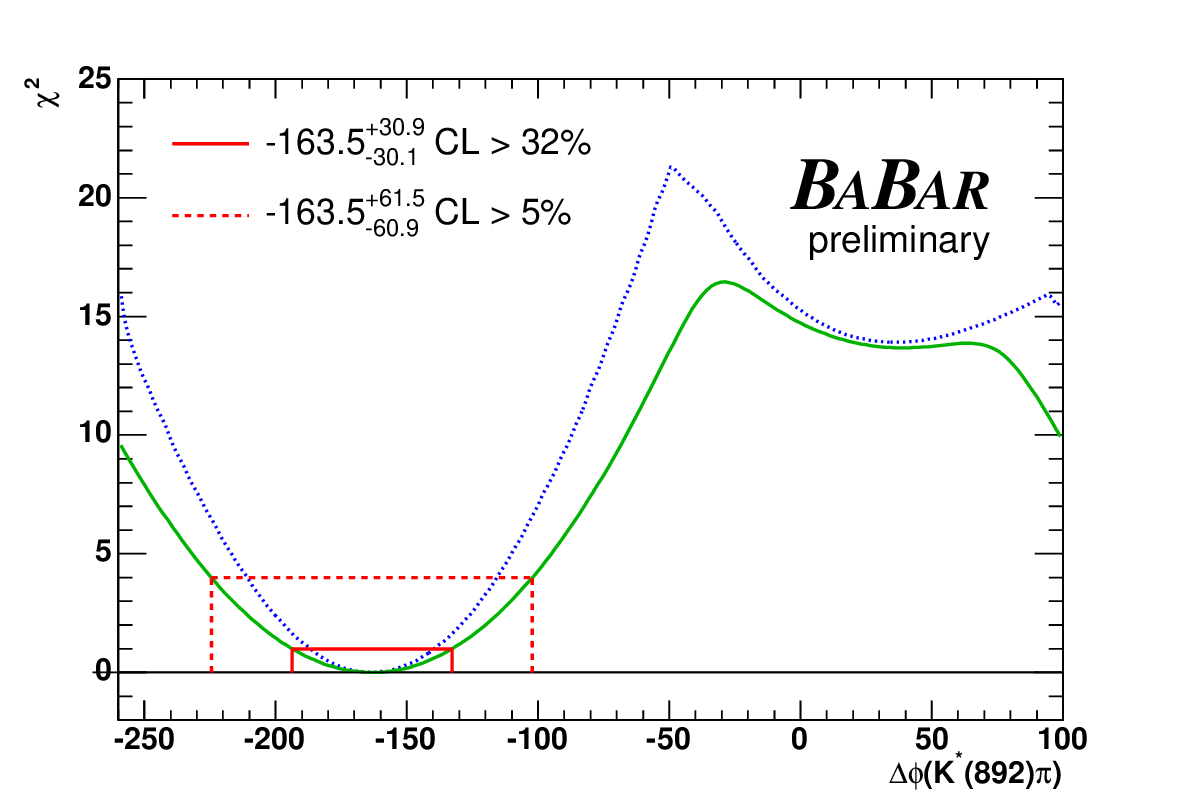}
\includegraphics[width=0.32\textwidth]{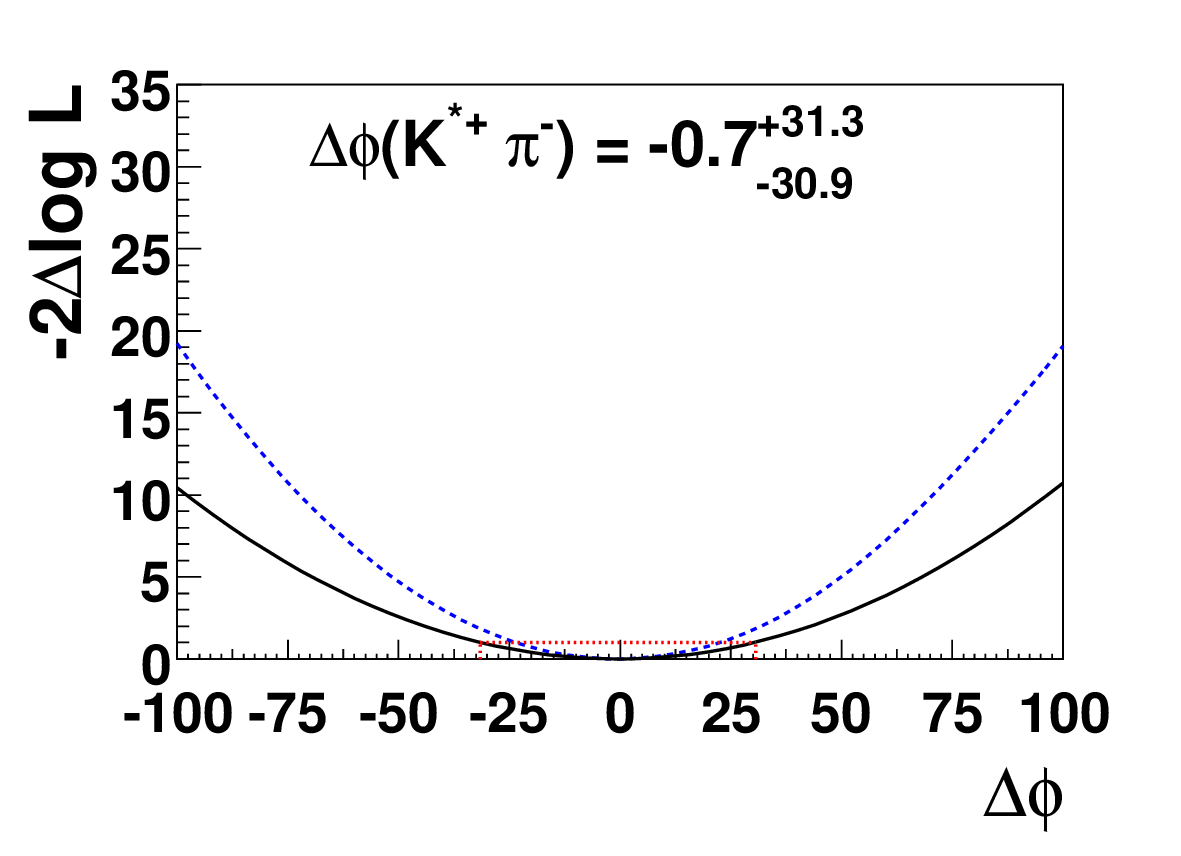}
\includegraphics[width=0.32\textwidth]{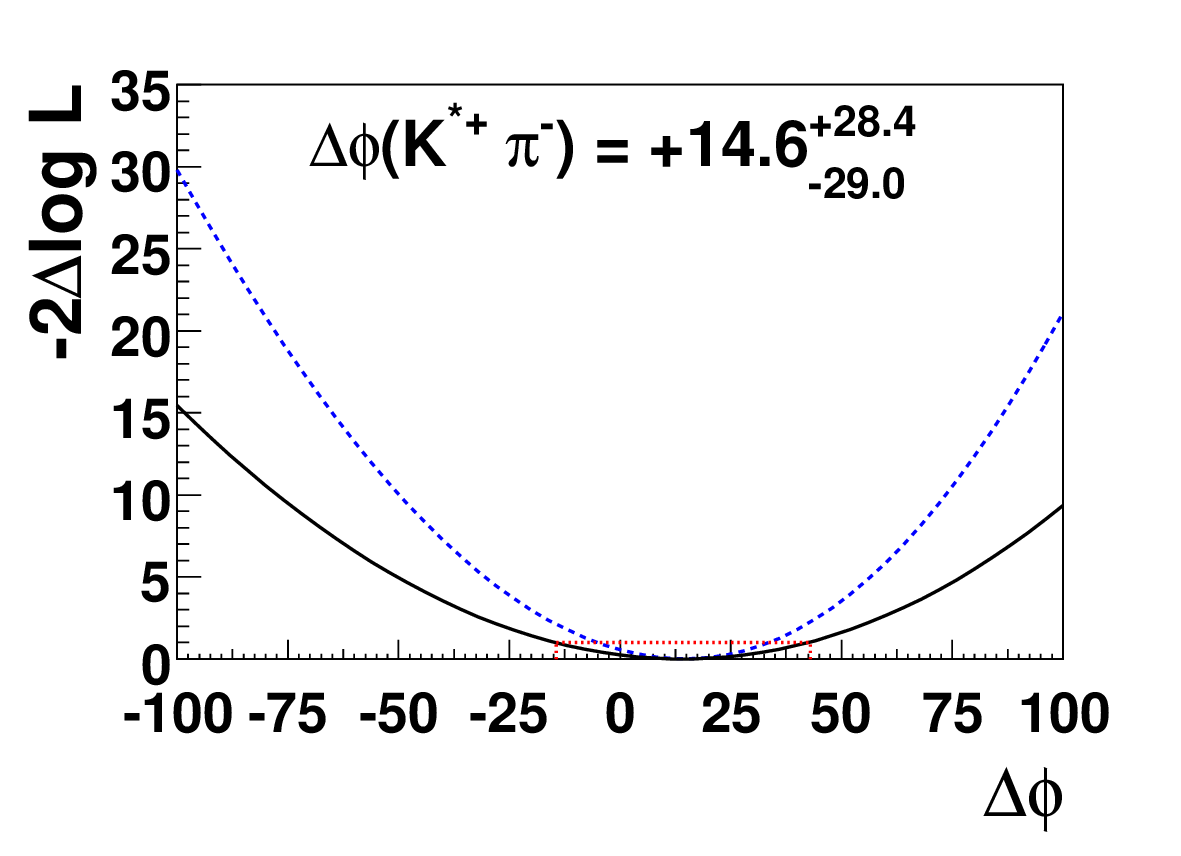}
\caption{Likelihood scans of $\Delta\phi$ from Dalitz-plot analyses of
$\Bz\to\KS\pipi$. The left plot is from BaBar, the middle and right plots
are from Belle and represent the scans of the two different solutions.}
\label{fig:angles:gamma-kspipi}
\end{figure}

\begin{table}[htb]
\centering
\caption{Summary of results for $\Delta\phi(\Kstarp\pim)$ from
         time-dependent Dalitz-plot analyses of $\Bz\to\KS\pipi$.
         The uncertainties are statistical, systematic and
	 model-dependent respectively.}
\label{tab:angles:gamma-kspipi}
\begin{tabular}{lc}
\hline
\hline
Experiment    & $\Delta\phi(\Kstarp\pim)$                    \\
\hline
BaBar         & $(-164 \pm 24 \pm 12 \pm 15)^{\circ}$        \\
Belle Soln. 1 & $(-1 \,^{+24}_{-23} \pm 11 \pm 18)^{\circ}$  \\
Belle Soln. 2 & $(+15 \,^{+19}_{-20} \pm 11 \pm 18)^{\circ}$ \\
\hline
\hline
\end{tabular}
\end{table}

The other two parameters required to determine \gm\ are $\phi$ and
$\bar{\phi}$.  These are the relative phases of $\Bz\to\Kstarp\pim$ and
$\Bz\to\Kstarz\piz$ and $\Bzb\to\Kstarm\pip$ and $\Bzb\to\Kstarzb\piz$,
respectively.  Both of these quantities can be determined from a
time-integrated Dalitz-plot analysis of $\Bz\to\Kp\pim\piz$ (and its charge
conjugate).  Such an analysis has not yet been performed by Belle but BaBar
has published results based on $232\times10^6$ \BB\ pairs~\cite{Aubert:2007bs}
and has preliminary results based on the full BaBar dataset of
$454\times10^6$ \BB\ pairs~\cite{Aubert:2008zu}.
The published analysis includes contributions from $\rho^-(770)\Kp$,
$\Kstarp(892)\pim$, $\Kstarz(892)\piz$, $\Kstarp_0(1430)\pim$ and
$\Kstarz_0(1430)\piz$.  The higher $\Kstar$ resonances are modeled by the
LASS parametrization, which also includes a slowly varying nonresonant term.
The fit exhibits multiple solutions that are not well separated.  This can
be seen in the likelihood scans in Fig.~\ref{fig:angles:gamma-kpipi0} and
unfortunately leads to a weaker constraint on \gm.
BaBar's preliminary results on the larger data sample indicate much better
separation between solutions, however likelihood scans of $\phi$ and
$\bar{\phi}$ are not yet completed.

\begin{figure}[htb]
\centering
\includegraphics[width=0.45\textwidth]{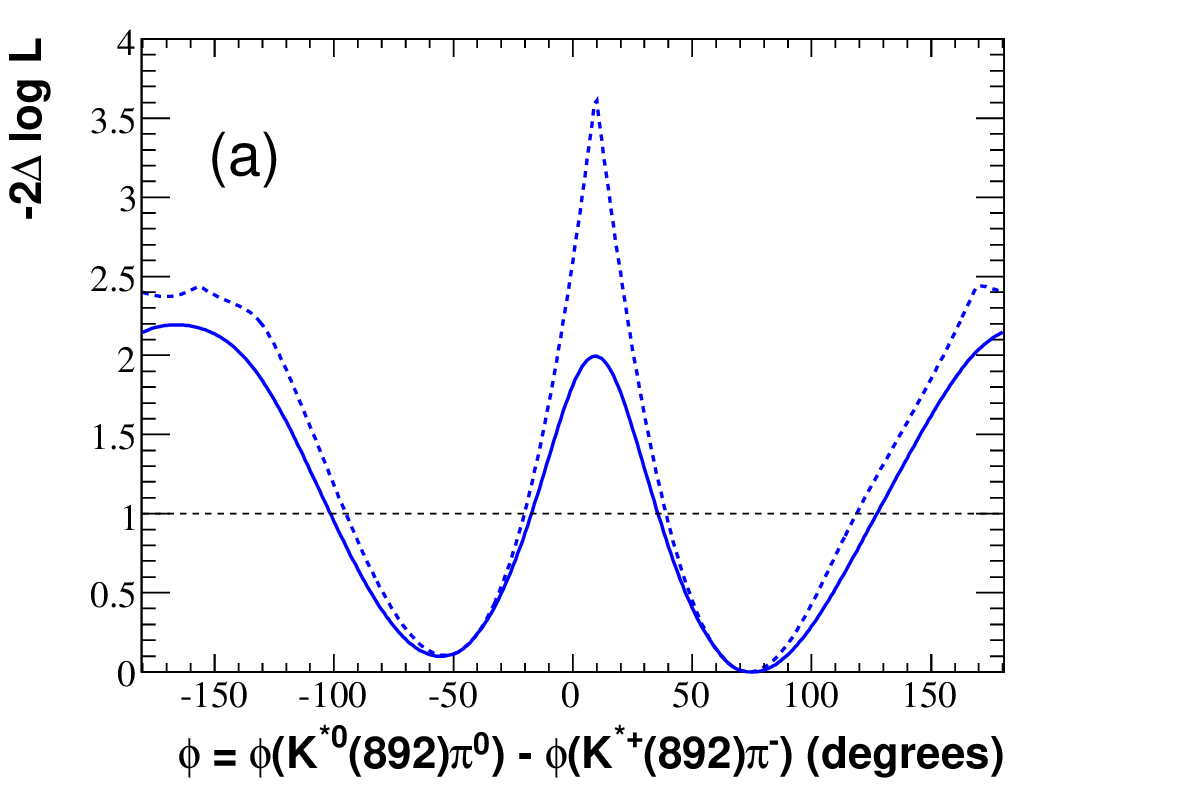}
\includegraphics[width=0.45\textwidth]{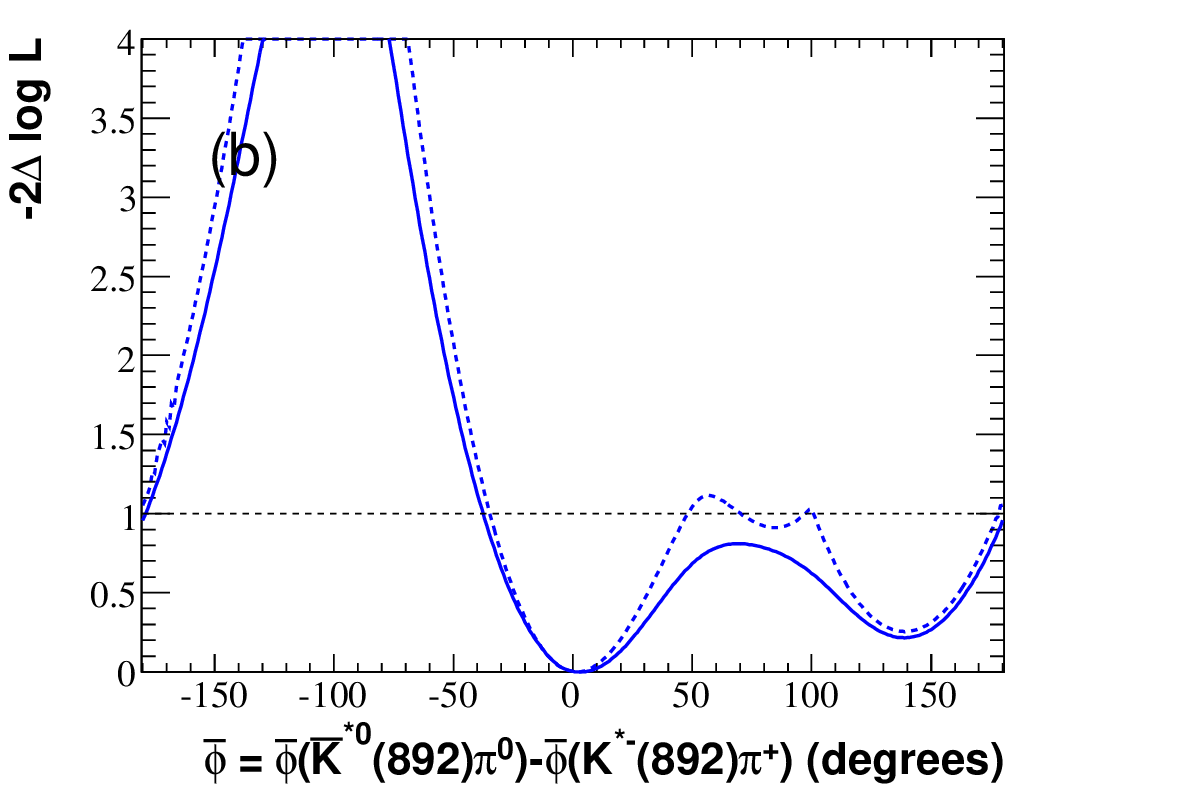}
\caption{Likelihood scans of $\phi$ (left) and $\bar{\phi}$ (right) from
BaBar Dalitz-plot analysis of $\Bz\to\Kp\pim\piz$.}
\label{fig:angles:gamma-kpipi0}
\end{figure}

The BaBar results on $\Delta\phi$~\cite{Aubert:2007vi}, $\phi$ and
$\bar{\phi}$~\cite{Aubert:2007bs} have been combined~\cite{Gronau:2007vr}
to create a constraint
\begin{equation}
39^{\circ} < \Phi_{3/2} < 112^{\circ} \; (68\% {\rm CL}) \, ,
\end{equation}
which can be converted to a constraint on the $\rhobar-\etabar$ plane,
using the relation \eqref{eq:angles:constraint}.
Both of these constraints are shown in
Fig.~\ref{fig:angles:gamma-constraint}.
%
%

\begin{figure}[htb]
  \centering
  \includegraphics[width=0.41\textwidth]{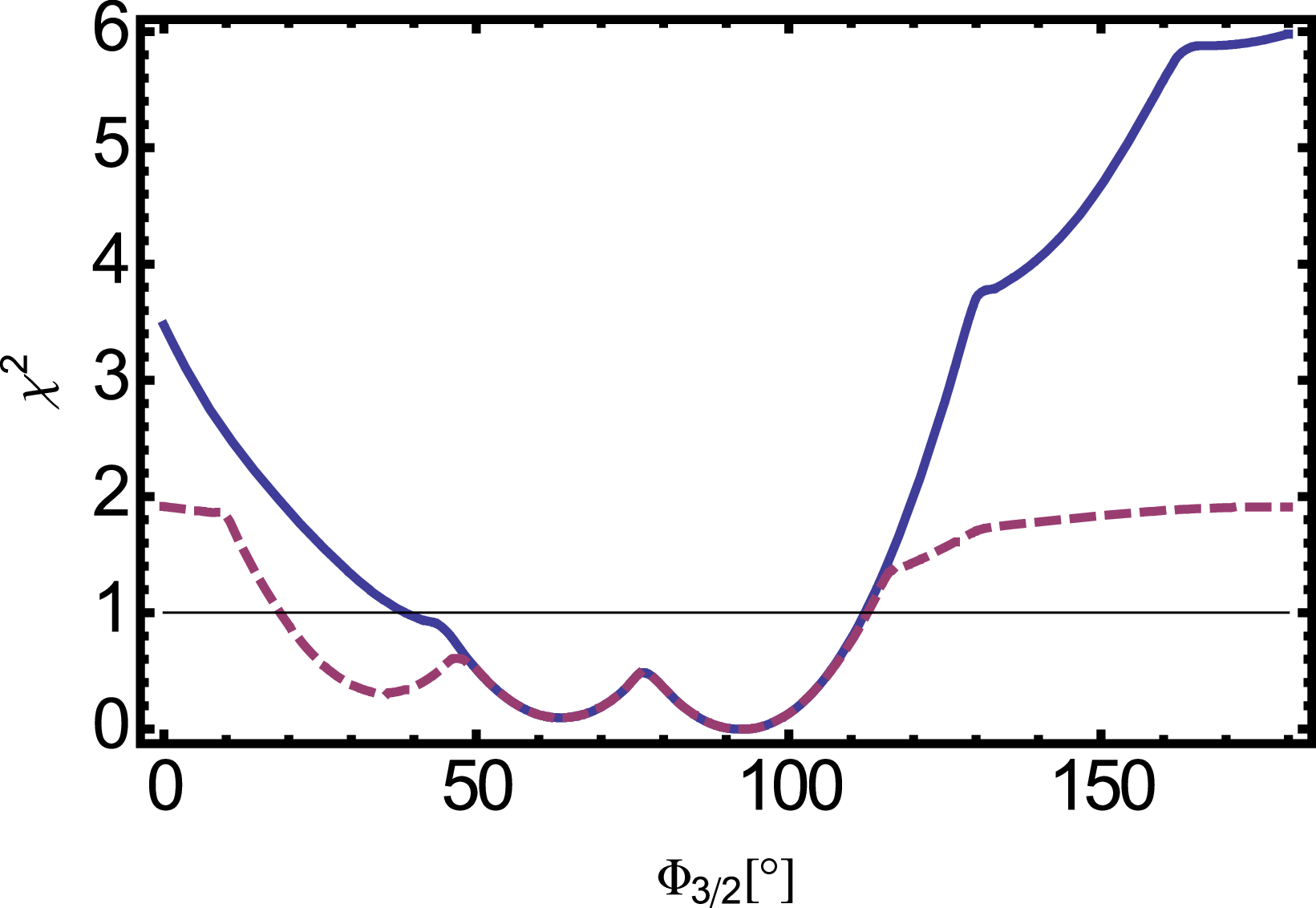}
  \hfill
  \includegraphics[width=0.58\textwidth]{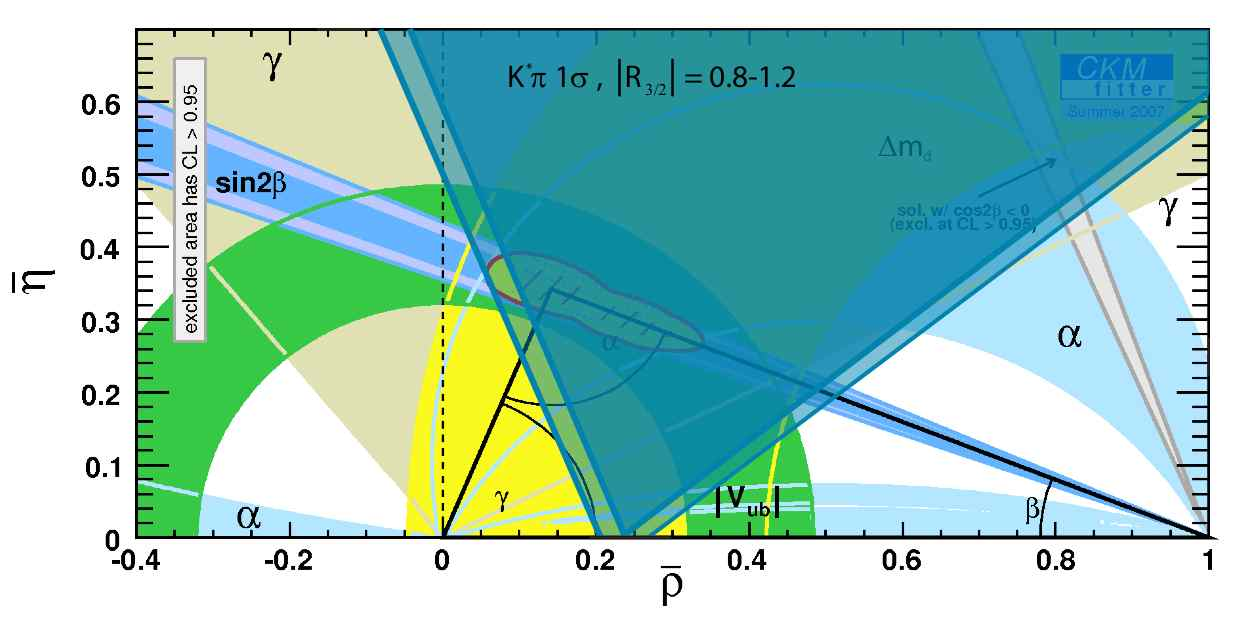}
  \caption{
    Constraint on the angle $\Phi_{3/2}$ (left) from combined
    information from $K\pi\pi$ Dalitz plot analyses. The dashed purple line is
    for the case when $\left|R_{3/2}\right|$ is unconstrained while the solid
    blue line is for the case when $0.8 < \left|R_{3/2}\right| < 1.2$.
    Constraint on the $\rhobar-\etabar$ plane (right) from combined
    information from $K\pi\pi$ Dalitz plot analyses. The dark shaded region
    corresponds to the experimental $1\sigma$ range while the light shaded
    region includes the theoretical error on the contributions from electroweak
    penguin processes.
  }
  \label{fig:angles:gamma-constraint}
\end{figure}

\section{Global Fits to the Unitarity Triangle and Constraints on New 
Physics}
\label{section:globalfits}
The large variety of precise measurements reported so far can be used to place constraints on theoretical models of flavor particles and their interactions. The impact of these constraints has been studied using global fits to the predictions of the Standard Model and other theoretical models. 

In this section, results of such studies will be presented. First, the results described in this report are interpreted within the Standard Model 
 (Sec.~\ref{sec:globalfits:SM}). 
Next, Sec.~\ref{sect:gfitsNP} summarizes the 
 constraints imposed by these measurements on deviations from the Standard Model.  Discussions of constraints, first in a model independent approach, then 
for Grand Unified Theories, and for models with Extra Dimensions conclude the report.

\subsection{Constraints on the Unitarity Triangle Parameters }
\label{sec:globalfits:SM}

The  measured quantities reported so far are sensitive to different combinations of the parameters of the CKM matrix 
(see Sec.~\ref{sec:intro:UT} for details). 
Their relations to the angles and sides of the Unitarity Triangle (UT) place constraints on the coordinates of its apex $(\overline{\rho},\overline{\eta})$ 
and thus can be used to test the predictions of the Standard Model or any other theory describing flavor physics. 
The most powerful way to make such tests is to perform global fits comparing the data to theoretical predictions.

To combine a large number of measurements of different quantities performed with different methods and data samples,
 widely different in size and composition, and 
thus with different statistical and systematic errors, not all of them Gaussian in nature, is a non-trivial task. 
Theoretical predictions have uncertainties that are very difficult to assess and that are usually not expected to adhere to Gaussian distributions.  Two 
groups have independently developed global analysis tools to determine the CKM parameters in the framework of the Standard Model and its extensions. The two 
approaches differ significantly, in particular in the treatment of uncertainties of data and of the theory predictions. The results of this report have been 
analyzed by the \utfit\ group. Most of this section therefore discusses their Bayesian approach in detail,  the work of the CKMfitter group being summarized 
for comparison in Sec.~\ref{sec:fitckmfitter}.

\subsubsection{Fitting technique }
\label{sec:statmethods}

The Unitarity Triangle analysis developed by the \utfit\ group relies on the Bayes Theorem.  Its specific application    
is briefly described in the following, more details can be found in elsewhere~\cite{Ciuchini:2000de}.

A given constraint $c_j$ relates the coordinates of the apex of the Unitarity Triangle $(\overline{\rho},\overline{\eta})$  to quantities that have been experimentally determined or theoretically calculated ($\vec{x}=\{x_1, x_2, ...,x_N\}$), through functional dependencies that are prescribed by the theory that is being tested,
$c_j = c_j(\bar\rho, \bar\eta, \vec{x})$.

In the case of perfect knowledge of $c_j$ and $x_i$, each of the constraints would represent a well defined curve in the $(\overline{\rho},\overline{\eta})$ plane.
In the presence of uncertainties, the constraints are represented by distributions of curves, each weighted according to the probability density derived from the error distributions.  
Based on Bayes Theorem, the \utfit\ group derives for $M$ constraints $c_j$ and $N$ free parameters $x_i$, a $pdf$ or probability density function, 
\begin{eqnarray}
f(\bar\rho, \bar\eta, \vec{x}|\hat{c_1},...,\hat{c_M}) \propto 
\prod_{j=1,{\rm M}}f_j(\hat{c}_j\,|\,\rhobar,\etabar,{\vec x})
\times \prod_{i=1,{\rm N}}f_i(x_i) \times f_\circ(\rhobar,\etabar) \label{eq:bayes_f} .
\end{eqnarray}
By integrating Eq.~(\ref{eq:bayes_f}) over ${\vec x}$, one obtains,
\begin{equation}
f(\rhobar,\etabar\,|\,{\bf \hat{c}},{\bf f})
\propto {\cal L}({\bf \hat{c}}\,|\,\rhobar,\etabar, {\bf f})
\times f_\circ(\rhobar,\etabar)\,,
\end{equation} 
where ${\bf \hat{c}}$ stands for the set of measured constraints,
 and
\begin{equation}
{\cal L}({\bf \hat{c}}\,|\,\rhobar,\etabar,{\bf f})
= \int 
\prod_{j=1,{\rm M}}f_j(\hat{c}_j\,|\,\rhobar,\etabar,{\vec x})
\prod_{i=1,{\rm N}}f_i(x_i)\, \mbox{d}{ x_i}
\label{eq:lik_int}
\end{equation}
is the effective overall likelihood function which takes into account 
all possible values of $x_j$ and their weights based on their associated error distributions. This expression underlines the dependence of the 
likelihood on the best knowledge of all $x_i$, described by $f({\vec x})$.
Assuming a flat  {\it a~priori} distribution for 
$\rhobar$ and $\etabar$, \ie\ all values are equally likely, the final (unnormalized) $pdf$ is,
\begin{equation}
f(\rhobar,\etabar) \propto 
 \int 
\prod_{j=1,{\rm M}}f_j(\hat{c}_j\,|\,\rhobar,\etabar,{\mathbf x})
\prod_{i=1,{\rm N}}f_i(x_i)\,\mbox{d}{x_i}\, .
\label{eq:flat_inf}
\end{equation}   
The integration is done by Monte Carlo methods, in which a large sample is 
extracted for the free parameters and a weight is assigned for each extraction . 
In this way an {\it a~posteriori} {\it pdf}\ for each parameter is obtained, 
generally different from the {\it a~priori} one, because of the weighting procedure.  The result of each 
extraction is considered more or less likely, depending on the agreement of the 
corresponding measured quantities with the actual experimental results or 
theoretical calculation. 
The \utfit\ group treats theoretical and experimental 
parameters in a uniform way, adopting the error distributions, 
Gaussian or non-Gaussian, directly as $a~priori$ probability density functions. 

The {\it a~posteriori} {\it pdf}s depend by construction on the choice 
of the {\it a~priori} ones, which are based on - to a certain degree subjective - assessments of systematic uncertainties, experimental and theoretical, on theoretical approximations and assumptions. In many Unitarity Triangle analyses, the precise and abundant measurements and theoretical inputs represent very stringent constraints and the results are not very sensitive to the particular choice of the 
{\it a~priori} distributions for the parameters. If this is not the case, an assessment of the sensitivity of the result to variations of the prior is required. 

As part of the \utfit\ analysis, the agreement of the measured quantities is quantified 
in the so-called {\it compatibility plots} \cite{Bona:2005vz}.  
An indirect determination of a particular quantity is obtained from a global fit including all the available constraints, except those from the direct measurement 
of the quantity of interest.  This indirectly determined value represents the  prediction by the Standard Model or any other theory from which the constraints 
are derived. 
The comparison of the prediction and the direct measurement, including their respective uncertainties, can be used to assess the compatibility with the underlying theoretical calculations or model.

Specifically, if $f(x_{th})$ and $f(x_{fit})$ are the $pdf$s for the predicted and the measured values, respectively,  
their compatibility is evaluated by constructing the $pdf$ for the difference, $x_{th}-x_{fit}$, and by estimating the distance of
its most probable value from zero, in units of standard deviations.
In the {\it compatibility plots}, contours of constant distance are shown 
in two dimensions, $\sigma(x_{fit})$ versus $\bar{x}_{fit}$. 
The compatibility between $x_{th}$ and $x_{fit}$ can be directly estimated, 
for any central value and error on $x_{fit}$. 
In this way, the compatibility of constraints with the measurements is simply assessed by comparing two different $pdf$s, without any assumption about their shapes. 
Examples of {\it compatibility plots} are shown in 
section \ref{sec:fitresults}.

\subsubsection{Inputs to the Unitarity Triangle Analysis}

\label{sec:inputs}
Not all measurements have sensitivity to the Unitarity Triangle parameters and  there are determinations of the same observable that are equivalent but not identical. A choice has to be made.
The best selection of experimental results discussed in this report has been used as input to the CKM analysis using UT fits. They are summarized in 
Tab.~\ref{tab:inputs_utfit}.
\begin{table}[htb]
\caption{Most relevant experimental inputs to the UT fits. Internal references with the details of the choice of the inputs are also included.  }
\label{tab:inputs_utfit}
\begin{center}
\begin{tabular}{lccc}
\hline
Input & Source & Value & Reference \\
\hline
$|V_{ud}|$ & Nuclear decays & $0.97425 \pm 0.00022$ & Eq.~\ref{Vud} \\
$|V_{us}|$ & SL Kaon decays & $0.2259 \pm 0.0009$ & Eq~\ref{eq:VusVud} \\
$|V_{cb}| incl.$ & SL charmed $B$ decays & $(41.54 \pm 0.73) \times 10^{-3}$ & Eq.~\ref{slep:incVcb:eq:Vcb} \\
$|V_{cb}| excl.$ & SL charmed $B$ decays & $(38.6 \pm 1.1) \times 10^{-3}$ & Eq.~\ref{slep:eq:exVcb} \\
$|V_{ub}|$ incl.& SL charmless $B$ decays & $(4.11^{+0.27}_{-0.28}) \times 10^{-3}$ &  Eq.~\ref{eq:slep:incVub} \\
$|V_{ub}|$ excl.& SL charmless $B$ decays & $(3.38\pm0.36) \times 10^{-3}$ &  Eq.~\ref{eq:slep:Vub} \\
${\cal B}(B^+ \to \tau^+ \nu)$ & Leptonic $B$ decays & $(1.51 \pm 0.33) \times 10^{-4}$ & Tab.~\ref{tab:BtoEllNuBF} \\
$\Delta m_d$ &  $B_d \bar{B}_d$ mixing & ($0.507 \pm 0.005$) ps$^{-1}$ & Fig.~\ref{fig:deltamd} \\
$\Delta m_s$ &  $B_s \bar{B}_s$ mixing & ($17.77 \pm 0.12$) ps$^{-1}$ & Sec.~\ref{sec:bmixingExp}   \\
\hline
$|\epsilon_K|$ & $K \bar{K}$ mixing & $(2.229 \pm 0.012) \times 10^{-3}$ & Eq.~\ref{eq:resultepsK}\\
$\sin 2\beta$ & Charmonium $B$ decays  & $0.671 \pm 0.023$ & Fig.~\ref{fig:btoccsS_CP} \\
${\cal B}$ \& ${CP}$ parameters \,\,& $B \to \pi \pi,\, \rho \rho,\, \rho \pi$ decays  & & Sec.~\ref{sec:angles} \\
($x^{\pm}, y^{\pm}$), ${\cal B}$ \& $A$ & $B \to D^{(*)0}K^{(*)\pm}$\; (GGSZ, GLW, ADS)  &  & Sec.~\ref{sec:tree} \\
\hline
\end{tabular}
\end{center}
\end{table} 
%
%
%
The set of lattice inputs (see Ref.~\cite{Lubicz:2008am} for details) chosen for UT fits is summarized in Tab.~\ref{tab:lattice_inputs_utfit}.

\begin{table*}[htb]
\caption{Phenomenological inputs obtained from Lattice QCD calculations}
\label{tab:lattice_inputs_utfit}
\begin{center}
\begin{tabular}{lc}
\hline
Input & Value \\
\hline
$f_{B_s}$ (MeV)                     & $245 \pm 25$     \\
$\hat{B}_{B_s}$               & $1.22 \pm 0.12$   \\
$f_{B_s}/f_{B_d}$              & $1.21 \pm 0.04$ \\
$\hat{B}_{B_s}/\hat{B}_{B_d}$  & $1.00 \pm 0.03$ \\
$B_K$                         & $0.75 \pm 0.07$  \\
\hline
\end{tabular}
\end{center}
\end{table*} 
\subsubsection{Results of Global Fits}
\label{sec:fitresults}

Figure~\ref{fig:global_fit} displays the results of the global fit in the ($\bar{\rho},\bar{\eta}$) plane. A 
summary of the fitted parameters of the CKM matrix  (see Sec.~\ref{sec:intro} for definitions) is presented in Table~\ref{tab:utfitresults}.
\begin{table*}[htb]
\caption{Results of the global fit for the parameters of the CKM matrix. Parameters obtained with the CKMfitter approach (see Sec.~\ref{sec:fitckmfitter}) are also shown for comparison.}
\label{tab:utfitresults}
\begin{center}
\begin{tabular}{lcc}
\hline\hline
Parameter     & Result          & CKMfitter \\\hline
$\bar{\rho}$  & $0.158 \pm 0.021$ &$0.139^{+0.025}_{-0.027}$\\
$\bar{\eta}$  & $0.343 \pm 0.013$ &$0.341^{+0.016}_{-0.015}$\\
$A$           & $0.802 \pm 0.015$                &$0.812^{+0.010}_{-0.024}$\\
$\lambda$     & $0.2259 \pm 0.0016$&$0.2252 \pm 0.0008$\\
\hline\hline
\end{tabular}
\end{center}
\end{table*} 
\begin{figure}[ht]
\begin{center}
\epsfig{file=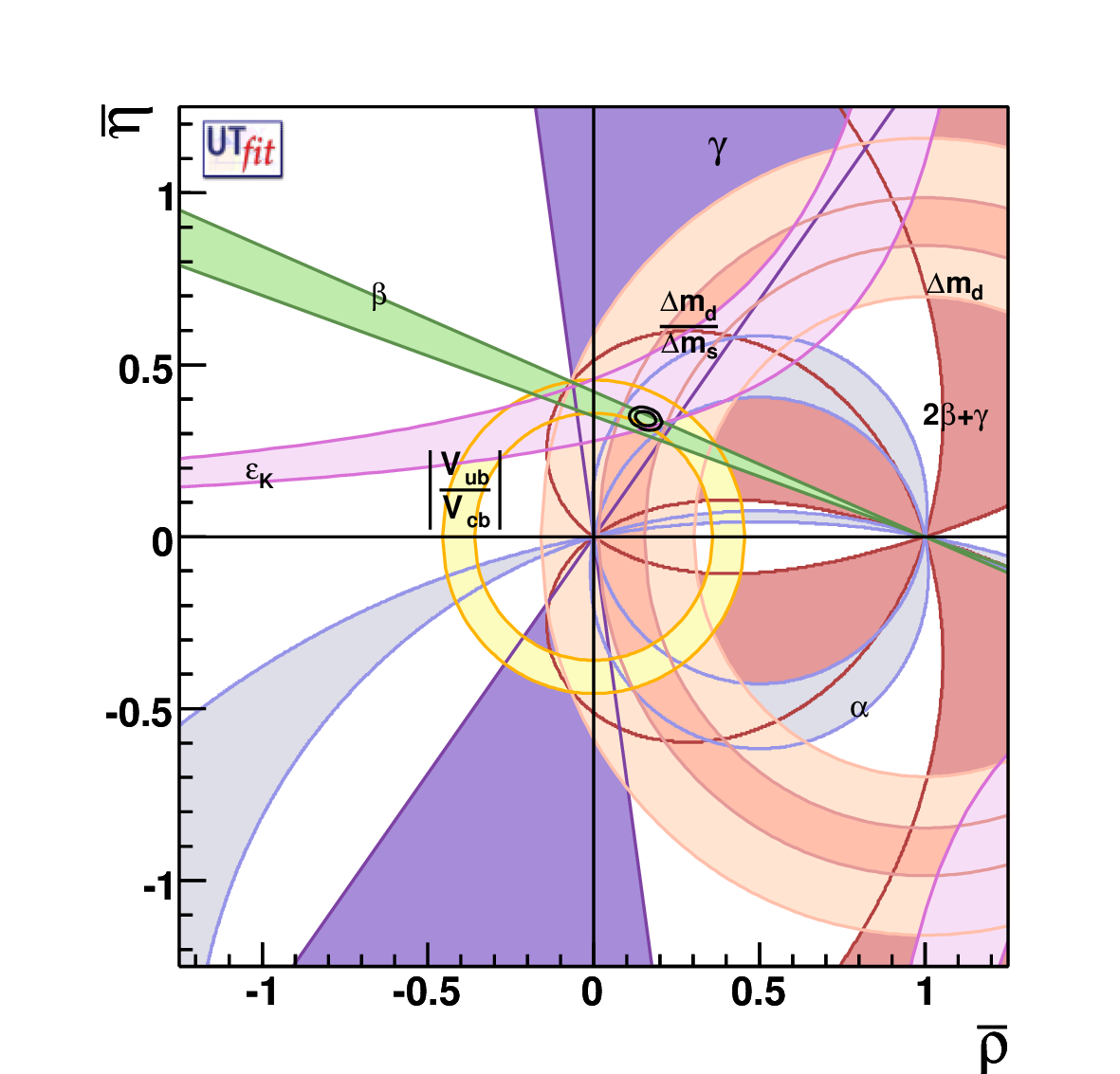,height=5cm}
\end{center}
\caption{Individual and global constraints in the ($\bar{\rho},\bar{\eta}$) plane from the global UT fits. 
The shaded areas indicate the individual constraints at 95\% CL. The contours of the overall constraints defining the apex of the UT triangle correspond to 68\% and 95\% C.L. .
}
\label{fig:global_fit}
\end{figure}
The global fits also result in improved determinations of the measured quantities, the angles and sides of the Unitarity Triangle which are 
listed in Tab.~\ref{tab:utfitresults_meas}.
\begin{table*}[htb]
\caption{Improved measurements of angles and sides of the Unitarity Triangle obtained from the global fits. Results obtained with the CKMfitter approach (see Sec.~\ref{sec:fitckmfitter}) are also shown for comparison.}
\label{tab:utfitresults_meas}
\begin{center}
\begin{tabular}{lcc}
\hline\hline
Parameter     & Results  &  CKMfitter               \\ \hline
$\alpha$($^o$)  &$92.6\pm3.2$& $90.6^{+3.8}_{-4.2}$\\ 
$sin2\beta$   &$0.698\pm0.019$& $0.684^{+0.023}_{-0.021}$\\ 
$\gamma$($^o$)&$65.4\pm3.1$& $67.8^{+4.2}_{-3.9}$\\ 
$|V_{ub}|$  &$0.00359\pm0.0012$& $0.00350^{+0.00015}_{-0.00014}$\\ 
$|V_{cb}|$  &$0.0409\pm0.0005$& $0.04117^{+0.00038}_{-0.00115}$\\ 
$|V_{td}|$  &$0.00842\pm0.00021$& $0.00859^{+0.00027}_{-0.00029}$\\ 
\hline\hline
\end{tabular}
\end{center}
\end{table*}

The increasing precision of the measurements and of the theoretical calculations have significantly improved the knowledge of the allowed region for the apex position $(\overline{\rho},\overline{\eta})$. 
Good overall consistency between the various measurements at 95 \% C.L. is observed, thus establishing the CKM mechanism as the dominant source of CP violation in $B$-meson decays. 
 
Furthermore, measurements of \CP-violating quantities from the $B$-factories are now so abundant and precise that the CKM parameters can be constrained by the angles of the Unitarity Triangle alone, as shown in Fig.~\ref{fig:Ut_angVSoth}.
In addition, $(\overline{\rho},\overline{\eta})$ can be determined independently using experimental 
information from \CP-conserving processes, $|V_{ub}|/|V_{cb}|$ 
from semileptonic $B$ decays, $\Delta m_d$ and $\Delta m_s$ from the $B_{d} - \bar{B}_{d}$ and 
$B_{s} - \bar{B}_{s}$ oscillations) and the direct \CP violation measurements in 
the Kaon sector, $\epsilon_K$ (see Fig.~\ref{fig:Ut_angVSoth}). 
Prior to the precise \babar and Belle measurements this was the strategy used to predict the value 
of $\sin2\beta$~\cite{Parodi:1999nr}.

\begin{figure}[h]
\begin{center}
\epsfig{file=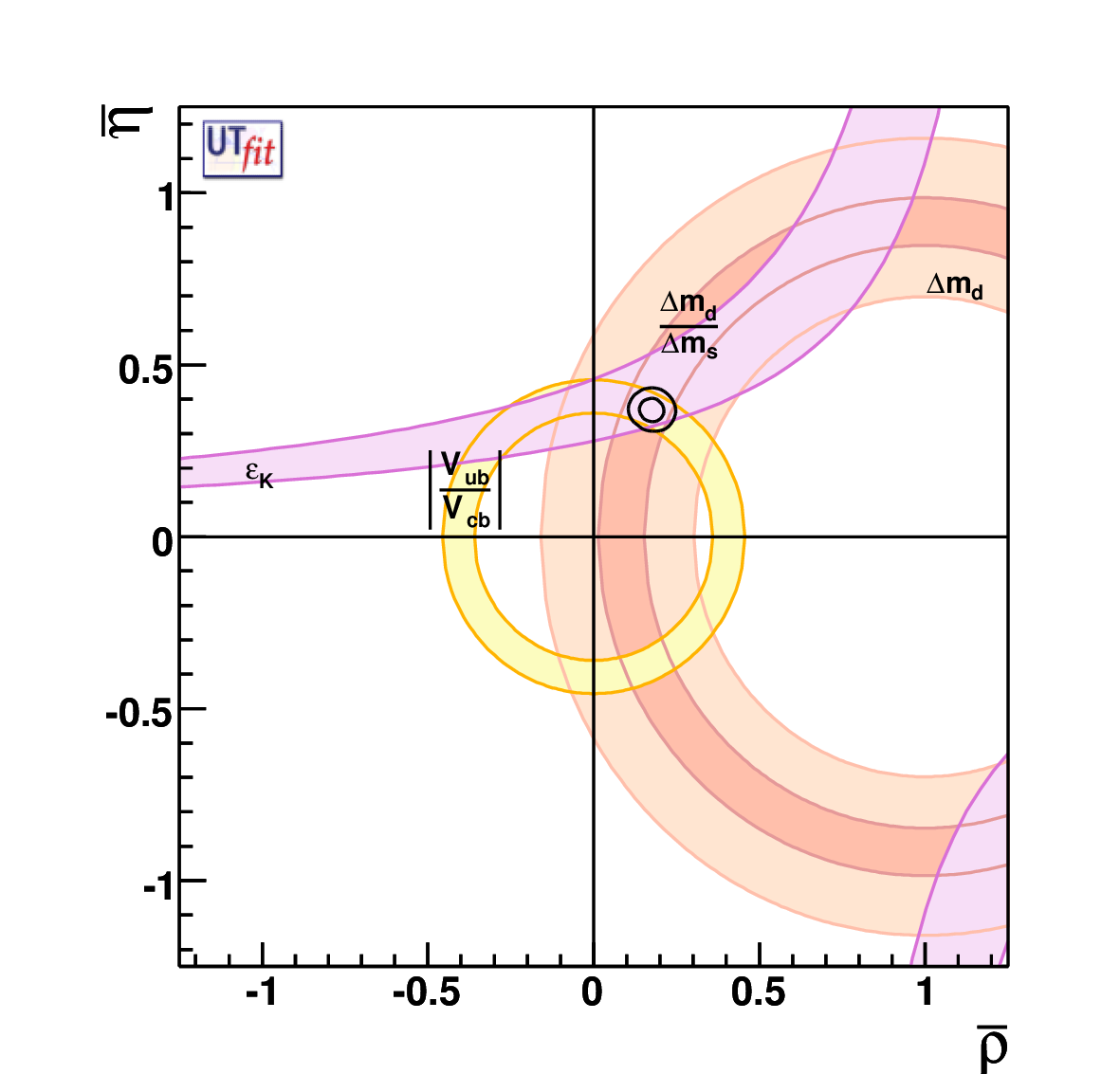,height=5.5cm}
\epsfig{file=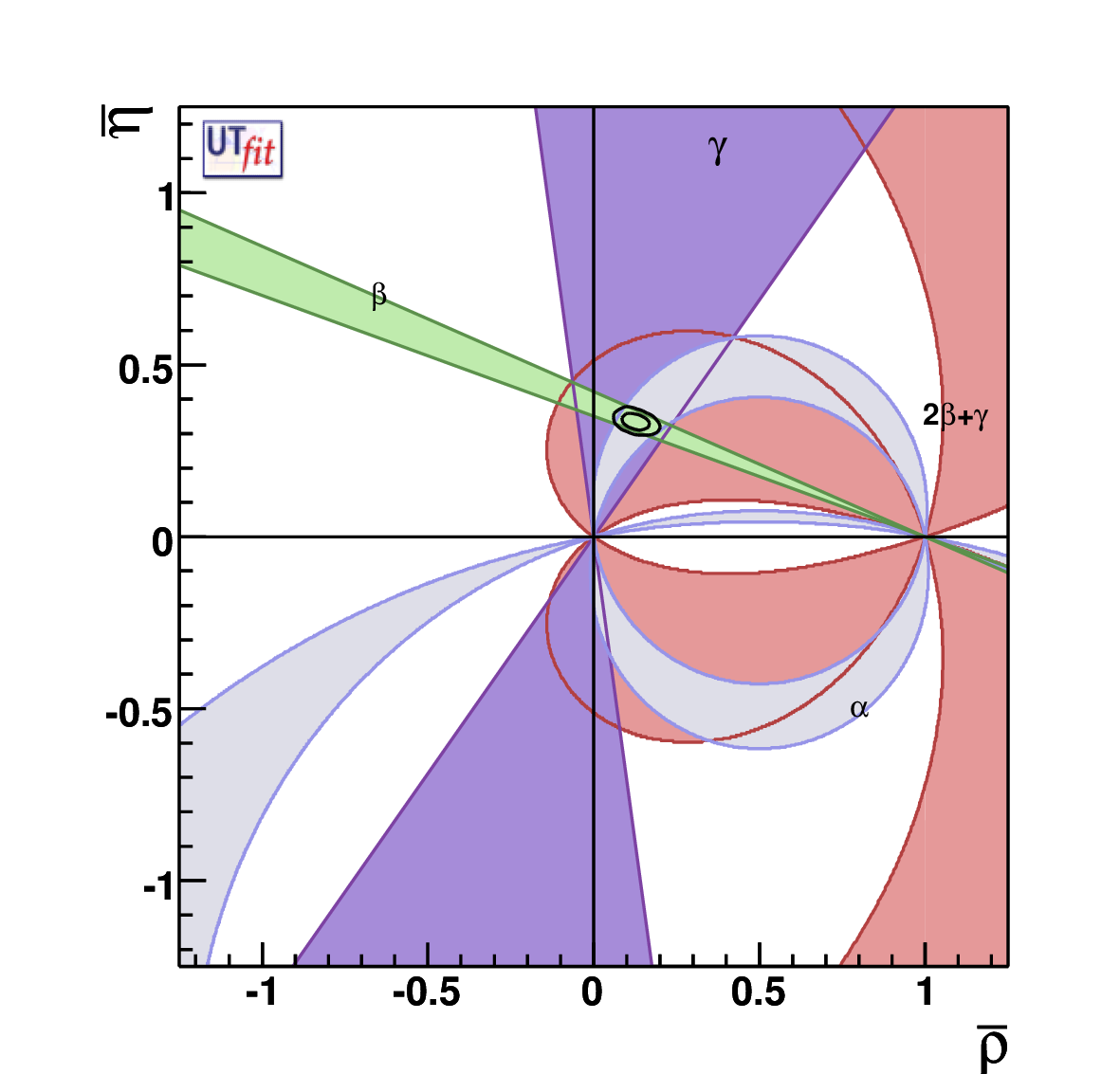,height=5.5cm}
\caption{ Allowed regions for $(\overline{\rho}, \overline{\eta})$, as constrained by the measurement 
of $|V_{ub}|/|V_{cb}|$, $\Delta m_d$, $\Delta m_s$ and $\epsilon_K$ (left)  and of the angles $\alpha$, $\sin 2\beta$, $\gamma$, $2\beta+\gamma$, $\beta$ 
and $\cos 2\beta$ (right) . 
The closed contours indicate the regions of $68\%$ and $95\%$ C.L. for the triangle apex, while the colored zones mark the $95\%$ C.L. for each constraint. \label{fig:Ut_angVSoth}}
\end{center}
\end{figure}

Although the global fits show very good agreement overall, there are some measurements for which the agreement is less convincing.
As described in Sec.~\ref{sec:statmethods}, \utfit\ quantifies the overall agreement of individual measurements with predictions of the global fit 
by means of {\it compatibility plots}.
Such plots for $\alpha$, $\sin 2\beta$, $\gamma$ and $\Delta {m}_s$ are shown 
in Fig. \ref{fig:Ut_various_compatibility}.
The direct measurements for $\alpha$ and $\Delta {m}_s$ are in excellent agreement with the indirect determination from the global fits, although for $\Delta {m}_s$ the effectiveness of the comparison 
is limited by the precision on the theoretical inputs, resulting in sizable uncertainties (compared to the experimental one) for the prediction extracted from the fit. 
The direct measurement of $\gamma$ yields a slightly 
higher value of $(78\pm 12)^{o}$ than the indirect one from the overall fit, $(65\pm 3)^{o}$, though they are compatible within 1$\sigma$.
The measurement of $\sin 2\beta$ based on 
the \CP\ asymmetry in $B^0 \to J/\psi K^0$ is slightly shifted with respect 
to the indirect determination, but compatible to within 2$\sigma$.

It has been observed for several years that the direct measurement of 
$\sin 2\beta$ favors a value of $|V_{ub}|$ that is more compatible with the direct 
determination of $|V_{ub}|$ based on exclusive rather than inclusive charmless semileptonic decays.  The problem is illustrated in  
Fig.~\ref{fig:Ut_Vub_compatibility} and reflects the great challenge that the extraction of $|V_{ub}|$ from charmless semileptonic decays represents. Experimentally, these charmless decays are impacted by very large backgrounds which are difficult to understand in detail and difficult to suppress due to the presence of a neutrino in the final state.  Theoretically,  the required normalization and corrections for hadronic effects based on QCD calculations are dominating the uncertainties. 
The QCD calculations and models are different for the two processes and their uncertainties are impacted by the selection of the experimental data. While there are now several calculations available, it remains very difficult to assess the overall theoretical uncertainties for the extraction of $|V_{ub}|$.  The fact that the current values of $|V_{ub}|$ from exclusive and inclusive decays are only marginally consistent could be taken as an indication that the uncertainties are larger than stated.

\begin{figure}[h]
\begin{center}
\epsfig{file=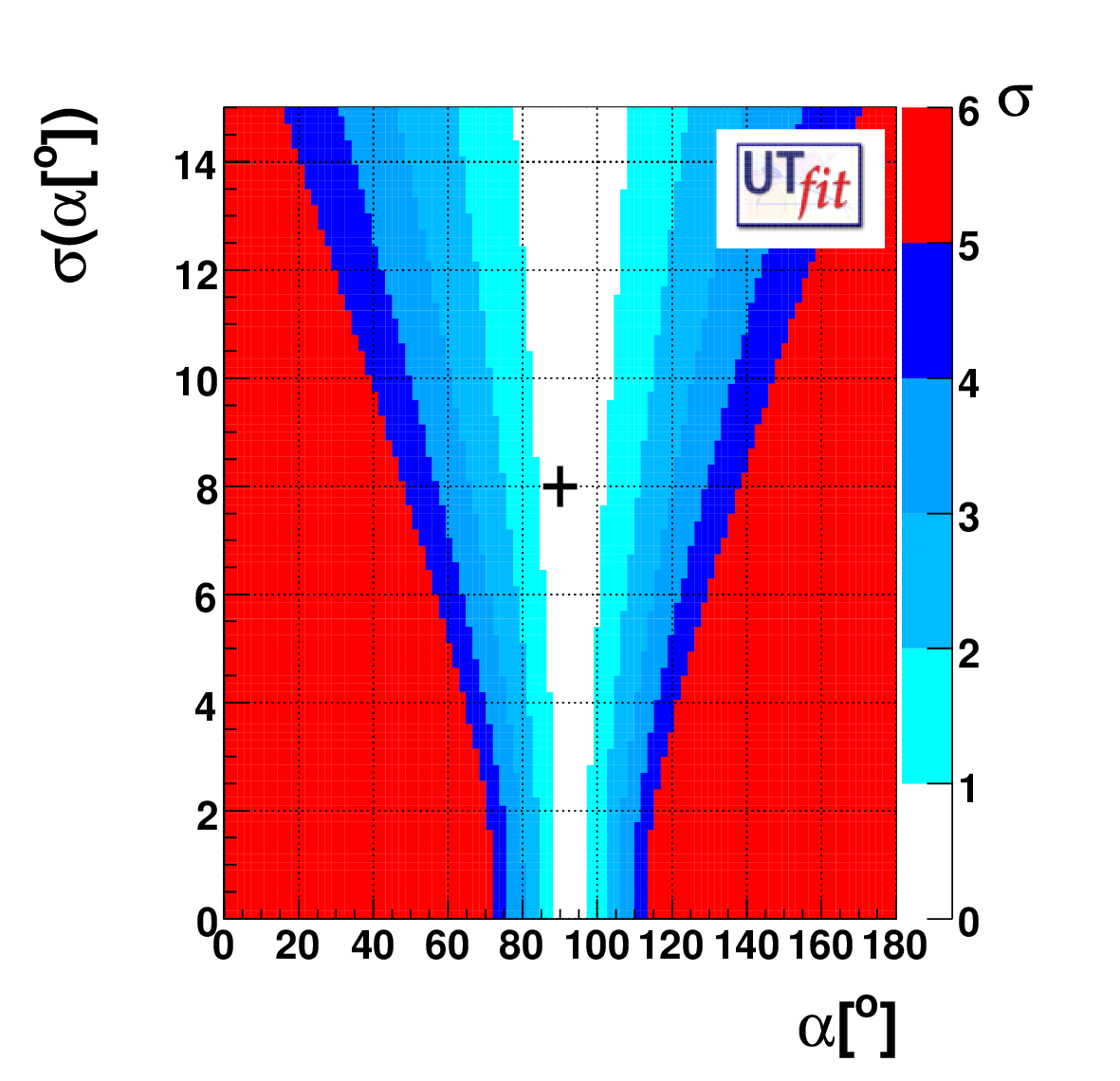,height=5.0cm}
\epsfig{file=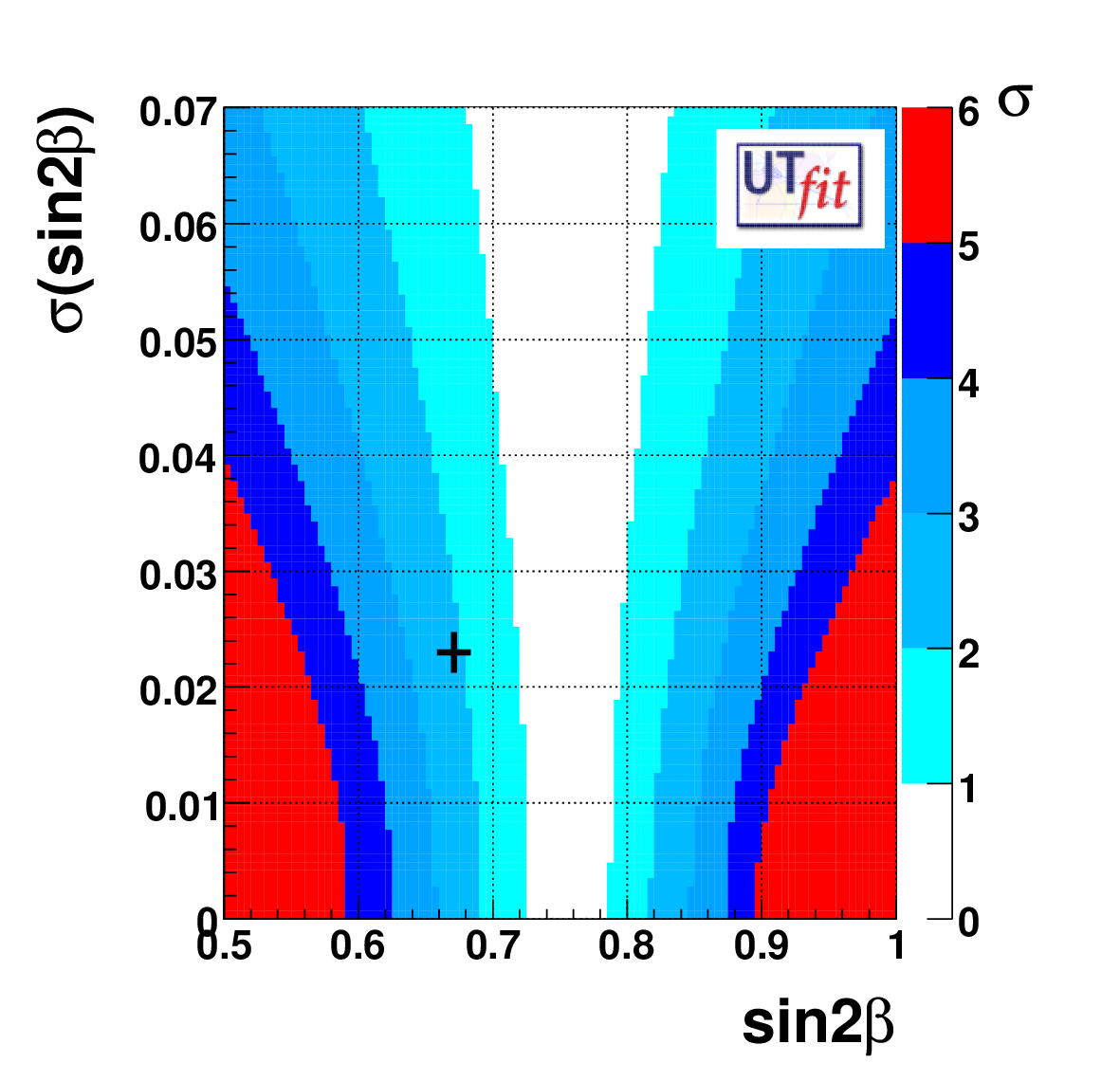,height=5.0cm} \\
\epsfig{file=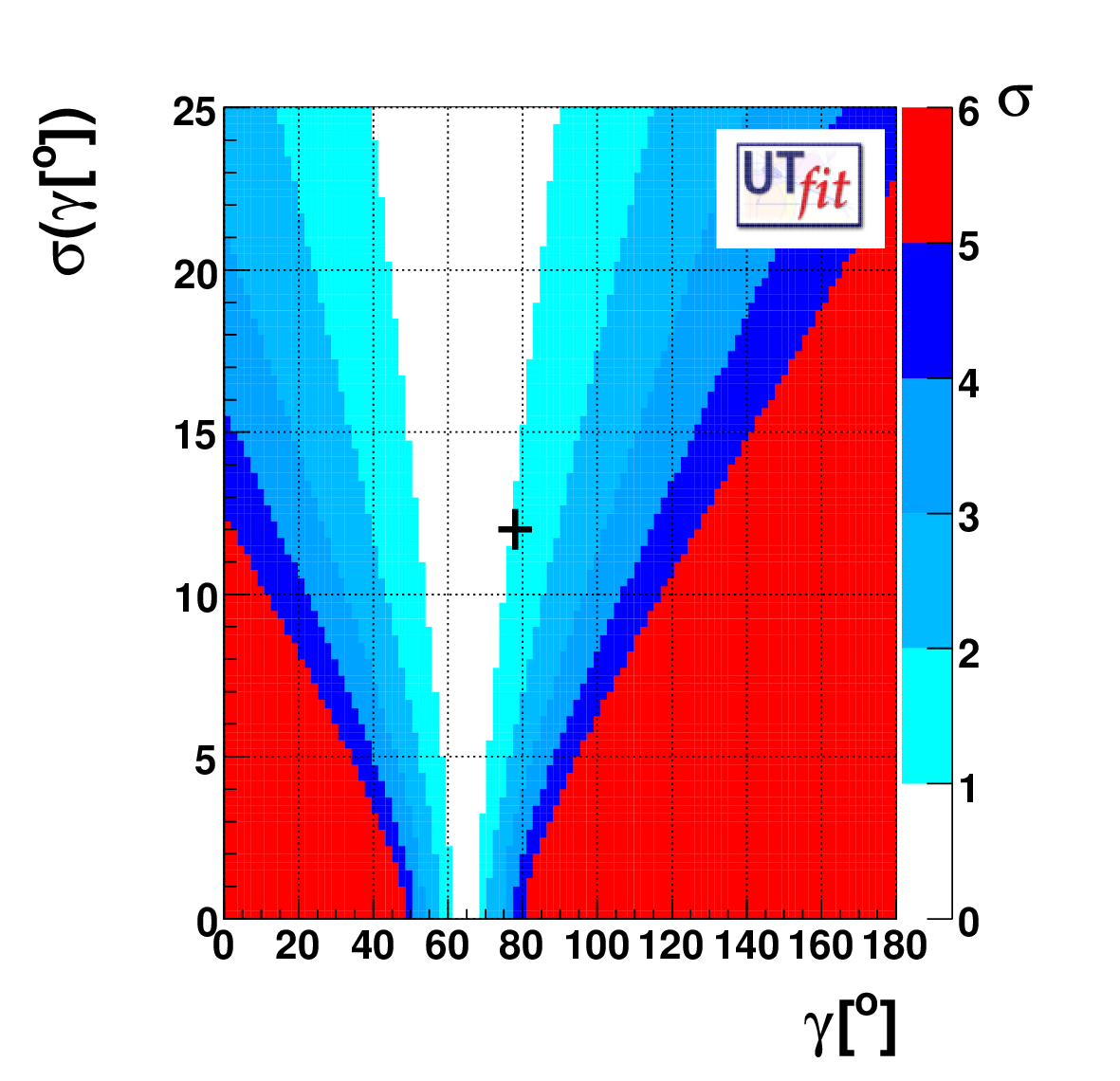,height=5.0cm}
\epsfig{file=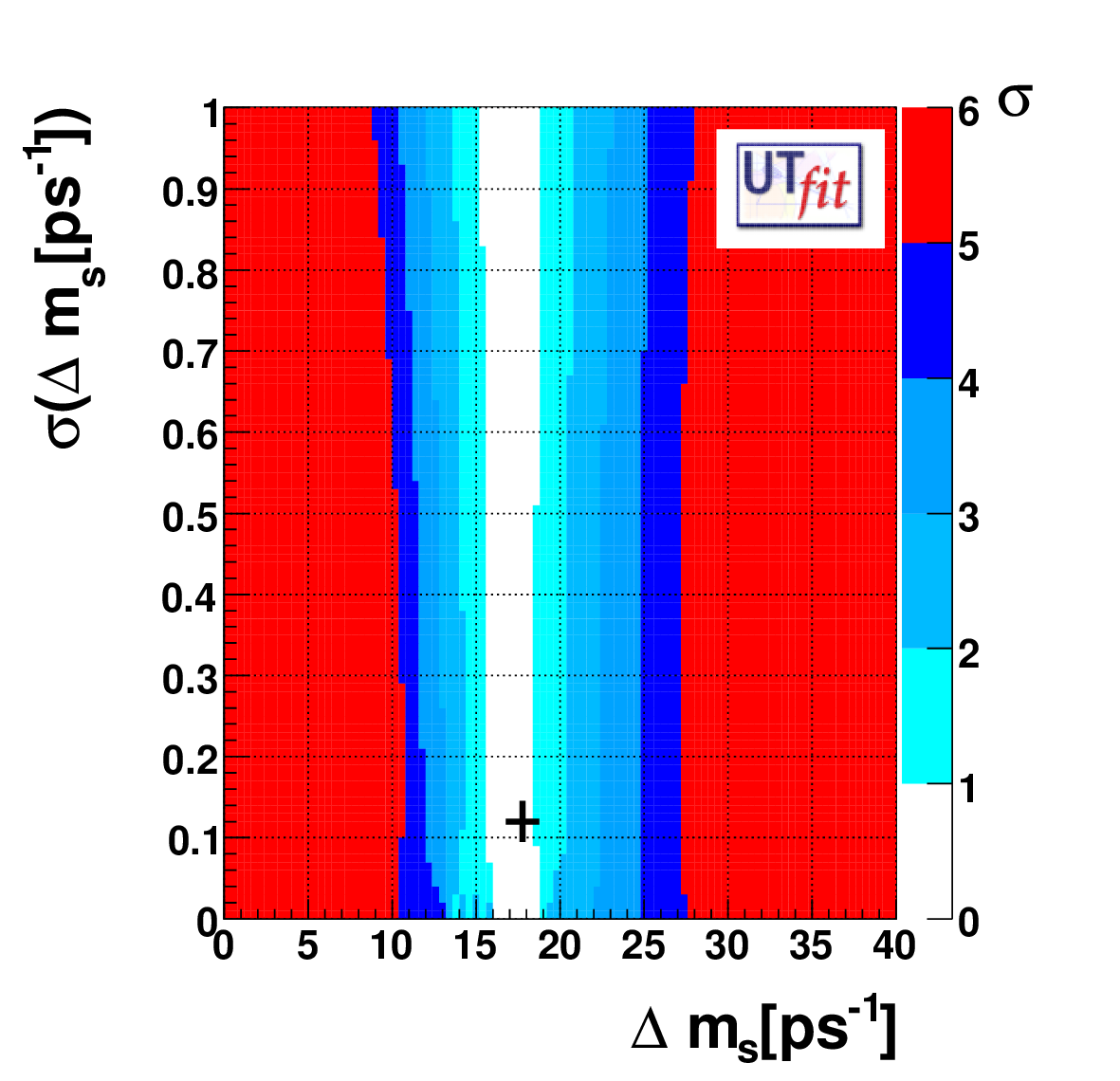,height=5.0cm}
\caption{ 
Compatibility plots for $\alpha$, $\sin 2\beta$ $\sin 2\beta$ from the measurement of the \CP asymmetry in $B^0 \to J/\psi K^0$, $\gamma$ and $\Delta {m}_s$. 
The color code indicates the compatibility between direct 
and indirect determinations, given in terms of standard deviations, as a function of the measured value and the experimental uncertainty. The crosses indicate 
the world averages and errors of the direct measurements.
\label{fig:Ut_various_compatibility}}
\end{center}
\end{figure}

\begin{figure}[h]
\begin{center}ls 
\epsfig{file=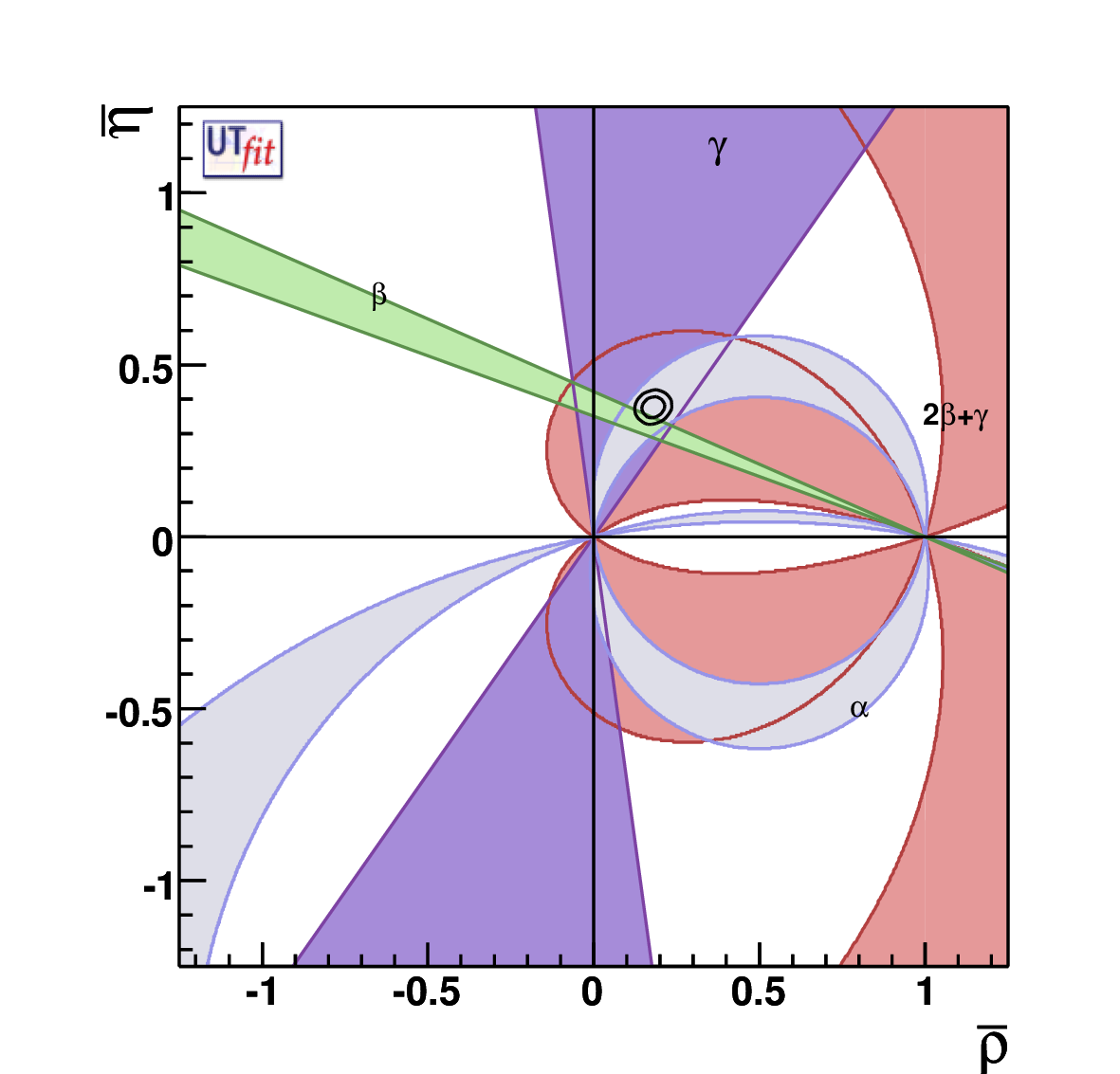,height=5.5cm}
\epsfig{file=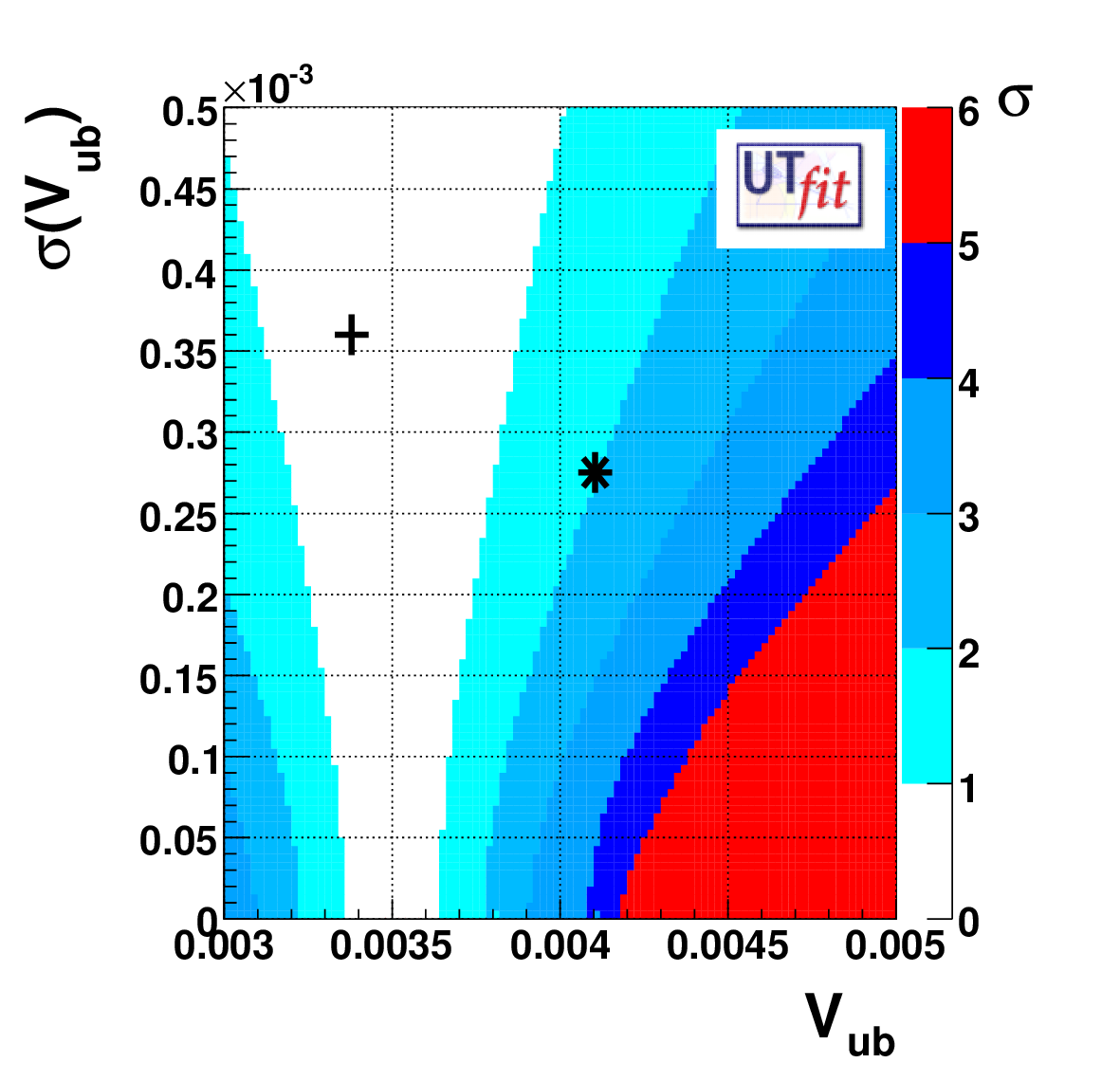,height=5.5cm}
\caption{ 
Left: Allowed regions for $\bar\rho$ and $\bar\eta$ obtained by 
using the measurements of $|V_{ub}|/|V_{cb}|$, $\Delta m_d$, $\Delta m_s$, $\epsilon_K$, 
The colored zones 
indicate the 68$\%$ and 95$\%$ probability regions for the angle measurements, 
which are not included in the fit.  
Right: Compatibility plot for $V_{ub}$. 
The cross and the star indicate the exclusive and inclusive measurements,
respectively.
\label{fig:Ut_Vub_compatibility}}
\end{center}
\end{figure}

Given the present experimental measurements, no significant deviation from the 
CKM picture has been observed.  Of course, this statement does not apply to observables that have no or very small impact on 
$\bar\rho$ and $\bar\eta$ (for instance the $B_s$ mixing phase). 
\subsubsection{Impact of the Uncertainties on Theoretical Quantities}
\label{sec:theo}

Given the abundance of constraints now available for the determination 
of the UT Triangle, $\bar\rho$ and $\bar\eta$, one can perform the global fit without 
the hadronic parameters derived from lattice calculations as input.
In this way, one can quantify the impact that future improvements in the lattice QCD calculation will have on the UT analysis. 

Figure~\ref{fig:Ut_vs_lattice}, shows the 68$\%$ and 95$\%$ probability 
regions for different lattice quantities, obtained from a UT fit using 
the measurements of angles and the constraints from semileptonic 
$B$ decays. 
The relations between observables and theoretical quantities used in this 
fit are obtained assuming the validity of the SM. 
Numerical results are given in Table \ref{tab:Ut_vs_lattice}. 

\begin{figure}[h]
\begin{center}
\epsfig{file=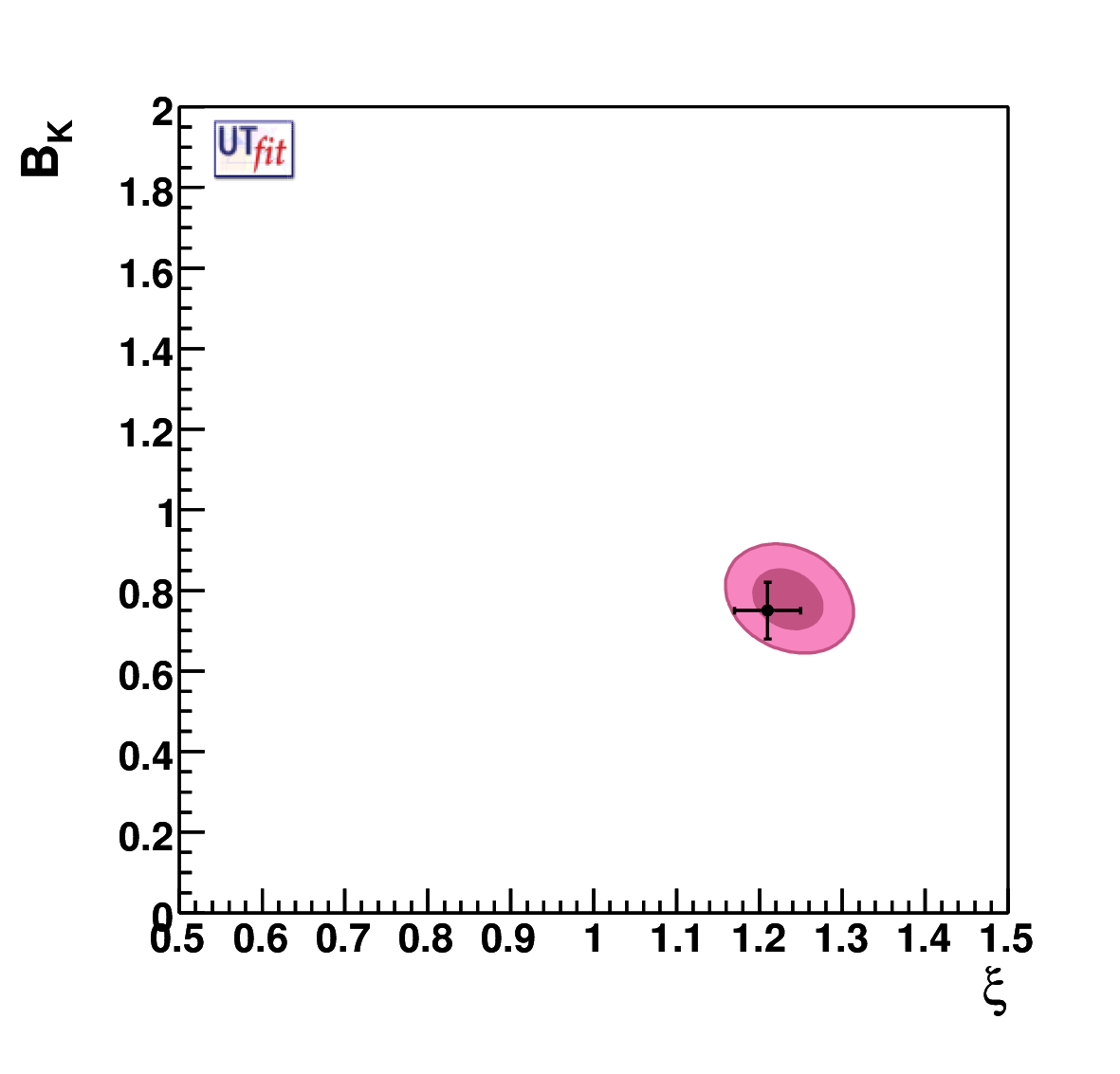,height=5.cm}
\epsfig{file=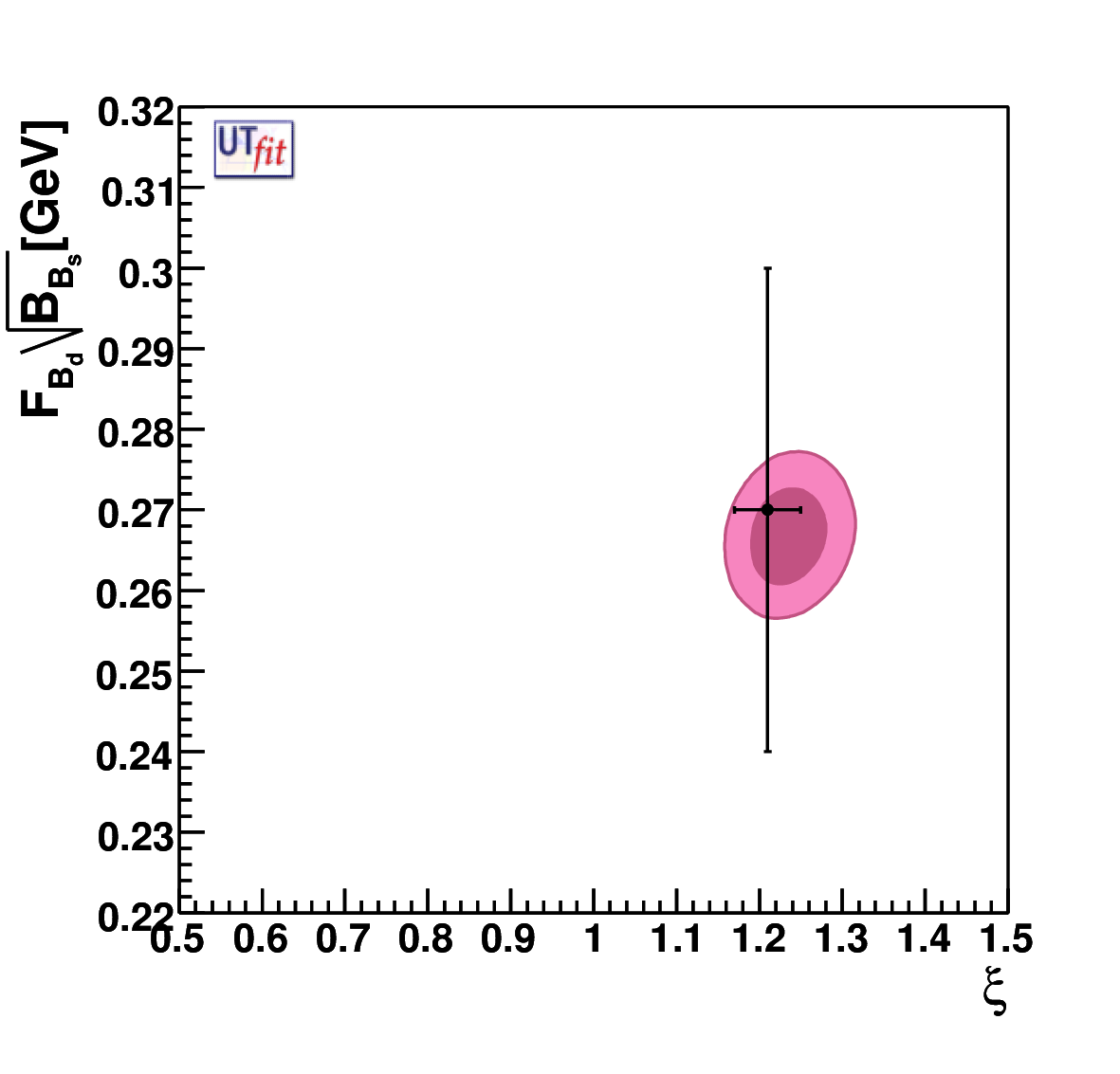,height=5.cm}
\caption{Comparison of the current lattice calculations (data points) 
with prediction of the global fit, left: $\xi$ versus $B_{K}$ and right:
$f_{B_{s}}\sqrt{B_{B_{s}}}$ versus $\xi$. 
The dark and light colored areas show the 68$\%$ and 95$\%$ probability 
regions. 
\label{fig:Ut_vs_lattice}}
\end{center}
\end{figure}

\begin{table*}[h]
\begin{center}
\caption{The values obtained for the theoretical parameters from a UT analysis 
using the angles and $V_{ub}/V_{cb}$ measurements are compared with the results 
of lattice calculations.}
\begin{tabular}{lcc}
\hline\hline
Parameter             & UT (angles+$V_{ub}/V_{cb}$) & Lattice QCD results \\
\hline
$B_{K}$                           & $0.78 \pm 0.05$ & $0.75 \pm 0.07$\\
$f_{B_{s}}\sqrt{B_{B_{s}}}$ [MeV] & $266.8 \pm 4.1$  & $270 \pm 30$   \\
$\xi=\frac{f_{B_{s}}\sqrt{B_{B_{s}}}}{f_{B_{d}}\sqrt{B_{B_{d}}}}$                           
                                  & $1.23 \pm 0.03$ & $1.21 \pm 0.04$ \\
$f_{B_{d}}$ [MeV]                 & $195 \pm 11$     & $200 \pm 20$ \\
\hline\hline
\end{tabular}
\label{tab:Ut_vs_lattice}
\end{center}
\end{table*} 
\subsubsection{Comparison with the Results of CKMfitter}
\label{sec:fitckmfitter}
Extracting Standard Model best values of parameters from the very large number of different measurements is difficult. It is not trivial to combine measurements with very different statistical errors and extract the best the information. However, it is much more difficult to combine measurements with widely different sources and estimations of the systematic errors, in many cases there is need for 
case-to-case judgment and margin for interpretation.

For these reasons, it has been extremely important to have more than one approach
to fits of the UT Triangle. In this section the results obtained by the \utfit\ group are compared with the most recent results of the CKMfitter group as summarized in http://ckmfitter.in2p3.fr/.
The inputs to the two fitting methods are different, and the choice of the lattice parameters differs and experimental inputs are taken from a slightly different sets of measurements, some of them taken from earlier publications.  Nonetheless, the comparison is important, because it shows that different approaches lead to somewhat different results.

\subsubsubsection{Statistical method}

The CKMfitter was developed in parallel to  \utfit\ to perform global UT analyses. 
The most significant difference to the \utfit\ approach is the treatment of non-Gaussian errors.  In particular, the CKMFitter group introduced 
$Range Fit$~\cite{Charles:2004jd}, a special procedure to deal with uncertainties of theoretical predictions.

The CKMfitter method is described briefly as follows. The experimental input is a set of $N_{exp}$ measurements,
$\vec{x}_{exp} $, related to a set of 
theoretical expressions or constraints, $\vec{x}_{theo} $.
The theoretical expressions are model-dependent functions of 
of $N_{mod}$ parameters $\vec{y}_{mod} $.
A subset of $N_{theo}$ parameters in  $\vec{y}_{mod}$  are 
considered fundamental and free parameters of the theory,
{\it e.g.} the four Wolfenstein parameters in the SM or the top quark mass.
These parameters are denoted as $\vec{y}_{theo}$. The remaining 
$N_{QCD} = N_{mod} - N_{theo}$ input parameters, which currently are 
less well known due to the difficulty of computing strong interaction 
effects, {\it e.g.} $f_{b_d}$, $B_d$,... are denoted as 
$\vec{y_{QCD}}$.

The fit is set up to minimize the quantity, 
$\chi^2 = -2 \ln {\cal L}(\vec{y_}{mod})$,
with the likelihood function ${\cal L}(\vec{y}_{mod})$, defined as a product 
two types of contributions,
\begin{eqnarray}
{\cal L}(\vec{y}_{mod}) = {\cal L}_{exp}(\vec{x}_{exp}-\vec{x}_{theo}(\vec{y}_{mod})) \times 
{\cal L}_{theo}(\vec{y}_{QCD}).
\label{eq_ckm}
\end{eqnarray}
${\cal L}_{exp}$ depends on the experimental 
measurements $\vec{x}_{exp}$, with errors that are Gaussian distributed in general
(and correlations, if known, are taken into account), 
and their theoretical predictions $\vec{x}_{theo}$, which are functions of 
the model parameters $\vec{y}_{mod}$. 
In the case of a non-Gaussian experimental errors, the exact 
description of the associated likelihood is used in the fit.
${\cal L}_{theo}$
describes the imperfect knowledge of the QCD parameters $\vec{y}_{QCD} \in {\vec{y}_{mod}}$,
where the theoretical uncertainties $\vec{\sigma}_{syst}$ are considered
to be bound by a range, 
$[\vec{y}_{QCD} - \vec{\sigma}_{syst}, \vec{y}_{QCD} + \vec{\sigma}_{syst}]$.
In $Range Fit$ the theoretical likelihood functions ${\cal L}_{theo}(i)$ 
do not contribute to the $\chi^2$ of the fit, as long as the $\vec{y}_{QCD}$ 
values are within their range.  With these constraints, all results should be
understood as valid only if the allowed ranges contain the true values of the  
$\vec{y}_{mod}$.

The minimization is performed in two steps. First, the global minimum, 
$\chi^2_{min, global}$, is determined with respect to all $N_{mod}$ 
parameters. Due to the systematic uncertainties from experiment and theory,
this minimum does in general not correspond to a unique $\vec{y}_{mod}$ .
Second, a selected subspace of the parameter
space, {\it e.g.} $a = ({\bar{\rho}, \bar{\eta}})$ is scanned, to determine
the local minimum $\chi^2$, $\chi^2_{min, local}(a)$, for each fixed
point on a grid in the parameter space $a$, with respect to the 
remaining parameters. The offset-corrected $\chi^2$ is calculated
as, 
$\Delta \chi^2(a) = \chi^2_{min, local}(a) - \chi^2_{min, global}$,
where its minimum is equal to zero by construction.

Finally, a confidence level (C.L.) for $a$ is obtained, assuming Gaussian distributions, by using the cumulative \chisq\ distribution:
\begin{align}
1- \rm CL & = \rm Prob(\Delta \chi^2(a), N_{dof})\\
& = \frac{1}{\sqrt{2^{N_{dof}}} \Gamma(N_{dof}/2) } 
\int_{\chi^2(y_{mod})}^{\infty} e^{-t/2}t^{N_{dof}/2-1} dt.
\end{align}

\subsubsubsection{Inputs}

The inputs to the fits performed by the CKMfitter group differ slightly from the results of this report and different choices of parameters estimated with lattice QCD calculations have been made. The latter difference is mostly due to the difference in the treatment of systematic errors. These differences are presented in Tables~\ref{tab:inputs_ckmfitter} and~\ref{tab:lattice_inputs_ckmfitter} to be compared with   Tables~\ref{tab:inputs_utfit} and~\ref{tab:lattice_inputs_utfit}. Identical input values are not included in these tables.
\begin{table*}[htb]
\caption{Most relevant experimental inputs used by CKMfitter 
for the global UT fit that are different from those used by UT fit. 
The numbers marked in bold are theoretical
uncertainties treated using $R$fit (flat likelihood).}
\label{tab:inputs_ckmfitter}
\begin{center}
\begin{tabular}{lccc}
\hline\hline
Input & Source & Value & Reference \\
\hline
$|V_{ud}|$ & Nuclear decays & $0.97418 \pm 0.00026$ & \cite{Towner:2007np} \\
$|V_{us}|$ & SL Kaon decays & $0.2246 \pm 0.0012$ & \cite{Antonelli:2008jg} \\
$|V_{cb}|$ & SL charmed $B$ decays & 
$(40.59 \pm 0.38 \pm {\bf 0.58}) \times 10^{-3}$ & \cite{Barberio:2008fa} 
\footnote{
For the average of inclusive measurements, the result based on BLNP was chosen.  
The  result based on DGE is very similar, but there is no simple method 
for combining two results based on the same measurements.
The theoretical error on the average for the inclusive decays is obtained by 
adding linearly the contributions from weak annihilation, subleading
shape functions, and the HQE uncertainty on $m_b$.
For exclusive decays, only the branching fractions measured for
$B \to \pi l \nu$ are used. The results obtained from two 
unquenched Lattice calculations and the LCSR calculation for 
the form factor are averaged in such a way that 
the smallest theoretical error is kept.
Also, when averaging the inclusive and exclusive result, the smallest theoretical error
is taken as the overall theoretical uncertainty.}\\
$|V_{ub}|$ & SL charmless $B$ decays & 
$(3.87 \pm 0.09 \pm {\bf 0.46}) \times 10^{-3}$ & \cite{Barberio:2008fa} \\
${\cal B}(B^+ \to \tau^+ \nu)$ & Leptonic $B$ decays & 
$(1.73 \pm 0.35) \times 10^{-4}$ &Tab.~\ref{tab:BtoEllNuBF} combined with \cite{:2008gx} \\
\hline\hline
\end{tabular}
\end{center}
\end{table*} 
\begin{table*}[htb]
\caption{Phenomenological inputs from Lattice QCD calculations as adopted by the CKMfitter group. The errors treated  
according to the $R$fit (see text) prescription are highlighted in bold.}
\label{tab:lattice_inputs_ckmfitter}
\begin{center}
\begin{tabular}{lc}
\hline\hline
$f_{B_s}$ (MeV)                     &  $228 \pm 3 \pm {\bf 17}$    \\
$\hat{B}_{B_s}$               &  $1.196 \pm 0.008 \pm {\bf 0.023}$ \\
$f_{B_s}/f_{B_d}$              & $1.23 \pm 0.03 \pm {\bf 0.05}$ \\
$\hat{B}_{B_s}/\hat{B}_{B_d}$  &  $1.05 \pm 0.02 \pm {\bf 0.05}$\\
$B_K$                        & $0.721 \pm 0.005 \pm {\bf 0.040}$\\
\hline\hline
\end{tabular}
\end{center}
\end{table*}
The sources of the experimental inputs are given in Tables~\ref{tab:inputs_ckmfitter}, the 
different choices of lattice parameter are justified in the following.

First of all, only unquenched lattice calculations with 2 or 2+1 
dynamical fermions, published in journals or proceedings are taken into account. 
The Gaussian and flat components of the errors are separated and the latter 
is treated according to $R$fit prescription~\cite{Charles:2004jd}. 

The Gaussian errors comprise purely statistical errors  as well as
systematic uncertainties that are expected to also have normal error distributions
({\it e.g.}interpolation errors). 
The remaining systematic uncertainties are handled as $R$fit errors. 
If there are several error sources in the $R$fit category, they are added linearly.

If $R$fit is taken {\it stricto~sensu} and the individual likelihood
functions are combined by multiplication, the resulting overall uncertainty might
be underestimated.
This effect is corrected for by adopting the following procedure:
first, the likelihood functions for the Gaussian uncertainties are combined; 
next this combination is assigned 
the smallest of the individual $R$fit errors. The underlying idea 
is as follows: 
The estimated error should not be smaller than the best of all estimates, 
but this best estimate should not be impacted by less precise methods,
as would be the case if one took the dispersion of the 
individual central values as a guess of the combined theoretical uncertainty.
All this underlines the fact that theoretical uncertainties are often ill-defined, 
and procedures to combine such errors should be judged critically. 
The CKMFitter approach is only one among several the alternatives that 
can be found in the literature. 

\subsubsubsection{Results of CKM Fitter}

Figure~\ref{fig:global_ckmfitter} displays the result of global fits performed with the CKMfitter, together with the 96\% C.L. contours of the individual constraints.   For comparison with UT fit, the  results  are listed in Tables~\ref{tab:utfitresults} and~\ref{tab:utfitresults_meas}).
The two global fit procedures give comparable results, although they arrive at somewhat different contours for the individual constraints. The results are an excellent proof of 
robustness of the fit methods, indicating that at present precision the different choices for the treatment of errors do not impact the conclusions significantly.

\begin{figure}[ht]
\begin{center}
\epsfig{file=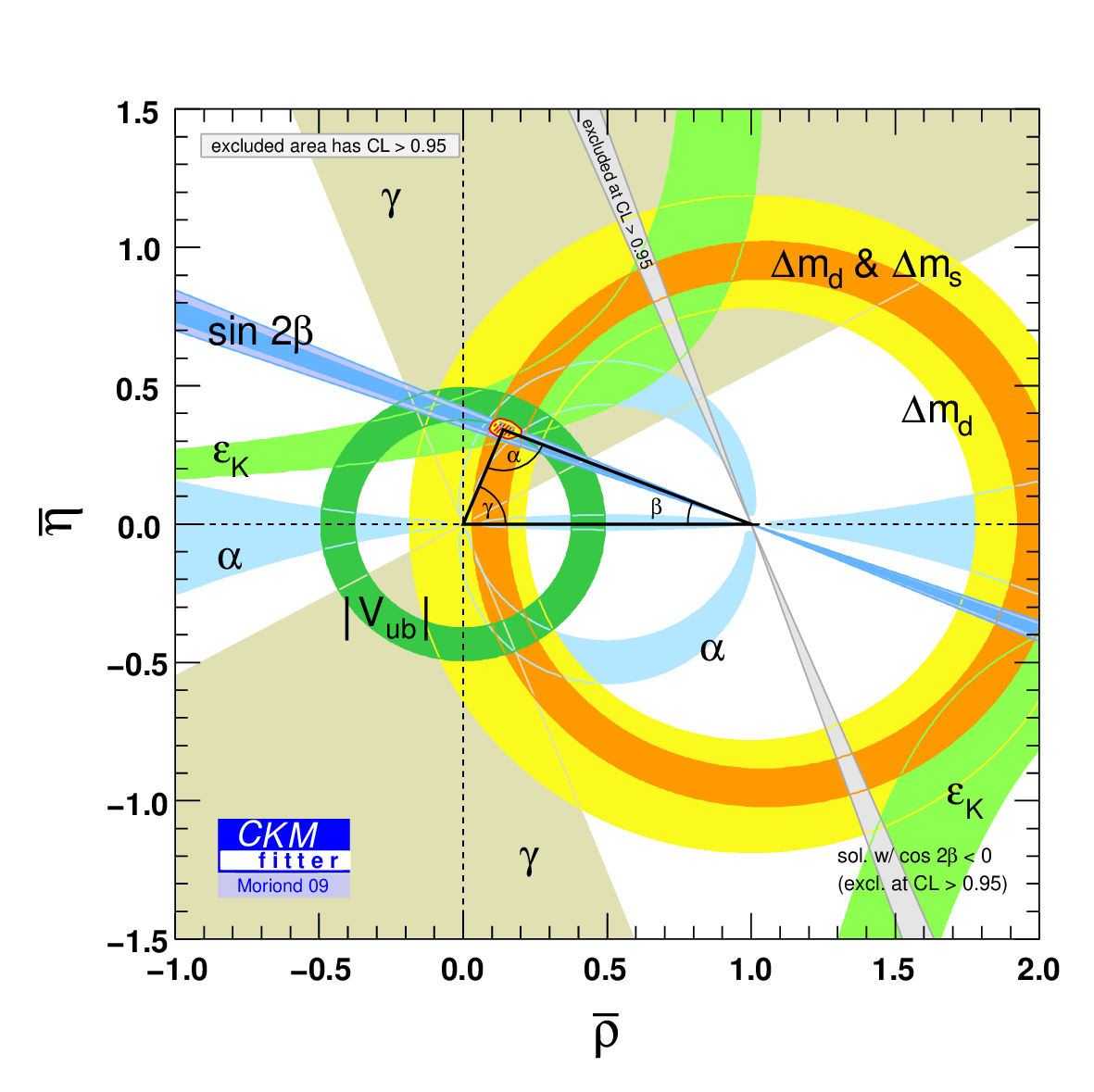,height=5.cm}
\end{center}
\caption{Individual and global constraints in the ($\bar{\rho},\bar{\eta}$) plane obtained by the CKMfitter group.
The shaded areas indicate the individual constraints at 95\% CL. The contours of the overall constraints defining the apex of the UT triangle correspond to 68\% and 95\% C.L. .
}
\label{fig:global_ckmfitter}
\end{figure}
\subsection{CKM angles in the presence of New Physics }
\label{sect:gfitsNP}
\subsubsection{Model independent constraints on New Physics from global fits}
\label{sect:gfitsMI}
The Standard Model (SM) of electroweak and strong interactions works
beautifully up to the highest energies presently explored at
colliders. However, there are several indications that it must be
embedded as an effective theory into a more complete model that
should, among other things, contain gravity, allow for gauge coupling
unification and provide a dark matter candidate and an efficient
mechanism for baryogenesis. As discussed in Sec.~\ref{sect:BSMprimer}, 
this effective theory can be described by a Lagrangian of the form
\begin{displaymath}
  \mathcal{L}(M_W)=\Lambda^2 H^\dagger H + \mathcal{L}_{\mathrm{SM}} +
  \frac{1}{\Lambda} \mathcal{L}^5 +
  \frac{1}{\Lambda^2} \mathcal{L}^6 + \ldots\,,
\end{displaymath}
where the logarithmic dependence on the cutoff $\Lambda$ has been
neglected. Barring the possibility of a conspiracy between physics at
scales below and above $\Lambda$ to give an electroweak symmetry
breaking scale $M_w \ll \Lambda$, we assume that the cutoff lies close
to $M_w$. Then the power suppression of higher dimensional operators
is not too severe for $\mathcal{L}^{5,6}$ to produce sizable effects
in low-energy processes, provided that they do not compete with
tree-level SM contributions.
Therefore, we should look for new physics effects in quantities that
are zero at the tree level in the SM and are finite and calculable
at the quantum level. Within the SM, such quantities fall in two
categories: i) electroweak precision observables (protected by the
electroweak symmetry) and ii) Flavor Changing Neutral Currents (FCNC)
(protected by the GIM mechanism).
In the SM, all FCNC and CP violating processes are computable in terms
of quark masses and of the elements of the Cabibbo-Kobayashi-Maskawa (CKM)
matrix~\cite{Cabibbo:1963yz, Kobayashi:1973fv}.
This implies very strong correlations among observables in the
flavor sector. NP contributions, or equivalently the operators in
$\mathcal{L}^{5,6}$, violate in general these correlations, so that NP
can be strongly constrained by combining all the available
experimental information on flavor and CP violation. 

A very useful tool to combine the available experimental data in the
quark sector is the Unitarity Triangle (UT)
analysis~\cite{Bona:2005vz,Charles:2004jd}.
Thanks to the measurements of the Unitarity Triangle (UT) angles recently
performed at $B$ factories, the UT fit is over-constrained.
Therefore, it has become possible to add NP contributions to all
quantities entering the UT analysis and to perform a combined fit of
both NP and SM parameters. In general, NP models introduce a
large number of new parameters: flavor changing couplings, short
distance coefficients and matrix elements of new local operators. The
specific list and the actual values of these parameters can only be
determined within a given model. Nevertheless, each of the
meson-antimeson mixing processes is described by a single amplitude
and can be parametrized, without loss of generality, in terms of two
parameters, which quantify the difference between the full amplitude
and the SM one.
Thus, for instance, in the case of $B^0_q-\bar{B}^0_q$ mixing
we define~\cite{Bona:2007vi}:

\begin{equation} 
  C_{B_q} \, e^{2 i \phi_{B_q}} = \frac{\langle
    B^0_q|H_\mathrm{eff}^\mathrm{full}|\bar{B}^0_q\rangle} {\langle
    B^0_q|H_\mathrm{eff}^\mathrm{SM}|\bar{B}^0_q\rangle}\,;
  \hspace{0.4cm}
  C_{\Delta m_K} = \frac{\mathrm{Re}[\langle
    K^0|H_{\mathrm{eff}}^{\mathrm{full}}|\bar{K}^0\rangle]}
  {\mathrm{Re}[\langle
    K^0|H_{\mathrm{eff}}^{\mathrm{SM}}|\bar{K}^0\rangle]}\,;
  \hspace{0.4cm}
  C_{\epsilon_K} = \frac{\mathrm{Im}[\langle
    K^0|H_{\mathrm{eff}}^{\mathrm{full}}|\bar{K}^0\rangle]}
  {\mathrm{Im}[\langle
    K^0|H_{\mathrm{eff}}^{\mathrm{SM}}|\bar{K}^0\rangle]}
  \label{eq:paranp}
\end{equation} 

\noindent
where $q=d,s$, 
$H_\mathrm{eff}^\mathrm{SM}$ includes only the SM box diagrams, while
$H_\mathrm{eff}^\mathrm{full}$ includes also the NP contributions.
For the $K^0-\bar{K}^0$ mixing, we find it convenient
to introduce two parameters related to the real and 
imaginary parts of the total amplitude to the SM one.
In summary, all NP effects in $\Delta F=2$ transitions are
parametrized in terms of six real quantities, $C_{\epsilon_K}$, 
$C_{\Delta m_K}$, $C_{B_d}$, $\phi_{B_d}$, $C_{B_s}$
and $\phi_{B_s}$~\cite{Ciuchini:2000de}.

To further improve the NP parameter determination in the $B_s$ sector,
mainly unconstrained in the classical UT analysis, we include
in the NP fit recent results from the Tevatron.
We use the following experimental inputs: the semileptonic asymmetry
in $B_s$ decays $A_\mathrm{SL}^{s}$, the dimuon charge
asymmetry $A_\mathrm{SL}^{\mu\mu}$ from CDF and 
D0, the measurement of the $B_s$ lifetime from
flavor-specific final states, and the
two-dimensional likelihood ratio for $\Delta \Gamma_s$ and
$\phi_s=2(\beta_s-\phi_{B_s})$ from the time-dependent tagged angular
analysis of $B_s\to J/\psi \phi$ decays by CDF and
D0\footnote{We use the latest D0 results without assumptions on the strong phases}. 
The new input parameters used in our analysis are given in
Ref.~\cite{Bona:2007vi} and continuously updated in~\cite{utfitorg}.
The relevant NLO formulas for $\Delta \Gamma_s$
and for the semileptonic asymmetries in the presence of NP
have been discussed in Refs.~\cite{Bona:2005eu,Bona:2006sa,Bona:2007vi}.

\begin{figure*}[htb!]
  \begin{tabular}{cccc}
    \includegraphics[width=0.25\textwidth]{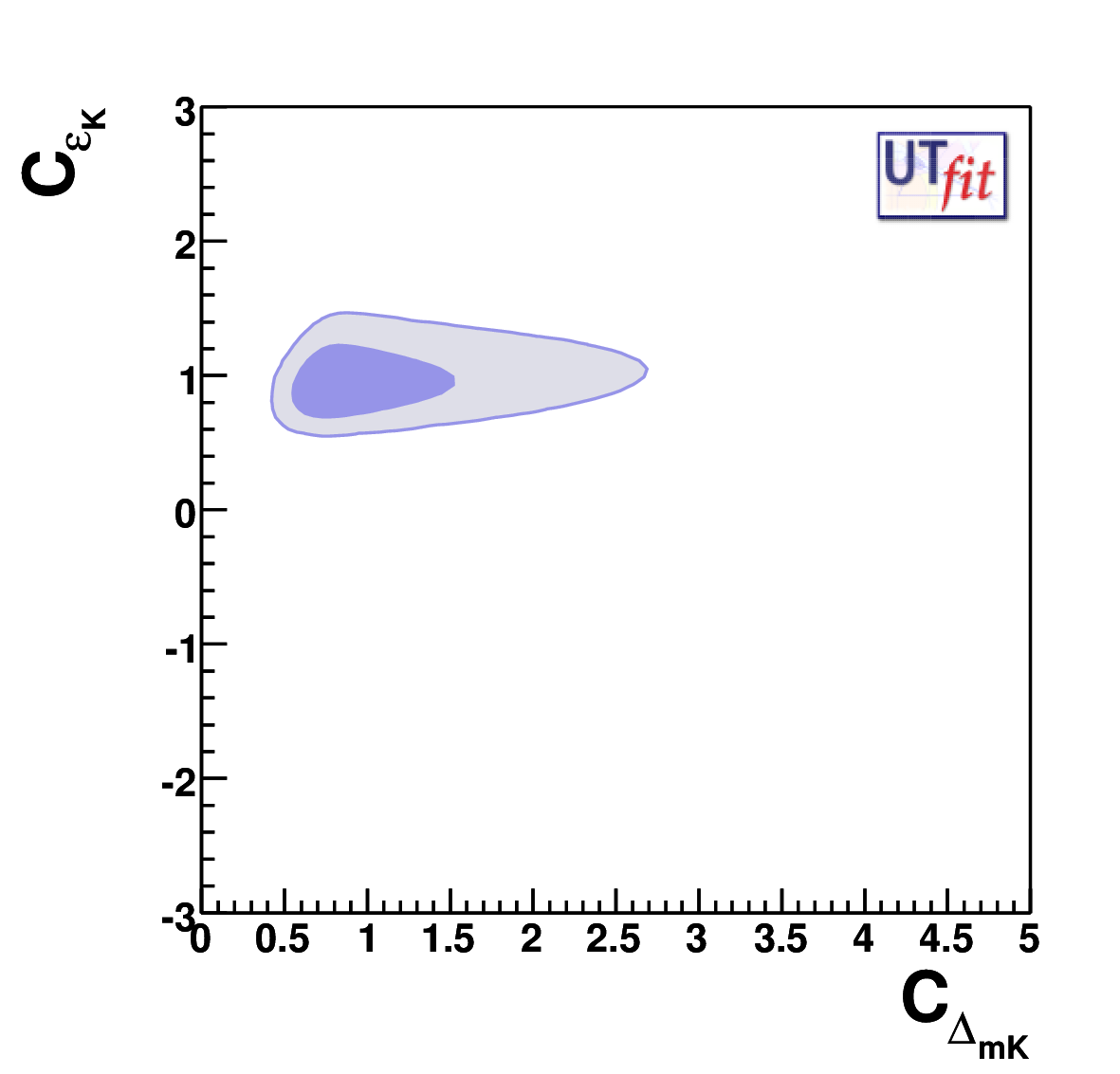} &
    \hspace*{-.5cm}
    \includegraphics[width=0.25\textwidth]{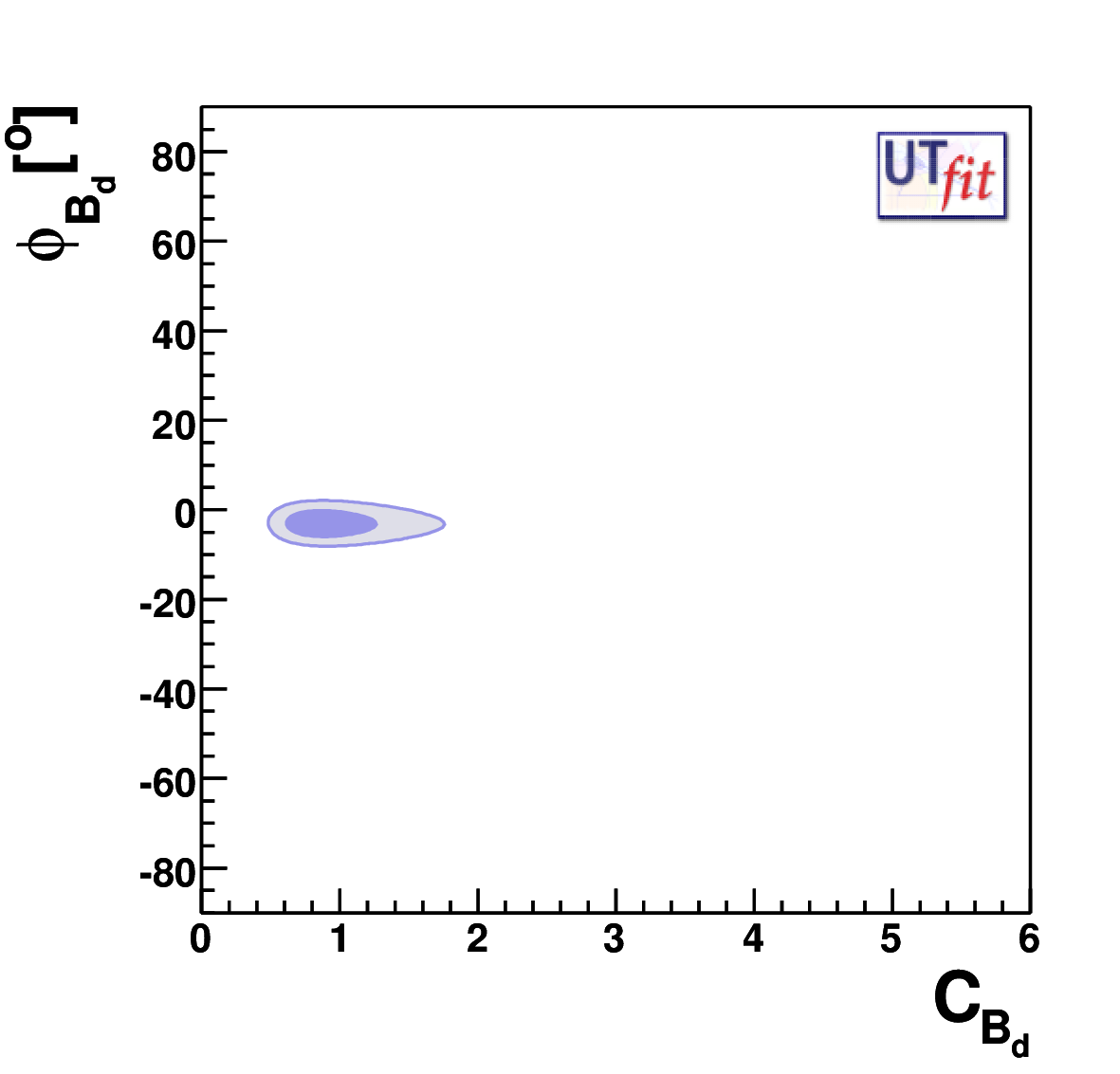} &
    \hspace*{-.5cm}
    \includegraphics[width=0.25\textwidth]{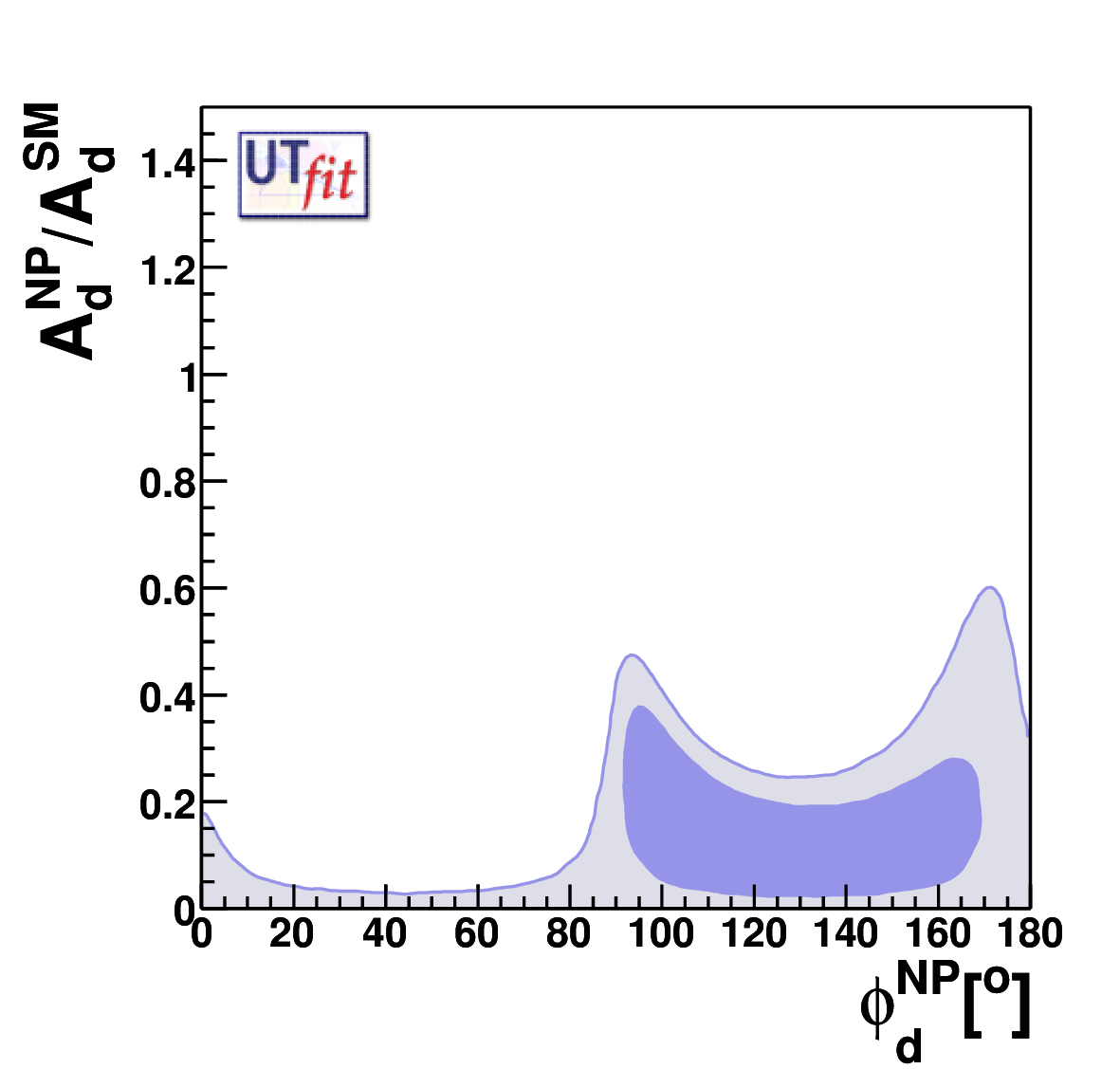} &
    \hspace*{-.5cm}
    \includegraphics[width=0.25\textwidth]{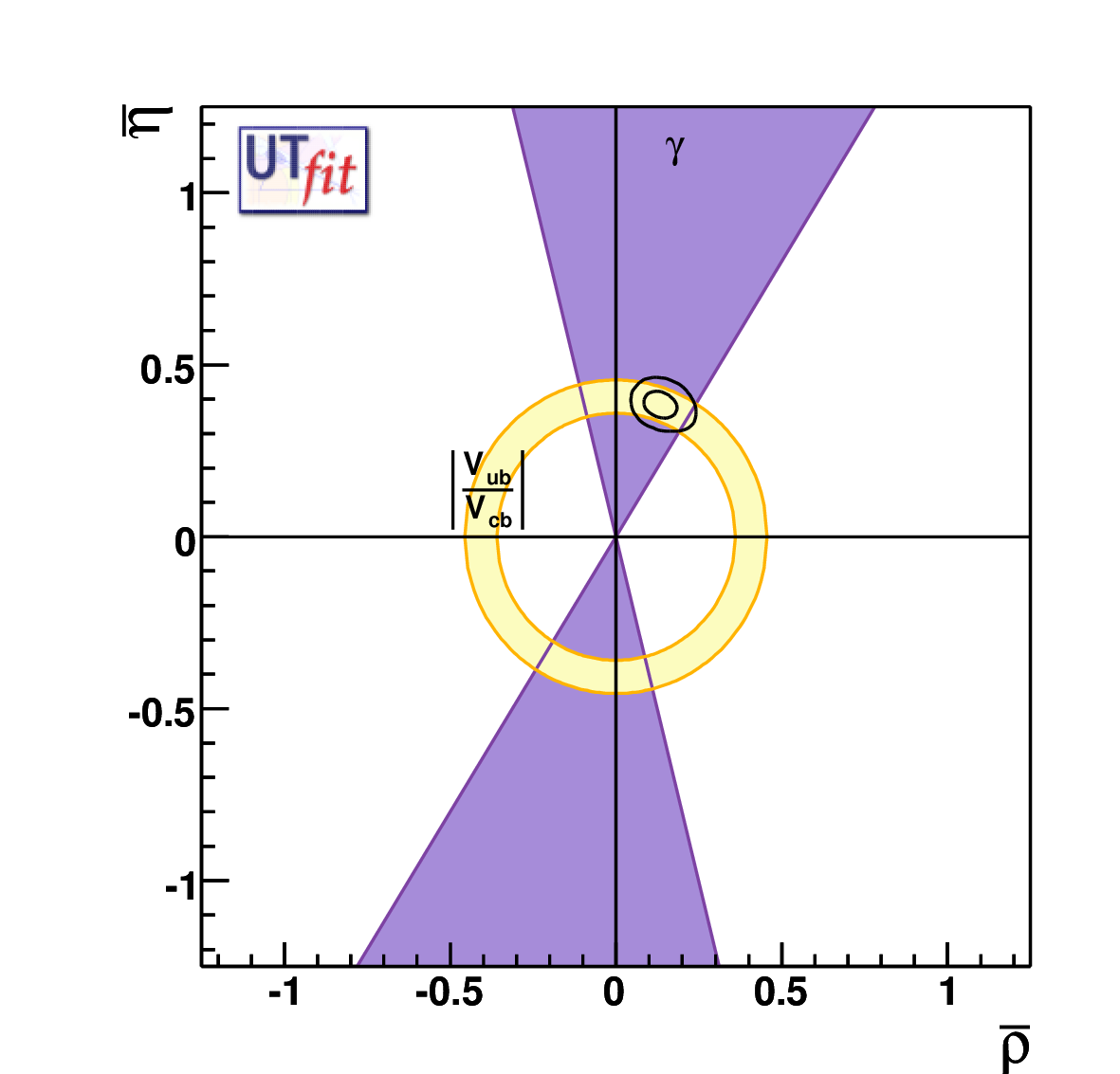} \\
\end{tabular}
  \caption{From left to right, $C_{\epsilon_K}$ vs. $C_{\Delta m_K}$,
    $\phi_{B_d}$ vs. $C_{B_d}$, $(A_\mathrm{NP}/A_\mathrm{SM})$ vs.
    $\phi_\mathrm{NP}$ for NP in the $B_d$ sector and the resulting
    selected region on the $\bar\rho-\bar\eta$ plane obtained
    from the NP analysis~\cite{Bona:2007vi}.}
  \label{fig:NP}
\end{figure*}

\begin{figure*}[htb!]
  \begin{tabular}{cccc}
    \includegraphics[width=0.25\textwidth]{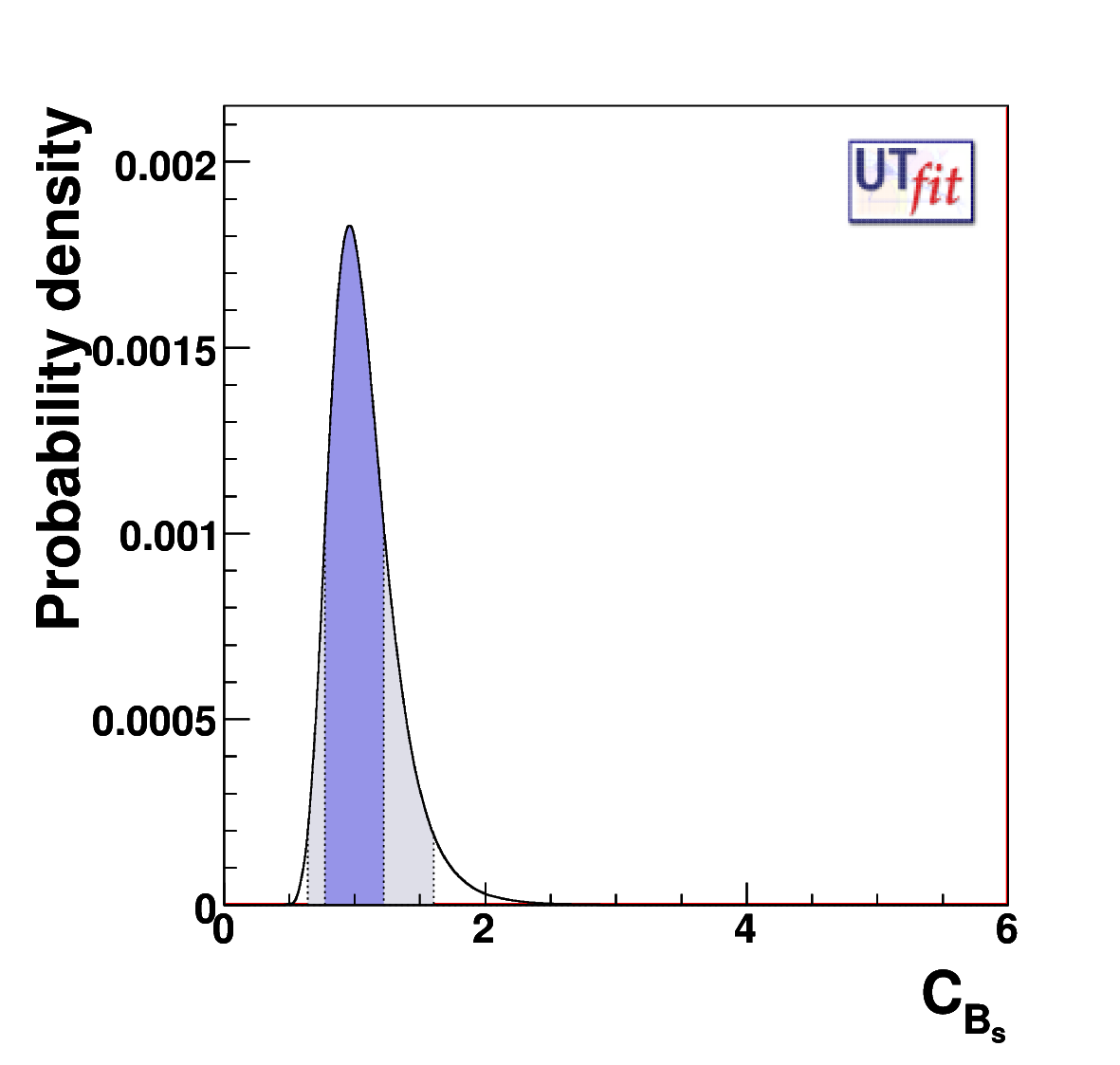} &
    \hspace*{-.5cm}
    \includegraphics[width=0.25\textwidth]{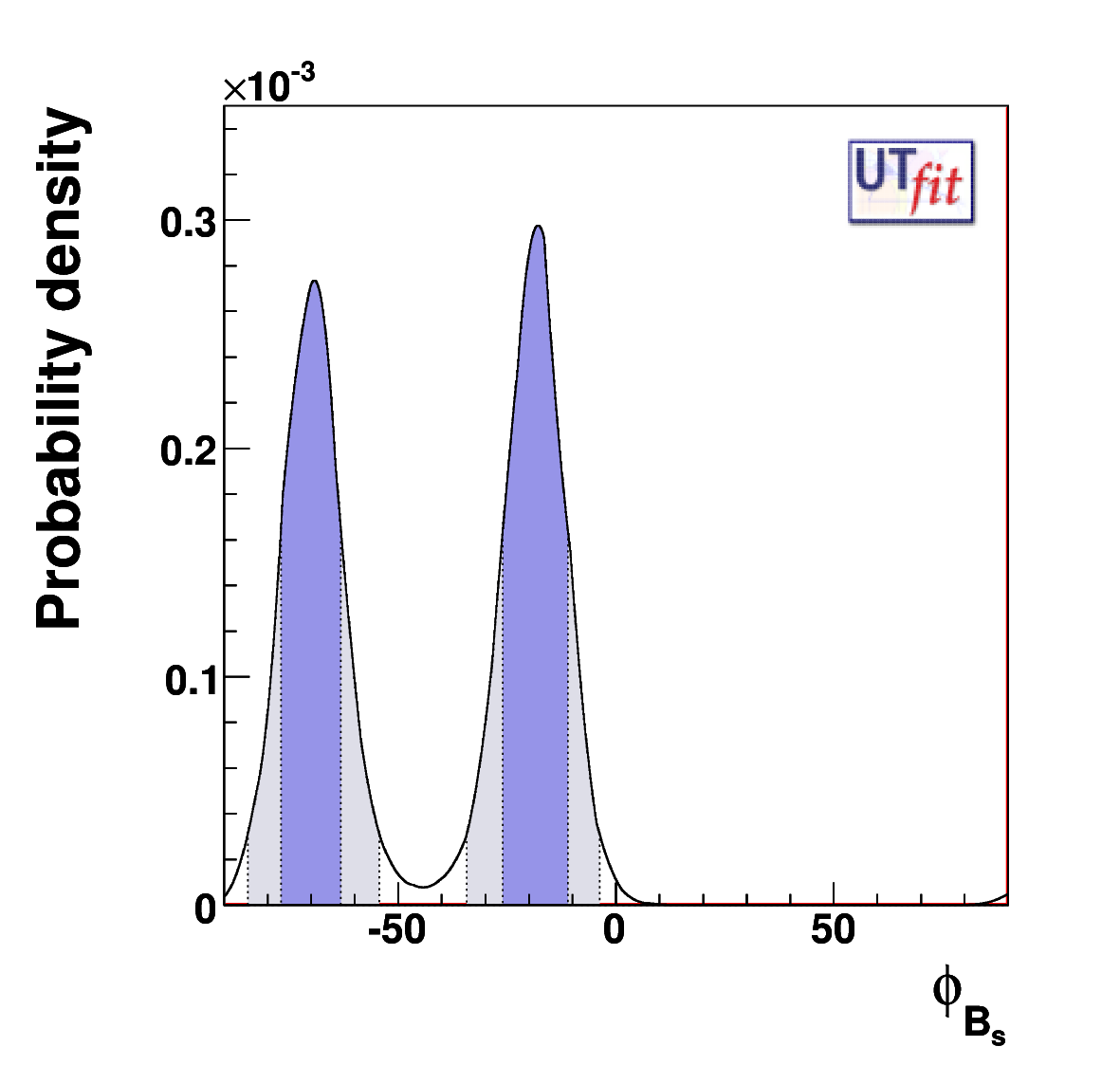} &
    \hspace*{-.5cm}
    \includegraphics[width=0.25\textwidth]{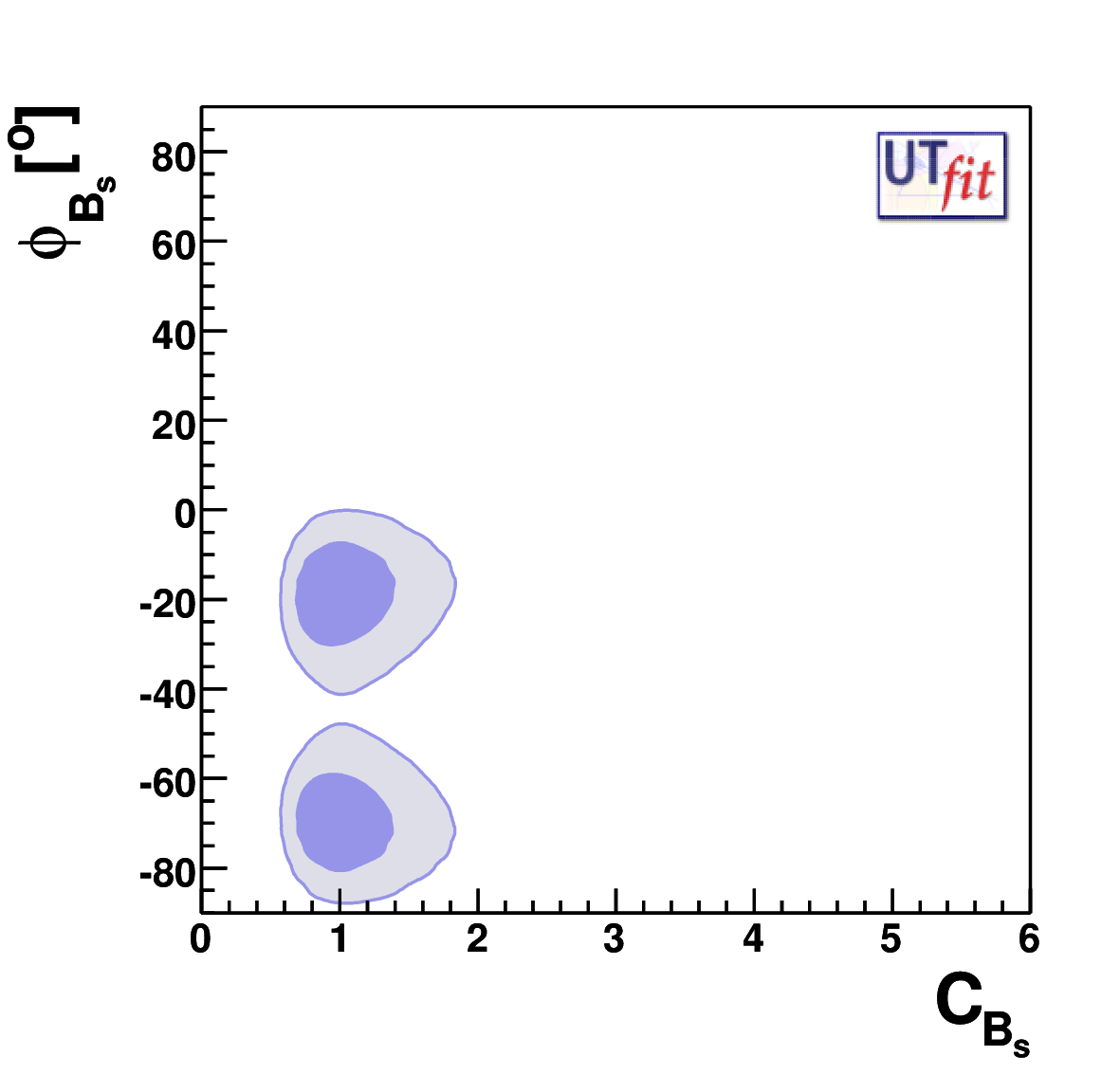} &
    \hspace*{-.5cm}
    \includegraphics[width=0.25\textwidth]{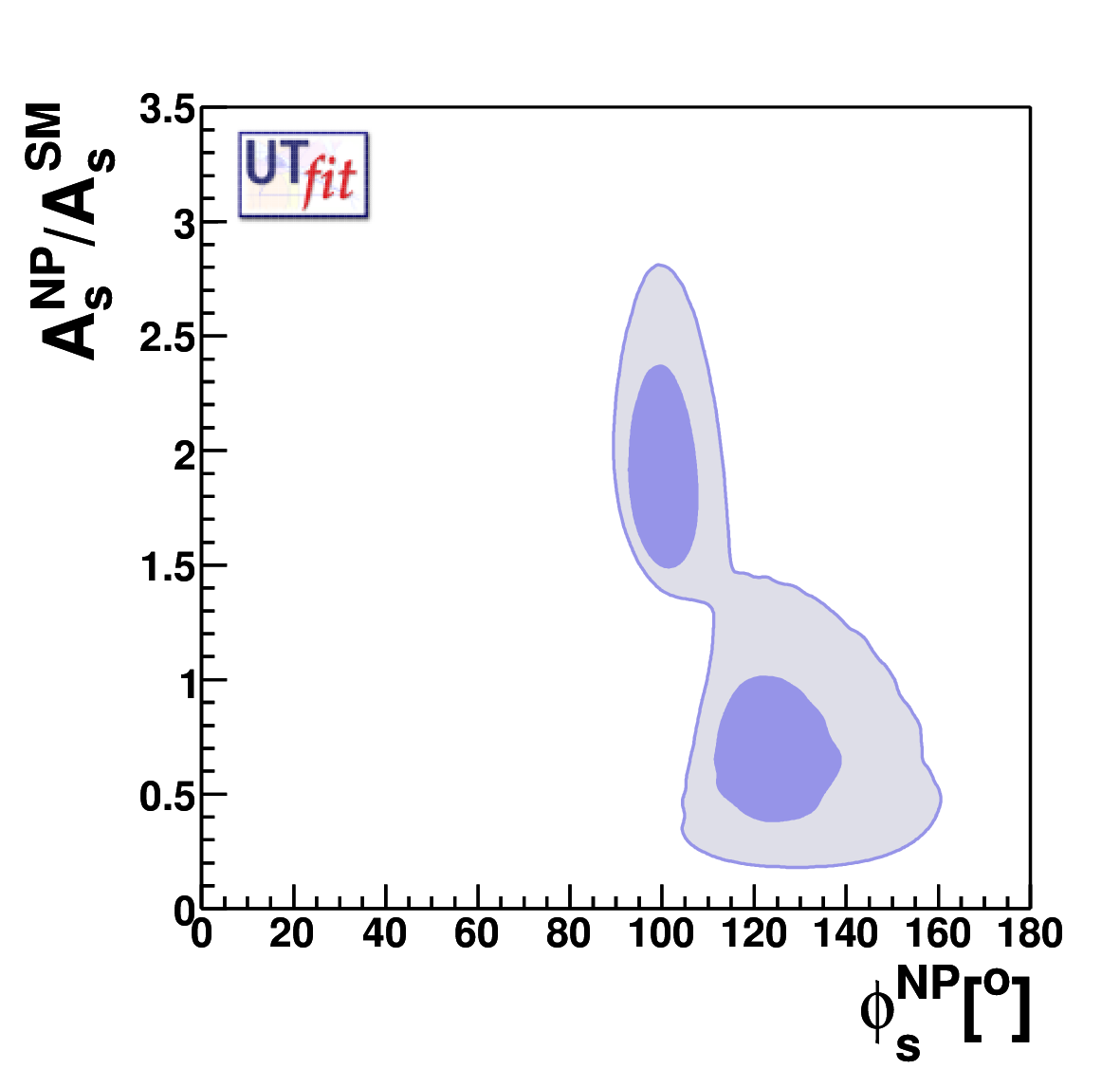} \\
    \hspace*{-.5cm}
\end{tabular}
  \caption{From left to right, p.d.f.'s for
    $C_{B_s}$, $\phi_{B_s}$, $\phi_{B_s}\,vs.\,C_{B_s}$ and
    $(A_\mathrm{NP}/A_\mathrm{SM})$ vs. $\phi_\mathrm{NP}$
    for NP in the $B_s$ sector~\cite{Bona:2008jn}.}
  \label{fig:BsNP}
\end{figure*}

We also include in the fit NP effects in $\Delta B=1$ transitions that
can also affect some of the measurements entering the UT analysis,
in particular the measurements of $\alpha$, $A_\mathrm{SL}$ and 
$\Delta \Gamma_s$\cite{Bona:2005eu, Bona:2006sa, Bona:2007vi}.

The results obtained in a global fit for the six NP parameters are
shown in Fig.~\ref{fig:NP}, together with the corresponding regions
in the $\bar\rho$--$\bar\eta$ plane.
More details on the analysis can be found in Ref.~\cite{Bona:2006sa} (see
Ref.~\cite{Charles:2004jd,Bona:2005eu} for previous analyses).

Writing $C_{B_q}e^{2 i \phi_{B_q}}$ $=$
$(A_\mathrm{SM} e^{2 i \beta_q} + A_\mathrm{NP} e^{2 i (\beta_q + \phi_\mathrm{NP})})/
(A_\mathrm{SM} e^{2 i \beta_q})$
and given the p.d.f. for $C_{B_q}$ and $\phi_{B_q}$, we can derive the
p.d.f. in the $(A_\mathrm{NP}/A_\mathrm{SM})$ vs. $\phi_\mathrm{NP}$
plane as seen in Fig.~\ref{fig:NP}.
We see that in the $B_d$ system,
the NP contribution can be substantial if its phase is close to
the SM phase, while for arbitrary phases its magnitude has to be much
smaller than the SM one. Notice that, with the latest data, the SM
($\phi_{B_d}=0$) is disfavored at $68\%$ probability due to the slight
disagreement between $\sin 2\beta$ and $|V_{ub}/V_{cb}|$.  This
requires $A_\mathrm{NP}\neq 0$ and $\phi_\mathrm{NP}\neq 0$. For the
same reason, $\phi_\mathrm{NP}>90^\circ$ at $68\%$ probability and the
plot is not symmetric around $\phi_\mathrm{NP}=90^\circ$. 
Assuming that the small but non-vanishing value for $\phi_{B_d}$ we
obtained is just due to a statistical fluctuation, the result of our
analysis points either towards models with no new source of flavor
and CP violation beyond the ones present in the SM (Minimal Flavor
Violation, MFV), or towards models in which new sources of flavor and
CP violation are only present in $b \to s$ transitions.

Conversely, from the results of our analysis in the $B_s$ system,
we see that the phase $\phi_{B_s}$ deviates from
zero at $\sim 3.0\sigma$. The solution around $\phi_{B_s}
\sim -20^\circ$ corresponds to $\phi^\mathrm{NP}_s \sim -50^\circ$ and
$A^\mathrm{NP}_s/A^\mathrm{SM}_s \sim 75\%$.  The second solution is
much more distant from the SM and it requires a dominant NP
contribution ($A^\mathrm{NP}_s/A^\mathrm{SM}_s \sim 190\%$). In this
case the NP phase is thus very well determined. The strong phase
ambiguity affects the sign of $\cos \phi_s$ and thus Re
$A^\mathrm{NP}_s/A^\mathrm{SM}_s$, while Im
$A^\mathrm{NP}_s/A^\mathrm{SM}_s \sim -0.74$ in any case.

This result shows an hint of discrepancy with respect to the SM
expectation in the $B_s$ CP-violating phase. We are eager
to see updated measurements using larger data sets from both the
Tevatron experiments in order to strengthen the present evidence,
waiting for the advent of LHCb for a high-precision measurement of the
NP phase.

It is remarkable that to explain the result obtained for $\phi_s$, new
sources of CP violation beyond the CKM phase are required, strongly
disfavoring the MFV hypothesis. These new phases will in general
produce correlated effects in $\Delta B=2$ processes and in $b\to s$
decays.  These correlations cannot be studied in a model-independent
way, but it will be interesting to analyze them in specific extensions
of the SM. In this respect, improving the results on CP violation in
$b\to s$ penguins at present and future experimental facilities is of
the utmost importance.

If we now consider the most general effective Hamiltonian for $\Delta F=2$
processes ($\mathcal{H}_\mathrm{eff}^{\Delta F=2}$~\cite{Bona:2007vi}),
we can translate the experimental constraints into allowed ranges
for the Wilson coefficients of $\mathcal{H}_\mathrm{eff}^{\Delta F=2}$.
These coefficients in general have the form
\begin{equation}
  \label{eq:cgenstruct}
  C_i (\Lambda) = \frac{F_i L_i}{\Lambda^2}\,
\end{equation}
where $F_i$ is a function of the (complex) NP flavor couplings, $L_i$
is a loop factor that is present in models with no tree-level Flavor
Changing Neutral Currents (FCNC), and $\Lambda$ is the scale of NP,
\emph{i.e.}  the typical mass of the new particles mediating
$\Delta F=2$ transitions. For a generic strongly-interacting theory
with arbitrary flavor structure, one expects $F_i \sim L_i \sim 1$ so
that the allowed range for each of the $C_i(\Lambda)$ can be
immediately translated into a lower bound on $\Lambda$. Specific
assumptions on the flavor structure of NP, for example Minimal or
Next-to-Minimal Flavor Violation (see Sec.~\ref{sect:BSMprimer}), correspond to
particular choices of the $F_i$ functions.
To obtain the p.d.f.~for the Wilson coefficients at the NP scale
$\Lambda$, we switch on one coefficient at a time in each sector and
calculate its value from the result of the NP analysis presented above.

The connection between the $C_i(\Lambda)$ and the NP scale $\Lambda$
depends on the general properties of the NP model, and in particular
on the flavor structure of the $F_i$. 
Assuming strongly interacting new particles, we have
\begin{equation}
  \label{eq:lambdagen}
  \Lambda=\sqrt{\frac{F_i}{C_i}}\,.
\end{equation}
In deriving the lower bounds on the NP
scale $\Lambda$, we assume $L_i = 1$, corresponding to
strongly-interacting and/or tree-level NP. Two other interesting
possibilities are given by loop-mediated NP contributions proportional
to $\alpha_s^2$ or $\alpha_W^2$.

Assuming strongly interacting and/or tree-level NP contributions with
generic flavor structure (\emph{i.e.} $L_i=\vert F_i\vert=1$), we can
translate the upper bounds on $C_i$ into the lower bounds on the NP
scale $\Lambda$. Conversely, in case of hints of NP effects, an upper
bounds on the NP scale $\Lambda$ is extracted.

\begin{table}[h!]
\begin{minipage}[c]{0.5\linewidth}
\caption{$95 \%$ probability lower bounds on the NP scale $\Lambda$ (in TeV)
  for several possible flavor structures and loop suppressions from the
  $K$ and $B_d$ systems.}
\begin{center}
  \begin{tabular}{|c|c|c|c|}
    \hline
    Scenario &~ strong/tree ~&~~ $\alpha_s$ loop ~~& ~~$\alpha_W$ loop~~ \\
    MFV  & 5.5  & 0.5  & 0.2  \\
    NMFV  & 62  & 6.2  & 2  \\
    General  & 240000  & 24000  & 8000  \\
    \hline
  \end{tabular}
\end{center}
\label{tab:NPscalesL}
\end{minipage}
\hspace{0.5cm}
\begin{minipage}[c]{0.5\linewidth}
\caption{$95 \%$ probability upper bounds on the NP scale $\Lambda$ (in TeV)
  for several possible flavor structures and loop suppressions from
  the $B_s$ system.}
\label{tab:NPscalesU}
\begin{center}
  \begin{tabular}{|c|c|c|c|}
    \hline
    Scenario &~ strong/tree ~&~~ $\alpha_s$ loop ~~& ~~$\alpha_W$ loop~~ \\
    NMFV     & 35  & 4  & 2  \\
    General  & 800   & 80  & 30  \\
    \hline
  \end{tabular}
\end{center}
\end{minipage}
\end{table}

From the lower bound Tab.~\ref{tab:NPscalesL}, we could conclude
that any model with strongly interacting NP and/or tree-level
contributions is beyond the reach of direct searches at the
LHC. Flavor and CP violation remain the main tool to constrain (or
detect) such NP models. Weakly-interacting extensions of the SM can be
accessible at the LHC provided that they enjoy a MFV-like suppression
of $\Delta F=2$ processes, or at least a NMFV-like suppression with an
additional depletion of the NP contribution to $\epsilon_K$.

If we consider the current effect in the $B_s$ mixing, we obtain
the upper bound Tab.~\ref{tab:NPscalesU} and we notice that
the general model is strongly problematic being the upper bound
at a much lower scale with respect to the corresponding lower bound
resulting from the $K$ and $B_d$ systems.
NMFV models are less problematic, but they can hardly reproduce
with the current size of the NP effect in the $B_s$ system while
keeping small effects in the $B_d$ and even smaller effects
in the $K$ system. Finally, MFV models would have possible solutions
in this scheme but they cannot generate the effect in the $B_s$
phase. So the current hint suggests some hierarchy in NP mixing
which is stronger that the SM one.

\subsubsection{Impact of flavor physics measurements on grand unified }
\label{sect:gfitsGU}
In a model of physics beyond the Standard Model, it is expected that
observables in the flavor physics are affected by the contributions from
new particles which couple to the quarks and leptons.
Comparing measured values of flavor observables with the Standard Model
predictions enables us to obtain information on the new physics
contributions.
If the measured value of certain observable differs from the Standard
Model prediction, the difference shows the magnitude of the new physics
contribution.
If the measured value is consistent with the Standard Model prediction,
that measurement is still useful as a constraint on the new physics.

Here we focus on the cases of supersymmetric grand unified models.
For general reviews of supersymmetric models, see
Refs.~\cite{Nilles:1983ge, Martin:1997ns, Chung:2003fi} and references
therein.

In supersymmetric extensions of the Standard Model, there exist
superpartners of the Standard Model particles, namely squarks, sleptons,
gauginos and higgsinos.
Supersymmetrized interactions include
quark-squark-gaugino and quark-squark-higgsino couplings.
The mass matrices of the superparticles are different from corresponding
ones of the Standard Model particles because of the supersymmetry
breaking.
Therefore the flavor mixing among the squarks depend on flavor structure
of the supersymmetry breaking mechanism.
The mismatch between the flavor bases of quarks and squarks generates
mixing matrices at the quark-squark-gaugino(higgsino) interactions.
These mixing matrices are not necessarily the same as the CKM matrix and
affect the flavor changing amplitudes through loop diagrams with squarks
in the internal lines.

Importance of the flavor physics in supersymmetric models have been
recognized since early 1980's \cite{Ellis:1981ts, Barbieri:1981gn}.
It was pointed out that squarks of the first and second generations must
be almost degenerate in mass, since otherwise too large contribution to
the $K-\bar{K}$ mixing would be given by squark-gaugino loops.
This requirement, which is known as ``SUSY flavor problem'', motivates
us to build a model of supersymmetry breaking mechanism that controls
the squark mass matrices.
The minimal supergravity (mSUGRA) model is one of those models.
In mSUGRA, it is assumed that the supersymmetry breaking occurs in a
hidden sector and its effect is transferred to the observable sector by
(super-)gravitational interactions.
Consequently, the supersymmetry breaking masses and interactions of
the superparticles are generated near the Planck scale and flavor-blind.
Mass differences and flavor mixings of the squarks are induced by the
(supersymmetrized) Yukawa interactions through radiative corrections.
Therefore the degeneracy of the first and the second generation squarks
is explained by the smallness of the Yukawa couplings of the light
quarks.
On the other hand, masses of the third generation squarks, particularly
stop, receive significant corrections from large top Yukawa coupling.
Squark flavor mixing occurs in left-handed squarks and the mixing matrix
is approximately the same as the CKM matrix \cite{Donoghue:1983mx,
  Bouquet:1984pp, Bouquet:1985qm}.

Effects of the superparticles on flavor observables have been studied in
the past decades \cite{Gabbiani:1988rb, Bigi:1991nt, Bertolini:1990if},
and it turns out that deviations from the
Standard Model predictions are small in the simplest mSUGRA scenario,
under the improved constraints from direct searches for the
superparticles and the Higgs bosons at LEP and Tevatron experiments.
An exception is the $b\to s\,\gamma$ decay.
$b\to s\,\gamma$ in supersymmetric models has been studied intensively
in 1990's \cite{Oshimo:1992zd, Hewett:1992is, Barger:1992dy,
  Barbieri:1993av}.
It is shown that the contributions from the suparparticle loops can be
as large as the Standard Model one, thus the agreement between the
measured value of $B(b\to s\,\gamma)$ and its Standard Model prediction
gives us an important constraint on the parameter space of a
supersymmetric model.

After the existence of the neutrino masses is established by neutrino
oscillation experiments, flavor mixing in the lepton sector has been
also taken into account.
Although the the neutrino masses are very small compared to the quark
and charged lepton masses, Yukawa couplings of the neutrinos need not to
be small.
If the see-saw mechanism \cite{Minkowski:1977sc, Yanagida:1979as,
  GellMann:1980vs} works and the Majorana
masses of the right-handed neutrinos are sufficiently large, the Yukawa
coupling constants of the neutrinos may be $O(1)$.
In the mSUGRA scenario, the neutrino Yukawa coupling generates the flavor
mixing in the slepton mass matrices through radiative corrections.
The flavor mixing in the sleptons eventually induces the lepton flavor
violating processes such as $\mu\to e\,\gamma$ \cite{Hisano:1997tc,
  Hisano:1998fj}.

In the supersymmetric grand unified models, the Yukawa interactions of
quarks and leptons are unified at the energy scale of the grand
unification.
Therefore both squark and slepton mass matrices receive flavor
off-diagonal contributions due to the unified Yukawa interactions above
the GUT scale.
In a $SU(5)$ unification model, flavor mixings of the left-handed squarks
and the right-handed charged sleptons are governed by the Yukawa
coupling matrix of the up-type quarks that consists of the top Yukawa
coupling \cite{Barbieri:1994pv, Barbieri:1995tw}.
On the other hand, the right-handed down-type squarks and the
left-handed sleptons receive contributions form the neutrino Yukawa
coupling matrix, which is related to the Maki-Nakagawa-Sakata neutrino
mixing matrix \cite{Maki:1962mu}.
Since the neutrino mixing angle between the second and the third
generations is known to be large, it is expected that significant
$\tilde{b}_R-\tilde{s}_R$ mixing is induced when the magnitudes of the
neutrino Yukawa couplings are sufficiently large \cite{Baek:2000sj,
  Moroi:2000mr, Moroi:2000tk}.

\begin{figure}
\begin{center}
\includegraphics[scale=.5]{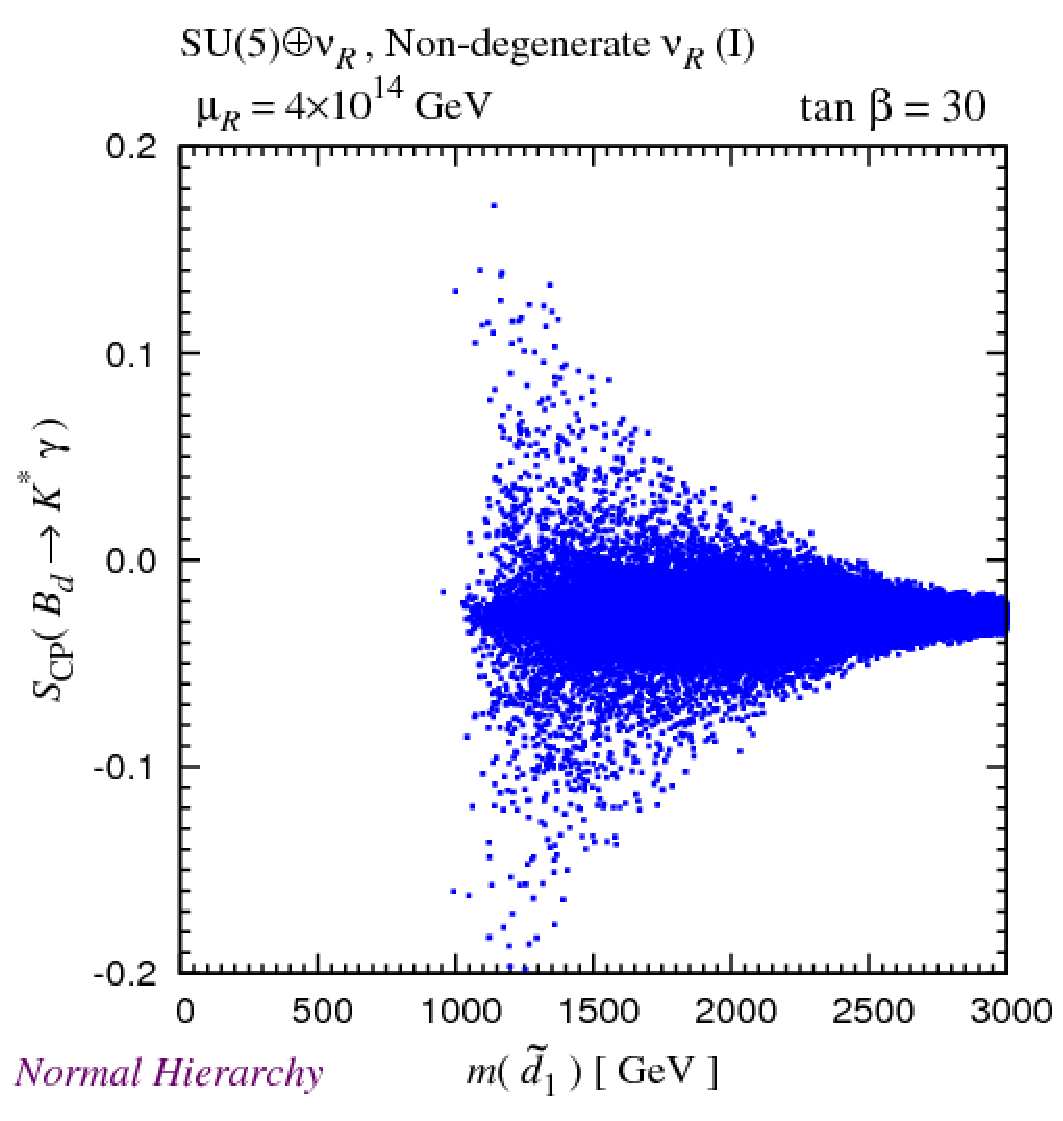}
\end{center}
\caption{Mixing-induced CP asymmetry in
$B_d\to K^*\,\gamma$ decay as a function of the lightest down-type
squark mass in the $SU(5)$ SUSY GUT with right-handed neutrinos
\cite{Goto:2007ee}.
}
\label{fig:SKsgm-msd1}
\end{figure}

The squark flavor mixings, which are generated by the (grand-unified)
neutrino Yukawa interactions, contribute to the quark flavor changing
amplitudes.
Since these additional contributions are independent of the CKM matrix, 
it is possible that deviations from the Standard Model predictions of
the flavor observables in the $B$ decays are sizably large while those
in $K$ decays are suppressed.
Fig.~\ref{fig:SKsgm-msd1} \cite{Goto:2007ee} shows the mixing-induced CP
asymmetry in $B_d\to K^*\,\gamma$ decay as a function of the lightest
down-type squark mass in the $SU(5)$ SUSY GUT with right-handed
neutrinos.
Each dot in the plot corresponds to a different choice of supersymmetry
breaking parameters in the mSUGRA scenario.
CKM matrix elements and neutrino parameters are fixed.
The neutrino Yukawa coupling matrix is chosen so that the flavor mixing
between the second and the third generations is large, whereas the
mixing between the first and the second generations is suppressed.
With this choice, SUSY contributions to the $K-\bar{K}$ mixing
($\varepsilon_K$) and $\mu\to e\,\gamma$ are small enough.
It is seen that there exist parameter regions where the asymmetry is as
large as $\pm 20$\% for the squark mass $\sim 1$TeV satisfying other
experimental constraints.
Other observables in $b\to s$ transition, such as the time-dependent CP
asymmetries in $B_d\to \phi\,K_S$ and $B_s\to J/\psi\,\phi$ are also
affected significantly in the same parameter region.

Another characteristic feature is that the SUSY flavor signals in the
quark and lepton sectors are correlated with each other
\cite{Goto:2007ee}.
As can be seen in Fig.~\ref{fig:SKsgm-tmg}, the branching fraction of
$\tau\to\mu\,\gamma$ can be as large as $10^{-8}$ in the parameter
region with large corrections to $b\to s$ observables.

The pattern of deviations from the Standard Model predictions depends on
the flavor structure of the masses and interactions of the squarks and
sleptons.
Therefore a combined analysis of many flavor observables provides us
with important clue on physics determining the structure of the SUSY
breaking sector.

\begin{figure}
\begin{center}
\includegraphics[scale=.5]{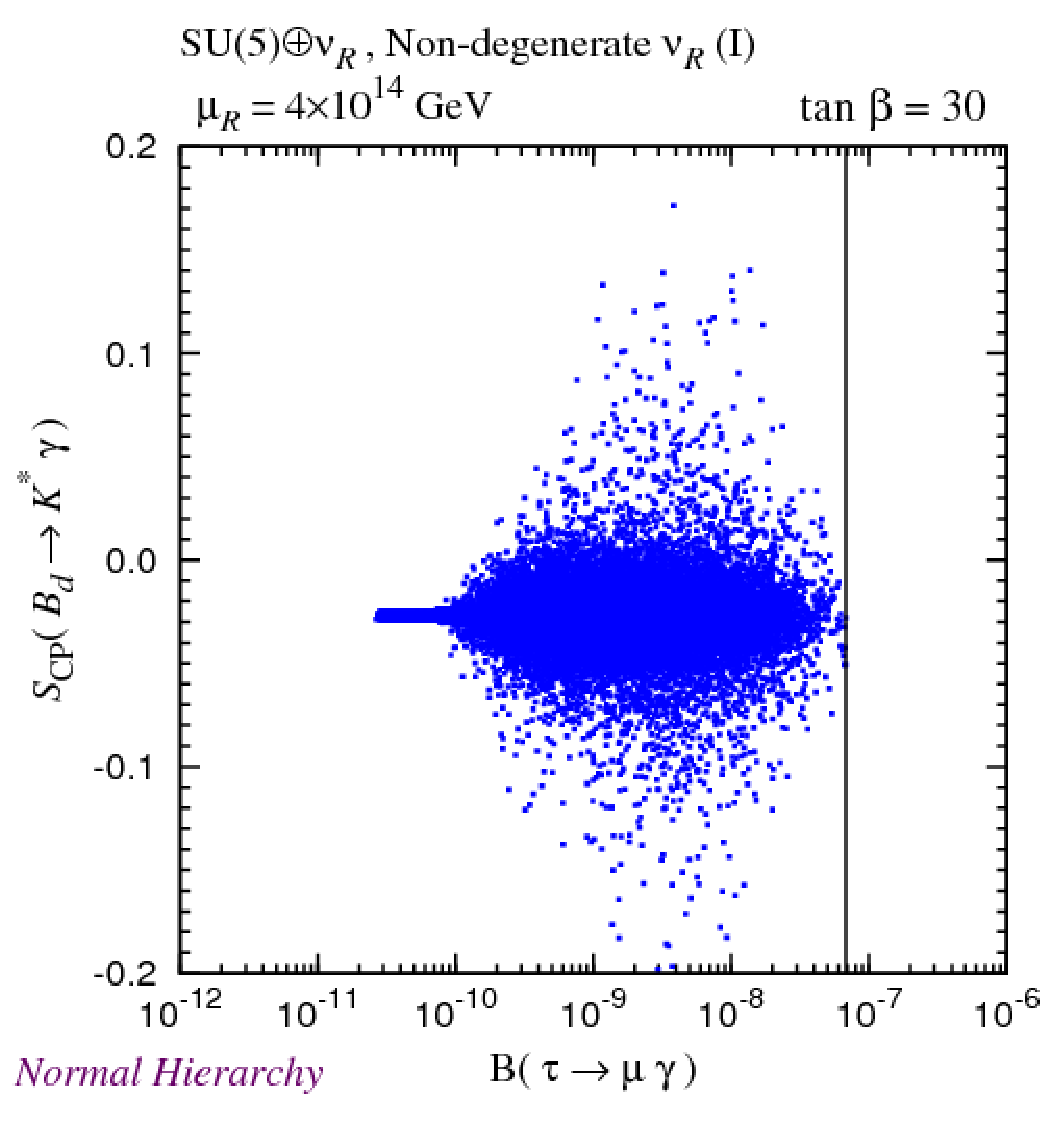}
\end{center}
\caption{Correlation between the mixing-induced CP asymmetry in
$B_d\to K^*\,\gamma$ and the branching fraction of $\tau\to\mu\,\gamma$
\cite{Goto:2007ee}.
}
\label{fig:SKsgm-tmg}
\end{figure}

\subsubsection{New physics in extra-dimension models}
\label{sect:gfitsED}
\label{sec:fits:NPinEDmodels}

In recent years a lot of interest was dedicated to extensions of the Standard Model involving 
one or more extra dimensions (ED), motivated by the possibility to find a  {\em natural}  
solution, in this context, of the hierarchy between the electroweak and the Planck scale. 
ED models can be grouped basically into three classes according to the space-time geometry of the ED 
and the localization properties of SM fields. 
In ADD~\cite{ArkaniHamed:1998rs,ArkaniHamed:1998nn} models the space-time is extended by one 
or more large (sub-millimeter) EDs with flat geometry. Only 
gravity is allowed to propagate in the higher-dimensional bulk, while all gauge and matter fields are confined to a 4d brane. In a different class of models, dubbed Universal Extra Dimensions (UED)~\cite{Appelquist:2000nn}, the EDs have flat geometry and are compactified, but now the SM fields are free to propagate in the bulk. Finally, in RS \cite{Randall:1999ee,Randall:1999vf} models, a 5d warped space-time is considered. Nowadays, in most phenomenological applications modifications of the original RS1 setup  \cite{Randall:1999ee} are considered, where gauge and matter fields propagate in the 5d bulk \cite{Chang:1999nh,Grossman:1999ra,Gherghetta:2000qt} and only the Higgs boson is confined on or near the IR brane.
In the following we will summarize flavor physics constraints on UED and warped models.


\paragraph{Universal extra dimensions (UED).}

\begin{figure}
\begin{center}
\includegraphics[width=0.37 \linewidth,height=0.36 \linewidth]{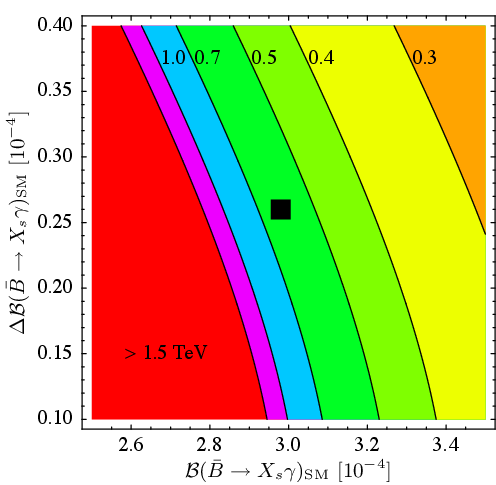}
\hspace{1.5cm}
\includegraphics[width=0.37 \linewidth,height=0.36 \linewidth]{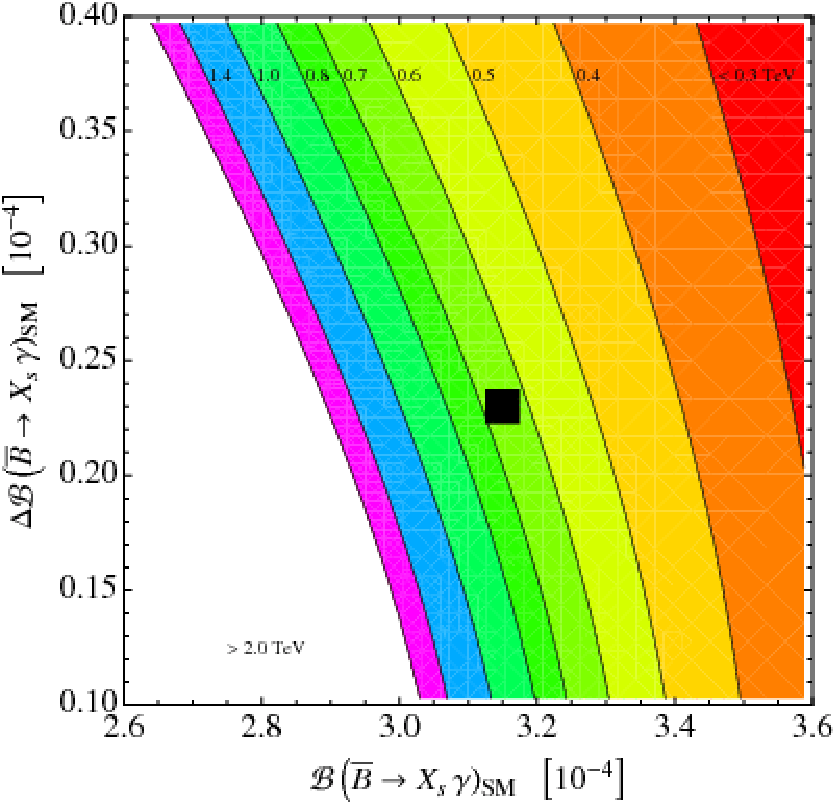}
\end{center}
\caption{95 \% C.L. limits on $1/R$ as a function of the SM central value and error on ${\cal B} (\bar B \to X_s \gamma)$ for the minimal UED5~\cite{Haisch:2007vb} and UED6~\cite{Freitas:2008vh} models. \label{fig:fits:UEDconstraints}}
\end{figure}

For what concerns UEDs we consider the so called minimal UED5~\cite{Appelquist:2000nn} and minimal UED6~\cite{Dobrescu:2004zi,Hashimoto:2004xz,Dobrescu:2007ec} models, characterized by one ED compactified on $S^1/\mathbb{Z}_2$ and two EDs compactified on $T^2/\mathbb{Z}_2$, respectively. The minimality refers to the absence of flavor non-universal boundary terms that would lead to unacceptably large flavor changing neutral currents. With these assumptions the Kaluza-Klein (KK) modes of the SM fields induce new contributions to flavor violating processes. As the models are minimal flavor violating (see Sec.~\ref{sect:BSMprimer}), 
those interactions are entirely controlled by the CKM matrix and the relevant 
free parameters of the models are the compactification radius $R$ and the cut-off scale 
$\Lambda$ at which the full (5d/6d) theory becomes non-perturbative. 
Detailed analyses of FCNC processes in UED5 and UED6 have been presented in \cite{Agashe:2001xt,Buras:2002ej,Buras:2003mk,Chakraverty:2002qk,Devidze:2005ua,Colangelo:2006vm,Colangelo:2006gv,Mohanta:2006ae,Colangelo:2007jy,Ferrandes:2007vn,Ahmed:2008ti,Bashiry:2008en,Bigi:2008bg,Haisch:2007vb} and \cite{Agashe:2001xt,Freitas:2008vh}, respectively. Lower bounds on $1/R$ come from oblique corrections, $Z\to b\bar b$, $(g-2)_\mu$ and  $b\to s \gamma$, with the latter providing by far the strongest constraint. It is interesting to note that UED contributions to $b\to s \gamma$ tend always to decrease the branching ratio and, within the 5d (6d) theory, have a negligible (logarithmic) dependence on the cut-off $\Lambda$. Utilizing the world average ${\cal B} (\bar B \to X_s \gamma)_{\rm exp} = (3.55 \pm 0.24^{+0.09}_{-0.10}\pm 0.03 ) \cdot 10^{-4}$ the authors of Refs.~\cite{Haisch:2007vb,Freitas:2008vh} find that in both models the inverse compactification radius has to be larger than about $600\gev$, with the exact bound depending quite sensitively on the theoretical prediction for the central value. Their findings are summarized in Fig.~\ref{fig:fits:UEDconstraints}. A discussion of these models in the context of dark matter and accelerator searches can be found, for instance, in Ref.~\cite{Hooper:2007qk}.

\paragraph{Bulk fermions in a warped ED.}

The case of bulk fermions in a warped ED is more interesting from the flavor physics point of view, as the localization of fermion zero modes along the 5$^{th}$ dimension provides an intrinsic explanation of the observed hierarchies in fermion masses and mixings \cite{Grossman:1999ra,Gherghetta:2000qt,Huber:2003tu}. Due to the absence of KK parity, here new physics contributions to FCNC observables appear already at the tree level, however they are strongly suppressed thanks to the so-called RS-GIM mechanism \cite{Agashe:2004cp}. In contrast to the UED models, this class of models goes beyond MFV and many new flavor violating parameters and CP phases are present, in addition to new flavor violating operators beyond the SM ones.

In order to obtain agreement with the electroweak $T$ parameter, usually an enlarged gauge sector $SU(2)_L\times SU(2)_R \times U(1)_X$ is considered 
\cite{Agashe:2003zs,Csaki:2003zu}, together with custodially protected fermion representations that avoid large anomalous $Z b_L \bar b_L$ 
\cite{Agashe:2006at,Contino:2006qr,Cacciapaglia:2006gp,Carena:2007ua} and at the same time also $Z d^i_L \bar d^j_L$ \cite{Blanke:2008zb,Blanke:2008yr,Buras:2009ka} 
couplings. Consequently the KK mass scale can be as low as $M_\text{KK}\simeq(2-3)\,\text{TeV}$ and therefore in the reach of direct LHC searches.

The impact of RS bulk matter on quark flavor violating 
observables has been discussed extensively in the literature, see 
e.g.~\cite{Burdman:2002gr,Burdman:2003nt,Agashe:2004ay,Agashe:2004cp,Moreau:2006np,Chang:2006si,Csaki:2008zd,Blanke:2008zb,Agashe:2008uz,Blanke:2008yr,Albrecht:2009xr} for details. Here we 
focus only on the most stringent constraint, coming from the $\varepsilon_K$ observable, that measures indirect CP-violation in the neutral $K$ meson sector, and on implications for flavor observables that have not yet been measured with high precision.

\begin{figure}
\begin{center}
\includegraphics[width=0.40 \linewidth,height=0.30 \linewidth]{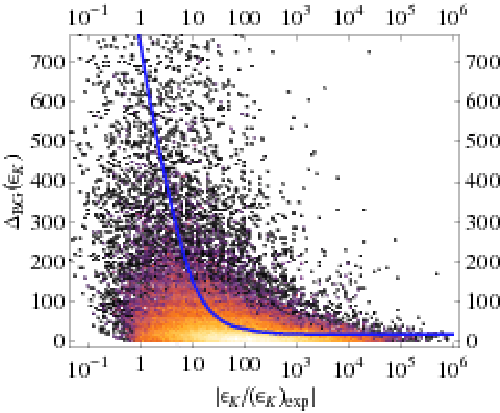}
\hspace{1.5cm}
\includegraphics[width=0.40 \linewidth,height=0.30 \linewidth]{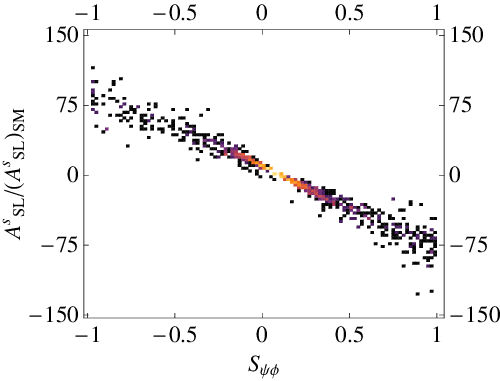}
\end{center}
\caption{{\it left: } Required Barbieri-Giudice \cite{Barbieri:1987fn} fine-tuning $\Delta_\text{BG}(\varepsilon_K)$ as a function of $\varepsilon_K$ in the custodially protected RS model. The blue curve displays the average fine-tuning \cite{Blanke:2008zb}.
{\it right: } Correlation between the CP-asymmetries $A^s_\text{SL}$ and $S_{\psi\phi}$ in the custodially protected RS model, fulfilling all available $\Delta F=2$ constraints \cite{Blanke:2008zb}. 
 \label{fig:fits:RS-DelF2}}
\end{figure}

\begin{figure}
\begin{center}
\includegraphics[width=0.40 \linewidth,height=0.27 \linewidth]{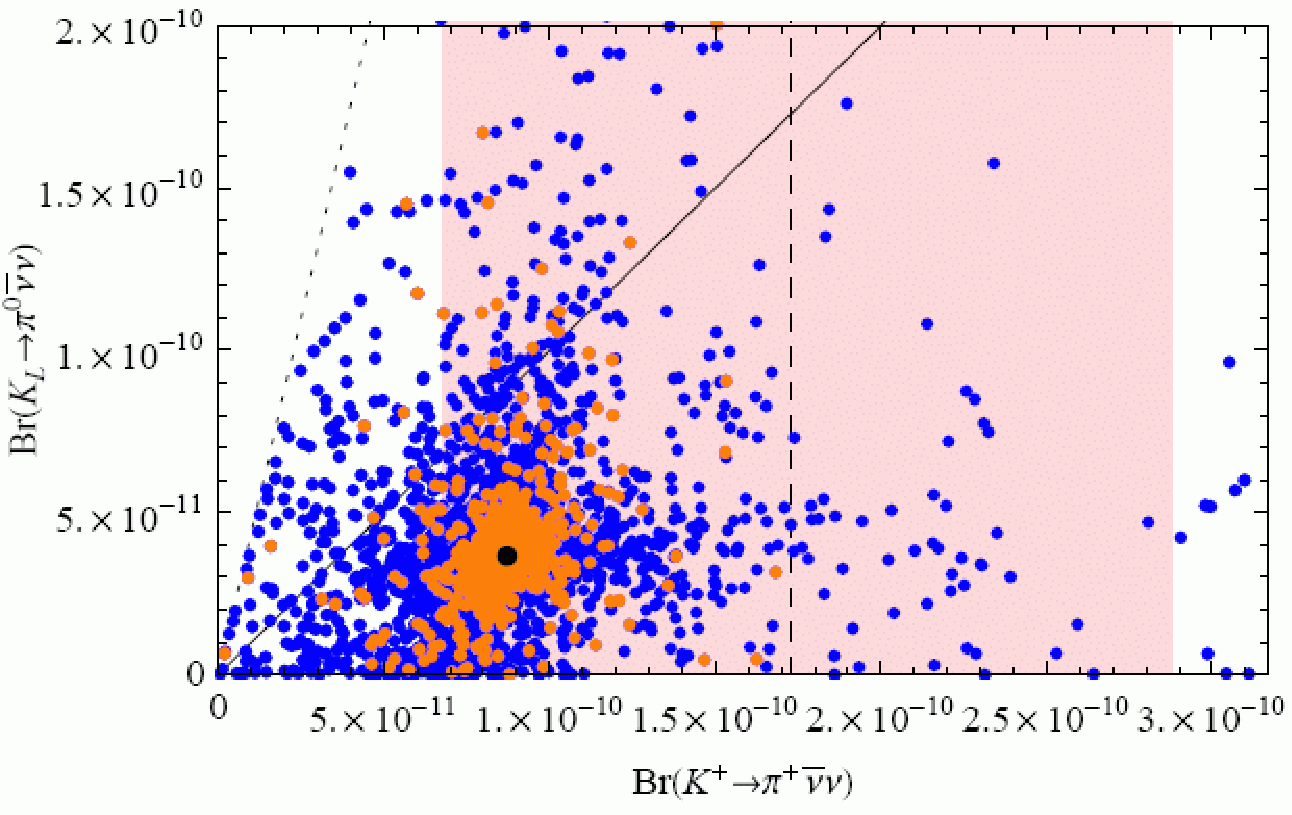}
\hspace{1.5cm}
\includegraphics[width=0.40 \linewidth,height=0.27 \linewidth]{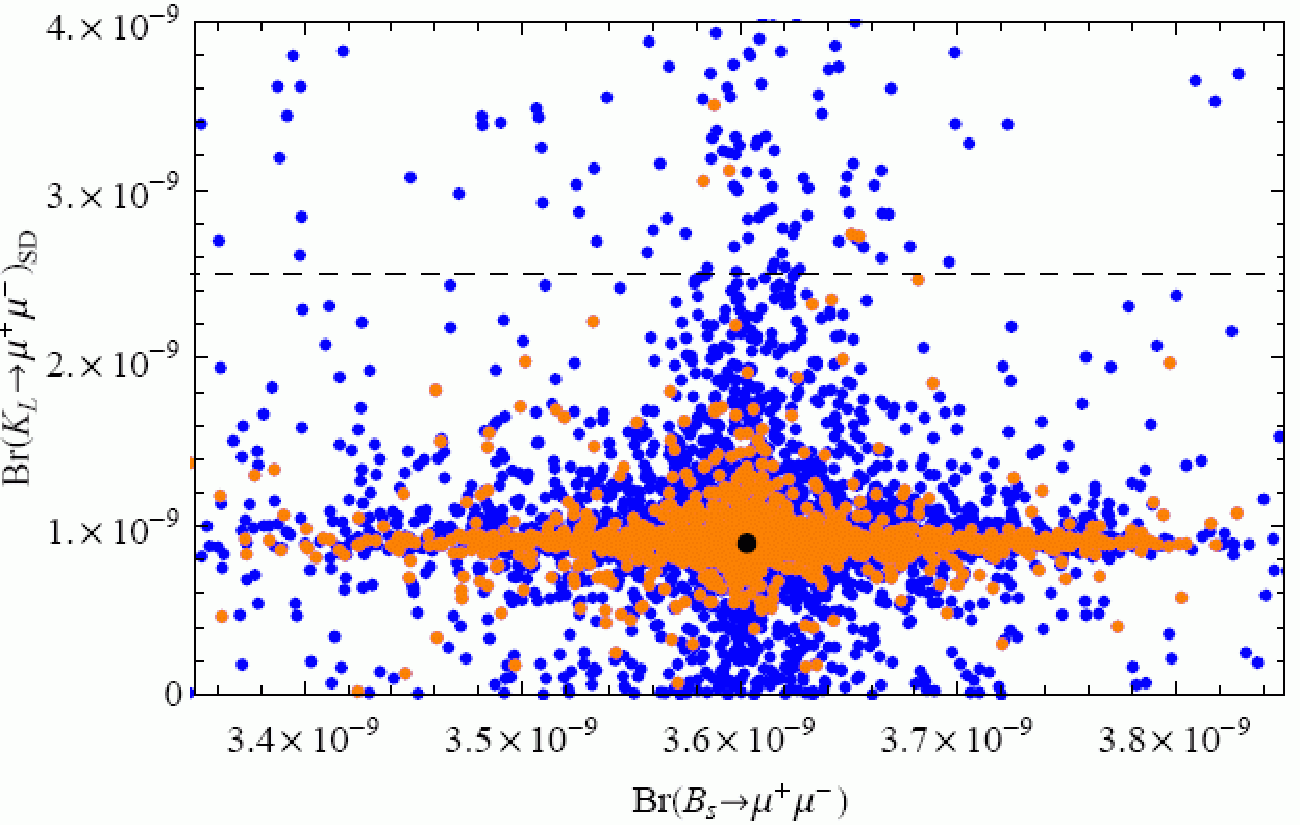}
\end{center}
\caption{{\it left: } $Br(K_L\to\pi^0\nu\bar\nu)$ as a function of $Br(K^+\to\pi^+\nu\bar\nu)$ in the custodially protected RS model \cite{Blanke:2008yr}. The shaded area represents the experimental $1\sigma$-range for $Br(K^+\to\pi^+\nu\bar\nu)$.
{\it right: } $Br(K_L\to\mu^+\mu^-)$ versus $Br(B_s\to\mu^+\mu^-)$ in the custodially protected RS model \cite{Blanke:2008yr}.
 \label{fig:fits:RSrare}}
\end{figure}

In contrast to the SM, the tree level exchange of KK gauge bosons induces 
amongst others the presence of left-right operators $\mathcal{Q}_{LR}$ 
contributing to $\Delta F=2$ processes. These operators receive large 
renormalisation group corrections and are in the case of $K^0-\bar K^0$ 
mixing in addition chirally enhanced. It turns out then that the otherwise 
so powerful RS-GIM mechanism in this case is not sufficient to suppress 
the new physics contribution below the experimental limits, so that 
assuming completely anarchic 5d Yukawa couplings a lower bound on the KK mass scale $M_\text{KK}\simeq 20\,\text{TeV}$ is obtained 
\cite{Csaki:2008zd,Blanke:2008zb}. In \cite{Blanke:2008zb} it has been shown however that allowing for modest hierarchies in the 5d Yukawas agreement with 
$\varepsilon_K$ can be obtained even without significant fine-tuning (see Fig.~\ref{fig:fits:RS-DelF2}), so that a natural solution to the ``$\varepsilon_K$-problem'' even for low KK scales cannot be excluded. Imposing then all available $\Delta F=2$ constraints on the RS parameter space, large new CP-violating effects in $B_s-\bar B_s$ mixing can still be found in this model \cite{Blanke:2008zb}, offering a neat explanation of the recent  CDF and \Dzero\ data \cite{Aaltonen:2007he,:2008fj,Brooijmans:2008nt}. In addition slight tensions between CP-violation in $K$ and $B_d$ observables \cite{Buras:2008nn,Lunghi:2008aa} could easily be resolved thanks to the presence of non-MFV interactions \cite{Blanke:2008zb}. An interesting pattern of deviations from the SM can also be found in the case of rare $K$ and $B_{d,s}$ decays (see Fig.~\ref{fig:fits:RSrare}) \cite{Blanke:2008yr}. As the dominant contribution stems from tree level flavor changing couplings of the $Z$ boson to right-handed down-type quarks, generally larger effects are to be expected in rare $K$ decays, e.\,g. $Br(K_L\to\pi^0\nu\bar\nu)$ can be enhanced by up to a factor 5. While the effects in $B_{d,s}$ decays are much more modest (e.\,g. $\pm 20\%$ in $Br(B_{d,s}\to\mu^+\mu^-)$), flavor universality can be strongly violated, so that interesting deviations from the MFV predictions appear. Striking correlations arise not only between various rare $K$ decays, but also between $K$ and $B_{d,s}$ physics observables, thus allowing to distinguish this framework from other new physics scenarios.

Alternative solutions to solve the ``$\varepsilon_K$-problem'', based on flavor symmetries, have as well been discussed in the recent literature. One 
approach is to protect the model from all tree
level FCNCs by incorporating a full 5d GIM mechanism~\cite{Cacciapaglia:2007fw}, in which the bulk respects a full $U(3)^3$ flavor symmetry. Although this model 
is safe, since its effective theory is MFV, it leaves the origin 
of the large hierarchies in the flavor sector unanswered.
More recent proposals therefore seek to suppress dangerous FCNCs and  simultaneously try to explain the hierarchical structure of the flavor sector.
One of them is the so called ``5d MFV'' model~\cite{Fitzpatrick:2007sa}. Here one postulates that the only sources of flavor breaking are two anarchic Yukawa spurions. The low-energy limit is not MFV, and
the additional assumption,
that brane and bulk terms in the down sector are effectively aligned, is needed to suppress dangerous FCNCs. Recently, an economical model has been proposed~\cite{Santiago:2008vq} in which one assumes a $U(3)$ flavor symmetry for the 5D fields containing the right handed down quarks. Dangerous contributions to $\mathcal{Q}_{LR}$ are then only generated by suppressed mass insertions on the IR brane where the symmetry is necessarily  broken.
Another recent approach \cite{Csaki:2008eh} presents a simple model where the key ingredient are two horizontal $U(1)$ symmetries which induce an alignment of bulk masses and down Yukawas, thus strongly suppressing FCNCs in the down sector. FCNCs in the up sector, however, can be close to experimental limits.

\section{Acknowledgements}
We would like to thank the Universit\'a "Sapienza" of Rome and in particular its Department of Physics  for the hospitality during the days 
of the workshop (9-13 September 2008). For the finantial and organizational support to the workshop itself we would like to thank INFN and in particular 
its Roma1 Section  and the Local Organizing Committee (D. Anzellotti, C. Bulfon, 
G. Bucci, E. Di Silvestro, R. Faccini, M. Mancini, G. Piredda, and R. Soldatelli) respectively.

The program of the workshop was elaborated by the Programm Committee (P. Ball, G. Cavoto, M. Ciuchini, R. 
Faccini -- chair, R. Forty, S. Giagu, P. Gambino, 
B. Grinstein, S. Hashimoto, T. Iijima, G. Isidori, V. Luth, G. Piredda, M. 
Rescigno, and A. Stocchi) under consultation of the International Advisory Committee (I. I. Bigi, C. Bloise, A.
Buras, N. Cabibbo, A. Ceccucci, P. Chang, F. Ferroni, A. 
Golutvin, A. Jawahery, A. S. Kronfeld, Y. Kwon, M. Mangano, W. J. 
Marciano, G. Martinelli, A. Masiero, T. Nakada, M. Neubert, P. Roudeau, A. I. Sanda, M. 
D. Shapiro, I. P.J. Shipsey, A. Soni, W. J. Taylor, N. G. 
Uraltsev, and M. Yamauchi).

This work is supported by 
Australian Research Council and the Australian Department of Industry Innovation, Science and Research,
the Natural Sciences and Engineering Research Council (Canada),
the National Science Fundation of China,
the Commissariat \`a l'Energie Atomique and Institut National de Physique Nucl\'eaire et de Physique des Particules (France), 
the Bundesministerium f\"ur Bildung und Forschung and Deutsche Forschungsgemeinschaft (Germany), 
the Department of Science and Technology of India
the Istituto Nazionale di Fisica Nucleare (Italy),
the Ministry of Education Culture, Sports, Science, and Technology (Japan), the Japan Society of Promotion of Science, 
the BK21 program of the Ministry of Education of Korea,
the Research Council of Norway,
 the Ministry of Education and Science of the Russian Federation,
the Slovenian Research Agency, 
Ministerio de Educaci\'on y Ciencia (Spain), 
the Science and Technology Facilities Council (United Kingdom),
and the US Department of Energy and National Science Foundation .

Individuals have received support from
 European Community's Marie-Curie Research Training Networks under contracts MRTN-CT-2006-035505 (`Tools and Precision Calculations for Physics Discoveries
at Colliders') and MRTN-CT-2006-035482 ('FLAVIAnet'),
from the National Science Fundation of China (grants 10735080 and 10625525),
the DFG Cluster of Excellence 'Origin and Structure of the Universe' (grant BU 706/2-1),
the Jpan Society for the Promotion of Science (grant 20244037),
 and the A. von Humboldt Stiftung,
from MICINN, Spain (grant FPA2007-60323), from the ITP at  University of Zurich,
from the US National Science Fundation (grant PHY-0555304) and Department of Energy (grants DE-FG02-96ER41005 and DE-AC02-07CH11359 -- Fermi Research
Alliance, LLC), and from Generalitat Valenciana (grant PROMETEO/2008/069).

\bibliographystyle{elsarticle-num}
\bibliography{CKMbook}

\end{document}